\RequirePackage{snapshot}
\documentclass[%
nofootinbib,
aip,
amsmath,amssymb,
reprint,%
floatfix,%
]{revtex4-2}

\usepackage{siunitx}
\DeclareSIUnit\pixel{pixel}

\usepackage{graphicx}

\usepackage[utf8]{inputenc}
\usepackage[T1]{fontenc}
\usepackage[T2A]{fontenc}
\DeclareUnicodeCharacter{017D}{\v{Z}}
\usepackage[usenames,dvipsnames]{xcolor}

\usepackage{longtable}
\usepackage{booktabs}
\usepackage{multirow}
\usepackage{tabularx} 
\usepackage{array}
\usepackage{makecell}
\newcolumntype{x}[1]{>{\centering\arraybackslash\hspace{0pt}}p{#1}}
\newcommand{\PreserveBackslash}[1]{\let\temp=\\#1\let\\=\temp}
\newcolumntype{P}[1]{>{\PreserveBackslash\raggedright}p{#1}}

\definecolor{cream}{RGB}{222,217,201}
\usepackage[normalem]{ulem}
\usepackage{amsmath,amssymb}

\usepackage{enumitem}
\setlist[enumerate]{itemsep=0pt}
\setlist[itemize]{itemsep=0pt}

\usepackage{hyperxmp}
\usepackage[type={CC},
            modifier={by},
            version={4.0}]{doclicense}
\usepackage[pdfa,unicode]{hyperref}
\hypersetup{pdftitle={Roadmap for focused ion beam technologies},
            pdfauthor={Katja Höflich, Gerhard Hobler, Frances I. Allen, Tom Wirtz, Gemma Rius, Lisa McElwee-White, Arkady V. Krasheninnikov, Matthias Schmidt, Ivo Utke, Nico Klingner, Markus Osenberg, Rosa Córdoba, Flyura Djurabekova, Ingo Manke, Philip Moll, Mariachiara Manoccio, José Maria De Teresa, Lothar Bischoff, Johann Michler, Olivier De Castro, Anne Delobbe, Peter Dunne, Oleksandr V. Dobrovolskiy, Natalie Frese, Armin Gölzhäuser, Paul Mazarov, Dieter Koelle, Wolfhard Möller, Francesc Pérez-Murano, Patrick Philipp, Florian Vollnhals, Gregor Hlawacek},
            pdfsubject={State-of-the-art and required future developments of advanced FIB technology.},
            pdfcaptionwriter={Gregor Hlawacek},
            pdfcontactemail={katja.hoeflich@fbh-berlin.de, gerhard.hobler@tuwien.ac.at, francesallen@berkeley.edu, tom.wirtz@list.lu, gemma.rius@csic.es, g.hlawacek@hzdr.de},
            pdfcontacturl={https://www.fit4nano.eu/, https://www.cost.eu/Actions/CA19140/},
            pdfpubstatus={AO},
            pdflang={en},
            pdfmetalang={en},
            pdfapart=3,
            pdfaconformance=B,
            keeppdfinfo,
            colorlinks=true, 
            citecolor=blue, 
            urlcolor=blue, 
            linkcolor=black}

\AtBeginDocument{%
  \hypersetup{%
    pdfpagerange={1-\ref*{TotPages}}
  }%
}
\pdfcatalog{%
  /OutputIntents [
  <<
  /Type /OutputIntent
  /S /GTS_PDFA1
  /DestOutputProfile \the\pdflastobj\space 0 R
  /OutputConditionIdentifier (sRGB)
  /Info (sRGB)
  >>
  ]
}

\usepackage{upgreek,textcomp}
\usepackage[printonlyused]{acronym}
\usepackage{lineno}

\usepackage[english]{babel}

\LetLtxMacro{\ORIGselectlanguage}{\selectlanguage}
\makeatletter
\DeclareRobustCommand{\selectlanguage}[1]{%
  \@ifundefined{alias@\string#1}
    {\ORIGselectlanguage{#1}}
    {\begingroup\edef\x{\endgroup
       \noexpand\ORIGselectlanguage{\@nameuse{alias@#1}}}\x}%
}
\newcommand{\definelanguagealias}[2]{%
  \@namedef{alias@#1}{#2}%
}
\makeatother

\definelanguagealias{en}{english}

\begin{document}
\linenumbersep4pt
\pagewiselinenumbers%
\nolinenumbers%

\title{Roadmap for \acl*{fib} technologies}

\author{Katja Höflich}
\email{katja.hoeflich@fbh-berlin.de}
\affiliation{Ferdinand-Braun-Institut gGmbH, Leibniz-Institut für Höchstfrequenztechnik, 12489 Berlin, Germany}
\author{Gerhard Hobler}
\email{gerhard.hobler@tuwien.ac.at}
\affiliation{Institute of Solid-State Electronics, TU Wien, Gu{\ss}hausstra{\ss}e 25--25a, 1040 Vienna, Austria.}
\author{Frances I. Allen}
\email{francesallen@berkeley.edu}
\affiliation{Department of Materials Science and Engineering, University of California, Berkeley, CA 94720, USA}
\author{Tom Wirtz}
\email{tom.wirtz@list.lu}
\affiliation{Advanced Instrumentation for Nano-Analytics (AINA), MRT Department, Luxembourg Institute of Science and Technology (LIST), 4422 Belvaux, Luxembourg}
\author{Gemma Rius} 
\email{gemma.rius@csic.es}
\affiliation{Institute of Microelectronics of Barcelona, IMB-CNM-CSIC, Carrer dels Til.lers s/n, E-08193, Cerdanyola, Spain}
\author{Lisa McElwee-White}
\affiliation{Department of Chemistry, University of Florida, Gainesville, Florida 32611--7200, USA}
\author{Arkady V. Krasheninnikov}
\affiliation{Institute of Ion Beam Physics and Materials Research, Helmholtz-Zentrum Dresden-Rossendorf, Bautzner~Landstr.~400, 01328 Dresden, Germany}
\author{Matthias Schmidt}
\affiliation{Helmholtz-Centre for Environmental Research GmbH---UFZ, Department of Isotope Biogeochemistry, Permoserstraße 15, 04318 Leipzig, Germany}
\author{Ivo Utke}
\affiliation{Laboratory for Mechanics of Materials and Nanostructures, EMPA Swiss Federal Laboratories for Material Science and Technology, Feuerwerkerstrasse 39, CH 3602 Thun, Switzerland}
\author{Nico Klingner}
\affiliation{Institute of Ion Beam Physics and Materials Research, Helmholtz-Zentrum Dresden-Rossendorf, Bautzner~Landstr.~400, 01328 Dresden, Germany}
\author{Markus Osenberg}
\affiliation{Institute of Applied Materials, Helmholtz-Zentrum Berlin für Materialien und Energie, Hahn Meitner Platz 1, 14109 Berlin, Germany}
\author{Rosa Córdoba}
\affiliation{Institute of Molecular Science, University of Valencia, Catedrático José Beltrán 2, Paterna 46980, Spain}
\author{Flyura Djurabekova}
\affiliation{Helsinki Institute of Physics and Department of Physics, University of Helsinki, P.O.\ Box 43, Helsinki FIB-00014, Finland}
\author{Ingo Manke}
\affiliation{Institute of Applied Materials, Helmholtz-Zentrum Berlin für Materialien und Energie, Hahn Meitner Platz 1, 14109 Berlin, Germany}
\author{Philip Moll}
\affiliation{Max Planck Institute for the Structure and Dynamics of Matter, Hamburg, Germany}
\author{Mariachiara Manoccio}
\affiliation{CNR-NANOTEC Institute of Nanotechnology, Via Monteroni, 73100 Lecce, Italy}
\author{José María De Teresa}
\affiliation{Instituto de Nanociencia y Materiales de Aragon (INMA), Universidad de Zaragoza-CSIC, 50009 Zaragoza, Spain}
\author{Lothar Bischoff}
\affiliation{Institute of Ion Beam Physics and Materials Research, Helmholtz-Zentrum Dresden-Rossendorf, Bautzner~Landstr.~400, 01328 Dresden, Germany}
\author{Johann Michler}
\affiliation{Laboratory for Mechanics of Materials and Nanostructures, EMPA Swiss Federal Laboratories for Material Science and Technology, Feuerwerkerstrasse 39, CH 3602 Thun, Switzerland}
\author{Olivier De Castro}
\affiliation{Advanced Instrumentation for Nano-Analytics (AINA), MRT Department, Luxembourg Institute of Science and Technology (LIST), 4422 Belvaux, Luxembourg}
\author{Anne Delobbe}
\affiliation{Orsay Physics S. A., 13710 Fuveau, France}
\author{Peter Dunne}
\affiliation{Université de Strasbourg, CNRS, IPCMS UMR 7504, 23 rue du Loess, F-67034 Strasbourg, France}
\author{Oleksandr V. Dobrovolskiy}
\affiliation{University of Vienna, Faculty of Physics, Boltzmanngasse 5, 1090 Vienna, Austria}
\author{Natalie Frese} 
\author{Armin Gölzhäuser} 
\affiliation{Faculty of Physics, Physics of Supramolecular Systems, University of Bielefeld, 33615 Bielefeld, Germany}
\author{Paul Mazarov}
\affiliation{Raith GmbH, Konrad-Adenauer-Allee 8, 44263 Dortmund, Germany}
\author{Dieter Koelle}
\affiliation{Physikalisches Institut, Center for Quantum Science (CQ) and LISA$^{+}$, Universität Tübingen, 72076 Tübingen, Germany}
\author{Wolfhard Möller}
\affiliation{Institute of Ion Beam Physics and Materials Research, Helmholtz-Zentrum Dresden-Rossendorf, Bautzner~Landstr.~400, 01328 Dresden, Germany}
\author{Francesc Pérez-Murano}
\affiliation{Institute of Microelectronics of Barcelona, IMB-CNM-CSIC, Carrer dels Til.lers s/n, E-08193, Cerdanyola, Spain}
\author{Patrick Philipp}
\affiliation{Advanced Instrumentation for Nano-Analytics (AINA), MRT Department, Luxembourg Institute of Science and Technology (LIST), 4422 Belvaux, Luxembourg}
\author{Florian Vollnhals} 
\affiliation{Institute for Nanotechnology and Correlative Microscopy gGmbH, Forchheim, Germany}
\author{Gregor Hlawacek}
\email{g.hlawacek@hzdr.de}
\affiliation{Institute of Ion Beam Physics and Materials Research, Helmholtz-Zentrum Dresden-Rossendorf, Bautzner~Landstr.~400, 01328 Dresden, Germany}

\begin{abstract}
The \ac{fib} is a powerful tool for the fabrication, modification and characterization of materials down to the nanoscale. 
Starting with the gallium \ac{fib}, which was originally intended for photomask repair in the semiconductor industry, there are now many different types of \ac{fib} that are commercially available. 
These instruments use a range of ion species and are applied broadly in materials science, physics, chemistry, biology, medicine, and even archaeology. 
The goal of this roadmap is to provide an overview of \ac{fib} instrumentation, theory, techniques and applications. 
By viewing \ac{fib} developments through the lens of the various research communities, we aim to identify future pathways for ion source and instrumentation development, as well as emerging applications and opportunities for improved understanding of the complex interplay of ion-solid interactions. 
We intend to provide a guide for all scientists in the field that identifies common research interests and will support future fruitful interactions connecting tool development, experiment and theory. 
While a comprehensive overview of the field is sought, it is not possible to cover all research related to \ac{fib} technologies in detail. 
We give examples of specific projects within the broader context, referencing original works and previous review articles throughout.
\end{abstract}

\maketitle


\tableofcontents

\listoftables%

\section*{Acronyms}
\begin{acronym}[HRTEM]
\itemsep0pt
	\acro{aes}[AES]{Auger electron spectroscopy}
	\acro{afm}[AFM]{atomic force microscopy}
  \acro{api}[API]{application programming interface}
	\acro{apt}[APT]{atom probe tomography}
  \acro{asis}[ASIS]{atomic-size ion source}
	\acro{bca}[BCA]{binary collision approximation}
  \acro{bi}[BI]{backscattered ion}
 	\acro{cad}[CAD]{computer-aided design}
  \acro{cmnt}[CMNT]{colloidal micro newton thrusters}
	\acro{cmos}[CMOS]{complementary metal-oxide-semiconductor}
	\acro{cnt}[CNT]{carbon nanotube}
	\acro{comb}[COMB]{charge optimized many body}
  \acro{dac}[DAC]{digital-to-analog converter}
	\acro{dc}[DC]{direct current}
	\acro{dft}[DFT]{density functional theory}
	\acro{eam}[EAM]{embedded atom method}
	\acro{ebit}[EBIT]{electron beam ion trap}
	\acro{ebl}[EBL]{electron beam lithography}
	\acro{ebsd}[EBSD]{electron backscatter diffraction}	
	\acro{ecr}[ECR]{electron cyclotron resonance}
	\acro{eds}[EDS]{energy-dispersive X-ray spectroscopy}
  \acro{esi}[ESI]{electrospray ionization source}
	\acro{eels}[EELS]{electron energy loss spectroscopy}
  \acro{esem}[ESEM]{environmental \acl{sem}}
	\acro{et}[ETD]{Everhardt-Thornley detector}
	\acro{euv}[EUV]{extreme ultraviolet}
	\acro{esa}[ESA]{excited surface atom}
	\acro{feb}[FEB]{focused electron beam}
	\acro{febid}[FEBID]{focused electron beam induced deposition}
	\acro{febie}[FEBIE]{focused electron beam induced etching}
  \acro{feep}[FEEP]{field emission electric propulsion}
	\acro{fib}[FIB]{focused ion beam}
	\acro{fibid}[FIBID]{focused ion beam induced deposition}
	\acro{fibie}[FIBIE]{focused ion beam induced etching}
    \acro{fim}[FIM]{field ion microscopy}
	\acro{finfet}[FinFET]{fin field-effect transistor}
  \acro{fmr}[FMR]{ferromagnetic resonance}
	\acro{fov}[FOV]{field of view}
	\acro{gap}[GAP]{Gaussian approximation potential}
    \acro{gui}[GUI]{graphical user interface} 
	\acro{fwhm}[FWHM]{full width at half maximum}
	\acro{gfis}[GFIS]{gas field-ionization source}
	\acro{gis}[GIS]{gas injection system}
  \acro{hbn}[hBN]{hexagonal boron nitride}
	\acro{hibl}[HIBL]{helium ion beam lithography}
	\acro{him}[HIM]{helium ion microscope}
	\acro{hrtem}[HRTEM]{high resolution transmission electron microscopy}
	\acro{hsq}[HSQ]{hydrogen silsesquioxane}
  \acro{ibic}[IBIC]{ion beam induced charge}
  \acro{ic}[IC]{integrated circuit}
  \acro{icd}[ICD]{image charge detector}
	\acro{icp}[ICP]{inductively coupled plasma}
	\acro{iiaes}[IIAES]{ion induced Auger electron spectroscopy}
	\acro{il}[IL]{ionoluminescence}
	\acro{ilis}[ILIS]{ionic liquid ion source}
	\acro{kmc}[kMC]{kinetic Monte Carlo}
	\acro{lefib}[LE-FIB]{low energy focused ion beam}
	\acro{lj}[LJ]{Lennard-Jones-type}
	\acro{lmais}[LMAIS]{liquid metal alloy ion source}
	\acro{lmis}[LMIS]{liquid metal ion source}
	\acro{lotis}[LoTIS]{low temperature ion source}
  \acro{mc}[MC]{Monte Carlo}
	\acro{mcp}[MCP]{micro channel plate}
	\acro{md}[MD]{molecular dynamics}
 	\acro{mfm}[MFM]{magnetic force microscopy}
	\acro{ml}[ML]{machine learning}
	\acro{ms}[MS]{molecular statics}
	\acro{mems}[MEMS]{micro-electro-mechanical systems}
	\acro{motis}[MOTIS]{magneto-optical trap ion source}
  \acro{mram}[MRAM]{magnetic random access memory}
  \acroplural{mram}[MRAMs]{magnetic random access memories}
	\acro{nais}[NAIS]{nano-aperture ion source}
	\acro{nems}[NEMS]{nano-electro-mechanical systems}
	\acro{nil}[NIL]{nanoimprint lithography}
	\acro{nsom}[NSOM]{near-field optical microscopy}
  \acro{nv}[NV]{nitrogen vacancy}
  \acroplural{nv}[NV]{nitrogen vacancies}
  \acro{pemfc}[PEMFC]{proton exchange membrane fuel cell}
	\acro{pfib}[PFIB]{plasma focused ion beam}
	\acro{pi}[PI]{primary ion}
	\acro{pixe}[PIXE]{particle induced X-ray emission}
    \acro{pmma}[PMMA]{poly(methyl methacrylate)}
	\acro{qms}[QMS]{quadrupole mass spectrometer}
	\acro{rbs}[RBS]{Rutherford backscattering spectrometry}
  \acro{sdd}[SDD]{silicon drift detector}
	\acro{se}[SE]{secondary electron}
	\acro{sem}[SEM]{scanning electron microscopy}
	\acro{sfim}[SFIM]{scanning field ion microscope}
  \acro{si}[SI]{secondary ion}
	\acro{sii}[SII]{single ion implantation}
	\acro{sims}[SIMS]{secondary ion mass spectrometry}
	\acro{snms}[SNMS]{secondary neutral mass spectrometry}
  \acro{snr}[SNR]{signal to noise ratio}
  \acro{spe}[SPE]{single photon emitter}
	\acro{spm}[SPM]{scanning probe microscopy}
  \acro{sspd}[SSPD]{superconducting single-photon detector}
	\acro{stem}[STEM]{scanning transmission electron microscopy}
	\acro{stim}[STIM]{scanning transmission ion microscopy}
	\acro{squid}[SQUID]{superconducting quantum interference device}
	\acro{tddft}[TDDFT]{time-dependent \acl*{dft}}
	\acro{tem}[TEM]{transmission electron microscopy}
  \acro{tic}[TIC]{total ion counter}
	\acro{tof}[TOF]{time-of-flight}
	\acro{uhv}[UHV]{ultra-high vacuum}
  \acro{ybco}[YBCO]{yttrium barium copper oxide}
  \acro{yig}[YIG]{yttrium iron garnet}
  \acro{yvo}[YVO]{yttrium orthovanadate}
  \acro{ysz}[YSZ]{yttrium stabilized zirconia}
	\acro{zbl}[ZBL]{Ziegler-Biersack-Littmark}
	\acro{zpl}[ZPL]{zero-phonon line}
\end{acronym}


\section{Introduction}%
\label{sec_intro}
\acresetall{}

The technological origin of the \ac{fib} instruments we use today lies in outer space, or more precisely, in the application of ion beams for spacecraft propulsion.
In space, thrust can only be generated by ejecting matter, the so-called reaction mass, which must be carried along with the spacecraft.
In addition to chemical thrusters based on combustion, ion thrusters have emerged as an important tool for high-precision movement.
The positively charged ions that are generated by field ionization or by electrospraying, are accelerated by electric or magnetic fields, and then neutralized before being ejected in the opposite direction to that of the intended motion\footnote{Neutralization is important here because otherwise the spacecraft would accumulate negative charge and thus attract the ejected positive ions.}.
Different types of thrusters based on liquid metal ion sources are currently being tested in space for ultra-precise position control of satellites, e.g.\ for the LISA gravitational wave interferometer~\cite{LISA}.
One of these thruster technologies, the electric field emission propulsion system (part of the LISA Pathfinder mission~\cite{Tajmar_2004} and the more recent CubeSat launches~\cite{Bock_2018}), is in fact very similar to the heart of many of our ground-based \ac{fib} instruments.

Whereas ion thrusters enable the exertion of forces in the micronewton range for the navigation of space objects, our ground-based \ac{fib} instruments enable the fabrication, modification and characterization of micrometer- to nanometer-sized objects.

The leading example of \ac{fib} processing is still the site-selective preparation of samples for high-resolution imaging techniques, in particular, for \ac{tem} and \ac{apt} and for cross-section imaging using \ac{sem}.
Several reviews have already been devoted to these important, well-established applications~\cite{Mayer_2007,Prosa_2017,Berger_2023}.  

However, since focused ion beams can be used to modify any material down to the nanoscale in a variety of ways, from targeted doping to structural modification and geometric shaping, the \ac{fib} is a powerful tool in all areas from basic research to technology.
Our focus is therefore on these novel and advanced applications of the \ac{fib}.

This document presents the state of the art of \ac{fib} research and development today, and discusses future perspectives. It is organized as follows:
Section~\ref{sec_instru} gives an overview of \ac{fib} instrumentation, starting with the generation and control of the focused ion beam, followed by detectors and other complementary tools and accessories.
Section~\ref{sec_theo} summarizes the theoretical approaches that can be used to describe various aspects of ion-matter interactions, including the \ac{bca}, \ac{md}, \ac{kmc} techniques, \ac{dft}, and continuum modeling. 
The last of these enables modeling over the longest length- and timescales, including treatment of ion- and electron-induced surface chemistry.
The wide range of applications of the \ac{fib} is discussed in Section~\ref{sec_appl}, which is organized according to the various experimental techniques (subtractive processing; defect engineering; imaging and tomography; elemental analysis; gas-assisted processing; and several other emerging directions).
For each application field, a selection of examples is discussed.
As a service to the community, summary tables with references have been compiled in order to provide a more comprehensive (albeit non-exhaustive) literature review of each of these application fields.
These tables can be used as a starting point to help the reader identify new \ac{fib} opportunities for their own research.

Central to this document is the roadmap in Section~\ref{sec_roadmap}, which highlights future perspectives for \ac{fib} research and development based on the state of the art in instrumentation, theory, and applications previously described. 
Here, key drivers for the \ac{fib} in different areas of science and technology have been identified and linked to \ac{fib}-specific challenges and the steps needed for future developments. 
We hope that this roadmap can serve as an incubator for future developments and will provide inspiration for scientific and technological breakthroughs, as well as serve as a unique resource for funding agencies and industry.

\acresetall{}
\section{Instrumentation}%
\label{sec_instru}

The creation of a finely focused beam of ions presents several engineering challenges.
Moreover, the beam requirements can vary widely depending on the application, sometimes with conflicting specifications.
For example, for many applications, a high-current beam is desired to enable efficient milling (i.e.\ material removal by sputtering). 
However, high-current sources tend to deliver ions with a large energy spread,  resulting in strong chromatic aberrations. 
Because of this (and other reasons, such as spherical aberrations from the lenses, as discussed later) it is challenging to build a high-current high-resolution source. 
Similar conflicting scenarios present themselves in many other areas of \ac{fib} instrumentation.

In the following, the various components of the \ac{fib} instrument are discussed.
We start by describing the different ion sources that are used (Section~\ref{subs_sources}). 
The source defines many properties of the final instrument, including the achievable spot size.
Then we address beam transport via the ion optical column (Section~\ref{subs_optics}), and subsequently detectors and analytics (Section~\ref{subs_detectors}); these elements, which are used to steer, shape and detect the ions, must be designed in such a way as to ensure the best possible end performance of the particular source being used.
In the sample chamber, there are several other components that can be incorporated, including specialized sample stages for in-situ or in-operando experiments, micro-/nanomanipulators, and gas injectors.
These are discusssed in Section~\ref{subs_accessories}.
Next, we address considerations concerning experiments with radiaoactive samples (Section~\ref{sec:subs_radioactive}).
Finally, we outline software needs and correlative approaches for beam control, automation, and multi-modal analysis (Section~\ref{subs_software}).
An overview of these topics can also be found in the book chapter by~\citet{Bassim2019}.

\subsection{Ion sources}%
\label{subs_sources}

In order to achieve high spatial resolution in a \ac{fib} instrument, an ion source with high brightness is required~\cite{Tondare_2005}.
Analogous to the definition in electron microscopy, the brightness $B$ of an ion source is a measure of the compactness and directionality of the ion beam, according to: 
\begin{equation}
    B = \frac{I}{A \Omega},
\end{equation}
where the emission current $I$ from a source area $A$ is emitted into a solid angle $\Omega$~\cite{Langmuir_1937,Wolf1995}.
In practice, this means that ions are suitable for microscopy and nanofabrication if they are emitted from a highly localized area into a well-defined direction.
The reduced brightness is derived from the above definition by taking into account the acceleration voltages of the \ac{fib} instrument; representative values for the different ion sources are given in Table~\ref{table:ion_sources}.
\begin{table*}[btp]
\caption[Common \ac{fib} sources]{\ac{fib} sources compared according to key parameters. 
Only common ion species are listed. 
All published or commercially available ion species can be found in Fig.~\ref{fig:pse_sources}. 
Superscript $q$ marks different charge states (from 1 to 3) obtained from \ac{lmis}/\ac{lmais}; for plasma sources even higher charge states are obtained.
Subscript $n$ for \ac{lmis}/\ac{lmais} indicates the possible emission of polyatomic clusters comprising up to 10 ions (or even more).
Parameters for \ac{ilis} indicated by * are estimated values from reference~\cite{Zorzos_2008}.
For better comparison, the lateral spatial resolutions have been converted to \ac{fwhm} values assuming a Gaussian beam profile resulting in an error-function-like edge profile.}%
\label{table:ion_sources}
\begin{center}
\sisetup{range-phrase=--, range-exponents = combine, tight-spacing=true}
\begin{tabular}{P{1.2cm}P{3.3cm}P{2.5cm}P{2cm}P{1.4cm}P{2cm}P{4cm}}
 \toprule
 Source		& Common ion species							& Reduced brightness (\si{\ampere\per\square\meter\per\steradian\per\volt})	& Min.\ energy spread (eV)					& Max.\ current (nA)  				& Lateral \newline resolution FWHM~(nm)					& Key applications\\
 \midrule

 \acs*{lmis}/ \acs*{lmais}	& (Li, Si, Ga, Ge, In, Sn, Sb, Au, Pb, Bi, \ldots{})$_{1..n}^{q+}$			& \num{1e6}~\cite{Bischoff_2016}					& \numrange[range-phrase=-]{2}{40}~\cite{Bischoff_2016, Swanson_1994}	& 100						& \numrange[range-phrase=-]{2}{2.8}~\cite{Chroma_2021, Bauerdick_2013}			& surface patterning and cross-sectioning, volume imaging, local doping and implantation,  \acs*{sims}, mask edit, ion thruster \\
 
\acs*{ilis}		& EMI$^+$, BMI$^+$, BF$_4^-$, I$^-$, \ldots{}, cluster ions 			& \num[]{5e6}*~\cite{Zorzos_2008}					& \numrange[range-phrase=-]{7}{10}~\cite{Lozano_2005b, Fedkiw_2009}	&  0.75*~\cite{Zorzos_2008}				& \numrange[range-phrase=*-]{50}{30000}~\cite{Zorzos_2008} 			& ion thruster, reactive ion etching, \acs*{sims} \\

\acs*{pfib}		& (Xe, Ar, Kr, N, O \ldots{})$^{q+}$						& \numrange{1e3}{1e4}~\cite{Smith_2006}			& \numrange[range-phrase=-]{7}{10}~\cite{Smith_2006}		& 2500~\cite{Leer_2019}				& \numrange[range-phrase=-]{20}{100}~\cite{Smith_2006, iFIB_2021}				& high rate sputtering, volume imaging, implantation of gaseous elements\\

\acs*{gfis}		& (He, Ne)$^+$							& \numrange{1e9}{4e9}~\cite{Ward_2006, Hill_2011, Notte_2016}	& \numrange[range-phrase=-]{0.25}{1}~\cite{Ward_2006, Ernst_1993}	& 0.15~\cite{Notte_2016}				& \numrange[range-phrase=/]{0.5}{4.0}~\cite{Hlawacek_2016, Livengood_2011, Klingner_2020}	& high resolution imaging and nanostructuring, mask repair, implantation of gaseous elements\\

\acs*{lotis}/ \acs*{motis}& (Li, Cs, Cr, Rb)$^+$ & \numrange{1e5}{2.4e7}~\cite{Steele_2017}			& \numrange[range-phrase=-]{0.2}{2}~\cite{Haaf_2018, Moufarej_2017}	& 25~\cite{Knuffman_2013}				& \num{4.9}~\cite{Steele_2017}						& \acs*{sims}, implantation\\

\acs*{nais} & H$^{q+}$, Ar$^{q+}$ 							& \num{4e5}~\cite{Kouwen_2018}						& \numrange{0.9}{2.3}~\cite{Kan_2020}				& \numrange{0.2}{20}~\cite{Kouwen_2018}						& \num{\approx2000}~\cite{Kouwen_2018}								& SEM as an ion source, proton beam writing\\
 \bottomrule
\end{tabular}
\end{center}
\end{table*}
In general, a smaller (virtual) source size means a higher brightness and consequently a higher achievable lateral resolution.
However, for the final resolution of the instrument additional parameters such as the extractable ion current, energy spread, and of course the performance of the ion optics, are also of importance.

In the field of ion sources, the first breakthroughs came in the 1970s, with the development of the \acf{lmis} and the \ac{gfis}.
While the latter evolved from \ac{fim} dating back to the 1950s~\cite{Mueller_1951} and was first employed for \ac{stim} of biological specimens~\cite{LeviSetti_1974,Escovitz_1975}, the \ac{lmis} was originally developed for space thruster applications~\cite{Krohn_1975,Clampitt_1975,Seliger_1979}.

The \ac{lmis} consists of a capillary or a sharp needle tip, wetted with a molten metal.
Through a combination of surface tension and applied electric field, the so-called Taylor cone~\cite{Taylor_1964} is formed, resulting in the extraction of ions from the source apex by field evaporation. 
The most common form of \ac{lmis} is based on gallium, due to its low melting temperature and low vapor pressure.
In fact, the Ga-\ac{lmis} is still the most used source for \ac{fib} instruments, not only because of its high brightness, but also because of its high source stability. 

Directly related to the \ac{lmis} is the \acf{lmais}, which through the use of a variety of alloys enables access to a much wider range of ion species~\cite{Bischoff_2016, Mazarov_2009, Bischoff_2010, Bauerdick_2013, Bruchhaus_2017, Pilz_2017, Gierak_2018, Pilz_2019, Bischoff_2020, Mazarov_2020,Bischoff_2022}, but with strongly varying source lifetimes from minutes to months.
Based on the same source principle, the \acf{ilis} is wetted with a compound that dissociates into molecular anions and cations, such as the ionic liquids C$_6$N$_2$H$_{11}$-BF$_4$ (EMI-BF$_4$), C$_6$N$_2$H$_{11}$-GaCl$_4$ (EMI-GaCl$_4$) or C$_8$N$_2$H$_{15}$-I (BMI-I)~\cite{Lozano_2004, Lozano_2005, Lozano_2005b, Krejci_2017, PerezMartinez_2010, Guilet_2011, Tolstogouzov_2015}. 
The \ac{ilis} is thus capable of producing beams of molecular ions, both positively and negatively charged, but has not yet been implemented commercially. 

The \Ac{gfis} operates at low temperature, producing ions from the field ionization of adsorbed gas atoms, and is characterized by the highest brightness and hence the highest deliverable spatial resolution of all sources developed to date~\cite{Ward_2006, Hlawacek_2014, Hlawacek_2016}.
This is a consequence of the atomically sharp emitter, which consists of only three atoms at its tip, the so-called trimer, whereby each atom emits a beamlet and thus forms a virtual source with a size in the Angstrom range. 
The trimer must be formed by the user and has a typical lifetime between a few days and a few weeks.
The source gases that are widely used for high-resolution imaging and high-resolution (metal-free) milling are He and Ne. 
Operation of the \ac{gfis} with H$_2$~\cite{Aramaki_2011}, N$_2$~\cite{Schmidt_2017, Schmidt_2018b}, Xe~\cite{Lai_2017}, and Kr~\cite{Shichi_2017} has also been demonstrated.
A closely related source technology is the \ac{asis} source~\cite{Purcell1999,Purcell2001}, which uses field emission of adatoms (Au and Ag have been demonstrated) deposited onto a tip made from refractory metal or another inert material with a high field ionization strength.
High resolution is to be expected due to the single atomic ionization site employed in this technology.
However, among other issues, it is the limited lifetime of the \ac{asis} which has hindered its application in actual \ac{fib} instruments so far.

Another source option offering non-metallic ions is the plasma ion source of the so-called \ac{pfib} instrument. 
Plasma ion sources achieve high currents of up to \qty{2}{\micro\ampere} and have a long lifetime, but lower brightness than \ac{lmis} sources.
The plasma can be generated by electron impacts, as in a duoplasmatron~\cite{Lejeune_1968, Smith_2006}, by inductive coupling of alternating currents in a radiofrequency antenna (an \ac{icp})~\cite{Boswell_2014}, or by microwaves in an \ac{ecr} ion source~\cite{Sortais_2010}.

\begin{figure*}[tbp]
    \centering
    \includegraphics[width=0.8\textwidth]{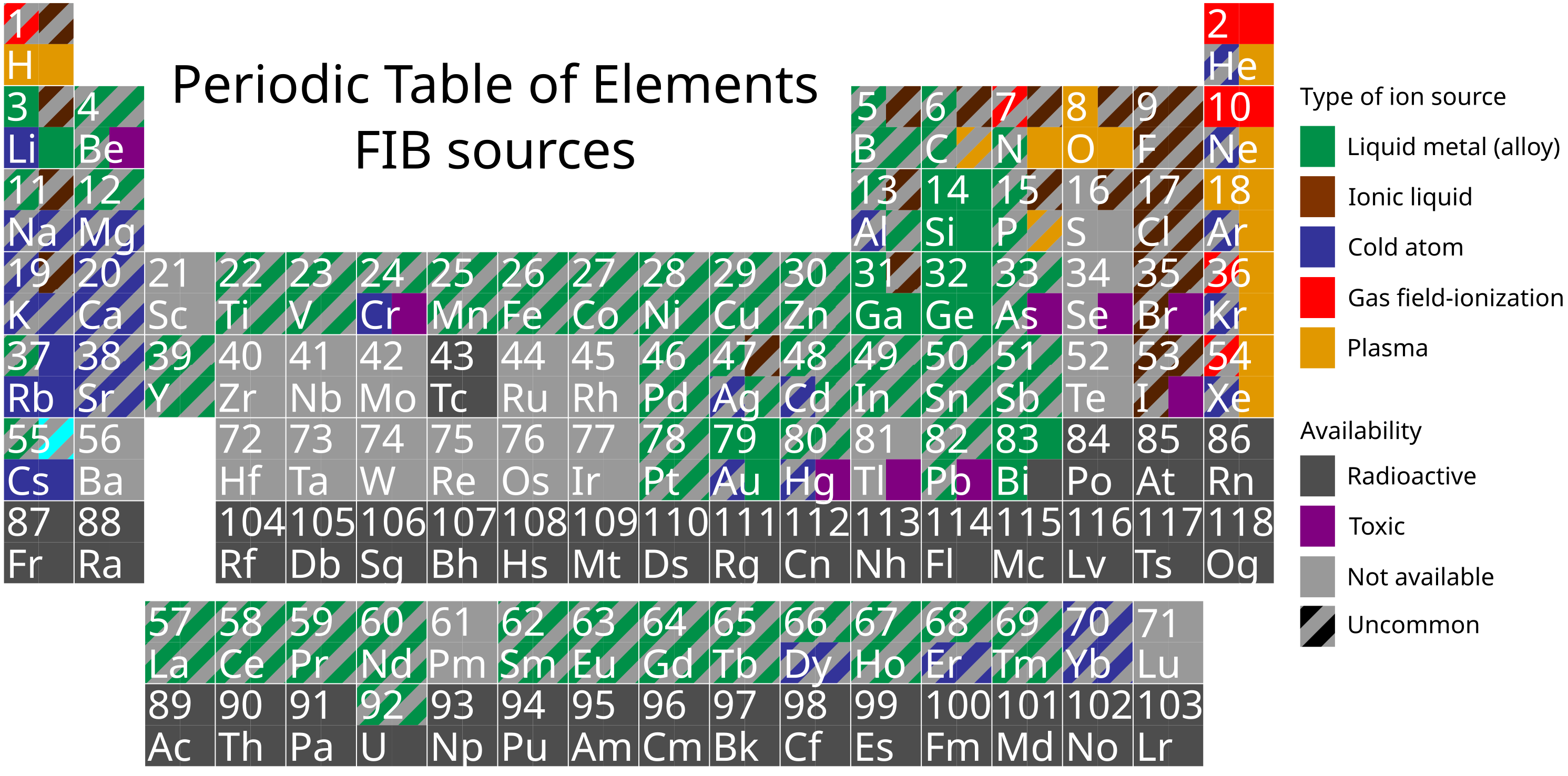}
    \caption{Periodic table of elements of available ion species for \acp{fib}.
    The color of each element shows the corresponding ion source type.
    Ions of elements that are not available, toxic or only available as radioactive isotopes are also indicated by color.
    If ions from a particular element are available using multiple types of ion source, this is shown as labeled.
    Diagonal stripes indicate that the ions of those elements are not widely used and/or are produced with a low yield.}%
    \label{fig:pse_sources}
\end{figure*}

In addition to the systems mentioned above, most of which are commercially available, there are other less common types of ion source. One example is the so-called cold atom ion source, of which there are two varieties: the \ac{motis} and the \ac{lotis}. These use magneto-optical trapping (in combination with laser-cooling in the case of the \ac{lotis}) to generate a trapped cloud of atoms or an intense atomic beam with a (transverse) temperature in the microkelvin regime~\cite{Hanssen_2006, Claessens_2007, Hanssen_2008, Knuffman2011, Kime_2013, Knuffman_2013, Gardner2019}.
The atoms in the trap or beam are then field- or photoionized to produce an isotopically pure, singly-charged ion beam of high brightness and low energy spread~\cite{Geer2007, Haaf_2014, McClelland_2016, Moufarej_2017, Haaf_2018}.
\ac{motis}/\ac{lotis} achieve good resolution at low energies and are suitable for a wide variety of ions~\cite{McClelland_2016}, but require a high degree of sophistication in their design and handling.
Consequently, these sources have so far only been demonstrated for a few ion species (see Table~\ref{table:ion_sources} and Fig.~\ref{fig:pse_sources}).

Another example is the \ac{nais} that uses an electron-impact gas ion source~\cite{Jun_2011,Xu_2017} and can be installed on standard \ac{sem} instruments to deliver ions from all noble gases, or protons.
Further less common source types are Paul traps~\cite{Schnitzler_2010}, the \ac{ebit}~\cite{Levine_1988,Raecke_2020}, the multicusp plasma ion sources~\cite{Ji_2002}, the solid electrolyte ion source (SEIS)~\cite{Tolstoguzov_2018} and the \ac{esi}~\cite{Nojima_2020}.

A systematic overview of all ion species currently available for \ac{fib} instruments is displayed in the form of the periodic table of elements in Figure~\ref{fig:pse_sources}.
The most common types of ion sources that are used for \acp{fib} and their key parameters are compared in Table~\ref{table:ion_sources}.

Crucial factors to be optimized for the routine use of new ion sources are source lifetime and stability.
Lifetime issues are usually related to contamination and can often be handled by working in cleaner conditions.
However, in other cases, lifetime and stability issues can be traced to the source itself.
For example, the formation of stable beams from certain elements, such as Al and P extracted from a \ac{lmais}, remains challenging.
The exact reasons for these complications are not yet fully understood. 
Source stability is also of particular importance for large volume sputtering applications, as well as for \ac{fibid} and resist-based \ac{fib} lithography.
For maximum patterning fidelity, these applications require \acp{fib} with highly stable emission currents and beam positioning.
Examples of poor stability are the short term current fluctuations and long time current drift of the Ne-\ac{gfis}~\cite{Klingner_2020}, and effects such as pulsation, droplet and globule emission from the source tip in \ac{lmis} and \ac{lmais}~\cite{Orloff_2009}.

\subsection{Beam transport}%
\label{subs_optics}

\begin{figure}[tbp]
   \centering
   \includegraphics[width=\columnwidth]{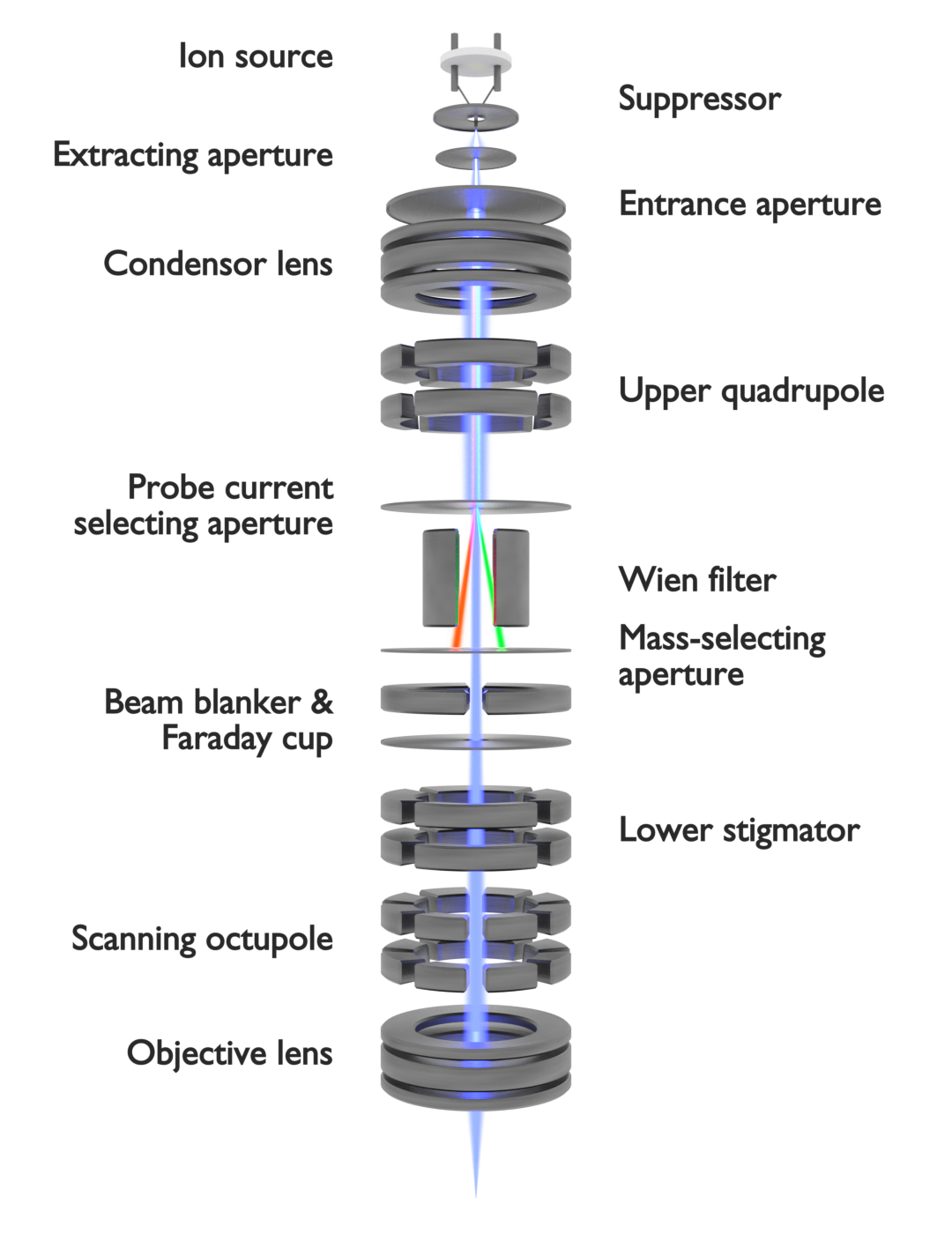}
   \caption{Principal components for beam transport in a \ac{fib} column (not to scale). Components labeled on the left are found in standard Ga-\ac{fib} columns, whereas labels on the right indicate components usually only found in special-purpose columns.}%
   \label{fig:fib_column}
\end{figure}

The main components responsible for transporting the beam in a \ac{fib} column from the source to the sample are shown in Fig.~\ref{fig:fib_column}.
The labels on the left correspond to components found in more standard instruments and the labels on the right correspond to additional components found in more specialized instruments.
To summarize: After extraction from the source, the ions are formed into a beam by the condenser lens (often after passing through an entrance aperture) and then guided toward the probe current selecting aperture (often via a quadrupole lens).
The current selecting aperture defines the probe current by selecting a small portion of the beam and thereby also reducing its angular dispersion.
To monitor the beam current, the beam is deflected into a Faraday cup.
The lower part of the column often houses a stigmator for final beam shaping, plus an octupole for beam scanning and a second electrostatic lens (objective) for beam focusing.
In more specialized instruments, a Wien filter is used to select isotopes, i.e.\ ions of a specific mass-to-charge ratio.

\begin{figure}[tbp]
   \centering
   \includegraphics[width=\columnwidth, clip, trim=100 0 200 0]{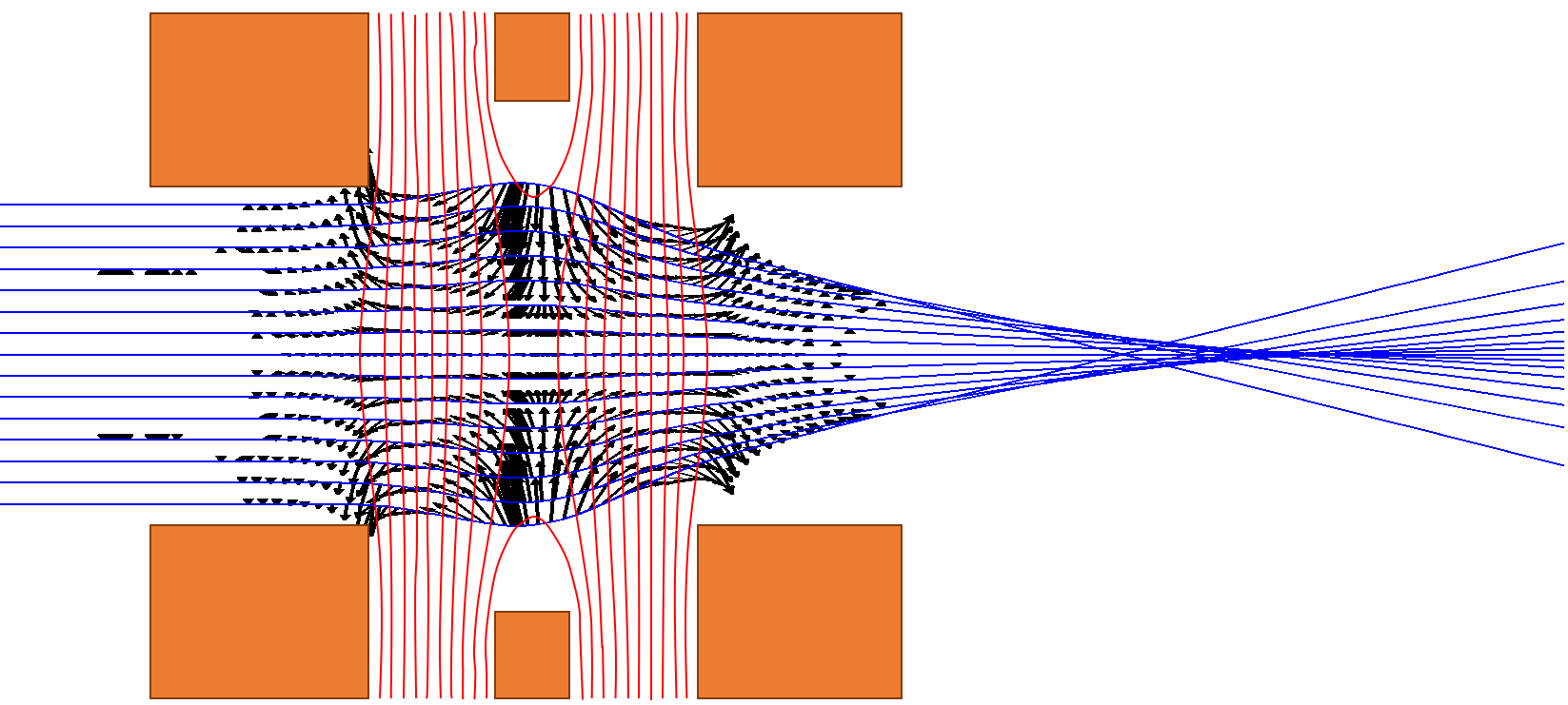}
   \caption{\label{fig:Einzel_lens} Example of an electrostatic (Einzel) lens, showing three cylindrical electrodes (in orange) where the central one is biased to focus the ions onto a given position. Equipotential lines are drawn in red, ion trajectories in blue, and the black arrows indicate the direction and magnitude of the electrostatic force.}%
\end{figure}

The condenser and objective lenses are convergent electrostatic lenses with a cylindrical geometry and are used to steer and focus the ion beam.
They are composed of several electrodes (typically three, i.e.\ an Einzel lens), which are electrically biased to generate electric fields that change the trajectories of the transiting ions through the electric force $\vec{F}=q\vec{E}$ (see Fig.~\ref{fig:Einzel_lens}).
Electrostatic lenses can be operated in retarding mode where the ion energy is reduced inside the lens, or in accelerating mode where the ion energy is increased inside the lens. 
Figure~\ref{fig:Einzel_lens} shows an example of a lens operated in retarding mode.  
Given the curved shape of the field lines inside the lens, the electric forces imposed on the ions have a radial component, in addition to the deceleration and acceleration axial components of these forces. 
The radial force components result in the overall focusing effect of the lens~\cite{Orloff_2009}. 
The focal length of an electrostatic lens depends on the ion energy, the charge state of the ions, the voltage applied to the lens, and the lens geometry (and not on the ion mass, as would be the case for a magnetic lens). 
For a lens operated at a given voltage, the focal length will be smaller if retarding mode is implemented compared to accelerating mode.

The beam blanker comprises an electrostatic deflector that diverts the ion beam into a Faraday cup. This allows the user to both blank the beam to prevent ion impingement on the specimen as needed, and also to measure the probe current. 
Beam blanking is an important feature, not only for standard applications such as nanostructuring by \ac{fib} milling or \ac{fibid}, but also for ion implantation tasks. 
In the most extreme case, beam blanking must be fast enough so as to allow implantation down to the single ion level from a non-deterministic ion source. 
In this context, fast means that for an assumed primary ion current of \qty{1}{pA} it is necessary to deliver a \qty{10}{ns} ion pulse in order to achieve the correct probability, following Poisson statistics, of no more than one ion per pulse. 
Fast blanking is achieved using dedicated electronics and electrodes located at a specific position in the column such that the beam is blanked in a really short time without creating any beam tail artifacts. 
Detection on the sample must also be able to count the arrival of each ion, one by one (see Section~\ref{subs_detectors} on detectors).

Other optical elements present in nearly all \ac{fib} columns are quadrupoles and octupoles, which are used to steer the beam and are key for proper alignment of the beam with respect to the major optical elements in the column (such as the lenses and apertures). 
These elements are comprised of two sets of either four or eight electrostatic deflectors, which allow the user to tilt and shift the beam (by applying two counteracting deflections) onto the vertical axis of the subsequent optical element.
In the lower part of the column, close to the final lens, a stigmator is often implemented, which allows the user to correct the shape of the beam by applying electric fields that compress (or expand) the beam along directions perpendicular to the optical axis.
All three of these elements (the quadru-/octupoles and stigmator) enable some compensation for the aberrations introduced by the imperfections of the lenses as well as correction for the mechanical alignment of the column.
In the \ac{fib} instruments commercially available today, chromatic or spherical aberrations are not routinely corrected. 
This is mostly because the magnitude of both is strongly defined by the source.
However, theoretical papers have shown the possible benefits of such corrections~\cite{Zhukov2007,Martin2014} and prototypes have been built using an electrostatic corrector~\cite{Nagata_2019, CEOSGmbH2023}.

\begin{figure}[tbp]
   \centering
   \includegraphics[width=\columnwidth]{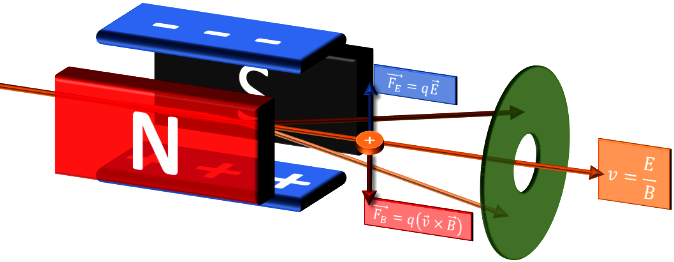}
   \caption{Principle of a Wien filter, with orthogonal electric and magnetic fields and an aperture to select the desired ion velocity.}%
   \label{fig:Wien_filter}
\end{figure}

More specialized columns are equipped with a Wien filter~\cite{Wien_1898,Galejs_1978}, by which perpendicular electric and magnetic fields separate the ions in the primary beam according to their mass-to-charge ratio (see Fig.~\ref{fig:Wien_filter}). 
This is necessary when the ion source produces ions from several elements and/or ions of different charge states.
The electromagnetic selection of ion species has of course no effect on any neutral atoms that may be present in the beam.
To prevent neutral atoms reaching the specimen, the beam can be sent through an electrostatic chicane blanker~\cite{Cassidy2019}, whereby all non-deflected particles (neutrals) are blocked such that only the charged particles (ions) reach the sample.

As mentioned previously, the performance of a given \ac{fib} instrument ultimately depends on the specifications of the ion source.
Most of the optical elements typically used in today's \ac{fib} columns do not correct for any beam aberrations, although they might create some due to inherent/manufacturing imperfections.
If the source has a poor performance, this will inevitably be propagated to the sample.
Consequently, adjusting the design of the beam transport elements on a state-of-the-art \ac{fib} instrument will typically only result in small improvements in the overall \ac{fib} performance~\cite{Liebl_2008, Orloff_2009, Rose2009}.

The main gap between the desired performance of a \ac{fib} instrument and its actual performance is currently in the area of low energy beams (\qty{<2}{keV}).
With the ongoing trend to reduce the dimensions and increase the complexity of (3D) device and sample architectures, the interactions between the ion beam and the sample become ever more critical. 
Thus, the penetration depth and straggle of the ions inside the material must be minimized. 
The key way to reduce these effects is to decrease the energy of the beam, but the consequence is an increase in beam spot size. 
This increase in beam spot size is due to chromatic aberrations from the energy spread of the ions in the beam, whereby the relative energy spread and hence aberrations become more pronounced when the beam is retarded.

Finally, the performance of current \ac{fib} systems is also limited in terms of the attainable processing speeds. Using a single focused beam requires scanning with varying exposure and move/blank sequences, and is inherently serial. 
In contrast, template masking of a broad ion beam on the sample can provide much higher throughput, but at the expense of flexibility and spatial resolution.
Therefore, it has been theoretically proposed to combine masking of a broad beam with full control of single beams in a multibeam approach that significantly improves throughput for nano-applications~\cite{Lischke1989}.
A working proof-of-concept demonstration of such a multibeam \ac{fib} column has been built around a controllable array of ion beamlets~\cite{Kapl2011}.
In brief, an aperture plate splits a broad parallel ion beam into a large number of \qty{2.5}{\micro\meter}-wide beamlets. 
Each of these 43,000 beamlets can be individually deflected by an array of apertures with adjacent electrostatic electrodes, fabricated using \ac{cmos} technology. 
All the beamlets are then passed through 200x reduction optics, which blocks any deflected beamlets. 
The remaining beamlets irradiate a user-defined pattern of pixels with a resolution of less than \qty{20}{nm}. 
This multibeam technique has been optimized for electron irradiation~\cite{Platzgummer2013} and is commercially available as a mask writer~\cite{Klein2016}, but further development of the concept for different ion species is highly desirable.

\subsection{Analytical tools and detectors}%
\label{subs_detectors}

Irradiating a sample with an ion beam to trigger and measure a response is a very common analytical technique, and 
\ac{fib} instruments that perform imaging, local irradiation, milling, etc., are typically equipped with various accessories to confer a range of analytical capabilities. 
For imaging purposes, the \acp{se} emitted from the sample surface are routinely detected using an \ac{et}~\cite{Everhart_1960}. The \ac{et} is scintillator-based, converting \ac{se} strikes to photons inside the sample chamber, that then travel via a light guide to a photomultiplier outside the chamber.
As a complementary imaging channel, \acp{si} can be detected using a \ac{tic} (typically in positive \ac{si} detection mode using a Faraday cup  or a channeltron).
In addition to these imaging modes, various other analytical techniques are implemented on \ac{fib} platforms; a general overview is given in Table~\ref{tab:analytic}.

\begin{table*}[btp]
   \caption[\ac{fib} based analytic methods]{Overview of \ac{fib} based analytical methods.}%
   \label{tab:analytic}
   \begin{tabular}{@{}m{4.5cm}m{2.5cm}m{3.0cm}m{2cm}m{2cm}m{2.0cm}@{}}
      \toprule
      Technique & Primary ion & Detected signal & \multicolumn{2}{p{4cm}}{\centering Spatial resolution} & Refs. \\ 
                &  & & \centering lateral & \centering depth & \\
      \midrule
      \acs*{qms}-\acs*{sims} &\multirow[c]{3}{=}{He$^+$, Ne$^+$, Ga$^+$, Cs$^+$, Xe$^+$, O$^+$, Bi$^+$, Bi$^{n+}$, Au$^+$, Au$^{n+}$, O$^-$}&\multirow[c]{3}{=}[-1ex]{Secondary ions}& \qty{100}{nm}& \qty{<20}{nm} & \citenum{LeviSetti1980,LeviSetti1986,Waugh_1986,McMahon_2002,Kubis_2004,Chater_2014,Chater_2016,Brugge_2021}\\ 
      Magnetic sector \acs*{sims} & & & \qty{<15}{nm} & \qty{<2}{nm} & \citenum{Nihei1987,LEVISETTI1994, Wirtz_2012,Vollnhals_2017,Wirtz_2019,Kim_2019,Ball_2020,Ovchinnikova_2021,Audinot_2021}\\ 
      \acs*{tof}-\acs*{sims}/orthogonal \acs*{tof} & & & \makecell[l]{\qty{<50}{nm}\\ (\qty{<30}{nm})}& \qty{<10}{nm} & \citenum{Waugh_1986,HAYASHI_2011,Whitby_2012,Kollmer_2013,Stevie_2014,Klingner_2016,Klingner_2019,Pillatsch_2019,Priebe_2019,Rickard_2020,Priebe_2020, Priebe_2021a, Priebe_2021b}\\ \addlinespace
     
      \acs*{snms} & Ga$^+$ & Post-ionized neutrals & \qty{20}{nm} & & \citenum{Nagata_2019, Bajo_2016, Ebata_2012,Draude_2012, Franzmann_2018} \\ \addlinespace

     \acs*{stim} & He$^+$ & \makecell[l]{Transmitted ions/neutrals,\\backside \acsp*{se}} & \qty{<5}{nm} & & \citenum{Hall_2013, Emmrich_2016, Woehl_2016, Kavanagh_2017, Wang_2018, Kavanagh_2020, Serralta_2020, Notte_2010,DeCastro_2021}\\ \addlinespace

      Backscattering spectroscopy & He$^+$ & \makecell[l]{Backscattered\\ions/neutrals} & \qty{\approx50}{nm} & & \citenum{Klingner_2016, Sijbrandij_2008, Sijbrandij_2010, Heller_2016, Bell_2009, Rabalais_2002, Abo_2012, Gastel_2015, Hlawacek_2012, Veligura_2012, Hijazi_2019} \\ \addlinespace

      Backscattering yield & He$^+$ & \makecell[l]{Backscattered\\ions/neutrals} & \qty{5}{nm} & & \citenum{Gastel_2012, Hlawacek_2016a} \\ \addlinespace

      Ion induced \acs*{se} spectroscopy & He$^+$, Ar$^+$ & \Aclp{se} & several nm & & \citenum{Stehling_2018, Masters_2015, Anikeva_2019, Suzuki_2008, Farr_2021, Petrov_2018, Petrov_2011, Ramachandra_2009, Pearson_2011} \\ \addlinespace

      \acs*{iiaes} & \makecell[l]{Si$^+$, Ne$^+$, Ar$^+$,\\ Au$^+$, Kr$^+$, Ga$^+$} & Auger electrons & \qty{10}{nm} & & \citenum{Polak_1995, Whaley_1984, Xu_1993, Gallon_1995, Parvaneh_2014, Pereyaslavtsev_2019} \\ \addlinespace

      \acs*{pixe} & He$^+$, Ne$^+$, Ga$^+$ & X-rays & \qty{100}{\nano\meter} & & \citenum{Giannuzzi_2006, Giannuzzi_2009} \\ \addlinespace

      \acs*{il} & He$^+$ & Photons & \qty{20}{nm} & & ~\citenum{Scipioni_2007,Boden_2012, Franklin_2012, Veligura_2014a, Veligura_2014, Veligura_2016, Veligura_2014b, Ogawa_2012}\\
      \bottomrule
   \end{tabular}
\end{table*}

\Acf{sims} enables the mapping of elemental/chemical compositions in the form of 2D/3D images or depth profiles with high sensitivity (down to the ppm level) in combination with a high dynamic range (i.e.\ a given element can be measured over a concentration range of several orders of magnitude). 
In principle, all elements including their isotopes can be measured.
However, for a truly quantitative analysis, reference samples are required since the ionization yields strongly depend on the local environment in the sample (known as the matrix effect)~\cite{Stevie_2005} 
Several implementations of \ac{fib}-\ac{sims} systems have been explored, with three main system types described in the literature for units installed on both \ac{fib} and \ac{fib}-\ac{sem} instruments:

(1) Historically, \acf{qms} systems have been used on \ac{fib} platforms due to their simple design, low weight, and reduced costs~\cite{Waugh_1986, Chater_2014}. 
In a \ac{qms}, only ions with a specific mass-to-charge ratio are able to travel through the applied quadrupole field.
Ions with different mass-to-charge ratios travel on unstable trajectories, thus leaving the mass spectrometer before the final exit aperture and so are not counted.
\acp{qms} therefore have the disadvantage of not allowing parallel detection (i.e.\ only one mass can be detected, the detection of several masses requiring sequential analyses and hence a duty cycle), and also have a lower performance in terms of sensitivity.

(2) More recently, various orthogonal and linear \acf{tof}-based mass spectrometers have been introduced~\cite{Klingner_2019, Pillatsch_2019}.
Here, mass separation is achieved because ions with different mass-to-charge ratios will reach different velocities;
measuring their time-of-flight allows the mass of the ions to be inferred.
The \ac{tof} systems offer the advantage of parallel detection, but since pulsing of the primary or secondary ion beam (or both) is required, a duty cycle results.

(3) Magnetic sector \ac{sims} systems can offer parallel detection if using so-called continuous focal plane detectors~\cite{Nihei1987,Wirtz2020}, and the highest sensitivity enabling high-resolution imaging applications~\cite{DeCastro_2021}. 
In a magnetic sector mass separator, ions are forced onto circular trajectories by a perpendicular magnetic field.
The radius of curvature of this trajectory depends again on the mass-to-charge ratio of the ion.
Parallel detection of all masses can be achieved using focal plane detectors and the appropriate spectrometer geometry.
Magnetic sector systems are operated in \ac{dc} mode i.e.\ they feature a 100\% duty cycle.
In terms of disadvantages, these systems typically have a larger footprint and are heavier because of the integrated electromagnet.

By combining a \ac{gis} with \ac{fib}-\ac{sims}, further improvements in analytical output have been demonstrated.
For example, in work combining \ac{fib}-\ac{tof}-\ac{sims} with a \ac{gis}, ionization probabilities and hence \ac{si} signals were found to significantly increase by 2--3 orders of magnitude, thereby improving the quality of 2D and 3D chemical maps~\cite{Priebe_2019}. 
Furthermore, it was observed that co-injection of XeF$_2$ during the ion bombardment can reduce mass interference~\cite{Priebe_2020, Priebe_2021a} and invert the polarity of the negatively-charged \acp{si} to positive~\cite{Priebe_2021b}, thus allowing the collection of more complete chemical information.
Enhancements in \ac{si} yields have also been obtained with magnetic sector \ac{sims} instruments using Cs deposition and O$_2$ flooding to boost the yield of negative and positive secondary ions, respectively~\cite{Frache_2010,Philipp_2006}.

Related to \ac{sims}, \acf{snms}~\cite{Draude_2012,Franzmann_2018} can obtain similar information, analyzing sputtered neutrals through laser-based post-ionization, and has the advantage of matrix-independent ion yields. 
However, while resonant post-ionization is an efficient process, non-resonant ionization yields are low.
\Ac{snms} requires significant experimental effort and is therefore associated with high costs.

\Acf{stim} is performed using the ions of light elements. It can provide mass-thickness contrast and also crystal structure information due to ion channeling~\cite{Serralta_2020}.
Several different \ac{stim} implementations have been investigated, most based on the collection of secondary electrons generated from impact of the transmitted ions on a conversion plate~\cite{Emmrich_2016,Notte_2010,Hall_2013}, but also some using direct detection of the transmitted ions~\cite{Serralta_2020,Kavanagh_2017,Woehl_2016}.
\ac{stim} is a quasi-nondestructive imaging technique, and for biological specimens has been shown to deliver structural contrast comparable to \ac{stem}~\cite{DeCastro_2021}.
Similar to \ac{stem}, \ac{stim} demands thin samples (\qty{<100}{nm}).
\ac{stim} can also be used to obtain high-resolution images that can be correlated with elemental/chemical maps determined by \ac{sims}~\cite{Vollnhals_2017}.

Backscattering spectroscopy in a \ac{fib} instrument is similar to \ac{rbs}, but is performed at much lower energies of around several tens of keV. 
This results in multiple scattering events due to the relatively large nuclear scattering cross-sections.
Even when performing time-consuming simulations of the ion spectra, quantitative results from the raw energy spectra are difficult to obtain.
However, it has been shown that when coupled with a \ac{tof} setup, backscattering spectroscopy can deliver damage-free depth-resolved elemental compositions~\cite{Klingner_2016}. 
Signal levels are good, since in addition to the \acp{si}, the backscattering method is also sensitive to the large number of neutrals generated during the \acl{pi} impact.

One can also analyze the yield of the backscattered primary particles, for which larger solid angles and higher signals can be achieved. This is accomplished using an annular \ac{mcp} located beneath the objective lens of the microscope.
While the backscatter yield approach is feasible regardless of \ac{fib} instrument type, it is most useful in a \ac{him}.
In this case, the backscatter yield of the primary He ions is high enough (in particular upon interaction with heavier elements), that damage from the primary beam can be neglected~\cite{Klingner_2016}.
Backscatter yields depend strongly on the atomic number of the target~\cite{Sijbrandij_2008}. To a certain extent elemental analysis is also possible, but requires prior knowledge of the elements in question~\cite{Gastel_2015}.
The backscatter approach has been shown to reveal elemental contrast from buried layers~\cite{Gastel_2012} and has also been applied to biological specimens~\cite{Ma_2017,Bidlack2014}.
Since backscatter yields depend strongly on ion channeling effects, this approach can be used to map crystal orientations~\cite{Magnuson1963,Magnuson1963a,Briggs2001,Veligura_2012,Hlawacek_2016a}. 
\Ac{se} yields are in fact also influenced by ion channeling and provide stronger signals, hence crystal orientation mapping based on the detection of \acp{se} offers a simple alternative~\cite{Veligura_2012} and has been shown to enable the visualization of interfacial nanoscale dislocation networks in thin-film alloys~\cite{Hlawacek_2015}.
The ability to infer crystallographic information from the \ac{se} yield has been extended to other ions and automated~\cite{Langlois_2015}.

Ion induced \ac{se} spectroscopy (using He ions) draws on variations in \ac{se} energy to map chemical variations on a sample surface~\cite{Rodenburg_2014,Anikeva_2019,Petrov_2016,Petrov_2010}.
The approach allows the user to maximize imaging contrast, thus permitting short beam exposure times, which is beneficial for beam-sensitive samples.
However, a quantitative application of the \ac{se} spectroscopy method is currently limited by the complicated nature of the data obtained and the lack of a suitable reference database. 

\Acf{iiaes} allows the chemical identification of surface layers, including bond structure, via the energy-resolved detection of Auger electrons emitted following an ion induced inner-shell electronic transition~\cite{Parvaneh_2014,Parvaneh_2014a}.
For elements in the third row of the periodic table, \ac{iiaes} has a superior signal-to-noise ratio compared to electron beam based \ac{aes}. 

Ion/\acf{pixe} in a \ac{fib} instrument is essentially the equivalent of \ac{eds} in the \ac{sem}, whereby characteristic X-rays (used for elemental mapping) are generated as a result of particle bombardment. However, in the \ac{fib} case, \ac{pixe} suffers from extremely low X-ray yields and hence has not proved practical thus far. 

The final entry in Table~\ref{tab:analytic}, \acf{il}, has been tested on various materials~\cite{Franklin_2012}. However, the authors of these works concluded that the merits of this technique only really emerge in the case of in-situ characterization of beam induced defects~\cite{Veligura_2014a}, and for the localization of rare earth elements~\cite{Veligura_2016}.

Specialized ion detection methods for \ac{sii} also deserve mention (see Section~\ref{subs_defects} for more detail).
These methods can be categorized into pre- and post-detection techniques.
In the pre-detection category, one example is the use of an \ac{icd} incorporated into the \ac{fib} column that registers the passage of a single ion (or ion bunch) and is coupled with fast blanking electronics~\cite{Raecke_2019,Raecke_2022}.
Post-detection can involve standard \ac{se} detection~\cite{Schenkel_2003} or more complex sample-integrated detection schemes, such as \ac{ibic} or source-drain current measurements~\cite{McCallum_2012,Hopf_2008,Johnson_2010,Robson_2022}.

\subsection[Other \acs*{fib} accessories]{Other \acs*{fib} accessories}%
\label{subs_accessories}

In addition to the above-described key components, a number of other accessories have been developed that extend the capabilities of \ac{fib} instruments even further. A summary of these is given below.

The most common addition to a \ac{fib} instrument is an electron column (\ac{sem}). Vice-versa, \ac{fib} columns are often added to \acp{sem}.
The electron column enables correlative imaging, in-situ monitoring of milling processes, \ac{fib}-\ac{sem} volumetric reconstructions, and SEM-based analytics such as \ac{eds}. The \ac{sem} addition facilitates sample navigation, since by imaging with electrons rather than ions, beam damage to the sample can be rendered negligible. 
Certain \ac{fib}-\ac{sem} implementations also allow automated metrology and analysis~\cite{Giannuzzi_2005, Yao_2007}.

For high-resolution, large-area direct patterning applications, laser interferometer stages become critical, since the positioning accuracy and stability of standard mechanical stages are not sufficient for these high-end applications.
In conjunction with specialized sample holders, laser interferometer stages enable a sample positioning accuracy in the nanometer range over a lateral distance of several \qty{100}{mm}~\cite{Laserstage_2021}.

A further rather common addition is a \acf{gis}, which enables \acf{fibid} and, in combination with an electron column, also \acf{febid}~\cite{Utke_2012}.

\ac{fib} columns are also sometimes operated in conjunction with an electron flood gun, 
whereby in-situ charge neutralization is achieved by illuminating an insulating specimen with low energy electrons. 
In particular for \ac{him}, this enables imaging of insulating samples at high spatial resolution without the need for conductive coatings~\cite{Golzhauser2016a}(see Section~\ref{sss:him-bio-imaging}), and more generally, \ac{fib} milling of insulating samples~\cite{Stokes_2007}.
A low resolution approach to allow charge-free macroscopic sample navigation is to use an optical camera.
The latter has been extended to enable in-situ fluorescence~\cite{Boltje_2022} and Raman spectroscopy~\cite{Timmermans_2015}.

Various types of micro/nanomanipulators have been developed for intuitive control and mechanical manipulation of micro- to nanoscale objects, including in-situ lift-out of lamella specimens for \ac{tem}~\cite{Langford_2004a}, and to allow local electrical connections for various in-situ experiments as demonstrated for \ac{sem}~\cite{Pawley_1972a,Shi_2016,Jiang_2017a}.

Through the addition of a femtosecond laser ablation system, large amounts of material can be removed very efficiently from around the final target area for subsequent finer milling with the \ac{fib}. This enables higher throughput for applications such as advanced package failure analysis and process optimization in the semiconductor industry~\cite{Stegmann_2011a}.

Numerous add-ons for a wide range of in-situ characterization experiments have also been developed, both by manufacturers and by researchers. 
These solutions include electrical probing stations, systems using Peltier elements for sample heating or cooling (not to be confused with more complex cryo-FIB add-ons), in-situ mechanical testing, plasma-based sample cleaners, and automatic laser-based height sensing.
Other often sought-after add-ons are inert gas transfer boxes to allow oxygen-free loading (and unloading) of air-sensitive samples~\cite{Stephant_2018}. 
Various solutions for incorporating \ac{afm} into \ac{fib} and \ac{fib}-\ac{sem} instruments have also been developed, e.g. Ref.~\citenum{Andany_2020}, which can be implemented in  combination with \ac{fib}-\ac{sims} to allow an assessment of surface roughness and thus improve the accuracy of the 3D reconstructions~\cite{Whitby_2012}.
A recent review article summarizing state-of-the-art solutions for in-situ characterization and micro-/nanomanipulation has been published by~\citet{Shi_2016}. 

\subsection{\acs*{fib} processing of radioactive samples}%
\label{sec:subs_radioactive}

The investigation of radioactive samples using a \ac{fib} system raises a number of critical issues (legal and technical) that need to be addressed.
The reason the \ac{fib} is such a useful instrument in the analysis of these samples is that it allows the preparation of small-scale specimens (by \ac{fib} milling), which by default have significantly reduced radioactivity levels. 
These small-scale samples can then be handled much more easily for a wide range of characterization experiments, both in-situ and ex-situ, including specimens for small-scale mechanical testing~\cite{Colldeweih2019}, 3D volume imaging~\cite{Teague2014,Baris2018}, lamellae for \ac{tem} or synchrotron investigations~\cite{Miller2012,Teague2014,Kuri2018}, and needle-shaped specimens for \ac{apt}~\cite{Kautz2021}.

The exact protocols to follow may vary widely depending on the national guidelines, individual facility, and the local safety regulations.
For example, individual sub-samples prepared from a larger radioactive sample may still be subject to radiation protection guidelines even in cases where radioactivities are below a clearance limit for which the sample can be considered quasi non-radioactive. 
The type of material, form of radioactivity, and isotopes involved, are all key factors to take into account.

\Ac{fib} instruments that are used for radioactive samples can be categorized as follows, depending on the laboratories where they are installed and on the shielding implemented:
\begin{enumerate}
  \item Instruments classed for ``low---medium'' radioactivity samples. These instruments are typically also used for experiments with non-radioactive samples, and are installed in labs with radiation protection guidelines and moderate shielding. 
  \item ``Hot cell'' instruments that are highly shielded (lead wall around the whole instrument) and may even allow spent fuel samples.
\end{enumerate}
Irrespective of the particular regulations, the main concerns when working with radioactive samples in \ac{fib} instruments are: 
\begin{itemize}
  \item Dose limits for the operator when loading and unloading the sample
  \item Contamination of the instrument to the extent that system maintenance becomes difficult
  \item Cross contamination between samples in the case of work with radioactive and non-radioactive samples
  \item Lifetime of detectors and other add-ons installed on the instrument (\ac{eds}, \acs{ebsd}, etc.).
\end{itemize}
Regarding contamination of the instrument itself, it has been determined that most of the milled material that does not redeposit on the sample is deposited on the pole piece and surrounding areas~\cite{Evelan2013,Aitkaliyeva2014}.
In the case of materials that generate dust, the sample stage can also easily become contaminated.
One approach to address contamination of the pole piece is to use a dedicated pole piece insert for experiments with radioactive samples.
\ac{fib}-\ac{sem} instruments with \ac{esem} capabilities can also be helpful here, as these provide another aperture for low-pressure mode, and the water vapor that is typically injected helps to volatilize the radioactive sputtered material. 
Frequent cleaning of areas prone to contamination is another effective strategy to increase the lifetime of the instrument.
The inside of the chamber can also be covered to protect from contamination.
Some approaches use sample shields to reduce the spread of the sputtered material by capturing most of it inside the shield. 
However, sooner or later, the entire \ac{fib} chamber will need to be treated as radioactively contaminated equipment and handled appropriately.

\subsection{Software and correlative approaches}%
\label{subs_software}
 
The software infrastructure of a \ac{fib} instrument needs to control various elementary subsystems, e.g.\ the vacuum system and sample stage, as well as the ion source and column functions (beam extraction parameters, aperture selection, column alignments, etc.). 
This is implemented by the manufacturer using a \ac{gui} that allows control of all \ac{fib} system components and integrated add-ons. 
Customized \ac{fib} setups employing e.g.\ specialized, non-commercial \ac{fib} columns or non-standard add-ons, usually use separate control software developed for these specific use cases, which are often difficult to set up and maintain due to closed hardware interfaces and patent issues.  
Regardless of the particular configuration, software control for the following applications is required:
\begin{itemize}
  \item Imaging (including 3D volume imaging) and elemental analysis using various detectors; see Sections~\ref{subs_detectors},~\ref{subs_imaging}, and~\ref{subs_analysis}
  \item Milling, ion irradiation and implantation; see Sections~\ref{subs_topdown} and~\ref{subs_defects}
  \item Gas-assisted processing and resist-based lithography; see Sections~\ref{subs_fibid} and~\ref{subs_recent_adv}
\end{itemize}
All of the above demand a scan generator that controls the path of the ion beam on the sample with high precision and repeatability. 

Beam scanning is realized using a \ac{dac} patterning board that determines the number of addressable pixels in each direction (e.g. 65,000 for 16 bits) and sets the voltages of the beam deflection system accordingly.  
The beam path is then a sequence of pixels with a certain dwell time and spacing depending on the field of view.
Simple line-by-line scanning routines are used for imaging and elemental analysis, and the respective detector output is synchronized to assign a signal (e.g.\ a gray value or a mass spectrum) to each pixel position. 

For subtractive processing, deposition, resist lithography, etc., special beam paths are needed.
Here, manufacturers typically offer software tools that enable the creation of geometric shapes that are rastered according to a chosen set of parameters.
Grayscale patterning offers an even easier way to realize complex shapes and the milling of 3D profiles, by loading an image in which the grayscale values of the individual pixels correspond to the local relative ion doses to be applied.
The latter is an appealing plug-and-play solution for markers, text, and other non-quantitative designs, yet it is difficult to optimize systematically and is therefore not commonly used for actual structures.
In the case of more advanced patterning needs, design files can be imported from lithographic software, assigned with parameters, and rasterized, thus enabling complex and large-scale structuring tasks~\cite{Wanzenboeck_2011}.
Standalone lithography systems from third-party manufacturers can also be directly integrated as accessories via the use of an external scan control unit for the \ac{fib} column.
Such systems typically also have access to stage control, allowing multi-step processing with marker-based registration or stitching, the latter requiring a laser interferometer stage for the highest level of positioning accuracy.

The capabilities of these various patterning systems are sufficient for many applications, but reach their limits when more complex geometric shapes and the highest spatial resolution are required, and/or when the position of the beam must be known with certainty at all times.
This is because with these systems, the user has limited control over the actual raster process and possible auxiliary routines can be hidden in the proprietary software (e.g.\ for stabilization of the beam position). 
The most robust solution is to address pixels directly on the patterning board using a point cloud, which can be encoded in so-called stream files or deflection lists. 
This allows arbitrarily rasterized beam paths to be executed by the integrated patterning software. 
Various groups have developed codes to generate such beam paths and the corresponding point clouds for specific tasks~\cite{Jamali_2014,Cui_2017}. 
For arbitrary geometric shapes, an open-source Python-based solution is available, which allows the generation of patterns and geometry-adapted beam paths that can be both variably rasterized and optimized~\cite{Deinhart_2021}.
For the creation of 3D profiles, approaches based on \ac{cad} have also been developed~\cite{Lalev_2008,svintsov_fib_2009,kim_multiple_2012,Felicis_2017,Niessen_2019}. 
These solutions implement material removal via thin slices plane-parallel to the sample surface, which allows systematic optimization since the amount of material removed per slice remains constant (a significant advantage over the grayscale patterning mentioned above).
Modeling of the milling process can also be used to take into account angle-dependent sputtering~\cite{svintsov_fib_2009} and redeposition~\cite{lindsey_inverse_2014}, or more generally, non-isotropic surface erosion~\cite{kim_multiple_2012}. For more detail see Sections~\ref{subs_bca} and~\ref{subs_cont}. 

In addition to beam control, another software need focuses on enhancing the analytical capabilities of the instrument. 
Commercial \acp{fib} are often coupled to a \ac{sem} column and/or equipped with analytical instrumentation such as \ac{sims}.
Most \ac{fib} instruments are thus furnished with electron detectors for the collection of \acp{se} to allow for \ac{sem}-like imaging (albeit the physics of the signal formation is different)~\cite{Ohya_2009}. 
Additional detectors offer a variety of other imaging possibilities via the detection of transmitted ions~\cite{Serralta_2020}, backscattered ions~\cite{Klingner_2016}, secondary ions~\cite{DeCastro_2021}, photons~\cite{Scipioni_2007,Boden_2012}, Auger electrons~\cite{Parvaneh_2014, Pereyaslavtsev_2019}, etc.; see Table~\ref{tab:analytic}.
These \ac{fib} instruments thus need to be equipped with the necessary interfaces to allow for either direct control of the beam position, or synchronization between the detector signal and internal scan generator.
Such interfaces are standard for the commercial \ac{fib} platforms offering these analytical modalities.

For common repetitive \ac{fib} tasks, instrument users can employ various forms of software automation.
Examples of automated tasks include column alignments and focusing, trench milling (including for \ac{tem} lamella preparation), large-scale sequential lithography projects, serial sectioning, and complex analytical tasks such as 3D tomography using multiple detectors. 
In the past, the technical challenges of many of these processes were handled largely without automation by expert \ac{fib} operators following intensive training.
However, over the last decade this situation has been significantly relieved by the introduction of various semi and fully automated processes by the major \ac{fib} manufacturers.
It must be stressed that there is still room for improvement and further automation is highly desired.
Here, automation routines that are more customizable or user-programmable with well documented \acp{api} or scripting environments within the \ac{fib} software stack would be of great benefit. This would allow standard automation routines to be set up and executed by lab technicians after just a short training period.

Finally, the correlative microscopy and spectroscopy functionalities of modern \ac{fib} instruments (both for 2D~\cite{Vollnhals_2017} and 3D~\cite{Burnett_2014,Holzer_2012,Andersen_2023} analysis; see Sections~\ref{subs_imaging} and~\ref{subs_analysis}) require a combination of the aforementioned software controls as well as advanced software tools for correlation, analysis, and visualization~\cite{Parlanti_2022,Ando_2018}. 
In the simplest case, a correlative data set is acquired using multiple detectors and a single scan of a surface (or in the case of 3D analysis, a stack thereof) such that multichannel data is available for every beam location (``pixel''). 
If this is not possible, data sets from separate experiments have to be combined post-acquisition under the condition that some form of alignment is possible, e.g.\ via fiducial markers. 
For the correlation of 2D images, open-source software is available, e.g.\ in the form of the Fiji/ImageJ image processing package~\cite{Schindelin_2012} and its plug-ins such as Correlia~\cite{ROHDE_2020,Schmidt_2021}.
Additional resources can be found on the webpages of the COMULIS project~\cite{COMULIS} and the BioImage Informatics Index~\cite{BIII}. 
For the most complex cases (e.g.\ the correlation of X-ray and \ac{fib} tomography data), dedicated analytical software~\cite{Lutter_2021} and novel approaches for data storage and management are required; see Section~\ref{subs_roadmap_data}.

\section{Simulation approaches for FIB processing}%
\label{sec_theo}%

Simulations carried out at various levels of sophistication have the potential to provide insights into the interaction of the \ac{fib} with the target. These calculations can  help rationalize the experimental results, optimize the beam parameters (ion energy, incidence angle, etc.), and guide the experimental work.
Various theoretical methods have been developed to describe the interaction of the impinging ions with the target and to assess the amount and type of sample modification produced by a given ion irradiation.\cite{Smith-1997,Krasheninnikov2010b, Nordlund-2019}
Among them, atomistic computer simulations, which describe the system as a collection of interacting atoms, have provided much insight into the behavior of materials under the impact of energetic ions.
These approaches can give precise information on ion ranges and energy losses and on the types of irradiation induced defects. Defect stability and long-term evolution under ambient conditions or at elevated temperatures during annealing can also be simulated. 
Continuum models, on the other hand, can be computationally more efficient and eliminate the statistical noise inherent to many atomistic methods. 
They describe the system under investigation by continuum quantities, such as concentrations or surface contours, which conceptually require averaging over finite volumes.
In practice, these quantities are often discretized on a mesh of elements with sizes larger than atomic dimensions.
Continuum models also require parameters to be provided by experiments.

The choice of simulation method is dictated not only by the main goal of the simulation (e.g.\ to assess the electronic stopping power or to calculate ion ranges, amount and type of damage, atomic mixing, surface evolution, etc.), but also by the system size, the required level of sophistication, and the computational costs.
A typical approach is to find a suitable compromise between the necessary computational resources and accuracy.
Multiscale simulations are frequently used, in which more accurate but computationally 
demanding approaches are employed as tests, or to provide parameters for lower spatial or temporal resolution techniques. 

\begin{figure}[tbp]
   \centering
   \includegraphics[width=\columnwidth]{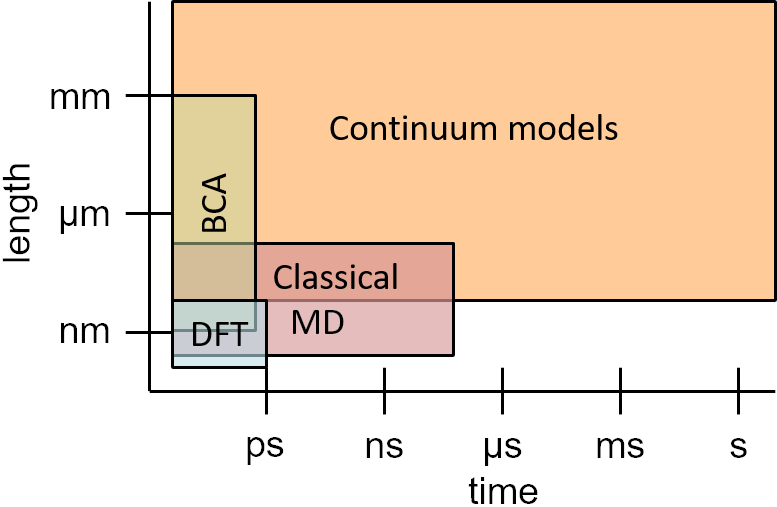}
   \caption{\label{fig:methods_overview}Time and length scales accessible via simulation methods. 
   The smallest time scales are accessible only using certain types of continuum models. 
 \Ac{kmc} can cover the entire length-time space displayed depending on the details of the system studied and on the sub-method used, and is therefore not included in the graph. 
Modified after Ref.~\citenum{Nordlund_2014}.}
\end{figure}

An overview of the time and length scales accessible via the various simulation methods is given in Fig.~\ref{fig:methods_overview}.
The \ac{bca} method, discussed in Section~\ref{subs_bca}, describes the ballistic phase of the slowing down of the ions and of the development of collision cascades. 
Since the ballistic phase is over in less than a picosecond, application of the \ac{bca} method makes sense only on a sub-ps time scale. 
Ab-initio methods, in practice mostly \ac{dft}, are the most accurate techniques, but are restricted to picosecond and nanometer ranges because of their computational cost. 
Classical \ac{md} may be used for system dimensions of roughly up to \qty{100}{ns} and \qty{100}{nm}. 
These methods are described in Section~\ref{subs_md}.
Longer time scales are accessible by the \acf{kmc} method (Section~\ref{subs_kmc}), which describes the evolution of a system given the probabilities of events. 
The actual limits of \ac{kmc} are determined by the event frequencies and the density of objects in the system to be studied.
Finally, continuum models (Section~\ref{subs_cont}) usually study systems beyond the single digit nanometer range, as they do not resolve materials on the atomic level. 

In the following subsections, a concise review of the available computational techniques, with examples of their use, is presented. It is hoped that this overview provides a clear picture of the opportunities simulation provides for \ac{fib} processing.

\subsection{Simulation techniques based on the \acf*{bca}}%
\label{subs_bca}

The \ac{bca} approach is the most widely used method to assess the scattering and slowing  of energetic ions in matter and the effects of the associated collision cascades, i.e.\ sputtering, atomic mixing, and the  formation of point defect damage. 
The motion of each energetic atom is described as a sequence of asymptotic trajectories between binary collisions with target atoms that are at rest before the collision. Detailed information on the associated algorithms is provided in the book by Eckstein~\cite{eckstein_computer_1991}.

\subsubsection{Principles and limitations}%
\label{subsubs_bca_principles}

The binary collisions are normally treated as elastic with repulsive-only screened Coulomb interaction potentials, as e.g.\ in the ``universal''\cite{ziegler_stopping_1985} or ``Kr-C''\cite{wilson_calculation_1977} parametrization. 
Energy transfer to the target electrons is included either nonlocally along the free paths between the collisions, (e.g.\ using a universal analytical description~\cite{lindhard_energy_1961} or specific semiempirical data~\cite{jf_ziegler_particle_2013}), or locally dependent on the collisional impact parameter~\cite{oen_computer_1976}, or as a combination of both. 

Simulations based on the \ac{bca} method are often denoted as \ac{mc} simulations due to the random choice of collisional parameters, such as the impact parameter and the azimuthal position of the target atoms relative to the trajectory of the moving atom. 
In random and amorphous media, this applies to all binary collisions during the simulation. 
In contrast, in crystalline media only the initial position of the incident projectile relative to the lattice atoms is treated randomly, while the subsequent slowing down, including the generation and termination of a collision cascade, is largely deterministic due to the positions of the lattice atoms being fixed except for thermal vibrations. 

A prominent advantage of \ac{bca} simulations is their low computational cost and very fast performance, which allows modeling even of large systems up to the macroscale and the treatment of incident ion energies up to the MeV regime. 
However, in contrast to \ac{md} simulations (see Sect.~\ref{subs_md}), the \ac{bca} model fails if a significant fraction of the collisions occurs between moving atoms. 
Such so-called ``collisional spikes'' or ``elastic thermal spikes''~\cite{sigmund_energy_1974} form in dense cascades generated under heavy-ion bombardment in the energy regime centered around the maximum of the nuclear stopping force~\cite{sigmund_theory_1969} (typically around \qty{100}{keV}). 
However, any collision cascade also dissipates into a collisional spike before its final thermalization, which sets a low-energy limit on the validity of the \ac{bca}. 
Depending on the material, estimates~\cite{eckstein_computer_1991} and comparisons with molecular dynamics simulations~\cite{hobler_useful_2001,hobler_channeling_2019} indicate a failure at energies below several tens of eV. 
However, experience shows that sputtering, which is dominated by cascade atom energies slightly above the surface binding energies of \qtyrange{2}{8}{eV}~\cite{sigmund_theory_1969}, is rather well described~\cite{w._eckstein_computer_2007,schlueter_absence_2020}. 
This suggests a practical lower energy limit of validity. 
Undoubtedly, \ac{bca} breaks down below the bulk binding energy of typically a few eV, where the purely repulsive interaction potential is no longer appropriate and many-body interaction has to be taken into account. 
This excludes e.g.\ the derivation of local atomic configurations after thermalization. 

It must also be noted that \ac{bca} simulations require predefinition of a number of options and parameters, such as the choice of interaction potential, the addition of ‘soft’ collisions with more distant target atoms, the type of electronic interaction, and the surface binding and atomic relocation threshold energies, which are often difficult to identify and unavailable in the literature, particularly for compounds~
\footnote{Note that except for the electronic interaction, these predefinitions are unnecessary in MD simulations because proper choice of the interaction potential ideally covers all bulk and surface interactions down to thermal energies.}. 
Whereas ion ranges derived from \ac{bca} simulations are mostly reliable within \SIrange{5}{10}{\percent}, sputtering and defect results are significantly influenced by these preselections. 
As a result of parameter variations within physically sound limits, calculated sputtering yields may easily change by more than 50\%~\cite{biersack_sputtering_1984}. 
For the different \ac{bca} codes, parameters are often not chosen from  theoretical or experimental information, but justified based on prior experience  comparing the simulation results to experimental data.

Originally, \ac{bca} codes were designed for ``static'' simulations during which the system is assumed not to be altered by the irradiation. This allows the prediction of range and damage distributions as well as sputtering yields with high statistical precision when a sufficiently large number of incident projectiles is employed. 
In the simulation of ion implantation into single-crystalline targets, ``dynamic'' consideration of damage buildup is required if the fluence exceeds a certain threshold,\cite{posselt_dynamic_1995,hobler_boron_1995} as crystal damage reduces or suppresses  channeling. 
This is achieved by choosing collision partners randomly instead of from the crystal lattice, with a probability proportional to the damage. 
More often, the term ``dynamic \ac{bca} simulation'' refers to the modification of the chemical composition and geometry of the target, which occurs experimentally for a sufficiently large number of incident ions. 
Specific \ac{bca} codes track implanted ions as well as relocated and sputtered recoil atoms in order to continuously update the local composition in multicomponent systems and the surface position or the surface contour in 1D or 2D/3D systems, respectively. 
Simultaneously, the composition-dependent local atomic density must be updated by volume relaxation and/or material transport~\footnote{Again, this dynamic modification is straightforward in MD simulations, but in systems exceeding the nanoscale and for higher ion energies it is often impeded by excessive computational cost.}. 

\subsubsection{Specific \acs*{bca} codes}%
\label{subsubs_specific_bca}

Among the large variety of \ac{bca}-based simulation codes that have been described in the literature, we will in the following only address a few selected ones, in particular those that are presently in broader use and/or offer new features with respect to nanosystems, specifically for \ac{fib} applications. 
For a rough classification, see Table~\ref{tab:BCA_codes}; access information for most of the codes can be found in Ref.~\citenum{noauthor_ion_nodate}.

\begin{table*}[btp]
    \caption[\acs*{bca} simulation codes]{Classification of selected \ac{bca} simulation codes. Note that the individual codes contain specific algorithms, use different parameters, and/or are designed for specific application areas, which are not detailed here.}%
    \label{tab:BCA_codes}
    \centering
    \begin{tabular}{lcccc}
	    \toprule
        \multirow{2}{*}{Code} 
         & \multirow{2}{*}{Geometry} 
         & Atomic 
         & \multirow{2}{*}{Surface} 
         & Dynamic  \\
         &  & structure &  & relaxation  \\
        \midrule
        TRIM\cite{jf_ziegler_particle_2013,biersack_monte_1980,ziegler_srim_2015} 
         & 1D & amorphous & planar & --- \\
        TRIDYN\cite{moller_tridyn_1988,moller_tridyn_1984} 
         & 1D & amorphous & planar & 1D \\
        SDTrimSP\cite{eckstein_sdtrimsp:_2007} 
         & 1D & amorphous & planar & 1D \\
        IMINTDYN\cite{hofsass_binary_2022}
         & 1D & amorphous & planar & 1D \\
        Crystal-TRIM\cite{posselt_computer_1993,posselt_dynamic_1995} 
         & 1D & crystalline & planar & --- \\
        IMSIL\cite{hobler_boron_1995,hobler_monte_1995,hobler_dynamic_2011,hobler_combined_2015} 
         & 1D/2D &~amorph./cryst.~& polygons/pixeled (adjusted) & --/1D/2D \\
        SDTrimSP-2D\cite{mutzke_sdtrimsp-2d:_2009} 
         & 2D & amorphous & pixeled & 2D \\
        CORTEO\cite{schiettekatte_fast_2008,schiettekatte_spectrum_2016} 
         & 3D & amorphous & voxeled & --- \\
        TRI3DST\cite{nietiadi_sputtering_2014} 
         & 3D & amorphous & analytical & --- \\
        IRADINA\cite{borschel_ion_2011} 
         & 3D & amorphous & voxeled & --- \\
        IM3D\cite{li_im3d:_2015} 
         & 3D & amorphous & analytical/triangularized & --- \\
        SDTrimSP-3D\cite{toussaint_sputtering_2017} 
         & 3D & amorphous & voxeled & 2D \\
        TRI3DYN\cite{w._moller_tri3dyn_2014,holland-moritz_dynamics_2018,moller_shaping_2016} 
         & 3D & amorphous & voxeled (planarized) & 3D \\
        EnvizION\cite{smith_monte_2010,timilsina_monte_2013,timilsina_comparison_2013,timilsina_monte_2014,mahady_monte_2017,mahady_simulating_2018,mahady_monte_2019}
         & 3D & amorphous & voxeled & 3D, near-surface \\
        \bottomrule
    \end{tabular}
\end{table*}

TRIM~\cite{biersack_monte_1980,ziegler_srim_2015} has been the most widely used \ac{bca} code for several decades. It offers a very convenient graphical user interface~\cite{jf_ziegler_particle_2013} for fast generation of range and damage statistics, and rough estimates of sputtering yields in amorphous semi-infinite or thin-film systems with a flat surface. 
The restriction to amorphous media is often not a severe limitation, since many materials become highly damaged or even amorphized under irradiation and, depending on the conditions,  most ion or recoil atom trajectories are random in nature, even in crystalline materials. 
TRIM partly fails in details, such as for sputtered atom angular/energy distributions~\cite{hofsass_simulation_2014}.

Sometimes the crystal structure of the irradiated material does matter. 
Typical cases include low-fluence or single-ion implantation aligned with a low-index crystallographic direction~\cite{Jakob_2021a} and the sputtering of elemental metals~\cite{schlueter_absence_2020}. 
Codes that consider the crystal structure are Crystal-TRIM~\cite{posselt_computer_1993,posselt_dynamic_1995}, which is restricted to certain materials, and IMSIL~\cite{hobler_boron_1995,hobler_monte_1995,hobler_dynamic_2011,hobler_combined_2015}, which allows more general crystal systems and sputtering simulations.

\begin{figure}[tbp]
  \centering
  \includegraphics[width=\columnwidth]{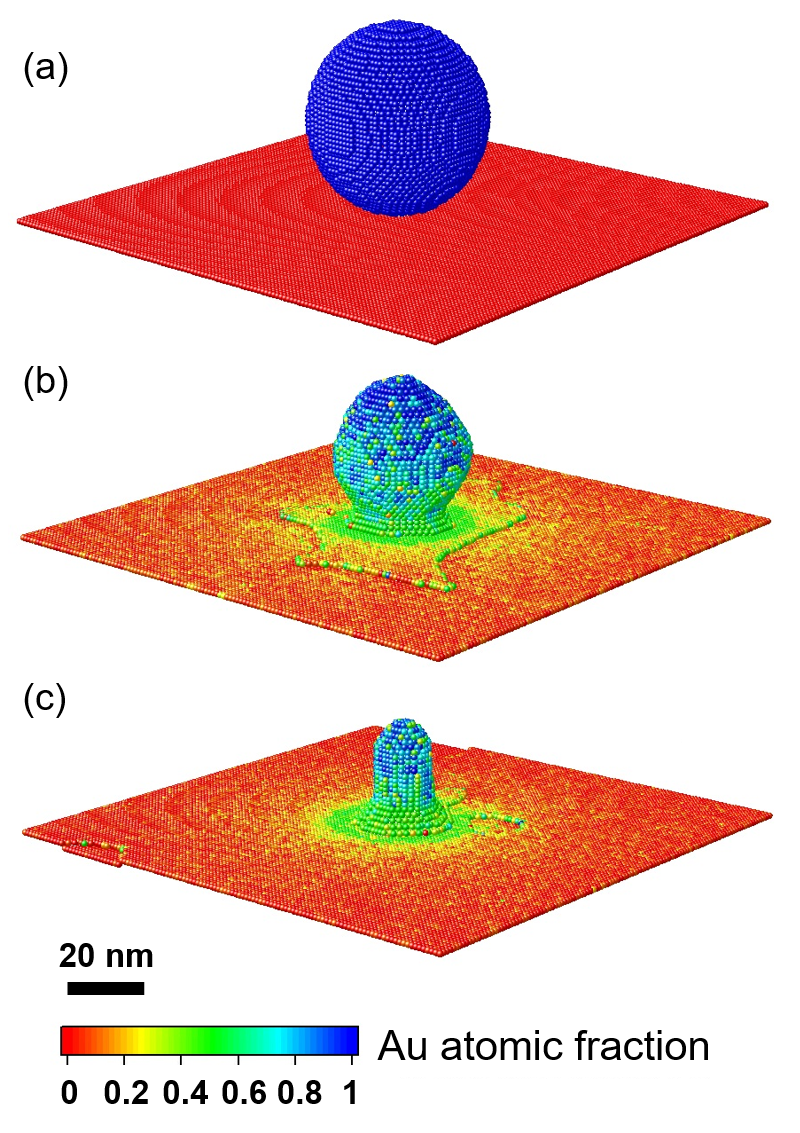}
  \caption{Sputter shaping of Au nanosphere of diameter \qty{50}{nm} on Si, (a) initially  and (b) after irradiation with 30~keV Ga ions under normal incidence for fluences of 4$\times$10$^{15}$ cm$^{-2}$ and (c) 1$\times$10$^{16}$ cm$^{-2}$.
    The TRI3DYN simulation (with periodic lateral boundary conditions) demonstrates size reduction and shaping by sputtering, substrate surface erosion, and contamination of the original sphere with sputtered substrate material and vice versa.
    The incorporation of Ga is small and not shown.
  For more information, see Ref.~\citenum{holland-moritz_dynamics_2018}}%
  \label{fig:BCA_Au_nanosphere}
\end{figure}

Based on the early TRIM sputtering version TRIM.SP~\cite{biersack_sputtering_1984}, 1D dynamic relaxation has been implemented in TRIDYN~\cite{moller_tridyn_1988,moller_tridyn_1984} and in  SDTrimSP~\cite{eckstein_sdtrimsp:_2007}. 
In connection with the recent broad interest in nanostructures, \ac{bca} codes have been extended to treat 2D and 3D systems using pixel and voxel grids, respectively. 
For static operation only, the surface for simple bodies may be defined analytically, such as in TRI3DST~\cite{nietiadi_sputtering_2014} and IM3D~\cite{li_im3d:_2015}, or triangularized as in a second option of IM3D. 
With a pure voxel approach, the stepped surface contour resulting from pixel or voxel grids may cause artifacts by e.g.\ capturing atoms at glancing incidence (which could in fact occur experimentally).
This may be overcome by a local adjustment of the near-surface nodes~\cite{hobler_combined_2015} (IMSIL), or by local planarization of the surface~\cite{moller_shaping_2016} (TRI3DYN). 
In IMSIL for 2D dynamic simulation~\cite{hobler_dynamic_2011}, pixels that are affected by collisional transport become distorted based on an algorithm which optimizes their volumes, and are subsequently projected onto the original pixel grid. 
In contrast, SDTrimSP-2D~\cite{mutzke_sdtrimsp-2d:_2009} relaxes the pixel volumes along one dimension only toward a specified surface. 
3D dynamic simulations have been demonstrated using TRI3DYN~\cite{w._moller_tri3dyn_2014,holland-moritz_dynamics_2018,moller_shaping_2016} and EnvizION~\cite{smith_monte_2010,timilsina_monte_2013,timilsina_comparison_2013,timilsina_monte_2014,mahady_monte_2017,mahady_simulating_2018,mahady_monte_2019}. 
For overall volume and surface relaxation, TRI3DYN makes use of material exchange with nearby voxels or the surface, whereas in EnvizION the relaxation is limited to near-surface regimes. 
TRI3DYN  even works for macroscopic systems, while EnvizION is limited to the nanoscale, since each voxel contains only one atom. 
A characteristic result of a TRI3DYN simulation is presented in Fig.~\ref{fig:BCA_Au_nanosphere}, demonstrating surface erosion and contamination during irradiation of a Au nanosphere on a Si surface.

\subsubsection{\acs*{bca}-based simulations of \acs*{fib} processing}%
\label{subsubs_BCA_FIB}
1D static \ac{bca} simulations have often been employed for understanding experimental findings in \ac{fib} processing, such as in recent studies of \ac{fibid}~\cite{DeTeresa_2019,Teresa_2020,SalvadorPorroche_2020,SalvadorPorroche_2021}. 
These have also been used to generate, e.g.\ angle-dependent sputtering yields and angle-energy distributions of sputtered atoms to describe erosion and re-deposition, respectively, in 2D and 3D ``level set''~\cite{kim_level_2007} or ``segment based''~\cite{ebm_assessment_2009,ebm_quantitative_2011,lindsey_significance_2012} simulation models of the surface contour development during \ac{fib} induced erosion, see Section~\ref{subs_cont}.

\begin{figure}[tbp]
    \centering
    \includegraphics[width=\columnwidth]{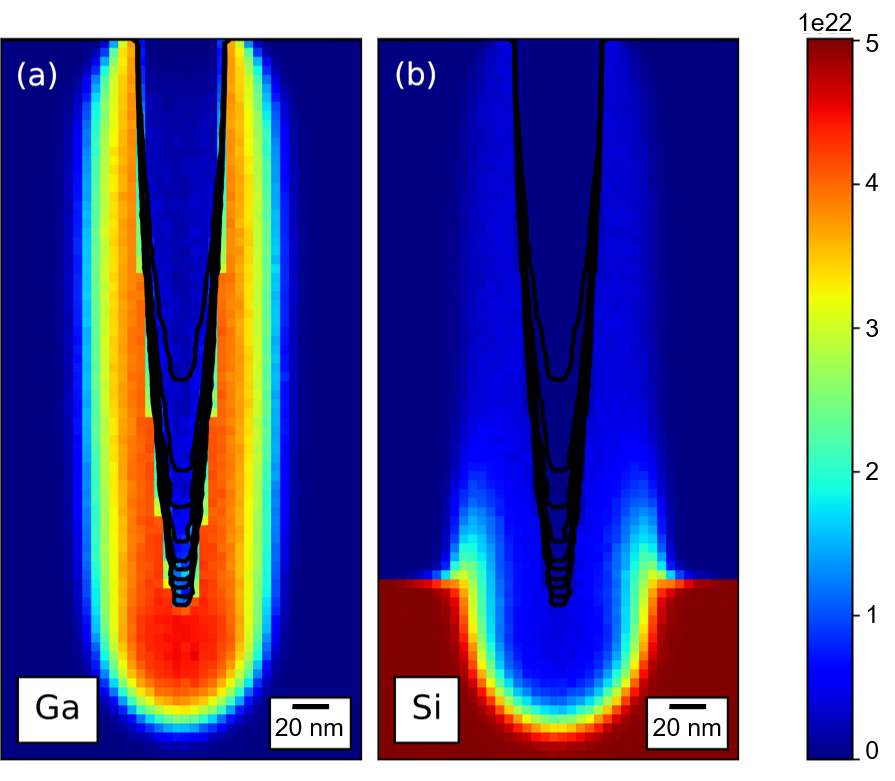}
    \caption{Bombardment of Si target with a sharp \qty{50}{nm}-wide, \qty{50}{keV} Ga beam with a ﬂuence of 5$\times$10$^{18}$~cm$^{-2}$. Final target composition displaying concentrations (cm$^{-3}$) of (a) Ga atoms, and (b) Si atoms that originate from a depth greater than \qty{30}{nm} in the virgin sample. Surface contours track fluence increments of 5$\times$10$^{17}$~cm$^{-2}$. The scan area of the beam is in the center of the top boundary. The simulations have been performed with IMSIL (modified from Ref.~\citenum{hobler_dynamic_2011}).}%
    \label{fig:BCA_trench}
\end{figure}

A direct 2D dynamic \ac{bca} simulation of \ac{fib} induced erosion is described in Ref.~\citenum{hobler_dynamic_2011}, where the milling of trenches is exemplified. 
The results shown in Fig.~\ref{fig:BCA_trench} demonstrate the slowing of the milling process by atoms sputtered from the bottom and redeposited to the sidewalls. 
In addition, a high concentration of implanted atoms at the bottom of the trench emerges, which pushes material upward along the trench sidewalls via volume relaxation. 
The simulation of the milling of a slit in a membrane is described in Ref.~\citenum{hobler_combined_2015}.

With particular focus on \ac{fib} processing, Rack and coworkers have published a series of 3D dynamic results obtained with EnvizION\cite{smith_monte_2010,timilsina_monte_2013,timilsina_comparison_2013,timilsina_monte_2014,mahady_monte_2017,mahady_simulating_2018,mahady_monte_2019}. 
In addition to the collisional \ac{bca} simulation and relaxation algorithms used, simplified models have been implemented for reactive-gas assisted erosion and deposition, including secondary-electron mechanisms~\cite{Ramachandra_2009}. Selected results have recently been reviewed in Ref.~\citenum{Li_2021}. For an introduction to gas-assisted processes, see also Section~\ref{subsubs_continuum_gas}.

\begin{figure}[tbp]
   \centering
   \includegraphics[width=0.9\columnwidth]{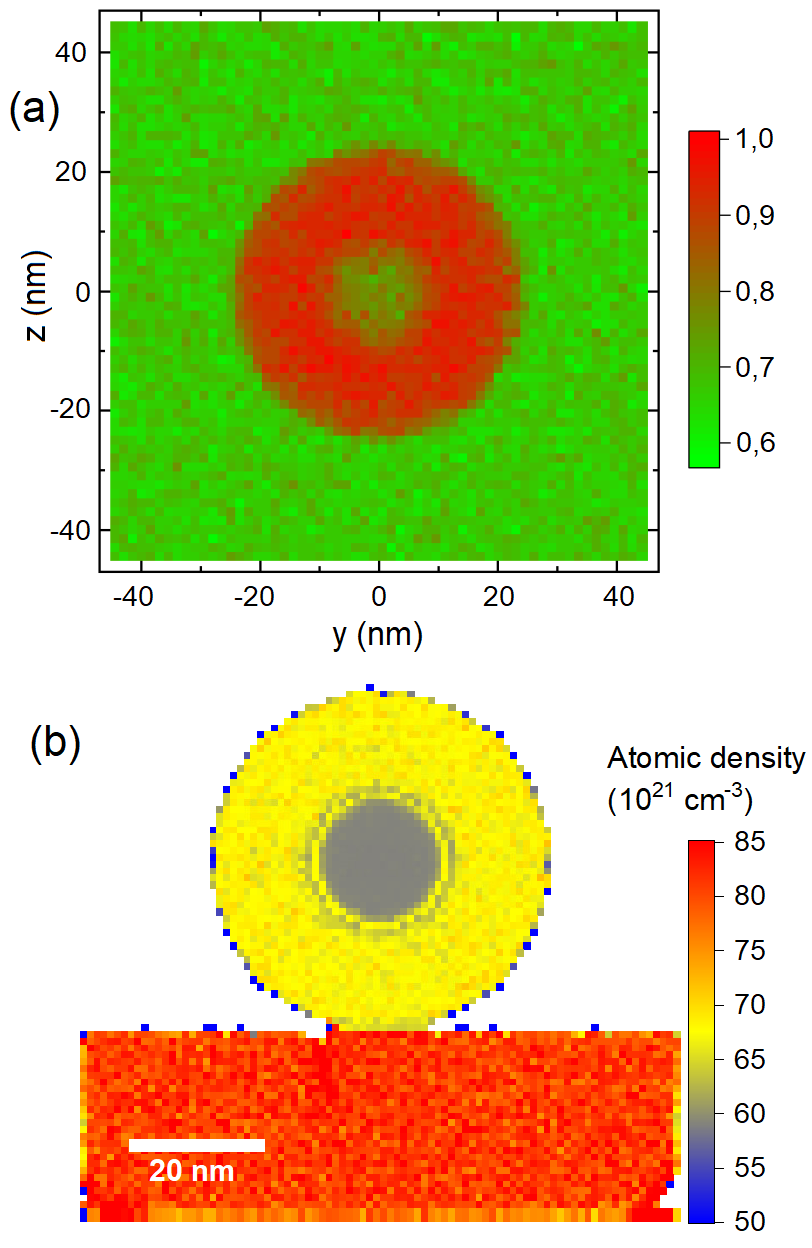} 
   \caption{\label{fig:BCA_TRI3DYN}TRI3DYN simulation of \ac{stim} in a \ac{him}. 
      (a) Ion transmission relative intensity image for Au (\qty{20}{nm} diam.) / SiO$_2$ (\qty{50}{nm} diam.) core/shell nanosphere on  Si$_3$N$_4$ membrane of thickness \qty{20}{nm}. 
    A \qty{20}{keV} He$^+$ beam with a Gaussian beam profile of \qty{0.5}{nm} \ac{fwhm} is scanned across an area of \qtyproduct{90x90}{nm} up to a fluence of \qty{2e16}{\per\centi\meter\squared}, with the transmitted ions being registered in an annular dark-field detector of \qty{0.6}{cm} and \qty{5}{cm} inner and outer radius, respectively, mounted \qty{2.5}{cm} behind the sample. 
      Custom evaluation software correlates the detection of the transmitted ions with the actual $y$-$z$ scan position of the beam. 
   (b) Cross-sectional image of the total atomic density after the irradiation, averaged over a central slice of thickness \qty{2}{nm}.} 
\end{figure}

Dynamic \ac{bca} simulations may also be employed to characterize specific \ac{fib} based nanoanalytical methods. 
An example is a 1D TRIDYN simulation of damage buildup during \ac{fib} milling of nanomembranes for \ac{tem} analysis~\cite{huang_si_2018}.  
Another example is the 3D TRI3DYN simulation of \ac{stim} in a \ac{him} device, as shown in Fig.~\ref{fig:BCA_TRI3DYN}. 
Simulations such as those presented in Fig.~\ref{fig:BCA_TRI3DYN}(a) help to identify the detector geometry for optimum image contrast and resolution.
Furthermore, Fig.~\ref{fig:BCA_TRI3DYN}(b) indicates that there is only minor degradation of the sample by surface sputtering and recoil atom transport between the core and the shell, and between the sphere and the substrate.

\subsection{\Acl*{md} and \acl*{ms}}%
\label{subs_md}

Although \ac{bca} methods are able to provide insights into processes developing during the ballistic (sub-ps) phase of ion irradiation, they cannot predict details of the permanent modification of the target's atomic structure, such as defect clusters or dislocations including their strain fields. Particularly with heavy ions, heat spikes may develop even in keV energy cascades, and the local melting of the target affects the amount of damage formation and atomic mixing~\cite{rubia_role_1987}.
The accuracy of the BCA approach diminishes~\cite{hobler_useful_2001,hobler_channeling_2019} at energies roughly below \qty{100}{eV} and it completely fails in the low single-digit eV energy range. 
Hence, \Ac{md} is often the method of choice for the simulation of phenomena associated with atomic motion at eV and sub-eV kinetic energies. 
\subsubsection{\label{subsubs_md_principle}Principles and simulation codes}

Most approaches in practical use are based on the Born-Oppenheimer approximation, which assumes that the electronic subsystem relaxes instantaneously to its ground state at any time as the atomic nuclei change their positions. 
Consequently, the effect of the electrons can be embodied in a potential energy function that depends only on the nuclear positions~\cite{eb_tadmor_modeling_2011}. 
Analysis of this energy function allows determination of many material properties as well as the time evolution of the system, e.g.\ in an irradiation experiment. 
This is even true when deposition of energy into electronic degrees of freedom  dominates, such as for 30 keV He ions impinging on Au clusters~\cite{Ghaderzadeh2020n} where defect production is still governed by ballistic energy transfer as excitations are quickly delocalized.

In the \ac{md} method, Newton's equations of motion are solved for all atoms in the system, using forces calculated from the potential energy function~\cite{Allen-Tildesley1989,Leach2001,eb_tadmor_modeling_2011}. This is done iteratively 
with time steps on the order of \qty{1}{fs}, with the time step being dependent on the ion/atom 
velocities and their masses. At the end of the simulation, when the energy is equilibrated everywhere in the system, the final structure can be analyzed. Due to the short time step required, total simulation times are limited to the nanosecond or at most microsecond range for classical \ac{md} and to picosecond for ab-initio \ac{md}. Effects corresponding to longer timescales are sometimes mimicked by increasing the temperature in the system, which then allows observation of annealing (recombination) of defects and the relaxation of strain~\cite{Nordlund_2018}. Detailed descriptions of the principles of these \ac{md} methods can be found in Refs.~\citenum{Nordlund-2019,Nordlund2021}.

\Ac{ms} methods are concerned with finding minima and saddle points in the potential energy function. The former identify (meta-)stable atomic configurations, while the latter allow the determination of energy barriers for diffusion and chemical reactions.
Finding global minima, corresponding to stable as opposed to metastable configurations, can be tricky. Various methods have been developed, see Ref.~\citenum{eb_tadmor_modeling_2011} for details.

Simulation codes can be classified into classical and ab-initio codes. In the first group, the potential energy and forces are calculated by evaluating mathematical expressions with empirical parameters, while in the second group, they are determined by solving approximations to the Schr{\"o}dinger equation. The most widely used classical \ac{md} code is LAMMPS~\cite{Plimpton1995,Thompson_2022}. Another general purpose \ac{md} code is DL\_POLY~\cite{Smith1996}. Codes specifically designed for the simulation of radiation effects include PARCAS~\cite{Granberg2016}, DYMOCA~\cite{Becquart1997}, and MOLDYCASC~\cite{delaRubia1990}. GROMACS~\cite{Berendsen1995} and NAMD\cite{Phillips2005} mainly target biomolecular systems. Popular ab-initio codes capable of performing MD simulations are VASP\cite{kresse_efficiency_1996} and QUANTUM ESPRESSO\cite{giannozzi_quantum_2009}.

\subsubsection{\label{subsubs_md_potentials}
Models for the potential energy function}

Among the ab-initio methods, \ac{dft}~\cite{sholl_density_2009} with local and semilocal exchange-correlation (XC) functionals provides a good compromise between accuracy and computational expense for the calculation of the potential energy function. 
Some phenomena such as van der Waals interactions, magnetism (especially when spin-orbit coupling is taken into account), or accurate evaluation of the bandgap in semiconductors, require one to go beyond ``vanilla'' \ac{dft} by either using more sophisticated XC functionals~\cite{doi:10.1021/cr200107z} or post-\ac{dft} methods, such as GW/RPA (many body perturbation theory / random phase approximation)~\cite{10.3389/fchem.2019.00377}. But otherwise, \ac{dft} simulations normally provide reliable results with respect to the atomic coordinates and energetics on the scales relevant for the modeling of ion irradiation effects. 
\Ac{dft} is capable of predicting charge states of defects and static charge transfer~\cite{gao_defect-enhanced_2009}, while \ac{tddft} can describe charge transfer dynamics in ion collisions.
The advantage of ab-initio simulations is that they do not require any material-dependent parameters from the user. 
The disadvantage is their high computational cost, which scales as $n^{2\ldots3}$, where $n$ denotes the number of valence electrons or atoms in the simulation. 
This imposes a practical limit on the system size that can be handled by \ac{dft}, which using supercomputers is currently on the order of 1000\,atoms.

If larger systems need to be handled, empirical interatomic potentials (sometimes called force fields) must be used to describe the potential energy function.
The computational expense of these simulations scales nearly linearly with the number of atoms.
On the downside, the mathematical expressions describing the interatomic potentials are only approximations of the true potential energy function, and the parameters in these models must be fitted for each combination of chemical elements in order to reproduce the physical properties of the material (e.g.\ lattice parameters, elastic properties, bond energies, energetics of defects, stability of sputtered fragments, etc.).
Care is needed if the simulation results in atomic arrangements that are outside the range of atomic arrangements used for the fit.
Moreover, the availability of force field parameters can be a limiting factor when looking for new applications, since the development of new parameter sets is a tedious task~\cite{Shchygol2019}.
To overcome this bottleneck, various methods have been developed in recent years~\cite{Angibaud2011,Jaramillo-Botero2014,Dittner2015,Larsson2013,Larentzos2015,Furman2018,Nakata2019,Sengul2021}.

Collections of 1000+ interatomic potentials together with their parameters for individual systems are available in the NIST Interatomic Potentials Repository~\cite{becker_nist_2010} and the Knowledgebase of Interatomic Models (OpenKIM)~\cite{noauthor_openkim_nodate}. Educated choice of an appropriate interatomic potential is essential for the success of a simulation. In Table~\ref{tab:MD_potentials}, the most popular options are listed together with the types of atomic interactions they can describe and their computational expense. 
\begin{table}[btp]
   \caption[Useful interatomic potentials]{Some popular interatomic potentials listed with the type of interaction they are able to describe and the execution times per atom relative to the Lennard-Jones potential taken from Refs.~\citenum{noauthor_lammps_nodate} and \citenum{schneider_atk-forcefield:_2017} using LAMMPS\@. Mean values are given in cases where both of these references list values.
   Execution times for \ac{ml} potentials vary widely, depending on which formalism they are implemented in (see main text), but still scale linearly with the number of atoms $n$. 
 \Ac{dft} times are usually orders of magnitude larger than those for empirical interatomic potentials. Ratios strongly depend on the system size, since \ac{dft} times scale as $n^{2\ldots3}$, hence it is not reasonable to give a scaling factor for \ac{dft}.}
    \label{tab:MD_potentials}
    \centering
    \begin{tabular}{lcc}
	    \toprule
        \multirow{2}{*}{Interatomic potential} & Type of & Execution \\
         & interaction & time  \\ 
        \midrule
        Lennard-Jones\cite{Lennard1931} & van der Waals & 1 \\
        EAM\cite{Daw1984} & metallic & 1.8 \\
        Stillinger-Weber\cite{SW-orig} & covalent & 3.4 \\
        Tersoff\cite{Tersoff-orig} & covalent & 4.4 \\
        Coulomb-Buckingham~\cite{vanBeest1990} & ionic & 4.7 \\ 
        CHARMM\cite{brooks_charmm_2009} & biomolecules & 14.1 \\
        MEAM\cite{baskes_modified_1992} & metallic/covalent & 16.5 \\
        COMB\cite{Yu2007} & chem.\ reactions & 188 \\
        ReaxFF\cite{VanDuin2001} & chem.\ reactions & 197 \\
        Kieffer\cite{Huang2003} & chem.\ reactions & --- \\
        machine learning (ML) & all & $\approx$ 10--10000 \\ 
        DFT & all & $\gg$ \\
        \bottomrule
    \end{tabular}
\end{table}
All but the CHARMM force field, which is mainly used to describe biochemical molecules such as proteins, lipids and nucleic acids, allow for bond breaking, a prerequisite for the simulation of almost all irradiation effects. For an introductory text on interatomic potentials, see Ref.~\citenum{eb_tadmor_modeling_2011}. 

Overall fitting of a single functional form with a limited number of parameters to reproduce a vast range of material properties has frequently proven to be an unmanageable task.
To overcome this limitation, \acf{ml} potentials have recently been developed~\cite{Behler2016,mishin_machine-learning_2021}. 
Common to all flavors of \ac{ml} potential is their description of  
the potential energy using multiparameter functions of the local environment around each atom.
To determine the parameters, a large data set of potential energies and corresponding interatomic forces for a specific set of chemical elements in different configurations is generated by \ac{dft} calculations. 
Hence, the flexibility of these potentials is much greater than that obtained using empirical interatomic potentials, and if carefully developed, they are able to provide results with an accuracy comparable to \ac{dft}. 
There have also been attempts to train \ac{ml} potentials on-the-fly to handle situations that had not originally been anticipated, such as large surface reconstructions or rare chemical events~\cite{Jacobsen2018,Vandermause2020}. 

\begin{figure}[tbp]
   \centering
\includegraphics[width=\columnwidth]{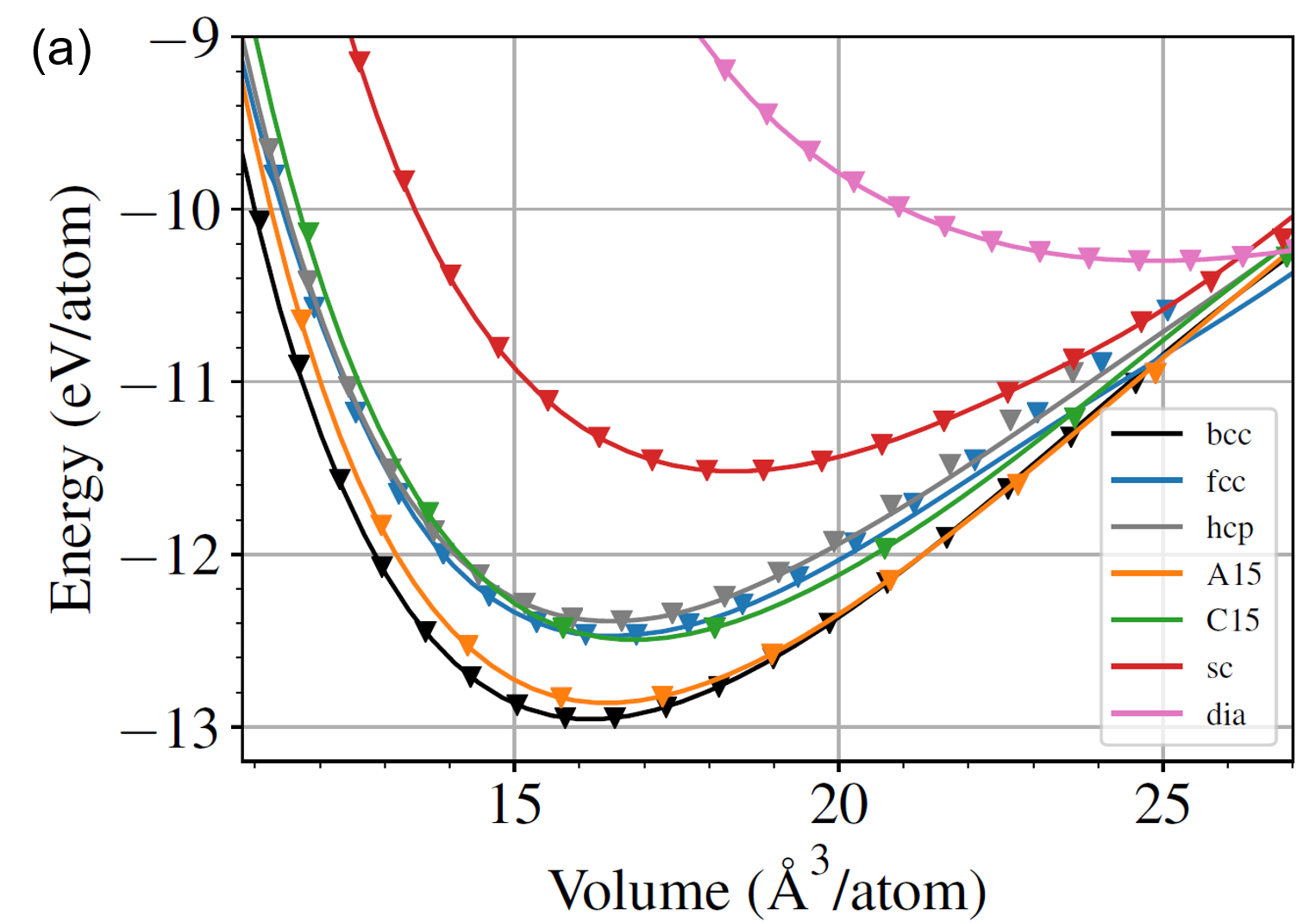}\newline
\includegraphics[width=\columnwidth]{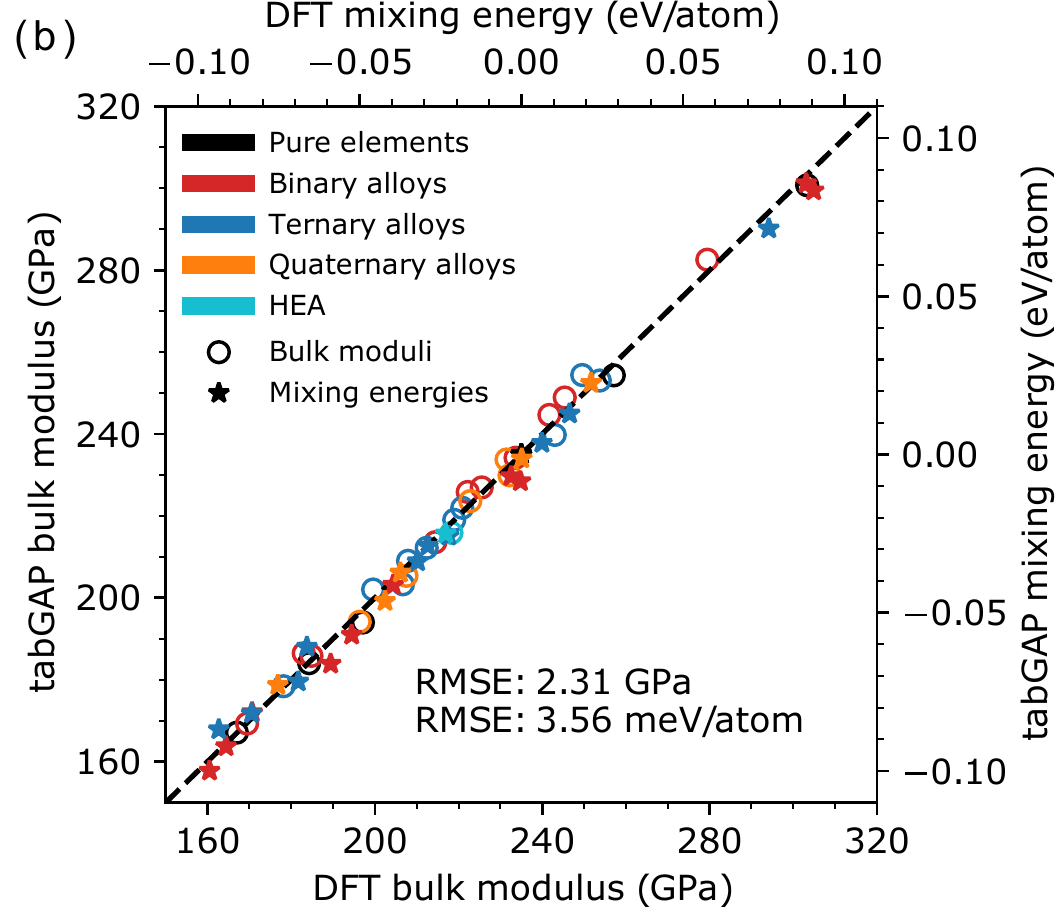}
   \caption{Example of the agreement between \ac{dft} training data sets and machine-learned interatomic potentials using (a) GAP and (b) tabGAP for computing the cohesive energy of pure W in different phases, and the mixing energies and bulk moduli of various alloy combinations, respectively. The symbols are the \ac{dft} data points and the lines show the GAP and tabGAP results. Reprinted from Refs.~\citenum{Byggmastar2019,Byggmastar2021}, Copyright (2019,2021) by the American Physical Society.}%
   \label{fig:DFT_GAP}
\end{figure}

As an example, Fig.~\ref{fig:DFT_GAP}(a) demonstrates the ability of the \ac{gap}~\cite{Bartok2010} to reproduce the \ac{dft} results for the cohesive energy of pure W as a function of atomic volume in different phases~\cite{Byggmastar2019}. 
Similar agreement between predictions of the tabGAP potential (a more efficient version of \ac{gap}~\cite{Byggmastar2021}) and the DFT calculations of mixing energies and bulk moduli is shown in Fig.~\ref{fig:DFT_GAP}(b) for multiple combinations of binary, ternary and quaternary alloys.

The interatomic potentials discussed so far describe the potential energy function near equilibrium or a few eV above equilibrium. During irradiation simulations, much higher potential energies occur in close collisions. 
In this regime, purely repulsive potentials such as also used in \ac{bca} simulations, e.g.\ the universal \ac{zbl} potential~\cite{ZBL} or the potential obtained with the all-electron \ac{dft} code DMol~\cite{Nordlund1997}, provide accurate results. 
These potentials need to be smoothly joined with the material-specific empirical potential at small distances~\cite{Nordlund1996}.
For \ac{ml} potentials, it is advantageous to already include the repulsive potential in the training phase~\cite{Byggmastar2019}.

\subsubsection{\label{subsubs_md_applications}Applications of \acl*{md} to \ac*{fib} irradiation}

\begin{figure}[tbp]
   \centering
   \includegraphics[width=\columnwidth]{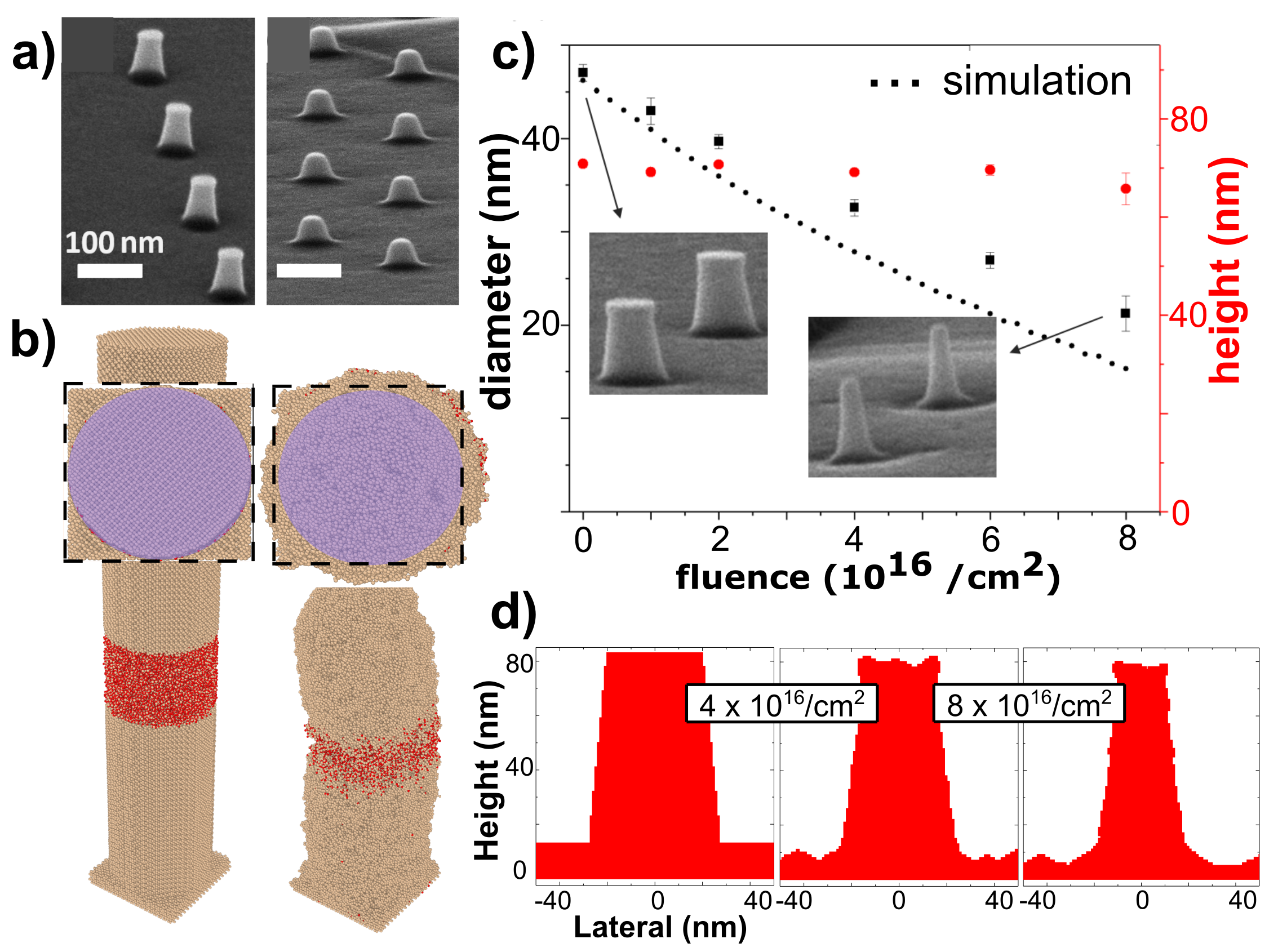}
   \caption{Comparison of the deformation of Si nanopillars under high fluence \qty{25}{keV} Ne or Si ion irradiation. 
   (a) \ac{sem} images before (left) and after (right) Ne ion fluence of \qty{2e16}{\per\centi\meter\squared} at room temperature~\cite{Fridlund2021}. 
   Note the short, but still wide shape of the pillars after irradiation.
   (b) Results of MD simulations for the Si/SiO$_2$/Si pillar ($r = \qty{5}{nm}$, $h = \qty{50}{nm}$) before and after Si fluence of \qty{7.6e15}{cm^{-2}}. 
   In the top-down views, the dashed squares outline the bottom pedestal and the purple circles show the size of the ion beam irradiation pattern used throughout the simulation. 
   (c) Experimental images of the evolution of Si nanopillar under prolonged Ne ion irradiation at \qty{400}{\celsius}~\cite{Xu2019}. 
   Since the Si does not amorphize, the shape is modified only due to erosion from the top and the sides of the pillar, which is well captured by \ac{bca} simulations. 
   (d) Corresponding nanopillar images from \ac{bca} simulation~\cite{Xu2019}.}%
   \label{fig:nanopillar_application}
\end{figure}

Large-scale \ac{md} simulations have been employed to gain insight into nanostructure formation and the modification of nanostructures. 
\citet{das_mechanisms_2015} investigated the Ga \ac{fib}-induced formation of nanopores in Si membranes, which are of interest for biological applications. 
They identified a local boiling mechanism that could be explosive at high enough ion fluxes~\cite{das_mechanisms_2015}. 
Moreover, they found recirculating material flow, which they explained as being driven by temperature-gradient induced surface tension gradients~\cite{das_atomic-scale_2015}.
Another interesting phenomenon occurs during ion irradiation of Si nanopillars. At room temperature, the pillars shorten (Fig.~\ref{fig:nanopillar_application}a), while at elevated temperature, they become thinner (Fig.~\ref{fig:nanopillar_application}c). 
In the \ac{md} simulation performed at room temperature (Fig.~\ref{fig:nanopillar_application}b) the Si amorphizes, which initiates viscous flow due to the ion hammering effect~\cite{Fridlund2021} resulting in a shortening of the nanopillar as seen in the experiment. 
In contrast, in the \ac{bca} simulations shown in Fig.~\ref{fig:nanopillar_application}d, the nanopillar mainly shrinks in diameter, in agreement with the experimental results at elevated temperature. 
This is explained by the neglect of viscous flow in the \ac{bca} simulations, which in fact does not occur when the target remains crystalline at the higher temperature.
As a third example, \ac{md} simulations have examined how a balance between beam spreading, erosion and defect-induced strain limits the thinning of \ac{fib} lamellae~\cite{holmstrom_atomic-scale_2012}.

\begin{figure}[tbp]
   \centering
   \includegraphics[width=\columnwidth]{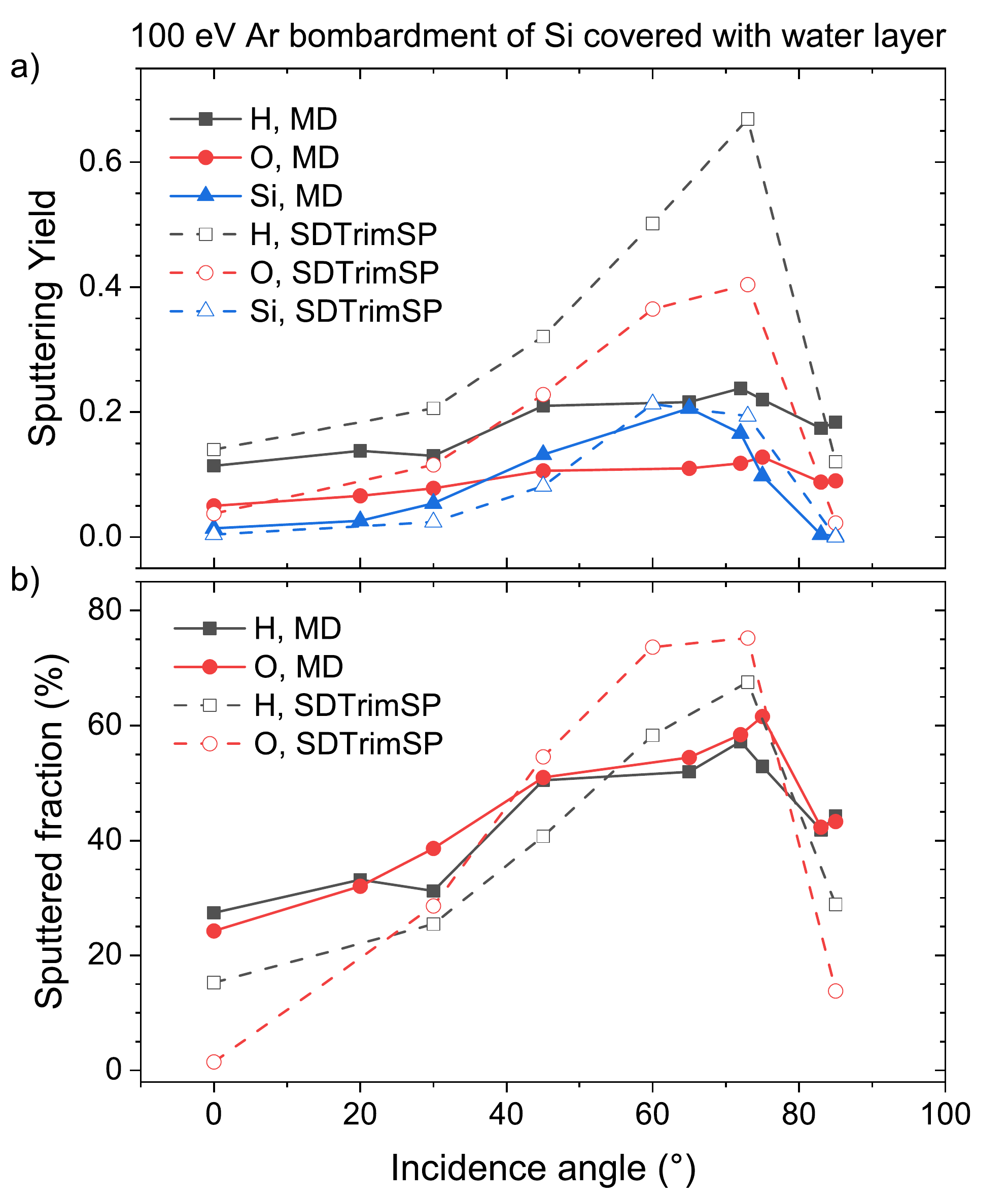}
   \caption{Results obtained by \ac{md} and \ac{bca} (SDTrimSP) simulations  after \qty{100}{eV} Ar irradiation of Si covered with a monolayer of water molecules to a fluence of \qty{5e15}{ions/cm^2}.
     For the MD Si(100) was used. The SDTrimSP sample was amorphous.
     The evolution of 
     (a) the sputtering yields and 
     (b) the fraction of sputtered H and O atoms, are shown as a function of incidence angle.
   }%
   \label{fig:MD_water_Si}
\end{figure}

Chemical effects in sputtering cannot be described by \ac{bca} simulations, but can be using \ac{md} simulations employing reactive force fields. 
For instance, the sputtering of Si by low-energy O and Si ions using the Kieffer force field~\cite{Huang2003} showed that two mechanisms are mainly responsible for sputtering: (i) emission of atoms forming clusters a few Angstroms above the surface, and (ii) ejection of clusters that are loosely bound to the surface~\cite{Barry2014}.
As another example, the impact of residual water molecules on sputter yields of Si bombarded with Ar has been studied (cf.\ Fig.~\ref{fig:MD_water_Si}). 
Here, \ac{md} simulations using the ReaxFF potential optimized for the Si-C-O-H system~\cite{Newsome2012} are compared to SDTrimSP \ac{bca} simulations. 
For the \ac{md} simulations, a much weaker dependence of the O and H partial sputter yields on the incidence angle is found.
The difference might be explained by the random positions of the H and O atoms in the SDTrimSP simulations, presenting more efficient scattering centers than the more regular atomic configurations in the \ac{md} simulations at incidence angles between about 50\textdegree\ and 80\textdegree. 
At angles closer to perpendicular and grazing incidence, the \ac{md} results show higher sputter yields than \ac{bca}.
This may be explained by the desorption of water molecules, which is not included in the \ac{bca} model.
Note that the maximum Si sputtering yield is found at around 65\textdegree, whereas the maxima for H and O are found between 70\textdegree\ and 80\textdegree.
As a practical conclusion, when the mixing of contaminants into the sample should be minimized, angles closer to grazing incidence should be favored.

\begin{figure}[tbp]
 \centering
 \includegraphics[width=0.95\columnwidth]{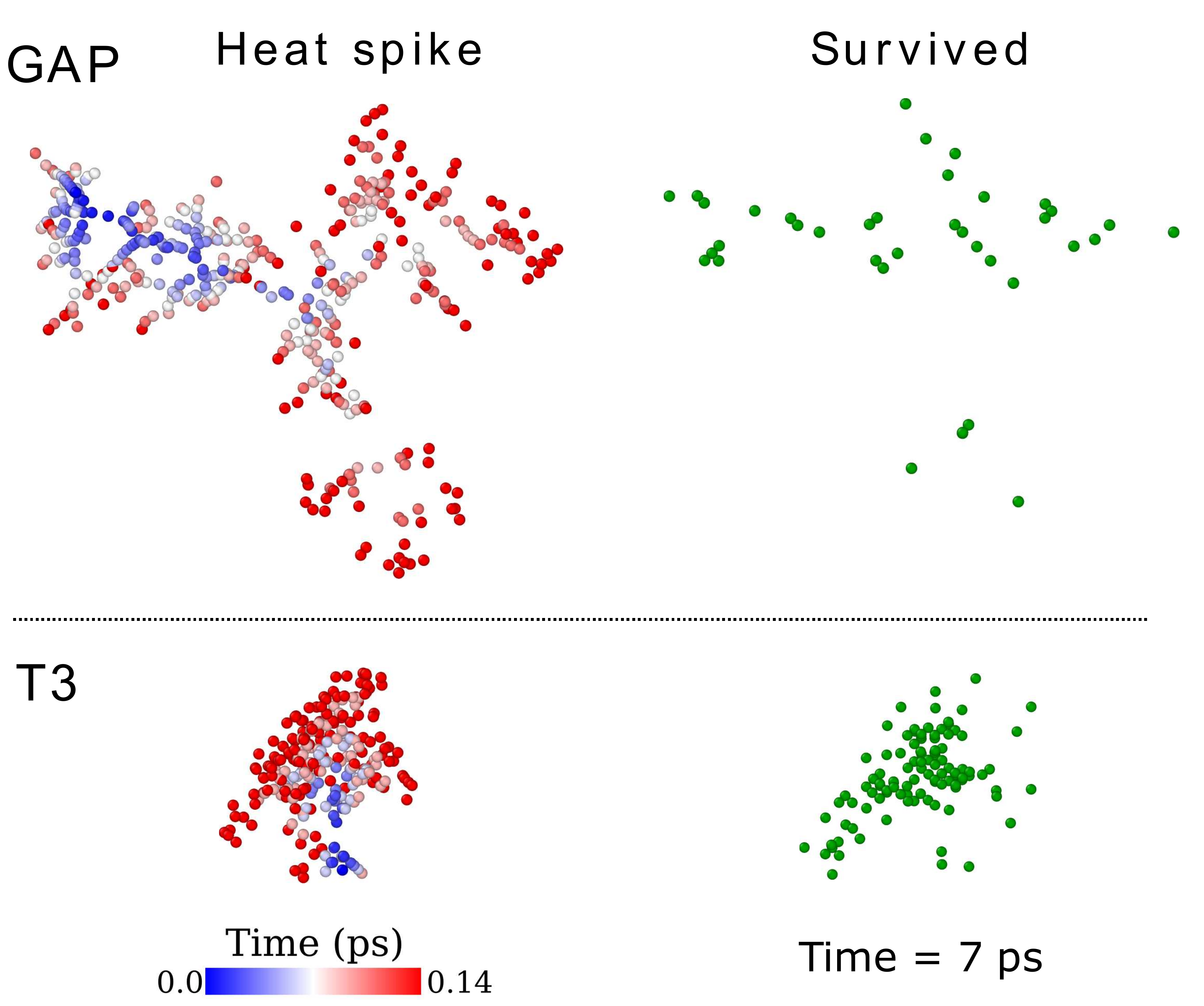}
 \caption{Typical snapshots of \qty{2}{keV} collision cascades at the heat spike phase and the resultant defects in Si. 
 Heat spike snapshots (left panel) present the atoms with kinetic energy above \qty{1}{eV}. 
 Atoms have been color-coded based on the time of generation after firing the primary knock-on atom. 
 Surviving defects (right panel) are acquired from Wigner–Seitz analysis of the cell after cooling to \qty{300}{K} ambient temperature. 
 The comparison visualizes the overestimated defect formation in Si predicted by the classical potential Tersoff (T3)\cite{Tersoff-orig} compared to \ac{gap}~\cite{Hamedani20}. Modified after Ref.~\citenum{Hamedani20}.}%
 \label{fig:survived_defects_comparison}
\end{figure}

Defect formation in all kinds of materials has been a traditional application field for \ac{md} simulations~\cite{Nordlund-2019}. 
Although much insight has been gained, quantitative and sometimes also qualitative results are in doubt, due to the inaccuracies of empirical interatomic potentials. 
For instance, the threshold displacement energy, the essential parameter for damage calculations in the \ac{bca} method,  turned out to be substantially different for Si when calculated by \ac{dft}-\ac{md} compared to using the Tersoff or Stillinger-Weber interatomic potentials~\cite{holmstrom_threshold_2008}.
On the other hand, when using a \ac{ml} potential, essentially the same values were obtained as with \ac{dft}~\cite{Hamedani20}.
Likewise, sputter yields for low-energy Ar irradiation of Si obtained with the \ac{ml} potential were found to be close to experimental values, while the empirical interatomic potentials largely underestimate the yields~\cite{Hamedani20}. 
Moreover, it was shown that damage buildup in keV energy cascades is strongly overestimated by the empirical potentials (see Fig.~\ref{fig:survived_defects_comparison}).

\ac{md} simulations have also provided good understanding of ion induced defect formation in free-standing~\cite{Ghorbani-Asl-2017,Ghaderzadeh-2021,Lehtinen_2010} and supported~\cite{Kretschmer-2018,Kalbac-2013,Standop13} 2D materials, which is of interest for developing ultrathin membranes to encapsulate gases and other volatile systems for ion beam analysis~\cite{Lehtinen_2011}, and also for the development of nanoporous membranes for advanced nanofiltration applications. As with bulk materials, new insight is gained for 2D materials using \ac{dft}-\ac{md}~\cite{Li-2015,Tripathi-2018,kretschmer_threshold_2022}. As an example, in the study by \citet{kretschmer_threshold_2022} the importance of accounting for chemical interactions when evaluating the threshold displacement energy in graphene and \ac{hbn} has been demonstrated.

\subsubsection{Limitations and extensions}

While in the \ac{bca} method, trajectories are only followed for atoms with energies exceeding a few eV, \ac{md} simulates all atoms in a region encompassing the collision cascade. 
Therefore, \ac{md} simulations are much more computationally demanding than \ac{bca} simulations.
Taking into account that actual \ac{fib} processes often use rather high fluences, it is clear that the computational cost of this may be a challenge.
Moreover, due to the limitation of the \ac{md} time step to values that resolve lattice vibrations, the time between collision cascades cannot be fully simulated.
Thermally activated events occurring during these time spans may be covered by \ac{kmc} simulations as needed (see Section~\ref{subs_kmc}).
Often, however, such thermally activated events are neglected, and in addition, the system is artificially  quenched to ambient temperature at the end of each cascade in order to shorten the time required to dissipate the energy deposited by the collision cascade~\cite{Barry2014}.
This comes at the risk of possibly missing some relevant defect annealing.

The limitations of empirical interatomic potentials have already been discussed above.
Another drawback of Born-Oppenheimer \ac{md} is the absence of an inherent description of electrons in the system. 
Electronic stopping can be included in \ac{md} as a frictional force~\cite{Allen-Tildesley1989}, and the same stopping powers may be used as in \ac{bca} simulations.
Modeling of electronic stopping of channeled ions, however, cannot be directly transferred from BCA to MD\@. Here, attempts have been carried out on a more sophisticated level~\cite{cai_phenomenological_1996,chan_molecular_2006}.
In addition, there is uncertainty about the low-energy limit to electronic stopping, which has been shown to be relevant for predicting sputter yields~\cite{sandoval_influence_2009} and damage formation~\cite{zarkadoula_electronic_2014}. 
A related problem is electron-phonon coupling, which describes the exchange of energy between the electronic and ionic subsystems. 
Electron-phonon coupling may be included in \ac{md} simulations by the so-called two-temperature model~\cite{Meftah-1994}.
However, so far no general consensus on the model for electron-phonon coupling nor its significance at \ac{fib}-relevant energies has been reached~\cite{jarrin_integration_2021}.
Recent insights from \ac{tddft}~\cite{runge_density-functional_1984} show promise for enabling a physically motivated parametrization of how electronic stopping and electron-phonon coupling relate to one another~\cite{Lim_2016,Tamm_2016,Sand_2019}. 
Other possible effects to consider include the modification of atomic interactions due to electron excitations in the system~\cite{Darkins2018}.

Born-Oppenheimer \ac{md} also fails to describe e.g.\ the transitions between electronic states or the neutralization of ions in the vicinity of a 2D target.
These situations may be treated by \ac{tddft}-\ac{md} (Ehrenfest dynamics). 
In irradiation simulations using Ehrenfest dynamics, the nuclei are usually approximated as classical particles moving in the time-dependent average potential of the electrons, while the electron subsystem evolves according to \ac{tddft}.
Due to their complexity and high  computational cost, Ehrenfest dynamics simulations using massively parallel computers are currently only feasible for systems comprising tens of atoms and for times up to picoseconds.
The method has been successfully applied to electronic stopping power calculation in various targets~\cite{Krasheninnikov-2007, Ojanpera-2014, Correa-2012, Ullah-2018}, excitation mediated diffusion modeling~\cite{Lee-2019}, and simulations of the behavior of inorganic 2D materials under electron beam irradiation~\cite{Kretschmer-2020}, but its applicability is limited by the mean-field approximation and the resulting propagation of mixed states.
For example, the sputtered atoms may have fractional charges that are unphysical.
It is also worth noting that this approach has been demonstrated to work for describing the formation of \ac{him} images by assessing the spatial distribution of excited electrons immediately after the ion impact~\cite{Zhang_2012}.

\subsection{\label{subs_kmc}\Acf*{kmc}} 

The time scale accessible by \ac{md} simulations is in the nanosecond range, which is sufficient for the system to reach thermal equilibrium after an ion impact. 
After this time, diffusion and interaction of defects take place by thermally activated events.
Examples of such events are diffusion hops, attachment of an atom to a cluster, or recombination of two defects.
The rate constant of each event may be determined by transition state theory using the potential energy function of the system, e.g.\ using \ac{dft} and the nudged elastic band (NEB) method~\cite{NEB-JCP}.
Sometimes, this parametrization can be achieved directly from experimental data, e.g.\ by fitting to a known temperature-dependent diffusion coefficient.
\ac{kmc} is a method that evolves systems described by such events stochastically.
For a basic introduction to the method see Ref.~\citenum{voter_introduction_2007}. Another review is found in Ref.~\citenum{Chatterjee-2007}, and more recent ones with special emphasis on radiation effects and semiconductor processing can be found in Refs.~\citenum{BECQUART2020} and \citenum{martin-bragado_kinetic_2018}, respectively. 
Other approaches for extending the time scale of \ac{md} are discussed in Ref.~\citenum{af_voter_extending_2002}.

Several variants of the \ac{kmc} method have been developed. On-the-fly \ac{kmc} determines energy barriers of possible events during the course of a simulation. 
This technique has been used to simulate defect diffusion in Fe~\cite{SEAKMC2012}, defect evolution in graphite~\cite{Zhou2019} as well as deposition and sputtering~\cite{Scott2011}.
On-the-fly \ac{kmc} can easily be combined with \ac{md}, since both methods can use the same potential energy function, and neither needs prior knowledge of possible events. 
However, the search for possible events makes the on-the-fly variant of \ac{kmc} the most computationally expensive.

Setting up a catalogue of events a priori can be approached in two ways.
In lattice (or atomistic) \ac{kmc}, a lattice is defined where each site can be  empty or occupied by an atom.
Events are hops from one lattice site to another with rates modeled based on the local environment of each site. 
Use cases include vacancy diffusion~\cite{malerba_comparison_2005}, solid-phase epitaxial regrowth~\cite{martin-bragado_facet_2009,Klingner_2022}, sponge structure formation in Ge~\cite{Boettger-2013}, self-organized surface pattern formation~\cite{Yang-2015}, knock-on damage in \ac{hbn}~\cite{Kotakoski-2010}, and phase decomposition of substoichiometric SiO$_2$ into Si nanoclusters in an amorphous SiO$_2$ matrix~\cite{xu_site-controlled_2018}.

In object \ac{kmc}, in addition to vacancies and impurity atoms, other defects including clusters are treated as individual objects that are tracked as they move and/or react with each other. 
The atoms of the ``background'' material are not explicitly represented. 
Therefore, the system size accessible by object \ac{kmc} is determined by the number of objects in the simulation, rather than by some fixed volume. 
The object \ac{kmc} approach has mainly been used in simulations of radiation effects in nuclear materials~\cite{Domain2004} and semiconductor processing~\cite{martin-bragado_kinetic_2018,Aboy2011}.

Lattice and object \ac{kmc} can be combined not only with \ac{md}~\cite{Zhou2019,Scott2011,Yang-2015,Domain2004}, but also with \ac{bca}~\cite{malerba_comparison_2005,xu_site-controlled_2018,Domain2004,Aboy2011}.
In addition, \ac{kmc} schemes are occasionally also used to describe system evolution due to non-thermally activated events. 
For instance, in Ref.~\citenum{Lehtinen_2011} the accumulation of defects in a graphene sheet is modeled based on defect production probabilities obtained by \ac{md} simulations of ion impacts.

\subsection{Continuum modeling}%
\label{subs_cont}

Unlike the methods described in the previous subsections, continuum models do not use individual atoms as the objects of simulation, but instead use quantities that are continuous functions of space and time. For example, in a continuum approach the ion beam is described by its spatially dependent flux (number of ions per area and time increment), the adsorbed precursor molecules are described by their concentration (number of molecules per surface area), and the boundaries of solids are described by mathematical surfaces. %
\subsubsection{Ion implantation and milling}%
\label{subsubs_continuum_principles}

In the absence of ion beam mixing and other secondary effects, the distribution of implanted ions may be obtained by the convolution of the beam flux $F_\mathrm{beam}(x,y)$ and a ``point response'' $f(x,y,z)$. The point response is the probability density function for the stop position of an ion entering a planar target at the origin of the coordinate system. For the purpose of this discussion, we assume the ions to be incident along the negative $z$ axis; $x$ and $y$ denote the other two Cartesian coordinates. If the surface is described by a height function $z\!=\!h(x,y)$, the increase in the concentration of implanted atoms $C(x,y,z)$ with time $t$ is given by~\cite{runge_distribution_1977}
\begin{equation}
	\frac{\partial C}{\partial t} = \iint F_\mathrm{beam}(\xi,\eta,t) \,
 		f(x-\xi,y-\eta,z-h(\xi,\eta))\,d\xi\,d\eta.
	\label{eq:Continuum_concentration}
\end{equation}
For practical reasons, $f(x,y,z)$ is often approximated by a Gaussian function whose parameters can be determined by a \ac{bca} simulation. For an unscanned beam, the beam flux $F_\mathrm{beam}(x,y,t)$ is constant in time, for a digitally scanned beam it is piecewise constant, and for an analog scanned beam it is a continuous function of time. 

Displaced atom and deposited energy distributions can be calculated in a similar way. They may be of interest for estimating the damage to the target, and as input for solving the heat transport equation, respectively.

Sputtering causes the surface to recede. The resulting surface velocity in the direction of the surface normal is given by 
\begin{equation}
	v_\perp = \Omega \, F ,
	\label{eq:Continuum_velocity}
\end{equation}
where $\Omega$ denotes the atomic volume of the target atoms, and $F$ the flux of atoms through the surface element around the point under consideration. In the absence of secondary effects, $F$ is given by the flux of sputtered atoms, $F = F_\mathrm{beam} Y(\theta) \cos\theta$, with $\theta$ the local incidence angle of the ions and $Y(\theta)$ the sputter yield. The sputter yield may be determined by \ac{bca} simulations or experiment.

As long as an initially flat surface, $h(x,y,t\!=\!0) = 0$, develops only small slopes, the surface contour reflects the beam profile: 
\begin{equation}
    h(x,y,t) = - \!\int_0^t v_\perp dt 
             = - \Omega Y(0)  \!\int_0^t F_\mathrm{beam}(x,y,t) \, dt. 
\end{equation}
Once larger slopes occur, further prediction of the target topography usually requires iteration~\cite{vasile_focused_1998} 
or more refined numerical techniques. One option is to move surface marker points along the surface normals~\cite{j.a._sethian_level_1999}. The drawback of this method is that as the nodes move, they tend to form loops which have to be removed. A simple technique to avoid loops, which works for surfaces that remain as height functions $h(x,y)$ throughout the simulation (i.e.\ the surface has no overhanging structures), is to interpolate the surface back to an $(x,y)$ grid after each time step~\cite{g_hobler_t_zahel_s_lindsey_pytopsim_2011}. More general solutions are cell-based and level-set methods. The former represent the target by filled or partially filled voxels, while the latter describe the surface as the zero level-set of a distance function and solve a differential equation for this quantity~\cite{j.a._sethian_level_1999}. Implementations of the marker and level-set methods for \ac{fib} milling have been presented in Refs.~\citenum{muller_simulation_1986,itoh_two-dimensional_1990,katardjiev_three-dimensional_1994,platzgummer_simulation_2006,kim_simulation_2007,kim_full_2007} and Refs.~\citenum{kim_level_2007,ertl_three-dimensional_2010,sun_investigation_2012,borgardt_simulation_2015,rumyantsev_study_2022}, respectively.

\begin{figure}[tbp]
  \centering
  \includegraphics[width=0.9\columnwidth]{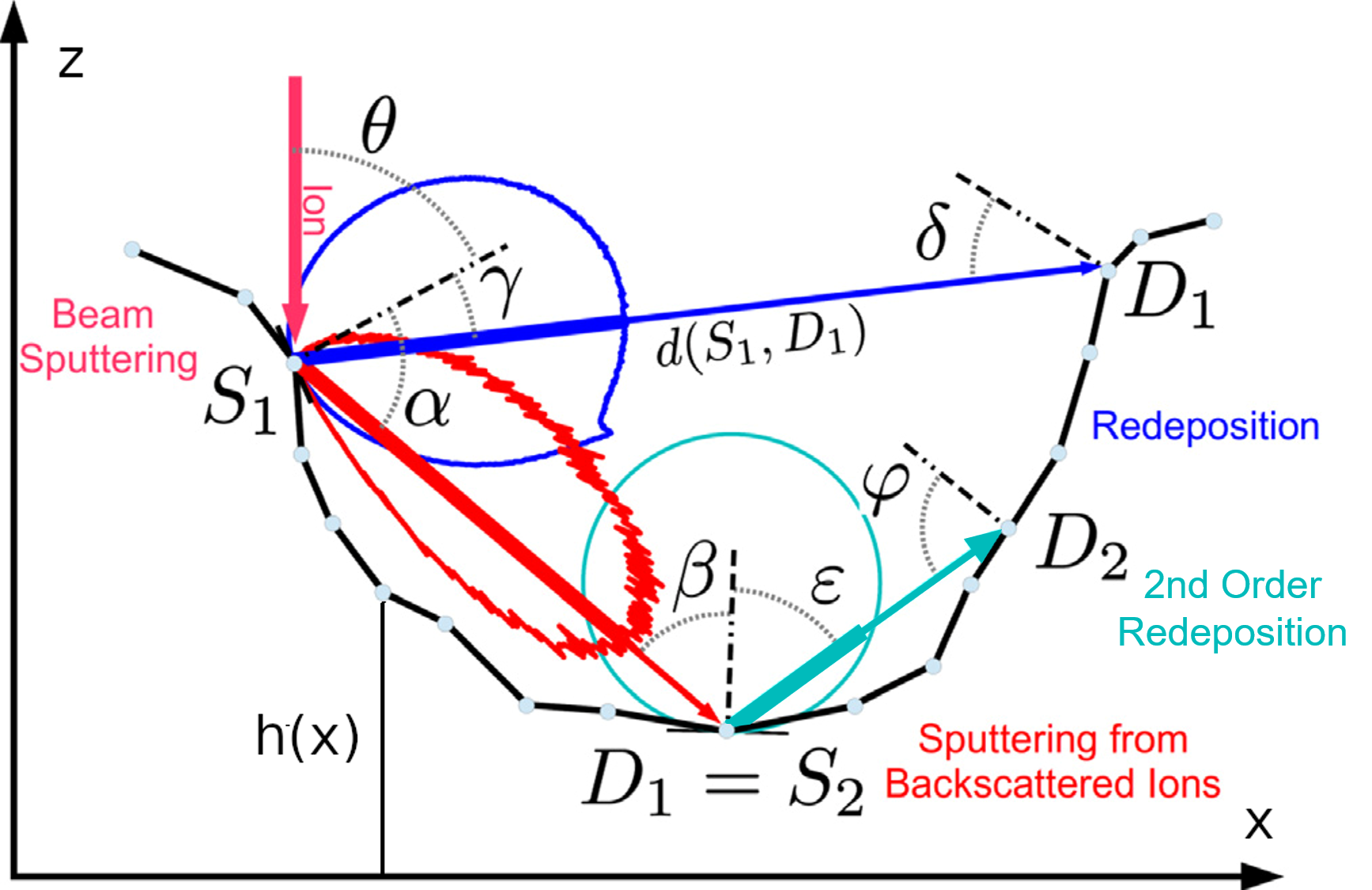}
  \caption{\label{fig:Continuum_redep_backsc}Illustration of ion backscattering (red lines) and redeposition of sputtered atoms (dark and light blue lines) during \ac{fib} milling of a trench.
    The ellipse-like curves represent angular distributions.
    In this case, the surface may be described by a height function $z\!=\!h(x)$.
  For clarity, no variation in $y$ direction is assumed.
  Reproduced from Ref.~\citenum{lindsey_inverse_2014}, Copyright Elsevier (2014).}
\end{figure}

When the surface develops larger slopes, two additional effects come into play (see Fig.~\ref{fig:Continuum_redep_backsc}). First, sputtered atoms may be redeposited on other regions of the surface (dark blue arrow). 
Second, ions may be backscattered from their primary incidence point and impact the target somewhere else (red arrow), where they cause sputtering. The secondary sputtered atoms may then be redeposited (light blue arrow). These processes add significantly to the complexity of the simulation: The angular (and possibly energy) distributions of the sputtered and backscattered atoms (blue and red elliptical lines, respectively) have to be described in some way, and the computation of the fluxes at each marker point (D$_1$ and D$_2$) requires the summation over the contributions of all other marker points (S$_1$ and S$_2$)~\cite{kim_simulation_2007,kim_full_2007,lindsey_inverse_2014}. This summation may be substituted by randomly starting individual rays at the source points, which adds a \ac{mc} aspect to the simulation~\cite{ertl_three-dimensional_2010,cupak_sputter_2021}. In contrast to a \ac{bca} simulation, however, the atom trajectories are not followed within the target, but their exit characteristics are chosen according to precalculated angular (and energy) distributions. Either way, redeposition and backscattering add contributions to the flux $F$ of Eq.~(\ref{eq:Continuum_velocity}). The same applies to contributions from etching and deposition in the case of gas-assisted processes (see Section~\ref{subsubs_continuum_gas}).

In principle, many physical and chemical effects may be considered in a continuum approach. However, when a new ion-material combination is considered, each model requires input from more fundamental simulations or experiments. This may be straightforward for the parameters of the most basic models, such as the mean values and standard deviations of implanted ions or the sputter yields. However, as the model complexity increases, parameter determination becomes more cumbersome.

Our discussion so far has neglected dynamic changes to the target composition. As the irradiation proceeds and ions are implanted, however, the stopping power for the ions changes, and so does the point response $f(x,y,z)$ in  Eq.~\ref{eq:Continuum_concentration}. The same applies to sputter yields and angular distributions. Consideration of dynamic effects would also require modeling of atomic mixing. This has not been attempted so far in a multidimensional continuum approach. Moreover, sputtering is a nonlocal effect, i.e.\ the sputtered atoms are not emitted exactly at the impact point of the ion, which may be important in nanofabrication. In addition, ion implantation and atom relocation may cause topography changes through volume relaxation (cf.\ Section~\ref{subs_bca}). These effects are building blocks of the theory of spontaneous pattern formation~\cite{bradley_nanoscale_2016}, but have not been considered in continuum simulations of \ac{fib} processes so far. 

Another effect that can be treated within a continuum approach is beam induced heating. Temperature distributions can be calculated by solving the heat transport equation with heat sources given by the energy deposition of the collision cascades, using finite element or finite difference methods~\cite{mutunga_impact_2019,Wolff_2018}. The estimates are important because elevated temperatures may reduce deposition rates in \ac{fibid} (see Section~\ref{subsubs_continuum_gas}) and may lead to heat damage in biological samples (see Section~\ref{subsubs_continuum_applications}).

\subsubsection{Gas-assisted deposition and etching}%
\label{subsubs_continuum_gas}

Gas assistance enables selective 3D nanoprinting and enhanced etch rates in \ac{fib} processing.  Continuum models attempt to quantify the local deposition or etch rate caused by the chemical surface reaction between the volatile adsorbates and the \acp{pi}, their locally generated \acp{se}, and \acp{esa}, as follows. 
~\\
\begin{minipage}{\columnwidth}
\vspace{1\baselineskip}
Deposition reaction:\vspace{-0.5\baselineskip}
\begin{multline}
    \text{adsorbate} + \text{(\acs*{pi},\acs*{se},\acs*{esa})} \stackrel{\text{\acs*{fibid}}}{\longrightarrow} \\[-0.25\baselineskip]
    \text{deposit fragments} + \text{volatile fragments}
    \label{eq:deposit}
\end{multline}
\end{minipage}
\begin{minipage}{\columnwidth}
\vspace*{\baselineskip}Etch reactions:\vspace{-0.5\baselineskip}
\begin{subequations}
\begin{multline}
    \text{adsorbate} + \text{(\acs*{pi},\acs*{se},\acs*{esa})} \stackrel{\text{\acs*{fibie}}}{\longrightarrow} \\[-0.25\baselineskip]
    \text{reactive fragments} + \text{volatile fragments}
    \label{eq:etch_a}
\end{multline}
\begin{multline}
    \text{reactive fragments} + \text{substrate atoms} \longrightarrow \\[-0.25\baselineskip]
    \text{volatile compound}
    \label{eq:etch_b}
\end{multline}%
\label{eq:etch}
\end{subequations}
\end{minipage}
\ac{fibid} typically employs volatile metalorganic compounds to deposit metal-containing nanostructures\cite{Utke_2008}. For \ac{fibie}, the adsorbed precursor  needs to dissociate upon irradiation into a reactive species that removes atoms from the underlying substrate by forming a volatile compound, but does not leave traces of other dissociated precursor elements. 

\begin{figure}[tbp]
  \centering
  \includegraphics[width=\columnwidth]{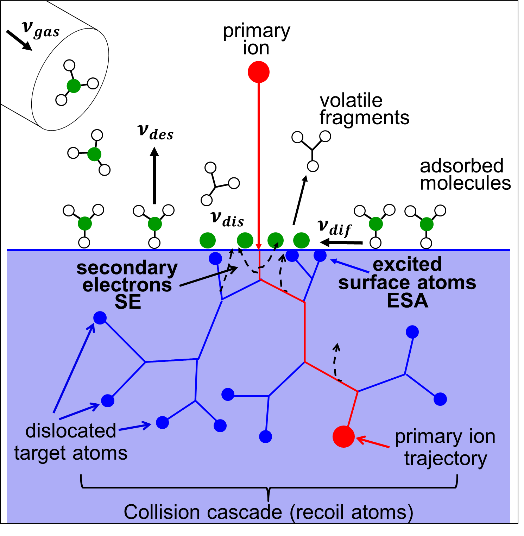}
  \caption{\label{fig:fibid}Concept of selective (local) gas assisted \ac{fibid}. 
    Volatile physisorbed molecules are dissociated by irradiation into volatile and non-volatile fragments, the latter forming the deposit (green spheres). 
    Primary ions typically arrive at keV energy while \acp{esa} and \acp{se} have an eV energy spectrum. 
    \Acp{esa} are generated by the collision cascade (recoil atoms) and have energies below the surface binding energy (not sputtered). 
  Rates ($\nu$) are explained in the text. Note that ion sputtering, ligand codeposition, and incomplete adsorbate dissociation have been omitted for clarity.
  Modified from Ref.~\citenum{Shorubalko_2016}.}
\end{figure}

Present \ac{fibid} and \ac{fibie} models treat the desorption of volatile fragments in reactions (\ref{eq:deposit}) and (\ref{eq:etch}) as instantaneous and assume that the etch reaction (\ref{eq:etch_b}) is fast compared to reaction (\ref{eq:etch_a}), see Ref.~\citenum{i_utke_continuum_2012} for a review. Both processes share the same formalism to calculate the surface coverage $\Theta$ (normalized concentration) of adsorbates, which is defined by the rate equation (in units of inverse time):
\begin{equation}
    \frac{\partial\Theta}{\partial t} = \nu_\mathrm{gas}(1-\Theta) - \nu_\mathrm{des} \Theta - \nu_\mathrm{dis} \Theta + D\,\nabla^2 \Theta.
    \label{eq:rate-equation}
\end{equation}
The terms $\nu_\mathrm{gas}(1-\Theta)$ and $\nu_\mathrm{dif} = D\,\nabla^2 \Theta$ are adsorbate source terms describing gas and surface-diffusion transport, respectively, while the other two terms are adsorbate sink terms related to the rates of desorption and dissociation, $\nu_\mathrm{des}$ and $\nu_\mathrm{dis}$, respectively, as schematically depicted in Fig.~\ref{fig:fibid}. This approach could be extended to multiple chemical pathways, as in models developed for \ac{febid} and \ac{febie}~\cite{toth_continuum_2015}.

\acp{pi}, \acp{se}, and \acp{esa} are the co-reactants. The concentration of impinging \acp{pi} is given by the ion beam spatial profile $F_\mathrm{beam}(x,y)$ and the \ac{fib} exposure time. Energy and spatial distributions $f(E,\vec{x})$ of \acp{se} and \acp{esa} result from the interaction of the \acp{pi} and the recoils with the etched substrate or growing deposit. $f(E,\vec{x})$ must be determined from more fundamental simulations. For \acp{se}, \ac{bca} simulations of \acp{pi} and recoils combined with \ac{mc} simulations of electron transport~\cite{mahady_monte_2017-1} have been performed using a simple model for \ac{se} generation. More refined theories of \ac{se} generation also exist~\cite{hagstrum_theory_1954,hagstrum_theory_1961,bercx_quantitative_2019}. For \acp{esa}, \ac{bca} simulations reach their low-energy limit of applicability, since atom energies of around 1~eV can already dissociate the precursors. An alternative for simulations of \acp{esa} would be \ac{md}. Recently, first attempts of \ac{md} were undertaken for \ac{febid} where atomistic detail on metal grain size, distribution, and composition in the deposited  metal-carbon material could be obtained\cite{Prosvetov_2023,Verkhovtsev_2021}. The treatment of  adsorbate dissociation by electrons relied on experimentally measured fragmentation mass spectra and fragment ion yields for structurally similar molecules, which needed smart appropriate scaling. Desorption of the adsorbate fragments was approached using an external force field. Both approaches limit the predictive power of the simulations at present, but show promise if appropriately refined in the future. Modeling of \ac{fibid} would require inclusion of the \ac{esa} and milling mechanisms into the simulation.

Finally, the chemical fluxes of deposition $F_\mathrm{dep}$ or etching $F_\mathrm{etch}$ (in units of atoms per area per time) are  proportional to the dissociation rate $\nu_\mathrm{dis}$ and the coverage of adsorbates $\Theta$ as shown
\begin{equation}
    F_\mathrm{dep,etch} = m \, \nu_\mathrm{dis} \Theta \, n_\mathrm{s} ,
    \label{eq:nu_etch}
\end{equation}
where $m$ denotes the number of atoms deposited or etched per dissociated precursor molecule and $n_\mathrm{s}$ is the number of adsorbate sites per unit surface area. Note that the product $\nu_\mathrm{dis} \Theta$ defines the rate of deposition or etching for a single site in units of inverse time. The flux according to Eq.~(\ref{eq:nu_etch}) must be added to the flux of sputtered and redeposited atoms discussed in Section~\ref{subsubs_continuum_principles}.

Modeling of the four terms on the right-hand side of Eq.~(\ref{eq:rate-equation}) is key to the successful simulation of \ac{fibid} and \ac{fibie}. The first term $\nu_\mathrm{gas}(1\!-\!\Theta)$ represents a widely used pragmatic simulation approach. It assumes non-dissociative Langmuir adsorption, which treats adsorption as physisorption with a maximum coverage of one monolayer of adsorbates. Other adsorption schemes include dissociative adsorption~\cite{utke_coordination_2022}, multilayer adsorption~\cite{sanz-hernandez_modelling_2017}, and chemisorption\cite{bishop_role_2012}. Open source MC codes are available for the calculation of the gas impingement rate $\nu_\mathrm{gas}$ on the irradiated area for a given geometrical arrangement of nozzle and substrate~\cite{toth_continuum_2015,i_utke_notitle_nodate,friedli_optimized_2009}. Typical values are in the range $0.1 \lesssim \nu_\mathrm{gas} \lesssim 10^3$~s$^{-1}$. 

The desorption rate $\nu_\mathrm{des}$, contained in the second term on the right-hand side of Eq.~(\ref{eq:rate-equation}), is the inverse of the surface residence time. It depends exponentially on the temperature as well as on the type of adsorption occuring on the growing/etched surface being irradiated. The desorption rate can cover several orders of magnitude depending on the molecule-surface interaction.

The third term, the adsorbate dissociation by irradiation, is key for selective \ac{fibid} and \ac{fibie} as it captures the crucial non-thermal irradiation-driven chemistry of the process. Different contributions involving \acp{pi}, \acp{se}, and \acp{esa} are involved in adsorbate dissociation, see Fig.~\ref{fig:fibid}. In principle, the dissociation rate can be calculated from the energy dependent dissociation cross section $\sigma(E)$ and the energy distribution $f(E,\vec{x})$ of the \acp{se} or \acp{esa} via 
\begin{equation}
    \nu_\mathrm{dis}(\vec{x}) = \int_0^\infty \sigma(E) f(E,\vec{x}) \, dE.    
    \label{eq:nu_dis}
\end{equation}
$\sigma(E)$ describes the efficiency of reactions (\ref{eq:deposit}) and (\ref{eq:etch}) as functions of the energy of the \acp{pi}, \acp{se}, and \acp{esa}. Such energy dependent calculation is hampered by the fact that so far reliable $\sigma(E)$ only exist for electron induced dissociation reactions of molecules in the gas phase, see for example Ref.~\citenum{thorman_role_2015}. A pragmatic approach assumes that such gas phase $\sigma(E)$ can be linearly scaled to the net deposition (etch) cross section (see below). This approach omits the quenching of certain dissociation channels (fragment production) due to the adsorbate state situation~\cite{i_utke_continuum_2012,o_ingolfson_low-energy_2019,spencer_understanding_2014,bohler_control_2013}. 
Simulations will thus not be accurate with respect to deposit composition, but the deposit (etch hole) shape can probably be adequately modeled. The only established energy dependent cross section for deposition conditions exists for the molecule (CH$_3$)$_3$Pt-CpMe under electron irradiation~\cite{dorp_role_2012}, describing the observed dissociation of one volatile CH$_3$ ligand from the adsorbed molecule leading to a deposit of atomic composition Pt:C=1:8.

So far, the \ac{esa} induced adsorbate dissociation reactions have not been studied, probably due to experimental issues in deconvolution of the naturally involved (secondary) electron contributions. Recent condensed phase studies using Ar-ion irradiation~\cite{bilgilisoy_surface_2020} point to the importance of the additional mechanical sputtering effect of the \ac{pi}s, which can help remove more ligand fragments than \ac{se}s alone and thus achieve a higher metal content material. The elemental species, charge, and energy of the ions open up a wide field of fragmentation chemistry with metalorganic or “etch” adsorbates, which is still unexplored, particularly with respect to missing $\sigma(E)$. Moreover, the energy distributions $f(E,\vec{x})$ in Eq.~(\ref{eq:nu_dis}) are not easily obtained in a continuum approach. Therefore, while using Eq.~(\ref{eq:nu_dis}) with a \ac{mc} approach for \ac{se} generation and transport is a promising approach, for the \ac{esa} mechanism one has to resort to simpler approaches, as is also often undertaken to describe the \ac{se} mechanism.

In a simpler approach, dissociation is generically modeled via the dissociation rate $\nu_\mathrm{dis}(\vec{x})=\sigma f(\vec{x})$, where $f(\vec{x})$  is the local flux of the \acp{pi}, and $\sigma$ is the so-called net deposition or etching cross section, which is deduced from fitting the shape of the deposit or etch hole. Such cross sections do not distinguish between the different contributions of \acp{pi}, \acp{se}, and \acp{esa}, nor do they distinguish between the fragmentation reactions. The modeled shape or rate can sometimes retroactively be attributed to a certain mechanism, when the yield and/or spatial distribution of either \acp{pi}, \acp{se}, or \acp{esa} matches the deposit or etch hole dimensions. In this way, the \ac{esa} dissociation mechanism was identified to be dominant for \ac{fibid} using noble gas ions heavier than He~\cite{dubner_role_1991,Shorubalko_2016} and the \ac{se} dissociation mechanism was identified as important for He-\ac{fibid} nanopillar growth.\cite{Alkemade_2014} In those cases, $f(\vec{x})$ may be interpreted as the flux of \acp{esa} or \acp{se}. Experimental net cross sections (yields) obtained from ion and electron induced deposition are available for a few precursor molecules~\cite{Utke_2008}, and deposit compositions upon electron irradiation were recently reviewed in Refs.~\citenum{barth_precursors_2020,utke_coordination_2022}. 

The fourth term in Eq.~\ref{eq:rate-equation}, the surface diffusion rate, is proportional to the surface diffusion coefficient $D$ and the second spatial derivative $\nabla^2 \Theta$ of the adsorbate coverage $\Theta$. Thus, it becomes largest for intense and finely focused ion beams and can result in the delivery of considerable amounts of adsorbates to non-irradiated surface areas. Increasing temperature increases the surface mobility of the adsorbates exponentially.

The generation of local heat during ion irradiation can be an important factor influencing the outcome of a deposition process. Thermal contributions can alter the nonthermal irradiative dissociation pathways and exponentially change the diffusion constant (mobility) and residence times of the adsorbates on the surface. As an example, thermal effects are thought to be responsible for the bending of lateral nanowires as they grow\cite{belianinov_direct_2020}. As mentioned in Section~\ref{subsubs_continuum_principles}, the temperature distribution needed as input for temperature dependent models of mobility and residence time can be obtained by numerically solving the heat transport equation.\cite{mutunga_impact_2019,Wolff_2018}

\subsubsection{Applications to \acs*{fib} processing}%
\label{subsubs_continuum_applications}

Simulations of \ac{fib} milling may be used to investigate the influence of beam shape and scan strategy on the final topography of the target~\cite{muller_simulation_1986}, and to better understand the milling process. For instance, Lindsey et al.~\cite{lindsey_significance_2012} investigated microtrench formation in trench milling and clarified the roles of redeposition and backscattering. In practice, often the inverse problem of determining beam and scan parameters for a desired geometry must be solved. In the simplest case, for low aspect ratio structures, this can be achieved approximately by choosing the pixel dwell times proportional to the desired depth at the respective pixel positions~\cite{svintsov_fib_2009}. For higher accuracy, a system of linear equations for the pixel dwell times must be solved in order to consider the effect of beam overlap~\cite{vasile_focused_1998,qin_large-scale_2013}. For higher aspect ratio structures, when redeposition and/or backscattering need to be considered, repeated numerical simulations of the milling process must be performed until convergence to the desired structure is achieved. In practice, such structures are fabricated by applying variable dwell times across the scan area in multiple passes. Here one can either choose to apply the same pattern in each pass~\cite{lindsey_inverse_2014}, or define layers to be milled in each pass and determine individual dwell time profiles~\cite{kim_multiple_2012}.

Finite element and finite difference simulations of heat transport have been performed by Wolff et al.~\cite{Wolff_2018}, who used them to verify a simple model for the estimation of target heating.
Schmied et al.~\cite{schmied_new_2012} employed a similar numerical approach to find beam scanning strategies that minimize local thermal stress in soft matter. 

Gas-assisted \ac{fib} processes have been treated in a continuum approach mostly to help identify in which regime the surface reactions (\ref{eq:deposit}) and (\ref{eq:etch_a}) take place, i.e.\ whether they are mass-transport-limited (adsorbate-limited), or reaction-rate limited (\ac{pi}, \ac{esa}, and/or \ac{se} limited), or driven by desorption or surface diffusion. These regimes strongly influence shape fidelity, homogeneity, and resolution, all of which are of utmost importance for 3D nanoprinting/nanoetching approaches. A review of both time dependent and steady-state analytical solutions of Eqs.~\ref{eq:rate-equation} and~\ref{eq:nu_etch}, using simplifying assumptions, is found in Ref.~\citenum{i_utke_continuum_2012}. This  theoretical framework closely resembles the one used for gas-assisted focused electron beam processing (\ac{febid}/\ac{febie}), with the exception of physical sputtering and redeposition. Hence, analytical solutions established for the analogue electron-based \ac{fibid}/\ac{fibie} variants straightforwardly apply. These include expressions obtained for steady-state surface diffusion and flat-top focused beam profiles~\cite{muller_elektronen-mikroschreiber_1971,c___1968}, long freestanding rods~\cite{aristov_direct_1991}, 1D time dependent solutions~\cite{hirsch_contamination_1994}, and expressions for lateral resolution~\cite{szkudlarek_lateral_2014} and etch/deposition gas mixtures~\cite{lobo_high_2007,toth_electron_2007}. Analytical solutions without surface diffusion can be used in the \ac{pi}, \ac{esa}, \ac{se} limited processing regimes (no concentration gradient of adsorbates due to high supply from the gas phase), or approximately when adsorbates cannot diffuse into the center of the \ac{fib} spot profile due their deposition at the perimeter. These conditions translate into very small \ac{fib} current and exposure (dwell) time~\cite{winkler_nanoscale_2014}. Without the surface diffusion term, analytical solutions are more easily derived and thus cover a broader range of adsorbate-related dissociation mechanisms. These include the resulting deposit composition upon injection of two precursor species~\cite{bernau_tunable_2010}, multistep reactions involved in gas enhanced etching~\cite{toth_continuum_2015,c_lobo_continuum_2012}, and experimental parameter maps~\cite{sanz-hernandez_modelling_2017}. 

\begin{figure}[tbp]
  \centering
  \includegraphics[width=\columnwidth]{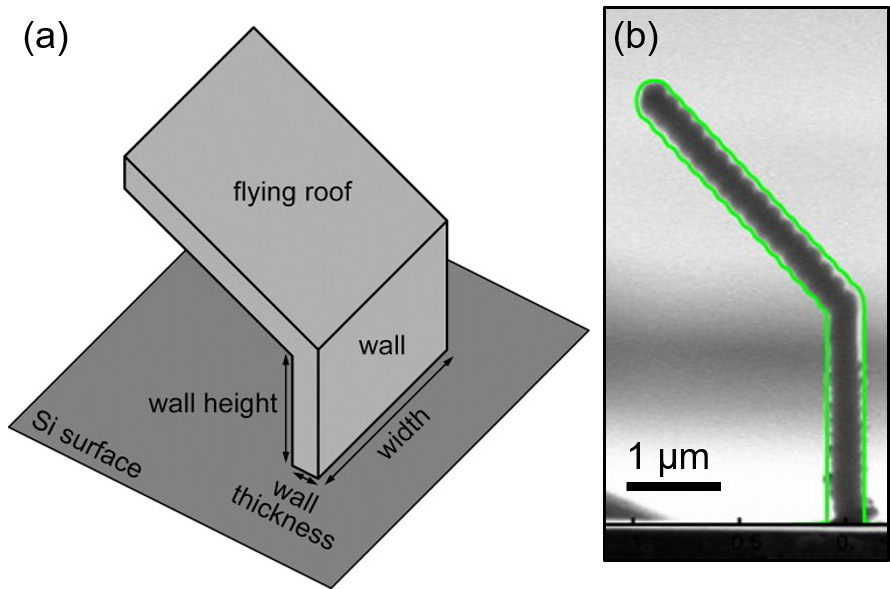}
  \caption{\label{fig:flying_roof}Flying roof structure fabricated by \ac{fibid}. 
    (a) Schematic illustration of the structure. 
    (b) Side view \ac{sem} image and simulation (green line) of the structure obtained using a \qty{30}{keV} Ga ion beam with a \ac{fwhm} of \qty{60}{nm} and pentamethylcyclopentasiloxane as the precursor. 
  After wall formation, the ion beam was moved to the left in 16 steps of spacing \qty{60}{nm}. Reproduced from Ref.~\citenum{ebm_quantitative_2011} with permission from the American Vacuum Society (AVS).}
\end{figure}

Full numerical continuum simulations of a \ac{fibid} process using Eqs.~(\ref{eq:rate-equation}) and (\ref{eq:nu_etch}) have been performed by Ebm et al.,\cite{ebm_quantitative_2011} see Fig.~\ref{fig:flying_roof}.
In these simulations, an \ac{esa}-based model was used, assuming the concentration of \acp{esa} to be proportional to the flux of sputtered atoms.
The authors considered the nonlocal nature of sputtering, a prerequisite for obtaining parallel sidewalls in the lower part of the structure and the flying roof in the upper part.
Simulations of pillar growth using voxel grids has been reported in Ref.~\citenum{fang_structural_2020}.

Presently, simulations of \ac{fibid}/\ac{fibie} are more advanced in the \ac{bca}-based approaches (cf.~Section~\ref{subsubs_BCA_FIB}). One reason for this might be that pure continuum approaches require the modeling of the \ac{se}/\ac{esa} intensities $f(\vec{x})$ as an additional step. Also, \ac{bca}-based approaches include several collision cascade related effects, which cannot easily be modeled in a continuum approach (see Section~\ref{subsubs_continuum_principles}). On the other hand, uncertainties in some of the assumptions of both the BCA and the continuum approaches might be more significant in practice.
These include the manner in which the energy and spatial distributions of the \acp{esa} and \acp{se} are determined, and assumptions about the fragmentation paths. Further development of either approach holds promise for model-assisted 3D nanoprinting, which has already been demonstrated for \ac{febid}\cite{Fowlkes_2018,winkler_high-fidelity_2018} and has been started for \ac{fibid}\cite{belianinov_direct_2020,timilsina_comparison_2013,mahady_monte_2019}.

\section{Applications of focused ion beams}%
\label{sec_appl}

\begin{figure*}[tbp]
   \centering
   \includegraphics[width=0.75\textwidth]{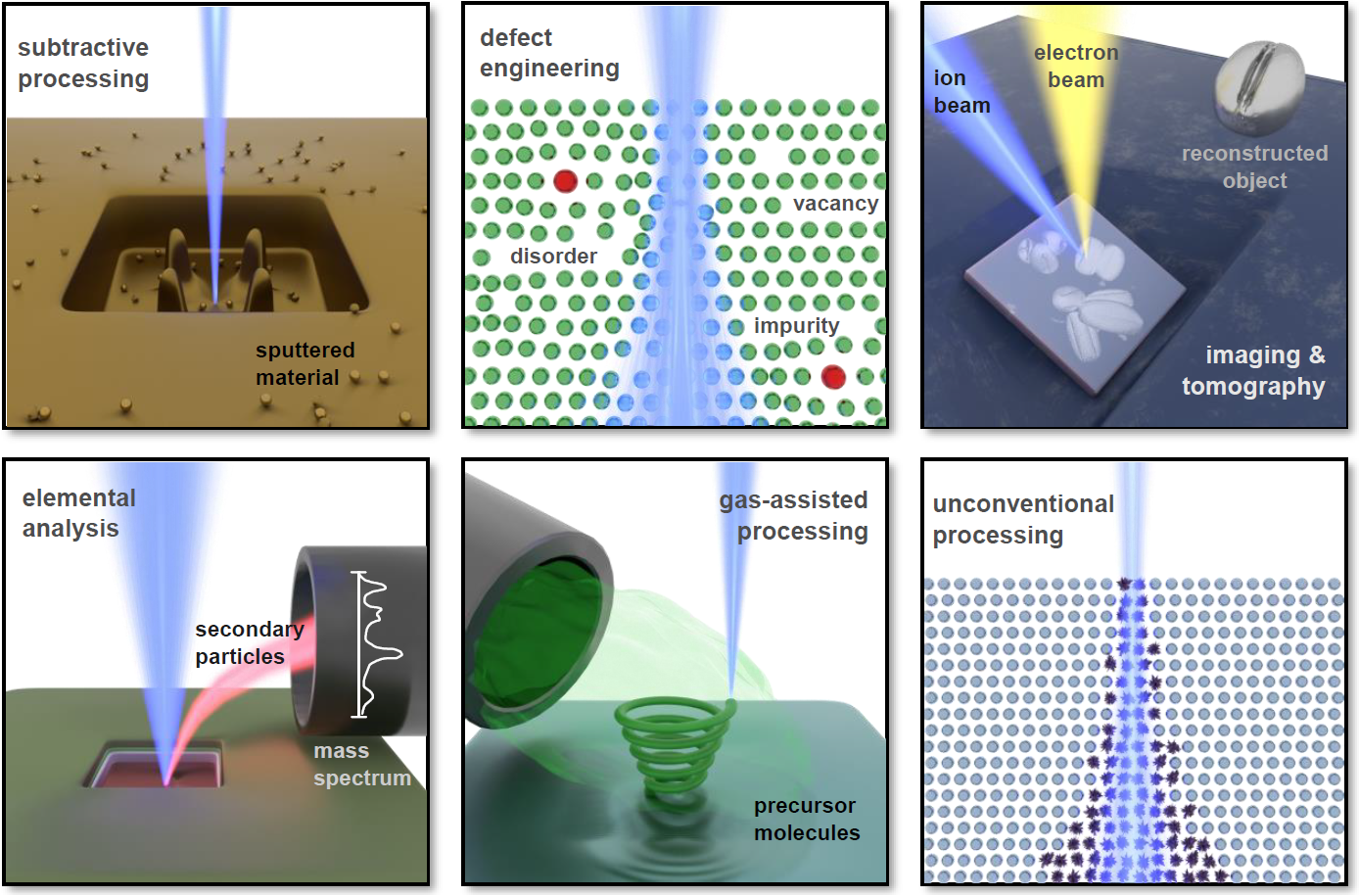}
   \caption{\label{fig:FIB_Overview_Alternative-Update} Schematic overview of the different \ac{fib} techniques}%
\end{figure*}

The various processes that are triggered when a beam of ions interacts with a sample form the basis of the multitude of \ac{fib} applications that are routinely used today. Accordingly, this section is intended to provide a comprehensive overview of this wide variety of experimental applications and is structured according to the different \ac{fib} processing techniques involved (see Fig.~\ref{fig:FIB_Overview_Alternative-Update} for an overview).

In the following paragraphs, the breadth of \ac{fib} techniques is briefly outlined, to be highlighted in more detail in the subsequent subsections. Further reviews of these topics can be found in a number of recent articles~\cite{Li_2021, Allen2021, Baglin_2020, Bruchhaus_2017, Bassim2019}.

\paragraph*{\bf Subtractive \acs*{fib} processing:}
Local material removal by a focused ion beam is based on physical sputtering. The incident projectiles knock out substrate atoms by momentum transfer, but also penetrate the material in a collision cascade. While the ejection of atoms is desired here, all other interactions with the substrate can cause unwanted side effects, such as amorphization, mixing, doping, dealloying, and others.
Conventional Ga-\ac{fib} tools provide spatial resolution and sputtering yields high enough for multiple prototyping purposes, but other ion species may offer advantages in terms of increased spatial resolution or sputter yield, and reduced damage or chemical reactivity. Section~\ref{subs_topdown} provides an overview of the rich variety of subtractive \Ac{fib} milling applications.

\paragraph*{\bf Defect engineering:}
\ac{fib} irradiation of materials at doses significantly lower than those used for milling has been implemented in a range of applications to locally tune material properties through the introduction of defects (vacancies and implanted ions). 
For example, by varying the dose, different concentrations of vacancy defects can be introduced into a crystalline material, thus modulating properties such as electrical, magnetic and thermal behavior in the patterned regions. 
In the case of ion implantation, this can be directed at \acl{sii}, or can be used to modify the chemical and structural properties of surfaces for resistless patterning. 
An overview of defect engineering applications is given in Section~\ref{subs_defects}.

\paragraph*{\bf Imaging and tomography:}
Just as is the case with the \ac{sem}, scanning the ion beam over a surface generates secondary electrons that can be detected to form an image.
In this respect, imaging using a focused ion or a focused electron beam is equivalent.
Even the same secondary electron detectors can be used.
However, what strongly differs, is the mass of the probe.
The mass of the ions is more than three orders of magnitude larger than that of the electrons, which significantly influences the contrast mechanisms and interaction volumes, and can also inflict significant damage on the sample.
In this respect, the light ions of the He-\ac{fib} bring great benefit. Section~\ref{subs_imaging} is dedicated to the application of the He-\ac{fib} specifically for the imaging of biological specimens.
The \ac{fib} can further be used in a subtractive manner as described above, to slice the sample and image the resulting cross-sections by \ac{sem}.
The images are then used for reconstruction of the full 3D information via so-called \ac{fib}-\ac{sem} tomography (see
subsections~\ref{sss:bio_tomo} and~\ref{sss:mat_tomo}).

\paragraph*{\bf Elemental analysis:}
During ion beam induced physical sputtering, atoms and ions originating from the sample are ejected and can be used to study its chemical composition. In \ac{sims}, the generated secondary ions are collected and separated by their mass in a spectrometer as detailed in Section~\ref{subs_analysis}. Other analytical techniques, such as \ac{eds} or \ac{ebsd}, are of minor importance in conventional \ac{fib} systems because the generation rate of X-ray quanta is negligible and the typical penetration depth of the conventional Ga ions is too low to generate sufficient information on  lattice orientation.

\paragraph*{\bf Gas-assisted \acs*{fib} processing:}
Gases can also be locally introduced into the \ac{fib} chamber to take advantage of other processing opportunities. For example, a reactive gas can be introduced to enhance sputtering yields in subtractive processing by milling (see, for example, Section \ref{subs_industry}). Alternatively, precursor gases can be selected to enable the highly localized deposition of material, which is referred to as \acf{fibid}. 
The flexibility of \ac{fibid} in terms of shape and spatial resolution (\SIrange{20}{30}{\nano\meter} for Ga-\ac{fibid}, and \qty{10}{\nano\meter} for He-\ac{fibid}) make it highly desirable for applications that require unique and precise planar or 3D designs. Top-down
\Ac{fibid} applications are discussed further in Section~\ref{subs_fibid}.

\paragraph*{\bf Unconventional \acs*{fib} processing}
Similar to gas-assisted processing, 
\acp{fib} can also be used for the local chemical modification of lithographic resists and condensed or spin-coated metalorganic precursors. Such `unconventional' \ac{fib} processing can increase processing speeds and provide new chemical reaction pathways for ion induced deposition of high-purity metals.
Focused ions can also be used to form self-assembled structures or combined with other micro-/nanofabrication techniques for advanced multiscale processing. Recent advances in the area of \ac{fib} based resist lithography and other non-standard \ac{fib} approaches are described in Section~\ref{subs_recent_adv}.

\subsection{Applications of subtractive \acs*{fib} processing}%
\label{subs_topdown} 

As material removal using a focused ion beam is local and direct, it offers extraordinary flexibility in terms of the target geometry to be realized, as well as the substrate materials and their geometries. Often \ac{fib} processing can be monitored in-operando using \ac{sem} for direct alignment and quality control.

\begin{longtable*}{P{2.6cm}P{4.6cm}P{1.3cm}P{8.2cm}}
   \caption[Subtractive \ac{fib} processing]{Overview of applications of subtractive \ac{fib} processing, in which material is locally removed by \ac{fib} based physical sputtering. Applications of FIB-SEM tomography are discussed in Section~\ref{sss:bio_tomo}.}%
   \label{table:subtractive}\\
	 \toprule
   Research area & Application & Ion & Material/Geometry/Device \\
	 \midrule
   \endfirsthead%
   \caption[]{continued}\\
	 \toprule
   Research area & Application & Ion & Material/Geometry/Device \\
   \midrule
   \endhead%
   \midrule\\
   \caption*{continued on the next page.}\\
   \endfoot%
   \bottomrule\\
   \endlastfoot%
	 Materials science
	 & TEM sample preparation
  	        & Ne 
	            & Al alloy and Si~\cite{Pekin2016}\\
	        & & Ga 
	            & wide range of materials~\cite{Mayer_2007}\\
	        & & Xe 
	            & Al and Si~\cite{Vitale_2022}, Al alloy~\cite{Zhong_2020}\\
	 & APT sample preparation
  	        & Ne 
	            & Al alloy and AlO$_x$ thin-film structure~\cite{Allen_2023}\\
	        & & Ga
	            & wide range of materials~\cite{Miller_2007}\\
	        & & Xe 
	            & Al alloy~\cite{Famelton_2021}, Ti alloy~\cite{Halpin_2019}\\

	 & mechanical testing
	       & Ga
         & micro/nanopillars~\cite{Uchic_2004,Uchic_2005, Greer_2005, Greer_2009,Shim_2009,Oestlund_2009, Huetsch_2014, Schwiedrzik_2014,  Xiao_2017,Bor_2021}, notches and micro- cantilevers~\cite{Armstrong_2009,Kiener_2009, Legros_2010,Kiener_2011,Ast_2014, Schoengrundner_2014, Best_2016, Kalacska_2020,Bor_2021}, `dogbone' tensile bars~\cite{Kiener_2008, Kiener_2009, Casari_2019}, suspended membrane beams~\cite{Soleimani_2022}, other geometries~\cite{Korsunsky_2010, Heyer_2014,Preiss_2021}
         \\
    & & Xe, Ne
         & micro/nanopillars~\cite{Xiao_2017}, trenches~\cite{Preiss_2021}\\

    \midrule
    Semiconductor technology
    & mask repair  
          & He, H, N
          & molybdenum silicide photomasks~\cite{Aramaki_2011, Aramaki_2013, Aramaki_2014, Edinger_2014, Tan_2016} \\
      &    & Ga
          & Cr-on-glass photomask~\cite{Heard_1985,Prewett_1987} including deposition for repair of clear defects~\cite{Cleaver_1985} and gas-assisted etching for opaque defects in TiN \acs{euv} masks~\cite{Liang_2000}\\
         & & Au
          & Cr-on-glass photomask~\cite{Heard_1985}, Au-on-Si X-ray mask including deposition and gas-assisted etching~\cite{Wagner_1990}\\
    & circuit edit  
          & He, H, N
          & \ac{cmos} circuits~\cite{Tan_2016, Tan_2015} \\
      &    & Ga
          & decomposition of Pd acetate for conductive lines~\cite{Harriott_1986a}\\

    \midrule
    Quantum \& magnetic materials
    & superconductivity
          & Ga
            & unconventional superconductivity in Sr$_2$RuO$_4$~\cite{Jang_2011,Yasui_2020}, 
            Josephson-vortices in strongly layered high-Tc superconductors~\cite{Moll_2014,Ooi_2002},
            vortex pinning~\cite{Pautrat_2004}, 
            directional vortex motion~\cite{Dobrovolskiy_2017,Dobrovolskiy_2015,Dobrovolskiy_2012},
            high-T$_c$ superconductivity \cite{Moll_2010,Jaroszynski_2008,Yang_2018}\\
    & superconductive circuitry 
	        & Ne 
		        & nanoconstrictions and nano\acsp*{squid} in Nb~\cite{Kennedy_2019,Burnett2017,Uhl2023}\\
    &
        & Ga
        & nanoconstrictions and nano\acsp*{squid}
        in Nb~\cite{Troeman2007,Granata2013,Schmelz2017,Wyss2022,Hao_2009a,Hao_2011,Hao_2008,Cox_2014}, NbN~\cite{Faley2022} and YBCO~\cite{Nagel2010,Schwarz2012,Schwarz2015,Lam2017}\\
    &      & Xe
            & nano\acsp*{squid} in Nb~\cite{Godfrey_2018}\\
    & spintronics and magnonics
        & Ne
        & permalloy microstrip trimming~\cite{Iurchuk_2023}, environment dependent \acs*{fmr}~\cite{Cansever_2022}\\ 
    &  & Ga
      & magnetic reversal~\cite{Lara_2014}, spin wave phase inverter~\cite{Dobrovolskiy_2019}\\ 
    & magnetism
          & Ga
            & surface time-reversal symmetry breaking in delafossites~\cite{Putzke_2020}, \\

    &topology
          & Ga
            & topological semi-metals~\cite{Geishendorf_2019,Gooth_2017,Liu_2021a},
            Fermi-arc surface states~\cite{Moll_2016},
            Kagome conductors~\cite{Guo2022}\\

    & strain engineering
          & Ga
          & Sr$_2$RuO$_4$~\cite{Maeno_1998}, CeIrIn$_5$~\cite{Delft_2022,Bachmann_2019}\\

 \midrule
	 
	 Optics
	 & photonic components 
		    & O 
		        & optical bridge in diamond~\cite{Martin_2015}\\
		&    & Ga 
    & aspherical microlens in Si~\cite{Langridge_2014}, solid immersion lenses in Si~\cite{Scholz_2010} and around pre-localized quantum emitters in diamond~\cite{Hadden_2010,Marseglia_2011,Jamali_2014}, photonic crystal cavity around quantum emitters in diamond~\cite{RiedrichMoeller_2014}, triangular nanobeam cavities in diamond~\cite{Bayn_2011} and in \acs*{yvo}~\cite{Kindem_2020,Zhong_2017,Zhong_2016}, Fresnel zone plates for X-ray microscopy~\cite{Keskinbora_2013}, Fresnel lenses milled through hard mask~\cite{Madison_2022}, metasurfaces for polarization and phase engineering~\cite{Li_2018}, tapered nanoantenna for single photon emission of CdSe quantum dots~\cite{Rakhlin_2021}\\
     & fiber-based optical components 
		    & Ga 
        & lab-on-a-fiber~\cite{Sloyan_2021}, Bragg gratings and microlenses in silica nanofibers~\cite{Romagnoli_2020a}, \ac{sspd} on optical fiber~\cite{Korneeva_2020,Bachar_2012}\\
     & plasmonic components 
         & He
        & Al bowtie antennae~\cite{Simeone_2018}, Au dimer antennae of different shapes~\cite{Kollmann_2014, Wang_2013, Scholder_2013, Seitl2019, Chen_2020a, Gittinger_2020, Smirnov2021}, Au coaxial~\cite{Melli_2013} and split-ball~\cite{Kuznetsov2014} and nanoslit~\cite{Chen2018} resonators, pores in Au pyramidal~\cite{Grant-Jacob_2016} and microsphere~\cite{Oo2016} cavities, Au oligomers~\cite{Deinhart_2021}, inverse Au trimers~\cite{Singh2018} and oligomers~\cite{Hahn_2020, Hajebifard_2021}\\
         &   & Ga 
	    & Au groove waveguides~\cite{Kriesch_2013} forming logic gates~\cite{Fu_2012}, Au bowtie antennae for EUV light generation~\cite{Sivis_2013}, 
	    inverse Ag bowties~\cite{Ding_2016}, 
	    ring-shaped beam patterning of Au antenna arrays~\cite{Lv_2015}, 
	    Yagi-Uda-type antennae~\cite{Kullock_2020, Kriesch_2013}, 
	    $\lambda/2$ antennae~\cite{Muehlschlegel_2005}, 
	    asymmetric Ag oligomers~\cite{Cherqui_2016}, two-wire waveguides in Au~\cite{Rewitz_2014,Geisler_2013} forming spin-optical nanodevice~\cite{Krauss_2019}\\

 \midrule	 
	 Custom probes and sensors
	& near-field microscopy 
 	          & He
	            & square nanopores in pyramidal Au nanocavities~\cite{Grant-Jacob_2016}, Au nanopyramid antenna~\cite{Archanjo_2018}\\
            & & Ga 
            & circular~\cite{Veerman_1998} and bowtie~\cite{Mivelle_2012} nanoaperture, tip-on-aperture geometry~\cite{Taminiau_2006,Zanten_2010},  Au bowtie~\cite{Farahani_2007} and nanocone antenna~\cite{Hoffmann_2015}, nanoslit grating in Au conical taper~\cite{Schmidt_2012}\\
             & & Xe
             & probe shaping for near-field scanning microwave microscope~\cite{Geaney_2019}\\
    & atomic force microscopy
            & Ga
		        & nanoscale apertures for liquid and ion transport~\cite{Meister_2004,Guillaume-Gentil_2014,Persaud_2005}, tip sharpening~\cite{Wang_2014}, modified mechanical response~\cite{Kunicki_2019}\\
    & magnetic force microscopy
            & Ga
            & V-shaped magnetic nanostructure at probe tip~\cite{CorteLeon_2019}\\
  & SQUID microscopy
            & Ga
            & \acsp*{squid} on AFM cantilever~\cite{Wyss2022}\\
    & mechanical high-Q resonators 
            & He 
	            & phononic crystal~\cite{Kirchhof_2021a}\\
	        & & Ga 
	            & graphene trampoline~\cite{Miller_2018a}\\
	 %
\end{longtable*}

\subsubsection{\acs*{fib} for materials science}%
\label{sss_subtract_mat}

Arguably the most widespread application of the \ac{fib} has been the site-selective fabrication of cross-section samples for high-resolution imaging using \ac{tem}~\cite{Volkert_2007, Mayer_2007, Lenrick_2013,Schaffer_2012, Berger_2023}.
Due to the prevalence of Ga-\ac{fib} instruments, Ga ions are most typically employed and various strategies have been developed to mill and extract cross-sections from bulk materials and to thin samples to electron transparency. 
This includes low-energy \ac{fib} milling under glancing incidence to minimize surface amorphization, and the use of material specific capping layers and rocking stages to avoid `curtaining' artifacts. 
More recently, the use of other \ac{fib} species for \ac{tem} lamella preparation has been demonstrated, such as the Ne-\ac{fib}~{\cite{Pekin2016}} and the Xe-\ac{fib}~\cite{Vitale_2022}. 
The advantage of a noble gas ion species is that it does not chemically react with the sample or accumulate along grain boundaries, as can be the case with Ga~\cite{Unocic2010}. 

In the same manner, \ac{fib} milling has emerged as an important tool for the creation of needle-shaped specimens for \ac{apt}~\cite{Thompson2007,Miller_2007, Prosa_2017}.
The required needle geometry is achieved via annular milling with different currents and successively smaller milling areas.
Similar considerations regarding beam damage apply as for \ac{tem} sample preparation.
For example, minimizing the damage layer is important and possible indiffusion of e.g.\ Ga atoms~\cite{Hugo_1998} has to be taken into account. 
In this respect, a final polish with Ne-\ac{fib} has recently been shown to remove all residual Ga contamination from Ga-\ac{fib}-milled \ac{apt} needle specimens~\cite{Allen2023}. Performing the Ga-\ac{fib} milling at cryogenic temperatures has also been shown to limit Ga diffusion into the specimen~\cite{Lilensten_2020}.

Another important application is the site-specific preparation of cross-section cuts that enable in-situ and ex-situ imaging~\cite{Uchic_2007} and/or other analysis such as \ac{eds} or \ac{ebsd} (e.g.~\citet{Kalacska_2020}).
By automating this process, full (destructive) tomographies can be performed by slicing the sample, extracting the desired information from each slice, and reconstructing the data in three dimensions (see for example~\citet{Kubis_2004, Burnett_2016}). Tomography and its applications for the study of energy materials and, in particular, complex organic systems in the life sciences, are discussed in more detail in Section~\ref{sss:bio_tomo}. 

All of these techniques are part of the standard materials science repertoire used to study the microstructure, composition, and properties of materials and devices at specific target locations. 
In depth overviews on these topics can be found, for example, in \citet{Giannuzzi_2005a}. 
One specific 
area in materials science that particularly benefits from the 3D sculpting capabilities of \ac{fib} concerns small-scale mechanical testing and is discussed in more detail in the following subsection.

\subsubsection{\acs*{fib} for mechanical testing}

As technology miniaturizes, the external dimensions of device components approach critical internal length scales such as dislocation interspacings.
The mechanical properties of such small material volumes are still not fully understood. 
In 2004, \citet{Uchic_2004} suggested the use of \ac{fib} milling to machine micropillars to study size dependent mechanical properties of metals, exploiting conventional nanoindentation equipment for microcompression experiments. 
In contrast to conventional cleanroom-based lithography and etching, \ac{fib} machining allows for virtually any material to be shaped into a micropillar at a specific target location.  
To achieve uniform pillar circumferences with minimal taper, lathe milling in combination with annular milling is often applied~\cite{Uchic_2005}.
Here, lathe milling refers to a particular \ac{fib} approach in which the sample is rotated stepwise and the ion beam performs a line scan along the rim of the pillar.
While lathe milling can introduce severe levels of stress requiring careful interpretation of the micromechanical testing results~\cite{Huetsch_2014}, 
it is well suited for insulating or very inhomogeneous samples comprising layers with very different sputter yields.

Today, most small-scale mechanical testing experiments are performed by \ac{sem} or \ac{tem} to capture changes from the atomic to micron scale (e.g.\ formation of dislocations, stacking faults, twins, grain orientations, contraction and expansion of volume). 
Complex workflows include some or all of the following steps:
(a) machining of micro-/nanopillars and other geometries,
(b) in-situ mechanical testing under \ac{sem} observation,
(c) deposition of a regular pattern of points by \ac{fibid} (or \ac{febid}) to follow deformations via digital image correlation algorithms, and 
(d) post-mortem \ac{fib} slicing for 3D tomographic or 3D \ac{ebsd} analysis of the deformed microstructure or the fracture pattern inside the sample~\cite{Kalacska_2020}. 
\begin{figure}[tbp]
   \centering
   \includegraphics[width=0.9\linewidth]{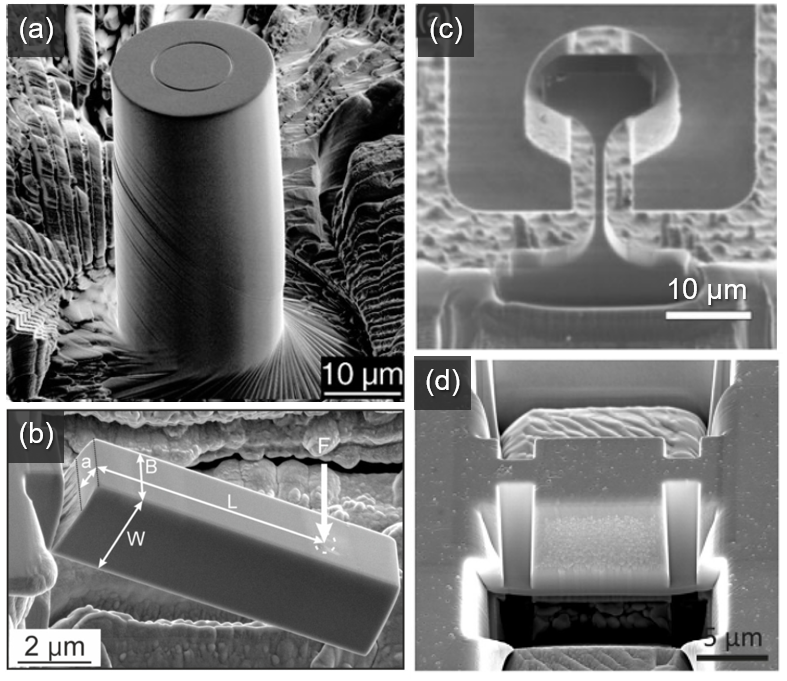}  
   \caption{\label{fig:fib-mechanics}Popular FIB-machined sample geometries for micromechanical testing: 
      (a) micropillars~\cite{Uchic_2004},
      (b) notched micro cantilvers~\cite{Ast_2019}, 
      (c) micro-tensile bars~\cite{Casari_2019} and 
   (d) micro-shear tests~\cite{Heyer_2014}.}
\end{figure}
\ac{fib} machining of sharp notches with a specified orientation to the loading axis has enabled a number of fracture mechanics studies~\cite{Ast_2019}, and 
for in-situ \ac{tem} tensile testing, various milling protocols have been developed to produce electron transparent specimens~\cite{Kiener_2011}. 

To reduce ion beam induced surface damage defects, a multistep milling strategy is typically employed. In this approach, the beam voltage and current are sequentially stepped down to the last polishing step~\cite{Mayer_2007}.
Among the still open questions are how much damage (amorphization, implantation) is induced by \ac{fib} milling into the structures and how the damaged zones affect the mechanical behaviour of the milled structure~\cite{Shim_2009}. 
Key factors are the penetration (implantation) depth of ions and the collision cascade (amorphization) they trigger (see Sections~\ref{subsubs_BCA_FIB},~\ref{subsubs_md_applications}, and~\ref{subsubs_continuum_principles}).

Once other ion sources, in particular He, Ne, and Xe ion sources, became available, the hope was to be able to circumvent some of the Ga damage issues. 
\citet{Best_2016} found, however, that using either He, Xe or Ga for the machining of notches affected the measured fracture toughness value in all cases. 
While He ions were shown to yield suitable sharp notches with radii below \qty{10}{nm}, Xe ion beams were resolution limited. 
It was unclear how the penetration depth of the different ions affects the fracture toughness. 
Currently, combinations of \ac{fib} milling with other technologies are becoming more and more popular. For example, the option of laser machining of large volumes followed by \ac{fib} fine sculpting has recently become available from several manufacturers in a single setup. 

\subsubsection{\acs*{fib} in the semiconductor industry}
\label{subs_industry}

In the field of semiconductor processing, \acp{fib} are most relevant during the debugging of new products and for identifying production errors in the context of circuit edit and fault isolation, as well as for the repair of photomasks.
Among other applications, \acp{fib} have been used for failure analysis, fault isolation, \ac{tem} sample preparation, mask repair and backside circuit edit~\cite{Orloff_2003, Giannuzzi_2005a,Gandhi_2019}.
The ongoing scaling of the transistor fin, gate, and interconnect via linear and volumetric shrinking, as well as the introduction of new transistor technologies, has led to a reduction of the critical length scale from \qty{22}{nm} in 2012 to the current \qty{3}{nm} technology node to be introduced by all major manufacturers between 2022 and 2024~\cite{TSMC_2022,Intel_2022, Samsung_2022}. 

\acp{fib} have been used since the 1980s for mask repair, initially 
using Ga and Au ions from \ac{lmis} and \ac{lmais}~\cite{Heard_1985}. However, staining from implanted ions was quickly identified as a problem limiting the application~\cite{Prewett_1987}.
Later, the application of \ac{gfis} based H$^+$, H$_2^+$, N$_2^+$, and He$^+$ for phase-shift mask and \ac{euv} mask repair was investigated~\cite{Aramaki_2011,Edinger_2014,Aramaki_2014}.
For H$_2^+$, the minimal repairable dimension is reported to be \qty{11}{nm}, which is half of what could be achieved using electron beams~\cite{Aramaki_2013}.
The application of \ac{fibid} was investigated early on for adding missing material to masks and for repairing pin holes using Ga-\ac{fib} induced carbon deposition~\cite{Cleaver_1985}.
This has been extended using other deposition precursors.
The addition of reactive gases such as Cl$_2$~\cite{Young_1990,Liang_2000},\ Br$_2$~\cite{Liang_2000}, XeF\textsubscript{2}~\cite{Xu_1989,Gamo_1988}, and O\textsubscript{2}~\cite{Gamo_1988} has also been investigated for mask repair. 
These gases can significantly increase etch rates, remove carbon contamination and trigger resist development.
In particular, XeF\textsubscript{2} is now readily available in commercial \ac{fib} systems.

For circuit edit and fault isolation applications, Ga-\ac{fib} was used initially~\cite{Harriott_1986a}, but later  Ne ions from \ac{gfis} instruments were employed. The latter yielded minimal trench widths of \qty{14}{nm} for high aspect ratio vias~\cite{Tan_2016}, and comparable/less collateral damage compared to Ga-\ac{fib} as measured from changes to ring oscillator timing~\cite{Tan_2015}.
Recently, \ac{gfis}-based H ions have been demonstrated for successful delayering~\cite{Matsubara_2019}.
Other source technologies,  e.g.\ Cr ions from \ac{motis}~\cite{Steele_2010}, also show potential for circuit edit applications~\cite{Livengood_2011a}.
In general it is found that lower beam energies are be required while still maintaining a suitably high spatial resolution.

To better understand the effects of the beam profile on the achievable mill resolution, approaches have been developed to elucidate the probe current distribution~\cite{Tan_2012, Tan_2012a}.
This is of particular importance for via fabrication and also for the fabrication of high aspect ratio holes/pores created by milling with a stationary beam.
For these applications a good signal to noise ratio is needed for reliable endpoint detection, and a new figure of merit---\ac{se} yield per sputtered atom---can be defined. 
\ac{gfis}-based He and Ne milling of these structures promises superior spatial and temporal precision compared to Ga-based \ac{fib} processing~\cite{Tan_2016a}.
However, \ac{gfis}-based milling suffers from the formation of gas bubbles and the creation of defects far below the milled surface of bulk samples~\cite{Livengood2009}.

\subsubsection{\acs*{fib} for quantum materials}
``Quantum materials'' are materials in which quantum phenomena manifest in non-trivial ways, leading to response functions that are qualitatively distinct from those of standard electronic materials such as Si or Cu.
These functions generally arise from the non-locality of quantum systems, with topology, frustration, and correlation all playing a central role~\cite{Basov_2017,Giustino_2020}.
The key goal is to refine our microscopic or phenomenological understanding of these exotic quantum states, and develop predictors to find promising materials candidates within the vast space of possible compounds. A second main driver of quantum materials research are novel electronic applications that harness this exotic behavior for solving technological challenges. Most famous is quantum information technology, including quantum computation, which directly utilizes the entanglement and vast configurational space of quantum coherent systems. The search for the most effective materials platform to realize systems like qubits remains an open and very active field, i.e.\ the search for the "silicon" of quantum technology. Beyond coherent information processing, a large materials space has developed in sensing and data communication that utilizes quantum coherence indirectly, such as in strongly correlated oxides. The ubiquity of quantum mechanics is the greatest challenge, in that interdisciplinary efforts in materials science, chemistry, physics, engineering and informatics are required to realize these concepts in a useful way. In this search for new physics and new quantum platforms, the \ac{fib} excels as a versatile patterning tool that is ideal for rapid processing of non-standard materials~\cite{Moll_2018}, as it (a) avoids exposure to wet chemistry as compared to resist-based lithography, (b) can operate with crystallites and microparticles, i.e.\ non-standard samples that do not resemble polished wafers, and (c) enables 3D structure control that is critical to many electronic 3D quantum materials.
The operational goal of this research is to fabricate hybrid structures that join micro-/nanostructures of a quantum material to a Si chip of conventional architecture, both to demonstrate advanced functionalities and to probe the underlying novel physics.

Quantum systems generically respond strongly to changes in their boundary conditions, in this case, their physical shape.
\acp{fib} enable the sculpting of non-trivial boundary conditions in crystals, which has been used to study and functionalize superconductivity, e.g.\ investigating the unconventional superconductivity of Sr$_2$RuO$_4$~\cite{Jang_2011,Yasui_2020}, and studying Josephson vortices in strongly layered high-T$_c$ superconductors and their excitations~\cite{Moll_2014,Ooi_2002}. Controlled patterns guide the motion of Abrikosov lattices by surface structuring~\cite{Dobrovolskiy_2017} towards non-trivial electronic responses such as vortex diodes~\cite{Dobrovolskiy_2015,Dobrovolskiy_2012}. Further towards superconducting applications on small dimensions, \ac{fib} milling also enables direct nanopatterning of Josephson junctions, nanowires and nanoSQUIDs~\cite{Granata2016,MartinezPerez2017}. 
The nanoSQUIDs are used for magnetization reversal detection of individual magnetic nanoparticles~\cite{Troeman2007, Schwarz2015} and for magnetic and thermal sensing in scanning SQUID microscopy~\cite{Wyss2022}. In microwave circuits, nanoSQUIDS can be used as tunable inductors, e.g. to enable tuning of the resonance frequency of microwave resonators~\cite{Kennedy_2019}.

Beyond superconductors, other exotic materials and properties can be probed, such as quasiparticle transport in ultra-clean metals~\cite{Putzke_2020}, and the magnetic exchange bias in time-reversal symmetry breaking topological semi-metals~\cite{Noah_2022}. Reducing the sample size to the micron scale further enhances the contribution of surface properties to the overall physics. This is particularly important in the quest to probe the contribution of topologically protected surface states, which have been proposed to enable novel electronic and spintronic applications in topological semi-metals~\cite{Geishendorf_2019,Moll_2016,Gooth_2017,Liu_2021a}. Precise \ac{fib} machining also enables optimization of the edge quality of structures, which is a prerequisite for the fabrication of efficient \acfp{sspd}~\cite{Korneeva_2020}.
In addition, the flexibility of \ac{fib} processing on highly 3D substrates has enabled \ac{fib} milling on top of an optical fiber~\cite{Bachar_2012}.

Subtractive \ac{fib} processing has also been applied in the fields of spintronics and magnonics.
Examples include guiding and control of propagating spin waves, which has been enabled using the Ga-\ac{fib}~\cite{Dobrovolskiy_2019}. 
High resolution metal free \ac{fib} processing using \ac{gfis} ion species has also enabled fine tuning of the interaction of spin waves across gaps~\cite{Iurchuk_2023}. 

Not all applications of \ac{fib} aim to alter the material.
In addition, there is a wide class of \ac{fib} milling experiments aimed at probing the bulk properties of novel materials that can be difficult to access when the material is in its native form, e.g.\ microcrystals that are too small for conventional measurements.
This has allowed insights into microscopic effects on critical charge transport in Fe-based superconductors~\cite{Moll_2010,Jaroszynski_2008}, the determination of topological bandstructures in microcrystals~\cite{Yang_2018}, and probing of the switchability of anti-ferromagnetic states under the high current densities achievable in \ac{fib}-machined crystals~\cite{Nair_2019,Nair_2020}.
The rich physical responses of quantum materials tend to be rooted in the competition between multiple electronic ground states, which leads to a strong (and potentially discontinuous) change in their properties under weak applied stimuli~\cite{Ronning_2017}.
One can hardly overstate the importance of residual stress in the samples, with a famous example being the enigma of the ``3K phase'' of superconductivity in Sr$_2$RuO$_4$, which originated from strain mismatch~\cite{Maeno_1998}.
Here a crucial advantage of \ac{fib} milling over mechanical sample preparation techniques, such as polishing or wire sawing, is its kinetic nature, which results in samples of low residual stress.
Precision control over the physical shape of a sample has also enabled microscopic control over desired strain fields and their gradients, which imprint correlation landscapes into quantum materials~\cite{Delft_2022,Bachmann_2019}.

\subsubsection{\acs*{fib} for photonics and plasmonics}

Miniaturized on-chip photonic components are designed to guide or focus light while having a footprint as small as possible. The fabrication of these relies mostly on conventional large-scale lithographic techniques, but there are applications where a maskless direct-write strategy has significant advantages. 
Prime examples are localized light sources, with the extreme case of single photon sources, where the optical component often has to be built around a pre-localized (quantum) emitter.
Such quantum emitters may be defects in wide-bandgap materials (see Section~\ref{sss:4_D_3-optical_appli}), where out-coupling of the generated quantum light is typically hindered by the large refractive index of the host material, e.g.\ diamond. 
Here, \Ac{fib} milling either using Ga ions~\cite{Hadden_2010, Marseglia_2011, Jamali_2014} or, to reduce Ga-induced damage, O ions~\cite{Martin_2015}, has been successfully used to engrave solid immersion lenses that enhance photon collection efficiencies. 
In the same manner, photonic crystal cavities with small mode volumes have been milled into a suspended diamond layer around pre-localized quantum emitters and proven to enhance spontaneous emission~\cite{RiedrichMoeller_2014}.  
An even further reduction of optical mode volume can be achieved using metallic nanostructures, since these have the ability to concentrate light below the diffraction limit by collective excitations of the free electron gas~\cite{Novotny_2012}. 
Since such plasmon-polariton modes are strongly dependent on geometry, \ac{fib} processing is a powerful tool for rapid prototyping of nanoscale resonators, sub-wavelength apertures~\cite{GarciaVidal_2010} and optical antennae~\cite{Novotny_2011}. 

Furthermore, extreme plasmonic field confinement allows the detection of molecular species down to the single molecule level by surface enhanced Raman scattering~\cite{Kneipp_1997} (the prime example), enhanced fluorescence~\cite{Punj_2013}, or enhanced F\"orster resonant energy transfer~\cite{Ghenuche_2015}. 
The latter two examples employ Ga-\ac{fib} milling to fabricate an antenna-in-a-box geometry, where the gap region of a dimer antenna provides the local electric field enhancement and correspondingly an enhanced local density of optical states.  The remaining metallic film around the antenna blocks fluorescence from contributing to the background. 
In general, Ga-FIB top down patterning, especially of single-crystalline Au or Ag nanoflakes, is often employed for high-fidelity rapid prototyping of plasmonic components~\cite{Huang_2010,Geisler_2013,Kern_2015,Schoerner_2019,Ochs_2021a}. 
When comparing the performance and material properties of plasmonic antennae, \ac{ebl} may lead to better results than \ac{fib} milling, depending on the actual geometry and processing conditions, especially in case of Ga-\ac{fib}~\cite{Horak_2018}.
However, \Ac{ebl} fails in fabricating truly 3D profiles.
In contrast, \ac{fib} milling has been used to e.g.\ mill out 3D Janus plasmonic helical nanoapertures for polarization-encrypted data storage~\cite{Chen_2019}.

A particularly beautiful example of 3D \ac{fib}-milling capabilities combines local material removal by Ga-\ac{fib} milling of a suspended gold film with low dose Ga ion irradiation at selected edges. The latter induces a stress distribution that leads to up- or downward bending of the predefined structures~\cite{Xia_2006}. This nano-kirigami approach has been demonstrated for various designs realizing optical chirality, polarization and phase engineering, and showing potential for opto-mechanical applications~\cite{Li_2018}. 

The resolution and shape fidelity of \Ac{fib} milling with light ions stands out compared to any other lithographic techniques~\cite{Allen2021,Deinhart_2021}, but comes at the cost of the slow speed of the inherently serial process.
To resolve this issue, lithographic large area processing by \Ac{ebl} can be combined with e.g.\ He-\ac{fib} for the ultimate in resolution. 
Extremely narrow gaps between large areas of lithographically defined plasmonic Au~\cite{Scholder_2013} and Al~\cite{Simeone_2018} antennae have been realized using this approach. 
By combining Ga-\ac{fib} patterning with He-\ac{fib} patterning, the achievable spatial resolution can be significantly improved while keeping a reasonable processing speed~\cite{Kollmann_2014}. 
Another approach to increase the speed of top-down FIB processing relies on milling only the outlines of desired plasmonic features, which are then subsequently obtained by peeling off the surrounding material~\cite{Chen_2016a,Laible_2019,Chen_2020a}.
This ``sketch and peel technique'', combining Ga-\ac{fib} patterning for the outlines with He-\ac{fib} for the definition of ultra-narrow gaps, was used to reproducibly demonstrate strong coupling in plasmonic dimer antennae~\cite{Gittinger_2020}.

\subsubsection{\label{sss:fab_afm}\acs*{fib} fabrication of custom probes}

Since \ac{fib} milling is not restricted to planar sample geometries, a vast body of research exists on the modification of optical fibers~\cite{Sloyan_2021}, nanofibers~\cite{Romagnoli_2020a} and nano-probes. 
The comprehensive review by \citet{Sloyan_2021} presents examples for optical near-field imaging, plasmonics,  beam shaping, fiber-based photonic cavities, fiber-chip coupling, and many more. He-\ac{fib} manufacturing was reviewed in detail by \citet{Allen2021}. 
Direct control over precise nanostructure geometries by \ac{fib} milling is especially interesting for \ac{nsom}, where a metal coated glass fiber tip localizes the excitation and/or collection of light by making use of a nanoaperture or antenna for plasmonic field concentration.
In applications concerned with \ac{nsom}, \ac{fib} milling has been employed to fabricate high-definition plasmonic circular nanoapertures at the apex of a metal coated glass fiber tip~\cite{Veerman_1998}. Further signal enhancement and fluorescent background reduction can be achieved by adding a tip at the aperture rim,  combining the concepts of scattering and aperture \ac{nsom}~\cite{Frey_2002,Frey_2004}. 
While the tip was grown by FEBID in these works and metallized afterwards, later work engraved the tip-on-aperture geometry into the metallic (Al) fiber coating by Ga-FIB milling and tuned the tip resonance to behave as a monopole antenna, thus achieving single molecule imaging~\cite{Taminiau_2006} and imaging of proteins on cell membranes~\cite{Zanten_2010}.
Additional geometries realized in this way are bowtie nanoapertures on the end of metal-coated fiber tips~\cite{Mivelle_2012} and square plasmonic nanoapertures in pyramidal cavity structures~\cite{Grant-Jacob_2016}, with the latter example employing He-FIB milling. 

Further specific geometries of metallic tips and metal-coated fiber tips have been fabricated to control the plasmonic response for other purposes.   
These include nanocones for resonance and radiation pattern tuning by Ga-FIB milling~\cite{Hoffmann_2015}, nanopyramids for optimizing local electric near-fields by He-FIB milling~\cite{Archanjo_2018}, and nanoslit gratings into the conical taper of metallic tips for nanofocusing upon far-field illumination, again by Ga-FIB milling~\cite{Schmidt_2012}.
\ac{fib} milling has also been been employed to modify \ac{afm} probes for specialized applications. 
In one example, \ac{fib}-milling was used to partially remove the magnetic layer from a commercial \ac{mfm}-probe. 
The V-shaped nanostructure remaining on one side of the triangular pyramid results in enhanced magnetic phase contrast without compromising spatial resolution~\cite{CorteLeon_2019}.  
Another example employs nanoscale apertures milled into hollow pyramidal \ac{afm} tips, that can be loaded with liquid to dispense attoliter volumes and create droplet arrays~\cite{Meister_2004}. Such tips can also be be used for the manipulation and analysis of single cells via so-called fluidic \ac{afm}~\cite{Guillaume-Gentil_2014}. 
Tips can be sharpened using \ac{fib} milling and can also be detached from their initial substrate and mounted onto a cantilever of choice with the help of a micro-manipulator. 
This technique has, for example, been used to produce customized tips with high aspect ratios~\cite{Wang_2014}. 
In other work, the \ac{afm} cantilevers themselves have been fashioned with \ac{fib}-milled apertures to enable ion implantation through those apertures aided by scanning probe alignment~\cite{Persaud_2005}. 
\Ac{fib} milling of larger patterns into piezoresitive cantilevers has been used to increase deflection sensitivity to optimize cantilever performance~\cite{Kunicki_2019}.
Recently, Ga-FIB milling has also been used for shaping AFM tips to produce mesa structures on the cantilever onto which Nb nano\acsp*{squid} have been fabricated. Such structures simultaneously enable topographic, magnetic and thermal imaging~\cite{Wyss2022}.

\subsubsection{Outlook}

As an all-in-one processing tool, the \ac{fib} provides a versatile platform for flexible and fast top-down fabrication and prototyping of nanostructures and devices.

To further advance these developments, \ac{fib} technology needs improvement not only in spatial resolution but also in the in-situ control of the material removal process.  
Critical issues include material sputtering selectivity and increasing the signal to noise ratio of \ac{se} monitoring during the milling of high aspect ratio structures.
However, \ac{fib} top-down nanofabrication is intrinsically limited by the fact that the energetic particles and the chemical species involved cause structural and chemical damage or changes, such as amorphization or poisoning.
While traditional Ga-\ac{fib} still represents the \ac{fib} standard, some of the aforementioned fundamental limitations may be resolved by using other ion species, such as those offered by \ac{gfis}, \ac{motis}, \ac{lotis}, \ac{pfib} or \ac{lmais} (see e.g.\ Section~\ref{subs_sources}).

Additionally, scale-up of \ac{fib} based top-down nanofabrication will only be achieved when increased throughput and reproducibility are realized.
Here, advances in software and hardware tools for e.g.\ modeling-informed automated operation will be beneficial, also allowing in-line integration of \ac{fib} processing into larger (nano)-fabrication pipelines.
Theoretical knowledge, modeling and simulation tools that provide precise descriptions of \ac{fib} modifications to materials, from the atomic to the microscopic scale, are a prerequisite for predictive design and for the realization of tailored properties and functionalities of nanostructured surfaces and operating devices.

There are also opportunities for coordinated \ac{fib} system development based on specific experimental needs. 
For example, strong interest in emerging fields such as quantum technology, as well as novel miniaturized platforms in the fields of microfluidics for lab-on-a-chip or organ-on-a-chip developments, would benefit from synergies with theory groups and tool makers in the \ac{fib} application community.

\subsection{Engineering materials properties through defects}%
\label{subs_defects}

In addition to the subtractive nanofabrication applications described above, site-specific and dose-controlled irradiation of samples with the \ac{fib} is used to tune material properties in spatially defined regions down to the nanoscale and even down to the single-atom level, without surface erosion or deposition. These applications are based on the local introduction of varying concentrations of defects (vacancies, dopants, and dopant-vacancy complexes), controlled by the dose.

\subsubsection{\label{sss:irra_defects}Basics of irradiation induced defect creation}

A common technique used to create defects in crystals relies on bombardment with high energy ions or electrons~\cite{Bernas2010,Krasheninnikov2010b,Zhang2020},
and in some cases photons, protons or neutrons~\cite{Kinchin1955}. For example, tuning the fluorescence properties or the refractive index of a material  for optical applications by bombardment with ions is a well-established technique~\cite{Morimoto1996,Chen_2016}.
Ion implanters are often employed for such applications. 
Since these deliver broad beams with spot sizes in the millimeter regime, a lithographically defined mask is needed when higher spatial resolution is required. 
However, the large collision cascades that are inherent to both high energy ion beams and light ion bombardment make mask design rather challenging. 
This is where \ac{fib} processing comes into play.

As described in Section~\ref{sec_theo}, the incident ion induces a collision cascade in the solid with a size that is determined by the target material and the ion's kinetic energy, as defined by the primary beam voltage.
In a simplified picture, when a recoil is generated, it leaves behind a vacancy and forms an interstitial when it comes to rest\cite{Ziegler2010}.
In reality, defect clusters and amorphous pockets may form depending on the target material\cite{nordlund_defect_1998} and on the ion mass and energy~\cite{pelaz_ion-beam-induced_2004}. 
Larger defect structures such as these are usually undesirable for defect engineering, as their properties are less well defined than those of simple defects. 
They are best avoided by using light ions, which cause more dilute collision cascades than heavier ions due to the lower nuclear stopping power. 
The recoils in these cases are also less energetic because of the unequal masses of the ion and target atoms. 
In addition, light ions cause less sputtering, which is generally an unwanted effect in defect engineering.
Regardless, the applied ion doses and beam parameters still have to be carefully adjusted. 
Considering the example of He ion irradiation of Si and Cu substrates, \citet{Livengood2009} differentiate the various defect regimes according to dose, ranging from individual point defects at low fluence, to extended subsurface microbubbles at very high fluence. 
While the picture will vary somewhat for other ion-substrate combinations, the same overall trends apply.

Table~\ref{table:property_engineering} gives the reader a flavor of the breadth of \ac{fib} based defect engineering applications, categorized according to the specific materials property engineered (electrical/electronic, magnetic, optical, quantum, chemical, thermal, and mechanical). Further examples can be found in the review articles by~\citet{Gierak_2009}, \citet{Orus_2020}, and~\citet{Allen2021}. 
\begin{longtable*}{P{1.8cm}P{5.4cm}P{1.3cm}P{8.2cm}}
 \caption[\ac{fib} based materials property engineering]{Overview of \ac{fib} based property engineering applications, in which fine control over the ion dose, beam energy, and irradiated regions enables local tuning of material properties.}%
 \label{table:property_engineering}\\
	 \toprule
	 Engineered Property & Experiment & Ion & Material/Device\\ 
	 \midrule
   \endfirsthead%
   \caption[]{continued}\\
	 \toprule
	 Engineered Property & Experiment & Ion & Material/Device\\ 
   \midrule
   \endhead%
   \midrule\\
   \caption*{continued on the next page.}\\
   \endfoot%
   \bottomrule\\
   \endlastfoot%
	 Electrical \&~electronic
	    & conductivity tuning 
	        & He 
          & graphene~\cite{Nakaharai2013,Moktadir2015,Araujo2015,Zhou2016,Naitou2017,Nanda2017}, few-layer/monolayer MoS$_2$~\cite{Fox-2015,Stanford2017,Jadwiszczak-2019}, epitaxial MoS$_2$~\cite{Parida_2022}, few-layer/monolayer WSe$_2$~\cite{Stanford-2016,Stanford2017}, monolayer WSe$_2$~\cite{Stanford2017}, InGaZnO thin film~\cite{Stanford2017b}\\
	     & & Ne 
          & Si nanocrystals (self-assembly of) in buried silica layer~\cite{xu_site-controlled_2018}\\   
         & & Ga 
          & AlGaAs/GaAs quantum wires~\cite{Hirayama_1988a, Itoh_1992,Wieck_1990,Nakata1990}, GaAs quantum constrictions~\cite{Hirayama_1989} GaN nanowires~\cite{Dhara_2010},
          2D MoSe$_2$/graphene heterostructure~\cite{Liu2021}, MoSi superconducting thin film~\cite{Mykkanen2020}, In$_2$O$_3$ thin film~\cite{Sosa2010}, Al$_2$O$_3$ film~\cite{Bianconi2019}, ZnO and VO$_2$~\cite{Mei_2022}\\
          & & Si
          & AlGaAs/GaAs quantum wires~\cite{Hiramoto1989}\\
	    & formation of barriers and nanoconstrictions 
	        & He 
          & superconducting electronic circuitry including Josephson junctions and nano\acsp*{squid} in \acs*{ybco}~\cite{Cybart2015,Cho2018,Muller2019,Couedo2020,LeFebvre_2022,Chen2022,Uhl2023,Karrer2023} and MgB$_2$~\cite{Kasaei2018}, nanoconstrictions in BSCCO~\cite{Seifert2021,Merino2023}\\
	       & & Ga 
	       & quantum dots in single-walled \acs*{cnt}~\cite{Maehashi_2007}\\
       & tuning thermoelectric properties
        & He
        & MoSe$_2$ thin films~\cite{Kim_2022}\\
      & tuning superconductivity
        & He
        & periodic arrays of artificial pinning sites in YBCO ~\cite{Aichner_2019,Backmeister2022}\\
	    & pinning ferroelectric domains 
	        & He 
	            & PbZr$_{0.2}$Ti$_{0.8}$O$_3$ thin film~\cite{Saremi2018}, multilayer MoTe$_2$ flake~\cite{Seol_2021}\\
        & ferroelectric phase change
	        & He 
          & induced ferroelectricity in HfO$_2$~\cite{Kang_2022}, conversion to paraelectric phase~\cite{Belianinov_2017}\\
        &   & Ga
        & HfO$_2$ thin film~\cite{Guberna_2021}, damage in ferroelectric capacitors~\cite{Stanishevsky_2002,Ganpule_1999}\\
	 \midrule
	 Magnetic 
	    & tuning magnetic anisotropy 
	        & He
	            & Co/Pt multilayers~\cite{Franken2011}, Ir$_{17}$Mn$_{83}$/Co$_{70}$Fe$_{30}$ bilayer~\cite{Huckfeldt2017}, Co-based thin-film stack~\cite{Dunne2020,Kurian_2023}, Co/Pt/Ru multilayer system~\cite{Samad_2021}\\
      &   &    & skyrmions in Pt/Co/MgO thin film~\cite{Juge_2021}, and in Pt/Co$_{60}$Fe$_{25}$B$_{15}$/MgO thin film~\cite{Kern_2022}\\
       &
          & Si/Ge
          & Bi and Ga substituted \acs*{yig}~\cite{Ishii_1989}\\
      & 
          & Ga
          & CoPt multilayers~\cite{Aign_1998}, Fe$_{78}$Ni$_{22}$ thin film~\cite{Flajsman_2020}, Ir/Co/Pt multilayers~\cite{Jong_2022}\\
      &   &    & skyrmions in FeGe thin plate~\cite{Shibata_2015}, and in Co/Pt multilayers~\cite{Sapozhnikov2019,Fallon2020,Hu_2022}\\

        & tuning metamagnetic transition temperature 
            & He 
                & FeRh thin film~\cite{Cress2021}\\
        & inducing ferromagnetism 
            & Ne 
                & Fe$_{60}$Al$_{40}$ thin film~\cite{Roder2015,Cansever_2022}\\
            & & Ga
                & Fe$_{60}$Al$_{40}$ alloy~\cite{Menendez_2008} and Fe$_{78}$Ni$_{22}$ on copper~\cite{Urbanek_2018}\\    
            
        & magnonics
            & Ga
            & patterned Fe$_{78}$Ni$_{22}$ films~\cite{Flajsman_2020}, spin wave optical elements in \acs*{yig}~\cite{Kiechle_2023}\\ 
	 \midrule
	 Optical \& quantum
		 & tuning photoluminescence 
		    & He 
        & SiN$_x$ membrane~\cite{Sawafta2014}, monolayer MoSe$_2$~\cite{Iberi2016a}, InGaAs/GaAs quantum well~\cite{Kapitonov_2016}, reducing parasitic fluorescence in \acs*{hbn}~\cite{Grosso_2017}\\
		    & & Ga 
        & AlGaAs/GaAs and GaInAsP/InP quantum wells~\cite{Reithmaier_1998}\\
		 & tuning of passive optical components 
		    & He 
		        & plasmonic resonance in graphene nanostructures~\cite{Jiang2018}, Bragg grating in silica nanofiber~\cite{Takashima_2019a} \\
		    & & Li 
		        & Si microdisc nanophotonic resonator~\cite{McGehee_2017a} \\
           & & Ga 
           & ZnSe quantum dot micropillar cavitities~\cite{Lohmeyer_2008}, periodic poling of LiNb waveguide~\cite{Chezganov_2023}\\
	  & creation of photoluminescence, quantum emitters and qubits 
	        & He 
          & \acs*{nv} centers in diamond with substitutional nitrogen~\cite{Huang2013,McCloskey2014}, Si vacancy centers in SiC~\cite{Song_2020,He_2022}, luminescent defects in monolayer MoS$_2$~\cite{Klein2017,Klein-2019}\\
	        & & Li 
              & Si vacancy centers in SiC~\cite{Pavunny_2021}\\
	        & & Si 
          & Si vacancy centers in diamond~\cite{Tamura_2014} and diamond nanostructures~\cite{Schroeder_2017}, W and G centers in silicon~\cite{Hollenbach_2022}, isotopic enrichment of $^{28}$Si to increase spin qubit coherence times~\cite{Acharya2023}\\
          	& & Ar 
		        & NV centers in diamond~\cite{Raecke_2021}\\
          & & P 
		        & deterministic single-atom implantation of $^{31}$P into Si~\cite{Jakob_2021a}\\
          & & Ga 
          & luminescent defects in \acs*{hbn}~\cite{Ziegler_2019,Klaiss_2022}, monolayer WSe$_2$~\cite{Qian_2020a}, MoS$_2$ and WS$_2$~\cite{Thiruraman_2019a}\\
	        & & Ge 
		        & Ge vacancy centers in diamond~\cite{Zhou_2018a}\\
	        & & Sb, Si 
		        & deterministic single-atom implantation into Si~\cite{Pacheco_2017a}\\
		    & & Xe
        & recoil implantation of color centers in diamond~\cite{Froech_2020}, spin defects in \ac*{hbn}~\cite{Kianinia_2020}, water assisted fabrication of extended defects in \ac*{hbn}~\cite{Glushkov_2022}\\
            & & Bi 
            & qubits in silicon and germanium~\cite{Cassidy_2020}\\
    \midrule
	Chemical 
	    & enhancing chemical etch rates 
	        & He 
	            & Si$_3$N$_4$ thin film~\cite{Petrov2015}, SiO$_2$ thin film~\cite{Petrov2020}, mono-/bilayer MoS$_2$~\cite{Maguire2019}\\
        & enhancing catalytic activity 
            & He 
                & MoS$_2$ and MoSe$_2$ flakes~\cite{Mitterreiter2019}\\
        & nucleation of growth 
            & He 
                & epitaxial graphene on SiC~\cite{Heilmann2021}\\
        & & Co
          & CoSi$_2$ nanowires in Co-saturated Si~\cite{Akhmadaliev_2006}\\
        & 
	        & Ga 
		        & InAs dots on GaAs~\cite{Lee_2011a}, Ge dots on Si~\cite{Portavoce_2007a,Gherasimova_2008a}\\
        & functionalization 
            & Ga 
                & graphene (fluorination using XeF$_2$)~\cite{Li-2016}\\
        & tuning surface wettability 
            & Ga 
	            & molecular self-assembled monolayer~\cite{Yamada_2015a}\\
	 \midrule
	 Thermal 
	    & reducing thermal conductivity 
	        & He 
	            & Si nanowires~\cite{Zhao2017}, Si membrane~\cite{Choe2019}, VO$_2$ nanowires~\cite{Jin2020}, Bi$_2$Te$_3$ nanoribbons~\cite{Choe2019b}\\
	 \midrule
	 Mechanical 
	    & tuning Young's modulus 
	        & He 
	            & monolayer MoSe$_2$~\cite{Iberi2016a} \\
	        & & Ga 
	            & polymer~\cite{Liu2015}, hydrogels~\cite{Kim2014}\\
        & gas bubble effects 
            & He 
                & nanobubble superlattice in Cu~\cite{Wang_2016}, nanobubbles in Si~\cite{Wang_2017b}, diamond~\cite{Chen_2020}, Eurofer steel~\cite{Bergner_2018}, W~\cite{Allen_2020}, ultrafine-grained W and W-Cu nanocomposite~\cite{Wurmshuber_2022}\\
      \end{longtable*}
In the remainder of this section we focus on those applications in Table~\ref{table:property_engineering} that are concerned with engineering the electronic, magnetic, and optical/quantum properties of materials, with additional detail on property engineering of low-dimensional materials (in particular, thin films and 2D materials). The study of irradiation damage mechanisms (e.g.\ the formation of gas nanobubbles) is also discussed.

\subsubsection{Engineering electronic properties}

From the outset, \acp{fib} have been used to fabricate electronic devices at the nanoscale.
Early examples, using Ga-\ac{fib} milling, are the fabrication of GaAs quantum wires~\cite{Hirayama_1988a} and a quantum wire transistor~\cite{Wieck_1990}.
More recently, however, Ga-\ac{fib} irradiation at lower dose was employed for the fabrication of a lateral diode based on a 2D heterostructure of MoSe$_2$ and graphene~\cite{Liu2021}.
Here, the preferential sputtering of Se locally transformed the top MoSe$_2$ layer into a quasimetallic state, generating a pn junction with rectification performance comparable to that of a vertical diode. 
These examples illustrate a development from the prevalence of subtractive materials modification by \ac{fib} milling, to increased interest in engineering the electrical and electronic properties of materials at lower dose through defects.

Here, light (and often preferably inert) ion beams are of relevance due to their high resolution defect engineering capabilities.
Ion beams of electronically active elements (e.g.\ P, Sb, As) also have the potential for interesting applications, such as local doping with high spatial resolution or the generation of qubits by \ac{sii}.
For applications in superconductivity, Josephson junctions, i.e.\ weak links between two superconducting electrodes, are key elements for superconducting electronic devices. 
For Josephson junctions based on conventional metallic superconductors, such as Nb or Al, thin-film devices with grown insulating (I) or normal conducting (N) barriers, sandwiched between superconducting (S) electrodes, are well established. 
However, in the case of the high-T$_C$ cuprate superconductors this as-grown approach does not produce well-defined SIS or SNS junctions, due to the complex nature of these materials and their strong sensitivity to defects on the atomic scale. 
As an alternative approach, the local \ac{fib}-enabled modification of superconducting thin films in order to ``write" thin N or I layers as Josephson barriers in prepatterned thin film structures has been successfully demonstrated. 
Here, the high spatial resolution offered by He-\ac{fib} has been shown to be of great importance. 
Prime examples are the fabrication of insulating or normal-conducting barriers in \ac{ybco} or MgB$_2$ thin films by He-\ac{fib} irradiation that locally suppress superconductivity on the nanometer scale~\cite{Cybart2015, Kasaei2018, Muller2019}, and the creation of He-\ac{fib}-induced nanoconstrictions in Bi$_2$Sr$_2$CaCu$_2$O$_8$ (BSCCO) single crystal flakes. 
The latter form the basis of two-dimensional cuprate nanodetectors with single telecom photon sensitivity~\cite{Seifert2021,Merino2023}. 
Moreover, He-\ac{fib} irradiation of \ac{ybco} films has produced artificial pinning arrays with unprecedented lattice constants down to \qty{30}{nm}~\cite{Aichner_2019,Backmeister2022}, which opens a new regime for motion control of Abrikosov vortices in superconducting fluxonic devices.

Due to the nanoscale beam size that is achievable with \acp{fib}, in particular with \ac{gfis}-based \acp{fib}, these are ideal tools for the fabrication of single quantum dots, tunneling devices and single electron transistors.
Examples include the fabrication of single electron transistors in Si/SiO$_2$ stacks~\cite{xu_site-controlled_2018} and the fabrication of single quantum dots in \acp{cnt}~\cite{Maehashi_2007}.

\subsubsection{Engineering magnetic properties} 

Ion irradiation in general has been used for many decades for the modification of magnetic materials~\cite{Fassbender_2004}.
To achieve this sort of modification, complex and expensive masking processes have to be used unless one employs \acp{fib}, which can modify the magnetic properties locally with nanometer precision.
In 1989, \citet{Ishii_1989} used Si and Ge ions from a \ac{lmais}-\ac{fib} to create nucleation sites for controlled switching of the magnetization in Bi- and Ga-substituted \ac{yig} films.
Ga-\ac{fib} processing has also been used for creating read/write heads~\cite{Pan_1999} for magnetic hard disks, and for modern approaches such as \acp{mram} composed of magnetic multilayer structures~\cite{Nishimura_2002}.
FIB-induced local changes in magnetic anisotropy have been used to create isolated tracks of single domain islands in recording media~\cite{Albrecht_2003} and to control the injection location and field dependency of magnetic domain walls in multilayers with perpendicular magnetic anisotropy (using Ga- and He-\ac{fib})~\cite{Franken2011}.
In the case of Co nanowires fabricated by \ac{febid}, irradiation with Ga-\ac{fib} has allowed modification of the propagation speed of the magnetic domain wall~\cite{SerranoRamon_2013}. 
In some materials, \acp{fib} can be used to generate ferromagnetism in selected regions, such as in paramagnetic FeAl alloys.
For example, the spatially resolved creation of disorder-induced ferromagnetism using Ga-\ac{fib} modification has been demonstrated by \citet{Menendez_2008}, 
achieving a spatial resolution of below \qty{100}{nm} with the possibility of erasing the pattern by annealing. 
Later, Fe$_{60}$Al$_{40}$ was studied in depth for order-disorder phase transitions and \ac{fmr} in magnetically written single structures induced by Ne-\ac{fib} irradiation~\cite{Roder2015, Cansever_2022}. 
Paramagnetic regions in non-magnetic materials can also be created by \ac{fib} irradiation, as has been done for InGaAs using Ga-\ac{fib} to create a spin filter device~\cite{Nguyen_2013}.

The continued quest for an increase in storage density in combination with novel data processing approaches (i.e.\ neuromorphic computing) has triggered the development of advanced patterning of complex magnetic alloys and multilayer systems~\cite{Devolder_1999,Warin_2001,Gierak_2005}.
In these systems various interactions due to the intrinsic anisotropy, or lattice and system geometry, may lead to topologically protected spin configurations such as skyrmions or the controlled formation and transport of spin waves.
Here the \ac{fib} stands out, as it enables tuning of magnetic anisotropy landscapes without any topographical changes to the devices.
This was demonstrated for the nucleation control of skyrmions in Co/Pt multilayers using Ga-\ac{fib}~\cite{Fallon2020} and He-\ac{fib}~\cite{Sapozhnikov2019}, and through the patterning of skyrmion racetracks in a Pt/Co/MgO thin film using the He-\ac{fib}~\cite{Juge_2021}. 
The patterning of dots as nucleation sites within a low-dose racetrack using He-\ac{fib} realized both the deterministic generation and movement of skyrmions with nanometric precision~\cite{Kern_2022}.
Finally, spin-wave optical components such as graded-index lenses for applications in magnonics can be created by ion induced modification of the local magnetization in \ac{yig}, which corresponds to an effective refractive index for spin wave propagation. Here, \qty{50}{keV} Ga-\ac{fib} was employed and the penetration depth of about \qty{100}{nm} allowed for an effective refractive index of up to 1.8, comparable to light-based optics. Since this depth was only one third of the actual device thickness, the much lighter He ions are expected to significantly improve the optical device performance~\cite{Kiechle_2023}.

\subsubsection{Engineering optical/quantum properties\label{sss:4_D_3-optical_appli}}

Typical applications of ion irradiation in optics include the local change of refractive index and the creation of fluorescence~\cite{Townsend_1987}.
However, since \ac{fib} processing is serial and the penetration depth is limited, the use of the \ac{fib} in this area is mostly of interest in cases where a small number of ions needs to be placed with high spatial accuracy.
A particularly beautiful example is the tuning and imaging of resonant optical modes in a silicon disc resonator using Li-\ac{fib}~\cite{McGehee_2017a}.
A low beam current of \qty{1}{pA} at \qty{3.9}{kV} acceleration voltage was used to radially raster scan in pulses of $\approx3000$\,ions while the transmitted light passing through a critically coupled waveguide was monitored.
The optical shifts observed could be attributed to ion damage and a rapid thermal (and thus reversible) response.
Knowledge of this time-varying response was used to enable imaging of different optical modes with minimum device degradation.

The spatial accuracy of defect creation using the mask-less \ac{fib} approach becomes especially beneficial in the field of quantum technology. 
Here, a defect in an otherwise ideally perfect crystal can take the role of an artificial atom emitting quantum light.
Furthermore, the defect may exhibit electronic and nuclear spin degrees of freedom that may be coherently coupled to the emitted quantum light, turning the defect into a solid state quantum bit (qubit).
The most popular example of a defect-based quantum emitter, or qubit, is the \ac{nv} center in diamond. Diamond \ac{nv} centers are routinely fabricated using ion implanters~\cite{Lagomarsino_2021}, but have  also been realized using He-\ac{fib} processing~\cite{Huang2013,McCloskey2014}.
\ac{fib} implantation of Si ions has been used to realize nearly lifetime-limited single Si vacancy (SiV) quantum emitters in diamond nanostructures~\cite{Schroeder_2017}.
This maskless approach achieved \qty{\approx32}{nm} lateral precision and \qty{<50}{nm} positioning accuracy,  and is thus promising for the development of scalable solid-state quantum information processors.
Focused Si ions at \qty{35}{keV} with beam diameters of \SIrange{5}{10}{nm} have also been used to fabricate negatively charged silicon vacancies (V$_{Si}^-$) in 4H-SiC, with single emitters observed starting from 40 ions per irradiated spot~\cite{Wang_2017a}.
A comparative study by the same author using H, He and C ions showed that He ions produced the highest defect concentration, while the lowest brightness emitters were observed for C ions~\cite{Wang_2019b}. 
In addition, Li-\ac{fib} has been employed for the creation of Si vacancies in SiC~\cite{Pavunny_2021}. Recently, focused Ar$^{8+}$ ions from an \ac{ebit} were also used for \ac{nv}-creation in diamond~\cite{Raecke_2021}.
A particularly interesting study made use of Xe-\ac{fib} for recoil implantation of targeted group IV dopants in a deposited surface layer~\cite{Froech_2020}.
In this work, Pb, Sn, Ge and Si color centers in diamond were fabricated with a spatial accuracy of better than \qty{50}{nm} in combination with an ultra-shallow doping profile.
So-called G and W centers in Si have recently been demonstrated using Si-\ac{fib} to create spatially resolved single defect centers of this type for application as quantum emitters in the telecom band~\cite{Hollenbach_2022}.
And in a very recent approach, a dedicated \ac{fib} system equipped with a \ac{uhv} sample chamber~\cite{Adshead2023} was used to suppress unwanted isotopes in Si that would otherwise limit coherence times and therewith the performance of solid-state spin qubits. 
This was achieved using a \ac{lmais} source to implant $^{28}$Si$^{++}$ into natural Si, lowering the level of unwanted $^{29}$Si to the single-digit ppm range without introducing contamination from the \ac{fib} processing~\cite{Acharya2023}.
The deterministic placement of single atoms (e.g.\ Sb or P) into Si to act as qubits is another challenge where progress has recently been made~\cite{Pacheco_2017a, Jakob_2021a}.
While the two cited works are a clear demonstration of the placement accuracy of single (or few) atoms using a \ac{fib}, the approaches rely on implantation into a host material that can also double as an active detector of the ion impacts.

In parallel, 2D materials have been heavily explored as hosts for usable quantum emitters e.g.\ sources of single photons~\cite{Kianinia_2022}. Here, ion irradiation can be used to both tailor the optical
properties of the 2D system~\cite{Ma2017aNS} as well as to produce defects that introduce novel optical activity~\cite{Choi2016acs,Fischer2021}. 
He-\ac{fib} processing in particular has been used for the site-selective creation of defects in e.g.\ 2D MoS$_2$~\cite{Klein-2019}. While the exact type of defects being introduced is still under investigation (true for most luminescent defects in 2D materials), they are expected to cause highly localized trapping potentials for exciton emission. 
Encapsulation with \ac{hbn} leads to very narrow linewidths of the emission lines in the visible range, \SIrange[range-phrase= -- ]{100}{220}{meV} below the neutral exciton~\cite{Klein-2019}. 
He irradiation has also been shown to result in a five-fold reduction in background and parasitic fluorescence, enabling the isolation of bright and stable quantum emitters in \ac{hbn}~\cite{Grosso_2017}.
Recent results using Ga-\ac{fib} patterning indicate that the formation of luminescent defects is positively correlated to material swelling rather than material removal by milling~\cite{Klaiss_2022}.
This helps to explain earlier results on the generation of defects by Ga-\ac{fib}, where it was not clear whether the single photon emission was actually generated inside the milled holes, at their edges, or in the damaged region near the edges~\cite{Ziegler_2019}. 

\subsubsection{Engineering of low-dimensional materials} 

2D materials such as graphene and atomically thin transition-metal dichalcogenides are expected to play an important role in the realization of future electronic and optoelectronic devices and there exists a significant body of work in which \ac{fib} based defect engineering has been used to tune the unique properties of these materials. In fact, as can be seen in Table~\ref{table:property_engineering}, the majority of materials that are used in \ac{fib} based defect engineering work are indeed low-dimensional (thin films, 2D materials, and nanowires).
The simple reason for this is that typical \ac{fib} instruments deliver beams with energies of several tens of keV, which limits the ion penetration depth. For example, in most materials the maximum penetration depth of the ubiquitous \qty{30}{keV} Ga ion beam is around \qty{60}{nm}, which severely limits possibilities for property engineering of bulk materials.
While lighter ions of the same energy penetrate more deeply, for many defect engineering applications implantation of the ion into the material is not actually desired. 
Rather, it is the local damage created by the ion in transit that is leveraged. 
Therefore, low-dimensional materials are often the ideal testbed material for defect engineering using the \ac{fib}. 
Notably, an increase in ion energy typically causes a decreasein defect production in 2D materials~\cite{Krasheninnikov2020}, such that lower beam energies are in fact better suited for efficient materials modification in these cases. 

An example of \ac{fib} based defect engineering of a low-dimensional material is the use of the He-\ac{fib} to tune the electrical resistivity of 2D MoS$_2$~\cite{Fox-2015}.
In the cited work, the electrical resistivity of the MoS$_2$ monolayer was tuned from semiconducting to insulating depending on the vacancy defect concentration, as controlled by the irradiation dose.
Above a certain threshold dose, total amorphization (i.e.\ disordering) of the material occurs.
By irradiating samples locally in specific patterns with fluences below the total amorphization threshold, the \ac{fib} can be used to tune the defect density and hence the electrical conductivity in spatially defined regions at the nanoscale. 
For example, using the He-\ac{fib} and a line irradiation strategy on 2D MoS$_2$, a neuromorphic memtransistor device has been demonstrated~\cite{Jadwiszczak-2019}.
Similarly, Ga-FIB has been used for defect creation in supported graphene, with MD simulations providing insight into the effect of irradiation angle on the lateral damage distribution~\cite{Wu-2018-1}.
He-\ac{fib} line irradiation of graphene has also been used to create graphene nanoribbons~\cite{Nanda2017}, which opens up a band gap in graphene due to quantum confinement.
Tuning of the ribbon width and hence the size of the band gap is a key step toward the use of graphene in electronic applications.
As discussed in Section~\ref{subsubs_md_applications}, for supported graphene the minimum achievable nanoribbon width is limited by the influence of the backscattered ions from the substrate~\cite{Kretschmer-2018}.

Other defect engineering applications concerning 2D materials include the use of Ga-\ac{fib} in combination with the simultaneous flow of XeF$_2$ gas for the localized fluorination of graphene~\cite{Li-2016}, and the use of the He-\ac{fib} to create localized defects in graphene to serve as nucleation points for the epitaxial growth of \ac{hbn}~\cite{Heilmann2021}. An example of defect engineering of a 1D structure is the employment of localized He-\ac{fib} irradiation to induce regions of disorder in a silicon nanowire in order to tune the thermal conductivity along its length~\cite{Zhao2017}.

\subsubsection{Study of irradiation damage mechanisms}

Fundamental curiosity and the need for experimentally straightforward replacements for long-duration reactor-based irradiation experiments have resulted in \ac{fib} instruments being used for the systematic study of irradiation effects on structural materials.
In order to simulate irradiation damage caused by $\alpha$ particles, for example, He$^+$ ions are a convenient species to use.
A further advantage of using an inert species such as He is that fundamental studies of irradiation effects can be made while avoiding alloying effects and grain boundary embrittlement, as can occur with a metallic ion species such as gallium~\cite{Unocic2010}.

A phenomenon that was discovered several decades ago is the formation of He nanobubbles upon He ion irradiation of various metal targets.
These nanobubbles form as a result of vacancy defects and He interstitials that diffuse and recombine to produce gas clusters.
However, fundamental experimental studies of the resulting changes in the mechanical properties of the target material were for a long time confined to the use of large-scale ion accelerators and plasma devices.
The introduction of the He-\ac{fib} changed this, enabling the first systematic dose-controlled irradiation studies.
Using the He-\ac{fib}, gas bubble phenomena that had been observed in large-scale experiments have been reproduced with much greater control over the experimental conditions.
Initial He-\ac{fib} studies in this area used Si targets, mostly driven by the need of the semiconductor industry to evaluate the invasiveness of He-\ac{fib} processing~\cite{Livengood2009}.
Later studies included the investigation of He gas-bubble superlattices, using single-crystal Cu as a simple model system and probing the effect of the superlattice on the mechanical properties of the host material using in-situ \ac{tem} nanoindentation~\cite{Wang_2016}.
Since samples are irradiated with the He-\ac{fib} on the nanoscale, individual grains and grain boundaries can be targeted in polycrystalline samples~\cite{Chen_2018}.
By increasing the dose, the coalescence of nanobubbles forming crack planes that ultimately result in blistering and delamination can also be studied with a level of control that was previously unattainable~\cite{Allen_2020}.
Furthermore, irradiation induced mechanical property changes of a number of materials that are directly relevant for nuclear applications have been investigated using the He-\ac{fib} combined with nanoindentation, e.g.\ in studies of Eurofer steel~\cite{Bergner_2018}, W~\cite{Balooch_2022}, ultrafine-grained W and a W-Cu nanocomposite~\cite{Wurmshuber_2022}.

Finally, He gas bubble effects have also been employed for nanofabrication tasks, using patterning with the He-\ac{fib} at high dose for localized subsurface swelling to deform surfaces and thereby create 3D nanostructures, such as nanopyramids and nanohemispheres~\cite{Fox2012,Zhang_2015,Kim_2020b,Wen_2022}.

\subsubsection{Outlook}

The use of the \ac{fib} for property engineering down to the nanoscale is now well established. 
Recently, strong efforts have been underway to increase the spatial resolution and control of the technique even further, with the ultimate goal of creating single defect centers or implanting single impurity atoms for applications in quantum technology.
Here, a significant challenge is posed by the intrinsically stochastic nature of both the production of ions from most \ac{fib} sources and of the ion-solid interaction itself.
The deterministic placement of single ions either into the bulk or at a surface thus pushes the \ac{fib} technique to its fundamental limits.

There are two distinct approaches that are being pursued for realizing deterministic \ac{sii} using the \ac{fib}.  
One is to count the ions as they travel to the sample and the other is to detect individual ion impact events, ideally in an universal manner.
The first approach can be realized by using an Paul ion trap as a deterministic source of single ions~\cite{GrootBerning_2021,GrootBerning_2019}. 
While the presence of the ion can be unambiguously proven, ejecting the ion and re-loading the trap is time-consuming and the spatial localization of the eventual impact event using ion-optical elements causes significant losses. 
Given that the probabilities for the creation of specific defects are usually low and highly material dependent  (see also Section~\ref{subs_md}) this deterministic ion source approach is appealing, but technologically extremely demanding. 
Alternatively, ion columns can be equipped with image-charge detectors that count ions produced from different types of ion sources. 
The beam is directed toward the sample only if an ion is detected. 
In this way, imperfect detection efficiencies can be mitigated since all non-detected ions are automatically blanked~\cite{Raecke_2019}.
The drawback of this technique is the large charge state (at least $10^+$) that is necessary for detection of the image charge, requiring ion sources such as an \ac{ebit}.
Here, Ar$^{18+}$ ions created by an \ac{ebit} have been successfully used~\cite{Raecke_2022}. 
Light ions are excluded from this technique, unless the sensitivity of the image-charge detectors can be significantly increased, and this method also presents significant technological challenges for column design.

The second approach, i.e.\ the detection of individual ion impacts on the sample, could in principle work for any ion source.
Promising attempts combine an on-chip \ac{ibic} detector~\cite{Robson_2022,Collins2023} with a plasma ion source and a near-surface aperture in an \ac{afm} cantilever~\cite{Jakob_2021a}. 
Here, scanning of the aperture realizes the spatial localization~\cite{Rangelow_2006, Meijer_2006, Meijer_2008} and thus allows use of various types of ion source, including broad beam implanters. 
While this is an important step for all quantum devices that tolerate a nearby pn-junction and amplifier, other device concepts e.g.\ in photonic quantum technology, require truly agnostic sample detector concepts.
Conventional \ac{se} detectors (Everhart-Thornley, see also Section~\ref{subs_detectors}) or novel detector types (e.g.\ channeltrons~\cite{Cassidy_2020}) may work as such, but suffer from their strong dependence on the \ac{se} yield, which is unknown for most materials and is very surface dependent.
There is also a need for a better understanding and prediction of \ac{se} generation and expected detection yields, and for the development of advanced \ac{se} detection schemes with detection efficiencies approaching \qty{100}{\%}~\cite{Murdin_2021,Cassidy_2020}.
Instead of detecting all ion impact events (that may each come with a small probability of creating the target defect), one may instead detect device function in-situ while delivering ions to the target locations. 
A combination of short dwell times at extremely low current results in a Poissonian probability of either one or zero ions per shot and the process may be stopped once the desired device operation is achieved. 
Here, the technological challenge of these implementations involves the co-integration of an extremely sensitive fluorescence detection setup~\cite{Budnik_2022, Chandrasekaran_2023}.

In conclusion, several promising development routes for \ac{sii} are currently being pursued that are showing first promising results. 
Whether modifying the ion supply toward being deterministic or enhancing the detection capabilities toward single-ion sensitivity (or a combination of both) will finally lead to success is still an open question.

\subsection{\label{ss:tomo}\acs*{fib}-based imaging and tomography}%
\label{subs_imaging}
Multiscale imaging of complex systems is becoming increasingly important. For example, to understand data transmission in the brain or chemical energy storage in a battery, one needs to capture the ``big picture'', ideally in three dimensions, as well as resolve features down to individual atoms.
This is where \ac{fib} based techniques play a central role, by direct imaging using light ions such as He, by 3D volume imaging combining \ac{fib} slicing with \ac{sem}, and by preparing thinned cross-sections and specialized geometries for \ac{tem} and \ac{apt}.

\subsubsection{A brief history of \acs*{fib}-based imaging and analysis}%
\label{ss:history_FIB_imag}

The enormous potential of electron microscopy for imaging the structure of condensed matter became clear with the invention of the \ac{tem} in \citeyear{Knoll_1932}~\cite{Knoll_1932} and soon thereafter first images of bacteria~\cite{Krause_1937, Brueche_1939} and virus particles~\cite{Ruska_1939} proved the same for the life sciences. 
The introduction of ion microscopy for imaging, however, took much longer. 
Pioneering work was carried out in the 1970s by Levi-Setti and co-workers, who experimented with a scanning transmission ion microscope using up to \qty{65}{keV} H$_{2}^{+}$ ions from a \ac{gfis} and published images of myofibrils from rabbit muscle~\cite{Escovitz_1975}.
However, since the achievable lateral resolution of ion microscopy had always been inferior to that of electron microscopy, ion microscopy was not used for nanoscale imaging until the early 2000s with the invention of the \ac{him}~\cite{Ward_2006,Hlawacek_2016} (see Section~\ref{sss:him-bio-imaging}).

The breakthrough in the use of ion microscopy for sub-cellular imaging in biology, or ``the revolution in ultra-microscopy'' as stated by~\citet{Ballerini_2001}, occurred once Ga ion beams from \ac{lmis}-based sources could be focused down to better than \qty{10}{nm}~\cite{Gierak_1999}.
It was recognized that the ion beam could be used ``like a tiny scalpel which slices away layers of the sample [and that one could] exploit this destruction to gain high-resolution three-dimensional structural, morphological, and chemical information''~\cite{Ballerini_2001}. Thus, the \ac{fib} based microscopy and machining that had until the turn of the century mainly been used in the semiconductor industry was brought to the life sciences~\cite{Ballerini_2001,HEKKING_2009}. Various other fields also adopted the technique.
For example, one of the first reports of \ac{fib} tomography was presented in 2001 by \citet{Inkson_2001} on the 3D characterization of a metallic nanocomposite, and \ac{fib} based tomography also became increasingly common in e.g.\ metallurgy~\cite{Lasagni_2008, Kubis_2004, Uchic_2007}, energy research~\cite{Ostadi_2010}, and geology~\cite{WINTER_2009}. 
Recently, \ac{fib} tomography has even been applied to paleontology samples~\cite{M_Yara_2021}.
While early work sometimes used the Ga-\ac{fib} for the imaging as well as the sectioning, the most common implementation of \ac{fib} tomography uses the dual beam \ac{fib}-\ac{sem}, comprising a Ga-\ac{fib} column for the milling coupled with a \ac{sem} for high-resolution surface imaging of the exposed faces.
Significant progress in 3D image segmentation and reconstruction~\cite{Pham_2000} was also key to enabling 3D structural imaging of complex specimens by the \ac{fib}-\ac{sem} technique. 
Furthermore, the development of cryo-workflows for cryo-FIB-SEM tomography was a major step forward allowing preservation of cell ultra-structure~\cite{Heymann_2006,Hayles_2007}. Such cryo workflows are now used for other beam-sensitive samples as well (see Section~\ref{sss:mat_tomo}).

An overview of the various \ac{fib} based imaging and analysis modalities used in the life sciences is given in Table~\ref{table:bioimaging}.
In all of the examples given, the imaging and analysis was primarily performed in-situ, i.e.\ inside the \ac{fib}-(\ac{sem}) instrument.
Key works on cryo-\ac{fib} milling for the site-selective preparation of electron transparent biological specimens for analysis by cryo-\ac{tem} can be found in the following references~\cite{Rigort_2015,Schaffer_2019}.

\begin{table*}[btp]
\caption[\ac{fib} based imaging of biological structures]{Overview of \ac{fib} based imaging and analysis in the life sciences.}%
\label{table:bioimaging}
  \begin{center}
   \begin{tabularx}{\linewidth}{%
        >{\centering} m{2em} 
        >{\hsize=.3\hsize}X 
        >{\hsize=.2\hsize}X 
        >{\hsize=.8\hsize}X} 
    \toprule
     & Application   & Technique & Experiment \\
    \midrule
    \multirow{8}{2em}{\rotatebox{90}{\makecell{molecules}}}
    & Localization of proteins 
    & cryo-\ac{fib}-\ac{sem} 
    & protein detection in vitrified HeLa-cells using Au-conjugated antibodies against RNA polymerase II~\cite{Spehner_2020}
    \\    
    &
    & \ac{him} \& fluorescence microscopy 
    & detection of quantum-dot labeled proteins~\cite{Lim_2020}
    \\
    & 
    & ToF-\ac{sims} 
    & detection of proteins on DNA bio-sensor chip~\cite{Arlinghaus_2008}
    \\
    & Metabolic imaging 
    & Orbi\ac{sims} 
    & imaging of the distribution of neurotransmitters in mouse hippocampus~\cite{Passarelli_2017a}
    \\
    & 
    & 
    & 3D metabolic imaging of Paramecium caudatum using orbitrap-MS coupled to \acs*{tof}-\acs*{sims}~\cite{Gilmore_2019}
    \\ 
    \midrule
    \multirow{6}{2em}{\rotatebox[origin=r]{90}{\makecell{viruses \&\\ nanoparticles}}}
    & Virus imaging
    & HIM
    & imaging of bacteriophage infection~\cite{Leppaenen_2017,Sharma_2018a}, and transport by bacteria~\cite{You_2021a} 
    \\
    & 
    & 
    & imaging of SARS-CoV-2 infected Vero cells~\cite{Frese2021,Merolli_2022} 
    \\ 
    & 
    & FIB-SEM
    & investigations into SARS-CoV-2 viral density at cell-cell contacts~~\cite{Baena_2021} 
    \\ 
    & Nanoparticle -- 
    & HIM
    & 
    \\ 
    & cell interactions
    & HIM-SIMS
    & investigation of biogenic nanoparticles by \ac{him}-\ac{sims}~\cite{Guillermier_2019}
    \\
    &
    & ToF-\ac{sims}
    & 3D-imaging of nanoparticles in an algal biofilm~\cite{Benettoni_2019}
    \\
    \midrule
    \multirow{16}{2em}{\rotatebox{90}{\makecell{cell-level}}}
    & Imaging of microbial cells
    & HIM
    & imaging of nitrate-reducing bacteria with spikey minerals~\cite{Joens_2013, Byrne_2018}, archea~\cite{Chen_2019a}, predatory bacteria~\cite{Said2018}
    \\
    & Imaging of biofilms
    & HIM
    & Geobacter sulfurreducens~\cite{Belianinov_2017a}, microbial mats from Himalayan hot springs~\cite{Sharma_2018a}, chlorella micro-algae~\cite{MorenoOsorio_2019}
    \\ 
    & Imaging of mammalian cells and tissues
    & \ac{him}
    & imaging of mitotic HeLa cells~\cite{Joens_2013}, visualization of lipid nanodomains in human neural stem-cells~~\cite{Schuermann_2015}, imaging of kidney~\cite{Paunescu_2014}
    \\ 
    & 3D imaging and remodeling of microbial cells 
    & (cryo-)\ac{fib}-\ac{sem}
    & FIB-SEM of yeast cells~\cite{Ballerini_2001, Heymann_2006}, 3D remodeling of Acantharia-microalgae~\cite{Decelle_2019} 
    \\
    & 3D imaging of cells and tissues
    & (cryo-)FIB-SEM
    & imaging of FIB-sectioned gland cells~\cite{Drobne_2005}, lymphoid tumor tissue~\cite{Heymann_2006}, \acp{cnt} in mouse lung tissue~\cite{Kobler_2014}, tobacco petal epidermis~\cite{Hayles_2007}, human bone~\cite{Binkley_2020}
    \\ 
    & Sub-cellular\goodbreak{} elemental mapping
    & nanoSIMS, \goodbreak{} ToF-SIMS
    & element distribution in E.coli exposed to natural antibacterial clay\cite{Morrisson_2016}, mapping of N/P ratio in a photosymbiotic Acantharia-microalgae system~\cite{Decelle_2019}, identification of TiO$_2$ in an algal biofilm~\cite{Benettoni_2019}
    \\
    &
    & PIXE
    & elemental analysis of hair, skin, nervous tissue~\cite{Forslind_1991} 
    \\
    &
    &
    & identification of Ni-U-phosphate deposits in bacteria~\cite{Basnakova_1998}
    \\ 
    & Metabolic activity\goodbreak{} measurements
    & nanoSIMS
    & nanoSIMS and stable isotope labeling to measure metabolic activity of microbial cells~\cite{Musat_2012, McNichol_2018} and heterogeneity in isogenic microbial populations~\cite{Stryhanyuk_2018, Calabrese_2019}
    \\ 
    & Biomarker imaging of \goodbreak{} single cells
    & ToF-SIMS
    & biomarker imaging of diatom cells in a microbial mat~\cite{Leefmann_2013}
    \\ 

    & Chemical analysis of \goodbreak{} biofilms
    & ToF-SIMS
    & localization of P in micro-algae biofilms~\cite{Moreno_2018}
    \\ 
    & Bacterial predators
    & HIM
    & imaging of the different stages of the life-cycle of Bdellovibrio bacteriovorus~\cite{Said2018}
    \\ 
    & Geomicrobiology
    & \ac{him}
    & imaging of iron-oxidizing bacteria~\cite{Joens_2013,Byrne_2018}, structure of Geobacter sulfurreducens biofilm\cite{Belianinov_2017a}
    \\
    &
    & ToF-\ac{sims}
    & microaerophilic Fe(II) oxidizing Zetaproteobacteria~\cite{Laufer_2017}, iron corrosion by methanogens~\cite{Tamisier_2022}
    \\ 
    & Anti-microbial properties of surfaces
    & HIM milling
    & investigation of the anti-bacterial properties of the nanostructures on dragonfly wings~\cite{Bandara_2020a} 
    \\ 
    \bottomrule
   \end{tabularx}
  \end{center}
\end{table*}

\subsubsection{\label{sss:him-bio-imaging}HIM---A tool for imaging of the tiniest living objects}

\ac{sem} is a standard method for imaging biological specimens, in which  samples are typically coated with a conductive layer to prevent charging under the beam. 
\ac{sem} imaging of uncoated samples without charging becomes possible if the accelerating voltage of the electron beam is reduced below \qty{1}{kV}, but at the expense of increased chromatic aberration and a reduced depth of field~\cite{Joens_2013}.
In contrast, with \ac{him}, in-situ charge compensation with an electron flood gun allows the investigation of non-conductive biological samples without conductive coatings while maintaining good spatial resolution. 
This is important, as metallic coatings can alter and conceal nanoscopic samples and features, such as virus particles or the cell membrane topography.
Furthermore, compared to \ac{sem}, \ac{him} delivers a larger depth of field, enhanced topographic contrast, and greater surface sensitivity~\cite{Hlawacek_2013, Hlawacek_2015}. 
Typically \ac{him} imaging relies on the detection of \acp{se}, but \acfp{bi} can also be detected, which enhances material contrast (this is comparable to \ac{se} versus backscattered electron detection in \ac{sem}).
Since backscatter detection in \ac{him} is also sensitive to the neutrals that are generated, detection yields are relatively high meaning that images can be collected at lower dose (see also Section~\ref{subs_detectors}).
\citet{Bidlack2014} pointed out that \ac{bi} imaging could improve the visualization of immuno-gold labels in cells and
\citet{Ma_2017} used \ac{bi} imaging to detect different concentrations of colloidal nanoparticles in cells. However, in general the use of \ac{bi} detection for \ac{him} in the life sciences is still underexplored.

\paragraph{Preparation of the specimen}

Since the sample requirements for room-temperature \ac{him} (dry and vacuum-compatible) are almost identical to those of electron microscopy, established protocols from the \ac{sem} community provide a good basis for \ac{him} sample preparation. 
In general the sample preparation steps involve:
\begin{enumerate}
  \item Chemical fixation with aldehydes (e.g.\ formaldehyde or glutaraldehyde)
  \item Optional post-fixation with strong oxidizing agents (e.g.\ osmium tetroxide)
  \item Dehydration in a graded series of an organic solvent (e.g.\ methanol, ethanol or acetone)
  \item Drying (e.g.\ critical point drying).
\end{enumerate}
Instead of the final dehydration and drying of the sample, the application of ionic liquids has been proposed as an alternative~\cite{Golding_2016a} and successfully applied for imaging bacterial biofilms using \ac{him}~\cite{Schmidt_2021a}. 
In all cases, no final metallization of the sample is required (due to the in-situ charge compensation mentioned previously), hence samples can be visualized in a more pristine state.

\paragraph{Viruses}

\begin{figure}[tbp]
  \centering
  \includegraphics[width=\linewidth]{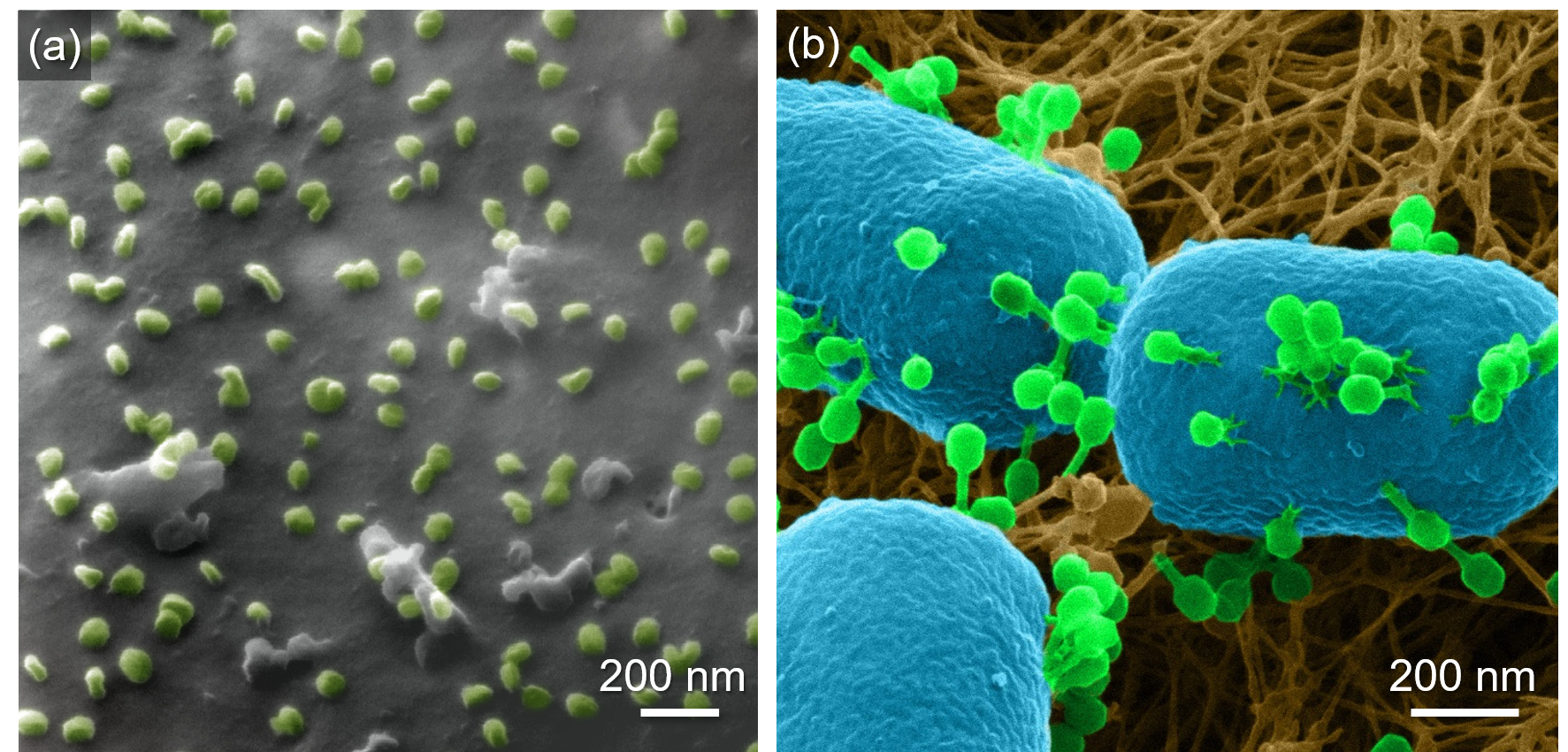}
  \caption{\label{fig:bioimagingViruses}
    (a) SARS-CoV-2 virus particles on the cell membrane of a Vero E6 cell. 
    (b) T4-phage infected E.\ coli bacteria on a fibrous agar surface. 
    (a) Copyright \textcopyright{}\citeyear{Frese2021}~\citet{Frese2021}, distributed under the terms and conditions of the Creative Commons Attribution License. 
 } 
\end{figure}

\citet{Sharma_2018a} used \ac{him} in investigations of the interactions of viruses and their hosts in sediment and microbial mat samples.
In the same manner, \ac{him} has been shown to be a powerful tool to investigate bacteriophages (i.e.\ viruses that attack bacteria) and their interactions with their host cells.
Both \citet{Leppaenen_2017} and \citet{You_2021a} published high-resolution \ac{him} images of T4-phage infected E. coli bacteria, and \citet{Vinner_2019a} visualized phage Felix O1 infected Salmonella bacteria.
The contributions that \ac{him} and other microscopies have made to our understanding of bacteriophages has been reviewed in more detail by \citet{Almeida_2018a}. 
Furthermore, \citet{Frese2021} and \citet{BarretoVieira_2022} recently reported the first visualization of SARS-CoV-2 virus particles by \ac{him}, demonstrating significant topographical differences compared to previously published \ac{sem} images of sputter-coated samples.
Examples of \ac{him} imaging of virus are shown in Fig.~\ref{fig:bioimagingViruses}.
In the future, \ac{him} could be used, for example, to examine the internal structure of phage-infected cells by milling cross-sections. 
In addition, \ac{him} could be used to better understand the spread of viral particles through other organisms, since it is possible to remove lightly attached phages during a washing step in the sample preparation and image what remains.

\paragraph{Single-celled organisms}%
\label{subs_life-sci_HIM_single_celled_organisms}

\ac{him} is an excellent tool for viewing single-celled organisms such as bacteria, archea, yeast, micro-algea, protozoa, etc.\, at a resolution approaching that of \ac{tem} without needing to embed and section the samples.
Here it is important to note that the \ac{him} technique primarily enables imaging of the cell surface.
In their 2013 paper, \citet{Joens_2013} employed \ac{him} for the first time for the imaging of microbes, namely iron-oxidizing bacteria.
The ability to view highly texturized biological specimens on non-conductive surfaces with strong topographic contrast, charge-compensation and large depth-of-field renders the \ac{him} an excellent tool for this work.
The first \ac{him} images of a culture of archea were published by \citet{Chen_2019a}, where \ac{him} proved particularly useful for investigation of the alleged cell budding of Ca.\ A.\ ethanivorans. 
Furthermore, \ac{him} has been utilized by \citet{Said2018} to image bacterial infections by the bacterial predator Bdellovibrio bacteriovorus. 
\ac{him} of bacterial biofilms (also often highly texturized) was published by~\citet{LeTourneau_2015a} (rhizobacteria), \citet{Belianinov_2017a} (Geobacter sulfurreducens) and \citet{Schmidt_2021a} (Pseudomonas putida).
Further \ac{him} studies of biofilms include imaging of microbial mats from hot springs in the Indian Himalayans by \citet{Sharma_2018a} and investigations of micro-algal biofilms with regard to architecture and P distribution by \citet{MorenoOsorio_2019}.
Other potential applications of \ac{him} imaging of microbiological objects include investigations of biofilms in different environments, e.g.\ on medical implants or in bio-reactors.
\ac{him} also holds great promise for studying phage-infection or predation by protists or bacteria in environmental bacterial biofilms.
Furthermore, one can conceive of \ac{him} investigations into the complex interplay of microbes with plants or fungi.
A recent example in this area is given by \citet{You_2021a}, where \ac{him} was used to analyze phage co-transport on hyphal-riding bacteria.

\paragraph{Cells, tissues, immuno-gold labelling}

The HIM has been used in many studies to investigate whole cells of multicellular organisms, including plant~\cite{MorenoOsorio_2019}, animal~\cite{Rice_2013a} and human cells~\cite{Chen_2011a, Schuermann_2018}.
In the study of nanoscopic cell structures, which are usually distorted or obscured by metallic coatings, charge compensation allows visualization of topographical structures that would not be observable by \ac{sem}.
For example, \citet{Schuermann_2015} visualized lipid nanodomains in the cell membranes of neurons, which were demonstrated to otherwise be concealed under a \qty{10}{nm} metallic coating.
\citet{Sato_2018a} have shown that immuno-gold-tagged antibodies can be resolved within cells to identify specific organelles, demonstrating the detection of ionoluminescence from ZnO nanoparticles in biological specimens using \ac{him}.
When imaging animal nanostructures such as insect wings, the superior depth of field of the \ac{him} has proven advantageous~\cite{Takei_2021a,Bandara_2020a}.
In addition, \citet{Joens_2013} demonstrated that the mouth cavity of a nematode can be dissected using Ne-\ac{fib} milling with subsequent imaging by \ac{him}, with minimal thermal damage during the milling process.

\paragraph{\acs*{him} in correlative workflows}

Besides structural analysis, the ability to investigate chemical compositions on the subcellular level is also needed.
In order to achieve this with a \ac{him}-based platform, detectors such as for \ac{sims} (cf.\ Subsection~\ref{subs_detectors}) can be installed and used for in-situ chemical investigations, as has been shown for E.\ coli cells incubated with TiO$\mathrm{_2}$ nanoparticles~\cite{DeCastro_2021}.
Alternatively, \ac{him} imaging can be integrated into correlative workflows ex-situ with \ac{sims}, \ac{eds}, or fluorescence microscopy using fluorescent markers (see also Section~\ref{subs_software}).
A recently published example shows how \ac{him} in correlation with the aforementioned techniques was used to analyze a resin-embedded soil sample with regard to roots, sediments, soil-organic matter and bacteria~\cite{Bandara_2021a}.
Further correlative studies that make use of the unique charge-free high-resolution imaging capabilities of \ac{him} can be expected, e.g.\ to analyze biomineralization processes, and to address ecological questions in environmental microbiology, plant and soil science, etc.

\subsubsection{\acs*{fib} tomography in the life sciences}%
\label{sss:bio_tomo}

Three-dimensional nano-imaging of biological samples has become increasingly important, e.g.\ for investigations of cell architecture, the ultra-structure of cell components, cell-cell interactions, structural analysis of tissues and biofilms, etc. 
Until the early 2000s, investigations into the 3D ultra-structure of samples such as resin-embedded tissue or cells were carried out by serial block-face \ac{sem}\cite{Denk_2004} or by serial sectioning for \ac{tem}\cite{Harris_2006}\@.
Whilst a very high lateral resolution can be achieved by these approaches, depth resolution is limited by the achievable thickness of the slices.
Alternatively, serial block-face \acp{sem} with an ultramicrotome fitted into the analysis chamber can be used for tomographic imaging of resin-embedded samples.
However, the depth resolution is still limited by the minimum about of material that can be removed in each slice.

\begin{figure*}[tbp]
  \centering
  \includegraphics[width=0.8\linewidth]{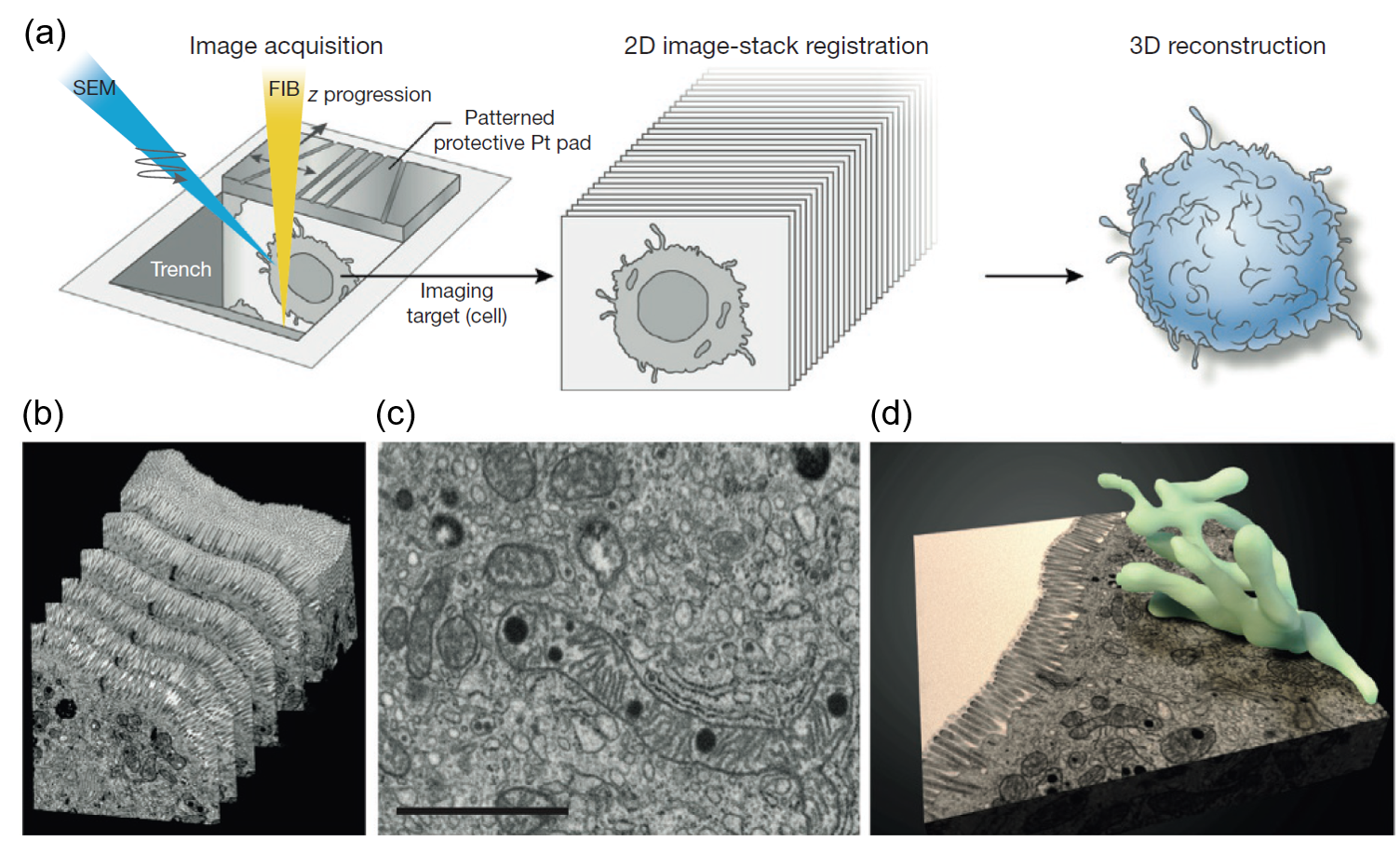}
  \caption{\label{fig:bioimagingFIB-SEMprinciple} Illustration of the basic workflow for volume imaging by \ac{fib}-\ac{sem} tomography: 
    (a) Repeated \ac{fib} milling followed by \ac{sem} imaging, image-stack registration and reconstruction, reproduced from \citet{Narayan_2015}.
  (b-d) Example of a mouse intestine sample, reproduced from \citet{Hartnell_2016}}
\end{figure*}

The advent of \acp{fib} (principally the Ga-\ac{fib}), propelled the field of biological micro-tomography, allowing much better control and much finer removal of material~\cite{Narayan_2015}.
In one of the first examples of \ac{fib} based tomography in the life sciences, \citet{Ballerini_2001} investigated the internal structure of yeast cells using Ga-\ac{fib} sectioning followed by a sample tilt to allow \ac{fib} based imaging of the cross section via secondary electrons and (negatively charged) secondary ions.
Soon thereafter, the image resolution and efficiency of the technique were improved using dual-beam \ac{fib}-\ac{sem} systems where the specimen is Ga-\ac{fib}-sectioned and imaged by \ac{sem}. 
The use of oxygen ions from a \ac{pfib} appears to reduce curtaining and other artifacts, allowing faster preparation of smoother surfaces for the imaging step~\cite{Sergey_2018}.
The basic workflow for \ac{fib}-\ac{sem} analysis of a resin-embedded biological specimen consists of repeated \ac{fib}-milling and \ac{sem} imaging of the exposed face followed by 2D image-stack registration and finally 3D reconstruction, as summarized in Figure~\ref{fig:bioimagingFIB-SEMprinciple}.

In order to obtain sufficient contrast in the \ac{sem} images, the sample is usually stained with heavy metals. Pioneering work was carried out by \citet{Drobne_2005}, who investigated the digestive system of a crustacean by \ac{fib}-\ac{sem} and compared imaging modes. In order to achieve better contrast, \citet{Leser_2009} systematically investigated different $\mathrm{OsO}_4$ and uranylacetate staining techniques and concluded that for plastic resin embedded samples, the OTOTO method ($\mathrm{OsO}_4$/thiocarbohydrazide/$\mathrm{OsO}_4$/thiocarbohydrazide/\\$\mathrm{OsO}_4$) in combination with backscattered electron imaging provides best results.
Other protocols involve a combination of potassium ferrocyanide, $\mathrm{OsO}_4$ and uranylacetate, and were successfully applied to brain~\cite{Knott_2011} and bone tissue~\cite{Kizilyaprak_2014}, as well as for the investigation of the interaction of \acp{cnt} with lung tissue~\cite{Kobler_2014}.
A detailed description of standard protocols for \ac{fib}-\ac{sem} of dried as well as plastic resin embedded (tissue) samples can be found in the work of \citet{Drobne_2013}. An alternative preparation method for \ac{fib}-\ac{sem} investigation of cells on surfaces uses thin layer plastification of the samples in place of resin embedding and full impregnation~\cite{Kizilyaprak_2014}. 

It is well known that dehydration and embedding in plastic resins introduces artifacts and may destroy the ultra-structure of cell components. 
These effects can be partially avoided by high-pressure or plunge freezing of the sample followed by cryo-substitution of water and slow infiltration with an epoxy-resin. 
However, \Ac{fib}-\ac{sem} imaging of the embedded sample is then done at room temperature, as described e.g.\ by~\citet{Schmid_2014} for samples of the nitrate-reducing and Fe-oxidizing bacteria BoFeN1\footnote{This is the same cell culture as used by~\citet{Joens_2013}, see Section~\ref{subs_life-sci_HIM_single_celled_organisms}}.
By the early 2000s it became clear that the ultra-structure of cells could be preserved even better if full cryo-workflows are employed, where a plunge- or high-pressure frozen specimen (still hydrated) is imaged at cryogenic temperatures.
\citet{Marko_2006} explored the possibilities of thinning vitrified biological specimens by cryo-\ac{fib} milling plunge-frozen water and concluded that (a) cryo \ac{fib}-milling of vitreous ice does not induce heating sufficient to cause devitrification, and (b) subsequent experiments should be performed below \qty{-135}{\celsius} to maintain a vitreous state.
In 2006, \citet{Heymann_2006} published \ac{sem} data of plunge-frozen and \ac{fib}-milled yeast cells, as well as \ac{sem} and \ac{tem} images of \ac{fib}-milled lamellae of plunge-frozen tumor tissue. Although the frozen samples were not stained, sufficient contrast for \ac{sem} imaging with \acp{se} was achieved through a combination of sublimation and metal deposition on the exposed surface. Similar work was performed by \citet{Hayles_2007}, who used a \ac{gis} system to deposit platinum onto frozen yeast cells and onto a petal epidermis sample of Nicotinia tabaccum. A recent overview of cryo-\ac{fib}-\ac{sem} techniques for frozen, hydrated biological samples is given by \citet{Hayles_2021}.

Another leap forward was the development of correlative confocal laser scanning and \ac{fib}-\ac{sem} tomography using fluorescent dyes, high-pressure freezing and cryo-substitution embedding\cite{Lucas_2014}. 
For example, \citet{Gorelick_2019} developed a photon-ion-electron microscope (PIE-scope), which employs correlative cryogenic fluorescent light microscopy and \ac{fib}-\ac{sem} enabling the rapid and precise location of protein complexes in cells.
Another example is the recently published (non-cryo) 3D-correlative workflow combining super-resolution light microscopy and \ac{fib}-\ac{sem} using reactive oxygen ion \ac{pfib} milling and a rotating sample stage to study embedded cultured cells~\cite{Wang_2021b}.

Driven by questions in cell biology and connectomics, strategies have also been developed to enable \ac{fib}-\ac{sem} analysis of large volumes at high spatial resolution. Pivotal work in this area by \citet{Xu_2017a}, involving automated recovery from beam/system malfunctions, enabled enhanced large-volume 3D \ac{fib}-\ac{sem} analysis of one third of a fruit fly brain at a depth resolution of \qty{8}{nm}. In order to analyze large volumes of hard materials such as bone, \ac{fib}-\ac{sem} using high-current beams of Xe ions available from \ac{pfib} sources has been demonstrated~\cite{Binkley_2020}.

\begin{figure}[tbp]
  \centering
  \includegraphics[width=\linewidth]{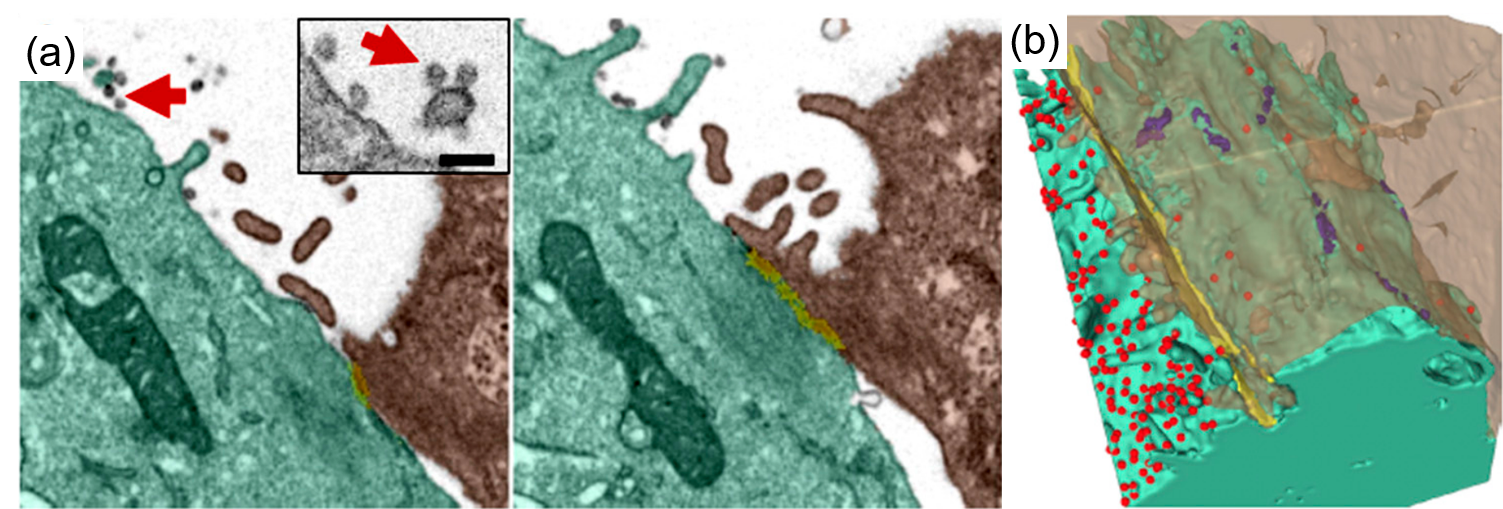}
  \caption{\label{fig:bioimagingFIBSEM_SARSCoV2} 
    (a) \ac{fib}-\ac{sem} images of two \qty{8}{nm} slices showing two contacted Vero E6 cells (green, brown) and SARS-CoV-2 particles in the extracellular space (arrows). 
    Inset scale bar: \qty{200}{nm}. 
    (b) 3D reconstruction of the plasma membranes of the two cells shown in (a). 
    The junction-mediated contact between the cells is coloured in yellow. 
    The virus particles are depicted in red and show a different density on either side of the junction. 
  Data reproduced from \citet{Baena_2021}.}
\end{figure}

In the context of the COVID-19 pandemic, \ac{fib}-\ac{sem} tomography proved valuable for gaining insight into the microscopic mechanisms of the disease.
\citet{Baena_2021} employed \ac{fib}-\ac{sem} to analyze how SARS-CoV-2 spreads from infected cells to new hosts, as shown in Figure~\ref{fig:bioimagingFIBSEM_SARSCoV2}. 
Using a combination of \ac{tem}-tomography and \ac{fib}-\ac{sem}, \citet{Cortese_2020} were able to visualize morphological changes in SARS-CoV-2-infected lung cells. In a recent study on the frequently observed infection of the human kidney by SARS-CoV-2, \ac{fib}-\ac{sem} was also used to image virus particles in vacuoles\cite{Jansen_2022}.

\subsubsection{\acs*{fib} tomography of energy materials}%
\label{sss:mat_tomo}

One of the very first applications of \ac{fib}-\ac{sem} tomography in energy materials research was presented by \citet{Wilson_2006} in \citeyear{Wilson_2006}.
The authors investigated a solid oxide fuel cell anode based on \ac{ysz} and used 3D reconstruction to differentiate the  \ac{ysz}, Ni and pore phases.
This enabled quantification of important structural features, such as volume fractions, surfaces and even information on the so-called three-phase boundary.

In \citeyear{Ostadi_2010}, \citet{Ostadi_2010} investigated a microporous layer used in a hydrogen-powered \ac{pemfc}.
\ac{fib} tomography was used to complement the X-ray tomography investigations of the fuel cell electrode structure in order to capture the nanostructure of the microporous layer, which cannot be resolved using standard X-ray imaging equipment due to the limited spatial resolution of \qty{\approx1}{\micro\meter} compared to the \ac{fib}-\ac{sem} tomography resolution of \qty{\approx10}{nm}.
The porous structure was reconstructed in 3D and the porosity and pore size distribution analyzed.

\begin{figure}[tbp]
   \centering
   \includegraphics[width=\linewidth]{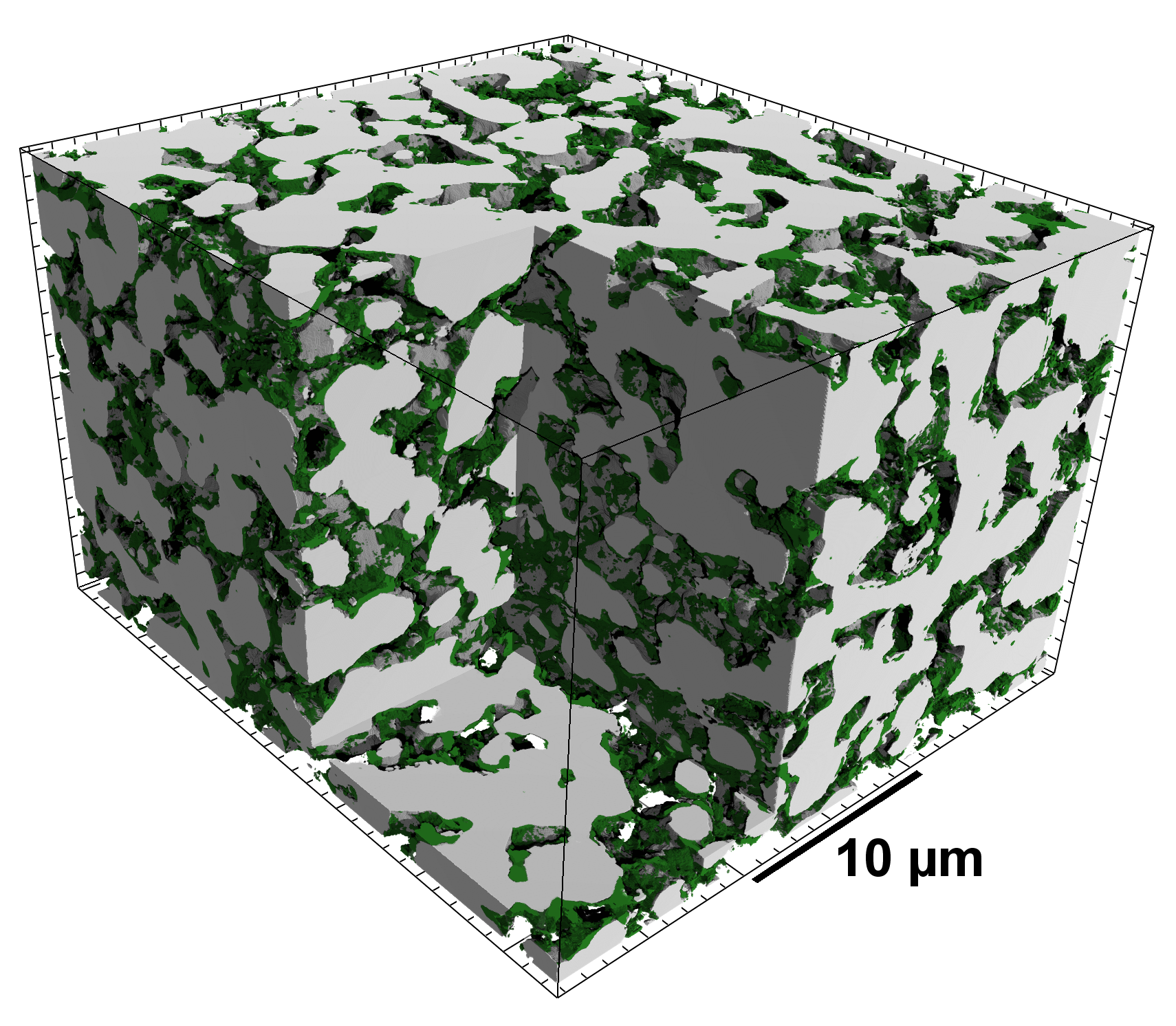}
   \caption{\label{fig:FOR2397_Silber} \ac{fib}-\ac{sem} tomography of Ag-based gas diffusion electrode. Gray: Ag grains, green: binder/PTFE within the pore system. Only the 3D image can reveal the complex pathways inside the pore system (\citet{Paulisch_2021})}
\end{figure}

In the same year, \citet{Zils_2010} demonstrated how \ac{fib} tomography can be used to study the effects of  manufacturing parameters on the 3D structure and resulting properties of the fabricated material.
For various catalyst electrode materials intended for use in \acp{pemfc}, the authors compared porosity and pore size distributions 
made with airbrushing and a fast spray coating technique, 
and related the 3D structural features to the performance of the materials. 
\citet{Schulenburg_2011} used \ac{fib} tomography to study degradation effects in the 3D structure of carbon-based fuel cell electrodes after many start-stop cycles and discovered breakdown of the entire nanoporous structure leading to a significant reduction in the supply of reaction gases to the catalyst.
In combination with other (X-ray based) tomography techniques and structural modeling, \ac{fib} tomography can also be used for correlative tomographic imaging~\cite{Moroni_2016} to fill in missing information at the nanometer scale and to produce multiscale 3D data covering structural features over a broad length scale. 
For example, \citet{Zielke_2014} have shown how \ac{fib} tomography can be used to supplement the information missing from synchrotron tomography data due to the limited spatial resolution. 
Here, high-resolution 3D data obtained using the \ac{fib} method was used to create a structural model
at the nanometer scale, to enhance the synchrotron data set with nanometer detail.
In a recent study, \citet{Paulisch_2021} used \ac{fib} tomography to analyze the complex 3D path network inside Ag-based gas diffusion electrodes (Figure~\ref{fig:FOR2397_Silber}). 
3D data acquired with \ac{fib} tomography has also been used for mathematical modeling of the 3D structure of energy materials~\cite{Neumann_2019}, whereby the computer-generated 3D structures can later be used for the development of digital twins and for future virtual materials design.

\ac{fib} tomography is thus now well established for high-resolution 3D structural studies of materials for energy-technology research, including battery materials~\cite{Vierrath_2015, Etiemble_2016, Billaud_2016, Liu_2016, Shearing_2012,Liu_2015,Lagadec_2018,Shi_2020,Houx_2020,Kroll_2021,Ghadban_2023}, fuel cell materials~\cite{Star_2017,Bevilacqua_2021,Netzeband_2016,YakalKremski_2012}, electrolyzers~\cite{Etzold_2021,Wang_2021,Hegge_2018,McLaughlin_2020a} and catalytic materials~\cite{Gebhard_2020,Grothausmann_2012,Franzen_2019,McLaughlin_2020}.
In the case of beam- and air-sensitive materials, cryogenic \ac{fib}-\ac{sem} implementations are becoming increasingly popular, since radiation damage can be significantly reduced at low temperature and cryotransfer workflows are by nature air free. For example, cryo \ac{fib}-\ac{sem} techniques have enabled 3D nanoscale insights into Li dendrite formation and the so-called solid-electrolyte interface~\cite{Fang_2019,Lee_2019,Cui_2020, Ren_2020,Dong_2021,Xu_2022}.

\subsubsection{Data processing of \acs*{fib} tomographies}

With the growing volume of data, the analysis of large tensors containing mostly grayscale information is becoming increasingly challenging. 
Sophisticated analysis requires data classification, i.e.\ assigning a class to each point.
This is particularly challenging for \ac{fib} based tomography, where a measurement contains a large number of classes, but in addition, artifacts that need to be detected and addressed.
Here, \ac{ml} approaches, especially easy-to-use mainly random forest based tools such as Ilastik~\cite{berg2019} or the trainable Weka segmentation~\cite{M_ArgandaC_2017}, are becoming increasingly established for image annotation and classification.

One of the major issues in tomographic imaging of porous materials is the problem of separating the pore foreground from the background.
In order to address this issue, infiltration of the pores with an appropriate resin may be carried out.
Silicone-based resins are frequently used to infiltrate the pore system of energy materials due to the generally high carbon content of the latter~\cite{Ender_2011}. 
However, due to sample alteration upon this treatment, or inaccessibility of the pore system, pore infiltration is often either impossible or undesirable.
Thus when the \ac{fib} mills through a pore, the \ac{sem} image will also collect signal from the back of the pore.
In such cases the 3D reconstruction becomes much more complex, because only the foreground around the pore edges represents the actual material distribution corresponding to the exposed surface.
To address this issue, various optimized segmentation algorithms have been developed~\cite{Moroni_2020,Salzer_2012,Salzer_2014,Salzer_2014a}, as well as some \ac{ml} tools~\cite{Furat_2019,Arlt_2020,Bailey_2023}.
Recently, \citet{M_Osenberg_2023} demonstrated a technique combining two different \ac{se} signals (from an in-lens detector and from a detector sensitive to \acp{se} from backscattering electrons) together with a \ac{ml} algorithm to improve results for the separation of pore foreground and background in the case of a lithium battery cathode material.
However, despite much progress, the general porosity problem is still not completely solved for all material structures.

\subsubsection{Outlook}

Over the past 20 years, \ac{fib}-\ac{sem} has become a mainstream technique for investigating the 3D structure of a range of complex specimens, often with fully established cryo-workflows.

Further improvements involving optimized milling strategies to reduce local heating effects due to the \ac{fib} (in non-cryo implementations) would be beneficial for nanoscale milling to reveal the internal structure of biological samples, and to reduce artifacts when milling energy materials, as these may also involve heat-sensitive components such as polymers~\cite{Orthacker2014}.

In parallel, the ever growing and complex data volumes demand many more automated analysis pipelines.
To make this possible, standardized data and metadata formats are needed~\cite{Scheffler2022}.
The combination of standardized data sets and high quality individual image classifications generated based on e.g.\ random forest models, would then allow for the compilation of diverse training sets to train deep learning models.
With sufficient training data, neural networks such as the U-Nets~\cite{ronneberger2015u} can be trained and applied to data that is very different from the original training set~\cite{M_Yara_2021}.

On the instrument side, much progress has been made in increasing the sample volumes accessible to \ac{fib} based tomographies through the use of high-current \ac{pfib} sources for heavy noble gas ions, especially Xe. 
Due also to advances in automated data acquisition and analysis, \ac{fib} tomography of cubic millimeter volumes with a spatial resolution on the order of tens of nanometers is now possible. 
Complex data from different experimental sources with different lateral resolutions can be made available to the research community in large multiscale atlases.
This is celebrated particularly in the life sciences, where Ga-\ac{fib}-\ac{sem} was used to map and reconstruct the connectome of the Drosophila fly brain\cite{Xu_2017a,Scheffer_2020}.
In terms of open data, life science offers prime examples: the virtual fly brain~\cite{Court_2023} and the fly brain atlas~\cite{Scheffer_2019}, a digital twin of Drosophila's brain combining huge amounts of data from different sources and methods in an easily accessible format.
Creating and analyzing such massive data in an automated way has relied on the rapid development of \ac{ml}, which will be one of the key factors for the future evolution of \ac{fib} technologies.

\subsection{Elemental analysis using \acs*{fib}-\acs*{sims}}%
\label{subs_analysis}

While many correlative \ac{fib} imaging and analysis approaches exist (mainly using \ac{fib}--\ac{sem} instruments), \ac{fib}-\ac{sims} is by far the leading method for achieving ion beam based analysis of elemental distributions.
Dedicated \ac{sims} instruments often use one or two \ac{fib} columns in combination with a highly optimized mass spectrometer to provide high mass resolution and sensitivity~\cite{AgueiGonzalez_2019,McPhail_2006a}, but a dedicated \ac{fib} instrument equipped with a \ac{sims} add-on can provide greater versatility.
So far, \ac{gfis}-based \ac{fib} instruments provide the highest spatial resolution for \ac{sims}~\cite{Wirtz_2016,Klingner_2019}.
Recently, \ac{lotis}-based \ac{fib}-\ac{sims} has also been introduced~\cite{NanoK}.

\subsubsection{Overview}

\Ac{sims} is based on irradiating the sample with a focused beam of ions (the so-called primary ions) and extracting, mass-filtering and detecting the ions sputtered from the sample surface (the so-called secondary ions).
It is a powerful technique for analyzing surfaces owing in particular to its ability to detect all elements from hydrogen (H) to uranium (U), its excellent sensitivity to low concentrations, its high dynamic range and its ability to differentiate between isotopes. 
\Ac{sims} offers four main analysis modes, namely mass spectrum acquisition, depth profiling, 2D imaging and 3D imaging.
In addition, \ac{sims} does not require any complex sample preparation (in contrast to \ac{tem} based analysis techniques such as \ac{eds} or \ac{eels}, or \ac{apt}).

Therefore, equipping a \ac{fib} instrument with a \ac{sims} system adds unique analytical capabilities for 3D chemical characterization. 
Furthermore, this add-on enables unique workflows for in-situ multimodal analytics by correlating the \ac{sims} data with that obtained from other imaging/analytical techniques available using the same \ac{fib} platform (e.g.\ \ac{se}-imaging, \ac{eds}, \ac{ebsd}, etc.).

\subsubsection{\acs*{fib}-\acs*{sims} in materials science}

\begin{figure}[tbp]
   \centering
   \includegraphics[width=\linewidth]{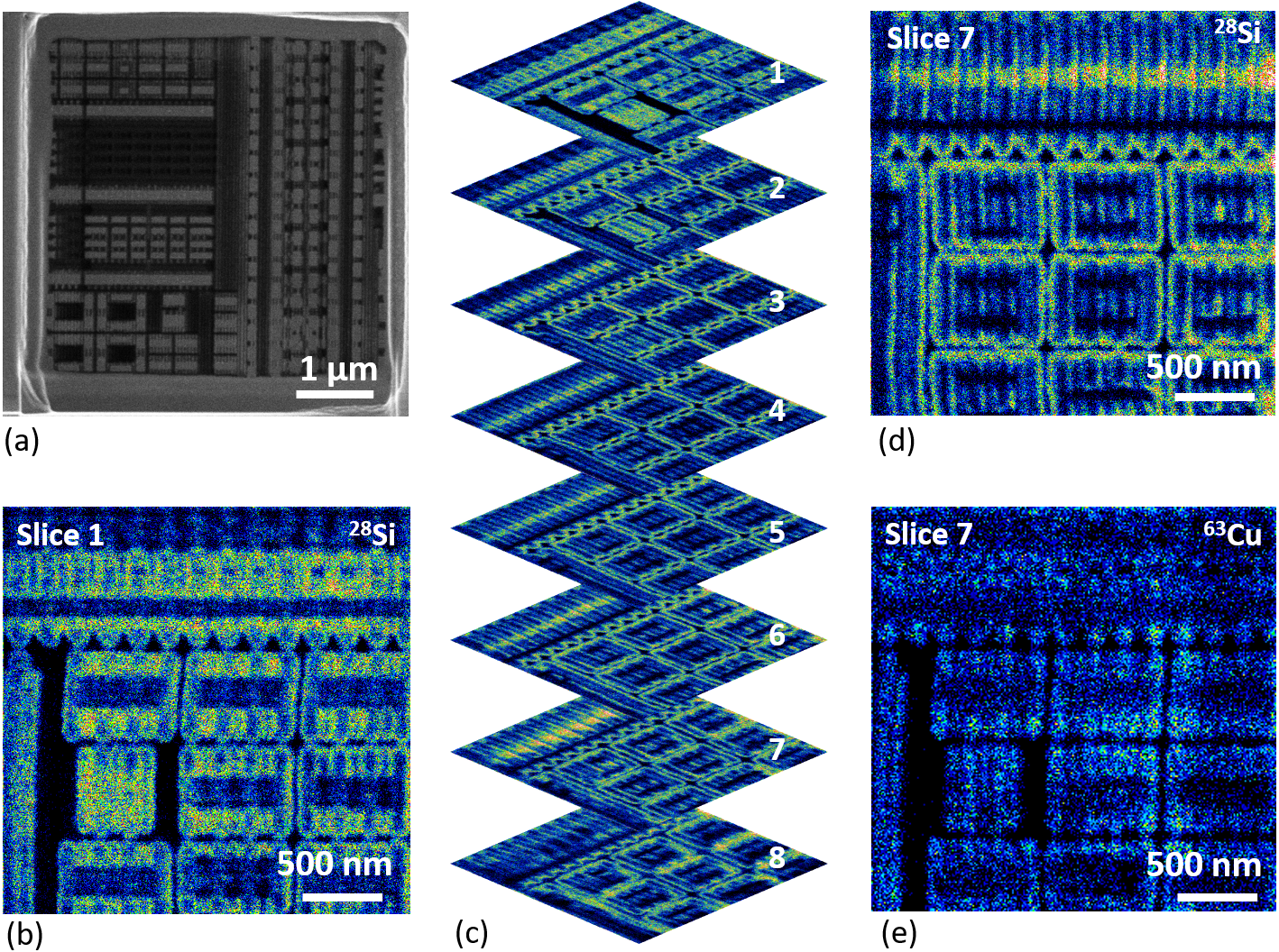}
   \caption{\ac{se} and \ac{sims} images of a \ac{finfet} device obtained using a \ac{him} instrument equipped with a magnetic sector \ac{sims} system:
   (a) \ac{se} image acquired using \qty{25}{\kilo\electronvolt} He$^{+}$ beam. 
   (b--e) \ac{sims} images for $^{28}$Si and $^{63}$Cu acquired using \qty{20}{\kilo\electronvolt} Ne$^{+}$ beam corresponding to a zoomed in region in the \ac{se} image.
   (c) Stack of eight successive \ac{sims} images recorded for the same area (\ac{fov} \qtyproduct{3x3}{\micro\meter}) allowing determination of the elemental composition in 3D. 
   (d) and (e) Individual $^{28}$Si and $^{63}$Cu \ac{sims} maps corresponding to the 7$^\text{th}$ slice. 
   Modified after~\cite{Audinot_2021}}.%
   \label{fig:finfet}
\end{figure}

\Ac{sims} on \ac{fib}(-\ac{sem}) instruments has been applied to a number of topics in materials science, in particular for the nano-characterization of complex 3D architectures and compositions. 
Microelectronic devices are a typical example, where 2D and 3D \ac{sims} imaging provides analysis of elemental distributions with high lateral and depth resolution~\cite{Stevie_1999}.
Figure~\ref{fig:finfet} shows an example of the 3D analysis of a \ac{finfet} using \ac{fib}-\ac{sims}~\cite{Audinot_2021}.
This data was obtained using a magnetic sector \ac{sims} system installed on a \ac{him} instrument, achieving a spatial resolution of better than \qty{15}{\nano\meter}.
3D maps such as these can be used for end pointing and circuit edit applications.

There are also a number of examples where \ac{fib} based \ac{sims} has been a major asset in the development of photovoltaic and battery materials.
The latest advances in photovoltaic conversion technologies are based on thin films, whether second generation, such as copper indium gallium selenide (CIGS), or third generation dye-based, such as lead halide perovskites. 
Their development requires a very comprehensive understanding of composition in order to correlate structure and chemical composition with performance and ageing.
\ac{fib} based \ac{sims} has allowed investigation of many questions related to the chemical heterogeneity of grains~\cite{Gratia_2016} and grain boundaries~\cite{Pillatsch_2019, Usiobo_2020}, precipitation of secondary phases (e.g.\ PbI$_2$), and chemical/structural modification at the interfaces with charge-collecting electrodes~\cite{Gratia_2017, Liu_2021, Zimmermann_2019}.
Optimizations based on these results have helped realize new photovoltaic cells with efficiencies approaching 25\%. 

In battery materials, \ac{sims} technology allows the detection of Li (in addition to any other element of interest) with a very high sensitivity, which is not possible using e.g.\ \ac{eds}.
Therefore, with \ac{fib}-\ac{sims} the Li concentration gradient can be mapped over several orders of magnitude at high spatial resolution across different surfaces and interfaces~\cite{Pillatsch_2019, Sarbada_2020, Wheatcroft_2020, Taeye_2021, Priebe_2021b, Sastre_2020a, Dubey_2021}.

Further examples of the application of \ac{fib}-\ac{sims} in materials science include studies of petroleum-based polyolefin plastics~\cite{Ovchinnikova_2021}, polyester and melamine resins~\cite{Nishinomiya_2016a}, the investigation of alloys and superalloys~\cite{Gaderbauer_2020, Kosari_2020, Wieczerzak_2021a}, the study of nanoparticles~\cite{Kennedy_2020, Fizesan_2019,Priebe_2019a, Priebe_2020a}, the investigation of chemical inhomogeneities in grain sizes between \qty{50}{\nano\meter} and \qty{200}{\nano\meter} in Delafossite materials~\cite{LuncaPopa_2020}, oxygen diffusion in grain boundaries~\cite{Ooi_2006} the distribution of rare earths in optical fibres~\cite{Sedlacek_2014, Lorincik_2014}, and the characterization of thin films~\cite{Priebe_2021a, Priebe_2021c}. 

\subsubsection{\acs*{fib}-\acs*{sims} in the life sciences}

Ion beam analysis has been used for chemical micro-analytics of biological specimens since the 1970s when high-energy \ac{pixe} was found to be valuable for determining trace elements in bulk tissue sections~\cite{Jundt_1974}. 
By the beginning of the 1990s accelerator-based proton beams could be focused down to \qty{1}{\micro\meter} and could thus be employed for the spatial mapping of trace elements in biological samples~\cite{Forslind_1991}. 
The development of high-brightness \ac{lmis} in the second-half of the 1970s~\cite{Krohn_1975} also facilitated the introduction of medium-energy (\qty{<100}{keV}) ion beam microprobe techniques for bio-imaging and micro-analysis, in particular in the field of \ac{sims}~\cite{LeviSetti_1986, LeviSetti_1988}.
Already at that time it turned out that \ac{sims} is advantageous for biological applications because of its low detection limits but also its sensitivity to different isotopes.
Hence in the life sciences, numerous studies based on imaging with \ac{sims} exist~\cite{Boxer_2009}. 
For example, \ac{fib}-\ac{sims} has been used to spatially resolve chemical signatures from the surface and interior of mammalian individual cells~\cite{Szakal_2011} and bone tissue~\cite{Malmberg_2008}.
In this area, the Cameca nanoSIMS (a sector-field mass spectrometer with about \qty{50}{\nano\meter} lateral resolution) has become an important tool~\cite{Nunez_2017}. 
It was designed to achieve a mass resolution high enough to resolve the stable isotopes of elements such as carbon and nitrogen ($^{13}\mathrm{C}$ and $^{15}\mathrm{N}$), which are commonly used for isotope-labeling experiments~\cite{Musat_2012}. 
In addition, \ac{tof}-\ac{sims} approaches for 2D and 3D chemical imaging in the life sciences are becoming increasingly important~\cite{Boxer_2009, MorenoOsorio_2019}. 
For example, metabolic imaging on a sub-cellular level with a mass resolution of 240,000 was recently achieved by combining \ac{tof}-\ac{sims} with an orbitrap mass analyzer~\cite{Passarelli_2017a}.

\begin{figure}[tbp]
   \centering
   \includegraphics[width=\linewidth]{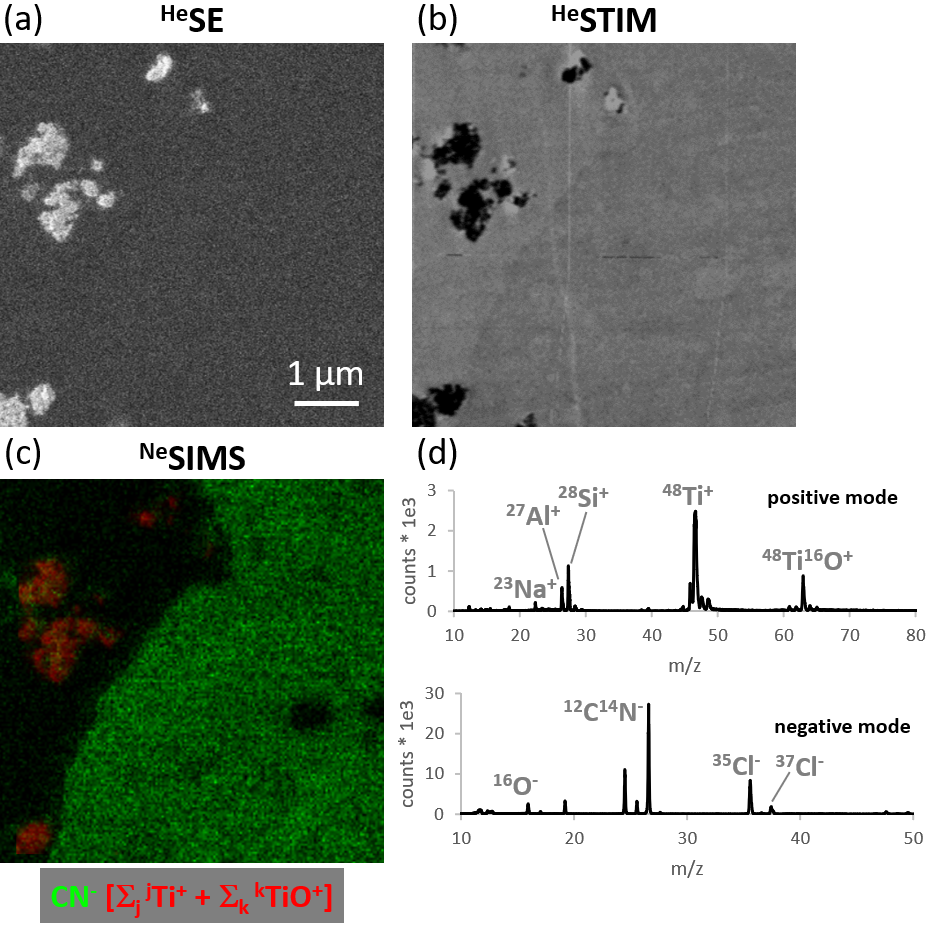}
   \caption{npSCOPE \ac{him}-\ac{stim}-\ac{sims} correlative high-resolution imaging and nano-analytics of keratinocyte cell cultures exposed to \qty{30}{\nano\meter} Si-Al-TiO$_2$ nanoparticles: 
   (a) He-\ac{fib} \ac{se} image. 
   (b) He-\ac{fib} \ac{stim} bright field image. 
   (c) Ne-\ac{fib}-\ac{sims} false color image of CN (green) and Ti species (red). 
   (d) Ne-\ac{fib}-\ac{sims} spectra.
   He-\ac{fib} \ac{se} image: \ac{fov} \qtyproduct[product-units=single]{15x15}{\micro\meter}, \qtyproduct[product-units=brackets]{1024x1024}{\pixel}, dwell time \qty{1}{\micro\second}; 
   He-\ac{fib} \ac{stim}: \ac{fov} \qtyproduct[product-units=single]{8x8}{\micro\meter}, beam current \qty{2}{\pico\ampere}, dwell time \qty{400}{\micro\second}, \qtyproduct[product-units=brackets]{512x512}{\pixel}. 
   Ne-\ac{fib}-\ac{sims} negative mode: \ac{fov} \qtyproduct[product-units=single]{8x8}{\micro\meter}, magnetic field \qty{300}{\milli\tesla}, dwell time \qty{9000}{\micro\second}, \qtyproduct[product-units=brackets]{256x256}{\pixel}. 
 Ne-\ac{fib}-\ac{sims} positive mode: \ac{fov} \qtyproduct[product-units=single]{8x8}{\micro\meter}, magnetic field \qty{325}{\milli\tesla}, dwell time \qty{4000}{\micro\second}, \qtyproduct[product-units=brackets]{256x256}{\pixel}.}%
   \label{fig:npScope}
\end{figure}

\ac{fib}-\ac{sims} at the highest spatial resolution has been enabled using the \ac{gfis} source, which can produce  sub-nanometer spot sizes. 
Here, \ac{him} instruments have been equipped with \ac{sims} spectrometers and used for high-resolution chemical analysis of a range of life science samples including micro-organisms, viruses, cells, small animals, plants, etc.~\cite{Schmidt_2021a}.  
Recent specific \ac{him}-\ac{sims} applications include the localization and identification of silver nanoparticles and nanowires in intestinal or alveolar cell cultures~\cite{Gillois_2021, Fizesan_2019} and the characterization of biogenic nanoparticles produced by bacteria~\cite{Guillermier_2019}.
Finally, correlative \ac{him} based \ac{se} imaging, \ac{stim}~\cite{Serralta_2020} and \ac{sims} (all in the same instrument) has been used in investigations of thin tissue sections of Daphnia magna exposed to TiO$_2$ nanoparticles (see Figure~\ref{fig:npScope})~\cite{DeCastro_2021}.
In this correlative nano-toxicological test study, the nanoparticles were shown to accumulate in the gut without penetration of the epithelial barrier.

Beyond materials science and life science applications, \ac{fib}-\ac{sims} is also proving valuable for high-resolution structural and chemical analysis in other fields, such as soil science~\cite{Ost_2021}, geology~\cite{Ball_2020}, and cosmochemistry~\cite{Ball_2019, McPhail_2006}, and even for the analysis of cultural heritage objects~\cite{Keune_2004, Adriaens_2006, Voras_2016}.

\subsubsection{Outlook}

In order to further increase the spatial resolution of \ac{fib}-\ac{sims}, we must now look beyond reducing the beam probe size. 
In \ac{him}-\ac{sims}, the lateral resolution is in fact mainly limited by the low signal yield and destructive nature of \ac{sims}.
In order to achieve the smallest possible detection limits, as much signal as possible must be extracted from the sputtered volume. 
One possibility here is to use other primary ions that produce a higher useful yield of sputtered ions.
However, other primary ion species typically only achieve a higher yield of either positive or negative ions (not both).
Higher yields of positive ions, which are particularly suitable for the detection of metal elements, are achieved using primary ions with high electronegativities, such as oxygen or halogen ions.
Enhancement of positive ion yields could thus be achieved using smaller spot sizes from \acp{pfib} or through the use of \ac{ilis}.
Higher yields of negative ions are achieved by using primary ions that lower the work function of the sample surface, which is where developments using \ac{lotis} can bring benefit by e.g.\ providing Cs ions~\cite{NanoK}.
Alternatively, the use of precursor gases can also increase the ion yields (see Section~\ref{subs_detectors}).

Another approach that could be explored to increase yields and therefore also spatial resolution involves further development of \ac{snms} schemes for \acp{fib}, in which sputtered neutrals are analyzed. 
Through the use of lasers for post-ionization of these neutrals, the dependence of the ion yields on the surrounding material in the sample (the so-called matrix effect) is reduced.
However, since high laser intensity is required, \ac{snms} is only possible in pulsed operation, which inevitably requires pulsing of the primary ion beam. 
Here, the use of \ac{tof} spectrometers becomes an obvious choice.
Post-ionization with focused electron beams is also conceivable, although preliminary experimental work in this area is scarce. 

A further issue in \ac{fib}-\ac{sims} that needs to be addressed is material intermixing induced by the primary ions.
In order to achieve the smallest beam probe sizes, comparatively high beam energies in the range of \SIrange{20}{50}{\kilo\electronvolt} are typically used.
However, the disadvantage is the resulting long range of the ions in the target (compared e.g.\ to that of the low energy Ar ions commonly used for \ac{sims} depth profiling).
Long ion ranges can quickly lead to atom mixing of deeper layers and therefore to a decrease in depth resolution with increasing sputter depth~\cite{Philipp_2016}.
In the future, primary ions with high atomic number and/or small kinetic energy per nucleon could be used.
For example, Cs ions from \ac{lotis} or Xe ions from \ac{gfis} could provide both small beam sizes and good depth resolutions. 
Alternatively, the energy per nucleon can also be reduced by using cluster ions, e.g.\  Bi or Au clusters from \ac{lmais} (see Table~\ref{table:ion_sources}).

Small cluster ions~\cite{Kollmer_2004}, reactive primary ions e.g.\ Cs~\cite{Knuffman_2013, Pilz_2019},
or laser post-ionization~\cite{Draude_2012} can also reduce the fragmentation of molecules, which is another aim for future developments (see references in Table~\ref{tab:analytic}). 
Molecule fragmentation needs to be understood both theoretically and experimentally in order to deduce the molecular composition of the sample from the measured spectra.
In this regard, the creation of molecular fragmentation databases for typically-used ion energies and types would enable fingerprinting in future work.

Continued spectrometer development is also needed to enable complete detection of all sputtered ions. Here the \ac{tof} spectrometers and magnetic sector spectrometers with large area \acp{mcp} lead the way.
Development efforts should also be invested into spectrometers that allow higher mass resolution in order to enable the separation of isobars.
In this area, results obtained using Orbitraps are very promising~\cite{Passarelli_2017a}. 

In the perfect \ac{fib}-\ac{sims} instrument, all the key parameters (spatial resolution, mass resolution, detection limit, measurement time) would be optimized simultaneously. 
In reality one has to compromise between different priorities.
Furthermore, in-situ combinations of \ac{fib}-\ac{sims} with other analytical modes, e.g.\ \ac{sem} based analysis, are highly desirable.
The integration of an electron flood gun for in-situ charge neutralization (as is implemented in \ac{him}) would be universally beneficial for the analysis of insulating sample surfaces, and cryo-\ac{fib}-\ac{sims} would be of great interest for life science and other beam-sensitive samples.
In-situ \ac{afm} would allow correlation of the \ac{sims} data with the actual sputtering depths enabling more accurate 3D reconstructions.

Last but not least, most elemental analysis methods, including \ac{sims}, benefit from vacuum levels that are better than  the usual high vacuum (HV) conditions of typical \ac{fib} chambers in order to avoid background signals originating from residual gas.
\Ac{fib} instruments with  \ac{uhv} or near \ac{uhv} chambers have been built by researchers in the past~\cite{Gastel2011,DeCastro_2021} and the benefits for normal imaging as well as analysis have been demonstrated~\cite{Veligura_2012}. Manufacturers should similarly consider the merits of \ac{uhv} in future instrument development.

\subsection{Applications of gas-assisted \acs*{fib} processing}%
\label{subs_fibid}
Gas-assisted bottom-up prototyping of nanostructures using the \ac{fib} is known as \acf{fibid}.
In this technique, the focused ion beam 
induces the partial decomposition of precursor molecules, dissociating them into volatile and non-volatile components. 
The volatile products are pumped away, while the non-volatile products are deposited on the surface, thus creating a nanostructure in a single-step process. 
This direct writing of nanostructures using the \ac{fib} is powerful, flexible, and conceptually simple, but at the same time extremely complex in the chemical and physical processes involved. In addition to the surface processes and electron-driven chemistry that also occur in direct writing with electron beams, competition with physical sputtering, and thermal or ion induced structural changes, contribute here to the shape and microstructure of the deposit. 
Hence, \Ac{fibid} strongly depends on multiple growth conditions, such as ion species, acceleration voltage, ion beam current, ion dose, beam dwell time, scanning strategy, substrate type (insulating or conductive), \ac{gis} nozzle position, precursor molecule (vapor pressure, sticking coefficient to the surface, residence time, dissociation cross-section), precursor temperature, base and process pressure in the vacuum chamber, among others.
This makes \Ac{fibid} of functional nanostructures highly dependent on the experience of the particular user and calls for better theoretical understanding and empirical modeling tools that generate patterns based on few calibration routines, thus also enabling less experienced users to fabricate complex nanostructures.
Table~\ref{table:fibid} provides an overview of the fields where \ac{fibid} processing has been employed so far.

\begin{table*}[btp]
   \caption[Applications of \acs*{fibid}]{Overview of applications of gas-assisted \ac{fib} processing, in which gaseous precursor compounds are locally dissociated by the ion beam to grow nanostructures.}%
   \label{table:fibid}
   \begin{center}
      \begin{tabular}{P{2cm}P{5.4cm}P{1.1cm}P{8.2cm}}
	 \toprule
	 Property & Target Application & Ion & Deposit Geometry \& Precursor\\ 
	 \midrule
	 Electrical\ \& electronic
	    & electrical connections
	        & He 
	            & semiconducting nanowires from CpPtMe$_3$~\cite{Wu_2013}, metallic Co lines from Co$_2$(CO)$_8$~\cite{Wu_2014}\\
	        & & Ne 
	            & metallic nanowires from 
             CpPtMe\textsubscript{3}~\cite{Wu_2013}\\
	        & & Ga 
          & conducting nanowires from CpPtMe\textsubscript{3}~\cite{Lin_2003,Lin_2003a}, MeCpPtMe\textsubscript{3}~\cite{Tao_1990, FernandezPacheco_2009, DeTeresa_2009}, Au(hfac)Me\textsubscript{2}~\cite{Blauner_1989,Shedd_1986}, Co(CO)\textsubscript{3}NO~\cite{Gazzadi_2011}, Cu(hfac)(VTMS)~\cite{DellaRatta_1993}, W(CO)$_6$ contacts to Co and Cu nanowires~\cite{Wang_2010,Kompaniiets_2014,Kompaniiets_2014a}, Pt-based sidewall contacts using MeCpPtMe\textsubscript{3} to MoSi layer stack~\cite{Dobrovolskiy_2018}\\
	   & insulating barriers
	            & Si
		    & thin layer from TMOS (Si(OCH\textsubscript{3})\textsubscript{4})~\cite{Komano_1989}\\
		    & & Ga & thin layer from TMOS~\cite{Young_1995,Lipp_1996} and PMCPS ((CH\textsubscript{3}SiHO)\textsubscript{5})~\cite{Edinger_1998}\\
	    & superconductivity
	        & He 
	            & planar and 3D superconducting wires from W(CO)\textsubscript{6}~\cite{Cordoba_2018, Cordoba_2019a, Cordoba_2020, Orus_2021a} \\
	       & & Ga 
         & planar nanowires from W(CO)\textsubscript{6}~\cite{Sadki_2004, Sadki_2005, Spoddig_2007, Luxmoore_2007, Guillamon_2008, Li_2008,Sun_2013,Cordoba_2013, CordobaCastillo_2014, Cordoba_2019b, Orus_2021} and Nb(NMe\textsubscript{2})\textsubscript{3}(N-\textit{t}-Bu)\cite{Porrati_2019,Dobrovolskiy_2020}, arrays of Josephson junctions from Nb(NMe\textsubscript{2})\textsubscript{3}(N-\textit{t}-Bu)~\cite{Porrati_2022}, 3D \acs*{squid} from W(CO)\textsubscript{6}~\cite{Romans_2010}, nano\acs*{squid} from W(CO)\textsubscript{6}~\cite{Sigloch_2022}\\

	 \midrule
	 Magnetic 
	    & nanomagnets 
	     & He
	            &  nanowires from Co\textsubscript{2}(CO)\textsubscript{8}\cite{Wu_2014}\\
          &  & Ga
	  & nanoparticles from Co\textsubscript{2}(CO)\textsubscript{8}~\cite{Lapicki_2002}, nanowires from Co(CO)\textsubscript{3}NO~\cite{Gazzadi_2011} and  Co\textsubscript{2}(CO)\textsubscript{8}~\cite{SanzMartin_2019}\\
        & nanomagnetic probes 
            & Ga 
	    & filling FIB milled trenches with W~\cite{Khizroev_2002}, \ac{mfm} tips via deposition of thin layer by milling Co\textsubscript{71}Cr\textsubscript{17}Pt\textsubscript{12} target~\cite{Hyun_2006a}, Hall-probes from Co\textsubscript{2}(CO)\textsubscript{8}~\cite{Gabureac_2010a}\\
	 \midrule
	 Optical
	  & photonic components 
		    & Ga
        & deposition of defects in etch mask for photonic crystal~\cite{Vogelaar_2001}, deposition of the central Pt beamstop in Fresnel zone plates~\cite{Keskinbora_2013}, chiral nano-antennae i.e.\ helices from MeCpPtMe\textsubscript{3}~\cite{Esposito_2014a,Manoccio_2020,Esposito_2015} and from phenanthrene~\cite{Esposito_2016a}\\
		  & plasmonic components 
	         & Ga 
		 & chiral nano-antennae i.e.\ helices from phenanthrene covered with Au~\cite{Wang_2019a, Wang_2019c,Wang_2021a}\\
		 & chiral biosensing 
		    & Ga 
		        & helix arrays from MeCpPtMe\textsubscript{3} coated with poly o-phenylediamine~\cite{Manoccio_2021} 
		        \\
	 \midrule
	 Mechanical 
	    & \ac{afm} tips	        
	        & He 
	            & hammerhead probes from MeCpPtMe\textsubscript{3}~\cite{Nanda2015}\\
	   & tunable elastic moduli 
	        & Ga 
	            & DLC (diamond-like carbon) core-shell pillars and springs from phenanthrene~\cite{Fujita_2007,Nakamatsu_2009},
	        metal containing pillars and springs from phenanthrene and ferrocene~\cite{Nakai_2010}, MeCpPtMe\textsubscript{3}~\cite{Reiser_2020}, W(CO)\textsubscript{6}~\cite{Cordoba_2017}, TMCTS ((HSiCH\textsubscript{3}O)\textsubscript{4}) and  O\textsubscript{2}~\cite{Reyntjens_2000}\\
	 \bottomrule
      \end{tabular}
   \end{center}
\end{table*}

\subsubsection{Basics of ion induced chemistry\label{subs_fibid_chem}}

The ion-molecule reactions that have been studied in the context of \ac{fibid} involve decomposition of the precursor on the substrate surface during ion bombardment.  
In \ac{fibid} of metal nanostructures from organometallic precursors, the deposited material is composed of the metal(s) and contamination from nonvolatile ligand-derived decomposition byproducts.  
Ion induced chemistry during \ac{fibid} is only partially understood, but has been postulated to involve ion-molecule collisions with the precursor, reactions with low energy \acp{se} produced by impact of ions with the substrate, and thermal processes driven by dissipation of the collision energy (see also Figure~\ref{fig:fibid})~\cite{Utke_2008,Yu_2021}.  
An additional feature of \ac{fibid} is the concomitant sputtering of material from the surface that competes with deposition.  
Although preferential sputtering of ligand-derived atoms can purify the material under appropriate conditions~\cite{Thorman_2020a, bilgilisoy_surface_2020}, etching of the metal will also occur and the choice of reaction conditions is critical to obtaining deposits with high metal content. 

Broad application of \ac{fibid} will require development of custom precursors that undergo well-understood and efficient ion induced decomposition to metal atoms and volatile byproducts.
In addition to the standard physical properties common to all deposition methods that involve gas phase delivery of reactants to the substrate (volatility, thermal stability, ready availability, etc.), mechanism-based design~\cite{Yu_2021, Carden_2018} of \ac{fibid} precursors will require a fundamental understanding of ion-molecule reactions on the surface.
Much of the current understanding of these processes comes from model studies of \ac{fibid} carried out by ion bombardment of a few monolayers of precursor molecules on a substrate in \ac{uhv}~\cite{Yu_2021,Carden_2018}.
Although there are differences between \ac{fibid} and its \ac{uhv} models (the \ac{uhv} environment is cleaner than standard HV \ac{fibid} deposition conditions, \ac{uhv} studies are carried out at lower temperatures, and the ion source is more diffuse), \ac{uhv} studies can be used to identify ligands that dissociate from the complex and ultimately desorb cleanly from the surface during ion-molecule reactions.
Purification of deposits by sputtering away ligand-derived contaminants can also be modeled in \ac{uhv}.
Additional mechanistic information has been obtained in gas phase studies of ion-molecule collisions between organometallic precursors and ions of different masses and energies~\cite{Indrajith_2019}.

As an example, a study of the reactions of Ru(CO)\textsubscript{4}I\textsubscript{2} with \qty{860}{eV} Ar\textsuperscript{+} ions on a surface in \ac{uhv} demonstrated that the initial reaction was loss of all four CO ligands to form adsorbed RuI\textsubscript{2}~\cite{bilgilisoy_surface_2020}.
Additional ion exposure resulted in sputtering of iodine from the surface at a rate five times faster than sputtering of Ru.
These results suggest that proper optimization of \ac{fibid} conditions could result in deposition of high metal content Ru from this precursor.
This study also highlights the difference in the charged-particle induced reactivity between \ac{fibid} and \ac{febid}, as modeling of the \ac{febid} process in \ac{uhv} with \qty{500}{eV} electrons results in loss of the CO ligands, but not the iodide, to afford RuI\textsubscript{2} as the final product~\cite{Thorman_2020}.
Rapid loss of all of the carbonyl ligands was also observed upon ion bombardment (\qty{860}{eV}) in a \ac{uhv} model study of the heterobimetallic precursor ($\eta^5$-C\textsubscript{5}H\textsubscript{5})Fe(CO)\textsubscript{2}Re(CO)\textsubscript{5}~\cite{Thorman_2020a}. 
In contrast to the electron-induced chemistry of this complex under the same conditions, the ion beam slowly removed the remaining carbon from the cyclopentadienyl ligand by sputtering, leaving behind non-stoichiometric Fe/Re material.

Additional information on ion-molecule reactions was obtained during studies of \ac{fibid} of Au from (Me)\textsubscript{2}Au(hfac) (hfac = hexafluoroacetylacetonate) with He\textsuperscript{+}, Ne\textsuperscript{+}, Ar\textsuperscript{+}, Kr\textsuperscript{+}, or Xe\textsuperscript{+} at energies of \SIrange{2}{100}{keV}~\cite{dubner_role_1991, Ro_1994}.  
The decomposition yield from the precursor increased with increasing ion mass and energy, consistent with a \ac{bca} model for decomposition.  
The authors concluded that the \ac{fibid} process was driven by energy deposition from ion-surface collisions.  
These results have implications for ion source choice in \ac{fibid}, because the variation in ion mass impacts the deposition chemistry through differences in \ac{se} yields, sputter yields and ion implantation yields.
Ar\textsuperscript{+} and Ne\textsuperscript{+} generate sufficient \acp{se} at low beam energies (\qty{30}{eV}) but have a shallower beam penetration depth than He\textsuperscript{+}, leading to lower ion-substrate interactions.  
Heavier ions have higher sputtering yields, which can remove light atom impurities but at the cost of deposition rates, as the target material will also be sputtered away~\cite{Joy_2013,Hlawacek_2014}.

Gas phase studies of reactions of Fe(CO)\textsubscript{5} with He, Ne, Ar and Kr ions of varying charge and energy~\cite{Indrajith_2019} also demonstrated facile loss of CO ligands upon ion-molecule collision.  
The distribution of fragment ions from CO loss Fe(CO)\textsubscript{5-x}, where x=\numrange{1}{5}, was dependent on the ion mass and the energy transfer in the collision.  
Under these conditions, light ions such as He$^+$ resulted in electronic excitation with low fragmentation efficiency.  
Heavier ions yielded more extensive CO dissociation from energy transfer and nuclear stopping.  
The highest extent of CO loss occurred upon collision with Ne$^+$, where both electron transfer and energy transfer were significant.  
It was pointed out that the gas phase results may not accurately predict behavior on a surface during \ac{fibid} due to the heat sink behavior of the substrate.  
However, more efficient fragmentation upon impact with heavier ions is to be expected.
Although there is not yet an extensive set of privileged ligands for \ac{fibid}, some preliminary conclusions about ion-molecule chemistry of organometallic precursors on substrate surfaces can be drawn from the mechanistic studies described above~\cite{Yu_2021}. Namely:
\begin{enumerate}
  \item To minimize ligand-derived contamination in deposits, ligands should be small and few in number 
  \item CO ligands can be removed quite easily by ions
  \item Halide ligands can be removed by sputtering, but more slowly than CO\@;
  \item Carbon derived from anionic $\pi$-facial ligands such as cyclopentadienyl and allyl can be removed by sputtering, but this is also slow and competes with removal of metal.
\end{enumerate}
  
\subsubsection{\acs*{fibid} nanostructures for electronics\label{subs_fibid_elec}}

The ease to directly write functional structures with different conductivities renders \Ac{fibid} a powerful tool in electronic applications, with
at least three different electrical conductivity regimes reported: semiconducting, metallic, and superconducting at low temperature (see Table~\ref{tab:precursors}).  
\begin{table*}[btp]
\caption[Common \acs*{fibid} precursors]{Overview of various precursor ion combinations and the resulting electrical properties. Room temperature resistivity values $\rho_{RT}$ are given for the respective deposits and for the pure bulk metal from literature.}%
\label{tab:precursors}
\begin{tabular}{@{}llllccllll@{}}
\toprule
Material &
  \begin{tabular}[t]{@{}l@{}}Precursor\\ compound\end{tabular} &
  Ion &
  Shape &
  \begin{tabular}[t]{@{}c@{}}Composition\\ (\si{at. \%})  \end{tabular} &
  Microstructure &
  \begin{tabular}[t]{@{}l@{}}Electrical\\ behavior\end{tabular}&
  \begin{tabular}[t]{@{}c@{}}$\rho_{RT}$\\ (\si{\micro\ohm\centi\meter})  \end{tabular} &
  \begin{tabular}[t]{@{}c@{}}$\rho_{RT}$\\ bulk \\ (\si{\micro\ohm\centi\meter})\end{tabular} &
  Ref.\\ \midrule
Platinum 
  & MeCpPtMe$_3$ &
  Ga$^+$ &
  planar &
  \begin{tabular}[c]{@{}c@{}}Pt:C:Ga:O\\ 46:24:28:2\vspace{3pt}\end{tabular} &
  amorphous &
  metal &
  70--700 &
    10.4 &
   \citenum{Tao_1990}
   \\
  & MeCpPtMe$_3$ &
  Ga$^+$ &
  planar &
  \begin{tabular}[c]{@{}c@{}}Pt:C:Ga:O\\ 22:60:11:7\vspace{3pt}\end{tabular} &
  \begin{tabular}[c]{@{}c@{}}\SIrange[range-units = single,range-phrase = --]{5}{10}{nm} \\ nanocrystals\end{tabular} &
  metal &
  700 &
 & 
  \citenum{FernandezPacheco_2009}
   \\

  & MeCpPtMe$_3$ &
  Ga$^+$ &
  planar &
  \begin{tabular}[c]{@{}c@{}}Pt:C:Ga\\ 17:72:11\vspace{3pt}\end{tabular} &
  \begin{tabular}[c]{@{}c@{}}\qty{\approx3}{nm}\\ nanocrystals\end{tabular} &
  metal &
  800 &
   &
   \citenum{DeTeresa_2009}
   \\
 & MeCpPtMe$_3$ &
  Ne$^+$ &
  planar &
  \begin{tabular}[c]{@{}c@{}}Pt:C\\ 17:83\vspace{3pt}\end{tabular} &
   \begin{tabular}[c]{@{}c@{}}\qty{\approx4.5}{nm} \\ nanocrystals\end{tabular}&
  metal &
  600 &
   &
   \citenum{Wu_2013}
   \\
  & MeCpPtMe$_3$ &
  He$^+$ &
  planar &
  \begin{tabular}[c]{@{}c@{}}Pt:C\\ 16:84\vspace{3pt}\end{tabular} &
  \begin{tabular}[c]{@{}c@{}}\qty{\approx3}{nm} \\ nanocrystals\end{tabular} &
  semiconductor &
  180000 &
   &
   \citenum{Wu_2013}
   \\ \midrule
Gold &
  Au(hfac)Me\textsubscript{2} &
  Ga$^+$ &
  planar &
  \begin{tabular}[c]{@{}c@{}}Au:C:Ga\\ 50:35:15\vspace{3pt}\end{tabular} &
  --- &
  metal &
  500--1500 &
  2.4 &
  \citenum{Blauner_1989}
   \\
&
  Au(hfac)Me\textsubscript{2} &
  Ga$^+$ &
  planar &
  \begin{tabular}[c]{@{}c@{}}Au:C:Ga:O\\ 75:\textless{}5:20:\textless{}5\vspace{3pt}\end{tabular} &
  --- &
  metal &
  500--1300 &
   &
   \citenum{Shedd_1986}
   \\ \midrule
Cobalt &
  Co$_2$(CO)$_8$ &
  Ga$^+$ &
  planar &
  --- &
  --- &
  metal &
  189 &
  6 &
  \citenum{Lapicki_2002}
   \\
  &Co$_2$(CO)$_8$ &
  Ga$^+$ &
  planar &
  --- &
  amorphous &
  metal &
  19--38 &
   &
   \citenum{SanzMartin_2019}
   \\
  &Co(CO)$_3$NO &
  Ga$^+$ &
  planar &
  \begin{tabular}[c]{@{}c@{}}Co:C:Ga:N:O\\ 54:7:9:13:17\vspace{3pt}\end{tabular} &
   \begin{tabular}[c]{@{}c@{}}\qty{>20}{nm}\\ nanocrystals\end{tabular} &
  metal &
  189 &
   &
   \citenum{Gazzadi_2011}
   \\
  &Co$_2$(CO)$_8$ &
  He$^+$ &
  planar &
  --- &
  \begin{tabular}[c]{@{}c@{}}\qty{6}{nm} \\ nanocrystals\end{tabular} &
  metal &
  64--116 &
   &
   \citenum{Wu_2014}
   \\ \midrule
Copper &
Cu(hfac)(VTMS) &
  Ga$^+$ &
  planar &
  \begin{tabular}[c]{@{}c@{}}Cu:C\\ 55:45\vspace{3pt}\end{tabular} &
  \begin{tabular}[c]{@{}c@{}}\qty{20}{nm}\\ nanocrystals\end{tabular} &
  metal &
  50 &
  1.7 &
  \citenum{DellaRatta_1993}
   \\ \midrule
Tungsten &
  W(CO)$_6$ &
  Ga$^+$ &
  planar &
  \begin{tabular}[c]{@{}c@{}}W:C:Ga\\ 40:40:20\vspace{3pt}\end{tabular} &
  amorphous &
  \begin{tabular}[c]{@{}c@{}}superconductor\\ below \qty{5.2}{K}\end{tabular} &
  200 &
  5.6 &
  \citenum{Sadki_2004,Sadki_2005}
   \\
  & W(CO)$_6$ &
  Ga$^+$ &
  planar &
  \begin{tabular}[c]{@{}c@{}}W:C:Ga:O\\ 40:43:10:7\vspace{3pt}\end{tabular} &
  \begin{tabular}[c]{@{}c@{}}\SIrange[range-units = single,range-phrase= -- ]{1}{2}{nm}\\ nanocrystals\end{tabular} &
  \begin{tabular}[c]{@{}c@{}}superconductor\\ below \SIrange[range-units = single,range-phrase= --]{4.2}{5.1}{K}\end{tabular} &
  275 &
   &
   \citenum{CordobaCastillo_2014}
   \\
  & W(CO)$_6$ &
  Ga$^+$ &
  planar &
  \begin{tabular}[c]{@{}c@{}}W:C:Ga:O\\ 53:34:11:2\vspace{3pt}\end{tabular} &
  \begin{tabular}[c]{@{}c@{}}\qty{\approx1}{nm} \\ nanocrystals\end{tabular} &
  \begin{tabular}[c]{@{}c@{}}superconductor\\ below \SIrange[range-units = single,range-phrase= --]{5}{5.5}{K}\end{tabular} &
  100--350 &
   &
   \citenum{Li_2008}
   \\
  & W(CO)$_6$ &
  He$^+$ &
  3D &
  \begin{tabular}[c]{@{}c@{}}W:C:O\\ 72:20:8\vspace{3pt}\end{tabular} &
  \begin{tabular}[c]{@{}c@{}}\SIrange[range-units = single,range-phrase= -- ]{20}{30}{nm} \\ nanocrystals\end{tabular} &
  \begin{tabular}[c]{@{}c@{}}superconductor\\ below \SIrange[range-units = single,range-phrase= --]{6.2}{7.1}{K}\end{tabular} &
  398 &
   &
   \citenum{Cordoba_2018, Cordoba_2019a, Cordoba_2020}
   \\ \midrule
Niobium &
  Nb(NMe$_2$)$_3$(N-t-Bu) &
  Ga$^+$ &
  planar &
  \begin{tabular}[c]{@{}c@{}}Nb:C:Ga:N\\ 29:43:15:13\vspace{3pt}\end{tabular} &
  \begin{tabular}[c]{@{}c@{}}1\SIrange[range-units = single,range-phrase= --]{5}{20}{nm} \\ nanocrystals\end{tabular} &
  \begin{tabular}[c]{@{}c@{}}superconductor\\ below \qty{5.0}{K}\end{tabular} &
  550 &
  15 &
  \citenum{Porrati_2019}
   \\
  & Nb(NMe$_2$)$_3$(N-t-Bu) &
  Ga$^+$ &
  3D &
  --- &
  \begin{tabular}[c]{@{}c@{}}\SIrange[range-units = single,range-phrase= --]{15}{20}{nm} \\ nanocrystals\end{tabular} &
  \begin{tabular}[c]{@{}c@{}}superconductor\\ below \qty{8.1}K\end{tabular} &
  380 &
   &
   \citenum{Porrati_2019}
   \\
   \bottomrule 
\end{tabular}
\end{table*}

Key parameters here are the carbon content of the deposit as well as its microstructure, which depend on the growth conditions (as well as on the precursor selected). Modulating the growth conditions thus offers an efficient way to tune electrical and electronic behavior.
Examples include Pt~\cite{FernandezPacheco_2009} and Au~\cite{Blauner_1989,Shedd_1986} deposition using Ga-\ac{fibid}, for which the conductivity increases with the thickness of the as-grown material.
This dependence is the result of a gradient in metal content through the film thickness, with a higher carbon content in the first deposited layers versus at the bottom.
Pt-based deposits below \qty{50}{\nano\meter} in thickness behave as semiconductors, but display metallic behavior~\cite{Tao_1990} for thicknesses above \qty{150}{\nano\meter}~\cite{FernandezPacheco_2009}. The metal content of these deposits is less than \qty{33}{at.\%}, consisting of fcc Pt nanocrystals with a diameter of \SIrange{5}{10}{\nano\meter} embedded in an amorphous carbon matrix~\cite{DeTeresa_2009}.
The observed conductivity regimes are expected to be strongly influenced by the disorder and doping introduced by Ga ion impact~\cite{DeTeresa_2009}.
In the case of Cu deposits (copper nanocrystals embedded in amorphous carbon), it has been found that the ion beam current also directly controls the metal content and thus the deposit conductivity~\cite{DellaRatta_1993}.
Futhermore, the ion species used for the deposition can have a great effect. For example, using Ne-\ac{fibid}, Pt deposits exhibit metallic behavior, whereas they behave as semiconductors using He-\ac{fibid}~\cite{Wu_2013}.
In the case of Ga-\ac{fibid} of Au-based deposits, 
metallic behavior is easily obtained, with significant Au contents of \qty{75}{at.\%}  and resisitivities exceeding the bulk resistivity of Au by two orders of magnitude~\cite{Shedd_1986}.

A number of works have investigated \ac{fibid} of Co deposits due to their magnetic properties, for example, as-grown~\cite{Lapicki_2002, Lapicki_2002a} and annealed~\cite{Gazzadi_2011} Ga-\ac{fibid} Co deposits (containing \qty{50}{at.\%} Co) exhibit ferromagnetic characteristics and resistivity values at room temperature that exceed the resistivity of bulk Co by several orders of magnitude.
Using He-\ac{fibid}, nanostructures can be fabricated with higher lateral resolution, such as metallic Co lines of \qty{10}{\nano\meter} in width displaying room temperature resistivities comparable to those grown using Ga-\ac{fibid}~\cite{Wu_2014}.

Intriguing \ac{fibid}-grown materials that exhibit metallic behavior at room temperature and transition into the superconducting state at low temperature include W and Nb based deposits, which become superconducting below \SIrange{4.2}{5.1}{\kelvin}~\cite{Sadki_2004, Sadki_2005, CordobaCastillo_2014, Li_2008} and \qty{5}{\kelvin}~\cite{Porrati_2019}, respectively.
In the case of freestanding 3D NbC nanowires, a broadened superconducting transition between \SIrange[range-phrase =\ and\ ]{4}{11}{\kelvin} has been reported~\cite{Porrati_2019}.

The WC nanostructures, in particular, exhibit remarkable superconducting properties, such as upper critical magnetic field values of B\textsubscript{c2}(2K)=\SIrange[range-phrase=--]{7}{8.5}{\tesla}~\cite{CordobaCastillo_2014, Sun_2013} and critical current density values of J\textsubscript{c}(2K)=\SIrange[range-phrase=--]{0.01}{0.1}{\mega\ampere\per\centi\meter\squared}~\cite{Sadki_2004, CordobaCastillo_2014, Luxmoore_2007, Spoddig_2007, Guillamon_2008}. 
For very narrow nanostructures (\qty{\leq50}{\nano\meter} in width), finite size effects that are key to their application in superconducting circuits, have been studied. This includes phenomena such as a magnetic field-driven re-entrance of the superconductivity~\cite{Cordoba_2013}, the occurrence of non-local voltages~\cite{Cordoba_2019b, Orus_2021}, and electric field-induced control of superconductivity~\cite{Orus_2021a}.

\begin{figure}[tbp]
   \centering
   \includegraphics[width=\linewidth]{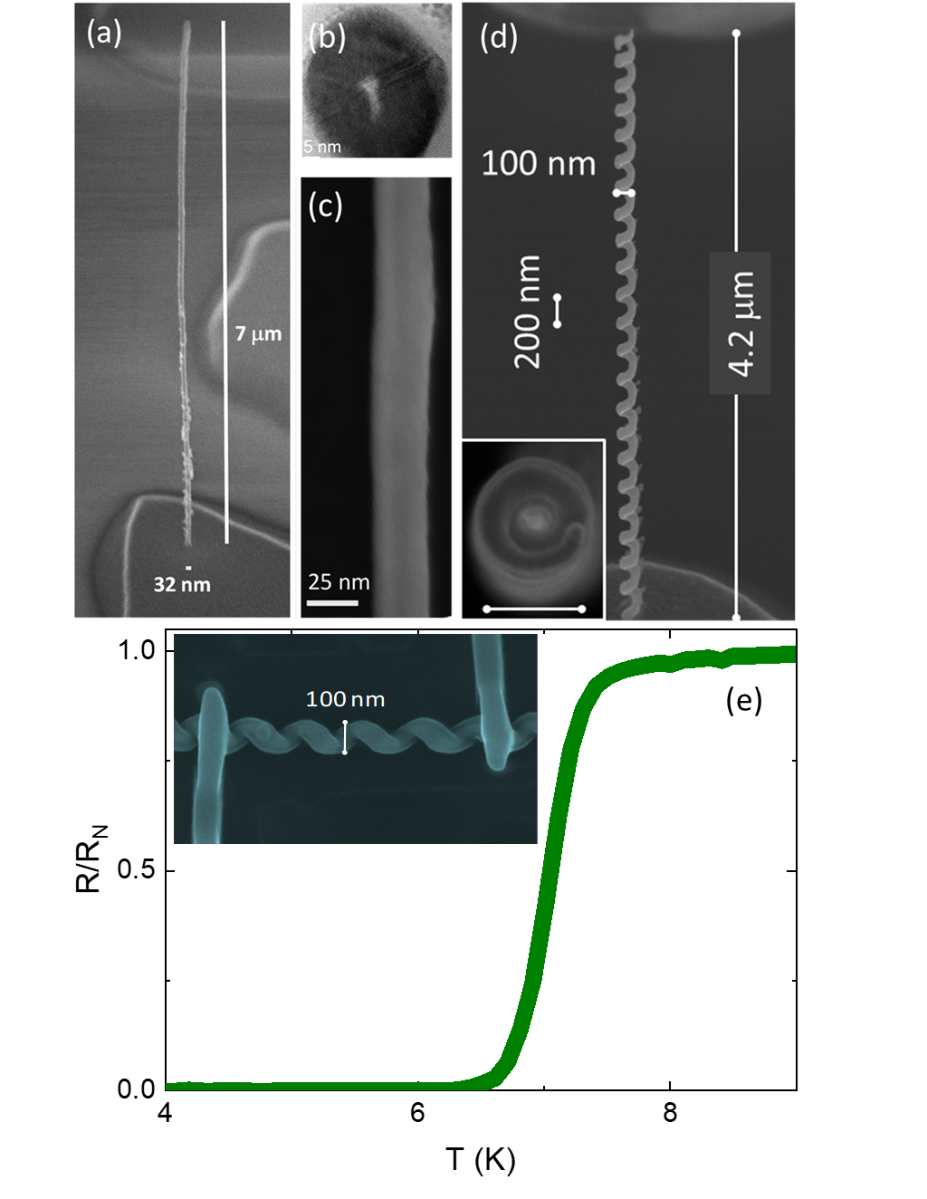} 
   \caption{3D \Ac{fibid} examples.
    (a)	\ac{sem} image of a hollow crystalline WC superconducting nanopillar grown by He-\ac{fibid}.
    (b)	Cross-sectional \ac{tem} view of the hollow nanowire.
    (c)	Higher magnification \ac{sem} image of (a).
    Adapted with permission from \citeauthor{Cordoba_2018}~\cite{Cordoba_2018} \copyright{}\citeyear{Cordoba_2018} American Chemical Society.
    (d) Superconducting WC nano-helix grown by He-\ac{fibid}. Inset shows its top view. 
    (e)	Superconducting transition of the nano-helix. Inset shows the nano-helix contacted for the electrical transport measurements. 
  Adapted with permission from \citet{Cordoba_2019a} \copyright{}\citeyear{Cordoba_2019a} American Chemical Society.}%
        \label{fig:FIBID_2}
\end{figure}

Furthermore, He-\ac{fibid} has facilitated the fabrication of sophisticated 3D WC nanosuperconductors with additional outstanding characteristics. These include a critical temperature ranging from T\textsubscript{c}=\SIrange[range-phrase=--]{6.2}{7.1}{\kelvin} and an upper critical magnetic field value extending from B\textsubscript{c2}(0)=\SIrange[range-phrase=--]{12}{15}{\tesla}. Noteworthy among these structures are hollow nanowires with an outer diameter of \qty{32}{\nano\meter} and aspect ratios of about \num{\approx200}~\cite{Cordoba_2018, Cordoba_2020}, as well as nano-helices with a diameter of \qty{100}{\nano\meter} and an aspect ratio of about \num{\approx65}~\cite{Cordoba_2019a} (see Figure~\ref{fig:FIBID_2}). 
These nanostructures exhibit a metallic content of \qty{70}{at.\%} W and consist of \SIrange{20}{30}{\nano\meter}-sized fcc WC\textsubscript{1$-$x } nanocrystals.

\subsubsection{\acs*{fibid} nanostructures for mechanics\label{subs_fibid_mech}} 

One of the main advantages of \ac{fibid} over standard lithographic techniques is the capability to directly write high aspect ratio structures onto any substrate topography. 
When it comes to mechanical applications, this allows e.g.\ placement of dedicated tips onto \ac{afm} cantilevers and spring structures into prefabricated mechanical devices.
A recent review by \citet{Utke_2020} summarizes the mechanical properties of \ac{fibid} pillars and springs. All structures were deposited using Ga-\ac{fibid} (ion energies mainly \SIrange{10}{30}{\kilo\electronvolt}, beam currents \SIrange{0.5}{9}{\pico\ampere}).
An interesting finding was that the Ga-\ac{fibid} pillars tended to grow as core-shell structures.
Using the precursor phenanthrene (C\textsubscript{14}H\textsubscript{10}), elastic moduli of the hydrogenated carbonaceous material in the core and shell were \qty{\approx300}{\giga\pascal} (stiff core with Ga implantation, low H content) and \qty{\approx30}{\giga\pascal} (soft shell, presumably high H content)~\cite{Fujita_2007,Nakamatsu_2009}.
Depending on the inner diameter of the core and the thickness of the shell, the elastic modulus of the structure ranges between these boundary values.
Metallic pillars with metal contents \qty{<40}{at.\%} (i.e.\ below the percolation threshold) have mechanical properties dominated by the co-deposited carbonaceous matrix~\cite{Utke_2020}.
Increasing metal contents yield elastic moduli as follows (metal at.\%~/~elastic modulus normalized to that of pure metal): Fe (21 {at.\%} / 0.4)~\cite{Nakai_2010}, Pt (60 {at.\%} / 0.8)~\cite{Reiser_2020}, W (80 {at.\%} / 0.4)~\cite{Cordoba_2017}, SiO\textsubscript{2} (95 {at.\%} / 0.8)~\cite{Reyntjens_2000}.
The non-correlation between the metal content and elastic modulus of the pillars may be due to Ga implantation, which softens the metals. 
Furthermore, the carbonaceous matrix stiffens or softens the pillar depending on its sp\textsuperscript{2} vs.\ sp\textsuperscript{3} hybridization ratio and its hydrogen content. 

So far, \Ac{fibid} is typically not used to fabricate high aspect ratio scanning probe tips, in contrast to \ac{febid}; see review by \citet{Plank_2019}.
This can be attributed to the strong tendency of the Ga ions to mill the cantilever substrate rather than induce tip (pillar) growth. Implantation of Ga into the cantilever can also be a problem. Here, the enhancement of deposition yields using pulsed irradiation can bring benefit~\cite{Utke_2008}.
Since the sputter yield of He ions is much lower, He-\ac{fibid} also shows promise in this area, and has been used for the fabrication of 3D hammerhead \ac{afm} probes~\cite{Nanda2015}.
In the case of strain sensors based on piezoresistive granular material, \ac{febid} is also still the method of choice~\cite{Dukic_2016}. 
However, here too changing to He or another light ion species for \ac{fibid} could prove fruitful~\cite{Shorubalko_2016}.

\subsubsection{\acs*{fibid} nanostructures for photonics and plasmonics\label{subs_fibid_photo}}

\Ac{fibid} is also a powerful nanofabrication tool for the exploration of novel optical properties thanks to its unique design flexibility~\cite{Li_2021, Matsui_2007, DeRidder_2007}.
The site specific and local nature of \ac{fibid} allows the growth of photonic structures on different material substrates~\cite{Peinado_2015}, providing a path to the fabrication of a number of optical elements, such as chiral optical antennae~\cite{Esposito_2014a, Wang_2019c}, Fresnel lenses~\cite{Keskinbora_2013}, and nanosensors~\cite{Manoccio_2021}. 
Combined with large area laser interference lithography to fabricate a large area photonic crystal,  \ac{fibid} has been used to define defects to form a multimode waveguide~\cite{Vogelaar_2001}.

\begin{figure*}[tbp]
   \centering
   \includegraphics[width=.95\linewidth]{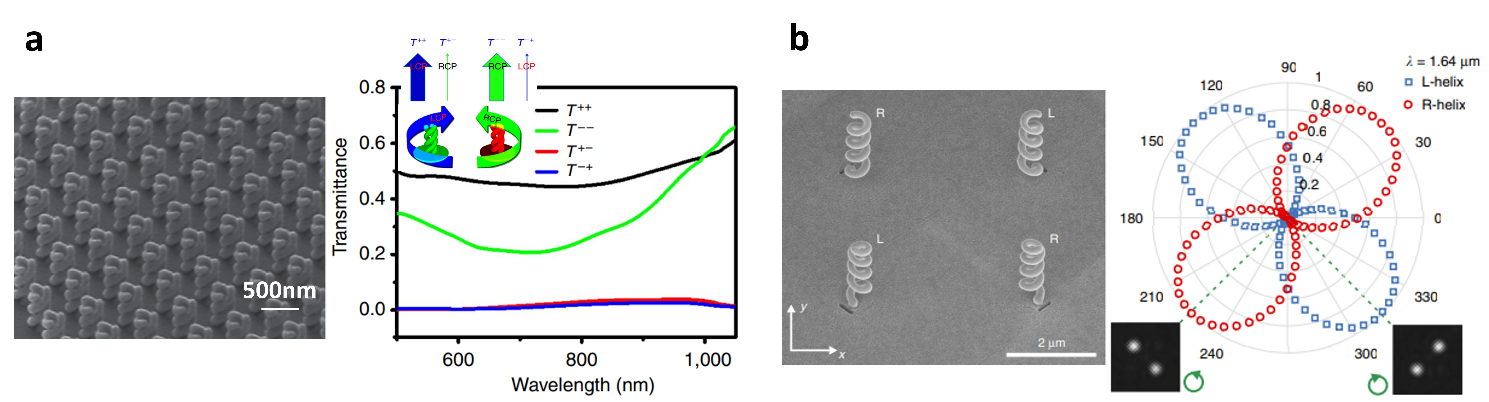}
   \caption{Examples of photonic nanostructures realized using \ac{fibid} and their related optical properties.
      (a) \ac{sem} image of an array of intertwined nano-helices and their transmission spectra T++, T--, T+-, T-+ for right and left circularly polarized incident light (RCP and LCP, respectively).
      The inset in the plot presents a scheme of the handedness of incident and transmitted circular polarization.
      Adapted from~\citet{Esposito_2015}.
      (b) \ac{sem} image of subwavelength optical device made of helices with opposite handedness (L and R) coupled to orthogonal apertures.
      This system enables conversion of an incoming light beam into four beams of tunable polarization and intensity.
      The plot shows helicity analysis of the device, recording the transmitted power after placing a rotating quarter-wave plate and a fixed polarizer in front of the detector.
      The insets show the far-field optical images of the device when the quarter-wave plate rotates by \qty{\pm45}{degrees}.
   Adapted from~\citet{Wang_2019c}}%
   \label{fig:FIBID_3}
\end{figure*}

Under optimized parameter conditions, it is possible to evolve planar structures along the third dimension, thus manufacturing free-standing nanostructures with more exotic optical properties.
For example, 3D chiral nano-helices, prototyped by \ac{fibid}, have shown tunable chiroptical response at optical frequencies~\cite{Manoccio_2020, Esposito_2015}, (Figure~\ref{fig:FIBID_3}(a)), been used as subwavelength circular polarization optics (Figure~\ref{fig:FIBID_3}(b))~\cite{Wang_2019c}, exploited in compact and miniaturized high-sensitivity biosensors~\cite{Manoccio_2021}, and integrated onto the apex of background-free and broadband chiroptical probes~\cite{Wang_2019a}. 

The wide range of available gas precursors enables the deposition of metal and dielectric materials that can be applied in the emerging field of photonic resonant nanostructures~\cite{Utke_2008}.
Pure and homogeneous compositions are, however, intrinsically hindered due to various aspects of the deposition process, such as incomplete dissociation of the precursor molecules and the residual contamination from hydrocarbons present in the vacuum chamber, both of which result in  carbon content in the nanostructure, and the inevitability of implanted ions, which cause absorption losses, and in some cases, material damage.
As a result, \ac{fibid} materials exhibit either a nanocrystalline composite structure~\cite{DeTeresa_2009, Tao_1990, Ro_1995} or a core/shell architecture, depending on the precursor properties and on the \ac{fib} irradiation conditions~\cite{Matsui_2012}.
Apart from purification approaches, which have mostly been used for \ac{febid} (and which might alter the nanostructure shape and dimensions)~\cite{PerezRoldan_2017, Shawrav_2016, Botman_2009, Hoeflich_2011}, or reduced implantation effects due to enhanced deposition yields (which require the employment of cryo-\ac{fibid}~\cite{Cordoba_2019b, DeTeresa_2019,Teresa_2020}), other promising strategies to control the final optical response are post-processing coating methods.
In the case of complex 3D shapes, conformal coverage can be obtained by combining \ac{fibid} with glancing angle deposition of metals~\cite{Wang_2019c, Wang_2019a}, cyclic voltammetry coating of polymers~\cite{Manoccio_2021} and atomic layer deposition of oxides~\cite{Cremers_2019}.
Also, the core-shell architecture (resulting from the scattering of Ga into a lower molecular weight matrix) can be controlled to engineer specific optical effects~\cite{Cuscuna_2021, Kometani_2010}.
Other approaches may emerge from the exploitation of other ion beam sources for \ac{fibid}, such as \ac{lmais}~\cite{Bischoff_2020,Hentschel_2020}.

\subsubsection{Outlook} 
Focused ion beams are a flexible and versatile tool for gas-assisted processing, but so far have been much less used than focused electron beams for this purpose. 
The reason for this may be the dominance of Ga-\ac{fib}, whereby Ga ions are not only implanted into the deposits but can also cause significant physical sputtering and trigger undesirable chemical and physical side effects.
In addition, there is still no precursor development specifically for \ac{fibid}. 
Deposition with ions can in fact deliver purer and denser materials due to their much higher mass compared to electrons. 
In this regard, ion beams from noble gases, which are chemically inert, are of particular interest. 
While several studies have already presented promising results for \Ac{fibid} using He and Ne ions (cf.\ Table~\ref{table:fibid}), \Ac{fibid} using beams of Ar, Kr or Xe ions from \acp{pfib} has not yet been reported and should be explored.

\subsection{Novel and unconventional approaches in \acs*{fib} processing}%
\label{subs_recent_adv}

Apart from removing or depositing material, introducing dopants or creating defects, the interaction between ions and solids can be used to change materials chemically or morphologically.
Such `non-standard' approaches include \ac{fib}-induced cross-linking or dissociation in different types of organic and metal-organic surface layers and also the triggering of self-organization processes.
Furthermore, `standard' \ac{fib} procedures can be combined with other nano- and microfabrication techniques to mitigate specific limitations and broaden applicability.

\subsubsection{Resist patterning by \acs*{fib}}

In resist based lithography the goal is to induce a chemical change in the exposed resist layer to render it either soluble (positive tone resist) or insoluble (negative tone resist) to the subsequent development step. Resist based nanopatterning techniques using photons (optical lithography) and electrons (\ac{ebl}) are common in industry, with the main properties being, respectively, high throughput and high resolution~\cite{DeTeresaNogueras_2020}. 
With the aim of improving the throughput of \ac{ebl} while maintaining a high resolution, pioneering studies demonstrated the use of Ga-\ac{fib} to expose resists~\cite{Seliger_1979, Atkinson_1992a,Hartney_1991, Stratton_1998}. 
Ions have the advantage over electrons that due to their mass they do not induce backscattering from the underlying substrate and therefore reduce unintended exposure of neighboring pixels (known as the proximity effect).
For example, line widths of less than \qty{100}{nm} and a sensitivity of \qty{8e11}{ions\per\centi\meter\squared} were found using a commercial negative resist and \qty{30}{kV} Ga-\ac{fib} irradiation~\cite{Hartney_1991}. 
Line widths of less than \qty{30}{nm} and a sensitivity of \qty{1e13}{ions\per\centi\meter\squared} were achieved using the positive \ac{pmma} resist and \qty{50}{kV} Ga-\ac{fib} irradiation~\cite{Atkinson_1992a}. 
Furthermore, using \qty{50}{kV} Ga-\ac{fib} irradiation of the negative poly(phenyl silsesquioxane) (PPSQ) resist, sub-\qty{100}{nm} tunneling tip structures were fabricated for a sensing device~\cite{Stratton_1998}.
Yet the drawbacks of Ga-\ac{fib} include sputtering of the resist and a relatively large beam spot size (several nanometers) with significant beam tails.
Compared to the significant developments of \ac{ebl} over the following years, these drawbacks hampered further use of \acp{fib} for resist exposure until the advent of the \ac{him}. 
\Ac{hibl} combines a smaller probe size with a low sputter yield, high resist sensitivity, and a reduced proximity effect~\cite{Kiran_2020}.

\begin{figure}[tbp]
   \centering
   \includegraphics[width=\columnwidth]{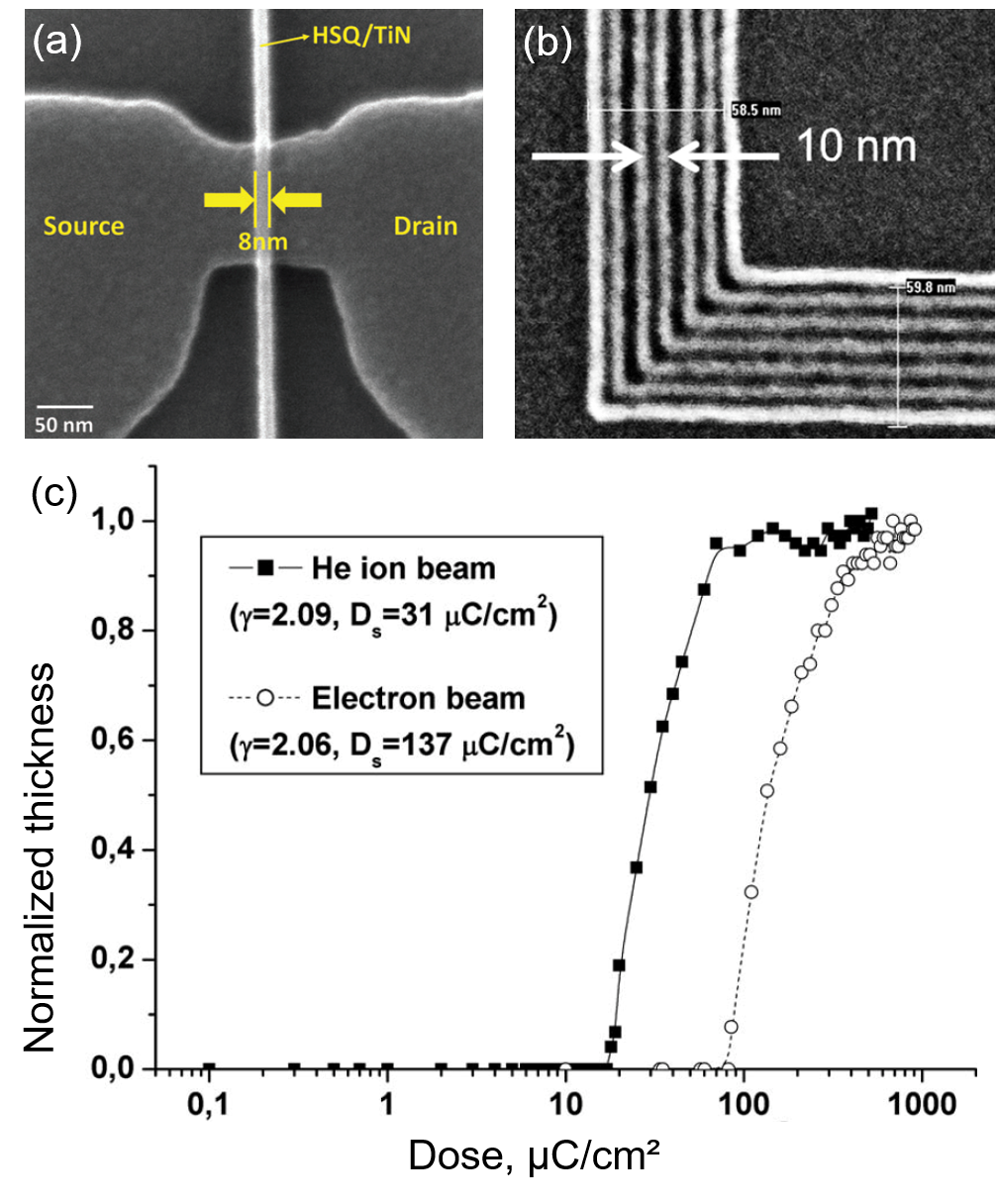}
   \caption{\label{fig:FigFIBHeLitho}(a) \qty{8}{nm}-wide \ac{hsq} lines patterned by \ac{hibl} and used as hard masks for subsequent etching of TiN gate. Reprinted from~\citet{Chang_2021a}. 
     (b) High-resolution dense nanopatterns in \ac{hsq} written using \ac{hibl}\@. Reprinted from~\citet{Li_2012}. 
   (c) Comparison of the charge dose required for \ac{hibl} vs.\ \ac{ebl} using \ac{hsq} resist. Reprinted from~\citet{Sidorkin_2009b}.}%
\end{figure}

First studies published in 2009 already highlighted the potential of \ac{hibl} for sub-10 nm patterning ~\cite{Winston_2009a,Sidorkin_2009b}. 
In both cases, \ac{hsq} negative resist was used.
An ion dose of only \qty{31}{\micro\coulomb\per\centi\meter\squared} was required for complete exposure~\cite{Sidorkin_2009b} (see Fig.~\ref{fig:FigFIBHeLitho}(c)). 
%
Compared to \ac{ebl} using the same resist, the dose required for \ac{hibl} is lower by a factor of four. 
\ac{hibl} can therefore simultaneously provide a similar resolution to \ac{ebl} with a better throughput. 
\ac{hibl} with \ac{hsq} has also been employed to fabricate a hard mask for subsequent metal etching~\cite{Chang_2021a} (Fig.~\ref{fig:FigFIBHeLitho}(a)) and as a mold for nanoimprint~\cite{Li_2012} (Fig.~\ref{fig:FigFIBHeLitho}(b)). 

The 3D nanofabrication of complex crosslinked \ac{hsq} nanostructures, such as embedded nanochannels and suspended grids, was demonstrated using  different exposure configurations~\cite{Cai_2018a}. 
Furthermore, due to the large depth of field of \ac{hibl}, large-area dense patterns have been achieved on \ac{hsq} as well as exposures on tilted surfaces~\cite{Flataboe_2018}. 
The use of the Ne-\ac{fib} for \ac{hsq} exposure has proven interesting, providing a trade-off between high resolution and dose efficiency~\cite{Winston_2011a}.

\ac{hibl} using other resists has also been investigated. For example, sub-\qty{10}{nm} resolution has been obtained on a fullerene based resist using a dose of \qty{40}{\micro\coulomb\per\centi\meter\squared}, which is three orders of magnitude lower than that of a similar process using \ac{ebl}~\cite{Shi_2016a}. 
Likewise, sub-\qty{10}{nm} resolution was achieved on an alumina-based resist, and \qty{10}{nm} lines (with an aspect ratio of 1:10) were then transferred to Si by means of low-bias reactive ion etching ~\cite{Cattoni_2018a}. 
\ac{hibl} using a new metalorganic negative resist, Cr$_8$F$_8$(pivalate)$_{16}$, has been shown to enable sub-\qty{10}{nm} resolution patterning using a line irradiation dose of \qty{20}{pC/cm} (three orders of magnitude lower compared to \ac{ebl}), demonstrated for Si and W substrates with good resistance to subsequent dry etching~\cite{Lewis_2019a}. 
New hybrid organic-inorganic resists based on Ni have also been demonstrated to perform well in \ac{hibl}, requiring an irradiation dose of \qty{22}{\micro\coulomb/cm^2}, producing sub-\qty{10}{nm} resolution with low line edge/width roughness~\cite{Kumar_2020a}. 
Notably, \ac{hibl} using a new resist composed of a radiation-sensitive monomer (MAPDST) with an organic functionality (TIPMA) only requires a very low exposure dose of \qty{6}{\micro\coulomb/cm^2}~\cite{Yogesh_2021a}. 
The exposure of PMMA to He and heavier ions has also been investigated~\cite{Zhang_2020a}.

Additionally, \ac{hibl} has shown good performance in combination with lift-off processes, in particular with Au, which has been used to fabricate plasmonic nanoantennae~\cite{Rashid_2021a}. 
A different approach consists of the direct He- and Ne-\ac{fib} milling of \ac{hsq} or PMMA resists that have been previously patterned using standard \ac{ebl}~\cite{Xia_2020a}. 
This can for instance be used for pattern repair. 
\ac{hibl} has also been proposed as a method for the (pre-) screening of \ac{euv} resists to help evaluate their performance~\cite{Maas_2014}.%

In general, it can be stated that \ac{hibl} is very promising for high-resolution resist based nanopatterning due to the low proximity effect and the lower ion dose required in comparison with \ac{ebl}. 
Thus broader use of \ac{hibl} in the coming years can be anticipated.
More detailed discussions on this topic can be found in previous review articles~\cite{Kalhor_2016,Kiran_2020,Allen2021}. 

\subsubsection{Cryo-\acs*{fibid}}

In the past, it was found that electron or ion irradiation of Sn-based precursors condensed on a substrate at \qty{120}{K} produced deposits with metallic behaviour~\cite{Funsten_1992a}. 
The topic was recently revived through the use of Ga-\ac{fib} to dissociate condensed W(CO)$_6$ precursor at \qty{-100}{\celsius}, giving rise to WC nanodeposits that exhibit metallic resistivity~\cite{Cordoba_2019b}. The particular advantage of the cryo method is the low ion dose required (600 times lower) compared to the standard Ga-\ac{fibid} process at room temperature. 
The new technique, coined cryo-\ac{fibid}, requires that the thickness of the condensed precursor layer is comparable to the ion penetration depth, so that the deposit does not lift off when heated back to room temperature~\cite{DeTeresa_2019,Teresa_2020}. 
Cryo-\ac{fibid} has also been carried out using the precursor MeCpPtMe\textsubscript{3}~\cite{SalvadorPorroche_2020}, although the resulting PtC deposits were less conductive than their WC counterparts.

\begin{figure}[tbp]
  \centering
  \includegraphics[width=\columnwidth]{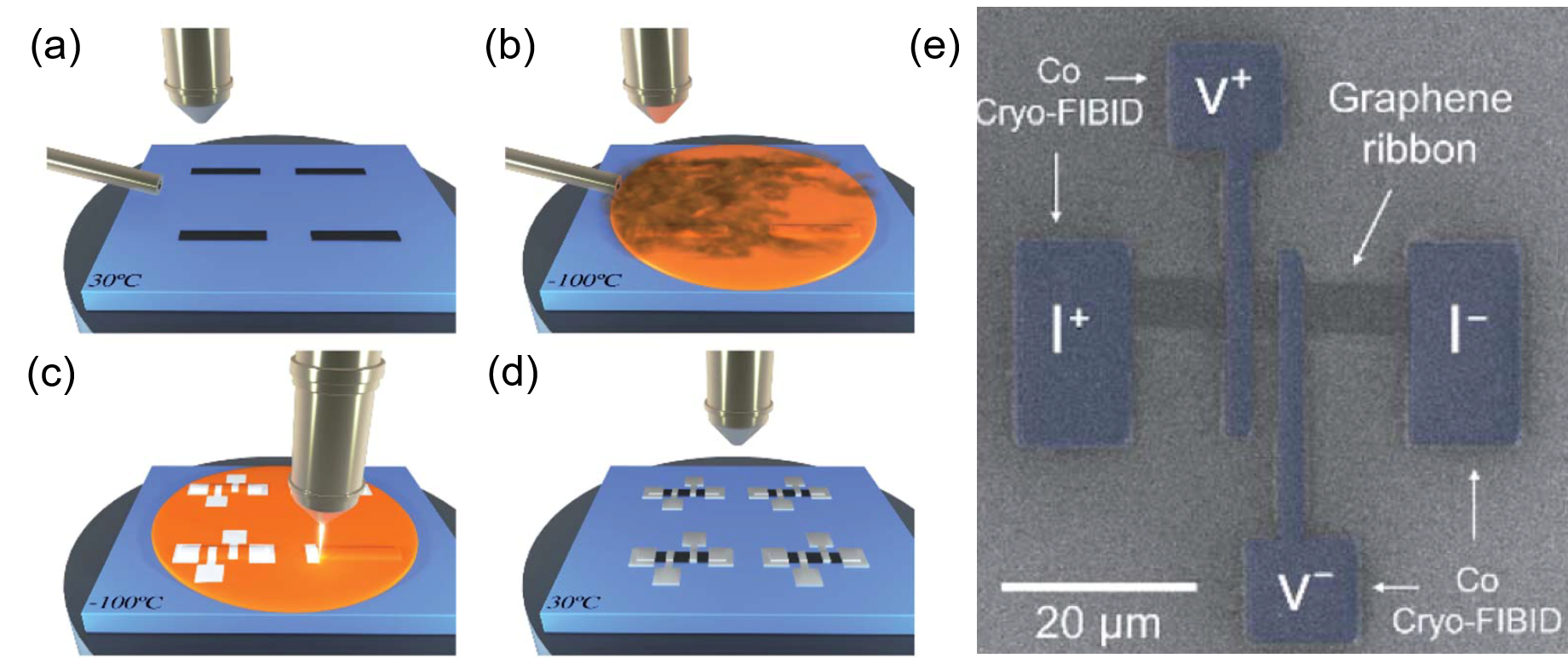}
  \caption{\label{fig:FigCryoFIBID}Cryo-\ac{fibid} process used to grow electrical contacts on 2D materials (e.g.\ graphene):
    (a) Graphene ribbons  on Si/SiO$_2$ substrate loaded into chamber. 
    (b) Substrate is cooled and precursor from \ac{gis} delivered  giving rise to a condensed precursor layer. 
    (c) \ac{fib} irradiation of the condensed layer using predefined pattern. 
    (d) Substrate heated up to \qty{30}{\celsius} thus sublimating the non-irradiated condensed layer. 
    (e) False color \ac{sem} image of a graphene ribbon with four electrical contacts grown by cryo-\ac{fibid} using Co precursor. 
  Reprinted from \citet{SalvadorPorroche_2021}.}%
\end{figure}

So far, the best cryo-\ac{fibid} results have been achieved using the precursor Co$_2$(CO)$_8$~\cite{SalvadorPorroche_2021}. 
In this case, the Ga ion dose required is only \qty{15}{\micro\coulomb\per\centi\meter\squared} and the obtained metallic resistivity is low (\qty{200}{\micro\ohm\centi\meter}). 
Using cryo-\ac{fibid} it is thus possible to write electrical contacts with greater efficiency directly onto a micro-/nanostructure, as outlined in Fig.~\ref{fig:FigCryoFIBID}(a--d), and applied to grow electrical contacts onto materials such as graphene, as shown in Figure~\ref{fig:FigCryoFIBID}(e).

Conveniently, the need for a cryogenic module in order to decrease the substrate temperature for the cryo-\ac{fibid} experiment can be circumvented by means of a cheap, vibration-free and compact thermoelectric plate. 
Recent process optimization along these lines has enabled thermoelectric cooling to \qty{-60}{\celsius} for cryo-\ac{fibid} with W(CO)$_6$~\cite{Orus_2021b}. 
The resistivity of the deposits (\qty{620}{\micro\ohm\centi\meter}), is even lower than that obtained for cryo-\ac{fibid} performed at \qty{-100}{\celsius}. 
Cooling via the thermoelectric plate method thus provides a promising route for the widespread use of cryo-\ac{fibid} in any \ac{fib} chamber.
Further cryo-\ac{fibid} studies will benefit from ongoing process optimization of the thermoelectric cooling method and from the use of alternative and novel precursors for metallic deposits in condensed form~\cite{Bilgilisoy_2021a,Thorman_2020a}. Future application of cryo-\ac{fibid} to directly and efficiently create electrical contacts onto 2D~\cite{SalvadorPorroche_2021} and other ion-sensitive materials is to be expected.

\subsubsection{\acs*{fib} irradiation of metalorganic spin-coated layers}

Although spin-coated layers such as resists are typically used as sacrificial material in a multi-step patterning process, it had also been proposed that metalorganic spin-coated layers could be directly patterned using ion irradiation, producing metallic structures based on Pd~\cite{Harriott_1986a, Gross_1989a}, Ir~\cite{Hoffmann_1991a} and Au~\cite{Hoffmann_1993a}. 
However, post-annealing treatments were required to achieve the desired metallic resistivity values, which limited interest in this approach. 
Similarly, electron irradiation of Pd-based spin-coated layers has been used to create nanopatterned structures, but there too, a post-annealing step is required to achieve metallic resistivity~\cite{Bhuvana_2008a}. 
However, recent process optimization using Pd based spin-coated layers in combination with either focused electron~\cite{SalvadorPorroche_2022} or focused ion irradiation~\cite{SalvadorPorroche_2022a} has led to the development of an efficient process that allows the direct patterning of metallic micro- and nanostructures at the wafer level without the need for annealing.
Due to the simplicity of the process, together with its high resolution and scalability, it is tempting to foresee its wide application for the patterning of electrical contacts or other conductive structures in nanoelectronics.

\subsubsection{\acs*{fib} induced mass transport}

Many effects known from broad beam irradiation studies are also observed during \ac{fib} processing.
This includes the observation of ion hammering and the viscous flow of nanostructures under \ac{fib} irradiation~\cite{Aramesh2017, Xu2019}.
Spatially resolved ion beam mixing using \ac{fib} with subsequent rapid thermal annealing has been used to create nanometer sized Si crystals at freely defined positions for single-electron transistor applications~\cite{xu_site-controlled_2018}.
Furthermore,  localized ion beam induced growth of nanostructures is often observed and can result in epitaxial overgrowth~\cite{Klingner_2022} or the selective growth of nanopillars~\cite{Aramesh2017}.
An overview of these and other effects related to mass transport along surfaces or between layers, including variations such as atomic mixing, can be found in~\citet{Allen2021}. 

A further example of such irradiation effects is the formation of self-organized patterns (such as ``ripples'') on the surface of semiconductors, depending on the angle of incidence and the energy of the ions. 
The broad beam process using noble gas ions is qualitatively understood with mass redistribution playing an important role~\cite{Keller2010,Ou2013,Hofsaess2013}.
\ac{fib} induced self organized pattern formation has mostly been studied for other (mostly metallic) ion species~\cite{Datta2001,Habenicht2002,Gnaser2012,Bischoff2012,Boettger2012,Boettger2013,Bischoff2014,Smirnova2022}. Here, the underlying physics is more complex due to the unavoidable change in stoichiometry of the sample as a result of the high fluence of primary ions. 
The situation could be simplified by using inert gas ions from \ac{gfis} or \ac{pfib}.
In combination with spatially resolved surface structuring, these mass transport effects have the potential for interesting applications, e.g.\ controlling the wettability of \ac{fib} milled trenches for microfluidic applications. Here, understanding and control of the wavelength of the surface ripples will be key~\cite{Ichim2005,Kramczynski2014}.

\subsubsection{Hybrid fabrication and heterogeneous integration} 

Despite the flexibility of \ac{fib} based processing, the intrinsically low throughput of the serial technique means that large-scale patterning remains elusive. A way to overcome this limitation is to combine the \ac{fib} step with other lithography methods, using the so-called `mix and match' approach. For example, in a method combining \ac{fib} patterning with \ac{nil}, \ac{hsq} resist was patterned by He-\ac{fib} and used as a nanoimprint template after development~\cite{Li_2012}. Remarkably, this allowed the transfer of lines with a half-pitch of \qty{4}{nm} into the UV-curable nanoimprint resist.
Similarly, miniaturized fluidic channels for single molecule analysis consisting of complex, multilevel, multiscale fluidic circuits have been patterned using \ac{fib} and photolithography to make a silicon master stamp followed by UV-\ac{nil} to fabricate a negative replica that was used as the working stamp to imprint the fluidic devices quickly and with high fidelity~\cite{Esmek_2019}.
The fabrication of arrays of complex plasmonic nanostructures has been achieved by combining \ac{ebl} and Au-\ac{fib}, demonstrating both planar and 3D nanostructures with high patterning accuracy over large areas~\cite{Hentschel_2020}.
In further examples, optical lithography has been combined with \ac{fib} milling to allow the fabrication of nanochannels in Si-glass microfluidic chips for studies of ion and molecule transport~\cite{Lebedev_2021,Postigo_2016} and for the separation of nanoparticles by size exclusion~\cite{Liao_2017}.
\ac{fib} techniques have also been combined with other processing methods, offering improved versatility in terms of the material to be processed and reducing total processing times.
For example, \ac{fib} patterning of molds into Si was applied for subsequent hot-filament CVD growth of diamond nanocone arrays on a free-standing diamond membrane formed after silicon removal~\cite{Wang_2005}. 
Similarly, direct patterning of inorganic hardmask layers has been demonstrated using \ac{fib} milling and \ac{fibid} combined with reactive ion etching, comparing favorably with the much more complex and material-limited resist based lithography techniques that are conventionally used in this area~\cite{Waid_2014}.
Instead of using the hard mask for later etching steps, it can also serve as a sacrificial layer that enables the use of high ion beam currents. For example, milling through a chromium oxide mask about \qty{63}{nm} thick provided the same spatial resolution with 100-fold increase in ion beam current (\qty{2.6}{nA} instead of \qty{26}{pA}). This approach has been used to fabricate Fresnel lenses from silicon dioxide, increasing throughput by a factor of 75 for functionally identical structures~\cite{Madison_2022}. 

A further promising application is the use of the \ac{fib} as a post-processing tool in \ac{cmos} chip manufacture. This extends the semiconductor industry's conventional use of the \ac{fib} as a debugging and inspection tool, to the top-down \ac{fib} based integration of novel semiconductor circuitry, such as nanometer-scale mechanical devices~\cite{Rius_2009,Llobet_2018}.
In connection with new (nano)-electronic materials, such as single walled \acp{cnt}, it has also been shown that there is low, amendable damage to electrical circuits when a \ac{fib} patterning process is implemented after fabrication of the basic structure~\cite{Rius_2007,Rius_2009a}.

\subsubsection{Outlook}

In terms of applications, machining using \acp{fib} goes far beyond a purely top-down approach, since it can also be used to trigger various chemical and physical processes, and combined with various other micro-/nanofabrication techniques.
Device fabrication based on unconventional processing steps, as well as unique combinations or alternative advanced sequences, may further gain in importance since these methods can mitigate typical limitations in either \ac{fib} processing or the other fabrication techniques.
It is hard to foresee in which way such unconventional processing techniques will develop, but they will certainly benefit from progress made in many other areas,  e.g.\ new insights into ion induced chemical processes, novel precursor materials, as well as novel ion species and newly accessible energy ranges. 

\section{\label{sec_roadmap}Roadmap}

Focused ion beams can be employed as an all-in-one tool for studying and modifying matter down to the nanoscale.  A showcase example of these diverse applications is given by the He-\ac{fib}, which enables imaging of corona virus particles and the fabrication of Josephson tunnel junctions (and much in between), all using the same instrument. 

Building on the previous sections of this article describing the current state of the art of \ac{fib} instrumentation, theory and applications, we can now ask where to go next with \ac{fib} technology. There are many avenues to pursue, so the answer to this question is multifaceted and complex. The present section is devoted to trying to answer this question in some detail.

\begin{figure*}[tbp]
   \centering
   \includegraphics[width=0.9\textheight, angle = 90]{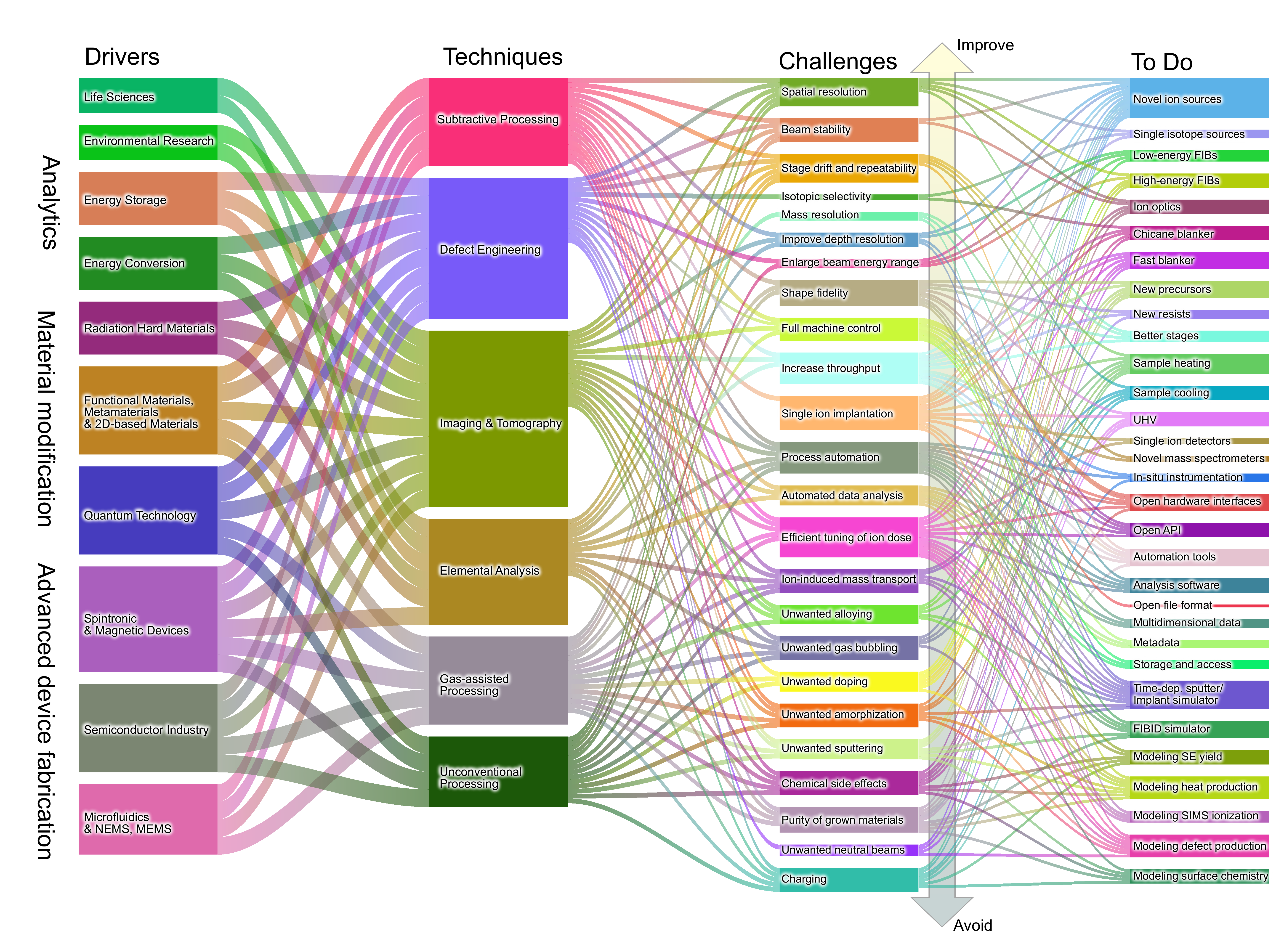}
   \caption[angle = 90]{\label{fig:sankey_main}Sankey diagram showing an overview of the interrelated drivers, techniques, challenges and to-do items for future development of \ac{fib} science and technology.}%
\end{figure*}

Figure~\ref{fig:sankey_main} provides an overview of the fields that drive various \ac{fib} techniques and the resulting challenges and demands for the further development of instrumentation, hardware/software interfaces, accessories, process modeling, and fundamental understanding. 
As the figure outlines, all of these required developments are driven by established or newly emerging research fields that use the \ac{fib} as a tool for any or all of the following: localized and comprehensive analytics, material modification, and advanced device fabrication.

To account for the complexity of the situation without getting buried in details, this section is organized around the important future developments that we have identified as necessary in order to meet the needs of the drivers. Each driver requires different FIB techniques, brings its own challenges, and demands different actions to move forward. 
In the following, a short description of these drivers is given and an evaluation of the needs of the user community is then presented. The latter is based on the results of a survey completed by over 50 scientists from across all \ac{fib} research fields.
The developments necessary for future progress are then derived and described in terms of their challenges and overall potential.

\subsection{Drivers for future applications}%

The drivers for \ac{fib} technologies span a wide range of fields in science and technology.
Currently, the semiconductor industry is the largest driver, demanding process automation, modeling and the analysis of big data. 

In general, the drivers can be separated into two interrelated areas. 
In the first, the \ac{fib} is mainly used for analytical purposes, providing access to a wealth of structural information and physical properties. 
In the second, the \ac{fib} enables rapid, flexible device prototyping and stimulates novel device development. 
In both of these areas material modification using the \ac{fib} is key, be it to enable a particular form or analysis, or to fabricate novel structures and devices. 

A leading example of \ac{fib} based analysis is in the life sciences, where it often involves large-volume 3D tomographic analysis of cells and tissues under cryogenic conditions (see Section~\ref{sss:bio_tomo}). This drives the need for increased processing speed, advanced cryo capabilities and automation.
One potential way to increase throughput is through cryo plasma-\ac{fib} milling, which enables faster removal of bulk material, as discussed in Section~\ref{subs_analysis}.
However, cryogenic processing is not straightforward and is further complicated by insufficient vacuum levels in the sample chamber of current dual beam instruments. A better vacuum, as has been demonstrated in the past, can reduce contamination, and in the case of cryo-\ac{fib}, can also lead to less water condensation on the target structure.

In both the life sciences and materials science, as well as in environmental research, \ac{fib} techniques such as cross sectioning, tomography and sample fabrication (e.g.\ for \ac{tem} and \ac{apt} studies, or ex-situ characterization of physical properties) have become established standards that would now benefit from powerful process automation. 
Materials science in particular can benefit from other ion types, such as ions from noble gas elements that (a) may not accumulate in the material, since they can either pass right through if the sample is thin enough or leave by out-diffusion, and (b) will not undergo chemical reactions. 
This also applies to radiation hardness testing of materials, where \ac{fib} is used as a means to locally mimic the effects of a plasma or solar radiation.
Increasingly, cryo-\ac{fib} techniques are also being applied to beam-sensitive samples, such as novel battery materials that are relevant for energy-technology applications, which also calls for the cryo-related developments mentioned above. 
In the case of multimodal imaging approaches and the identification of target areas for e.g.\ \ac{tem} lamella preparation, workflows are currently hindered by the very different contrast mechanisms in e.g.\ the fluorescence light microscope versus \ac{sem}. Addressing this will require improved correlative techniques as well as more advanced data management and analysis (cf. Section~\ref{subs_software}).

In a broader context, \ac{fib} research into quantum and 2D materials, as well as other novel materials, can largely be subsumed under materials science, but should be recognized as drivers of ``non-conventional'' \ac{fib} techniques in their own right.
For example, the growing interest in quantum materials is driving the need for rapid and high-precision patterning of large 3D structures from novel materials.
This could be achieved by combining process modeling with an easy-to-use interface for 3D shape definition and calculation of the required beam and patterning parameters.
In addition, chemical functionalities can be integrated into patterning through the development of novel ion sources; oxygen in particular is an important tuning parameter for the electronic structure of many quantum materials.

At a more fundamental level, the maskless approach of \ac{fib} processing renders it an ideal technique for the precise creation of defects at the exact location where they are needed. 
This is driven by applications in quantum technology, where defects in an otherwise perfect crystal can act as quantum emitters or, if they carry spin information, as qubits.

On the theoretical side, the creation of specific defects and especially single ion implantation (SII) require a deeper understanding of the mechanisms of defect formation and the evolution of the defect's local environment, which in turn determines the physical properties. Such challenges can be addressed by modeling at different levels of sophistication, as described in Section~\ref{sec_theo}. 
\begin{figure*}[tbp]
    \centering
    \includegraphics[width=\textwidth]{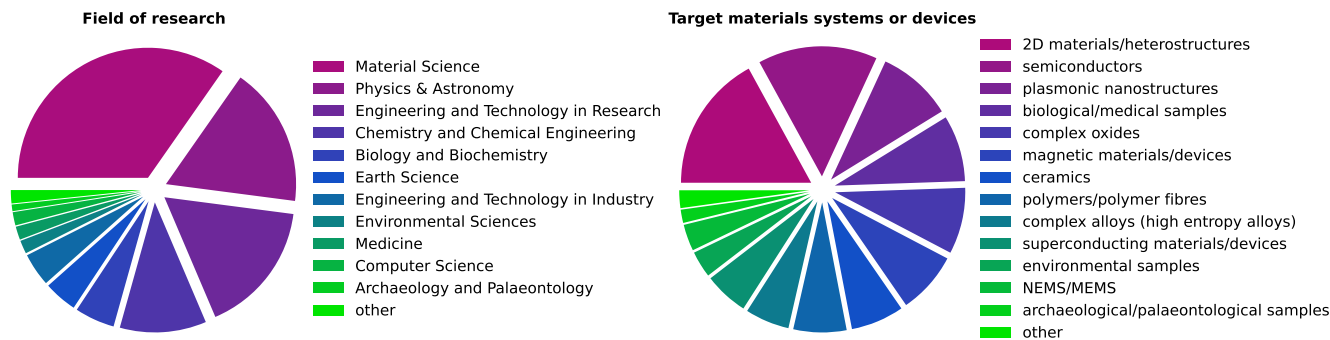}
    \caption{Results of a survey of the fields of research (left) and the target materials systems or devices (right) being investigated or addressed.}%
    \label{fig:pie2}
\end{figure*}

In terms of instrumentation, a key technological challenge for quantum applications is to be able to place individual ions in a (quasi-)deterministic manner and detect them with very high efficiency.
Certain defect types may require the implantation of a specific ion for which stable ion sources still need to be developed.
The implantation depth is controlled by the mass and kinetic energy of the incident ion, and in order to reduce dephasing, quantum emitters in photonic chip architectures are usually desired to have a depth of \qtyrange{10}{100}{nm} beneath the surface.
As such, the ion energy ranges of tens of keV that are currently available are mostly well suited. 
However, in the case of spatially resolved heavy element implantation, a higher primary beam energy would be required, which poses challenges for the power supplies and for electrical isolation in the ion optical column.
If instead non-native atoms are to be implanted into 2D materials, e.g.\ for high resolution flexible area doping, lower energies are needed. 
In this context, low energy means less than \qty{100}{eV} to ensure a significant number of replacement collisions in which the ion exchanges position with a target atom and the latter is removed from the 2D material~\cite{kretschmer_threshold_2022}.
Methods to reduce the primary energy of the beam include using deceleration lenses in the final stage of the beam focusing or implementing reverse sample biasing (see also Section~\ref{subs_future}). 
For the detection of the transiting primary ion itself or the resulting interaction with the substrate
, various approaches are under development (see the Outlook of Section~\ref{subs_defects}). 

Another strong driver for \ac{fib} technologies are \ac{mems}/\ac{nems} and microfluidic devices, which are ubiquitous in our modern world, be it in sensors, displays, printers, transducers or microphones.
Such devices often combine various active and passive components consisting of electronic and moving parts, and therefore typically rely on a whole chain of thin film lithographic, deposition and etching  techniques. 
However, these fabrication techniques require that the devices are built up layer by layer. 
As such, true 3D geometries are difficult to achieve.

\ac{fib} processing by milling and deposition, on the other hand, enables the flexible fabrication of complex multi-level geometric structures and smooth 3D shapes, as it permits direct top-down and bottom-up structuring in the same instrument.
What has limited the default implementation of \ac{fib} technologies in device manufacturing thus far is the slow processing speed of the inherently serial process. 
Here, the availability of a multibeam system comprising thousands of individually addressable focused beams would represent a major leap forward (see also Section~\ref{subs_optics} ).

\subsection{The generic needs of the community}%
\label{subs_needs}

\ac{fib} systems have become a standard in scientific research and industrial processing, but the market is still heavily dominated by the Ga-\ac{fib}. While Ga represents a good compromise between spatial resolution and sputtering yield for many material systems, other applications require different beam parameters and/or suffer from Ga contamination. 
In our survey, the participants were therefore asked in which areas they conduct research and upon which material systems and/or devices they focus. 
The corresponding results and percentages thus reflect the perspectives of this particular group of users. 
Figure~\ref{fig:pie2} shows that material science, physics and engineering are the most prominent research areas in the \ac{fib} community surveyed. Fields such as biology, medicine and earth science may be underrepresented in these statistics, since fewer scientists from those areas participated in the survey. 
It would be interesting to repeat the survey with a larger number of participants from a wider range of fields in order to be more statistically representative. 
The material systems most commonly studied by the respondents are 2D materials and semiconductors, followed by various other systems including biological samples, ceramics, polymers and environmental samples. This reflects current research trends, but also underscores the universality of the \ac{fib} as a tool for modifying matter at the nanoscale.

\begin{figure*}[tbp]
    \centering
    \includegraphics[width=\textwidth]{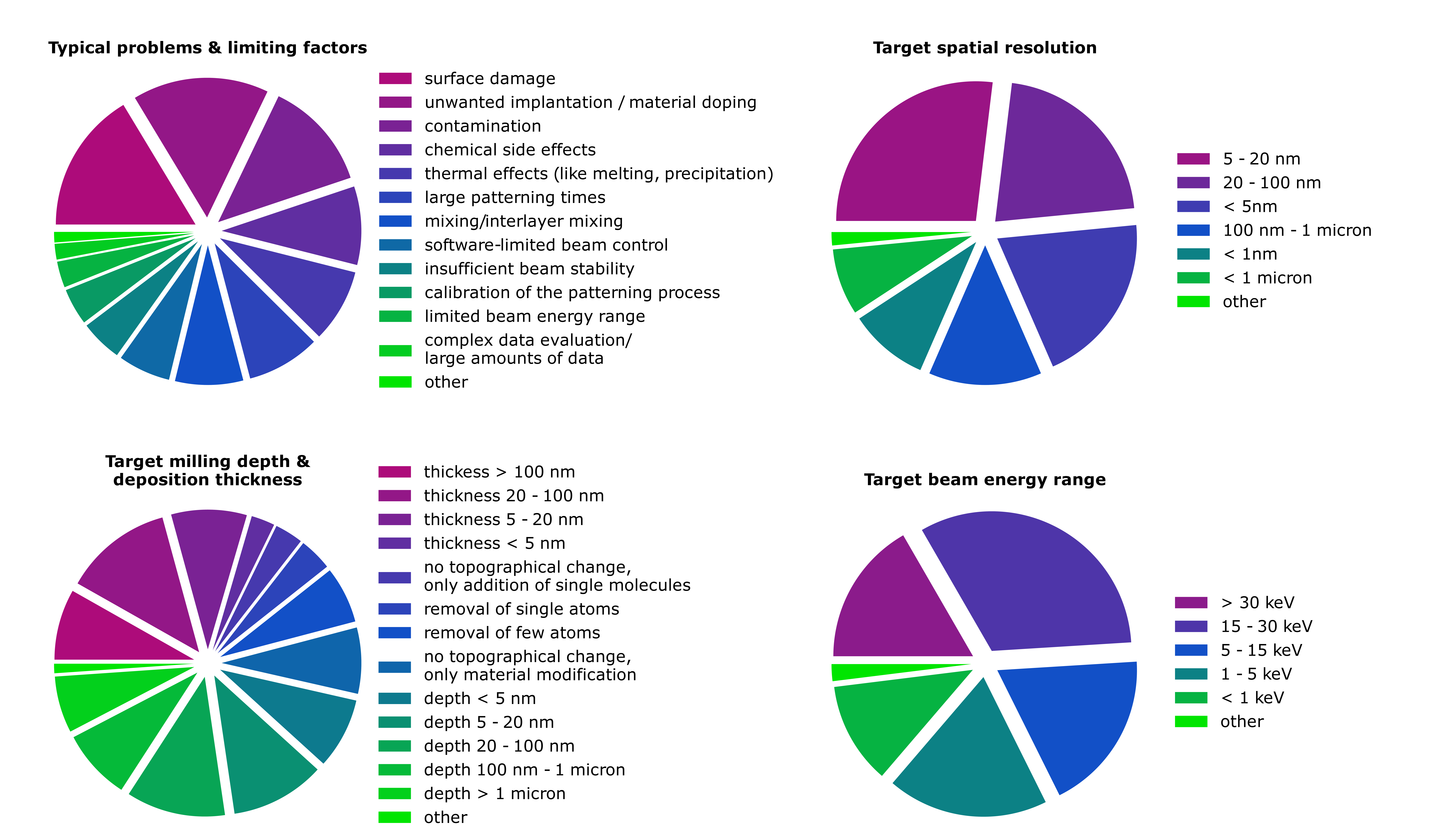}
    \caption{Results of a survey of the typical problems and limiting factors in \ac{fib} processing (top left), the target spatial resolution (top right), the target milling depth and deposition thickness (bottom left), and the target ion beam energy range (bottom right).}%
    \label{fig:pie4}
\end{figure*}

The typical problems encountered by the \ac{fib} users participating in our survey (see Fig.~\ref{fig:pie4}) show a clear trend, with surface damage, unwanted implantation/doping and contamination being the most dominant problems. 
This last point on contamination is mostly related to the typical vacuum operating range of around $10^{-6} - 10^{-7}$ mbar and the associated residual gases in the vacuum chamber. 
In this pressure range, a monolayer of molecules still forms on the sample roughly every second, and with a typical water content of around 60$\%$ in the residual gas, most surfaces will also be covered with a double layer of water molecules.
The hydrocarbons from the residual gas are also decomposed by the ion beam and, depending on the ratio of the sputtering to deposition yield, can form a carbonaceous layer on the sample. 
Especially in the case of \ac{fib} processing with light ions, even a small amount of contamination can result in greatly reduced net sputtering rates or in the extreme, net contamination deposition.
Generally, both water and hydrocarbons can add various, mostly unwanted, chemical reaction channels (cf. Section~\ref{subs_fibid_chem}).
Here, a significant improvement of the vacuum level toward \ac{uhv} would be an important development. 
Since some of the contamination is introduced with the sample itself, a dedicated load lock for plasma cleaning of samples prior to \ac{fib} processing would be an ideal complement. 
Other unwanted chemical side-effects may occur when the ion interacts inside the sample and can be mitigated using inert ion species. 
However, unwanted physical side-effects such as interlayer mixing or dealloying are difficult to avoid as this would require a significant reduction in primary ion energies; 
in the shorter term these effects can be better addressed through process modeling (cf.\ Sections~\ref{subs_bca} and~\ref{subs_md}).

Problems with long patterning times are either due to beam currents not being high enough or to stability issues of the system. 
\ac{pfib} columns address the first issue by proving higher beam currents, but at the expense of a larger probe size.
One especially intriguing way of increasing processing speed and thereby reducing patterning times would be the implementation of a multibeam system. 
Thousands of parallel beams could in principle be available while maintaining small probe sizes (cf. Section~\ref{subs_optics}).
System stability is mainly limited by the use of mechanical stages, which suffer from drift over time and from inaccuracies in positioning. 
In principle, laser interferometric stages solve this problem. 
However, since most \ac{fib} systems use a tilted column, the sample must also be tilted in order to allow \ac{fib} processing under normal incidence. 
Thus novel system designs with a vertical \ac{fib} column are of great value.
Finally, instabilities in the beam current itself (more of a problem for some of the novel ion sources) can also limit patterning performance, so the development of novel and stable emitters is needed (cf. Section~\ref{subs_sources}).

Interestingly, the survey answers to questions about desired spatial resolution, milling depths, deposition thicknesses and beam energy ranges more or less correspond to what is currently technologically possible. 
This makes sense, in that users often aim for what is feasible with their instruments. 
Only a few users were interested in depositing single molecules or removing/implanting single atoms. Currently, such approaches are beyond the technological capabilities of commercial systems and would require substantial developments. However, with these developments the impact on areas such as quantum technology research would be substantial.

\subsection{Important future developments}%
\label{subs_future}

Having discussed the needs and wants of the \ac{fib} research community in terms of future applications and user experiences, we now turn to specific technologies that need to be developed.
For this purpose, Figure~\ref{fig:sankey_main} has been divided into subplots based on different groups of to-do items. 
This allows us to trace the impact that certain developments will have on specific challenges, to explore the techniques that are then enabled, and ultimately link back to the drivers.

\subsubsection{Ion beam generation and transport}

\begin{figure*}[tbp]
   \centering
   \includegraphics[width=\textwidth]{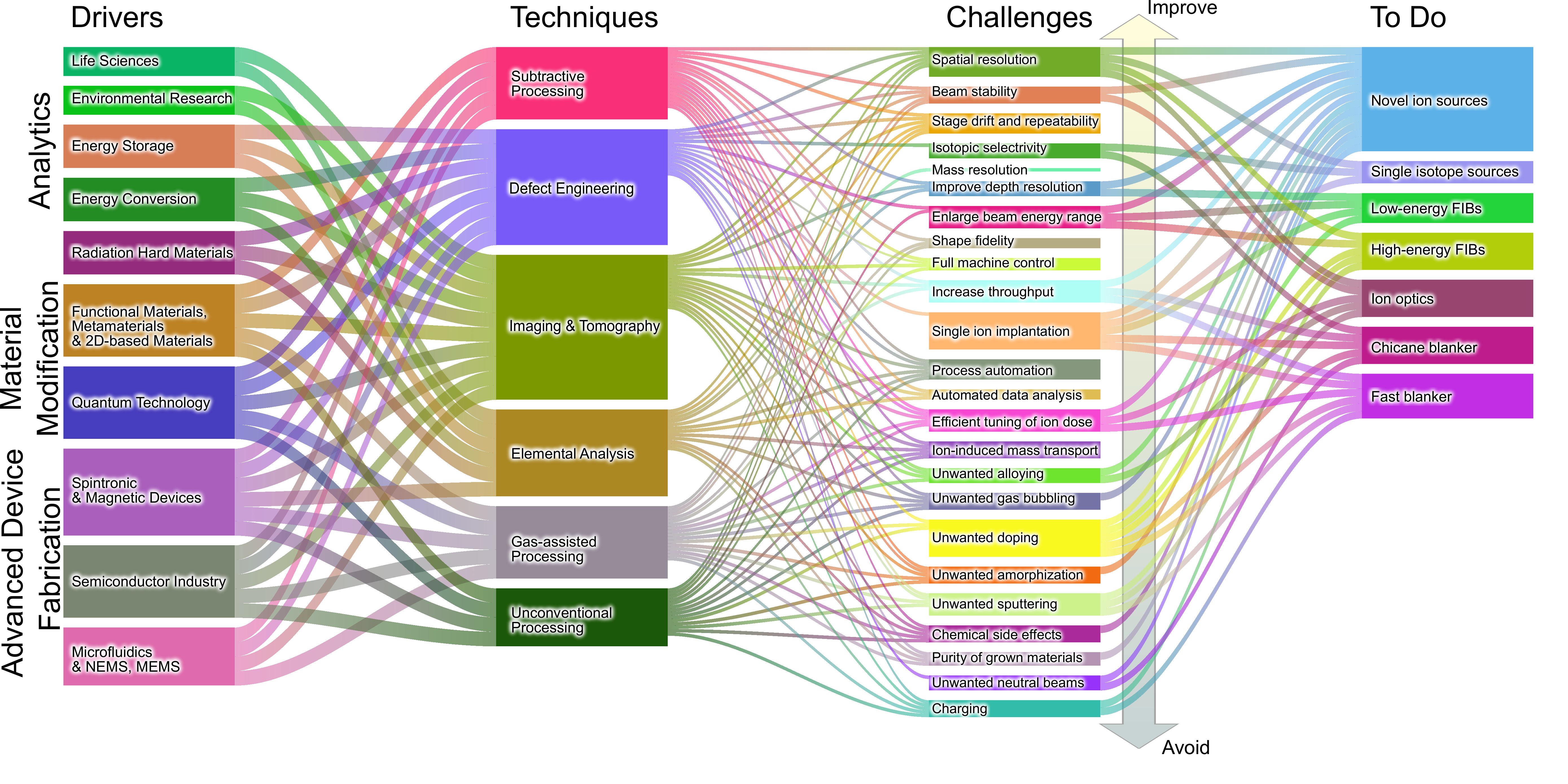}
   \caption{\label{fig:sankey_sources-beams}Developments that are required with respect to the generation and transport of focused ion beams.}%
\end{figure*}

Figure~\ref{fig:sankey_sources-beams} displays the required developments (i.e.\ to-do items) and the associated challenges, techniques and drivers in the area of next-generation \ac{fib} sources and ion beam transport. Specific needs for novel ion sources, ion optics for both low and high energy beams, and fast beam control/blanking have been identified.

\paragraph{Novel source types including single isotope sources}
A key area requiring development concerns novel ion sources, in particular, high brightness and light element sources, which are crucial for  high-resolution imaging and nanostructuring applications. 
Here, further development of \ac{gfis} based sources to operate with gases other than He, Ne, and N$_2$ is desired.
The first steps in this direction have already been taken by various researchers testing \ac{gfis} based emission of H~\cite{Matsubara_2016}, Xe~\cite{Lai_2017}, and Kr~\cite{Shichi_2017}. 
However, the heavier noble gas elements have numerous isotopes, sometimes in nearly equal proportions;
if the ion optical column is not well shielded from magnetic fields, this can affect spatial resolution, since magnetic fields present in the column can split the ion beam according to the mass of the isotopes.
Solutions to this problem, other than shielding, include the addition of a Wien filter.
Again, these developments will require new ion optical columns that can reduce or correct the aberration induced by the Wien filter.
Alternatively, isotopically pure source gases can be used, similar to the already successful implemention of isotopically pure source materials by manufacturers of Ga-\ac{lmis} and AuGeSi-\ac{lmais}\@.
Further \ac{lmais} single isotope sources might also be of interest in the context of \ac{sii}~\cite{Jakob_2021a, Raecke_2022, Collins2023, Chandrasekaran_2023}, local isotope enrichment~\cite{Acharya2023} and for a wide range of other applications e.g.\ \ac{sims}, also in combination with reactive gas flooding~\cite{Priebe_2019, Priebe_2020, Priebe_2021c, Wieczerzak_2021a}. 
The GaBiLi-source~\cite{Bischoff_2010a, Pilz_2019}, for example, provides heavy Bi atoms and small Bi clusters for milling, and a reactive Li ion beam for analysis by \ac{sims}.
With the \ac{lmais}, switching between ion beams for these different species is fast (only takes a few seconds) and can be automated.
For the future, new alloy concepts that include heavier reactive species are desired; to date only Au-Si, Au-Ge, Au-Ge-Si and Ga-Bi-Li \ac{lmais} are commercially available.

Ion sources for heavy ions are desirable for applications such as \ac{sims} in order to increase sputter yields and also for device debugging in the semiconductor industry, where the shorter penetration depth of the heavier ion is an advantage.
Furthermore, ions from group 1 and group 17 in the periodic table are of interest due to their chemical reactivity, which enhances sputter yields.
While successful attempts have been made in the past with group 1 elements~\cite{Knuffman_2013, Pilz_2019,Haaf_2018, Viteau_2016}, efforts for the development of sources based on group 17 elements such as fluorine or iodine are more limited.
A possible approach for reactive ion sources using group 17 elements could be the \ac{ilis} if implementing the appropriate salts~\cite{Guilet_2011}.
Such \ac{ilis} prototypes have been developed, but they did not yet enter the market.
One of the intrinsic issues when operating an \ac{ilis} is the need to switch the polarity of the extracted ion beam in order to keep the source composition in balance. 
While this is usually considered a nuisance it can become an advantage, since by using clever chemistry, ion beams suitable for performing \ac{sims} with both polarities could be extracted.
However, collateral damage by the reactive ions to the ion optical elements needs to be considered and avoided as much as possible.
We note that the first commercial \ac{fib} systems combining reactive ion species with \ac{sims} are currently entering the market~\cite{NanoK,RAITH2023}.

With regard to source brightness, the recently developed \ac{motis}/\ac{lotis} have been shown to achieve high brightness for a number of ion types~\cite{Knuffman_2013, Haaf_2018,Viteau_2016}.
Furthermore, fundamentally new concepts based on e.g.\ insulator-on-conductor ion sources~\cite{Laurencich_2022} and other ideas that currently have a very low technology readiness level (TRL) may also have potential for application as high brightness sources, since they partially exploit similar concepts to the \ac{gfis} technology.

An aspect that is becoming more and more important for users is instrument uptime, which can be increased by minimizing source maintenance needs. For example, \ac{lmis} and \ac{lmais} lifetimes could be increased by increasing the reservoir size of the liquid metal/alloy. The lifetimes of the apertures could be increased by adjusting the material and thickness of the aperture strip, as well as the number of apertures incorporated. 

\paragraph{Ion optics for low and high energy ion beams}
Access to low energy ion beams (\qty{<1}{keV})
is becoming increasingly important for a range of \ac{fib} applications in which the penetration depth and ion straggle in the material must be minimized. 
However, the transport of a low-energy beam without sacrificing on resolution is a challenge. This is mainly due to chromatic aberrations (see also Section~\ref{subs_optics}). To address this issue, specialized beam optics such as energy filters and chromatic aberration correctors are needed, which are in general complicated to produce, align and tune~\cite{Hawkes_2019} thus posing extra challenges for system automation. Low-energy beams also pose challenges in terms of the stability and precision of the associated electronics.

Key areas where low-energy high-resolution \acp{fib} will bring benefit are in \ac{fib} based debugging and mask repair in the semiconductor industry~\cite{Tan_2016}, where minimized ion penetration and thus minimal damage to underlying layers and structures is required. If these low-energy \acp{fib} employ heavy ions, the beam-sample interactions can be confined even more closely to the surface. A further application that requires ultra-low energy beams (down to a few tens of eV) is the controlled and spatially-resolved doping of 2D materials or the ion-beam induced modification of surfaces (see Section~\ref{subs_bca}). For the latter, the energy spread of the ions emitted from the source must be exceptionally small to begin with, smaller even than the \ac{lotis} and \ac{gfis} can deliver. Indeed, in order to reduce chromatic aberrations for all of the above, it makes sense to start with new types of ion sources that have a reduced energy spread and then employ subsequent beam elements for chromatic aberration correction accordingly.

For other applications, higher energy beams are required, such as for the deep placement of (heavy) atoms for quantum technology applications, and for applications requiring even higher spatial resolution.
Since the ion collision cascade expands the \ac{fib} beam in the material and most damage occurs at the end (cf.\ again Section~\ref{subs_bca}), it may be useful to use high energies for devices based on (potentially suspended) thin films. 
The target defects are then generated close to the surface with high spatial precision, but low yield, and the beam damage occurs spatially separated from this in the substrate (or not at all in the case of suspended membranes).
While the technological challenges for increasing the beam energy are moderate, the increased size of the required electrical isolation and power supplies will actually limit the achievable resolution, since they may act as additional noise sources.

\paragraph{Fast beam control and blanking}
Standard beam blankers installed on \ac{fib} instruments typically achieve blanking times (known as rise times) on the order of \qtyrange{20}{100}{ns}. 
However, for a number of applications, rise times below \qty{10}{ns} are desirable and have been demonstrated in isolated cases~\cite{Klingner_2016}.
Applications requiring faster blanking times include resist patterning, high-end \ac{fibid} and \ac{sii}, in all cases to avoid unintentional exposure.

Another need for specialized beam blanking arises from the fact that ion beam sources typically also create a certain fraction of neutrals in the beam, which is generally unwanted.
These neutrals result from charge exchange collisions between energetic ions  and neutral gas atoms.
The latter are generally present due to insufficient vacuum at the source and in the ion optical column, and in the case of the \ac{gfis}, as a direct result of the gas based source technology. 
In the case of the \ac{gfis}, \qty{0.29}{\percent} He$^0$ have been reported under conditions used for medium resolution imaging and for \ac{tof}-\ac{sims} (\qty{6.3e-6}{mbar} He pressure, \qty{20}{\micro\meter} aperture)~\cite{Klingner2016}.
However, these numbers depend heavily on the gas pressure and aperture used.
Since neutral species are not influenced by electric or magnetic fields and thus cannot be deflected away using ion optical elements or the standard electrostatic blankers, they can result in broad background exposure, especially in the case of long \ac{fib} processing times.
This is detrimental in applications concerned with few or single-ion impacts at specific target locations, which often also involve high-dose irradiation (and hence even more unwanted neutrals reaching the sample) to make fiducial markers for subsequent correlative analysis.
In order to eliminate the neutral beam, a physical obstacle can be introduced into the central portion of the column, which only the ions can bypass by means of a series of electrostatic deflectors.
This is known as a chicane blanker.
Similar solutions, such as a curved column, could also be explored but come with their own challenges.

\subsubsection{Accessories and complementary instrumentation}

\begin{figure*}[tbp]
   \centering
   \includegraphics[width=\textwidth]{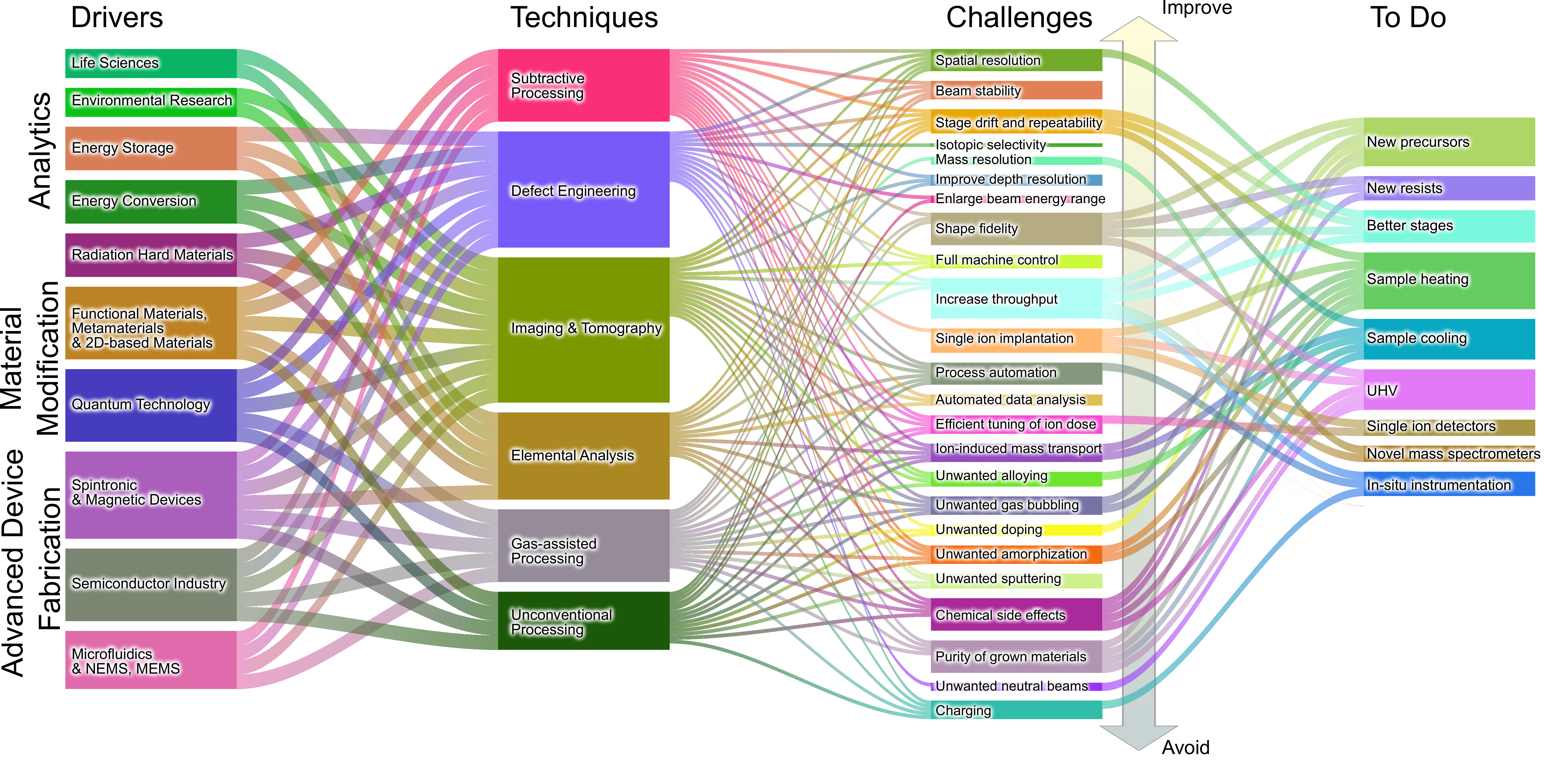}
   \caption{\label{fig:sankey_accessories}Developments that are required with respect to accessories and complementary instrumentation.}%
\end{figure*}

In Fig.~\ref{fig:sankey_accessories}, the required developments related to new accessories (add-ons) for \acp{fib} and complementary instrumentation are outlined. These to-do items can be broken down into categories focused on ion induced chemistry, adaptable sample environments and analytical add-ons.

\paragraph{Ion induced chemistry}
Applications using \ac{fibid}, gas-assisted etching, and resist-based lithography could all be greatly enhanced by the development of more advanced modeling tools, new gaseous precursors and new resists. 

In the case of \ac{fibid}, new precursor chemistries that enable higher deposit purity (e.g.\ for plasmonic applications) and superior resistivity (e.g.\ for circuit edit applications) are needed.
The prerequisite for the development of these precursors is a deeper understanding of the ion-molecule reactions that occur during the deposition process~\cite{Utke_2008}. 
For high-purity metallic deposits, the precursor molecule will ideally be fully decomposed so that only the desired metal component is deposited and the ligand fragments are pumped away. 
Similarly, in the case of gas-assisted etching, the precursor molecules should undergo ligand loss to generate empty coordination sites that bind with the substrate to enhance the rate of removal of substrate atoms without forming a residue of nonvolatile reaction products. 
Once the fundamentals of beam induced chemical reactions are better understood (including aspects such as the effect of beam induced heating), researchers will be better poised to develop simulation tools to enable the prediction and design of more complex deposition and etching scenarios. 
The development of these simulation tools will also rely on input from systematic parameter optimization, where automated beam control and data analysis will play a key role. 

A common but generally undesired effect from ion induced chemistry is the deposition of a thin carbon film over ion-irradiated regions, arising due to residual hydrocarbon species present on the sample surface and in the microscope chamber~\cite{Hlawacek_2013}.
This effect can be minimized by e.g.\ plasma cleaning samples before loading into the microscope or working in \ac{uhv}.
Other unwanted beam induced effects that can occur during intentional deposition processes include implantation of the ions themselves into the deposit during the growth process, or concurrent milling of the deposit.
To minimize these effects, it is desirable to increase the deposition rate, so that a lower total ion dose is needed for the process.
Recently it has been shown that cryogenic cooling of the substrate is one way to increase deposition rates, which could be investigated further~\cite{Cordoba_2019b}.
However, it is worth mentioning at least one application where a balance between deposition and concurrent milling by the primary ion beam is actually desired, namely the formation of high-aspect-ratio hollow nanopillar deposits with unique electronic properties using He-\ac{fibid}~\cite{Cordoba_2018}.
Overall, fine control over the rate of deposition should prove highly beneficial for future applications.

For resist based lithography, further development of resists specifically for lithography using ion beams is needed.
It has already been shown that resists tend to be much more sensitive to ions compared to electrons, but so far most resist development has still been for \ac{ebl}.
In order to optimize resists for lithography using different ion species, a deeper understanding of the ion-resist interaction process is required~\cite{Zhang_2020a}.
Consequently, the full potential of \ac{fib} resist lithography has yet to be realized, even though \ac{hibl} has already demonstrated high resolution patterning with dramatically reduced proximity effects compared with \ac{ebl}~\cite{Sidorkin_2009b,Winston_2009a}.

\paragraph{Adapted sample environments}

In electron microscopy, sample environments are now routinely adapted through the use of liquid cells that separate e.g.\ chemical reagents, from the surrounding vacuum in the chamber.
This principle does not readily translate to the \ac{fib}, since the typically used Ga ions can easily damage the cell membrane by sputtering and also suffer from limited penetration depths.
Accordingly, only He-\ac{fib} has been used in a prototype liquid cell~\cite{Belianinov2018}, because He ions combine negligible sputtering of the cell membrane with a sufficient penetration depth to reach the reaction volume.  
In contrast, heating of samples during \ac{fib} processing has been used for decades, including  more recently for the reduction of ion beam induced amorphization and blistering during He- and Ne-\ac{fib} processing.
Quasi-local heating with an infrared laser~\cite{Stanford_2016} and homogeneous heating using a heating stage~\cite{Xu2019} have both been tested.
However, the latter approach in particular suffers from sample drift induced by the high thermal mass of the sample heater.
Low drift implementations are thus needed that also have lower power requirements in order to protect the rest of the system (e.g.\ the piezo motors in the stage and ion optical components in the pole piece) from overheating and to allow faster heating ramps~\cite{Mele_2016}.
\Ac{fibid}-related applications could also benefit from new simple heating stages for exploring in-situ annealing of the deposited structures.

Sample cooling can also be of interest to reduce beam damage, de-alloying, and surface microstructure formation, and for applications using \ac{fibid} or \ac{hibl}.
For example, often temperatures slightly below \qty{0}{^{\circ}C} are sufficient to prevent the formation of eutectic phases or phase separation. 
In the same manner, molecular desorption rates may be tuned by either decreasing (or increasing) the temperature by only few tens of Kelvin.
Hence, for many experiments with a relatively small temperature range to bridge, Peltier based solutions become possible~\cite{Orus_2021b}.

A pressing requirement for substantial cooling comes from biological applications, where cryo-stages and cryo-transfer units have been developed to enable complete cryo workflows (see Section~\ref{subs_imaging} for examples).
However, due to the stage cooling mechanism and shielding, these systems can suffer from reduced imaging performance as well as limitations regarding sample geometry and compatibility with certain accessories in the \ac{fib} chamber.
Further development of specialized \ac{fib} instruments dedicated to the handling of biological and other sensitive samples is expected. 
These systems require special load locks for vacuum and/or cryo transfer, and better vacuum conditions in general will result in less condensation on the specimens. 

By adding fluorescence light microscopy to a \ac{fib}-\ac{sem} system, a dedicated and time-efficient platform for the correlative preparation of \ac{tem} lamellae for biology, medicine and related fields becomes possible, since well-established fluorescence light microscopy protocols can be used to locate regions of interest for \ac{fib} milling in-situ.
Indeed, a system including fluorescence light microscopy has recently been demonstrated in a cryo triple-beam configuration~\cite{Boltje_2022}. 
Other implementations for fluorescent light detection either employ an ellipsoidal mirror in conventional Ga and plasma \ac{fib}-\ac{sem} systems for fluorescence spectroscopy of the atoms sputtered by the ion beam~\cite{Budnik_2022}, or a full confocal microscope setup integrated into the vacuum chamber of a \qty{100}{kV} \ac{lmais}-\ac{fib} with a high-precision stage to move between working positions~\cite{Chandrasekaran_2023}.

The latter system demonstrated the in-situ detection of fluorescent light from arrays of optically active defects created by the \ac{fib}, thus representing an important step toward realizing the deterministic creation of quantum emitters (cf. Outlook of Section~\ref{subs_defects}). 
For this application, it is of utmost importance to reduce all kinds of residual contamination (fluorescent and chemically active), both on the sample and in the vacuum chamber using \ac{uhv} chambers and ideally inert sample transfer.
Dedicated \ac{uhv} \ac{fib} systems for quantum technology do not yet belong to the commercial standard, but have recently been demonstrated~\cite{Cassidy2019,Adshead2023}.

\paragraph{Novel detectors and analytical add-ons} 

Quantum technology and therewith the deterministic implantation of ions is one of the main drivers in science and technology today, and correspondingly, there is an urgent need for single ion detection.
As outlined in the Outlook of Section~\ref{subs_defects}, specialized detectors~\cite{Raecke_2019, Raecke_2022} or specific types of ion sources like a Paul trap~\cite{GrootBerning_2021,GrootBerning_2019}, can either count the ions that impinge on the sample or measure the ion impact event.
The latter can be realized using an on-chip \ac{ibic} measurement~\cite{Robson_2022,Collins2023}, which can be implemented for many \ac{cmos} compatible quantum devices that are not affected if there is an amplifier nearby.
For all other quantum devices, truly agnostic \ac{sii} detectors that detect e.g.\ \acp{se} will constitute an especially important development. 
Besides very high detection efficiencies approaching \qty{100}{\%}~\cite{Murdin_2021,Cassidy_2020}, appropriate theory support is needed to calculate the expected \ac{se} yields. 
If such detectors are combined with a complementary system that detects device function, e.g.\ fluorescence~\cite{Chandrasekaran_2023} associated with single-photon emission from the defects generated, truly deterministic implantation can be realized for an industry-ready commercial platform.

Several types of detection systems exist, which can often be combined to enable the acquisition of unique multimodal data sets. The general trend is toward better spatial resolution in 3D, higher sensitivity (and hence lower detection limits), and better quantification of results. A promising approach in this area is \ac{fib}-\ac{sims}.
Resolution and sensitivity are directly linked, which must be considered in efforts to improve performance.
For example, in the case of \ac{gfis} and \ac{lotis} based \ac{sims},  simultaneous improvement in both resolution and sensitivity is not possible, since further reducing the sampling volume renders the number of target species in that volume statistically insignificant.
Another analytical approach that is complementary to \ac{sims} uses fluorescence spectroscopy to detect and identify atoms sputtered by the \ac{fib} and has been shown to be feasible for certain applications~\cite{Budnik_2022}.

Various other ``classic'' ion beam analytical methods have been implemented in \ac{fib} instruments or are interesting candidates for future implementations.
These include \ac{il}~\cite{Veligura_2016} and \ac{rbs}~\cite{Heller_2016}, which have been explored by a number of researchers but are not available as commercial solutions.
In \ac{il}, the luminescence of excited target atoms, defects or recoiled ions is detected using parabolic mirrors and a combination of gratings and lenses to obtain spectral information that allows identification of the emission source. 
\ac{il} was recently demonstrated for semiconductor heterostructures and Li-ion batteries~\cite{Budnik_2022}, and also has potential for various insulating samples in fields such as geology, mineralogy, biology and medicine, including in nanoparticle research such as nanotoxicology.
In contrast, \ac{rbs} measures the energy and scattering angle of backscattered ions and allows standard-free and depth-resolved elemental analysis.
However, \ac{rbs} in \ac{fib} systems has largely been abandoned due to fundamental problems related to the high number of multiple scattering events. 
Implementing a low energy ion scattering (LEIS) configuration may be a good alternative, as it only measures the energy of the charged particles that are primarily scattered at or close to the sample surface.
Nevertheless, due to the low yield of charged scattered particles, high fluences are required to obtain good statistics.

There has been some debate around the potential to use X-rays as a signal for ion induced processes.
Some reports indicate that this could indeed be possible~\cite{Giannuzzi_2009}.
However, in that particular case the observed X-rays were not generated by the interaction of the primary ion beam with the sample. 
Instead, non-conductive samples became positively charged under ion impact and backscattered ions caused the emission of \acp{se} from components in the chamber. These \acp{se} were then accelerated toward the positively charged sample surface and generated X-ray peaks due to random interactions with the chamber components. 
It turns out that because the inelastic interactions that are responsible for X-ray generation depend on velocity and not on energy, typical primary beam energies in \ac{fib} systems are not sufficient for X-ray production. 
As stated in \citet{Joy_2007}, a beam energy of about \qty{53}{MeV} for He ions, for example, would be required in order to excite the same X-ray lines as can be obtained using a \qty{6}{keV} electron beam. 
Such high beam energies are out of reach for conventional \ac{fib} systems and thus the high-energy \ac{pixe} technique remains elusive to \ac{fib}.

In contrast to X-ray spectroscopy, the generation and detection of Auger electrons is indeed possible using the beam energies of current \ac{fib} systems. 
Auger electrons are generated due to the creation of core-holes in both the projectile ions and the substrate atoms, thus Auger signals emanate from both colliding particles, resulting in spectra with rich features that can be quantitatively evaluated~\cite{Parvaneh_2014}.
 
Another analytical approach is based on in-situ monitoring of ion beam-induced property changes, such as in-situ electrical measurements to test nanostructured samples or to probe e.g.\ magnetization via the measurement of a Hall voltage.
Various other analytical techniques can also be incorporated into the \ac{fib} chamber, including \ac{afm} for in-situ surface characterization or a nanoindenter for in-situ mechanical testing.
The benefit of performing these techniques in-situ is not only control of the sample environment (e.g.\ no exposure to air), but also the possibility of real-time imaging (usually with the \ac{sem}) enabling immediate quality control of the measurement.
For improved implementation of all these approaches, open hard- and software \acp{api} are required.

Further developments that will facilitate work with analytical add-ons are related to the continued integration of sample cleaning options based on UV light or remote plasma, as well as the combination of \ac{fib} processing with direct laser milling~\cite{Sugioka_2014}. 
Systems integrating femtosecond laser milling with \ac{fib} have recently become available and the associated workflows and possibilities are currently being explored.

\subsubsection{Automation and data management}%
\label{subs_roadmap_data}

\begin{figure*}[tbp]
  \centering
  \includegraphics[width=\textwidth]{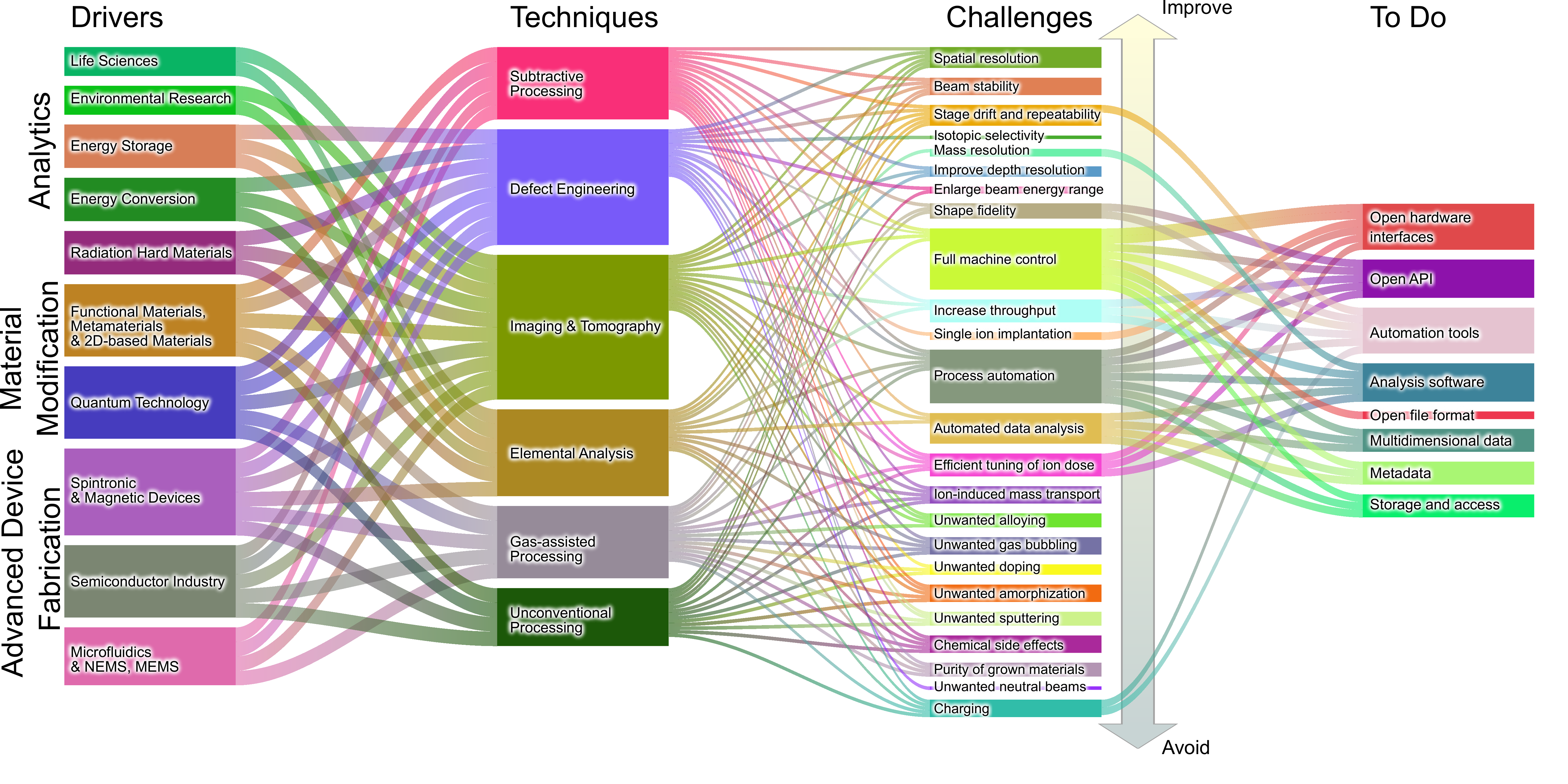}
  \caption{\label{fig:sankey_sub3}Developments that are required with respect to automation and data management.}%
\end{figure*}

Figure~\ref{fig:sankey_sub3} shows the technological developments required in terms of automation and data management. Both of these areas are rapidly growing in importance due to the sheer volume of samples and data.
There is enormous potential to facilitate the daily work of technologists and researchers using recent developments in \ac{ml} and artificial intelligence. However, the black-box nature of such routines requires awareness and regular cross-checking of both the training data used and the results obtained. In the following, open interfaces for hardware and software, and then needs related to data access and management, are discussed in more detail.

\paragraph{Open hardware and software interfaces}
In recent decades, science has increasingly moved toward openness and transparency. Today, this is generally a prerequisite for receiving taxpayer money for scientific research~\cite{EREA2023}. The idea of sharing not only raw data (as discussed in the next subsection) but also task-specific developments in terms of machine operation and data analysis holds great potential. Instead of developing isolated solutions specific to particular research groups, which often become obsolete or are lost when the original developers move on, a larger community can maintain and further develop methods much more efficiently. In addition, the overall speed of development is significantly increased, as existing data and tools can be reviewed much more easily before ``reinventing the wheel''. 

Open access to non-proprietary development is already standard in many scientific fields, but the more technologically sophisticated and commercially valuable scientific work becomes, the more conflicts over intellectual property, licenses and potential patents can arise. Nevertheless, the major \ac{fib} instrument manufacturers have recognized the benefits of opening up their machines for modification by the users.
Everyday users seek the ultimate in machine performance.
Therefore a key resource for driving future developments comes from enabling those users to make the most efficient use of instrument time and gain the most insight from the data generated.
The larger the active user community is, the more far-reaching the benefit, which in turn translates to increased sales for the manufacturers even if new technological developments do not have immediate impact on a patent or a price tag.
What users need are open hardware interfaces to give them real-time access to detector signals and to enable them to add their own custom components.
Even more important is a full-fledged \ac{api} that allows control of all system parameters accessible in the \ac{gui} and ideally all others in developer mode. 
In this respect, software development on the manufacturer side must be a compromise between the needs of those who require easy-to-use software yet with more limited access, and those who require full access. 

For process automation, certain tools are already included in some \ac{fib} systems. For example, routines that have already been executed can be stored and modified in a quasi-graphical approach, removing the hurdle of programming in a text-based interface. 
A key area needing a high degree of automation is in the efficient acquisition of tomographic data sets with imaging based on \ac{se} detection, as well as \ac{eds} and \ac{ebsd}\@. \ac{fib} based tomography is particularly used in the life sciences and in environmental research, and increasingly in all branches of materials science, including for battery research and energy conversion. In all cases high precision and repeatability are required.

In the future, automation tools should be complemented by \ac{ml} algorithms that help with experimental control and real-time data analysis. Integration of an active learning user interface (human in the loop)~\cite{Settles_2009} for in-operando improvement of the supporting \ac{ml} algorithms would then allow easy continuous training and thus automatic control and/or optimization of processing based on predefined quality parameters or figures of merit.
Such feedback control would also be beneficial for advanced patterning tasks, where modeling-informed pattern generation and optimization could be coupled with \ac{fib}-processing in a fully automatic way.

Open software interfaces are also required for data analysis, which varies greatly depending on the application typically requiring a tailored approach. 
\ac{ml}-based solutions are similarly very diverse, with application to problems including simple data noise reduction, the broad field of feature recognition and data reduction, super resolution imaging, and full quantification. 
The integration or linking of different methods and data streams thus needs open interfaces.
Therefore, in place of closed all-in-one tools for data analysis and automation, manufacturers should instead provide the infrastructures needed for users to efficiently develop solutions for the broader community.

\paragraph{Data access and management}
Likewise, all data generated should be open and ideally FAIR.
The acronym FAIR, coined by~\citet{Wilkinson2016} in 2016, stands for Findable, Accessible, Interoperable, and Reusable.
The FAIR principles have led to global initiatives in all scientific disciplines~\cite{Wittenburg2020, WorldFAIRproject2024,DGC2023,NationaleForschungsdateninfrastruktur2023}.
In particular, research areas that generate large amounts of data, such as the life sciences and materials science, are driving the development of data harmonization including a comprehensive set of metadata, ontologies, and workflows that are accessible along with the data~\cite{Scheffler2022}. 
For experimentally-derived data, this includes sample tagging, e.g., with QR codes linked to each resulting data set, along with the corresponding experimental workflows documented in electronic lab journals.
One of the biggest challenges with open and FAIR data is the diversity of data sets from different sources. 
The Novel Materials Discovery (NOMAD) laboratory~\cite{NOMAD2023,Draxl2019} can serve as a blueprint here, as it maintains a collection of data repositories (called oases) for data from virtually all codes in computational materials science.

Several factors are crucial for the successful implementation of such data repositories:
\begin{itemize}
    \item The data must be easy to upload and integrate into the repository.
    \item The data must be presented in a common format that allows easy access and efficient searching.
    \item Ideally, additional toolkits such as metadata crawlers or \ac{ml} tools for data analysis or prediction should be offered.
    \item There should be an embargo period option to restrict access to collaborators and/or industry partners before publication or patent.
    \item The data sets themselves, as well as scripts used for data processing and analysis, should be given DOIs to make them citable.
\end{itemize}

With respect to data from (\ac{fib}-related) microscopy and spectroscopy, proprietary file formats are used by various manufacturers. 
While this can be useful when providing additional functionality or easier access to certain data analysis routines, it poses a major problem if there is no way to access the raw unprocessed data to analyze it and make it available. 
In order to integrate such data into a FAIR infrastructure, a standardized and multidimensional open file format specifically for correlative image and spectroscopy data is highly desirable. 
HDF5 would be such an option (where HDF stands for hierarchical data format), allowing raw data to be stored in image or spectroscopy file formats (such as tiff or binary files) together with metadata and all other types of information in a structured way~\cite{HDF}.
However, this would require that instrument manufacturers agree on such a uniform standard, possibly following the lead of the neutron, X-ray and muon research communities where the HDF5-based data format neXus is used~\cite{K_nnecke_2015}.

In general, open and FAIR data enables platform- and tool-independent analysis as well as correlation with data from other sources. 
Furthermore, image processing tools originally developed for other purposes (which may be based on completely different data generation platforms) can be applied and further developed for \ac{fib}-specific analysis. 
A prime example is the open source and platform-independent image processing software ImageJ and its derivatives~\cite{Rasband_1997,Schneider_2012,Schindelin_2012}, which has already been extended through a variety of community-developed plug-ins as well as further adapted and integrated into other programs via its \ac{api}\@.
Here again, \ac{ml} can be leveraged as an extremely powerful tool for automated data evaluation, especially for large and complex data sets from correlative measurements or tomography. 
For example, external programming libraries for \ac{ml}-based image classification and analysis such as scikit-learn~\cite{Pedregosa_2011}, Tensorflow~\cite{abadi2016tensorflow},  PyTorch~\cite{paszke2019pytorch}, dragonfy~\cite{ORSI_2020}%
and Ilastik~\cite{berg2019} are already available, actively developed and are becoming easier to use. 

Ultimately, protocols will need to be established for the pre-processing of raw data into data packages for storage while others parts are to be discarded. 
In this regard, bottom-up harmonization of FAIR data including hierarchical and structured data repositories with full metadata can help with data management and sorting in order to provide ``useful'' raw data for further processing both internally and externally, while avoiding or at least minimizing the typical data graveyards.

\subsubsection{Process modeling}%

\begin{figure*}[tbp]
   \centering
   \includegraphics[width=\textwidth]{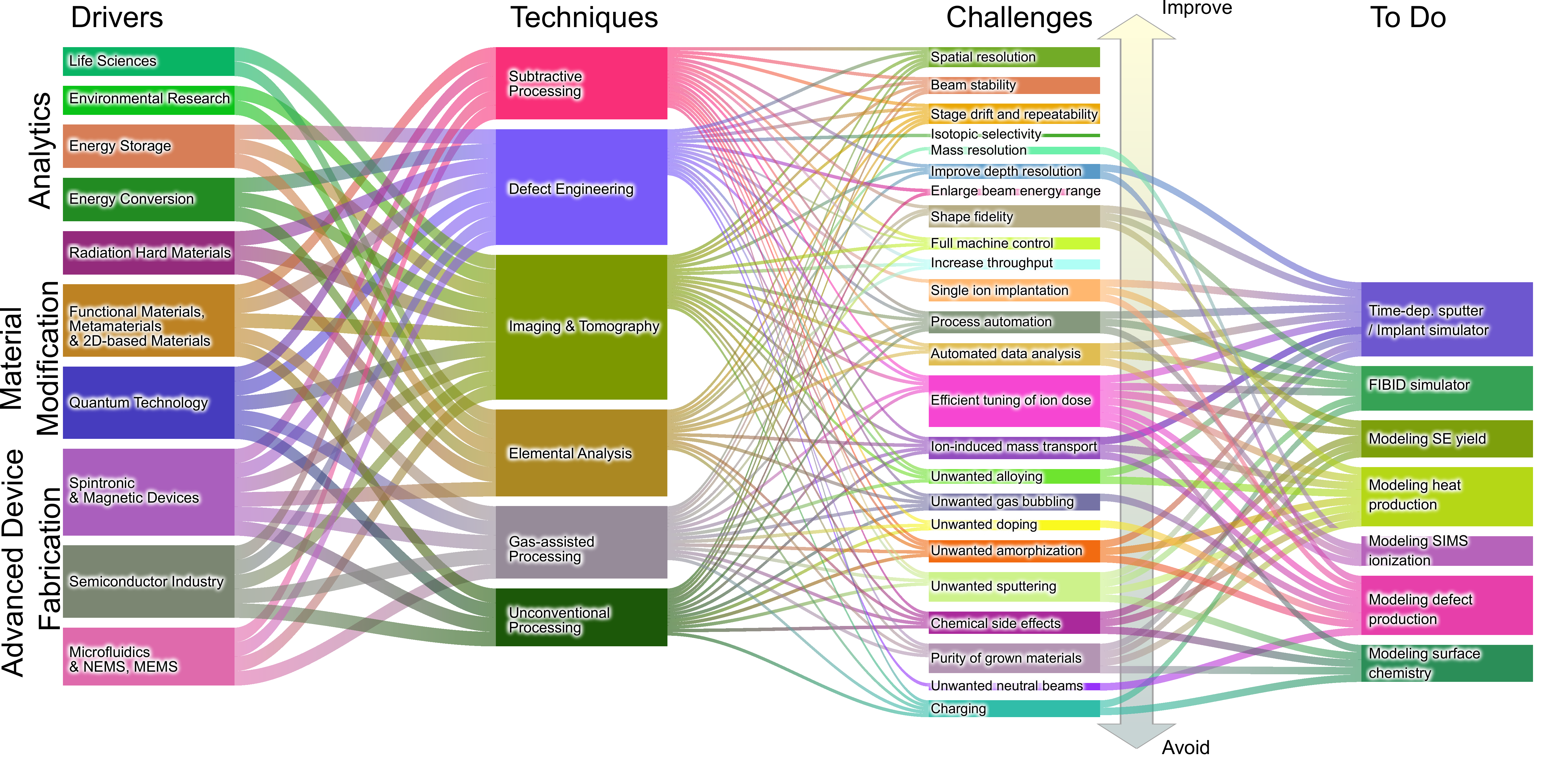}
\caption{\label{fig:sankey_sub4}Developments that are required with respect to process modeling.}%
\end{figure*}

The required developments concerning process modeling are outlined in Figure~\ref{fig:sankey_sub4}. 
Process modeling plays a key role in technology development for \acp{ic}\cite{cea_process_2013,lorenz_challenges_2013}
and together with other types of simulation, is considered one of the drivers of Moore's law.
It is also heavily used beyond traditional device scaling (``Beyond CMOS'' and ``More than Moore''\cite{irds_2022}).
However, in spite of \ac{fib} processes sharing a lot of physics and chemistry with \ac{ic} processes, process modeling is much less used in the \ac{fib} community. 
This is due to the hefty price tag for commercial simulation tools from the \ac{ic} business and the much smaller industry behind \ac{fib} activities.
Nevertheless, the success of process modeling for \ac{ic} development demonstrates its potential for the further development of related \ac{fib} processes.

In the following, process modeling needs for \ac{fib} applications are discussed in terms of new \ac{bca} simulation tools, continuum modeling and approaches for modeling \ac{fibid}, as well as further areas where more advanced computational techniques are critical for progress.

\paragraph{BCA simulations beyond TRIM}
The simulation program most widely used by the \ac{fib} community is
TRIM~\cite{jf_ziegler_particle_2013},
which allows the calculation of range and damage statistics, in addition to rough estimates of sputtering yields in 1D amorphous targets (see Section~\ref{subsubs_specific_bca}). 
Its success is based on the fact that it provides reasonable estimates for any ion--target combination, and the fact that it incorporates a convenient user interface. 
Other \ac{bca} simulation tools that in addition allow consideration of dynamic changes to the target and other target types (2D and 3D, and crystalline targets) would certainly benefit from more intuitive user interfaces. 
Work toward this goal has recently been reported for one particular code\cite{szabo_graphical_2022} and could probably be translated to other codes with moderate effort. 
This would promote adoption of a wider range of \ac{bca} simulation codes by the community.

A further desired feature is the implementation of \ac{se} generation in \ac{bca} codes. 
This will allow more accurate interpretation of \ac{fib} images, will foster model development for precursor dissociation in \ac{fibid} processes, and may also aid  \ac{sii} when detecting single ion impacts using an \ac{se} detector.

\paragraph{Continuum modeling and approaches toward FIBID modeling}
Continuum simulations of surface evolution are hampered by the need for various input parameters and the fact that they neglect processes taking place beneath the surface (see Sections~\ref{subsubs_continuum_principles} and~\ref{subsubs_continuum_applications}). 
However, these simulations are still useful for qualitative investigations. 
Wider applicability is expected from coupling \ac{bca} with continuum simulations, which could enable modeling of surface diffusion and reactions, heat transport and/or relaxation of density changes. 
This could pave the way for accurate and general process simulation of \ac{fib} processes, although taking into account
physical and chemical parameters will still be an issue. 
Here, input from more fundamental methods, or strategies for experimental parameter determination, will be required.
The former is already standard in \ac{ic} process simulation, while an example of the latter in the context of \ac{febid} 
can be found in the use of tilted nanowires on a pedestal for model calibration~\cite{Fowlkes_2018}.

As an alternative to physico-chemical modeling, \ac{ml} has opened up new possibilities in many fields of research.
In \ac{fib} technology, an example of this is a neural network that has been trained on a set of binary design patterns, the corresponding \ac{sem} images, and the ion beam parameters used in their production~\cite{buchnev_deep-learning-assisted_2022}.  
The trained network was then able to accurately predict the patterning outcome for previously unseen designs in nearly real time.
Such a neural network approach holds promise for process control and, in a more general setup, for reducing the number of trial and error iterations in the fabrication of nanostructures. 

\paragraph{Addressing the gap in understanding}%
\label{subs_understanding}

The most important fundamental issues which cannot be addressed at the \ac{bca} or continuum level are:  
\begin{itemize}
  \item Accurate description of the energy loss of energetic ions and recoils accounting for their charge state and the electronic structure of the target.
  \item Production of defects in the irradiated material accounting for electronic excitations.
  \item Evolution of the atomic structure of the target over macroscopic time scales at room or elevated temperatures during annealing.
  \item Effects of defects and impurities on the mechanical, electronic, optical and magnetic properties of the irradiated system.
\end{itemize}

The simulations needed here can be divided into three groups: modeling of the collision, evolution of the system after the ion impact, and assessment of the influence of defects on materials properties.

Future progress in these fields is expected simply due to increasingly powerful computers, which allow for the simulation of larger systems and longer times.  These advances will lead to more widespread use of \ac{md} and \ac{kmc} simulations for specific \ac{fib} applications and to explore the physics of ion-target interaction in general. Currently these research areas are only accessible using \ac{bca}-type simulations. In addition, new developments in computational techniques widen the range of applicability of these advanced methods. For example, \ac{md} simulations with \ac{ml}-derived interatomic potentials trained by \ac{dft} calculations (see Section~\ref{subsubs_md_potentials}) raise the accuracy and reliability of \ac{md} predictions to \ac{dft} level while being applicable to much larger systems.

Combinations of various techniques in a single run (e.g.\ first-principles approaches combined with \ac{ml}-potential \ac{md}, or analytical-potential \ac{md} combined with \ac{bca} methods) can also help elucidate irradiation effects in various situations, such as in strongly non-uniform systems when the main focus is on the description of damage production in one of the subsystems, or over longer time scales. Examples are the response of 2D materials on substrates to ion bombardment~\cite{Kretschmer-2018,Jain2022} or the evolution of the target under high-dose \ac{fib} irradiation. 

Computers with massively parallel architectures and recent development of Ehrenfest dynamics allow prediction of the energy transfer from the ion to the target through ballistic collisions and electronic excitations, and of the energy dissipation in the target involving defect formation without any empirical parameters.
In particular, the direct juxtaposition of experimental data and corresponding simulation results concerning ion transit through atomically thin targets (e.g.\ graphene and other 2D materials) should shed light on the mechanisms of ion neutralization and energy transfer.
Indeed, if the initial/final charge states of the ion and its kinetic energy are known, along with the scattering angle, this information can be correlated with impact parameter-dependent results of atomistic simulations carried out using \ac{tddft}-based Ehrenfest dynamics with the assumption that the nuclei behave as classical particles.
\Ac{tddft}-based Ehrenfest dynamics can also improve parametrization of two-temperature models, bridging the gap between different approaches in multiscale atomistic simulations. Though very useful, their application will remain limited to small systems for quite some time. 

While significant progress in process modeling has been made, there are still challenges for the development of more advanced methods and their application to irradiation simulations. 
One of the challenges is simulation of more complex scenarios of ions passing through 2D materials (touched on above), and interpretation of the theoretical results. After passing through the target, the ion may be in a mixed charge state because the nuclei move in an average potential created by the electron subsystem, which can be in an excited state.  For example, in the experiment, a hydrogen atom (proton) which has passed through the target can be in charge state zero or plus one, while non-integral charge localized around the atom can be found in the simulations. This problem is related to the fundamental question of the connection between quantum-mechanical averages and observable quantities, and requires careful analysis in the context of ion-target interaction. 

There are also challenges in the methodology to be overcome. 
As an example, the common computational schemes of \ac{dft} codes fail at small separations between the nuclei, when atom cores overlap. Even for full-electron codes, numerical instabilities in the eigenvalue solvers are expected at small interatomic distances. There are also other challenges related to spin-polarization energies of atoms, as recent calculations have indicated~\cite{kretschmer_threshold_2022}.  

As for the long-term evolution of the system after irradiation (e.g.\ during post-treatments such as annealing at temperatures of several hundred Kelvin over hours), techniques other than \ac{md} with \ac{dft} or ML/analytical potentials are required. 
The bottleneck is the large number of computational steps to be performed. Parallelization can resolve the issue of system size (ultimately each processor calculates the forces acting on one specific atom) and thus reduce the cost of the time step down to a certain minimum.  However, the time step in \ac{md} simulations should still be at the atomic scale (normally a few femtoseconds), which indicates that $\sim 10^{15}$ steps must still be performed in order to simulate one second of the physical time. 
Even if each time step is $\sim 10^{-4}$ seconds (a typical value for a system comprised of less than 100 atoms), the total CPU time will be several years. In some cases (e.g.\ migration of point defects and their annealing in crystalline solids), \ac{kmc} simulations may provide insights into the evolution of the system on a macroscopic time scale (see Section~\ref{subs_kmc}). However, \ac{kmc} can hardly be applied to systems with a high degree of amorphization or when the atomic structure of the system changes substantially during annealing.

Assessment of the effects of defects on the properties of materials is in a much better situation. 
\ac{dft}-based approaches make it possible to calculate the electronic and magnetic properties of the defective system and directly juxtapose the
results of calculations with the experimental data. In particular, insights into the optical properties and exciton binding
energies of defective materials have recently been obtained\cite{Refaely-Abramson2018}, and photoluminescence spectra of materials with defects such as single-photon emitters have been calculated~\cite{Bertoldo2022,Fischer2021,Mendelson2021,tran2016quantum,Shang2020}.

\section{Executive summary and perspective}
\acresetall{}

A \ac{fib} instrument employs a finely focused beam of ions with an energy of typically \SIrange{2}{30}{keV} that is scanned across a sample to effect modifications down to the nanometer scale.
The technique relies on the transfer of energy from the energetic primary ion to target atoms, resulting in the displacement of the substrate atoms and potentially their removal from the sample.
As such the \ac{fib} instrument is a universal tool for all areas of research enabling comprehensive analyses, maskless local material changes and rapid prototyping of devices.

The aim of this document is to provide an overview of the current state of the art of \ac{fib} technology, its applications and important tool developments, all of which require attention by researchers and technologists developing new \ac{fib} based workflows and instrumentation. This document can thus serve as an important reference work for students, \ac{fib} users, academic and commercial developers of related technologies, and funding agencies. It includes overview tables, providing a bird's eye view of the relevant works in this field. New developments in the various driving fields of research pose new challenges for \ac{fib} techniques, for which we propose targeted solutions and provide a list of technical developments that will be required.

From an instrumentation point of view, the ubiquitous gallium \ac{fib} is an excellent tool for the preparation of both \ac{tem} samples and functional nanostructures. 
However, \ac{fib} sources producing ions of other elements are becoming increasingly common and provide additional degrees of freedom for the fabrication of new functional materials.
In fact, \acp{lmais} and \acp{gfis} were used even before the gallium \ac{lmis} started to control the market.  Newer developments include the \ac{pfib}, which has quickly gained relevance due to its ability to remove large volumes of material while maintaining a lateral resolution sufficient for many applications.
Combined with the very recent addition of the \ac{motis}, about two thirds of the elements of the periodic table can now be accessed for \ac{fib} applications (see Fig.~\ref{fig:pse_sources} and Tab.~\ref{table:ion_sources}).

New applications related to, e.g.\ life and environmental science, quantum technology, energy conversion and storage, present new challenges for existing techniques that will require innovations in ion sources, beam transport and detectors.
Low energy \acp{fib} (\qty{<1}{keV}), for example, are particularly relevant for the semiconductor industry, quantum technology, and applications based on low dimensional materials.
New developments related to heavy and unconventional ion beams are important for \ac{sii} and analytical applications such as \ac{sims}.

The theory of ion-solid interactions and all related processes that may occur in \ac{fib} processing must bridge many orders of magnitude in terms of time and length scales (see Fig.~\ref{fig:methods_overview}). 
Therefore, a tradeoff must always be found between accuracy and the size of the modeled system. 
The \ac{bca} provides insight into a range of \ac{fib}-relevant quantities that are mainly driven by the ballistic phase of the collision cascade, such as sputtering yield, ion penetration depth, and degree of amorphization (see Tab.~\ref{tab:BCA_codes} for an overview of available \ac{bca} codes). Other techniques such as \ac{md}, \ac{kmc}, \ac{dft}, and continuum modeling can address further important aspects. 
These include short-term defect evolution (\ac{md}), long-term system evolution at elevated temperatures (\ac{kmc}), charge state of defects and charge transfer (\ac{dft}), and ion induced chemistry as well as surface processes (continuum modeling). 

An important \ac{fib} application that needs support from theory and simulation in this respect is gas-assisted processing, usually referred to as \ac{fibid}.
This is the nanometer-scale counterpart to industrial and consumer-based 3D printing.
The variety of precursor gases (see Tab.~\ref{tab:precursors}) and the patterning flexibility of the \ac{fib} enable the fabrication of nanoscale devices with unique properties and geometries for emerging applications (see Tab.~\ref{table:fibid}).
Here, an in-depth understanding of the complex processes related to the non-uniform gas flow, the surface kinetics of all involved molecular species, and the secondary particle generation, together with the related energy-dependent dissociation cross-sections of molecular bonds, is needed for accurate process simulation. 

Process modeling is also relevant for other \ac{fib} applications, such as subtractive processing (see Tab.~\ref{table:subtractive}) and defect engineering (see Tab.~\ref{table:property_engineering}).
Applicability, however, will also require an extension of the existing knowledge on ion-solid interactions into the energy range relevant for \ac{fib} applications.
This knowledge transfer would be particularly helpful for calculations of the production rate of charged secondary particles, especially \acp{se}.
The yield of \acp{se} is relevant for several applications, including \ac{sii} and also general imaging and analysis. 
For analysis applications, robust models for the prediction of \ac{si} yields are needed.
In addition to addressing these fundamental questions, efforts are required in the areas of empirical process modeling and shape prediction for additive and subtractive patterning.
Here, open \acp{api} and open source software in combination with \ac{ml} will enable better control over the \aclp{fib}, which will in turn allow higher precision, resulting in faster and more reproducible device fabrication. 

New software and computing technologies enable the collection and evaluation of large data sets, e.g.\ for the \ac{fib} based analytical methods of \ac{fib}-\ac{sims} and serial sectioning (see Tab.~\ref{table:bioimaging}).
\Ac{ml} algorithms can assist with the acquisition, alignment, and segmentation of these typically multidimensional data sets, which should ideally be made available in a FAIR data infrastructure.
Currently, biological applications are driving this field by using data sets from various analytical modalities, including \ac{fib} based tomography, to generate nanoscale 3D representations of biological specimens that provide new insight into microscopic structures and how they function.
Other drivers are energy storage and conversion, meta- and functional materials, and the semiconductor industry.

As the applications of \ac{fib} processing are extremely diverse, it will not be a single technological development that brings a breakthrough. Instead, numerous advances will be needed in order to address the challenges that have been identified.
The resulting developments will often impact several scientific fields, as can be seen from the overview graph in Fig.~\ref{fig:sankey_main}.
Therefore, we look to an exciting future where \ac{fib} processing will grow in importance and open new avenues in many areas of science and technology, spanning quantum technology to the life sciences.

\section*{Acknowledgements}

This publication is based on work from the COST Action FIT4NANO CA19140, supported by COST (European Cooperation in Science and Technology) \url{https://www.cost.eu/} and \url{http://www.fit4nano.eu/}. 
The authors would like to thank the FIT4NANO members for participating in the survey that forms part of Section~\ref{subs_needs}, as well as for input provided during in-person meetings and online discussions.
Specifically we would like to thank (in alphabetical order) Richard J. Curry, Damjana Drobne, Ádám Gali, Hans Hofsäss, Peter Hosemann, Karen L. Kavanagh, Daniel Kiener, Wolfgang Lang, Elisabeth A. Müller Gubler, Pablo A. Postigo, Paul Räcke, Joakim Reuteler, Samuel M. Stavis, Iwona B. Szyma\'{n}ska, Edgar J.D. Vredenbregt, and Robert Winkler  
for their valuable feedback on the manuscript. 
The writing of this document and the implementation of FIT4NANO also benefitted greatly from the guidance and support of worldwide \ac{fib} manufacturers and add-on producers, including (in alphabetical order) Carl Zeiss, Ionoptika, Kleindiek Nanotechnik, Nanores, NenoVision, Orsay Physics, PlasmaSolve, Quantum Design Microscopy, RAITH, Tescan, ThermoFisher Scientific, TOFWERK, and X-Spektrum. 

\section*{Author contributions}

\textbf{Katja Höflich}: Conceptualization (equal); Data curation (equal); Writing---Section 1: Introduction (leading), Writing---Section 4: Applications (leading), Section 5: Roadmap (leading); Writing---review \& editing (equal);  Visualization (leading).
\textbf{Gregor Hlawacek}: Conceptualization (leading); Funding acquisition (leading); Writing---original draft (equal); Writing---review \& editing (equal); Visualization (equal);
\textbf{Gerhard Hobler}: Conceptualization (equal); Writing---Section 3: Simulation Approaches (leading); Writing---review \& editing (equal). 
\textbf{Frances Allen}: Writing---Section 4: Applications (leading); Writing---review \& editing (leading).
\textbf{Tom Wirtz}: Conceptualization (equal); Writing---Section 2: Instrumentation (leading); Writing---review \& editing (equal). 
\textbf{Gemma Rius}: Conceptualization (equal); Data curation (leading); Writing---original draft (equal); Writing---review \& editing (equal);  Visualization (equal). 
\textbf{Lisa McElwee-White}: Writing---original draft (equal); Writing---review \& editing (leading).
\textbf{Lothar Bischoff, Olivier De Castro, Rosa Córdoba, Anne Delobbe, Flyura Djurabekova, Peter Dunne, Natalie Frese, Armin Gölzhäuser, Nico Klingner, Dieter Koelle, Arkady V. Krasheninnikov, Ingo Manke, Mariachiara Manoccio, Paul Mazarov, Johann Michler, Philip Moll, Oleksandr V. Dobrovolskiy, Wolfhard Möller, Markus Osenberg, Francesc Pérez-Murano, Patrick Philip, Matthias Schmidt, José Maria De Teresa, Ivo Utke, Florian Vollnhals}:  Writing---original draft (equal).

\bibliography{merged} 

\begin{thebibliography}{1197}%
\makeatletter
\providecommand \@ifxundefined [1]{%
 \@ifx{#1\undefined}
}%
\providecommand \@ifnum [1]{%
 \ifnum #1\expandafter \@firstoftwo
 \else \expandafter \@secondoftwo
 \fi
}%
\providecommand \@ifx [1]{%
 \ifx #1\expandafter \@firstoftwo
 \else \expandafter \@secondoftwo
 \fi
}%
\providecommand \natexlab [1]{#1}%
\providecommand \enquote  [1]{``#1''}%
\providecommand \bibnamefont  [1]{#1}%
\providecommand \bibfnamefont [1]{#1}%
\providecommand \citenamefont [1]{#1}%
\providecommand \href@noop [0]{\@secondoftwo}%
\providecommand \href [0]{\begingroup \@sanitize@url \@href}%
\providecommand \@href[1]{\@@startlink{#1}\@@href}%
\providecommand \@@href[1]{\endgroup#1\@@endlink}%
\providecommand \@sanitize@url [0]{\catcode `\\12\catcode `\$12\catcode
  `\&12\catcode `\#12\catcode `\^12\catcode `\_12\catcode `\%12\relax}%
\providecommand \@@startlink[1]{}%
\providecommand \@@endlink[0]{}%
\providecommand \url  [0]{\begingroup\@sanitize@url \@url }%
\providecommand \@url [1]{\endgroup\@href {#1}{\urlprefix }}%
\providecommand \urlprefix  [0]{URL }%
\providecommand \Eprint [0]{\href }%
\providecommand \doibase [0]{https://doi.org/}%
\providecommand \selectlanguage [0]{\@gobble}%
\providecommand \bibinfo  [0]{\@secondoftwo}%
\providecommand \bibfield  [0]{\@secondoftwo}%
\providecommand \translation [1]{[#1]}%
\providecommand \BibitemOpen [0]{}%
\providecommand \bibitemStop [0]{}%
\providecommand \bibitemNoStop [0]{.\EOS\space}%
\providecommand \EOS [0]{\spacefactor3000\relax}%
\providecommand \BibitemShut  [1]{\csname bibitem#1\endcsname}%
\let\auto@bib@innerbib\@empty
\bibitem [{\citenamefont {ESA}(2023)}]{LISA}%
  \BibitemOpen
  \bibfield  {author} {\bibinfo {author} {\bibnamefont {ESA}},\ }\href
  {https://www.esa.int/Science_Exploration/Space_Science/LISA_factsheet}
  {\enquote {\bibinfo {title} {{LISA} factsheet},}\ }\bibinfo {howpublished}
  {Online} (\bibinfo {year} {2023}),\ \bibinfo {note}
  {\url{https://www.esa.int/Science_Exploration/Space_Science/LISA_factsheet}}\BibitemShut
  {NoStop}%
\bibitem [{\citenamefont {Tajmar}, \citenamefont {Genovese},\ and\
  \citenamefont {Steiger}(2004)}]{Tajmar_2004}%
  \BibitemOpen
  \bibfield  {author} {\bibinfo {author} {\bibfnamefont {M.}~\bibnamefont
  {Tajmar}}, \bibinfo {author} {\bibfnamefont {A.}~\bibnamefont {Genovese}},\
  and\ \bibinfo {author} {\bibfnamefont {W.}~\bibnamefont {Steiger}},\
  }\bibfield  {title} {\enquote {\bibinfo {title} {Indium field emission
  electric propulsion microthruster experimental characterization},}\ }\href
  {https://doi.org/10.2514/1.9247} {\bibfield  {journal} {\bibinfo  {journal}
  {J. Propul. Power}\ }\textbf {\bibinfo {volume} {20}},\ \bibinfo {pages}
  {211--218} (\bibinfo {year} {2004})}\BibitemShut {NoStop}%
\bibitem [{\citenamefont {Bock}\ and\ \citenamefont
  {Tajmar}(2018)}]{Bock_2018}%
  \BibitemOpen
  \bibfield  {author} {\bibinfo {author} {\bibfnamefont {D.}~\bibnamefont
  {Bock}}\ and\ \bibinfo {author} {\bibfnamefont {M.}~\bibnamefont {Tajmar}},\
  }\bibfield  {title} {\enquote {\bibinfo {title} {Highly miniaturized feep
  propulsion system (nanofeep) for attitude and orbit control of cubesats},}\
  }\href {https://doi.org/10.1016/j.actaastro.2018.01.012} {\bibfield
  {journal} {\bibinfo  {journal} {Acta Astronaut.}\ }\textbf {\bibinfo {volume}
  {144}},\ \bibinfo {pages} {422--428} (\bibinfo {year} {2018})}\BibitemShut
  {NoStop}%
\bibitem [{\citenamefont {Mayer}\ \emph {et~al.}(2007)\citenamefont {Mayer},
  \citenamefont {Giannuzzi}, \citenamefont {Kamino},\ and\ \citenamefont
  {Michael}}]{Mayer_2007}%
  \BibitemOpen
  \bibfield  {author} {\bibinfo {author} {\bibfnamefont {J.}~\bibnamefont
  {Mayer}}, \bibinfo {author} {\bibfnamefont {L.~A.}\ \bibnamefont
  {Giannuzzi}}, \bibinfo {author} {\bibfnamefont {T.}~\bibnamefont {Kamino}},\
  and\ \bibinfo {author} {\bibfnamefont {J.}~\bibnamefont {Michael}},\
  }\bibfield  {title} {\enquote {\bibinfo {title} {{TEM} sample preparation and
  {FIB}-induced damage},}\ }\href {https://doi.org/10.1557/mrs2007.63}
  {\bibfield  {journal} {\bibinfo  {journal} {{MRS} Bulletin}\ }\textbf
  {\bibinfo {volume} {32}},\ \bibinfo {pages} {400--407} (\bibinfo {year}
  {2007})}\BibitemShut {NoStop}%
\bibitem [{\citenamefont {Prosa}\ and\ \citenamefont
  {Larson}(2017)}]{Prosa_2017}%
  \BibitemOpen
  \bibfield  {author} {\bibinfo {author} {\bibfnamefont {T.~J.}\ \bibnamefont
  {Prosa}}\ and\ \bibinfo {author} {\bibfnamefont {D.~J.}\ \bibnamefont
  {Larson}},\ }\bibfield  {title} {\enquote {\bibinfo {title} {Modern
  focused-ion-beam-based site-specific specimen preparation for atom probe
  tomography},}\ }\href {https://doi.org/10.1017/S1431927616012642} {\bibfield
  {journal} {\bibinfo  {journal} {Microsc. Microanal.}\ }\textbf {\bibinfo
  {volume} {23}},\ \bibinfo {pages} {194--209} (\bibinfo {year}
  {2017})}\BibitemShut {NoStop}%
\bibitem [{\citenamefont {Berger}\ \emph {et~al.}(2023)\citenamefont {Berger},
  \citenamefont {Premaraj}, \citenamefont {Ravelli}, \citenamefont {Knoops},
  \citenamefont {L{\'o}pez-Iglesias},\ and\ \citenamefont
  {Peters}}]{Berger_2023}%
  \BibitemOpen
  \bibfield  {author} {\bibinfo {author} {\bibfnamefont {C.}~\bibnamefont
  {Berger}}, \bibinfo {author} {\bibfnamefont {N.}~\bibnamefont {Premaraj}},
  \bibinfo {author} {\bibfnamefont {R.~B.~G.}\ \bibnamefont {Ravelli}},
  \bibinfo {author} {\bibfnamefont {K.}~\bibnamefont {Knoops}}, \bibinfo
  {author} {\bibfnamefont {C.}~\bibnamefont {L{\'o}pez-Iglesias}},\ and\
  \bibinfo {author} {\bibfnamefont {P.~J.}\ \bibnamefont {Peters}},\ }\bibfield
   {title} {{\selectlanguage {en}\enquote {\bibinfo {title} {Cryo-electron
  tomography on focused ion beam lamellae transforms structural cell
  biology},}\ }}\href {https://doi.org/10.1038/s41592-023-01783-5} {\bibfield
  {journal} {\bibinfo  {journal} {Nat. Methods}\ }\textbf {\bibinfo {volume}
  {20}},\ \bibinfo {pages} {499--511} (\bibinfo {year} {2023})}\BibitemShut
  {NoStop}%
\bibitem [{\citenamefont {Bassim}\ and\ \citenamefont
  {Notte}(2019)}]{Bassim2019}%
  \BibitemOpen
  \bibfield  {author} {\bibinfo {author} {\bibfnamefont {N.}~\bibnamefont
  {Bassim}}\ and\ \bibinfo {author} {\bibfnamefont {J.~A.}\ \bibnamefont
  {Notte}},\ }\bibfield  {title} {\enquote {\bibinfo {title} {Focused ion beam
  instruments},}\ }in\ \href {https://doi.org/10.31399/asm.hb.v10.a0006677}
  {\emph {\bibinfo {booktitle} {Materials Characterization}}}\ (\bibinfo
  {publisher} {{ASM} International},\ \bibinfo {year} {2019})\ pp.\ \bibinfo
  {pages} {635--670}\BibitemShut {NoStop}%
\bibitem [{\citenamefont {Tondare}(2005)}]{Tondare_2005}%
  \BibitemOpen
  \bibfield  {author} {\bibinfo {author} {\bibfnamefont {V.~N.}\ \bibnamefont
  {Tondare}},\ }\bibfield  {title} {\enquote {\bibinfo {title} {Quest for high
  brightness, monochromatic noble gas ion sources},}\ }\href
  {https://doi.org/10.1116/1.2101792} {\bibfield  {journal} {\bibinfo
  {journal} {J. Vac. Sci. Technol. A}\ }\textbf {\bibinfo {volume} {23}},\
  \bibinfo {pages} {1498} (\bibinfo {year} {2005})}\BibitemShut {NoStop}%
\bibitem [{\citenamefont {Langmuir}(1937)}]{Langmuir_1937}%
  \BibitemOpen
  \bibfield  {author} {\bibinfo {author} {\bibfnamefont {D.~B.}\ \bibnamefont
  {Langmuir}},\ }\bibfield  {title} {\enquote {\bibinfo {title} {Theoretical
  limitations of cathode-ray tubes},}\ }in\ \href
  {https://doi.org/10.1109/jrproc.1937.228419} {\emph {\bibinfo {booktitle}
  {Proceedings of the IRE}}},\ Vol.~\bibinfo {volume} {25}\ (\bibinfo
  {publisher} {Institute of Electrical and Electronics Engineers ({IEEE})},\
  \bibinfo {year} {1937})\ pp.\ \bibinfo {pages} {977--991}\BibitemShut
  {NoStop}%
\bibitem [{\citenamefont {Wolf}(1995)}]{Wolf1995}%
  \BibitemOpen
  \bibfield  {author} {\bibinfo {author} {\bibfnamefont {B.}~\bibnamefont
  {Wolf}},\ }\href {https://doi.org/10.1201/9781315214054} {\emph {\bibinfo
  {title} {Handbook of ion sources}}}\ (\bibinfo  {publisher} {CRC Press},\
  \bibinfo {address} {Bosa Roca},\ \bibinfo {year} {1995})\BibitemShut
  {NoStop}%
\bibitem [{\citenamefont {Zorzos}\ and\ \citenamefont
  {Lozano}(2008)}]{Zorzos_2008}%
  \BibitemOpen
  \bibfield  {author} {\bibinfo {author} {\bibfnamefont {A.~N.}\ \bibnamefont
  {Zorzos}}\ and\ \bibinfo {author} {\bibfnamefont {P.~C.}\ \bibnamefont
  {Lozano}},\ }\bibfield  {title} {\enquote {\bibinfo {title} {The use of ionic
  liquid ion sources in focused ion beam applications},}\ }\href
  {https://doi.org/10.1116/1.2991619} {\bibfield  {journal} {\bibinfo
  {journal} {J. Vac. Sci. Technol. B}\ }\textbf {\bibinfo {volume} {26}},\
  \bibinfo {pages} {2097--2102} (\bibinfo {year} {2008})}\BibitemShut {NoStop}%
\bibitem [{\citenamefont {Bischoff}\ \emph {et~al.}(2016)\citenamefont
  {Bischoff}, \citenamefont {Mazarov}, \citenamefont {Bruchhaus},\ and\
  \citenamefont {Gierak}}]{Bischoff_2016}%
  \BibitemOpen
  \bibfield  {author} {\bibinfo {author} {\bibfnamefont {L.}~\bibnamefont
  {Bischoff}}, \bibinfo {author} {\bibfnamefont {P.}~\bibnamefont {Mazarov}},
  \bibinfo {author} {\bibfnamefont {L.}~\bibnamefont {Bruchhaus}},\ and\
  \bibinfo {author} {\bibfnamefont {J.}~\bibnamefont {Gierak}},\ }\bibfield
  {title} {\enquote {\bibinfo {title} {{Liquid metal alloy ion sources—An
  alternative for focussed ion beam technology}},}\ }\href
  {https://doi.org/10.1063/1.4947095} {\bibfield  {journal} {\bibinfo
  {journal} {Appl. Phys. Rev.}\ }\textbf {\bibinfo {volume} {3}},\ \bibinfo
  {pages} {021101} (\bibinfo {year} {2016})}\BibitemShut {NoStop}%
\bibitem [{\citenamefont {Swanson}(1994)}]{Swanson_1994}%
  \BibitemOpen
  \bibfield  {author} {\bibinfo {author} {\bibfnamefont {L.~W.}\ \bibnamefont
  {Swanson}},\ }\bibfield  {title} {\enquote {\bibinfo {title} {{Use of the
  liquid metal ion source for focused beam applications}},}\ }\href
  {https://doi.org/10.1016/0169-4332(94)90327-1} {\bibfield  {journal}
  {\bibinfo  {journal} {Appl. Surf. Sci.}\ }\textbf {\bibinfo {volume}
  {76-77}},\ \bibinfo {pages} {80--88} (\bibinfo {year} {1994})}\BibitemShut
  {NoStop}%
\bibitem [{\citenamefont {{Orsay physics}}(2021{\natexlab{a}})}]{Chroma_2021}%
  \BibitemOpen
  \bibfield  {author} {\bibinfo {author} {\bibnamefont {{Orsay physics}}},\
  }\href {https://www.orsayphysics.com/fib} {\enquote {\bibinfo {title} {Orsay
  physics datasheet chroma {ExB}},}\ }\bibinfo {howpublished} {Online}
  (\bibinfo {year} {2021}{\natexlab{a}}),\ \bibinfo {note}
  {\url{https://www.orsayphysics.com/fib}}\BibitemShut {NoStop}%
\bibitem [{\citenamefont {Bauerdick}\ \emph {et~al.}(2013)\citenamefont
  {Bauerdick}, \citenamefont {Bruchhaus}, \citenamefont {Mazarov},
  \citenamefont {Nadzeyka}, \citenamefont {Jede}, \citenamefont {Fridmann},
  \citenamefont {Sanabia}, \citenamefont {Gila},\ and\ \citenamefont
  {Appleton}}]{Bauerdick_2013}%
  \BibitemOpen
  \bibfield  {author} {\bibinfo {author} {\bibfnamefont {S.}~\bibnamefont
  {Bauerdick}}, \bibinfo {author} {\bibfnamefont {L.}~\bibnamefont
  {Bruchhaus}}, \bibinfo {author} {\bibfnamefont {P.}~\bibnamefont {Mazarov}},
  \bibinfo {author} {\bibfnamefont {A.}~\bibnamefont {Nadzeyka}}, \bibinfo
  {author} {\bibfnamefont {R.}~\bibnamefont {Jede}}, \bibinfo {author}
  {\bibfnamefont {J.}~\bibnamefont {Fridmann}}, \bibinfo {author}
  {\bibfnamefont {J.~E.}\ \bibnamefont {Sanabia}}, \bibinfo {author}
  {\bibfnamefont {B.}~\bibnamefont {Gila}},\ and\ \bibinfo {author}
  {\bibfnamefont {B.~R.}\ \bibnamefont {Appleton}},\ }\bibfield  {title}
  {\enquote {\bibinfo {title} {Multispecies focused ion beam lithography system
  and its applications},}\ }\href {https://doi.org/10.1116/1.4824327}
  {\bibfield  {journal} {\bibinfo  {journal} {J. Vac. Sci. Technol. B}\
  }\textbf {\bibinfo {volume} {31}},\ \bibinfo {pages} {06F404} (\bibinfo
  {year} {2013})}\BibitemShut {NoStop}%
\bibitem [{\citenamefont {Lozano}(2005)}]{Lozano_2005b}%
  \BibitemOpen
  \bibfield  {author} {\bibinfo {author} {\bibfnamefont {P.~C.}\ \bibnamefont
  {Lozano}},\ }\bibfield  {title} {\enquote {\bibinfo {title} {Energy
  properties of an {EMI}-im ionic liquid ion source},}\ }\href
  {https://doi.org/10.1088/0022-3727/39/1/020} {\bibfield  {journal} {\bibinfo
  {journal} {J. Phys. D: Appl. Phys.}\ }\textbf {\bibinfo {volume} {39}},\
  \bibinfo {pages} {126--134} (\bibinfo {year} {2005})}\BibitemShut {NoStop}%
\bibitem [{\citenamefont {Fedkiw}\ and\ \citenamefont
  {Lozano}(2009)}]{Fedkiw_2009}%
  \BibitemOpen
  \bibfield  {author} {\bibinfo {author} {\bibfnamefont {T.~P.}\ \bibnamefont
  {Fedkiw}}\ and\ \bibinfo {author} {\bibfnamefont {P.~C.}\ \bibnamefont
  {Lozano}},\ }\bibfield  {title} {\enquote {\bibinfo {title} {Development and
  characterization of an iodine field emission ion source for focused ion beam
  applications},}\ }\href {https://doi.org/10.1116/1.3253604} {\bibfield
  {journal} {\bibinfo  {journal} {J. Vac. Sci. Technol. B}\ }\textbf {\bibinfo
  {volume} {27}},\ \bibinfo {pages} {2648} (\bibinfo {year}
  {2009})}\BibitemShut {NoStop}%
\bibitem [{\citenamefont {Smith}\ \emph {et~al.}(2006)\citenamefont {Smith},
  \citenamefont {Skoczylas}, \citenamefont {Kellogg}, \citenamefont {Kinion},
  \citenamefont {Tesch}, \citenamefont {Sutherland}, \citenamefont
  {Aanesland},\ and\ \citenamefont {Boswell}}]{Smith_2006}%
  \BibitemOpen
  \bibfield  {author} {\bibinfo {author} {\bibfnamefont {N.~S.}\ \bibnamefont
  {Smith}}, \bibinfo {author} {\bibfnamefont {W.~P.}\ \bibnamefont
  {Skoczylas}}, \bibinfo {author} {\bibfnamefont {S.~M.}\ \bibnamefont
  {Kellogg}}, \bibinfo {author} {\bibfnamefont {D.~E.}\ \bibnamefont {Kinion}},
  \bibinfo {author} {\bibfnamefont {P.~P.}\ \bibnamefont {Tesch}}, \bibinfo
  {author} {\bibfnamefont {O.}~\bibnamefont {Sutherland}}, \bibinfo {author}
  {\bibfnamefont {A.}~\bibnamefont {Aanesland}},\ and\ \bibinfo {author}
  {\bibfnamefont {R.~W.}\ \bibnamefont {Boswell}},\ }\bibfield  {title}
  {\enquote {\bibinfo {title} {{High brightness inductively coupled plasma
  source for high current focused ion beam applications}},}\ }\href
  {https://doi.org/10.1116/1.2366617} {\bibfield  {journal} {\bibinfo
  {journal} {J. Vac. Sci. Technol. B}\ }\textbf {\bibinfo {volume} {24}},\
  \bibinfo {pages} {2902} (\bibinfo {year} {2006})}\BibitemShut {NoStop}%
\bibitem [{\citenamefont {Leer}\ and\ \citenamefont {Dutka}(2019)}]{Leer_2019}%
  \BibitemOpen
  \bibfield  {author} {\bibinfo {author} {\bibfnamefont {B.~V.}\ \bibnamefont
  {Leer}}\ and\ \bibinfo {author} {\bibfnamefont {M.}~\bibnamefont {Dutka}},\
  }\bibfield  {title} {\enquote {\bibinfo {title} {Latest developments in
  multiple ion species plasma {FIB} technology},}\ }\href
  {https://doi.org/10.1017/s1431927619003581} {\bibfield  {journal} {\bibinfo
  {journal} {Microsc. Microanal.}\ }\textbf {\bibinfo {volume} {25}},\ \bibinfo
  {pages} {570--571} (\bibinfo {year} {2019})}\BibitemShut {NoStop}%
\bibitem [{\citenamefont {{Orsay physics}}(2021{\natexlab{b}})}]{iFIB_2021}%
  \BibitemOpen
  \bibfield  {author} {\bibinfo {author} {\bibnamefont {{Orsay physics}}},\
  }\href {https://www.orsayphysics.com/fib} {\enquote {\bibinfo {title} {Orsay
  physics datasheet i{FIB}},}\ }\bibinfo {howpublished} {Online} (\bibinfo
  {year} {2021}{\natexlab{b}}),\ \bibinfo {note}
  {\url{https://www.orsayphysics.com/fib}}\BibitemShut {NoStop}%
\bibitem [{\citenamefont {Ward}, \citenamefont {Notte},\ and\ \citenamefont
  {Economou}(2006)}]{Ward_2006}%
  \BibitemOpen
  \bibfield  {author} {\bibinfo {author} {\bibfnamefont {B.~W.}\ \bibnamefont
  {Ward}}, \bibinfo {author} {\bibfnamefont {J.~A.}\ \bibnamefont {Notte}},\
  and\ \bibinfo {author} {\bibfnamefont {N.~P.}\ \bibnamefont {Economou}},\
  }\bibfield  {title} {\enquote {\bibinfo {title} {Helium ion microscope: a new
  tool for nanoscale microscopy and metrology},}\ }\href
  {https://doi.org/10.1116/1.2357967} {\bibfield  {journal} {\bibinfo
  {journal} {J. Vac. Sci. Technol. B}\ }\textbf {\bibinfo {volume} {24}},\
  \bibinfo {pages} {2871} (\bibinfo {year} {2006})}\BibitemShut {NoStop}%
\bibitem [{\citenamefont {Hill}\ and\ \citenamefont
  {Faridur~Rahman}(2011)}]{Hill_2011}%
  \BibitemOpen
  \bibfield  {author} {\bibinfo {author} {\bibfnamefont {R.}~\bibnamefont
  {Hill}}\ and\ \bibinfo {author} {\bibfnamefont {F.~H. M.~M.}\ \bibnamefont
  {Faridur~Rahman}},\ }\bibfield  {title} {\enquote {\bibinfo {title}
  {{Advances in helium ion microscopy}},}\ }\href
  {https://doi.org/10.1016/j.nima.2010.12.123} {\bibfield  {journal} {\bibinfo
  {journal} {Nucl. Instrum. Meth. Phys. Res., A}\ }\textbf {\bibinfo {volume}
  {645}},\ \bibinfo {pages} {96--101} (\bibinfo {year} {2011})}\BibitemShut
  {NoStop}%
\bibitem [{\citenamefont {Notte}\ and\ \citenamefont
  {Huang}(2016)}]{Notte_2016}%
  \BibitemOpen
  \bibfield  {author} {\bibinfo {author} {\bibfnamefont {J.}~\bibnamefont
  {Notte}}\ and\ \bibinfo {author} {\bibfnamefont {J.}~\bibnamefont {Huang}},\
  }\bibfield  {title} {\enquote {\bibinfo {title} {The helium ion
  microscope},}\ }in\  \cite{Hlawacek_2016},\ Chap.~\bibinfo {chapter} {1},
  pp.\ \bibinfo {pages} {3--30}\BibitemShut {NoStop}%
\bibitem [{\citenamefont {Ernst}\ \emph {et~al.}(1993)\citenamefont {Ernst},
  \citenamefont {Bozdech}, \citenamefont {Schmidt}, \citenamefont {Schmidt},\
  and\ \citenamefont {Larkins}}]{Ernst_1993}%
  \BibitemOpen
  \bibfield  {author} {\bibinfo {author} {\bibfnamefont {N.}~\bibnamefont
  {Ernst}}, \bibinfo {author} {\bibfnamefont {G.}~\bibnamefont {Bozdech}},
  \bibinfo {author} {\bibfnamefont {H.}~\bibnamefont {Schmidt}}, \bibinfo
  {author} {\bibfnamefont {W.~A.}\ \bibnamefont {Schmidt}},\ and\ \bibinfo
  {author} {\bibfnamefont {G.~L.}\ \bibnamefont {Larkins}},\ }\bibfield
  {title} {\enquote {\bibinfo {title} {On the full-width-at-half-maximum of
  field ion energy distributions},}\ }\href
  {https://doi.org/10.1016/0169-4332(93)90301-q} {\bibfield  {journal}
  {\bibinfo  {journal} {Appl. Surf. Sci.}\ }\textbf {\bibinfo {volume} {67}},\
  \bibinfo {pages} {111--117} (\bibinfo {year} {1993})}\BibitemShut {NoStop}%
\bibitem [{\citenamefont {Hlawacek}\ and\ \citenamefont
  {Gölzhäuser}(2016)}]{Hlawacek_2016}%
  \BibitemOpen
  \bibinfo {editor} {\bibfnamefont {G.}~\bibnamefont {Hlawacek}}\ and\ \bibinfo
  {editor} {\bibfnamefont {A.}~\bibnamefont {Gölzhäuser}},\ eds.,\ \href
  {https://doi.org/10.1007/978-3-319-41990-9} {\emph {\bibinfo {title} {Helium
  ion microscopy}}}\ (\bibinfo  {publisher} {Springer International
  Publishing},\ \bibinfo {year} {2016})\BibitemShut {NoStop}%
\bibitem [{\citenamefont {Livengood}\ \emph
  {et~al.}(2011{\natexlab{a}})\citenamefont {Livengood}, \citenamefont {Tan},
  \citenamefont {Hallstein}, \citenamefont {Notte}, \citenamefont {McVey},\
  and\ \citenamefont {Rahman}}]{Livengood_2011}%
  \BibitemOpen
  \bibfield  {author} {\bibinfo {author} {\bibfnamefont {R.~H.}\ \bibnamefont
  {Livengood}}, \bibinfo {author} {\bibfnamefont {S.}~\bibnamefont {Tan}},
  \bibinfo {author} {\bibfnamefont {R.}~\bibnamefont {Hallstein}}, \bibinfo
  {author} {\bibfnamefont {J.}~\bibnamefont {Notte}}, \bibinfo {author}
  {\bibfnamefont {S.}~\bibnamefont {McVey}},\ and\ \bibinfo {author}
  {\bibfnamefont {F.~H. M.~F.}\ \bibnamefont {Rahman}},\ }\bibfield  {title}
  {\enquote {\bibinfo {title} {The neon gas field ion source---a first
  characterization of neon nanomachining properties},}\ }\href
  {https://doi.org/10.1016/j.nima.2010.12.220} {\bibfield  {journal} {\bibinfo
  {journal} {Nucl. Instrum. Meth. Phys. Res. A}\ }\textbf {\bibinfo {volume}
  {645}},\ \bibinfo {pages} {136--140} (\bibinfo {year}
  {2011}{\natexlab{a}})}\BibitemShut {NoStop}%
\bibitem [{\citenamefont {Klingner}\ \emph {et~al.}(2020)\citenamefont
  {Klingner}, \citenamefont {Hlawacek}, \citenamefont {Mazarov}, \citenamefont
  {Pilz}, \citenamefont {Meyer},\ and\ \citenamefont
  {Bischoff}}]{Klingner_2020}%
  \BibitemOpen
  \bibfield  {author} {\bibinfo {author} {\bibfnamefont {N.}~\bibnamefont
  {Klingner}}, \bibinfo {author} {\bibfnamefont {G.}~\bibnamefont {Hlawacek}},
  \bibinfo {author} {\bibfnamefont {P.}~\bibnamefont {Mazarov}}, \bibinfo
  {author} {\bibfnamefont {W.}~\bibnamefont {Pilz}}, \bibinfo {author}
  {\bibfnamefont {F.}~\bibnamefont {Meyer}},\ and\ \bibinfo {author}
  {\bibfnamefont {L.}~\bibnamefont {Bischoff}},\ }\bibfield  {title} {\enquote
  {\bibinfo {title} {Imaging and milling resolution of light ion beams from
  helium ion microscopy and {FIBs} driven by liquid metal alloy ion sources},}\
  }\href {https://doi.org/10.3762/bjnano.11.156} {\bibfield  {journal}
  {\bibinfo  {journal} {Beilstein J. Nanotechnol.}\ }\textbf {\bibinfo {volume}
  {11}},\ \bibinfo {pages} {1742--1749} (\bibinfo {year} {2020})}\BibitemShut
  {NoStop}%
\bibitem [{\citenamefont {Steele}\ \emph {et~al.}(2017)\citenamefont {Steele},
  \citenamefont {Schwarzkopf}, \citenamefont {McClelland},\ and\ \citenamefont
  {Knuffman}}]{Steele_2017}%
  \BibitemOpen
  \bibfield  {author} {\bibinfo {author} {\bibfnamefont {A.~V.}\ \bibnamefont
  {Steele}}, \bibinfo {author} {\bibfnamefont {A.}~\bibnamefont {Schwarzkopf}},
  \bibinfo {author} {\bibfnamefont {J.~J.}\ \bibnamefont {McClelland}},\ and\
  \bibinfo {author} {\bibfnamefont {B.}~\bibnamefont {Knuffman}},\ }\bibfield
  {title} {\enquote {\bibinfo {title} {{High-brightness Cs focused ion beam
  from a cold-atomic-beam ion source}},}\ }\href
  {https://doi.org/10.1088/2399-1984/aa6a48} {\bibfield  {journal} {\bibinfo
  {journal} {Nano Futur.}\ }\textbf {\bibinfo {volume} {1}},\ \bibinfo {pages}
  {015005} (\bibinfo {year} {2017})}\BibitemShut {NoStop}%
\bibitem [{\citenamefont {ten Haaf}\ \emph {et~al.}(2018)\citenamefont {ten
  Haaf}, \citenamefont {Wouters}, \citenamefont {Nijhof}, \citenamefont
  {Mutsaers},\ and\ \citenamefont {Vredenbregt}}]{Haaf_2018}%
  \BibitemOpen
  \bibfield  {author} {\bibinfo {author} {\bibfnamefont {G.}~\bibnamefont {ten
  Haaf}}, \bibinfo {author} {\bibfnamefont {S.~H.~W.}\ \bibnamefont {Wouters}},
  \bibinfo {author} {\bibfnamefont {D.~F.~J.}\ \bibnamefont {Nijhof}}, \bibinfo
  {author} {\bibfnamefont {P.~H.~A.}\ \bibnamefont {Mutsaers}},\ and\ \bibinfo
  {author} {\bibfnamefont {E.~J.~D.}\ \bibnamefont {Vredenbregt}},\ }\bibfield
  {title} {\enquote {\bibinfo {title} {{Measurements of the energy distribution
  of a high brightness rubidium ion beam}},}\ }\href
  {https://doi.org/10.1016/j.ultramic.2018.03.014} {\bibfield  {journal}
  {\bibinfo  {journal} {Ultramicroscopy}\ }\textbf {\bibinfo {volume} {190}},\
  \bibinfo {pages} {12--20} (\bibinfo {year} {2018})}\BibitemShut {NoStop}%
\bibitem [{\citenamefont {Moufarej}\ \emph {et~al.}(2017)\citenamefont
  {Moufarej}, \citenamefont {Vielle-Grosjean}, \citenamefont {Khalili},
  \citenamefont {McCulloch}, \citenamefont {Robicheaux}, \citenamefont
  {Picard},\ and\ \citenamefont {Comparat}}]{Moufarej_2017}%
  \BibitemOpen
  \bibfield  {author} {\bibinfo {author} {\bibfnamefont {E.}~\bibnamefont
  {Moufarej}}, \bibinfo {author} {\bibfnamefont {M.}~\bibnamefont
  {Vielle-Grosjean}}, \bibinfo {author} {\bibfnamefont {G.}~\bibnamefont
  {Khalili}}, \bibinfo {author} {\bibfnamefont {A.~J.}\ \bibnamefont
  {McCulloch}}, \bibinfo {author} {\bibfnamefont {F.}~\bibnamefont
  {Robicheaux}}, \bibinfo {author} {\bibfnamefont {Y.~J.}\ \bibnamefont
  {Picard}},\ and\ \bibinfo {author} {\bibfnamefont {D.}~\bibnamefont
  {Comparat}},\ }\bibfield  {title} {\enquote {\bibinfo {title} {{Forced field
  ionization of Rydberg states for the production of monochromatic beams}},}\
  }\href {https://doi.org/10.1103/PhysRevA.95.043409} {\bibfield  {journal}
  {\bibinfo  {journal} {Phys. Rev. A}\ }\textbf {\bibinfo {volume} {95}},\
  \bibinfo {pages} {043409} (\bibinfo {year} {2017})}\BibitemShut {NoStop}%
\bibitem [{\citenamefont {Knuffman}, \citenamefont {Steele},\ and\
  \citenamefont {McClelland}(2013)}]{Knuffman_2013}%
  \BibitemOpen
  \bibfield  {author} {\bibinfo {author} {\bibfnamefont {B.}~\bibnamefont
  {Knuffman}}, \bibinfo {author} {\bibfnamefont {A.~V.}\ \bibnamefont
  {Steele}},\ and\ \bibinfo {author} {\bibfnamefont {J.~J.}\ \bibnamefont
  {McClelland}},\ }\bibfield  {title} {\enquote {\bibinfo {title} {{Cold atomic
  beam ion source for focused ion beam applications}},}\ }\href
  {https://doi.org/10.1063/1.4816248} {\bibfield  {journal} {\bibinfo
  {journal} {J. Appl. Phys.}\ }\textbf {\bibinfo {volume} {114}},\ \bibinfo
  {pages} {044303} (\bibinfo {year} {2013})}\BibitemShut {NoStop}%
\bibitem [{\citenamefont {van Kouwen}\ and\ \citenamefont
  {Kruit}(2018)}]{Kouwen_2018}%
  \BibitemOpen
  \bibfield  {author} {\bibinfo {author} {\bibfnamefont {L.}~\bibnamefont {van
  Kouwen}}\ and\ \bibinfo {author} {\bibfnamefont {P.}~\bibnamefont {Kruit}},\
  }\bibfield  {title} {\enquote {\bibinfo {title} {Brightness measurements of
  the nano-aperture ion source},}\ }\href {https://doi.org/10.1116/1.5048054}
  {\bibfield  {journal} {\bibinfo  {journal} {J. Vac. Sci. Technol. B}\
  }\textbf {\bibinfo {volume} {36}},\ \bibinfo {pages} {06J901} (\bibinfo
  {year} {2018})}\BibitemShut {NoStop}%
\bibitem [{\citenamefont {van Kan}\ \emph {et~al.}(2020)\citenamefont {van
  Kan}, \citenamefont {Pang}, \citenamefont {Basu}, \citenamefont {Dou},
  \citenamefont {Gokul}, \citenamefont {Tarino}, \citenamefont {Tregidga},
  \citenamefont {Roy},\ and\ \citenamefont {Tan}}]{Kan_2020}%
  \BibitemOpen
  \bibfield  {author} {\bibinfo {author} {\bibfnamefont {J.~A.}\ \bibnamefont
  {van Kan}}, \bibinfo {author} {\bibfnamefont {R.}~\bibnamefont {Pang}},
  \bibinfo {author} {\bibfnamefont {T.}~\bibnamefont {Basu}}, \bibinfo {author}
  {\bibfnamefont {Y.}~\bibnamefont {Dou}}, \bibinfo {author} {\bibnamefont
  {Gokul}}, \bibinfo {author} {\bibfnamefont {N.}~\bibnamefont {Tarino}},
  \bibinfo {author} {\bibfnamefont {J.}~\bibnamefont {Tregidga}}, \bibinfo
  {author} {\bibfnamefont {S.~C.}\ \bibnamefont {Roy}},\ and\ \bibinfo {author}
  {\bibfnamefont {H.~M.}\ \bibnamefont {Tan}},\ }\bibfield  {title} {\enquote
  {\bibinfo {title} {Considerations for the nano aperture ion source:
  geometrical design and electrical control},}\ }\href
  {https://doi.org/10.1063/1.5128657} {\bibfield  {journal} {\bibinfo
  {journal} {Rev. Sci. Instrum.}\ }\textbf {\bibinfo {volume} {91}},\ \bibinfo
  {pages} {013310} (\bibinfo {year} {2020})}\BibitemShut {NoStop}%
\bibitem [{\citenamefont {M{\"{u}}ller}(1951)}]{Mueller_1951}%
  \BibitemOpen
  \bibfield  {author} {\bibinfo {author} {\bibfnamefont {E.~W.}\ \bibnamefont
  {M{\"{u}}ller}},\ }\bibfield  {title} {\enquote {\bibinfo {title} {{Das
  Feldionenmikroskop}},}\ }\href {https://doi.org/10.1007/bf01329651}
  {\bibfield  {journal} {\bibinfo  {journal} {Zeitschrift f{\"{u}}r Phys.}\
  }\textbf {\bibinfo {volume} {131}},\ \bibinfo {pages} {136--142} (\bibinfo
  {year} {1951})}\BibitemShut {NoStop}%
\bibitem [{\citenamefont {Levi-Setti}(1974)}]{LeviSetti_1974}%
  \BibitemOpen
  \bibfield  {author} {\bibinfo {author} {\bibfnamefont {R.}~\bibnamefont
  {Levi-Setti}},\ }\bibfield  {title} {\enquote {\bibinfo {title} {{Proton
  scanning microscopy: feasibility and promise}},}\ }\href@noop {} {\bibfield
  {journal} {\bibinfo  {journal} {Scan. Electron Microsc.}\ }\textbf {\bibinfo
  {volume} {7}},\ \bibinfo {pages} {125--134} (\bibinfo {year}
  {1974})}\BibitemShut {NoStop}%
\bibitem [{\citenamefont {Escovitz}, \citenamefont {Fox},\ and\ \citenamefont
  {Levi-Setti}(1975)}]{Escovitz_1975}%
  \BibitemOpen
  \bibfield  {author} {\bibinfo {author} {\bibfnamefont {W.~H.}\ \bibnamefont
  {Escovitz}}, \bibinfo {author} {\bibfnamefont {T.~R.}\ \bibnamefont {Fox}},\
  and\ \bibinfo {author} {\bibfnamefont {R.}~\bibnamefont {Levi-Setti}},\
  }\bibfield  {title} {\enquote {\bibinfo {title} {Scanning transmission ion
  microscope with a field ion source.}}\ }\href
  {https://doi.org/10.1073/pnas.72.5.1826} {\bibfield  {journal} {\bibinfo
  {journal} {Proc. Natl. Acad. Sci.}\ }\textbf {\bibinfo {volume} {72}},\
  \bibinfo {pages} {1826--1828} (\bibinfo {year} {1975})}\BibitemShut {NoStop}%
\bibitem [{\citenamefont {Krohn}\ and\ \citenamefont
  {Ringo}(1975)}]{Krohn_1975}%
  \BibitemOpen
  \bibfield  {author} {\bibinfo {author} {\bibfnamefont {V.~E.}\ \bibnamefont
  {Krohn}}\ and\ \bibinfo {author} {\bibfnamefont {G.~R.}\ \bibnamefont
  {Ringo}},\ }\bibfield  {title} {\enquote {\bibinfo {title} {Ion source of
  high brightness using liquid metal},}\ }\href
  {https://doi.org/10.1063/1.88540} {\bibfield  {journal} {\bibinfo  {journal}
  {Appl. Phys. Lett.}\ }\textbf {\bibinfo {volume} {27}},\ \bibinfo {pages}
  {479--481} (\bibinfo {year} {1975})}\BibitemShut {NoStop}%
\bibitem [{\citenamefont {Clampitt}, \citenamefont {Aitken},\ and\
  \citenamefont {Jefferies}(1975)}]{Clampitt_1975}%
  \BibitemOpen
  \bibfield  {author} {\bibinfo {author} {\bibfnamefont {R.}~\bibnamefont
  {Clampitt}}, \bibinfo {author} {\bibfnamefont {K.~L.}\ \bibnamefont
  {Aitken}},\ and\ \bibinfo {author} {\bibfnamefont {D.~K.}\ \bibnamefont
  {Jefferies}},\ }\bibfield  {title} {\enquote {\bibinfo {title} {Abstract:
  intense field-emission ion source of liquid metals},}\ }\href
  {https://doi.org/10.1116/1.568496} {\bibfield  {journal} {\bibinfo  {journal}
  {J. Vac. Sci. Technol.}\ }\textbf {\bibinfo {volume} {12}},\ \bibinfo {pages}
  {1208--1208} (\bibinfo {year} {1975})}\BibitemShut {NoStop}%
\bibitem [{\citenamefont {Seliger}\ \emph {et~al.}(1979)\citenamefont
  {Seliger}, \citenamefont {Ward}, \citenamefont {Wang},\ and\ \citenamefont
  {Kubena}}]{Seliger_1979}%
  \BibitemOpen
  \bibfield  {author} {\bibinfo {author} {\bibfnamefont {R.~L.}\ \bibnamefont
  {Seliger}}, \bibinfo {author} {\bibfnamefont {J.~W.}\ \bibnamefont {Ward}},
  \bibinfo {author} {\bibfnamefont {V.}~\bibnamefont {Wang}},\ and\ \bibinfo
  {author} {\bibfnamefont {R.~L.}\ \bibnamefont {Kubena}},\ }\bibfield  {title}
  {\enquote {\bibinfo {title} {{A high‐intensity scanning ion probe with
  submicrometer spot size}},}\ }\href {https://doi.org/10.1063/1.90786}
  {\bibfield  {journal} {\bibinfo  {journal} {Appl. Phys. Lett.}\ }\textbf
  {\bibinfo {volume} {34}},\ \bibinfo {pages} {310--312} (\bibinfo {year}
  {1979})}\BibitemShut {NoStop}%
\bibitem [{\citenamefont {Taylor}(1964)}]{Taylor_1964}%
  \BibitemOpen
  \bibfield  {author} {\bibinfo {author} {\bibfnamefont {G.~I.}\ \bibnamefont
  {Taylor}},\ }\bibfield  {title} {\enquote {\bibinfo {title} {Disintegration
  of water drops in an electric field},}\ }\href
  {https://doi.org/10.1098/rspa.1964.0151} {\bibfield  {journal} {\bibinfo
  {journal} {Proc. R. Soc. Lon. Ser-A.}\ }\textbf {\bibinfo {volume} {280}},\
  \bibinfo {pages} {383--397} (\bibinfo {year} {1964})}\BibitemShut {NoStop}%
\bibitem [{\citenamefont {Mazarov}\ \emph {et~al.}(2009)\citenamefont
  {Mazarov}, \citenamefont {Wieck}, \citenamefont {Bischoff},\ and\
  \citenamefont {Pilz}}]{Mazarov_2009}%
  \BibitemOpen
  \bibfield  {author} {\bibinfo {author} {\bibfnamefont {P.}~\bibnamefont
  {Mazarov}}, \bibinfo {author} {\bibfnamefont {A.~D.}\ \bibnamefont {Wieck}},
  \bibinfo {author} {\bibfnamefont {L.}~\bibnamefont {Bischoff}},\ and\
  \bibinfo {author} {\bibfnamefont {W.}~\bibnamefont {Pilz}},\ }\bibfield
  {title} {\enquote {\bibinfo {title} {Alloy liquid metal ion source for carbon
  focused ion beams},}\ }\href {https://doi.org/10.1116/1.3253471} {\bibfield
  {journal} {\bibinfo  {journal} {J. Vac. Sci. Technol.}\ }\textbf {\bibinfo
  {volume} {27}},\ \bibinfo {pages} {L47} (\bibinfo {year} {2009})}\BibitemShut
  {NoStop}%
\bibitem [{\citenamefont {Bischoff}\ \emph {et~al.}(2010)\citenamefont
  {Bischoff}, \citenamefont {Pilz}, \citenamefont {Mazarov},\ and\
  \citenamefont {Wieck}}]{Bischoff_2010}%
  \BibitemOpen
  \bibfield  {author} {\bibinfo {author} {\bibfnamefont {L.}~\bibnamefont
  {Bischoff}}, \bibinfo {author} {\bibfnamefont {W.}~\bibnamefont {Pilz}},
  \bibinfo {author} {\bibfnamefont {P.}~\bibnamefont {Mazarov}},\ and\ \bibinfo
  {author} {\bibfnamefont {A.~D.}\ \bibnamefont {Wieck}},\ }\bibfield  {title}
  {\enquote {\bibinfo {title} {{Comparison of bismuth emitting liquid metal ion
  sources}},}\ }\href {https://doi.org/10.1007/s00339-010-5597-0} {\bibfield
  {journal} {\bibinfo  {journal} {Appl. Phys. A}\ }\textbf {\bibinfo {volume}
  {99}},\ \bibinfo {pages} {145--150} (\bibinfo {year} {2010})}\BibitemShut
  {NoStop}%
\bibitem [{\citenamefont {Bruchhaus}\ \emph {et~al.}(2017)\citenamefont
  {Bruchhaus}, \citenamefont {Mazarov}, \citenamefont {Bischoff}, \citenamefont
  {Gierak}, \citenamefont {Wieck},\ and\ \citenamefont
  {H{\"{o}}vel}}]{Bruchhaus_2017}%
  \BibitemOpen
  \bibfield  {author} {\bibinfo {author} {\bibfnamefont {L.}~\bibnamefont
  {Bruchhaus}}, \bibinfo {author} {\bibfnamefont {P.}~\bibnamefont {Mazarov}},
  \bibinfo {author} {\bibfnamefont {L.}~\bibnamefont {Bischoff}}, \bibinfo
  {author} {\bibfnamefont {J.}~\bibnamefont {Gierak}}, \bibinfo {author}
  {\bibfnamefont {A.~D.}\ \bibnamefont {Wieck}},\ and\ \bibinfo {author}
  {\bibfnamefont {H.}~\bibnamefont {H{\"{o}}vel}},\ }\bibfield  {title}
  {\enquote {\bibinfo {title} {{Comparison of technologies for nano device
  prototyping with a special focus on ion beams: A review}},}\ }\href
  {https://doi.org/10.1063/1.4972262} {\bibfield  {journal} {\bibinfo
  {journal} {Appl. Phys. Rev.}\ }\textbf {\bibinfo {volume} {4}},\ \bibinfo
  {pages} {011302} (\bibinfo {year} {2017})}\BibitemShut {NoStop}%
\bibitem [{\citenamefont {Pilz}\ \emph {et~al.}(2017)\citenamefont {Pilz},
  \citenamefont {Laufer}, \citenamefont {Tajmar}, \citenamefont
  {B{\"{o}}ttger},\ and\ \citenamefont {Bischoff}}]{Pilz_2017}%
  \BibitemOpen
  \bibfield  {author} {\bibinfo {author} {\bibfnamefont {W.}~\bibnamefont
  {Pilz}}, \bibinfo {author} {\bibfnamefont {P.}~\bibnamefont {Laufer}},
  \bibinfo {author} {\bibfnamefont {M.}~\bibnamefont {Tajmar}}, \bibinfo
  {author} {\bibfnamefont {R.}~\bibnamefont {B{\"{o}}ttger}},\ and\ \bibinfo
  {author} {\bibfnamefont {L.}~\bibnamefont {Bischoff}},\ }\bibfield  {title}
  {\enquote {\bibinfo {title} {Polyatomic ions from a high current ion
  implanter driven by a liquid metal ion source},}\ }\href
  {https://doi.org/10.1063/1.4995537} {\bibfield  {journal} {\bibinfo
  {journal} {Rev. Sci. Instrum.}\ }\textbf {\bibinfo {volume} {88}},\ \bibinfo
  {pages} {123302} (\bibinfo {year} {2017})}\BibitemShut {NoStop}%
\bibitem [{\citenamefont {Gierak}\ \emph {et~al.}(2018)\citenamefont {Gierak},
  \citenamefont {Mazarov}, \citenamefont {Bruchhaus}, \citenamefont {Jede},\
  and\ \citenamefont {Bischoff}}]{Gierak_2018}%
  \BibitemOpen
  \bibfield  {author} {\bibinfo {author} {\bibfnamefont {J.}~\bibnamefont
  {Gierak}}, \bibinfo {author} {\bibfnamefont {P.}~\bibnamefont {Mazarov}},
  \bibinfo {author} {\bibfnamefont {L.}~\bibnamefont {Bruchhaus}}, \bibinfo
  {author} {\bibfnamefont {R.}~\bibnamefont {Jede}},\ and\ \bibinfo {author}
  {\bibfnamefont {L.}~\bibnamefont {Bischoff}},\ }\bibfield  {title} {\enquote
  {\bibinfo {title} {Review article: review of electrohydrodynamical ion
  sources and their applications to focused ion beam technology},}\ }\href
  {https://doi.org/10.1116/1.5047150} {\bibfield  {journal} {\bibinfo
  {journal} {J. Vac. Sci. Technol. B}\ }\textbf {\bibinfo {volume} {36}},\
  \bibinfo {pages} {06J101} (\bibinfo {year} {2018})}\BibitemShut {NoStop}%
\bibitem [{\citenamefont {Pilz}\ \emph {et~al.}(2019)\citenamefont {Pilz},
  \citenamefont {Klingner}, \citenamefont {Bischoff}, \citenamefont {Mazarov},\
  and\ \citenamefont {Bauerdick}}]{Pilz_2019}%
  \BibitemOpen
  \bibfield  {author} {\bibinfo {author} {\bibfnamefont {W.}~\bibnamefont
  {Pilz}}, \bibinfo {author} {\bibfnamefont {N.}~\bibnamefont {Klingner}},
  \bibinfo {author} {\bibfnamefont {L.}~\bibnamefont {Bischoff}}, \bibinfo
  {author} {\bibfnamefont {P.}~\bibnamefont {Mazarov}},\ and\ \bibinfo {author}
  {\bibfnamefont {S.}~\bibnamefont {Bauerdick}},\ }\bibfield  {title} {\enquote
  {\bibinfo {title} {Lithium ion beams from liquid metal alloy ion sources},}\
  }\href {https://doi.org/10.1116/1.5086271} {\bibfield  {journal} {\bibinfo
  {journal} {J. Vac. Sci. Technol. B}\ }\textbf {\bibinfo {volume} {37}},\
  \bibinfo {pages} {021802} (\bibinfo {year} {2019})}\BibitemShut {NoStop}%
\bibitem [{\citenamefont {Bischoff}\ \emph {et~al.}(2020)\citenamefont
  {Bischoff}, \citenamefont {Klingner}, \citenamefont {Mazarov}, \citenamefont
  {Pilz},\ and\ \citenamefont {Meyer}}]{Bischoff_2020}%
  \BibitemOpen
  \bibfield  {author} {\bibinfo {author} {\bibfnamefont {L.}~\bibnamefont
  {Bischoff}}, \bibinfo {author} {\bibfnamefont {N.}~\bibnamefont {Klingner}},
  \bibinfo {author} {\bibfnamefont {P.}~\bibnamefont {Mazarov}}, \bibinfo
  {author} {\bibfnamefont {W.}~\bibnamefont {Pilz}},\ and\ \bibinfo {author}
  {\bibfnamefont {F.}~\bibnamefont {Meyer}},\ }\bibfield  {title} {\enquote
  {\bibinfo {title} {Boron liquid metal alloy ion sources for special focused
  ion beam applications},}\ }\href {https://doi.org/10.1116/6.0000073}
  {\bibfield  {journal} {\bibinfo  {journal} {J. Vac. Sci. Technol. B}\
  }\textbf {\bibinfo {volume} {38}},\ \bibinfo {pages} {042801} (\bibinfo
  {year} {2020})},\ \Eprint {https://arxiv.org/abs/2004.12722}
  {arXiv:2004.12722} \BibitemShut {NoStop}%
\bibitem [{\citenamefont {Mazarov}, \citenamefont {Dudnikov},\ and\
  \citenamefont {Tolstoguzov}(2020)}]{Mazarov_2020}%
  \BibitemOpen
  \bibfield  {author} {\bibinfo {author} {\bibfnamefont {P.~A.}\ \bibnamefont
  {Mazarov}}, \bibinfo {author} {\bibfnamefont {V.~G.}\ \bibnamefont
  {Dudnikov}},\ and\ \bibinfo {author} {\bibfnamefont {A.~B.}\ \bibnamefont
  {Tolstoguzov}},\ }\bibfield  {title} {\enquote {\bibinfo {title}
  {{Electrohydrodynamic emitters of ion beams}},}\ }\href
  {https://doi.org/10.3367/UFNr.2020.09.038845} {\bibfield  {journal} {\bibinfo
   {journal} {Uspekhi Fizicheskih Nauk}\ }\textbf {\bibinfo {volume} {190}},\
  \bibinfo {pages} {1293--1333} (\bibinfo {year} {2020})}\BibitemShut {NoStop}%
\bibitem [{\citenamefont {Bischoff}\ \emph {et~al.}(2022)\citenamefont
  {Bischoff}, \citenamefont {Klingner}, \citenamefont {Mazarov}, \citenamefont
  {Lenz}, \citenamefont {Narkowicz}, \citenamefont {Pilz},\ and\ \citenamefont
  {Meyer}}]{Bischoff_2022}%
  \BibitemOpen
  \bibfield  {author} {\bibinfo {author} {\bibfnamefont {L.}~\bibnamefont
  {Bischoff}}, \bibinfo {author} {\bibfnamefont {N.}~\bibnamefont {Klingner}},
  \bibinfo {author} {\bibfnamefont {P.}~\bibnamefont {Mazarov}}, \bibinfo
  {author} {\bibfnamefont {K.}~\bibnamefont {Lenz}}, \bibinfo {author}
  {\bibfnamefont {R.}~\bibnamefont {Narkowicz}}, \bibinfo {author}
  {\bibfnamefont {W.}~\bibnamefont {Pilz}},\ and\ \bibinfo {author}
  {\bibfnamefont {F.}~\bibnamefont {Meyer}},\ }\bibfield  {title} {\enquote
  {\bibinfo {title} {Dysprosium liquid metal alloy ion source for magnetic
  nanostructures},}\ }\href {https://doi.org/10.1116/6.0001837} {\bibfield
  {journal} {\bibinfo  {journal} {J. Vac. Sci. Technol. B}\ }\textbf {\bibinfo
  {volume} {40}},\ \bibinfo {pages} {052802} (\bibinfo {year}
  {2022})}\BibitemShut {NoStop}%
\bibitem [{\citenamefont {Lozano}\ and\ \citenamefont
  {Mart{\'\i}nez-S{\'a}nchez}(2004)}]{Lozano_2004}%
  \BibitemOpen
  \bibfield  {author} {\bibinfo {author} {\bibfnamefont {P.}~\bibnamefont
  {Lozano}}\ and\ \bibinfo {author} {\bibfnamefont {M.}~\bibnamefont
  {Mart{\'\i}nez-S{\'a}nchez}},\ }\bibfield  {title} {\enquote {\bibinfo
  {title} {Ionic liquid ion sources: suppression of electrochemical reactions
  using voltage alternation},}\ }\href
  {https://doi.org/10.1016/j.jcis.2004.07.037} {\bibfield  {journal} {\bibinfo
  {journal} {J. Colloid Interface Sci.}\ }\textbf {\bibinfo {volume} {280}},\
  \bibinfo {pages} {149--154} (\bibinfo {year} {2004})}\BibitemShut {NoStop}%
\bibitem [{\citenamefont {Lozano}\ and\ \citenamefont
  {Mart{\'\i}nez-S{\'a}nchez}(2005)}]{Lozano_2005}%
  \BibitemOpen
  \bibfield  {author} {\bibinfo {author} {\bibfnamefont {P.}~\bibnamefont
  {Lozano}}\ and\ \bibinfo {author} {\bibfnamefont {M.}~\bibnamefont
  {Mart{\'\i}nez-S{\'a}nchez}},\ }\bibfield  {title} {\enquote {\bibinfo
  {title} {{Ionic liquid ion sources: Characterization of externally wetted
  emitters}},}\ }\href {https://doi.org/10.1016/j.jcis.2004.08.132} {\bibfield
  {journal} {\bibinfo  {journal} {J. Colloid Interface Sci.}\ }\textbf
  {\bibinfo {volume} {282}},\ \bibinfo {pages} {415--421} (\bibinfo {year}
  {2005})}\BibitemShut {NoStop}%
\bibitem [{\citenamefont {Krejci}\ \emph {et~al.}(2017)\citenamefont {Krejci},
  \citenamefont {Mier-Hicks}, \citenamefont {Thomas}, \citenamefont {Haag},\
  and\ \citenamefont {Lozano}}]{Krejci_2017}%
  \BibitemOpen
  \bibfield  {author} {\bibinfo {author} {\bibfnamefont {D.}~\bibnamefont
  {Krejci}}, \bibinfo {author} {\bibfnamefont {F.}~\bibnamefont {Mier-Hicks}},
  \bibinfo {author} {\bibfnamefont {R.}~\bibnamefont {Thomas}}, \bibinfo
  {author} {\bibfnamefont {T.}~\bibnamefont {Haag}},\ and\ \bibinfo {author}
  {\bibfnamefont {P.}~\bibnamefont {Lozano}},\ }\bibfield  {title} {\enquote
  {\bibinfo {title} {Emission characteristics of passively fed electrospray
  microthrusters with propellant reservoirs},}\ }\href
  {https://doi.org/10.2514/1.a33531} {\bibfield  {journal} {\bibinfo  {journal}
  {J. Spacecraft Rockets}\ }\textbf {\bibinfo {volume} {54}},\ \bibinfo {pages}
  {447--458} (\bibinfo {year} {2017})}\BibitemShut {NoStop}%
\bibitem [{\citenamefont {Perez-Martinez}\ \emph {et~al.}(2010)\citenamefont
  {Perez-Martinez}, \citenamefont {Guilet}, \citenamefont {Gogneau},
  \citenamefont {Jegou}, \citenamefont {Gierak},\ and\ \citenamefont
  {Lozano}}]{PerezMartinez_2010}%
  \BibitemOpen
  \bibfield  {author} {\bibinfo {author} {\bibfnamefont {C.}~\bibnamefont
  {Perez-Martinez}}, \bibinfo {author} {\bibfnamefont {S.}~\bibnamefont
  {Guilet}}, \bibinfo {author} {\bibfnamefont {N.}~\bibnamefont {Gogneau}},
  \bibinfo {author} {\bibfnamefont {P.}~\bibnamefont {Jegou}}, \bibinfo
  {author} {\bibfnamefont {J.}~\bibnamefont {Gierak}},\ and\ \bibinfo {author}
  {\bibfnamefont {P.}~\bibnamefont {Lozano}},\ }\bibfield  {title} {\enquote
  {\bibinfo {title} {Development of ion sources from ionic liquids for
  microfabrication},}\ }\href {https://doi.org/10.1116/1.3432125} {\bibfield
  {journal} {\bibinfo  {journal} {J. Vac. Sci. Technol. B}\ }\textbf {\bibinfo
  {volume} {28}},\ \bibinfo {pages} {L25--L27} (\bibinfo {year}
  {2010})}\BibitemShut {NoStop}%
\bibitem [{\citenamefont {Guilet}\ \emph {et~al.}(2011)\citenamefont {Guilet},
  \citenamefont {Perez-Martinez}, \citenamefont {Jegou}, \citenamefont
  {Lozano},\ and\ \citenamefont {Gierak}}]{Guilet_2011}%
  \BibitemOpen
  \bibfield  {author} {\bibinfo {author} {\bibfnamefont {S.}~\bibnamefont
  {Guilet}}, \bibinfo {author} {\bibfnamefont {C.}~\bibnamefont
  {Perez-Martinez}}, \bibinfo {author} {\bibfnamefont {P.}~\bibnamefont
  {Jegou}}, \bibinfo {author} {\bibfnamefont {P.}~\bibnamefont {Lozano}},\ and\
  \bibinfo {author} {\bibfnamefont {J.}~\bibnamefont {Gierak}},\ }\bibfield
  {title} {\enquote {\bibinfo {title} {{Ionic liquid ion sources for silicon
  reactive machining}},}\ }\href {https://doi.org/10.1016/j.mee.2010.12.037}
  {\bibfield  {journal} {\bibinfo  {journal} {Microelectron. Eng.}\ }\textbf
  {\bibinfo {volume} {88}},\ \bibinfo {pages} {1968--1971} (\bibinfo {year}
  {2011})}\BibitemShut {NoStop}%
\bibitem [{\citenamefont {Tolstogouzov}\ \emph {et~al.}(2015)\citenamefont
  {Tolstogouzov}, \citenamefont {Belykh}, \citenamefont {Gurov}, \citenamefont
  {Lozovan}, \citenamefont {Taganov}, \citenamefont {Teodoro}, \citenamefont
  {Trubitsyn},\ and\ \citenamefont {Chenakin}}]{Tolstogouzov_2015}%
  \BibitemOpen
  \bibfield  {author} {\bibinfo {author} {\bibfnamefont {A.~B.}\ \bibnamefont
  {Tolstogouzov}}, \bibinfo {author} {\bibfnamefont {S.~F.}\ \bibnamefont
  {Belykh}}, \bibinfo {author} {\bibfnamefont {V.~S.}\ \bibnamefont {Gurov}},
  \bibinfo {author} {\bibfnamefont {A.~A.}\ \bibnamefont {Lozovan}}, \bibinfo
  {author} {\bibfnamefont {A.~I.}\ \bibnamefont {Taganov}}, \bibinfo {author}
  {\bibfnamefont {O.~M. N.~D.}\ \bibnamefont {Teodoro}}, \bibinfo {author}
  {\bibfnamefont {A.~A.}\ \bibnamefont {Trubitsyn}},\ and\ \bibinfo {author}
  {\bibfnamefont {S.~P.}\ \bibnamefont {Chenakin}},\ }\bibfield  {title}
  {\enquote {\bibinfo {title} {Ion-beam sources based on room-temperature ionic
  liquids for aerospace applications, nanotechnology, and microprobe analysis
  (review)},}\ }\href {https://doi.org/10.1134/s002044121501011x} {\bibfield
  {journal} {\bibinfo  {journal} {Instrum. Exp. Tech.}\ }\textbf {\bibinfo
  {volume} {58}},\ \bibinfo {pages} {1--14} (\bibinfo {year}
  {2015})}\BibitemShut {NoStop}%
\bibitem [{\citenamefont {Hlawacek}\ \emph {et~al.}(2014)\citenamefont
  {Hlawacek}, \citenamefont {Veligura}, \citenamefont {van Gastel},\ and\
  \citenamefont {Poelsema}}]{Hlawacek_2014}%
  \BibitemOpen
  \bibfield  {author} {\bibinfo {author} {\bibfnamefont {G.}~\bibnamefont
  {Hlawacek}}, \bibinfo {author} {\bibfnamefont {V.}~\bibnamefont {Veligura}},
  \bibinfo {author} {\bibfnamefont {R.}~\bibnamefont {van Gastel}},\ and\
  \bibinfo {author} {\bibfnamefont {B.}~\bibnamefont {Poelsema}},\ }\bibfield
  {title} {\enquote {\bibinfo {title} {Helium ion microscopy},}\ }\href
  {https://doi.org/10.1116/1.4863676} {\bibfield  {journal} {\bibinfo
  {journal} {J. Vac. Sci. Technol. B}\ }\textbf {\bibinfo {volume} {32}},\
  \bibinfo {pages} {020801} (\bibinfo {year} {2014})}\BibitemShut {NoStop}%
\bibitem [{\citenamefont {Aramaki}\ \emph {et~al.}(2011)\citenamefont
  {Aramaki}, \citenamefont {Ogawa}, \citenamefont {Matsuda}, \citenamefont
  {Kozakai},\ and\ \citenamefont {Sugiyama}}]{Aramaki_2011}%
  \BibitemOpen
  \bibfield  {author} {\bibinfo {author} {\bibfnamefont {F.}~\bibnamefont
  {Aramaki}}, \bibinfo {author} {\bibfnamefont {T.}~\bibnamefont {Ogawa}},
  \bibinfo {author} {\bibfnamefont {O.}~\bibnamefont {Matsuda}}, \bibinfo
  {author} {\bibfnamefont {T.}~\bibnamefont {Kozakai}},\ and\ \bibinfo {author}
  {\bibfnamefont {Y.}~\bibnamefont {Sugiyama}},\ }\bibfield  {title} {\enquote
  {\bibinfo {title} {Development of new {FIB} technology for {EUVL} mask
  repair},}\ }in\ \href {https://doi.org/10.1117/12.879609} {\emph {\bibinfo
  {booktitle} {{SPIE} Proceedings}}},\ \bibinfo {editor} {edited by\ \bibinfo
  {editor} {\bibfnamefont {B.~M.~L.}\ \bibnamefont {Fontaine}}\ and\ \bibinfo
  {editor} {\bibfnamefont {P.~P.}\ \bibnamefont {Naulleau}}}\ (\bibinfo
  {publisher} {{SPIE}},\ \bibinfo {year} {2011})\ p.\ \bibinfo {pages}
  {7969}\BibitemShut {NoStop}%
\bibitem [{\citenamefont {Schmidt}\ \emph {et~al.}(2017)\citenamefont
  {Schmidt}, \citenamefont {Yasaka}, \citenamefont {Akabori},\ and\
  \citenamefont {Mizuta}}]{Schmidt_2017}%
  \BibitemOpen
  \bibfield  {author} {\bibinfo {author} {\bibfnamefont {M.~E.}\ \bibnamefont
  {Schmidt}}, \bibinfo {author} {\bibfnamefont {A.}~\bibnamefont {Yasaka}},
  \bibinfo {author} {\bibfnamefont {M.}~\bibnamefont {Akabori}},\ and\ \bibinfo
  {author} {\bibfnamefont {H.}~\bibnamefont {Mizuta}},\ }\bibfield  {title}
  {\enquote {\bibinfo {title} {Nitrogen gas field ion source ({GFIS}) focused
  ion beam ({FIB}) secondary electron imaging: a first look},}\ }\href
  {https://doi.org/10.1017/s1431927617000502} {\bibfield  {journal} {\bibinfo
  {journal} {Microsc. Microanal.}\ }\textbf {\bibinfo {volume} {23}},\ \bibinfo
  {pages} {758--768} (\bibinfo {year} {2017})}\BibitemShut {NoStop}%
\bibitem [{\citenamefont {Schmidt}, \citenamefont {Akabori},\ and\
  \citenamefont {Mizuta}(2018)}]{Schmidt_2018b}%
  \BibitemOpen
  \bibfield  {author} {\bibinfo {author} {\bibfnamefont {M.~E.}\ \bibnamefont
  {Schmidt}}, \bibinfo {author} {\bibfnamefont {M.}~\bibnamefont {Akabori}},\
  and\ \bibinfo {author} {\bibfnamefont {H.}~\bibnamefont {Mizuta}},\
  }\bibfield  {title} {\enquote {\bibinfo {title} {Nitrogen ion microscopy},}\
  }in\ \href {https://doi.org/10.5772/intechopen.76383} {\emph {\bibinfo
  {booktitle} {Ion Beam Applications}}}\ (\bibinfo  {publisher} {{InTech}},\
  \bibinfo {year} {2018})\BibitemShut {NoStop}%
\bibitem [{\citenamefont {Lai}\ \emph {et~al.}(2017)\citenamefont {Lai},
  \citenamefont {Lin}, \citenamefont {Chang}, \citenamefont {Li}, \citenamefont
  {Fu}, \citenamefont {Chang}, \citenamefont {Tsong},\ and\ \citenamefont
  {Hwang}}]{Lai_2017}%
  \BibitemOpen
  \bibfield  {author} {\bibinfo {author} {\bibfnamefont {W.-C.}\ \bibnamefont
  {Lai}}, \bibinfo {author} {\bibfnamefont {C.-Y.}\ \bibnamefont {Lin}},
  \bibinfo {author} {\bibfnamefont {W.-T.}\ \bibnamefont {Chang}}, \bibinfo
  {author} {\bibfnamefont {P.-C.}\ \bibnamefont {Li}}, \bibinfo {author}
  {\bibfnamefont {T.-Y.}\ \bibnamefont {Fu}}, \bibinfo {author} {\bibfnamefont
  {C.-S.}\ \bibnamefont {Chang}}, \bibinfo {author} {\bibfnamefont {T.~T.}\
  \bibnamefont {Tsong}},\ and\ \bibinfo {author} {\bibfnamefont {I.-S.}\
  \bibnamefont {Hwang}},\ }\bibfield  {title} {\enquote {\bibinfo {title}
  {Xenon gas field ion source from a single-atom tip},}\ }\href
  {https://doi.org/10.1088/1361-6528/aa6ed3} {\bibfield  {journal} {\bibinfo
  {journal} {Nanotechnology}\ }\textbf {\bibinfo {volume} {28}},\ \bibinfo
  {pages} {255301} (\bibinfo {year} {2017})}\BibitemShut {NoStop}%
\bibitem [{\citenamefont {Shichi}, \citenamefont {Matsubara},\ and\
  \citenamefont {Hashizume}(2017)}]{Shichi_2017}%
  \BibitemOpen
  \bibfield  {author} {\bibinfo {author} {\bibfnamefont {H.}~\bibnamefont
  {Shichi}}, \bibinfo {author} {\bibfnamefont {S.}~\bibnamefont {Matsubara}},\
  and\ \bibinfo {author} {\bibfnamefont {T.}~\bibnamefont {Hashizume}},\
  }\bibfield  {title} {\enquote {\bibinfo {title} {{Characteristics of krypton
  ion emission from a gas field ionization source with a single atom tip}},}\
  }\href {https://doi.org/10.7567/JJAP.56.06GC01} {\bibfield  {journal}
  {\bibinfo  {journal} {Jpn. J. Appl. Phys.}\ }\textbf {\bibinfo {volume} {56}}
  (\bibinfo {year} {2017}),\ 10.7567/JJAP.56.06GC01}\BibitemShut {NoStop}%
\bibitem [{\citenamefont {Purcell}\ and\ \citenamefont
  {Binh}(1999)}]{Purcell1999}%
  \BibitemOpen
  \bibfield  {author} {\bibinfo {author} {\bibfnamefont {S.~T.}\ \bibnamefont
  {Purcell}}\ and\ \bibinfo {author} {\bibfnamefont {V.~T.}\ \bibnamefont
  {Binh}},\ }\bibfield  {title} {\enquote {\bibinfo {title} {Realization of an
  axially aligned au-ion source of atomic size},}\ }\href
  {https://doi.org/10.1063/1.124684} {\bibfield  {journal} {\bibinfo  {journal}
  {Appl. Phys. Lett.}\ }\textbf {\bibinfo {volume} {75}},\ \bibinfo {pages}
  {1332--1334} (\bibinfo {year} {1999})}\BibitemShut {NoStop}%
\bibitem [{\citenamefont {Purcell}, \citenamefont {Binh},\ and\ \citenamefont
  {Thevenard}(2001)}]{Purcell2001}%
  \BibitemOpen
  \bibfield  {author} {\bibinfo {author} {\bibfnamefont {S.~T.}\ \bibnamefont
  {Purcell}}, \bibinfo {author} {\bibfnamefont {V.~T.}\ \bibnamefont {Binh}},\
  and\ \bibinfo {author} {\bibfnamefont {P.}~\bibnamefont {Thevenard}},\
  }\bibfield  {title} {\enquote {\bibinfo {title} {Atomic-size metal ion
  sources: principles and use},}\ }\href
  {https://doi.org/10.1088/0957-4484/12/2/320} {\bibfield  {journal} {\bibinfo
  {journal} {Nanotechnology}\ }\textbf {\bibinfo {volume} {12}},\ \bibinfo
  {pages} {168--172} (\bibinfo {year} {2001})}\BibitemShut {NoStop}%
\bibitem [{\citenamefont {Lejeune}\ and\ \citenamefont
  {Prangere}(1968)}]{Lejeune_1968}%
  \BibitemOpen
  \bibfield  {author} {\bibinfo {author} {\bibfnamefont {C.}~\bibnamefont
  {Lejeune}}\ and\ \bibinfo {author} {\bibfnamefont {F.}~\bibnamefont
  {Prangere}},\ }\bibfield  {title} {\enquote {\bibinfo {title} {Un
  mod{\`{e}}le simple d{\'{e}}crivant les propri{\'{e}}t{\'{e}}s de la
  d{\'{e}}charge d'une source du type duoplasmatron},}\ }\href
  {https://doi.org/10.1051/jphyscol:1968322} {\bibfield  {journal} {\bibinfo
  {journal} {Le Journal de Physique Colloques}\ }\textbf {\bibinfo {volume}
  {29}},\ \bibinfo {pages} {90--92} (\bibinfo {year} {1968})}\BibitemShut
  {NoStop}%
\bibitem [{\citenamefont {Boswell}\ \emph {et~al.}(2014)\citenamefont
  {Boswell}, \citenamefont {Smith}, \citenamefont {Tesch},\ and\ \citenamefont
  {Martin}}]{Boswell_2014}%
  \BibitemOpen
  \bibfield  {author} {\bibinfo {author} {\bibfnamefont {R.~W.}\ \bibnamefont
  {Boswell}}, \bibinfo {author} {\bibfnamefont {N.~S.}\ \bibnamefont {Smith}},
  \bibinfo {author} {\bibfnamefont {P.~P.}\ \bibnamefont {Tesch}},\ and\
  \bibinfo {author} {\bibfnamefont {N.~P.}\ \bibnamefont {Martin}},\ }\href
  {https://patents.google.com/patent/US20140077699} {\enquote {\bibinfo {title}
  {{RF} system, magnetic filter, and high voltage isolation for an inductively
  coupled plasma ion source},}\ }\bibinfo {type} {patent}\ \bibinfo {number}
  {{US}20140077699}\ (\bibinfo  {institution} {Oregon Physics LLC},\ \bibinfo
  {year} {2014})\BibitemShut {NoStop}%
\bibitem [{\citenamefont {Sortais}\ \emph {et~al.}(2010)\citenamefont
  {Sortais}, \citenamefont {Lamy}, \citenamefont {M{\'{e}}dard}, \citenamefont
  {Angot}, \citenamefont {Latrasse},\ and\ \citenamefont
  {Thuillier}}]{Sortais_2010}%
  \BibitemOpen
  \bibfield  {author} {\bibinfo {author} {\bibfnamefont {P.}~\bibnamefont
  {Sortais}}, \bibinfo {author} {\bibfnamefont {T.}~\bibnamefont {Lamy}},
  \bibinfo {author} {\bibfnamefont {J.}~\bibnamefont {M{\'{e}}dard}}, \bibinfo
  {author} {\bibfnamefont {J.}~\bibnamefont {Angot}}, \bibinfo {author}
  {\bibfnamefont {L.}~\bibnamefont {Latrasse}},\ and\ \bibinfo {author}
  {\bibfnamefont {T.}~\bibnamefont {Thuillier}},\ }\bibfield  {title} {\enquote
  {\bibinfo {title} {Ultracompact/ultralow power electron cyclotron resonance
  ion source for multipurpose applications},}\ }\href
  {https://doi.org/10.1063/1.3272878} {\bibfield  {journal} {\bibinfo
  {journal} {Rev. Sci. Instrum.}\ }\textbf {\bibinfo {volume} {81}},\ \bibinfo
  {pages} {02B314} (\bibinfo {year} {2010})}\BibitemShut {NoStop}%
\bibitem [{\citenamefont {Hanssen}\ \emph {et~al.}(2006)\citenamefont
  {Hanssen}, \citenamefont {McClelland}, \citenamefont {Dakin},\ and\
  \citenamefont {Jacka}}]{Hanssen_2006}%
  \BibitemOpen
  \bibfield  {author} {\bibinfo {author} {\bibfnamefont {J.~L.}\ \bibnamefont
  {Hanssen}}, \bibinfo {author} {\bibfnamefont {J.~J.}\ \bibnamefont
  {McClelland}}, \bibinfo {author} {\bibfnamefont {E.~A.}\ \bibnamefont
  {Dakin}},\ and\ \bibinfo {author} {\bibfnamefont {M.}~\bibnamefont {Jacka}},\
  }\bibfield  {title} {\enquote {\bibinfo {title} {{Laser-cooled atoms as a
  focused ion-beam source}},}\ }\href
  {https://doi.org/10.1103/PhysRevA.74.063416} {\bibfield  {journal} {\bibinfo
  {journal} {Phys. Rev. A}\ }\textbf {\bibinfo {volume} {74}},\ \bibinfo
  {pages} {063416} (\bibinfo {year} {2006})}\BibitemShut {NoStop}%
\bibitem [{\citenamefont {Claessens}\ \emph {et~al.}(2007)\citenamefont
  {Claessens}, \citenamefont {Reijnders}, \citenamefont {Taban}, \citenamefont
  {Luiten},\ and\ \citenamefont {Vredenbregt}}]{Claessens_2007}%
  \BibitemOpen
  \bibfield  {author} {\bibinfo {author} {\bibfnamefont {B.~J.}\ \bibnamefont
  {Claessens}}, \bibinfo {author} {\bibfnamefont {M.~P.}\ \bibnamefont
  {Reijnders}}, \bibinfo {author} {\bibfnamefont {G.}~\bibnamefont {Taban}},
  \bibinfo {author} {\bibfnamefont {O.~J.}\ \bibnamefont {Luiten}},\ and\
  \bibinfo {author} {\bibfnamefont {E.~J.~D.}\ \bibnamefont {Vredenbregt}},\
  }\bibfield  {title} {\enquote {\bibinfo {title} {{Cold electron and ion beams
  generated from trapped atoms}},}\ }\href {https://doi.org/10.1063/1.2771518}
  {\bibfield  {journal} {\bibinfo  {journal} {Phys. Plasmas}\ }\textbf
  {\bibinfo {volume} {14}},\ \bibinfo {pages} {093101} (\bibinfo {year}
  {2007})}\BibitemShut {NoStop}%
\bibitem [{\citenamefont {Hanssen}\ \emph {et~al.}(2008)\citenamefont
  {Hanssen}, \citenamefont {Hill}, \citenamefont {Orloff},\ and\ \citenamefont
  {McClelland}}]{Hanssen_2008}%
  \BibitemOpen
  \bibfield  {author} {\bibinfo {author} {\bibfnamefont {J.~L.}\ \bibnamefont
  {Hanssen}}, \bibinfo {author} {\bibfnamefont {S.~B.}\ \bibnamefont {Hill}},
  \bibinfo {author} {\bibfnamefont {J.}~\bibnamefont {Orloff}},\ and\ \bibinfo
  {author} {\bibfnamefont {J.~J.}\ \bibnamefont {McClelland}},\ }\bibfield
  {title} {\enquote {\bibinfo {title} {{Magneto-Optical-Trap-Based, High
  Brightness Ion Source for Use as a Nanoscale Probe}},}\ }\href
  {https://doi.org/10.1021/nl801472n} {\bibfield  {journal} {\bibinfo
  {journal} {Nano Lett.}\ }\textbf {\bibinfo {volume} {8}},\ \bibinfo {pages}
  {2844--2850} (\bibinfo {year} {2008})}\BibitemShut {NoStop}%
\bibitem [{\citenamefont {Knuffman}\ \emph {et~al.}(2011)\citenamefont
  {Knuffman}, \citenamefont {Steele}, \citenamefont {Orloff}, \citenamefont
  {Maazouz}, \citenamefont {McClelland}, \citenamefont {Seiler}, \citenamefont
  {Diebold}, \citenamefont {McDonald}, \citenamefont {Chabli},\ and\
  \citenamefont {Secula}}]{Knuffman2011}%
  \BibitemOpen
  \bibfield  {author} {\bibinfo {author} {\bibfnamefont {B.}~\bibnamefont
  {Knuffman}}, \bibinfo {author} {\bibfnamefont {A.~V.}\ \bibnamefont
  {Steele}}, \bibinfo {author} {\bibfnamefont {J.}~\bibnamefont {Orloff}},
  \bibinfo {author} {\bibfnamefont {M.}~\bibnamefont {Maazouz}}, \bibinfo
  {author} {\bibfnamefont {J.~J.}\ \bibnamefont {McClelland}}, \bibinfo
  {author} {\bibfnamefont {D.~G.}\ \bibnamefont {Seiler}}, \bibinfo {author}
  {\bibfnamefont {A.~C.}\ \bibnamefont {Diebold}}, \bibinfo {author}
  {\bibfnamefont {R.}~\bibnamefont {McDonald}}, \bibinfo {author}
  {\bibfnamefont {A.}~\bibnamefont {Chabli}},\ and\ \bibinfo {author}
  {\bibfnamefont {E.~M.}\ \bibnamefont {Secula}},\ }\bibfield  {title}
  {\enquote {\bibinfo {title} {{MOTIS}: A focused ion beam source based on
  laser-cooled atoms},}\ }in\ \href {https://doi.org/10.1063/1.3657870} {\emph
  {\bibinfo {booktitle} {{AIP} Conference Proceedings}}},\ Vol.~\bibinfo
  {volume} {85}\ (\bibinfo  {publisher} {{AIP}},\ \bibinfo {year} {2011})\ pp.\
  \bibinfo {pages} {85--89}\BibitemShut {NoStop}%
\bibitem [{\citenamefont {Kime}\ \emph {et~al.}(2013)\citenamefont {Kime},
  \citenamefont {Fioretti}, \citenamefont {Bruneau}, \citenamefont {Porfido},
  \citenamefont {Fuso}, \citenamefont {Viteau}, \citenamefont {Khalili},
  \citenamefont {{\v{S}}anti{\'{c}}}, \citenamefont {Gloter}, \citenamefont
  {Rasser}, \citenamefont {Sudraud}, \citenamefont {Pillet},\ and\
  \citenamefont {Comparat}}]{Kime_2013}%
  \BibitemOpen
  \bibfield  {author} {\bibinfo {author} {\bibfnamefont {L.}~\bibnamefont
  {Kime}}, \bibinfo {author} {\bibfnamefont {A.}~\bibnamefont {Fioretti}},
  \bibinfo {author} {\bibfnamefont {Y.}~\bibnamefont {Bruneau}}, \bibinfo
  {author} {\bibfnamefont {N.}~\bibnamefont {Porfido}}, \bibinfo {author}
  {\bibfnamefont {F.}~\bibnamefont {Fuso}}, \bibinfo {author} {\bibfnamefont
  {M.}~\bibnamefont {Viteau}}, \bibinfo {author} {\bibfnamefont
  {G.}~\bibnamefont {Khalili}}, \bibinfo {author} {\bibfnamefont
  {N.}~\bibnamefont {{\v{S}}anti{\'{c}}}}, \bibinfo {author} {\bibfnamefont
  {A.}~\bibnamefont {Gloter}}, \bibinfo {author} {\bibfnamefont
  {B.}~\bibnamefont {Rasser}}, \bibinfo {author} {\bibfnamefont
  {P.}~\bibnamefont {Sudraud}}, \bibinfo {author} {\bibfnamefont
  {P.}~\bibnamefont {Pillet}},\ and\ \bibinfo {author} {\bibfnamefont
  {D.}~\bibnamefont {Comparat}},\ }\bibfield  {title} {\enquote {\bibinfo
  {title} {{High-flux monochromatic ion and electron beams based on
  laser-cooled atoms}},}\ }\href {https://doi.org/10.1103/PhysRevA.88.033424}
  {\bibfield  {journal} {\bibinfo  {journal} {Phys. Rev. A}\ }\textbf {\bibinfo
  {volume} {88}},\ \bibinfo {pages} {033424} (\bibinfo {year}
  {2013})}\BibitemShut {NoStop}%
\bibitem [{\citenamefont {Gardner}, \citenamefont {McGehee},\ and\
  \citenamefont {McClelland}(2019)}]{Gardner2019}%
  \BibitemOpen
  \bibfield  {author} {\bibinfo {author} {\bibfnamefont {J.~R.}\ \bibnamefont
  {Gardner}}, \bibinfo {author} {\bibfnamefont {W.~R.}\ \bibnamefont
  {McGehee}},\ and\ \bibinfo {author} {\bibfnamefont {J.~J.}\ \bibnamefont
  {McClelland}},\ }\bibfield  {title} {\enquote {\bibinfo {title}
  {Characterization of a high-brightness, laser-cooled {L}i${+}$ ion source},}\
  }\href {https://doi.org/10.1063/1.5085068} {\bibfield  {journal} {\bibinfo
  {journal} {J. Appl. Phys.}\ }\textbf {\bibinfo {volume} {125}},\ \bibinfo
  {pages} {074904} (\bibinfo {year} {2019})}\BibitemShut {NoStop}%
\bibitem [{\citenamefont {van~der Geer}\ \emph {et~al.}(2007)\citenamefont
  {van~der Geer}, \citenamefont {Reijnders}, \citenamefont {de~Loos},
  \citenamefont {Vredenbregt}, \citenamefont {Mutsaers},\ and\ \citenamefont
  {Luiten}}]{Geer2007}%
  \BibitemOpen
  \bibfield  {author} {\bibinfo {author} {\bibfnamefont {S.~B.}\ \bibnamefont
  {van~der Geer}}, \bibinfo {author} {\bibfnamefont {M.~P.}\ \bibnamefont
  {Reijnders}}, \bibinfo {author} {\bibfnamefont {M.~J.}\ \bibnamefont
  {de~Loos}}, \bibinfo {author} {\bibfnamefont {E.~J.~D.}\ \bibnamefont
  {Vredenbregt}}, \bibinfo {author} {\bibfnamefont {P.~H.~A.}\ \bibnamefont
  {Mutsaers}},\ and\ \bibinfo {author} {\bibfnamefont {O.~J.}\ \bibnamefont
  {Luiten}},\ }\bibfield  {title} {\enquote {\bibinfo {title} {Simulated
  performance of an ultracold ion source},}\ }\href
  {https://doi.org/10.1063/1.2804287} {\bibfield  {journal} {\bibinfo
  {journal} {J. Appl. Phys.}\ }\textbf {\bibinfo {volume} {102}} (\bibinfo
  {year} {2007}),\ 10.1063/1.2804287}\BibitemShut {NoStop}%
\bibitem [{\citenamefont {ten Haaf}\ \emph {et~al.}(2014)\citenamefont {ten
  Haaf}, \citenamefont {Wouters}, \citenamefont {van~der Geer}, \citenamefont
  {Vredenbregt},\ and\ \citenamefont {Mutsaers}}]{Haaf_2014}%
  \BibitemOpen
  \bibfield  {author} {\bibinfo {author} {\bibfnamefont {G.}~\bibnamefont {ten
  Haaf}}, \bibinfo {author} {\bibfnamefont {S.~H.~W.}\ \bibnamefont {Wouters}},
  \bibinfo {author} {\bibfnamefont {S.~B.}\ \bibnamefont {van~der Geer}},
  \bibinfo {author} {\bibfnamefont {E.~J.~D.}\ \bibnamefont {Vredenbregt}},\
  and\ \bibinfo {author} {\bibfnamefont {P.~H.~a.}\ \bibnamefont {Mutsaers}},\
  }\bibfield  {title} {\enquote {\bibinfo {title} {{Performance predictions of
  a focused ion beam from a laser cooled and compressed atomic beam}},}\ }\href
  {https://doi.org/10.1063/1.4905022} {\bibfield  {journal} {\bibinfo
  {journal} {J. Appl. Phys.}\ }\textbf {\bibinfo {volume} {116}},\ \bibinfo
  {pages} {244301} (\bibinfo {year} {2014})},\ \Eprint
  {https://arxiv.org/abs/1410.4362} {arXiv:1410.4362} \BibitemShut {NoStop}%
\bibitem [{\citenamefont {McClelland}\ \emph {et~al.}(2016)\citenamefont
  {McClelland}, \citenamefont {Steele}, \citenamefont {Knuffman}, \citenamefont
  {Twedt}, \citenamefont {Schwarzkopf},\ and\ \citenamefont
  {Wilson}}]{McClelland_2016}%
  \BibitemOpen
  \bibfield  {author} {\bibinfo {author} {\bibfnamefont {J.~J.}\ \bibnamefont
  {McClelland}}, \bibinfo {author} {\bibfnamefont {A.~V.}\ \bibnamefont
  {Steele}}, \bibinfo {author} {\bibfnamefont {B.}~\bibnamefont {Knuffman}},
  \bibinfo {author} {\bibfnamefont {K.~A.}\ \bibnamefont {Twedt}}, \bibinfo
  {author} {\bibfnamefont {A.}~\bibnamefont {Schwarzkopf}},\ and\ \bibinfo
  {author} {\bibfnamefont {T.~M.}\ \bibnamefont {Wilson}},\ }\bibfield  {title}
  {\enquote {\bibinfo {title} {{Bright focused ion beam sources based on
  laser-cooled atoms}},}\ }\href {https://doi.org/10.1063/1.4944491} {\bibfield
   {journal} {\bibinfo  {journal} {Appl. Phys. Rev.}\ }\textbf {\bibinfo
  {volume} {3}},\ \bibinfo {pages} {011302} (\bibinfo {year}
  {2016})}\BibitemShut {NoStop}%
\bibitem [{\citenamefont {Jun}, \citenamefont {Kutchoukov},\ and\ \citenamefont
  {Kruit}(2011)}]{Jun_2011}%
  \BibitemOpen
  \bibfield  {author} {\bibinfo {author} {\bibfnamefont {D.~S.}\ \bibnamefont
  {Jun}}, \bibinfo {author} {\bibfnamefont {V.~G.}\ \bibnamefont
  {Kutchoukov}},\ and\ \bibinfo {author} {\bibfnamefont {P.}~\bibnamefont
  {Kruit}},\ }\bibfield  {title} {\enquote {\bibinfo {title} {Ion beams in
  {SEM}: an experiment towards a high brightness low energy spread electron
  impact gas ion source},}\ }\href {https://doi.org/10.1116/1.3660390}
  {\bibfield  {journal} {\bibinfo  {journal} {J. Vac. Sci. Technol. B}\
  }\textbf {\bibinfo {volume} {29}},\ \bibinfo {pages} {06F603} (\bibinfo
  {year} {2011})}\BibitemShut {NoStop}%
\bibitem [{\citenamefont {Xu}\ \emph {et~al.}(2017{\natexlab{a}})\citenamefont
  {Xu}, \citenamefont {Pang}, \citenamefont {Raman}, \citenamefont {Mariappan},
  \citenamefont {Khursheed},\ and\ \citenamefont {van Kan}}]{Xu_2017}%
  \BibitemOpen
  \bibfield  {author} {\bibinfo {author} {\bibfnamefont {X.}~\bibnamefont
  {Xu}}, \bibinfo {author} {\bibfnamefont {R.}~\bibnamefont {Pang}}, \bibinfo
  {author} {\bibfnamefont {P.~S.}\ \bibnamefont {Raman}}, \bibinfo {author}
  {\bibfnamefont {R.}~\bibnamefont {Mariappan}}, \bibinfo {author}
  {\bibfnamefont {A.}~\bibnamefont {Khursheed}},\ and\ \bibinfo {author}
  {\bibfnamefont {J.~A.}\ \bibnamefont {van Kan}},\ }\bibfield  {title}
  {\enquote {\bibinfo {title} {{Fabrication and development of high brightness
  nano-aperture ion source}},}\ }\href
  {https://doi.org/10.1016/j.mee.2016.12.009} {\bibfield  {journal} {\bibinfo
  {journal} {Microelectron. Eng.}\ }\textbf {\bibinfo {volume} {174}},\
  \bibinfo {pages} {20--23} (\bibinfo {year} {2017}{\natexlab{a}})}\BibitemShut
  {NoStop}%
\bibitem [{\citenamefont {Schnitzler}\ \emph {et~al.}(2010)\citenamefont
  {Schnitzler}, \citenamefont {Jacob}, \citenamefont {Fickler}, \citenamefont
  {Schmidt-Kaler},\ and\ \citenamefont {Singer}}]{Schnitzler_2010}%
  \BibitemOpen
  \bibfield  {author} {\bibinfo {author} {\bibfnamefont {W.}~\bibnamefont
  {Schnitzler}}, \bibinfo {author} {\bibfnamefont {G.}~\bibnamefont {Jacob}},
  \bibinfo {author} {\bibfnamefont {R.}~\bibnamefont {Fickler}}, \bibinfo
  {author} {\bibfnamefont {F.}~\bibnamefont {Schmidt-Kaler}},\ and\ \bibinfo
  {author} {\bibfnamefont {K.}~\bibnamefont {Singer}},\ }\bibfield  {title}
  {\enquote {\bibinfo {title} {Focusing a deterministic single-ion beam},}\
  }\href {https://doi.org/10.1088/1367-2630/12/6/065023} {\bibfield  {journal}
  {\bibinfo  {journal} {New J. Phys.}\ }\textbf {\bibinfo {volume} {12}},\
  \bibinfo {pages} {065023} (\bibinfo {year} {2010})}\BibitemShut {NoStop}%
\bibitem [{\citenamefont {Levine}\ \emph {et~al.}(1988)\citenamefont {Levine},
  \citenamefont {Marrs}, \citenamefont {Henderson}, \citenamefont {Knapp},\
  and\ \citenamefont {Schneider}}]{Levine_1988}%
  \BibitemOpen
  \bibfield  {author} {\bibinfo {author} {\bibfnamefont {M.~A.}\ \bibnamefont
  {Levine}}, \bibinfo {author} {\bibfnamefont {R.~E.}\ \bibnamefont {Marrs}},
  \bibinfo {author} {\bibfnamefont {J.~R.}\ \bibnamefont {Henderson}}, \bibinfo
  {author} {\bibfnamefont {D.~A.}\ \bibnamefont {Knapp}},\ and\ \bibinfo
  {author} {\bibfnamefont {M.~B.}\ \bibnamefont {Schneider}},\ }\bibfield
  {title} {\enquote {\bibinfo {title} {The electron beam ion trap: a new
  instrument for atomic physics measurements},}\ }\href
  {https://doi.org/10.1088/0031-8949/1988/t22/024} {\bibfield  {journal}
  {\bibinfo  {journal} {Phys. Scripta}\ }\textbf {\bibinfo {volume} {T22}},\
  \bibinfo {pages} {157--163} (\bibinfo {year} {1988})}\BibitemShut {NoStop}%
\bibitem [{\citenamefont {R{\"a}cke}\ \emph {et~al.}(2020)\citenamefont
  {R{\"a}cke}, \citenamefont {Wunderlich}, \citenamefont {Gerlach},
  \citenamefont {Meijer},\ and\ \citenamefont {Spemann}}]{Raecke_2020}%
  \BibitemOpen
  \bibfield  {author} {\bibinfo {author} {\bibfnamefont {P.}~\bibnamefont
  {R{\"a}cke}}, \bibinfo {author} {\bibfnamefont {R.}~\bibnamefont
  {Wunderlich}}, \bibinfo {author} {\bibfnamefont {J.~W.}\ \bibnamefont
  {Gerlach}}, \bibinfo {author} {\bibfnamefont {J.}~\bibnamefont {Meijer}},\
  and\ \bibinfo {author} {\bibfnamefont {D.}~\bibnamefont {Spemann}},\
  }\bibfield  {title} {{\selectlanguage {en}\enquote {\bibinfo {title}
  {Nanoscale ion implantation using focussed highly charged ions},}\ }}\href
  {https://doi.org/10.1088/1367-2630/aba0e6} {\bibfield  {journal} {\bibinfo
  {journal} {New J. Phys.}\ }\textbf {\bibinfo {volume} {22}},\ \bibinfo
  {pages} {083028} (\bibinfo {year} {2020})}\BibitemShut {NoStop}%
\bibitem [{\citenamefont {Ji}\ \emph {et~al.}(2002)\citenamefont {Ji},
  \citenamefont {Jiang}, \citenamefont {King}, \citenamefont {Leung},
  \citenamefont {Standiford},\ and\ \citenamefont {Wilde}}]{Ji_2002}%
  \BibitemOpen
  \bibfield  {author} {\bibinfo {author} {\bibfnamefont {Q.}~\bibnamefont
  {Ji}}, \bibinfo {author} {\bibfnamefont {X.}~\bibnamefont {Jiang}}, \bibinfo
  {author} {\bibfnamefont {T.-J.}\ \bibnamefont {King}}, \bibinfo {author}
  {\bibfnamefont {K.-N.}\ \bibnamefont {Leung}}, \bibinfo {author}
  {\bibfnamefont {K.}~\bibnamefont {Standiford}},\ and\ \bibinfo {author}
  {\bibfnamefont {S.~B.}\ \bibnamefont {Wilde}},\ }\bibfield  {title} {\enquote
  {\bibinfo {title} {{Improvement in brightness of multicusp-plasma ion
  source}},}\ }\href {https://doi.org/10.1116/1.1526694} {\bibfield  {journal}
  {\bibinfo  {journal} {J. Vac. Sci. Technol. B}\ }\textbf {\bibinfo {volume}
  {20}},\ \bibinfo {pages} {2717} (\bibinfo {year} {2002})}\BibitemShut
  {NoStop}%
\bibitem [{\citenamefont {Tolstoguzov}\ \emph {et~al.}(2018)\citenamefont
  {Tolstoguzov}, \citenamefont {Belykh}, \citenamefont {Gololobov},
  \citenamefont {Gurov}, \citenamefont {Gusev}, \citenamefont {Suvorov},
  \citenamefont {Taganov}, \citenamefont {Fud}, \citenamefont {Ai},\ and\
  \citenamefont {Liu}}]{Tolstoguzov_2018}%
  \BibitemOpen
  \bibfield  {author} {\bibinfo {author} {\bibfnamefont {A.~B.}\ \bibnamefont
  {Tolstoguzov}}, \bibinfo {author} {\bibfnamefont {S.~F.}\ \bibnamefont
  {Belykh}}, \bibinfo {author} {\bibfnamefont {G.~P.}\ \bibnamefont
  {Gololobov}}, \bibinfo {author} {\bibfnamefont {V.~S.}\ \bibnamefont
  {Gurov}}, \bibinfo {author} {\bibfnamefont {S.~I.}\ \bibnamefont {Gusev}},
  \bibinfo {author} {\bibfnamefont {D.~V.}\ \bibnamefont {Suvorov}}, \bibinfo
  {author} {\bibfnamefont {A.~I.}\ \bibnamefont {Taganov}}, \bibinfo {author}
  {\bibfnamefont {D.~J.}\ \bibnamefont {Fud}}, \bibinfo {author} {\bibfnamefont
  {Z.}~\bibnamefont {Ai}},\ and\ \bibinfo {author} {\bibfnamefont {C.~S.}\
  \bibnamefont {Liu}},\ }\bibfield  {title} {\enquote {\bibinfo {title}
  {Ion-beam sources based on solid electrolytes for aerospace applications and
  ion-beam technologies (review)},}\ }\href
  {https://doi.org/10.1134/s0020441218020100} {\bibfield  {journal} {\bibinfo
  {journal} {Instrum. Exp. Tech.}\ }\textbf {\bibinfo {volume} {61}},\ \bibinfo
  {pages} {159--172} (\bibinfo {year} {2018})}\BibitemShut {NoStop}%
\bibitem [{\citenamefont {Nojima}, \citenamefont {Mita},\ and\ \citenamefont
  {Hara}(2020)}]{Nojima_2020}%
  \BibitemOpen
  \bibfield  {author} {\bibinfo {author} {\bibfnamefont {M.}~\bibnamefont
  {Nojima}}, \bibinfo {author} {\bibfnamefont {H.}~\bibnamefont {Mita}},\ and\
  \bibinfo {author} {\bibfnamefont {K.}~\bibnamefont {Hara}},\ }\bibfield
  {title} {\enquote {\bibinfo {title} {Mass separation of water cluster ion
  beam using two rotating electric fields and sputtering of a polymer thin
  film},}\ }\href {https://doi.org/10.1380/ejssnt.2020.101} {\bibfield
  {journal} {\bibinfo  {journal} {E-J. Surf. Sci. Nanotechnol.}\ }\textbf
  {\bibinfo {volume} {18}},\ \bibinfo {pages} {101--105} (\bibinfo {year}
  {2020})}\BibitemShut {NoStop}%
\bibitem [{\citenamefont {Orloff}(2009)}]{Orloff_2009}%
  \BibitemOpen
  \bibfield  {author} {\bibinfo {author} {\bibfnamefont {J.}~\bibnamefont
  {Orloff}},\ }\href@noop {} {\emph {\bibinfo {title} {Handbook of charged
  particle optics}}},\ \bibinfo {edition} {second edition}\ ed.,\ edited by\
  \bibinfo {editor} {\bibfnamefont {J.}~\bibnamefont {Orloff}}\ (\bibinfo
  {publisher} {CRC Press},\ \bibinfo {address} {Boca Raton},\ \bibinfo {year}
  {2009})\BibitemShut {NoStop}%
\bibitem [{\citenamefont {Zhukov}, \citenamefont {Titov},\ and\ \citenamefont
  {Zav'yalova}(2007)}]{Zhukov2007}%
  \BibitemOpen
  \bibfield  {author} {\bibinfo {author} {\bibfnamefont {V.~A.}\ \bibnamefont
  {Zhukov}}, \bibinfo {author} {\bibfnamefont {A.~I.}\ \bibnamefont {Titov}},\
  and\ \bibinfo {author} {\bibfnamefont {A.~V.}\ \bibnamefont {Zav'yalova}},\
  }\bibfield  {title} {\enquote {\bibinfo {title} {Using a chromatic-aberration
  correction system to achieve sub-1.6-nm resolutions of a focused-ion-beam
  microscope designed for characterization and processing},}\ }\href
  {https://doi.org/10.1134/s1063739707050010} {\bibfield  {journal} {\bibinfo
  {journal} {Russian Microelectronics}\ }\textbf {\bibinfo {volume} {36}},\
  \bibinfo {pages} {279--287} (\bibinfo {year} {2007})}\BibitemShut {NoStop}%
\bibitem [{\citenamefont {Martin}(2014)}]{Martin2014}%
  \BibitemOpen
  \bibfield  {author} {\bibinfo {author} {\bibfnamefont {F.~W.}\ \bibnamefont
  {Martin}},\ }\bibfield  {title} {\enquote {\bibinfo {title} {A proposal for
  improved helium microscopy},}\ }\href
  {https://doi.org/10.1017/s1431927614000555} {\bibfield  {journal} {\bibinfo
  {journal} {Microsc. Microanal.}\ }\textbf {\bibinfo {volume} {20}},\ \bibinfo
  {pages} {1619--1622} (\bibinfo {year} {2014})}\BibitemShut {NoStop}%
\bibitem [{\citenamefont {Nagata}\ \emph {et~al.}(2019)\citenamefont {Nagata},
  \citenamefont {Bajo}, \citenamefont {Itose}, \citenamefont {Matsuya},
  \citenamefont {Ishihara}, \citenamefont {Uchino},\ and\ \citenamefont
  {Yurimoto}}]{Nagata_2019}%
  \BibitemOpen
  \bibfield  {author} {\bibinfo {author} {\bibfnamefont {K.}~\bibnamefont
  {Nagata}}, \bibinfo {author} {\bibfnamefont {K.-i.}\ \bibnamefont {Bajo}},
  \bibinfo {author} {\bibfnamefont {S.}~\bibnamefont {Itose}}, \bibinfo
  {author} {\bibfnamefont {M.}~\bibnamefont {Matsuya}}, \bibinfo {author}
  {\bibfnamefont {M.}~\bibnamefont {Ishihara}}, \bibinfo {author}
  {\bibfnamefont {K.}~\bibnamefont {Uchino}},\ and\ \bibinfo {author}
  {\bibfnamefont {H.}~\bibnamefont {Yurimoto}},\ }\bibfield  {title} {\enquote
  {\bibinfo {title} {{Aberration-corrected focused ion beam for time-of-flight
  secondary neutral mass spectrometry}},}\ }\href
  {https://doi.org/10.7567/1882-0786/ab30e4} {\bibfield  {journal} {\bibinfo
  {journal} {Appl. Phys. Express}\ }\textbf {\bibinfo {volume} {12}},\ \bibinfo
  {pages} {085005} (\bibinfo {year} {2019})}\BibitemShut {NoStop}%
\bibitem [{\citenamefont {{CEOS GmbH}}(2023)}]{CEOSGmbH2023}%
  \BibitemOpen
  \bibfield  {author} {\bibinfo {author} {\bibnamefont {{CEOS GmbH}}},\ }\href
  {https://www.ceos-gmbh.de/en/research/electrostat} {\enquote {\bibinfo
  {title} {Electrostatic corrector},}\ }\bibinfo {howpublished} {Online}
  (\bibinfo {year} {2023}),\ \bibinfo {note}
  {\url{https://www.ceos-gmbh.de/en/research/electrostat}}\BibitemShut
  {NoStop}%
\bibitem [{\citenamefont {Wien}(1898)}]{Wien_1898}%
  \BibitemOpen
  \bibfield  {author} {\bibinfo {author} {\bibfnamefont {W.}~\bibnamefont
  {Wien}},\ }\bibfield  {title} {\enquote {\bibinfo {title} {Untersuchungen
  {\"{u}}ber die electrische {E}ntladung in verd{\"{u}}nnten {G}asen},}\ }\href
  {https://doi.org/10.1002/andp.18983010618} {\bibfield  {journal} {\bibinfo
  {journal} {Ann. Phys. (Berlin)}\ }\textbf {\bibinfo {volume} {301}},\
  \bibinfo {pages} {440--452} (\bibinfo {year} {1898})}\BibitemShut {NoStop}%
\bibitem [{\citenamefont {Galejs}\ and\ \citenamefont
  {Kuyatt}(1978)}]{Galejs_1978}%
  \BibitemOpen
  \bibfield  {author} {\bibinfo {author} {\bibfnamefont {A.}~\bibnamefont
  {Galejs}}\ and\ \bibinfo {author} {\bibfnamefont {C.~E.}\ \bibnamefont
  {Kuyatt}},\ }\bibfield  {title} {\enquote {\bibinfo {title} {Focusing and
  dispersing properties of a stigmatic crossed-field energy analyzer},}\ }\href
  {https://doi.org/10.1116/1.569615} {\bibfield  {journal} {\bibinfo  {journal}
  {J. Vac. Sci. Technol.}\ }\textbf {\bibinfo {volume} {15}},\ \bibinfo {pages}
  {865--867} (\bibinfo {year} {1978})}\BibitemShut {NoStop}%
\bibitem [{\citenamefont {Cassidy}\ \emph {et~al.}(2019)\citenamefont
  {Cassidy}, \citenamefont {Cox}, \citenamefont {Webb}, \citenamefont {Murdin},
  \citenamefont {Blenkinsopp}, \citenamefont {Brown},\ and\ \citenamefont
  {Curry}}]{Cassidy2019}%
  \BibitemOpen
  \bibfield  {author} {\bibinfo {author} {\bibfnamefont {N.}~\bibnamefont
  {Cassidy}}, \bibinfo {author} {\bibfnamefont {D.}~\bibnamefont {Cox}},
  \bibinfo {author} {\bibfnamefont {R.}~\bibnamefont {Webb}}, \bibinfo {author}
  {\bibfnamefont {B.}~\bibnamefont {Murdin}}, \bibinfo {author} {\bibfnamefont
  {P.}~\bibnamefont {Blenkinsopp}}, \bibinfo {author} {\bibfnamefont
  {I.}~\bibnamefont {Brown}},\ and\ \bibinfo {author} {\bibfnamefont
  {R.}~\bibnamefont {Curry}},\ }\bibfield  {title} {\enquote {\bibinfo {title}
  {New fib for deterministic single ion implantation},}\ }in\ \href
  {https://www.avssymposium.org/AVS2019/Sessions/Schedule/33365} {\emph
  {\bibinfo {booktitle} {AVS66}}}\ (\bibinfo {organization} {AVS},\ \bibinfo
  {year} {2019})\BibitemShut {NoStop}%
\bibitem [{\citenamefont {Liebl}(2008)}]{Liebl_2008}%
  \BibitemOpen
  \bibfield  {author} {\bibinfo {author} {\bibfnamefont {H.}~\bibnamefont
  {Liebl}},\ }\href {https://doi.org/10.1007/978-3-540-71925-0} {\emph
  {\bibinfo {title} {Applied charged particle optics}}}\ (\bibinfo  {publisher}
  {Springer},\ \bibinfo {address} {Berlin, Heidelberg},\ \bibinfo {year}
  {2008})\BibitemShut {NoStop}%
\bibitem [{\citenamefont {Rose}(2009)}]{Rose2009}%
  \BibitemOpen
  \bibfield  {author} {\bibinfo {author} {\bibfnamefont {H.}~\bibnamefont
  {Rose}},\ }\href@noop {} {\emph {\bibinfo {title} {Geometrical
  charged-particle optics}}},\ \bibinfo {edition} {second edition}\ ed.\
  (\bibinfo  {publisher} {Springer},\ \bibinfo {year} {2009})\BibitemShut
  {NoStop}%
\bibitem [{\citenamefont {Lischke}\ \emph {et~al.}(1989)\citenamefont
  {Lischke}, \citenamefont {Bennecke}, \citenamefont {Brunner}, \citenamefont
  {Herrmann}, \citenamefont {Heuberger}, \citenamefont {Knapek}, \citenamefont
  {Schäffer},\ and\ \citenamefont {Schnakenberg}}]{Lischke1989}%
  \BibitemOpen
  \bibfield  {author} {\bibinfo {author} {\bibfnamefont {B.}~\bibnamefont
  {Lischke}}, \bibinfo {author} {\bibfnamefont {W.}~\bibnamefont {Bennecke}},
  \bibinfo {author} {\bibfnamefont {M.}~\bibnamefont {Brunner}}, \bibinfo
  {author} {\bibfnamefont {K.~H.}\ \bibnamefont {Herrmann}}, \bibinfo {author}
  {\bibfnamefont {A.}~\bibnamefont {Heuberger}}, \bibinfo {author}
  {\bibfnamefont {E.}~\bibnamefont {Knapek}}, \bibinfo {author} {\bibfnamefont
  {P.}~\bibnamefont {Schäffer}},\ and\ \bibinfo {author} {\bibfnamefont
  {U.}~\bibnamefont {Schnakenberg}},\ }\bibfield  {title} {\enquote {\bibinfo
  {title} {Multi-beam concepts for nanometer devices},}\ }\href
  {https://doi.org/10.1143/jjap.28.2058} {\bibfield  {journal} {\bibinfo
  {journal} {Jpn. J. Appl. Phys.}\ }\textbf {\bibinfo {volume} {28}},\ \bibinfo
  {pages} {2058} (\bibinfo {year} {1989})}\BibitemShut {NoStop}%
\bibitem [{\citenamefont {Kapl}\ \emph {et~al.}(2011)\citenamefont {Kapl},
  \citenamefont {Loeschner}, \citenamefont {Piller}, \citenamefont {Witt},
  \citenamefont {Pilz}, \citenamefont {Letzkus}, \citenamefont {Jurisch},
  \citenamefont {Irmscher},\ and\ \citenamefont {Platzgummer}}]{Kapl2011}%
  \BibitemOpen
  \bibfield  {author} {\bibinfo {author} {\bibfnamefont {S.~E.}\ \bibnamefont
  {Kapl}}, \bibinfo {author} {\bibfnamefont {H.}~\bibnamefont {Loeschner}},
  \bibinfo {author} {\bibfnamefont {W.}~\bibnamefont {Piller}}, \bibinfo
  {author} {\bibfnamefont {M.}~\bibnamefont {Witt}}, \bibinfo {author}
  {\bibfnamefont {W.}~\bibnamefont {Pilz}}, \bibinfo {author} {\bibfnamefont
  {F.}~\bibnamefont {Letzkus}}, \bibinfo {author} {\bibfnamefont
  {M.}~\bibnamefont {Jurisch}}, \bibinfo {author} {\bibfnamefont
  {M.}~\bibnamefont {Irmscher}},\ and\ \bibinfo {author} {\bibfnamefont
  {E.}~\bibnamefont {Platzgummer}},\ }\bibfield  {title} {\enquote {\bibinfo
  {title} {Characterization of {CMOS} programmable multi-beam blanking arrays
  as used for programmable multi-beam projection lithography and resistless
  nanopatterning},}\ }\href {https://doi.org/10.1088/0960-1317/21/4/045038}
  {\bibfield  {journal} {\bibinfo  {journal} {J. Micromech. Microeng.}\
  }\textbf {\bibinfo {volume} {21}},\ \bibinfo {pages} {045038} (\bibinfo
  {year} {2011})}\BibitemShut {NoStop}%
\bibitem [{\citenamefont {Platzgummer}, \citenamefont {Klein},\ and\
  \citenamefont {Loeschner}(2013)}]{Platzgummer2013}%
  \BibitemOpen
  \bibfield  {author} {\bibinfo {author} {\bibfnamefont {E.}~\bibnamefont
  {Platzgummer}}, \bibinfo {author} {\bibfnamefont {C.}~\bibnamefont {Klein}},\
  and\ \bibinfo {author} {\bibfnamefont {H.}~\bibnamefont {Loeschner}},\
  }\bibfield  {title} {\enquote {\bibinfo {title} {Electron multibeam
  technology for mask and wafer writing at 0.1~nm address grid},}\ }\href
  {https://doi.org/10.1117/1.jmm.12.3.031108} {\bibfield  {journal} {\bibinfo
  {journal} {J. Micro/Nanolithogr. MEMS MOEMS}\ }\textbf {\bibinfo {volume}
  {12}},\ \bibinfo {pages} {031108} (\bibinfo {year} {2013})}\BibitemShut
  {NoStop}%
\bibitem [{\citenamefont {Klein}\ and\ \citenamefont
  {Platzgummer}(2016)}]{Klein2016}%
  \BibitemOpen
  \bibfield  {author} {\bibinfo {author} {\bibfnamefont {C.}~\bibnamefont
  {Klein}}\ and\ \bibinfo {author} {\bibfnamefont {E.}~\bibnamefont
  {Platzgummer}},\ }\bibfield  {title} {\enquote {\bibinfo {title} {{MBMW}-101:
  {W}orld's 1st high-throughput multi-beam mask writer},}\ }in\ \href
  {https://doi.org/10.1117/12.2243638} {\emph {\bibinfo {booktitle} {{SPIE}
  Proceedings}}},\ \bibinfo {editor} {edited by\ \bibinfo {editor}
  {\bibfnamefont {B.~S.}\ \bibnamefont {Kasprowicz}}\ and\ \bibinfo {editor}
  {\bibfnamefont {P.~D.}\ \bibnamefont {Buck}}}\ (\bibinfo  {publisher}
  {{SPIE}},\ \bibinfo {year} {2016})\BibitemShut {NoStop}%
\bibitem [{\citenamefont {Everhart}\ and\ \citenamefont
  {Thornley}(1960)}]{Everhart_1960}%
  \BibitemOpen
  \bibfield  {author} {\bibinfo {author} {\bibfnamefont {T.~E.}\ \bibnamefont
  {Everhart}}\ and\ \bibinfo {author} {\bibfnamefont {R.~F.~M.}\ \bibnamefont
  {Thornley}},\ }\bibfield  {title} {\enquote {\bibinfo {title} {Wide-band
  detector for micro-microampere low-energy electron currents},}\ }\href
  {https://doi.org/10.1088/0950-7671/37/7/307} {\bibfield  {journal} {\bibinfo
  {journal} {J. Sci. Instrum.}\ }\textbf {\bibinfo {volume} {37}},\ \bibinfo
  {pages} {246--248} (\bibinfo {year} {1960})}\BibitemShut {NoStop}%
\bibitem [{\citenamefont {Levi-Setti}\ and\ \citenamefont
  {Fox}(1980)}]{LeviSetti1980}%
  \BibitemOpen
  \bibfield  {author} {\bibinfo {author} {\bibfnamefont {R.}~\bibnamefont
  {Levi-Setti}}\ and\ \bibinfo {author} {\bibfnamefont {T.~R.}\ \bibnamefont
  {Fox}},\ }\bibfield  {title} {\enquote {\bibinfo {title} {High resolution
  scanning ion probes: Applications to physics and biology},}\ }\href
  {https://doi.org/10.1016/0029-554x(80)91244-6} {\bibfield  {journal}
  {\bibinfo  {journal} {Nucl. Instr. Meth.}\ }\textbf {\bibinfo {volume}
  {168}},\ \bibinfo {pages} {139--149} (\bibinfo {year} {1980})}\BibitemShut
  {NoStop}%
\bibitem [{\citenamefont {Levi-Setti}, \citenamefont {Crow},\ and\
  \citenamefont {Wang}(1986)}]{LeviSetti1986}%
  \BibitemOpen
  \bibfield  {author} {\bibinfo {author} {\bibfnamefont {R.}~\bibnamefont
  {Levi-Setti}}, \bibinfo {author} {\bibfnamefont {G.}~\bibnamefont {Crow}},\
  and\ \bibinfo {author} {\bibfnamefont {Y.~L.}\ \bibnamefont {Wang}},\
  }\bibfield  {title} {\enquote {\bibinfo {title} {Imaging {SIMS} at 20 nm
  lateral resolution: exploratory research applications},}\ }in\ \href
  {https://doi.org/10.1007/978-3-642-82724-2_31} {\emph {\bibinfo {booktitle}
  {Springer Series in Chemical Physics}}}\ (\bibinfo  {publisher} {Springer
  Berlin Heidelberg},\ \bibinfo {year} {1986})\ pp.\ \bibinfo {pages}
  {132--138}\BibitemShut {NoStop}%
\bibitem [{\citenamefont {Waugh}\ \emph {et~al.}(1986)\citenamefont {Waugh},
  \citenamefont {Bayly}, \citenamefont {Walls}, \citenamefont {Vohralik},\ and\
  \citenamefont {Fathers}}]{Waugh_1986}%
  \BibitemOpen
  \bibfield  {author} {\bibinfo {author} {\bibfnamefont {A.~R.}\ \bibnamefont
  {Waugh}}, \bibinfo {author} {\bibfnamefont {A.~R.}\ \bibnamefont {Bayly}},
  \bibinfo {author} {\bibfnamefont {M.}~\bibnamefont {Walls}}, \bibinfo
  {author} {\bibfnamefont {P.}~\bibnamefont {Vohralik}},\ and\ \bibinfo
  {author} {\bibfnamefont {D.}~\bibnamefont {Fathers}},\ }\bibfield  {title}
  {\enquote {\bibinfo {title} {Recent developments in the application of liquid
  metal ion sources to {SIMS}},}\ }\href
  {https://doi.org/10.1051/jphyscol:1986219} {\bibfield  {journal} {\bibinfo
  {journal} {Le Journal de Physique Colloques}\ }\textbf {\bibinfo {volume}
  {47}},\ \bibinfo {pages} {133--136} (\bibinfo {year} {1986})}\BibitemShut
  {NoStop}%
\bibitem [{\citenamefont {McMahon}, \citenamefont {Nxumalo},\ and\
  \citenamefont {Phaneuf}(2002)}]{McMahon_2002}%
  \BibitemOpen
  \bibfield  {author} {\bibinfo {author} {\bibfnamefont {G.}~\bibnamefont
  {McMahon}}, \bibinfo {author} {\bibfnamefont {J.}~\bibnamefont {Nxumalo}},\
  and\ \bibinfo {author} {\bibfnamefont {M.~W.}\ \bibnamefont {Phaneuf}},\
  }\bibfield  {title} {\enquote {\bibinfo {title} {Applications of a novel
  {FIB-SIMS} instrument in {SIMS} image depth profiling},}\ }\href
  {https://doi.org/10.1017/S1431927602107938} {\bibfield  {journal} {\bibinfo
  {journal} {Microsc. Microanal.}\ }\textbf {\bibinfo {volume} {8}},\ \bibinfo
  {pages} {1212--1213} (\bibinfo {year} {2002})}\BibitemShut {NoStop}%
\bibitem [{\citenamefont {Kubis}\ \emph {et~al.}(2004)\citenamefont {Kubis},
  \citenamefont {Shiflet}, \citenamefont {Hull},\ and\ \citenamefont
  {Dunn}}]{Kubis_2004}%
  \BibitemOpen
  \bibfield  {author} {\bibinfo {author} {\bibfnamefont {A.~J.}\ \bibnamefont
  {Kubis}}, \bibinfo {author} {\bibfnamefont {G.~J.}\ \bibnamefont {Shiflet}},
  \bibinfo {author} {\bibfnamefont {R.}~\bibnamefont {Hull}},\ and\ \bibinfo
  {author} {\bibfnamefont {D.~N.}\ \bibnamefont {Dunn}},\ }\bibfield  {title}
  {\enquote {\bibinfo {title} {Focused ion-beam tomography},}\ }\href
  {https://doi.org/10.1007/s11661-004-0142-4} {\bibfield  {journal} {\bibinfo
  {journal} {Metall. Mater. Trans. A.}\ }\textbf {\bibinfo {volume} {35}},\
  \bibinfo {pages} {1935--1943} (\bibinfo {year} {2004})}\BibitemShut {NoStop}%
\bibitem [{\citenamefont {Chater}\ \emph {et~al.}(2014)\citenamefont {Chater},
  \citenamefont {Shollock}, \citenamefont {McPhail}, \citenamefont {Smith},\
  and\ \citenamefont {Cooke}}]{Chater_2014}%
  \BibitemOpen
  \bibfield  {author} {\bibinfo {author} {\bibfnamefont {R.~J.}\ \bibnamefont
  {Chater}}, \bibinfo {author} {\bibfnamefont {B.~A.}\ \bibnamefont
  {Shollock}}, \bibinfo {author} {\bibfnamefont {D.~S.}\ \bibnamefont
  {McPhail}}, \bibinfo {author} {\bibfnamefont {A.~J.}\ \bibnamefont {Smith}},\
  and\ \bibinfo {author} {\bibfnamefont {G.}~\bibnamefont {Cooke}},\ }\bibfield
   {title} {\enquote {\bibinfo {title} {Differentially pumped quadrupole {SIMS}
  probe on {FIB}-based and two-beam microscopes},}\ }\href
  {https://doi.org/10.1002/sia.5665} {\bibfield  {journal} {\bibinfo  {journal}
  {Surf. Interface Anal.}\ }\textbf {\bibinfo {volume} {46}},\ \bibinfo {pages}
  {372--374} (\bibinfo {year} {2014})}\BibitemShut {NoStop}%
\bibitem [{\citenamefont {Chater}, \citenamefont {Smith},\ and\ \citenamefont
  {Cooke}(2016)}]{Chater_2016}%
  \BibitemOpen
  \bibfield  {author} {\bibinfo {author} {\bibfnamefont {R.~J.}\ \bibnamefont
  {Chater}}, \bibinfo {author} {\bibfnamefont {A.~J.}\ \bibnamefont {Smith}},\
  and\ \bibinfo {author} {\bibfnamefont {G.}~\bibnamefont {Cooke}},\ }\bibfield
   {title} {\enquote {\bibinfo {title} {{Simultaneous detection of positive and
  negative secondary ions}},}\ }\href {https://doi.org/10.1116/1.4943531}
  {\bibfield  {journal} {\bibinfo  {journal} {J. Vac. Sci. Technol. B}\
  }\textbf {\bibinfo {volume} {34}},\ \bibinfo {pages} {03H122} (\bibinfo
  {year} {2016})}\BibitemShut {NoStop}%
\bibitem [{\citenamefont {Brugge}\ \emph {et~al.}(2021)\citenamefont {Brugge},
  \citenamefont {Chater}, \citenamefont {Kilner},\ and\ \citenamefont
  {Aguadero}}]{Brugge_2021}%
  \BibitemOpen
  \bibfield  {author} {\bibinfo {author} {\bibfnamefont {R.~H.}\ \bibnamefont
  {Brugge}}, \bibinfo {author} {\bibfnamefont {R.~J.}\ \bibnamefont {Chater}},
  \bibinfo {author} {\bibfnamefont {J.~A.}\ \bibnamefont {Kilner}},\ and\
  \bibinfo {author} {\bibfnamefont {A.}~\bibnamefont {Aguadero}},\ }\bibfield
  {title} {\enquote {\bibinfo {title} {Experimental determination of {L}i
  diffusivity in {LLZO} using isotopic exchange and {FIB-SIMS}},}\ }\href
  {https://doi.org/10.1088/2515-7655/abe2f7} {\bibfield  {journal} {\bibinfo
  {journal} {J. Phys. Energy}\ }\textbf {\bibinfo {volume} {3}},\ \bibinfo
  {pages} {034001} (\bibinfo {year} {2021})}\BibitemShut {NoStop}%
\bibitem [{\citenamefont {Nihei}\ \emph {et~al.}(1987)\citenamefont {Nihei},
  \citenamefont {Satoh}, \citenamefont {Tatsuzawa}, \citenamefont {Owari},
  \citenamefont {Ataka}, \citenamefont {Aihara}, \citenamefont {Azuma},\ and\
  \citenamefont {Kammei}}]{Nihei1987}%
  \BibitemOpen
  \bibfield  {author} {\bibinfo {author} {\bibfnamefont {Y.}~\bibnamefont
  {Nihei}}, \bibinfo {author} {\bibfnamefont {H.}~\bibnamefont {Satoh}},
  \bibinfo {author} {\bibfnamefont {S.}~\bibnamefont {Tatsuzawa}}, \bibinfo
  {author} {\bibfnamefont {M.}~\bibnamefont {Owari}}, \bibinfo {author}
  {\bibfnamefont {M.}~\bibnamefont {Ataka}}, \bibinfo {author} {\bibfnamefont
  {R.}~\bibnamefont {Aihara}}, \bibinfo {author} {\bibfnamefont
  {K.}~\bibnamefont {Azuma}},\ and\ \bibinfo {author} {\bibfnamefont
  {Y.}~\bibnamefont {Kammei}},\ }\bibfield  {title} {\enquote {\bibinfo {title}
  {High spatial resolution secondary ion mass spectrometry with parallel
  detection system},}\ }\href {https://doi.org/10.1116/1.574784} {\bibfield
  {journal} {\bibinfo  {journal} {J. Vac. Sci. Technol. A.}\ }\textbf {\bibinfo
  {volume} {5}},\ \bibinfo {pages} {1254--1257} (\bibinfo {year}
  {1987})}\BibitemShut {NoStop}%
\bibitem [{\citenamefont {Levi-Setti}, \citenamefont {Chabala},\ and\
  \citenamefont {Smolik}(1994)}]{LEVISETTI1994}%
  \BibitemOpen
  \bibfield  {author} {\bibinfo {author} {\bibfnamefont {R.}~\bibnamefont
  {Levi-Setti}}, \bibinfo {author} {\bibfnamefont {J.~M.}\ \bibnamefont
  {Chabala}},\ and\ \bibinfo {author} {\bibfnamefont {S.}~\bibnamefont
  {Smolik}},\ }\bibfield  {title} {\enquote {\bibinfo {title} {Nucleotide and
  protein distribution in {BrdU}-labelled polytene chromosomes revealed by ion
  probe mass spectrometry},}\ }\href
  {https://doi.org/10.1111/j.1365-2818.1994.tb04786.x} {\bibfield  {journal}
  {\bibinfo  {journal} {J. Microsc.}\ }\textbf {\bibinfo {volume} {175}},\
  \bibinfo {pages} {44--53} (\bibinfo {year} {1994})}\BibitemShut {NoStop}%
\bibitem [{\citenamefont {Wirtz}\ \emph {et~al.}(2012)\citenamefont {Wirtz},
  \citenamefont {Vanhove}, \citenamefont {Pillatsch}, \citenamefont {Dowsett},
  \citenamefont {Sijbrandij},\ and\ \citenamefont {Notte}}]{Wirtz_2012}%
  \BibitemOpen
  \bibfield  {author} {\bibinfo {author} {\bibfnamefont {T.}~\bibnamefont
  {Wirtz}}, \bibinfo {author} {\bibfnamefont {N.}~\bibnamefont {Vanhove}},
  \bibinfo {author} {\bibfnamefont {L.}~\bibnamefont {Pillatsch}}, \bibinfo
  {author} {\bibfnamefont {D.}~\bibnamefont {Dowsett}}, \bibinfo {author}
  {\bibfnamefont {S.}~\bibnamefont {Sijbrandij}},\ and\ \bibinfo {author}
  {\bibfnamefont {J.}~\bibnamefont {Notte}},\ }\bibfield  {title} {\enquote
  {\bibinfo {title} {Towards secondary ion mass spectrometry on the helium ion
  microscope: an experimental and simulation based feasibility study with
  he$^{+}$ and ne$^{+}$ bombardment},}\ }\href
  {https://doi.org/10.1063/1.4739240} {\bibfield  {journal} {\bibinfo
  {journal} {Appl. Phys. Lett.}\ }\textbf {\bibinfo {volume} {101}},\ \bibinfo
  {pages} {1--5} (\bibinfo {year} {2012})}\BibitemShut {NoStop}%
\bibitem [{\citenamefont {Vollnhals}\ \emph {et~al.}(2017)\citenamefont
  {Vollnhals}, \citenamefont {Audinot}, \citenamefont {Wirtz}, \citenamefont
  {Mercier-Bonin}, \citenamefont {Fourquaux}, \citenamefont {Schroeppel},
  \citenamefont {Kraushaar}, \citenamefont {Lev-Ram}, \citenamefont
  {Ellisman},\ and\ \citenamefont {Eswara}}]{Vollnhals_2017}%
  \BibitemOpen
  \bibfield  {author} {\bibinfo {author} {\bibfnamefont {F.}~\bibnamefont
  {Vollnhals}}, \bibinfo {author} {\bibfnamefont {J.-N.}\ \bibnamefont
  {Audinot}}, \bibinfo {author} {\bibfnamefont {T.}~\bibnamefont {Wirtz}},
  \bibinfo {author} {\bibfnamefont {M.}~\bibnamefont {Mercier-Bonin}}, \bibinfo
  {author} {\bibfnamefont {I.}~\bibnamefont {Fourquaux}}, \bibinfo {author}
  {\bibfnamefont {B.}~\bibnamefont {Schroeppel}}, \bibinfo {author}
  {\bibfnamefont {U.}~\bibnamefont {Kraushaar}}, \bibinfo {author}
  {\bibfnamefont {V.}~\bibnamefont {Lev-Ram}}, \bibinfo {author} {\bibfnamefont
  {M.~H.}\ \bibnamefont {Ellisman}},\ and\ \bibinfo {author} {\bibfnamefont
  {S.}~\bibnamefont {Eswara}},\ }\bibfield  {title} {{\selectlanguage
  {en}\enquote {\bibinfo {title} {Correlative microscopy combining secondary
  ion mass spectrometry and electron microscopy: comparison of
  {Intensity-Hue-Saturation} and laplacian pyramid methods for image fusion},}\
  }}\href {https://doi.org/10.1021/acs.analchem.7b01256} {\bibfield  {journal}
  {\bibinfo  {journal} {Anal. Chem.}\ }\textbf {\bibinfo {volume} {89}},\
  \bibinfo {pages} {10702--10710} (\bibinfo {year} {2017})}\BibitemShut
  {NoStop}%
\bibitem [{\citenamefont {Wirtz}\ \emph {et~al.}(2019)\citenamefont {Wirtz},
  \citenamefont {De~Castro}, \citenamefont {Audinot},\ and\ \citenamefont
  {Philipp}}]{Wirtz_2019}%
  \BibitemOpen
  \bibfield  {author} {\bibinfo {author} {\bibfnamefont {T.}~\bibnamefont
  {Wirtz}}, \bibinfo {author} {\bibfnamefont {O.}~\bibnamefont {De~Castro}},
  \bibinfo {author} {\bibfnamefont {J.~N.}\ \bibnamefont {Audinot}},\ and\
  \bibinfo {author} {\bibfnamefont {P.}~\bibnamefont {Philipp}},\ }\bibfield
  {title} {\enquote {\bibinfo {title} {{Imaging and Analytics on the Helium Ion
  Microscope}},}\ }\href {https://doi.org/10.1146/annurev-anchem-061318-115457}
  {\bibfield  {journal} {\bibinfo  {journal} {Annu. Rev. Anal. Chem.}\ }\textbf
  {\bibinfo {volume} {12}},\ \bibinfo {pages} {523--543} (\bibinfo {year}
  {2019})}\BibitemShut {NoStop}%
\bibitem [{\citenamefont {Kim}\ \emph {et~al.}(2019{\natexlab{a}})\citenamefont
  {Kim}, \citenamefont {Trofimov}, \citenamefont {Khanom}, \citenamefont
  {Stern}, \citenamefont {Lamberti}, \citenamefont {Colby}, \citenamefont
  {Abmayr}, \citenamefont {Belianinov},\ and\ \citenamefont
  {Ovchinnikova}}]{Kim_2019}%
  \BibitemOpen
  \bibfield  {author} {\bibinfo {author} {\bibfnamefont {S.}~\bibnamefont
  {Kim}}, \bibinfo {author} {\bibfnamefont {A.}~\bibnamefont {Trofimov}},
  \bibinfo {author} {\bibfnamefont {F.}~\bibnamefont {Khanom}}, \bibinfo
  {author} {\bibfnamefont {L.}~\bibnamefont {Stern}}, \bibinfo {author}
  {\bibfnamefont {W.}~\bibnamefont {Lamberti}}, \bibinfo {author}
  {\bibfnamefont {R.}~\bibnamefont {Colby}}, \bibinfo {author} {\bibfnamefont
  {D.}~\bibnamefont {Abmayr}}, \bibinfo {author} {\bibfnamefont
  {A.}~\bibnamefont {Belianinov}},\ and\ \bibinfo {author} {\bibfnamefont
  {O.~S.}\ \bibnamefont {Ovchinnikova}},\ }\bibfield  {title} {\enquote
  {\bibinfo {title} {High resolution multimodal chemical imaging platform for
  organics and inorganics},}\ }\href
  {https://doi.org/10.1021/acs.analchem.9b03377} {\bibfield  {journal}
  {\bibinfo  {journal} {Anal. Chem.}\ }\textbf {\bibinfo {volume} {91}},\
  \bibinfo {pages} {12142--12148} (\bibinfo {year}
  {2019}{\natexlab{a}})}\BibitemShut {NoStop}%
\bibitem [{\citenamefont {Ball}\ \emph {et~al.}(2020)\citenamefont {Ball},
  \citenamefont {Taylor}, \citenamefont {Einsle}, \citenamefont {Khanom},
  \citenamefont {Guillermier},\ and\ \citenamefont {Harrison}}]{Ball_2020}%
  \BibitemOpen
  \bibfield  {author} {\bibinfo {author} {\bibfnamefont {M.~R.}\ \bibnamefont
  {Ball}}, \bibinfo {author} {\bibfnamefont {R.~J.~M.}\ \bibnamefont {Taylor}},
  \bibinfo {author} {\bibfnamefont {J.~F.}\ \bibnamefont {Einsle}}, \bibinfo
  {author} {\bibfnamefont {F.}~\bibnamefont {Khanom}}, \bibinfo {author}
  {\bibfnamefont {C.}~\bibnamefont {Guillermier}},\ and\ \bibinfo {author}
  {\bibfnamefont {R.~J.}\ \bibnamefont {Harrison}},\ }\bibfield  {title}
  {\enquote {\bibinfo {title} {Helium ion microscope {\textendash} secondary
  ion mass spectrometry for geological materials},}\ }\href
  {https://doi.org/10.3762/bjnano.11.133} {\bibfield  {journal} {\bibinfo
  {journal} {Beilstein J. Nanotechnol.}\ }\textbf {\bibinfo {volume} {11}},\
  \bibinfo {pages} {1504--1515} (\bibinfo {year} {2020})}\BibitemShut {NoStop}%
\bibitem [{\citenamefont {Ovchinnikova}\ \emph {et~al.}(2021)\citenamefont
  {Ovchinnikova}, \citenamefont {Borodinov}, \citenamefont {Trofimov},
  \citenamefont {King}, \citenamefont {Lorenz}, \citenamefont {Lamberti},
  \citenamefont {Abmayr},\ and\ \citenamefont {Ievlev}}]{Ovchinnikova_2021}%
  \BibitemOpen
  \bibfield  {author} {\bibinfo {author} {\bibfnamefont {O.~S.}\ \bibnamefont
  {Ovchinnikova}}, \bibinfo {author} {\bibfnamefont {N.}~\bibnamefont
  {Borodinov}}, \bibinfo {author} {\bibfnamefont {A.~A.}\ \bibnamefont
  {Trofimov}}, \bibinfo {author} {\bibfnamefont {S.~T.}\ \bibnamefont {King}},
  \bibinfo {author} {\bibfnamefont {M.}~\bibnamefont {Lorenz}}, \bibinfo
  {author} {\bibfnamefont {W.}~\bibnamefont {Lamberti}}, \bibinfo {author}
  {\bibfnamefont {D.}~\bibnamefont {Abmayr}},\ and\ \bibinfo {author}
  {\bibfnamefont {A.~V.}\ \bibnamefont {Ievlev}},\ }\bibfield  {title}
  {\enquote {\bibinfo {title} {Helium ion microscopy with secondary ion mass
  spectrometry for nanoscale chemical imaging and analysis of polyolefins},}\
  }\href {https://doi.org/10.1021/acsapm.1c00407} {\bibfield  {journal}
  {\bibinfo  {journal} {{ACS} Applied Polymer Materials}\ }\textbf {\bibinfo
  {volume} {3}},\ \bibinfo {pages} {3478--3484} (\bibinfo {year}
  {2021})}\BibitemShut {NoStop}%
\bibitem [{\citenamefont {Audinot}\ \emph {et~al.}(2021)\citenamefont
  {Audinot}, \citenamefont {Philipp}, \citenamefont {Castro}, \citenamefont
  {Biesemeier}, \citenamefont {Hoang},\ and\ \citenamefont
  {Wirtz}}]{Audinot_2021}%
  \BibitemOpen
  \bibfield  {author} {\bibinfo {author} {\bibfnamefont {J.-N.}\ \bibnamefont
  {Audinot}}, \bibinfo {author} {\bibfnamefont {P.}~\bibnamefont {Philipp}},
  \bibinfo {author} {\bibfnamefont {O.~D.}\ \bibnamefont {Castro}}, \bibinfo
  {author} {\bibfnamefont {A.}~\bibnamefont {Biesemeier}}, \bibinfo {author}
  {\bibfnamefont {Q.~H.}\ \bibnamefont {Hoang}},\ and\ \bibinfo {author}
  {\bibfnamefont {T.}~\bibnamefont {Wirtz}},\ }\bibfield  {title} {\enquote
  {\bibinfo {title} {Highest resolution chemical imaging based on secondary ion
  mass spectrometry performed on the helium ion microscope},}\ }\href
  {https://doi.org/10.1088/1361-6633/ac1e32} {\bibfield  {journal} {\bibinfo
  {journal} {Rep. Prog. Phys.}\ }\textbf {\bibinfo {volume} {84}},\ \bibinfo
  {pages} {105901} (\bibinfo {year} {2021})}\BibitemShut {NoStop}%
\bibitem [{\citenamefont {Hayashi}\ \emph {et~al.}(2011)\citenamefont
  {Hayashi}, \citenamefont {Tsuji}, \citenamefont {Nishinomiya},\ and\
  \citenamefont {Kubota}}]{HAYASHI_2011}%
  \BibitemOpen
  \bibfield  {author} {\bibinfo {author} {\bibfnamefont {S.}~\bibnamefont
  {Hayashi}}, \bibinfo {author} {\bibfnamefont {N.}~\bibnamefont {Tsuji}},
  \bibinfo {author} {\bibfnamefont {S.}~\bibnamefont {Nishinomiya}},\ and\
  \bibinfo {author} {\bibfnamefont {N.}~\bibnamefont {Kubota}},\ }\href
  {https://www.nipponsteel.com/en/tech/report/nsc/pdf/n10007.pdf} {\enquote
  {\bibinfo {title} {Development of {FIB-TOF-SIMS} apparatus to evaluate
  suspended particulate matters},}\ }\bibinfo {type} {Tech. Rep.}\ (\bibinfo
  {institution} {Nippon Steel},\ \bibinfo {year} {2011})\BibitemShut {NoStop}%
\bibitem [{\citenamefont {Whitby}\ \emph {et~al.}(2012)\citenamefont {Whitby},
  \citenamefont {{\"{O}}stlund}, \citenamefont {Horvath}, \citenamefont
  {Gabureac}, \citenamefont {Riesterer}, \citenamefont {Utke}, \citenamefont
  {Hohl}, \citenamefont {Sedl{\'{a}}{\v{c}}ek}, \citenamefont {Jiru{\v{s}}e},
  \citenamefont {Friedli}, \citenamefont {Bechelany},\ and\ \citenamefont
  {Michler}}]{Whitby_2012}%
  \BibitemOpen
  \bibfield  {author} {\bibinfo {author} {\bibfnamefont {J.~A.}\ \bibnamefont
  {Whitby}}, \bibinfo {author} {\bibfnamefont {F.}~\bibnamefont
  {{\"{O}}stlund}}, \bibinfo {author} {\bibfnamefont {P.}~\bibnamefont
  {Horvath}}, \bibinfo {author} {\bibfnamefont {M.}~\bibnamefont {Gabureac}},
  \bibinfo {author} {\bibfnamefont {J.~L.}\ \bibnamefont {Riesterer}}, \bibinfo
  {author} {\bibfnamefont {I.}~\bibnamefont {Utke}}, \bibinfo {author}
  {\bibfnamefont {M.}~\bibnamefont {Hohl}}, \bibinfo {author} {\bibfnamefont
  {L.}~\bibnamefont {Sedl{\'{a}}{\v{c}}ek}}, \bibinfo {author} {\bibfnamefont
  {J.}~\bibnamefont {Jiru{\v{s}}e}}, \bibinfo {author} {\bibfnamefont
  {V.}~\bibnamefont {Friedli}}, \bibinfo {author} {\bibfnamefont
  {M.}~\bibnamefont {Bechelany}},\ and\ \bibinfo {author} {\bibfnamefont
  {J.}~\bibnamefont {Michler}},\ }\bibfield  {title} {\enquote {\bibinfo
  {title} {High spatial resolution time-of-flight secondary ion mass
  spectrometry for the masses: a novel orthogonal {ToF FIB-SIMS} instrument
  with in situ {AFM}},}\ }\href {https://doi.org/10.1155/2012/180437}
  {\bibfield  {journal} {\bibinfo  {journal} {Adv. Mater. Sci. Eng.}\ }\textbf
  {\bibinfo {volume} {2012}},\ \bibinfo {pages} {1--13} (\bibinfo {year}
  {2012})}\BibitemShut {NoStop}%
\bibitem [{\citenamefont {Kollmer}\ \emph {et~al.}(2013)\citenamefont
  {Kollmer}, \citenamefont {Paul}, \citenamefont {Krehl},\ and\ \citenamefont
  {Niehuis}}]{Kollmer_2013}%
  \BibitemOpen
  \bibfield  {author} {\bibinfo {author} {\bibfnamefont {F.}~\bibnamefont
  {Kollmer}}, \bibinfo {author} {\bibfnamefont {W.}~\bibnamefont {Paul}},
  \bibinfo {author} {\bibfnamefont {M.}~\bibnamefont {Krehl}},\ and\ \bibinfo
  {author} {\bibfnamefont {E.}~\bibnamefont {Niehuis}},\ }\bibfield  {title}
  {\enquote {\bibinfo {title} {{Ultra high spatial resolution SIMS with cluster
  ions - approaching the physical limits}},}\ }\href
  {https://doi.org/10.1002/sia.5093} {\bibfield  {journal} {\bibinfo  {journal}
  {Surf. Interface Anal.}\ }\textbf {\bibinfo {volume} {45}},\ \bibinfo {pages}
  {312--314} (\bibinfo {year} {2013})}\BibitemShut {NoStop}%
\bibitem [{\citenamefont {Stevie}\ \emph {et~al.}(2014)\citenamefont {Stevie},
  \citenamefont {Sedlacek}, \citenamefont {Babor}, \citenamefont {Jiruse},
  \citenamefont {Principe},\ and\ \citenamefont {Klosova}}]{Stevie_2014}%
  \BibitemOpen
  \bibfield  {author} {\bibinfo {author} {\bibfnamefont {F.~A.}\ \bibnamefont
  {Stevie}}, \bibinfo {author} {\bibfnamefont {L.}~\bibnamefont {Sedlacek}},
  \bibinfo {author} {\bibfnamefont {P.}~\bibnamefont {Babor}}, \bibinfo
  {author} {\bibfnamefont {J.}~\bibnamefont {Jiruse}}, \bibinfo {author}
  {\bibfnamefont {E.}~\bibnamefont {Principe}},\ and\ \bibinfo {author}
  {\bibfnamefont {K.}~\bibnamefont {Klosova}},\ }\bibfield  {title} {\enquote
  {\bibinfo {title} {{FIB-SIMS} quantification using {TOF-SIMS} with {Ar} and
  {Xe} plasma sources},}\ }\href {https://doi.org/10.1002/sia.5483} {\bibfield
  {journal} {\bibinfo  {journal} {Surf. Interface Anal.}\ }\textbf {\bibinfo
  {volume} {46}},\ \bibinfo {pages} {285--287} (\bibinfo {year}
  {2014})}\BibitemShut {NoStop}%
\bibitem [{\citenamefont {Klingner}\ \emph {et~al.}(2016)\citenamefont
  {Klingner}, \citenamefont {Heller}, \citenamefont {Hlawacek}, \citenamefont
  {von Borany}, \citenamefont {Notte}, \citenamefont {Huang},\ and\
  \citenamefont {Facsko}}]{Klingner_2016}%
  \BibitemOpen
  \bibfield  {author} {\bibinfo {author} {\bibfnamefont {N.}~\bibnamefont
  {Klingner}}, \bibinfo {author} {\bibfnamefont {R.}~\bibnamefont {Heller}},
  \bibinfo {author} {\bibfnamefont {G.}~\bibnamefont {Hlawacek}}, \bibinfo
  {author} {\bibfnamefont {J.}~\bibnamefont {von Borany}}, \bibinfo {author}
  {\bibfnamefont {J.}~\bibnamefont {Notte}}, \bibinfo {author} {\bibfnamefont
  {J.}~\bibnamefont {Huang}},\ and\ \bibinfo {author} {\bibfnamefont
  {S.}~\bibnamefont {Facsko}},\ }\bibfield  {title} {\enquote {\bibinfo {title}
  {{Nanometer scale elemental analysis in the helium ion microscope using time
  of flight spectrometry}},}\ }\href
  {https://doi.org/10.1016/j.ultramic.2015.12.005} {\bibfield  {journal}
  {\bibinfo  {journal} {Ultramicroscopy}\ }\textbf {\bibinfo {volume} {162}},\
  \bibinfo {pages} {91--97} (\bibinfo {year} {2016})},\ \Eprint
  {https://arxiv.org/abs/1510.04594} {arXiv:1510.04594} \BibitemShut {NoStop}%
\bibitem [{\citenamefont {Klingner}\ \emph {et~al.}(2019)\citenamefont
  {Klingner}, \citenamefont {Heller}, \citenamefont {Hlawacek}, \citenamefont
  {Facsko},\ and\ \citenamefont {von Borany}}]{Klingner_2019}%
  \BibitemOpen
  \bibfield  {author} {\bibinfo {author} {\bibfnamefont {N.}~\bibnamefont
  {Klingner}}, \bibinfo {author} {\bibfnamefont {R.}~\bibnamefont {Heller}},
  \bibinfo {author} {\bibfnamefont {G.}~\bibnamefont {Hlawacek}}, \bibinfo
  {author} {\bibfnamefont {S.}~\bibnamefont {Facsko}},\ and\ \bibinfo {author}
  {\bibfnamefont {J.}~\bibnamefont {von Borany}},\ }\bibfield  {title}
  {\enquote {\bibinfo {title} {{Time-of-flight secondary ion mass spectrometry
  in the helium ion microscope}},}\ }\href
  {https://doi.org/10.1016/j.ultramic.2018.12.014} {\bibfield  {journal}
  {\bibinfo  {journal} {Ultramicroscopy}\ }\textbf {\bibinfo {volume} {198}},\
  \bibinfo {pages} {10--17} (\bibinfo {year} {2019})}\BibitemShut {NoStop}%
\bibitem [{\citenamefont {Pillatsch}, \citenamefont {{\"{O}}stlund},\ and\
  \citenamefont {Michler}(2019)}]{Pillatsch_2019}%
  \BibitemOpen
  \bibfield  {author} {\bibinfo {author} {\bibfnamefont {L.}~\bibnamefont
  {Pillatsch}}, \bibinfo {author} {\bibfnamefont {F.}~\bibnamefont
  {{\"{O}}stlund}},\ and\ \bibinfo {author} {\bibfnamefont {J.}~\bibnamefont
  {Michler}},\ }\bibfield  {title} {\enquote {\bibinfo {title} {{FIBSIMS}: a
  review of secondary ion mass spectrometry for analytical dual beam focussed
  ion beam instruments},}\ }\href
  {https://doi.org/10.1016/j.pcrysgrow.2018.10.001} {\bibfield  {journal}
  {\bibinfo  {journal} {Prog. Cryst. Growth Ch.}\ }\textbf {\bibinfo {volume}
  {65}},\ \bibinfo {pages} {1--19} (\bibinfo {year} {2019})}\BibitemShut
  {NoStop}%
\bibitem [{\citenamefont {Priebe}\ \emph
  {et~al.}(2019{\natexlab{a}})\citenamefont {Priebe}, \citenamefont {Utke},
  \citenamefont {Peth{\"{o}}},\ and\ \citenamefont {Michler}}]{Priebe_2019}%
  \BibitemOpen
  \bibfield  {author} {\bibinfo {author} {\bibfnamefont {A.}~\bibnamefont
  {Priebe}}, \bibinfo {author} {\bibfnamefont {I.}~\bibnamefont {Utke}},
  \bibinfo {author} {\bibfnamefont {L.}~\bibnamefont {Peth{\"{o}}}},\ and\
  \bibinfo {author} {\bibfnamefont {J.}~\bibnamefont {Michler}},\ }\bibfield
  {title} {\enquote {\bibinfo {title} {Application of a gas-injection system
  during the {FIB-TOF-SIMS} analysis—influence of water vapor and fluorine
  gas on secondary ion signals and sputtering rates},}\ }\href
  {https://doi.org/10.1021/acs.analchem.9b02287} {\bibfield  {journal}
  {\bibinfo  {journal} {Anal. Chem.}\ }\textbf {\bibinfo {volume} {91}},\
  \bibinfo {pages} {11712--11722} (\bibinfo {year}
  {2019}{\natexlab{a}})}\BibitemShut {NoStop}%
\bibitem [{\citenamefont {Rickard}\ \emph {et~al.}(2020)\citenamefont
  {Rickard}, \citenamefont {Reddy}, \citenamefont {Saxey}, \citenamefont
  {Fougerouse}, \citenamefont {Timms}, \citenamefont {Daly}, \citenamefont
  {Peterman}, \citenamefont {Cavosie},\ and\ \citenamefont
  {Jourdan}}]{Rickard_2020}%
  \BibitemOpen
  \bibfield  {author} {\bibinfo {author} {\bibfnamefont {W.~D.~A.}\
  \bibnamefont {Rickard}}, \bibinfo {author} {\bibfnamefont {S.~M.}\
  \bibnamefont {Reddy}}, \bibinfo {author} {\bibfnamefont {D.~W.}\ \bibnamefont
  {Saxey}}, \bibinfo {author} {\bibfnamefont {D.}~\bibnamefont {Fougerouse}},
  \bibinfo {author} {\bibfnamefont {N.~E.}\ \bibnamefont {Timms}}, \bibinfo
  {author} {\bibfnamefont {L.}~\bibnamefont {Daly}}, \bibinfo {author}
  {\bibfnamefont {E.}~\bibnamefont {Peterman}}, \bibinfo {author}
  {\bibfnamefont {A.~J.}\ \bibnamefont {Cavosie}},\ and\ \bibinfo {author}
  {\bibfnamefont {F.}~\bibnamefont {Jourdan}},\ }\bibfield  {title} {\enquote
  {\bibinfo {title} {Novel applications of {FIB-SEM}-based {ToF-SIMS} in atom
  probe tomography workflows},}\ }\href
  {https://doi.org/10.1017/S1431927620000136} {\bibfield  {journal} {\bibinfo
  {journal} {Microsc. Microanal.}\ }\textbf {\bibinfo {volume} {26}},\ \bibinfo
  {pages} {750--757} (\bibinfo {year} {2020})}\BibitemShut {NoStop}%
\bibitem [{\citenamefont {Priebe}, \citenamefont {Peth{\"{o}}},\ and\
  \citenamefont {Michler}(2020)}]{Priebe_2020}%
  \BibitemOpen
  \bibfield  {author} {\bibinfo {author} {\bibfnamefont {A.}~\bibnamefont
  {Priebe}}, \bibinfo {author} {\bibfnamefont {L.}~\bibnamefont
  {Peth{\"{o}}}},\ and\ \bibinfo {author} {\bibfnamefont {J.}~\bibnamefont
  {Michler}},\ }\bibfield  {title} {\enquote {\bibinfo {title} {Fluorine gas
  coinjection as a solution for enhancing spatial resolution of time-of-flight
  secondary ion mass spectrometry and separating mass interference},}\ }\href
  {https://doi.org/10.1021/acs.analchem.9b04647} {\bibfield  {journal}
  {\bibinfo  {journal} {Anal. Chem.}\ }\textbf {\bibinfo {volume} {92}},\
  \bibinfo {pages} {2121--2129} (\bibinfo {year} {2020})}\BibitemShut {NoStop}%
\bibitem [{\citenamefont {Priebe}\ \emph
  {et~al.}(2021{\natexlab{a}})\citenamefont {Priebe}, \citenamefont {Huszar},
  \citenamefont {Nowicki}, \citenamefont {Peth{\"{o}}},\ and\ \citenamefont
  {Michler}}]{Priebe_2021a}%
  \BibitemOpen
  \bibfield  {author} {\bibinfo {author} {\bibfnamefont {A.}~\bibnamefont
  {Priebe}}, \bibinfo {author} {\bibfnamefont {E.}~\bibnamefont {Huszar}},
  \bibinfo {author} {\bibfnamefont {M.}~\bibnamefont {Nowicki}}, \bibinfo
  {author} {\bibfnamefont {L.}~\bibnamefont {Peth{\"{o}}}},\ and\ \bibinfo
  {author} {\bibfnamefont {J.}~\bibnamefont {Michler}},\ }\bibfield  {title}
  {\enquote {\bibinfo {title} {Mechanisms of fluorine-induced separation of
  mass interference during {TOF}-{SIMS} analysis},}\ }\href
  {https://doi.org/10.1021/acs.analchem.1c01661} {\bibfield  {journal}
  {\bibinfo  {journal} {Anal. Chem.}\ }\textbf {\bibinfo {volume} {93}},\
  \bibinfo {pages} {10261--10271} (\bibinfo {year}
  {2021}{\natexlab{a}})}\BibitemShut {NoStop}%
\bibitem [{\citenamefont {Priebe}\ \emph
  {et~al.}(2021{\natexlab{b}})\citenamefont {Priebe}, \citenamefont {Sastre},
  \citenamefont {Futscher}, \citenamefont {Jurczyk}, \citenamefont {dos
  Santos}, \citenamefont {Romanyuk},\ and\ \citenamefont
  {Michler}}]{Priebe_2021b}%
  \BibitemOpen
  \bibfield  {author} {\bibinfo {author} {\bibfnamefont {A.}~\bibnamefont
  {Priebe}}, \bibinfo {author} {\bibfnamefont {J.}~\bibnamefont {Sastre}},
  \bibinfo {author} {\bibfnamefont {M.~H.}\ \bibnamefont {Futscher}}, \bibinfo
  {author} {\bibfnamefont {J.}~\bibnamefont {Jurczyk}}, \bibinfo {author}
  {\bibfnamefont {M.~V.~P.}\ \bibnamefont {dos Santos}}, \bibinfo {author}
  {\bibfnamefont {Y.~E.}\ \bibnamefont {Romanyuk}},\ and\ \bibinfo {author}
  {\bibfnamefont {J.}~\bibnamefont {Michler}},\ }\bibfield  {title} {\enquote
  {\bibinfo {title} {Detection of {A}u$^+$ ions during fluorine gas-assisted
  time-of-flight secondary ion mass spectrometry ({TOF}-{SIMS}) for the
  complete elemental characterization of microbatteries},}\ }\href
  {https://doi.org/10.1021/acsami.1c10352} {\bibfield  {journal} {\bibinfo
  {journal} {ACS Appl. Mater. Interfaces}\ }\textbf {\bibinfo {volume} {13}},\
  \bibinfo {pages} {41262--41274} (\bibinfo {year}
  {2021}{\natexlab{b}})}\BibitemShut {NoStop}%
\bibitem [{\citenamefont {Bajo}\ \emph {et~al.}(2016)\citenamefont {Bajo},
  \citenamefont {Itose}, \citenamefont {Matsuya}, \citenamefont {Ishihara},
  \citenamefont {Uchino}, \citenamefont {Kudo}, \citenamefont {Sakaguchi},\
  and\ \citenamefont {Yurimoto}}]{Bajo_2016}%
  \BibitemOpen
  \bibfield  {author} {\bibinfo {author} {\bibfnamefont {K.-i.}\ \bibnamefont
  {Bajo}}, \bibinfo {author} {\bibfnamefont {S.}~\bibnamefont {Itose}},
  \bibinfo {author} {\bibfnamefont {M.}~\bibnamefont {Matsuya}}, \bibinfo
  {author} {\bibfnamefont {M.}~\bibnamefont {Ishihara}}, \bibinfo {author}
  {\bibfnamefont {K.}~\bibnamefont {Uchino}}, \bibinfo {author} {\bibfnamefont
  {M.}~\bibnamefont {Kudo}}, \bibinfo {author} {\bibfnamefont {I.}~\bibnamefont
  {Sakaguchi}},\ and\ \bibinfo {author} {\bibfnamefont {H.}~\bibnamefont
  {Yurimoto}},\ }\bibfield  {title} {\enquote {\bibinfo {title} {{High spatial
  resolution imaging of helium isotope by TOF-SNMS}},}\ }\href
  {https://doi.org/10.1002/sia.6085} {\bibfield  {journal} {\bibinfo  {journal}
  {Surf. Interface Anal.}\ }\textbf {\bibinfo {volume} {48}},\ \bibinfo {pages}
  {1190--1193} (\bibinfo {year} {2016})}\BibitemShut {NoStop}%
\bibitem [{\citenamefont {Ebata}\ \emph {et~al.}(2012)\citenamefont {Ebata},
  \citenamefont {Ishihara}, \citenamefont {Uchino}, \citenamefont {Itose},
  \citenamefont {Matsuya}, \citenamefont {Kudo}, \citenamefont {Bajo},\ and\
  \citenamefont {Yurimoto}}]{Ebata_2012}%
  \BibitemOpen
  \bibfield  {author} {\bibinfo {author} {\bibfnamefont {S.}~\bibnamefont
  {Ebata}}, \bibinfo {author} {\bibfnamefont {M.}~\bibnamefont {Ishihara}},
  \bibinfo {author} {\bibfnamefont {K.}~\bibnamefont {Uchino}}, \bibinfo
  {author} {\bibfnamefont {S.}~\bibnamefont {Itose}}, \bibinfo {author}
  {\bibfnamefont {M.}~\bibnamefont {Matsuya}}, \bibinfo {author} {\bibfnamefont
  {M.}~\bibnamefont {Kudo}}, \bibinfo {author} {\bibfnamefont {K.-i.}\
  \bibnamefont {Bajo}},\ and\ \bibinfo {author} {\bibfnamefont
  {H.}~\bibnamefont {Yurimoto}},\ }\bibfield  {title} {\enquote {\bibinfo
  {title} {{Development of laser ionization mass nanoscope (LIMAS)}},}\ }\href
  {https://doi.org/10.1002/sia.4857} {\bibfield  {journal} {\bibinfo  {journal}
  {Surf. Interface Anal.}\ }\textbf {\bibinfo {volume} {44}},\ \bibinfo {pages}
  {635--640} (\bibinfo {year} {2012})}\BibitemShut {NoStop}%
\bibitem [{\citenamefont {Draude}\ \emph {et~al.}(2012)\citenamefont {Draude},
  \citenamefont {Galla}, \citenamefont {Pelster}, \citenamefont {Tentschert},
  \citenamefont {Jungnickel}, \citenamefont {Haase}, \citenamefont {Mantion},
  \citenamefont {Th{\"{u}}nemann}, \citenamefont {Taubert}, \citenamefont
  {Luch},\ and\ \citenamefont {Arlinghaus}}]{Draude_2012}%
  \BibitemOpen
  \bibfield  {author} {\bibinfo {author} {\bibfnamefont {F.}~\bibnamefont
  {Draude}}, \bibinfo {author} {\bibfnamefont {S.}~\bibnamefont {Galla}},
  \bibinfo {author} {\bibfnamefont {A.}~\bibnamefont {Pelster}}, \bibinfo
  {author} {\bibfnamefont {J.}~\bibnamefont {Tentschert}}, \bibinfo {author}
  {\bibfnamefont {H.}~\bibnamefont {Jungnickel}}, \bibinfo {author}
  {\bibfnamefont {A.}~\bibnamefont {Haase}}, \bibinfo {author} {\bibfnamefont
  {A.}~\bibnamefont {Mantion}}, \bibinfo {author} {\bibfnamefont {A.~F.}\
  \bibnamefont {Th{\"{u}}nemann}}, \bibinfo {author} {\bibfnamefont
  {A.}~\bibnamefont {Taubert}}, \bibinfo {author} {\bibfnamefont
  {A.}~\bibnamefont {Luch}},\ and\ \bibinfo {author} {\bibfnamefont {H.~F.}\
  \bibnamefont {Arlinghaus}},\ }\bibfield  {title} {\enquote {\bibinfo {title}
  {{ToF}-{SIMS} and laser-{SNMS} analysis of macrophages after exposure to
  silver nanoparticles},}\ }\href {https://doi.org/10.1002/sia.4902} {\bibfield
   {journal} {\bibinfo  {journal} {Surf. Interface Anal.}\ }\textbf {\bibinfo
  {volume} {45}},\ \bibinfo {pages} {286--289} (\bibinfo {year}
  {2012})}\BibitemShut {NoStop}%
\bibitem [{\citenamefont {Franzmann}\ \emph {et~al.}(2018)\citenamefont
  {Franzmann}, \citenamefont {Bosco}, \citenamefont {Hamann}, \citenamefont
  {Walther},\ and\ \citenamefont {Wendt}}]{Franzmann_2018}%
  \BibitemOpen
  \bibfield  {author} {\bibinfo {author} {\bibfnamefont {M.}~\bibnamefont
  {Franzmann}}, \bibinfo {author} {\bibfnamefont {H.}~\bibnamefont {Bosco}},
  \bibinfo {author} {\bibfnamefont {L.}~\bibnamefont {Hamann}}, \bibinfo
  {author} {\bibfnamefont {C.}~\bibnamefont {Walther}},\ and\ \bibinfo {author}
  {\bibfnamefont {K.}~\bibnamefont {Wendt}},\ }\bibfield  {title} {\enquote
  {\bibinfo {title} {Resonant laser{\textendash}{SNMS} for spatially resolved
  and element selective ultra-trace analysis of radionuclides},}\ }\href
  {https://doi.org/10.1039/c7ja00423k} {\bibfield  {journal} {\bibinfo
  {journal} {J. Anal. Atom. Spectrom.}\ }\textbf {\bibinfo {volume} {33}},\
  \bibinfo {pages} {730--737} (\bibinfo {year} {2018})}\BibitemShut {NoStop}%
\bibitem [{\citenamefont {Hall}(2013)}]{Hall_2013}%
  \BibitemOpen
  \bibfield  {author} {\bibinfo {author} {\bibfnamefont {A.~R.}\ \bibnamefont
  {Hall}},\ }\bibfield  {title} {\enquote {\bibinfo {title} {{In Situ Thickness
  Assessment During Ion Milling of a Free-Standing Membrane Using Transmission
  Helium Ion Microscopy}},}\ }\href {https://doi.org/10.1017/S1431927613000500}
  {\bibfield  {journal} {\bibinfo  {journal} {Microsc. Microanal.}\ }\textbf
  {\bibinfo {volume} {19}},\ \bibinfo {pages} {740--744} (\bibinfo {year}
  {2013})}\BibitemShut {NoStop}%
\bibitem [{\citenamefont {Emmrich}\ \emph {et~al.}(2016)\citenamefont
  {Emmrich}, \citenamefont {Beyer}, \citenamefont {Nadzeyka}, \citenamefont
  {Bauerdick}, \citenamefont {Meyer}, \citenamefont {Kotakoski},\ and\
  \citenamefont {G{\"{o}}lzh{\"{a}}user}}]{Emmrich_2016}%
  \BibitemOpen
  \bibfield  {author} {\bibinfo {author} {\bibfnamefont {D.}~\bibnamefont
  {Emmrich}}, \bibinfo {author} {\bibfnamefont {A.}~\bibnamefont {Beyer}},
  \bibinfo {author} {\bibfnamefont {A.}~\bibnamefont {Nadzeyka}}, \bibinfo
  {author} {\bibfnamefont {S.}~\bibnamefont {Bauerdick}}, \bibinfo {author}
  {\bibfnamefont {J.~C.}\ \bibnamefont {Meyer}}, \bibinfo {author}
  {\bibfnamefont {J.}~\bibnamefont {Kotakoski}},\ and\ \bibinfo {author}
  {\bibfnamefont {A.}~\bibnamefont {G{\"{o}}lzh{\"{a}}user}},\ }\bibfield
  {title} {\enquote {\bibinfo {title} {{Nanopore fabrication and
  characterization by helium ion microscopy}},}\ }\href
  {https://doi.org/10.1063/1.4947277} {\bibfield  {journal} {\bibinfo
  {journal} {Appl. Phys. Lett.}\ }\textbf {\bibinfo {volume} {108}},\ \bibinfo
  {pages} {163103} (\bibinfo {year} {2016})}\BibitemShut {NoStop}%
\bibitem [{\citenamefont {Woehl}, \citenamefont {White},\ and\ \citenamefont
  {Keller}(2016)}]{Woehl_2016}%
  \BibitemOpen
  \bibfield  {author} {\bibinfo {author} {\bibfnamefont {T.~J.}\ \bibnamefont
  {Woehl}}, \bibinfo {author} {\bibfnamefont {R.~M.}\ \bibnamefont {White}},\
  and\ \bibinfo {author} {\bibfnamefont {R.~R.}\ \bibnamefont {Keller}},\
  }\bibfield  {title} {\enquote {\bibinfo {title} {Dark-field scanning
  transmission ion microscopy via detection of forward-scattered helium ions
  with a microchannel plate},}\ }\href
  {https://doi.org/10.1017/S1431927616000775} {\bibfield  {journal} {\bibinfo
  {journal} {Microsc. Microanal.}\ }\textbf {\bibinfo {volume} {22}},\ \bibinfo
  {pages} {544--550} (\bibinfo {year} {2016})}\BibitemShut {NoStop}%
\bibitem [{\citenamefont {Kavanagh}, \citenamefont {Herrmann},\ and\
  \citenamefont {J.Notte}(2017)}]{Kavanagh_2017}%
  \BibitemOpen
  \bibfield  {author} {\bibinfo {author} {\bibfnamefont {K.~L.}\ \bibnamefont
  {Kavanagh}}, \bibinfo {author} {\bibfnamefont {C.}~\bibnamefont {Herrmann}},\
  and\ \bibinfo {author} {\bibnamefont {J.Notte}},\ }\bibfield  {title}
  {\enquote {\bibinfo {title} {{Camera for transmission He+ ion microscopy}},}\
  }\href {https://doi.org/10.1116/1.4991898} {\bibfield  {journal} {\bibinfo
  {journal} {J. Vac. Sci. Technol. B}\ }\textbf {\bibinfo {volume} {35}},\
  \bibinfo {pages} {1--5} (\bibinfo {year} {2017})}\BibitemShut {NoStop}%
\bibitem [{\citenamefont {Wang}\ \emph {et~al.}(2018)\citenamefont {Wang},
  \citenamefont {Huang}, \citenamefont {Herrmann}, \citenamefont {Scott},
  \citenamefont {Schiettekatte},\ and\ \citenamefont {Kavanagh}}]{Wang_2018}%
  \BibitemOpen
  \bibfield  {author} {\bibinfo {author} {\bibfnamefont {J.}~\bibnamefont
  {Wang}}, \bibinfo {author} {\bibfnamefont {S.~H.}\ \bibnamefont {Huang}},
  \bibinfo {author} {\bibfnamefont {C.}~\bibnamefont {Herrmann}}, \bibinfo
  {author} {\bibfnamefont {S.~A.}\ \bibnamefont {Scott}}, \bibinfo {author}
  {\bibfnamefont {F.}~\bibnamefont {Schiettekatte}},\ and\ \bibinfo {author}
  {\bibfnamefont {K.~L.}\ \bibnamefont {Kavanagh}},\ }\bibfield  {title}
  {\enquote {\bibinfo {title} {Focussed helium ion channeling through {S}i
  nanomembranes},}\ }\href {https://doi.org/10.1116/1.5020667} {\bibfield
  {journal} {\bibinfo  {journal} {J. Vac. Sci. Technol. B}\ }\textbf {\bibinfo
  {volume} {36}},\ \bibinfo {pages} {1--5} (\bibinfo {year} {2018})},\ \Eprint
  {https://arxiv.org/abs/2004.01682} {arxiv:2004.01682} \BibitemShut {NoStop}%
\bibitem [{\citenamefont {Kavanagh}, \citenamefont {Bunevich},\ and\
  \citenamefont {Motapothula}(2020)}]{Kavanagh_2020}%
  \BibitemOpen
  \bibfield  {author} {\bibinfo {author} {\bibfnamefont {K.~L.}\ \bibnamefont
  {Kavanagh}}, \bibinfo {author} {\bibfnamefont {A.}~\bibnamefont {Bunevich}},\
  and\ \bibinfo {author} {\bibfnamefont {M.~R.}\ \bibnamefont {Motapothula}},\
  }\href {https://doi.org/10.48550/ARXIV.2004.01682} {\enquote {\bibinfo
  {title} {Transmission helium ion microscopy of graphene},}\ } (\bibinfo
  {year} {2020}),\ \bibinfo {note} {preprint},\ \Eprint
  {https://arxiv.org/abs/2004.01682} {arXiv:2004.01682 [physics.app-ph]}
  \BibitemShut {NoStop}%
\bibitem [{\citenamefont {Serralta}\ \emph {et~al.}(2020)\citenamefont
  {Serralta}, \citenamefont {Klingner}, \citenamefont {De~Castro},
  \citenamefont {Mousley}, \citenamefont {Eswara}, \citenamefont
  {Duarte~Pinto}, \citenamefont {Wirtz},\ and\ \citenamefont
  {Hlawacek}}]{Serralta_2020}%
  \BibitemOpen
  \bibfield  {author} {\bibinfo {author} {\bibfnamefont {E.}~\bibnamefont
  {Serralta}}, \bibinfo {author} {\bibfnamefont {N.}~\bibnamefont {Klingner}},
  \bibinfo {author} {\bibfnamefont {O.}~\bibnamefont {De~Castro}}, \bibinfo
  {author} {\bibfnamefont {M.}~\bibnamefont {Mousley}}, \bibinfo {author}
  {\bibfnamefont {S.}~\bibnamefont {Eswara}}, \bibinfo {author} {\bibfnamefont
  {S.}~\bibnamefont {Duarte~Pinto}}, \bibinfo {author} {\bibfnamefont
  {T.}~\bibnamefont {Wirtz}},\ and\ \bibinfo {author} {\bibfnamefont
  {G.}~\bibnamefont {Hlawacek}},\ }\bibfield  {title} {\enquote {\bibinfo
  {title} {{Scanning transmission imaging in the helium ion microscope using a
  microchannel plate with a delay line detector}},}\ }\href
  {https://doi.org/10.3762/bjnano.11.167} {\bibfield  {journal} {\bibinfo
  {journal} {Beilstein J. Nanotechnol.}\ }\textbf {\bibinfo {volume} {11}},\
  \bibinfo {pages} {1854--1864} (\bibinfo {year} {2020})}\BibitemShut {NoStop}%
\bibitem [{\citenamefont {Notte}\ \emph {et~al.}(2010)\citenamefont {Notte},
  \citenamefont {Hill}, \citenamefont {McVey}, \citenamefont {Ramachandra},
  \citenamefont {Griffin},\ and\ \citenamefont {Joy}}]{Notte_2010}%
  \BibitemOpen
  \bibfield  {author} {\bibinfo {author} {\bibfnamefont {J.}~\bibnamefont
  {Notte}}, \bibinfo {author} {\bibfnamefont {R.}~\bibnamefont {Hill}},
  \bibinfo {author} {\bibfnamefont {S.~M.}\ \bibnamefont {McVey}}, \bibinfo
  {author} {\bibfnamefont {R.}~\bibnamefont {Ramachandra}}, \bibinfo {author}
  {\bibfnamefont {B.}~\bibnamefont {Griffin}},\ and\ \bibinfo {author}
  {\bibfnamefont {D.}~\bibnamefont {Joy}},\ }\bibfield  {title} {\enquote
  {\bibinfo {title} {Diffraction imaging in a he$^{+}$ ion beam scanning
  transmission microscope},}\ }\href
  {https://doi.org/10.1017/S1431927610093682} {\bibfield  {journal} {\bibinfo
  {journal} {Microsc. Microanal.}\ }\textbf {\bibinfo {volume} {16}},\ \bibinfo
  {pages} {599--603} (\bibinfo {year} {2010})}\BibitemShut {NoStop}%
\bibitem [{\citenamefont {Castro}\ \emph {et~al.}(2021)\citenamefont {Castro},
  \citenamefont {Biesemeier}, \citenamefont {Serralta}, \citenamefont {Bouton},
  \citenamefont {Barrahma}, \citenamefont {Hoang}, \citenamefont {Cambier},
  \citenamefont {Taubitz}, \citenamefont {Klingner}, \citenamefont {Hlawacek},
  \citenamefont {Pinto}, \citenamefont {Gnauck}, \citenamefont {Lucas},
  \citenamefont {Bebeacua},\ and\ \citenamefont {Wirtz}}]{DeCastro_2021}%
  \BibitemOpen
  \bibfield  {author} {\bibinfo {author} {\bibfnamefont {O.~D.}\ \bibnamefont
  {Castro}}, \bibinfo {author} {\bibfnamefont {A.}~\bibnamefont {Biesemeier}},
  \bibinfo {author} {\bibfnamefont {E.}~\bibnamefont {Serralta}}, \bibinfo
  {author} {\bibfnamefont {O.}~\bibnamefont {Bouton}}, \bibinfo {author}
  {\bibfnamefont {R.}~\bibnamefont {Barrahma}}, \bibinfo {author}
  {\bibfnamefont {Q.~H.}\ \bibnamefont {Hoang}}, \bibinfo {author}
  {\bibfnamefont {S.}~\bibnamefont {Cambier}}, \bibinfo {author} {\bibfnamefont
  {T.}~\bibnamefont {Taubitz}}, \bibinfo {author} {\bibfnamefont
  {N.}~\bibnamefont {Klingner}}, \bibinfo {author} {\bibfnamefont
  {G.}~\bibnamefont {Hlawacek}}, \bibinfo {author} {\bibfnamefont {S.~D.}\
  \bibnamefont {Pinto}}, \bibinfo {author} {\bibfnamefont {P.}~\bibnamefont
  {Gnauck}}, \bibinfo {author} {\bibfnamefont {F.}~\bibnamefont {Lucas}},
  \bibinfo {author} {\bibfnamefont {C.}~\bibnamefont {Bebeacua}},\ and\
  \bibinfo {author} {\bibfnamefont {T.}~\bibnamefont {Wirtz}},\ }\bibfield
  {title} {\enquote {\bibinfo {title} {{npSCOPE}: a new multimodal instrument
  for in situ correlative analysis of nanoparticles},}\ }\href
  {https://doi.org/10.1021/acs.analchem.1c02337} {\bibfield  {journal}
  {\bibinfo  {journal} {Anal. Chem.}\ }\textbf {\bibinfo {volume} {93}},\
  \bibinfo {pages} {14417--14424} (\bibinfo {year} {2021})}\BibitemShut
  {NoStop}%
\bibitem [{\citenamefont {Sijbrandij}\ \emph {et~al.}(2008)\citenamefont
  {Sijbrandij}, \citenamefont {Thompson}, \citenamefont {Notte}, \citenamefont
  {Ward},\ and\ \citenamefont {Economou}}]{Sijbrandij_2008}%
  \BibitemOpen
  \bibfield  {author} {\bibinfo {author} {\bibfnamefont {S.}~\bibnamefont
  {Sijbrandij}}, \bibinfo {author} {\bibfnamefont {B.}~\bibnamefont
  {Thompson}}, \bibinfo {author} {\bibfnamefont {J.}~\bibnamefont {Notte}},
  \bibinfo {author} {\bibfnamefont {B.~W.}\ \bibnamefont {Ward}},\ and\
  \bibinfo {author} {\bibfnamefont {N.~P.}\ \bibnamefont {Economou}},\
  }\bibfield  {title} {\enquote {\bibinfo {title} {{Elemental analysis with the
  helium ion microscope}},}\ }\href {https://doi.org/10.1116/1.2993262}
  {\bibfield  {journal} {\bibinfo  {journal} {J. Vac. Sci. Technol. B}\
  }\textbf {\bibinfo {volume} {26}},\ \bibinfo {pages} {2103--2106} (\bibinfo
  {year} {2008})}\BibitemShut {NoStop}%
\bibitem [{\citenamefont {Sijbrandij}\ \emph {et~al.}(2010)\citenamefont
  {Sijbrandij}, \citenamefont {Notte}, \citenamefont {Scipioni}, \citenamefont
  {Huynh},\ and\ \citenamefont {Sanford}}]{Sijbrandij_2010}%
  \BibitemOpen
  \bibfield  {author} {\bibinfo {author} {\bibfnamefont {S.}~\bibnamefont
  {Sijbrandij}}, \bibinfo {author} {\bibfnamefont {J.}~\bibnamefont {Notte}},
  \bibinfo {author} {\bibfnamefont {L.}~\bibnamefont {Scipioni}}, \bibinfo
  {author} {\bibfnamefont {C.}~\bibnamefont {Huynh}},\ and\ \bibinfo {author}
  {\bibfnamefont {C.}~\bibnamefont {Sanford}},\ }\bibfield  {title} {\enquote
  {\bibinfo {title} {{Analysis and metrology with a focused helium ion
  beam}},}\ }\href {https://doi.org/10.1116/1.3271254} {\bibfield  {journal}
  {\bibinfo  {journal} {J. Vac. Sci. Technol. B}\ }\textbf {\bibinfo {volume}
  {28}},\ \bibinfo {pages} {73--77} (\bibinfo {year} {2010})}\BibitemShut
  {NoStop}%
\bibitem [{\citenamefont {Heller}, \citenamefont {Klingner},\ and\
  \citenamefont {Hlawacek}(2016)}]{Heller_2016}%
  \BibitemOpen
  \bibfield  {author} {\bibinfo {author} {\bibfnamefont {R.}~\bibnamefont
  {Heller}}, \bibinfo {author} {\bibfnamefont {N.}~\bibnamefont {Klingner}},\
  and\ \bibinfo {author} {\bibfnamefont {G.}~\bibnamefont {Hlawacek}},\
  }\bibfield  {title} {\enquote {\bibinfo {title} {{Backscattering Spectrometry
  in the Helium Ion Microscope: Imaging Elemental Compositions on the nm
  Scale}},}\ }in\  \cite{Hlawacek_2016},\ Chap.~\bibinfo {chapter} {12}, pp.\
  \bibinfo {pages} {265--295}\BibitemShut {NoStop}%
\bibitem [{\citenamefont {Bell}(2009)}]{Bell_2009}%
  \BibitemOpen
  \bibfield  {author} {\bibinfo {author} {\bibfnamefont {D.~C.}\ \bibnamefont
  {Bell}},\ }\bibfield  {title} {\enquote {\bibinfo {title} {{Contrast
  Mechanisms and Image Formation in Helium Ion Microscopy}},}\ }\href
  {https://doi.org/10.1017/S1431927609090138} {\bibfield  {journal} {\bibinfo
  {journal} {Microsc. Microanal.}\ }\textbf {\bibinfo {volume} {15}},\ \bibinfo
  {pages} {147--153} (\bibinfo {year} {2009})}\BibitemShut {NoStop}%
\bibitem [{\citenamefont {Rabalais}(2002)}]{Rabalais_2002}%
  \BibitemOpen
  \bibfield  {author} {\bibinfo {author} {\bibfnamefont {J.~W.}\ \bibnamefont
  {Rabalais}},\ }\href
  {https://www.ebook.de/de/product/4242732/j_wayne_rabalais_principles_and_applications_of_ion_scattering_spectrometry_surface_chemical_and_structural_analysis.html}
  {\emph {\bibinfo {title} {Principles and applications of ion scattering
  spectrometry: surface chemical and structural analysis}}}\ (\bibinfo
  {publisher} {Wiley},\ \bibinfo {year} {2002})\ p.\ \bibinfo {pages}
  {344}\BibitemShut {NoStop}%
\bibitem [{\citenamefont {Abo}\ \emph {et~al.}(2012)\citenamefont {Abo},
  \citenamefont {Azuma}, \citenamefont {Lohner}, \citenamefont {Wakaya},\ and\
  \citenamefont {Takai}}]{Abo_2012}%
  \BibitemOpen
  \bibfield  {author} {\bibinfo {author} {\bibfnamefont {S.}~\bibnamefont
  {Abo}}, \bibinfo {author} {\bibfnamefont {T.}~\bibnamefont {Azuma}}, \bibinfo
  {author} {\bibfnamefont {T.}~\bibnamefont {Lohner}}, \bibinfo {author}
  {\bibfnamefont {F.}~\bibnamefont {Wakaya}},\ and\ \bibinfo {author}
  {\bibfnamefont {M.}~\bibnamefont {Takai}},\ }\bibfield  {title} {\enquote
  {\bibinfo {title} {{Study on spatial resolution of three-dimensional analysis
  by full count TOF-RBS with beryllium nanoprobe}},}\ }\href
  {https://doi.org/10.1016/j.nimb.2011.07.091} {\bibfield  {journal} {\bibinfo
  {journal} {Nucl. Instrum. Meth. Phys. Res. B}\ }\textbf {\bibinfo {volume}
  {273}},\ \bibinfo {pages} {266--269} (\bibinfo {year} {2012})}\BibitemShut
  {NoStop}%
\bibitem [{\citenamefont {van Gastel}\ \emph {et~al.}(2015)\citenamefont {van
  Gastel}, \citenamefont {Hlawacek}, \citenamefont {Dutta},\ and\ \citenamefont
  {Poelsema}}]{Gastel_2015}%
  \BibitemOpen
  \bibfield  {author} {\bibinfo {author} {\bibfnamefont {R.}~\bibnamefont {van
  Gastel}}, \bibinfo {author} {\bibfnamefont {G.}~\bibnamefont {Hlawacek}},
  \bibinfo {author} {\bibfnamefont {S.}~\bibnamefont {Dutta}},\ and\ \bibinfo
  {author} {\bibfnamefont {B.}~\bibnamefont {Poelsema}},\ }\bibfield  {title}
  {\enquote {\bibinfo {title} {{Backscattered helium spectroscopy in the helium
  ion microscope: Principles, resolution and applications}},}\ }\href
  {https://doi.org/10.1016/j.nimb.2014.11.073} {\bibfield  {journal} {\bibinfo
  {journal} {Nucl. Instrum. Meth. Phys. Res. B}\ }\textbf {\bibinfo {volume}
  {344}},\ \bibinfo {pages} {44--49} (\bibinfo {year} {2015})}\BibitemShut
  {NoStop}%
\bibitem [{\citenamefont {Hlawacek}\ \emph {et~al.}(2012)\citenamefont
  {Hlawacek}, \citenamefont {Veligura}, \citenamefont {Lorbek}, \citenamefont
  {Mocking}, \citenamefont {George}, \citenamefont {van Gastel}, \citenamefont
  {Zandvliet},\ and\ \citenamefont {Poelsema}}]{Hlawacek_2012}%
  \BibitemOpen
  \bibfield  {author} {\bibinfo {author} {\bibfnamefont {G.}~\bibnamefont
  {Hlawacek}}, \bibinfo {author} {\bibfnamefont {V.}~\bibnamefont {Veligura}},
  \bibinfo {author} {\bibfnamefont {S.}~\bibnamefont {Lorbek}}, \bibinfo
  {author} {\bibfnamefont {T.~F.}\ \bibnamefont {Mocking}}, \bibinfo {author}
  {\bibfnamefont {A.}~\bibnamefont {George}}, \bibinfo {author} {\bibfnamefont
  {R.}~\bibnamefont {van Gastel}}, \bibinfo {author} {\bibfnamefont {H.~J.~W.}\
  \bibnamefont {Zandvliet}},\ and\ \bibinfo {author} {\bibfnamefont
  {B.}~\bibnamefont {Poelsema}},\ }\bibfield  {title} {\enquote {\bibinfo
  {title} {{Imaging ultra thin layers with helium ion microscopy: Utilizing the
  channeling contrast mechanism}},}\ }\href
  {https://doi.org/10.3762/bjnano.3.58} {\bibfield  {journal} {\bibinfo
  {journal} {Beilstein J. Nanotechnol.}\ }\textbf {\bibinfo {volume} {3}},\
  \bibinfo {pages} {507--512} (\bibinfo {year} {2012})}\BibitemShut {NoStop}%
\bibitem [{\citenamefont {Veligura}\ \emph {et~al.}(2012)\citenamefont
  {Veligura}, \citenamefont {Hlawacek}, \citenamefont {van Gastel},
  \citenamefont {Zandvliet},\ and\ \citenamefont {Poelsema}}]{Veligura_2012}%
  \BibitemOpen
  \bibfield  {author} {\bibinfo {author} {\bibfnamefont {V.}~\bibnamefont
  {Veligura}}, \bibinfo {author} {\bibfnamefont {G.}~\bibnamefont {Hlawacek}},
  \bibinfo {author} {\bibfnamefont {R.}~\bibnamefont {van Gastel}}, \bibinfo
  {author} {\bibfnamefont {H.~J.~W.}\ \bibnamefont {Zandvliet}},\ and\ \bibinfo
  {author} {\bibfnamefont {B.}~\bibnamefont {Poelsema}},\ }\bibfield  {title}
  {\enquote {\bibinfo {title} {{Channeling in helium ion microscopy: Mapping of
  crystal orientation}},}\ }\href {https://doi.org/10.3762/bjnano.3.57}
  {\bibfield  {journal} {\bibinfo  {journal} {Beilstein J. Nanotechnol.}\
  }\textbf {\bibinfo {volume} {3}},\ \bibinfo {pages} {501--506} (\bibinfo
  {year} {2012})}\BibitemShut {NoStop}%
\bibitem [{\citenamefont {Hijazi}\ \emph {et~al.}(2019)\citenamefont {Hijazi},
  \citenamefont {Li}, \citenamefont {Barbacci}, \citenamefont {Schultz},
  \citenamefont {Thorpe}, \citenamefont {Gustafsson},\ and\ \citenamefont
  {Feldman}}]{Hijazi_2019}%
  \BibitemOpen
  \bibfield  {author} {\bibinfo {author} {\bibfnamefont {H.}~\bibnamefont
  {Hijazi}}, \bibinfo {author} {\bibfnamefont {M.}~\bibnamefont {Li}}, \bibinfo
  {author} {\bibfnamefont {D.}~\bibnamefont {Barbacci}}, \bibinfo {author}
  {\bibfnamefont {A.}~\bibnamefont {Schultz}}, \bibinfo {author} {\bibfnamefont
  {R.}~\bibnamefont {Thorpe}}, \bibinfo {author} {\bibfnamefont
  {T.}~\bibnamefont {Gustafsson}},\ and\ \bibinfo {author} {\bibfnamefont
  {L.~C.}\ \bibnamefont {Feldman}},\ }\bibfield  {title} {\enquote {\bibinfo
  {title} {{Channeling in the helium ion microscope}},}\ }\href
  {https://doi.org/10.1016/j.nimb.2019.07.002} {\bibfield  {journal} {\bibinfo
  {journal} {Nucl. Instrum. Meth. Phys. Res. B}\ }\textbf {\bibinfo {volume}
  {456}},\ \bibinfo {pages} {92--96} (\bibinfo {year} {2019})}\BibitemShut
  {NoStop}%
\bibitem [{\citenamefont {van Gastel}\ \emph {et~al.}(2012)\citenamefont {van
  Gastel}, \citenamefont {Hlawacek}, \citenamefont {Zandvliet},\ and\
  \citenamefont {Poelsema}}]{Gastel_2012}%
  \BibitemOpen
  \bibfield  {author} {\bibinfo {author} {\bibfnamefont {R.}~\bibnamefont {van
  Gastel}}, \bibinfo {author} {\bibfnamefont {G.}~\bibnamefont {Hlawacek}},
  \bibinfo {author} {\bibfnamefont {H.~J.~W.}\ \bibnamefont {Zandvliet}},\ and\
  \bibinfo {author} {\bibfnamefont {B.}~\bibnamefont {Poelsema}},\ }\bibfield
  {title} {\enquote {\bibinfo {title} {{Subsurface analysis of semiconductor
  structures with helium ion microscopy}},}\ }\href
  {https://doi.org/10.1016/j.microrel.2012.06.130} {\bibfield  {journal}
  {\bibinfo  {journal} {Microelectron. Reliab.}\ }\textbf {\bibinfo {volume}
  {52}},\ \bibinfo {pages} {2104--2109} (\bibinfo {year} {2012})}\BibitemShut
  {NoStop}%
\bibitem [{\citenamefont {Hlawacek}\ \emph {et~al.}(2016)\citenamefont
  {Hlawacek}, \citenamefont {Veligura}, \citenamefont {van Gastel},\ and\
  \citenamefont {Poelsema}}]{Hlawacek_2016a}%
  \BibitemOpen
  \bibfield  {author} {\bibinfo {author} {\bibfnamefont {G.}~\bibnamefont
  {Hlawacek}}, \bibinfo {author} {\bibfnamefont {V.}~\bibnamefont {Veligura}},
  \bibinfo {author} {\bibfnamefont {R.}~\bibnamefont {van Gastel}},\ and\
  \bibinfo {author} {\bibfnamefont {B.}~\bibnamefont {Poelsema}},\ }\bibfield
  {title} {\enquote {\bibinfo {title} {Channeling and backscatter imaging},}\
  }in\  \cite{Hlawacek_2016},\ Chap.~\bibinfo {chapter} {9}, pp.\ \bibinfo
  {pages} {205--224}\BibitemShut {NoStop}%
\bibitem [{\citenamefont {Stehling}\ \emph {et~al.}(2018)\citenamefont
  {Stehling}, \citenamefont {Masters}, \citenamefont {Zhou}, \citenamefont
  {O'Connell}, \citenamefont {Holland}, \citenamefont {Zhang},\ and\
  \citenamefont {Rodenburg}}]{Stehling_2018}%
  \BibitemOpen
  \bibfield  {author} {\bibinfo {author} {\bibfnamefont {N.}~\bibnamefont
  {Stehling}}, \bibinfo {author} {\bibfnamefont {R.}~\bibnamefont {Masters}},
  \bibinfo {author} {\bibfnamefont {Y.}~\bibnamefont {Zhou}}, \bibinfo {author}
  {\bibfnamefont {R.}~\bibnamefont {O'Connell}}, \bibinfo {author}
  {\bibfnamefont {C.}~\bibnamefont {Holland}}, \bibinfo {author} {\bibfnamefont
  {H.}~\bibnamefont {Zhang}},\ and\ \bibinfo {author} {\bibfnamefont
  {C.}~\bibnamefont {Rodenburg}},\ }\bibfield  {title} {\enquote {\bibinfo
  {title} {{New perspectives on nano-engineering by secondary electron
  spectroscopy in the helium ion and scanning electron microscope}},}\ }\href
  {https://doi.org/10.1557/mrc.2018.75} {\bibfield  {journal} {\bibinfo
  {journal} {MRS Commun.}\ }\textbf {\bibinfo {volume} {8}},\ \bibinfo {pages}
  {226--240} (\bibinfo {year} {2018})}\BibitemShut {NoStop}%
\bibitem [{\citenamefont {Masters}\ \emph {et~al.}(2015)\citenamefont
  {Masters}, \citenamefont {Pearson}, \citenamefont {Glen}, \citenamefont
  {Sasam}, \citenamefont {Li}, \citenamefont {Dapor}, \citenamefont {Donald},
  \citenamefont {Lidzey},\ and\ \citenamefont {Rodenburg}}]{Masters_2015}%
  \BibitemOpen
  \bibfield  {author} {\bibinfo {author} {\bibfnamefont {R.~C.}\ \bibnamefont
  {Masters}}, \bibinfo {author} {\bibfnamefont {A.~J.}\ \bibnamefont
  {Pearson}}, \bibinfo {author} {\bibfnamefont {T.~S.}\ \bibnamefont {Glen}},
  \bibinfo {author} {\bibfnamefont {F.-C.}\ \bibnamefont {Sasam}}, \bibinfo
  {author} {\bibfnamefont {L.}~\bibnamefont {Li}}, \bibinfo {author}
  {\bibfnamefont {M.}~\bibnamefont {Dapor}}, \bibinfo {author} {\bibfnamefont
  {A.~M.}\ \bibnamefont {Donald}}, \bibinfo {author} {\bibfnamefont {D.~G.}\
  \bibnamefont {Lidzey}},\ and\ \bibinfo {author} {\bibfnamefont
  {C.}~\bibnamefont {Rodenburg}},\ }\bibfield  {title} {\enquote {\bibinfo
  {title} {Sub-nanometre resolution imaging of polymer--fullerene photovoltaic
  blends using energy-filtered scanning electron microscopy},}\ }\href
  {https://doi.org/10.1038/ncomms7928} {\bibfield  {journal} {\bibinfo
  {journal} {Nat. Commun.}\ }\textbf {\bibinfo {volume} {6}},\ \bibinfo {pages}
  {6928} (\bibinfo {year} {2015})}\BibitemShut {NoStop}%
\bibitem [{\citenamefont {Anikeva}, \citenamefont {Petrov},\ and\ \citenamefont
  {Vyvenko}(2019)}]{Anikeva_2019}%
  \BibitemOpen
  \bibfield  {author} {\bibinfo {author} {\bibfnamefont {A.~E.}\ \bibnamefont
  {Anikeva}}, \bibinfo {author} {\bibfnamefont {Y.~V.}\ \bibnamefont
  {Petrov}},\ and\ \bibinfo {author} {\bibfnamefont {O.~F.}\ \bibnamefont
  {Vyvenko}},\ }\bibfield  {title} {\enquote {\bibinfo {title} {Secondary
  electron energy distribution from insulators in helium ion microscope},}\
  }in\ \href {https://doi.org/10.1063/1.5087657} {\emph {\bibinfo {booktitle}
  {AIP Conference Proceedings}}}\ (\bibinfo {year} {2019})\ p.\ \bibinfo
  {pages} {020001}\BibitemShut {NoStop}%
\bibitem [{\citenamefont {Suzuki}\ \emph {et~al.}(2008)\citenamefont {Suzuki},
  \citenamefont {Kudo}, \citenamefont {Sakai},\ and\ \citenamefont
  {Ichinokawa}}]{Suzuki_2008}%
  \BibitemOpen
  \bibfield  {author} {\bibinfo {author} {\bibfnamefont {T.}~\bibnamefont
  {Suzuki}}, \bibinfo {author} {\bibfnamefont {M.}~\bibnamefont {Kudo}},
  \bibinfo {author} {\bibfnamefont {Y.}~\bibnamefont {Sakai}},\ and\ \bibinfo
  {author} {\bibfnamefont {T.}~\bibnamefont {Ichinokawa}},\ }\bibfield  {title}
  {\enquote {\bibinfo {title} {Material contrast of scanning electron and ion
  microscope images of metals},}\ }\href
  {https://doi.org/10.1017/s1551929500054250} {\bibfield  {journal} {\bibinfo
  {journal} {Microscopy Today}\ }\textbf {\bibinfo {volume} {16}},\ \bibinfo
  {pages} {6--10} (\bibinfo {year} {2008})}\BibitemShut {NoStop}%
\bibitem [{\citenamefont {Farr}, \citenamefont {Hughes},\ and\ \citenamefont
  {Rodenburg}(2021)}]{Farr_2021}%
  \BibitemOpen
  \bibfield  {author} {\bibinfo {author} {\bibfnamefont {N.~T.~H.}\
  \bibnamefont {Farr}}, \bibinfo {author} {\bibfnamefont {G.~M.}\ \bibnamefont
  {Hughes}},\ and\ \bibinfo {author} {\bibfnamefont {C.}~\bibnamefont
  {Rodenburg}},\ }\bibfield  {title} {\enquote {\bibinfo {title} {Monitoring
  carbon in electron and ion beam deposition within {FIB-SEM}},}\ }\href
  {https://doi.org/10.3390/ma14113034} {\bibfield  {journal} {\bibinfo
  {journal} {Materials}\ }\textbf {\bibinfo {volume} {14}},\ \bibinfo {pages}
  {3034} (\bibinfo {year} {2021})}\BibitemShut {NoStop}%
\bibitem [{\citenamefont {Petrov}, \citenamefont {Anikeva},\ and\ \citenamefont
  {Vyvenko}(2018)}]{Petrov_2018}%
  \BibitemOpen
  \bibfield  {author} {\bibinfo {author} {\bibfnamefont {Y.~V.}\ \bibnamefont
  {Petrov}}, \bibinfo {author} {\bibfnamefont {A.~E.}\ \bibnamefont
  {Anikeva}},\ and\ \bibinfo {author} {\bibfnamefont {O.~F.}\ \bibnamefont
  {Vyvenko}},\ }\bibfield  {title} {\enquote {\bibinfo {title} {{Helium ion
  beam induced electron emission from insulating silicon nitride films under
  charging conditions}},}\ }\href {https://doi.org/10.1016/j.nimb.2018.04.001}
  {\bibfield  {journal} {\bibinfo  {journal} {Nucl. Instrum. Meth. Phys. Res.
  B}\ }\textbf {\bibinfo {volume} {425}},\ \bibinfo {pages} {11--17} (\bibinfo
  {year} {2018})}\BibitemShut {NoStop}%
\bibitem [{\citenamefont {Petrov}\ and\ \citenamefont
  {Vyvenko}(2011)}]{Petrov_2011}%
  \BibitemOpen
  \bibfield  {author} {\bibinfo {author} {\bibfnamefont {Y.}~\bibnamefont
  {Petrov}}\ and\ \bibinfo {author} {\bibfnamefont {O.}~\bibnamefont
  {Vyvenko}},\ }\bibfield  {title} {\enquote {\bibinfo {title} {Secondary
  electron emission spectra and energy selective imaging in helium ion
  microscope},}\ }in\ \href {https://doi.org/10.1117/12.886347} {\emph
  {\bibinfo {booktitle} {{SPIE} Proceedings}}},\ \bibinfo {editor} {edited by\
  \bibinfo {editor} {\bibfnamefont {M.~T.}\ \bibnamefont {Postek}}, \bibinfo
  {editor} {\bibfnamefont {D.~E.}\ \bibnamefont {Newbury}}, \bibinfo {editor}
  {\bibfnamefont {S.~F.}\ \bibnamefont {Platek}}, \bibinfo {editor}
  {\bibfnamefont {D.~C.}\ \bibnamefont {Joy}},\ and\ \bibinfo {editor}
  {\bibfnamefont {T.~K.}\ \bibnamefont {Maugel}}}\ (\bibinfo  {publisher}
  {{SPIE}},\ \bibinfo {year} {2011})\ p.\ \bibinfo {pages} {80360O}\BibitemShut
  {NoStop}%
\bibitem [{\citenamefont {Ramachandra}, \citenamefont {Griffin},\ and\
  \citenamefont {Joy}(2009)}]{Ramachandra_2009}%
  \BibitemOpen
  \bibfield  {author} {\bibinfo {author} {\bibfnamefont {R.}~\bibnamefont
  {Ramachandra}}, \bibinfo {author} {\bibfnamefont {B.}~\bibnamefont
  {Griffin}},\ and\ \bibinfo {author} {\bibfnamefont {D.}~\bibnamefont {Joy}},\
  }\bibfield  {title} {\enquote {\bibinfo {title} {A model of secondary
  electron imaging in the helium ion scanning microscope},}\ }\href
  {https://doi.org/10.1016/j.ultramic.2009.01.013} {\bibfield  {journal}
  {\bibinfo  {journal} {Ultramicroscopy}\ }\textbf {\bibinfo {volume} {109}},\
  \bibinfo {pages} {748--757} (\bibinfo {year} {2009})}\BibitemShut {NoStop}%
\bibitem [{\citenamefont {Pearson}\ \emph {et~al.}(2011)\citenamefont
  {Pearson}, \citenamefont {Boden}, \citenamefont {Bagnall}, \citenamefont
  {Lidzey},\ and\ \citenamefont {Rodenburg}}]{Pearson_2011}%
  \BibitemOpen
  \bibfield  {author} {\bibinfo {author} {\bibfnamefont {A.~J.}\ \bibnamefont
  {Pearson}}, \bibinfo {author} {\bibfnamefont {S.~A.}\ \bibnamefont {Boden}},
  \bibinfo {author} {\bibfnamefont {D.~M.}\ \bibnamefont {Bagnall}}, \bibinfo
  {author} {\bibfnamefont {D.~G.}\ \bibnamefont {Lidzey}},\ and\ \bibinfo
  {author} {\bibfnamefont {C.}~\bibnamefont {Rodenburg}},\ }\bibfield  {title}
  {\enquote {\bibinfo {title} {{Imaging the bulk nanoscale morphology of
  organic solar cell blends using helium ion microscopy.}}}\ }\href
  {https://doi.org/10.1021/nl202269n} {\bibfield  {journal} {\bibinfo
  {journal} {Nano Lett.}\ }\textbf {\bibinfo {volume} {11}},\ \bibinfo {pages}
  {4275--4281} (\bibinfo {year} {2011})}\BibitemShut {NoStop}%
\bibitem [{\citenamefont {Polak}(1995)}]{Polak_1995}%
  \BibitemOpen
  \bibfield  {author} {\bibinfo {author} {\bibfnamefont {M.}~\bibnamefont
  {Polak}},\ }\bibfield  {title} {\enquote {\bibinfo {title} {Ion-excited
  low-energy {A}uger electron emission from {Ti} and {TiNi}},}\ }\href
  {https://doi.org/10.1088/0953-8984/7/27/012} {\bibfield  {journal} {\bibinfo
  {journal} {J. Phys.: Condens. Matter}\ }\textbf {\bibinfo {volume} {7}},\
  \bibinfo {pages} {5275--5279} (\bibinfo {year} {1995})}\BibitemShut {NoStop}%
\bibitem [{\citenamefont {Whaley}\ and\ \citenamefont
  {Thomas}(1984)}]{Whaley_1984}%
  \BibitemOpen
  \bibfield  {author} {\bibinfo {author} {\bibfnamefont {R.}~\bibnamefont
  {Whaley}}\ and\ \bibinfo {author} {\bibfnamefont {E.~W.}\ \bibnamefont
  {Thomas}},\ }\bibfield  {title} {\enquote {\bibinfo {title} {{Auger spectra
  induced by Ne + and Ar + impact on Mg, Al, and Si}},}\ }\href
  {https://doi.org/10.1063/1.334106} {\bibfield  {journal} {\bibinfo  {journal}
  {J. Appl. Phys.}\ }\textbf {\bibinfo {volume} {56}},\ \bibinfo {pages}
  {1505--1513} (\bibinfo {year} {1984})}\BibitemShut {NoStop}%
\bibitem [{\citenamefont {Xu}\ \emph {et~al.}(1993)\citenamefont {Xu},
  \citenamefont {Riccardi}, \citenamefont {Oliva},\ and\ \citenamefont
  {Bonanno}}]{Xu_1993}%
  \BibitemOpen
  \bibfield  {author} {\bibinfo {author} {\bibfnamefont {F.}~\bibnamefont
  {Xu}}, \bibinfo {author} {\bibfnamefont {P.}~\bibnamefont {Riccardi}},
  \bibinfo {author} {\bibfnamefont {A.}~\bibnamefont {Oliva}},\ and\ \bibinfo
  {author} {\bibfnamefont {A.}~\bibnamefont {Bonanno}},\ }\bibfield  {title}
  {\enquote {\bibinfo {title} {{Ar L-shell and metal M-shell Auger electron
  emission for 14 keV Ar+ ion impact on Ca, Sc, Ti, V, Cr, Fe, Co, Ni, and
  Cu}},}\ }\href {https://doi.org/10.1016/0168-583X(93)95806-G} {\bibfield
  {journal} {\bibinfo  {journal} {Nucl. Instrum. Meth. Phys. Res. B}\ }\textbf
  {\bibinfo {volume} {78}},\ \bibinfo {pages} {251--254} (\bibinfo {year}
  {1993})}\BibitemShut {NoStop}%
\bibitem [{\citenamefont {Gallon}, \citenamefont {Orgassa},\ and\ \citenamefont
  {Matthew}(1995)}]{Gallon_1995}%
  \BibitemOpen
  \bibfield  {author} {\bibinfo {author} {\bibfnamefont {T.~E.}\ \bibnamefont
  {Gallon}}, \bibinfo {author} {\bibfnamefont {K.}~\bibnamefont {Orgassa}},\
  and\ \bibinfo {author} {\bibfnamefont {J.~A.~D.}\ \bibnamefont {Matthew}},\
  }\bibfield  {title} {\enquote {\bibinfo {title} {{Low-energy Auger electron
  emission from titanium induced by ion bombardment}},}\ }\href
  {https://doi.org/10.1088/0953-8984/7/45/009} {\bibfield  {journal} {\bibinfo
  {journal} {J. Phys.: Condens. Matter}\ }\textbf {\bibinfo {volume} {7}},\
  \bibinfo {pages} {8539--8546} (\bibinfo {year} {1995})}\BibitemShut {NoStop}%
\bibitem [{\citenamefont {Parvaneh}\ and\ \citenamefont
  {Hull}(2014{\natexlab{a}})}]{Parvaneh_2014}%
  \BibitemOpen
  \bibfield  {author} {\bibinfo {author} {\bibfnamefont {H.}~\bibnamefont
  {Parvaneh}}\ and\ \bibinfo {author} {\bibfnamefont {R.}~\bibnamefont
  {Hull}},\ }\bibfield  {title} {\enquote {\bibinfo {title} {Examination of
  ion-induced {A}uger electron spectra of {Ti}, {Cr} and {Co} in a
  mass-selecting focused ion beam with a gold–silicon liquid metal source},}\
  }\href {https://doi.org/10.1016/j.vacuum.2014.08.012} {\bibfield  {journal}
  {\bibinfo  {journal} {Vacuum}\ }\textbf {\bibinfo {volume} {110}},\ \bibinfo
  {pages} {69--73} (\bibinfo {year} {2014}{\natexlab{a}})}\BibitemShut
  {NoStop}%
\bibitem [{\citenamefont {Pereyaslavtsev}\ and\ \citenamefont
  {Naumkin}(2019)}]{Pereyaslavtsev_2019}%
  \BibitemOpen
  \bibfield  {author} {\bibinfo {author} {\bibfnamefont {A.~Y.}\ \bibnamefont
  {Pereyaslavtsev}}\ and\ \bibinfo {author} {\bibfnamefont {A.~V.}\
  \bibnamefont {Naumkin}},\ }\bibfield  {title} {\enquote {\bibinfo {title}
  {{Ion-induced Auger electrons contrast on cross-beam systems}},}\ }\href
  {https://doi.org/10.1088/1361-6641/ab541f} {\bibfield  {journal} {\bibinfo
  {journal} {Semicond. Sci. Tech.}\ }\textbf {\bibinfo {volume} {34}},\
  \bibinfo {pages} {124005} (\bibinfo {year} {2019})}\BibitemShut {NoStop}%
\bibitem [{\citenamefont {Giannuzzi}(2006)}]{Giannuzzi_2006}%
  \BibitemOpen
  \bibfield  {author} {\bibinfo {author} {\bibfnamefont {L.~A.}\ \bibnamefont
  {Giannuzzi}},\ }\bibfield  {title} {\enquote {\bibinfo {title}
  {{Particle-induced x-ray analysis using focused ion beams}},}\ }\href
  {https://doi.org/10.1002/sca.4950270402} {\bibfield  {journal} {\bibinfo
  {journal} {Scanning}\ }\textbf {\bibinfo {volume} {27}},\ \bibinfo {pages}
  {165--169} (\bibinfo {year} {2006})}\BibitemShut {NoStop}%
\bibitem [{\citenamefont {Giannuzzi}\ and\ \citenamefont
  {Gorman}(2009)}]{Giannuzzi_2009}%
  \BibitemOpen
  \bibfield  {author} {\bibinfo {author} {\bibfnamefont {L.~A.}\ \bibnamefont
  {Giannuzzi}}\ and\ \bibinfo {author} {\bibfnamefont {B.~P.}\ \bibnamefont
  {Gorman}},\ }\bibfield  {title} {\enquote {\bibinfo {title}
  {{Particle-induced x-ray emission in stainless steel using 30keV Ga+ focused
  ion beams}},}\ }\href {https://doi.org/10.1116/1.3136852} {\bibfield
  {journal} {\bibinfo  {journal} {J. Vac. Sci. Technol. A}\ }\textbf {\bibinfo
  {volume} {27}},\ \bibinfo {pages} {668--671} (\bibinfo {year}
  {2009})}\BibitemShut {NoStop}%
\bibitem [{\citenamefont {Scipioni}, \citenamefont {Stern},\ and\ \citenamefont
  {Notte}(2007)}]{Scipioni_2007}%
  \BibitemOpen
  \bibfield  {author} {\bibinfo {author} {\bibfnamefont {L.}~\bibnamefont
  {Scipioni}}, \bibinfo {author} {\bibfnamefont {L.}~\bibnamefont {Stern}},\
  and\ \bibinfo {author} {\bibfnamefont {J.}~\bibnamefont {Notte}},\ }\bibfield
   {title} {\enquote {\bibinfo {title} {Applications of the helium ion
  microscope},}\ }\href {https://doi.org/10.1017/s1551929500061897} {\bibfield
  {journal} {\bibinfo  {journal} {Microscopy Today}\ }\textbf {\bibinfo
  {volume} {15}},\ \bibinfo {pages} {12--15} (\bibinfo {year}
  {2007})}\BibitemShut {NoStop}%
\bibitem [{\citenamefont {Boden}\ \emph {et~al.}(2012)\citenamefont {Boden},
  \citenamefont {Franklin}, \citenamefont {Scipioni}, \citenamefont {Bagnall},\
  and\ \citenamefont {Rutt}}]{Boden_2012}%
  \BibitemOpen
  \bibfield  {author} {\bibinfo {author} {\bibfnamefont {S.~A.}\ \bibnamefont
  {Boden}}, \bibinfo {author} {\bibfnamefont {T.~M.~W.}\ \bibnamefont
  {Franklin}}, \bibinfo {author} {\bibfnamefont {L.}~\bibnamefont {Scipioni}},
  \bibinfo {author} {\bibfnamefont {D.~M.}\ \bibnamefont {Bagnall}},\ and\
  \bibinfo {author} {\bibfnamefont {H.~N.}\ \bibnamefont {Rutt}},\ }\bibfield
  {title} {\enquote {\bibinfo {title} {{Ionoluminescence in the Helium Ion
  Microscope}},}\ }\href {https://doi.org/10.1017/S1431927612013463} {\bibfield
   {journal} {\bibinfo  {journal} {Microsc. Microanal.}\ }\textbf {\bibinfo
  {volume} {18}},\ \bibinfo {pages} {1253--1262} (\bibinfo {year}
  {2012})}\BibitemShut {NoStop}%
\bibitem [{\citenamefont {Franklin}(2012)}]{Franklin_2012}%
  \BibitemOpen
  \bibfield  {author} {\bibinfo {author} {\bibfnamefont {T.~M.~W.}\
  \bibnamefont {Franklin}},\ }\emph {\bibinfo {title} {{Scanning
  Ionoluminescence Microscopy with a Helium Ion Microscope}}},\ \href
  {http://eprints.soton.ac.uk/id/eprint/352281} {Ph.D. thesis},\ \bibinfo
  {school} {University of Southampton} (\bibinfo {year} {2012})\BibitemShut
  {NoStop}%
\bibitem [{\citenamefont {Veligura}\ \emph
  {et~al.}(2014{\natexlab{a}})\citenamefont {Veligura}, \citenamefont
  {Hlawacek}, \citenamefont {van Gastel}, \citenamefont {Zandvliet},\ and\
  \citenamefont {Poelsema}}]{Veligura_2014a}%
  \BibitemOpen
  \bibfield  {author} {\bibinfo {author} {\bibfnamefont {V.}~\bibnamefont
  {Veligura}}, \bibinfo {author} {\bibfnamefont {G.}~\bibnamefont {Hlawacek}},
  \bibinfo {author} {\bibfnamefont {R.}~\bibnamefont {van Gastel}}, \bibinfo
  {author} {\bibfnamefont {H.~J.~W.}\ \bibnamefont {Zandvliet}},\ and\ \bibinfo
  {author} {\bibfnamefont {B.}~\bibnamefont {Poelsema}},\ }\bibfield  {title}
  {\enquote {\bibinfo {title} {{A high resolution ionoluminescence study of
  defect creation and interaction}},}\ }\href
  {https://doi.org/10.1088/0953-8984/26/16/165401} {\bibfield  {journal}
  {\bibinfo  {journal} {J. Phys.: Condens. Matter}\ }\textbf {\bibinfo {volume}
  {26}},\ \bibinfo {pages} {165401} (\bibinfo {year}
  {2014}{\natexlab{a}})}\BibitemShut {NoStop}%
\bibitem [{\citenamefont {Veligura}(2014)}]{Veligura_2014}%
  \BibitemOpen
  \bibfield  {author} {\bibinfo {author} {\bibfnamefont {V.}~\bibnamefont
  {Veligura}},\ }\emph {\bibinfo {title} {Material characterization and
  modification using helium ion microscopy: various examples}},\ \href
  {https://doi.org/10.3990/1.9789036536271} {\bibinfo {type} {phdthesis}},\
  \bibinfo  {school} {University Twente, Enschede, Netherlands} (\bibinfo
  {year} {2014})\BibitemShut {NoStop}%
\bibitem [{\citenamefont {Veligura}\ and\ \citenamefont
  {Hlawacek}(2016)}]{Veligura_2016}%
  \BibitemOpen
  \bibfield  {author} {\bibinfo {author} {\bibfnamefont {V.}~\bibnamefont
  {Veligura}}\ and\ \bibinfo {author} {\bibfnamefont {G.}~\bibnamefont
  {Hlawacek}},\ }\bibfield  {title} {\enquote {\bibinfo {title}
  {{Ionoluminescence}},}\ }in\  \cite{Hlawacek_2016},\ Chap.~\bibinfo {chapter}
  {14}, pp.\ \bibinfo {pages} {325--351}\BibitemShut {NoStop}%
\bibitem [{\citenamefont {Veligura}\ \emph
  {et~al.}(2014{\natexlab{b}})\citenamefont {Veligura}, \citenamefont
  {Hlawacek}, \citenamefont {Jahn}, \citenamefont {van Gastel}, \citenamefont
  {Zandvliet},\ and\ \citenamefont {Poelsema}}]{Veligura_2014b}%
  \BibitemOpen
  \bibfield  {author} {\bibinfo {author} {\bibfnamefont {V.}~\bibnamefont
  {Veligura}}, \bibinfo {author} {\bibfnamefont {G.}~\bibnamefont {Hlawacek}},
  \bibinfo {author} {\bibfnamefont {U.}~\bibnamefont {Jahn}}, \bibinfo {author}
  {\bibfnamefont {R.}~\bibnamefont {van Gastel}}, \bibinfo {author}
  {\bibfnamefont {H.~J.~W.}\ \bibnamefont {Zandvliet}},\ and\ \bibinfo {author}
  {\bibfnamefont {B.}~\bibnamefont {Poelsema}},\ }\bibfield  {title} {\enquote
  {\bibinfo {title} {{Creation and physical aspects of luminescent patterns
  using helium ion microscopy}},}\ }\href {https://doi.org/10.1063/1.4875480}
  {\bibfield  {journal} {\bibinfo  {journal} {J. Appl. Phys.}\ }\textbf
  {\bibinfo {volume} {115}},\ \bibinfo {pages} {183502} (\bibinfo {year}
  {2014}{\natexlab{b}})}\BibitemShut {NoStop}%
\bibitem [{\citenamefont {Ogawa}\ \emph {et~al.}(2012)\citenamefont {Ogawa},
  \citenamefont {Iijima}, \citenamefont {Awata}, \citenamefont {Sugie},
  \citenamefont {Kawasaki},\ and\ \citenamefont {Otsuka}}]{Ogawa_2012}%
  \BibitemOpen
  \bibfield  {author} {\bibinfo {author} {\bibfnamefont {S.}~\bibnamefont
  {Ogawa}}, \bibinfo {author} {\bibfnamefont {T.}~\bibnamefont {Iijima}},
  \bibinfo {author} {\bibfnamefont {S.}~\bibnamefont {Awata}}, \bibinfo
  {author} {\bibfnamefont {R.}~\bibnamefont {Sugie}}, \bibinfo {author}
  {\bibfnamefont {N.}~\bibnamefont {Kawasaki}},\ and\ \bibinfo {author}
  {\bibfnamefont {Y.}~\bibnamefont {Otsuka}},\ }\bibfield  {title} {\enquote
  {\bibinfo {title} {{Characterization of Damage in SiO2 during Helium Ion
  Microscope Observation by Luminescence and TEM-EELS}},}\ }\href
  {https://doi.org/10.1017/S1431927612005922} {\bibfield  {journal} {\bibinfo
  {journal} {Microsc. Microanal.}\ }\textbf {\bibinfo {volume} {18}},\ \bibinfo
  {pages} {814--815} (\bibinfo {year} {2012})}\BibitemShut {NoStop}%
\bibitem [{\citenamefont {Stevie}(2005)}]{Stevie_2005}%
  \BibitemOpen
  \bibfield  {author} {\bibinfo {author} {\bibfnamefont {F.~A.}\ \bibnamefont
  {Stevie}},\ }\bibfield  {title} {\enquote {\bibinfo {title} {Focused ion beam
  secondary ion mass spectrometry {(FIB-SIMS)}},}\ }in\
  \cite{Giannuzzi_2005a},\ pp.\ \bibinfo {pages} {269--280}\BibitemShut
  {NoStop}%
\bibitem [{\citenamefont {Wirtz}\ \emph {et~al.}(2020)\citenamefont {Wirtz},
  \citenamefont {Castro}, \citenamefont {Biesemeier}, \citenamefont {Hoang},\
  and\ \citenamefont {Audinot}}]{Wirtz2020}%
  \BibitemOpen
  \bibfield  {author} {\bibinfo {author} {\bibfnamefont {T.}~\bibnamefont
  {Wirtz}}, \bibinfo {author} {\bibfnamefont {O.~D.}\ \bibnamefont {Castro}},
  \bibinfo {author} {\bibfnamefont {A.}~\bibnamefont {Biesemeier}}, \bibinfo
  {author} {\bibfnamefont {H.~Q.}\ \bibnamefont {Hoang}},\ and\ \bibinfo
  {author} {\bibfnamefont {J.-N.}\ \bibnamefont {Audinot}},\ }\bibfield
  {title} {\enquote {\bibinfo {title} {Magnetic sector {SIMS} system with
  continuous focal plane detector for advanced analytical capabilities on {FIB}
  instruments},}\ }\href {https://doi.org/10.1017/s1431927620019984} {\bibfield
   {journal} {\bibinfo  {journal} {Microsc. Microanal.}\ }\textbf {\bibinfo
  {volume} {26}},\ \bibinfo {pages} {1972--1974} (\bibinfo {year}
  {2020})}\BibitemShut {NoStop}%
\bibitem [{\citenamefont {Frache}\ \emph {et~al.}(2010)\citenamefont {Frache},
  \citenamefont {Adib}, \citenamefont {Audinot},\ and\ \citenamefont
  {Migeon}}]{Frache_2010}%
  \BibitemOpen
  \bibfield  {author} {\bibinfo {author} {\bibfnamefont {G.}~\bibnamefont
  {Frache}}, \bibinfo {author} {\bibfnamefont {B.~E.}\ \bibnamefont {Adib}},
  \bibinfo {author} {\bibfnamefont {J.-N.}\ \bibnamefont {Audinot}},\ and\
  \bibinfo {author} {\bibfnamefont {H.-N.}\ \bibnamefont {Migeon}},\ }\bibfield
   {title} {\enquote {\bibinfo {title} {Evaluation of ionization yields under
  gallium bombardment},}\ }\href {https://doi.org/10.1002/sia.3418} {\bibfield
  {journal} {\bibinfo  {journal} {Surf. Interface Anal.}\ }\textbf {\bibinfo
  {volume} {43}},\ \bibinfo {pages} {639--642} (\bibinfo {year}
  {2010})}\BibitemShut {NoStop}%
\bibitem [{\citenamefont {Philipp}\ \emph {et~al.}(2006)\citenamefont
  {Philipp}, \citenamefont {Wirtz}, \citenamefont {Migeon},\ and\ \citenamefont
  {Scherrer}}]{Philipp_2006}%
  \BibitemOpen
  \bibfield  {author} {\bibinfo {author} {\bibfnamefont {P.}~\bibnamefont
  {Philipp}}, \bibinfo {author} {\bibfnamefont {T.}~\bibnamefont {Wirtz}},
  \bibinfo {author} {\bibfnamefont {H.-N.}\ \bibnamefont {Migeon}},\ and\
  \bibinfo {author} {\bibfnamefont {H.}~\bibnamefont {Scherrer}},\ }\bibfield
  {title} {\enquote {\bibinfo {title} {{SIMS} analysis with neutral cesium
  deposition: negative secondary ion sensitivity increase and quantification
  aspects},}\ }\href {https://doi.org/10.1016/j.ijms.2006.02.018} {\bibfield
  {journal} {\bibinfo  {journal} {Int. J. Mass Spectrom.}\ }\textbf {\bibinfo
  {volume} {253}},\ \bibinfo {pages} {71--78} (\bibinfo {year}
  {2006})}\BibitemShut {NoStop}%
\bibitem [{\citenamefont {Ma}\ \emph {et~al.}(2017{\natexlab{a}})\citenamefont
  {Ma}, \citenamefont {Hartmann}, \citenamefont {de~Aberasturi}, \citenamefont
  {Yang}, \citenamefont {Soenen}, \citenamefont {Manshian}, \citenamefont
  {Franz}, \citenamefont {Valdeperez}, \citenamefont {Pelaz}, \citenamefont
  {Feliu}, \citenamefont {Hampp}, \citenamefont {Riethm{\"{u}}ller},
  \citenamefont {Vieker}, \citenamefont {Frese}, \citenamefont
  {G{\"{o}}lzh{\"{a}}user}, \citenamefont {Simonich}, \citenamefont {Tanguay},
  \citenamefont {Liang},\ and\ \citenamefont {Parak}}]{Ma_2017}%
  \BibitemOpen
  \bibfield  {author} {\bibinfo {author} {\bibfnamefont {X.}~\bibnamefont
  {Ma}}, \bibinfo {author} {\bibfnamefont {R.}~\bibnamefont {Hartmann}},
  \bibinfo {author} {\bibfnamefont {D.~J.}\ \bibnamefont {de~Aberasturi}},
  \bibinfo {author} {\bibfnamefont {F.}~\bibnamefont {Yang}}, \bibinfo {author}
  {\bibfnamefont {S.~J.~H.}\ \bibnamefont {Soenen}}, \bibinfo {author}
  {\bibfnamefont {B.~B.}\ \bibnamefont {Manshian}}, \bibinfo {author}
  {\bibfnamefont {J.}~\bibnamefont {Franz}}, \bibinfo {author} {\bibfnamefont
  {D.}~\bibnamefont {Valdeperez}}, \bibinfo {author} {\bibfnamefont
  {B.}~\bibnamefont {Pelaz}}, \bibinfo {author} {\bibfnamefont
  {N.}~\bibnamefont {Feliu}}, \bibinfo {author} {\bibfnamefont
  {N.}~\bibnamefont {Hampp}}, \bibinfo {author} {\bibfnamefont
  {C.}~\bibnamefont {Riethm{\"{u}}ller}}, \bibinfo {author} {\bibfnamefont
  {H.}~\bibnamefont {Vieker}}, \bibinfo {author} {\bibfnamefont
  {N.}~\bibnamefont {Frese}}, \bibinfo {author} {\bibfnamefont
  {A.}~\bibnamefont {G{\"{o}}lzh{\"{a}}user}}, \bibinfo {author} {\bibfnamefont
  {M.}~\bibnamefont {Simonich}}, \bibinfo {author} {\bibfnamefont {R.~L.}\
  \bibnamefont {Tanguay}}, \bibinfo {author} {\bibfnamefont {X.-J.}\
  \bibnamefont {Liang}},\ and\ \bibinfo {author} {\bibfnamefont {W.~J.}\
  \bibnamefont {Parak}},\ }\bibfield  {title} {\enquote {\bibinfo {title}
  {Colloidal gold nanoparticles induce changes in cellular and subcellular
  morphology},}\ }\href {https://doi.org/10.1021/acsnano.7b01760} {\bibfield
  {journal} {\bibinfo  {journal} {{ACS} Nano}\ }\textbf {\bibinfo {volume}
  {11}},\ \bibinfo {pages} {7807--7820} (\bibinfo {year}
  {2017}{\natexlab{a}})}\BibitemShut {NoStop}%
\bibitem [{\citenamefont {Bidlack}\ \emph {et~al.}(2014)\citenamefont
  {Bidlack}, \citenamefont {Huynh}, \citenamefont {Marshman},\ and\
  \citenamefont {Goetze}}]{Bidlack2014}%
  \BibitemOpen
  \bibfield  {author} {\bibinfo {author} {\bibfnamefont {F.~B.}\ \bibnamefont
  {Bidlack}}, \bibinfo {author} {\bibfnamefont {C.}~\bibnamefont {Huynh}},
  \bibinfo {author} {\bibfnamefont {J.}~\bibnamefont {Marshman}},\ and\
  \bibinfo {author} {\bibfnamefont {B.}~\bibnamefont {Goetze}},\ }\bibfield
  {title} {\enquote {\bibinfo {title} {{Helium ion microscopy of enamel
  crystallites and extracellular tooth enamel matrix}},}\ }\href
  {https://doi.org/10.3389/fphys.2014.00395} {\bibfield  {journal} {\bibinfo
  {journal} {Front. Physiol.}\ }\textbf {\bibinfo {volume} {5}},\ \bibinfo
  {pages} {1--10} (\bibinfo {year} {2014})}\BibitemShut {NoStop}%
\bibitem [{\citenamefont {Magnuson}\ and\ \citenamefont
  {Carlston}(1963{\natexlab{a}})}]{Magnuson1963}%
  \BibitemOpen
  \bibfield  {author} {\bibinfo {author} {\bibfnamefont {G.~D.}\ \bibnamefont
  {Magnuson}}\ and\ \bibinfo {author} {\bibfnamefont {C.~E.}\ \bibnamefont
  {Carlston}},\ }\bibfield  {title} {\enquote {\bibinfo {title} {Electron
  ejection from metals due to 1- to 10-ke{V} noble gas ion bombardment. {I}.
  {P}olycrystalline materials},}\ }\href
  {https://doi.org/10.1103/PhysRev.129.2403} {\bibfield  {journal} {\bibinfo
  {journal} {Phys. Rev.}\ }\textbf {\bibinfo {volume} {129}},\ \bibinfo {pages}
  {2403--2408} (\bibinfo {year} {1963}{\natexlab{a}})}\BibitemShut {NoStop}%
\bibitem [{\citenamefont {Magnuson}\ and\ \citenamefont
  {Carlston}(1963{\natexlab{b}})}]{Magnuson1963a}%
  \BibitemOpen
  \bibfield  {author} {\bibinfo {author} {\bibfnamefont {G.~D.}\ \bibnamefont
  {Magnuson}}\ and\ \bibinfo {author} {\bibfnamefont {C.~E.}\ \bibnamefont
  {Carlston}},\ }\bibfield  {title} {\enquote {\bibinfo {title} {Electron
  ejection from metals due to 1- to 10-ke{v} noble gas ion bombardment. {II.}
  {S}ingle crystal},}\ }\href {https://doi.org/10.1103/PhysRev.129.2409}
  {\bibfield  {journal} {\bibinfo  {journal} {Phys. Rev.}\ }\textbf {\bibinfo
  {volume} {129}},\ \bibinfo {pages} {2409--2412} (\bibinfo {year}
  {1963}{\natexlab{b}})}\BibitemShut {NoStop}%
\bibitem [{\citenamefont {Briggs}\ and\ \citenamefont
  {Pathak}(2001)}]{Briggs2001}%
  \BibitemOpen
  \bibfield  {author} {\bibinfo {author} {\bibfnamefont {J.~S.}\ \bibnamefont
  {Briggs}}\ and\ \bibinfo {author} {\bibfnamefont {a.~P.}\ \bibnamefont
  {Pathak}},\ }\bibfield  {title} {\enquote {\bibinfo {title} {{The stopping
  power of solids for low-velocity channelled heavy ions}},}\ }\href
  {https://doi.org/10.1088/0022-3719/7/11/009} {\bibfield  {journal} {\bibinfo
  {journal} {J. Phys. C Solid State Phys.}\ }\textbf {\bibinfo {volume} {7}},\
  \bibinfo {pages} {1929--1936} (\bibinfo {year} {2001})}\BibitemShut {NoStop}%
\bibitem [{\citenamefont {Hlawacek}\ \emph {et~al.}(2015)\citenamefont
  {Hlawacek}, \citenamefont {Jankowski}, \citenamefont {Wormeester},
  \citenamefont {van Gastel}, \citenamefont {Zandvliet},\ and\ \citenamefont
  {Poelsema}}]{Hlawacek_2015}%
  \BibitemOpen
  \bibfield  {author} {\bibinfo {author} {\bibfnamefont {G.}~\bibnamefont
  {Hlawacek}}, \bibinfo {author} {\bibfnamefont {M.}~\bibnamefont {Jankowski}},
  \bibinfo {author} {\bibfnamefont {H.}~\bibnamefont {Wormeester}}, \bibinfo
  {author} {\bibfnamefont {R.}~\bibnamefont {van Gastel}}, \bibinfo {author}
  {\bibfnamefont {H.~J.~W.}\ \bibnamefont {Zandvliet}},\ and\ \bibinfo {author}
  {\bibfnamefont {B.}~\bibnamefont {Poelsema}},\ }\bibfield  {title} {\enquote
  {\bibinfo {title} {{Visualization of steps and surface reconstructions in
  Helium Ion Microscopy with atomic precision}},}\ }\href
  {https://doi.org/10.1016/j.ultramic.2015.11.009} {\bibfield  {journal}
  {\bibinfo  {journal} {Ultramicroscopy}\ }\textbf {\bibinfo {volume} {162}},\
  \bibinfo {pages} {17--24} (\bibinfo {year} {2015})},\ \Eprint
  {https://arxiv.org/abs/1505.00020} {arXiv:1505.00020} \BibitemShut {NoStop}%
\bibitem [{\citenamefont {Langlois}\ \emph {et~al.}(2015)\citenamefont
  {Langlois}, \citenamefont {Douillard}, \citenamefont {Yuan}, \citenamefont
  {Blanchard}, \citenamefont {Descamps-Mandine}, \citenamefont {Van~de
  Moort{\`{e}}le}, \citenamefont {Rigotti},\ and\ \citenamefont
  {Epicier}}]{Langlois_2015}%
  \BibitemOpen
  \bibfield  {author} {\bibinfo {author} {\bibfnamefont {C.}~\bibnamefont
  {Langlois}}, \bibinfo {author} {\bibfnamefont {T.}~\bibnamefont {Douillard}},
  \bibinfo {author} {\bibfnamefont {H.}~\bibnamefont {Yuan}}, \bibinfo {author}
  {\bibfnamefont {N.~P.}\ \bibnamefont {Blanchard}}, \bibinfo {author}
  {\bibfnamefont {A.}~\bibnamefont {Descamps-Mandine}}, \bibinfo {author}
  {\bibfnamefont {B.}~\bibnamefont {Van~de Moort{\`{e}}le}}, \bibinfo {author}
  {\bibfnamefont {C.}~\bibnamefont {Rigotti}},\ and\ \bibinfo {author}
  {\bibfnamefont {T.}~\bibnamefont {Epicier}},\ }\bibfield  {title} {\enquote
  {\bibinfo {title} {Crystal orientation mapping via ion channeling: an
  alternative to {EBSD}},}\ }\href
  {https://doi.org/10.1016/j.ultramic.2015.05.023} {\bibfield  {journal}
  {\bibinfo  {journal} {Ultramicroscopy}\ }\textbf {\bibinfo {volume} {157}},\
  \bibinfo {pages} {65--72} (\bibinfo {year} {2015})}\BibitemShut {NoStop}%
\bibitem [{\citenamefont {Rodenburg}\ \emph {et~al.}(2014)\citenamefont
  {Rodenburg}, \citenamefont {Jepson}, \citenamefont {Boden},\ and\
  \citenamefont {Bagnall}}]{Rodenburg_2014}%
  \BibitemOpen
  \bibfield  {author} {\bibinfo {author} {\bibfnamefont {C.}~\bibnamefont
  {Rodenburg}}, \bibinfo {author} {\bibfnamefont {M.~A.~E.}\ \bibnamefont
  {Jepson}}, \bibinfo {author} {\bibfnamefont {S.~A.}\ \bibnamefont {Boden}},\
  and\ \bibinfo {author} {\bibfnamefont {D.~M.}\ \bibnamefont {Bagnall}},\
  }\bibfield  {title} {\enquote {\bibinfo {title} {Helium ion microscopy and
  energy selective scanning electron microscopy -- two advanced microscopy
  techniques with complementary applications},}\ }\href
  {https://doi.org/10.1088/1742-6596/522/1/012049} {\bibfield  {journal}
  {\bibinfo  {journal} {J. Phys.: Conf. Ser.}\ }\textbf {\bibinfo {volume}
  {522}},\ \bibinfo {pages} {012049} (\bibinfo {year} {2014})}\BibitemShut
  {NoStop}%
\bibitem [{\citenamefont {Petrov}\ and\ \citenamefont
  {Vyvenko}(2016)}]{Petrov_2016}%
  \BibitemOpen
  \bibfield  {author} {\bibinfo {author} {\bibfnamefont {Y.~V.}\ \bibnamefont
  {Petrov}}\ and\ \bibinfo {author} {\bibfnamefont {O.~F.}\ \bibnamefont
  {Vyvenko}},\ }\bibfield  {title} {\enquote {\bibinfo {title} {{Secondary
  Electron Generation in the Helium Ion Microscope: Basics and Imaging}},}\
  }in\  \cite{Hlawacek_2016},\ Chap.~\bibinfo {chapter} {5}, pp.\ \bibinfo
  {pages} {119--146}\BibitemShut {NoStop}%
\bibitem [{\citenamefont {Petrov}, \citenamefont {Vyvenko},\ and\ \citenamefont
  {Bondarenko}(2010)}]{Petrov_2010}%
  \BibitemOpen
  \bibfield  {author} {\bibinfo {author} {\bibfnamefont {Y.~V.}\ \bibnamefont
  {Petrov}}, \bibinfo {author} {\bibfnamefont {O.~F.}\ \bibnamefont
  {Vyvenko}},\ and\ \bibinfo {author} {\bibfnamefont {A.~S.}\ \bibnamefont
  {Bondarenko}},\ }\bibfield  {title} {\enquote {\bibinfo {title} {{Scanning
  helium ion microscope: Distribution of secondary electrons and ion
  channeling}},}\ }\href {https://doi.org/10.1134/S1027451010050186} {\bibfield
   {journal} {\bibinfo  {journal} {J. Surf. Investig.}\ }\textbf {\bibinfo
  {volume} {4}},\ \bibinfo {pages} {792--795} (\bibinfo {year}
  {2010})}\BibitemShut {NoStop}%
\bibitem [{\citenamefont {Parvaneh}\ and\ \citenamefont
  {Hull}(2014{\natexlab{b}})}]{Parvaneh_2014a}%
  \BibitemOpen
  \bibfield  {author} {\bibinfo {author} {\bibfnamefont {H.}~\bibnamefont
  {Parvaneh}}\ and\ \bibinfo {author} {\bibfnamefont {R.}~\bibnamefont
  {Hull}},\ }\bibfield  {title} {\enquote {\bibinfo {title} {{Ion-induced Auger
  Electron Spectroscopy as a Potential Route to Chemical Focused-Ion Beam
  Tomography}},}\ }\href {https://doi.org/10.1017/S1431927614003274} {\bibfield
   {journal} {\bibinfo  {journal} {Microsc. Microanal.}\ }\textbf {\bibinfo
  {volume} {20}},\ \bibinfo {pages} {310--311} (\bibinfo {year}
  {2014}{\natexlab{b}})}\BibitemShut {NoStop}%
\bibitem [{\citenamefont {R{\"{a}}cke}\ \emph {et~al.}(2019)\citenamefont
  {R{\"{a}}cke}, \citenamefont {Staacke}, \citenamefont {Gerlach},
  \citenamefont {Meijer},\ and\ \citenamefont {Spemann}}]{Raecke_2019}%
  \BibitemOpen
  \bibfield  {author} {\bibinfo {author} {\bibfnamefont {P.}~\bibnamefont
  {R{\"{a}}cke}}, \bibinfo {author} {\bibfnamefont {R.}~\bibnamefont
  {Staacke}}, \bibinfo {author} {\bibfnamefont {J.~W.}\ \bibnamefont
  {Gerlach}}, \bibinfo {author} {\bibfnamefont {J.}~\bibnamefont {Meijer}},\
  and\ \bibinfo {author} {\bibfnamefont {D.}~\bibnamefont {Spemann}},\
  }\bibfield  {title} {\enquote {\bibinfo {title} {Image charge detection
  statistics relevant for deterministic ion implantation},}\ }\href
  {https://doi.org/10.1088/1361-6463/ab1d04} {\bibfield  {journal} {\bibinfo
  {journal} {J. Phys. D: Appl. Phys.}\ }\textbf {\bibinfo {volume} {52}},\
  \bibinfo {pages} {305103} (\bibinfo {year} {2019})}\BibitemShut {NoStop}%
\bibitem [{\citenamefont {R{\"{a}}cke}, \citenamefont {Meijer},\ and\
  \citenamefont {Spemann}(2022)}]{Raecke_2022}%
  \BibitemOpen
  \bibfield  {author} {\bibinfo {author} {\bibfnamefont {P.}~\bibnamefont
  {R{\"{a}}cke}}, \bibinfo {author} {\bibfnamefont {J.}~\bibnamefont
  {Meijer}},\ and\ \bibinfo {author} {\bibfnamefont {D.}~\bibnamefont
  {Spemann}},\ }\bibfield  {title} {\enquote {\bibinfo {title} {Image charge
  detection of ion bunches using a segmented, cryogenic detector},}\ }\href
  {https://doi.org/10.1063/5.0096094} {\bibfield  {journal} {\bibinfo
  {journal} {J. Appl. Phys.}\ }\textbf {\bibinfo {volume} {131}},\ \bibinfo
  {pages} {204502} (\bibinfo {year} {2022})}\BibitemShut {NoStop}%
\bibitem [{\citenamefont {Schenkel}\ \emph {et~al.}(2003)\citenamefont
  {Schenkel}, \citenamefont {Persaud}, \citenamefont {Park}, \citenamefont
  {Nilsson}, \citenamefont {Bokor}, \citenamefont {Liddle}, \citenamefont
  {Keller}, \citenamefont {Schneider}, \citenamefont {Cheng},\ and\
  \citenamefont {Humphries}}]{Schenkel_2003}%
  \BibitemOpen
  \bibfield  {author} {\bibinfo {author} {\bibfnamefont {T.}~\bibnamefont
  {Schenkel}}, \bibinfo {author} {\bibfnamefont {A.}~\bibnamefont {Persaud}},
  \bibinfo {author} {\bibfnamefont {S.~J.}\ \bibnamefont {Park}}, \bibinfo
  {author} {\bibfnamefont {J.}~\bibnamefont {Nilsson}}, \bibinfo {author}
  {\bibfnamefont {J.}~\bibnamefont {Bokor}}, \bibinfo {author} {\bibfnamefont
  {J.~A.}\ \bibnamefont {Liddle}}, \bibinfo {author} {\bibfnamefont
  {R.}~\bibnamefont {Keller}}, \bibinfo {author} {\bibfnamefont {D.~H.}\
  \bibnamefont {Schneider}}, \bibinfo {author} {\bibfnamefont {D.~W.}\
  \bibnamefont {Cheng}},\ and\ \bibinfo {author} {\bibfnamefont {D.~E.}\
  \bibnamefont {Humphries}},\ }\bibfield  {title} {\enquote {\bibinfo {title}
  {Solid state quantum computer development in silicon with single ion
  implantation},}\ }\href {https://doi.org/10.1063/1.1622109} {\bibfield
  {journal} {\bibinfo  {journal} {J. Appl. Phys.}\ }\textbf {\bibinfo {volume}
  {94}},\ \bibinfo {pages} {7017--7024} (\bibinfo {year} {2003})}\BibitemShut
  {NoStop}%
\bibitem [{\citenamefont {McCallum}\ \emph {et~al.}(2012)\citenamefont
  {McCallum}, \citenamefont {Jamieson}, \citenamefont {Yang}, \citenamefont
  {Alves}, \citenamefont {Johnson}, \citenamefont {Hopf}, \citenamefont
  {Thompson},\ and\ \citenamefont {van Donkelaar}}]{McCallum_2012}%
  \BibitemOpen
  \bibfield  {author} {\bibinfo {author} {\bibfnamefont {J.~C.}\ \bibnamefont
  {McCallum}}, \bibinfo {author} {\bibfnamefont {D.~N.}\ \bibnamefont
  {Jamieson}}, \bibinfo {author} {\bibfnamefont {C.}~\bibnamefont {Yang}},
  \bibinfo {author} {\bibfnamefont {A.~D.}\ \bibnamefont {Alves}}, \bibinfo
  {author} {\bibfnamefont {B.~C.}\ \bibnamefont {Johnson}}, \bibinfo {author}
  {\bibfnamefont {T.}~\bibnamefont {Hopf}}, \bibinfo {author} {\bibfnamefont
  {S.~C.}\ \bibnamefont {Thompson}},\ and\ \bibinfo {author} {\bibfnamefont
  {J.~A.}\ \bibnamefont {van Donkelaar}},\ }\bibfield  {title} {\enquote
  {\bibinfo {title} {Single-ion implantation for the development of si-based
  {MOSFET} devices with quantum functionalities},}\ }\href
  {https://doi.org/10.1155/2012/272694} {\bibfield  {journal} {\bibinfo
  {journal} {Adv. Mater. Sci. Eng.}\ }\textbf {\bibinfo {volume} {2012}},\
  \bibinfo {pages} {1--10} (\bibinfo {year} {2012})}\BibitemShut {NoStop}%
\bibitem [{\citenamefont {Hopf}\ \emph {et~al.}(2008)\citenamefont {Hopf},
  \citenamefont {Yang}, \citenamefont {Hearne}, \citenamefont {Jamieson},
  \citenamefont {Gauja}, \citenamefont {Andresen},\ and\ \citenamefont
  {Dzurak}}]{Hopf_2008}%
  \BibitemOpen
  \bibfield  {author} {\bibinfo {author} {\bibfnamefont {T.}~\bibnamefont
  {Hopf}}, \bibinfo {author} {\bibfnamefont {C.}~\bibnamefont {Yang}}, \bibinfo
  {author} {\bibfnamefont {S.~M.}\ \bibnamefont {Hearne}}, \bibinfo {author}
  {\bibfnamefont {D.~N.}\ \bibnamefont {Jamieson}}, \bibinfo {author}
  {\bibfnamefont {E.}~\bibnamefont {Gauja}}, \bibinfo {author} {\bibfnamefont
  {S.~E.}\ \bibnamefont {Andresen}},\ and\ \bibinfo {author} {\bibfnamefont
  {A.~S.}\ \bibnamefont {Dzurak}},\ }\bibfield  {title} {\enquote {\bibinfo
  {title} {Low-noise detection system for the counted implantation of single
  ions in silicon},}\ }\href {https://doi.org/10.1109/tns.2008.918529}
  {\bibfield  {journal} {\bibinfo  {journal} {IEEE Trans. Nucl. Sci.}\ }\textbf
  {\bibinfo {volume} {55}},\ \bibinfo {pages} {812--816} (\bibinfo {year}
  {2008})}\BibitemShut {NoStop}%
\bibitem [{\citenamefont {Johnson}\ \emph {et~al.}(2010)\citenamefont
  {Johnson}, \citenamefont {Tettamanzi}, \citenamefont {Alves}, \citenamefont
  {Thompson}, \citenamefont {Yang}, \citenamefont {Verduijn}, \citenamefont
  {Mol}, \citenamefont {Wacquez}, \citenamefont {Vinet}, \citenamefont
  {Sanquer}, \citenamefont {Rogge},\ and\ \citenamefont
  {Jamieson}}]{Johnson_2010}%
  \BibitemOpen
  \bibfield  {author} {\bibinfo {author} {\bibfnamefont {B.~C.}\ \bibnamefont
  {Johnson}}, \bibinfo {author} {\bibfnamefont {G.~C.}\ \bibnamefont
  {Tettamanzi}}, \bibinfo {author} {\bibfnamefont {A.~D.~C.}\ \bibnamefont
  {Alves}}, \bibinfo {author} {\bibfnamefont {S.}~\bibnamefont {Thompson}},
  \bibinfo {author} {\bibfnamefont {C.}~\bibnamefont {Yang}}, \bibinfo {author}
  {\bibfnamefont {J.}~\bibnamefont {Verduijn}}, \bibinfo {author}
  {\bibfnamefont {J.~A.}\ \bibnamefont {Mol}}, \bibinfo {author} {\bibfnamefont
  {R.}~\bibnamefont {Wacquez}}, \bibinfo {author} {\bibfnamefont
  {M.}~\bibnamefont {Vinet}}, \bibinfo {author} {\bibfnamefont
  {M.}~\bibnamefont {Sanquer}}, \bibinfo {author} {\bibfnamefont
  {S.}~\bibnamefont {Rogge}},\ and\ \bibinfo {author} {\bibfnamefont {D.~N.}\
  \bibnamefont {Jamieson}},\ }\bibfield  {title} {\enquote {\bibinfo {title}
  {Drain current modulation in a nanoscale field-effect-transistor channel by
  single dopant implantation},}\ }\href {https://doi.org/10.1063/1.3458783}
  {\bibfield  {journal} {\bibinfo  {journal} {Appl. Phys. Lett.}\ }\textbf
  {\bibinfo {volume} {96}},\ \bibinfo {pages} {264102} (\bibinfo {year}
  {2010})}\BibitemShut {NoStop}%
\bibitem [{\citenamefont {Robson}\ \emph {et~al.}(2022)\citenamefont {Robson},
  \citenamefont {R{\"{a}}cke}, \citenamefont {Jakob}, \citenamefont {Collins},
  \citenamefont {Firgau}, \citenamefont {Schmitt}, \citenamefont {Mourik},
  \citenamefont {Morello}, \citenamefont {Mayes}, \citenamefont {Spemann},\
  and\ \citenamefont {Jamieson}}]{Robson_2022}%
  \BibitemOpen
  \bibfield  {author} {\bibinfo {author} {\bibfnamefont {S.~G.}\ \bibnamefont
  {Robson}}, \bibinfo {author} {\bibfnamefont {P.}~\bibnamefont {R{\"{a}}cke}},
  \bibinfo {author} {\bibfnamefont {A.~M.}\ \bibnamefont {Jakob}}, \bibinfo
  {author} {\bibfnamefont {N.}~\bibnamefont {Collins}}, \bibinfo {author}
  {\bibfnamefont {H.~R.}\ \bibnamefont {Firgau}}, \bibinfo {author}
  {\bibfnamefont {V.}~\bibnamefont {Schmitt}}, \bibinfo {author} {\bibfnamefont
  {V.}~\bibnamefont {Mourik}}, \bibinfo {author} {\bibfnamefont
  {A.}~\bibnamefont {Morello}}, \bibinfo {author} {\bibfnamefont
  {E.}~\bibnamefont {Mayes}}, \bibinfo {author} {\bibfnamefont
  {D.}~\bibnamefont {Spemann}},\ and\ \bibinfo {author} {\bibfnamefont {D.~N.}\
  \bibnamefont {Jamieson}},\ }\bibfield  {title} {\enquote {\bibinfo {title}
  {Near-surface electrical characterization of silicon electronic devices using
  focused {keV}-range ions},}\ }\href
  {https://doi.org/10.1103/physrevapplied.18.034037} {\bibfield  {journal}
  {\bibinfo  {journal} {Phys. Rev. Applied}\ }\textbf {\bibinfo {volume}
  {18}},\ \bibinfo {pages} {034037} (\bibinfo {year} {2022})}\BibitemShut
  {NoStop}%
\bibitem [{\citenamefont {Giannuzzi}, \citenamefont {Prenitzer},\ and\
  \citenamefont {Kempshall}(2005)}]{Giannuzzi_2005}%
  \BibitemOpen
  \bibfield  {author} {\bibinfo {author} {\bibfnamefont {L.~A.}\ \bibnamefont
  {Giannuzzi}}, \bibinfo {author} {\bibfnamefont {B.~I.}\ \bibnamefont
  {Prenitzer}},\ and\ \bibinfo {author} {\bibfnamefont {B.~W.}\ \bibnamefont
  {Kempshall}},\ }\bibfield  {title} {\enquote {\bibinfo {title} {Ion - solid
  interactions},}\ }in\  \cite{Giannuzzi_2005a},\ Chap.~\bibinfo {chapter} {2},
  pp.\ \bibinfo {pages} {13--52}\BibitemShut {NoStop}%
\bibitem [{\citenamefont {Yao}(2007)}]{Yao_2007}%
  \BibitemOpen
  \bibfield  {author} {\bibinfo {author} {\bibfnamefont {N.}~\bibnamefont
  {Yao}},\ }\href {https://doi.org/10.1017/CBO9780511600302} {\emph {\bibinfo
  {title} {Focused ion beam systems}}},\ edited by\ \bibinfo {editor}
  {\bibfnamefont {N.}~\bibnamefont {Yao}}\ (\bibinfo  {publisher} {Cambridge
  University Press},\ \bibinfo {address} {Cambridge},\ \bibinfo {year} {2007})\
  p.\ \bibinfo {pages} {395}\BibitemShut {NoStop}%
\bibitem [{\citenamefont {{RAITH GmbH}}(2021)}]{Laserstage_2021}%
  \BibitemOpen
  \bibfield  {author} {\bibinfo {author} {\bibnamefont {{RAITH GmbH}}},\ }\href
  {https://www.raith.com/technology/laser-interferometer-stage/} {\enquote
  {\bibinfo {title} {Raith laser interferometer stage},}\ }\bibinfo
  {howpublished} {Online} (\bibinfo {year} {2021}),\ \bibinfo {note}
  {\url{https://www.raith.com/technology/laser-interferometer-stage/}}\BibitemShut
  {NoStop}%
\bibitem [{\citenamefont {Utke}, \citenamefont {Moshkalev},\ and\ \citenamefont
  {Russell}(2012)}]{Utke_2012}%
  \BibitemOpen
  \bibinfo {editor} {\bibfnamefont {I.}~\bibnamefont {Utke}}, \bibinfo {editor}
  {\bibfnamefont {S.}~\bibnamefont {Moshkalev}},\ and\ \bibinfo {editor}
  {\bibfnamefont {P.}~\bibnamefont {Russell}},\ eds.,\ \href@noop {} {\emph
  {\bibinfo {title} {Nanofabrication using focused ion and electron beams}}}\
  (\bibinfo  {publisher} {Oxford University Press},\ \bibinfo {year}
  {2012})\BibitemShut {NoStop}%
\bibitem [{\citenamefont {G{\"{o}}lzh{\"{a}}user}\ and\ \citenamefont
  {Hlawacek}(2016)}]{Golzhauser2016a}%
  \BibitemOpen
  \bibfield  {author} {\bibinfo {author} {\bibfnamefont {A.}~\bibnamefont
  {G{\"{o}}lzh{\"{a}}user}}\ and\ \bibinfo {author} {\bibfnamefont
  {G.}~\bibnamefont {Hlawacek}},\ }\bibfield  {title} {\enquote {\bibinfo
  {title} {{HIM of Biological Samples}},}\ }in\  \cite{Hlawacek_2016},\
  Chap.~\bibinfo {chapter} {7}, pp.\ \bibinfo {pages} {173--185}\BibitemShut
  {NoStop}%
\bibitem [{\citenamefont {Stokes}, \citenamefont {Vystavel},\ and\
  \citenamefont {Morrissey}(2007)}]{Stokes_2007}%
  \BibitemOpen
  \bibfield  {author} {\bibinfo {author} {\bibfnamefont {D.~J.}\ \bibnamefont
  {Stokes}}, \bibinfo {author} {\bibfnamefont {T.}~\bibnamefont {Vystavel}},\
  and\ \bibinfo {author} {\bibfnamefont {F.}~\bibnamefont {Morrissey}},\
  }\bibfield  {title} {\enquote {\bibinfo {title} {Focused ion beam ({FIB})
  milling of electrically insulating specimens using simultaneous primary
  electron and ion beam irradiation},}\ }\href
  {https://doi.org/10.1088/0022-3727/40/3/028} {\bibfield  {journal} {\bibinfo
  {journal} {J. Phys. D: Appl. Phys.}\ }\textbf {\bibinfo {volume} {40}},\
  \bibinfo {pages} {874--877} (\bibinfo {year} {2007})}\BibitemShut {NoStop}%
\bibitem [{\citenamefont {Boltje}\ \emph {et~al.}(2022)\citenamefont {Boltje},
  \citenamefont {Hoogenboom}, \citenamefont {Jakobi}, \citenamefont {Jensen},
  \citenamefont {Jonker}, \citenamefont {Kaag}, \citenamefont {Koster},
  \citenamefont {Last}, \citenamefont {de~Agrela~Pinto}, \citenamefont
  {Plitzko}, \citenamefont {Raunser}, \citenamefont {Tacke}, \citenamefont
  {Wang}, \citenamefont {van~der Wee}, \citenamefont {Wepf},\ and\
  \citenamefont {den Hoedt}}]{Boltje_2022}%
  \BibitemOpen
  \bibfield  {author} {\bibinfo {author} {\bibfnamefont {D.~B.}\ \bibnamefont
  {Boltje}}, \bibinfo {author} {\bibfnamefont {J.~P.}\ \bibnamefont
  {Hoogenboom}}, \bibinfo {author} {\bibfnamefont {A.~J.}\ \bibnamefont
  {Jakobi}}, \bibinfo {author} {\bibfnamefont {G.~J.}\ \bibnamefont {Jensen}},
  \bibinfo {author} {\bibfnamefont {C.~T.~H.}\ \bibnamefont {Jonker}}, \bibinfo
  {author} {\bibfnamefont {M.~J.}\ \bibnamefont {Kaag}}, \bibinfo {author}
  {\bibfnamefont {A.~J.}\ \bibnamefont {Koster}}, \bibinfo {author}
  {\bibfnamefont {M.~G.~F.}\ \bibnamefont {Last}}, \bibinfo {author}
  {\bibfnamefont {C.}~\bibnamefont {de~Agrela~Pinto}}, \bibinfo {author}
  {\bibfnamefont {J.~M.}\ \bibnamefont {Plitzko}}, \bibinfo {author}
  {\bibfnamefont {S.}~\bibnamefont {Raunser}}, \bibinfo {author} {\bibfnamefont
  {S.}~\bibnamefont {Tacke}}, \bibinfo {author} {\bibfnamefont
  {Z.}~\bibnamefont {Wang}}, \bibinfo {author} {\bibfnamefont {E.~B.}\
  \bibnamefont {van~der Wee}}, \bibinfo {author} {\bibfnamefont
  {R.}~\bibnamefont {Wepf}},\ and\ \bibinfo {author} {\bibfnamefont
  {S.}~\bibnamefont {den Hoedt}},\ }\bibfield  {title} {\enquote {\bibinfo
  {title} {A cryogenic, coincident fluorescence, electron, and ion beam
  microscope},}\ }\href {https://doi.org/10.7554/elife.82891} {\bibfield
  {journal} {\bibinfo  {journal} {{eLife}}\ }\textbf {\bibinfo {volume} {11}}
  (\bibinfo {year} {2022}),\ 10.7554/elife.82891}\BibitemShut {NoStop}%
\bibitem [{\citenamefont {Timmermans}\ and\ \citenamefont
  {Otto}(2015)}]{Timmermans_2015}%
  \BibitemOpen
  \bibfield  {author} {\bibinfo {author} {\bibfnamefont {F.~J.}\ \bibnamefont
  {Timmermans}}\ and\ \bibinfo {author} {\bibfnamefont {C.}~\bibnamefont
  {Otto}},\ }\bibfield  {title} {\enquote {\bibinfo {title} {Contributed
  review: review of integrated correlative light and electron microscopy},}\
  }\href {https://doi.org/10.1063/1.4905434} {\bibfield  {journal} {\bibinfo
  {journal} {Rev. Sci. Instrum.}\ }\textbf {\bibinfo {volume} {86}},\ \bibinfo
  {pages} {011501} (\bibinfo {year} {2015})}\BibitemShut {NoStop}%
\bibitem [{\citenamefont {Langford}\ and\ \citenamefont
  {Clinton}(2004)}]{Langford_2004a}%
  \BibitemOpen
  \bibfield  {author} {\bibinfo {author} {\bibfnamefont {R.~M.}\ \bibnamefont
  {Langford}}\ and\ \bibinfo {author} {\bibfnamefont {C.}~\bibnamefont
  {Clinton}},\ }\bibfield  {title} {\enquote {\bibinfo {title} {In situ
  lift-out using a {FIB}-{SEM} system},}\ }\href
  {https://doi.org/10.1016/j.micron.2004.03.002} {\bibfield  {journal}
  {\bibinfo  {journal} {Micron}\ }\textbf {\bibinfo {volume} {35}},\ \bibinfo
  {pages} {607--611} (\bibinfo {year} {2004})}\BibitemShut {NoStop}%
\bibitem [{\citenamefont {Pawley}(1972)}]{Pawley_1972a}%
  \BibitemOpen
  \bibfield  {author} {\bibinfo {author} {\bibfnamefont {J.~B.}\ \bibnamefont
  {Pawley}},\ }\bibfield  {title} {\enquote {\bibinfo {title} {A dual needle
  piezoelectric micromanipulator for the scanning electron microscope},}\
  }\href {https://doi.org/10.1063/1.1685701} {\bibfield  {journal} {\bibinfo
  {journal} {Rev. Sci. Instrum.}\ }\textbf {\bibinfo {volume} {43}},\ \bibinfo
  {pages} {600--602} (\bibinfo {year} {1972})}\BibitemShut {NoStop}%
\bibitem [{\citenamefont {Shi}\ \emph {et~al.}(2016)\citenamefont {Shi},
  \citenamefont {Luu}, \citenamefont {Yang}, \citenamefont {Liu}, \citenamefont
  {Chen}, \citenamefont {Ru}, \citenamefont {Xie}, \citenamefont {Luo},
  \citenamefont {Ge},\ and\ \citenamefont {Sun}}]{Shi_2016}%
  \BibitemOpen
  \bibfield  {author} {\bibinfo {author} {\bibfnamefont {C.}~\bibnamefont
  {Shi}}, \bibinfo {author} {\bibfnamefont {D.~K.}\ \bibnamefont {Luu}},
  \bibinfo {author} {\bibfnamefont {Q.}~\bibnamefont {Yang}}, \bibinfo {author}
  {\bibfnamefont {J.}~\bibnamefont {Liu}}, \bibinfo {author} {\bibfnamefont
  {J.}~\bibnamefont {Chen}}, \bibinfo {author} {\bibfnamefont {C.}~\bibnamefont
  {Ru}}, \bibinfo {author} {\bibfnamefont {S.}~\bibnamefont {Xie}}, \bibinfo
  {author} {\bibfnamefont {J.}~\bibnamefont {Luo}}, \bibinfo {author}
  {\bibfnamefont {J.}~\bibnamefont {Ge}},\ and\ \bibinfo {author}
  {\bibfnamefont {Y.}~\bibnamefont {Sun}},\ }\bibfield  {title} {\enquote
  {\bibinfo {title} {Recent advances in nanorobotic manipulation inside
  scanning electron microscopes},}\ }\href
  {https://doi.org/10.1038/micronano.2016.24} {\bibfield  {journal} {\bibinfo
  {journal} {Microsystems \& Nanoengineering}\ }\textbf {\bibinfo {volume} {2}}
  (\bibinfo {year} {2016}),\ 10.1038/micronano.2016.24}\BibitemShut {NoStop}%
\bibitem [{\citenamefont {Jiang}\ \emph {et~al.}(2017)\citenamefont {Jiang},
  \citenamefont {Lu}, \citenamefont {Zhang}, \citenamefont {Shen},\ and\
  \citenamefont {Lu}}]{Jiang_2017a}%
  \BibitemOpen
  \bibfield  {author} {\bibinfo {author} {\bibfnamefont {C.}~\bibnamefont
  {Jiang}}, \bibinfo {author} {\bibfnamefont {H.}~\bibnamefont {Lu}}, \bibinfo
  {author} {\bibfnamefont {H.}~\bibnamefont {Zhang}}, \bibinfo {author}
  {\bibfnamefont {Y.}~\bibnamefont {Shen}},\ and\ \bibinfo {author}
  {\bibfnamefont {Y.}~\bibnamefont {Lu}},\ }\bibfield  {title} {\enquote
  {\bibinfo {title} {Recent advances on in situ {SEM} mechanical and electrical
  characterization of low-dimensional nanomaterials},}\ }\href
  {https://doi.org/10.1155/2017/1985149} {\bibfield  {journal} {\bibinfo
  {journal} {Scanning}\ }\textbf {\bibinfo {volume} {2017}},\ \bibinfo {pages}
  {1--11} (\bibinfo {year} {2017})}\BibitemShut {NoStop}%
\bibitem [{\citenamefont {Stegmann}\ \emph {et~al.}(2011)\citenamefont
  {Stegmann}, \citenamefont {D{\"{o}}mer}, \citenamefont {Rosenkranz},\ and\
  \citenamefont {Zschech}}]{Stegmann_2011a}%
  \BibitemOpen
  \bibfield  {author} {\bibinfo {author} {\bibfnamefont {H.}~\bibnamefont
  {Stegmann}}, \bibinfo {author} {\bibfnamefont {H.}~\bibnamefont
  {D{\"{o}}mer}}, \bibinfo {author} {\bibfnamefont {R.}~\bibnamefont
  {Rosenkranz}},\ and\ \bibinfo {author} {\bibfnamefont {E.}~\bibnamefont
  {Zschech}},\ }\bibfield  {title} {\enquote {\bibinfo {title} {Efficient
  target preparation by combined pulsed laser ablation and {FIB} milling},}\
  }\href {https://doi.org/10.1017/s1431927611004168} {\bibfield  {journal}
  {\bibinfo  {journal} {Microsc. Microanal.}\ }\textbf {\bibinfo {volume}
  {17}},\ \bibinfo {pages} {658--659} (\bibinfo {year} {2011})}\BibitemShut
  {NoStop}%
\bibitem [{\citenamefont {Stephant}\ \emph {et~al.}(2018)\citenamefont
  {Stephant}, \citenamefont {Grissa}, \citenamefont {Guillou}, \citenamefont
  {Bretaudeau}, \citenamefont {Borjon-Piron}, \citenamefont {Guillet},\ and\
  \citenamefont {Moreau}}]{Stephant_2018}%
  \BibitemOpen
  \bibfield  {author} {\bibinfo {author} {\bibfnamefont {N.}~\bibnamefont
  {Stephant}}, \bibinfo {author} {\bibfnamefont {R.}~\bibnamefont {Grissa}},
  \bibinfo {author} {\bibfnamefont {F.}~\bibnamefont {Guillou}}, \bibinfo
  {author} {\bibfnamefont {M.}~\bibnamefont {Bretaudeau}}, \bibinfo {author}
  {\bibfnamefont {Y.}~\bibnamefont {Borjon-Piron}}, \bibinfo {author}
  {\bibfnamefont {J.}~\bibnamefont {Guillet}},\ and\ \bibinfo {author}
  {\bibfnamefont {P.}~\bibnamefont {Moreau}},\ }\bibfield  {title} {\enquote
  {\bibinfo {title} {New airtight transfer box for {SEM} experiments:
  application to lithium and sodium metals observation and analyses},}\ }\href
  {https://doi.org/10.1016/j.micron.2018.04.003} {\bibfield  {journal}
  {\bibinfo  {journal} {Micron}\ }\textbf {\bibinfo {volume} {110}},\ \bibinfo
  {pages} {10--17} (\bibinfo {year} {2018})}\BibitemShut {NoStop}%
\bibitem [{\citenamefont {Andany}\ \emph {et~al.}(2020)\citenamefont {Andany},
  \citenamefont {Hlawacek}, \citenamefont {Hummel}, \citenamefont {Brillard},
  \citenamefont {Kang{\"{u}}l},\ and\ \citenamefont {Fantner}}]{Andany_2020}%
  \BibitemOpen
  \bibfield  {author} {\bibinfo {author} {\bibfnamefont {S.~H.}\ \bibnamefont
  {Andany}}, \bibinfo {author} {\bibfnamefont {G.}~\bibnamefont {Hlawacek}},
  \bibinfo {author} {\bibfnamefont {S.}~\bibnamefont {Hummel}}, \bibinfo
  {author} {\bibfnamefont {C.}~\bibnamefont {Brillard}}, \bibinfo {author}
  {\bibfnamefont {M.}~\bibnamefont {Kang{\"{u}}l}},\ and\ \bibinfo {author}
  {\bibfnamefont {G.~E.}\ \bibnamefont {Fantner}},\ }\bibfield  {title}
  {\enquote {\bibinfo {title} {An atomic force microscope integrated with a
  helium ion microscope for correlative nanoscale characterization},}\ }\href
  {https://doi.org/10.3762/bjnano.11.111} {\bibfield  {journal} {\bibinfo
  {journal} {Beilstein J. Nanotechnol.}\ }\textbf {\bibinfo {volume} {11}},\
  \bibinfo {pages} {1272--1279} (\bibinfo {year} {2020})},\ \Eprint
  {https://arxiv.org/abs/2004.00536} {arXiv:2004.00536 [physics.ins-det]}
  \BibitemShut {NoStop}%
\bibitem [{\citenamefont {Colldeweih}\ \emph {et~al.}(2019)\citenamefont
  {Colldeweih}, \citenamefont {Baris}, \citenamefont {Sp{\"{a}}tig},\ and\
  \citenamefont {Abolhassani}}]{Colldeweih2019}%
  \BibitemOpen
  \bibfield  {author} {\bibinfo {author} {\bibfnamefont {A.~W.}\ \bibnamefont
  {Colldeweih}}, \bibinfo {author} {\bibfnamefont {A.}~\bibnamefont {Baris}},
  \bibinfo {author} {\bibfnamefont {P.}~\bibnamefont {Sp{\"{a}}tig}},\ and\
  \bibinfo {author} {\bibfnamefont {S.}~\bibnamefont {Abolhassani}},\
  }\bibfield  {title} {\enquote {\bibinfo {title} {Evaluation of mechanical
  properties of irradiated zirconium alloys in the vicinity of the metal-oxide
  interface},}\ }\href {https://doi.org/10.1016/j.msea.2018.09.107} {\bibfield
  {journal} {\bibinfo  {journal} {Mat. Sci. Eng. A-struct.}\ }\textbf {\bibinfo
  {volume} {742}},\ \bibinfo {pages} {842--850} (\bibinfo {year}
  {2019})}\BibitemShut {NoStop}%
\bibitem [{\citenamefont {Teague}\ \emph {et~al.}(2014)\citenamefont {Teague},
  \citenamefont {Gorman}, \citenamefont {Miller},\ and\ \citenamefont
  {King}}]{Teague2014}%
  \BibitemOpen
  \bibfield  {author} {\bibinfo {author} {\bibfnamefont {M.}~\bibnamefont
  {Teague}}, \bibinfo {author} {\bibfnamefont {B.}~\bibnamefont {Gorman}},
  \bibinfo {author} {\bibfnamefont {B.}~\bibnamefont {Miller}},\ and\ \bibinfo
  {author} {\bibfnamefont {J.}~\bibnamefont {King}},\ }\bibfield  {title}
  {\enquote {\bibinfo {title} {{EBSD} and {TEM} characterization of high
  burn-up mixed oxide fuel},}\ }\href
  {https://doi.org/10.1016/j.jnucmat.2013.10.037} {\bibfield  {journal}
  {\bibinfo  {journal} {J. Nucl. Mater.}\ }\textbf {\bibinfo {volume} {444}},\
  \bibinfo {pages} {475--480} (\bibinfo {year} {2014})}\BibitemShut {NoStop}%
\bibitem [{\citenamefont {Baris}\ \emph {et~al.}(2018)\citenamefont {Baris},
  \citenamefont {Restani}, \citenamefont {Grabherr}, \citenamefont {Chiu},
  \citenamefont {Evans}, \citenamefont {Ammon}, \citenamefont {Limb{\"{a}}ck},\
  and\ \citenamefont {Abolhassani}}]{Baris2018}%
  \BibitemOpen
  \bibfield  {author} {\bibinfo {author} {\bibfnamefont {A.}~\bibnamefont
  {Baris}}, \bibinfo {author} {\bibfnamefont {R.}~\bibnamefont {Restani}},
  \bibinfo {author} {\bibfnamefont {R.}~\bibnamefont {Grabherr}}, \bibinfo
  {author} {\bibfnamefont {Y.-L.}\ \bibnamefont {Chiu}}, \bibinfo {author}
  {\bibfnamefont {H.~E.}\ \bibnamefont {Evans}}, \bibinfo {author}
  {\bibfnamefont {K.}~\bibnamefont {Ammon}}, \bibinfo {author} {\bibfnamefont
  {M.}~\bibnamefont {Limb{\"{a}}ck}},\ and\ \bibinfo {author} {\bibfnamefont
  {S.}~\bibnamefont {Abolhassani}},\ }\bibfield  {title} {\enquote {\bibinfo
  {title} {Chemical and microstructural characterization of a 9 cycle
  {Z}ircaloy-2 cladding using {EPMA} and {FIB} tomography},}\ }\href
  {https://doi.org/10.1016/j.jnucmat.2018.01.065} {\bibfield  {journal}
  {\bibinfo  {journal} {J. Nucl. Mater.}\ }\textbf {\bibinfo {volume} {504}},\
  \bibinfo {pages} {144--160} (\bibinfo {year} {2018})}\BibitemShut {NoStop}%
\bibitem [{\citenamefont {Miller}\ \emph {et~al.}(2012)\citenamefont {Miller},
  \citenamefont {Gan}, \citenamefont {Madden}, \citenamefont {Jue},
  \citenamefont {Robinson},\ and\ \citenamefont {Keiser}}]{Miller2012}%
  \BibitemOpen
  \bibfield  {author} {\bibinfo {author} {\bibfnamefont {B.~D.}\ \bibnamefont
  {Miller}}, \bibinfo {author} {\bibfnamefont {J.}~\bibnamefont {Gan}},
  \bibinfo {author} {\bibfnamefont {J.}~\bibnamefont {Madden}}, \bibinfo
  {author} {\bibfnamefont {J.~F.}\ \bibnamefont {Jue}}, \bibinfo {author}
  {\bibfnamefont {A.}~\bibnamefont {Robinson}},\ and\ \bibinfo {author}
  {\bibfnamefont {D.~D.}\ \bibnamefont {Keiser}},\ }\bibfield  {title}
  {\enquote {\bibinfo {title} {Advantages and disadvantages of using a focused
  ion beam to prepare {TEM} samples from irradiated {U-10Mo} monolithic nuclear
  fuel},}\ }\href {https://doi.org/10.1016/j.jnucmat.2012.01.022} {\bibfield
  {journal} {\bibinfo  {journal} {J. Nucl. Mater.}\ }\textbf {\bibinfo {volume}
  {424}},\ \bibinfo {pages} {38--42} (\bibinfo {year} {2012})}\BibitemShut
  {NoStop}%
\bibitem [{\citenamefont {Kuri}\ \emph {et~al.}(2018)\citenamefont {Kuri},
  \citenamefont {Ramanantoanina}, \citenamefont {Bertsch}, \citenamefont
  {Martin},\ and\ \citenamefont {Panas}}]{Kuri2018}%
  \BibitemOpen
  \bibfield  {author} {\bibinfo {author} {\bibfnamefont {G.}~\bibnamefont
  {Kuri}}, \bibinfo {author} {\bibfnamefont {H.}~\bibnamefont
  {Ramanantoanina}}, \bibinfo {author} {\bibfnamefont {J.}~\bibnamefont
  {Bertsch}}, \bibinfo {author} {\bibfnamefont {M.}~\bibnamefont {Martin}},\
  and\ \bibinfo {author} {\bibfnamefont {I.}~\bibnamefont {Panas}},\ }\bibfield
   {title} {\enquote {\bibinfo {title} {Chemical state and atomic scale
  environment of nickel in the corrosion layer of irradiated {Z}ircaloy-2 at a
  burn-up around 45 {MW}d/kg},}\ }\href
  {https://doi.org/10.1016/j.corsci.2018.08.032} {\bibfield  {journal}
  {\bibinfo  {journal} {Corros. Sci.}\ }\textbf {\bibinfo {volume} {143}},\
  \bibinfo {pages} {200--211} (\bibinfo {year} {2018})}\BibitemShut {NoStop}%
\bibitem [{\citenamefont {Kautz}\ \emph {et~al.}(2021)\citenamefont {Kautz},
  \citenamefont {Cliff}, \citenamefont {Lach}, \citenamefont {Reilly},\ and\
  \citenamefont {Devaraj}}]{Kautz2021}%
  \BibitemOpen
  \bibfield  {author} {\bibinfo {author} {\bibfnamefont {E.}~\bibnamefont
  {Kautz}}, \bibinfo {author} {\bibfnamefont {J.}~\bibnamefont {Cliff}},
  \bibinfo {author} {\bibfnamefont {T.}~\bibnamefont {Lach}}, \bibinfo {author}
  {\bibfnamefont {D.}~\bibnamefont {Reilly}},\ and\ \bibinfo {author}
  {\bibfnamefont {A.}~\bibnamefont {Devaraj}},\ }\bibfield  {title} {\enquote
  {\bibinfo {title} {Correlating nanoscale secondary ion mass spectrometry and
  atom probe tomography analysis of uranium enrichment in metallic nuclear
  fuel},}\ }\href {https://doi.org/10.1039/d0an01831g} {\bibfield  {journal}
  {\bibinfo  {journal} {The Analyst}\ }\textbf {\bibinfo {volume} {146}},\
  \bibinfo {pages} {69--74} (\bibinfo {year} {2021})}\BibitemShut {NoStop}%
\bibitem [{\citenamefont {Evelan}\ and\ \citenamefont
  {Brey}(2013)}]{Evelan2013}%
  \BibitemOpen
  \bibfield  {author} {\bibinfo {author} {\bibfnamefont {A.~R.}\ \bibnamefont
  {Evelan}}\ and\ \bibinfo {author} {\bibfnamefont {R.~R.}\ \bibnamefont
  {Brey}},\ }\bibfield  {title} {\enquote {\bibinfo {title} {Contamination
  analysis of radioactive samples in focused ion beam instruments},}\ }\href
  {https://doi.org/10.1097/hp.0b013e3182732979} {\bibfield  {journal} {\bibinfo
   {journal} {Health Phys.}\ }\textbf {\bibinfo {volume} {104}},\ \bibinfo
  {pages} {S23--S30} (\bibinfo {year} {2013})}\BibitemShut {NoStop}%
\bibitem [{\citenamefont {Aitkaliyeva}\ \emph {et~al.}(2014)\citenamefont
  {Aitkaliyeva}, \citenamefont {Madden}, \citenamefont {Miller},\ and\
  \citenamefont {Cole}}]{Aitkaliyeva2014}%
  \BibitemOpen
  \bibfield  {author} {\bibinfo {author} {\bibfnamefont {A.}~\bibnamefont
  {Aitkaliyeva}}, \bibinfo {author} {\bibfnamefont {J.~W.}\ \bibnamefont
  {Madden}}, \bibinfo {author} {\bibfnamefont {B.~D.}\ \bibnamefont {Miller}},\
  and\ \bibinfo {author} {\bibfnamefont {J.~I.}\ \bibnamefont {Cole}},\
  }\bibfield  {title} {\enquote {\bibinfo {title} {Implementation of focused
  ion beam ({FIB}) system in characterization of nuclear fuels and
  materials},}\ }\href {https://doi.org/10.1016/j.micron.2014.06.010}
  {\bibfield  {journal} {\bibinfo  {journal} {Micron}\ }\textbf {\bibinfo
  {volume} {67}},\ \bibinfo {pages} {65--73} (\bibinfo {year}
  {2014})}\BibitemShut {NoStop}%
\bibitem [{\citenamefont {Wanzenb{\"{o}}ck}\ and\ \citenamefont
  {Wai}(2011)}]{Wanzenboeck_2011}%
  \BibitemOpen
  \bibfield  {author} {\bibinfo {author} {\bibfnamefont {H.~D.}\ \bibnamefont
  {Wanzenb{\"{o}}ck}}\ and\ \bibinfo {author} {\bibfnamefont {S.}~\bibnamefont
  {Wai}},\ }\bibfield  {title} {\enquote {\bibinfo {title} {Focused ion beam
  lithography},}\ }in\ \href {https://doi.org/10.5772/22075} {\emph {\bibinfo
  {booktitle} {Recent Advances in Nanofabrication Techniques and
  Applications}}}\ (\bibinfo  {publisher} {{InTech}},\ \bibinfo {year}
  {2011})\BibitemShut {NoStop}%
\bibitem [{\citenamefont {Jamali}\ \emph {et~al.}(2014)\citenamefont {Jamali},
  \citenamefont {Gerhardt}, \citenamefont {Rezai}, \citenamefont {Frenner},
  \citenamefont {Fedder},\ and\ \citenamefont {Wrachtrup}}]{Jamali_2014}%
  \BibitemOpen
  \bibfield  {author} {\bibinfo {author} {\bibfnamefont {M.}~\bibnamefont
  {Jamali}}, \bibinfo {author} {\bibfnamefont {I.}~\bibnamefont {Gerhardt}},
  \bibinfo {author} {\bibfnamefont {M.}~\bibnamefont {Rezai}}, \bibinfo
  {author} {\bibfnamefont {K.}~\bibnamefont {Frenner}}, \bibinfo {author}
  {\bibfnamefont {H.}~\bibnamefont {Fedder}},\ and\ \bibinfo {author}
  {\bibfnamefont {J.}~\bibnamefont {Wrachtrup}},\ }\bibfield  {title} {\enquote
  {\bibinfo {title} {Microscopic diamond solid-immersion-lenses fabricated
  around single defect centers by focused ion beam milling},}\ }\href
  {https://doi.org/10.1063/1.4902818} {\bibfield  {journal} {\bibinfo
  {journal} {Rev. Sci. Instrum.}\ }\textbf {\bibinfo {volume} {85}},\ \bibinfo
  {pages} {123703} (\bibinfo {year} {2014})}\BibitemShut {NoStop}%
\bibitem [{\citenamefont {Cui}, \citenamefont {Shalaginov},\ and\ \citenamefont
  {Kildishev}(2017)}]{Cui_2017}%
  \BibitemOpen
  \bibfield  {author} {\bibinfo {author} {\bibfnamefont {R.}~\bibnamefont
  {Cui}}, \bibinfo {author} {\bibfnamefont {M.}~\bibnamefont {Shalaginov}},\
  and\ \bibinfo {author} {\bibfnamefont {A.~V.}\ \bibnamefont {Kildishev}},\
  }\href {https://doi.org/10.4231/D3DN3ZX26} {\enquote {\bibinfo {title}
  {{NanoCraft-FIBstream}: {F}ocused ion beam stream file generator},}\ }
  (\bibinfo {year} {2017})\BibitemShut {NoStop}%
\bibitem [{\citenamefont {Deinhart}\ \emph {et~al.}(2021)\citenamefont
  {Deinhart}, \citenamefont {Kern}, \citenamefont {Kirchhof}, \citenamefont
  {Juergensen}, \citenamefont {Sturm}, \citenamefont {Krauss}, \citenamefont
  {Feichtner}, \citenamefont {Kovalchuk}, \citenamefont {Schneider},
  \citenamefont {Engel}, \citenamefont {Pfau}, \citenamefont {Hecht},
  \citenamefont {Bolotin}, \citenamefont {Reich},\ and\ \citenamefont
  {H{\"{o}}flich}}]{Deinhart_2021}%
  \BibitemOpen
  \bibfield  {author} {\bibinfo {author} {\bibfnamefont {V.}~\bibnamefont
  {Deinhart}}, \bibinfo {author} {\bibfnamefont {L.-M.}\ \bibnamefont {Kern}},
  \bibinfo {author} {\bibfnamefont {J.~N.}\ \bibnamefont {Kirchhof}}, \bibinfo
  {author} {\bibfnamefont {S.}~\bibnamefont {Juergensen}}, \bibinfo {author}
  {\bibfnamefont {J.}~\bibnamefont {Sturm}}, \bibinfo {author} {\bibfnamefont
  {E.}~\bibnamefont {Krauss}}, \bibinfo {author} {\bibfnamefont
  {T.}~\bibnamefont {Feichtner}}, \bibinfo {author} {\bibfnamefont
  {S.}~\bibnamefont {Kovalchuk}}, \bibinfo {author} {\bibfnamefont
  {M.}~\bibnamefont {Schneider}}, \bibinfo {author} {\bibfnamefont
  {D.}~\bibnamefont {Engel}}, \bibinfo {author} {\bibfnamefont
  {B.}~\bibnamefont {Pfau}}, \bibinfo {author} {\bibfnamefont {B.}~\bibnamefont
  {Hecht}}, \bibinfo {author} {\bibfnamefont {K.~I.}\ \bibnamefont {Bolotin}},
  \bibinfo {author} {\bibfnamefont {S.}~\bibnamefont {Reich}},\ and\ \bibinfo
  {author} {\bibfnamefont {K.}~\bibnamefont {H{\"{o}}flich}},\ }\bibfield
  {title} {\enquote {\bibinfo {title} {The patterning toolbox {FIB}-o-mat:
  exploiting the full potential of focused helium ions for nanofabrication},}\
  }\href {https://doi.org/10.3762/bjnano.12.25} {\bibfield  {journal} {\bibinfo
   {journal} {Beilstein J. Nanotechnol.}\ }\textbf {\bibinfo {volume} {12}},\
  \bibinfo {pages} {304--318} (\bibinfo {year} {2021})}\BibitemShut {NoStop}%
\bibitem [{\citenamefont {Lalev}\ \emph {et~al.}(2008)\citenamefont {Lalev},
  \citenamefont {Dimov}, \citenamefont {Kettle}, \citenamefont {van Delft},\
  and\ \citenamefont {Minev}}]{Lalev_2008}%
  \BibitemOpen
  \bibfield  {author} {\bibinfo {author} {\bibfnamefont {G.}~\bibnamefont
  {Lalev}}, \bibinfo {author} {\bibfnamefont {S.}~\bibnamefont {Dimov}},
  \bibinfo {author} {\bibfnamefont {J.}~\bibnamefont {Kettle}}, \bibinfo
  {author} {\bibfnamefont {F.}~\bibnamefont {van Delft}},\ and\ \bibinfo
  {author} {\bibfnamefont {R.}~\bibnamefont {Minev}},\ }\bibfield  {title}
  {\enquote {\bibinfo {title} {Data preparation for focused ion beam machining
  of complex three-dimensional structures},}\ }in\ \href
  {https://doi.org/10.1243/09544054jem864} {\emph {\bibinfo {booktitle}
  {Proceedings of the Institution of Mechanical Engineers, Part B: Journal of
  Engineering Manufacture}}},\ Vol.\ \bibinfo {volume} {222}\ (\bibinfo
  {publisher} {{SAGE} Publications},\ \bibinfo {year} {2008})\ pp.\ \bibinfo
  {pages} {67--76}\BibitemShut {NoStop}%
\bibitem [{\citenamefont {Svintsov}\ \emph {et~al.}(2009)\citenamefont
  {Svintsov}, \citenamefont {Zaitsev}, \citenamefont {Lalev}, \citenamefont
  {Dimov}, \citenamefont {Velkova},\ and\ \citenamefont
  {Hirshy}}]{svintsov_fib_2009}%
  \BibitemOpen
  \bibfield  {author} {\bibinfo {author} {\bibfnamefont {A.}~\bibnamefont
  {Svintsov}}, \bibinfo {author} {\bibfnamefont {S.}~\bibnamefont {Zaitsev}},
  \bibinfo {author} {\bibfnamefont {G.}~\bibnamefont {Lalev}}, \bibinfo
  {author} {\bibfnamefont {S.}~\bibnamefont {Dimov}}, \bibinfo {author}
  {\bibfnamefont {V.}~\bibnamefont {Velkova}},\ and\ \bibinfo {author}
  {\bibfnamefont {H.}~\bibnamefont {Hirshy}},\ }\bibfield  {title} {\enquote
  {\bibinfo {title} {{FIB} sputtering optimization using ion reverse
  software},}\ }\href {https://doi.org/10.1016/j.mee.2009.01.073} {\bibfield
  {journal} {\bibinfo  {journal} {Microelectron. Eng.}\ }\textbf {\bibinfo
  {volume} {86}},\ \bibinfo {pages} {544--547} (\bibinfo {year}
  {2009})}\BibitemShut {NoStop}%
\bibitem [{\citenamefont {Kim}(2012)}]{kim_multiple_2012}%
  \BibitemOpen
  \bibfield  {author} {\bibinfo {author} {\bibfnamefont {H.-B.}\ \bibnamefont
  {Kim}},\ }\bibfield  {title} {\enquote {\bibinfo {title} {Multiple surface
  driving method for the accurate sculpting of predefined arbitrary surfaces in
  the micro/nano regime},}\ }\href {https://doi.org/10.1016/j.mee.2011.11.015}
  {\bibfield  {journal} {\bibinfo  {journal} {Microelectron. Eng.}\ }\textbf
  {\bibinfo {volume} {91}},\ \bibinfo {pages} {14--18} (\bibinfo {year}
  {2012})}\BibitemShut {NoStop}%
\bibitem [{\citenamefont {Felicis}, \citenamefont {Mughal},\ and\ \citenamefont
  {Bemporad}(2017)}]{Felicis_2017}%
  \BibitemOpen
  \bibfield  {author} {\bibinfo {author} {\bibfnamefont {D.~D.}\ \bibnamefont
  {Felicis}}, \bibinfo {author} {\bibfnamefont {M.~Z.}\ \bibnamefont
  {Mughal}},\ and\ \bibinfo {author} {\bibfnamefont {E.}~\bibnamefont
  {Bemporad}},\ }\bibfield  {title} {\enquote {\bibinfo {title} {A method to
  improve the quality of 2.5 dimensional micro-and nano-structures produced by
  focused ion beam machining},}\ }\href
  {https://doi.org/10.1016/j.micron.2017.05.005} {\bibfield  {journal}
  {\bibinfo  {journal} {Micron}\ }\textbf {\bibinfo {volume} {101}},\ \bibinfo
  {pages} {8--15} (\bibinfo {year} {2017})}\BibitemShut {NoStop}%
\bibitem [{\citenamefont {Niessen}\ and\ \citenamefont
  {Nancarrow}(2019)}]{Niessen_2019}%
  \BibitemOpen
  \bibfield  {author} {\bibinfo {author} {\bibfnamefont {F.}~\bibnamefont
  {Niessen}}\ and\ \bibinfo {author} {\bibfnamefont {M.~J.~B.}\ \bibnamefont
  {Nancarrow}},\ }\bibfield  {title} {\enquote {\bibinfo {title}
  {Computer-aided manufacturing and focused ion beam technology enable
  machining of complex micro- and nano-structures},}\ }\href
  {https://doi.org/10.1088/1361-6528/ab329d} {\bibfield  {journal} {\bibinfo
  {journal} {Nanotechnology}\ }\textbf {\bibinfo {volume} {30}},\ \bibinfo
  {pages} {435301} (\bibinfo {year} {2019})}\BibitemShut {NoStop}%
\bibitem [{\citenamefont {Lindsey}\ \emph {et~al.}(2014)\citenamefont
  {Lindsey}, \citenamefont {Waid}, \citenamefont {Hobler}, \citenamefont
  {Wanzenb{\"{o}}ck},\ and\ \citenamefont
  {Bertagnolli}}]{lindsey_inverse_2014}%
  \BibitemOpen
  \bibfield  {author} {\bibinfo {author} {\bibfnamefont {S.}~\bibnamefont
  {Lindsey}}, \bibinfo {author} {\bibfnamefont {S.}~\bibnamefont {Waid}},
  \bibinfo {author} {\bibfnamefont {G.}~\bibnamefont {Hobler}}, \bibinfo
  {author} {\bibfnamefont {H.~D.}\ \bibnamefont {Wanzenb{\"{o}}ck}},\ and\
  \bibinfo {author} {\bibfnamefont {E.}~\bibnamefont {Bertagnolli}},\
  }\bibfield  {title} {\enquote {\bibinfo {title} {Inverse modeling of {FIB}
  milling by dose profile optimization},}\ }\href
  {https://doi.org/10.1016/j.nimb.2014.09.006} {\bibfield  {journal} {\bibinfo
  {journal} {Nucl. Instr. Meth. Phys. Res. B}\ }\textbf {\bibinfo {volume}
  {341}},\ \bibinfo {pages} {77--83} (\bibinfo {year} {2014})}\BibitemShut
  {NoStop}%
\bibitem [{\citenamefont {Ohya}\ \emph {et~al.}(2009)\citenamefont {Ohya},
  \citenamefont {Yamanaka}, \citenamefont {Inai},\ and\ \citenamefont
  {Ishitani}}]{Ohya_2009}%
  \BibitemOpen
  \bibfield  {author} {\bibinfo {author} {\bibfnamefont {K.}~\bibnamefont
  {Ohya}}, \bibinfo {author} {\bibfnamefont {T.}~\bibnamefont {Yamanaka}},
  \bibinfo {author} {\bibfnamefont {K.}~\bibnamefont {Inai}},\ and\ \bibinfo
  {author} {\bibfnamefont {T.}~\bibnamefont {Ishitani}},\ }\bibfield  {title}
  {\enquote {\bibinfo {title} {{Comparison of secondary electron emission in
  helium ion microscope with gallium ion and electron microscopes}},}\ }\href
  {https://doi.org/10.1016/j.nimb.2008.11.003} {\bibfield  {journal} {\bibinfo
  {journal} {Nucl. Instrum. Methods Phys. Res. B}\ }\textbf {\bibinfo {volume}
  {267}},\ \bibinfo {pages} {584--589} (\bibinfo {year} {2009})}\BibitemShut
  {NoStop}%
\bibitem [{\citenamefont {Burnett}\ \emph {et~al.}(2014)\citenamefont
  {Burnett}, \citenamefont {McDonald}, \citenamefont {Gholinia}, \citenamefont
  {Geurts}, \citenamefont {Janus}, \citenamefont {Slater}, \citenamefont
  {Haigh}, \citenamefont {Ornek}, \citenamefont {Almuaili}, \citenamefont
  {Engelberg}, \citenamefont {Thompson},\ and\ \citenamefont
  {Withers}}]{Burnett_2014}%
  \BibitemOpen
  \bibfield  {author} {\bibinfo {author} {\bibfnamefont {T.~L.}\ \bibnamefont
  {Burnett}}, \bibinfo {author} {\bibfnamefont {S.~A.}\ \bibnamefont
  {McDonald}}, \bibinfo {author} {\bibfnamefont {A.}~\bibnamefont {Gholinia}},
  \bibinfo {author} {\bibfnamefont {R.}~\bibnamefont {Geurts}}, \bibinfo
  {author} {\bibfnamefont {M.}~\bibnamefont {Janus}}, \bibinfo {author}
  {\bibfnamefont {T.}~\bibnamefont {Slater}}, \bibinfo {author} {\bibfnamefont
  {S.~J.}\ \bibnamefont {Haigh}}, \bibinfo {author} {\bibfnamefont
  {C.}~\bibnamefont {Ornek}}, \bibinfo {author} {\bibfnamefont
  {F.}~\bibnamefont {Almuaili}}, \bibinfo {author} {\bibfnamefont {D.~L.}\
  \bibnamefont {Engelberg}}, \bibinfo {author} {\bibfnamefont {G.~E.}\
  \bibnamefont {Thompson}},\ and\ \bibinfo {author} {\bibfnamefont {P.~J.}\
  \bibnamefont {Withers}},\ }\bibfield  {title} {\enquote {\bibinfo {title}
  {Correlative tomography},}\ }\href {https://doi.org/10.1038/srep04711}
  {\bibfield  {journal} {\bibinfo  {journal} {Sci. Rep.}\ }\textbf {\bibinfo
  {volume} {4}} (\bibinfo {year} {2014}),\ 10.1038/srep04711}\BibitemShut
  {NoStop}%
\bibitem [{\citenamefont {Holzer}\ and\ \citenamefont
  {Cantoni}(2012)}]{Holzer_2012}%
  \BibitemOpen
  \bibfield  {author} {\bibinfo {author} {\bibfnamefont {L.}~\bibnamefont
  {Holzer}}\ and\ \bibinfo {author} {\bibfnamefont {M.}~\bibnamefont
  {Cantoni}},\ }\bibfield  {title} {\enquote {\bibinfo {title} {Review of {FIB}
  tomography},}\ }in\  \cite{Utke_2012},\ Chap.~\bibinfo {chapter} {11}, pp.\
  \bibinfo {pages} {410--435}\BibitemShut {NoStop}%
\bibitem [{\citenamefont {Andersen}\ \emph {et~al.}(2023)\citenamefont
  {Andersen}, \citenamefont {Chen}, \citenamefont {Pal}, \citenamefont
  {Cressa}, \citenamefont {Castro}, \citenamefont {Wirtz}, \citenamefont
  {Schmitz},\ and\ \citenamefont {Eswara}}]{Andersen_2023}%
  \BibitemOpen
  \bibfield  {author} {\bibinfo {author} {\bibfnamefont {D.}~\bibnamefont
  {Andersen}}, \bibinfo {author} {\bibfnamefont {H.}~\bibnamefont {Chen}},
  \bibinfo {author} {\bibfnamefont {S.}~\bibnamefont {Pal}}, \bibinfo {author}
  {\bibfnamefont {L.}~\bibnamefont {Cressa}}, \bibinfo {author} {\bibfnamefont
  {O.~D.}\ \bibnamefont {Castro}}, \bibinfo {author} {\bibfnamefont
  {T.}~\bibnamefont {Wirtz}}, \bibinfo {author} {\bibfnamefont
  {G.}~\bibnamefont {Schmitz}},\ and\ \bibinfo {author} {\bibfnamefont
  {S.}~\bibnamefont {Eswara}},\ }\bibfield  {title} {\enquote {\bibinfo {title}
  {Correlative high-resolution imaging of hydrogen in mg$_2$ni hydrogen storage
  thin films},}\ }\href {https://doi.org/10.1016/j.ijhydene.2022.12.216}
  {\bibfield  {journal} {\bibinfo  {journal} {Int. J. Hydrogen Energ.}\
  }\textbf {\bibinfo {volume} {48}},\ \bibinfo {pages} {13943--13954} (\bibinfo
  {year} {2023})}\BibitemShut {NoStop}%
\bibitem [{\citenamefont {Parlanti}\ and\ \citenamefont
  {Cappello}(2022)}]{Parlanti_2022}%
  \BibitemOpen
  \bibfield  {author} {\bibinfo {author} {\bibfnamefont {P.}~\bibnamefont
  {Parlanti}}\ and\ \bibinfo {author} {\bibfnamefont {V.}~\bibnamefont
  {Cappello}},\ }\bibfield  {title} {\enquote {\bibinfo {title} {Microscopes,
  tools, probes, and protocols: a guide in the route of correlative microscopy
  for biomedical investigation},}\ }\href
  {https://doi.org/10.1016/j.micron.2021.103182} {\bibfield  {journal}
  {\bibinfo  {journal} {Micron}\ }\textbf {\bibinfo {volume} {152}},\ \bibinfo
  {pages} {103182} (\bibinfo {year} {2022})}\BibitemShut {NoStop}%
\bibitem [{\citenamefont {Ando}\ \emph {et~al.}(2018)\citenamefont {Ando},
  \citenamefont {Bhamidimarri}, \citenamefont {Brending}, \citenamefont
  {Colin-York}, \citenamefont {Collinson}, \citenamefont {Jonge}, \citenamefont
  {de~Pablo}, \citenamefont {Debroye}, \citenamefont {Eggeling}, \citenamefont
  {Franck}, \citenamefont {Fritzsche}, \citenamefont {Gerritsen}, \citenamefont
  {Giepmans}, \citenamefont {Grunewald}, \citenamefont {Hofkens}, \citenamefont
  {Hoogenboom}, \citenamefont {Janssen}, \citenamefont {Kaufmann},
  \citenamefont {Klumperman}, \citenamefont {Kurniawan}, \citenamefont {Kusch},
  \citenamefont {Liv}, \citenamefont {Parekh}, \citenamefont {Peckys},
  \citenamefont {Rehfeldt}, \citenamefont {Reutens}, \citenamefont {Roeffaers},
  \citenamefont {Salditt}, \citenamefont {Schaap}, \citenamefont {Schwarz},
  \citenamefont {Verkade}, \citenamefont {Vogel}, \citenamefont {Wagner},
  \citenamefont {Winterhalter}, \citenamefont {Yuan},\ and\ \citenamefont
  {Zifarelli}}]{Ando_2018}%
  \BibitemOpen
  \bibfield  {author} {\bibinfo {author} {\bibfnamefont {T.}~\bibnamefont
  {Ando}}, \bibinfo {author} {\bibfnamefont {S.~P.}\ \bibnamefont
  {Bhamidimarri}}, \bibinfo {author} {\bibfnamefont {N.}~\bibnamefont
  {Brending}}, \bibinfo {author} {\bibfnamefont {H.}~\bibnamefont
  {Colin-York}}, \bibinfo {author} {\bibfnamefont {L.}~\bibnamefont
  {Collinson}}, \bibinfo {author} {\bibfnamefont {N.~D.}\ \bibnamefont
  {Jonge}}, \bibinfo {author} {\bibfnamefont {P.~J.}\ \bibnamefont {de~Pablo}},
  \bibinfo {author} {\bibfnamefont {E.}~\bibnamefont {Debroye}}, \bibinfo
  {author} {\bibfnamefont {C.}~\bibnamefont {Eggeling}}, \bibinfo {author}
  {\bibfnamefont {C.}~\bibnamefont {Franck}}, \bibinfo {author} {\bibfnamefont
  {M.}~\bibnamefont {Fritzsche}}, \bibinfo {author} {\bibfnamefont
  {H.}~\bibnamefont {Gerritsen}}, \bibinfo {author} {\bibfnamefont {B.~N.~G.}\
  \bibnamefont {Giepmans}}, \bibinfo {author} {\bibfnamefont {K.}~\bibnamefont
  {Grunewald}}, \bibinfo {author} {\bibfnamefont {J.}~\bibnamefont {Hofkens}},
  \bibinfo {author} {\bibfnamefont {J.~P.}\ \bibnamefont {Hoogenboom}},
  \bibinfo {author} {\bibfnamefont {K.~P.~F.}\ \bibnamefont {Janssen}},
  \bibinfo {author} {\bibfnamefont {R.}~\bibnamefont {Kaufmann}}, \bibinfo
  {author} {\bibfnamefont {J.}~\bibnamefont {Klumperman}}, \bibinfo {author}
  {\bibfnamefont {N.}~\bibnamefont {Kurniawan}}, \bibinfo {author}
  {\bibfnamefont {J.}~\bibnamefont {Kusch}}, \bibinfo {author} {\bibfnamefont
  {N.}~\bibnamefont {Liv}}, \bibinfo {author} {\bibfnamefont {V.}~\bibnamefont
  {Parekh}}, \bibinfo {author} {\bibfnamefont {D.~B.}\ \bibnamefont {Peckys}},
  \bibinfo {author} {\bibfnamefont {F.}~\bibnamefont {Rehfeldt}}, \bibinfo
  {author} {\bibfnamefont {D.~C.}\ \bibnamefont {Reutens}}, \bibinfo {author}
  {\bibfnamefont {M.~B.~J.}\ \bibnamefont {Roeffaers}}, \bibinfo {author}
  {\bibfnamefont {T.}~\bibnamefont {Salditt}}, \bibinfo {author} {\bibfnamefont
  {I.~A.~T.}\ \bibnamefont {Schaap}}, \bibinfo {author} {\bibfnamefont {U.~S.}\
  \bibnamefont {Schwarz}}, \bibinfo {author} {\bibfnamefont {P.}~\bibnamefont
  {Verkade}}, \bibinfo {author} {\bibfnamefont {M.~W.}\ \bibnamefont {Vogel}},
  \bibinfo {author} {\bibfnamefont {R.}~\bibnamefont {Wagner}}, \bibinfo
  {author} {\bibfnamefont {M.}~\bibnamefont {Winterhalter}}, \bibinfo {author}
  {\bibfnamefont {H.}~\bibnamefont {Yuan}},\ and\ \bibinfo {author}
  {\bibfnamefont {G.}~\bibnamefont {Zifarelli}},\ }\bibfield  {title} {\enquote
  {\bibinfo {title} {The 2018 correlative microscopy techniques roadmap},}\
  }\href {https://doi.org/10.1088/1361-6463/aad055} {\bibfield  {journal}
  {\bibinfo  {journal} {J. Phys. D: Appl. Phys.}\ }\textbf {\bibinfo {volume}
  {51}},\ \bibinfo {pages} {443001} (\bibinfo {year} {2018})}\BibitemShut
  {NoStop}%
\bibitem [{\citenamefont {Schindelin}\ \emph {et~al.}(2012)\citenamefont
  {Schindelin}, \citenamefont {Arganda-Carreras}, \citenamefont {Frise},
  \citenamefont {Kaynig}, \citenamefont {Longair}, \citenamefont {Pietzsch},
  \citenamefont {Preibisch}, \citenamefont {Rueden}, \citenamefont {Saalfeld},
  \citenamefont {Schmid}, \citenamefont {Tinevez}, \citenamefont {White},
  \citenamefont {Hartenstein}, \citenamefont {Eliceiri}, \citenamefont
  {Tomancak},\ and\ \citenamefont {Cardona}}]{Schindelin_2012}%
  \BibitemOpen
  \bibfield  {author} {\bibinfo {author} {\bibfnamefont {J.}~\bibnamefont
  {Schindelin}}, \bibinfo {author} {\bibfnamefont {I.}~\bibnamefont
  {Arganda-Carreras}}, \bibinfo {author} {\bibfnamefont {E.}~\bibnamefont
  {Frise}}, \bibinfo {author} {\bibfnamefont {V.}~\bibnamefont {Kaynig}},
  \bibinfo {author} {\bibfnamefont {M.}~\bibnamefont {Longair}}, \bibinfo
  {author} {\bibfnamefont {T.}~\bibnamefont {Pietzsch}}, \bibinfo {author}
  {\bibfnamefont {S.}~\bibnamefont {Preibisch}}, \bibinfo {author}
  {\bibfnamefont {C.}~\bibnamefont {Rueden}}, \bibinfo {author} {\bibfnamefont
  {S.}~\bibnamefont {Saalfeld}}, \bibinfo {author} {\bibfnamefont
  {B.}~\bibnamefont {Schmid}}, \bibinfo {author} {\bibfnamefont {J.-Y.}\
  \bibnamefont {Tinevez}}, \bibinfo {author} {\bibfnamefont {D.~J.}\
  \bibnamefont {White}}, \bibinfo {author} {\bibfnamefont {V.}~\bibnamefont
  {Hartenstein}}, \bibinfo {author} {\bibfnamefont {K.}~\bibnamefont
  {Eliceiri}}, \bibinfo {author} {\bibfnamefont {P.}~\bibnamefont {Tomancak}},\
  and\ \bibinfo {author} {\bibfnamefont {A.}~\bibnamefont {Cardona}},\
  }\bibfield  {title} {\enquote {\bibinfo {title} {Fiji: an open-source
  platform for biological-image analysis},}\ }\href
  {https://doi.org/10.1038/nmeth.2019} {\bibfield  {journal} {\bibinfo
  {journal} {Nat. Methods}\ }\textbf {\bibinfo {volume} {9}},\ \bibinfo {pages}
  {676--682} (\bibinfo {year} {2012})}\BibitemShut {NoStop}%
\bibitem [{\citenamefont {Rohde}, \citenamefont {Braumann},\ and\ \citenamefont
  {Schmidt}(2020)}]{ROHDE_2020}%
  \BibitemOpen
  \bibfield  {author} {\bibinfo {author} {\bibfnamefont {F.}~\bibnamefont
  {Rohde}}, \bibinfo {author} {\bibfnamefont {U.-D.}\ \bibnamefont
  {Braumann}},\ and\ \bibinfo {author} {\bibfnamefont {M.}~\bibnamefont
  {Schmidt}},\ }\bibfield  {title} {\enquote {\bibinfo {title} {Correlia: an
  {ImageJ} plug-in to co-register and visualise multimodal correlative
  micrographs},}\ }\href {https://doi.org/10.1111/jmi.12928} {\bibfield
  {journal} {\bibinfo  {journal} {J. Microsc.}\ }\textbf {\bibinfo {volume}
  {280}},\ \bibinfo {pages} {3--11} (\bibinfo {year} {2020})}\BibitemShut
  {NoStop}%
\bibitem [{\citenamefont {Schmidt}, \citenamefont {Rohde},\ and\ \citenamefont
  {Braumann}(2021)}]{Schmidt_2021}%
  \BibitemOpen
  \bibfield  {author} {\bibinfo {author} {\bibfnamefont {M.}~\bibnamefont
  {Schmidt}}, \bibinfo {author} {\bibfnamefont {F.}~\bibnamefont {Rohde}},\
  and\ \bibinfo {author} {\bibfnamefont {U.-D.}\ \bibnamefont {Braumann}},\
  }\bibfield  {title} {\enquote {\bibinfo {title} {Visualization and
  co-registration of correlative microscopy data with the {ImageJ} plug-in
  correlia},}\ }in\ \href {https://doi.org/10.1016/bs.mcb.2020.10.001} {\emph
  {\bibinfo {booktitle} {Methods in Cell Biology}}}\ (\bibinfo  {publisher}
  {Elsevier},\ \bibinfo {year} {2021})\ pp.\ \bibinfo {pages}
  {353--388}\BibitemShut {NoStop}%
\bibitem [{\citenamefont {COMULIS}(2023)}]{COMULIS}%
  \BibitemOpen
  \bibfield  {author} {\bibinfo {author} {\bibnamefont {COMULIS}},\ }\href
  {https://www.comulis.eu/correlation-software} {\enquote {\bibinfo {title}
  {Correlated multimodal imaging in life sciences},}\ }\bibinfo {howpublished}
  {Online} (\bibinfo {year} {2023}),\ \bibinfo {note}
  {\url{https://www.comulis.eu/correlation-software}}\BibitemShut {NoStop}%
\bibitem [{\citenamefont {BIII}(2023)}]{BIII}%
  \BibitemOpen
  \bibfield  {author} {\bibinfo {author} {\bibnamefont {BIII}},\ }\href
  {https://biii.eu} {\enquote {\bibinfo {title} {Bioimage informatics index},}\
  }\bibinfo {howpublished} {Online} (\bibinfo {year} {2023}),\ \bibinfo {note}
  {\url{https://biii.eu}}\BibitemShut {NoStop}%
\bibitem [{\citenamefont {Lutter}\ \emph {et~al.}(2021)\citenamefont {Lutter},
  \citenamefont {Stahlhut}, \citenamefont {Dremel}, \citenamefont {Zabler},
  \citenamefont {Fell}, \citenamefont {Herrmann},\ and\ \citenamefont
  {Hanke}}]{Lutter_2021}%
  \BibitemOpen
  \bibfield  {author} {\bibinfo {author} {\bibfnamefont {F.}~\bibnamefont
  {Lutter}}, \bibinfo {author} {\bibfnamefont {P.}~\bibnamefont {Stahlhut}},
  \bibinfo {author} {\bibfnamefont {K.}~\bibnamefont {Dremel}}, \bibinfo
  {author} {\bibfnamefont {S.}~\bibnamefont {Zabler}}, \bibinfo {author}
  {\bibfnamefont {J.}~\bibnamefont {Fell}}, \bibinfo {author} {\bibfnamefont
  {H.-G.}\ \bibnamefont {Herrmann}},\ and\ \bibinfo {author} {\bibfnamefont
  {R.}~\bibnamefont {Hanke}},\ }\bibfield  {title} {\enquote {\bibinfo {title}
  {{Combining X-ray Nano Tomography with focused ion beam serial section
  imaging — Application of correlative tomography to integrated circuits}},}\
  }\href {https://doi.org/10.1016/j.nimb.2021.05.006} {\bibfield  {journal}
  {\bibinfo  {journal} {Nucl. Instrum. Meth. Phys. Res. B}\ }\textbf {\bibinfo
  {volume} {500-501}},\ \bibinfo {pages} {10--17} (\bibinfo {year}
  {2021})}\BibitemShut {NoStop}%
\bibitem [{\citenamefont {Smith}(1997)}]{Smith-1997}%
  \BibitemOpen
  \bibinfo {editor} {\bibfnamefont {R.}~\bibnamefont {Smith}},\ ed.,\ \href
  {https://doi.org/10.1017/CBO9780511524325} {\emph {\bibinfo {title} {{Atomic
  and Ion Collisions in Solids and at Surfaces: Theory, Simulation and
  Applications}}}}\ (\bibinfo  {publisher} {Cambridge University Press},\
  \bibinfo {address} {Cambridge},\ \bibinfo {year} {1997})\BibitemShut
  {NoStop}%
\bibitem [{\citenamefont {Krasheninnikov}\ and\ \citenamefont
  {Nordlund}(2010)}]{Krasheninnikov2010b}%
  \BibitemOpen
  \bibfield  {author} {\bibinfo {author} {\bibfnamefont {A.~V.}\ \bibnamefont
  {Krasheninnikov}}\ and\ \bibinfo {author} {\bibfnamefont {K.}~\bibnamefont
  {Nordlund}},\ }\bibfield  {title} {\enquote {\bibinfo {title} {Ion and
  electron irradiation-induced effects in nanostructured materials},}\ }\href
  {https://doi.org/10.1063/1.3318261} {\bibfield  {journal} {\bibinfo
  {journal} {J. Appl. Phys.}\ }\textbf {\bibinfo {volume} {107}},\ \bibinfo
  {pages} {071301} (\bibinfo {year} {2010})}\BibitemShut {NoStop}%
\bibitem [{\citenamefont {Nordlund}(2019)}]{Nordlund-2019}%
  \BibitemOpen
  \bibfield  {author} {\bibinfo {author} {\bibfnamefont {K.}~\bibnamefont
  {Nordlund}},\ }\bibfield  {title} {\enquote {\bibinfo {title} {{Historical
  review of computer simulation of radiation effects in materials}},}\ }\href
  {https://doi.org/10.1016/j.jnucmat.2019.04.028} {\bibfield  {journal}
  {\bibinfo  {journal} {J. Nucl. Mater.}\ }\textbf {\bibinfo {volume} {520}},\
  \bibinfo {pages} {273--295} (\bibinfo {year} {2019})}\BibitemShut {NoStop}%
\bibitem [{\citenamefont {Nordlund}\ and\ \citenamefont
  {Djurabekova}(2014)}]{Nordlund_2014}%
  \BibitemOpen
  \bibfield  {author} {\bibinfo {author} {\bibfnamefont {K.}~\bibnamefont
  {Nordlund}}\ and\ \bibinfo {author} {\bibfnamefont {F.}~\bibnamefont
  {Djurabekova}},\ }\bibfield  {title} {{\selectlanguage {english}\enquote
  {\bibinfo {title} {{Multiscale modelling of irradiation in
  nanostructures}},}\ }}\href {https://doi.org/10.1007/s10825-013-0542-z}
  {\bibfield  {journal} {\bibinfo  {journal} {J. Comput. Electron.}\ }\textbf
  {\bibinfo {volume} {13}},\ \bibinfo {pages} {122--141} (\bibinfo {year}
  {2014})}\BibitemShut {NoStop}%
\bibitem [{\citenamefont {Eckstein}(1991)}]{eckstein_computer_1991}%
  \BibitemOpen
  \bibfield  {author} {\bibinfo {author} {\bibfnamefont {W.}~\bibnamefont
  {Eckstein}},\ }\href {https://doi.org/10.1007/978-3-642-73513-4} {\emph
  {\bibinfo {title} {Computer Simulation of Ion-Solid Interactions}}}\
  (\bibinfo  {publisher} {Springer Berlin Heidelberg},\ \bibinfo {year}
  {1991})\BibitemShut {NoStop}%
\bibitem [{\citenamefont {Ziegler}, \citenamefont {Biersack},\ and\
  \citenamefont {Littmark}(1985)}]{ziegler_stopping_1985}%
  \BibitemOpen
  \bibfield  {author} {\bibinfo {author} {\bibfnamefont {J.~F.}\ \bibnamefont
  {Ziegler}}, \bibinfo {author} {\bibfnamefont {J.~P.}\ \bibnamefont
  {Biersack}},\ and\ \bibinfo {author} {\bibfnamefont {U.}~\bibnamefont
  {Littmark}},\ }\href@noop {} {\emph {\bibinfo {title} {The {Stopping} and
  {Range} of {Ions} in {Solids}}}}\ (\bibinfo  {publisher} {Pergamon Press, New
  York},\ \bibinfo {year} {1985})\BibitemShut {NoStop}%
\bibitem [{\citenamefont {Wilson}, \citenamefont {Haggmark},\ and\
  \citenamefont {Biersack}(1977)}]{wilson_calculation_1977}%
  \BibitemOpen
  \bibfield  {author} {\bibinfo {author} {\bibfnamefont {W.~D.}\ \bibnamefont
  {Wilson}}, \bibinfo {author} {\bibfnamefont {L.~G.}\ \bibnamefont
  {Haggmark}},\ and\ \bibinfo {author} {\bibfnamefont {J.~P.}\ \bibnamefont
  {Biersack}},\ }\bibfield  {title} {\enquote {\bibinfo {title} {Calculation of
  {Nuclear} {Stopping}, {Ranges}, and {Straggling} in the {Low}-{Energy}
  {Region}},}\ }\href {https://doi.org/10.1103/PhysRevB.15.2458} {\bibfield
  {journal} {\bibinfo  {journal} {Phys. Rev.}\ }\textbf {\bibinfo {volume} {B
  15}},\ \bibinfo {pages} {2458--2468} (\bibinfo {year} {1977})}\BibitemShut
  {NoStop}%
\bibitem [{\citenamefont {Lindhard}\ and\ \citenamefont
  {Scharff}(1961)}]{lindhard_energy_1961}%
  \BibitemOpen
  \bibfield  {author} {\bibinfo {author} {\bibfnamefont {J.}~\bibnamefont
  {Lindhard}}\ and\ \bibinfo {author} {\bibfnamefont {M.}~\bibnamefont
  {Scharff}},\ }\bibfield  {title} {\enquote {\bibinfo {title} {Energy
  {Dissipation} by {Ions} in the {keV} {Region}},}\ }\href
  {https://doi.org/10.1103/PhysRev.124.128} {\bibfield  {journal} {\bibinfo
  {journal} {Phys. Rev.}\ }\textbf {\bibinfo {volume} {124}},\ \bibinfo {pages}
  {128--130} (\bibinfo {year} {1961})}\BibitemShut {NoStop}%
\bibitem [{\citenamefont {Ziegler}(2013)}]{jf_ziegler_particle_2013}%
  \BibitemOpen
  \bibfield  {author} {\bibinfo {author} {\bibfnamefont {J.~F.}\ \bibnamefont
  {Ziegler}},\ }\href {http://www.srim.org} {\enquote {\bibinfo {title}
  {Particle interactions with matter},}\ }\bibinfo {type} {software}\ (\bibinfo
   {institution} {SRIM},\ \bibinfo {year} {2013})\BibitemShut {NoStop}%
\bibitem [{\citenamefont {Oen}\ and\ \citenamefont
  {Robinson}(1976)}]{oen_computer_1976}%
  \BibitemOpen
  \bibfield  {author} {\bibinfo {author} {\bibfnamefont {O.~S.}\ \bibnamefont
  {Oen}}\ and\ \bibinfo {author} {\bibfnamefont {M.~T.}\ \bibnamefont
  {Robinson}},\ }\bibfield  {title} {\enquote {\bibinfo {title} {Computer
  studies of the reflection of light ions from solids},}\ }\href
  {https://doi.org/10.1016/0029-554x(76)90806-5} {\bibfield  {journal}
  {\bibinfo  {journal} {Nucl. Instr. Meth.}\ }\textbf {\bibinfo {volume}
  {132}},\ \bibinfo {pages} {647--653} (\bibinfo {year} {1976})}\BibitemShut
  {NoStop}%
\bibitem [{\citenamefont {Sigmund}(1974)}]{sigmund_energy_1974}%
  \BibitemOpen
  \bibfield  {author} {\bibinfo {author} {\bibfnamefont {P.}~\bibnamefont
  {Sigmund}},\ }\bibfield  {title} {\enquote {\bibinfo {title} {Energy density
  and time constant of heavy-ion-induced elastic-collision spikes in solids},}\
  }\href {https://doi.org/10.1063/1.1655425} {\bibfield  {journal} {\bibinfo
  {journal} {Appl. Phys. Lett.}\ }\textbf {\bibinfo {volume} {25}},\ \bibinfo
  {pages} {169--171} (\bibinfo {year} {1974})}\BibitemShut {NoStop}%
\bibitem [{\citenamefont {Sigmund}(1969)}]{sigmund_theory_1969}%
  \BibitemOpen
  \bibfield  {author} {\bibinfo {author} {\bibfnamefont {P.}~\bibnamefont
  {Sigmund}},\ }\bibfield  {title} {\enquote {\bibinfo {title} {Theory of
  {Sputtering}. {I}. {Sputtering} {Yield} of {Amorphous} and {Polycrystalline}
  {Targets}},}\ }\href {https://doi.org/10.1103/PhysRev.184.383} {\bibfield
  {journal} {\bibinfo  {journal} {Phys. Rev.}\ }\textbf {\bibinfo {volume}
  {184}},\ \bibinfo {pages} {383--416} (\bibinfo {year} {1969})}\BibitemShut
  {NoStop}%
\bibitem [{\citenamefont {Hobler}\ and\ \citenamefont
  {Betz}(2001)}]{hobler_useful_2001}%
  \BibitemOpen
  \bibfield  {author} {\bibinfo {author} {\bibfnamefont {G.}~\bibnamefont
  {Hobler}}\ and\ \bibinfo {author} {\bibfnamefont {G.}~\bibnamefont {Betz}},\
  }\bibfield  {title} {\enquote {\bibinfo {title} {On the useful range of
  application of molecular dynamics simulations in the recoil interaction
  approximation},}\ }\href {https://doi.org/10.1016/S0168-583X(01)00418-9}
  {\bibfield  {journal} {\bibinfo  {journal} {Nucl. Instrum. Meth. Phys. Res.
  B}\ }\textbf {\bibinfo {volume} {180}},\ \bibinfo {pages} {203--208}
  (\bibinfo {year} {2001})}\BibitemShut {NoStop}%
\bibitem [{\citenamefont {Hobler}\ and\ \citenamefont
  {Nordlund}(2019)}]{hobler_channeling_2019}%
  \BibitemOpen
  \bibfield  {author} {\bibinfo {author} {\bibfnamefont {G.}~\bibnamefont
  {Hobler}}\ and\ \bibinfo {author} {\bibfnamefont {K.}~\bibnamefont
  {Nordlund}},\ }\bibfield  {title} {\enquote {\bibinfo {title} {Channeling
  maps for {Si} ions in {Si}: {Assessing} the binary collision
  approximation},}\ }\href {https://doi.org/10.1016/j.nimb.2019.04.029}
  {\bibfield  {journal} {\bibinfo  {journal} {Nucl. Instrum. Meth. Phys. Res.
  B}\ }\textbf {\bibinfo {volume} {449}},\ \bibinfo {pages} {17--21} (\bibinfo
  {year} {2019})}\BibitemShut {NoStop}%
\bibitem [{\citenamefont {Eckstein}\ and\ \citenamefont
  {Urbassek}(2007)}]{w._eckstein_computer_2007}%
  \BibitemOpen
  \bibfield  {author} {\bibinfo {author} {\bibfnamefont {W.}~\bibnamefont
  {Eckstein}}\ and\ \bibinfo {author} {\bibfnamefont {H.~M.}\ \bibnamefont
  {Urbassek}},\ }\bibfield  {title} {\enquote {\bibinfo {title} {Computer
  simulation of the sputtering process},}\ }in\ \href
  {https://doi.org/10.1007/978-3-540-44502-9_2} {\emph {\bibinfo {booktitle}
  {Sputtering by Particle Bombardment}}}\ (\bibinfo  {publisher} {Springer
  Berlin Heidelberg},\ \bibinfo {year} {2007})\ pp.\ \bibinfo {pages}
  {21--31}\BibitemShut {NoStop}%
\bibitem [{\citenamefont {Schlueter}\ \emph {et~al.}(2020)\citenamefont
  {Schlueter}, \citenamefont {Nordlund}, \citenamefont {Hobler}, \citenamefont
  {Balden}, \citenamefont {Granberg}, \citenamefont {Flinck}, \citenamefont
  {da~Silva},\ and\ \citenamefont {Neu}}]{schlueter_absence_2020}%
  \BibitemOpen
  \bibfield  {author} {\bibinfo {author} {\bibfnamefont {K.}~\bibnamefont
  {Schlueter}}, \bibinfo {author} {\bibfnamefont {K.}~\bibnamefont {Nordlund}},
  \bibinfo {author} {\bibfnamefont {G.}~\bibnamefont {Hobler}}, \bibinfo
  {author} {\bibfnamefont {M.}~\bibnamefont {Balden}}, \bibinfo {author}
  {\bibfnamefont {F.}~\bibnamefont {Granberg}}, \bibinfo {author}
  {\bibfnamefont {O.}~\bibnamefont {Flinck}}, \bibinfo {author} {\bibfnamefont
  {T.~F.}\ \bibnamefont {da~Silva}},\ and\ \bibinfo {author} {\bibfnamefont
  {R.}~\bibnamefont {Neu}},\ }\bibfield  {title} {\enquote {\bibinfo {title}
  {Absence of a {Crystal} {Direction} {Regime} in which {Sputtering}
  {Corresponds} to {Amorphous} {Material}},}\ }\href
  {https://doi.org/10.1103/PhysRevLett.125.225502} {\bibfield  {journal}
  {\bibinfo  {journal} {Phys. Rev. Lett.}\ }\textbf {\bibinfo {volume} {125}},\
  \bibinfo {pages} {225502} (\bibinfo {year} {2020})},\ \bibinfo {note}
  {publisher: American Physical Society}\BibitemShut {NoStop}%
\bibitem [{\citenamefont {Biersack}\ and\ \citenamefont
  {Eckstein}(1984)}]{biersack_sputtering_1984}%
  \BibitemOpen
  \bibfield  {author} {\bibinfo {author} {\bibfnamefont {J.~P.}\ \bibnamefont
  {Biersack}}\ and\ \bibinfo {author} {\bibfnamefont {W.}~\bibnamefont
  {Eckstein}},\ }\bibfield  {title} {\enquote {\bibinfo {title} {Sputtering
  {Studies} with the {Monte} {Carlo} {Program} {TRIM}.{SP}},}\ }\href
  {https://doi.org/10.1007/BF00614759} {\bibfield  {journal} {\bibinfo
  {journal} {Appl. Phys. A}\ }\textbf {\bibinfo {volume} {34}},\ \bibinfo
  {pages} {73--94} (\bibinfo {year} {1984})}\BibitemShut {NoStop}%
\bibitem [{\citenamefont {Posselt}(1995)}]{posselt_dynamic_1995}%
  \BibitemOpen
  \bibfield  {author} {\bibinfo {author} {\bibfnamefont {M.}~\bibnamefont
  {Posselt}},\ }\bibfield  {title} {\enquote {\bibinfo {title} {Dynamic
  simulation of damage accumulation during implantation of
  {BF}\$\_2{\textasciicircum}+\$ molecular ions into crystalline silicon},}\
  }\href {https://doi.org/10.1016/0168-583x(95)80136-a} {\bibfield  {journal}
  {\bibinfo  {journal} {Nucl. Instrum. Meth. Phys. Res. B}\ }\textbf {\bibinfo
  {volume} {B 102}},\ \bibinfo {pages} {167--172} (\bibinfo {year}
  {1995})}\BibitemShut {NoStop}%
\bibitem [{\citenamefont {Hobler}\ \emph {et~al.}(1995)\citenamefont {Hobler},
  \citenamefont {Simionescu}, \citenamefont {Palmetshofer}, \citenamefont
  {Tian},\ and\ \citenamefont {Stingeder}}]{hobler_boron_1995}%
  \BibitemOpen
  \bibfield  {author} {\bibinfo {author} {\bibfnamefont {G.}~\bibnamefont
  {Hobler}}, \bibinfo {author} {\bibfnamefont {A.}~\bibnamefont {Simionescu}},
  \bibinfo {author} {\bibfnamefont {L.}~\bibnamefont {Palmetshofer}}, \bibinfo
  {author} {\bibfnamefont {C.}~\bibnamefont {Tian}},\ and\ \bibinfo {author}
  {\bibfnamefont {G.}~\bibnamefont {Stingeder}},\ }\bibfield  {title} {\enquote
  {\bibinfo {title} {Boron channeling implantations in silicon: Modeling of
  electronic stopping and damage accumulation},}\ }\href
  {https://doi.org/10.1063/1.358608} {\bibfield  {journal} {\bibinfo  {journal}
  {J. Appl. Phys.}\ }\textbf {\bibinfo {volume} {77}},\ \bibinfo {pages}
  {3697--3703} (\bibinfo {year} {1995})}\BibitemShut {NoStop}%
\bibitem [{\citenamefont {{RADIATE}}(2023)}]{noauthor_ion_nodate}%
  \BibitemOpen
  \bibfield  {author} {\bibinfo {author} {\bibnamefont {{RADIATE}}},\ }\href
  {https://www.ionbeamcenters.eu/resources/ion-beam-software/}
  {{\selectlanguage {english}\enquote {\bibinfo {title} {Ion {Beam}
  {Software}},}\ }}\bibinfo {howpublished} {Online} (\bibinfo {year} {2023}),\
  \bibinfo {note}
  {\url{https://www.ionbeamcenters.eu/resources/ion-beam-software/}}\BibitemShut
  {NoStop}%
\bibitem [{\citenamefont {Biersack}\ and\ \citenamefont
  {Haggmark}(1980)}]{biersack_monte_1980}%
  \BibitemOpen
  \bibfield  {author} {\bibinfo {author} {\bibfnamefont {J.~P.}\ \bibnamefont
  {Biersack}}\ and\ \bibinfo {author} {\bibfnamefont {L.~G.}\ \bibnamefont
  {Haggmark}},\ }\bibfield  {title} {\enquote {\bibinfo {title} {A {Monte}
  {Carlo} {Computer} {Program} for the {Transport} of {Energetic} {Ions} in
  {Amorphous} {Targets}},}\ }\href
  {https://doi.org/10.1016/0029-554X(80)90440-1} {\bibfield  {journal}
  {\bibinfo  {journal} {Nucl. Instr. Meth.}\ }\textbf {\bibinfo {volume}
  {174}},\ \bibinfo {pages} {257--269} (\bibinfo {year} {1980})}\BibitemShut
  {NoStop}%
\bibitem [{\citenamefont {Ziegler}, \citenamefont {Biersack},\ and\
  \citenamefont {Ziegler}(2015)}]{ziegler_srim_2015}%
  \BibitemOpen
  \bibfield  {author} {\bibinfo {author} {\bibfnamefont {J.~F.}\ \bibnamefont
  {Ziegler}}, \bibinfo {author} {\bibfnamefont {J.}~\bibnamefont {Biersack}},\
  and\ \bibinfo {author} {\bibfnamefont {M.~D.}\ \bibnamefont {Ziegler}},\
  }\href@noop {} {{\selectlanguage {english}\emph {\bibinfo {title} {{SRIM} -
  {The} {Stopping} and {Range} of {Ions} in {Matter}}}}}\ (\bibinfo
  {publisher} {SRIM},\ \bibinfo {address} {Chester, Maryland},\ \bibinfo {year}
  {2015})\BibitemShut {NoStop}%
\bibitem [{\citenamefont {M{\"{o}}ller}, \citenamefont {Eckstein},\ and\
  \citenamefont {Biersack}(1988)}]{moller_tridyn_1988}%
  \BibitemOpen
  \bibfield  {author} {\bibinfo {author} {\bibfnamefont {W.}~\bibnamefont
  {M{\"{o}}ller}}, \bibinfo {author} {\bibfnamefont {W.}~\bibnamefont
  {Eckstein}},\ and\ \bibinfo {author} {\bibfnamefont {J.~P.}\ \bibnamefont
  {Biersack}},\ }\bibfield  {title} {\enquote {\bibinfo {title} {Tridyn -
  {Binary} collision simulation of atomic collisions and dynamic composition
  changes in solids},}\ }\href {https://doi.org/10.1016/0010-4655(88)90148-8}
  {\bibfield  {journal} {\bibinfo  {journal} {Comput. Phys. Commun.}\ }\textbf
  {\bibinfo {volume} {51}},\ \bibinfo {pages} {355--368} (\bibinfo {year}
  {1988})}\BibitemShut {NoStop}%
\bibitem [{\citenamefont {M{\"{o}}ller}\ and\ \citenamefont
  {Eckstein}(1984)}]{moller_tridyn_1984}%
  \BibitemOpen
  \bibfield  {author} {\bibinfo {author} {\bibfnamefont {W.}~\bibnamefont
  {M{\"{o}}ller}}\ and\ \bibinfo {author} {\bibfnamefont {W.}~\bibnamefont
  {Eckstein}},\ }\bibfield  {title} {\enquote {\bibinfo {title} {Tridyn — {A}
  {TRIM} simulation code including dynamic composition changes},}\ }\href
  {https://doi.org/10.1016/0168-583X(84)90321-5} {\bibfield  {journal}
  {\bibinfo  {journal} {Nucl. Instrum. Meth. Phys. Res. B}\ }\textbf {\bibinfo
  {volume} {2}},\ \bibinfo {pages} {814--818} (\bibinfo {year}
  {1984})}\BibitemShut {NoStop}%
\bibitem [{\citenamefont {Eckstein}\ \emph {et~al.}(2007)\citenamefont
  {Eckstein}, \citenamefont {Mutzke}, \citenamefont {Dohmen},\ and\
  \citenamefont {Schneider}}]{eckstein_sdtrimsp:_2007}%
  \BibitemOpen
  \bibfield  {author} {\bibinfo {author} {\bibfnamefont {W.}~\bibnamefont
  {Eckstein}}, \bibinfo {author} {\bibfnamefont {A.}~\bibnamefont {Mutzke}},
  \bibinfo {author} {\bibfnamefont {R.}~\bibnamefont {Dohmen}},\ and\ \bibinfo
  {author} {\bibfnamefont {R.}~\bibnamefont {Schneider}},\ }\href
  {https://hdl.handle.net/11858/00-001M-0000-0027-04E9-D} {\enquote {\bibinfo
  {title} {{SDTrimSP}: {A} {Monte}-{Carlo} {Code} for calculating collision
  phenomena in randomized targets},}\ }\bibinfo {type} {techreport}\ \bibinfo
  {number} {IPP 12/3}\ (\bibinfo  {institution} {Max-Planck-Institut für
  Plasmaphysik},\ \bibinfo {year} {2007})\BibitemShut {NoStop}%
\bibitem [{\citenamefont {Hofs{\"{a}}ss}\ and\ \citenamefont
  {Stegmaier}(2022)}]{hofsass_binary_2022}%
  \BibitemOpen
  \bibfield  {author} {\bibinfo {author} {\bibfnamefont {H.}~\bibnamefont
  {Hofs{\"{a}}ss}}\ and\ \bibinfo {author} {\bibfnamefont {A.}~\bibnamefont
  {Stegmaier}},\ }\bibfield  {title} {{\selectlanguage {english}\enquote
  {\bibinfo {title} {Binary collision approximation simulations of ion solid
  interaction without the concept of surface binding energies},}\ }}\href
  {https://doi.org/10.1016/j.nimb.2022.02.012} {\bibfield  {journal} {\bibinfo
  {journal} {Nucl. Instrum. Meth. Phys. Res. B}\ }\textbf {\bibinfo {volume}
  {517}},\ \bibinfo {pages} {49--62} (\bibinfo {year} {2022})}\BibitemShut
  {NoStop}%
\bibitem [{\citenamefont {Posselt}(1993)}]{posselt_computer_1993}%
  \BibitemOpen
  \bibfield  {author} {\bibinfo {author} {\bibfnamefont {M.}~\bibnamefont
  {Posselt}},\ }\bibfield  {title} {\enquote {\bibinfo {title} {Computer
  {Simulation} of {Channeling} {Implantation} at {High} and {Medium}
  {Energies}},}\ }\href {https://doi.org/10.1016/0168-583X(93)96069-O}
  {\bibfield  {journal} {\bibinfo  {journal} {Nucl. Instrum. Meth. Phys. Res.
  B}\ }\textbf {\bibinfo {volume} {80/81}},\ \bibinfo {pages} {28--32}
  (\bibinfo {year} {1993})}\BibitemShut {NoStop}%
\bibitem [{\citenamefont {Hobler}(1995)}]{hobler_monte_1995}%
  \BibitemOpen
  \bibfield  {author} {\bibinfo {author} {\bibfnamefont {G.}~\bibnamefont
  {Hobler}},\ }\bibfield  {title} {\enquote {\bibinfo {title} {Monte {Carlo}
  {Simulation} of {Two}-{Dimensional} {Implanted} {Dopant} {Distributions} at
  {Mask} {Edges}},}\ }\href {https://doi.org/10.1016/0168-583X(94)00476-5}
  {\bibfield  {journal} {\bibinfo  {journal} {Nucl. Instrum. Meth. Phys. Res.
  B}\ }\textbf {\bibinfo {volume} {96}},\ \bibinfo {pages} {155--162} (\bibinfo
  {year} {1995})}\BibitemShut {NoStop}%
\bibitem [{\citenamefont {Hobler}\ and\ \citenamefont
  {Kovac}(2011)}]{hobler_dynamic_2011}%
  \BibitemOpen
  \bibfield  {author} {\bibinfo {author} {\bibfnamefont {G.}~\bibnamefont
  {Hobler}}\ and\ \bibinfo {author} {\bibfnamefont {D.}~\bibnamefont {Kovac}},\
  }\bibfield  {title} {\enquote {\bibinfo {title} {Dynamic binary collision
  simulation of focused ion beam milling of deep trenches},}\ }\href
  {https://doi.org/10.1016/j.nimb.2010.12.076} {\bibfield  {journal} {\bibinfo
  {journal} {Nucl. Instrum. Meth. Phys. Res. B}\ }\textbf {\bibinfo {volume}
  {269}},\ \bibinfo {pages} {1609--1613} (\bibinfo {year} {2011})}\BibitemShut
  {NoStop}%
\bibitem [{\citenamefont {Hobler}(2015)}]{hobler_combined_2015}%
  \BibitemOpen
  \bibfield  {author} {\bibinfo {author} {\bibfnamefont {G.}~\bibnamefont
  {Hobler}},\ }\bibfield  {title} {\enquote {\bibinfo {title} {Combined binary
  collision and continuum mechanics model applied to focused ion beam milling
  of a silicon membrane},}\ }\href {https://doi.org/10.1016/j.nimb.2014.11.109}
  {\bibfield  {journal} {\bibinfo  {journal} {Nucl. Instrum. Meth. Phys. Res.
  B}\ }\bibinfo {series} {Proceedings of the 12th {International} {Conference}
  on {Computer} {Simulation} of {Radiation} {Effects} in {Solids}, {Alacant},
  {Spain}, 8-13 {June}, 2014},\ \textbf {\bibinfo {volume} {352}},\ \bibinfo
  {pages} {22--26} (\bibinfo {year} {2015})}\BibitemShut {NoStop}%
\bibitem [{\citenamefont {Mutzke}\ and\ \citenamefont
  {Schneider}(2009)}]{mutzke_sdtrimsp-2d:_2009}%
  \BibitemOpen
  \bibfield  {author} {\bibinfo {author} {\bibfnamefont {A.}~\bibnamefont
  {Mutzke}}\ and\ \bibinfo {author} {\bibfnamefont {R.}~\bibnamefont
  {Schneider}},\ }\href {https://hdl.handle.net/11858/00-001M-0000-0026-F4C2-4}
  {\enquote {\bibinfo {title} {{SDTrimSP}-{2D}: {Simulation} of particles
  bombarding on a two dimensional target, version 1.0},}\ }\bibinfo {type}
  {techreport}\ \bibinfo {number} {IPP 12/4}\ (\bibinfo  {institution}
  {Max-Planck-Institut für Plasmaphysik},\ \bibinfo {year} {2009})\BibitemShut
  {NoStop}%
\bibitem [{\citenamefont {Schiettekatte}(2008)}]{schiettekatte_fast_2008}%
  \BibitemOpen
  \bibfield  {author} {\bibinfo {author} {\bibfnamefont {F.}~\bibnamefont
  {Schiettekatte}},\ }\bibfield  {title} {\enquote {\bibinfo {title} {Fast
  {Monte} {Carlo} for ion beam analysis simulations},}\ }\href
  {https://doi.org/10.1016/j.nimb.2007.11.075} {\bibfield  {journal} {\bibinfo
  {journal} {Nucl. Instrum. Meth. Phys. Res. B}\ }\textbf {\bibinfo {volume} {B
  226}},\ \bibinfo {pages} {1880--1885} (\bibinfo {year} {2008})}\BibitemShut
  {NoStop}%
\bibitem [{\citenamefont {Schiettekatte}\ and\ \citenamefont
  {Chicoine}(2016)}]{schiettekatte_spectrum_2016}%
  \BibitemOpen
  \bibfield  {author} {\bibinfo {author} {\bibfnamefont {F.}~\bibnamefont
  {Schiettekatte}}\ and\ \bibinfo {author} {\bibfnamefont {M.}~\bibnamefont
  {Chicoine}},\ }\bibfield  {title} {\enquote {\bibinfo {title} {Spectrum
  simulation of rough and nanostructured targets from their {2D} and {3D} image
  by {Monte} {Carlo} methods},}\ }\href
  {https://doi.org/10.1016/j.nimb.2015.09.089} {\bibfield  {journal} {\bibinfo
  {journal} {Nucl. Instrum. Meth. Phys. Res. B}\ }\bibinfo {series} {The 22nd
  {International} {Conference} on {Ion} {Beam} {Analysis} ({IBA} 2015)},\
  \textbf {\bibinfo {volume} {371}},\ \bibinfo {pages} {106--110} (\bibinfo
  {year} {2016})}\BibitemShut {NoStop}%
\bibitem [{\citenamefont {Nietiadi}\ \emph {et~al.}(2014)\citenamefont
  {Nietiadi}, \citenamefont {Sandoval}, \citenamefont {Urbassek},\ and\
  \citenamefont {M{\"{o}}ller}}]{nietiadi_sputtering_2014}%
  \BibitemOpen
  \bibfield  {author} {\bibinfo {author} {\bibfnamefont {M.~L.}\ \bibnamefont
  {Nietiadi}}, \bibinfo {author} {\bibfnamefont {L.}~\bibnamefont {Sandoval}},
  \bibinfo {author} {\bibfnamefont {H.~M.}\ \bibnamefont {Urbassek}},\ and\
  \bibinfo {author} {\bibfnamefont {W.}~\bibnamefont {M{\"{o}}ller}},\
  }\bibfield  {title} {\enquote {\bibinfo {title} {Sputtering of {Si}
  nanospheres},}\ }\href {https://doi.org/10.1103/PhysRevB.90.045417}
  {\bibfield  {journal} {\bibinfo  {journal} {Phys. Rev. B}\ }\textbf {\bibinfo
  {volume} {90}},\ \bibinfo {pages} {045417} (\bibinfo {year}
  {2014})}\BibitemShut {NoStop}%
\bibitem [{\citenamefont {Borschel}\ and\ \citenamefont
  {Ronning}(2011)}]{borschel_ion_2011}%
  \BibitemOpen
  \bibfield  {author} {\bibinfo {author} {\bibfnamefont {C.}~\bibnamefont
  {Borschel}}\ and\ \bibinfo {author} {\bibfnamefont {C.}~\bibnamefont
  {Ronning}},\ }\bibfield  {title} {\enquote {\bibinfo {title} {Ion beam
  irradiation of nanostructures – {A} {3D} {Monte} {Carlo} simulation
  code},}\ }\href {https://doi.org/10.1016/j.nimb.2011.07.004} {\bibfield
  {journal} {\bibinfo  {journal} {Nucl. Instrum. Meth. Phys. Res. B}\ }\textbf
  {\bibinfo {volume} {269}},\ \bibinfo {pages} {2133--2138} (\bibinfo {year}
  {2011})}\BibitemShut {NoStop}%
\bibitem [{\citenamefont {Li}\ \emph {et~al.}(2015)\citenamefont {Li},
  \citenamefont {Yang}, \citenamefont {Short}, \citenamefont {Ding},
  \citenamefont {Zeng},\ and\ \citenamefont {Li}}]{li_im3d:_2015}%
  \BibitemOpen
  \bibfield  {author} {\bibinfo {author} {\bibfnamefont {Y.~G.}\ \bibnamefont
  {Li}}, \bibinfo {author} {\bibfnamefont {Y.}~\bibnamefont {Yang}}, \bibinfo
  {author} {\bibfnamefont {M.~P.}\ \bibnamefont {Short}}, \bibinfo {author}
  {\bibfnamefont {Z.~J.}\ \bibnamefont {Ding}}, \bibinfo {author}
  {\bibfnamefont {Z.}~\bibnamefont {Zeng}},\ and\ \bibinfo {author}
  {\bibfnamefont {J.}~\bibnamefont {Li}},\ }\bibfield  {title} {\enquote
  {\bibinfo {title} {{IM3D}: {A} parallel {Monte} {Carlo} code for efficient
  simulations of primary radiation displacements and damage in {3D}
  geometry},}\ }\href {https://doi.org/10.1038/srep18130} {\bibfield  {journal}
  {\bibinfo  {journal} {Sci. Rep.}\ }\textbf {\bibinfo {volume} {5}},\ \bibinfo
  {pages} {18130} (\bibinfo {year} {2015})}\BibitemShut {NoStop}%
\bibitem [{\citenamefont {Toussaint}, \citenamefont {Mutzke},\ and\
  \citenamefont {Manhard}(2017)}]{toussaint_sputtering_2017}%
  \BibitemOpen
  \bibfield  {author} {\bibinfo {author} {\bibfnamefont {U.~v.}\ \bibnamefont
  {Toussaint}}, \bibinfo {author} {\bibfnamefont {A.}~\bibnamefont {Mutzke}},\
  and\ \bibinfo {author} {\bibfnamefont {A.}~\bibnamefont {Manhard}},\
  }\bibfield  {title} {{\selectlanguage {english}\enquote {\bibinfo {title}
  {Sputtering of rough surfaces: a {3D} simulation study},}\ }}\href
  {https://doi.org/10.1088/1402-4896/aa90be} {\bibfield  {journal} {\bibinfo
  {journal} {Phys. Scr.}\ }\textbf {\bibinfo {volume} {2017}},\ \bibinfo
  {pages} {014056} (\bibinfo {year} {2017})}\BibitemShut {NoStop}%
\bibitem [{\citenamefont {M{\"{o}}ller}(2014)}]{w._moller_tri3dyn_2014}%
  \BibitemOpen
  \bibfield  {author} {\bibinfo {author} {\bibfnamefont {W.}~\bibnamefont
  {M{\"{o}}ller}},\ }\bibfield  {title} {\enquote {\bibinfo {title} {{TRI3DYN}
  - {C}ollisional computer simulation of the dynamic evolution of 3-dimensional
  nanostructures under ion irradiation},}\ }\href
  {https://doi.org/10.1016/j.nimb.2013.12.027} {\bibfield  {journal} {\bibinfo
  {journal} {Nucl. Instrum. Meth. Phys. Res. B}\ }\textbf {\bibinfo {volume}
  {322}},\ \bibinfo {pages} {23--33} (\bibinfo {year} {2014})}\BibitemShut
  {NoStop}%
\bibitem [{\citenamefont {Holland-Moritz}\ \emph {et~al.}(2018)\citenamefont
  {Holland-Moritz}, \citenamefont {Graupner}, \citenamefont {M{\"{o}}ller},
  \citenamefont {Pacholski},\ and\ \citenamefont
  {Ronning}}]{holland-moritz_dynamics_2018}%
  \BibitemOpen
  \bibfield  {author} {\bibinfo {author} {\bibfnamefont {H.}~\bibnamefont
  {Holland-Moritz}}, \bibinfo {author} {\bibfnamefont {J.}~\bibnamefont
  {Graupner}}, \bibinfo {author} {\bibfnamefont {W.}~\bibnamefont
  {M{\"{o}}ller}}, \bibinfo {author} {\bibfnamefont {C.}~\bibnamefont
  {Pacholski}},\ and\ \bibinfo {author} {\bibfnamefont {C.}~\bibnamefont
  {Ronning}},\ }\bibfield  {title} {{\selectlanguage {english}\enquote
  {\bibinfo {title} {Dynamics of nanoparticle morphology under low energy ion
  irradiation},}\ }}\href {https://doi.org/10.1088/1361-6528/aac36c} {\bibfield
   {journal} {\bibinfo  {journal} {Nanotechnology}\ }\textbf {\bibinfo {volume}
  {29}},\ \bibinfo {pages} {314002} (\bibinfo {year} {2018})}\BibitemShut
  {NoStop}%
\bibitem [{\citenamefont {M{\"{o}}ller}, \citenamefont {Johannes},\ and\
  \citenamefont {Ronning}(2016)}]{moller_shaping_2016}%
  \BibitemOpen
  \bibfield  {author} {\bibinfo {author} {\bibfnamefont {W.}~\bibnamefont
  {M{\"{o}}ller}}, \bibinfo {author} {\bibfnamefont {A.}~\bibnamefont
  {Johannes}},\ and\ \bibinfo {author} {\bibfnamefont {C.}~\bibnamefont
  {Ronning}},\ }\bibfield  {title} {{\selectlanguage {english}\enquote
  {\bibinfo {title} {Shaping and compositional modification of zinc oxide
  nanowires under energetic manganese ion irradiation},}\ }}\href
  {https://doi.org/10.1088/0957-4484/27/17/175301} {\bibfield  {journal}
  {\bibinfo  {journal} {Nanotechnology}\ }\textbf {\bibinfo {volume} {27}},\
  \bibinfo {pages} {175301} (\bibinfo {year} {2016})}\BibitemShut {NoStop}%
\bibitem [{\citenamefont {Smith}, \citenamefont {Joy},\ and\ \citenamefont
  {Rack}(2010)}]{smith_monte_2010}%
  \BibitemOpen
  \bibfield  {author} {\bibinfo {author} {\bibfnamefont {D.~A.}\ \bibnamefont
  {Smith}}, \bibinfo {author} {\bibfnamefont {D.~C.}\ \bibnamefont {Joy}},\
  and\ \bibinfo {author} {\bibfnamefont {P.~D.}\ \bibnamefont {Rack}},\
  }\bibfield  {title} {{\selectlanguage {english}\enquote {\bibinfo {title}
  {Monte {Carlo} simulation of focused helium ion beam induced deposition},}\
  }}\href {https://doi.org/10.1088/0957-4484/21/17/175302} {\bibfield
  {journal} {\bibinfo  {journal} {Nanotechnology}\ }\textbf {\bibinfo {volume}
  {21}},\ \bibinfo {pages} {175302} (\bibinfo {year} {2010})}\BibitemShut
  {NoStop}%
\bibitem [{\citenamefont {Timilsina}\ and\ \citenamefont
  {Rack}(2013)}]{timilsina_monte_2013}%
  \BibitemOpen
  \bibfield  {author} {\bibinfo {author} {\bibfnamefont {R.}~\bibnamefont
  {Timilsina}}\ and\ \bibinfo {author} {\bibfnamefont {P.~D.}\ \bibnamefont
  {Rack}},\ }\bibfield  {title} {\enquote {\bibinfo {title} {Monte {Carlo}
  simulations of nanoscale focused neon ion beam sputtering},}\ }\href
  {https://doi.org/10.1088/0957-4484/24/49/495303} {\bibfield  {journal}
  {\bibinfo  {journal} {Nanotechnology}\ }\textbf {\bibinfo {volume} {24}},\
  \bibinfo {pages} {495303} (\bibinfo {year} {2013})}\BibitemShut {NoStop}%
\bibitem [{\citenamefont {Timilsina}, \citenamefont {Smith},\ and\
  \citenamefont {Rack}(2013)}]{timilsina_comparison_2013}%
  \BibitemOpen
  \bibfield  {author} {\bibinfo {author} {\bibfnamefont {R.}~\bibnamefont
  {Timilsina}}, \bibinfo {author} {\bibfnamefont {D.~A.}\ \bibnamefont
  {Smith}},\ and\ \bibinfo {author} {\bibfnamefont {P.~D.}\ \bibnamefont
  {Rack}},\ }\bibfield  {title} {{\selectlanguage {english}\enquote {\bibinfo
  {title} {A comparison of neon versus helium ion beam induced deposition via
  {Monte} {Carlo} simulations},}\ }}\href
  {https://doi.org/10.1088/0957-4484/24/11/115302} {\bibfield  {journal}
  {\bibinfo  {journal} {Nanotechnology}\ }\textbf {\bibinfo {volume} {24}},\
  \bibinfo {pages} {115302} (\bibinfo {year} {2013})}\BibitemShut {NoStop}%
\bibitem [{\citenamefont {Timilsina}\ \emph {et~al.}(2014)\citenamefont
  {Timilsina}, \citenamefont {Tan}, \citenamefont {Livengood},\ and\
  \citenamefont {Rack}}]{timilsina_monte_2014}%
  \BibitemOpen
  \bibfield  {author} {\bibinfo {author} {\bibfnamefont {R.}~\bibnamefont
  {Timilsina}}, \bibinfo {author} {\bibfnamefont {S.}~\bibnamefont {Tan}},
  \bibinfo {author} {\bibfnamefont {R.}~\bibnamefont {Livengood}},\ and\
  \bibinfo {author} {\bibfnamefont {P.~D.}\ \bibnamefont {Rack}},\ }\bibfield
  {title} {{\selectlanguage {english}\enquote {\bibinfo {title} {Monte {Carlo}
  simulations of nanoscale focused neon ion beam sputtering of copper:
  elucidating resolution limits and sub-surface damage},}\ }}\href
  {https://doi.org/10.1088/0957-4484/25/48/485704} {\bibfield  {journal}
  {\bibinfo  {journal} {Nanotechnology}\ }\textbf {\bibinfo {volume} {25}},\
  \bibinfo {pages} {485704} (\bibinfo {year} {2014})}\BibitemShut {NoStop}%
\bibitem [{\citenamefont {Mahady}\ \emph {et~al.}(2016)\citenamefont {Mahady},
  \citenamefont {Tan}, \citenamefont {Greenzweig}, \citenamefont {Livengood},
  \citenamefont {Raveh},\ and\ \citenamefont {Rack}}]{mahady_monte_2017}%
  \BibitemOpen
  \bibfield  {author} {\bibinfo {author} {\bibfnamefont {K.}~\bibnamefont
  {Mahady}}, \bibinfo {author} {\bibfnamefont {S.}~\bibnamefont {Tan}},
  \bibinfo {author} {\bibfnamefont {Y.}~\bibnamefont {Greenzweig}}, \bibinfo
  {author} {\bibfnamefont {R.}~\bibnamefont {Livengood}}, \bibinfo {author}
  {\bibfnamefont {A.}~\bibnamefont {Raveh}},\ and\ \bibinfo {author}
  {\bibfnamefont {P.}~\bibnamefont {Rack}},\ }\bibfield  {title}
  {{\selectlanguage {english}\enquote {\bibinfo {title} {Monte {Carlo}
  simulations of nanoscale {Ne}$^+$ ion beam sputtering: investigating the
  influence of surface effects, interstitial formation, and the nanostructural
  evolution},}\ }}\href {https://doi.org/10.1088/1361-6528/28/4/045305}
  {\bibfield  {journal} {\bibinfo  {journal} {Nanotechnology}\ }\textbf
  {\bibinfo {volume} {28}},\ \bibinfo {pages} {045305} (\bibinfo {year}
  {2016})}\BibitemShut {NoStop}%
\bibitem [{\citenamefont {Mahady}\ \emph {et~al.}(2018)\citenamefont {Mahady},
  \citenamefont {Tan}, \citenamefont {Greenzweig}, \citenamefont {Raveh},\ and\
  \citenamefont {Rack}}]{mahady_simulating_2018}%
  \BibitemOpen
  \bibfield  {author} {\bibinfo {author} {\bibfnamefont {K.~T.}\ \bibnamefont
  {Mahady}}, \bibinfo {author} {\bibfnamefont {S.}~\bibnamefont {Tan}},
  \bibinfo {author} {\bibfnamefont {Y.}~\bibnamefont {Greenzweig}}, \bibinfo
  {author} {\bibfnamefont {A.}~\bibnamefont {Raveh}},\ and\ \bibinfo {author}
  {\bibfnamefont {P.~D.}\ \bibnamefont {Rack}},\ }\bibfield  {title}
  {{\selectlanguage {english}\enquote {\bibinfo {title} {Simulating advanced
  focused ion beam nanomachining: a quantitative comparison of simulation and
  experimental results},}\ }}\href {https://doi.org/10.1088/1361-6528/aae183}
  {\bibfield  {journal} {\bibinfo  {journal} {Nanotechnology}\ }\textbf
  {\bibinfo {volume} {29}},\ \bibinfo {pages} {495301} (\bibinfo {year}
  {2018})}\BibitemShut {NoStop}%
\bibitem [{\citenamefont {Mahady}\ \emph {et~al.}(2019)\citenamefont {Mahady},
  \citenamefont {Tan}, \citenamefont {Greenzweig}, \citenamefont {Raveh},\ and\
  \citenamefont {Rack}}]{mahady_monte_2019}%
  \BibitemOpen
  \bibfield  {author} {\bibinfo {author} {\bibfnamefont {K.}~\bibnamefont
  {Mahady}}, \bibinfo {author} {\bibfnamefont {S.}~\bibnamefont {Tan}},
  \bibinfo {author} {\bibfnamefont {Y.}~\bibnamefont {Greenzweig}}, \bibinfo
  {author} {\bibfnamefont {A.}~\bibnamefont {Raveh}},\ and\ \bibinfo {author}
  {\bibfnamefont {P.}~\bibnamefont {Rack}},\ }\bibfield  {title}
  {{\selectlanguage {english}\enquote {\bibinfo {title} {{Monte Carlo}
  simulation of nanoscale material focused ion beam gas-assisted etching:
  {Ga}$^+$ and {Ne}$^+$ etching of {SiO}$_2$ in the presence of a {XeF}$_2$
  precursor gas},}\ }}\href {https://doi.org/10.1039/C9NA00390H} {\bibfield
  {journal} {\bibinfo  {journal} {Nanoscale Adv.}\ }\textbf {\bibinfo {volume}
  {1}},\ \bibinfo {pages} {3584--3596} (\bibinfo {year} {2019})}\BibitemShut
  {NoStop}%
\bibitem [{\citenamefont {Hofs{\"{a}}ss}, \citenamefont {Zhang},\ and\
  \citenamefont {Mutzke}(2014)}]{hofsass_simulation_2014}%
  \BibitemOpen
  \bibfield  {author} {\bibinfo {author} {\bibfnamefont {H.}~\bibnamefont
  {Hofs{\"{a}}ss}}, \bibinfo {author} {\bibfnamefont {K.}~\bibnamefont
  {Zhang}},\ and\ \bibinfo {author} {\bibfnamefont {A.}~\bibnamefont
  {Mutzke}},\ }\bibfield  {title} {\enquote {\bibinfo {title} {Simulation of
  ion beam sputtering with {SDTrimSP}, {TRIDYN} and {SRIM}},}\ }\href
  {https://doi.org/10.1016/j.apsusc.2014.03.152} {\bibfield  {journal}
  {\bibinfo  {journal} {Appl. Surf. Sci.}\ }\textbf {\bibinfo {volume} {310}},\
  \bibinfo {pages} {134--141} (\bibinfo {year} {2014})}\BibitemShut {NoStop}%
\bibitem [{\citenamefont {Jakob}\ \emph {et~al.}(2021)\citenamefont {Jakob},
  \citenamefont {Robson}, \citenamefont {Schmitt}, \citenamefont {Mourik},
  \citenamefont {Posselt}, \citenamefont {Spemann}, \citenamefont {Johnson},
  \citenamefont {Firgau}, \citenamefont {Mayes}, \citenamefont {McCallum},
  \citenamefont {Morello},\ and\ \citenamefont {Jamieson}}]{Jakob_2021a}%
  \BibitemOpen
  \bibfield  {author} {\bibinfo {author} {\bibfnamefont {A.~M.}\ \bibnamefont
  {Jakob}}, \bibinfo {author} {\bibfnamefont {S.~G.}\ \bibnamefont {Robson}},
  \bibinfo {author} {\bibfnamefont {V.}~\bibnamefont {Schmitt}}, \bibinfo
  {author} {\bibfnamefont {V.}~\bibnamefont {Mourik}}, \bibinfo {author}
  {\bibfnamefont {M.}~\bibnamefont {Posselt}}, \bibinfo {author} {\bibfnamefont
  {D.}~\bibnamefont {Spemann}}, \bibinfo {author} {\bibfnamefont {B.~C.}\
  \bibnamefont {Johnson}}, \bibinfo {author} {\bibfnamefont {H.~R.}\
  \bibnamefont {Firgau}}, \bibinfo {author} {\bibfnamefont {E.}~\bibnamefont
  {Mayes}}, \bibinfo {author} {\bibfnamefont {J.~C.}\ \bibnamefont {McCallum}},
  \bibinfo {author} {\bibfnamefont {A.}~\bibnamefont {Morello}},\ and\ \bibinfo
  {author} {\bibfnamefont {D.~N.}\ \bibnamefont {Jamieson}},\ }\bibfield
  {title} {\enquote {\bibinfo {title} {Deterministic shallow dopant
  implantation in silicon with detection confidence upper-bound to 99.85{\%} by
  ion{\textendash}solid interactions},}\ }\href
  {https://doi.org/10.1002/adma.202103235} {\bibfield  {journal} {\bibinfo
  {journal} {Adv. Mater.}\ }\textbf {\bibinfo {volume} {34}},\ \bibinfo {pages}
  {2103235} (\bibinfo {year} {2021})}\BibitemShut {NoStop}%
\bibitem [{\citenamefont {De~Teresa}\ \emph {et~al.}(2019)\citenamefont
  {De~Teresa}, \citenamefont {Or{\'{u}}s}, \citenamefont {C{\'{o}}rdoba},\ and\
  \citenamefont {Philipp}}]{DeTeresa_2019}%
  \BibitemOpen
  \bibfield  {author} {\bibinfo {author} {\bibfnamefont {J.~M.}\ \bibnamefont
  {De~Teresa}}, \bibinfo {author} {\bibfnamefont {P.}~\bibnamefont
  {Or{\'{u}}s}}, \bibinfo {author} {\bibfnamefont {R.}~\bibnamefont
  {C{\'{o}}rdoba}},\ and\ \bibinfo {author} {\bibfnamefont {P.}~\bibnamefont
  {Philipp}},\ }\bibfield  {title} {{\selectlanguage {english}\enquote
  {\bibinfo {title} {Comparison between focused electron/ion beam-induced
  deposition at room temperature and under cryogenic conditions},}\ }}\href
  {https://doi.org/10.3390/mi10120799} {\bibfield  {journal} {\bibinfo
  {journal} {Micromachines}\ }\textbf {\bibinfo {volume} {10}},\ \bibinfo
  {pages} {799} (\bibinfo {year} {2019})},\ \bibinfo {note} {number: 12
  Publisher: Multidisciplinary Digital Publishing Institute}\BibitemShut
  {NoStop}%
\bibitem [{\citenamefont {De~Teresa}\ \emph {et~al.}(2020)\citenamefont
  {De~Teresa}, \citenamefont {Or{\'{u}}s}, \citenamefont {C{\'{o}}rdoba},\ and\
  \citenamefont {Philipp}}]{Teresa_2020}%
  \BibitemOpen
  \bibfield  {author} {\bibinfo {author} {\bibfnamefont {J.~M.}\ \bibnamefont
  {De~Teresa}}, \bibinfo {author} {\bibfnamefont {P.}~\bibnamefont
  {Or{\'{u}}s}}, \bibinfo {author} {\bibfnamefont {R.}~\bibnamefont
  {C{\'{o}}rdoba}},\ and\ \bibinfo {author} {\bibfnamefont {P.}~\bibnamefont
  {Philipp}},\ }\bibfield  {title} {\enquote {\bibinfo {title} {Erratum: de
  teresa, j.m. et al. comparison between focused electron/ion beam-induced
  deposition at room temperature and under cryogenic conditions. micromachines
  2019, 10, 799},}\ }\href {https://doi.org/10.3390/mi11020211} {\bibfield
  {journal} {\bibinfo  {journal} {Micromachines}\ }\textbf {\bibinfo {volume}
  {11}},\ \bibinfo {pages} {211} (\bibinfo {year} {2020})}\BibitemShut
  {NoStop}%
\bibitem [{\citenamefont {Salvador-Porroche}\ \emph {et~al.}(2020)\citenamefont
  {Salvador-Porroche}, \citenamefont {Sangiao}, \citenamefont {Philipp},
  \citenamefont {Cea},\ and\ \citenamefont
  {De~Teresa}}]{SalvadorPorroche_2020}%
  \BibitemOpen
  \bibfield  {author} {\bibinfo {author} {\bibfnamefont {A.}~\bibnamefont
  {Salvador-Porroche}}, \bibinfo {author} {\bibfnamefont {S.}~\bibnamefont
  {Sangiao}}, \bibinfo {author} {\bibfnamefont {P.}~\bibnamefont {Philipp}},
  \bibinfo {author} {\bibfnamefont {P.}~\bibnamefont {Cea}},\ and\ \bibinfo
  {author} {\bibfnamefont {J.~M.}\ \bibnamefont {De~Teresa}},\ }\bibfield
  {title} {{\selectlanguage {english}\enquote {\bibinfo {title} {Optimization
  of {Pt}-{C} {Deposits} by cryo-{FIBID}: substantial growth rate increase and
  quasi-metallic behaviour},}\ }}\href {https://doi.org/10.3390/nano10101906}
  {\bibfield  {journal} {\bibinfo  {journal} {Nanomaterials}\ }\textbf
  {\bibinfo {volume} {10}},\ \bibinfo {pages} {1906} (\bibinfo {year}
  {2020})},\ \bibinfo {note} {number: 10 Publisher: Multidisciplinary Digital
  Publishing Institute}\BibitemShut {NoStop}%
\bibitem [{\citenamefont {Salvador-Porroche}\ \emph {et~al.}(2021)\citenamefont
  {Salvador-Porroche}, \citenamefont {Sangiao}, \citenamefont {Mag{\'{e}}n},
  \citenamefont {Barrado}, \citenamefont {Philipp}, \citenamefont
  {Belotcerkovtceva}, \citenamefont {Kamalakar}, \citenamefont {Cea},\ and\
  \citenamefont {De~Teresa}}]{SalvadorPorroche_2021}%
  \BibitemOpen
  \bibfield  {author} {\bibinfo {author} {\bibfnamefont {A.}~\bibnamefont
  {Salvador-Porroche}}, \bibinfo {author} {\bibfnamefont {S.}~\bibnamefont
  {Sangiao}}, \bibinfo {author} {\bibfnamefont {C.}~\bibnamefont
  {Mag{\'{e}}n}}, \bibinfo {author} {\bibfnamefont {M.}~\bibnamefont
  {Barrado}}, \bibinfo {author} {\bibfnamefont {P.}~\bibnamefont {Philipp}},
  \bibinfo {author} {\bibfnamefont {D.}~\bibnamefont {Belotcerkovtceva}},
  \bibinfo {author} {\bibfnamefont {M.~V.}\ \bibnamefont {Kamalakar}}, \bibinfo
  {author} {\bibfnamefont {P.}~\bibnamefont {Cea}},\ and\ \bibinfo {author}
  {\bibfnamefont {J.~M.}\ \bibnamefont {De~Teresa}},\ }\bibfield  {title}
  {{\selectlanguage {english}\enquote {\bibinfo {title} {Highly-efficient
  growth of cobalt nanostructures using focused ion beam induced deposition
  under cryogenic conditions: application to electrical contacts on graphene,
  magnetism and hard masking},}\ }}\href {https://doi.org/10.1039/d1na00580d}
  {\bibfield  {journal} {\bibinfo  {journal} {Nanoscale Adv.}\ }\textbf
  {\bibinfo {volume} {3}},\ \bibinfo {pages} {5656--5662} (\bibinfo {year}
  {2021})},\ \bibinfo {note} {publisher: RSC}\BibitemShut {NoStop}%
\bibitem [{\citenamefont {Kim}\ \emph {et~al.}(2007{\natexlab{a}})\citenamefont
  {Kim}, \citenamefont {Hobler}, \citenamefont {Steiger}, \citenamefont
  {Lugstein},\ and\ \citenamefont {Bertagnolli}}]{kim_level_2007}%
  \BibitemOpen
  \bibfield  {author} {\bibinfo {author} {\bibfnamefont {H.-B.}\ \bibnamefont
  {Kim}}, \bibinfo {author} {\bibfnamefont {G.}~\bibnamefont {Hobler}},
  \bibinfo {author} {\bibfnamefont {A.}~\bibnamefont {Steiger}}, \bibinfo
  {author} {\bibfnamefont {A.}~\bibnamefont {Lugstein}},\ and\ \bibinfo
  {author} {\bibfnamefont {E.}~\bibnamefont {Bertagnolli}},\ }\bibfield
  {title} {\enquote {\bibinfo {title} {Level set approach for the simulation of
  focused ion beam processing on the micro/nano scale},}\ }\href
  {https://doi.org/10.1088/0957-4484/18/26/265307} {\bibfield  {journal}
  {\bibinfo  {journal} {Nanotechnology}\ }\textbf {\bibinfo {volume} {18}},\
  \bibinfo {pages} {265307} (\bibinfo {year} {2007}{\natexlab{a}})}\BibitemShut
  {NoStop}%
\bibitem [{\citenamefont {Ebm}\ and\ \citenamefont
  {Hobler}(2009)}]{ebm_assessment_2009}%
  \BibitemOpen
  \bibfield  {author} {\bibinfo {author} {\bibfnamefont {C.}~\bibnamefont
  {Ebm}}\ and\ \bibinfo {author} {\bibfnamefont {G.}~\bibnamefont {Hobler}},\
  }\bibfield  {title} {\enquote {\bibinfo {title} {Assessment of approximations
  for efficient topography simulation of ion beam processes: 10 {keV} {Ar} on
  {Si}},}\ }\href {https://doi.org/10.1016/j.nimb.2009.06.014} {\bibfield
  {journal} {\bibinfo  {journal} {Nucl. Instrum. Meth. Phys. Res. B}\ }\textbf
  {\bibinfo {volume} {267}},\ \bibinfo {pages} {2987--2990} (\bibinfo {year}
  {2009})}\BibitemShut {NoStop}%
\bibitem [{\citenamefont {Ebm}\ \emph {et~al.}(2011)\citenamefont {Ebm},
  \citenamefont {Hobler}, \citenamefont {Waid},\ and\ \citenamefont
  {Wanzenboeck}}]{ebm_quantitative_2011}%
  \BibitemOpen
  \bibfield  {author} {\bibinfo {author} {\bibfnamefont {C.}~\bibnamefont
  {Ebm}}, \bibinfo {author} {\bibfnamefont {G.}~\bibnamefont {Hobler}},
  \bibinfo {author} {\bibfnamefont {S.}~\bibnamefont {Waid}},\ and\ \bibinfo
  {author} {\bibfnamefont {H.~D.}\ \bibnamefont {Wanzenboeck}},\ }\bibfield
  {title} {\enquote {\bibinfo {title} {Quantitative simulation of ion-beam
  induced deposition of nanostructures},}\ }\href
  {https://doi.org/10.1116/1.3533951} {\bibfield  {journal} {\bibinfo
  {journal} {J. Vac. Sci. Technol. B}\ }\textbf {\bibinfo {volume} {29}},\
  \bibinfo {pages} {011031} (\bibinfo {year} {2011})}\BibitemShut {NoStop}%
\bibitem [{\citenamefont {Lindsey}\ and\ \citenamefont
  {Hobler}(2012)}]{lindsey_significance_2012}%
  \BibitemOpen
  \bibfield  {author} {\bibinfo {author} {\bibfnamefont {S.}~\bibnamefont
  {Lindsey}}\ and\ \bibinfo {author} {\bibfnamefont {G.}~\bibnamefont
  {Hobler}},\ }\bibfield  {title} {\enquote {\bibinfo {title} {The significance
  of redeposition and backscattering in nanostructure formation by focused ion
  beams},}\ }\href {https://doi.org/10.1016/j.nimb.2011.08.051} {\bibfield
  {journal} {\bibinfo  {journal} {Nucl. Instrum. Meth. Phys. Res. B}\ }\textbf
  {\bibinfo {volume} {282}},\ \bibinfo {pages} {12--16} (\bibinfo {year}
  {2012})}\BibitemShut {NoStop}%
\bibitem [{\citenamefont {Li}\ \emph {et~al.}(2021)\citenamefont {Li},
  \citenamefont {Chen}, \citenamefont {Dai}, \citenamefont {Yang},
  \citenamefont {Chen}, \citenamefont {Wang}, \citenamefont {Chen},
  \citenamefont {Peng}, \citenamefont {Shan},\ and\ \citenamefont
  {Duan}}]{Li_2021}%
  \BibitemOpen
  \bibfield  {author} {\bibinfo {author} {\bibfnamefont {P.}~\bibnamefont
  {Li}}, \bibinfo {author} {\bibfnamefont {S.}~\bibnamefont {Chen}}, \bibinfo
  {author} {\bibfnamefont {H.}~\bibnamefont {Dai}}, \bibinfo {author}
  {\bibfnamefont {Z.}~\bibnamefont {Yang}}, \bibinfo {author} {\bibfnamefont
  {Z.}~\bibnamefont {Chen}}, \bibinfo {author} {\bibfnamefont {Y.}~\bibnamefont
  {Wang}}, \bibinfo {author} {\bibfnamefont {Y.}~\bibnamefont {Chen}}, \bibinfo
  {author} {\bibfnamefont {W.}~\bibnamefont {Peng}}, \bibinfo {author}
  {\bibfnamefont {W.}~\bibnamefont {Shan}},\ and\ \bibinfo {author}
  {\bibfnamefont {H.}~\bibnamefont {Duan}},\ }\bibfield  {title}
  {{\selectlanguage {english}\enquote {\bibinfo {title} {Recent advances in
  focused ion beam nanofabrication for nanostructures and devices: fundamentals
  and applications},}\ }}\href {https://doi.org/10.1039/D0NR07539F} {\bibfield
  {journal} {\bibinfo  {journal} {Nanoscale}\ }\textbf {\bibinfo {volume}
  {13}},\ \bibinfo {pages} {1529--1565} (\bibinfo {year} {2021})},\ \bibinfo
  {note} {publisher: The Royal Society of Chemistry}\BibitemShut {NoStop}%
\bibitem [{\citenamefont {Huang}\ \emph {et~al.}(2018)\citenamefont {Huang},
  \citenamefont {Loeffler}, \citenamefont {Muehle}, \citenamefont {Moeller},
  \citenamefont {Mulders}, \citenamefont {Kwakman}, \citenamefont {Van~Dorp},\
  and\ \citenamefont {Zschech}}]{huang_si_2018}%
  \BibitemOpen
  \bibfield  {author} {\bibinfo {author} {\bibfnamefont {J.}~\bibnamefont
  {Huang}}, \bibinfo {author} {\bibfnamefont {M.}~\bibnamefont {Loeffler}},
  \bibinfo {author} {\bibfnamefont {U.}~\bibnamefont {Muehle}}, \bibinfo
  {author} {\bibfnamefont {W.}~\bibnamefont {Moeller}}, \bibinfo {author}
  {\bibfnamefont {J.~J.~L.}\ \bibnamefont {Mulders}}, \bibinfo {author}
  {\bibfnamefont {L.~F.~T.}\ \bibnamefont {Kwakman}}, \bibinfo {author}
  {\bibfnamefont {W.~F.}\ \bibnamefont {Van~Dorp}},\ and\ \bibinfo {author}
  {\bibfnamefont {E.}~\bibnamefont {Zschech}},\ }\bibfield  {title} {\enquote
  {\bibinfo {title} {{S}i amorphization by focused ion beam milling: {Point}
  defect model with dynamic {BCA} simulation and experimental validation},}\
  }\href {https://doi.org/10.1016/j.ultramic.2017.10.011} {\bibfield  {journal}
  {\bibinfo  {journal} {Ultramicroscopy}\ }\textbf {\bibinfo {volume} {184}},\
  \bibinfo {pages} {52--56} (\bibinfo {year} {2018})}\BibitemShut {NoStop}%
\bibitem [{\citenamefont {Rubia}\ \emph {et~al.}(1988)\citenamefont {Rubia},
  \citenamefont {Averback}, \citenamefont {Benedek},\ and\ \citenamefont
  {King}}]{rubia_role_1987}%
  \BibitemOpen
  \bibfield  {author} {\bibinfo {author} {\bibfnamefont {T.~D. d.~l.}\
  \bibnamefont {Rubia}}, \bibinfo {author} {\bibfnamefont {R.~S.}\ \bibnamefont
  {Averback}}, \bibinfo {author} {\bibfnamefont {R.}~\bibnamefont {Benedek}},\
  and\ \bibinfo {author} {\bibfnamefont {W.~E.}\ \bibnamefont {King}},\
  }\bibfield  {title} {\enquote {\bibinfo {title} {Role of thermal spikes in
  energetic displacement cascades},}\ }\href
  {https://doi.org/10.1103/physrevlett.60.76.3} {\bibfield  {journal} {\bibinfo
   {journal} {Phys. Rev. Lett.}\ }\textbf {\bibinfo {volume} {60}},\ \bibinfo
  {pages} {76--76} (\bibinfo {year} {1988})}\BibitemShut {NoStop}%
\bibitem [{\citenamefont {Tadmor}\ and\ \citenamefont
  {Miller}(2011)}]{eb_tadmor_modeling_2011}%
  \BibitemOpen
  \bibfield  {author} {\bibinfo {author} {\bibfnamefont {E.~B.}\ \bibnamefont
  {Tadmor}}\ and\ \bibinfo {author} {\bibfnamefont {R.~E.}\ \bibnamefont
  {Miller}},\ }\href@noop {} {\emph {\bibinfo {title} {Modeling {Materials}:
  {Continuum}, {Atomistic} and {Multiscale} {Techniques}}}}\ (\bibinfo
  {publisher} {Cambridge University Press},\ \bibinfo {address} {United
  Kingdom},\ \bibinfo {year} {2011})\BibitemShut {NoStop}%
\bibitem [{\citenamefont {Ghaderzadeh}\ \emph {et~al.}(2019)\citenamefont
  {Ghaderzadeh}, \citenamefont {Ghorbani-Asl}, \citenamefont {Kretschmer},
  \citenamefont {Hlawacek},\ and\ \citenamefont
  {Krasheninnikov}}]{Ghaderzadeh2020n}%
  \BibitemOpen
  \bibfield  {author} {\bibinfo {author} {\bibfnamefont {S.}~\bibnamefont
  {Ghaderzadeh}}, \bibinfo {author} {\bibfnamefont {M.}~\bibnamefont
  {Ghorbani-Asl}}, \bibinfo {author} {\bibfnamefont {S.}~\bibnamefont
  {Kretschmer}}, \bibinfo {author} {\bibfnamefont {G.}~\bibnamefont
  {Hlawacek}},\ and\ \bibinfo {author} {\bibfnamefont {A.~V.}\ \bibnamefont
  {Krasheninnikov}},\ }\bibfield  {title} {\enquote {\bibinfo {title}
  {{Channeling effects in gold nanoclusters under He ion irradiation: Insights
  from molecular dynamics simulations}},}\ }\href
  {https://doi.org/10.1088/1361-6528/ab4847} {\bibfield  {journal} {\bibinfo
  {journal} {Nanotechnology}\ }\textbf {\bibinfo {volume} {31}},\ \bibinfo
  {pages} {035302} (\bibinfo {year} {2019})}\BibitemShut {NoStop}%
\bibitem [{\citenamefont {Allen}\ and\ \citenamefont
  {Tildesley}(1989)}]{Allen-Tildesley1989}%
  \BibitemOpen
  \bibfield  {author} {\bibinfo {author} {\bibfnamefont {M.~P.}\ \bibnamefont
  {Allen}}\ and\ \bibinfo {author} {\bibfnamefont {D.~J.}\ \bibnamefont
  {Tildesley}},\ }\href@noop {} {\emph {\bibinfo {title} {{Computer Simulation
  of Liquids}}}}\ (\bibinfo  {publisher} {Oxford University Press},\ \bibinfo
  {address} {Oxford, England},\ \bibinfo {year} {1989})\BibitemShut {NoStop}%
\bibitem [{\citenamefont {Leach}(2001)}]{Leach2001}%
  \BibitemOpen
  \bibfield  {author} {\bibinfo {author} {\bibfnamefont {A.~R.}\ \bibnamefont
  {Leach}},\ }\href@noop {} {\emph {\bibinfo {title} {{Molecular modelling:
  principles and applications}}}},\ \bibinfo {edition} {second edition}\ ed.\
  (\bibinfo  {publisher} {Pearson Education},\ \bibinfo {address} {Harlow,
  England},\ \bibinfo {year} {2001})\BibitemShut {NoStop}%
\bibitem [{\citenamefont {Nordlund}\ \emph {et~al.}(2018)\citenamefont
  {Nordlund}, \citenamefont {Zinkle}, \citenamefont {Sand}, \citenamefont
  {Granberg}, \citenamefont {Averback}, \citenamefont {Stoller}, \citenamefont
  {Suzudo}, \citenamefont {Malerba}, \citenamefont {Banhart}, \citenamefont
  {Weber}, \citenamefont {Willaime}, \citenamefont {Dudarev},\ and\
  \citenamefont {Simeone}}]{Nordlund_2018}%
  \BibitemOpen
  \bibfield  {author} {\bibinfo {author} {\bibfnamefont {K.}~\bibnamefont
  {Nordlund}}, \bibinfo {author} {\bibfnamefont {S.~J.}\ \bibnamefont
  {Zinkle}}, \bibinfo {author} {\bibfnamefont {A.~E.}\ \bibnamefont {Sand}},
  \bibinfo {author} {\bibfnamefont {F.}~\bibnamefont {Granberg}}, \bibinfo
  {author} {\bibfnamefont {R.~S.}\ \bibnamefont {Averback}}, \bibinfo {author}
  {\bibfnamefont {R.~E.}\ \bibnamefont {Stoller}}, \bibinfo {author}
  {\bibfnamefont {T.}~\bibnamefont {Suzudo}}, \bibinfo {author} {\bibfnamefont
  {L.}~\bibnamefont {Malerba}}, \bibinfo {author} {\bibfnamefont
  {F.}~\bibnamefont {Banhart}}, \bibinfo {author} {\bibfnamefont {W.~J.}\
  \bibnamefont {Weber}}, \bibinfo {author} {\bibfnamefont {F.}~\bibnamefont
  {Willaime}}, \bibinfo {author} {\bibfnamefont {S.~L.}\ \bibnamefont
  {Dudarev}},\ and\ \bibinfo {author} {\bibfnamefont {D.}~\bibnamefont
  {Simeone}},\ }\bibfield  {title} {\enquote {\bibinfo {title} {{Primary
  radiation damage: A review of current understanding and models}},}\ }\href
  {https://doi.org/10.1016/j.jnucmat.2018.10.027} {\bibfield  {journal}
  {\bibinfo  {journal} {J. Nucl. Mater.}\ }\textbf {\bibinfo {volume} {512}},\
  \bibinfo {pages} {450--479} (\bibinfo {year} {2018})}\BibitemShut {NoStop}%
\bibitem [{\citenamefont {Nordlund}\ \emph {et~al.}(2021)\citenamefont
  {Nordlund}, \citenamefont {Djurabekova}, \citenamefont {Murphy},\ and\
  \citenamefont {Duffy}}]{Nordlund2021}%
  \BibitemOpen
  \bibfield  {author} {\bibinfo {author} {\bibfnamefont {K.}~\bibnamefont
  {Nordlund}}, \bibinfo {author} {\bibfnamefont {F.}~\bibnamefont
  {Djurabekova}}, \bibinfo {author} {\bibfnamefont {S.~T.}\ \bibnamefont
  {Murphy}},\ and\ \bibinfo {author} {\bibfnamefont {D.~M.}\ \bibnamefont
  {Duffy}},\ }\bibfield  {title} {\enquote {\bibinfo {title} {Tools for
  investigating electronic excitation: experiment and multi-scale modelling},}\
  }in\ \href {https://doi.org/10.20868/upm.book.69109} {\emph {\bibinfo
  {booktitle} {Tools for investigating electronic excitation: experiment and
  multi-scale modelling}}}\ (\bibinfo  {publisher} {COST},\ \bibinfo {address}
  {Brussels, Belgium},\ \bibinfo {year} {2021})\ Chap.\ \bibinfo {chapter}
  {Molecular dynamics for radiation effects}, pp.\ \bibinfo {pages}
  {227--259}\BibitemShut {NoStop}%
\bibitem [{\citenamefont {Plimpton}(1995)}]{Plimpton1995}%
  \BibitemOpen
  \bibfield  {author} {\bibinfo {author} {\bibfnamefont {S.}~\bibnamefont
  {Plimpton}},\ }\bibfield  {title} {\enquote {\bibinfo {title} {Fast parallel
  algorithms for short-range molecular dynamics},}\ }\href
  {https://doi.org/10.1006/jcph.1995.1039} {\bibfield  {journal} {\bibinfo
  {journal} {J. Comput. Phys.}\ }\textbf {\bibinfo {volume} {115}},\ \bibinfo
  {pages} {468--472} (\bibinfo {year} {1995})}\BibitemShut {NoStop}%
\bibitem [{\citenamefont {Thompson}\ \emph {et~al.}(2022)\citenamefont
  {Thompson}, \citenamefont {Aktulga}, \citenamefont {Berger}, \citenamefont
  {Bolintineanu}, \citenamefont {Brown}, \citenamefont {Crozier}, \citenamefont
  {in~'t Veld}, \citenamefont {Kohlmeyer}, \citenamefont {Moore}, \citenamefont
  {Nguyen}, \citenamefont {Shan}, \citenamefont {Stevens}, \citenamefont
  {Tranchida}, \citenamefont {Trott},\ and\ \citenamefont
  {Plimpton}}]{Thompson_2022}%
  \BibitemOpen
  \bibfield  {author} {\bibinfo {author} {\bibfnamefont {A.~P.}\ \bibnamefont
  {Thompson}}, \bibinfo {author} {\bibfnamefont {H.~M.}\ \bibnamefont
  {Aktulga}}, \bibinfo {author} {\bibfnamefont {R.}~\bibnamefont {Berger}},
  \bibinfo {author} {\bibfnamefont {D.~S.}\ \bibnamefont {Bolintineanu}},
  \bibinfo {author} {\bibfnamefont {W.~M.}\ \bibnamefont {Brown}}, \bibinfo
  {author} {\bibfnamefont {P.~S.}\ \bibnamefont {Crozier}}, \bibinfo {author}
  {\bibfnamefont {P.~J.}\ \bibnamefont {in~'t Veld}}, \bibinfo {author}
  {\bibfnamefont {A.}~\bibnamefont {Kohlmeyer}}, \bibinfo {author}
  {\bibfnamefont {S.~G.}\ \bibnamefont {Moore}}, \bibinfo {author}
  {\bibfnamefont {T.~D.}\ \bibnamefont {Nguyen}}, \bibinfo {author}
  {\bibfnamefont {R.}~\bibnamefont {Shan}}, \bibinfo {author} {\bibfnamefont
  {M.~J.}\ \bibnamefont {Stevens}}, \bibinfo {author} {\bibfnamefont
  {J.}~\bibnamefont {Tranchida}}, \bibinfo {author} {\bibfnamefont
  {C.}~\bibnamefont {Trott}},\ and\ \bibinfo {author} {\bibfnamefont {S.~J.}\
  \bibnamefont {Plimpton}},\ }\bibfield  {title} {{\selectlanguage
  {english}\enquote {\bibinfo {title} {{LAMMPS} - a flexible simulation tool
  for particle-based materials modeling at the atomic, meso, and continuum
  scales},}\ }}\href {https://doi.org/10.1016/j.cpc.2021.108171} {\bibfield
  {journal} {\bibinfo  {journal} {Comp. Phys. Comm.}\ }\textbf {\bibinfo
  {volume} {271}},\ \bibinfo {pages} {108171} (\bibinfo {year}
  {2022})}\BibitemShut {NoStop}%
\bibitem [{\citenamefont {Smith}\ and\ \citenamefont
  {Forester}(1996)}]{Smith1996}%
  \BibitemOpen
  \bibfield  {author} {\bibinfo {author} {\bibfnamefont {W.}~\bibnamefont
  {Smith}}\ and\ \bibinfo {author} {\bibfnamefont {T.~R.}\ \bibnamefont
  {Forester}},\ }\bibfield  {title} {\enquote {\bibinfo {title} {Dl\_poly\_2.0:
  a general-purpose parallel molecular dynamics simulation package},}\ }\href
  {https://doi.org/10.1016/S0263-7855(96)00043-4} {\bibfield  {journal}
  {\bibinfo  {journal} {J. Mol. Graph.}\ }\textbf {\bibinfo {volume} {14}},\
  \bibinfo {pages} {136--141} (\bibinfo {year} {1996})}\BibitemShut {NoStop}%
\bibitem [{\citenamefont {Granberg}\ \emph {et~al.}(2016)\citenamefont
  {Granberg}, \citenamefont {Nordlund}, \citenamefont {Ullah}, \citenamefont
  {Jin}, \citenamefont {Lu}, \citenamefont {Bei}, \citenamefont {Wang},
  \citenamefont {Djurabekova}, \citenamefont {Weber},\ and\ \citenamefont
  {Zhang}}]{Granberg2016}%
  \BibitemOpen
  \bibfield  {author} {\bibinfo {author} {\bibfnamefont {F.}~\bibnamefont
  {Granberg}}, \bibinfo {author} {\bibfnamefont {K.}~\bibnamefont {Nordlund}},
  \bibinfo {author} {\bibfnamefont {M.~W.}\ \bibnamefont {Ullah}}, \bibinfo
  {author} {\bibfnamefont {K.}~\bibnamefont {Jin}}, \bibinfo {author}
  {\bibfnamefont {C.}~\bibnamefont {Lu}}, \bibinfo {author} {\bibfnamefont
  {H.}~\bibnamefont {Bei}}, \bibinfo {author} {\bibfnamefont {L.~M.}\
  \bibnamefont {Wang}}, \bibinfo {author} {\bibfnamefont {F.}~\bibnamefont
  {Djurabekova}}, \bibinfo {author} {\bibfnamefont {W.~J.}\ \bibnamefont
  {Weber}},\ and\ \bibinfo {author} {\bibfnamefont {Y.}~\bibnamefont {Zhang}},\
  }\bibfield  {title} {\enquote {\bibinfo {title} {Mechanism of radiation
  damage reduction in equiatomic multicomponent single phase alloys},}\ }\href
  {https://doi.org/10.1103/PhysRevLett.116.135504} {\bibfield  {journal}
  {\bibinfo  {journal} {Phys. Rev. Lett.}\ }\textbf {\bibinfo {volume} {116}},\
  \bibinfo {pages} {135504} (\bibinfo {year} {2016})}\BibitemShut {NoStop}%
\bibitem [{\citenamefont {Becquart}\ \emph {et~al.}(1997)\citenamefont
  {Becquart}, \citenamefont {Decker}, \citenamefont {Domain}, \citenamefont
  {Ruste}, \citenamefont {Souffez}, \citenamefont {Turbatte},\ and\
  \citenamefont {Duysen}}]{Becquart1997}%
  \BibitemOpen
  \bibfield  {author} {\bibinfo {author} {\bibfnamefont {C.~S.}\ \bibnamefont
  {Becquart}}, \bibinfo {author} {\bibfnamefont {K.~M.}\ \bibnamefont
  {Decker}}, \bibinfo {author} {\bibfnamefont {C.}~\bibnamefont {Domain}},
  \bibinfo {author} {\bibfnamefont {J.}~\bibnamefont {Ruste}}, \bibinfo
  {author} {\bibfnamefont {Y.}~\bibnamefont {Souffez}}, \bibinfo {author}
  {\bibfnamefont {J.~C.}\ \bibnamefont {Turbatte}},\ and\ \bibinfo {author}
  {\bibfnamefont {J.~C.~V.}\ \bibnamefont {Duysen}},\ }\bibfield  {title}
  {\enquote {\bibinfo {title} {Massively parallel molecular dynamics
  simulations with eam potentials},}\ }\href
  {https://doi.org/10.1080/10420159708211592} {\bibfield  {journal} {\bibinfo
  {journal} {Radiat. Eff. Defect Solid}\ }\textbf {\bibinfo {volume} {142}},\
  \bibinfo {pages} {9--21} (\bibinfo {year} {1997})}\BibitemShut {NoStop}%
\bibitem [{\citenamefont {de~la Rubia}\ and\ \citenamefont
  {Guinan}(1990)}]{delaRubia1990}%
  \BibitemOpen
  \bibfield  {author} {\bibinfo {author} {\bibfnamefont {D.}~\bibnamefont
  {de~la Rubia}}\ and\ \bibinfo {author} {\bibfnamefont {M.~W.}\ \bibnamefont
  {Guinan}},\ }\bibfield  {title} {\enquote {\bibinfo {title} {Progress in the
  development of a molecular dynamics code for high-energy cascade studies},}\
  }\href {https://doi.org/10.1016/0022-3115(90)90229-g} {\bibfield  {journal}
  {\bibinfo  {journal} {J. Nucl. Mater.}\ }\textbf {\bibinfo {volume} {174}},\
  \bibinfo {pages} {151--157} (\bibinfo {year} {1990})}\BibitemShut {NoStop}%
\bibitem [{\citenamefont {Berendsen}, \citenamefont {van~der Spoel},\ and\
  \citenamefont {van Drunen}(1995)}]{Berendsen1995}%
  \BibitemOpen
  \bibfield  {author} {\bibinfo {author} {\bibfnamefont {H.~J.~C.}\
  \bibnamefont {Berendsen}}, \bibinfo {author} {\bibfnamefont {D.}~\bibnamefont
  {van~der Spoel}},\ and\ \bibinfo {author} {\bibfnamefont {R.}~\bibnamefont
  {van Drunen}},\ }\bibfield  {title} {\enquote {\bibinfo {title} {Gromacs: a
  message-passing parallel molecular dynamics implementation},}\ }\href
  {https://doi.org/10.1016/0010-4655(95)00042-E} {\bibfield  {journal}
  {\bibinfo  {journal} {Comput. Phys. Commun.}\ }\textbf {\bibinfo {volume}
  {91}},\ \bibinfo {pages} {43--56} (\bibinfo {year} {1995})}\BibitemShut
  {NoStop}%
\bibitem [{\citenamefont {Phillips}\ \emph {et~al.}(2005)\citenamefont
  {Phillips}, \citenamefont {Braun}, \citenamefont {Wang}, \citenamefont
  {Gumbart}, \citenamefont {Tajkhorshid}, \citenamefont {Villa}, \citenamefont
  {Chipot}, \citenamefont {Skeel}, \citenamefont {Kal{\'e}},\ and\
  \citenamefont {Schulten}}]{Phillips2005}%
  \BibitemOpen
  \bibfield  {author} {\bibinfo {author} {\bibfnamefont {J.~C.}\ \bibnamefont
  {Phillips}}, \bibinfo {author} {\bibfnamefont {R.}~\bibnamefont {Braun}},
  \bibinfo {author} {\bibfnamefont {W.}~\bibnamefont {Wang}}, \bibinfo {author}
  {\bibfnamefont {J.}~\bibnamefont {Gumbart}}, \bibinfo {author} {\bibfnamefont
  {E.}~\bibnamefont {Tajkhorshid}}, \bibinfo {author} {\bibfnamefont
  {E.}~\bibnamefont {Villa}}, \bibinfo {author} {\bibfnamefont
  {C.}~\bibnamefont {Chipot}}, \bibinfo {author} {\bibfnamefont {R.~D.}\
  \bibnamefont {Skeel}}, \bibinfo {author} {\bibfnamefont {L.}~\bibnamefont
  {Kal{\'e}}},\ and\ \bibinfo {author} {\bibfnamefont {K.}~\bibnamefont
  {Schulten}},\ }\bibfield  {title} {{\selectlanguage {english}\enquote
  {\bibinfo {title} {{Scalable molecular dynamics with NAMD}},}\ }}\href
  {https://doi.org/10.1002/jcc.20289} {\bibfield  {journal} {\bibinfo
  {journal} {J. Comput. Chem.}\ }\textbf {\bibinfo {volume} {26}},\ \bibinfo
  {pages} {1781--1802} (\bibinfo {year} {2005})},\ \bibinfo {note} {\_eprint:
  https://onlinelibrary.wiley.com/doi/pdf/10.1002/jcc.20289}\BibitemShut
  {NoStop}%
\bibitem [{\citenamefont {Kresse}\ and\ \citenamefont
  {Furthm{\"{u}}ller}(1996)}]{kresse_efficiency_1996}%
  \BibitemOpen
  \bibfield  {author} {\bibinfo {author} {\bibfnamefont {G.}~\bibnamefont
  {Kresse}}\ and\ \bibinfo {author} {\bibfnamefont {J.}~\bibnamefont
  {Furthm{\"{u}}ller}},\ }\bibfield  {title} {\enquote {\bibinfo {title}
  {Efficiency of ab-initio total energy calculations for metals and
  semiconductors using a plane-wave basis set},}\ }\href
  {https://doi.org/10.1016/0927-0256(96)00008-0} {\bibfield  {journal}
  {\bibinfo  {journal} {Comput. Mater. Sci.}\ }\textbf {\bibinfo {volume}
  {6}},\ \bibinfo {pages} {15--50} (\bibinfo {year} {1996})}\BibitemShut
  {NoStop}%
\bibitem [{\citenamefont {Giannozzi}\ \emph {et~al.}(2009)\citenamefont
  {Giannozzi}, \citenamefont {Baroni}, \citenamefont {Bonini}, \citenamefont
  {Calandra}, \citenamefont {Car}, \citenamefont {Cavazzoni}, \citenamefont
  {Ceresoli}, \citenamefont {Chiarotti}, \citenamefont {Cococcioni},
  \citenamefont {Dabo}, \citenamefont {Corso}, \citenamefont {Gironcoli},
  \citenamefont {Fabris}, \citenamefont {Fratesi}, \citenamefont {Gebauer},
  \citenamefont {Gerstmann}, \citenamefont {Gougoussis}, \citenamefont
  {Kokalj}, \citenamefont {Lazzeri}, \citenamefont {Martin-Samos},
  \citenamefont {Marzari}, \citenamefont {Mauri}, \citenamefont {Mazzarello},
  \citenamefont {Paolini}, \citenamefont {Pasquarello}, \citenamefont
  {Paulatto}, \citenamefont {Sbraccia}, \citenamefont {Scandolo}, \citenamefont
  {Sclauzero}, \citenamefont {Seitsonen}, \citenamefont {Smogunov},
  \citenamefont {Umari},\ and\ \citenamefont
  {Wentzcovitch}}]{giannozzi_quantum_2009}%
  \BibitemOpen
  \bibfield  {author} {\bibinfo {author} {\bibfnamefont {P.}~\bibnamefont
  {Giannozzi}}, \bibinfo {author} {\bibfnamefont {S.}~\bibnamefont {Baroni}},
  \bibinfo {author} {\bibfnamefont {N.}~\bibnamefont {Bonini}}, \bibinfo
  {author} {\bibfnamefont {M.}~\bibnamefont {Calandra}}, \bibinfo {author}
  {\bibfnamefont {R.}~\bibnamefont {Car}}, \bibinfo {author} {\bibfnamefont
  {C.}~\bibnamefont {Cavazzoni}}, \bibinfo {author} {\bibfnamefont
  {D.}~\bibnamefont {Ceresoli}}, \bibinfo {author} {\bibfnamefont {G.~L.}\
  \bibnamefont {Chiarotti}}, \bibinfo {author} {\bibfnamefont {M.}~\bibnamefont
  {Cococcioni}}, \bibinfo {author} {\bibfnamefont {I.}~\bibnamefont {Dabo}},
  \bibinfo {author} {\bibfnamefont {A.~D.}\ \bibnamefont {Corso}}, \bibinfo
  {author} {\bibfnamefont {S.~d.}\ \bibnamefont {Gironcoli}}, \bibinfo {author}
  {\bibfnamefont {S.}~\bibnamefont {Fabris}}, \bibinfo {author} {\bibfnamefont
  {G.}~\bibnamefont {Fratesi}}, \bibinfo {author} {\bibfnamefont
  {R.}~\bibnamefont {Gebauer}}, \bibinfo {author} {\bibfnamefont
  {U.}~\bibnamefont {Gerstmann}}, \bibinfo {author} {\bibfnamefont
  {C.}~\bibnamefont {Gougoussis}}, \bibinfo {author} {\bibfnamefont
  {A.}~\bibnamefont {Kokalj}}, \bibinfo {author} {\bibfnamefont
  {M.}~\bibnamefont {Lazzeri}}, \bibinfo {author} {\bibfnamefont
  {L.}~\bibnamefont {Martin-Samos}}, \bibinfo {author} {\bibfnamefont
  {N.}~\bibnamefont {Marzari}}, \bibinfo {author} {\bibfnamefont
  {F.}~\bibnamefont {Mauri}}, \bibinfo {author} {\bibfnamefont
  {R.}~\bibnamefont {Mazzarello}}, \bibinfo {author} {\bibfnamefont
  {S.}~\bibnamefont {Paolini}}, \bibinfo {author} {\bibfnamefont
  {A.}~\bibnamefont {Pasquarello}}, \bibinfo {author} {\bibfnamefont
  {L.}~\bibnamefont {Paulatto}}, \bibinfo {author} {\bibfnamefont
  {C.}~\bibnamefont {Sbraccia}}, \bibinfo {author} {\bibfnamefont
  {S.}~\bibnamefont {Scandolo}}, \bibinfo {author} {\bibfnamefont
  {G.}~\bibnamefont {Sclauzero}}, \bibinfo {author} {\bibfnamefont {A.~P.}\
  \bibnamefont {Seitsonen}}, \bibinfo {author} {\bibfnamefont {A.}~\bibnamefont
  {Smogunov}}, \bibinfo {author} {\bibfnamefont {P.}~\bibnamefont {Umari}},\
  and\ \bibinfo {author} {\bibfnamefont {R.~M.}\ \bibnamefont {Wentzcovitch}},\
  }\bibfield  {title} {{\selectlanguage {english}\enquote {\bibinfo {title}
  {{QUANTUM} {ESPRESSO}: a modular and open-source software project for quantum
  simulations of materials},}\ }}\href
  {https://doi.org/10.1088/0953-8984/21/39/395502} {\bibfield  {journal}
  {\bibinfo  {journal} {J. Phys.: Condens. Matter}\ }\textbf {\bibinfo {volume}
  {21}},\ \bibinfo {pages} {395502} (\bibinfo {year} {2009})}\BibitemShut
  {NoStop}%
\bibitem [{\citenamefont {Sholl}\ and\ \citenamefont
  {Steckel}(2009)}]{sholl_density_2009}%
  \BibitemOpen
  \bibfield  {author} {\bibinfo {author} {\bibfnamefont {D.~S.}\ \bibnamefont
  {Sholl}}\ and\ \bibinfo {author} {\bibfnamefont {J.~A.}\ \bibnamefont
  {Steckel}},\ }\href@noop {} {\emph {\bibinfo {title} {Density {Functional}
  {Theory}, {A} {Practical} {Introduction}}}}\ (\bibinfo  {publisher} {Wiley},\
  \bibinfo {address} {Hoboken, NJ},\ \bibinfo {year} {2009})\BibitemShut
  {NoStop}%
\bibitem [{\citenamefont {Cohen}, \citenamefont {Mori-S{\'{a}}nchez},\ and\
  \citenamefont {Yang}(2012)}]{doi:10.1021/cr200107z}%
  \BibitemOpen
  \bibfield  {author} {\bibinfo {author} {\bibfnamefont {A.~J.}\ \bibnamefont
  {Cohen}}, \bibinfo {author} {\bibfnamefont {P.}~\bibnamefont
  {Mori-S{\'{a}}nchez}},\ and\ \bibinfo {author} {\bibfnamefont
  {W.}~\bibnamefont {Yang}},\ }\bibfield  {title} {\enquote {\bibinfo {title}
  {Challenges for density functional theory},}\ }\href
  {https://doi.org/10.1021/cr200107z} {\bibfield  {journal} {\bibinfo
  {journal} {Chem. Rev.}\ }\textbf {\bibinfo {volume} {112}},\ \bibinfo {pages}
  {289--320} (\bibinfo {year} {2012})}\BibitemShut {NoStop}%
\bibitem [{\citenamefont {Golze}, \citenamefont {Dvorak},\ and\ \citenamefont
  {Rinke}(2019)}]{10.3389/fchem.2019.00377}%
  \BibitemOpen
  \bibfield  {author} {\bibinfo {author} {\bibfnamefont {D.}~\bibnamefont
  {Golze}}, \bibinfo {author} {\bibfnamefont {M.}~\bibnamefont {Dvorak}},\ and\
  \bibinfo {author} {\bibfnamefont {P.}~\bibnamefont {Rinke}},\ }\bibfield
  {title} {\enquote {\bibinfo {title} {The gw compendium: a practical guide to
  theoretical photoemission spectroscopy},}\ }\href
  {https://doi.org/10.3389/fchem.2019.00377} {\bibfield  {journal} {\bibinfo
  {journal} {Front. Chem.}\ }\textbf {\bibinfo {volume} {7}} (\bibinfo {year}
  {2019}),\ 10.3389/fchem.2019.00377}\BibitemShut {NoStop}%
\bibitem [{\citenamefont {Gao}\ \emph {et~al.}(2009)\citenamefont {Gao},
  \citenamefont {Xiao}, \citenamefont {Zu}, \citenamefont {Posselt},\ and\
  \citenamefont {Weber}}]{gao_defect-enhanced_2009}%
  \BibitemOpen
  \bibfield  {author} {\bibinfo {author} {\bibfnamefont {F.}~\bibnamefont
  {Gao}}, \bibinfo {author} {\bibfnamefont {H.}~\bibnamefont {Xiao}}, \bibinfo
  {author} {\bibfnamefont {X.}~\bibnamefont {Zu}}, \bibinfo {author}
  {\bibfnamefont {M.}~\bibnamefont {Posselt}},\ and\ \bibinfo {author}
  {\bibfnamefont {W.~J.}\ \bibnamefont {Weber}},\ }\bibfield  {title} {\enquote
  {\bibinfo {title} {Defect-enhanced charge transfer by ion-solid interactions
  in {SiC} using large-scale ab initio molecular dynamics simulations},}\
  }\href {https://doi.org/10.1103/PhysRevLett.103.027405} {\bibfield  {journal}
  {\bibinfo  {journal} {Phys. Rev. Lett.}\ }\textbf {\bibinfo {volume} {103}},\
  \bibinfo {pages} {027405} (\bibinfo {year} {2009})}\BibitemShut {NoStop}%
\bibitem [{\citenamefont {Shchygol}\ \emph {et~al.}(2019)\citenamefont
  {Shchygol}, \citenamefont {Yakovlev}, \citenamefont {Trnka}, \citenamefont
  {van Duin},\ and\ \citenamefont {Verstraelen}}]{Shchygol2019}%
  \BibitemOpen
  \bibfield  {author} {\bibinfo {author} {\bibfnamefont {G.}~\bibnamefont
  {Shchygol}}, \bibinfo {author} {\bibfnamefont {A.}~\bibnamefont {Yakovlev}},
  \bibinfo {author} {\bibfnamefont {T.}~\bibnamefont {Trnka}}, \bibinfo
  {author} {\bibfnamefont {A.~C.~T.}\ \bibnamefont {van Duin}},\ and\ \bibinfo
  {author} {\bibfnamefont {T.}~\bibnamefont {Verstraelen}},\ }\bibfield
  {title} {\enquote {\bibinfo {title} {{ReaxFF Parameter Optimization with
  Monte-Carlo and Evolutionary Algorithms: Guidelines and Insights}},}\ }\href
  {https://doi.org/10.1021/ACS.JCTC.9B00769} {\bibfield  {journal} {\bibinfo
  {journal} {J. Chem. Theory Comput.}\ }\textbf {\bibinfo {volume} {15}},\
  \bibinfo {pages} {6799--6812} (\bibinfo {year} {2019})}\BibitemShut {NoStop}%
\bibitem [{\citenamefont {Angibaud}\ \emph {et~al.}(2011)\citenamefont
  {Angibaud}, \citenamefont {Briquet}, \citenamefont {Philipp}, \citenamefont
  {Wirtz},\ and\ \citenamefont {Kieffer}}]{Angibaud2011}%
  \BibitemOpen
  \bibfield  {author} {\bibinfo {author} {\bibfnamefont {L.}~\bibnamefont
  {Angibaud}}, \bibinfo {author} {\bibfnamefont {L.}~\bibnamefont {Briquet}},
  \bibinfo {author} {\bibfnamefont {P.}~\bibnamefont {Philipp}}, \bibinfo
  {author} {\bibfnamefont {T.}~\bibnamefont {Wirtz}},\ and\ \bibinfo {author}
  {\bibfnamefont {J.}~\bibnamefont {Kieffer}},\ }\bibfield  {title} {\enquote
  {\bibinfo {title} {{Parameter optimization in molecular dynamics simulations
  using a genetic algorithm}},}\ }\href
  {https://doi.org/10.1016/j.nimb.2010.11.024} {\bibfield  {journal} {\bibinfo
  {journal} {Nucl. Instrum. Meth. Phys. Res. B}\ }\textbf {\bibinfo {volume}
  {269}},\ \bibinfo {pages} {1559--1563} (\bibinfo {year} {2011})}\BibitemShut
  {NoStop}%
\bibitem [{\citenamefont {Jaramillo-Botero}, \citenamefont {Naserifar},\ and\
  \citenamefont {Goddard}(2014)}]{Jaramillo-Botero2014}%
  \BibitemOpen
  \bibfield  {author} {\bibinfo {author} {\bibfnamefont {A.}~\bibnamefont
  {Jaramillo-Botero}}, \bibinfo {author} {\bibfnamefont {S.}~\bibnamefont
  {Naserifar}},\ and\ \bibinfo {author} {\bibfnamefont {W.~A.}\ \bibnamefont
  {Goddard}},\ }\bibfield  {title} {\enquote {\bibinfo {title} {{General
  Multiobjective Force Field Optimization Framework, with Application to
  Reactive Force Fields for Silicon Carbide}},}\ }\href
  {https://doi.org/10.1021/ct5001044} {\bibfield  {journal} {\bibinfo
  {journal} {J. Chem. Theory Comput.}\ }\textbf {\bibinfo {volume} {10}},\
  \bibinfo {pages} {1426--1439} (\bibinfo {year} {2014})}\BibitemShut {NoStop}%
\bibitem [{\citenamefont {Dittner}\ \emph {et~al.}(2015)\citenamefont
  {Dittner}, \citenamefont {M{\"{u}}ller}, \citenamefont {Aktulga},\ and\
  \citenamefont {Hartke}}]{Dittner2015}%
  \BibitemOpen
  \bibfield  {author} {\bibinfo {author} {\bibfnamefont {M.}~\bibnamefont
  {Dittner}}, \bibinfo {author} {\bibfnamefont {J.}~\bibnamefont
  {M{\"{u}}ller}}, \bibinfo {author} {\bibfnamefont {H.~M.}\ \bibnamefont
  {Aktulga}},\ and\ \bibinfo {author} {\bibfnamefont {B.}~\bibnamefont
  {Hartke}},\ }\bibfield  {title} {\enquote {\bibinfo {title} {{Efficient
  global optimization of reactive force-field parameters}},}\ }\href
  {https://doi.org/10.1002/JCC.23966} {\bibfield  {journal} {\bibinfo
  {journal} {J. Comput. Chem.}\ }\textbf {\bibinfo {volume} {36}},\ \bibinfo
  {pages} {1550--1561} (\bibinfo {year} {2015})}\BibitemShut {NoStop}%
\bibitem [{\citenamefont {Larsson}, \citenamefont {van Duin},\ and\
  \citenamefont {Hartke}(2013)}]{Larsson2013}%
  \BibitemOpen
  \bibfield  {author} {\bibinfo {author} {\bibfnamefont {H.~R.}\ \bibnamefont
  {Larsson}}, \bibinfo {author} {\bibfnamefont {A.~C.~T.}\ \bibnamefont {van
  Duin}},\ and\ \bibinfo {author} {\bibfnamefont {B.}~\bibnamefont {Hartke}},\
  }\bibfield  {title} {\enquote {\bibinfo {title} {{Global optimization of
  parameters in the reactive force field ReaxFF for SiOH}},}\ }\href
  {https://doi.org/10.1002/JCC.23382} {\bibfield  {journal} {\bibinfo
  {journal} {J. Comput. Chem.}\ }\textbf {\bibinfo {volume} {34}},\ \bibinfo
  {pages} {2178--2189} (\bibinfo {year} {2013})}\BibitemShut {NoStop}%
\bibitem [{\citenamefont {Larentzos}\ \emph {et~al.}(2015)\citenamefont
  {Larentzos}, \citenamefont {Rice}, \citenamefont {Byrd}, \citenamefont
  {Weingarten},\ and\ \citenamefont {Lill}}]{Larentzos2015}%
  \BibitemOpen
  \bibfield  {author} {\bibinfo {author} {\bibfnamefont {J.~P.}\ \bibnamefont
  {Larentzos}}, \bibinfo {author} {\bibfnamefont {B.~M.}\ \bibnamefont {Rice}},
  \bibinfo {author} {\bibfnamefont {E.~F.~C.}\ \bibnamefont {Byrd}}, \bibinfo
  {author} {\bibfnamefont {N.~S.}\ \bibnamefont {Weingarten}},\ and\ \bibinfo
  {author} {\bibfnamefont {J.~V.}\ \bibnamefont {Lill}},\ }\bibfield  {title}
  {\enquote {\bibinfo {title} {{Parameterizing Complex Reactive Force Fields
  Using Multiple Objective Evolutionary Strategies (MOES). Part 1: reaxff
  models for cyclotrimethylene trinitramine (rdx) and
  1,1-diamino-2,2-dinitroethene (fox-7)}},}\ }\href
  {https://doi.org/10.1021/CT500788C} {\bibfield  {journal} {\bibinfo
  {journal} {J. Chem. Theory Comput.}\ }\textbf {\bibinfo {volume} {11}},\
  \bibinfo {pages} {381--391} (\bibinfo {year} {2015})}\BibitemShut {NoStop}%
\bibitem [{\citenamefont {Furman}\ \emph {et~al.}(2018)\citenamefont {Furman},
  \citenamefont {Carmeli}, \citenamefont {Zeiri},\ and\ \citenamefont
  {Kosloff}}]{Furman2018}%
  \BibitemOpen
  \bibfield  {author} {\bibinfo {author} {\bibfnamefont {D.}~\bibnamefont
  {Furman}}, \bibinfo {author} {\bibfnamefont {B.}~\bibnamefont {Carmeli}},
  \bibinfo {author} {\bibfnamefont {Y.}~\bibnamefont {Zeiri}},\ and\ \bibinfo
  {author} {\bibfnamefont {R.}~\bibnamefont {Kosloff}},\ }\bibfield  {title}
  {\enquote {\bibinfo {title} {{Enhanced Particle Swarm Optimization Algorithm:
  Efficient Training of ReaxFF Reactive Force Fields}},}\ }\href
  {https://doi.org/10.1021/ACS.JCTC.7B01272} {\bibfield  {journal} {\bibinfo
  {journal} {J. Chem. Theory Comput.}\ }\textbf {\bibinfo {volume} {14}},\
  \bibinfo {pages} {3100--3112} (\bibinfo {year} {2018})}\BibitemShut {NoStop}%
\bibitem [{\citenamefont {Nakata}\ and\ \citenamefont
  {Bai}(2019)}]{Nakata2019}%
  \BibitemOpen
  \bibfield  {author} {\bibinfo {author} {\bibfnamefont {H.}~\bibnamefont
  {Nakata}}\ and\ \bibinfo {author} {\bibfnamefont {S.}~\bibnamefont {Bai}},\
  }\bibfield  {title} {\enquote {\bibinfo {title} {{Development of a new
  parameter optimization scheme for a reactive force field based on a machine
  learning approach}},}\ }\href {https://doi.org/10.1002/JCC.25841} {\bibfield
  {journal} {\bibinfo  {journal} {J. Comput. Chem.}\ }\textbf {\bibinfo
  {volume} {40}},\ \bibinfo {pages} {2000--2012} (\bibinfo {year}
  {2019})}\BibitemShut {NoStop}%
\bibitem [{\citenamefont {Sengul}\ \emph {et~al.}(2021)\citenamefont {Sengul},
  \citenamefont {Song}, \citenamefont {Nayir}, \citenamefont {Gao},
  \citenamefont {Hung}, \citenamefont {Dasgupta},\ and\ \citenamefont {van
  Duin}}]{Sengul2021}%
  \BibitemOpen
  \bibfield  {author} {\bibinfo {author} {\bibfnamefont {M.~Y.}\ \bibnamefont
  {Sengul}}, \bibinfo {author} {\bibfnamefont {Y.}~\bibnamefont {Song}},
  \bibinfo {author} {\bibfnamefont {N.}~\bibnamefont {Nayir}}, \bibinfo
  {author} {\bibfnamefont {Y.}~\bibnamefont {Gao}}, \bibinfo {author}
  {\bibfnamefont {Y.}~\bibnamefont {Hung}}, \bibinfo {author} {\bibfnamefont
  {T.}~\bibnamefont {Dasgupta}},\ and\ \bibinfo {author} {\bibfnamefont
  {A.~C.~T.}\ \bibnamefont {van Duin}},\ }\bibfield  {title} {\enquote
  {\bibinfo {title} {{INDEEDopt}: a deep learning-based {ReaxFF}
  parameterization framework},}\ }\href
  {https://doi.org/10.1038/s41524-021-00534-4} {\bibfield  {journal} {\bibinfo
  {journal} {npj Comput. Mater.}\ }\textbf {\bibinfo {volume} {7}},\ \bibinfo
  {pages} {1--9} (\bibinfo {year} {2021})}\BibitemShut {NoStop}%
\bibitem [{\citenamefont {Becker}, \citenamefont {Trautt},\ and\ \citenamefont
  {Hale}(2010)}]{becker_nist_2010}%
  \BibitemOpen
  \bibfield  {author} {\bibinfo {author} {\bibfnamefont {C.}~\bibnamefont
  {Becker}}, \bibinfo {author} {\bibfnamefont {Z.}~\bibnamefont {Trautt}},\
  and\ \bibinfo {author} {\bibfnamefont {L.}~\bibnamefont {Hale}},\ }\bibfield
  {title} {{\selectlanguage {english}\enquote {\bibinfo {title} {{NIST}
  interatomic potentials repository},}\ }}\href {https://doi.org/10.18434/M37}
  {10.18434/M37} (\bibinfo {year} {2010})\BibitemShut {NoStop}%
\bibitem [{\citenamefont {Tadmor}\ \emph {et~al.}(2011)\citenamefont {Tadmor},
  \citenamefont {Elliott}, \citenamefont {Sethna}, \citenamefont {Miller},\
  and\ \citenamefont {Becker}}]{noauthor_openkim_nodate}%
  \BibitemOpen
  \bibfield  {author} {\bibinfo {author} {\bibfnamefont {E.~B.}\ \bibnamefont
  {Tadmor}}, \bibinfo {author} {\bibfnamefont {R.~S.}\ \bibnamefont {Elliott}},
  \bibinfo {author} {\bibfnamefont {J.~P.}\ \bibnamefont {Sethna}}, \bibinfo
  {author} {\bibfnamefont {R.~E.}\ \bibnamefont {Miller}},\ and\ \bibinfo
  {author} {\bibfnamefont {C.~A.}\ \bibnamefont {Becker}},\ }\bibfield  {title}
  {\enquote {\bibinfo {title} {{OpenKIM} - {Knowledgebase} of {Interatomic}
  {Models}},}\ }\href {https://doi.org/10.1007/s11837-011-0102-6} {\bibfield
  {journal} {\bibinfo  {journal} {{JOM}}\ }\textbf {\bibinfo {volume} {63}},\
  \bibinfo {pages} {17--17} (\bibinfo {year} {2011})},\ \bibinfo {note}
  {\url{https://openkim.org/}}\BibitemShut {NoStop}%
\bibitem [{\citenamefont {LAMMPS}(2023)}]{noauthor_lammps_nodate}%
  \BibitemOpen
  \bibfield  {author} {\bibinfo {author} {\bibnamefont {LAMMPS}},\ }\href
  {https://www.lammps.org/bench.html} {\enquote {\bibinfo {title} {{LAMMPS}
  {Benchmarks}},}\ }\bibinfo {howpublished} {Online} (\bibinfo {year} {2023}),\
  \bibinfo {note} {\url{https://www.lammps.org/bench.html}}\BibitemShut
  {NoStop}%
\bibitem [{\citenamefont {Schneider}\ \emph {et~al.}(2017)\citenamefont
  {Schneider}, \citenamefont {Hamaekers}, \citenamefont {Chill}, \citenamefont
  {Smidstrup}, \citenamefont {Bulin}, \citenamefont {Thesen}, \citenamefont
  {Blom},\ and\ \citenamefont {Stokbro}}]{schneider_atk-forcefield:_2017}%
  \BibitemOpen
  \bibfield  {author} {\bibinfo {author} {\bibfnamefont {J.}~\bibnamefont
  {Schneider}}, \bibinfo {author} {\bibfnamefont {J.}~\bibnamefont
  {Hamaekers}}, \bibinfo {author} {\bibfnamefont {S.~T.}\ \bibnamefont
  {Chill}}, \bibinfo {author} {\bibfnamefont {S.}~\bibnamefont {Smidstrup}},
  \bibinfo {author} {\bibfnamefont {J.}~\bibnamefont {Bulin}}, \bibinfo
  {author} {\bibfnamefont {R.}~\bibnamefont {Thesen}}, \bibinfo {author}
  {\bibfnamefont {A.}~\bibnamefont {Blom}},\ and\ \bibinfo {author}
  {\bibfnamefont {K.}~\bibnamefont {Stokbro}},\ }\bibfield  {title}
  {{\selectlanguage {english}\enquote {\bibinfo {title} {{ATK}-{ForceField}: a
  new generation molecular dynamics software package},}\ }}\href
  {https://doi.org/10.1088/1361-651X/aa8ff0} {\bibfield  {journal} {\bibinfo
  {journal} {Modelling Simul. Mater. Sci. Eng.}\ }\textbf {\bibinfo {volume}
  {25}},\ \bibinfo {pages} {085007} (\bibinfo {year} {2017})}\BibitemShut
  {NoStop}%
\bibitem [{\citenamefont {Lennard-Jones}(1931)}]{Lennard1931}%
  \BibitemOpen
  \bibfield  {author} {\bibinfo {author} {\bibfnamefont {J.~E.}\ \bibnamefont
  {Lennard-Jones}},\ }\bibfield  {title} {\enquote {\bibinfo {title}
  {Cohesion},}\ }\href {https://doi.org/10.1088/0959-5309/43/5/301} {\bibfield
  {journal} {\bibinfo  {journal} {Proc. Phys. Soc.}\ }\textbf {\bibinfo
  {volume} {43}},\ \bibinfo {pages} {461--482} (\bibinfo {year}
  {1931})}\BibitemShut {NoStop}%
\bibitem [{\citenamefont {Daw}\ and\ \citenamefont {Baskes}(1984)}]{Daw1984}%
  \BibitemOpen
  \bibfield  {author} {\bibinfo {author} {\bibfnamefont {M.~S.}\ \bibnamefont
  {Daw}}\ and\ \bibinfo {author} {\bibfnamefont {M.~I.}\ \bibnamefont
  {Baskes}},\ }\bibfield  {title} {\enquote {\bibinfo {title} {Embedded-atom
  method: derivation and application to impurities, surfaces, and other defects
  in metals},}\ }\href {https://doi.org/10.1103/PhysRevB.29.6443} {\bibfield
  {journal} {\bibinfo  {journal} {Phys. Rev. B}\ }\textbf {\bibinfo {volume}
  {29}},\ \bibinfo {pages} {6443--6453} (\bibinfo {year} {1984})}\BibitemShut
  {NoStop}%
\bibitem [{\citenamefont {Stillinger}\ and\ \citenamefont
  {Weber}(1985)}]{SW-orig}%
  \BibitemOpen
  \bibfield  {author} {\bibinfo {author} {\bibfnamefont {F.~H.}\ \bibnamefont
  {Stillinger}}\ and\ \bibinfo {author} {\bibfnamefont {T.~A.}\ \bibnamefont
  {Weber}},\ }\bibfield  {title} {\enquote {\bibinfo {title} {Computer
  simulation of local order in condensed phases of silicon},}\ }\href
  {https://doi.org/10.1103/PhysRevB.31.5262} {\bibfield  {journal} {\bibinfo
  {journal} {Phys. Rev. B}\ }\textbf {\bibinfo {volume} {31}},\ \bibinfo
  {pages} {5262--5271} (\bibinfo {year} {1985})}\BibitemShut {NoStop}%
\bibitem [{\citenamefont {Tersoff}(1988)}]{Tersoff-orig}%
  \BibitemOpen
  \bibfield  {author} {\bibinfo {author} {\bibfnamefont {J.}~\bibnamefont
  {Tersoff}},\ }\bibfield  {title} {\enquote {\bibinfo {title} {New empirical
  approach for the structure and energy of covalent systems},}\ }\href
  {https://doi.org/10.1103/PhysRevB.37.6991} {\bibfield  {journal} {\bibinfo
  {journal} {Phys. Rev. B}\ }\textbf {\bibinfo {volume} {37}},\ \bibinfo
  {pages} {6991--7000} (\bibinfo {year} {1988})}\BibitemShut {NoStop}%
\bibitem [{\citenamefont {van Beest}, \citenamefont {Kramer},\ and\
  \citenamefont {van Santen}(1990)}]{vanBeest1990}%
  \BibitemOpen
  \bibfield  {author} {\bibinfo {author} {\bibfnamefont {B.~W.~H.}\
  \bibnamefont {van Beest}}, \bibinfo {author} {\bibfnamefont {G.~J.}\
  \bibnamefont {Kramer}},\ and\ \bibinfo {author} {\bibfnamefont {R.~A.}\
  \bibnamefont {van Santen}},\ }\bibfield  {title} {\enquote {\bibinfo {title}
  {Force fields for silicas and aluminophosphates based on ab initio
  calculations},}\ }\href {https://doi.org/10.1103/physrevlett.64.1955}
  {\bibfield  {journal} {\bibinfo  {journal} {Phys. Rev. Lett.}\ }\textbf
  {\bibinfo {volume} {64}},\ \bibinfo {pages} {1955--1958} (\bibinfo {year}
  {1990})}\BibitemShut {NoStop}%
\bibitem [{\citenamefont {Brooks}\ \emph {et~al.}(2009)\citenamefont {Brooks},
  \citenamefont {Brooks~III}, \citenamefont {Mackerell~Jr.}, \citenamefont
  {Nilsson}, \citenamefont {Petrella}, \citenamefont {Roux}, \citenamefont
  {Won}, \citenamefont {Archontis}, \citenamefont {Bartels}, \citenamefont
  {Boresch}, \citenamefont {Caflisch}, \citenamefont {Caves}, \citenamefont
  {Cui}, \citenamefont {Dinner}, \citenamefont {Feig}, \citenamefont {Fischer},
  \citenamefont {Gao}, \citenamefont {Hodoscek}, \citenamefont {Im},
  \citenamefont {Kuczera}, \citenamefont {Lazaridis}, \citenamefont {Ma},
  \citenamefont {Ovchinnikov}, \citenamefont {Paci}, \citenamefont {Pastor},
  \citenamefont {Post}, \citenamefont {Pu}, \citenamefont {Schaefer},
  \citenamefont {Tidor}, \citenamefont {Venable}, \citenamefont {Woodcock},
  \citenamefont {Wu}, \citenamefont {Yang}, \citenamefont {York},\ and\
  \citenamefont {Karplus}}]{brooks_charmm_2009}%
  \BibitemOpen
  \bibfield  {author} {\bibinfo {author} {\bibfnamefont {B.~R.}\ \bibnamefont
  {Brooks}}, \bibinfo {author} {\bibfnamefont {C.~L.}\ \bibnamefont
  {Brooks~III}}, \bibinfo {author} {\bibfnamefont {A.~D.}\ \bibnamefont
  {Mackerell~Jr.}}, \bibinfo {author} {\bibfnamefont {L.}~\bibnamefont
  {Nilsson}}, \bibinfo {author} {\bibfnamefont {R.~J.}\ \bibnamefont
  {Petrella}}, \bibinfo {author} {\bibfnamefont {B.}~\bibnamefont {Roux}},
  \bibinfo {author} {\bibfnamefont {Y.}~\bibnamefont {Won}}, \bibinfo {author}
  {\bibfnamefont {G.}~\bibnamefont {Archontis}}, \bibinfo {author}
  {\bibfnamefont {C.}~\bibnamefont {Bartels}}, \bibinfo {author} {\bibfnamefont
  {S.}~\bibnamefont {Boresch}}, \bibinfo {author} {\bibfnamefont
  {A.}~\bibnamefont {Caflisch}}, \bibinfo {author} {\bibfnamefont
  {L.}~\bibnamefont {Caves}}, \bibinfo {author} {\bibfnamefont
  {Q.}~\bibnamefont {Cui}}, \bibinfo {author} {\bibfnamefont {A.~R.}\
  \bibnamefont {Dinner}}, \bibinfo {author} {\bibfnamefont {M.}~\bibnamefont
  {Feig}}, \bibinfo {author} {\bibfnamefont {S.}~\bibnamefont {Fischer}},
  \bibinfo {author} {\bibfnamefont {J.}~\bibnamefont {Gao}}, \bibinfo {author}
  {\bibfnamefont {M.}~\bibnamefont {Hodoscek}}, \bibinfo {author}
  {\bibfnamefont {W.}~\bibnamefont {Im}}, \bibinfo {author} {\bibfnamefont
  {K.}~\bibnamefont {Kuczera}}, \bibinfo {author} {\bibfnamefont
  {T.}~\bibnamefont {Lazaridis}}, \bibinfo {author} {\bibfnamefont
  {J.}~\bibnamefont {Ma}}, \bibinfo {author} {\bibfnamefont {V.}~\bibnamefont
  {Ovchinnikov}}, \bibinfo {author} {\bibfnamefont {E.}~\bibnamefont {Paci}},
  \bibinfo {author} {\bibfnamefont {R.~W.}\ \bibnamefont {Pastor}}, \bibinfo
  {author} {\bibfnamefont {C.~B.}\ \bibnamefont {Post}}, \bibinfo {author}
  {\bibfnamefont {J.~Z.}\ \bibnamefont {Pu}}, \bibinfo {author} {\bibfnamefont
  {M.}~\bibnamefont {Schaefer}}, \bibinfo {author} {\bibfnamefont
  {B.}~\bibnamefont {Tidor}}, \bibinfo {author} {\bibfnamefont {R.~M.}\
  \bibnamefont {Venable}}, \bibinfo {author} {\bibfnamefont {H.~L.}\
  \bibnamefont {Woodcock}}, \bibinfo {author} {\bibfnamefont {X.}~\bibnamefont
  {Wu}}, \bibinfo {author} {\bibfnamefont {W.}~\bibnamefont {Yang}}, \bibinfo
  {author} {\bibfnamefont {D.~M.}\ \bibnamefont {York}},\ and\ \bibinfo
  {author} {\bibfnamefont {M.}~\bibnamefont {Karplus}},\ }\bibfield  {title}
  {{\selectlanguage {english}\enquote {\bibinfo {title} {{CHARMM}: {The}
  biomolecular simulation program},}\ }}\href
  {https://doi.org/10.1002/jcc.21287} {\bibfield  {journal} {\bibinfo
  {journal} {J. Comput. Chem.}\ }\textbf {\bibinfo {volume} {30}},\ \bibinfo
  {pages} {1545--1614} (\bibinfo {year} {2009})},\ \bibinfo {note} {\_eprint:
  https://onlinelibrary.wiley.com/doi/pdf/10.1002/jcc.21287}\BibitemShut
  {NoStop}%
\bibitem [{\citenamefont {Baskes}(1992)}]{baskes_modified_1992}%
  \BibitemOpen
  \bibfield  {author} {\bibinfo {author} {\bibfnamefont {M.~I.}\ \bibnamefont
  {Baskes}},\ }\bibfield  {title} {\enquote {\bibinfo {title} {Modified
  embedded-atom potentials for cubic materials and impurities},}\ }\href
  {https://doi.org/10.1103/PhysRevB.46.2727} {\bibfield  {journal} {\bibinfo
  {journal} {Phys. Rev. B}\ }\textbf {\bibinfo {volume} {46}},\ \bibinfo
  {pages} {2727--2742} (\bibinfo {year} {1992})}\BibitemShut {NoStop}%
\bibitem [{\citenamefont {Yu}, \citenamefont {Sinnott},\ and\ \citenamefont
  {Phillpot}(2007)}]{Yu2007}%
  \BibitemOpen
  \bibfield  {author} {\bibinfo {author} {\bibfnamefont {J.}~\bibnamefont
  {Yu}}, \bibinfo {author} {\bibfnamefont {S.~B.}\ \bibnamefont {Sinnott}},\
  and\ \bibinfo {author} {\bibfnamefont {S.~R.}\ \bibnamefont {Phillpot}},\
  }\bibfield  {title} {\enquote {\bibinfo {title} {{Charge optimized many-body
  potential for the Si/SiO2 system}},}\ }\href
  {https://doi.org/10.1103/PhysRevB.75.085311} {\bibfield  {journal} {\bibinfo
  {journal} {Phys. Rev. B}\ }\textbf {\bibinfo {volume} {75}},\ \bibinfo
  {pages} {1--13} (\bibinfo {year} {2007})}\BibitemShut {NoStop}%
\bibitem [{\citenamefont {Van~Duin}\ \emph {et~al.}(2001)\citenamefont
  {Van~Duin}, \citenamefont {Dasgupta}, \citenamefont {Lorant},\ and\
  \citenamefont {Goddard}}]{VanDuin2001}%
  \BibitemOpen
  \bibfield  {author} {\bibinfo {author} {\bibfnamefont {A.~C.~T.}\
  \bibnamefont {Van~Duin}}, \bibinfo {author} {\bibfnamefont {S.}~\bibnamefont
  {Dasgupta}}, \bibinfo {author} {\bibfnamefont {F.}~\bibnamefont {Lorant}},\
  and\ \bibinfo {author} {\bibfnamefont {W.~A.}\ \bibnamefont {Goddard}},\
  }\bibfield  {title} {\enquote {\bibinfo {title} {{ReaxFF: A reactive force
  field for hydrocarbons}},}\ }\href {https://doi.org/10.1021/jp004368u}
  {\bibfield  {journal} {\bibinfo  {journal} {J. Phys. Chem. A}\ }\textbf
  {\bibinfo {volume} {105}},\ \bibinfo {pages} {9396--9409} (\bibinfo {year}
  {2001})},\ \Eprint {https://arxiv.org/abs/1011.1669v3} {arXiv:1011.1669v3}
  \BibitemShut {NoStop}%
\bibitem [{\citenamefont {Huang}\ and\ \citenamefont
  {Kieffer}(2003)}]{Huang2003}%
  \BibitemOpen
  \bibfield  {author} {\bibinfo {author} {\bibfnamefont {L.}~\bibnamefont
  {Huang}}\ and\ \bibinfo {author} {\bibfnamefont {J.}~\bibnamefont
  {Kieffer}},\ }\bibfield  {title} {\enquote {\bibinfo {title} {{Molecular
  dynamics study of cristobalite silica using a charge transfer three-body
  potential: Phase transformation and structural disorder}},}\ }\href
  {https://doi.org/10.1063/1.1529684} {\bibfield  {journal} {\bibinfo
  {journal} {J. Chem. Phys.}\ }\textbf {\bibinfo {volume} {118}},\ \bibinfo
  {pages} {1487--1498} (\bibinfo {year} {2003})}\BibitemShut {NoStop}%
\bibitem [{\citenamefont {Behler}(2016)}]{Behler2016}%
  \BibitemOpen
  \bibfield  {author} {\bibinfo {author} {\bibfnamefont {J.}~\bibnamefont
  {Behler}},\ }\bibfield  {title} {\enquote {\bibinfo {title} {{Perspective:
  Machine learning potentials for atomistic simulations}},}\ }\href
  {https://doi.org/10.1063/1.4966192} {\bibfield  {journal} {\bibinfo
  {journal} {J. Chem. Phys.}\ }\textbf {\bibinfo {volume} {145}},\ \bibinfo
  {pages} {170901} (\bibinfo {year} {2016})},\ \bibinfo {note} {publisher:
  American Institute of Physics}\BibitemShut {NoStop}%
\bibitem [{\citenamefont {Mishin}(2021)}]{mishin_machine-learning_2021}%
  \BibitemOpen
  \bibfield  {author} {\bibinfo {author} {\bibfnamefont {Y.}~\bibnamefont
  {Mishin}},\ }\bibfield  {title} {{\selectlanguage {english}\enquote {\bibinfo
  {title} {Machine-learning interatomic potentials for materials science},}\
  }}\href {https://doi.org/10.1016/j.actamat.2021.116980} {\bibfield  {journal}
  {\bibinfo  {journal} {Acta Mater.}\ }\textbf {\bibinfo {volume} {214}},\
  \bibinfo {pages} {116980} (\bibinfo {year} {2021})}\BibitemShut {NoStop}%
\bibitem [{\citenamefont {Jacobsen}, \citenamefont {Jorgensen},\ and\
  \citenamefont {Hammer}(2018)}]{Jacobsen2018}%
  \BibitemOpen
  \bibfield  {author} {\bibinfo {author} {\bibfnamefont {T.~L.}\ \bibnamefont
  {Jacobsen}}, \bibinfo {author} {\bibfnamefont {M.~S.}\ \bibnamefont
  {Jorgensen}},\ and\ \bibinfo {author} {\bibfnamefont {B.}~\bibnamefont
  {Hammer}},\ }\bibfield  {title} {\enquote {\bibinfo {title} {On-the-fly
  machine learning of atomic potential in density functional theory structure
  optimization},}\ }\href {https://doi.org/10.1103/PhysRevLett.120.026102}
  {\bibfield  {journal} {\bibinfo  {journal} {Phys. Rev. Lett.}\ }\textbf
  {\bibinfo {volume} {120}},\ \bibinfo {pages} {026102} (\bibinfo {year}
  {2018})},\ \bibinfo {note} {publisher: American Physical Society}\BibitemShut
  {NoStop}%
\bibitem [{\citenamefont {Vandermause}\ \emph {et~al.}(2020)\citenamefont
  {Vandermause}, \citenamefont {Torrisi}, \citenamefont {Batzner},
  \citenamefont {Xie}, \citenamefont {Sun}, \citenamefont {Kolpak},\ and\
  \citenamefont {Kozinsky}}]{Vandermause2020}%
  \BibitemOpen
  \bibfield  {author} {\bibinfo {author} {\bibfnamefont {J.}~\bibnamefont
  {Vandermause}}, \bibinfo {author} {\bibfnamefont {S.~B.}\ \bibnamefont
  {Torrisi}}, \bibinfo {author} {\bibfnamefont {S.}~\bibnamefont {Batzner}},
  \bibinfo {author} {\bibfnamefont {Y.}~\bibnamefont {Xie}}, \bibinfo {author}
  {\bibfnamefont {L.}~\bibnamefont {Sun}}, \bibinfo {author} {\bibfnamefont
  {A.~M.}\ \bibnamefont {Kolpak}},\ and\ \bibinfo {author} {\bibfnamefont
  {B.}~\bibnamefont {Kozinsky}},\ }\bibfield  {title} {\enquote {\bibinfo
  {title} {On-the-fly active learning of interpretable bayesian force fields
  for atomistic rare events},}\ }\href
  {https://doi.org/10.1038/s41524-020-0283-z} {\bibfield  {journal} {\bibinfo
  {journal} {npj Comput. Mater.}\ }\textbf {\bibinfo {volume} {6}} (\bibinfo
  {year} {2020}),\ 10.1038/s41524-020-0283-z}\BibitemShut {NoStop}%
\bibitem [{\citenamefont {Bartok}, \citenamefont {Kondor},\ and\ \citenamefont
  {Csanyi}(2010)}]{Bartok2010}%
  \BibitemOpen
  \bibfield  {author} {\bibinfo {author} {\bibfnamefont {A.~P.}\ \bibnamefont
  {Bartok}}, \bibinfo {author} {\bibfnamefont {R.}~\bibnamefont {Kondor}},\
  and\ \bibinfo {author} {\bibfnamefont {G.}~\bibnamefont {Csanyi}},\
  }\bibfield  {title} {\enquote {\bibinfo {title} {Gaussian approximation
  potentials: the accuracy of quantum mechanics, without the electrons},}\
  }\href {https://doi.org/10.1103/PhysRevLett.104.136403} {\bibfield  {journal}
  {\bibinfo  {journal} {Phys. Rev. Lett.}\ }\textbf {\bibinfo {volume} {104}},\
  \bibinfo {pages} {136403} (\bibinfo {year} {2010})}\BibitemShut {NoStop}%
\bibitem [{\citenamefont {Byggm\"astar}\ \emph {et~al.}(2019)\citenamefont
  {Byggm\"astar}, \citenamefont {Hamedani}, \citenamefont {Nordlund},\ and\
  \citenamefont {Djurabekova}}]{Byggmastar2019}%
  \BibitemOpen
  \bibfield  {author} {\bibinfo {author} {\bibfnamefont {J.}~\bibnamefont
  {Byggm\"astar}}, \bibinfo {author} {\bibfnamefont {A.}~\bibnamefont
  {Hamedani}}, \bibinfo {author} {\bibfnamefont {K.}~\bibnamefont {Nordlund}},\
  and\ \bibinfo {author} {\bibfnamefont {F.}~\bibnamefont {Djurabekova}},\
  }\bibfield  {title} {\enquote {\bibinfo {title} {Machine-learning interatomic
  potential for radiation damage and defects in tungsten},}\ }\href
  {https://doi.org/10.1103/PhysRevB.100.144105} {\bibfield  {journal} {\bibinfo
   {journal} {Phys. Rev. B}\ }\textbf {\bibinfo {volume} {100}},\ \bibinfo
  {pages} {144105} (\bibinfo {year} {2019})}\BibitemShut {NoStop}%
\bibitem [{\citenamefont {Byggmastar}, \citenamefont {Nordlund},\ and\
  \citenamefont {Djurabekova}(2021)}]{Byggmastar2021}%
  \BibitemOpen
  \bibfield  {author} {\bibinfo {author} {\bibfnamefont {J.}~\bibnamefont
  {Byggmastar}}, \bibinfo {author} {\bibfnamefont {K.}~\bibnamefont
  {Nordlund}},\ and\ \bibinfo {author} {\bibfnamefont {F.}~\bibnamefont
  {Djurabekova}},\ }\bibfield  {title} {\enquote {\bibinfo {title} {{Modeling
  refractory high-entropy alloys with efficient machine-learned interatomic
  potentials: Defects and segregation}},}\ }\href
  {https://doi.org/10.1103/PhysRevB.104.104101} {\bibfield  {journal} {\bibinfo
   {journal} {Phys. Rev. B}\ }\textbf {\bibinfo {volume} {104}},\ \bibinfo
  {pages} {104101} (\bibinfo {year} {2021})}\BibitemShut {NoStop}%
\bibitem [{\citenamefont {Ziegler}\ and\ \citenamefont {Biersack}(1985)}]{ZBL}%
  \BibitemOpen
  \bibfield  {author} {\bibinfo {author} {\bibfnamefont {J.~F.}\ \bibnamefont
  {Ziegler}}\ and\ \bibinfo {author} {\bibfnamefont {J.~P.}\ \bibnamefont
  {Biersack}},\ }\bibfield  {title} {\enquote {\bibinfo {title} {The stopping
  and range of ions in matter},}\ \ }(\bibinfo  {publisher} {Springer {US}},\
  \bibinfo {address} {Boston, MA},\ \bibinfo {year} {1985})\ pp.\ \bibinfo
  {pages} {93--129}\BibitemShut {NoStop}%
\bibitem [{\citenamefont {Nordlund}, \citenamefont {Runeberg},\ and\
  \citenamefont {Sundholm}(1997)}]{Nordlund1997}%
  \BibitemOpen
  \bibfield  {author} {\bibinfo {author} {\bibfnamefont {K.}~\bibnamefont
  {Nordlund}}, \bibinfo {author} {\bibfnamefont {N.}~\bibnamefont {Runeberg}},\
  and\ \bibinfo {author} {\bibfnamefont {D.}~\bibnamefont {Sundholm}},\
  }\bibfield  {title} {\enquote {\bibinfo {title} {Repulsive interatomic
  potentials calculated using hartree-fock and density-functional theory
  methods},}\ }\href {https://doi.org/10.1016/s0168-583x(97)00447-3} {\bibfield
   {journal} {\bibinfo  {journal} {Nucl. Instr. Meth. Phys. Res. B}\ }\textbf
  {\bibinfo {volume} {132}},\ \bibinfo {pages} {45--54} (\bibinfo {year}
  {1997})}\BibitemShut {NoStop}%
\bibitem [{\citenamefont {Nordlund}, \citenamefont {Keinonen},\ and\
  \citenamefont {Mattila}(1996)}]{Nordlund1996}%
  \BibitemOpen
  \bibfield  {author} {\bibinfo {author} {\bibfnamefont {K.}~\bibnamefont
  {Nordlund}}, \bibinfo {author} {\bibfnamefont {J.}~\bibnamefont {Keinonen}},\
  and\ \bibinfo {author} {\bibfnamefont {T.}~\bibnamefont {Mattila}},\
  }\bibfield  {title} {\enquote {\bibinfo {title} {Formation of ion irradiation
  induced small-scale defects on graphite surfaces},}\ }\href
  {https://doi.org/10.1103/physrevlett.77.699} {\bibfield  {journal} {\bibinfo
  {journal} {Phys. Rev. Lett.}\ }\textbf {\bibinfo {volume} {77}},\ \bibinfo
  {pages} {699--702} (\bibinfo {year} {1996})}\BibitemShut {NoStop}%
\bibitem [{\citenamefont {Das}, \citenamefont {Freund},\ and\ \citenamefont
  {Johnson}(2015)}]{das_mechanisms_2015}%
  \BibitemOpen
  \bibfield  {author} {\bibinfo {author} {\bibfnamefont {K.}~\bibnamefont
  {Das}}, \bibinfo {author} {\bibfnamefont {J.~B.}\ \bibnamefont {Freund}},\
  and\ \bibinfo {author} {\bibfnamefont {H.~T.}\ \bibnamefont {Johnson}},\
  }\bibfield  {title} {\enquote {\bibinfo {title} {Mechanisms of material
  removal and mass transport in focused ion beam nanopore formation},}\ }\href
  {https://doi.org/10.1063/1.4913449} {\bibfield  {journal} {\bibinfo
  {journal} {J. Appl. Phys.}\ }\textbf {\bibinfo {volume} {117}},\ \bibinfo
  {pages} {085304} (\bibinfo {year} {2015})}\BibitemShut {NoStop}%
\bibitem [{\citenamefont {Das}, \citenamefont {Johnson},\ and\ \citenamefont
  {Freund}(2015)}]{das_atomic-scale_2015}%
  \BibitemOpen
  \bibfield  {author} {\bibinfo {author} {\bibfnamefont {K.}~\bibnamefont
  {Das}}, \bibinfo {author} {\bibfnamefont {H.~T.}\ \bibnamefont {Johnson}},\
  and\ \bibinfo {author} {\bibfnamefont {J.~B.}\ \bibnamefont {Freund}},\
  }\bibfield  {title} {\enquote {\bibinfo {title} {Atomic-scale thermocapillary
  flow in focused ion beam milling},}\ }\href
  {https://doi.org/10.1063/1.4919782} {\bibfield  {journal} {\bibinfo
  {journal} {Phys. Fluids}\ }\textbf {\bibinfo {volume} {27}},\ \bibinfo
  {pages} {052003} (\bibinfo {year} {2015})}\BibitemShut {NoStop}%
\bibitem [{\citenamefont {Fridlund}\ \emph {et~al.}(2020)\citenamefont
  {Fridlund}, \citenamefont {Lopez-Cazalilla}, \citenamefont {Nordlund},\ and\
  \citenamefont {Djurabekova}}]{Fridlund2021}%
  \BibitemOpen
  \bibfield  {author} {\bibinfo {author} {\bibfnamefont {C.}~\bibnamefont
  {Fridlund}}, \bibinfo {author} {\bibfnamefont {A.}~\bibnamefont
  {Lopez-Cazalilla}}, \bibinfo {author} {\bibfnamefont {K.}~\bibnamefont
  {Nordlund}},\ and\ \bibinfo {author} {\bibfnamefont {F.}~\bibnamefont
  {Djurabekova}},\ }\bibfield  {title} {\enquote {\bibinfo {title}
  {{Deformations related to atom mixing in Si/SiO2/Si nanopillars under
  high-fluence broad-beam irradiation}},}\ }\href
  {https://doi.org/10.1103/physrevmaterials.4.013601} {\bibfield  {journal}
  {\bibinfo  {journal} {Phys. Rev. Materials}\ }\textbf {\bibinfo {volume}
  {4}},\ \bibinfo {pages} {013601} (\bibinfo {year} {2020})}\BibitemShut
  {NoStop}%
\bibitem [{\citenamefont {Xu}\ \emph {et~al.}(2019)\citenamefont {Xu},
  \citenamefont {Heinig}, \citenamefont {M{\"{o}}ller}, \citenamefont
  {Engelmann}, \citenamefont {Klingner}, \citenamefont {Gharbi}, \citenamefont
  {Tiron}, \citenamefont {von Borany},\ and\ \citenamefont
  {Hlawacek}}]{Xu2019}%
  \BibitemOpen
  \bibfield  {author} {\bibinfo {author} {\bibfnamefont {X.}~\bibnamefont
  {Xu}}, \bibinfo {author} {\bibfnamefont {K.-H.}\ \bibnamefont {Heinig}},
  \bibinfo {author} {\bibfnamefont {W.}~\bibnamefont {M{\"{o}}ller}}, \bibinfo
  {author} {\bibfnamefont {H.-J.}\ \bibnamefont {Engelmann}}, \bibinfo {author}
  {\bibfnamefont {N.}~\bibnamefont {Klingner}}, \bibinfo {author}
  {\bibfnamefont {A.}~\bibnamefont {Gharbi}}, \bibinfo {author} {\bibfnamefont
  {R.}~\bibnamefont {Tiron}}, \bibinfo {author} {\bibfnamefont
  {J.}~\bibnamefont {von Borany}},\ and\ \bibinfo {author} {\bibfnamefont
  {G.}~\bibnamefont {Hlawacek}},\ }\bibfield  {title} {\enquote {\bibinfo
  {title} {Morphology modification of si nanopillars under ion irradiation at
  elevated temperatures: plastic deformation and controlled thinning to 10
  nm},}\ }\href {https://doi.org/10.1088/1361-6641/ab57ba} {\bibfield
  {journal} {\bibinfo  {journal} {Semicond. Sci. Tech.}\ }\textbf {\bibinfo
  {volume} {35}},\ \bibinfo {pages} {015021} (\bibinfo {year}
  {2019})}\BibitemShut {NoStop}%
\bibitem [{\citenamefont {Holmstr{\"{o}}m}\ \emph {et~al.}(2012)\citenamefont
  {Holmstr{\"{o}}m}, \citenamefont {Kotakoski}, \citenamefont {Lechner},
  \citenamefont {Kaiser},\ and\ \citenamefont
  {Nordlund}}]{holmstrom_atomic-scale_2012}%
  \BibitemOpen
  \bibfield  {author} {\bibinfo {author} {\bibfnamefont {E.}~\bibnamefont
  {Holmstr{\"{o}}m}}, \bibinfo {author} {\bibfnamefont {J.}~\bibnamefont
  {Kotakoski}}, \bibinfo {author} {\bibfnamefont {L.}~\bibnamefont {Lechner}},
  \bibinfo {author} {\bibfnamefont {U.}~\bibnamefont {Kaiser}},\ and\ \bibinfo
  {author} {\bibfnamefont {K.}~\bibnamefont {Nordlund}},\ }\bibfield  {title}
  {\enquote {\bibinfo {title} {Atomic-scale effects behind structural
  instabilities in {Si} lamellae during ion beam thinning},}\ }\href
  {https://doi.org/10.1063/1.3698411} {\bibfield  {journal} {\bibinfo
  {journal} {AIP Adv.}\ }\textbf {\bibinfo {volume} {2}},\ \bibinfo {pages}
  {012186--012186--13} (\bibinfo {year} {2012})}\BibitemShut {NoStop}%
\bibitem [{\citenamefont {Barry}\ \emph {et~al.}(2014)\citenamefont {Barry},
  \citenamefont {Philipp}, \citenamefont {Wirtz},\ and\ \citenamefont
  {Kieffer}}]{Barry2014}%
  \BibitemOpen
  \bibfield  {author} {\bibinfo {author} {\bibfnamefont {P.~R.}\ \bibnamefont
  {Barry}}, \bibinfo {author} {\bibfnamefont {P.}~\bibnamefont {Philipp}},
  \bibinfo {author} {\bibfnamefont {T.}~\bibnamefont {Wirtz}},\ and\ \bibinfo
  {author} {\bibfnamefont {J.}~\bibnamefont {Kieffer}},\ }\bibfield  {title}
  {\enquote {\bibinfo {title} {{Mechanisms of silicon sputtering and cluster
  formation explained by atomic level simulations}},}\ }\href
  {https://doi.org/10.1002/jms.3317} {\bibfield  {journal} {\bibinfo  {journal}
  {J. Mass Spectrom.}\ }\textbf {\bibinfo {volume} {49}},\ \bibinfo {pages}
  {185--194} (\bibinfo {year} {2014})}\BibitemShut {NoStop}%
\bibitem [{\citenamefont {Newsome}\ \emph {et~al.}(2012)\citenamefont
  {Newsome}, \citenamefont {Sengupta}, \citenamefont {Foroutan}, \citenamefont
  {Russo},\ and\ \citenamefont {Van~Duin}}]{Newsome2012}%
  \BibitemOpen
  \bibfield  {author} {\bibinfo {author} {\bibfnamefont {D.~A.}\ \bibnamefont
  {Newsome}}, \bibinfo {author} {\bibfnamefont {D.}~\bibnamefont {Sengupta}},
  \bibinfo {author} {\bibfnamefont {H.}~\bibnamefont {Foroutan}}, \bibinfo
  {author} {\bibfnamefont {M.~F.}\ \bibnamefont {Russo}},\ and\ \bibinfo
  {author} {\bibfnamefont {A.~C.~T.}\ \bibnamefont {Van~Duin}},\ }\bibfield
  {title} {\enquote {\bibinfo {title} {{Oxidation of silicon carbide by O 2 and
  H 2O: A ReaxFF reactive molecular dynamics study, part i}},}\ }\href
  {https://doi.org/10.1021/jp306391p} {\bibfield  {journal} {\bibinfo
  {journal} {J. Phys. Chem. C}\ }\textbf {\bibinfo {volume} {116}},\ \bibinfo
  {pages} {16111--16121} (\bibinfo {year} {2012})}\BibitemShut {NoStop}%
\bibitem [{\citenamefont {Holmstr{\"{o}}m}, \citenamefont {Kuronen},\ and\
  \citenamefont {Nordlund}(2008)}]{holmstrom_threshold_2008}%
  \BibitemOpen
  \bibfield  {author} {\bibinfo {author} {\bibfnamefont {E.}~\bibnamefont
  {Holmstr{\"{o}}m}}, \bibinfo {author} {\bibfnamefont {A.}~\bibnamefont
  {Kuronen}},\ and\ \bibinfo {author} {\bibfnamefont {K.}~\bibnamefont
  {Nordlund}},\ }\bibfield  {title} {\enquote {\bibinfo {title} {Threshold
  defect production in silicon determined by density functional theory
  molecular dynamics simulations},}\ }\href
  {https://doi.org/10.1103/physrevb.78.045202} {\bibfield  {journal} {\bibinfo
  {journal} {Phys. Rev. B}\ }\textbf {\bibinfo {volume} {78}},\ \bibinfo
  {pages} {045202} (\bibinfo {year} {2008})}\BibitemShut {NoStop}%
\bibitem [{\citenamefont {Hamedani}\ \emph {et~al.}(2020)\citenamefont
  {Hamedani}, \citenamefont {Byggma{\"a}star}, \citenamefont {Djurabekova},
  \citenamefont {Alahyarizadeh}, \citenamefont {Ghaderi}, \citenamefont
  {Minuchehr},\ and\ \citenamefont {Nordlund}}]{Hamedani20}%
  \BibitemOpen
  \bibfield  {author} {\bibinfo {author} {\bibfnamefont {A.}~\bibnamefont
  {Hamedani}}, \bibinfo {author} {\bibfnamefont {J.}~\bibnamefont
  {Byggma{\"a}star}}, \bibinfo {author} {\bibfnamefont {F.}~\bibnamefont
  {Djurabekova}}, \bibinfo {author} {\bibfnamefont {G.}~\bibnamefont
  {Alahyarizadeh}}, \bibinfo {author} {\bibfnamefont {R.}~\bibnamefont
  {Ghaderi}}, \bibinfo {author} {\bibfnamefont {A.}~\bibnamefont {Minuchehr}},\
  and\ \bibinfo {author} {\bibfnamefont {K.}~\bibnamefont {Nordlund}},\
  }\bibfield  {title} {\enquote {\bibinfo {title} {Insights into the primary
  radiation damage in silicon by machine learning interatomic potential},}\
  }\href {https://doi.org/10.1080/21663831.2020.1771451} {\bibfield  {journal}
  {\bibinfo  {journal} {Mater. Res. Lett.}\ }\textbf {\bibinfo {volume} {8}},\
  \bibinfo {pages} {364--372} (\bibinfo {year} {2020})}\BibitemShut {NoStop}%
\bibitem [{\citenamefont {Ghorbani-Asl}\ \emph {et~al.}(2017)\citenamefont
  {Ghorbani-Asl}, \citenamefont {Kretschmer}, \citenamefont {Spearot},\ and\
  \citenamefont {Krasheninnikov}}]{Ghorbani-Asl-2017}%
  \BibitemOpen
  \bibfield  {author} {\bibinfo {author} {\bibfnamefont {M.}~\bibnamefont
  {Ghorbani-Asl}}, \bibinfo {author} {\bibfnamefont {S.}~\bibnamefont
  {Kretschmer}}, \bibinfo {author} {\bibfnamefont {D.~E.}\ \bibnamefont
  {Spearot}},\ and\ \bibinfo {author} {\bibfnamefont {A.~V.}\ \bibnamefont
  {Krasheninnikov}},\ }\bibfield  {title} {\enquote {\bibinfo {title}
  {{Two-dimensional MoS2 under ion irradiation: from controlled defect
  production to electronic structure engineering}},}\ }\href
  {https://doi.org/10.1088/2053-1583/aa6b17} {\bibfield  {journal} {\bibinfo
  {journal} {2D Mater.}\ }\textbf {\bibinfo {volume} {4}},\ \bibinfo {pages}
  {25078} (\bibinfo {year} {2017})}\BibitemShut {NoStop}%
\bibitem [{\citenamefont {Ghaderzadeh}\ \emph {et~al.}(2021)\citenamefont
  {Ghaderzadeh}, \citenamefont {Kretschmer}, \citenamefont {Ghorbani-Asl},
  \citenamefont {Hlawacek},\ and\ \citenamefont
  {Krasheninnikov}}]{Ghaderzadeh-2021}%
  \BibitemOpen
  \bibfield  {author} {\bibinfo {author} {\bibfnamefont {S.}~\bibnamefont
  {Ghaderzadeh}}, \bibinfo {author} {\bibfnamefont {S.}~\bibnamefont
  {Kretschmer}}, \bibinfo {author} {\bibfnamefont {M.}~\bibnamefont
  {Ghorbani-Asl}}, \bibinfo {author} {\bibfnamefont {G.}~\bibnamefont
  {Hlawacek}},\ and\ \bibinfo {author} {\bibfnamefont {A.~V.}\ \bibnamefont
  {Krasheninnikov}},\ }\bibfield  {title} {\enquote {\bibinfo {title}
  {Atomistic simulations of defect production in monolayer and bulk hexagonal
  boron nitride under low- and high-fluence ion irradiation},}\ }\href
  {https://doi.org/10.3390/nano11051214} {\bibfield  {journal} {\bibinfo
  {journal} {Nanomaterials}\ }\textbf {\bibinfo {volume} {11}} (\bibinfo {year}
  {2021}),\ 10.3390/nano11051214}\BibitemShut {NoStop}%
\bibitem [{\citenamefont {Lehtinen}\ \emph {et~al.}(2010)\citenamefont
  {Lehtinen}, \citenamefont {Kotakoski}, \citenamefont {Krasheninnikov},
  \citenamefont {Tolvanen}, \citenamefont {Nordlund},\ and\ \citenamefont
  {Keinonen}}]{Lehtinen_2010}%
  \BibitemOpen
  \bibfield  {author} {\bibinfo {author} {\bibfnamefont {O.}~\bibnamefont
  {Lehtinen}}, \bibinfo {author} {\bibfnamefont {J.}~\bibnamefont {Kotakoski}},
  \bibinfo {author} {\bibfnamefont {A.~V.}\ \bibnamefont {Krasheninnikov}},
  \bibinfo {author} {\bibfnamefont {A.}~\bibnamefont {Tolvanen}}, \bibinfo
  {author} {\bibfnamefont {K.}~\bibnamefont {Nordlund}},\ and\ \bibinfo
  {author} {\bibfnamefont {J.}~\bibnamefont {Keinonen}},\ }\bibfield  {title}
  {\enquote {\bibinfo {title} {{Effects of ion bombardment on a two-dimensional
  target: Atomistic simulations of graphene irradiation}},}\ }\href
  {https://doi.org/10.1103/PhysRevB.81.153401} {\bibfield  {journal} {\bibinfo
  {journal} {Phys. Rev. B}\ }\textbf {\bibinfo {volume} {81}},\ \bibinfo
  {pages} {153401} (\bibinfo {year} {2010})},\ \bibinfo {note} {publisher:
  American Physical Society}\BibitemShut {NoStop}%
\bibitem [{\citenamefont {Kretschmer}\ \emph {et~al.}(2018)\citenamefont
  {Kretschmer}, \citenamefont {Maslov}, \citenamefont {Ghaderzadeh},
  \citenamefont {Ghorbani-Asl}, \citenamefont {Hlawacek},\ and\ \citenamefont
  {Krasheninnikov}}]{Kretschmer-2018}%
  \BibitemOpen
  \bibfield  {author} {\bibinfo {author} {\bibfnamefont {S.}~\bibnamefont
  {Kretschmer}}, \bibinfo {author} {\bibfnamefont {M.}~\bibnamefont {Maslov}},
  \bibinfo {author} {\bibfnamefont {S.}~\bibnamefont {Ghaderzadeh}}, \bibinfo
  {author} {\bibfnamefont {M.}~\bibnamefont {Ghorbani-Asl}}, \bibinfo {author}
  {\bibfnamefont {G.}~\bibnamefont {Hlawacek}},\ and\ \bibinfo {author}
  {\bibfnamefont {A.~V.}\ \bibnamefont {Krasheninnikov}},\ }\bibfield  {title}
  {\enquote {\bibinfo {title} {{Supported Two-Dimensional Materials under Ion
  Irradiation: The Substrate Governs Defect Production}},}\ }\href
  {https://doi.org/10.1021/acsami.8b08471} {\bibfield  {journal} {\bibinfo
  {journal} {ACS Appl. Mater. Interfaces}\ }\textbf {\bibinfo {volume} {10}},\
  \bibinfo {pages} {30827--30836} (\bibinfo {year} {2018})}\BibitemShut
  {NoStop}%
\bibitem [{\citenamefont {Kalbac}\ \emph {et~al.}(2012)\citenamefont {Kalbac},
  \citenamefont {Lehtinen}, \citenamefont {Krasheninnikov},\ and\ \citenamefont
  {Keinonen}}]{Kalbac-2013}%
  \BibitemOpen
  \bibfield  {author} {\bibinfo {author} {\bibfnamefont {M.}~\bibnamefont
  {Kalbac}}, \bibinfo {author} {\bibfnamefont {O.}~\bibnamefont {Lehtinen}},
  \bibinfo {author} {\bibfnamefont {A.~V.}\ \bibnamefont {Krasheninnikov}},\
  and\ \bibinfo {author} {\bibfnamefont {J.}~\bibnamefont {Keinonen}},\
  }\bibfield  {title} {\enquote {\bibinfo {title} {Ion-irradiation-induced
  defects in isotopically-labeled two layered graphene: enhanced in-situ
  annealing of the damage},}\ }\href {https://doi.org/10.1002/adma.201203807}
  {\bibfield  {journal} {\bibinfo  {journal} {Adv. Mater.}\ }\textbf {\bibinfo
  {volume} {25}},\ \bibinfo {pages} {1004--1009} (\bibinfo {year}
  {2012})}\BibitemShut {NoStop}%
\bibitem [{\citenamefont {Standop}\ \emph {et~al.}(2013)\citenamefont
  {Standop}, \citenamefont {Lehtinen}, \citenamefont {Herbig}, \citenamefont
  {Lewes-Malandrakis}, \citenamefont {Craes}, \citenamefont {Kotakoski},
  \citenamefont {Michely}, \citenamefont {Krasheninnikov},\ and\ \citenamefont
  {Busse}}]{Standop13}%
  \BibitemOpen
  \bibfield  {author} {\bibinfo {author} {\bibfnamefont {S.}~\bibnamefont
  {Standop}}, \bibinfo {author} {\bibfnamefont {O.}~\bibnamefont {Lehtinen}},
  \bibinfo {author} {\bibfnamefont {C.}~\bibnamefont {Herbig}}, \bibinfo
  {author} {\bibfnamefont {G.}~\bibnamefont {Lewes-Malandrakis}}, \bibinfo
  {author} {\bibfnamefont {F.}~\bibnamefont {Craes}}, \bibinfo {author}
  {\bibfnamefont {J.}~\bibnamefont {Kotakoski}}, \bibinfo {author}
  {\bibfnamefont {T.}~\bibnamefont {Michely}}, \bibinfo {author} {\bibfnamefont
  {A.~V.}\ \bibnamefont {Krasheninnikov}},\ and\ \bibinfo {author}
  {\bibfnamefont {C.}~\bibnamefont {Busse}},\ }\bibfield  {title} {\enquote
  {\bibinfo {title} {Ion impacts on graphene/ir(111): interface channeling,
  vacancy funnels, and a nanomesh},}\ }\href
  {https://doi.org/10.1021/nl304659n} {\bibfield  {journal} {\bibinfo
  {journal} {Nano Lett.}\ }\textbf {\bibinfo {volume} {13}},\ \bibinfo {pages}
  {1948--1955} (\bibinfo {year} {2013})}\BibitemShut {NoStop}%
\bibitem [{\citenamefont {Lehtinen}\ \emph {et~al.}(2011)\citenamefont
  {Lehtinen}, \citenamefont {Kotakoski}, \citenamefont {Krasheninnikov},\ and\
  \citenamefont {Keinonen}}]{Lehtinen_2011}%
  \BibitemOpen
  \bibfield  {author} {\bibinfo {author} {\bibfnamefont {O.}~\bibnamefont
  {Lehtinen}}, \bibinfo {author} {\bibfnamefont {J.}~\bibnamefont {Kotakoski}},
  \bibinfo {author} {\bibfnamefont {A.~V.}\ \bibnamefont {Krasheninnikov}},\
  and\ \bibinfo {author} {\bibfnamefont {J.}~\bibnamefont {Keinonen}},\
  }\bibfield  {title} {{\selectlanguage {english}\enquote {\bibinfo {title}
  {Cutting and controlled modification of graphene with ion beams},}\ }}\href
  {https://doi.org/10.1088/0957-4484/22/17/175306} {\bibfield  {journal}
  {\bibinfo  {journal} {Nanotechnology}\ }\textbf {\bibinfo {volume} {22}},\
  \bibinfo {pages} {175306} (\bibinfo {year} {2011})}\BibitemShut {NoStop}%
\bibitem [{\citenamefont {Li}\ and\ \citenamefont {Xue}(2015)}]{Li-2015}%
  \BibitemOpen
  \bibfield  {author} {\bibinfo {author} {\bibfnamefont {W.}~\bibnamefont
  {Li}}\ and\ \bibinfo {author} {\bibfnamefont {J.}~\bibnamefont {Xue}},\
  }\bibfield  {title} {\enquote {\bibinfo {title} {{Ion implantation of low
  energy Si into graphene: Insight from computational studies}},}\ }\href
  {https://doi.org/10.1039/c5ra17250k} {\bibfield  {journal} {\bibinfo
  {journal} {RSC Adv.}\ }\textbf {\bibinfo {volume} {5}},\ \bibinfo {pages}
  {99920--99926} (\bibinfo {year} {2015})}\BibitemShut {NoStop}%
\bibitem [{\citenamefont {Tripathi}\ \emph {et~al.}(2018)\citenamefont
  {Tripathi}, \citenamefont {Markevich}, \citenamefont {B{\"{o}}ttger},
  \citenamefont {Facsko}, \citenamefont {Besley}, \citenamefont {Kotakoski},\
  and\ \citenamefont {Susi}}]{Tripathi-2018}%
  \BibitemOpen
  \bibfield  {author} {\bibinfo {author} {\bibfnamefont {M.}~\bibnamefont
  {Tripathi}}, \bibinfo {author} {\bibfnamefont {A.}~\bibnamefont {Markevich}},
  \bibinfo {author} {\bibfnamefont {R.}~\bibnamefont {B{\"{o}}ttger}}, \bibinfo
  {author} {\bibfnamefont {S.}~\bibnamefont {Facsko}}, \bibinfo {author}
  {\bibfnamefont {E.}~\bibnamefont {Besley}}, \bibinfo {author} {\bibfnamefont
  {J.}~\bibnamefont {Kotakoski}},\ and\ \bibinfo {author} {\bibfnamefont
  {T.}~\bibnamefont {Susi}},\ }\bibfield  {title} {\enquote {\bibinfo {title}
  {{Implanting Germanium into Graphene}},}\ }\href
  {https://doi.org/10.1021/acsnano.8b01191} {\bibfield  {journal} {\bibinfo
  {journal} {ACS Nano}\ }\textbf {\bibinfo {volume} {12}},\ \bibinfo {pages}
  {4641--4647} (\bibinfo {year} {2018})}\BibitemShut {NoStop}%
\bibitem [{\citenamefont {Kretschmer}\ \emph {et~al.}(2022)\citenamefont
  {Kretschmer}, \citenamefont {Ghaderzadeh}, \citenamefont {Facsko},\ and\
  \citenamefont {Krasheninnikov}}]{kretschmer_threshold_2022}%
  \BibitemOpen
  \bibfield  {author} {\bibinfo {author} {\bibfnamefont {S.}~\bibnamefont
  {Kretschmer}}, \bibinfo {author} {\bibfnamefont {S.}~\bibnamefont
  {Ghaderzadeh}}, \bibinfo {author} {\bibfnamefont {S.}~\bibnamefont
  {Facsko}},\ and\ \bibinfo {author} {\bibfnamefont {A.~V.}\ \bibnamefont
  {Krasheninnikov}},\ }\bibfield  {title} {\enquote {\bibinfo {title}
  {Threshold {Ion} {Energies} for {Creating} {Defects} in {2D} {Materials} from
  {First}-{Principles} {Calculations}: {Chemical} {Interactions} {Are}
  {Important}},}\ }\href {https://doi.org/10.1021/acs.jpclett.1c03995}
  {\bibfield  {journal} {\bibinfo  {journal} {J. Phys. Chem. Lett.}\ }\textbf
  {\bibinfo {volume} {13}},\ \bibinfo {pages} {514--519} (\bibinfo {year}
  {2022})},\ \bibinfo {note} {publisher: American Chemical Society}\BibitemShut
  {NoStop}%
\bibitem [{\citenamefont {Cai}\ \emph {et~al.}(1996)\citenamefont {Cai},
  \citenamefont {Gr{\textbackslash}onsbech-Jensen}, \citenamefont {Snell},\
  and\ \citenamefont {Beardmore}}]{cai_phenomenological_1996}%
  \BibitemOpen
  \bibfield  {author} {\bibinfo {author} {\bibfnamefont {D.}~\bibnamefont
  {Cai}}, \bibinfo {author} {\bibfnamefont {N.}~\bibnamefont
  {Gr{\textbackslash}onsbech-Jensen}}, \bibinfo {author} {\bibfnamefont
  {C.~M.}\ \bibnamefont {Snell}},\ and\ \bibinfo {author} {\bibfnamefont
  {K.~M.}\ \bibnamefont {Beardmore}},\ }\bibfield  {title} {\enquote {\bibinfo
  {title} {Phenomenological electronic stopping-power model for molecular
  dynamics and {Monte} {Carlo} simulation of ion implantation into silicon},}\
  }\href {https://doi.org/10.1103/physrevb.54.17147} {\bibfield  {journal}
  {\bibinfo  {journal} {Phys. Rev. B}\ }\textbf {\bibinfo {volume} {54}},\
  \bibinfo {pages} {17147--17157} (\bibinfo {year} {1996})}\BibitemShut
  {NoStop}%
\bibitem [{\citenamefont {Chan}\ \emph {et~al.}(2006)\citenamefont {Chan},
  \citenamefont {Nordlund}, \citenamefont {Gossmann}, \citenamefont {Harris},
  \citenamefont {Montgomery}, \citenamefont {Mulcahy}, \citenamefont {Biswas},
  \citenamefont {Srinivasan}, \citenamefont {Benistant}, \citenamefont {Ng},\
  and\ \citenamefont {Chan}}]{chan_molecular_2006}%
  \BibitemOpen
  \bibfield  {author} {\bibinfo {author} {\bibfnamefont {H.~Y.}\ \bibnamefont
  {Chan}}, \bibinfo {author} {\bibfnamefont {K.}~\bibnamefont {Nordlund}},
  \bibinfo {author} {\bibfnamefont {H.~J.~L.}\ \bibnamefont {Gossmann}},
  \bibinfo {author} {\bibfnamefont {M.}~\bibnamefont {Harris}}, \bibinfo
  {author} {\bibfnamefont {N.~J.}\ \bibnamefont {Montgomery}}, \bibinfo
  {author} {\bibfnamefont {C.~P.~A.}\ \bibnamefont {Mulcahy}}, \bibinfo
  {author} {\bibfnamefont {S.}~\bibnamefont {Biswas}}, \bibinfo {author}
  {\bibfnamefont {M.~P.}\ \bibnamefont {Srinivasan}}, \bibinfo {author}
  {\bibfnamefont {F.}~\bibnamefont {Benistant}}, \bibinfo {author}
  {\bibfnamefont {C.~M.}\ \bibnamefont {Ng}},\ and\ \bibinfo {author}
  {\bibfnamefont {L.}~\bibnamefont {Chan}},\ }\bibfield  {title} {\enquote
  {\bibinfo {title} {Molecular dynamics with phase-shift-based electronic
  stopping for calibration of ion implantation profiles in crystalline
  silicon},}\ }\href {https://doi.org/10.1016/j.tsf.2005.09.153} {\bibfield
  {journal} {\bibinfo  {journal} {Thin Solid Films}\ }\textbf {\bibinfo
  {volume} {504}},\ \bibinfo {pages} {121--125} (\bibinfo {year}
  {2006})}\BibitemShut {NoStop}%
\bibitem [{\citenamefont {Sandoval}\ and\ \citenamefont
  {Urbassek}(2009)}]{sandoval_influence_2009}%
  \BibitemOpen
  \bibfield  {author} {\bibinfo {author} {\bibfnamefont {L.}~\bibnamefont
  {Sandoval}}\ and\ \bibinfo {author} {\bibfnamefont {H.~M.}\ \bibnamefont
  {Urbassek}},\ }\bibfield  {title} {\enquote {\bibinfo {title} {Influence of
  electronic stopping on sputtering induced by cluster impact on metallic
  targets},}\ }\href {https://doi.org/10.1103/PhysRevB.79.144115} {\bibfield
  {journal} {\bibinfo  {journal} {Phys. Rev. B}\ }\textbf {\bibinfo {volume}
  {79}},\ \bibinfo {pages} {144115} (\bibinfo {year} {2009})}\BibitemShut
  {NoStop}%
\bibitem [{\citenamefont {Zarkadoula}\ \emph {et~al.}(2014)\citenamefont
  {Zarkadoula}, \citenamefont {Daraszewicz}, \citenamefont {Duffy},
  \citenamefont {Seaton}, \citenamefont {Todorov}, \citenamefont {Nordlund},
  \citenamefont {Dove},\ and\ \citenamefont
  {Trachenko}}]{zarkadoula_electronic_2014}%
  \BibitemOpen
  \bibfield  {author} {\bibinfo {author} {\bibfnamefont {E.}~\bibnamefont
  {Zarkadoula}}, \bibinfo {author} {\bibfnamefont {S.~L.}\ \bibnamefont
  {Daraszewicz}}, \bibinfo {author} {\bibfnamefont {D.~M.}\ \bibnamefont
  {Duffy}}, \bibinfo {author} {\bibfnamefont {M.~A.}\ \bibnamefont {Seaton}},
  \bibinfo {author} {\bibfnamefont {I.~T.}\ \bibnamefont {Todorov}}, \bibinfo
  {author} {\bibfnamefont {K.}~\bibnamefont {Nordlund}}, \bibinfo {author}
  {\bibfnamefont {M.~T.}\ \bibnamefont {Dove}},\ and\ \bibinfo {author}
  {\bibfnamefont {K.}~\bibnamefont {Trachenko}},\ }\bibfield  {title}
  {{\selectlanguage {english}\enquote {\bibinfo {title} {Electronic effects in
  high-energy radiation damage in iron},}\ }}\href
  {https://doi.org/10.1088/0953-8984/26/8/085401} {\bibfield  {journal}
  {\bibinfo  {journal} {J. Phys.: Condens. Matter}\ }\textbf {\bibinfo {volume}
  {26}},\ \bibinfo {pages} {085401} (\bibinfo {year} {2014})},\ \bibinfo {note}
  {publisher: IOP Publishing}\BibitemShut {NoStop}%
\bibitem [{\citenamefont {Meftah}\ \emph {et~al.}(1994)\citenamefont {Meftah},
  \citenamefont {Brisard}, \citenamefont {Costantini}, \citenamefont
  {Dooryhee}, \citenamefont {Hage-Ali}, \citenamefont {Hervieu}, \citenamefont
  {Stoquert}, \citenamefont {Studer},\ and\ \citenamefont
  {Toulemonde}}]{Meftah-1994}%
  \BibitemOpen
  \bibfield  {author} {\bibinfo {author} {\bibfnamefont {A.}~\bibnamefont
  {Meftah}}, \bibinfo {author} {\bibfnamefont {F.}~\bibnamefont {Brisard}},
  \bibinfo {author} {\bibfnamefont {J.~M.}\ \bibnamefont {Costantini}},
  \bibinfo {author} {\bibfnamefont {E.}~\bibnamefont {Dooryhee}}, \bibinfo
  {author} {\bibfnamefont {M.}~\bibnamefont {Hage-Ali}}, \bibinfo {author}
  {\bibfnamefont {M.}~\bibnamefont {Hervieu}}, \bibinfo {author} {\bibfnamefont
  {J.~P.}\ \bibnamefont {Stoquert}}, \bibinfo {author} {\bibfnamefont
  {F.}~\bibnamefont {Studer}},\ and\ \bibinfo {author} {\bibfnamefont
  {M.}~\bibnamefont {Toulemonde}},\ }\bibfield  {title} {\enquote {\bibinfo
  {title} {Track formation in ${\mathrm{sio}}_{2}$ quartz and the thermal-spike
  mechanism},}\ }\href {https://doi.org/10.1103/PhysRevB.49.12457} {\bibfield
  {journal} {\bibinfo  {journal} {Phys. Rev. B}\ }\textbf {\bibinfo {volume}
  {49}},\ \bibinfo {pages} {12457--12463} (\bibinfo {year} {1994})}\BibitemShut
  {NoStop}%
\bibitem [{\citenamefont {Jarrin}\ \emph {et~al.}(2021)\citenamefont {Jarrin},
  \citenamefont {Richard}, \citenamefont {Teunissen}, \citenamefont
  {Da~Pieve},\ and\ \citenamefont {H{\'{e}}meryck}}]{jarrin_integration_2021}%
  \BibitemOpen
  \bibfield  {author} {\bibinfo {author} {\bibfnamefont {T.}~\bibnamefont
  {Jarrin}}, \bibinfo {author} {\bibfnamefont {N.}~\bibnamefont {Richard}},
  \bibinfo {author} {\bibfnamefont {J.}~\bibnamefont {Teunissen}}, \bibinfo
  {author} {\bibfnamefont {F.}~\bibnamefont {Da~Pieve}},\ and\ \bibinfo
  {author} {\bibfnamefont {A.}~\bibnamefont {H{\'{e}}meryck}},\ }\bibfield
  {title} {\enquote {\bibinfo {title} {Integration of electronic effects into
  molecular dynamics simulations of collision cascades in silicon from
  first-principles calculations},}\ }\href
  {https://doi.org/10.1103/PhysRevB.104.195203} {\bibfield  {journal} {\bibinfo
   {journal} {Phys. Rev. B}\ }\textbf {\bibinfo {volume} {104}},\ \bibinfo
  {pages} {195203} (\bibinfo {year} {2021})},\ \bibinfo {note} {publisher:
  American Physical Society}\BibitemShut {NoStop}%
\bibitem [{\citenamefont {Runge}\ and\ \citenamefont
  {Gross}(1984)}]{runge_density-functional_1984}%
  \BibitemOpen
  \bibfield  {author} {\bibinfo {author} {\bibfnamefont {E.}~\bibnamefont
  {Runge}}\ and\ \bibinfo {author} {\bibfnamefont {E.~K.~U.}\ \bibnamefont
  {Gross}},\ }\bibfield  {title} {\enquote {\bibinfo {title}
  {Density-{Functional} {Theory} for {Time}-{Dependent} {Systems}},}\ }\href
  {https://doi.org/10.1103/PhysRevLett.52.997} {\bibfield  {journal} {\bibinfo
  {journal} {Phys. Rev. Lett.}\ }\textbf {\bibinfo {volume} {52}},\ \bibinfo
  {pages} {997--1000} (\bibinfo {year} {1984})},\ \bibinfo {note} {publisher:
  American Physical Society}\BibitemShut {NoStop}%
\bibitem [{\citenamefont {Lim}\ \emph {et~al.}(2016)\citenamefont {Lim},
  \citenamefont {Foulkes}, \citenamefont {Horsfield}, \citenamefont {Mason},
  \citenamefont {Schleife}, \citenamefont {Draeger},\ and\ \citenamefont
  {Correa}}]{Lim_2016}%
  \BibitemOpen
  \bibfield  {author} {\bibinfo {author} {\bibfnamefont {A.}~\bibnamefont
  {Lim}}, \bibinfo {author} {\bibfnamefont {W.~M.~C.}\ \bibnamefont {Foulkes}},
  \bibinfo {author} {\bibfnamefont {A.~P.}\ \bibnamefont {Horsfield}}, \bibinfo
  {author} {\bibfnamefont {D.~R.}\ \bibnamefont {Mason}}, \bibinfo {author}
  {\bibfnamefont {A.}~\bibnamefont {Schleife}}, \bibinfo {author}
  {\bibfnamefont {E.~W.}\ \bibnamefont {Draeger}},\ and\ \bibinfo {author}
  {\bibfnamefont {A.~A.}\ \bibnamefont {Correa}},\ }\bibfield  {title}
  {\enquote {\bibinfo {title} {{The electron elevator: excitations across the
  band gap via a dynamical gap state}},}\ }\href
  {https://doi.org/10.1103/physrevlett.116.043201} {\bibfield  {journal}
  {\bibinfo  {journal} {Phys. Rev. Lett.}\ }\textbf {\bibinfo {volume} {116}},\
  \bibinfo {pages} {043201} (\bibinfo {year} {2016})}\BibitemShut {NoStop}%
\bibitem [{\citenamefont {Tamm}\ \emph {et~al.}(2016)\citenamefont {Tamm},
  \citenamefont {Samolyuk}, \citenamefont {Correa}, \citenamefont
  {Klintenberg}, \citenamefont {Aabloo},\ and\ \citenamefont
  {Caro}}]{Tamm_2016}%
  \BibitemOpen
  \bibfield  {author} {\bibinfo {author} {\bibfnamefont {A.}~\bibnamefont
  {Tamm}}, \bibinfo {author} {\bibfnamefont {G.}~\bibnamefont {Samolyuk}},
  \bibinfo {author} {\bibfnamefont {A.~A.}\ \bibnamefont {Correa}}, \bibinfo
  {author} {\bibfnamefont {M.}~\bibnamefont {Klintenberg}}, \bibinfo {author}
  {\bibfnamefont {A.}~\bibnamefont {Aabloo}},\ and\ \bibinfo {author}
  {\bibfnamefont {A.}~\bibnamefont {Caro}},\ }\bibfield  {title} {\enquote
  {\bibinfo {title} {Electron-phonon interaction within classical molecular
  dynamics},}\ }\href {https://doi.org/10.1103/PhysRevB.94.024305} {\bibfield
  {journal} {\bibinfo  {journal} {Phys. Rev. B}\ }\textbf {\bibinfo {volume}
  {94}},\ \bibinfo {pages} {024305} (\bibinfo {year} {2016})}\BibitemShut
  {NoStop}%
\bibitem [{\citenamefont {Sand~Andrea}, \citenamefont {Rafi},\ and\
  \citenamefont {Correa~Alfredo}(2019)}]{Sand_2019}%
  \BibitemOpen
  \bibfield  {author} {\bibinfo {author} {\bibfnamefont {E.}~\bibnamefont
  {Sand~Andrea}}, \bibinfo {author} {\bibfnamefont {U.}~\bibnamefont {Rafi}},\
  and\ \bibinfo {author} {\bibfnamefont {A.}~\bibnamefont {Correa~Alfredo}},\
  }\bibfield  {title} {\enquote {\bibinfo {title} {{Heavy ion ranges from
  first-principles electron dynamics}},}\ }\href
  {https://doi.org/10.1038/s41524-019-0180-5} {\bibfield  {journal} {\bibinfo
  {journal} {npj Comput. Mater.}\ }\textbf {\bibinfo {volume} {5}},\ \bibinfo
  {pages} {43} (\bibinfo {year} {2019})}\BibitemShut {NoStop}%
\bibitem [{\citenamefont {Darkins}\ \emph {et~al.}(2018)\citenamefont
  {Darkins}, \citenamefont {Ma}, \citenamefont {Murphy},\ and\ \citenamefont
  {Duffy}}]{Darkins2018}%
  \BibitemOpen
  \bibfield  {author} {\bibinfo {author} {\bibfnamefont {R.}~\bibnamefont
  {Darkins}}, \bibinfo {author} {\bibfnamefont {P.-W.}\ \bibnamefont {Ma}},
  \bibinfo {author} {\bibfnamefont {S.}~\bibnamefont {Murphy}},\ and\ \bibinfo
  {author} {\bibfnamefont {D.}~\bibnamefont {Duffy}},\ }\bibfield  {title}
  {\enquote {\bibinfo {title} {Simulating electronically driven structural
  changes in silicon with two-temperature molecular dynamics},}\ }\href
  {https://doi.org/10.1103/PhysRevB.98.024304} {\bibfield  {journal} {\bibinfo
  {journal} {Phys. Rev. B}\ }\textbf {\bibinfo {volume} {98}},\ \bibinfo
  {pages} {024304} (\bibinfo {year} {2018})}\BibitemShut {NoStop}%
\bibitem [{\citenamefont {Krasheninnikov}, \citenamefont {Miyamoto},\ and\
  \citenamefont {Tom\'anek}(2007)}]{Krasheninnikov-2007}%
  \BibitemOpen
  \bibfield  {author} {\bibinfo {author} {\bibfnamefont {A.~V.}\ \bibnamefont
  {Krasheninnikov}}, \bibinfo {author} {\bibfnamefont {Y.}~\bibnamefont
  {Miyamoto}},\ and\ \bibinfo {author} {\bibfnamefont {D.}~\bibnamefont
  {Tom\'anek}},\ }\bibfield  {title} {\enquote {\bibinfo {title} {Role of
  electronic excitations in ion collisions with carbon nanostructures},}\
  }\href {https://doi.org/10.1103/PhysRevLett.99.016104} {\bibfield  {journal}
  {\bibinfo  {journal} {Phys. Rev. Lett.}\ }\textbf {\bibinfo {volume} {99}},\
  \bibinfo {pages} {016104} (\bibinfo {year} {2007})}\BibitemShut {NoStop}%
\bibitem [{\citenamefont {Ojanpera}, \citenamefont {Krasheninnikov},\ and\
  \citenamefont {Puska}(2014)}]{Ojanpera-2014}%
  \BibitemOpen
  \bibfield  {author} {\bibinfo {author} {\bibfnamefont {A.}~\bibnamefont
  {Ojanpera}}, \bibinfo {author} {\bibfnamefont {A.~V.}\ \bibnamefont
  {Krasheninnikov}},\ and\ \bibinfo {author} {\bibfnamefont {M.}~\bibnamefont
  {Puska}},\ }\bibfield  {title} {\enquote {\bibinfo {title} {Electronic
  stopping power from first-principles calculations with account for core
  electron excitations and projectile ionization},}\ }\href
  {https://doi.org/10.1103/PhysRevB.89.035120} {\bibfield  {journal} {\bibinfo
  {journal} {Phys. Rev. B}\ }\textbf {\bibinfo {volume} {89}},\ \bibinfo
  {pages} {035120} (\bibinfo {year} {2014})}\BibitemShut {NoStop}%
\bibitem [{\citenamefont {Correa}\ \emph {et~al.}(2012)\citenamefont {Correa},
  \citenamefont {Kohanoff}, \citenamefont {Artacho}, \citenamefont
  {S\'anchez-Portal},\ and\ \citenamefont {Caro}}]{Correa-2012}%
  \BibitemOpen
  \bibfield  {author} {\bibinfo {author} {\bibfnamefont {A.~A.}\ \bibnamefont
  {Correa}}, \bibinfo {author} {\bibfnamefont {J.}~\bibnamefont {Kohanoff}},
  \bibinfo {author} {\bibfnamefont {E.}~\bibnamefont {Artacho}}, \bibinfo
  {author} {\bibfnamefont {D.}~\bibnamefont {S\'anchez-Portal}},\ and\ \bibinfo
  {author} {\bibfnamefont {A.}~\bibnamefont {Caro}},\ }\bibfield  {title}
  {\enquote {\bibinfo {title} {Nonadiabatic forces in ion-solid interactions:
  the initial stages of radiation damage},}\ }\href
  {https://doi.org/10.1103/PhysRevLett.108.213201} {\bibfield  {journal}
  {\bibinfo  {journal} {Phys. Rev. Lett.}\ }\textbf {\bibinfo {volume} {108}},\
  \bibinfo {pages} {213201} (\bibinfo {year} {2012})}\BibitemShut {NoStop}%
\bibitem [{\citenamefont {Ullah}, \citenamefont {Artacho},\ and\ \citenamefont
  {Correa}(2018)}]{Ullah-2018}%
  \BibitemOpen
  \bibfield  {author} {\bibinfo {author} {\bibfnamefont {R.}~\bibnamefont
  {Ullah}}, \bibinfo {author} {\bibfnamefont {E.}~\bibnamefont {Artacho}},\
  and\ \bibinfo {author} {\bibfnamefont {A.~A.}\ \bibnamefont {Correa}},\
  }\bibfield  {title} {\enquote {\bibinfo {title} {Core electrons in the
  electronic stopping of heavy ions},}\ }\href
  {https://doi.org/10.1103/PhysRevLett.121.116401} {\bibfield  {journal}
  {\bibinfo  {journal} {Phys. Rev. Lett.}\ }\textbf {\bibinfo {volume} {121}},\
  \bibinfo {pages} {116401} (\bibinfo {year} {2018})}\BibitemShut {NoStop}%
\bibitem [{\citenamefont {Lee}\ and\ \citenamefont
  {Schleife}(2019)}]{Lee-2019}%
  \BibitemOpen
  \bibfield  {author} {\bibinfo {author} {\bibfnamefont {C.-W.}\ \bibnamefont
  {Lee}}\ and\ \bibinfo {author} {\bibfnamefont {A.}~\bibnamefont {Schleife}},\
  }\bibfield  {title} {\enquote {\bibinfo {title} {Hot-electron-mediated ion
  diffusion in semiconductors for ion-beam nanostructuring},}\ }\href
  {https://doi.org/10.1021/acs.nanolett.9b01214} {\bibfield  {journal}
  {\bibinfo  {journal} {Nano Lett.}\ }\textbf {\bibinfo {volume} {19}},\
  \bibinfo {pages} {3939--3947} (\bibinfo {year} {2019})}\BibitemShut {NoStop}%
\bibitem [{\citenamefont {Kretschmer}\ \emph {et~al.}(2020)\citenamefont
  {Kretschmer}, \citenamefont {Lehnert}, \citenamefont {Kaiser},\ and\
  \citenamefont {Krasheninnikov}}]{Kretschmer-2020}%
  \BibitemOpen
  \bibfield  {author} {\bibinfo {author} {\bibfnamefont {S.}~\bibnamefont
  {Kretschmer}}, \bibinfo {author} {\bibfnamefont {T.}~\bibnamefont {Lehnert}},
  \bibinfo {author} {\bibfnamefont {U.}~\bibnamefont {Kaiser}},\ and\ \bibinfo
  {author} {\bibfnamefont {A.~V.}\ \bibnamefont {Krasheninnikov}},\ }\bibfield
  {title} {\enquote {\bibinfo {title} {{Formation of Defects in Two-Dimensional
  MoS 2 in the Transmission Electron Microscope at Electron Energies below the
  Knock-on Threshold: The Role of Electronic Excitations}},}\ }\href
  {https://doi.org/10.1021/acs.nanolett.0c00670} {\bibfield  {journal}
  {\bibinfo  {journal} {Nano Lett.}\ }\textbf {\bibinfo {volume} {20}},\
  \bibinfo {pages} {2865--2870} (\bibinfo {year} {2020})}\BibitemShut {NoStop}%
\bibitem [{\citenamefont {Zhang}, \citenamefont {Miyamoto},\ and\ \citenamefont
  {Rubio}(2012)}]{Zhang_2012}%
  \BibitemOpen
  \bibfield  {author} {\bibinfo {author} {\bibfnamefont {H.}~\bibnamefont
  {Zhang}}, \bibinfo {author} {\bibfnamefont {Y.}~\bibnamefont {Miyamoto}},\
  and\ \bibinfo {author} {\bibfnamefont {A.}~\bibnamefont {Rubio}},\ }\bibfield
   {title} {\enquote {\bibinfo {title} {{Ab-initio Simulation of Helium-Ion
  Microscopy Images: The Case of Suspended Graphene}},}\ }\href
  {https://doi.org/10.1103/PhysRevLett.109.265505} {\bibfield  {journal}
  {\bibinfo  {journal} {Phys. Rev. Lett.}\ }\textbf {\bibinfo {volume} {109}},\
  \bibinfo {pages} {265505} (\bibinfo {year} {2012})}\BibitemShut {NoStop}%
\bibitem [{\citenamefont {Henkelman}, \citenamefont {Uberuaga},\ and\
  \citenamefont {Jonsson}(2000)}]{NEB-JCP}%
  \BibitemOpen
  \bibfield  {author} {\bibinfo {author} {\bibfnamefont {G.}~\bibnamefont
  {Henkelman}}, \bibinfo {author} {\bibfnamefont {B.~P.}\ \bibnamefont
  {Uberuaga}},\ and\ \bibinfo {author} {\bibfnamefont {H.}~\bibnamefont
  {Jonsson}},\ }\bibfield  {title} {\enquote {\bibinfo {title} {A climbing
  image nudged elastic band method for finding saddle points and minimum energy
  paths},}\ }\href {https://doi.org/10.1063/1.1329672} {\bibfield  {journal}
  {\bibinfo  {journal} {J. Chem. Phys.}\ }\textbf {\bibinfo {volume} {113}},\
  \bibinfo {pages} {9901--9904} (\bibinfo {year} {2000})}\BibitemShut {NoStop}%
\bibitem [{\citenamefont {Voter}(2007)}]{voter_introduction_2007}%
  \BibitemOpen
  \bibfield  {author} {\bibinfo {author} {\bibfnamefont {A.~F.}\ \bibnamefont
  {Voter}},\ }\bibfield  {title} {{\selectlanguage {english}\enquote {\bibinfo
  {title} {Introduction to the kinetic {M}onte {C}arlo method},}\ }}in\ \href
  {https://doi.org/10.1007/978-1-4020-5295-8_1} {{\selectlanguage
  {english}\emph {\bibinfo {booktitle} {Radiation {Effects} in {Solids}}}}},\
  \bibinfo {series and number} {\bibinfo {series} {{NATO} {Science} {Series}}\
  No.\ \bibinfo {number} {235}},\ \bibinfo {editor} {edited by\ \bibinfo
  {editor} {\bibfnamefont {K.~E.}\ \bibnamefont {Sickafus}}, \bibinfo {editor}
  {\bibfnamefont {E.~A.}\ \bibnamefont {Kotomin}},\ and\ \bibinfo {editor}
  {\bibfnamefont {B.~P.}\ \bibnamefont {Uberuaga}}}\ (\bibinfo  {publisher}
  {Springer Netherlands},\ \bibinfo {year} {2007})\ pp.\ \bibinfo {pages}
  {1--23}\BibitemShut {NoStop}%
\bibitem [{\citenamefont {Chatterjee}\ and\ \citenamefont
  {Vlachos}(2007)}]{Chatterjee-2007}%
  \BibitemOpen
  \bibfield  {author} {\bibinfo {author} {\bibfnamefont {A.}~\bibnamefont
  {Chatterjee}}\ and\ \bibinfo {author} {\bibfnamefont {D.~G.}\ \bibnamefont
  {Vlachos}},\ }\bibfield  {title} {\enquote {\bibinfo {title} {{An overview of
  spatial microscopic and accelerated kinetic Monte Carlo methods}},}\ }\href
  {https://doi.org/10.1007/s10820-006-9042-9} {\bibfield  {journal} {\bibinfo
  {journal} {J. Comput.-Aided Mater. Des.}\ }\textbf {\bibinfo {volume} {14}},\
  \bibinfo {pages} {253--308} (\bibinfo {year} {2007})}\BibitemShut {NoStop}%
\bibitem [{\citenamefont {Becquart}, \citenamefont {Mousseau},\ and\
  \citenamefont {Domain}(2020)}]{BECQUART2020}%
  \BibitemOpen
  \bibfield  {author} {\bibinfo {author} {\bibfnamefont {C.~S.}\ \bibnamefont
  {Becquart}}, \bibinfo {author} {\bibfnamefont {N.}~\bibnamefont {Mousseau}},\
  and\ \bibinfo {author} {\bibfnamefont {C.}~\bibnamefont {Domain}},\
  }\bibfield  {title} {\enquote {\bibinfo {title} {1.24 - kinetic monte carlo
  simulations of irradiation effects},}\ }in\ \href
  {https://doi.org/10.1016/b978-0-12-803581-8.11685-6} {\emph {\bibinfo
  {booktitle} {Comprehensive Nuclear Materials}}},\ \bibinfo {editor} {edited
  by\ \bibinfo {editor} {\bibfnamefont {R.~J.~M.}\ \bibnamefont {Konings}}\
  and\ \bibinfo {editor} {\bibfnamefont {R.~E.}\ \bibnamefont {Stoller}}}\
  (\bibinfo  {publisher} {Elsevier},\ \bibinfo {address} {Oxford},\ \bibinfo
  {year} {2020})\ \bibinfo {edition} {second edition}\ ed.,\ pp.\ \bibinfo
  {pages} {754--778}\BibitemShut {NoStop}%
\bibitem [{\citenamefont {Martin-Bragado}\ \emph {et~al.}(2018)\citenamefont
  {Martin-Bragado}, \citenamefont {Borges}, \citenamefont {Balbuena},\ and\
  \citenamefont {Jaraiz}}]{martin-bragado_kinetic_2018}%
  \BibitemOpen
  \bibfield  {author} {\bibinfo {author} {\bibfnamefont {I.}~\bibnamefont
  {Martin-Bragado}}, \bibinfo {author} {\bibfnamefont {R.}~\bibnamefont
  {Borges}}, \bibinfo {author} {\bibfnamefont {J.~P.}\ \bibnamefont
  {Balbuena}},\ and\ \bibinfo {author} {\bibfnamefont {M.}~\bibnamefont
  {Jaraiz}},\ }\bibfield  {title} {{\selectlanguage {english}\enquote {\bibinfo
  {title} {Kinetic {Monte} {Carlo} simulation for semiconductor processing: {A}
  review},}\ }}\href {https://doi.org/10.1016/j.pmatsci.2017.09.003} {\bibfield
   {journal} {\bibinfo  {journal} {Prog. Mater. Sci.}\ }\textbf {\bibinfo
  {volume} {92}},\ \bibinfo {pages} {1--32} (\bibinfo {year}
  {2018})}\BibitemShut {NoStop}%
\bibitem [{\citenamefont {Voter}, \citenamefont {Montalenti},\ and\
  \citenamefont {Germann}(2002)}]{af_voter_extending_2002}%
  \BibitemOpen
  \bibfield  {author} {\bibinfo {author} {\bibfnamefont {A.~F.}\ \bibnamefont
  {Voter}}, \bibinfo {author} {\bibfnamefont {F.}~\bibnamefont {Montalenti}},\
  and\ \bibinfo {author} {\bibfnamefont {T.~C.}\ \bibnamefont {Germann}},\
  }\bibfield  {title} {\enquote {\bibinfo {title} {Extending the time scale in
  atomistic simulation of materials},}\ }\href
  {https://doi.org/10.1146/annurev.matsci.32.112601.141541} {\bibfield
  {journal} {\bibinfo  {journal} {Ann. Rev. Mater. Res.}\ }\textbf {\bibinfo
  {volume} {32}},\ \bibinfo {pages} {321--346} (\bibinfo {year}
  {2002})}\BibitemShut {NoStop}%
\bibitem [{\citenamefont {Xu}, \citenamefont {Osetsky},\ and\ \citenamefont
  {Stoller}(2012)}]{SEAKMC2012}%
  \BibitemOpen
  \bibfield  {author} {\bibinfo {author} {\bibfnamefont {H.}~\bibnamefont
  {Xu}}, \bibinfo {author} {\bibfnamefont {Y.~N.}\ \bibnamefont {Osetsky}},\
  and\ \bibinfo {author} {\bibfnamefont {R.~E.}\ \bibnamefont {Stoller}},\
  }\bibfield  {title} {\enquote {\bibinfo {title} {Self-evolving atomistic
  kinetic {Monte Carlo}: fundamentals and applications},}\ }\href
  {https://doi.org/10.1088/0953-8984/24/37/375402} {\bibfield  {journal}
  {\bibinfo  {journal} {J. Phys.: Condens. Matter}\ }\textbf {\bibinfo {volume}
  {24}},\ \bibinfo {pages} {375402} (\bibinfo {year} {2012})}\BibitemShut
  {NoStop}%
\bibitem [{\citenamefont {Zhou}\ \emph {et~al.}(2019)\citenamefont {Zhou},
  \citenamefont {Jolley}, \citenamefont {Phillips}, \citenamefont {Smith},\
  and\ \citenamefont {Wu}}]{Zhou2019}%
  \BibitemOpen
  \bibfield  {author} {\bibinfo {author} {\bibfnamefont {Y.}~\bibnamefont
  {Zhou}}, \bibinfo {author} {\bibfnamefont {K.}~\bibnamefont {Jolley}},
  \bibinfo {author} {\bibfnamefont {R.}~\bibnamefont {Phillips}}, \bibinfo
  {author} {\bibfnamefont {R.}~\bibnamefont {Smith}},\ and\ \bibinfo {author}
  {\bibfnamefont {H.}~\bibnamefont {Wu}},\ }\bibfield  {title} {\enquote
  {\bibinfo {title} {{Modelling defect evolution in irradiated graphite}},}\
  }\href {https://doi.org/10.1016/J.CARBON.2019.07.092} {\bibfield  {journal}
  {\bibinfo  {journal} {Carbon}\ }\textbf {\bibinfo {volume} {154}},\ \bibinfo
  {pages} {192--202} (\bibinfo {year} {2019})}\BibitemShut {NoStop}%
\bibitem [{\citenamefont {Scott}\ \emph {et~al.}(2011)\citenamefont {Scott},
  \citenamefont {Blackwell}, \citenamefont {Vernon}, \citenamefont {Kenny},
  \citenamefont {Walls},\ and\ \citenamefont {Smith}}]{Scott2011}%
  \BibitemOpen
  \bibfield  {author} {\bibinfo {author} {\bibfnamefont {C.}~\bibnamefont
  {Scott}}, \bibinfo {author} {\bibfnamefont {S.}~\bibnamefont {Blackwell}},
  \bibinfo {author} {\bibfnamefont {L.}~\bibnamefont {Vernon}}, \bibinfo
  {author} {\bibfnamefont {S.}~\bibnamefont {Kenny}}, \bibinfo {author}
  {\bibfnamefont {M.}~\bibnamefont {Walls}},\ and\ \bibinfo {author}
  {\bibfnamefont {R.}~\bibnamefont {Smith}},\ }\bibfield  {title} {\enquote
  {\bibinfo {title} {{Atomistic surface erosion and thin film growth modelled
  over realistic time scales}},}\ }\href {https://doi.org/10.1063/1.3657436}
  {\bibfield  {journal} {\bibinfo  {journal} {J. Chem. Phys.}\ }\textbf
  {\bibinfo {volume} {135}},\ \bibinfo {pages} {174706} (\bibinfo {year}
  {2011})},\ \bibinfo {note} {publisher: American Institute of
  Physics}\BibitemShut {NoStop}%
\bibitem [{\citenamefont {Malerba}\ \emph {et~al.}(2005)\citenamefont
  {Malerba}, \citenamefont {Becquart}, \citenamefont {Hou},\ and\ \citenamefont
  {Domain}}]{malerba_comparison_2005}%
  \BibitemOpen
  \bibfield  {author} {\bibinfo {author} {\bibfnamefont {L.}~\bibnamefont
  {Malerba}}, \bibinfo {author} {\bibfnamefont {C.~S.}\ \bibnamefont
  {Becquart}}, \bibinfo {author} {\bibfnamefont {M.}~\bibnamefont {Hou}},\ and\
  \bibinfo {author} {\bibfnamefont {C.}~\bibnamefont {Domain}},\ }\bibfield
  {title} {\enquote {\bibinfo {title} {Comparison of algorithms for multiscale
  modelling of radiation damage in {Fe}-{Cu} alloys},}\ }\href
  {https://doi.org/10.1080/02678370412331320242} {\bibfield  {journal}
  {\bibinfo  {journal} {Philos. Mag.}\ }\textbf {\bibinfo {volume} {85}},\
  \bibinfo {pages} {417--428} (\bibinfo {year} {2005})},\ \bibinfo {note}
  {publisher: Taylor \& Francis \_eprint:
  https://doi.org/10.1080/02678370412331320242}\BibitemShut {NoStop}%
\bibitem [{\citenamefont {Martin-Bragado}\ and\ \citenamefont
  {Moroz}(2009)}]{martin-bragado_facet_2009}%
  \BibitemOpen
  \bibfield  {author} {\bibinfo {author} {\bibfnamefont {I.}~\bibnamefont
  {Martin-Bragado}}\ and\ \bibinfo {author} {\bibfnamefont {V.}~\bibnamefont
  {Moroz}},\ }\bibfield  {title} {\enquote {\bibinfo {title} {Facet formation
  during solid phase epitaxy regrowth: {A} lattice kinetic {Monte} {Carlo}
  model},}\ }\href {https://doi.org/10.1063/1.3236535} {\bibfield  {journal}
  {\bibinfo  {journal} {Appl. Phys. Lett.}\ }\textbf {\bibinfo {volume} {95}},\
  \bibinfo {pages} {123123--123123--3} (\bibinfo {year} {2009})}\BibitemShut
  {NoStop}%
\bibitem [{\citenamefont {Klingner}\ \emph {et~al.}(2022)\citenamefont
  {Klingner}, \citenamefont {Heinig}, \citenamefont {Tucholski}, \citenamefont
  {M{\"{o}}ller}, \citenamefont {H{\"{u}}bner}, \citenamefont {Bischoff},
  \citenamefont {Hlawacek},\ and\ \citenamefont {Facsko}}]{Klingner_2022}%
  \BibitemOpen
  \bibfield  {author} {\bibinfo {author} {\bibfnamefont {N.}~\bibnamefont
  {Klingner}}, \bibinfo {author} {\bibfnamefont {K.-H.}\ \bibnamefont
  {Heinig}}, \bibinfo {author} {\bibfnamefont {D.}~\bibnamefont {Tucholski}},
  \bibinfo {author} {\bibfnamefont {W.}~\bibnamefont {M{\"{o}}ller}}, \bibinfo
  {author} {\bibfnamefont {R.}~\bibnamefont {H{\"{u}}bner}}, \bibinfo {author}
  {\bibfnamefont {L.}~\bibnamefont {Bischoff}}, \bibinfo {author}
  {\bibfnamefont {G.}~\bibnamefont {Hlawacek}},\ and\ \bibinfo {author}
  {\bibfnamefont {S.}~\bibnamefont {Facsko}},\ }\bibfield  {title} {\enquote
  {\bibinfo {title} {Epitaxial lateral overgrowth of tin spheres driven and
  directly observed by helium ion microscopy},}\ }\href
  {https://doi.org/10.1021/acs.jpcc.2c03707} {\bibfield  {journal} {\bibinfo
  {journal} {J. Phys. Chem. C}\ }\textbf {\bibinfo {volume} {126}},\ \bibinfo
  {pages} {16332--16340} (\bibinfo {year} {2022})}\BibitemShut {NoStop}%
\bibitem [{\citenamefont {B{\"o}ttger}\ \emph {et~al.}(2013)\citenamefont
  {B{\"o}ttger}, \citenamefont {Heinig}, \citenamefont {Bischoff},
  \citenamefont {Liedke},\ and\ \citenamefont {Facsko}}]{Boettger-2013}%
  \BibitemOpen
  \bibfield  {author} {\bibinfo {author} {\bibfnamefont {R.}~\bibnamefont
  {B{\"o}ttger}}, \bibinfo {author} {\bibfnamefont {K.-H.}\ \bibnamefont
  {Heinig}}, \bibinfo {author} {\bibfnamefont {L.}~\bibnamefont {Bischoff}},
  \bibinfo {author} {\bibfnamefont {B.}~\bibnamefont {Liedke}},\ and\ \bibinfo
  {author} {\bibfnamefont {S.}~\bibnamefont {Facsko}},\ }\bibfield  {title}
  {\enquote {\bibinfo {title} {From holes to sponge at irradiated {G}e surfaces
  with increasing ion energy - an effect of defect kinetics?}}\ }\href
  {https://doi.org/10.1007/s00339-013-7911-0} {\bibfield  {journal} {\bibinfo
  {journal} {Appl. Phys. A}\ }\textbf {\bibinfo {volume} {113}},\ \bibinfo
  {pages} {53--59} (\bibinfo {year} {2013})}\BibitemShut {NoStop}%
\bibitem [{\citenamefont {Yang}, \citenamefont {Lively},\ and\ \citenamefont
  {Allain}(2015)}]{Yang-2015}%
  \BibitemOpen
  \bibfield  {author} {\bibinfo {author} {\bibfnamefont {Z.}~\bibnamefont
  {Yang}}, \bibinfo {author} {\bibfnamefont {M.~A.}\ \bibnamefont {Lively}},\
  and\ \bibinfo {author} {\bibfnamefont {J.~P.}\ \bibnamefont {Allain}},\
  }\bibfield  {title} {\enquote {\bibinfo {title} {Kinetic monte carlo
  simulation of self-organized pattern formation induced by ion beam sputtering
  using crater functions},}\ }\href
  {https://doi.org/10.1103/PhysRevB.91.075427} {\bibfield  {journal} {\bibinfo
  {journal} {Phys. Rev. B}\ }\textbf {\bibinfo {volume} {91}},\ \bibinfo
  {pages} {075427} (\bibinfo {year} {2015})}\BibitemShut {NoStop}%
\bibitem [{\citenamefont {Kotakoski}\ \emph {et~al.}(2010)\citenamefont
  {Kotakoski}, \citenamefont {Jin}, \citenamefont {Lehtinen}, \citenamefont
  {Suenaga},\ and\ \citenamefont {Krasheninnikov}}]{Kotakoski-2010}%
  \BibitemOpen
  \bibfield  {author} {\bibinfo {author} {\bibfnamefont {J.}~\bibnamefont
  {Kotakoski}}, \bibinfo {author} {\bibfnamefont {C.~H.}\ \bibnamefont {Jin}},
  \bibinfo {author} {\bibfnamefont {O.}~\bibnamefont {Lehtinen}}, \bibinfo
  {author} {\bibfnamefont {K.}~\bibnamefont {Suenaga}},\ and\ \bibinfo {author}
  {\bibfnamefont {A.~V.}\ \bibnamefont {Krasheninnikov}},\ }\bibfield  {title}
  {\enquote {\bibinfo {title} {Electron knock-on damage in hexagonal boron
  nitride monolayers},}\ }\href {https://doi.org/10.1103/PhysRevB.82.113404}
  {\bibfield  {journal} {\bibinfo  {journal} {Phys. Rev. B}\ }\textbf {\bibinfo
  {volume} {82}},\ \bibinfo {pages} {113404} (\bibinfo {year}
  {2010})}\BibitemShut {NoStop}%
\bibitem [{\citenamefont {Xu}\ \emph {et~al.}(2018)\citenamefont {Xu},
  \citenamefont {Pr{\"{u}}fer}, \citenamefont {Wolf}, \citenamefont
  {Engelmann}, \citenamefont {Bischoff}, \citenamefont {H{\"{u}}bner},
  \citenamefont {Heinig}, \citenamefont {M{\"{o}}ller}, \citenamefont {Facsko},
  \citenamefont {Borany},\ and\ \citenamefont
  {Hlawacek}}]{xu_site-controlled_2018}%
  \BibitemOpen
  \bibfield  {author} {\bibinfo {author} {\bibfnamefont {X.}~\bibnamefont
  {Xu}}, \bibinfo {author} {\bibfnamefont {T.}~\bibnamefont {Pr{\"{u}}fer}},
  \bibinfo {author} {\bibfnamefont {D.}~\bibnamefont {Wolf}}, \bibinfo {author}
  {\bibfnamefont {H.-J.}\ \bibnamefont {Engelmann}}, \bibinfo {author}
  {\bibfnamefont {L.}~\bibnamefont {Bischoff}}, \bibinfo {author}
  {\bibfnamefont {R.}~\bibnamefont {H{\"{u}}bner}}, \bibinfo {author}
  {\bibfnamefont {K.-H.}\ \bibnamefont {Heinig}}, \bibinfo {author}
  {\bibfnamefont {W.}~\bibnamefont {M{\"{o}}ller}}, \bibinfo {author}
  {\bibfnamefont {S.}~\bibnamefont {Facsko}}, \bibinfo {author} {\bibfnamefont
  {J.~v.}\ \bibnamefont {Borany}},\ and\ \bibinfo {author} {\bibfnamefont
  {G.}~\bibnamefont {Hlawacek}},\ }\bibfield  {title} {{\selectlanguage
  {english}\enquote {\bibinfo {title} {Site-controlled formation of single {Si}
  nanocrystals in a buried {SiO2} matrix using ion beam mixing},}\ }}\href
  {https://doi.org/10.3762/bjnano.9.267} {\bibfield  {journal} {\bibinfo
  {journal} {Beilstein J. Nanotechnol.}\ }\textbf {\bibinfo {volume} {9}},\
  \bibinfo {pages} {2883--2892} (\bibinfo {year} {2018})},\ \bibinfo {note}
  {publisher: Beilstein-Institut}\BibitemShut {NoStop}%
\bibitem [{\citenamefont {Domain}, \citenamefont {Becquart},\ and\
  \citenamefont {Malerba}(2004)}]{Domain2004}%
  \BibitemOpen
  \bibfield  {author} {\bibinfo {author} {\bibfnamefont {C.}~\bibnamefont
  {Domain}}, \bibinfo {author} {\bibfnamefont {C.}~\bibnamefont {Becquart}},\
  and\ \bibinfo {author} {\bibfnamefont {L.}~\bibnamefont {Malerba}},\
  }\bibfield  {title} {\enquote {\bibinfo {title} {Simulation of radiation
  damage in fe alloys: an object kinetic monte carlo approach},}\ }\href
  {https://doi.org/10.1016/j.jnucmat.2004.07.037} {\bibfield  {journal}
  {\bibinfo  {journal} {J. Nucl. Mater.}\ }\textbf {\bibinfo {volume} {335}},\
  \bibinfo {pages} {121--145} (\bibinfo {year} {2004})}\BibitemShut {NoStop}%
\bibitem [{\citenamefont {Aboy}\ \emph {et~al.}(2011)\citenamefont {Aboy},
  \citenamefont {Pelaz}, \citenamefont {Bruno}, \citenamefont {Mirabella},\
  and\ \citenamefont {Boninelli}}]{Aboy2011}%
  \BibitemOpen
  \bibfield  {author} {\bibinfo {author} {\bibfnamefont {M.}~\bibnamefont
  {Aboy}}, \bibinfo {author} {\bibfnamefont {L.}~\bibnamefont {Pelaz}},
  \bibinfo {author} {\bibfnamefont {E.}~\bibnamefont {Bruno}}, \bibinfo
  {author} {\bibfnamefont {S.}~\bibnamefont {Mirabella}},\ and\ \bibinfo
  {author} {\bibfnamefont {S.}~\bibnamefont {Boninelli}},\ }\bibfield  {title}
  {\enquote {\bibinfo {title} {Kinetics of large {B} clusters in crystalline
  and preamorphized silicon},}\ }\href {https://doi.org/10.1063/1.3639280}
  {\bibfield  {journal} {\bibinfo  {journal} {J. Appl. Phys.}\ }\textbf
  {\bibinfo {volume} {110}} (\bibinfo {year} {2011}),\
  10.1063/1.3639280}\BibitemShut {NoStop}%
\bibitem [{\citenamefont {Runge}(1977)}]{runge_distribution_1977}%
  \BibitemOpen
  \bibfield  {author} {\bibinfo {author} {\bibfnamefont {H.}~\bibnamefont
  {Runge}},\ }\bibfield  {title} {\enquote {\bibinfo {title} {Distribution of
  implanted ions under arbitrarily shaped mask edges},}\ }\href
  {https://doi.org/10.1002/pssa.2210390228} {\bibfield  {journal} {\bibinfo
  {journal} {Phys. Status Solidi A}\ }\textbf {\bibinfo {volume} {39}},\
  \bibinfo {pages} {595--599} (\bibinfo {year} {1977})}\BibitemShut {NoStop}%
\bibitem [{\citenamefont {Vasile}, \citenamefont {Nassar},\ and\ \citenamefont
  {Xie}(1998)}]{vasile_focused_1998}%
  \BibitemOpen
  \bibfield  {author} {\bibinfo {author} {\bibfnamefont {M.~J.}\ \bibnamefont
  {Vasile}}, \bibinfo {author} {\bibfnamefont {R.}~\bibnamefont {Nassar}},\
  and\ \bibinfo {author} {\bibfnamefont {J.}~\bibnamefont {Xie}},\ }\bibfield
  {title} {\enquote {\bibinfo {title} {Focused ion beam technology applied to
  microstructure fabrication},}\ }in\ \href {https://doi.org/10.1116/1.590198}
  {\emph {\bibinfo {booktitle} {Papers from the {Japan}/{U}.{S}. workshop on
  formation of ion nanobeams}}},\ Vol.~\bibinfo {volume} {16}\ (\bibinfo
  {publisher} {AVS},\ \bibinfo {address} {Osaka (Japan)},\ \bibinfo {year}
  {1998})\ pp.\ \bibinfo {pages} {2499--2505}\BibitemShut {NoStop}%
\bibitem [{\citenamefont {Sethian}(1999)}]{j.a._sethian_level_1999}%
  \BibitemOpen
  \bibfield  {author} {\bibinfo {author} {\bibfnamefont {J.~A.}\ \bibnamefont
  {Sethian}},\ }\href@noop {} {\emph {\bibinfo {title} {Level {Set} {Methods}
  and {Fast} {Marching} {Methods}}}},\ Cambridge {Monographs} on {Applied} and
  {Computational} {Methematics}\ (\bibinfo  {publisher} {Cambridge University
  Press},\ \bibinfo {year} {1999})\BibitemShut {NoStop}%
\bibitem [{\citenamefont {Hobler}, \citenamefont {Zahel},\ and\ \citenamefont
  {Lindsey}(2011)}]{g_hobler_t_zahel_s_lindsey_pytopsim_2011}%
  \BibitemOpen
  \bibfield  {author} {\bibinfo {author} {\bibfnamefont {G.}~\bibnamefont
  {Hobler}}, \bibinfo {author} {\bibfnamefont {T.}~\bibnamefont {Zahel}},\ and\
  \bibinfo {author} {\bibfnamefont {S.}~\bibnamefont {Lindsey}},\ }\href
  {https://github.com/hobler/PyTopSim} {\enquote {\bibinfo {title}
  {{PyTopSim}},}\ }\bibinfo {type} {software}\ (\bibinfo  {institution} {TU
  Wien},\ \bibinfo {year} {2011})\BibitemShut {NoStop}%
\bibitem [{\citenamefont {M{\"{u}}ller}, \citenamefont {Weigmann},\ and\
  \citenamefont {Burghause}(1986)}]{muller_simulation_1986}%
  \BibitemOpen
  \bibfield  {author} {\bibinfo {author} {\bibfnamefont {K.~P.}\ \bibnamefont
  {M{\"{u}}ller}}, \bibinfo {author} {\bibfnamefont {U.}~\bibnamefont
  {Weigmann}},\ and\ \bibinfo {author} {\bibfnamefont {H.}~\bibnamefont
  {Burghause}},\ }\bibfield  {title} {{\selectlanguage {english}\enquote
  {\bibinfo {title} {Simulation of focused ion beam milling},}\ }}\href
  {https://doi.org/10.1016/0167-9317(86)90081-X} {\bibfield  {journal}
  {\bibinfo  {journal} {Microelectron. Eng.}\ }\textbf {\bibinfo {volume}
  {5}},\ \bibinfo {pages} {481--489} (\bibinfo {year} {1986})}\BibitemShut
  {NoStop}%
\bibitem [{\citenamefont {Itoh}, \citenamefont {Shimase},\ and\ \citenamefont
  {Haraichi}(1990)}]{itoh_two-dimensional_1990}%
  \BibitemOpen
  \bibfield  {author} {\bibinfo {author} {\bibfnamefont {F.}~\bibnamefont
  {Itoh}}, \bibinfo {author} {\bibfnamefont {A.}~\bibnamefont {Shimase}},\ and\
  \bibinfo {author} {\bibfnamefont {S.}~\bibnamefont {Haraichi}},\ }\bibfield
  {title} {\enquote {\bibinfo {title} {Two-dimensional profile simulation of
  focused ion-beam milling of {LSI}},}\ }\href
  {https://doi.org/10.1149/1.2086592} {\bibfield  {journal} {\bibinfo
  {journal} {J. Electrochem. Soc.}\ }\textbf {\bibinfo {volume} {137}},\
  \bibinfo {pages} {983--988} (\bibinfo {year} {1990})}\BibitemShut {NoStop}%
\bibitem [{\citenamefont {Katardjiev}\ \emph {et~al.}(1994)\citenamefont
  {Katardjiev}, \citenamefont {Carter}, \citenamefont {Nobes}, \citenamefont
  {Berg},\ and\ \citenamefont {Blom}}]{katardjiev_three-dimensional_1994}%
  \BibitemOpen
  \bibfield  {author} {\bibinfo {author} {\bibfnamefont {I.~V.}\ \bibnamefont
  {Katardjiev}}, \bibinfo {author} {\bibfnamefont {G.}~\bibnamefont {Carter}},
  \bibinfo {author} {\bibfnamefont {M.~J.}\ \bibnamefont {Nobes}}, \bibinfo
  {author} {\bibfnamefont {S.}~\bibnamefont {Berg}},\ and\ \bibinfo {author}
  {\bibfnamefont {H.-O.}\ \bibnamefont {Blom}},\ }\bibfield  {title} {\enquote
  {\bibinfo {title} {Three-dimensional simulation of surface evolution during
  growth and erosion},}\ }\href {https://doi.org/10.1116/1.578859} {\bibfield
  {journal} {\bibinfo  {journal} {J. Vac. Sci. Technol. A}\ }\textbf {\bibinfo
  {volume} {12}},\ \bibinfo {pages} {61--68} (\bibinfo {year}
  {1994})}\BibitemShut {NoStop}%
\bibitem [{\citenamefont {Platzgummer}\ \emph {et~al.}(2006)\citenamefont
  {Platzgummer}, \citenamefont {Biedermann}, \citenamefont {Langfischer},
  \citenamefont {Eder-Kapl}, \citenamefont {Kuemmel}, \citenamefont {Cernusca},
  \citenamefont {Loeschner}, \citenamefont {Lehrer}, \citenamefont {Frey},
  \citenamefont {Lugstein},\ and\ \citenamefont
  {Bertagnolli}}]{platzgummer_simulation_2006}%
  \BibitemOpen
  \bibfield  {author} {\bibinfo {author} {\bibfnamefont {E.}~\bibnamefont
  {Platzgummer}}, \bibinfo {author} {\bibfnamefont {A.}~\bibnamefont
  {Biedermann}}, \bibinfo {author} {\bibfnamefont {H.}~\bibnamefont
  {Langfischer}}, \bibinfo {author} {\bibfnamefont {S.}~\bibnamefont
  {Eder-Kapl}}, \bibinfo {author} {\bibfnamefont {M.}~\bibnamefont {Kuemmel}},
  \bibinfo {author} {\bibfnamefont {S.}~\bibnamefont {Cernusca}}, \bibinfo
  {author} {\bibfnamefont {H.}~\bibnamefont {Loeschner}}, \bibinfo {author}
  {\bibfnamefont {C.}~\bibnamefont {Lehrer}}, \bibinfo {author} {\bibfnamefont
  {L.}~\bibnamefont {Frey}}, \bibinfo {author} {\bibfnamefont {A.}~\bibnamefont
  {Lugstein}},\ and\ \bibinfo {author} {\bibfnamefont {E.}~\bibnamefont
  {Bertagnolli}},\ }\bibfield  {title} {\enquote {\bibinfo {title} {Simulation
  of ion beam direct structuring for {3D} nanoimprint template fabrication},}\
  }\href {https://doi.org/10.1016/j.mee.2006.01.140} {\bibfield  {journal}
  {\bibinfo  {journal} {Microelectron. Eng.}\ }\textbf {\bibinfo {volume}
  {83}},\ \bibinfo {pages} {936--939} (\bibinfo {year} {2006})}\BibitemShut
  {NoStop}%
\bibitem [{\citenamefont {Kim}\ \emph {et~al.}(2007{\natexlab{b}})\citenamefont
  {Kim}, \citenamefont {Hobler}, \citenamefont {Lugstein},\ and\ \citenamefont
  {Bertagnolli}}]{kim_simulation_2007}%
  \BibitemOpen
  \bibfield  {author} {\bibinfo {author} {\bibfnamefont {H.-B.}\ \bibnamefont
  {Kim}}, \bibinfo {author} {\bibfnamefont {G.}~\bibnamefont {Hobler}},
  \bibinfo {author} {\bibfnamefont {A.}~\bibnamefont {Lugstein}},\ and\
  \bibinfo {author} {\bibfnamefont {E.}~\bibnamefont {Bertagnolli}},\
  }\bibfield  {title} {\enquote {\bibinfo {title} {Simulation of ion beam
  induced micro/nano fabrication},}\ }\href
  {https://doi.org/10.1088/0960-1317/17/6/011} {\bibfield  {journal} {\bibinfo
  {journal} {J. Micromech. Microeng.}\ }\textbf {\bibinfo {volume} {17}},\
  \bibinfo {pages} {1178--1183} (\bibinfo {year}
  {2007}{\natexlab{b}})}\BibitemShut {NoStop}%
\bibitem [{\citenamefont {Kim}\ \emph {et~al.}(2007{\natexlab{c}})\citenamefont
  {Kim}, \citenamefont {Hobler}, \citenamefont {Steiger}, \citenamefont
  {Lugstein},\ and\ \citenamefont {Bertagnolli}}]{kim_full_2007}%
  \BibitemOpen
  \bibfield  {author} {\bibinfo {author} {\bibfnamefont {H.-B.}\ \bibnamefont
  {Kim}}, \bibinfo {author} {\bibfnamefont {G.}~\bibnamefont {Hobler}},
  \bibinfo {author} {\bibfnamefont {A.}~\bibnamefont {Steiger}}, \bibinfo
  {author} {\bibfnamefont {A.}~\bibnamefont {Lugstein}},\ and\ \bibinfo
  {author} {\bibfnamefont {E.}~\bibnamefont {Bertagnolli}},\ }\bibfield
  {title} {\enquote {\bibinfo {title} {Full three-dimensional simulation of
  focused ion beam micro/nanofabrication},}\ }\href
  {https://doi.org/10.1088/0957-4484/18/24/245303} {\bibfield  {journal}
  {\bibinfo  {journal} {Nanotechnology}\ }\textbf {\bibinfo {volume} {18}},\
  \bibinfo {pages} {245303} (\bibinfo {year} {2007}{\natexlab{c}})}\BibitemShut
  {NoStop}%
\bibitem [{\citenamefont {Ertl}, \citenamefont {Filipovi{\'{c}}},\ and\
  \citenamefont {Selberherr}(2010)}]{ertl_three-dimensional_2010}%
  \BibitemOpen
  \bibfield  {author} {\bibinfo {author} {\bibfnamefont {O.}~\bibnamefont
  {Ertl}}, \bibinfo {author} {\bibfnamefont {L.}~\bibnamefont
  {Filipovi{\'{c}}}},\ and\ \bibinfo {author} {\bibfnamefont {S.}~\bibnamefont
  {Selberherr}},\ }\bibfield  {title} {\enquote {\bibinfo {title}
  {Three-{Dimensional} {Simulation} of {Focused} {Ion} {Beam} {Processing}
  {Using} the {Level} {Set} {Method}},}\ }in\ \href
  {https://doi.org/10.1109/SISPAD.2010.5604573} {\emph {\bibinfo {booktitle}
  {Proc. {SISPAD} 2010}}}\ (\bibinfo  {publisher} {IEEE},\ \bibinfo {year}
  {2010})\ pp.\ \bibinfo {pages} {49--52}\BibitemShut {NoStop}%
\bibitem [{\citenamefont {Sun}\ \emph {et~al.}(2012)\citenamefont {Sun},
  \citenamefont {Luo}, \citenamefont {Ritchie}, \citenamefont {Chang},\ and\
  \citenamefont {Wang}}]{sun_investigation_2012}%
  \BibitemOpen
  \bibfield  {author} {\bibinfo {author} {\bibfnamefont {J.}~\bibnamefont
  {Sun}}, \bibinfo {author} {\bibfnamefont {X.}~\bibnamefont {Luo}}, \bibinfo
  {author} {\bibfnamefont {J.~M.}\ \bibnamefont {Ritchie}}, \bibinfo {author}
  {\bibfnamefont {W.}~\bibnamefont {Chang}},\ and\ \bibinfo {author}
  {\bibfnamefont {W.}~\bibnamefont {Wang}},\ }\bibfield  {title} {\enquote
  {\bibinfo {title} {An investigation of redeposition effect for deterministic
  fabrication of nanodots by focused ion beam},}\ }\href
  {https://doi.org/10.1016/j.precisioneng.2011.07.001} {\bibfield  {journal}
  {\bibinfo  {journal} {Precision Engineering}\ }\textbf {\bibinfo {volume}
  {36}},\ \bibinfo {pages} {31--36} (\bibinfo {year} {2012})}\BibitemShut
  {NoStop}%
\bibitem [{\citenamefont {Borgardt}\ \emph {et~al.}(2015)\citenamefont
  {Borgardt}, \citenamefont {Volkov}, \citenamefont {Rumyantsev},\ and\
  \citenamefont {Chaplygin}}]{borgardt_simulation_2015}%
  \BibitemOpen
  \bibfield  {author} {\bibinfo {author} {\bibfnamefont {N.~I.}\ \bibnamefont
  {Borgardt}}, \bibinfo {author} {\bibfnamefont {R.~L.}\ \bibnamefont
  {Volkov}}, \bibinfo {author} {\bibfnamefont {A.~V.}\ \bibnamefont
  {Rumyantsev}},\ and\ \bibinfo {author} {\bibfnamefont {Y.~A.}\ \bibnamefont
  {Chaplygin}},\ }\bibfield  {title} {{\selectlanguage {english}\enquote
  {\bibinfo {title} {Simulation of material sputtering with a focused ion
  beam},}\ }}\href {https://doi.org/10.1134/S106378501506019X} {\bibfield
  {journal} {\bibinfo  {journal} {Tech. Phys. Lett.}\ }\textbf {\bibinfo
  {volume} {41}},\ \bibinfo {pages} {610--613} (\bibinfo {year}
  {2015})}\BibitemShut {NoStop}%
\bibitem [{\citenamefont {Rumyantsev}\ \emph {et~al.}(2022)\citenamefont
  {Rumyantsev}, \citenamefont {Borgardt}, \citenamefont {Volkov},\ and\
  \citenamefont {Chaplygin}}]{rumyantsev_study_2022}%
  \BibitemOpen
  \bibfield  {author} {\bibinfo {author} {\bibfnamefont {A.~V.}\ \bibnamefont
  {Rumyantsev}}, \bibinfo {author} {\bibfnamefont {N.~I.}\ \bibnamefont
  {Borgardt}}, \bibinfo {author} {\bibfnamefont {R.~L.}\ \bibnamefont
  {Volkov}},\ and\ \bibinfo {author} {\bibfnamefont {Y.~A.}\ \bibnamefont
  {Chaplygin}},\ }\bibfield  {title} {{\selectlanguage {english}\enquote
  {\bibinfo {title} {Study of silicon dioxide focused ion beam sputtering using
  electron microscopy imaging and level set simulation},}\ }}\href
  {https://doi.org/10.1016/j.vacuum.2022.111128} {\bibfield  {journal}
  {\bibinfo  {journal} {Vacuum}\ }\textbf {\bibinfo {volume} {202}},\ \bibinfo
  {pages} {111128} (\bibinfo {year} {2022})}\BibitemShut {NoStop}%
\bibitem [{\citenamefont {Cupak}\ \emph {et~al.}(2021)\citenamefont {Cupak},
  \citenamefont {Szabo}, \citenamefont {Biber}, \citenamefont {Stadlmayr},
  \citenamefont {Grave}, \citenamefont {Fellinger}, \citenamefont
  {Br{\"{o}}tzner}, \citenamefont {Wilhelm}, \citenamefont {M{\"{o}}ller},
  \citenamefont {Mutzke}, \citenamefont {Moro},\ and\ \citenamefont
  {Aumayr}}]{cupak_sputter_2021}%
  \BibitemOpen
  \bibfield  {author} {\bibinfo {author} {\bibfnamefont {C.}~\bibnamefont
  {Cupak}}, \bibinfo {author} {\bibfnamefont {P.~S.}\ \bibnamefont {Szabo}},
  \bibinfo {author} {\bibfnamefont {H.}~\bibnamefont {Biber}}, \bibinfo
  {author} {\bibfnamefont {R.}~\bibnamefont {Stadlmayr}}, \bibinfo {author}
  {\bibfnamefont {C.}~\bibnamefont {Grave}}, \bibinfo {author} {\bibfnamefont
  {M.}~\bibnamefont {Fellinger}}, \bibinfo {author} {\bibfnamefont
  {J.}~\bibnamefont {Br{\"{o}}tzner}}, \bibinfo {author} {\bibfnamefont
  {R.~A.}\ \bibnamefont {Wilhelm}}, \bibinfo {author} {\bibfnamefont
  {W.}~\bibnamefont {M{\"{o}}ller}}, \bibinfo {author} {\bibfnamefont
  {A.}~\bibnamefont {Mutzke}}, \bibinfo {author} {\bibfnamefont {M.~V.}\
  \bibnamefont {Moro}},\ and\ \bibinfo {author} {\bibfnamefont
  {F.}~\bibnamefont {Aumayr}},\ }\bibfield  {title} {{\selectlanguage
  {english}\enquote {\bibinfo {title} {Sputter yields of rough surfaces:
  {Importance} of the mean surface inclination angle from nano- to microscopic
  rough regimes},}\ }}\href {https://doi.org/10.1016/j.apsusc.2021.151204}
  {\bibfield  {journal} {\bibinfo  {journal} {Appl. Surf. Sci.}\ ,\ \bibinfo
  {pages} {151204}} (\bibinfo {year} {2021})}\BibitemShut {NoStop}%
\bibitem [{\citenamefont {Bradley}\ and\ \citenamefont
  {Hofs{\"{a}}ss}(2016)}]{bradley_nanoscale_2016}%
  \BibitemOpen
  \bibfield  {author} {\bibinfo {author} {\bibfnamefont {R.~M.}\ \bibnamefont
  {Bradley}}\ and\ \bibinfo {author} {\bibfnamefont {H.}~\bibnamefont
  {Hofs{\"{a}}ss}},\ }\bibfield  {title} {\enquote {\bibinfo {title} {Nanoscale
  patterns produced by self-sputtering of solid surfaces: {The} effect of ion
  implantation},}\ }\href {https://doi.org/10.1063/1.4960807} {\bibfield
  {journal} {\bibinfo  {journal} {J. Appl. Phys.}\ }\textbf {\bibinfo {volume}
  {120}},\ \bibinfo {pages} {074302} (\bibinfo {year} {2016})}\BibitemShut
  {NoStop}%
\bibitem [{\citenamefont {Mutunga}\ \emph {et~al.}(2019)\citenamefont
  {Mutunga}, \citenamefont {Winkler}, \citenamefont {Sattelkow}, \citenamefont
  {Rack}, \citenamefont {Plank},\ and\ \citenamefont
  {Fowlkes}}]{mutunga_impact_2019}%
  \BibitemOpen
  \bibfield  {author} {\bibinfo {author} {\bibfnamefont {E.}~\bibnamefont
  {Mutunga}}, \bibinfo {author} {\bibfnamefont {R.}~\bibnamefont {Winkler}},
  \bibinfo {author} {\bibfnamefont {J.}~\bibnamefont {Sattelkow}}, \bibinfo
  {author} {\bibfnamefont {P.~D.}\ \bibnamefont {Rack}}, \bibinfo {author}
  {\bibfnamefont {H.}~\bibnamefont {Plank}},\ and\ \bibinfo {author}
  {\bibfnamefont {J.~D.}\ \bibnamefont {Fowlkes}},\ }\bibfield  {title}
  {\enquote {\bibinfo {title} {Impact of {Electron}-{Beam} {Heating} during
  {3D} {Nanoprinting}},}\ }\href {https://doi.org/10.1021/acsnano.8b09341}
  {\bibfield  {journal} {\bibinfo  {journal} {ACS Nano}\ }\textbf {\bibinfo
  {volume} {13}},\ \bibinfo {pages} {5198--5213} (\bibinfo {year} {2019})},\
  \bibinfo {note} {publisher: American Chemical Society}\BibitemShut {NoStop}%
\bibitem [{\citenamefont {Wolff}\ \emph {et~al.}(2018)\citenamefont {Wolff},
  \citenamefont {Klingner}, \citenamefont {Thompson}, \citenamefont {Zhou},
  \citenamefont {Lin}, \citenamefont {Peng}, \citenamefont {Ramshaw},\ and\
  \citenamefont {Xiao}}]{Wolff_2018}%
  \BibitemOpen
  \bibfield  {author} {\bibinfo {author} {\bibfnamefont {A.}~\bibnamefont
  {Wolff}}, \bibinfo {author} {\bibfnamefont {N.}~\bibnamefont {Klingner}},
  \bibinfo {author} {\bibfnamefont {W.}~\bibnamefont {Thompson}}, \bibinfo
  {author} {\bibfnamefont {Y.}~\bibnamefont {Zhou}}, \bibinfo {author}
  {\bibfnamefont {J.}~\bibnamefont {Lin}}, \bibinfo {author} {\bibfnamefont
  {Y.~Y.}\ \bibnamefont {Peng}}, \bibinfo {author} {\bibfnamefont {J.~A.~M.}\
  \bibnamefont {Ramshaw}},\ and\ \bibinfo {author} {\bibfnamefont
  {Y.}~\bibnamefont {Xiao}},\ }\bibfield  {title} {{\selectlanguage
  {english}\enquote {\bibinfo {title} {Modelling of focused ion beam induced
  increases in sample temperature: a case study of heat damage in biological
  samples},}\ }}\href {https://doi.org/10.1111/jmi.12731} {\bibfield  {journal}
  {\bibinfo  {journal} {J. Microsc.}\ }\textbf {\bibinfo {volume} {272}},\
  \bibinfo {pages} {47--59} (\bibinfo {year} {2018})},\ \bibinfo {note}
  {\_eprint:
  https://onlinelibrary.wiley.com/doi/pdf/10.1111/jmi.12731}\BibitemShut
  {NoStop}%
\bibitem [{\citenamefont {Utke}, \citenamefont {Hoffmann},\ and\ \citenamefont
  {Melngailis}(2008)}]{Utke_2008}%
  \BibitemOpen
  \bibfield  {author} {\bibinfo {author} {\bibfnamefont {I.}~\bibnamefont
  {Utke}}, \bibinfo {author} {\bibfnamefont {P.}~\bibnamefont {Hoffmann}},\
  and\ \bibinfo {author} {\bibfnamefont {J.}~\bibnamefont {Melngailis}},\
  }\bibfield  {title} {\enquote {\bibinfo {title} {Gas-assisted focused
  electron beam and ion beam processing and fabrication},}\ }\href
  {https://doi.org/10.1116/1.2955728} {\bibfield  {journal} {\bibinfo
  {journal} {J. Vac. Sci. Technol. B}\ }\textbf {\bibinfo {volume} {26}},\
  \bibinfo {pages} {1197} (\bibinfo {year} {2008})}\BibitemShut {NoStop}%
\bibitem [{\citenamefont {Utke}(2012)}]{i_utke_continuum_2012}%
  \BibitemOpen
  \bibfield  {author} {\bibinfo {author} {\bibfnamefont {I.}~\bibnamefont
  {Utke}},\ }\bibfield  {title} {\enquote {\bibinfo {title} {Continuum {Models}
  for {One} {Adsorbate} {Species}},}\ }in\ \href@noop {} {\emph {\bibinfo
  {booktitle} {Nanofabrication {Using} {Focused} {Ion} and {Electron} {Beams}:
  {Principles} and {Applications}}}},\ \bibinfo {editor} {edited by\ \bibinfo
  {editor} {\bibfnamefont {I.}~\bibnamefont {Utke}}, \bibinfo {editor}
  {\bibfnamefont {S.}~\bibnamefont {Moshkalev}},\ and\ \bibinfo {editor}
  {\bibfnamefont {P.}~\bibnamefont {Russell}}}\ (\bibinfo  {publisher} {Oxford
  University Press},\ \bibinfo {year} {2012})\ pp.\ \bibinfo {pages}
  {248--285}\BibitemShut {NoStop}%
\bibitem [{\citenamefont {Toth}\ \emph {et~al.}(2015)\citenamefont {Toth},
  \citenamefont {Lobo}, \citenamefont {Friedli}, \citenamefont {Szkudlarek},\
  and\ \citenamefont {Utke}}]{toth_continuum_2015}%
  \BibitemOpen
  \bibfield  {author} {\bibinfo {author} {\bibfnamefont {M.}~\bibnamefont
  {Toth}}, \bibinfo {author} {\bibfnamefont {C.}~\bibnamefont {Lobo}}, \bibinfo
  {author} {\bibfnamefont {V.}~\bibnamefont {Friedli}}, \bibinfo {author}
  {\bibfnamefont {A.}~\bibnamefont {Szkudlarek}},\ and\ \bibinfo {author}
  {\bibfnamefont {I.}~\bibnamefont {Utke}},\ }\bibfield  {title}
  {{\selectlanguage {english}\enquote {\bibinfo {title} {Continuum models of
  focused electron beam induced processing},}\ }}\href
  {https://doi.org/10.3762/bjnano.6.157} {\bibfield  {journal} {\bibinfo
  {journal} {Beilstein J. Nanotechnol.}\ }\textbf {\bibinfo {volume} {6}},\
  \bibinfo {pages} {1518--1540} (\bibinfo {year} {2015})},\ \bibinfo {note}
  {publisher: Beilstein-Institut}\BibitemShut {NoStop}%
\bibitem [{\citenamefont {Shorubalko}, \citenamefont {Pillatsch},\ and\
  \citenamefont {Utke}(2016)}]{Shorubalko_2016}%
  \BibitemOpen
  \bibfield  {author} {\bibinfo {author} {\bibfnamefont {I.}~\bibnamefont
  {Shorubalko}}, \bibinfo {author} {\bibfnamefont {L.}~\bibnamefont
  {Pillatsch}},\ and\ \bibinfo {author} {\bibfnamefont {I.}~\bibnamefont
  {Utke}},\ }\bibfield  {title} {\enquote {\bibinfo {title} {Direct-write
  milling and deposition with noble gases},}\ }in\  \cite{Hlawacek_2016},\ pp.\
  \bibinfo {pages} {355--393}\BibitemShut {NoStop}%
\bibitem [{\citenamefont {Mahady}\ \emph {et~al.}(2017)\citenamefont {Mahady},
  \citenamefont {Tan}, \citenamefont {Greenzweig}, \citenamefont {Livengood},
  \citenamefont {Raveh}, \citenamefont {Fowlkes},\ and\ \citenamefont
  {Rack}}]{mahady_monte_2017-1}%
  \BibitemOpen
  \bibfield  {author} {\bibinfo {author} {\bibfnamefont {K.}~\bibnamefont
  {Mahady}}, \bibinfo {author} {\bibfnamefont {S.}~\bibnamefont {Tan}},
  \bibinfo {author} {\bibfnamefont {Y.}~\bibnamefont {Greenzweig}}, \bibinfo
  {author} {\bibfnamefont {R.}~\bibnamefont {Livengood}}, \bibinfo {author}
  {\bibfnamefont {A.}~\bibnamefont {Raveh}}, \bibinfo {author} {\bibfnamefont
  {J.~D.}\ \bibnamefont {Fowlkes}},\ and\ \bibinfo {author} {\bibfnamefont
  {P.}~\bibnamefont {Rack}},\ }\bibfield  {title} {\enquote {\bibinfo {title}
  {Monte {Carlo} simulations of secondary electron emission due to ion beam
  milling},}\ }\href {https://doi.org/10.1116/1.4994801} {\bibfield  {journal}
  {\bibinfo  {journal} {J. Vac. Sci. Technol. B}\ }\textbf {\bibinfo {volume}
  {35}},\ \bibinfo {pages} {041805} (\bibinfo {year} {2017})}\BibitemShut
  {NoStop}%
\bibitem [{\citenamefont {Hagstrum}(1954)}]{hagstrum_theory_1954}%
  \BibitemOpen
  \bibfield  {author} {\bibinfo {author} {\bibfnamefont {H.~D.}\ \bibnamefont
  {Hagstrum}},\ }\bibfield  {title} {\enquote {\bibinfo {title} {Theory of
  {Auger} {Ejection} of {Electrons} from {Metals} by {Ions}},}\ }\href
  {https://doi.org/10.1103/PhysRev.96.336} {\bibfield  {journal} {\bibinfo
  {journal} {Phys. Rev.}\ }\textbf {\bibinfo {volume} {96}},\ \bibinfo {pages}
  {336--365} (\bibinfo {year} {1954})},\ \bibinfo {note} {publisher: American
  Physical Society}\BibitemShut {NoStop}%
\bibitem [{\citenamefont {Hagstrum}(1961)}]{hagstrum_theory_1961}%
  \BibitemOpen
  \bibfield  {author} {\bibinfo {author} {\bibfnamefont {H.~D.}\ \bibnamefont
  {Hagstrum}},\ }\bibfield  {title} {\enquote {\bibinfo {title} {Theory of
  {Auger} {Neutralization} of {Ions} at the {Surface} of a {Diamond}-{Type}
  {Semiconductor}},}\ }\href {https://doi.org/10.1103/PhysRev.122.83}
  {\bibfield  {journal} {\bibinfo  {journal} {Phys. Rev.}\ }\textbf {\bibinfo
  {volume} {122}},\ \bibinfo {pages} {83--113} (\bibinfo {year} {1961})},\
  \bibinfo {note} {publisher: American Physical Society}\BibitemShut {NoStop}%
\bibitem [{\citenamefont {Bercx}, \citenamefont {Partoens},\ and\ \citenamefont
  {Lamoen}(2019)}]{bercx_quantitative_2019}%
  \BibitemOpen
  \bibfield  {author} {\bibinfo {author} {\bibfnamefont {M.}~\bibnamefont
  {Bercx}}, \bibinfo {author} {\bibfnamefont {B.}~\bibnamefont {Partoens}},\
  and\ \bibinfo {author} {\bibfnamefont {D.}~\bibnamefont {Lamoen}},\
  }\bibfield  {title} {\enquote {\bibinfo {title} {Quantitative modeling of
  secondary electron emission from slow-ion bombardment on semiconductors},}\
  }\href {https://doi.org/10.1103/PhysRevB.99.085413} {\bibfield  {journal}
  {\bibinfo  {journal} {Phys. Rev. B}\ }\textbf {\bibinfo {volume} {99}},\
  \bibinfo {pages} {085413} (\bibinfo {year} {2019})},\ \bibinfo {note}
  {publisher: American Physical Society}\BibitemShut {NoStop}%
\bibitem [{\citenamefont {Prosvetov}\ \emph {et~al.}(2023)\citenamefont
  {Prosvetov}, \citenamefont {Verkhovtsev}, \citenamefont {Sushko},\ and\
  \citenamefont {Solov{\textquoteright}yov}}]{Prosvetov_2023}%
  \BibitemOpen
  \bibfield  {author} {\bibinfo {author} {\bibfnamefont {A.}~\bibnamefont
  {Prosvetov}}, \bibinfo {author} {\bibfnamefont {A.~V.}\ \bibnamefont
  {Verkhovtsev}}, \bibinfo {author} {\bibfnamefont {G.}~\bibnamefont
  {Sushko}},\ and\ \bibinfo {author} {\bibfnamefont {A.~V.}\ \bibnamefont
  {Solov{\textquoteright}yov}},\ }\bibfield  {title} {{\selectlanguage
  {en}\enquote {\bibinfo {title} {Atomistic modeling of thermal effects in
  focused electron beam-induced deposition of {Me}$_2${Au}(tfac)},}\ }}\href
  {https://doi.org/10.1140/epjd/s10053-023-00598-5} {\bibfield  {journal}
  {\bibinfo  {journal} {Eur. Phys. J. D}\ }\textbf {\bibinfo {volume} {77}},\
  \bibinfo {pages} {15} (\bibinfo {year} {2023})}\BibitemShut {NoStop}%
\bibitem [{\citenamefont {Verkhovtsev}, \citenamefont
  {Solov{\textquoteright}yov},\ and\ \citenamefont
  {Solov{\textquoteright}yov}(2021)}]{Verkhovtsev_2021}%
  \BibitemOpen
  \bibfield  {author} {\bibinfo {author} {\bibfnamefont {A.~V.}\ \bibnamefont
  {Verkhovtsev}}, \bibinfo {author} {\bibfnamefont {I.~A.}\ \bibnamefont
  {Solov{\textquoteright}yov}},\ and\ \bibinfo {author} {\bibfnamefont {A.~V.}\
  \bibnamefont {Solov{\textquoteright}yov}},\ }\bibfield  {title}
  {{\selectlanguage {en}\enquote {\bibinfo {title} {Irradiation-driven
  molecular dynamics: a review},}\ }}\href
  {https://doi.org/10.1140/epjd/s10053-021-00223-3} {\bibfield  {journal}
  {\bibinfo  {journal} {Eur. Phys. J. D}\ }\textbf {\bibinfo {volume} {75}},\
  \bibinfo {pages} {213} (\bibinfo {year} {2021})}\BibitemShut {NoStop}%
\bibitem [{\citenamefont {Utke}\ \emph {et~al.}(2022)\citenamefont {Utke},
  \citenamefont {Swiderek}, \citenamefont {H{\"{o}}flich}, \citenamefont
  {Madajska}, \citenamefont {Jurczyk}, \citenamefont {Martinovi{\'{c}}},\ and\
  \citenamefont {Szyma{{\'{n}}}ska}}]{utke_coordination_2022}%
  \BibitemOpen
  \bibfield  {author} {\bibinfo {author} {\bibfnamefont {I.}~\bibnamefont
  {Utke}}, \bibinfo {author} {\bibfnamefont {P.}~\bibnamefont {Swiderek}},
  \bibinfo {author} {\bibfnamefont {K.}~\bibnamefont {H{\"{o}}flich}}, \bibinfo
  {author} {\bibfnamefont {K.}~\bibnamefont {Madajska}}, \bibinfo {author}
  {\bibfnamefont {J.}~\bibnamefont {Jurczyk}}, \bibinfo {author} {\bibfnamefont
  {P.}~\bibnamefont {Martinovi{\'{c}}}},\ and\ \bibinfo {author} {\bibfnamefont
  {I.~B.}\ \bibnamefont {Szyma{{\'{n}}}ska}},\ }\bibfield  {title}
  {{\selectlanguage {english}\enquote {\bibinfo {title} {Coordination and
  organometallic precursors of group 10 and 11: {Focused} electron beam induced
  deposition of metals and insight gained from chemical vapour deposition,
  atomic layer deposition, and fundamental surface and gas phase studies},}\
  }}\href {https://doi.org/10.1016/j.ccr.2021.213851} {\bibfield  {journal}
  {\bibinfo  {journal} {Coordin. Chem. Rev.}\ ,\ \bibinfo {pages} {213851}}
  (\bibinfo {year} {2022})}\BibitemShut {NoStop}%
\bibitem [{\citenamefont {Sanz-Hern{\'{a}}ndez}\ and\ \citenamefont
  {Fern{\'{a}}ndez-Pacheco}(2017)}]{sanz-hernandez_modelling_2017}%
  \BibitemOpen
  \bibfield  {author} {\bibinfo {author} {\bibfnamefont {D.}~\bibnamefont
  {Sanz-Hern{\'{a}}ndez}}\ and\ \bibinfo {author} {\bibfnamefont
  {A.}~\bibnamefont {Fern{\'{a}}ndez-Pacheco}},\ }\bibfield  {title}
  {{\selectlanguage {english}\enquote {\bibinfo {title} {Modelling focused
  electron beam induced deposition beyond {Langmuir} adsorption},}\ }}\href
  {https://doi.org/10.3762/bjnano.8.214} {\bibfield  {journal} {\bibinfo
  {journal} {Beilstein J. Nanotechnol.}\ }\textbf {\bibinfo {volume} {8}},\
  \bibinfo {pages} {2151--2161} (\bibinfo {year} {2017})},\ \bibinfo {note}
  {publisher: Beilstein-Institut}\BibitemShut {NoStop}%
\bibitem [{\citenamefont {Bishop}\ \emph {et~al.}(2012)\citenamefont {Bishop},
  \citenamefont {Lobo}, \citenamefont {Martin}, \citenamefont {Ford},
  \citenamefont {Phillips},\ and\ \citenamefont {Toth}}]{bishop_role_2012}%
  \BibitemOpen
  \bibfield  {author} {\bibinfo {author} {\bibfnamefont {J.}~\bibnamefont
  {Bishop}}, \bibinfo {author} {\bibfnamefont {C.~J.}\ \bibnamefont {Lobo}},
  \bibinfo {author} {\bibfnamefont {A.}~\bibnamefont {Martin}}, \bibinfo
  {author} {\bibfnamefont {M.}~\bibnamefont {Ford}}, \bibinfo {author}
  {\bibfnamefont {M.}~\bibnamefont {Phillips}},\ and\ \bibinfo {author}
  {\bibfnamefont {M.}~\bibnamefont {Toth}},\ }\bibfield  {title} {\enquote
  {\bibinfo {title} {Role of {Activated} {Chemisorption} in {Gas}-{Mediated}
  {Electron} {Beam} {Induced} {Deposition}},}\ }\href
  {https://doi.org/10.1103/PhysRevLett.109.146103} {\bibfield  {journal}
  {\bibinfo  {journal} {Phys. Rev. Lett.}\ }\textbf {\bibinfo {volume} {109}},\
  \bibinfo {pages} {146103} (\bibinfo {year} {2012})},\ \bibinfo {note}
  {publisher: American Physical Society}\BibitemShut {NoStop}%
\bibitem [{\citenamefont {Utke}(2023)}]{i_utke_notitle_nodate}%
  \BibitemOpen
  \bibfield  {author} {\bibinfo {author} {\bibfnamefont {I.}~\bibnamefont
  {Utke}},\ }\href {https://www.empa.ch/web/s206/febip-codes} {\enquote
  {\bibinfo {title} {{FEBID} codes},}\ }\bibinfo {howpublished} {Online}
  (\bibinfo {year} {2023}),\ \bibinfo {note}
  {\url{https://www.empa.ch/web/s206/febip-codes}}\BibitemShut {NoStop}%
\bibitem [{\citenamefont {Friedli}\ and\ \citenamefont
  {Utke}(2009)}]{friedli_optimized_2009}%
  \BibitemOpen
  \bibfield  {author} {\bibinfo {author} {\bibfnamefont {V.}~\bibnamefont
  {Friedli}}\ and\ \bibinfo {author} {\bibfnamefont {I.}~\bibnamefont {Utke}},\
  }\bibfield  {title} {\enquote {\bibinfo {title} {Optimized molecule supply
  from nozzle-based gas injection systems for focused electron- and ion-beam
  induced deposition and etching: simulation and experiment},}\ }\href
  {https://doi.org/10.1088/0022-3727/42/12/125305} {\bibfield  {journal}
  {\bibinfo  {journal} {J. Phys. D Appl. Phys.}\ }\textbf {\bibinfo {volume}
  {42}},\ \bibinfo {pages} {125305} (\bibinfo {year} {2009})}\BibitemShut
  {NoStop}%
\bibitem [{\citenamefont {Thorman}\ \emph {et~al.}(2015)\citenamefont
  {Thorman}, \citenamefont {Ragesh Kumar~T.}, \citenamefont {Fairbrother},\
  and\ \citenamefont {Ing{\'{o}}lfsson}}]{thorman_role_2015}%
  \BibitemOpen
  \bibfield  {author} {\bibinfo {author} {\bibfnamefont {R.~M.}\ \bibnamefont
  {Thorman}}, \bibinfo {author} {\bibfnamefont {P.}~\bibnamefont {Ragesh
  Kumar~T.}}, \bibinfo {author} {\bibfnamefont {D.~H.}\ \bibnamefont
  {Fairbrother}},\ and\ \bibinfo {author} {\bibfnamefont {O.}~\bibnamefont
  {Ing{\'{o}}lfsson}},\ }\bibfield  {title} {{\selectlanguage {english}\enquote
  {\bibinfo {title} {The role of low-energy electrons in focused electron beam
  induced deposition: four case studies of representative precursors},}\
  }}\href {https://doi.org/10.3762/bjnano.6.194} {\bibfield  {journal}
  {\bibinfo  {journal} {Beilstein J. Nanotechnol.}\ }\textbf {\bibinfo {volume}
  {6}},\ \bibinfo {pages} {1904--1926} (\bibinfo {year} {2015})},\ \bibinfo
  {note} {publisher: Beilstein-Institut}\BibitemShut {NoStop}%
\bibitem [{\citenamefont {Ingolfson}(2019)}]{o_ingolfson_low-energy_2019}%
  \BibitemOpen
  \bibfield  {author} {\bibinfo {author} {\bibfnamefont {O.}~\bibnamefont
  {Ingolfson}},\ }\href@noop {} {\emph {\bibinfo {title} {Low-energy electrons:
  {Fundamentals} and applications}}}\ (\bibinfo  {publisher} {Pan Stanford
  Publishing},\ \bibinfo {address} {Singapore},\ \bibinfo {year}
  {2019})\BibitemShut {NoStop}%
\bibitem [{\citenamefont {Spencer}\ \emph {et~al.}(2014)\citenamefont
  {Spencer}, \citenamefont {Rosenberg}, \citenamefont {Barclay}, \citenamefont
  {Wu}, \citenamefont {McElwee-White},\ and\ \citenamefont
  {Howard~Fairbrother}}]{spencer_understanding_2014}%
  \BibitemOpen
  \bibfield  {author} {\bibinfo {author} {\bibfnamefont {J.~A.}\ \bibnamefont
  {Spencer}}, \bibinfo {author} {\bibfnamefont {S.~G.}\ \bibnamefont
  {Rosenberg}}, \bibinfo {author} {\bibfnamefont {M.}~\bibnamefont {Barclay}},
  \bibinfo {author} {\bibfnamefont {Y.-C.}\ \bibnamefont {Wu}}, \bibinfo
  {author} {\bibfnamefont {L.}~\bibnamefont {McElwee-White}},\ and\ \bibinfo
  {author} {\bibfnamefont {D.}~\bibnamefont {Howard~Fairbrother}},\ }\bibfield
  {title} {{\selectlanguage {english}\enquote {\bibinfo {title} {Understanding
  the electron-stimulated surface reactions of organometallic complexes to
  enable design of precursors for electron beam-induced deposition},}\ }}\href
  {https://doi.org/10.1007/s00339-014-8570-5} {\bibfield  {journal} {\bibinfo
  {journal} {Appl. Phys. A}\ }\textbf {\bibinfo {volume} {117}},\ \bibinfo
  {pages} {1631--1644} (\bibinfo {year} {2014})}\BibitemShut {NoStop}%
\bibitem [{\citenamefont {B{\"{o}}hler}, \citenamefont {Warneke},\ and\
  \citenamefont {Swiderek}(2013)}]{bohler_control_2013}%
  \BibitemOpen
  \bibfield  {author} {\bibinfo {author} {\bibfnamefont {E.}~\bibnamefont
  {B{\"{o}}hler}}, \bibinfo {author} {\bibfnamefont {J.}~\bibnamefont
  {Warneke}},\ and\ \bibinfo {author} {\bibfnamefont {P.}~\bibnamefont
  {Swiderek}},\ }\bibfield  {title} {{\selectlanguage {english}\enquote
  {\bibinfo {title} {Control of chemical reactions and synthesis by low-energy
  electrons},}\ }}\href {https://doi.org/10.1039/C3CS60180C} {\bibfield
  {journal} {\bibinfo  {journal} {Chem. Soc. Rev.}\ }\textbf {\bibinfo {volume}
  {42}},\ \bibinfo {pages} {9219--9231} (\bibinfo {year} {2013})},\ \bibinfo
  {note} {publisher: The Royal Society of Chemistry}\BibitemShut {NoStop}%
\bibitem [{\citenamefont {Dorp}(2012)}]{dorp_role_2012}%
  \BibitemOpen
  \bibfield  {author} {\bibinfo {author} {\bibfnamefont {W.~F.~v.}\
  \bibnamefont {Dorp}},\ }\bibfield  {title} {{\selectlanguage
  {english}\enquote {\bibinfo {title} {The role of electron scattering in
  electron-induced surface chemistry},}\ }}\href
  {https://doi.org/10.1039/C2CP42275A} {\bibfield  {journal} {\bibinfo
  {journal} {Phys. Chem. Chem. Phys.}\ }\textbf {\bibinfo {volume} {14}},\
  \bibinfo {pages} {16753--16759} (\bibinfo {year} {2012})},\ \bibinfo {note}
  {publisher: The Royal Society of Chemistry}\BibitemShut {NoStop}%
\bibitem [{\citenamefont {Bilgilisoy}\ \emph {et~al.}(2020)\citenamefont
  {Bilgilisoy}, \citenamefont {Thorman}, \citenamefont {Yu}, \citenamefont
  {Dunn}, \citenamefont {Marbach}, \citenamefont {McElwee-White},\ and\
  \citenamefont {Fairbrother}}]{bilgilisoy_surface_2020}%
  \BibitemOpen
  \bibfield  {author} {\bibinfo {author} {\bibfnamefont {E.}~\bibnamefont
  {Bilgilisoy}}, \bibinfo {author} {\bibfnamefont {R.~M.}\ \bibnamefont
  {Thorman}}, \bibinfo {author} {\bibfnamefont {J.-C.}\ \bibnamefont {Yu}},
  \bibinfo {author} {\bibfnamefont {T.~B.}\ \bibnamefont {Dunn}}, \bibinfo
  {author} {\bibfnamefont {H.}~\bibnamefont {Marbach}}, \bibinfo {author}
  {\bibfnamefont {L.}~\bibnamefont {McElwee-White}},\ and\ \bibinfo {author}
  {\bibfnamefont {D.~H.}\ \bibnamefont {Fairbrother}},\ }\bibfield  {title}
  {\enquote {\bibinfo {title} {Surface {Reactions} of {Low}-{Energy} {Argon}
  {Ions} with {Organometallic} {Precursors}},}\ }\href
  {https://doi.org/10.1021/acs.jpcc.0c07269} {\bibfield  {journal} {\bibinfo
  {journal} {J. Phys. Chem. C}\ }\textbf {\bibinfo {volume} {124}},\ \bibinfo
  {pages} {24795--24808} (\bibinfo {year} {2020})},\ \bibinfo {note}
  {publisher: American Chemical Society}\BibitemShut {NoStop}%
\bibitem [{\citenamefont {Dubner}\ \emph {et~al.}(1991)\citenamefont {Dubner},
  \citenamefont {Wagner}, \citenamefont {Melngailis},\ and\ \citenamefont
  {Thompson}}]{dubner_role_1991}%
  \BibitemOpen
  \bibfield  {author} {\bibinfo {author} {\bibfnamefont {A.~D.}\ \bibnamefont
  {Dubner}}, \bibinfo {author} {\bibfnamefont {A.}~\bibnamefont {Wagner}},
  \bibinfo {author} {\bibfnamefont {J.}~\bibnamefont {Melngailis}},\ and\
  \bibinfo {author} {\bibfnamefont {C.~V.}\ \bibnamefont {Thompson}},\
  }\bibfield  {title} {\enquote {\bibinfo {title} {The role of the ion-solid
  interaction in ion-beam-induced deposition of gold},}\ }\href
  {https://doi.org/10.1063/1.349671} {\bibfield  {journal} {\bibinfo  {journal}
  {J. Appl. Phys.}\ }\textbf {\bibinfo {volume} {70}},\ \bibinfo {pages}
  {665--673} (\bibinfo {year} {1991})}\BibitemShut {NoStop}%
\bibitem [{\citenamefont {Alkemade}\ and\ \citenamefont
  {Miro}(2014)}]{Alkemade_2014}%
  \BibitemOpen
  \bibfield  {author} {\bibinfo {author} {\bibfnamefont {P.~F.~A.}\
  \bibnamefont {Alkemade}}\ and\ \bibinfo {author} {\bibfnamefont
  {H.}~\bibnamefont {Miro}},\ }\bibfield  {title} {{\selectlanguage
  {english}\enquote {\bibinfo {title} {Focused helium-ion-beam-induced
  deposition},}\ }}\href {https://doi.org/10.1007/s00339-014-8763-y} {\bibfield
   {journal} {\bibinfo  {journal} {Appl. Phys. A}\ }\textbf {\bibinfo {volume}
  {117}},\ \bibinfo {pages} {1727--1747} (\bibinfo {year} {2014})}\BibitemShut
  {NoStop}%
\bibitem [{\citenamefont {Barth}, \citenamefont {Huth},\ and\ \citenamefont
  {Jungwirth}(2020)}]{barth_precursors_2020}%
  \BibitemOpen
  \bibfield  {author} {\bibinfo {author} {\bibfnamefont {S.}~\bibnamefont
  {Barth}}, \bibinfo {author} {\bibfnamefont {M.}~\bibnamefont {Huth}},\ and\
  \bibinfo {author} {\bibfnamefont {F.}~\bibnamefont {Jungwirth}},\ }\bibfield
  {title} {{\selectlanguage {english}\enquote {\bibinfo {title} {Precursors for
  direct-write nanofabrication with electrons},}\ }}\href
  {https://doi.org/10.1039/D0TC03689G} {\bibfield  {journal} {\bibinfo
  {journal} {J. Mater. Chem. C}\ }\textbf {\bibinfo {volume} {8}},\ \bibinfo
  {pages} {15884--15919} (\bibinfo {year} {2020})},\ \bibinfo {note}
  {publisher: The Royal Society of Chemistry}\BibitemShut {NoStop}%
\bibitem [{\citenamefont {Belianinov}\ \emph {et~al.}(2020)\citenamefont
  {Belianinov}, \citenamefont {Burch}, \citenamefont {Ievlev}, \citenamefont
  {Kim}, \citenamefont {Stanford}, \citenamefont {Mahady}, \citenamefont
  {Lewis}, \citenamefont {Fowlkes}, \citenamefont {Rack},\ and\ \citenamefont
  {Ovchinnikova}}]{belianinov_direct_2020}%
  \BibitemOpen
  \bibfield  {author} {\bibinfo {author} {\bibfnamefont {A.}~\bibnamefont
  {Belianinov}}, \bibinfo {author} {\bibfnamefont {M.~J.}\ \bibnamefont
  {Burch}}, \bibinfo {author} {\bibfnamefont {A.}~\bibnamefont {Ievlev}},
  \bibinfo {author} {\bibfnamefont {S.}~\bibnamefont {Kim}}, \bibinfo {author}
  {\bibfnamefont {M.~G.}\ \bibnamefont {Stanford}}, \bibinfo {author}
  {\bibfnamefont {K.}~\bibnamefont {Mahady}}, \bibinfo {author} {\bibfnamefont
  {B.~B.}\ \bibnamefont {Lewis}}, \bibinfo {author} {\bibfnamefont {J.~D.}\
  \bibnamefont {Fowlkes}}, \bibinfo {author} {\bibfnamefont {P.~D.}\
  \bibnamefont {Rack}},\ and\ \bibinfo {author} {\bibfnamefont {O.~S.}\
  \bibnamefont {Ovchinnikova}},\ }\bibfield  {title} {{\selectlanguage
  {english}\enquote {\bibinfo {title} {Direct write of {3D} nanoscale mesh
  objects with platinum precursor via focused helium ion beam induced
  deposition},}\ }}\href {https://doi.org/10.3390/mi11050527} {\bibfield
  {journal} {\bibinfo  {journal} {Micromachines}\ }\textbf {\bibinfo {volume}
  {11}},\ \bibinfo {pages} {527} (\bibinfo {year} {2020})},\ \bibinfo {note}
  {number: 5 Publisher: Multidisciplinary Digital Publishing
  Institute}\BibitemShut {NoStop}%
\bibitem [{\citenamefont {Qin}\ \emph {et~al.}(2013)\citenamefont {Qin},
  \citenamefont {Fu}, \citenamefont {Yin},\ and\ \citenamefont
  {Zheng}}]{qin_large-scale_2013}%
  \BibitemOpen
  \bibfield  {author} {\bibinfo {author} {\bibfnamefont {R.}~\bibnamefont
  {Qin}}, \bibinfo {author} {\bibfnamefont {J.}~\bibnamefont {Fu}}, \bibinfo
  {author} {\bibfnamefont {Z.}~\bibnamefont {Yin}},\ and\ \bibinfo {author}
  {\bibfnamefont {C.}~\bibnamefont {Zheng}},\ }\bibfield  {title}
  {{\selectlanguage {english}\enquote {\bibinfo {title} {Large-scale process
  optimization for focused ion beam 3-{D} nanofabrication},}\ }}\href
  {https://doi.org/10.1007/s00170-012-4429-0} {\bibfield  {journal} {\bibinfo
  {journal} {Int J Adv Manuf Technol}\ }\textbf {\bibinfo {volume} {64}},\
  \bibinfo {pages} {587--600} (\bibinfo {year} {2013})}\BibitemShut {NoStop}%
\bibitem [{\citenamefont {Schmied}\ \emph {et~al.}(2012)\citenamefont
  {Schmied}, \citenamefont {Chernev}, \citenamefont {Trimmel},\ and\
  \citenamefont {Plank}}]{schmied_new_2012}%
  \BibitemOpen
  \bibfield  {author} {\bibinfo {author} {\bibfnamefont {R.}~\bibnamefont
  {Schmied}}, \bibinfo {author} {\bibfnamefont {B.}~\bibnamefont {Chernev}},
  \bibinfo {author} {\bibfnamefont {G.}~\bibnamefont {Trimmel}},\ and\ \bibinfo
  {author} {\bibfnamefont {H.}~\bibnamefont {Plank}},\ }\bibfield  {title}
  {{\selectlanguage {english}\enquote {\bibinfo {title} {New possibilities for
  soft matter applications: eliminating technically induced thermal stress
  during {FIB} processing},}\ }}\href {https://doi.org/10.1039/C2RA21025H}
  {\bibfield  {journal} {\bibinfo  {journal} {RSC Adv.}\ }\textbf {\bibinfo
  {volume} {2}},\ \bibinfo {pages} {6932--6938} (\bibinfo {year}
  {2012})}\BibitemShut {NoStop}%
\bibitem [{\citenamefont
  {M{\"{u}}ller}(1971)}]{muller_elektronen-mikroschreiber_1971}%
  \BibitemOpen
  \bibfield  {author} {\bibinfo {author} {\bibfnamefont {K.-H.}\ \bibnamefont
  {M{\"{u}}ller}},\ }\bibfield  {title} {\enquote {\bibinfo {title}
  {Elektronen-{Mikroschreiber} mit geschwindigkeitsgesteuerter
  {Strahlführung}. {I}},}\ }\href@noop {} {\bibfield  {journal} {\bibinfo
  {journal} {Optik}\ }\textbf {\bibinfo {volume} {33}},\ \bibinfo {pages}
  {296--311} (\bibinfo {year} {1971})}\BibitemShut {NoStop}%
\bibitem [{\citenamefont {{ЖДАНОВ, C.}}\ and\ \citenamefont
  {{ВЕРЦНЕР, В. Н.}}(1968)}]{c___1968}%
  \BibitemOpen
  \bibfield  {author} {\bibinfo {author} {\bibnamefont {{ЖДАНОВ, C.}}}\
  and\ \bibinfo {author} {\bibnamefont {{ВЕРЦНЕР, В. Н.}}},\
  }\bibfield  {title} {\enquote {\bibinfo {title} {{ИССЛЕДОВАНИЕ}
  механизма образования углеводородных
  загрязнений на объе{RТАХ}, облучаемых
  эле{KТРОНАМИ}},}\ }\href@noop {} {\bibfield  {journal} {\bibinfo
  {journal} {ИЗВЕСТИЯ АКАДЕМИИ НАУК СССР}\ }\textbf
  {\bibinfo {volume} {32}},\ \bibinfo {pages} {1091--1094} (\bibinfo {year}
  {1968})}\BibitemShut {NoStop}%
\bibitem [{\citenamefont {Aristov}, \citenamefont {Kislov},\ and\ \citenamefont
  {Khodos}(1991)}]{aristov_direct_1991}%
  \BibitemOpen
  \bibfield  {author} {\bibinfo {author} {\bibfnamefont {V.~V.}\ \bibnamefont
  {Aristov}}, \bibinfo {author} {\bibfnamefont {N.~A.}\ \bibnamefont
  {Kislov}},\ and\ \bibinfo {author} {\bibfnamefont {I.~I.}\ \bibnamefont
  {Khodos}},\ }\bibfield  {title} {{\selectlanguage {english}\enquote {\bibinfo
  {title} {Direct electron beam-induced formation of nanometer carbon
  structures in {STEM}},}\ }}\href@noop {} {\bibfield  {journal} {\bibinfo
  {journal} {Inst. Phys. Conf. Ser.}\ }\textbf {\bibinfo {volume} {117}},\
  \bibinfo {pages} {775--780} (\bibinfo {year} {1991})},\ \bibinfo {note}
  {iSSN: 0373-0751 Issue: 117}\BibitemShut {NoStop}%
\bibitem [{\citenamefont {Hirsch}\ \emph {et~al.}(1994)\citenamefont {Hirsch},
  \citenamefont {K{\"{a}}ssens}, \citenamefont {P{\"{u}}ttmann},\ and\
  \citenamefont {Reimer}}]{hirsch_contamination_1994}%
  \BibitemOpen
  \bibfield  {author} {\bibinfo {author} {\bibfnamefont {P.}~\bibnamefont
  {Hirsch}}, \bibinfo {author} {\bibfnamefont {M.}~\bibnamefont
  {K{\"{a}}ssens}}, \bibinfo {author} {\bibfnamefont {M.}~\bibnamefont
  {P{\"{u}}ttmann}},\ and\ \bibinfo {author} {\bibfnamefont {L.}~\bibnamefont
  {Reimer}},\ }\bibfield  {title} {\enquote {\bibinfo {title} {Contamination in
  a scanning electron microscope and the influence of specimen cooling},}\
  }\href {https://doi.org/10.1002/sca.4950160207} {\bibfield  {journal}
  {\bibinfo  {journal} {Scanning}\ }\textbf {\bibinfo {volume} {16}},\ \bibinfo
  {pages} {101--110} (\bibinfo {year} {1994})}\BibitemShut {NoStop}%
\bibitem [{\citenamefont {Szkudlarek}\ \emph {et~al.}(2014)\citenamefont
  {Szkudlarek}, \citenamefont {Szmyt}, \citenamefont {Kapusta},\ and\
  \citenamefont {Utke}}]{szkudlarek_lateral_2014}%
  \BibitemOpen
  \bibfield  {author} {\bibinfo {author} {\bibfnamefont {A.}~\bibnamefont
  {Szkudlarek}}, \bibinfo {author} {\bibfnamefont {W.}~\bibnamefont {Szmyt}},
  \bibinfo {author} {\bibfnamefont {C.}~\bibnamefont {Kapusta}},\ and\ \bibinfo
  {author} {\bibfnamefont {I.}~\bibnamefont {Utke}},\ }\bibfield  {title}
  {{\selectlanguage {english}\enquote {\bibinfo {title} {Lateral resolution in
  focused electron beam-induced deposition: scaling laws for pulsed and static
  exposure},}\ }}\href {https://doi.org/10.1007/s00339-014-8751-2} {\bibfield
  {journal} {\bibinfo  {journal} {Appl. Phys. A}\ }\textbf {\bibinfo {volume}
  {117}},\ \bibinfo {pages} {1715--1726} (\bibinfo {year} {2014})}\BibitemShut
  {NoStop}%
\bibitem [{\citenamefont {Lobo}\ \emph {et~al.}(2007)\citenamefont {Lobo},
  \citenamefont {Toth}, \citenamefont {Wagner}, \citenamefont {Thiel},\ and\
  \citenamefont {Lysaght}}]{lobo_high_2007}%
  \BibitemOpen
  \bibfield  {author} {\bibinfo {author} {\bibfnamefont {C.~J.}\ \bibnamefont
  {Lobo}}, \bibinfo {author} {\bibfnamefont {M.}~\bibnamefont {Toth}}, \bibinfo
  {author} {\bibfnamefont {R.}~\bibnamefont {Wagner}}, \bibinfo {author}
  {\bibfnamefont {B.~L.}\ \bibnamefont {Thiel}},\ and\ \bibinfo {author}
  {\bibfnamefont {M.}~\bibnamefont {Lysaght}},\ }\bibfield  {title}
  {{\selectlanguage {english}\enquote {\bibinfo {title} {High resolution
  radially symmetric nanostructures from simultaneous electron beam induced
  etching and deposition},}\ }}\href
  {https://doi.org/10.1088/0957-4484/19/02/025303} {\bibfield  {journal}
  {\bibinfo  {journal} {Nanotechnology}\ }\textbf {\bibinfo {volume} {19}},\
  \bibinfo {pages} {025303} (\bibinfo {year} {2007})},\ \bibinfo {note}
  {publisher: IOP Publishing}\BibitemShut {NoStop}%
\bibitem [{\citenamefont {Toth}\ \emph {et~al.}(2007)\citenamefont {Toth},
  \citenamefont {Lobo}, \citenamefont {Hartigan},\ and\ \citenamefont
  {Ralph~Knowles}}]{toth_electron_2007}%
  \BibitemOpen
  \bibfield  {author} {\bibinfo {author} {\bibfnamefont {M.}~\bibnamefont
  {Toth}}, \bibinfo {author} {\bibfnamefont {C.~J.}\ \bibnamefont {Lobo}},
  \bibinfo {author} {\bibfnamefont {G.}~\bibnamefont {Hartigan}},\ and\
  \bibinfo {author} {\bibfnamefont {W.}~\bibnamefont {Ralph~Knowles}},\
  }\bibfield  {title} {\enquote {\bibinfo {title} {Electron flux controlled
  switching between electron beam induced etching and deposition},}\ }\href
  {https://doi.org/10.1063/1.2437667} {\bibfield  {journal} {\bibinfo
  {journal} {J. Appl. Phys.}\ }\textbf {\bibinfo {volume} {101}},\ \bibinfo
  {pages} {054309} (\bibinfo {year} {2007})},\ \bibinfo {note} {publisher:
  American Institute of Physics}\BibitemShut {NoStop}%
\bibitem [{\citenamefont {Winkler}\ \emph {et~al.}(2014)\citenamefont
  {Winkler}, \citenamefont {Fowlkes}, \citenamefont {Szkudlarek}, \citenamefont
  {Utke}, \citenamefont {Rack},\ and\ \citenamefont
  {Plank}}]{winkler_nanoscale_2014}%
  \BibitemOpen
  \bibfield  {author} {\bibinfo {author} {\bibfnamefont {R.}~\bibnamefont
  {Winkler}}, \bibinfo {author} {\bibfnamefont {J.}~\bibnamefont {Fowlkes}},
  \bibinfo {author} {\bibfnamefont {A.}~\bibnamefont {Szkudlarek}}, \bibinfo
  {author} {\bibfnamefont {I.}~\bibnamefont {Utke}}, \bibinfo {author}
  {\bibfnamefont {P.~D.}\ \bibnamefont {Rack}},\ and\ \bibinfo {author}
  {\bibfnamefont {H.}~\bibnamefont {Plank}},\ }\bibfield  {title} {\enquote
  {\bibinfo {title} {The {Nanoscale} {Implications} of a {Molecular} {Gas}
  {Beam} during {Electron} {Beam} {Induced} {Deposition}},}\ }\href
  {https://doi.org/10.1021/am405591d} {\bibfield  {journal} {\bibinfo
  {journal} {ACS Appl. Mater. Interfaces}\ }\textbf {\bibinfo {volume} {6}},\
  \bibinfo {pages} {2987--2995} (\bibinfo {year} {2014})},\ \bibinfo {note}
  {publisher: American Chemical Society}\BibitemShut {NoStop}%
\bibitem [{\citenamefont {Bernau}\ \emph {et~al.}(2010)\citenamefont {Bernau},
  \citenamefont {Gabureac}, \citenamefont {Erni},\ and\ \citenamefont
  {Utke}}]{bernau_tunable_2010}%
  \BibitemOpen
  \bibfield  {author} {\bibinfo {author} {\bibfnamefont {L.}~\bibnamefont
  {Bernau}}, \bibinfo {author} {\bibfnamefont {M.}~\bibnamefont {Gabureac}},
  \bibinfo {author} {\bibfnamefont {R.}~\bibnamefont {Erni}},\ and\ \bibinfo
  {author} {\bibfnamefont {I.}~\bibnamefont {Utke}},\ }\bibfield  {title}
  {\enquote {\bibinfo {title} {Tunable nanosynthesis of composite materials by
  electron-impact reaction},}\ }\href {https://doi.org/10.1002/anie.201004220}
  {\bibfield  {journal} {\bibinfo  {journal} {Angew. Chem. Int. Ed.}\ }\textbf
  {\bibinfo {volume} {49}},\ \bibinfo {pages} {8880--8884} (\bibinfo {year}
  {2010})}\BibitemShut {NoStop}%
\bibitem [{\citenamefont {Lobo}\ and\ \citenamefont
  {Toth}(2012)}]{c_lobo_continuum_2012}%
  \BibitemOpen
  \bibfield  {author} {\bibinfo {author} {\bibfnamefont {C.}~\bibnamefont
  {Lobo}}\ and\ \bibinfo {author} {\bibfnamefont {M.}~\bibnamefont {Toth}},\
  }\bibfield  {title} {\enquote {\bibinfo {title} {Continuum {Modeling} of
  {Electron} {Beam} {Induced} {Process}},}\ }in\ \href@noop {} {\emph {\bibinfo
  {booktitle} {Nanofabrication {Using} {Focused} {Ion} and {Electron} {Beams}:
  {Principles} and {Applications} ({I}. {Utke}, {S}. {Moshkalev}, {Ph}.
  {Russell}, ed.)}}}\ (\bibinfo  {publisher} {Oxford University Press},\
  \bibinfo {year} {2012})\ pp.\ \bibinfo {pages} {286--320}\BibitemShut
  {NoStop}%
\bibitem [{\citenamefont {Fang}\ \emph {et~al.}(2020)\citenamefont {Fang},
  \citenamefont {Xing}, \citenamefont {Zhou}, \citenamefont {Chai},\ and\
  \citenamefont {Chen}}]{fang_structural_2020}%
  \BibitemOpen
  \bibfield  {author} {\bibinfo {author} {\bibfnamefont {C.}~\bibnamefont
  {Fang}}, \bibinfo {author} {\bibfnamefont {Y.}~\bibnamefont {Xing}}, \bibinfo
  {author} {\bibfnamefont {Z.}~\bibnamefont {Zhou}}, \bibinfo {author}
  {\bibfnamefont {Q.}~\bibnamefont {Chai}},\ and\ \bibinfo {author}
  {\bibfnamefont {Q.}~\bibnamefont {Chen}},\ }\bibfield  {title}
  {{\selectlanguage {english}\enquote {\bibinfo {title} {Structural
  characterization and continuous cellular automata simulation of micro
  structures by focused {Ga} ion beam induced deposition process},}\ }}\href
  {https://doi.org/10.1016/j.sna.2020.112150} {\bibfield  {journal} {\bibinfo
  {journal} {Sens. Actuators Phys.}\ }\textbf {\bibinfo {volume} {312}},\
  \bibinfo {pages} {112150} (\bibinfo {year} {2020})}\BibitemShut {NoStop}%
\bibitem [{\citenamefont {Fowlkes}\ \emph {et~al.}(2018)\citenamefont
  {Fowlkes}, \citenamefont {Winkler}, \citenamefont {Lewis}, \citenamefont
  {Fern{\'{a}}ndez-Pacheco}, \citenamefont {Skoric}, \citenamefont
  {Sanz-Hern{\'{a}}ndez}, \citenamefont {Stanford}, \citenamefont {Mutunga},
  \citenamefont {Rack},\ and\ \citenamefont {Plank}}]{Fowlkes_2018}%
  \BibitemOpen
  \bibfield  {author} {\bibinfo {author} {\bibfnamefont {J.~D.}\ \bibnamefont
  {Fowlkes}}, \bibinfo {author} {\bibfnamefont {R.}~\bibnamefont {Winkler}},
  \bibinfo {author} {\bibfnamefont {B.~B.}\ \bibnamefont {Lewis}}, \bibinfo
  {author} {\bibfnamefont {A.}~\bibnamefont {Fern{\'{a}}ndez-Pacheco}},
  \bibinfo {author} {\bibfnamefont {L.}~\bibnamefont {Skoric}}, \bibinfo
  {author} {\bibfnamefont {D.}~\bibnamefont {Sanz-Hern{\'{a}}ndez}}, \bibinfo
  {author} {\bibfnamefont {M.~G.}\ \bibnamefont {Stanford}}, \bibinfo {author}
  {\bibfnamefont {E.}~\bibnamefont {Mutunga}}, \bibinfo {author} {\bibfnamefont
  {P.~D.}\ \bibnamefont {Rack}},\ and\ \bibinfo {author} {\bibfnamefont
  {H.}~\bibnamefont {Plank}},\ }\bibfield  {title} {\enquote {\bibinfo {title}
  {High-fidelity {3D}-nanoprinting via focused electron beams: computer-aided
  design ({3BID})},}\ }\href {https://doi.org/10.1021/acsanm.7b00342}
  {\bibfield  {journal} {\bibinfo  {journal} {ACS Appl. Nano Mater.}\ }\textbf
  {\bibinfo {volume} {1}},\ \bibinfo {pages} {1028--1041} (\bibinfo {year}
  {2018})},\ \bibinfo {note} {publisher: American Chemical Society}\BibitemShut
  {NoStop}%
\bibitem [{\citenamefont {Winkler}\ \emph {et~al.}(2018)\citenamefont
  {Winkler}, \citenamefont {Lewis}, \citenamefont {Fowlkes}, \citenamefont
  {Rack},\ and\ \citenamefont {Plank}}]{winkler_high-fidelity_2018}%
  \BibitemOpen
  \bibfield  {author} {\bibinfo {author} {\bibfnamefont {R.}~\bibnamefont
  {Winkler}}, \bibinfo {author} {\bibfnamefont {B.~B.}\ \bibnamefont {Lewis}},
  \bibinfo {author} {\bibfnamefont {J.~D.}\ \bibnamefont {Fowlkes}}, \bibinfo
  {author} {\bibfnamefont {P.~D.}\ \bibnamefont {Rack}},\ and\ \bibinfo
  {author} {\bibfnamefont {H.}~\bibnamefont {Plank}},\ }\bibfield  {title}
  {\enquote {\bibinfo {title} {High-fidelity {3D}-nanoprinting via focused
  electron beams: growth fundamentals},}\ }\href
  {https://doi.org/10.1021/acsanm.8b00158} {\bibfield  {journal} {\bibinfo
  {journal} {ACS Appl. Nano Mater.}\ }\textbf {\bibinfo {volume} {1}},\
  \bibinfo {pages} {1014--1027} (\bibinfo {year} {2018})},\ \bibinfo {note}
  {publisher: American Chemical Society}\BibitemShut {NoStop}%
\bibitem [{\citenamefont {Allen}(2021)}]{Allen2021}%
  \BibitemOpen
  \bibfield  {author} {\bibinfo {author} {\bibfnamefont {F.~I.}\ \bibnamefont
  {Allen}},\ }\bibfield  {title} {{\selectlanguage {english}\enquote {\bibinfo
  {title} {A review of defect engineering, ion implantation, and
  nanofabrication using the helium ion microscope},}\ }}\href
  {https://doi.org/10.3762/bjnano.12.52} {\bibfield  {journal} {\bibinfo
  {journal} {Beilstein J. Nanotechnol.}\ }\textbf {\bibinfo {volume} {12}},\
  \bibinfo {pages} {633--664} (\bibinfo {year} {2021})}\BibitemShut {NoStop}%
\bibitem [{\citenamefont {Baglin}(2020)}]{Baglin_2020}%
  \BibitemOpen
  \bibfield  {author} {\bibinfo {author} {\bibfnamefont {J.~E.~E.}\
  \bibnamefont {Baglin}},\ }\bibfield  {title} {\enquote {\bibinfo {title} {Ion
  beam enabled nanoscale fabrication, surface patterning, and self-assembly},}\
  }\href {https://doi.org/10.1063/1.5143650} {\bibfield  {journal} {\bibinfo
  {journal} {Appl. Phys. Rev.}\ }\textbf {\bibinfo {volume} {7}},\ \bibinfo
  {pages} {011601} (\bibinfo {year} {2020})}\BibitemShut {NoStop}%
\bibitem [{\citenamefont {Pekin}, \citenamefont {Allen},\ and\ \citenamefont
  {Minor}(2016)}]{Pekin2016}%
  \BibitemOpen
  \bibfield  {author} {\bibinfo {author} {\bibfnamefont {T.~C.}\ \bibnamefont
  {Pekin}}, \bibinfo {author} {\bibfnamefont {F.~I.}\ \bibnamefont {Allen}},\
  and\ \bibinfo {author} {\bibfnamefont {A.~M.}\ \bibnamefont {Minor}},\
  }\bibfield  {title} {\enquote {\bibinfo {title} {Evaluation of neon focused
  ion beam milling for {TEM} sample preparation},}\ }\href
  {https://doi.org/10.1111/jmi.12416} {\bibfield  {journal} {\bibinfo
  {journal} {J. Microsc.}\ }\textbf {\bibinfo {volume} {264}},\ \bibinfo
  {pages} {59--63} (\bibinfo {year} {2016})}\BibitemShut {NoStop}%
\bibitem [{\citenamefont {Vitale}\ and\ \citenamefont
  {Sugar}(2022)}]{Vitale_2022}%
  \BibitemOpen
  \bibfield  {author} {\bibinfo {author} {\bibfnamefont {S.~M.}\ \bibnamefont
  {Vitale}}\ and\ \bibinfo {author} {\bibfnamefont {J.~D.}\ \bibnamefont
  {Sugar}},\ }\bibfield  {title} {{\selectlanguage {english}\enquote {\bibinfo
  {title} {Using {Xe} plasma {FIB} for high-quality {TEM} sample
  preparation},}\ }}\href {https://doi.org/10.1017/S1431927622000344}
  {\bibfield  {journal} {\bibinfo  {journal} {Microsc. Microanal.}\ ,\ \bibinfo
  {pages} {1--13}} (\bibinfo {year} {2022})}\BibitemShut {NoStop}%
\bibitem [{\citenamefont {Zhong}\ \emph {et~al.}(2020)\citenamefont {Zhong},
  \citenamefont {Wade}, \citenamefont {Withers}, \citenamefont {Zhou},
  \citenamefont {Cai}, \citenamefont {Haigh},\ and\ \citenamefont
  {Burke}}]{Zhong_2020}%
  \BibitemOpen
  \bibfield  {author} {\bibinfo {author} {\bibfnamefont {X.~L.}\ \bibnamefont
  {Zhong}}, \bibinfo {author} {\bibfnamefont {C.~A.}\ \bibnamefont {Wade}},
  \bibinfo {author} {\bibfnamefont {P.~J.}\ \bibnamefont {Withers}}, \bibinfo
  {author} {\bibfnamefont {X.}~\bibnamefont {Zhou}}, \bibinfo {author}
  {\bibfnamefont {C.}~\bibnamefont {Cai}}, \bibinfo {author} {\bibfnamefont
  {S.~J.}\ \bibnamefont {Haigh}},\ and\ \bibinfo {author} {\bibfnamefont
  {M.~G.}\ \bibnamefont {Burke}},\ }\bibfield  {title} {{\selectlanguage
  {english}\enquote {\bibinfo {title} {Comparing {Xe}$^+$ {pFIB} and {G}a$^+$
  {FIB} for {TEM} sample preparation of al alloys: minimising {FIB-induced}
  artefacts},}\ }}\href {https://doi.org/10.1111/jmi.12983} {\bibfield
  {journal} {\bibinfo  {journal} {J. Microsc.}\ } (\bibinfo {year} {2020}),\
  10.1111/jmi.12983}\BibitemShut {NoStop}%
\bibitem [{\citenamefont {Allen}\ \emph
  {et~al.}(2023{\natexlab{a}})\citenamefont {Allen}, \citenamefont {Blanchard},
  \citenamefont {Lake}, \citenamefont {Pappas}, \citenamefont {Xia},
  \citenamefont {Notte}, \citenamefont {Zhang}, \citenamefont {Minor},\ and\
  \citenamefont {Sanford}}]{Allen_2023}%
  \BibitemOpen
  \bibfield  {author} {\bibinfo {author} {\bibfnamefont {F.~I.}\ \bibnamefont
  {Allen}}, \bibinfo {author} {\bibfnamefont {P.~T.}\ \bibnamefont
  {Blanchard}}, \bibinfo {author} {\bibfnamefont {R.}~\bibnamefont {Lake}},
  \bibinfo {author} {\bibfnamefont {D.}~\bibnamefont {Pappas}}, \bibinfo
  {author} {\bibfnamefont {D.}~\bibnamefont {Xia}}, \bibinfo {author}
  {\bibfnamefont {J.~A.}\ \bibnamefont {Notte}}, \bibinfo {author}
  {\bibfnamefont {R.}~\bibnamefont {Zhang}}, \bibinfo {author} {\bibfnamefont
  {A.~M.}\ \bibnamefont {Minor}},\ and\ \bibinfo {author} {\bibfnamefont
  {N.~A.}\ \bibnamefont {Sanford}},\ }\bibfield  {title} {{\selectlanguage
  {en}\enquote {\bibinfo {title} {Fabrication of specimens for atom probe
  tomography using a combined gallium and neon focused ion beam milling
  approach},}\ }}\href {https://doi.org/10.1093/micmic/ozad078} {\bibfield
  {journal} {\bibinfo  {journal} {Microsc. Microanal.}\ } (\bibinfo {year}
  {2023}{\natexlab{a}}),\ 10.1093/micmic/ozad078}\BibitemShut {NoStop}%
\bibitem [{\citenamefont {Miller}\ \emph {et~al.}(2007)\citenamefont {Miller},
  \citenamefont {Russell}, \citenamefont {Thompson}, \citenamefont {Alvis},\
  and\ \citenamefont {Larson}}]{Miller_2007}%
  \BibitemOpen
  \bibfield  {author} {\bibinfo {author} {\bibfnamefont {M.~K.}\ \bibnamefont
  {Miller}}, \bibinfo {author} {\bibfnamefont {K.~F.}\ \bibnamefont {Russell}},
  \bibinfo {author} {\bibfnamefont {K.}~\bibnamefont {Thompson}}, \bibinfo
  {author} {\bibfnamefont {R.}~\bibnamefont {Alvis}},\ and\ \bibinfo {author}
  {\bibfnamefont {D.~J.}\ \bibnamefont {Larson}},\ }\bibfield  {title}
  {{\selectlanguage {english}\enquote {\bibinfo {title} {Review of atom probe
  {FIB-based} specimen preparation methods},}\ }}\href
  {https://doi.org/10.1017/S1431927607070845} {\bibfield  {journal} {\bibinfo
  {journal} {Microsc. Microanal.}\ }\textbf {\bibinfo {volume} {13}},\ \bibinfo
  {pages} {428--436} (\bibinfo {year} {2007})}\BibitemShut {NoStop}%
\bibitem [{\citenamefont {Famelton}\ \emph {et~al.}(2021)\citenamefont
  {Famelton}, \citenamefont {Hughes}, \citenamefont {Williams}, \citenamefont
  {Barbatti}, \citenamefont {Moody},\ and\ \citenamefont
  {Bagot}}]{Famelton_2021}%
  \BibitemOpen
  \bibfield  {author} {\bibinfo {author} {\bibfnamefont {J.~R.}\ \bibnamefont
  {Famelton}}, \bibinfo {author} {\bibfnamefont {G.~M.}\ \bibnamefont
  {Hughes}}, \bibinfo {author} {\bibfnamefont {C.~A.}\ \bibnamefont
  {Williams}}, \bibinfo {author} {\bibfnamefont {C.}~\bibnamefont {Barbatti}},
  \bibinfo {author} {\bibfnamefont {M.~P.}\ \bibnamefont {Moody}},\ and\
  \bibinfo {author} {\bibfnamefont {P.~A.~J.}\ \bibnamefont {Bagot}},\
  }\bibfield  {title} {\enquote {\bibinfo {title} {Xenon plasma focussed ion
  beam preparation of an {Al-6XXX} alloy sample for atom probe tomography
  including analysis of an {$\alpha$-Al(Fe,Mn)Si} dispersoid},}\ }\href
  {https://doi.org/10.1016/j.matchar.2021.111194} {\bibfield  {journal}
  {\bibinfo  {journal} {Mater. Charact.}\ }\textbf {\bibinfo {volume} {178}},\
  \bibinfo {pages} {111194} (\bibinfo {year} {2021})}\BibitemShut {NoStop}%
\bibitem [{\citenamefont {Halpin}\ \emph {et~al.}(2019)\citenamefont {Halpin},
  \citenamefont {Webster}, \citenamefont {Gardner}, \citenamefont {Moody},
  \citenamefont {Bagot},\ and\ \citenamefont {MacLaren}}]{Halpin_2019}%
  \BibitemOpen
  \bibfield  {author} {\bibinfo {author} {\bibfnamefont {J.~E.}\ \bibnamefont
  {Halpin}}, \bibinfo {author} {\bibfnamefont {R.~W.~H.}\ \bibnamefont
  {Webster}}, \bibinfo {author} {\bibfnamefont {H.}~\bibnamefont {Gardner}},
  \bibinfo {author} {\bibfnamefont {M.~P.}\ \bibnamefont {Moody}}, \bibinfo
  {author} {\bibfnamefont {P.~A.~J.}\ \bibnamefont {Bagot}},\ and\ \bibinfo
  {author} {\bibfnamefont {D.~A.}\ \bibnamefont {MacLaren}},\ }\bibfield
  {title} {\enquote {\bibinfo {title} {An in-situ approach for preparing atom
  probe tomography specimens by xenon plasma-focussed ion beam},}\ }\href
  {https://doi.org/10.1016/j.ultramic.2019.04.005} {\bibfield  {journal}
  {\bibinfo  {journal} {Ultramicroscopy}\ }\textbf {\bibinfo {volume} {202}},\
  \bibinfo {pages} {121--127} (\bibinfo {year} {2019})}\BibitemShut {NoStop}%
\bibitem [{\citenamefont {Uchic}\ \emph {et~al.}(2004)\citenamefont {Uchic},
  \citenamefont {Dimiduk}, \citenamefont {Florando},\ and\ \citenamefont
  {Nix}}]{Uchic_2004}%
  \BibitemOpen
  \bibfield  {author} {\bibinfo {author} {\bibfnamefont {M.~D.}\ \bibnamefont
  {Uchic}}, \bibinfo {author} {\bibfnamefont {D.~M.}\ \bibnamefont {Dimiduk}},
  \bibinfo {author} {\bibfnamefont {J.~N.}\ \bibnamefont {Florando}},\ and\
  \bibinfo {author} {\bibfnamefont {W.~D.}\ \bibnamefont {Nix}},\ }\bibfield
  {title} {\enquote {\bibinfo {title} {Sample dimensions influence strength and
  crystal plasticity},}\ }\href {https://doi.org/10.1126/science.1098993}
  {\bibfield  {journal} {\bibinfo  {journal} {Science}\ }\textbf {\bibinfo
  {volume} {305}},\ \bibinfo {pages} {986--989} (\bibinfo {year}
  {2004})}\BibitemShut {NoStop}%
\bibitem [{\citenamefont {Uchic}\ and\ \citenamefont
  {Dimiduk}(2005)}]{Uchic_2005}%
  \BibitemOpen
  \bibfield  {author} {\bibinfo {author} {\bibfnamefont {M.~D.}\ \bibnamefont
  {Uchic}}\ and\ \bibinfo {author} {\bibfnamefont {D.~M.}\ \bibnamefont
  {Dimiduk}},\ }\bibfield  {title} {\enquote {\bibinfo {title} {A methodology
  to investigate size scale effects in crystalline plasticity using uniaxial
  compression testing},}\ }\href {https://doi.org/10.1016/j.msea.2005.03.082}
  {\bibfield  {journal} {\bibinfo  {journal} {Mater. Sci. Eng., A}\ }\textbf
  {\bibinfo {volume} {400-401}},\ \bibinfo {pages} {268--278} (\bibinfo {year}
  {2005})}\BibitemShut {NoStop}%
\bibitem [{\citenamefont {Greer}, \citenamefont {Oliver},\ and\ \citenamefont
  {Nix}(2005)}]{Greer_2005}%
  \BibitemOpen
  \bibfield  {author} {\bibinfo {author} {\bibfnamefont {J.~R.}\ \bibnamefont
  {Greer}}, \bibinfo {author} {\bibfnamefont {W.~C.}\ \bibnamefont {Oliver}},\
  and\ \bibinfo {author} {\bibfnamefont {W.~D.}\ \bibnamefont {Nix}},\
  }\bibfield  {title} {\enquote {\bibinfo {title} {Size dependence of
  mechanical properties of gold at the micron scale in the absence of strain
  gradients},}\ }\href {https://doi.org/10.1016/j.actamat.2004.12.031}
  {\bibfield  {journal} {\bibinfo  {journal} {Acta Mater.}\ }\textbf {\bibinfo
  {volume} {53}},\ \bibinfo {pages} {1821--1830} (\bibinfo {year}
  {2005})}\BibitemShut {NoStop}%
\bibitem [{\citenamefont {Greer}, \citenamefont {Kim},\ and\ \citenamefont
  {Burek}(2009)}]{Greer_2009}%
  \BibitemOpen
  \bibfield  {author} {\bibinfo {author} {\bibfnamefont {J.~R.}\ \bibnamefont
  {Greer}}, \bibinfo {author} {\bibfnamefont {J.-Y.}\ \bibnamefont {Kim}},\
  and\ \bibinfo {author} {\bibfnamefont {M.~J.}\ \bibnamefont {Burek}},\
  }\bibfield  {title} {\enquote {\bibinfo {title} {The in-situ mechanical
  testing of nanoscale single-crystalline nanopillars},}\ }\href
  {https://doi.org/10.1007/s11837-009-0174-8} {\bibfield  {journal} {\bibinfo
  {journal} {{JOM}}\ }\textbf {\bibinfo {volume} {61}},\ \bibinfo {pages}
  {19--25} (\bibinfo {year} {2009})}\BibitemShut {NoStop}%
\bibitem [{\citenamefont {Shim}\ \emph {et~al.}(2009)\citenamefont {Shim},
  \citenamefont {Bei}, \citenamefont {Miller}, \citenamefont {Pharr},\ and\
  \citenamefont {George}}]{Shim_2009}%
  \BibitemOpen
  \bibfield  {author} {\bibinfo {author} {\bibfnamefont {S.}~\bibnamefont
  {Shim}}, \bibinfo {author} {\bibfnamefont {H.}~\bibnamefont {Bei}}, \bibinfo
  {author} {\bibfnamefont {M.~K.}\ \bibnamefont {Miller}}, \bibinfo {author}
  {\bibfnamefont {G.~M.}\ \bibnamefont {Pharr}},\ and\ \bibinfo {author}
  {\bibfnamefont {E.~P.}\ \bibnamefont {George}},\ }\bibfield  {title}
  {\enquote {\bibinfo {title} {Effects of focused ion beam milling on the
  compressive behavior of directionally solidified micropillars and the
  nanoindentation response of an electropolished surface},}\ }\href
  {https://doi.org/10.1016/j.actamat.2008.09.033} {\bibfield  {journal}
  {\bibinfo  {journal} {Acta Mater.}\ }\textbf {\bibinfo {volume} {57}},\
  \bibinfo {pages} {503--510} (\bibinfo {year} {2009})}\BibitemShut {NoStop}%
\bibitem [{\citenamefont {{{\"{O}}}stlund}\ \emph {et~al.}(2009)\citenamefont
  {{{\"{O}}}stlund}, \citenamefont {Rzepiejewska-Malyska}, \citenamefont
  {Leifer}, \citenamefont {Hale}, \citenamefont {Tang}, \citenamefont
  {Ballarini}, \citenamefont {Gerberich},\ and\ \citenamefont
  {Michler}}]{Oestlund_2009}%
  \BibitemOpen
  \bibfield  {author} {\bibinfo {author} {\bibfnamefont {F.}~\bibnamefont
  {{{\"{O}}}stlund}}, \bibinfo {author} {\bibfnamefont {K.}~\bibnamefont
  {Rzepiejewska-Malyska}}, \bibinfo {author} {\bibfnamefont {K.}~\bibnamefont
  {Leifer}}, \bibinfo {author} {\bibfnamefont {L.~M.}\ \bibnamefont {Hale}},
  \bibinfo {author} {\bibfnamefont {Y.}~\bibnamefont {Tang}}, \bibinfo {author}
  {\bibfnamefont {R.}~\bibnamefont {Ballarini}}, \bibinfo {author}
  {\bibfnamefont {W.~W.}\ \bibnamefont {Gerberich}},\ and\ \bibinfo {author}
  {\bibfnamefont {J.}~\bibnamefont {Michler}},\ }\bibfield  {title} {\enquote
  {\bibinfo {title} {Brittle-to-ductile transition in uniaxial compression of
  silicon pillars at room temperature},}\ }\href
  {https://doi.org/10.1002/adfm.200900418} {\bibfield  {journal} {\bibinfo
  {journal} {Adv. Funct. Mater.}\ }\textbf {\bibinfo {volume} {19}},\ \bibinfo
  {pages} {2439--2444} (\bibinfo {year} {2009})}\BibitemShut {NoStop}%
\bibitem [{\citenamefont {H{\"{u}}tsch}\ and\ \citenamefont
  {Lilleodden}(2014)}]{Huetsch_2014}%
  \BibitemOpen
  \bibfield  {author} {\bibinfo {author} {\bibfnamefont {J.}~\bibnamefont
  {H{\"{u}}tsch}}\ and\ \bibinfo {author} {\bibfnamefont {E.~T.}\ \bibnamefont
  {Lilleodden}},\ }\bibfield  {title} {\enquote {\bibinfo {title} {The
  influence of focused-ion beam preparation technique on microcompression
  investigations: lathe vs. annular milling},}\ }\href
  {https://doi.org/10.1016/j.scriptamat.2014.01.016} {\bibfield  {journal}
  {\bibinfo  {journal} {Scripta Mater.}\ }\textbf {\bibinfo {volume} {77}},\
  \bibinfo {pages} {49--51} (\bibinfo {year} {2014})}\BibitemShut {NoStop}%
\bibitem [{\citenamefont {Schwiedrzik}\ \emph {et~al.}(2014)\citenamefont
  {Schwiedrzik}, \citenamefont {Raghavan}, \citenamefont {B{\"{u}}rki},
  \citenamefont {LeNader}, \citenamefont {Wolfram}, \citenamefont {Michler},\
  and\ \citenamefont {Zysset}}]{Schwiedrzik_2014}%
  \BibitemOpen
  \bibfield  {author} {\bibinfo {author} {\bibfnamefont {J.}~\bibnamefont
  {Schwiedrzik}}, \bibinfo {author} {\bibfnamefont {R.}~\bibnamefont
  {Raghavan}}, \bibinfo {author} {\bibfnamefont {A.}~\bibnamefont
  {B{\"{u}}rki}}, \bibinfo {author} {\bibfnamefont {V.}~\bibnamefont
  {LeNader}}, \bibinfo {author} {\bibfnamefont {U.}~\bibnamefont {Wolfram}},
  \bibinfo {author} {\bibfnamefont {J.}~\bibnamefont {Michler}},\ and\ \bibinfo
  {author} {\bibfnamefont {P.}~\bibnamefont {Zysset}},\ }\bibfield  {title}
  {\enquote {\bibinfo {title} {In situ micropillar compression reveals superior
  strength and ductility but an absence of{~}damage in{~}lamellar bone},}\
  }\href {https://doi.org/10.1038/nmat3959} {\bibfield  {journal} {\bibinfo
  {journal} {Nat. Mater.}\ }\textbf {\bibinfo {volume} {13}},\ \bibinfo {pages}
  {740--747} (\bibinfo {year} {2014})}\BibitemShut {NoStop}%
\bibitem [{\citenamefont {Xiao}\ \emph {et~al.}(2017)\citenamefont {Xiao},
  \citenamefont {Wehrs}, \citenamefont {Ma}, \citenamefont {Al-Samman},
  \citenamefont {Korte-Kerzel}, \citenamefont {G{\"{o}}ken}, \citenamefont
  {Michler}, \citenamefont {Spolenak},\ and\ \citenamefont
  {Wheeler}}]{Xiao_2017}%
  \BibitemOpen
  \bibfield  {author} {\bibinfo {author} {\bibfnamefont {Y.}~\bibnamefont
  {Xiao}}, \bibinfo {author} {\bibfnamefont {J.}~\bibnamefont {Wehrs}},
  \bibinfo {author} {\bibfnamefont {H.}~\bibnamefont {Ma}}, \bibinfo {author}
  {\bibfnamefont {T.}~\bibnamefont {Al-Samman}}, \bibinfo {author}
  {\bibfnamefont {S.}~\bibnamefont {Korte-Kerzel}}, \bibinfo {author}
  {\bibfnamefont {M.}~\bibnamefont {G{\"{o}}ken}}, \bibinfo {author}
  {\bibfnamefont {J.}~\bibnamefont {Michler}}, \bibinfo {author} {\bibfnamefont
  {R.}~\bibnamefont {Spolenak}},\ and\ \bibinfo {author} {\bibfnamefont
  {J.~M.}\ \bibnamefont {Wheeler}},\ }\bibfield  {title} {\enquote {\bibinfo
  {title} {Investigation of the deformation behavior of aluminum micropillars
  produced by focused ion beam machining using ga and xe ions},}\ }\href
  {https://doi.org/10.1016/j.scriptamat.2016.08.028} {\bibfield  {journal}
  {\bibinfo  {journal} {Scripta Mater.}\ }\textbf {\bibinfo {volume} {127}},\
  \bibinfo {pages} {191--194} (\bibinfo {year} {2017})}\BibitemShut {NoStop}%
\bibitem [{\citenamefont {Bor}\ \emph {et~al.}(2021)\citenamefont {Bor},
  \citenamefont {Giuntini}, \citenamefont {Dom{\`{e}}nech}, \citenamefont
  {Plunkett}, \citenamefont {Kampferbeck}, \citenamefont {Vossmeyer},
  \citenamefont {Weller}, \citenamefont {Scheider},\ and\ \citenamefont
  {Schneider}}]{Bor_2021}%
  \BibitemOpen
  \bibfield  {author} {\bibinfo {author} {\bibfnamefont {B.}~\bibnamefont
  {Bor}}, \bibinfo {author} {\bibfnamefont {D.}~\bibnamefont {Giuntini}},
  \bibinfo {author} {\bibfnamefont {B.}~\bibnamefont {Dom{\`{e}}nech}},
  \bibinfo {author} {\bibfnamefont {A.}~\bibnamefont {Plunkett}}, \bibinfo
  {author} {\bibfnamefont {M.}~\bibnamefont {Kampferbeck}}, \bibinfo {author}
  {\bibfnamefont {T.}~\bibnamefont {Vossmeyer}}, \bibinfo {author}
  {\bibfnamefont {H.}~\bibnamefont {Weller}}, \bibinfo {author} {\bibfnamefont
  {I.}~\bibnamefont {Scheider}},\ and\ \bibinfo {author} {\bibfnamefont
  {G.~A.}\ \bibnamefont {Schneider}},\ }\bibfield  {title} {\enquote {\bibinfo
  {title} {Constitutive and fracture behavior of ultra-strong supercrystalline
  nanocomposites},}\ }\href {https://doi.org/10.1063/5.0056616} {\bibfield
  {journal} {\bibinfo  {journal} {Appl. Phys. Rev.}\ }\textbf {\bibinfo
  {volume} {8}},\ \bibinfo {pages} {031414} (\bibinfo {year}
  {2021})}\BibitemShut {NoStop}%
\bibitem [{\citenamefont {Armstrong}, \citenamefont {Wilkinson},\ and\
  \citenamefont {Roberts}(2009)}]{Armstrong_2009}%
  \BibitemOpen
  \bibfield  {author} {\bibinfo {author} {\bibfnamefont {D.~E.~J.}\
  \bibnamefont {Armstrong}}, \bibinfo {author} {\bibfnamefont {A.~J.}\
  \bibnamefont {Wilkinson}},\ and\ \bibinfo {author} {\bibfnamefont {S.~G.}\
  \bibnamefont {Roberts}},\ }\bibfield  {title} {\enquote {\bibinfo {title}
  {Measuring anisotropy in young’s modulus of copper using microcantilever
  testing},}\ }\href {https://doi.org/10.1557/jmr.2009.0396} {\bibfield
  {journal} {\bibinfo  {journal} {J. Mater. Res.}\ }\textbf {\bibinfo {volume}
  {24}},\ \bibinfo {pages} {3268--3276} (\bibinfo {year} {2009})}\BibitemShut
  {NoStop}%
\bibitem [{\citenamefont {Kiener}\ \emph {et~al.}(2009)\citenamefont {Kiener},
  \citenamefont {Motz}, \citenamefont {Dehm},\ and\ \citenamefont
  {Pippan}}]{Kiener_2009}%
  \BibitemOpen
  \bibfield  {author} {\bibinfo {author} {\bibfnamefont {D.}~\bibnamefont
  {Kiener}}, \bibinfo {author} {\bibfnamefont {C.}~\bibnamefont {Motz}},
  \bibinfo {author} {\bibfnamefont {G.}~\bibnamefont {Dehm}},\ and\ \bibinfo
  {author} {\bibfnamefont {R.}~\bibnamefont {Pippan}},\ }\bibfield  {title}
  {\enquote {\bibinfo {title} {Overview on established and novel {FIB} based
  miniaturized mechanical testing using in-situ {SEM}},}\ }\href
  {https://doi.org/10.3139/146.110149} {\bibfield  {journal} {\bibinfo
  {journal} {Int. J. Mater. Res.}\ }\textbf {\bibinfo {volume} {100}},\
  \bibinfo {pages} {1074--1087} (\bibinfo {year} {2009})}\BibitemShut {NoStop}%
\bibitem [{\citenamefont {Legros}, \citenamefont {Gianola},\ and\ \citenamefont
  {Motz}(2010)}]{Legros_2010}%
  \BibitemOpen
  \bibfield  {author} {\bibinfo {author} {\bibfnamefont {M.}~\bibnamefont
  {Legros}}, \bibinfo {author} {\bibfnamefont {D.~S.}\ \bibnamefont
  {Gianola}},\ and\ \bibinfo {author} {\bibfnamefont {C.}~\bibnamefont
  {Motz}},\ }\bibfield  {title} {\enquote {\bibinfo {title} {Quantitative
  in-situ mechanical testing in electron microscopes},}\ }\href
  {https://doi.org/10.1557/mrs2010.567} {\bibfield  {journal} {\bibinfo
  {journal} {{MRS} Bulletin}\ }\textbf {\bibinfo {volume} {35}},\ \bibinfo
  {pages} {354--360} (\bibinfo {year} {2010})}\BibitemShut {NoStop}%
\bibitem [{\citenamefont {Kiener}\ and\ \citenamefont
  {Minor}(2011)}]{Kiener_2011}%
  \BibitemOpen
  \bibfield  {author} {\bibinfo {author} {\bibfnamefont {D.}~\bibnamefont
  {Kiener}}\ and\ \bibinfo {author} {\bibfnamefont {A.~M.}\ \bibnamefont
  {Minor}},\ }\bibfield  {title} {\enquote {\bibinfo {title} {Source truncation
  and exhaustion: insights from quantitative in situ tem tensile testing},}\
  }\href {https://doi.org/10.1021/nl201890s} {\bibfield  {journal} {\bibinfo
  {journal} {Nano Lett.}\ }\textbf {\bibinfo {volume} {11}},\ \bibinfo {pages}
  {3816--3820} (\bibinfo {year} {2011})}\BibitemShut {NoStop}%
\bibitem [{\citenamefont {Ast}\ \emph {et~al.}(2014)\citenamefont {Ast},
  \citenamefont {Przybilla}, \citenamefont {Maier}, \citenamefont {Durst},\
  and\ \citenamefont {G{\"{o}}ken}}]{Ast_2014}%
  \BibitemOpen
  \bibfield  {author} {\bibinfo {author} {\bibfnamefont {J.}~\bibnamefont
  {Ast}}, \bibinfo {author} {\bibfnamefont {T.}~\bibnamefont {Przybilla}},
  \bibinfo {author} {\bibfnamefont {V.}~\bibnamefont {Maier}}, \bibinfo
  {author} {\bibfnamefont {K.}~\bibnamefont {Durst}},\ and\ \bibinfo {author}
  {\bibfnamefont {M.}~\bibnamefont {G{\"{o}}ken}},\ }\bibfield  {title}
  {\enquote {\bibinfo {title} {Microcantilever bending experiments in nial –
  evaluation, size effects, and crack tip plasticity},}\ }\href
  {https://doi.org/10.1557/jmr.2014.240} {\bibfield  {journal} {\bibinfo
  {journal} {J. Mater. Res.}\ }\textbf {\bibinfo {volume} {29}},\ \bibinfo
  {pages} {2129--2140} (\bibinfo {year} {2014})}\BibitemShut {NoStop}%
\bibitem [{\citenamefont {Sch{\"{o}}ngrundner}\ \emph
  {et~al.}(2014)\citenamefont {Sch{\"{o}}ngrundner}, \citenamefont {Treml},
  \citenamefont {Antretter}, \citenamefont {Kozic}, \citenamefont {Ecker},
  \citenamefont {Kiener},\ and\ \citenamefont {Brunner}}]{Schoengrundner_2014}%
  \BibitemOpen
  \bibfield  {author} {\bibinfo {author} {\bibfnamefont {R.}~\bibnamefont
  {Sch{\"{o}}ngrundner}}, \bibinfo {author} {\bibfnamefont {R.}~\bibnamefont
  {Treml}}, \bibinfo {author} {\bibfnamefont {T.}~\bibnamefont {Antretter}},
  \bibinfo {author} {\bibfnamefont {D.}~\bibnamefont {Kozic}}, \bibinfo
  {author} {\bibfnamefont {W.}~\bibnamefont {Ecker}}, \bibinfo {author}
  {\bibfnamefont {D.}~\bibnamefont {Kiener}},\ and\ \bibinfo {author}
  {\bibfnamefont {R.}~\bibnamefont {Brunner}},\ }\bibfield  {title} {\enquote
  {\bibinfo {title} {Critical assessment of the determination of residual
  stress profiles in thin films by means of the ion beam layer removal
  method},}\ }\href {https://doi.org/10.1016/j.tsf.2014.06.003} {\bibfield
  {journal} {\bibinfo  {journal} {Thin Solid Films}\ }\textbf {\bibinfo
  {volume} {564}},\ \bibinfo {pages} {321--330} (\bibinfo {year}
  {2014})}\BibitemShut {NoStop}%
\bibitem [{\citenamefont {Best}\ \emph {et~al.}(2016)\citenamefont {Best},
  \citenamefont {Zechner}, \citenamefont {Shorubalko}, \citenamefont
  {Obo{\v{n}}a}, \citenamefont {Wehrs}, \citenamefont {Morstein},\ and\
  \citenamefont {Michler}}]{Best_2016}%
  \BibitemOpen
  \bibfield  {author} {\bibinfo {author} {\bibfnamefont {J.~P.}\ \bibnamefont
  {Best}}, \bibinfo {author} {\bibfnamefont {J.}~\bibnamefont {Zechner}},
  \bibinfo {author} {\bibfnamefont {I.}~\bibnamefont {Shorubalko}}, \bibinfo
  {author} {\bibfnamefont {J.~V.}\ \bibnamefont {Obo{\v{n}}a}}, \bibinfo
  {author} {\bibfnamefont {J.}~\bibnamefont {Wehrs}}, \bibinfo {author}
  {\bibfnamefont {M.}~\bibnamefont {Morstein}},\ and\ \bibinfo {author}
  {\bibfnamefont {J.}~\bibnamefont {Michler}},\ }\bibfield  {title} {\enquote
  {\bibinfo {title} {A comparison of three different notching ions for
  small-scale fracture toughness measurement},}\ }\href
  {https://doi.org/10.1016/j.scriptamat.2015.09.014} {\bibfield  {journal}
  {\bibinfo  {journal} {Scripta Mater.}\ }\textbf {\bibinfo {volume} {112}},\
  \bibinfo {pages} {71--74} (\bibinfo {year} {2016})}\BibitemShut {NoStop}%
\bibitem [{\citenamefont {Kal{\'{a}}cska}\ \emph {et~al.}(2020)\citenamefont
  {Kal{\'{a}}cska}, \citenamefont {Ast}, \citenamefont {Isp{\'{a}}novity},
  \citenamefont {Michler},\ and\ \citenamefont {Maeder}}]{Kalacska_2020}%
  \BibitemOpen
  \bibfield  {author} {\bibinfo {author} {\bibfnamefont {S.}~\bibnamefont
  {Kal{\'{a}}cska}}, \bibinfo {author} {\bibfnamefont {J.}~\bibnamefont {Ast}},
  \bibinfo {author} {\bibfnamefont {P.~D.}\ \bibnamefont {Isp{\'{a}}novity}},
  \bibinfo {author} {\bibfnamefont {J.}~\bibnamefont {Michler}},\ and\ \bibinfo
  {author} {\bibfnamefont {X.}~\bibnamefont {Maeder}},\ }\bibfield  {title}
  {\enquote {\bibinfo {title} {{3D HR-EBSD} characterization of the plastic
  zone around crack tips in tungsten single crystals at the micron scale},}\
  }\href {https://doi.org/10.1016/j.actamat.2020.09.009} {\bibfield  {journal}
  {\bibinfo  {journal} {Acta Mater.}\ }\textbf {\bibinfo {volume} {200}},\
  \bibinfo {pages} {211--222} (\bibinfo {year} {2020})}\BibitemShut {NoStop}%
\bibitem [{\citenamefont {Kiener}\ \emph {et~al.}(2008)\citenamefont {Kiener},
  \citenamefont {Grosinger}, \citenamefont {Dehm},\ and\ \citenamefont
  {Pippan}}]{Kiener_2008}%
  \BibitemOpen
  \bibfield  {author} {\bibinfo {author} {\bibfnamefont {D.}~\bibnamefont
  {Kiener}}, \bibinfo {author} {\bibfnamefont {W.}~\bibnamefont {Grosinger}},
  \bibinfo {author} {\bibfnamefont {G.}~\bibnamefont {Dehm}},\ and\ \bibinfo
  {author} {\bibfnamefont {R.}~\bibnamefont {Pippan}},\ }\bibfield  {title}
  {\enquote {\bibinfo {title} {A further step towards an understanding of
  size-dependent crystal plasticity: in situ tension experiments of
  miniaturized single-crystal copper samples},}\ }\href
  {https://doi.org/10.1016/j.actamat.2007.10.015} {\bibfield  {journal}
  {\bibinfo  {journal} {Acta Mater.}\ }\textbf {\bibinfo {volume} {56}},\
  \bibinfo {pages} {580--592} (\bibinfo {year} {2008})}\BibitemShut {NoStop}%
\bibitem [{\citenamefont {Casari}\ \emph {et~al.}(2019)\citenamefont {Casari},
  \citenamefont {Peth{\"{o}}}, \citenamefont {Sch{\"{u}}rch}, \citenamefont
  {Maeder}, \citenamefont {Philippe}, \citenamefont {Michler}, \citenamefont
  {Zysset},\ and\ \citenamefont {Schwiedrzik}}]{Casari_2019}%
  \BibitemOpen
  \bibfield  {author} {\bibinfo {author} {\bibfnamefont {D.}~\bibnamefont
  {Casari}}, \bibinfo {author} {\bibfnamefont {L.}~\bibnamefont {Peth{\"{o}}}},
  \bibinfo {author} {\bibfnamefont {P.}~\bibnamefont {Sch{\"{u}}rch}}, \bibinfo
  {author} {\bibfnamefont {X.}~\bibnamefont {Maeder}}, \bibinfo {author}
  {\bibfnamefont {L.}~\bibnamefont {Philippe}}, \bibinfo {author}
  {\bibfnamefont {J.}~\bibnamefont {Michler}}, \bibinfo {author} {\bibfnamefont
  {P.}~\bibnamefont {Zysset}},\ and\ \bibinfo {author} {\bibfnamefont
  {J.}~\bibnamefont {Schwiedrzik}},\ }\bibfield  {title} {\enquote {\bibinfo
  {title} {A self-aligning microtensile setup: application to single-crystal
  {GaAs} microscale tension--compression asymmetry},}\ }\href
  {https://doi.org/10.1557/jmr.2019.183} {\bibfield  {journal} {\bibinfo
  {journal} {J. Mater. Res.}\ }\textbf {\bibinfo {volume} {34}},\ \bibinfo
  {pages} {2517--2534} (\bibinfo {year} {2019})}\BibitemShut {NoStop}%
\bibitem [{\citenamefont {Soleimani}\ \emph {et~al.}(2022)\citenamefont
  {Soleimani}, \citenamefont {Maddala}, \citenamefont {Wismans}, \citenamefont
  {Liao}, \citenamefont {Breemen}, \citenamefont {Benthem},\ and\ \citenamefont
  {Friedrich}}]{Soleimani_2022}%
  \BibitemOpen
  \bibfield  {author} {\bibinfo {author} {\bibfnamefont {M.}~\bibnamefont
  {Soleimani}}, \bibinfo {author} {\bibfnamefont {S.~P.}\ \bibnamefont
  {Maddala}}, \bibinfo {author} {\bibfnamefont {M.}~\bibnamefont {Wismans}},
  \bibinfo {author} {\bibfnamefont {W.-C.}\ \bibnamefont {Liao}}, \bibinfo
  {author} {\bibfnamefont {L.~C.~A.}\ \bibnamefont {Breemen}}, \bibinfo
  {author} {\bibfnamefont {R.~A. T.~M.}\ \bibnamefont {Benthem}},\ and\
  \bibinfo {author} {\bibfnamefont {H.}~\bibnamefont {Friedrich}},\ }\bibfield
  {title} {\enquote {\bibinfo {title} {In situ fabrication, manipulation, and
  mechanical characterization of free-standing silica thin films using focused
  ion beam scanning electron microscopy},}\ }\href
  {https://doi.org/10.1002/admi.202102201} {\bibfield  {journal} {\bibinfo
  {journal} {Adv. Mater. Interfaces}\ }\textbf {\bibinfo {volume} {9}},\
  \bibinfo {pages} {2102201} (\bibinfo {year} {2022})}\BibitemShut {NoStop}%
\bibitem [{\citenamefont {Korsunsky}, \citenamefont {Sebastiani},\ and\
  \citenamefont {Bemporad}(2010)}]{Korsunsky_2010}%
  \BibitemOpen
  \bibfield  {author} {\bibinfo {author} {\bibfnamefont {A.~M.}\ \bibnamefont
  {Korsunsky}}, \bibinfo {author} {\bibfnamefont {M.}~\bibnamefont
  {Sebastiani}},\ and\ \bibinfo {author} {\bibfnamefont {E.}~\bibnamefont
  {Bemporad}},\ }\bibfield  {title} {\enquote {\bibinfo {title} {Residual
  stress evaluation at the micrometer scale: analysis of thin coatings by {FIB}
  milling and digital image correlation},}\ }\href
  {https://doi.org/10.1016/j.surfcoat.2010.09.033} {\bibfield  {journal}
  {\bibinfo  {journal} {Surf. Coat. Technol.}\ }\textbf {\bibinfo {volume}
  {205}},\ \bibinfo {pages} {2393--2403} (\bibinfo {year} {2010})}\BibitemShut
  {NoStop}%
\bibitem [{\citenamefont {Heyer}\ \emph {et~al.}(2014)\citenamefont {Heyer},
  \citenamefont {Brinckmann}, \citenamefont {Pfetzing-Micklich},\ and\
  \citenamefont {Eggeler}}]{Heyer_2014}%
  \BibitemOpen
  \bibfield  {author} {\bibinfo {author} {\bibfnamefont {J.-K.}\ \bibnamefont
  {Heyer}}, \bibinfo {author} {\bibfnamefont {S.}~\bibnamefont {Brinckmann}},
  \bibinfo {author} {\bibfnamefont {J.}~\bibnamefont {Pfetzing-Micklich}},\
  and\ \bibinfo {author} {\bibfnamefont {G.}~\bibnamefont {Eggeler}},\
  }\bibfield  {title} {\enquote {\bibinfo {title} {Microshear deformation of
  gold single crystals},}\ }\href
  {https://doi.org/10.1016/j.actamat.2013.10.002} {\bibfield  {journal}
  {\bibinfo  {journal} {Acta Mater.}\ }\textbf {\bibinfo {volume} {62}},\
  \bibinfo {pages} {225--238} (\bibinfo {year} {2014})}\BibitemShut {NoStop}%
\bibitem [{\citenamefont {Prei{\ss}}\ \emph {et~al.}(2021)\citenamefont
  {Prei{\ss}}, \citenamefont {Merle}, \citenamefont {Xiao}, \citenamefont
  {Gannott}, \citenamefont {Liebig}, \citenamefont {Wheeler},\ and\
  \citenamefont {G{\"{o}}ken}}]{Preiss_2021}%
  \BibitemOpen
  \bibfield  {author} {\bibinfo {author} {\bibfnamefont {E.~I.}\ \bibnamefont
  {Prei{\ss}}}, \bibinfo {author} {\bibfnamefont {B.}~\bibnamefont {Merle}},
  \bibinfo {author} {\bibfnamefont {Y.}~\bibnamefont {Xiao}}, \bibinfo {author}
  {\bibfnamefont {F.}~\bibnamefont {Gannott}}, \bibinfo {author} {\bibfnamefont
  {J.~P.}\ \bibnamefont {Liebig}}, \bibinfo {author} {\bibfnamefont {J.~M.}\
  \bibnamefont {Wheeler}},\ and\ \bibinfo {author} {\bibfnamefont
  {M.}~\bibnamefont {G{\"{o}}ken}},\ }\bibfield  {title} {\enquote {\bibinfo
  {title} {Applicability of focused ion beam ({FIB}) milling with gallium,
  neon, and xenon to the fracture toughness characterization of gold thin
  films},}\ }\href {https://doi.org/10.1557/s43578-020-00045-w} {\bibfield
  {journal} {\bibinfo  {journal} {J. Mater. Res.}\ }\textbf {\bibinfo {volume}
  {36}},\ \bibinfo {pages} {2505--2514} (\bibinfo {year} {2021})}\BibitemShut
  {NoStop}%
\bibitem [{\citenamefont {Aramaki}\ \emph {et~al.}(2013)\citenamefont
  {Aramaki}, \citenamefont {Kozakai}, \citenamefont {Matsuda}, \citenamefont
  {Takaoka}, \citenamefont {Sugiyama}, \citenamefont {Oba}, \citenamefont
  {Aita},\ and\ \citenamefont {Yasaka}}]{Aramaki_2013}%
  \BibitemOpen
  \bibfield  {author} {\bibinfo {author} {\bibfnamefont {F.}~\bibnamefont
  {Aramaki}}, \bibinfo {author} {\bibfnamefont {T.}~\bibnamefont {Kozakai}},
  \bibinfo {author} {\bibfnamefont {O.}~\bibnamefont {Matsuda}}, \bibinfo
  {author} {\bibfnamefont {O.}~\bibnamefont {Takaoka}}, \bibinfo {author}
  {\bibfnamefont {Y.}~\bibnamefont {Sugiyama}}, \bibinfo {author}
  {\bibfnamefont {H.}~\bibnamefont {Oba}}, \bibinfo {author} {\bibfnamefont
  {K.}~\bibnamefont {Aita}},\ and\ \bibinfo {author} {\bibfnamefont
  {A.}~\bibnamefont {Yasaka}},\ }\bibfield  {title} {\enquote {\bibinfo {title}
  {{Photomask Repair Technology by using Gas Field Ion Source}},}\ }\href@noop
  {} {\bibfield  {journal} {\bibinfo  {journal} {BACUS}\ }\textbf {\bibinfo
  {volume} {29}},\ \bibinfo {pages} {1--8} (\bibinfo {year}
  {2013})}\BibitemShut {NoStop}%
\bibitem [{\citenamefont {Aramaki}\ \emph {et~al.}(2014)\citenamefont
  {Aramaki}, \citenamefont {Kozakai}, \citenamefont {Matsuda}, \citenamefont
  {Yasaka}, \citenamefont {Yoshikawa}, \citenamefont {Kanno}, \citenamefont
  {Miyashita},\ and\ \citenamefont {Hayashi}}]{Aramaki_2014}%
  \BibitemOpen
  \bibfield  {author} {\bibinfo {author} {\bibfnamefont {F.}~\bibnamefont
  {Aramaki}}, \bibinfo {author} {\bibfnamefont {T.}~\bibnamefont {Kozakai}},
  \bibinfo {author} {\bibfnamefont {O.}~\bibnamefont {Matsuda}}, \bibinfo
  {author} {\bibfnamefont {A.}~\bibnamefont {Yasaka}}, \bibinfo {author}
  {\bibfnamefont {S.}~\bibnamefont {Yoshikawa}}, \bibinfo {author}
  {\bibfnamefont {K.}~\bibnamefont {Kanno}}, \bibinfo {author} {\bibfnamefont
  {H.}~\bibnamefont {Miyashita}},\ and\ \bibinfo {author} {\bibfnamefont
  {N.}~\bibnamefont {Hayashi}},\ }\bibfield  {title} {\enquote {\bibinfo
  {title} {Performance of {GFIS} mask repair system for various mask
  materials},}\ }in\ \href {https://doi.org/10.1117/12.2069435} {\emph
  {\bibinfo {booktitle} {{SPIE} Proceedings}}},\ \bibinfo {editor} {edited by\
  \bibinfo {editor} {\bibfnamefont {P.~W.}\ \bibnamefont {Ackmann}}\ and\
  \bibinfo {editor} {\bibfnamefont {N.}~\bibnamefont {Hayashi}}}\ (\bibinfo
  {publisher} {{SPIE}},\ \bibinfo {year} {2014})\BibitemShut {NoStop}%
\bibitem [{\citenamefont {Edinger}\ \emph {et~al.}(2014)\citenamefont
  {Edinger}, \citenamefont {Wolff}, \citenamefont {Steigerwald}, \citenamefont
  {Auth}, \citenamefont {Spies}, \citenamefont {Oster}, \citenamefont
  {Schneider}, \citenamefont {Budach}, \citenamefont {Hofmann},\ and\
  \citenamefont {Waiblinger}}]{Edinger_2014}%
  \BibitemOpen
  \bibfield  {author} {\bibinfo {author} {\bibfnamefont {K.}~\bibnamefont
  {Edinger}}, \bibinfo {author} {\bibfnamefont {K.}~\bibnamefont {Wolff}},
  \bibinfo {author} {\bibfnamefont {H.}~\bibnamefont {Steigerwald}}, \bibinfo
  {author} {\bibfnamefont {N.}~\bibnamefont {Auth}}, \bibinfo {author}
  {\bibfnamefont {P.}~\bibnamefont {Spies}}, \bibinfo {author} {\bibfnamefont
  {J.}~\bibnamefont {Oster}}, \bibinfo {author} {\bibfnamefont
  {H.}~\bibnamefont {Schneider}}, \bibinfo {author} {\bibfnamefont
  {M.}~\bibnamefont {Budach}}, \bibinfo {author} {\bibfnamefont
  {T.}~\bibnamefont {Hofmann}},\ and\ \bibinfo {author} {\bibfnamefont
  {M.}~\bibnamefont {Waiblinger}},\ }\bibfield  {title} {\enquote {\bibinfo
  {title} {Bringing mask repair to the next level},}\ }in\ \href
  {https://doi.org/10.1117/12.2072474} {\emph {\bibinfo {booktitle} {{SPIE}
  Proceedings}}},\ \bibinfo {editor} {edited by\ \bibinfo {editor}
  {\bibfnamefont {P.~W.}\ \bibnamefont {Ackmann}}\ and\ \bibinfo {editor}
  {\bibfnamefont {N.}~\bibnamefont {Hayashi}}}\ (\bibinfo  {publisher}
  {{SPIE}},\ \bibinfo {year} {2014})\BibitemShut {NoStop}%
\bibitem [{\citenamefont {Tan}\ and\ \citenamefont
  {Livengood}(2016)}]{Tan_2016}%
  \BibitemOpen
  \bibfield  {author} {\bibinfo {author} {\bibfnamefont {S.}~\bibnamefont
  {Tan}}\ and\ \bibinfo {author} {\bibfnamefont {R.~H.}\ \bibnamefont
  {Livengood}},\ }\bibfield  {title} {\enquote {\bibinfo {title} {Applications
  of {GFIS} in semiconductors},}\ }in\  \cite{Hlawacek_2016},\ Chap.~\bibinfo
  {chapter} {19}, pp.\ \bibinfo {pages} {471--498}\BibitemShut {NoStop}%
\bibitem [{\citenamefont {Heard}(1985)}]{Heard_1985}%
  \BibitemOpen
  \bibfield  {author} {\bibinfo {author} {\bibfnamefont {P.~J.}\ \bibnamefont
  {Heard}},\ }\bibfield  {title} {\enquote {\bibinfo {title} {Application of a
  focused ion beam system to defect repair of {VLSI} masks},}\ }\href
  {https://doi.org/10.1116/1.583297} {\bibfield  {journal} {\bibinfo  {journal}
  {J. Vac. Sci. Technol. B}\ }\textbf {\bibinfo {volume} {3}},\ \bibinfo
  {pages} {87} (\bibinfo {year} {1985})}\BibitemShut {NoStop}%
\bibitem [{\citenamefont {Prewett}\ and\ \citenamefont
  {Heard}(1987)}]{Prewett_1987}%
  \BibitemOpen
  \bibfield  {author} {\bibinfo {author} {\bibfnamefont {P.~D.}\ \bibnamefont
  {Prewett}}\ and\ \bibinfo {author} {\bibfnamefont {P.~J.}\ \bibnamefont
  {Heard}},\ }\bibfield  {title} {\enquote {\bibinfo {title} {Repair of opaque
  defects in photomasks using focused ion beams},}\ }\href
  {https://doi.org/10.1088/0022-3727/20/9/021} {\bibfield  {journal} {\bibinfo
  {journal} {J. Phys. D: Appl. Phys.}\ }\textbf {\bibinfo {volume} {20}},\
  \bibinfo {pages} {1207--1209} (\bibinfo {year} {1987})}\BibitemShut {NoStop}%
\bibitem [{\citenamefont {Cleaver}\ \emph {et~al.}(1985)\citenamefont
  {Cleaver}, \citenamefont {Ahmed}, \citenamefont {Heard}, \citenamefont
  {Prewett}, \citenamefont {Dunn},\ and\ \citenamefont
  {Kaufmann}}]{Cleaver_1985}%
  \BibitemOpen
  \bibfield  {author} {\bibinfo {author} {\bibfnamefont {J.~R.~A.}\
  \bibnamefont {Cleaver}}, \bibinfo {author} {\bibfnamefont {H.}~\bibnamefont
  {Ahmed}}, \bibinfo {author} {\bibfnamefont {P.~J.}\ \bibnamefont {Heard}},
  \bibinfo {author} {\bibfnamefont {P.~D.}\ \bibnamefont {Prewett}}, \bibinfo
  {author} {\bibfnamefont {G.~J.}\ \bibnamefont {Dunn}},\ and\ \bibinfo
  {author} {\bibfnamefont {H.}~\bibnamefont {Kaufmann}},\ }\bibfield  {title}
  {\enquote {\bibinfo {title} {Focused ion beam repair techniques for clear and
  opaque defects in masks},}\ }\href
  {https://doi.org/10.1016/0167-9317(85)90034-6} {\bibfield  {journal}
  {\bibinfo  {journal} {Microelectron. Eng.}\ }\textbf {\bibinfo {volume}
  {3}},\ \bibinfo {pages} {253--260} (\bibinfo {year} {1985})}\BibitemShut
  {NoStop}%
\bibitem [{\citenamefont {Liang}\ \emph {et~al.}(2000)\citenamefont {Liang},
  \citenamefont {Stivers}, \citenamefont {Livengood}, \citenamefont {Yan},
  \citenamefont {Zhang},\ and\ \citenamefont {Lo}}]{Liang_2000}%
  \BibitemOpen
  \bibfield  {author} {\bibinfo {author} {\bibfnamefont {T.}~\bibnamefont
  {Liang}}, \bibinfo {author} {\bibfnamefont {A.}~\bibnamefont {Stivers}},
  \bibinfo {author} {\bibfnamefont {R.}~\bibnamefont {Livengood}}, \bibinfo
  {author} {\bibfnamefont {P.-Y.}\ \bibnamefont {Yan}}, \bibinfo {author}
  {\bibfnamefont {G.}~\bibnamefont {Zhang}},\ and\ \bibinfo {author}
  {\bibfnamefont {F.-C.}\ \bibnamefont {Lo}},\ }\bibfield  {title} {\enquote
  {\bibinfo {title} {Progress in extreme ultraviolet mask repair using a
  focused ion beam},}\ }\href {https://doi.org/10.1116/1.1319687} {\bibfield
  {journal} {\bibinfo  {journal} {J. Vac. Sci. Technol. B}\ }\textbf {\bibinfo
  {volume} {18}},\ \bibinfo {pages} {3216} (\bibinfo {year}
  {2000})}\BibitemShut {NoStop}%
\bibitem [{\citenamefont {Wagner}(1990)}]{Wagner_1990}%
  \BibitemOpen
  \bibfield  {author} {\bibinfo {author} {\bibfnamefont {A.}~\bibnamefont
  {Wagner}},\ }\bibfield  {title} {\enquote {\bibinfo {title} {X-ray mask
  repair with focused ion beams},}\ }\href {https://doi.org/10.1116/1.585115}
  {\bibfield  {journal} {\bibinfo  {journal} {J. Vac. Sci. Technol. B}\
  }\textbf {\bibinfo {volume} {8}},\ \bibinfo {pages} {1557} (\bibinfo {year}
  {1990})}\BibitemShut {NoStop}%
\bibitem [{\citenamefont {Tan}\ \emph {et~al.}(2015)\citenamefont {Tan},
  \citenamefont {Hallstein}, \citenamefont {Livengood}, \citenamefont
  {Prakasam}, \citenamefont {Patel}, \citenamefont {Nastasescu},\ and\
  \citenamefont {Scheffler}}]{Tan_2015}%
  \BibitemOpen
  \bibfield  {author} {\bibinfo {author} {\bibfnamefont {S.}~\bibnamefont
  {Tan}}, \bibinfo {author} {\bibfnamefont {R.}~\bibnamefont {Hallstein}},
  \bibinfo {author} {\bibfnamefont {R.}~\bibnamefont {Livengood}}, \bibinfo
  {author} {\bibfnamefont {H.}~\bibnamefont {Prakasam}}, \bibinfo {author}
  {\bibfnamefont {L.}~\bibnamefont {Patel}}, \bibinfo {author} {\bibfnamefont
  {O.}~\bibnamefont {Nastasescu}},\ and\ \bibinfo {author} {\bibfnamefont
  {C.}~\bibnamefont {Scheffler}},\ }\bibfield  {title} {\enquote {\bibinfo
  {title} {Exploring neon gfis nano-machining applications incircuit edit},}\
  }in\ \href {https://eipbn.org/abstracts/2015/papers/2C1.pdf} {\emph {\bibinfo
  {booktitle} {EIPBN}}}\ (\bibinfo {year} {2015})\BibitemShut {NoStop}%
\bibitem [{\citenamefont {Harriott}\ \emph {et~al.}(1986)\citenamefont
  {Harriott}, \citenamefont {Cummings}, \citenamefont {Gross},\ and\
  \citenamefont {Brown}}]{Harriott_1986a}%
  \BibitemOpen
  \bibfield  {author} {\bibinfo {author} {\bibfnamefont {L.~R.}\ \bibnamefont
  {Harriott}}, \bibinfo {author} {\bibfnamefont {K.~D.}\ \bibnamefont
  {Cummings}}, \bibinfo {author} {\bibfnamefont {M.~E.}\ \bibnamefont
  {Gross}},\ and\ \bibinfo {author} {\bibfnamefont {W.~L.}\ \bibnamefont
  {Brown}},\ }\bibfield  {title} {\enquote {\bibinfo {title} {Decomposition of
  palladium acetate films with a microfocused ion beam},}\ }\href
  {https://doi.org/10.1063/1.97259} {\bibfield  {journal} {\bibinfo  {journal}
  {Appl. Phys. Lett.}\ }\textbf {\bibinfo {volume} {49}},\ \bibinfo {pages}
  {1661--1662} (\bibinfo {year} {1986})}\BibitemShut {NoStop}%
\bibitem [{\citenamefont {Jang}\ \emph {et~al.}(2011)\citenamefont {Jang},
  \citenamefont {Ferguson}, \citenamefont {Vakaryuk}, \citenamefont {Budakian},
  \citenamefont {Chung}, \citenamefont {Goldbart},\ and\ \citenamefont
  {Maeno}}]{Jang_2011}%
  \BibitemOpen
  \bibfield  {author} {\bibinfo {author} {\bibfnamefont {J.}~\bibnamefont
  {Jang}}, \bibinfo {author} {\bibfnamefont {D.~G.}\ \bibnamefont {Ferguson}},
  \bibinfo {author} {\bibfnamefont {V.}~\bibnamefont {Vakaryuk}}, \bibinfo
  {author} {\bibfnamefont {R.}~\bibnamefont {Budakian}}, \bibinfo {author}
  {\bibfnamefont {S.~B.}\ \bibnamefont {Chung}}, \bibinfo {author}
  {\bibfnamefont {P.~M.}\ \bibnamefont {Goldbart}},\ and\ \bibinfo {author}
  {\bibfnamefont {Y.}~\bibnamefont {Maeno}},\ }\bibfield  {title} {\enquote
  {\bibinfo {title} {Observation of half-height magnetization steps in
  {Sr$_2$RuO$_4$}},}\ }\href {https://doi.org/10.1126/science.1193839}
  {\bibfield  {journal} {\bibinfo  {journal} {Science}\ }\textbf {\bibinfo
  {volume} {331}},\ \bibinfo {pages} {186--188} (\bibinfo {year}
  {2011})}\BibitemShut {NoStop}%
\bibitem [{\citenamefont {Yasui}\ \emph {et~al.}(2020)\citenamefont {Yasui},
  \citenamefont {Lahabi}, \citenamefont {Becerra}, \citenamefont {Fermin},
  \citenamefont {Anwar}, \citenamefont {Yonezawa}, \citenamefont {Terashima},
  \citenamefont {Milo{\v{s}}evi{\'{c}}}, \citenamefont {Aarts},\ and\
  \citenamefont {Maeno}}]{Yasui_2020}%
  \BibitemOpen
  \bibfield  {author} {\bibinfo {author} {\bibfnamefont {Y.}~\bibnamefont
  {Yasui}}, \bibinfo {author} {\bibfnamefont {K.}~\bibnamefont {Lahabi}},
  \bibinfo {author} {\bibfnamefont {V.~F.}\ \bibnamefont {Becerra}}, \bibinfo
  {author} {\bibfnamefont {R.}~\bibnamefont {Fermin}}, \bibinfo {author}
  {\bibfnamefont {M.~S.}\ \bibnamefont {Anwar}}, \bibinfo {author}
  {\bibfnamefont {S.}~\bibnamefont {Yonezawa}}, \bibinfo {author}
  {\bibfnamefont {T.}~\bibnamefont {Terashima}}, \bibinfo {author}
  {\bibfnamefont {M.~V.}\ \bibnamefont {Milo{\v{s}}evi{\'{c}}}}, \bibinfo
  {author} {\bibfnamefont {J.}~\bibnamefont {Aarts}},\ and\ \bibinfo {author}
  {\bibfnamefont {Y.}~\bibnamefont {Maeno}},\ }\bibfield  {title} {\enquote
  {\bibinfo {title} {Spontaneous emergence of josephson junctions in
  homogeneous rings of single-crystal {Sr$_2$RuO$_4$}},}\ }\href
  {https://doi.org/10.1038/s41535-020-0223-7} {\bibfield  {journal} {\bibinfo
  {journal} {npj Quantum Materials}\ }\textbf {\bibinfo {volume} {5}} (\bibinfo
  {year} {2020}),\ 10.1038/s41535-020-0223-7}\BibitemShut {NoStop}%
\bibitem [{\citenamefont {Moll}\ \emph {et~al.}(2014)\citenamefont {Moll},
  \citenamefont {Zhu}, \citenamefont {Cheng}, \citenamefont {Wen},\ and\
  \citenamefont {Batlogg}}]{Moll_2014}%
  \BibitemOpen
  \bibfield  {author} {\bibinfo {author} {\bibfnamefont {P.~J.~W.}\
  \bibnamefont {Moll}}, \bibinfo {author} {\bibfnamefont {X.}~\bibnamefont
  {Zhu}}, \bibinfo {author} {\bibfnamefont {P.}~\bibnamefont {Cheng}}, \bibinfo
  {author} {\bibfnamefont {H.-H.}\ \bibnamefont {Wen}},\ and\ \bibinfo {author}
  {\bibfnamefont {B.}~\bibnamefont {Batlogg}},\ }\bibfield  {title} {\enquote
  {\bibinfo {title} {Intrinsic josephson junctions in the iron-based multi-band
  superconductor {(V$_2$Sr$_4$O$_6$)Fe$_2$As$_2$}},}\ }\href
  {https://doi.org/10.1038/nphys3034} {\bibfield  {journal} {\bibinfo
  {journal} {Nat. Phys.}\ }\textbf {\bibinfo {volume} {10}},\ \bibinfo {pages}
  {644--647} (\bibinfo {year} {2014})}\BibitemShut {NoStop}%
\bibitem [{\citenamefont {Ooi}, \citenamefont {Mochiku},\ and\ \citenamefont
  {Hirata}(2002)}]{Ooi_2002}%
  \BibitemOpen
  \bibfield  {author} {\bibinfo {author} {\bibfnamefont {S.}~\bibnamefont
  {Ooi}}, \bibinfo {author} {\bibfnamefont {T.}~\bibnamefont {Mochiku}},\ and\
  \bibinfo {author} {\bibfnamefont {K.}~\bibnamefont {Hirata}},\ }\bibfield
  {title} {\enquote {\bibinfo {title} {Periodic oscillations of
  josephson-vortex flow resistance in {Bi$_2$Sr$_2$CaCu$_2$O$_{8+y}$}},}\
  }\href {https://doi.org/10.1103/physrevlett.89.247002} {\bibfield  {journal}
  {\bibinfo  {journal} {Phys. Rev. Lett.}\ }\textbf {\bibinfo {volume} {89}},\
  \bibinfo {pages} {247002} (\bibinfo {year} {2002})}\BibitemShut {NoStop}%
\bibitem [{\citenamefont {Pautrat}\ \emph {et~al.}(2004)\citenamefont
  {Pautrat}, \citenamefont {Scola}, \citenamefont {Goupil}, \citenamefont
  {Simon}, \citenamefont {Villard}, \citenamefont {Domeng\`es}, \citenamefont
  {Simon}, \citenamefont {Guilpin},\ and\ \citenamefont
  {M\'echin}}]{Pautrat_2004}%
  \BibitemOpen
  \bibfield  {author} {\bibinfo {author} {\bibfnamefont {A.}~\bibnamefont
  {Pautrat}}, \bibinfo {author} {\bibfnamefont {J.}~\bibnamefont {Scola}},
  \bibinfo {author} {\bibfnamefont {C.}~\bibnamefont {Goupil}}, \bibinfo
  {author} {\bibfnamefont {C.}~\bibnamefont {Simon}}, \bibinfo {author}
  {\bibfnamefont {C.}~\bibnamefont {Villard}}, \bibinfo {author} {\bibfnamefont
  {B.}~\bibnamefont {Domeng\`es}}, \bibinfo {author} {\bibfnamefont
  {Y.}~\bibnamefont {Simon}}, \bibinfo {author} {\bibfnamefont
  {C.}~\bibnamefont {Guilpin}},\ and\ \bibinfo {author} {\bibfnamefont
  {L.}~\bibnamefont {M\'echin}},\ }\bibfield  {title} {\enquote {\bibinfo
  {title} {Quantitative analysis of the critical current due to vortex pinning
  by surface corrugation},}\ }\href
  {https://doi.org/10.1103/PhysRevB.69.224504} {\bibfield  {journal} {\bibinfo
  {journal} {Phys. Rev. B}\ }\textbf {\bibinfo {volume} {69}},\ \bibinfo
  {pages} {224504--1--5} (\bibinfo {year} {2004})}\BibitemShut {NoStop}%
\bibitem [{\citenamefont {Dobrovolskiy}(2017)}]{Dobrovolskiy_2017}%
  \BibitemOpen
  \bibfield  {author} {\bibinfo {author} {\bibfnamefont {O.~V.}\ \bibnamefont
  {Dobrovolskiy}},\ }\bibfield  {title} {\enquote {\bibinfo {title} {Abrikosov
  fluxonics in washboard nanolandscapes},}\ }\href
  {https://doi.org/10.1016/j.physc.2016.07.008} {\bibfield  {journal} {\bibinfo
   {journal} {Physica C}\ }\textbf {\bibinfo {volume} {533}},\ \bibinfo {pages}
  {80--90} (\bibinfo {year} {2017})}\BibitemShut {NoStop}%
\bibitem [{\citenamefont {Dobrovolskiy}\ and\ \citenamefont
  {Huth}(2015)}]{Dobrovolskiy_2015}%
  \BibitemOpen
  \bibfield  {author} {\bibinfo {author} {\bibfnamefont {O.~V.}\ \bibnamefont
  {Dobrovolskiy}}\ and\ \bibinfo {author} {\bibfnamefont {M.}~\bibnamefont
  {Huth}},\ }\bibfield  {title} {\enquote {\bibinfo {title} {Dual cut-off
  direct current-tunable microwave low-pass filter on superconducting {Nb}
  microstrips with asymmetric nanogrooves},}\ }\href
  {https://doi.org/10.1063/1.4917229} {\bibfield  {journal} {\bibinfo
  {journal} {Appl. Phys. Lett.}\ }\textbf {\bibinfo {volume} {106}},\ \bibinfo
  {pages} {142601--1--5} (\bibinfo {year} {2015})}\BibitemShut {NoStop}%
\bibitem [{\citenamefont {Dobrovolskiy}\ \emph {et~al.}(2012)\citenamefont
  {Dobrovolskiy}, \citenamefont {Begun}, \citenamefont {Huth},\ and\
  \citenamefont {Shklovskij}}]{Dobrovolskiy_2012}%
  \BibitemOpen
  \bibfield  {author} {\bibinfo {author} {\bibfnamefont {O.~V.}\ \bibnamefont
  {Dobrovolskiy}}, \bibinfo {author} {\bibfnamefont {E.}~\bibnamefont {Begun}},
  \bibinfo {author} {\bibfnamefont {M.}~\bibnamefont {Huth}},\ and\ \bibinfo
  {author} {\bibfnamefont {V.~A.}\ \bibnamefont {Shklovskij}},\ }\bibfield
  {title} {\enquote {\bibinfo {title} {Electrical transport and pinning
  properties of {Nb} thin films patterned with focused ion beam-milled
  washboard nanostructures},}\ }\href
  {https://doi.org/10.1088/1367-2630/14/11/113027} {\bibfield  {journal}
  {\bibinfo  {journal} {New J. Phys.}\ }\textbf {\bibinfo {volume} {14}},\
  \bibinfo {pages} {113027} (\bibinfo {year} {2012})}\BibitemShut {NoStop}%
\bibitem [{\citenamefont {Moll}\ \emph {et~al.}(2010)\citenamefont {Moll},
  \citenamefont {Puzniak}, \citenamefont {Balakirev}, \citenamefont {Rogacki},
  \citenamefont {Karpinski}, \citenamefont {Zhigadlo},\ and\ \citenamefont
  {Batlogg}}]{Moll_2010}%
  \BibitemOpen
  \bibfield  {author} {\bibinfo {author} {\bibfnamefont {P.~J.~W.}\
  \bibnamefont {Moll}}, \bibinfo {author} {\bibfnamefont {R.}~\bibnamefont
  {Puzniak}}, \bibinfo {author} {\bibfnamefont {F.}~\bibnamefont {Balakirev}},
  \bibinfo {author} {\bibfnamefont {K.}~\bibnamefont {Rogacki}}, \bibinfo
  {author} {\bibfnamefont {J.}~\bibnamefont {Karpinski}}, \bibinfo {author}
  {\bibfnamefont {N.~D.}\ \bibnamefont {Zhigadlo}},\ and\ \bibinfo {author}
  {\bibfnamefont {B.}~\bibnamefont {Batlogg}},\ }\bibfield  {title} {\enquote
  {\bibinfo {title} {High magnetic-field scales and critical currents in
  {SmFeAs}({O, F}) crystals},}\ }\href {https://doi.org/10.1038/nmat2795}
  {\bibfield  {journal} {\bibinfo  {journal} {Nat. Mater.}\ }\textbf {\bibinfo
  {volume} {9}},\ \bibinfo {pages} {628--633} (\bibinfo {year}
  {2010})}\BibitemShut {NoStop}%
\bibitem [{\citenamefont {Jaroszynski}\ \emph {et~al.}(2008)\citenamefont
  {Jaroszynski}, \citenamefont {Hunte}, \citenamefont {Balicas}, \citenamefont
  {Jo}, \citenamefont {Rai{\v{c}}evi{\'{c}}}, \citenamefont {Gurevich},
  \citenamefont {Larbalestier}, \citenamefont {Balakirev}, \citenamefont
  {Fang}, \citenamefont {Cheng}, \citenamefont {Jia},\ and\ \citenamefont
  {Wen}}]{Jaroszynski_2008}%
  \BibitemOpen
  \bibfield  {author} {\bibinfo {author} {\bibfnamefont {J.}~\bibnamefont
  {Jaroszynski}}, \bibinfo {author} {\bibfnamefont {F.}~\bibnamefont {Hunte}},
  \bibinfo {author} {\bibfnamefont {L.}~\bibnamefont {Balicas}}, \bibinfo
  {author} {\bibfnamefont {Y.-j.}\ \bibnamefont {Jo}}, \bibinfo {author}
  {\bibfnamefont {I.}~\bibnamefont {Rai{\v{c}}evi{\'{c}}}}, \bibinfo {author}
  {\bibfnamefont {A.}~\bibnamefont {Gurevich}}, \bibinfo {author}
  {\bibfnamefont {D.~C.}\ \bibnamefont {Larbalestier}}, \bibinfo {author}
  {\bibfnamefont {F.~F.}\ \bibnamefont {Balakirev}}, \bibinfo {author}
  {\bibfnamefont {L.}~\bibnamefont {Fang}}, \bibinfo {author} {\bibfnamefont
  {P.}~\bibnamefont {Cheng}}, \bibinfo {author} {\bibfnamefont
  {Y.}~\bibnamefont {Jia}},\ and\ \bibinfo {author} {\bibfnamefont {H.~H.}\
  \bibnamefont {Wen}},\ }\bibfield  {title} {\enquote {\bibinfo {title} {Upper
  critical fields and thermally-activated transport of {NdFeAsO$_0.7$F$_0.3$}
  single crystal},}\ }\href {https://doi.org/10.1103/physrevb.78.174523}
  {\bibfield  {journal} {\bibinfo  {journal} {Phys. Rev. B}\ }\textbf {\bibinfo
  {volume} {78}},\ \bibinfo {pages} {174523} (\bibinfo {year}
  {2008})}\BibitemShut {NoStop}%
\bibitem [{\citenamefont {Yang}\ \emph {et~al.}(2018)\citenamefont {Yang},
  \citenamefont {Schmidt}, \citenamefont {S{\"{u}}ss}, \citenamefont {Chan},
  \citenamefont {Balakirev}, \citenamefont {McDonald}, \citenamefont {Parkin},
  \citenamefont {Felser}, \citenamefont {Yan},\ and\ \citenamefont
  {Moll}}]{Yang_2018}%
  \BibitemOpen
  \bibfield  {author} {\bibinfo {author} {\bibfnamefont {H.}~\bibnamefont
  {Yang}}, \bibinfo {author} {\bibfnamefont {M.}~\bibnamefont {Schmidt}},
  \bibinfo {author} {\bibfnamefont {V.}~\bibnamefont {S{\"{u}}ss}}, \bibinfo
  {author} {\bibfnamefont {M.}~\bibnamefont {Chan}}, \bibinfo {author}
  {\bibfnamefont {F.~F.}\ \bibnamefont {Balakirev}}, \bibinfo {author}
  {\bibfnamefont {R.~D.}\ \bibnamefont {McDonald}}, \bibinfo {author}
  {\bibfnamefont {S.~S.~P.}\ \bibnamefont {Parkin}}, \bibinfo {author}
  {\bibfnamefont {C.}~\bibnamefont {Felser}}, \bibinfo {author} {\bibfnamefont
  {B.}~\bibnamefont {Yan}},\ and\ \bibinfo {author} {\bibfnamefont {P.~J.~W.}\
  \bibnamefont {Moll}},\ }\bibfield  {title} {\enquote {\bibinfo {title}
  {Quantum oscillations in the type-{II} dirac semi-metal candidate
  {PtSe$_2$}},}\ }\href {https://doi.org/10.1088/1367-2630/aab32f} {\bibfield
  {journal} {\bibinfo  {journal} {New J. Phys.}\ }\textbf {\bibinfo {volume}
  {20}},\ \bibinfo {pages} {043008} (\bibinfo {year} {2018})}\BibitemShut
  {NoStop}%
\bibitem [{\citenamefont {Kennedy}\ \emph {et~al.}(2019)\citenamefont
  {Kennedy}, \citenamefont {Burnett}, \citenamefont {Fenton}, \citenamefont
  {Constantino}, \citenamefont {Warburton}, \citenamefont {Morton},\ and\
  \citenamefont {Dupont-Ferrier}}]{Kennedy_2019}%
  \BibitemOpen
  \bibfield  {author} {\bibinfo {author} {\bibfnamefont {O.~W.}\ \bibnamefont
  {Kennedy}}, \bibinfo {author} {\bibfnamefont {J.}~\bibnamefont {Burnett}},
  \bibinfo {author} {\bibfnamefont {J.~C.}\ \bibnamefont {Fenton}}, \bibinfo
  {author} {\bibfnamefont {N.~G.~N.}\ \bibnamefont {Constantino}}, \bibinfo
  {author} {\bibfnamefont {P.~A.}\ \bibnamefont {Warburton}}, \bibinfo {author}
  {\bibfnamefont {J.~J.~L.}\ \bibnamefont {Morton}},\ and\ \bibinfo {author}
  {\bibfnamefont {E.}~\bibnamefont {Dupont-Ferrier}},\ }\bibfield  {title}
  {\enquote {\bibinfo {title} {Tunable {Nb} superconducting resonator based on
  a constriction nano-{SQUID} fabricated with a {Ne} focused ion beam},}\
  }\href {https://doi.org/10.1103/physrevapplied.11.014006} {\bibfield
  {journal} {\bibinfo  {journal} {Phys. Rev. Applied}\ }\textbf {\bibinfo
  {volume} {11}},\ \bibinfo {pages} {014006} (\bibinfo {year}
  {2019})}\BibitemShut {NoStop}%
\bibitem [{\citenamefont {Burnett}\ \emph {et~al.}(2017)\citenamefont
  {Burnett}, \citenamefont {Sagar}, \citenamefont {Kennedy}, \citenamefont
  {Warburton},\ and\ \citenamefont {Fenton}}]{Burnett2017}%
  \BibitemOpen
  \bibfield  {author} {\bibinfo {author} {\bibfnamefont {J.}~\bibnamefont
  {Burnett}}, \bibinfo {author} {\bibfnamefont {J.}~\bibnamefont {Sagar}},
  \bibinfo {author} {\bibfnamefont {O.}~\bibnamefont {Kennedy}}, \bibinfo
  {author} {\bibfnamefont {P.}~\bibnamefont {Warburton}},\ and\ \bibinfo
  {author} {\bibfnamefont {J.}~\bibnamefont {Fenton}},\ }\bibfield  {title}
  {\enquote {\bibinfo {title} {Low-loss superconducting nanowire circuits using
  a neon focused ion beam},}\ }\href
  {https://doi.org/10.1103/physrevapplied.8.014039} {\bibfield  {journal}
  {\bibinfo  {journal} {Phys. Rev. Applied}\ }\textbf {\bibinfo {volume} {8}},\
  \bibinfo {pages} {014039} (\bibinfo {year} {2017})}\BibitemShut {NoStop}%
\bibitem [{\citenamefont {Uhl}\ \emph {et~al.}(2023)\citenamefont {Uhl},
  \citenamefont {Hackenbeck}, \citenamefont {Peter}, \citenamefont {Kleiner},
  \citenamefont {Koelle},\ and\ \citenamefont {Bothner}}]{Uhl2023}%
  \BibitemOpen
  \bibfield  {author} {\bibinfo {author} {\bibfnamefont {K.}~\bibnamefont
  {Uhl}}, \bibinfo {author} {\bibfnamefont {D.}~\bibnamefont {Hackenbeck}},
  \bibinfo {author} {\bibfnamefont {J.}~\bibnamefont {Peter}}, \bibinfo
  {author} {\bibfnamefont {R.}~\bibnamefont {Kleiner}}, \bibinfo {author}
  {\bibfnamefont {D.}~\bibnamefont {Koelle}},\ and\ \bibinfo {author}
  {\bibfnamefont {D.}~\bibnamefont {Bothner}},\ }\href
  {https://doi.org/10.48550/ARXIV.2305.16276} {\enquote {\bibinfo {title}
  {Niobium quantum interference microwave circuits with monolithic
  three-dimensional ({3D}) nanobridge junctions},}\ } (\bibinfo {year}
  {2023}),\ \bibinfo {note} {preprint},\ \Eprint
  {https://arxiv.org/abs/2305.16276} {arXiv:2305.16276 [quant-ph]} \BibitemShut
  {NoStop}%
\bibitem [{\citenamefont {Troeman}\ \emph {et~al.}(2007)\citenamefont
  {Troeman}, \citenamefont {Derking}, \citenamefont {Borger}, \citenamefont
  {Pleikies}, \citenamefont {Veldhuis},\ and\ \citenamefont
  {Hilgenkamp}}]{Troeman2007}%
  \BibitemOpen
  \bibfield  {author} {\bibinfo {author} {\bibfnamefont {A.~G.~P.}\
  \bibnamefont {Troeman}}, \bibinfo {author} {\bibfnamefont {H.}~\bibnamefont
  {Derking}}, \bibinfo {author} {\bibfnamefont {B.}~\bibnamefont {Borger}},
  \bibinfo {author} {\bibfnamefont {J.}~\bibnamefont {Pleikies}}, \bibinfo
  {author} {\bibfnamefont {D.}~\bibnamefont {Veldhuis}},\ and\ \bibinfo
  {author} {\bibfnamefont {H.}~\bibnamefont {Hilgenkamp}},\ }\bibfield  {title}
  {\enquote {\bibinfo {title} {{NanoSQUIDs} based on niobium constrictions},}\
  }\href {https://doi.org/10.1021/nl070870f} {\bibfield  {journal} {\bibinfo
  {journal} {Nano Lett.}\ }\textbf {\bibinfo {volume} {7}},\ \bibinfo {pages}
  {2152--2156} (\bibinfo {year} {2007})}\BibitemShut {NoStop}%
\bibitem [{\citenamefont {Granata}\ \emph {et~al.}(2013)\citenamefont
  {Granata}, \citenamefont {Vettoliere}, \citenamefont {Russo}, \citenamefont
  {Fretto}, \citenamefont {Leo},\ and\ \citenamefont
  {Lacquaniti}}]{Granata2013}%
  \BibitemOpen
  \bibfield  {author} {\bibinfo {author} {\bibfnamefont {C.}~\bibnamefont
  {Granata}}, \bibinfo {author} {\bibfnamefont {A.}~\bibnamefont {Vettoliere}},
  \bibinfo {author} {\bibfnamefont {R.}~\bibnamefont {Russo}}, \bibinfo
  {author} {\bibfnamefont {M.}~\bibnamefont {Fretto}}, \bibinfo {author}
  {\bibfnamefont {N.~D.}\ \bibnamefont {Leo}},\ and\ \bibinfo {author}
  {\bibfnamefont {V.}~\bibnamefont {Lacquaniti}},\ }\bibfield  {title}
  {\enquote {\bibinfo {title} {Three-dimensional spin nanosensor based on
  reliable tunnel josephson nano-junctions for nanomagnetism investigations},}\
  }\href {https://doi.org/10.1063/1.4819463} {\bibfield  {journal} {\bibinfo
  {journal} {Appl. Phys. Lett.}\ }\textbf {\bibinfo {volume} {103}} (\bibinfo
  {year} {2013}),\ 10.1063/1.4819463}\BibitemShut {NoStop}%
\bibitem [{\citenamefont {Schmelz}\ \emph {et~al.}(2017)\citenamefont
  {Schmelz}, \citenamefont {Vettoliere}, \citenamefont {Zakosarenko},
  \citenamefont {Leo}, \citenamefont {Fretto}, \citenamefont {Stolz},\ and\
  \citenamefont {Granata}}]{Schmelz2017}%
  \BibitemOpen
  \bibfield  {author} {\bibinfo {author} {\bibfnamefont {M.}~\bibnamefont
  {Schmelz}}, \bibinfo {author} {\bibfnamefont {A.}~\bibnamefont {Vettoliere}},
  \bibinfo {author} {\bibfnamefont {V.}~\bibnamefont {Zakosarenko}}, \bibinfo
  {author} {\bibfnamefont {N.~D.}\ \bibnamefont {Leo}}, \bibinfo {author}
  {\bibfnamefont {M.}~\bibnamefont {Fretto}}, \bibinfo {author} {\bibfnamefont
  {R.}~\bibnamefont {Stolz}},\ and\ \bibinfo {author} {\bibfnamefont
  {C.}~\bibnamefont {Granata}},\ }\bibfield  {title} {\enquote {\bibinfo
  {title} {{3D} {nanoSQUID} based on tunnel nano-junctions with an energy
  sensitivity of 1.3 h at 4.2{K}},}\ }\href {https://doi.org/10.1063/1.4986655}
  {\bibfield  {journal} {\bibinfo  {journal} {Appl. Phys. Lett.}\ }\textbf
  {\bibinfo {volume} {111}} (\bibinfo {year} {2017}),\
  10.1063/1.4986655}\BibitemShut {NoStop}%
\bibitem [{\citenamefont {Wyss}\ \emph {et~al.}(2022)\citenamefont {Wyss},
  \citenamefont {Bagani}, \citenamefont {Jetter}, \citenamefont {Marchiori},
  \citenamefont {Vervelaki}, \citenamefont {Gross}, \citenamefont {Ridderbos},
  \citenamefont {Gliga}, \citenamefont {Sch{\"{o}}nenberger},\ and\
  \citenamefont {Poggio}}]{Wyss2022}%
  \BibitemOpen
  \bibfield  {author} {\bibinfo {author} {\bibfnamefont {M.}~\bibnamefont
  {Wyss}}, \bibinfo {author} {\bibfnamefont {K.}~\bibnamefont {Bagani}},
  \bibinfo {author} {\bibfnamefont {D.}~\bibnamefont {Jetter}}, \bibinfo
  {author} {\bibfnamefont {E.}~\bibnamefont {Marchiori}}, \bibinfo {author}
  {\bibfnamefont {A.}~\bibnamefont {Vervelaki}}, \bibinfo {author}
  {\bibfnamefont {B.}~\bibnamefont {Gross}}, \bibinfo {author} {\bibfnamefont
  {J.}~\bibnamefont {Ridderbos}}, \bibinfo {author} {\bibfnamefont
  {S.}~\bibnamefont {Gliga}}, \bibinfo {author} {\bibfnamefont
  {C.}~\bibnamefont {Sch{\"{o}}nenberger}},\ and\ \bibinfo {author}
  {\bibfnamefont {M.}~\bibnamefont {Poggio}},\ }\bibfield  {title} {\enquote
  {\bibinfo {title} {Magnetic, thermal, and topographic imaging with a
  nanometer-scale {SQUID}-on-lever scanning probe},}\ }\href
  {https://doi.org/10.1103/physrevapplied.17.034002} {\bibfield  {journal}
  {\bibinfo  {journal} {Phys. Rev. Applied}\ }\textbf {\bibinfo {volume}
  {17}},\ \bibinfo {pages} {034002} (\bibinfo {year} {2022})}\BibitemShut
  {NoStop}%
\bibitem [{\citenamefont {Hao}, \citenamefont {Cox},\ and\ \citenamefont
  {Gallop}(2009)}]{Hao_2009a}%
  \BibitemOpen
  \bibfield  {author} {\bibinfo {author} {\bibfnamefont {L.}~\bibnamefont
  {Hao}}, \bibinfo {author} {\bibfnamefont {D.~C.}\ \bibnamefont {Cox}},\ and\
  \bibinfo {author} {\bibfnamefont {J.~C.}\ \bibnamefont {Gallop}},\ }\bibfield
   {title} {\enquote {\bibinfo {title} {Characteristics of focused ion beam
  nanoscale josephson devices},}\ }\href
  {https://doi.org/10.1088/0953-2048/22/6/064011} {\bibfield  {journal}
  {\bibinfo  {journal} {Supercond. Sci. Technol.}\ }\textbf {\bibinfo {volume}
  {22}},\ \bibinfo {pages} {064011} (\bibinfo {year} {2009})}\BibitemShut
  {NoStop}%
\bibitem [{\citenamefont {Hao}\ \emph {et~al.}(2011)\citenamefont {Hao},
  \citenamefont {A{\ss}mann}, \citenamefont {Gallop}, \citenamefont {Cox},
  \citenamefont {Ruede}, \citenamefont {Kazakova}, \citenamefont
  {Josephs-Franks}, \citenamefont {Drung},\ and\ \citenamefont
  {Schurig}}]{Hao_2011}%
  \BibitemOpen
  \bibfield  {author} {\bibinfo {author} {\bibfnamefont {L.}~\bibnamefont
  {Hao}}, \bibinfo {author} {\bibfnamefont {C.}~\bibnamefont {A{\ss}mann}},
  \bibinfo {author} {\bibfnamefont {J.~C.}\ \bibnamefont {Gallop}}, \bibinfo
  {author} {\bibfnamefont {D.}~\bibnamefont {Cox}}, \bibinfo {author}
  {\bibfnamefont {F.}~\bibnamefont {Ruede}}, \bibinfo {author} {\bibfnamefont
  {O.}~\bibnamefont {Kazakova}}, \bibinfo {author} {\bibfnamefont
  {P.}~\bibnamefont {Josephs-Franks}}, \bibinfo {author} {\bibfnamefont
  {D.}~\bibnamefont {Drung}},\ and\ \bibinfo {author} {\bibfnamefont
  {T.}~\bibnamefont {Schurig}},\ }\bibfield  {title} {\enquote {\bibinfo
  {title} {Detection of single magnetic nanobead with a nano-superconducting
  quantum interference device},}\ }\href {https://doi.org/10.1063/1.3561743}
  {\bibfield  {journal} {\bibinfo  {journal} {Appl. Phys. Lett.}\ }\textbf
  {\bibinfo {volume} {98}},\ \bibinfo {pages} {092504} (\bibinfo {year}
  {2011})}\BibitemShut {NoStop}%
\bibitem [{\citenamefont {Hao}\ \emph {et~al.}(2008)\citenamefont {Hao},
  \citenamefont {Macfarlane}, \citenamefont {Gallop}, \citenamefont {Cox},
  \citenamefont {Beyer}, \citenamefont {Drung},\ and\ \citenamefont
  {Schurig}}]{Hao_2008}%
  \BibitemOpen
  \bibfield  {author} {\bibinfo {author} {\bibfnamefont {L.}~\bibnamefont
  {Hao}}, \bibinfo {author} {\bibfnamefont {J.~C.}\ \bibnamefont {Macfarlane}},
  \bibinfo {author} {\bibfnamefont {J.~C.}\ \bibnamefont {Gallop}}, \bibinfo
  {author} {\bibfnamefont {D.}~\bibnamefont {Cox}}, \bibinfo {author}
  {\bibfnamefont {J.}~\bibnamefont {Beyer}}, \bibinfo {author} {\bibfnamefont
  {D.}~\bibnamefont {Drung}},\ and\ \bibinfo {author} {\bibfnamefont
  {T.}~\bibnamefont {Schurig}},\ }\bibfield  {title} {\enquote {\bibinfo
  {title} {Measurement and noise performance of
  nano-superconducting-quantum-interference devices fabricated by focused ion
  beam},}\ }\href {https://doi.org/10.1063/1.2917580} {\bibfield  {journal}
  {\bibinfo  {journal} {Appl. Phys. Lett.}\ }\textbf {\bibinfo {volume} {92}},\
  \bibinfo {pages} {192507} (\bibinfo {year} {2008})}\BibitemShut {NoStop}%
\bibitem [{\citenamefont {Cox}, \citenamefont {Gallop},\ and\ \citenamefont
  {Hao}(2014)}]{Cox_2014}%
  \BibitemOpen
  \bibfield  {author} {\bibinfo {author} {\bibfnamefont {D.~C.}\ \bibnamefont
  {Cox}}, \bibinfo {author} {\bibfnamefont {J.~C.}\ \bibnamefont {Gallop}},\
  and\ \bibinfo {author} {\bibfnamefont {L.}~\bibnamefont {Hao}},\ }\bibfield
  {title} {\enquote {\bibinfo {title} {Focused ion beam processing of
  superconducting junctions and {SQUID} based devices},}\ }\href
  {https://doi.org/10.2478/nanofab-2014-0005} {\bibfield  {journal} {\bibinfo
  {journal} {Nanofabrication}\ }\textbf {\bibinfo {volume} {1}} (\bibinfo
  {year} {2014}),\ 10.2478/nanofab-2014-0005}\BibitemShut {NoStop}%
\bibitem [{\citenamefont {Faley}\ and\ \citenamefont
  {Dunin-Borkowski}(2022)}]{Faley2022}%
  \BibitemOpen
  \bibfield  {author} {\bibinfo {author} {\bibfnamefont {M.~I.}\ \bibnamefont
  {Faley}}\ and\ \bibinfo {author} {\bibfnamefont {R.~E.}\ \bibnamefont
  {Dunin-Borkowski}},\ }\bibfield  {title} {\enquote {\bibinfo {title} {A
  self-flux-biased {NanoSQUID} with four {NbN}-{TiN}-{NbN} nanobridge josephson
  junctions},}\ }\href {https://doi.org/10.3390/electronics11111704} {\bibfield
   {journal} {\bibinfo  {journal} {Electronics}\ }\textbf {\bibinfo {volume}
  {11}},\ \bibinfo {pages} {1704} (\bibinfo {year} {2022})}\BibitemShut
  {NoStop}%
\bibitem [{\citenamefont {Nagel}\ \emph {et~al.}(2010)\citenamefont {Nagel},
  \citenamefont {Konovalenko}, \citenamefont {Kemmler}, \citenamefont {Turad},
  \citenamefont {Werner}, \citenamefont {Kleisz}, \citenamefont {Menzel},
  \citenamefont {Klingeler}, \citenamefont {B{\"{u}}chner}, \citenamefont
  {Kleiner},\ and\ \citenamefont {Koelle}}]{Nagel2010}%
  \BibitemOpen
  \bibfield  {author} {\bibinfo {author} {\bibfnamefont {J.}~\bibnamefont
  {Nagel}}, \bibinfo {author} {\bibfnamefont {K.~B.}\ \bibnamefont
  {Konovalenko}}, \bibinfo {author} {\bibfnamefont {M.}~\bibnamefont
  {Kemmler}}, \bibinfo {author} {\bibfnamefont {M.}~\bibnamefont {Turad}},
  \bibinfo {author} {\bibfnamefont {R.}~\bibnamefont {Werner}}, \bibinfo
  {author} {\bibfnamefont {E.}~\bibnamefont {Kleisz}}, \bibinfo {author}
  {\bibfnamefont {S.}~\bibnamefont {Menzel}}, \bibinfo {author} {\bibfnamefont
  {R.}~\bibnamefont {Klingeler}}, \bibinfo {author} {\bibfnamefont
  {B.}~\bibnamefont {B{\"{u}}chner}}, \bibinfo {author} {\bibfnamefont
  {R.}~\bibnamefont {Kleiner}},\ and\ \bibinfo {author} {\bibfnamefont
  {D.}~\bibnamefont {Koelle}},\ }\bibfield  {title} {\enquote {\bibinfo {title}
  {Resistively shunted {YBa}$_{2}${C}u$_{3}${O}$_{7}$ grain boundary junctions
  and low-noise {SQUIDs} patterned by a focused ion beam down to 80 nm
  linewidth},}\ }\href {https://doi.org/10.1088/0953-2048/24/1/015015}
  {\bibfield  {journal} {\bibinfo  {journal} {Supercond. Sci. Technol.}\
  }\textbf {\bibinfo {volume} {24}},\ \bibinfo {pages} {015015} (\bibinfo
  {year} {2010})}\BibitemShut {NoStop}%
\bibitem [{\citenamefont {Schwarz}\ \emph {et~al.}(2012)\citenamefont
  {Schwarz}, \citenamefont {Nagel}, \citenamefont {W{\"{o}}lbing},
  \citenamefont {Kemmler}, \citenamefont {Kleiner},\ and\ \citenamefont
  {Koelle}}]{Schwarz2012}%
  \BibitemOpen
  \bibfield  {author} {\bibinfo {author} {\bibfnamefont {T.}~\bibnamefont
  {Schwarz}}, \bibinfo {author} {\bibfnamefont {J.}~\bibnamefont {Nagel}},
  \bibinfo {author} {\bibfnamefont {R.}~\bibnamefont {W{\"{o}}lbing}}, \bibinfo
  {author} {\bibfnamefont {M.}~\bibnamefont {Kemmler}}, \bibinfo {author}
  {\bibfnamefont {R.}~\bibnamefont {Kleiner}},\ and\ \bibinfo {author}
  {\bibfnamefont {D.}~\bibnamefont {Koelle}},\ }\bibfield  {title} {\enquote
  {\bibinfo {title} {Low-noise nano superconducting quantum interference device
  operating in {T}esla magnetic fields},}\ }\href
  {https://doi.org/10.1021/nn305431c} {\bibfield  {journal} {\bibinfo
  {journal} {{ACS} Nano}\ }\textbf {\bibinfo {volume} {7}},\ \bibinfo {pages}
  {844--850} (\bibinfo {year} {2012})}\BibitemShut {NoStop}%
\bibitem [{\citenamefont {Schwarz}\ \emph {et~al.}(2015)\citenamefont
  {Schwarz}, \citenamefont {W{\"{o}}lbing}, \citenamefont {Reiche},
  \citenamefont {M{\"{u}}ller}, \citenamefont {Mart{\'{\i}}nez-P{\'{e}}rez},
  \citenamefont {M{\"{u}}hl}, \citenamefont {B{\"{u}}chner}, \citenamefont
  {Kleiner},\ and\ \citenamefont {Koelle}}]{Schwarz2015}%
  \BibitemOpen
  \bibfield  {author} {\bibinfo {author} {\bibfnamefont {T.}~\bibnamefont
  {Schwarz}}, \bibinfo {author} {\bibfnamefont {R.}~\bibnamefont
  {W{\"{o}}lbing}}, \bibinfo {author} {\bibfnamefont {C.~F.}\ \bibnamefont
  {Reiche}}, \bibinfo {author} {\bibfnamefont {B.}~\bibnamefont
  {M{\"{u}}ller}}, \bibinfo {author} {\bibfnamefont {M.~J.}\ \bibnamefont
  {Mart{\'{\i}}nez-P{\'{e}}rez}}, \bibinfo {author} {\bibfnamefont
  {T.}~\bibnamefont {M{\"{u}}hl}}, \bibinfo {author} {\bibfnamefont
  {B.}~\bibnamefont {B{\"{u}}chner}}, \bibinfo {author} {\bibfnamefont
  {R.}~\bibnamefont {Kleiner}},\ and\ \bibinfo {author} {\bibfnamefont
  {D.}~\bibnamefont {Koelle}},\ }\bibfield  {title} {\enquote {\bibinfo {title}
  {Low-noise {YBa}$_{2}${C}u$_{3}${O}$_{7}$ nano-{SQUIDs} for performing
  magnetization-reversal measurements on magnetic nanoparticles},}\ }\href
  {https://doi.org/10.1103/physrevapplied.3.044011} {\bibfield  {journal}
  {\bibinfo  {journal} {Phys. Rev. Applied}\ }\textbf {\bibinfo {volume} {3}},\
  \bibinfo {pages} {044011} (\bibinfo {year} {2015})}\BibitemShut {NoStop}%
\bibitem [{\citenamefont {Lam}, \citenamefont {Bendavid},\ and\ \citenamefont
  {Du}(2017)}]{Lam2017}%
  \BibitemOpen
  \bibfield  {author} {\bibinfo {author} {\bibfnamefont {S.~K.~H.}\
  \bibnamefont {Lam}}, \bibinfo {author} {\bibfnamefont {A.}~\bibnamefont
  {Bendavid}},\ and\ \bibinfo {author} {\bibfnamefont {J.}~\bibnamefont {Du}},\
  }\bibfield  {title} {\enquote {\bibinfo {title} {Trimming the electrical
  properties on nanoscale {YBa}$_{2}${C}u$_{3}${O}$_{7-x}$ constrictions by
  focus ion beam technique},}\ }\href
  {https://doi.org/10.1016/j.physc.2017.07.009} {\bibfield  {journal} {\bibinfo
   {journal} {Phys. C: Supercond. Its Appl.}\ }\textbf {\bibinfo {volume}
  {540}},\ \bibinfo {pages} {38--43} (\bibinfo {year} {2017})}\BibitemShut
  {NoStop}%
\bibitem [{\citenamefont {Godfrey}\ \emph {et~al.}(2018)\citenamefont
  {Godfrey}, \citenamefont {Gallop}, \citenamefont {Cox}, \citenamefont
  {Romans}, \citenamefont {Chen},\ and\ \citenamefont {Hao}}]{Godfrey_2018}%
  \BibitemOpen
  \bibfield  {author} {\bibinfo {author} {\bibfnamefont {T.}~\bibnamefont
  {Godfrey}}, \bibinfo {author} {\bibfnamefont {J.~C.}\ \bibnamefont {Gallop}},
  \bibinfo {author} {\bibfnamefont {D.~C.}\ \bibnamefont {Cox}}, \bibinfo
  {author} {\bibfnamefont {E.~J.}\ \bibnamefont {Romans}}, \bibinfo {author}
  {\bibfnamefont {J.}~\bibnamefont {Chen}},\ and\ \bibinfo {author}
  {\bibfnamefont {L.}~\bibnamefont {Hao}},\ }\bibfield  {title} {\enquote
  {\bibinfo {title} {Investigation of dayem bridge {NanoSQUIDs} made by xe
  focused ion beam},}\ }\href {https://doi.org/10.1109/tasc.2018.2854624}
  {\bibfield  {journal} {\bibinfo  {journal} {IEEE Trans. Appl. Supercond.}\
  }\textbf {\bibinfo {volume} {28}},\ \bibinfo {pages} {1--5} (\bibinfo {year}
  {2018})}\BibitemShut {NoStop}%
\bibitem [{\citenamefont {Iurchuk}\ \emph {et~al.}(2023)\citenamefont
  {Iurchuk}, \citenamefont {Pablo-Navarro}, \citenamefont {Hula}, \citenamefont
  {Narkowicz}, \citenamefont {Hlawacek}, \citenamefont {K{\"{o}}rber},
  \citenamefont {K{\'{a}}kay}, \citenamefont {Schultheiss}, \citenamefont
  {Fassbender}, \citenamefont {Lenz},\ and\ \citenamefont
  {Lindner}}]{Iurchuk_2023}%
  \BibitemOpen
  \bibfield  {author} {\bibinfo {author} {\bibfnamefont {V.}~\bibnamefont
  {Iurchuk}}, \bibinfo {author} {\bibfnamefont {J.}~\bibnamefont
  {Pablo-Navarro}}, \bibinfo {author} {\bibfnamefont {T.}~\bibnamefont {Hula}},
  \bibinfo {author} {\bibfnamefont {R.}~\bibnamefont {Narkowicz}}, \bibinfo
  {author} {\bibfnamefont {G.}~\bibnamefont {Hlawacek}}, \bibinfo {author}
  {\bibfnamefont {L.}~\bibnamefont {K{\"{o}}rber}}, \bibinfo {author}
  {\bibfnamefont {A.}~\bibnamefont {K{\'{a}}kay}}, \bibinfo {author}
  {\bibfnamefont {H.}~\bibnamefont {Schultheiss}}, \bibinfo {author}
  {\bibfnamefont {J.}~\bibnamefont {Fassbender}}, \bibinfo {author}
  {\bibfnamefont {K.}~\bibnamefont {Lenz}},\ and\ \bibinfo {author}
  {\bibfnamefont {J.}~\bibnamefont {Lindner}},\ }\bibfield  {title} {\enquote
  {\bibinfo {title} {Tailoring crosstalk between localized 1d spin-wave
  nanochannels using focused ion beams},}\ }\href
  {https://doi.org/10.1038/s41598-022-27249-w} {\bibfield  {journal} {\bibinfo
  {journal} {Sci. Rep.}\ }\textbf {\bibinfo {volume} {13}} (\bibinfo {year}
  {2023}),\ 10.1038/s41598-022-27249-w},\ \Eprint
  {https://arxiv.org/abs/2209.13180} {arXiv:2209.13180 [cond-mat.mes-hall]}
  \BibitemShut {NoStop}%
\bibitem [{\citenamefont {Cansever}\ \emph {et~al.}(2022)\citenamefont
  {Cansever}, \citenamefont {Anwar}, \citenamefont {Stienen}, \citenamefont
  {Lenz}, \citenamefont {Narkowicz}, \citenamefont {Hlawacek}, \citenamefont
  {Potzger}, \citenamefont {Hellwig}, \citenamefont {Fassbender}, \citenamefont
  {Lindner},\ and\ \citenamefont {Bali}}]{Cansever_2022}%
  \BibitemOpen
  \bibfield  {author} {\bibinfo {author} {\bibfnamefont {H.}~\bibnamefont
  {Cansever}}, \bibinfo {author} {\bibfnamefont {M.~S.}\ \bibnamefont {Anwar}},
  \bibinfo {author} {\bibfnamefont {S.}~\bibnamefont {Stienen}}, \bibinfo
  {author} {\bibfnamefont {K.}~\bibnamefont {Lenz}}, \bibinfo {author}
  {\bibfnamefont {R.}~\bibnamefont {Narkowicz}}, \bibinfo {author}
  {\bibfnamefont {G.}~\bibnamefont {Hlawacek}}, \bibinfo {author}
  {\bibfnamefont {K.}~\bibnamefont {Potzger}}, \bibinfo {author} {\bibfnamefont
  {O.}~\bibnamefont {Hellwig}}, \bibinfo {author} {\bibfnamefont
  {J.}~\bibnamefont {Fassbender}}, \bibinfo {author} {\bibfnamefont
  {J.}~\bibnamefont {Lindner}},\ and\ \bibinfo {author} {\bibfnamefont
  {R.}~\bibnamefont {Bali}},\ }\bibfield  {title} {\enquote {\bibinfo {title}
  {Resonance behavior of embedded and freestanding microscale ferromagnets},}\
  }\href {https://doi.org/10.1038/s41598-022-15959-0} {\bibfield  {journal}
  {\bibinfo  {journal} {Sci. Rep.}\ }\textbf {\bibinfo {volume} {12}} (\bibinfo
  {year} {2022}),\ 10.1038/s41598-022-15959-0}\BibitemShut {NoStop}%
\bibitem [{\citenamefont {Lara}\ \emph {et~al.}(2014)\citenamefont {Lara},
  \citenamefont {Dobrovolskiy}, \citenamefont {Prieto}, \citenamefont {Huth},\
  and\ \citenamefont {Aliev}}]{Lara_2014}%
  \BibitemOpen
  \bibfield  {author} {\bibinfo {author} {\bibfnamefont {A.}~\bibnamefont
  {Lara}}, \bibinfo {author} {\bibfnamefont {O.~V.}\ \bibnamefont
  {Dobrovolskiy}}, \bibinfo {author} {\bibfnamefont {J.~L.}\ \bibnamefont
  {Prieto}}, \bibinfo {author} {\bibfnamefont {M.}~\bibnamefont {Huth}},\ and\
  \bibinfo {author} {\bibfnamefont {F.~G.}\ \bibnamefont {Aliev}},\ }\bibfield
  {title} {\enquote {\bibinfo {title} {Magnetization reversal assisted by half
  antivortex states in nanostructured circular cobalt disks},}\ }\href
  {https://doi.org/10.1063/1.4900789} {\bibfield  {journal} {\bibinfo
  {journal} {Appl. Phys. Lett.}\ }\textbf {\bibinfo {volume} {105}},\ \bibinfo
  {eid} {182402} (\bibinfo {year} {2014})}\BibitemShut {NoStop}%
\bibitem [{\citenamefont {Dobrovolskiy}\ \emph {et~al.}(2019)\citenamefont
  {Dobrovolskiy}, \citenamefont {Sachser}, \citenamefont {Bunyaev},
  \citenamefont {Navas}, \citenamefont {Bevz}, \citenamefont {Zelent},
  \citenamefont {{\'{S}}migaj}, \citenamefont {Rych{\l}y}, \citenamefont
  {Krawczyk}, \citenamefont {Vovk}, \citenamefont {Huth},\ and\ \citenamefont
  {Kakazei}}]{Dobrovolskiy_2019}%
  \BibitemOpen
  \bibfield  {author} {\bibinfo {author} {\bibfnamefont {O.~V.}\ \bibnamefont
  {Dobrovolskiy}}, \bibinfo {author} {\bibfnamefont {R.}~\bibnamefont
  {Sachser}}, \bibinfo {author} {\bibfnamefont {S.~A.}\ \bibnamefont
  {Bunyaev}}, \bibinfo {author} {\bibfnamefont {D.}~\bibnamefont {Navas}},
  \bibinfo {author} {\bibfnamefont {V.~M.}\ \bibnamefont {Bevz}}, \bibinfo
  {author} {\bibfnamefont {M.}~\bibnamefont {Zelent}}, \bibinfo {author}
  {\bibfnamefont {W.}~\bibnamefont {{\'{S}}migaj}}, \bibinfo {author}
  {\bibfnamefont {J.}~\bibnamefont {Rych{\l}y}}, \bibinfo {author}
  {\bibfnamefont {M.}~\bibnamefont {Krawczyk}}, \bibinfo {author}
  {\bibfnamefont {R.~V.}\ \bibnamefont {Vovk}}, \bibinfo {author}
  {\bibfnamefont {M.}~\bibnamefont {Huth}},\ and\ \bibinfo {author}
  {\bibfnamefont {G.~N.}\ \bibnamefont {Kakazei}},\ }\bibfield  {title}
  {\enquote {\bibinfo {title} {Spin-wave phase inverter upon a single
  nanodefect},}\ }\href {https://doi.org/10.1021/acsami.9b02717} {\bibfield
  {journal} {\bibinfo  {journal} {ACS Appl. Mater. Interfaces}\ }\textbf
  {\bibinfo {volume} {11}},\ \bibinfo {pages} {17654--17662} (\bibinfo {year}
  {2019})}\BibitemShut {NoStop}%
\bibitem [{\citenamefont {Putzke}\ \emph {et~al.}(2020)\citenamefont {Putzke},
  \citenamefont {Bachmann}, \citenamefont {McGuinness}, \citenamefont
  {Zhakina}, \citenamefont {Sunko}, \citenamefont {Konczykowski}, \citenamefont
  {Oka}, \citenamefont {Moessner}, \citenamefont {Stern}, \citenamefont
  {K{\"{o}}nig}, \citenamefont {Khim}, \citenamefont {Mackenzie},\ and\
  \citenamefont {Moll}}]{Putzke_2020}%
  \BibitemOpen
  \bibfield  {author} {\bibinfo {author} {\bibfnamefont {C.}~\bibnamefont
  {Putzke}}, \bibinfo {author} {\bibfnamefont {M.~D.}\ \bibnamefont
  {Bachmann}}, \bibinfo {author} {\bibfnamefont {P.}~\bibnamefont
  {McGuinness}}, \bibinfo {author} {\bibfnamefont {E.}~\bibnamefont {Zhakina}},
  \bibinfo {author} {\bibfnamefont {V.}~\bibnamefont {Sunko}}, \bibinfo
  {author} {\bibfnamefont {M.}~\bibnamefont {Konczykowski}}, \bibinfo {author}
  {\bibfnamefont {T.}~\bibnamefont {Oka}}, \bibinfo {author} {\bibfnamefont
  {R.}~\bibnamefont {Moessner}}, \bibinfo {author} {\bibfnamefont
  {A.}~\bibnamefont {Stern}}, \bibinfo {author} {\bibfnamefont
  {M.}~\bibnamefont {K{\"{o}}nig}}, \bibinfo {author} {\bibfnamefont
  {S.}~\bibnamefont {Khim}}, \bibinfo {author} {\bibfnamefont {A.~P.}\
  \bibnamefont {Mackenzie}},\ and\ \bibinfo {author} {\bibfnamefont {P.~J.~W.}\
  \bibnamefont {Moll}},\ }\bibfield  {title} {\enquote {\bibinfo {title} {H/e
  oscillations in interlayer transport of delafossites},}\ }\href
  {https://doi.org/10.1126/science.aay8413} {\bibfield  {journal} {\bibinfo
  {journal} {Science}\ }\textbf {\bibinfo {volume} {368}},\ \bibinfo {pages}
  {1234--1238} (\bibinfo {year} {2020})}\BibitemShut {NoStop}%
\bibitem [{\citenamefont {Geishendorf}\ \emph {et~al.}(2019)\citenamefont
  {Geishendorf}, \citenamefont {Vir}, \citenamefont {Shekhar}, \citenamefont
  {Felser}, \citenamefont {Facio}, \citenamefont {van~den Brink}, \citenamefont
  {Nielsch}, \citenamefont {Thomas},\ and\ \citenamefont
  {Goennenwein}}]{Geishendorf_2019}%
  \BibitemOpen
  \bibfield  {author} {\bibinfo {author} {\bibfnamefont {K.}~\bibnamefont
  {Geishendorf}}, \bibinfo {author} {\bibfnamefont {P.}~\bibnamefont {Vir}},
  \bibinfo {author} {\bibfnamefont {C.}~\bibnamefont {Shekhar}}, \bibinfo
  {author} {\bibfnamefont {C.}~\bibnamefont {Felser}}, \bibinfo {author}
  {\bibfnamefont {J.~I.}\ \bibnamefont {Facio}}, \bibinfo {author}
  {\bibfnamefont {J.}~\bibnamefont {van~den Brink}}, \bibinfo {author}
  {\bibfnamefont {K.}~\bibnamefont {Nielsch}}, \bibinfo {author} {\bibfnamefont
  {A.}~\bibnamefont {Thomas}},\ and\ \bibinfo {author} {\bibfnamefont
  {S.~T.~B.}\ \bibnamefont {Goennenwein}},\ }\bibfield  {title} {\enquote
  {\bibinfo {title} {Signatures of the magnetic entropy in the thermopower
  signals in nanoribbons of the magnetic {W}eyl semimetal
  {Co$_3$Sn$_2$S$_2$}},}\ }\href {https://doi.org/10.1021/acs.nanolett.9b03822}
  {\bibfield  {journal} {\bibinfo  {journal} {Nano Lett.}\ }\textbf {\bibinfo
  {volume} {20}},\ \bibinfo {pages} {300--305} (\bibinfo {year}
  {2019})}\BibitemShut {NoStop}%
\bibitem [{\citenamefont {Gooth}\ \emph {et~al.}(2017)\citenamefont {Gooth},
  \citenamefont {Niemann}, \citenamefont {Meng}, \citenamefont {Grushin},
  \citenamefont {Landsteiner}, \citenamefont {Gotsmann}, \citenamefont
  {Menges}, \citenamefont {Schmidt}, \citenamefont {Shekhar}, \citenamefont
  {S{\"{u}}{\ss}}, \citenamefont {H{\"{u}}hne}, \citenamefont {Rellinghaus},
  \citenamefont {Felser}, \citenamefont {Yan},\ and\ \citenamefont
  {Nielsch}}]{Gooth_2017}%
  \BibitemOpen
  \bibfield  {author} {\bibinfo {author} {\bibfnamefont {J.}~\bibnamefont
  {Gooth}}, \bibinfo {author} {\bibfnamefont {A.~C.}\ \bibnamefont {Niemann}},
  \bibinfo {author} {\bibfnamefont {T.}~\bibnamefont {Meng}}, \bibinfo {author}
  {\bibfnamefont {A.~G.}\ \bibnamefont {Grushin}}, \bibinfo {author}
  {\bibfnamefont {K.}~\bibnamefont {Landsteiner}}, \bibinfo {author}
  {\bibfnamefont {B.}~\bibnamefont {Gotsmann}}, \bibinfo {author}
  {\bibfnamefont {F.}~\bibnamefont {Menges}}, \bibinfo {author} {\bibfnamefont
  {M.}~\bibnamefont {Schmidt}}, \bibinfo {author} {\bibfnamefont
  {C.}~\bibnamefont {Shekhar}}, \bibinfo {author} {\bibfnamefont
  {V.}~\bibnamefont {S{\"{u}}{\ss}}}, \bibinfo {author} {\bibfnamefont
  {R.}~\bibnamefont {H{\"{u}}hne}}, \bibinfo {author} {\bibfnamefont
  {B.}~\bibnamefont {Rellinghaus}}, \bibinfo {author} {\bibfnamefont
  {C.}~\bibnamefont {Felser}}, \bibinfo {author} {\bibfnamefont
  {B.}~\bibnamefont {Yan}},\ and\ \bibinfo {author} {\bibfnamefont
  {K.}~\bibnamefont {Nielsch}},\ }\bibfield  {title} {\enquote {\bibinfo
  {title} {Experimental signatures of the mixed axial{\textendash}gravitational
  anomaly in the weyl semimetal {NbP}},}\ }\href
  {https://doi.org/10.1038/nature23005} {\bibfield  {journal} {\bibinfo
  {journal} {Nature}\ }\textbf {\bibinfo {volume} {547}},\ \bibinfo {pages}
  {324--327} (\bibinfo {year} {2017})}\BibitemShut {NoStop}%
\bibitem [{\citenamefont {Liu}\ \emph {et~al.}(2021{\natexlab{a}})\citenamefont
  {Liu}, \citenamefont {Yu}, \citenamefont {Ning}, \citenamefont {Yi},
  \citenamefont {Miao}, \citenamefont {Min}, \citenamefont {Zhao},
  \citenamefont {Ning}, \citenamefont {Lopez}, \citenamefont {Zhu},
  \citenamefont {Pillsbury}, \citenamefont {Zhang}, \citenamefont {Wang},
  \citenamefont {Hu}, \citenamefont {Cao}, \citenamefont {Chakoumakos},
  \citenamefont {Balakirev}, \citenamefont {Weickert}, \citenamefont {Jaime},
  \citenamefont {Lai}, \citenamefont {Yang}, \citenamefont {Sun}, \citenamefont
  {Alem}, \citenamefont {Gopalan}, \citenamefont {Chang}, \citenamefont
  {Samarth}, \citenamefont {Liu}, \citenamefont {McDonald},\ and\ \citenamefont
  {Mao}}]{Liu_2021a}%
  \BibitemOpen
  \bibfield  {author} {\bibinfo {author} {\bibfnamefont {J.~Y.}\ \bibnamefont
  {Liu}}, \bibinfo {author} {\bibfnamefont {J.}~\bibnamefont {Yu}}, \bibinfo
  {author} {\bibfnamefont {J.~L.}\ \bibnamefont {Ning}}, \bibinfo {author}
  {\bibfnamefont {H.~M.}\ \bibnamefont {Yi}}, \bibinfo {author} {\bibfnamefont
  {L.}~\bibnamefont {Miao}}, \bibinfo {author} {\bibfnamefont {L.~J.}\
  \bibnamefont {Min}}, \bibinfo {author} {\bibfnamefont {Y.~F.}\ \bibnamefont
  {Zhao}}, \bibinfo {author} {\bibfnamefont {W.}~\bibnamefont {Ning}}, \bibinfo
  {author} {\bibfnamefont {K.~A.}\ \bibnamefont {Lopez}}, \bibinfo {author}
  {\bibfnamefont {Y.~L.}\ \bibnamefont {Zhu}}, \bibinfo {author} {\bibfnamefont
  {T.}~\bibnamefont {Pillsbury}}, \bibinfo {author} {\bibfnamefont {Y.~B.}\
  \bibnamefont {Zhang}}, \bibinfo {author} {\bibfnamefont {Y.}~\bibnamefont
  {Wang}}, \bibinfo {author} {\bibfnamefont {J.}~\bibnamefont {Hu}}, \bibinfo
  {author} {\bibfnamefont {H.~B.}\ \bibnamefont {Cao}}, \bibinfo {author}
  {\bibfnamefont {B.~C.}\ \bibnamefont {Chakoumakos}}, \bibinfo {author}
  {\bibfnamefont {F.}~\bibnamefont {Balakirev}}, \bibinfo {author}
  {\bibfnamefont {F.}~\bibnamefont {Weickert}}, \bibinfo {author}
  {\bibfnamefont {M.}~\bibnamefont {Jaime}}, \bibinfo {author} {\bibfnamefont
  {Y.}~\bibnamefont {Lai}}, \bibinfo {author} {\bibfnamefont {K.}~\bibnamefont
  {Yang}}, \bibinfo {author} {\bibfnamefont {J.~W.}\ \bibnamefont {Sun}},
  \bibinfo {author} {\bibfnamefont {N.}~\bibnamefont {Alem}}, \bibinfo {author}
  {\bibfnamefont {V.}~\bibnamefont {Gopalan}}, \bibinfo {author} {\bibfnamefont
  {C.~Z.}\ \bibnamefont {Chang}}, \bibinfo {author} {\bibfnamefont
  {N.}~\bibnamefont {Samarth}}, \bibinfo {author} {\bibfnamefont {C.~X.}\
  \bibnamefont {Liu}}, \bibinfo {author} {\bibfnamefont {R.~D.}\ \bibnamefont
  {McDonald}},\ and\ \bibinfo {author} {\bibfnamefont {Z.~Q.}\ \bibnamefont
  {Mao}},\ }\bibfield  {title} {\enquote {\bibinfo {title} {Spin-valley locking
  and bulk quantum hall effect in a noncentrosymmetric dirac semimetal
  {BaMnSb}$_2$},}\ }\href {https://doi.org/10.1038/s41467-021-24369-1}
  {\bibfield  {journal} {\bibinfo  {journal} {Nat. Commun.}\ }\textbf {\bibinfo
  {volume} {12}} (\bibinfo {year} {2021}{\natexlab{a}}),\
  10.1038/s41467-021-24369-1}\BibitemShut {NoStop}%
\bibitem [{\citenamefont {Moll}\ \emph {et~al.}(2016)\citenamefont {Moll},
  \citenamefont {Nair}, \citenamefont {Helm}, \citenamefont {Potter},
  \citenamefont {Kimchi}, \citenamefont {Vishwanath},\ and\ \citenamefont
  {Analytis}}]{Moll_2016}%
  \BibitemOpen
  \bibfield  {author} {\bibinfo {author} {\bibfnamefont {P.~J.~W.}\
  \bibnamefont {Moll}}, \bibinfo {author} {\bibfnamefont {N.~L.}\ \bibnamefont
  {Nair}}, \bibinfo {author} {\bibfnamefont {T.}~\bibnamefont {Helm}}, \bibinfo
  {author} {\bibfnamefont {A.~C.}\ \bibnamefont {Potter}}, \bibinfo {author}
  {\bibfnamefont {I.}~\bibnamefont {Kimchi}}, \bibinfo {author} {\bibfnamefont
  {A.}~\bibnamefont {Vishwanath}},\ and\ \bibinfo {author} {\bibfnamefont
  {J.~G.}\ \bibnamefont {Analytis}},\ }\bibfield  {title} {\enquote {\bibinfo
  {title} {Transport evidence for {F}ermi-arc-mediated chirality transfer in
  the {D}irac semimetal {Cd$_3$As$_2$}},}\ }\href
  {https://doi.org/10.1038/nature18276} {\bibfield  {journal} {\bibinfo
  {journal} {Nature}\ }\textbf {\bibinfo {volume} {535}},\ \bibinfo {pages}
  {266--270} (\bibinfo {year} {2016})}\BibitemShut {NoStop}%
\bibitem [{\citenamefont {Guo}\ \emph {et~al.}(2022)\citenamefont {Guo},
  \citenamefont {Putzke}, \citenamefont {Konyzheva}, \citenamefont {Huang},
  \citenamefont {Gutierrez-Amigo}, \citenamefont {Errea}, \citenamefont {Chen},
  \citenamefont {Vergniory}, \citenamefont {Felser}, \citenamefont {Fischer},
  \citenamefont {Neupert},\ and\ \citenamefont {Moll}}]{Guo2022}%
  \BibitemOpen
  \bibfield  {author} {\bibinfo {author} {\bibfnamefont {C.}~\bibnamefont
  {Guo}}, \bibinfo {author} {\bibfnamefont {C.}~\bibnamefont {Putzke}},
  \bibinfo {author} {\bibfnamefont {S.}~\bibnamefont {Konyzheva}}, \bibinfo
  {author} {\bibfnamefont {X.}~\bibnamefont {Huang}}, \bibinfo {author}
  {\bibfnamefont {M.}~\bibnamefont {Gutierrez-Amigo}}, \bibinfo {author}
  {\bibfnamefont {I.}~\bibnamefont {Errea}}, \bibinfo {author} {\bibfnamefont
  {D.}~\bibnamefont {Chen}}, \bibinfo {author} {\bibfnamefont {M.~G.}\
  \bibnamefont {Vergniory}}, \bibinfo {author} {\bibfnamefont {C.}~\bibnamefont
  {Felser}}, \bibinfo {author} {\bibfnamefont {M.~H.}\ \bibnamefont {Fischer}},
  \bibinfo {author} {\bibfnamefont {T.}~\bibnamefont {Neupert}},\ and\ \bibinfo
  {author} {\bibfnamefont {P.~J.~W.}\ \bibnamefont {Moll}},\ }\bibfield
  {title} {\enquote {\bibinfo {title} {Switchable chiral transport in
  charge-ordered kagome metal {CsV}$_3${S}b$_5$},}\ }\href
  {https://doi.org/10.1038/s41586-022-05127-9} {\bibfield  {journal} {\bibinfo
  {journal} {Nature}\ }\textbf {\bibinfo {volume} {611}},\ \bibinfo {pages}
  {461--466} (\bibinfo {year} {2022})}\BibitemShut {NoStop}%
\bibitem [{\citenamefont {Maeno}\ \emph {et~al.}(1998)\citenamefont {Maeno},
  \citenamefont {Ando}, \citenamefont {Mori}, \citenamefont {Ohmichi},
  \citenamefont {Ikeda}, \citenamefont {NishiZaki},\ and\ \citenamefont
  {Nakatsuji}}]{Maeno_1998}%
  \BibitemOpen
  \bibfield  {author} {\bibinfo {author} {\bibfnamefont {Y.}~\bibnamefont
  {Maeno}}, \bibinfo {author} {\bibfnamefont {T.}~\bibnamefont {Ando}},
  \bibinfo {author} {\bibfnamefont {Y.}~\bibnamefont {Mori}}, \bibinfo {author}
  {\bibfnamefont {E.}~\bibnamefont {Ohmichi}}, \bibinfo {author} {\bibfnamefont
  {S.}~\bibnamefont {Ikeda}}, \bibinfo {author} {\bibfnamefont
  {S.}~\bibnamefont {NishiZaki}},\ and\ \bibinfo {author} {\bibfnamefont
  {S.}~\bibnamefont {Nakatsuji}},\ }\bibfield  {title} {\enquote {\bibinfo
  {title} {Enhancement of superconductivity of {Sr$_2$RuO$_4$} to 3\,{K} by
  embedded metallic microdomains},}\ }\href
  {https://doi.org/10.1103/physrevlett.81.3765} {\bibfield  {journal} {\bibinfo
   {journal} {Phys. Rev. Lett.}\ }\textbf {\bibinfo {volume} {81}},\ \bibinfo
  {pages} {3765--3768} (\bibinfo {year} {1998})}\BibitemShut {NoStop}%
\bibitem [{\citenamefont {van Delft}\ \emph {et~al.}(2022)\citenamefont {van
  Delft}, \citenamefont {Bachmann}, \citenamefont {Putzke}, \citenamefont
  {Guo}, \citenamefont {Straquadine}, \citenamefont {Bauer}, \citenamefont
  {Ronning},\ and\ \citenamefont {Moll}}]{Delft_2022}%
  \BibitemOpen
  \bibfield  {author} {\bibinfo {author} {\bibfnamefont {M.~R.}\ \bibnamefont
  {van Delft}}, \bibinfo {author} {\bibfnamefont {M.~D.}\ \bibnamefont
  {Bachmann}}, \bibinfo {author} {\bibfnamefont {C.}~\bibnamefont {Putzke}},
  \bibinfo {author} {\bibfnamefont {C.}~\bibnamefont {Guo}}, \bibinfo {author}
  {\bibfnamefont {J.~A.~W.}\ \bibnamefont {Straquadine}}, \bibinfo {author}
  {\bibfnamefont {E.~D.}\ \bibnamefont {Bauer}}, \bibinfo {author}
  {\bibfnamefont {F.}~\bibnamefont {Ronning}},\ and\ \bibinfo {author}
  {\bibfnamefont {P.~J.~W.}\ \bibnamefont {Moll}},\ }\bibfield  {title}
  {\enquote {\bibinfo {title} {Controlling superconductivity of {CeIrIn}$_5$
  microstructures by substrate selection},}\ }\href
  {https://doi.org/10.1063/5.0082561} {\bibfield  {journal} {\bibinfo
  {journal} {Appl. Phys. Lett.}\ }\textbf {\bibinfo {volume} {120}},\ \bibinfo
  {pages} {092601} (\bibinfo {year} {2022})}\BibitemShut {NoStop}%
\bibitem [{\citenamefont {Bachmann}\ \emph {et~al.}(2019)\citenamefont
  {Bachmann}, \citenamefont {Ferguson}, \citenamefont {Theuss}, \citenamefont
  {Meng}, \citenamefont {Putzke}, \citenamefont {Helm}, \citenamefont {Shirer},
  \citenamefont {Li}, \citenamefont {Modic}, \citenamefont {Nicklas},
  \citenamefont {K{\"{o}}nig}, \citenamefont {Low}, \citenamefont {Ghosh},
  \citenamefont {Mackenzie}, \citenamefont {Arnold}, \citenamefont {Hassinger},
  \citenamefont {McDonald}, \citenamefont {Winter}, \citenamefont {Bauer},
  \citenamefont {Ronning}, \citenamefont {Ramshaw}, \citenamefont {Nowack},\
  and\ \citenamefont {Moll}}]{Bachmann_2019}%
  \BibitemOpen
  \bibfield  {author} {\bibinfo {author} {\bibfnamefont {M.~D.}\ \bibnamefont
  {Bachmann}}, \bibinfo {author} {\bibfnamefont {G.~M.}\ \bibnamefont
  {Ferguson}}, \bibinfo {author} {\bibfnamefont {F.}~\bibnamefont {Theuss}},
  \bibinfo {author} {\bibfnamefont {T.}~\bibnamefont {Meng}}, \bibinfo {author}
  {\bibfnamefont {C.}~\bibnamefont {Putzke}}, \bibinfo {author} {\bibfnamefont
  {T.}~\bibnamefont {Helm}}, \bibinfo {author} {\bibfnamefont {K.~R.}\
  \bibnamefont {Shirer}}, \bibinfo {author} {\bibfnamefont {Y.-S.}\
  \bibnamefont {Li}}, \bibinfo {author} {\bibfnamefont {K.~A.}\ \bibnamefont
  {Modic}}, \bibinfo {author} {\bibfnamefont {M.}~\bibnamefont {Nicklas}},
  \bibinfo {author} {\bibfnamefont {M.}~\bibnamefont {K{\"{o}}nig}}, \bibinfo
  {author} {\bibfnamefont {D.}~\bibnamefont {Low}}, \bibinfo {author}
  {\bibfnamefont {S.}~\bibnamefont {Ghosh}}, \bibinfo {author} {\bibfnamefont
  {A.~P.}\ \bibnamefont {Mackenzie}}, \bibinfo {author} {\bibfnamefont
  {F.}~\bibnamefont {Arnold}}, \bibinfo {author} {\bibfnamefont
  {E.}~\bibnamefont {Hassinger}}, \bibinfo {author} {\bibfnamefont {R.~D.}\
  \bibnamefont {McDonald}}, \bibinfo {author} {\bibfnamefont {L.~E.}\
  \bibnamefont {Winter}}, \bibinfo {author} {\bibfnamefont {E.~D.}\
  \bibnamefont {Bauer}}, \bibinfo {author} {\bibfnamefont {F.}~\bibnamefont
  {Ronning}}, \bibinfo {author} {\bibfnamefont {B.~J.}\ \bibnamefont
  {Ramshaw}}, \bibinfo {author} {\bibfnamefont {K.~C.}\ \bibnamefont
  {Nowack}},\ and\ \bibinfo {author} {\bibfnamefont {P.~J.~W.}\ \bibnamefont
  {Moll}},\ }\bibfield  {title} {\enquote {\bibinfo {title} {Spatial control of
  heavy-fermion superconductivity in {CeIrIn}$_5$},}\ }\href
  {https://doi.org/10.1126/science.aao6640} {\bibfield  {journal} {\bibinfo
  {journal} {Science}\ }\textbf {\bibinfo {volume} {366}},\ \bibinfo {pages}
  {221--226} (\bibinfo {year} {2019})}\BibitemShut {NoStop}%
\bibitem [{\citenamefont {Martin}\ \emph {et~al.}(2015)\citenamefont {Martin},
  \citenamefont {Randolph}, \citenamefont {Botman}, \citenamefont {Toth},\ and\
  \citenamefont {Aharonovich}}]{Martin_2015}%
  \BibitemOpen
  \bibfield  {author} {\bibinfo {author} {\bibfnamefont {A.~A.}\ \bibnamefont
  {Martin}}, \bibinfo {author} {\bibfnamefont {S.}~\bibnamefont {Randolph}},
  \bibinfo {author} {\bibfnamefont {A.}~\bibnamefont {Botman}}, \bibinfo
  {author} {\bibfnamefont {M.}~\bibnamefont {Toth}},\ and\ \bibinfo {author}
  {\bibfnamefont {I.}~\bibnamefont {Aharonovich}},\ }\bibfield  {title}
  {\enquote {\bibinfo {title} {Maskless milling of diamond by a focused oxygen
  ion beam},}\ }\href {https://doi.org/10.1038/srep08958} {\bibfield  {journal}
  {\bibinfo  {journal} {Sci. Rep.}\ }\textbf {\bibinfo {volume} {5}} (\bibinfo
  {year} {2015}),\ 10.1038/srep08958}\BibitemShut {NoStop}%
\bibitem [{\citenamefont {Langridge}\ \emph {et~al.}(2014)\citenamefont
  {Langridge}, \citenamefont {Cox}, \citenamefont {Webb},\ and\ \citenamefont
  {Stolojan}}]{Langridge_2014}%
  \BibitemOpen
  \bibfield  {author} {\bibinfo {author} {\bibfnamefont {M.~T.}\ \bibnamefont
  {Langridge}}, \bibinfo {author} {\bibfnamefont {D.~C.}\ \bibnamefont {Cox}},
  \bibinfo {author} {\bibfnamefont {R.~P.}\ \bibnamefont {Webb}},\ and\
  \bibinfo {author} {\bibfnamefont {V.}~\bibnamefont {Stolojan}},\ }\bibfield
  {title} {\enquote {\bibinfo {title} {The fabrication of aspherical
  microlenses using focused ion-beam techniques},}\ }\href
  {https://doi.org/10.1016/j.micron.2013.10.013} {\bibfield  {journal}
  {\bibinfo  {journal} {Micron}\ }\textbf {\bibinfo {volume} {57}},\ \bibinfo
  {pages} {56--66} (\bibinfo {year} {2014})}\BibitemShut {NoStop}%
\bibitem [{\citenamefont {Scholz}\ \emph {et~al.}(2010)\citenamefont {Scholz},
  \citenamefont {Gallrapp}, \citenamefont {Kerst}, \citenamefont {Lundquist},\
  and\ \citenamefont {Boit}}]{Scholz_2010}%
  \BibitemOpen
  \bibfield  {author} {\bibinfo {author} {\bibfnamefont {P.}~\bibnamefont
  {Scholz}}, \bibinfo {author} {\bibfnamefont {C.}~\bibnamefont {Gallrapp}},
  \bibinfo {author} {\bibfnamefont {U.}~\bibnamefont {Kerst}}, \bibinfo
  {author} {\bibfnamefont {T.}~\bibnamefont {Lundquist}},\ and\ \bibinfo
  {author} {\bibfnamefont {C.}~\bibnamefont {Boit}},\ }\bibfield  {title}
  {\enquote {\bibinfo {title} {Optimizing focused ion beam created solid
  immersion lenses in bulk silicon using design of experiments},}\ }\href
  {https://doi.org/10.1016/j.microrel.2010.07.031} {\bibfield  {journal}
  {\bibinfo  {journal} {Microelectron. Reliab.}\ }\textbf {\bibinfo {volume}
  {50}},\ \bibinfo {pages} {1441--1445} (\bibinfo {year} {2010})}\BibitemShut
  {NoStop}%
\bibitem [{\citenamefont {Hadden}\ \emph {et~al.}(2010)\citenamefont {Hadden},
  \citenamefont {Harrison}, \citenamefont {Stanley-Clarke}, \citenamefont
  {Marseglia}, \citenamefont {Ho}, \citenamefont {Patton}, \citenamefont
  {O'Brien},\ and\ \citenamefont {Rarity}}]{Hadden_2010}%
  \BibitemOpen
  \bibfield  {author} {\bibinfo {author} {\bibfnamefont {J.~P.}\ \bibnamefont
  {Hadden}}, \bibinfo {author} {\bibfnamefont {J.~P.}\ \bibnamefont
  {Harrison}}, \bibinfo {author} {\bibfnamefont {A.~C.}\ \bibnamefont
  {Stanley-Clarke}}, \bibinfo {author} {\bibfnamefont {L.}~\bibnamefont
  {Marseglia}}, \bibinfo {author} {\bibfnamefont {Y.-L.~D.}\ \bibnamefont
  {Ho}}, \bibinfo {author} {\bibfnamefont {B.~R.}\ \bibnamefont {Patton}},
  \bibinfo {author} {\bibfnamefont {J.~L.}\ \bibnamefont {O'Brien}},\ and\
  \bibinfo {author} {\bibfnamefont {J.~G.}\ \bibnamefont {Rarity}},\ }\bibfield
   {title} {\enquote {\bibinfo {title} {Strongly enhanced photon collection
  from diamond defect centers under microfabricated integrated solid immersion
  lenses},}\ }\href {https://doi.org/10.1063/1.3519847} {\bibfield  {journal}
  {\bibinfo  {journal} {Appl. Phys. Lett.}\ }\textbf {\bibinfo {volume} {97}},\
  \bibinfo {pages} {241901} (\bibinfo {year} {2010})}\BibitemShut {NoStop}%
\bibitem [{\citenamefont {Marseglia}\ \emph {et~al.}(2011)\citenamefont
  {Marseglia}, \citenamefont {Hadden}, \citenamefont {Stanley-Clarke},
  \citenamefont {Harrison}, \citenamefont {Patton}, \citenamefont {Ho},
  \citenamefont {Naydenov}, \citenamefont {Jelezko}, \citenamefont {Meijer},
  \citenamefont {Dolan}, \citenamefont {Smith}, \citenamefont {Rarity},\ and\
  \citenamefont {O'Brien}}]{Marseglia_2011}%
  \BibitemOpen
  \bibfield  {author} {\bibinfo {author} {\bibfnamefont {L.}~\bibnamefont
  {Marseglia}}, \bibinfo {author} {\bibfnamefont {J.~P.}\ \bibnamefont
  {Hadden}}, \bibinfo {author} {\bibfnamefont {A.~C.}\ \bibnamefont
  {Stanley-Clarke}}, \bibinfo {author} {\bibfnamefont {J.~P.}\ \bibnamefont
  {Harrison}}, \bibinfo {author} {\bibfnamefont {B.}~\bibnamefont {Patton}},
  \bibinfo {author} {\bibfnamefont {Y.-L.~D.}\ \bibnamefont {Ho}}, \bibinfo
  {author} {\bibfnamefont {B.}~\bibnamefont {Naydenov}}, \bibinfo {author}
  {\bibfnamefont {F.}~\bibnamefont {Jelezko}}, \bibinfo {author} {\bibfnamefont
  {J.}~\bibnamefont {Meijer}}, \bibinfo {author} {\bibfnamefont {P.~R.}\
  \bibnamefont {Dolan}}, \bibinfo {author} {\bibfnamefont {J.~M.}\ \bibnamefont
  {Smith}}, \bibinfo {author} {\bibfnamefont {J.~G.}\ \bibnamefont {Rarity}},\
  and\ \bibinfo {author} {\bibfnamefont {J.~L.}\ \bibnamefont {O'Brien}},\
  }\bibfield  {title} {\enquote {\bibinfo {title} {Nanofabricated solid
  immersion lenses registered to single emitters in diamond},}\ }\href
  {https://doi.org/10.1063/1.3573870} {\bibfield  {journal} {\bibinfo
  {journal} {Appl. Phys. Lett.}\ }\textbf {\bibinfo {volume} {98}},\ \bibinfo
  {pages} {133107} (\bibinfo {year} {2011})}\BibitemShut {NoStop}%
\bibitem [{\citenamefont {Riedrich-M{\"{o}}ller}\ \emph
  {et~al.}(2014)\citenamefont {Riedrich-M{\"{o}}ller}, \citenamefont {Arend},
  \citenamefont {Pauly}, \citenamefont {M{\"{u}}cklich}, \citenamefont
  {Fischer}, \citenamefont {Gsell}, \citenamefont {Schreck},\ and\
  \citenamefont {Becher}}]{RiedrichMoeller_2014}%
  \BibitemOpen
  \bibfield  {author} {\bibinfo {author} {\bibfnamefont {J.}~\bibnamefont
  {Riedrich-M{\"{o}}ller}}, \bibinfo {author} {\bibfnamefont {C.}~\bibnamefont
  {Arend}}, \bibinfo {author} {\bibfnamefont {C.}~\bibnamefont {Pauly}},
  \bibinfo {author} {\bibfnamefont {F.}~\bibnamefont {M{\"{u}}cklich}},
  \bibinfo {author} {\bibfnamefont {M.}~\bibnamefont {Fischer}}, \bibinfo
  {author} {\bibfnamefont {S.}~\bibnamefont {Gsell}}, \bibinfo {author}
  {\bibfnamefont {M.}~\bibnamefont {Schreck}},\ and\ \bibinfo {author}
  {\bibfnamefont {C.}~\bibnamefont {Becher}},\ }\bibfield  {title} {\enquote
  {\bibinfo {title} {Deterministic coupling of a single silicon-vacancy color
  center to a photonic crystal cavity in diamond},}\ }\href
  {https://doi.org/10.1021/nl502327b} {\bibfield  {journal} {\bibinfo
  {journal} {Nano Lett.}\ }\textbf {\bibinfo {volume} {14}},\ \bibinfo {pages}
  {5281--5287} (\bibinfo {year} {2014})}\BibitemShut {NoStop}%
\bibitem [{\citenamefont {Bayn}\ \emph {et~al.}(2011)\citenamefont {Bayn},
  \citenamefont {Meyler}, \citenamefont {Salzman},\ and\ \citenamefont
  {Kalish}}]{Bayn_2011}%
  \BibitemOpen
  \bibfield  {author} {\bibinfo {author} {\bibfnamefont {I.}~\bibnamefont
  {Bayn}}, \bibinfo {author} {\bibfnamefont {B.}~\bibnamefont {Meyler}},
  \bibinfo {author} {\bibfnamefont {J.}~\bibnamefont {Salzman}},\ and\ \bibinfo
  {author} {\bibfnamefont {R.}~\bibnamefont {Kalish}},\ }\bibfield  {title}
  {\enquote {\bibinfo {title} {Triangular nanobeam photonic cavities in
  single-crystal diamond},}\ }\href
  {https://doi.org/10.1088/1367-2630/13/2/025018} {\bibfield  {journal}
  {\bibinfo  {journal} {New J. Phys.}\ }\textbf {\bibinfo {volume} {13}},\
  \bibinfo {pages} {025018} (\bibinfo {year} {2011})}\BibitemShut {NoStop}%
\bibitem [{\citenamefont {Kindem}\ \emph {et~al.}(2020)\citenamefont {Kindem},
  \citenamefont {Ruskuc}, \citenamefont {Bartholomew}, \citenamefont {Rochman},
  \citenamefont {Huan},\ and\ \citenamefont {Faraon}}]{Kindem_2020}%
  \BibitemOpen
  \bibfield  {author} {\bibinfo {author} {\bibfnamefont {J.~M.}\ \bibnamefont
  {Kindem}}, \bibinfo {author} {\bibfnamefont {A.}~\bibnamefont {Ruskuc}},
  \bibinfo {author} {\bibfnamefont {J.~G.}\ \bibnamefont {Bartholomew}},
  \bibinfo {author} {\bibfnamefont {J.}~\bibnamefont {Rochman}}, \bibinfo
  {author} {\bibfnamefont {Y.~Q.}\ \bibnamefont {Huan}},\ and\ \bibinfo
  {author} {\bibfnamefont {A.}~\bibnamefont {Faraon}},\ }\bibfield  {title}
  {\enquote {\bibinfo {title} {Control and single-shot readout of an ion
  embedded in a nanophotonic cavity},}\ }\href
  {https://doi.org/10.1038/s41586-020-2160-9} {\bibfield  {journal} {\bibinfo
  {journal} {Nature}\ }\textbf {\bibinfo {volume} {580}},\ \bibinfo {pages}
  {201--204} (\bibinfo {year} {2020})}\BibitemShut {NoStop}%
\bibitem [{\citenamefont {Zhong}\ \emph {et~al.}(2017)\citenamefont {Zhong},
  \citenamefont {Kindem}, \citenamefont {Bartholomew}, \citenamefont {Rochman},
  \citenamefont {Craiciu}, \citenamefont {Miyazono}, \citenamefont
  {Bettinelli}, \citenamefont {Cavalli}, \citenamefont {Verma}, \citenamefont
  {Nam}, \citenamefont {Marsili}, \citenamefont {Shaw}, \citenamefont {Beyer},\
  and\ \citenamefont {Faraon}}]{Zhong_2017}%
  \BibitemOpen
  \bibfield  {author} {\bibinfo {author} {\bibfnamefont {T.}~\bibnamefont
  {Zhong}}, \bibinfo {author} {\bibfnamefont {J.~M.}\ \bibnamefont {Kindem}},
  \bibinfo {author} {\bibfnamefont {J.~G.}\ \bibnamefont {Bartholomew}},
  \bibinfo {author} {\bibfnamefont {J.}~\bibnamefont {Rochman}}, \bibinfo
  {author} {\bibfnamefont {I.}~\bibnamefont {Craiciu}}, \bibinfo {author}
  {\bibfnamefont {E.}~\bibnamefont {Miyazono}}, \bibinfo {author}
  {\bibfnamefont {M.}~\bibnamefont {Bettinelli}}, \bibinfo {author}
  {\bibfnamefont {E.}~\bibnamefont {Cavalli}}, \bibinfo {author} {\bibfnamefont
  {V.}~\bibnamefont {Verma}}, \bibinfo {author} {\bibfnamefont {S.~W.}\
  \bibnamefont {Nam}}, \bibinfo {author} {\bibfnamefont {F.}~\bibnamefont
  {Marsili}}, \bibinfo {author} {\bibfnamefont {M.~D.}\ \bibnamefont {Shaw}},
  \bibinfo {author} {\bibfnamefont {A.~D.}\ \bibnamefont {Beyer}},\ and\
  \bibinfo {author} {\bibfnamefont {A.}~\bibnamefont {Faraon}},\ }\bibfield
  {title} {\enquote {\bibinfo {title} {Nanophotonic rare-earth quantum memory
  with optically controlled retrieval},}\ }\href
  {https://doi.org/10.1126/science.aan5959} {\bibfield  {journal} {\bibinfo
  {journal} {Science}\ }\textbf {\bibinfo {volume} {357}},\ \bibinfo {pages}
  {1392--1395} (\bibinfo {year} {2017})}\BibitemShut {NoStop}%
\bibitem [{\citenamefont {Zhong}\ \emph {et~al.}(2016)\citenamefont {Zhong},
  \citenamefont {Rochman}, \citenamefont {Kindem}, \citenamefont {Miyazono},\
  and\ \citenamefont {Faraon}}]{Zhong_2016}%
  \BibitemOpen
  \bibfield  {author} {\bibinfo {author} {\bibfnamefont {T.}~\bibnamefont
  {Zhong}}, \bibinfo {author} {\bibfnamefont {J.}~\bibnamefont {Rochman}},
  \bibinfo {author} {\bibfnamefont {J.~M.}\ \bibnamefont {Kindem}}, \bibinfo
  {author} {\bibfnamefont {E.}~\bibnamefont {Miyazono}},\ and\ \bibinfo
  {author} {\bibfnamefont {A.}~\bibnamefont {Faraon}},\ }\bibfield  {title}
  {\enquote {\bibinfo {title} {High quality factor nanophotonic resonators in
  bulk rare-earth doped crystals},}\ }\href
  {https://doi.org/10.1364/oe.24.000536} {\bibfield  {journal} {\bibinfo
  {journal} {Opt. Express}\ }\textbf {\bibinfo {volume} {24}},\ \bibinfo
  {pages} {536} (\bibinfo {year} {2016})}\BibitemShut {NoStop}%
\bibitem [{\citenamefont {Keskinbora}\ \emph {et~al.}(2013)\citenamefont
  {Keskinbora}, \citenamefont {Gr{\'{e}}vent}, \citenamefont {Eigenthaler},
  \citenamefont {Weigand},\ and\ \citenamefont
  {Sch{\"{u}}tz}}]{Keskinbora_2013}%
  \BibitemOpen
  \bibfield  {author} {\bibinfo {author} {\bibfnamefont {K.}~\bibnamefont
  {Keskinbora}}, \bibinfo {author} {\bibfnamefont {C.}~\bibnamefont
  {Gr{\'{e}}vent}}, \bibinfo {author} {\bibfnamefont {U.}~\bibnamefont
  {Eigenthaler}}, \bibinfo {author} {\bibfnamefont {M.}~\bibnamefont
  {Weigand}},\ and\ \bibinfo {author} {\bibfnamefont {G.}~\bibnamefont
  {Sch{\"{u}}tz}},\ }\bibfield  {title} {\enquote {\bibinfo {title} {Rapid
  prototyping of fresnel zone plates via direct ga$^+$ ion beam lithography for
  high-resolution x-ray imaging},}\ }\href {https://doi.org/10.1021/nn403295k}
  {\bibfield  {journal} {\bibinfo  {journal} {{ACS} Nano}\ }\textbf {\bibinfo
  {volume} {7}},\ \bibinfo {pages} {9788--9797} (\bibinfo {year}
  {2013})}\BibitemShut {NoStop}%
\bibitem [{\citenamefont {Madison}\ \emph {et~al.}(2022)\citenamefont
  {Madison}, \citenamefont {Villarrubia}, \citenamefont {Liao}, \citenamefont
  {Copeland}, \citenamefont {Schumacher}, \citenamefont {Siebein},
  \citenamefont {Ilic}, \citenamefont {Liddle},\ and\ \citenamefont
  {Stavis}}]{Madison_2022}%
  \BibitemOpen
  \bibfield  {author} {\bibinfo {author} {\bibfnamefont {A.~C.}\ \bibnamefont
  {Madison}}, \bibinfo {author} {\bibfnamefont {J.~S.}\ \bibnamefont
  {Villarrubia}}, \bibinfo {author} {\bibfnamefont {K.-T.}\ \bibnamefont
  {Liao}}, \bibinfo {author} {\bibfnamefont {C.~R.}\ \bibnamefont {Copeland}},
  \bibinfo {author} {\bibfnamefont {J.}~\bibnamefont {Schumacher}}, \bibinfo
  {author} {\bibfnamefont {K.}~\bibnamefont {Siebein}}, \bibinfo {author}
  {\bibfnamefont {B.~R.}\ \bibnamefont {Ilic}}, \bibinfo {author}
  {\bibfnamefont {J.~A.}\ \bibnamefont {Liddle}},\ and\ \bibinfo {author}
  {\bibfnamefont {S.~M.}\ \bibnamefont {Stavis}},\ }\bibfield  {title}
  {\enquote {\bibinfo {title} {Unmasking the resolution{\textendash}throughput
  tradespace of focused-ion-beam machining},}\ }\href
  {https://doi.org/10.1002/adfm.202111840} {\bibfield  {journal} {\bibinfo
  {journal} {Adv. Funct. Mater.}\ }\textbf {\bibinfo {volume} {32}},\ \bibinfo
  {pages} {2111840} (\bibinfo {year} {2022})}\BibitemShut {NoStop}%
\bibitem [{\citenamefont {Li}\ and\ \citenamefont {Liu}(2018)}]{Li_2018}%
  \BibitemOpen
  \bibfield  {author} {\bibinfo {author} {\bibfnamefont {J.}~\bibnamefont
  {Li}}\ and\ \bibinfo {author} {\bibfnamefont {Z.}~\bibnamefont {Liu}},\
  }\bibfield  {title} {\enquote {\bibinfo {title} {Focused-ion-beam-based
  nano-kirigami: from art to photonics},}\ }\href
  {https://doi.org/10.1515/nanoph-2018-0117} {\bibfield  {journal} {\bibinfo
  {journal} {Nanophotonics}\ }\textbf {\bibinfo {volume} {7}},\ \bibinfo
  {pages} {1637--1650} (\bibinfo {year} {2018})}\BibitemShut {NoStop}%
\bibitem [{\citenamefont {Rakhlin}\ \emph {et~al.}(2021)\citenamefont
  {Rakhlin}, \citenamefont {Sorokin}, \citenamefont {Kazanov}, \citenamefont
  {Sedova}, \citenamefont {Shubina}, \citenamefont {Ivanov}, \citenamefont
  {Mikhailovskii},\ and\ \citenamefont {Toropov}}]{Rakhlin_2021}%
  \BibitemOpen
  \bibfield  {author} {\bibinfo {author} {\bibfnamefont {M.}~\bibnamefont
  {Rakhlin}}, \bibinfo {author} {\bibfnamefont {S.}~\bibnamefont {Sorokin}},
  \bibinfo {author} {\bibfnamefont {D.}~\bibnamefont {Kazanov}}, \bibinfo
  {author} {\bibfnamefont {I.}~\bibnamefont {Sedova}}, \bibinfo {author}
  {\bibfnamefont {T.}~\bibnamefont {Shubina}}, \bibinfo {author} {\bibfnamefont
  {S.}~\bibnamefont {Ivanov}}, \bibinfo {author} {\bibfnamefont
  {V.}~\bibnamefont {Mikhailovskii}},\ and\ \bibinfo {author} {\bibfnamefont
  {A.}~\bibnamefont {Toropov}},\ }\bibfield  {title} {\enquote {\bibinfo
  {title} {Bright single-photon emitters with a {CdSe} quantum dot and
  multimode tapered nanoantenna for the visible spectral range},}\ }\href
  {https://doi.org/10.3390/nano11040916} {\bibfield  {journal} {\bibinfo
  {journal} {Nanomaterials}\ }\textbf {\bibinfo {volume} {11}},\ \bibinfo
  {pages} {916} (\bibinfo {year} {2021})}\BibitemShut {NoStop}%
\bibitem [{\citenamefont {Sloyan}\ \emph {et~al.}(2021)\citenamefont {Sloyan},
  \citenamefont {Melkonyan}, \citenamefont {Apostoleris}, \citenamefont
  {Dahlem}, \citenamefont {Chiesa},\ and\ \citenamefont
  {Ghaferi}}]{Sloyan_2021}%
  \BibitemOpen
  \bibfield  {author} {\bibinfo {author} {\bibfnamefont {K.}~\bibnamefont
  {Sloyan}}, \bibinfo {author} {\bibfnamefont {H.}~\bibnamefont {Melkonyan}},
  \bibinfo {author} {\bibfnamefont {H.}~\bibnamefont {Apostoleris}}, \bibinfo
  {author} {\bibfnamefont {M.~S.}\ \bibnamefont {Dahlem}}, \bibinfo {author}
  {\bibfnamefont {M.}~\bibnamefont {Chiesa}},\ and\ \bibinfo {author}
  {\bibfnamefont {A.~A.}\ \bibnamefont {Ghaferi}},\ }\bibfield  {title}
  {\enquote {\bibinfo {title} {A review of focused ion beam applications in
  optical fibers},}\ }\href {https://doi.org/10.1088/1361-6528/ac1d75}
  {\bibfield  {journal} {\bibinfo  {journal} {Nanotechnology}\ }\textbf
  {\bibinfo {volume} {32}},\ \bibinfo {pages} {472004} (\bibinfo {year}
  {2021})}\BibitemShut {NoStop}%
\bibitem [{\citenamefont {Romagnoli}\ \emph {et~al.}(2020)\citenamefont
  {Romagnoli}, \citenamefont {Maeda}, \citenamefont {Ward}, \citenamefont
  {Truong},\ and\ \citenamefont {Chormaic}}]{Romagnoli_2020a}%
  \BibitemOpen
  \bibfield  {author} {\bibinfo {author} {\bibfnamefont {P.}~\bibnamefont
  {Romagnoli}}, \bibinfo {author} {\bibfnamefont {M.}~\bibnamefont {Maeda}},
  \bibinfo {author} {\bibfnamefont {J.~M.}\ \bibnamefont {Ward}}, \bibinfo
  {author} {\bibfnamefont {V.~G.}\ \bibnamefont {Truong}},\ and\ \bibinfo
  {author} {\bibfnamefont {S.~N.}\ \bibnamefont {Chormaic}},\ }\bibfield
  {title} {\enquote {\bibinfo {title} {Fabrication of optical nanofibre-based
  cavities using focussed ion-beam milling: a review},}\ }\href
  {https://doi.org/10.1007/s00340-020-07456-x} {\bibfield  {journal} {\bibinfo
  {journal} {Appl. Phys. B}\ }\textbf {\bibinfo {volume} {126}} (\bibinfo
  {year} {2020}),\ 10.1007/s00340-020-07456-x}\BibitemShut {NoStop}%
\bibitem [{\citenamefont {Korneeva}\ \emph {et~al.}(2020)\citenamefont
  {Korneeva}, \citenamefont {Manova}, \citenamefont {Florya}, \citenamefont
  {Mikhailov}, \citenamefont {Dobrovolskiy}, \citenamefont {Korneev},\ and\
  \citenamefont {Vodolazov}}]{Korneeva_2020}%
  \BibitemOpen
  \bibfield  {author} {\bibinfo {author} {\bibfnamefont {Y.~P.}\ \bibnamefont
  {Korneeva}}, \bibinfo {author} {\bibfnamefont {N.~N.}\ \bibnamefont
  {Manova}}, \bibinfo {author} {\bibfnamefont {I.~N.}\ \bibnamefont {Florya}},
  \bibinfo {author} {\bibfnamefont {M.~Y.}\ \bibnamefont {Mikhailov}}, \bibinfo
  {author} {\bibfnamefont {O.~V.}\ \bibnamefont {Dobrovolskiy}}, \bibinfo
  {author} {\bibfnamefont {A.~A.}\ \bibnamefont {Korneev}},\ and\ \bibinfo
  {author} {\bibfnamefont {D.~Y.}\ \bibnamefont {Vodolazov}},\ }\bibfield
  {title} {\enquote {\bibinfo {title} {Different single-photon response of wide
  and narrow superconducting {Mo}$_{x}${Si}$_{1-x}$ strips},}\ }\href
  {https://doi.org/10.1103/physrevapplied.13.024011} {\bibfield  {journal}
  {\bibinfo  {journal} {Phys. Rev. Applied}\ }\textbf {\bibinfo {volume}
  {13}},\ \bibinfo {pages} {024011} (\bibinfo {year} {2020})}\BibitemShut
  {NoStop}%
\bibitem [{\citenamefont {Bachar}\ \emph {et~al.}(2012)\citenamefont {Bachar},
  \citenamefont {Baskin}, \citenamefont {Shtempluck},\ and\ \citenamefont
  {Buks}}]{Bachar_2012}%
  \BibitemOpen
  \bibfield  {author} {\bibinfo {author} {\bibfnamefont {G.}~\bibnamefont
  {Bachar}}, \bibinfo {author} {\bibfnamefont {I.}~\bibnamefont {Baskin}},
  \bibinfo {author} {\bibfnamefont {O.}~\bibnamefont {Shtempluck}},\ and\
  \bibinfo {author} {\bibfnamefont {E.}~\bibnamefont {Buks}},\ }\bibfield
  {title} {\enquote {\bibinfo {title} {Superconducting nanowire single photon
  detectors on-fiber},}\ }\href {https://doi.org/10.1063/1.4773305} {\bibfield
  {journal} {\bibinfo  {journal} {Appl. Phys. Lett.}\ }\textbf {\bibinfo
  {volume} {101}},\ \bibinfo {pages} {262601} (\bibinfo {year}
  {2012})}\BibitemShut {NoStop}%
\bibitem [{\citenamefont {Simeone}\ \emph {et~al.}(2018)\citenamefont
  {Simeone}, \citenamefont {Esposito}, \citenamefont {Scuderi}, \citenamefont
  {Calafiore}, \citenamefont {Palermo}, \citenamefont {Luca}, \citenamefont
  {Todisco}, \citenamefont {Sanvitto}, \citenamefont {Nicotra}, \citenamefont
  {Cabrini}, \citenamefont {Tasco}, \citenamefont {Passaseo},\ and\
  \citenamefont {Cuscun{\`{a}}}}]{Simeone_2018}%
  \BibitemOpen
  \bibfield  {author} {\bibinfo {author} {\bibfnamefont {D.}~\bibnamefont
  {Simeone}}, \bibinfo {author} {\bibfnamefont {M.}~\bibnamefont {Esposito}},
  \bibinfo {author} {\bibfnamefont {M.}~\bibnamefont {Scuderi}}, \bibinfo
  {author} {\bibfnamefont {G.}~\bibnamefont {Calafiore}}, \bibinfo {author}
  {\bibfnamefont {G.}~\bibnamefont {Palermo}}, \bibinfo {author} {\bibfnamefont
  {A.~D.}\ \bibnamefont {Luca}}, \bibinfo {author} {\bibfnamefont
  {F.}~\bibnamefont {Todisco}}, \bibinfo {author} {\bibfnamefont
  {D.}~\bibnamefont {Sanvitto}}, \bibinfo {author} {\bibfnamefont
  {G.}~\bibnamefont {Nicotra}}, \bibinfo {author} {\bibfnamefont
  {S.}~\bibnamefont {Cabrini}}, \bibinfo {author} {\bibfnamefont
  {V.}~\bibnamefont {Tasco}}, \bibinfo {author} {\bibfnamefont
  {A.}~\bibnamefont {Passaseo}},\ and\ \bibinfo {author} {\bibfnamefont
  {M.}~\bibnamefont {Cuscun{\`{a}}}},\ }\bibfield  {title} {\enquote {\bibinfo
  {title} {Tailoring electromagnetic hot spots toward visible frequencies in
  ultra-narrow gap {A}l/{A}l$_{2}${O}$_{3}$ bowtie nanoantennas},}\ }\href
  {https://doi.org/10.1021/acsphotonics.8b00665} {\bibfield  {journal}
  {\bibinfo  {journal} {{ACS} Photonics}\ }\textbf {\bibinfo {volume} {5}},\
  \bibinfo {pages} {3399--3407} (\bibinfo {year} {2018})}\BibitemShut {NoStop}%
\bibitem [{\citenamefont {Kollmann}\ \emph {et~al.}(2014)\citenamefont
  {Kollmann}, \citenamefont {Piao}, \citenamefont {Esmann}, \citenamefont
  {Becker}, \citenamefont {Hou}, \citenamefont {Huynh}, \citenamefont
  {Kautschor}, \citenamefont {B{\"{o}}sker}, \citenamefont {Vieker},
  \citenamefont {Beyer}, \citenamefont {G{\"{o}}lzh{\"{a}}user}, \citenamefont
  {Park}, \citenamefont {Vogelgesang}, \citenamefont {Silies},\ and\
  \citenamefont {Lienau}}]{Kollmann_2014}%
  \BibitemOpen
  \bibfield  {author} {\bibinfo {author} {\bibfnamefont {H.}~\bibnamefont
  {Kollmann}}, \bibinfo {author} {\bibfnamefont {X.}~\bibnamefont {Piao}},
  \bibinfo {author} {\bibfnamefont {M.}~\bibnamefont {Esmann}}, \bibinfo
  {author} {\bibfnamefont {S.~F.}\ \bibnamefont {Becker}}, \bibinfo {author}
  {\bibfnamefont {D.}~\bibnamefont {Hou}}, \bibinfo {author} {\bibfnamefont
  {C.}~\bibnamefont {Huynh}}, \bibinfo {author} {\bibfnamefont {L.-O.}\
  \bibnamefont {Kautschor}}, \bibinfo {author} {\bibfnamefont {G.}~\bibnamefont
  {B{\"{o}}sker}}, \bibinfo {author} {\bibfnamefont {H.}~\bibnamefont
  {Vieker}}, \bibinfo {author} {\bibfnamefont {A.}~\bibnamefont {Beyer}},
  \bibinfo {author} {\bibfnamefont {A.}~\bibnamefont {G{\"{o}}lzh{\"{a}}user}},
  \bibinfo {author} {\bibfnamefont {N.}~\bibnamefont {Park}}, \bibinfo {author}
  {\bibfnamefont {R.}~\bibnamefont {Vogelgesang}}, \bibinfo {author}
  {\bibfnamefont {M.}~\bibnamefont {Silies}},\ and\ \bibinfo {author}
  {\bibfnamefont {C.}~\bibnamefont {Lienau}},\ }\bibfield  {title} {\enquote
  {\bibinfo {title} {Toward plasmonics with nanometer precision: nonlinear
  optics of helium-ion milled gold nanoantennas},}\ }\href
  {https://doi.org/10.1021/nl5019589} {\bibfield  {journal} {\bibinfo
  {journal} {Nano Lett.}\ }\textbf {\bibinfo {volume} {14}},\ \bibinfo {pages}
  {4778--4784} (\bibinfo {year} {2014})}\BibitemShut {NoStop}%
\bibitem [{\citenamefont {Wang}\ \emph {et~al.}(2013)\citenamefont {Wang},
  \citenamefont {Abb}, \citenamefont {Boden}, \citenamefont {Aizpurua},
  \citenamefont {de~Groot},\ and\ \citenamefont {Muskens}}]{Wang_2013}%
  \BibitemOpen
  \bibfield  {author} {\bibinfo {author} {\bibfnamefont {Y.}~\bibnamefont
  {Wang}}, \bibinfo {author} {\bibfnamefont {M.}~\bibnamefont {Abb}}, \bibinfo
  {author} {\bibfnamefont {S.~A.}\ \bibnamefont {Boden}}, \bibinfo {author}
  {\bibfnamefont {J.}~\bibnamefont {Aizpurua}}, \bibinfo {author}
  {\bibfnamefont {C.~H.}\ \bibnamefont {de~Groot}},\ and\ \bibinfo {author}
  {\bibfnamefont {O.~L.}\ \bibnamefont {Muskens}},\ }\bibfield  {title}
  {\enquote {\bibinfo {title} {Ultrafast nonlinear control of progressively
  loaded, single plasmonic nanoantennas fabricated using helium ion milling},}\
  }\href {https://doi.org/10.1021/nl403316z} {\bibfield  {journal} {\bibinfo
  {journal} {Nano Lett.}\ }\textbf {\bibinfo {volume} {13}},\ \bibinfo {pages}
  {5647--5653} (\bibinfo {year} {2013})}\BibitemShut {NoStop}%
\bibitem [{\citenamefont {Scholder}\ \emph {et~al.}(2013)\citenamefont
  {Scholder}, \citenamefont {Jefimovs}, \citenamefont {Shorubalko},
  \citenamefont {Hafner}, \citenamefont {Sennhauser},\ and\ \citenamefont
  {Bona}}]{Scholder_2013}%
  \BibitemOpen
  \bibfield  {author} {\bibinfo {author} {\bibfnamefont {O.}~\bibnamefont
  {Scholder}}, \bibinfo {author} {\bibfnamefont {K.}~\bibnamefont {Jefimovs}},
  \bibinfo {author} {\bibfnamefont {I.}~\bibnamefont {Shorubalko}}, \bibinfo
  {author} {\bibfnamefont {C.}~\bibnamefont {Hafner}}, \bibinfo {author}
  {\bibfnamefont {U.}~\bibnamefont {Sennhauser}},\ and\ \bibinfo {author}
  {\bibfnamefont {G.-L.}\ \bibnamefont {Bona}},\ }\bibfield  {title} {\enquote
  {\bibinfo {title} {Helium focused ion beam fabricated plasmonic antennas with
  sub-5 nm gaps},}\ }\href {https://doi.org/10.1088/0957-4484/24/39/395301}
  {\bibfield  {journal} {\bibinfo  {journal} {Nanotechnology}\ }\textbf
  {\bibinfo {volume} {24}},\ \bibinfo {pages} {395301} (\bibinfo {year}
  {2013})}\BibitemShut {NoStop}%
\bibitem [{\citenamefont {Seitl}\ \emph {et~al.}(2019)\citenamefont {Seitl},
  \citenamefont {Laible}, \citenamefont {Dickreuter}, \citenamefont {Gollmer},
  \citenamefont {Kern},\ and\ \citenamefont {Fleischer}}]{Seitl2019}%
  \BibitemOpen
  \bibfield  {author} {\bibinfo {author} {\bibfnamefont {L.}~\bibnamefont
  {Seitl}}, \bibinfo {author} {\bibfnamefont {F.}~\bibnamefont {Laible}},
  \bibinfo {author} {\bibfnamefont {S.}~\bibnamefont {Dickreuter}}, \bibinfo
  {author} {\bibfnamefont {D.~A.}\ \bibnamefont {Gollmer}}, \bibinfo {author}
  {\bibfnamefont {D.~P.}\ \bibnamefont {Kern}},\ and\ \bibinfo {author}
  {\bibfnamefont {M.}~\bibnamefont {Fleischer}},\ }\bibfield  {title} {\enquote
  {\bibinfo {title} {Miniaturized fractal optical nanoantennas defined by
  focused helium ion beam milling},}\ }\href
  {https://doi.org/10.1088/1361-6528/ab5120} {\bibfield  {journal} {\bibinfo
  {journal} {Nanotechnology}\ }\textbf {\bibinfo {volume} {31}},\ \bibinfo
  {pages} {075301} (\bibinfo {year} {2019})}\BibitemShut {NoStop}%
\bibitem [{\citenamefont {Chen}\ \emph
  {et~al.}(2020{\natexlab{a}})\citenamefont {Chen}, \citenamefont {Hu},
  \citenamefont {Zhao}, \citenamefont {Deng}, \citenamefont {Wang},
  \citenamefont {Cheng}, \citenamefont {Lei}, \citenamefont {Deng},\ and\
  \citenamefont {Duan}}]{Chen_2020a}%
  \BibitemOpen
  \bibfield  {author} {\bibinfo {author} {\bibfnamefont {Y.}~\bibnamefont
  {Chen}}, \bibinfo {author} {\bibfnamefont {Y.}~\bibnamefont {Hu}}, \bibinfo
  {author} {\bibfnamefont {J.}~\bibnamefont {Zhao}}, \bibinfo {author}
  {\bibfnamefont {Y.}~\bibnamefont {Deng}}, \bibinfo {author} {\bibfnamefont
  {Z.}~\bibnamefont {Wang}}, \bibinfo {author} {\bibfnamefont {X.}~\bibnamefont
  {Cheng}}, \bibinfo {author} {\bibfnamefont {D.}~\bibnamefont {Lei}}, \bibinfo
  {author} {\bibfnamefont {Y.}~\bibnamefont {Deng}},\ and\ \bibinfo {author}
  {\bibfnamefont {H.}~\bibnamefont {Duan}},\ }\bibfield  {title} {\enquote
  {\bibinfo {title} {Topology optimization-based inverse design of plasmonic
  nanodimer with maximum near-field enhancement},}\ }\href
  {https://doi.org/10.1002/adfm.202000642} {\bibfield  {journal} {\bibinfo
  {journal} {Adv. Funct. Mater.}\ }\textbf {\bibinfo {volume} {30}},\ \bibinfo
  {pages} {2000642} (\bibinfo {year} {2020}{\natexlab{a}})}\BibitemShut
  {NoStop}%
\bibitem [{\citenamefont {Gittinger}\ \emph {et~al.}(2020)\citenamefont
  {Gittinger}, \citenamefont {H{\"{o}}flich}, \citenamefont {Smirnov},
  \citenamefont {Kollmann}, \citenamefont {Lienau},\ and\ \citenamefont
  {Silies}}]{Gittinger_2020}%
  \BibitemOpen
  \bibfield  {author} {\bibinfo {author} {\bibfnamefont {M.}~\bibnamefont
  {Gittinger}}, \bibinfo {author} {\bibfnamefont {K.}~\bibnamefont
  {H{\"{o}}flich}}, \bibinfo {author} {\bibfnamefont {V.}~\bibnamefont
  {Smirnov}}, \bibinfo {author} {\bibfnamefont {H.}~\bibnamefont {Kollmann}},
  \bibinfo {author} {\bibfnamefont {C.}~\bibnamefont {Lienau}},\ and\ \bibinfo
  {author} {\bibfnamefont {M.}~\bibnamefont {Silies}},\ }\bibfield  {title}
  {\enquote {\bibinfo {title} {Strongly coupled, high-quality plasmonic dimer
  antennas fabricated using a sketch-and-peel technique},}\ }\href
  {https://doi.org/10.1515/nanoph-2019-0379} {\bibfield  {journal} {\bibinfo
  {journal} {Nanophotonics}\ }\textbf {\bibinfo {volume} {9}},\ \bibinfo
  {pages} {401--412} (\bibinfo {year} {2020})}\BibitemShut {NoStop}%
\bibitem [{\citenamefont {Smirnov}\ \emph {et~al.}(2021)\citenamefont
  {Smirnov}, \citenamefont {Stephan}, \citenamefont {Westphal}, \citenamefont
  {Emmrich}, \citenamefont {Beyer}, \citenamefont {G{\"{o}}lzh{\"{a}}user},
  \citenamefont {Lienau},\ and\ \citenamefont {Silies}}]{Smirnov2021}%
  \BibitemOpen
  \bibfield  {author} {\bibinfo {author} {\bibfnamefont {V.}~\bibnamefont
  {Smirnov}}, \bibinfo {author} {\bibfnamefont {S.}~\bibnamefont {Stephan}},
  \bibinfo {author} {\bibfnamefont {M.}~\bibnamefont {Westphal}}, \bibinfo
  {author} {\bibfnamefont {D.}~\bibnamefont {Emmrich}}, \bibinfo {author}
  {\bibfnamefont {A.}~\bibnamefont {Beyer}}, \bibinfo {author} {\bibfnamefont
  {A.}~\bibnamefont {G{\"{o}}lzh{\"{a}}user}}, \bibinfo {author} {\bibfnamefont
  {C.}~\bibnamefont {Lienau}},\ and\ \bibinfo {author} {\bibfnamefont
  {M.}~\bibnamefont {Silies}},\ }\bibfield  {title} {\enquote {\bibinfo {title}
  {Transmitting surface plasmon polaritons across nanometer-sized gaps by
  optical near-field coupling},}\ }\href
  {https://doi.org/10.1021/acsphotonics.0c01797} {\bibfield  {journal}
  {\bibinfo  {journal} {{ACS} Photonics}\ }\textbf {\bibinfo {volume} {8}},\
  \bibinfo {pages} {832--840} (\bibinfo {year} {2021})}\BibitemShut {NoStop}%
\bibitem [{\citenamefont {Melli}\ \emph {et~al.}(2013)\citenamefont {Melli},
  \citenamefont {Polyakov}, \citenamefont {Gargas}, \citenamefont {Huynh},
  \citenamefont {Scipioni}, \citenamefont {Bao}, \citenamefont {Ogletree},
  \citenamefont {Schuck}, \citenamefont {Cabrini},\ and\ \citenamefont
  {Weber-Bargioni}}]{Melli_2013}%
  \BibitemOpen
  \bibfield  {author} {\bibinfo {author} {\bibfnamefont {M.}~\bibnamefont
  {Melli}}, \bibinfo {author} {\bibfnamefont {A.}~\bibnamefont {Polyakov}},
  \bibinfo {author} {\bibfnamefont {D.}~\bibnamefont {Gargas}}, \bibinfo
  {author} {\bibfnamefont {C.}~\bibnamefont {Huynh}}, \bibinfo {author}
  {\bibfnamefont {L.}~\bibnamefont {Scipioni}}, \bibinfo {author}
  {\bibfnamefont {W.}~\bibnamefont {Bao}}, \bibinfo {author} {\bibfnamefont
  {D.~F.}\ \bibnamefont {Ogletree}}, \bibinfo {author} {\bibfnamefont {P.~J.}\
  \bibnamefont {Schuck}}, \bibinfo {author} {\bibfnamefont {S.}~\bibnamefont
  {Cabrini}},\ and\ \bibinfo {author} {\bibfnamefont {A.}~\bibnamefont
  {Weber-Bargioni}},\ }\bibfield  {title} {\enquote {\bibinfo {title} {Reaching
  the theoretical resonance quality factor limit in coaxial plasmonic
  nanoresonators fabricated by helium ion lithography},}\ }\href
  {https://doi.org/10.1021/nl400844a} {\bibfield  {journal} {\bibinfo
  {journal} {Nano Lett.}\ }\textbf {\bibinfo {volume} {13}},\ \bibinfo {pages}
  {2687--2691} (\bibinfo {year} {2013})}\BibitemShut {NoStop}%
\bibitem [{\citenamefont {Kuznetsov}\ \emph {et~al.}(2014)\citenamefont
  {Kuznetsov}, \citenamefont {Miroshnichenko}, \citenamefont {Fu},
  \citenamefont {Viswanathan}, \citenamefont {Rahmani}, \citenamefont
  {Valuckas}, \citenamefont {Pan}, \citenamefont {Kivshar}, \citenamefont
  {Pickard},\ and\ \citenamefont {Luk'yanchuk}}]{Kuznetsov2014}%
  \BibitemOpen
  \bibfield  {author} {\bibinfo {author} {\bibfnamefont {A.~I.}\ \bibnamefont
  {Kuznetsov}}, \bibinfo {author} {\bibfnamefont {A.~E.}\ \bibnamefont
  {Miroshnichenko}}, \bibinfo {author} {\bibfnamefont {Y.~H.}\ \bibnamefont
  {Fu}}, \bibinfo {author} {\bibfnamefont {V.}~\bibnamefont {Viswanathan}},
  \bibinfo {author} {\bibfnamefont {M.}~\bibnamefont {Rahmani}}, \bibinfo
  {author} {\bibfnamefont {V.}~\bibnamefont {Valuckas}}, \bibinfo {author}
  {\bibfnamefont {Z.~Y.}\ \bibnamefont {Pan}}, \bibinfo {author} {\bibfnamefont
  {Y.}~\bibnamefont {Kivshar}}, \bibinfo {author} {\bibfnamefont {D.~S.}\
  \bibnamefont {Pickard}},\ and\ \bibinfo {author} {\bibfnamefont
  {B.}~\bibnamefont {Luk'yanchuk}},\ }\bibfield  {title} {\enquote {\bibinfo
  {title} {Split-ball resonator as a three-dimensional analogue of planar
  split-rings},}\ }\href {https://doi.org/10.1038/ncomms4104} {\bibfield
  {journal} {\bibinfo  {journal} {Nat. Commun.}\ }\textbf {\bibinfo {volume}
  {5}} (\bibinfo {year} {2014}),\ 10.1038/ncomms4104}\BibitemShut {NoStop}%
\bibitem [{\citenamefont {Chen}\ \emph
  {et~al.}(2018{\natexlab{a}})\citenamefont {Chen}, \citenamefont {Razinskas},
  \citenamefont {Vieker}, \citenamefont {Gross}, \citenamefont {Wu},
  \citenamefont {Beyer}, \citenamefont {G{\"{o}}lzh{\"{a}}user},\ and\
  \citenamefont {Hecht}}]{Chen2018}%
  \BibitemOpen
  \bibfield  {author} {\bibinfo {author} {\bibfnamefont {K.}~\bibnamefont
  {Chen}}, \bibinfo {author} {\bibfnamefont {G.}~\bibnamefont {Razinskas}},
  \bibinfo {author} {\bibfnamefont {H.}~\bibnamefont {Vieker}}, \bibinfo
  {author} {\bibfnamefont {H.}~\bibnamefont {Gross}}, \bibinfo {author}
  {\bibfnamefont {X.}~\bibnamefont {Wu}}, \bibinfo {author} {\bibfnamefont
  {A.}~\bibnamefont {Beyer}}, \bibinfo {author} {\bibfnamefont
  {A.}~\bibnamefont {G{\"{o}}lzh{\"{a}}user}},\ and\ \bibinfo {author}
  {\bibfnamefont {B.}~\bibnamefont {Hecht}},\ }\bibfield  {title} {\enquote
  {\bibinfo {title} {High-{Q}, low-mode-volume and multiresonant plasmonic
  nanoslit cavities fabricated by helium ion milling},}\ }\href
  {https://doi.org/10.1039/c8nr02160k} {\bibfield  {journal} {\bibinfo
  {journal} {Nanoscale}\ }\textbf {\bibinfo {volume} {10}},\ \bibinfo {pages}
  {17148--17155} (\bibinfo {year} {2018}{\natexlab{a}})}\BibitemShut {NoStop}%
\bibitem [{\citenamefont {Grant-Jacob}\ \emph {et~al.}(2016)\citenamefont
  {Grant-Jacob}, \citenamefont {Oo}, \citenamefont {Carpignano}, \citenamefont
  {Boden}, \citenamefont {Brocklesby}, \citenamefont {Charlton},\ and\
  \citenamefont {Melvin}}]{Grant-Jacob_2016}%
  \BibitemOpen
  \bibfield  {author} {\bibinfo {author} {\bibfnamefont {J.~A.}\ \bibnamefont
  {Grant-Jacob}}, \bibinfo {author} {\bibfnamefont {S.~Z.}\ \bibnamefont {Oo}},
  \bibinfo {author} {\bibfnamefont {F.}~\bibnamefont {Carpignano}}, \bibinfo
  {author} {\bibfnamefont {S.~A.}\ \bibnamefont {Boden}}, \bibinfo {author}
  {\bibfnamefont {W.~S.}\ \bibnamefont {Brocklesby}}, \bibinfo {author}
  {\bibfnamefont {M.~D.~B.}\ \bibnamefont {Charlton}},\ and\ \bibinfo {author}
  {\bibfnamefont {T.}~\bibnamefont {Melvin}},\ }\bibfield  {title}
  {{\selectlanguage {english}\enquote {\bibinfo {title} {Design and fabrication
  of a 3{D}-structured gold film with nanopores for local electric field
  enhancement in the pore},}\ }}\href
  {https://doi.org/10.1088/0957-4484/27/6/065302} {\bibfield  {journal}
  {\bibinfo  {journal} {Nanotechnology}\ }\textbf {\bibinfo {volume} {27}},\
  \bibinfo {pages} {065302} (\bibinfo {year} {2016})}\BibitemShut {NoStop}%
\bibitem [{\citenamefont {Oo}\ \emph {et~al.}(2016)\citenamefont {Oo},
  \citenamefont {Silva}, \citenamefont {Carpignano}, \citenamefont {Noual},
  \citenamefont {Pechstedt}, \citenamefont {Mateos}, \citenamefont
  {Grant-Jacob}, \citenamefont {Brocklesby}, \citenamefont {Horak},
  \citenamefont {Charlton}, \citenamefont {Boden},\ and\ \citenamefont
  {Melvin}}]{Oo2016}%
  \BibitemOpen
  \bibfield  {author} {\bibinfo {author} {\bibfnamefont {S.~Z.}\ \bibnamefont
  {Oo}}, \bibinfo {author} {\bibfnamefont {G.}~\bibnamefont {Silva}}, \bibinfo
  {author} {\bibfnamefont {F.}~\bibnamefont {Carpignano}}, \bibinfo {author}
  {\bibfnamefont {A.}~\bibnamefont {Noual}}, \bibinfo {author} {\bibfnamefont
  {K.}~\bibnamefont {Pechstedt}}, \bibinfo {author} {\bibfnamefont
  {L.}~\bibnamefont {Mateos}}, \bibinfo {author} {\bibfnamefont {J.~A.}\
  \bibnamefont {Grant-Jacob}}, \bibinfo {author} {\bibfnamefont
  {B.}~\bibnamefont {Brocklesby}}, \bibinfo {author} {\bibfnamefont
  {P.}~\bibnamefont {Horak}}, \bibinfo {author} {\bibfnamefont
  {M.}~\bibnamefont {Charlton}}, \bibinfo {author} {\bibfnamefont {S.~A.}\
  \bibnamefont {Boden}},\ and\ \bibinfo {author} {\bibfnamefont
  {T.}~\bibnamefont {Melvin}},\ }\bibfield  {title} {\enquote {\bibinfo {title}
  {A nanoporous gold membrane for sensing applications},}\ }\href
  {https://doi.org/10.1016/j.sbsr.2016.01.001} {\bibfield  {journal} {\bibinfo
  {journal} {Sens. Bio-Sens. Res.}\ }\textbf {\bibinfo {volume} {7}},\ \bibinfo
  {pages} {133--140} (\bibinfo {year} {2016})}\BibitemShut {NoStop}%
\bibitem [{\citenamefont {Singh}\ \emph {et~al.}(2018)\citenamefont {Singh},
  \citenamefont {Panchenko}, \citenamefont {Nasr}, \citenamefont {Liu},
  \citenamefont {Wesemann}, \citenamefont {Davis},\ and\ \citenamefont
  {Roberts}}]{Singh2018}%
  \BibitemOpen
  \bibfield  {author} {\bibinfo {author} {\bibfnamefont {K.}~\bibnamefont
  {Singh}}, \bibinfo {author} {\bibfnamefont {E.}~\bibnamefont {Panchenko}},
  \bibinfo {author} {\bibfnamefont {B.}~\bibnamefont {Nasr}}, \bibinfo {author}
  {\bibfnamefont {A.}~\bibnamefont {Liu}}, \bibinfo {author} {\bibfnamefont
  {L.}~\bibnamefont {Wesemann}}, \bibinfo {author} {\bibfnamefont {T.~J.}\
  \bibnamefont {Davis}},\ and\ \bibinfo {author} {\bibfnamefont
  {A.}~\bibnamefont {Roberts}},\ }\bibfield  {title} {\enquote {\bibinfo
  {title} {Cathodoluminescence as a probe of the optical properties of resonant
  apertures in a metallic film},}\ }\href
  {https://doi.org/10.3762/bjnano.9.140} {\bibfield  {journal} {\bibinfo
  {journal} {Beilstein J. Nanotechnol.}\ }\textbf {\bibinfo {volume} {9}},\
  \bibinfo {pages} {1491--1500} (\bibinfo {year} {2018})}\BibitemShut {NoStop}%
\bibitem [{\citenamefont {Hahn}, \citenamefont {Hajebifard},\ and\
  \citenamefont {Berini}(2020)}]{Hahn_2020}%
  \BibitemOpen
  \bibfield  {author} {\bibinfo {author} {\bibfnamefont {C.}~\bibnamefont
  {Hahn}}, \bibinfo {author} {\bibfnamefont {A.}~\bibnamefont {Hajebifard}},\
  and\ \bibinfo {author} {\bibfnamefont {P.}~\bibnamefont {Berini}},\
  }\bibfield  {title} {\enquote {\bibinfo {title} {Helium focused ion beam
  direct milling of plasmonic heptamer-arranged nanohole arrays},}\ }\href
  {https://doi.org/10.1515/nanoph-2019-0385} {\bibfield  {journal} {\bibinfo
  {journal} {Nanophotonics}\ }\textbf {\bibinfo {volume} {9}},\ \bibinfo
  {pages} {393--399} (\bibinfo {year} {2020})}\BibitemShut {NoStop}%
\bibitem [{\citenamefont {Hajebifard}\ \emph {et~al.}(2021)\citenamefont
  {Hajebifard}, \citenamefont {Hahn}, \citenamefont {Walia}, \citenamefont
  {Suleman}, \citenamefont {Variola}, \citenamefont {Weck},\ and\ \citenamefont
  {Berini}}]{Hajebifard_2021}%
  \BibitemOpen
  \bibfield  {author} {\bibinfo {author} {\bibfnamefont {A.}~\bibnamefont
  {Hajebifard}}, \bibinfo {author} {\bibfnamefont {C.}~\bibnamefont {Hahn}},
  \bibinfo {author} {\bibfnamefont {J.}~\bibnamefont {Walia}}, \bibinfo
  {author} {\bibfnamefont {H.}~\bibnamefont {Suleman}}, \bibinfo {author}
  {\bibfnamefont {F.}~\bibnamefont {Variola}}, \bibinfo {author} {\bibfnamefont
  {A.}~\bibnamefont {Weck}},\ and\ \bibinfo {author} {\bibfnamefont
  {P.}~\bibnamefont {Berini}},\ }\bibfield  {title} {\enquote {\bibinfo {title}
  {Fano resonances in nanohole oligomers in a gold film},}\ }\href
  {https://doi.org/10.1063/5.0038448} {\bibfield  {journal} {\bibinfo
  {journal} {J. Appl. Phys.}\ }\textbf {\bibinfo {volume} {129}},\ \bibinfo
  {pages} {033103} (\bibinfo {year} {2021})}\BibitemShut {NoStop}%
\bibitem [{\citenamefont {Kriesch}\ \emph {et~al.}(2013)\citenamefont
  {Kriesch}, \citenamefont {Burgos}, \citenamefont {Ploss}, \citenamefont
  {Pfeifer}, \citenamefont {Atwater},\ and\ \citenamefont
  {Peschel}}]{Kriesch_2013}%
  \BibitemOpen
  \bibfield  {author} {\bibinfo {author} {\bibfnamefont {A.}~\bibnamefont
  {Kriesch}}, \bibinfo {author} {\bibfnamefont {S.~P.}\ \bibnamefont {Burgos}},
  \bibinfo {author} {\bibfnamefont {D.}~\bibnamefont {Ploss}}, \bibinfo
  {author} {\bibfnamefont {H.}~\bibnamefont {Pfeifer}}, \bibinfo {author}
  {\bibfnamefont {H.~A.}\ \bibnamefont {Atwater}},\ and\ \bibinfo {author}
  {\bibfnamefont {U.}~\bibnamefont {Peschel}},\ }\bibfield  {title} {\enquote
  {\bibinfo {title} {Functional plasmonic nanocircuits with low insertion and
  propagation losses},}\ }\href {https://doi.org/10.1021/nl402580c} {\bibfield
  {journal} {\bibinfo  {journal} {Nano Lett.}\ }\textbf {\bibinfo {volume}
  {13}},\ \bibinfo {pages} {4539--4545} (\bibinfo {year} {2013})}\BibitemShut
  {NoStop}%
\bibitem [{\citenamefont {Fu}\ \emph {et~al.}(2012)\citenamefont {Fu},
  \citenamefont {Hu}, \citenamefont {Lu}, \citenamefont {Yue}, \citenamefont
  {Yang},\ and\ \citenamefont {Gong}}]{Fu_2012}%
  \BibitemOpen
  \bibfield  {author} {\bibinfo {author} {\bibfnamefont {Y.}~\bibnamefont
  {Fu}}, \bibinfo {author} {\bibfnamefont {X.}~\bibnamefont {Hu}}, \bibinfo
  {author} {\bibfnamefont {C.}~\bibnamefont {Lu}}, \bibinfo {author}
  {\bibfnamefont {S.}~\bibnamefont {Yue}}, \bibinfo {author} {\bibfnamefont
  {H.}~\bibnamefont {Yang}},\ and\ \bibinfo {author} {\bibfnamefont
  {Q.}~\bibnamefont {Gong}},\ }\bibfield  {title} {\enquote {\bibinfo {title}
  {All-optical logic gates based on nanoscale plasmonic slot waveguides},}\
  }\href {https://doi.org/10.1021/nl303095s} {\bibfield  {journal} {\bibinfo
  {journal} {Nano Lett.}\ }\textbf {\bibinfo {volume} {12}},\ \bibinfo {pages}
  {5784--5790} (\bibinfo {year} {2012})}\BibitemShut {NoStop}%
\bibitem [{\citenamefont {Sivis}\ \emph {et~al.}(2013)\citenamefont {Sivis},
  \citenamefont {Duwe}, \citenamefont {Abel},\ and\ \citenamefont
  {Ropers}}]{Sivis_2013}%
  \BibitemOpen
  \bibfield  {author} {\bibinfo {author} {\bibfnamefont {M.}~\bibnamefont
  {Sivis}}, \bibinfo {author} {\bibfnamefont {M.}~\bibnamefont {Duwe}},
  \bibinfo {author} {\bibfnamefont {B.}~\bibnamefont {Abel}},\ and\ \bibinfo
  {author} {\bibfnamefont {C.}~\bibnamefont {Ropers}},\ }\bibfield  {title}
  {\enquote {\bibinfo {title} {Extreme-ultraviolet light generation in
  plasmonic nanostructures},}\ }\href {https://doi.org/10.1038/nphys2590}
  {\bibfield  {journal} {\bibinfo  {journal} {Nat. Phys.}\ }\textbf {\bibinfo
  {volume} {9}},\ \bibinfo {pages} {304--309} (\bibinfo {year}
  {2013})}\BibitemShut {NoStop}%
\bibitem [{\citenamefont {Ding}\ \emph {et~al.}(2016)\citenamefont {Ding},
  \citenamefont {Qin}, \citenamefont {Guo}, \citenamefont {Liu}, \citenamefont
  {Kinzel},\ and\ \citenamefont {Wang}}]{Ding_2016}%
  \BibitemOpen
  \bibfield  {author} {\bibinfo {author} {\bibfnamefont {L.}~\bibnamefont
  {Ding}}, \bibinfo {author} {\bibfnamefont {J.}~\bibnamefont {Qin}}, \bibinfo
  {author} {\bibfnamefont {S.}~\bibnamefont {Guo}}, \bibinfo {author}
  {\bibfnamefont {T.}~\bibnamefont {Liu}}, \bibinfo {author} {\bibfnamefont
  {E.}~\bibnamefont {Kinzel}},\ and\ \bibinfo {author} {\bibfnamefont
  {L.}~\bibnamefont {Wang}},\ }\bibfield  {title} {\enquote {\bibinfo {title}
  {Resonant effects in nanoscale bowtie apertures},}\ }\href
  {https://doi.org/10.1038/srep27254} {\bibfield  {journal} {\bibinfo
  {journal} {Sci. Rep.}\ }\textbf {\bibinfo {volume} {6}} (\bibinfo {year}
  {2016}),\ 10.1038/srep27254}\BibitemShut {NoStop}%
\bibitem [{\citenamefont {Lv}\ \emph {et~al.}(2015)\citenamefont {Lv},
  \citenamefont {Yan}, \citenamefont {Zhang}, \citenamefont {Liu},
  \citenamefont {Jiang},\ and\ \citenamefont {Si}}]{Lv_2015}%
  \BibitemOpen
  \bibfield  {author} {\bibinfo {author} {\bibfnamefont {J.-T.}\ \bibnamefont
  {Lv}}, \bibinfo {author} {\bibfnamefont {Y.}~\bibnamefont {Yan}}, \bibinfo
  {author} {\bibfnamefont {W.-K.}\ \bibnamefont {Zhang}}, \bibinfo {author}
  {\bibfnamefont {Y.-H.}\ \bibnamefont {Liu}}, \bibinfo {author} {\bibfnamefont
  {Z.-Y.}\ \bibnamefont {Jiang}},\ and\ \bibinfo {author} {\bibfnamefont
  {G.-Y.}\ \bibnamefont {Si}},\ }\bibfield  {title} {\enquote {\bibinfo {title}
  {Plasmonic nanoantennae fabricated by focused ion beam milling},}\ }\href
  {https://doi.org/10.1007/s12541-015-0112-3} {\bibfield  {journal} {\bibinfo
  {journal} {Int. J. Precis. Eng. Man.}\ }\textbf {\bibinfo {volume} {16}},\
  \bibinfo {pages} {851--855} (\bibinfo {year} {2015})}\BibitemShut {NoStop}%
\bibitem [{\citenamefont {Kullock}\ \emph {et~al.}(2020)\citenamefont
  {Kullock}, \citenamefont {Ochs}, \citenamefont {Grimm}, \citenamefont
  {Emmerling},\ and\ \citenamefont {Hecht}}]{Kullock_2020}%
  \BibitemOpen
  \bibfield  {author} {\bibinfo {author} {\bibfnamefont {R.}~\bibnamefont
  {Kullock}}, \bibinfo {author} {\bibfnamefont {M.}~\bibnamefont {Ochs}},
  \bibinfo {author} {\bibfnamefont {P.}~\bibnamefont {Grimm}}, \bibinfo
  {author} {\bibfnamefont {M.}~\bibnamefont {Emmerling}},\ and\ \bibinfo
  {author} {\bibfnamefont {B.}~\bibnamefont {Hecht}},\ }\bibfield  {title}
  {\enquote {\bibinfo {title} {Electrically-driven yagi-uda antennas for
  light},}\ }\href {https://doi.org/10.1038/s41467-019-14011-6} {\bibfield
  {journal} {\bibinfo  {journal} {Nat. Commun.}\ }\textbf {\bibinfo {volume}
  {11}} (\bibinfo {year} {2020}),\ 10.1038/s41467-019-14011-6}\BibitemShut
  {NoStop}%
\bibitem [{\citenamefont {M\"uhlschlegel}\ \emph {et~al.}(2005)\citenamefont
  {M\"uhlschlegel}, \citenamefont {Eisler}, \citenamefont {Martin},
  \citenamefont {Hecht},\ and\ \citenamefont {Pohl}}]{Muehlschlegel_2005}%
  \BibitemOpen
  \bibfield  {author} {\bibinfo {author} {\bibfnamefont {P.}~\bibnamefont
  {M\"uhlschlegel}}, \bibinfo {author} {\bibfnamefont {H.-J.}\ \bibnamefont
  {Eisler}}, \bibinfo {author} {\bibfnamefont {O.~J.~F.}\ \bibnamefont
  {Martin}}, \bibinfo {author} {\bibfnamefont {B.}~\bibnamefont {Hecht}},\ and\
  \bibinfo {author} {\bibfnamefont {D.~W.}\ \bibnamefont {Pohl}},\ }\bibfield
  {title} {\enquote {\bibinfo {title} {Resonant optical antennas},}\ }\href
  {https://doi.org/10.1126/science.1111886} {\bibfield  {journal} {\bibinfo
  {journal} {Science}\ }\textbf {\bibinfo {volume} {308}},\ \bibinfo {pages}
  {1607--1609} (\bibinfo {year} {2005})}\BibitemShut {NoStop}%
\bibitem [{\citenamefont {Cherqui}\ \emph {et~al.}(2016)\citenamefont
  {Cherqui}, \citenamefont {Wu}, \citenamefont {Li}, \citenamefont {Quillin},
  \citenamefont {Busche}, \citenamefont {Thakkar}, \citenamefont {West},
  \citenamefont {Montoni}, \citenamefont {Rack}, \citenamefont {Camden},\ and\
  \citenamefont {Masiello}}]{Cherqui_2016}%
  \BibitemOpen
  \bibfield  {author} {\bibinfo {author} {\bibfnamefont {C.}~\bibnamefont
  {Cherqui}}, \bibinfo {author} {\bibfnamefont {Y.}~\bibnamefont {Wu}},
  \bibinfo {author} {\bibfnamefont {G.}~\bibnamefont {Li}}, \bibinfo {author}
  {\bibfnamefont {S.~C.}\ \bibnamefont {Quillin}}, \bibinfo {author}
  {\bibfnamefont {J.~A.}\ \bibnamefont {Busche}}, \bibinfo {author}
  {\bibfnamefont {N.}~\bibnamefont {Thakkar}}, \bibinfo {author} {\bibfnamefont
  {C.~A.}\ \bibnamefont {West}}, \bibinfo {author} {\bibfnamefont {N.~P.}\
  \bibnamefont {Montoni}}, \bibinfo {author} {\bibfnamefont {P.~D.}\
  \bibnamefont {Rack}}, \bibinfo {author} {\bibfnamefont {J.~P.}\ \bibnamefont
  {Camden}},\ and\ \bibinfo {author} {\bibfnamefont {D.~J.}\ \bibnamefont
  {Masiello}},\ }\bibfield  {title} {\enquote {\bibinfo {title} {{STEM}/{EELS}
  imaging of magnetic hybridization in symmetric and symmetry-broken plasmon
  oligomer dimers and all-magnetic fano interference},}\ }\href
  {https://doi.org/10.1021/acs.nanolett.6b03504} {\bibfield  {journal}
  {\bibinfo  {journal} {Nano Lett.}\ }\textbf {\bibinfo {volume} {16}},\
  \bibinfo {pages} {6668--6676} (\bibinfo {year} {2016})}\BibitemShut {NoStop}%
\bibitem [{\citenamefont {Rewitz}\ \emph {et~al.}(2014)\citenamefont {Rewitz},
  \citenamefont {Razinskas}, \citenamefont {Geisler}, \citenamefont {Krauss},
  \citenamefont {Goetz}, \citenamefont {Paw{\l}owska}, \citenamefont {Hecht},\
  and\ \citenamefont {Brixner}}]{Rewitz_2014}%
  \BibitemOpen
  \bibfield  {author} {\bibinfo {author} {\bibfnamefont {C.}~\bibnamefont
  {Rewitz}}, \bibinfo {author} {\bibfnamefont {G.}~\bibnamefont {Razinskas}},
  \bibinfo {author} {\bibfnamefont {P.}~\bibnamefont {Geisler}}, \bibinfo
  {author} {\bibfnamefont {E.}~\bibnamefont {Krauss}}, \bibinfo {author}
  {\bibfnamefont {S.}~\bibnamefont {Goetz}}, \bibinfo {author} {\bibfnamefont
  {M.}~\bibnamefont {Paw{\l}owska}}, \bibinfo {author} {\bibfnamefont
  {B.}~\bibnamefont {Hecht}},\ and\ \bibinfo {author} {\bibfnamefont
  {T.}~\bibnamefont {Brixner}},\ }\bibfield  {title} {\enquote {\bibinfo
  {title} {Coherent control of plasmon propagation in a nanocircuit},}\ }\href
  {https://doi.org/10.1103/physrevapplied.1.014007} {\bibfield  {journal}
  {\bibinfo  {journal} {Phys. Rev. Applied}\ }\textbf {\bibinfo {volume} {1}},\
  \bibinfo {pages} {014007} (\bibinfo {year} {2014})}\BibitemShut {NoStop}%
\bibitem [{\citenamefont {Geisler}\ \emph {et~al.}(2013)\citenamefont
  {Geisler}, \citenamefont {Razinskas}, \citenamefont {Krauss}, \citenamefont
  {Wu}, \citenamefont {Rewitz}, \citenamefont {Tuchscherer}, \citenamefont
  {Goetz}, \citenamefont {Huang}, \citenamefont {Brixner},\ and\ \citenamefont
  {Hecht}}]{Geisler_2013}%
  \BibitemOpen
  \bibfield  {author} {\bibinfo {author} {\bibfnamefont {P.}~\bibnamefont
  {Geisler}}, \bibinfo {author} {\bibfnamefont {G.}~\bibnamefont {Razinskas}},
  \bibinfo {author} {\bibfnamefont {E.}~\bibnamefont {Krauss}}, \bibinfo
  {author} {\bibfnamefont {X.-F.}\ \bibnamefont {Wu}}, \bibinfo {author}
  {\bibfnamefont {C.}~\bibnamefont {Rewitz}}, \bibinfo {author} {\bibfnamefont
  {P.}~\bibnamefont {Tuchscherer}}, \bibinfo {author} {\bibfnamefont
  {S.}~\bibnamefont {Goetz}}, \bibinfo {author} {\bibfnamefont {C.-B.}\
  \bibnamefont {Huang}}, \bibinfo {author} {\bibfnamefont {T.}~\bibnamefont
  {Brixner}},\ and\ \bibinfo {author} {\bibfnamefont {B.}~\bibnamefont
  {Hecht}},\ }\bibfield  {title} {\enquote {\bibinfo {title} {Multimode plasmon
  excitation and \textit{In{~}Situ} analysis in top-down fabricated
  nanocircuits},}\ }\href {https://doi.org/10.1103/physrevlett.111.183901}
  {\bibfield  {journal} {\bibinfo  {journal} {Phys. Rev. Lett.}\ }\textbf
  {\bibinfo {volume} {111}},\ \bibinfo {pages} {183901} (\bibinfo {year}
  {2013})}\BibitemShut {NoStop}%
\bibitem [{\citenamefont {Krauss}\ \emph {et~al.}(2019)\citenamefont {Krauss},
  \citenamefont {Razinskas}, \citenamefont {K{\"{o}}ck}, \citenamefont
  {Grossmann},\ and\ \citenamefont {Hecht}}]{Krauss_2019}%
  \BibitemOpen
  \bibfield  {author} {\bibinfo {author} {\bibfnamefont {E.}~\bibnamefont
  {Krauss}}, \bibinfo {author} {\bibfnamefont {G.}~\bibnamefont {Razinskas}},
  \bibinfo {author} {\bibfnamefont {D.}~\bibnamefont {K{\"{o}}ck}}, \bibinfo
  {author} {\bibfnamefont {S.}~\bibnamefont {Grossmann}},\ and\ \bibinfo
  {author} {\bibfnamefont {B.}~\bibnamefont {Hecht}},\ }\bibfield  {title}
  {\enquote {\bibinfo {title} {Reversible mapping and sorting the spin of
  photons on the nanoscale: a spin-optical nanodevice},}\ }\href
  {https://doi.org/10.1021/acs.nanolett.9b01162} {\bibfield  {journal}
  {\bibinfo  {journal} {Nano Lett.}\ }\textbf {\bibinfo {volume} {19}},\
  \bibinfo {pages} {3364--3369} (\bibinfo {year} {2019})}\BibitemShut {NoStop}%
\bibitem [{\citenamefont {Archanjo}\ \emph {et~al.}(2018)\citenamefont
  {Archanjo}, \citenamefont {Vasconcelos}, \citenamefont {Oliveira},
  \citenamefont {Song}, \citenamefont {Allen}, \citenamefont {Achete},\ and\
  \citenamefont {Ercius}}]{Archanjo_2018}%
  \BibitemOpen
  \bibfield  {author} {\bibinfo {author} {\bibfnamefont {B.~S.}\ \bibnamefont
  {Archanjo}}, \bibinfo {author} {\bibfnamefont {T.~L.}\ \bibnamefont
  {Vasconcelos}}, \bibinfo {author} {\bibfnamefont {B.~S.}\ \bibnamefont
  {Oliveira}}, \bibinfo {author} {\bibfnamefont {S.}~\bibnamefont {Song}},
  \bibinfo {author} {\bibfnamefont {F.~I.}\ \bibnamefont {Allen}}, \bibinfo
  {author} {\bibfnamefont {C.~A.}\ \bibnamefont {Achete}},\ and\ \bibinfo
  {author} {\bibfnamefont {P.}~\bibnamefont {Ercius}},\ }\bibfield  {title}
  {\enquote {\bibinfo {title} {Plasmon {3D} electron tomography and local
  {Electric-Field} enhancement of engineered plasmonic nanoantennas},}\ }\href
  {https://doi.org/10.1021/acsphotonics.8b00125} {\bibfield  {journal}
  {\bibinfo  {journal} {ACS Photonics}\ }\textbf {\bibinfo {volume} {7}},\
  \bibinfo {pages} {2834--2842} (\bibinfo {year} {2018})}\BibitemShut {NoStop}%
\bibitem [{\citenamefont {Veerman}\ \emph {et~al.}(1998)\citenamefont
  {Veerman}, \citenamefont {Otter}, \citenamefont {Kuipers},\ and\
  \citenamefont {van Hulst}}]{Veerman_1998}%
  \BibitemOpen
  \bibfield  {author} {\bibinfo {author} {\bibfnamefont {J.~A.}\ \bibnamefont
  {Veerman}}, \bibinfo {author} {\bibfnamefont {A.~M.}\ \bibnamefont {Otter}},
  \bibinfo {author} {\bibfnamefont {L.}~\bibnamefont {Kuipers}},\ and\ \bibinfo
  {author} {\bibfnamefont {N.~F.}\ \bibnamefont {van Hulst}},\ }\bibfield
  {title} {\enquote {\bibinfo {title} {High definition aperture probes for
  near-field optical microscopy fabricated by focused ion beam milling},}\
  }\href {https://doi.org/10.1063/1.121564} {\bibfield  {journal} {\bibinfo
  {journal} {Appl. Phys. Lett.}\ }\textbf {\bibinfo {volume} {72}},\ \bibinfo
  {pages} {3115--3117} (\bibinfo {year} {1998})}\BibitemShut {NoStop}%
\bibitem [{\citenamefont {Mivelle}\ \emph {et~al.}(2012)\citenamefont
  {Mivelle}, \citenamefont {van Zanten}, \citenamefont {Neumann}, \citenamefont
  {van Hulst},\ and\ \citenamefont {Garcia-Parajo}}]{Mivelle_2012}%
  \BibitemOpen
  \bibfield  {author} {\bibinfo {author} {\bibfnamefont {M.}~\bibnamefont
  {Mivelle}}, \bibinfo {author} {\bibfnamefont {T.~S.}\ \bibnamefont {van
  Zanten}}, \bibinfo {author} {\bibfnamefont {L.}~\bibnamefont {Neumann}},
  \bibinfo {author} {\bibfnamefont {N.~F.}\ \bibnamefont {van Hulst}},\ and\
  \bibinfo {author} {\bibfnamefont {M.~F.}\ \bibnamefont {Garcia-Parajo}},\
  }\bibfield  {title} {{\selectlanguage {english}\enquote {\bibinfo {title}
  {Ultrabright bowtie nanoaperture antenna probes studied by single molecule
  fluorescence},}\ }}\href {https://doi.org/10.1021/nl303440w} {\bibfield
  {journal} {\bibinfo  {journal} {Nano Lett.}\ }\textbf {\bibinfo {volume}
  {12}},\ \bibinfo {pages} {5972--5978} (\bibinfo {year} {2012})}\BibitemShut
  {NoStop}%
\bibitem [{\citenamefont {Taminiau}\ \emph {et~al.}(2006)\citenamefont
  {Taminiau}, \citenamefont {Moerland}, \citenamefont {Segerink}, \citenamefont
  {Kuipers},\ and\ \citenamefont {van Hulst}}]{Taminiau_2006}%
  \BibitemOpen
  \bibfield  {author} {\bibinfo {author} {\bibfnamefont {T.~H.}\ \bibnamefont
  {Taminiau}}, \bibinfo {author} {\bibfnamefont {R.~J.}\ \bibnamefont
  {Moerland}}, \bibinfo {author} {\bibfnamefont {F.~B.}\ \bibnamefont
  {Segerink}}, \bibinfo {author} {\bibfnamefont {L.}~\bibnamefont {Kuipers}},\
  and\ \bibinfo {author} {\bibfnamefont {N.~F.}\ \bibnamefont {van Hulst}},\
  }\bibfield  {title} {\enquote {\bibinfo {title} {$\uplambda$/4 resonance of
  an optical monopole antenna probed by single molecule fluorescence},}\ }\href
  {https://doi.org/10.1021/nl061726h} {\bibfield  {journal} {\bibinfo
  {journal} {Nano Lett.}\ }\textbf {\bibinfo {volume} {7}},\ \bibinfo {pages}
  {28--33} (\bibinfo {year} {2006})}\BibitemShut {NoStop}%
\bibitem [{\citenamefont {van Zanten}, \citenamefont {Lopez-Bosque},\ and\
  \citenamefont {Garcia-Parajo}(2010)}]{Zanten_2010}%
  \BibitemOpen
  \bibfield  {author} {\bibinfo {author} {\bibfnamefont {T.~S.}\ \bibnamefont
  {van Zanten}}, \bibinfo {author} {\bibfnamefont {M.~J.}\ \bibnamefont
  {Lopez-Bosque}},\ and\ \bibinfo {author} {\bibfnamefont {M.~F.}\ \bibnamefont
  {Garcia-Parajo}},\ }\bibfield  {title} {\enquote {\bibinfo {title} {Imaging
  individual proteins and nanodomains on intact cell membranes with a
  probe-based optical antenna},}\ }\href
  {https://doi.org/10.1002/smll.200901204} {\bibfield  {journal} {\bibinfo
  {journal} {Small}\ }\textbf {\bibinfo {volume} {6}},\ \bibinfo {pages}
  {270--275} (\bibinfo {year} {2010})}\BibitemShut {NoStop}%
\bibitem [{\citenamefont {Farahani}\ \emph {et~al.}(2007)\citenamefont
  {Farahani}, \citenamefont {Eisler}, \citenamefont {Pohl}, \citenamefont
  {Pavius}, \citenamefont {Fl{\"{u}}ckiger}, \citenamefont {Gasser},\ and\
  \citenamefont {Hecht}}]{Farahani_2007}%
  \BibitemOpen
  \bibfield  {author} {\bibinfo {author} {\bibfnamefont {J.~N.}\ \bibnamefont
  {Farahani}}, \bibinfo {author} {\bibfnamefont {H.-J.}\ \bibnamefont
  {Eisler}}, \bibinfo {author} {\bibfnamefont {D.~W.}\ \bibnamefont {Pohl}},
  \bibinfo {author} {\bibfnamefont {M.}~\bibnamefont {Pavius}}, \bibinfo
  {author} {\bibfnamefont {P.}~\bibnamefont {Fl{\"{u}}ckiger}}, \bibinfo
  {author} {\bibfnamefont {P.}~\bibnamefont {Gasser}},\ and\ \bibinfo {author}
  {\bibfnamefont {B.}~\bibnamefont {Hecht}},\ }\bibfield  {title} {\enquote
  {\bibinfo {title} {Bow-tie optical antenna probes for single-emitter scanning
  near-field optical microscopy},}\ }\href
  {https://doi.org/10.1088/0957-4484/18/12/125506} {\bibfield  {journal}
  {\bibinfo  {journal} {Nanotechnology}\ }\textbf {\bibinfo {volume} {18}},\
  \bibinfo {pages} {125506} (\bibinfo {year} {2007})}\BibitemShut {NoStop}%
\bibitem [{\citenamefont {Hoffmann}\ \emph {et~al.}(2015)\citenamefont
  {Hoffmann}, \citenamefont {Vassant}, \citenamefont {Chen}, \citenamefont
  {G{\"o}tzinger}, \citenamefont {Sandoghdar},\ and\ \citenamefont
  {Christiansen}}]{Hoffmann_2015}%
  \BibitemOpen
  \bibfield  {author} {\bibinfo {author} {\bibfnamefont {B.}~\bibnamefont
  {Hoffmann}}, \bibinfo {author} {\bibfnamefont {S.}~\bibnamefont {Vassant}},
  \bibinfo {author} {\bibfnamefont {X.-W.}\ \bibnamefont {Chen}}, \bibinfo
  {author} {\bibfnamefont {S.}~\bibnamefont {G{\"o}tzinger}}, \bibinfo {author}
  {\bibfnamefont {V.}~\bibnamefont {Sandoghdar}},\ and\ \bibinfo {author}
  {\bibfnamefont {S.}~\bibnamefont {Christiansen}},\ }\bibfield  {title}
  {{\selectlanguage {english}\enquote {\bibinfo {title} {Fabrication and
  characterization of plasmonic nanocone antennas for strong spontaneous
  emission enhancement},}\ }}\href
  {https://doi.org/10.1088/0957-4484/26/40/404001} {\bibfield  {journal}
  {\bibinfo  {journal} {Nanotechnology}\ }\textbf {\bibinfo {volume} {26}},\
  \bibinfo {pages} {404001} (\bibinfo {year} {2015})}\BibitemShut {NoStop}%
\bibitem [{\citenamefont {Schmidt}\ \emph {et~al.}(2012)\citenamefont
  {Schmidt}, \citenamefont {Piglosiewicz}, \citenamefont {Sadiq}, \citenamefont
  {Shirdel}, \citenamefont {Lee}, \citenamefont {Vasa}, \citenamefont {Park},
  \citenamefont {Kim},\ and\ \citenamefont {Lienau}}]{Schmidt_2012}%
  \BibitemOpen
  \bibfield  {author} {\bibinfo {author} {\bibfnamefont {S.}~\bibnamefont
  {Schmidt}}, \bibinfo {author} {\bibfnamefont {B.}~\bibnamefont
  {Piglosiewicz}}, \bibinfo {author} {\bibfnamefont {D.}~\bibnamefont {Sadiq}},
  \bibinfo {author} {\bibfnamefont {J.}~\bibnamefont {Shirdel}}, \bibinfo
  {author} {\bibfnamefont {J.~S.}\ \bibnamefont {Lee}}, \bibinfo {author}
  {\bibfnamefont {P.}~\bibnamefont {Vasa}}, \bibinfo {author} {\bibfnamefont
  {N.}~\bibnamefont {Park}}, \bibinfo {author} {\bibfnamefont {D.-S.}\
  \bibnamefont {Kim}},\ and\ \bibinfo {author} {\bibfnamefont {C.}~\bibnamefont
  {Lienau}},\ }\bibfield  {title} {{\selectlanguage {english}\enquote {\bibinfo
  {title} {Adiabatic nanofocusing on ultrasmooth single-crystalline gold tapers
  creates a 10-nm-sized light source with few-cycle time resolution},}\ }}\href
  {https://doi.org/10.1021/nn301121h} {\bibfield  {journal} {\bibinfo
  {journal} {ACS Nano}\ }\textbf {\bibinfo {volume} {6}},\ \bibinfo {pages}
  {6040--6048} (\bibinfo {year} {2012})}\BibitemShut {NoStop}%
\bibitem [{\citenamefont {Geaney}\ \emph {et~al.}(2019)\citenamefont {Geaney},
  \citenamefont {Cox}, \citenamefont {H{\"{o}}nigl-Decrinis}, \citenamefont
  {Shaikhaidarov}, \citenamefont {Kubatkin}, \citenamefont {Lindstr{\"{o}}m},
  \citenamefont {Danilov},\ and\ \citenamefont {de~Graaf}}]{Geaney_2019}%
  \BibitemOpen
  \bibfield  {author} {\bibinfo {author} {\bibfnamefont {S.}~\bibnamefont
  {Geaney}}, \bibinfo {author} {\bibfnamefont {D.}~\bibnamefont {Cox}},
  \bibinfo {author} {\bibfnamefont {T.}~\bibnamefont {H{\"{o}}nigl-Decrinis}},
  \bibinfo {author} {\bibfnamefont {R.}~\bibnamefont {Shaikhaidarov}}, \bibinfo
  {author} {\bibfnamefont {S.~E.}\ \bibnamefont {Kubatkin}}, \bibinfo {author}
  {\bibfnamefont {T.}~\bibnamefont {Lindstr{\"{o}}m}}, \bibinfo {author}
  {\bibfnamefont {A.~V.}\ \bibnamefont {Danilov}},\ and\ \bibinfo {author}
  {\bibfnamefont {S.~E.}\ \bibnamefont {de~Graaf}},\ }\bibfield  {title}
  {\enquote {\bibinfo {title} {Near-field scanning microwave microscopy in the
  single photon regime},}\ }\href {https://doi.org/10.1038/s41598-019-48780-3}
  {\bibfield  {journal} {\bibinfo  {journal} {Sci. Rep.}\ }\textbf {\bibinfo
  {volume} {9}} (\bibinfo {year} {2019}),\
  10.1038/s41598-019-48780-3}\BibitemShut {NoStop}%
\bibitem [{\citenamefont {Meister}\ \emph {et~al.}(2004)\citenamefont
  {Meister}, \citenamefont {Liley}, \citenamefont {Brugger}, \citenamefont
  {Pugin},\ and\ \citenamefont {Heinzelmann}}]{Meister_2004}%
  \BibitemOpen
  \bibfield  {author} {\bibinfo {author} {\bibfnamefont {A.}~\bibnamefont
  {Meister}}, \bibinfo {author} {\bibfnamefont {M.}~\bibnamefont {Liley}},
  \bibinfo {author} {\bibfnamefont {J.}~\bibnamefont {Brugger}}, \bibinfo
  {author} {\bibfnamefont {R.}~\bibnamefont {Pugin}},\ and\ \bibinfo {author}
  {\bibfnamefont {H.}~\bibnamefont {Heinzelmann}},\ }\bibfield  {title}
  {\enquote {\bibinfo {title} {Nanodispenser for attoliter volume deposition
  using atomic force microscopy probes modified by focused-ion-beam milling},}\
  }\href {https://doi.org/10.1063/1.1842352} {\bibfield  {journal} {\bibinfo
  {journal} {Appl. Phys. Lett.}\ }\textbf {\bibinfo {volume} {85}},\ \bibinfo
  {pages} {6260--6262} (\bibinfo {year} {2004})}\BibitemShut {NoStop}%
\bibitem [{\citenamefont {Guillaume-Gentil}\ \emph {et~al.}(2014)\citenamefont
  {Guillaume-Gentil}, \citenamefont {Potthoff}, \citenamefont {Ossola},
  \citenamefont {Franz}, \citenamefont {Zambelli},\ and\ \citenamefont
  {Vorholt}}]{Guillaume-Gentil_2014}%
  \BibitemOpen
  \bibfield  {author} {\bibinfo {author} {\bibfnamefont {O.}~\bibnamefont
  {Guillaume-Gentil}}, \bibinfo {author} {\bibfnamefont {E.}~\bibnamefont
  {Potthoff}}, \bibinfo {author} {\bibfnamefont {D.}~\bibnamefont {Ossola}},
  \bibinfo {author} {\bibfnamefont {C.~M.}\ \bibnamefont {Franz}}, \bibinfo
  {author} {\bibfnamefont {T.}~\bibnamefont {Zambelli}},\ and\ \bibinfo
  {author} {\bibfnamefont {J.~A.}\ \bibnamefont {Vorholt}},\ }\bibfield
  {title} {{\selectlanguage {english}\enquote {\bibinfo {title}
  {Force-controlled manipulation of single cells: from {AFM} to {FluidFM}},}\
  }}\href {https://doi.org/10.1016/j.tibtech.2014.04.008} {\bibfield  {journal}
  {\bibinfo  {journal} {Trends Biotechnol.}\ }\textbf {\bibinfo {volume}
  {32}},\ \bibinfo {pages} {381--388} (\bibinfo {year} {2014})}\BibitemShut
  {NoStop}%
\bibitem [{\citenamefont {Persaud}\ \emph {et~al.}(2005)\citenamefont
  {Persaud}, \citenamefont {Park}, \citenamefont {Liddle}, \citenamefont
  {Schenkel}, \citenamefont {Bokor},\ and\ \citenamefont
  {Rangelow}}]{Persaud_2005}%
  \BibitemOpen
  \bibfield  {author} {\bibinfo {author} {\bibfnamefont {A.}~\bibnamefont
  {Persaud}}, \bibinfo {author} {\bibfnamefont {S.~J.}\ \bibnamefont {Park}},
  \bibinfo {author} {\bibfnamefont {J.~A.}\ \bibnamefont {Liddle}}, \bibinfo
  {author} {\bibfnamefont {T.}~\bibnamefont {Schenkel}}, \bibinfo {author}
  {\bibfnamefont {J.}~\bibnamefont {Bokor}},\ and\ \bibinfo {author}
  {\bibfnamefont {I.~W.}\ \bibnamefont {Rangelow}},\ }\bibfield  {title}
  {{\selectlanguage {english}\enquote {\bibinfo {title} {Integration of
  scanning probes and ion beams},}\ }}\href {https://doi.org/10.1021/nl0506103}
  {\bibfield  {journal} {\bibinfo  {journal} {Nano Lett.}\ }\textbf {\bibinfo
  {volume} {5}},\ \bibinfo {pages} {1087--1091} (\bibinfo {year}
  {2005})}\BibitemShut {NoStop}%
\bibitem [{\citenamefont {Wang}\ and\ \citenamefont {Butte}(2014)}]{Wang_2014}%
  \BibitemOpen
  \bibfield  {author} {\bibinfo {author} {\bibfnamefont {A.}~\bibnamefont
  {Wang}}\ and\ \bibinfo {author} {\bibfnamefont {M.~J.}\ \bibnamefont
  {Butte}},\ }\bibfield  {title} {{\selectlanguage {english}\enquote {\bibinfo
  {title} {Customized atomic force microscopy probe by
  focused-ion-beam-assisted tip transfer},}\ }}\href
  {https://doi.org/10.1063/1.4892075} {\bibfield  {journal} {\bibinfo
  {journal} {Appl. Phys. Lett.}\ }\textbf {\bibinfo {volume} {105}},\ \bibinfo
  {pages} {053101} (\bibinfo {year} {2014})}\BibitemShut {NoStop}%
\bibitem [{\citenamefont {Kunicki}\ \emph {et~al.}(2019)\citenamefont
  {Kunicki}, \citenamefont {Angelov}, \citenamefont {Ivanov}, \citenamefont
  {Gotszalk},\ and\ \citenamefont {Rangelow}}]{Kunicki_2019}%
  \BibitemOpen
  \bibfield  {author} {\bibinfo {author} {\bibfnamefont {P.}~\bibnamefont
  {Kunicki}}, \bibinfo {author} {\bibfnamefont {T.}~\bibnamefont {Angelov}},
  \bibinfo {author} {\bibfnamefont {T.}~\bibnamefont {Ivanov}}, \bibinfo
  {author} {\bibfnamefont {T.}~\bibnamefont {Gotszalk}},\ and\ \bibinfo
  {author} {\bibfnamefont {I.}~\bibnamefont {Rangelow}},\ }\bibfield  {title}
  {{\selectlanguage {english}\enquote {\bibinfo {title} {Sensitivity
  improvement to active piezoresistive {AFM} probes using focused ion beam
  processing},}\ }}\href {https://doi.org/10.3390/s19204429} {\bibfield
  {journal} {\bibinfo  {journal} {Sensors}\ }\textbf {\bibinfo {volume} {19}}
  (\bibinfo {year} {2019}),\ 10.3390/s19204429}\BibitemShut {NoStop}%
\bibitem [{\citenamefont {Corte-Le{\'{o}}n}\ \emph {et~al.}(2019)\citenamefont
  {Corte-Le{\'{o}}n}, \citenamefont {Rodr{\'{\i}}guez}, \citenamefont
  {Pancaldi}, \citenamefont {Gatel}, \citenamefont {Cox}, \citenamefont
  {Snoeck}, \citenamefont {Antonov}, \citenamefont {Vavassori},\ and\
  \citenamefont {Kazakova}}]{CorteLeon_2019}%
  \BibitemOpen
  \bibfield  {author} {\bibinfo {author} {\bibfnamefont {H.}~\bibnamefont
  {Corte-Le{\'{o}}n}}, \bibinfo {author} {\bibfnamefont {L.~A.}\ \bibnamefont
  {Rodr{\'{\i}}guez}}, \bibinfo {author} {\bibfnamefont {M.}~\bibnamefont
  {Pancaldi}}, \bibinfo {author} {\bibfnamefont {C.}~\bibnamefont {Gatel}},
  \bibinfo {author} {\bibfnamefont {D.}~\bibnamefont {Cox}}, \bibinfo {author}
  {\bibfnamefont {E.}~\bibnamefont {Snoeck}}, \bibinfo {author} {\bibfnamefont
  {V.}~\bibnamefont {Antonov}}, \bibinfo {author} {\bibfnamefont
  {P.}~\bibnamefont {Vavassori}},\ and\ \bibinfo {author} {\bibfnamefont
  {O.}~\bibnamefont {Kazakova}},\ }\bibfield  {title} {\enquote {\bibinfo
  {title} {Magnetic imaging using geometrically constrained nano-domain
  walls},}\ }\href {https://doi.org/10.1039/c8nr07729k} {\bibfield  {journal}
  {\bibinfo  {journal} {Nanoscale}\ }\textbf {\bibinfo {volume} {11}},\
  \bibinfo {pages} {4478--4488} (\bibinfo {year} {2019})}\BibitemShut {NoStop}%
\bibitem [{\citenamefont {Kirchhof}\ \emph {et~al.}(2021)\citenamefont
  {Kirchhof}, \citenamefont {Weinel}, \citenamefont {Heeg}, \citenamefont
  {Deinhart}, \citenamefont {Kovalchuk}, \citenamefont {H{\"{o}}flich},\ and\
  \citenamefont {Bolotin}}]{Kirchhof_2021a}%
  \BibitemOpen
  \bibfield  {author} {\bibinfo {author} {\bibfnamefont {J.~N.}\ \bibnamefont
  {Kirchhof}}, \bibinfo {author} {\bibfnamefont {K.}~\bibnamefont {Weinel}},
  \bibinfo {author} {\bibfnamefont {S.}~\bibnamefont {Heeg}}, \bibinfo {author}
  {\bibfnamefont {V.}~\bibnamefont {Deinhart}}, \bibinfo {author}
  {\bibfnamefont {S.}~\bibnamefont {Kovalchuk}}, \bibinfo {author}
  {\bibfnamefont {K.}~\bibnamefont {H{\"{o}}flich}},\ and\ \bibinfo {author}
  {\bibfnamefont {K.~I.}\ \bibnamefont {Bolotin}},\ }\bibfield  {title}
  {\enquote {\bibinfo {title} {Tunable graphene phononic crystal},}\ }\href
  {https://doi.org/10.1021/acs.nanolett.0c04986} {\bibfield  {journal}
  {\bibinfo  {journal} {Nano Lett.}\ }\textbf {\bibinfo {volume} {21}},\
  \bibinfo {pages} {2174--2182} (\bibinfo {year} {2021})}\BibitemShut {NoStop}%
\bibitem [{\citenamefont {Miller}\ \emph {et~al.}(2018)\citenamefont {Miller},
  \citenamefont {Blaikie}, \citenamefont {Carter},\ and\ \citenamefont
  {Aleman}}]{Miller_2018a}%
  \BibitemOpen
  \bibfield  {author} {\bibinfo {author} {\bibfnamefont {D.}~\bibnamefont
  {Miller}}, \bibinfo {author} {\bibfnamefont {A.}~\bibnamefont {Blaikie}},
  \bibinfo {author} {\bibfnamefont {B.}~\bibnamefont {Carter}},\ and\ \bibinfo
  {author} {\bibfnamefont {B.}~\bibnamefont {Aleman}},\ }\bibfield  {title}
  {\enquote {\bibinfo {title} {Engineering the modal shape of graphene
  nanoelectromechanical systems using focused ion beam milling},}\ }in\ \href
  {https://doi.org/10.1109/nmdc.2018.8605924} {\emph {\bibinfo {booktitle}
  {2018 {IEEE} 13th Nanotechnology Materials and Devices Conference
  ({NMDC})}}}\ (\bibinfo  {publisher} {{IEEE}},\ \bibinfo {year}
  {2018})\BibitemShut {NoStop}%
\bibitem [{\citenamefont {Volkert}\ and\ \citenamefont
  {Minor}(2007)}]{Volkert_2007}%
  \BibitemOpen
  \bibfield  {author} {\bibinfo {author} {\bibfnamefont {C.~A.}\ \bibnamefont
  {Volkert}}\ and\ \bibinfo {author} {\bibfnamefont {A.~M.}\ \bibnamefont
  {Minor}},\ }\bibfield  {title} {\enquote {\bibinfo {title} {Focused ion beam
  microscopy and micromachining},}\ }\href {https://doi.org/10.1557/mrs2007.62}
  {\bibfield  {journal} {\bibinfo  {journal} {MRS Bull.}\ }\textbf {\bibinfo
  {volume} {32}},\ \bibinfo {pages} {389--399} (\bibinfo {year}
  {2007})}\BibitemShut {NoStop}%
\bibitem [{\citenamefont {Lenrick}\ \emph {et~al.}(2013)\citenamefont
  {Lenrick}, \citenamefont {Ek}, \citenamefont {Jacobsson}, \citenamefont
  {Borgstr{\"{o}}m},\ and\ \citenamefont {Wallenberg}}]{Lenrick_2013}%
  \BibitemOpen
  \bibfield  {author} {\bibinfo {author} {\bibfnamefont {F.}~\bibnamefont
  {Lenrick}}, \bibinfo {author} {\bibfnamefont {M.}~\bibnamefont {Ek}},
  \bibinfo {author} {\bibfnamefont {D.}~\bibnamefont {Jacobsson}}, \bibinfo
  {author} {\bibfnamefont {M.~T.}\ \bibnamefont {Borgstr{\"{o}}m}},\ and\
  \bibinfo {author} {\bibfnamefont {L.~R.}\ \bibnamefont {Wallenberg}},\
  }\bibfield  {title} {\enquote {\bibinfo {title} {{FIB} plan and side view
  cross-sectional {TEM} sample preparation of nanostructures},}\ }\href
  {https://doi.org/10.1017/s1431927613013780} {\bibfield  {journal} {\bibinfo
  {journal} {Microsc. Microanal.}\ }\textbf {\bibinfo {volume} {20}},\ \bibinfo
  {pages} {133--140} (\bibinfo {year} {2013})}\BibitemShut {NoStop}%
\bibitem [{\citenamefont {Schaffer}, \citenamefont {Schaffer},\ and\
  \citenamefont {Ramasse}(2012)}]{Schaffer_2012}%
  \BibitemOpen
  \bibfield  {author} {\bibinfo {author} {\bibfnamefont {M.}~\bibnamefont
  {Schaffer}}, \bibinfo {author} {\bibfnamefont {B.}~\bibnamefont {Schaffer}},\
  and\ \bibinfo {author} {\bibfnamefont {Q.}~\bibnamefont {Ramasse}},\
  }\bibfield  {title} {\enquote {\bibinfo {title} {Sample preparation for
  atomic-resolution {STEM} at low voltages by {FIB}},}\ }\href
  {https://doi.org/10.1016/j.ultramic.2012.01.005} {\bibfield  {journal}
  {\bibinfo  {journal} {Ultramicroscopy}\ }\textbf {\bibinfo {volume} {114}},\
  \bibinfo {pages} {62--71} (\bibinfo {year} {2012})}\BibitemShut {NoStop}%
\bibitem [{\citenamefont {Unocic}, \citenamefont {Mills},\ and\ \citenamefont
  {Daehn}(2010)}]{Unocic2010}%
  \BibitemOpen
  \bibfield  {author} {\bibinfo {author} {\bibfnamefont {K.~A.}\ \bibnamefont
  {Unocic}}, \bibinfo {author} {\bibfnamefont {M.~J.}\ \bibnamefont {Mills}},\
  and\ \bibinfo {author} {\bibfnamefont {G.~S.}\ \bibnamefont {Daehn}},\
  }\bibfield  {title} {\enquote {\bibinfo {title} {Effect of gallium focused
  ion beam milling on preparation of aluminium thin foils},}\ }\href
  {https://doi.org/10.1111/j.1365-2818.2010.03401.x} {\bibfield  {journal}
  {\bibinfo  {journal} {J. Microsc.}\ }\textbf {\bibinfo {volume} {240}},\
  \bibinfo {pages} {227--238} (\bibinfo {year} {2010})}\BibitemShut {NoStop}%
\bibitem [{\citenamefont {Thompson}\ \emph {et~al.}(2007)\citenamefont
  {Thompson}, \citenamefont {Lawrence}, \citenamefont {Larson}, \citenamefont
  {Olson}, \citenamefont {Kelly},\ and\ \citenamefont {Gorman}}]{Thompson2007}%
  \BibitemOpen
  \bibfield  {author} {\bibinfo {author} {\bibfnamefont {K.}~\bibnamefont
  {Thompson}}, \bibinfo {author} {\bibfnamefont {D.}~\bibnamefont {Lawrence}},
  \bibinfo {author} {\bibfnamefont {D.~J.}\ \bibnamefont {Larson}}, \bibinfo
  {author} {\bibfnamefont {J.~D.}\ \bibnamefont {Olson}}, \bibinfo {author}
  {\bibfnamefont {T.~F.}\ \bibnamefont {Kelly}},\ and\ \bibinfo {author}
  {\bibfnamefont {B.}~\bibnamefont {Gorman}},\ }\bibfield  {title} {\enquote
  {\bibinfo {title} {In situ site-specific specimen preparation for atom probe
  tomography},}\ }\href {https://doi.org/10.1016/j.ultramic.2006.06.008}
  {\bibfield  {journal} {\bibinfo  {journal} {Ultramicroscopy}\ }\textbf
  {\bibinfo {volume} {107}},\ \bibinfo {pages} {131--139} (\bibinfo {year}
  {2007})}\BibitemShut {NoStop}%
\bibitem [{\citenamefont {Hugo}\ and\ \citenamefont
  {Hoagland}(1998)}]{Hugo_1998}%
  \BibitemOpen
  \bibfield  {author} {\bibinfo {author} {\bibfnamefont {R.~C.}\ \bibnamefont
  {Hugo}}\ and\ \bibinfo {author} {\bibfnamefont {R.~G.}\ \bibnamefont
  {Hoagland}},\ }\bibfield  {title} {\enquote {\bibinfo {title} {In-situ {TEM}
  observation of aluminum embrittlement by liquid gallium},}\ }\href
  {https://doi.org/10.1016/s1359-6462(97)00464-8} {\bibfield  {journal}
  {\bibinfo  {journal} {Scripta Mater.}\ }\textbf {\bibinfo {volume} {38}},\
  \bibinfo {pages} {523--529} (\bibinfo {year} {1998})}\BibitemShut {NoStop}%
\bibitem [{\citenamefont {Allen}\ \emph
  {et~al.}(2023{\natexlab{b}})\citenamefont {Allen}, \citenamefont {Blanchard},
  \citenamefont {Lake}, \citenamefont {Pappas}, \citenamefont {Xia},
  \citenamefont {Notte}, \citenamefont {Zhang}, \citenamefont {Minor},\ and\
  \citenamefont {Sanford}}]{Allen2023}%
  \BibitemOpen
  \bibfield  {author} {\bibinfo {author} {\bibfnamefont {F.~I.}\ \bibnamefont
  {Allen}}, \bibinfo {author} {\bibfnamefont {P.~T.}\ \bibnamefont
  {Blanchard}}, \bibinfo {author} {\bibfnamefont {R.}~\bibnamefont {Lake}},
  \bibinfo {author} {\bibfnamefont {D.}~\bibnamefont {Pappas}}, \bibinfo
  {author} {\bibfnamefont {D.}~\bibnamefont {Xia}}, \bibinfo {author}
  {\bibfnamefont {J.~A.}\ \bibnamefont {Notte}}, \bibinfo {author}
  {\bibfnamefont {R.}~\bibnamefont {Zhang}}, \bibinfo {author} {\bibfnamefont
  {A.~M.}\ \bibnamefont {Minor}},\ and\ \bibinfo {author} {\bibfnamefont
  {N.~A.}\ \bibnamefont {Sanford}},\ }\href
  {https://doi.org/10.48550/ARXIV.2302.09254} {\enquote {\bibinfo {title}
  {Fabrication of specimens for atom probe tomography using a combined gallium
  and neon focused ion beam milling approach},}\ } (\bibinfo {year}
  {2023}{\natexlab{b}}),\ \bibinfo {note} {preprint},\ \Eprint
  {https://arxiv.org/abs/2302.09254} {arXiv:2302.09254 [cond-mat.mtrl-sci]}
  \BibitemShut {NoStop}%
\bibitem [{\citenamefont {Lilensten}\ and\ \citenamefont
  {Gault}(2020)}]{Lilensten_2020}%
  \BibitemOpen
  \bibfield  {author} {\bibinfo {author} {\bibfnamefont {L.}~\bibnamefont
  {Lilensten}}\ and\ \bibinfo {author} {\bibfnamefont {B.}~\bibnamefont
  {Gault}},\ }\bibfield  {title} {{\selectlanguage {en}\enquote {\bibinfo
  {title} {New approach for {FIB-preparation} of atom probe specimens for
  aluminum alloys},}\ }}\href {https://doi.org/10.1371/journal.pone.0231179}
  {\bibfield  {journal} {\bibinfo  {journal} {PLoS One}\ }\textbf {\bibinfo
  {volume} {15}},\ \bibinfo {pages} {e0231179} (\bibinfo {year}
  {2020})}\BibitemShut {NoStop}%
\bibitem [{\citenamefont {Uchic}\ \emph {et~al.}(2007)\citenamefont {Uchic},
  \citenamefont {Holzer}, \citenamefont {Inkson}, \citenamefont {Principe},\
  and\ \citenamefont {Munroe}}]{Uchic_2007}%
  \BibitemOpen
  \bibfield  {author} {\bibinfo {author} {\bibfnamefont {M.~D.}\ \bibnamefont
  {Uchic}}, \bibinfo {author} {\bibfnamefont {L.}~\bibnamefont {Holzer}},
  \bibinfo {author} {\bibfnamefont {B.~J.}\ \bibnamefont {Inkson}}, \bibinfo
  {author} {\bibfnamefont {E.~L.}\ \bibnamefont {Principe}},\ and\ \bibinfo
  {author} {\bibfnamefont {P.}~\bibnamefont {Munroe}},\ }\bibfield  {title}
  {\enquote {\bibinfo {title} {Three-dimensional microstructural
  characterization using focused ion beam tomography},}\ }\href
  {https://doi.org/10.1557/mrs2007.64} {\bibfield  {journal} {\bibinfo
  {journal} {{MRS} Bulletin}\ }\textbf {\bibinfo {volume} {32}},\ \bibinfo
  {pages} {408--416} (\bibinfo {year} {2007})}\BibitemShut {NoStop}%
\bibitem [{\citenamefont {Burnett}\ \emph {et~al.}(2016)\citenamefont
  {Burnett}, \citenamefont {Kelley}, \citenamefont {Winiarski}, \citenamefont
  {Contreras}, \citenamefont {Daly}, \citenamefont {Gholinia}, \citenamefont
  {Burke},\ and\ \citenamefont {Withers}}]{Burnett_2016}%
  \BibitemOpen
  \bibfield  {author} {\bibinfo {author} {\bibfnamefont {T.~L.}\ \bibnamefont
  {Burnett}}, \bibinfo {author} {\bibfnamefont {R.}~\bibnamefont {Kelley}},
  \bibinfo {author} {\bibfnamefont {B.}~\bibnamefont {Winiarski}}, \bibinfo
  {author} {\bibfnamefont {L.}~\bibnamefont {Contreras}}, \bibinfo {author}
  {\bibfnamefont {M.}~\bibnamefont {Daly}}, \bibinfo {author} {\bibfnamefont
  {A.}~\bibnamefont {Gholinia}}, \bibinfo {author} {\bibfnamefont {M.~G.}\
  \bibnamefont {Burke}},\ and\ \bibinfo {author} {\bibfnamefont {P.~J.}\
  \bibnamefont {Withers}},\ }\bibfield  {title} {{\selectlanguage
  {english}\enquote {\bibinfo {title} {Large volume serial section tomography
  by {X}e plasma {FIB} dual beam microscopy},}\ }}\href
  {https://doi.org/10.1016/j.ultramic.2015.11.001} {\bibfield  {journal}
  {\bibinfo  {journal} {Ultramicroscopy}\ }\textbf {\bibinfo {volume} {161}},\
  \bibinfo {pages} {119--129} (\bibinfo {year} {2016})}\BibitemShut {NoStop}%
\bibitem [{\citenamefont {Giannuzzi}\ and\ \citenamefont
  {Stevie}(2005)}]{Giannuzzi_2005a}%
  \BibitemOpen
  \bibfield  {author} {\bibinfo {author} {\bibfnamefont {L.~A.}\ \bibnamefont
  {Giannuzzi}}\ and\ \bibinfo {author} {\bibfnamefont {F.~A.}\ \bibnamefont
  {Stevie}},\ }\href {https://doi.org/10.1007/b101190} {\emph {\bibinfo {title}
  {{Introduction to Focused Ion Beams}}}},\ edited by\ \bibinfo {editor}
  {\bibfnamefont {L.~A.}\ \bibnamefont {Giannuzzi}}\ and\ \bibinfo {editor}
  {\bibfnamefont {F.~A.}\ \bibnamefont {Stevie}}\ (\bibinfo  {publisher}
  {Kluwer Academic Publishers},\ \bibinfo {address} {Boston},\ \bibinfo {year}
  {2005})\ pp.\ \bibinfo {pages} {1--357}\BibitemShut {NoStop}%
\bibitem [{\citenamefont {Ast}\ \emph {et~al.}(2019)\citenamefont {Ast},
  \citenamefont {Ghidelli}, \citenamefont {Durst}, \citenamefont {G{\"{o}}ken},
  \citenamefont {Sebastiani},\ and\ \citenamefont {Korsunsky}}]{Ast_2019}%
  \BibitemOpen
  \bibfield  {author} {\bibinfo {author} {\bibfnamefont {J.}~\bibnamefont
  {Ast}}, \bibinfo {author} {\bibfnamefont {M.}~\bibnamefont {Ghidelli}},
  \bibinfo {author} {\bibfnamefont {K.}~\bibnamefont {Durst}}, \bibinfo
  {author} {\bibfnamefont {M.}~\bibnamefont {G{\"{o}}ken}}, \bibinfo {author}
  {\bibfnamefont {M.}~\bibnamefont {Sebastiani}},\ and\ \bibinfo {author}
  {\bibfnamefont {A.~M.}\ \bibnamefont {Korsunsky}},\ }\bibfield  {title}
  {\enquote {\bibinfo {title} {A review of experimental approaches to fracture
  toughness evaluation at the micro-scale},}\ }\href
  {https://doi.org/10.1016/j.matdes.2019.107762} {\bibfield  {journal}
  {\bibinfo  {journal} {Mater. Design}\ }\textbf {\bibinfo {volume} {173}},\
  \bibinfo {pages} {107762} (\bibinfo {year} {2019})}\BibitemShut {NoStop}%
\bibitem [{\citenamefont {Orloff}, \citenamefont {Utlaut},\ and\ \citenamefont
  {Swanson}(2003)}]{Orloff_2003}%
  \BibitemOpen
  \bibfield  {author} {\bibinfo {author} {\bibfnamefont {J.~H.}\ \bibnamefont
  {Orloff}}, \bibinfo {author} {\bibfnamefont {M.}~\bibnamefont {Utlaut}},\
  and\ \bibinfo {author} {\bibfnamefont {L.~W.}\ \bibnamefont {Swanson}},\
  }\href {https://doi.org/10.1007/978-1-4615-0765-9} {\emph {\bibinfo {title}
  {High resolution focused ion beams: {FIB} and its applications}}}\ (\bibinfo
  {publisher} {Springer US},\ \bibinfo {address} {Boston, MA},\ \bibinfo {year}
  {2003})\BibitemShut {NoStop}%
\bibitem [{\citenamefont {Gandhi}(2019)}]{Gandhi_2019}%
  \BibitemOpen
  \bibinfo {editor} {\bibfnamefont {T.}~\bibnamefont {Gandhi}},\ ed.,\ \href
  {https://doi.org/10.31399/asm.mfadr7.9781627082471} {\emph {\bibinfo {title}
  {Microelectronics failure analysis}}}\ (\bibinfo  {publisher} {{ASM}
  {I}nternational},\ \bibinfo {year} {2019})\BibitemShut {NoStop}%
\bibitem [{\citenamefont {TSMC}(2022)}]{TSMC_2022}%
  \BibitemOpen
  \bibfield  {author} {\bibinfo {author} {\bibnamefont {TSMC}},\ }\href
  {https://www.tsmc.com/english/dedicatedFoundry/technology/logic/l_3nm}
  {\enquote {\bibinfo {title} {3nm technology},}\ }\bibinfo {howpublished}
  {Online} (\bibinfo {year} {2022}),\ \bibinfo {note}
  {\url{https://www.tsmc.com/english/dedicatedFoundry/technology/logic/l_3nm}}\BibitemShut
  {NoStop}%
\bibitem [{\citenamefont {Intel}(2022)}]{Intel_2022}%
  \BibitemOpen
  \bibfield  {author} {\bibinfo {author} {\bibnamefont {Intel}},\ }\href
  {https://www.intel.com/content/www/us/en/newsroom/news/intel-technology-roadmaps-milestones.html#gs.tuhd2s}
  {\enquote {\bibinfo {title} {Intel technology roadmaps and milestones},}\
  }\bibinfo {howpublished} {Online} (\bibinfo {year} {2022}),\ \bibinfo {note}
  {\url{https://www.intel.com/content/www/us/en/newsroom/news/intel-technology-roadmaps-milestones.html\#gs.tuhd2s}}\BibitemShut
  {NoStop}%
\bibitem [{\citenamefont {Samsung}(2022)}]{Samsung_2022}%
  \BibitemOpen
  \bibfield  {author} {\bibinfo {author} {\bibnamefont {Samsung}},\ }\href
  {https://news.samsung.com/global/samsung-electronics-announces-first-quarter-2022-results}
  {\enquote {\bibinfo {title} {Samsung electronics announces first quarter 2022
  results},}\ }\bibinfo {howpublished} {Online} (\bibinfo {year} {2022}),\
  \bibinfo {note}
  {\url{https://news.samsung.com/global/samsung-electronics-announces-first-quarter-2022-results}}\BibitemShut
  {NoStop}%
\bibitem [{\citenamefont {Young}, \citenamefont {Cleaver},\ and\ \citenamefont
  {Ahmed}(1990)}]{Young_1990}%
  \BibitemOpen
  \bibfield  {author} {\bibinfo {author} {\bibfnamefont {R.~J.}\ \bibnamefont
  {Young}}, \bibinfo {author} {\bibfnamefont {J.~R.~A.}\ \bibnamefont
  {Cleaver}},\ and\ \bibinfo {author} {\bibfnamefont {H.}~\bibnamefont
  {Ahmed}},\ }\bibfield  {title} {\enquote {\bibinfo {title} {Gas-assisted
  focused ion beam etching for microfabrication and inspection},}\ }\href
  {https://doi.org/10.1016/0167-9317(90)90140-O} {\bibfield  {journal}
  {\bibinfo  {journal} {Microelectron. Eng.}\ }\textbf {\bibinfo {volume}
  {11}},\ \bibinfo {pages} {409--412} (\bibinfo {year} {1990})}\BibitemShut
  {NoStop}%
\bibitem [{\citenamefont {Xu}\ \emph {et~al.}(1989)\citenamefont {Xu},
  \citenamefont {Gamo}, \citenamefont {Shiokawa},\ and\ \citenamefont
  {Namba}}]{Xu_1989}%
  \BibitemOpen
  \bibfield  {author} {\bibinfo {author} {\bibfnamefont {Z.}~\bibnamefont
  {Xu}}, \bibinfo {author} {\bibfnamefont {K.}~\bibnamefont {Gamo}}, \bibinfo
  {author} {\bibfnamefont {T.}~\bibnamefont {Shiokawa}},\ and\ \bibinfo
  {author} {\bibfnamefont {S.}~\bibnamefont {Namba}},\ }\bibfield  {title}
  {\enquote {\bibinfo {title} {In situ patterning of {Si3N4} by an
  ion-beam-induced gas surface reaction},}\ }\href
  {https://doi.org/10.1016/0168-583X(89)90890-2} {\bibfield  {journal}
  {\bibinfo  {journal} {Nucl. Instrum. Methods Phys. Res. B}\ }\textbf
  {\bibinfo {volume} {39}},\ \bibinfo {pages} {750--753} (\bibinfo {year}
  {1989})}\BibitemShut {NoStop}%
\bibitem [{\citenamefont {Gamo}\ \emph {et~al.}(1988)\citenamefont {Gamo},
  \citenamefont {Hamauzu}, \citenamefont {Xu},\ and\ \citenamefont
  {Namba}}]{Gamo_1988}%
  \BibitemOpen
  \bibfield  {author} {\bibinfo {author} {\bibfnamefont {K.}~\bibnamefont
  {Gamo}}, \bibinfo {author} {\bibfnamefont {H.}~\bibnamefont {Hamauzu}},
  \bibinfo {author} {\bibfnamefont {Z.}~\bibnamefont {Xu}},\ and\ \bibinfo
  {author} {\bibfnamefont {S.}~\bibnamefont {Namba}},\ }\bibfield  {title}
  {\enquote {\bibinfo {title} {In situ development of ion bombarded
  poly(methylmethacrylate) resist in a reactive gas ambient},}\ }\href
  {https://doi.org/10.1116/1.584294} {\bibfield  {journal} {\bibinfo  {journal}
  {J. Vac. Sci. Technol. B}\ }\textbf {\bibinfo {volume} {6}},\ \bibinfo
  {pages} {989--992} (\bibinfo {year} {1988})}\BibitemShut {NoStop}%
\bibitem [{\citenamefont {Matsubara}\ \emph {et~al.}(2019)\citenamefont
  {Matsubara}, \citenamefont {Shichi}, \citenamefont {Tanimoto}, \citenamefont
  {Ikota},\ and\ \citenamefont {Hashizume}}]{Matsubara_2019}%
  \BibitemOpen
  \bibfield  {author} {\bibinfo {author} {\bibfnamefont {S.}~\bibnamefont
  {Matsubara}}, \bibinfo {author} {\bibfnamefont {H.}~\bibnamefont {Shichi}},
  \bibinfo {author} {\bibfnamefont {S.}~\bibnamefont {Tanimoto}}, \bibinfo
  {author} {\bibfnamefont {M.}~\bibnamefont {Ikota}},\ and\ \bibinfo {author}
  {\bibfnamefont {T.}~\bibnamefont {Hashizume}},\ }\bibfield  {title} {\enquote
  {\bibinfo {title} {Visualization of 3{D} structure of semiconductor devices
  by "dig and see" using {GFIS}-{SIM}},}\ }in\ \href
  {https://doi.org/10.1117/12.2514934} {\emph {\bibinfo {booktitle} {Metrology,
  Inspection, and Process Control for Microlithography {XXXIII}}}},\ \bibinfo
  {editor} {edited by\ \bibinfo {editor} {\bibfnamefont {O.}~\bibnamefont
  {Adan}}\ and\ \bibinfo {editor} {\bibfnamefont {V.~A.}\ \bibnamefont
  {Ukraintsev}}}\ (\bibinfo  {publisher} {{SPIE}},\ \bibinfo {year}
  {2019})\BibitemShut {NoStop}%
\bibitem [{\citenamefont {Steele}\ \emph {et~al.}(2010)\citenamefont {Steele},
  \citenamefont {Knuffman}, \citenamefont {McClelland},\ and\ \citenamefont
  {Orloff}}]{Steele_2010}%
  \BibitemOpen
  \bibfield  {author} {\bibinfo {author} {\bibfnamefont {A.~V.}\ \bibnamefont
  {Steele}}, \bibinfo {author} {\bibfnamefont {B.}~\bibnamefont {Knuffman}},
  \bibinfo {author} {\bibfnamefont {J.~J.}\ \bibnamefont {McClelland}},\ and\
  \bibinfo {author} {\bibfnamefont {J.}~\bibnamefont {Orloff}},\ }\bibfield
  {title} {\enquote {\bibinfo {title} {{Focused chromium ion beam}},}\ }\href
  {https://doi.org/10.1116/1.3502668} {\bibfield  {journal} {\bibinfo
  {journal} {J. Vac. Sci. Technol. B}\ }\textbf {\bibinfo {volume} {28}},\
  \bibinfo {pages} {C6F1--C6F5} (\bibinfo {year} {2010})}\BibitemShut {NoStop}%
\bibitem [{\citenamefont {Livengood}\ \emph
  {et~al.}(2011{\natexlab{b}})\citenamefont {Livengood}, \citenamefont {Tan},
  \citenamefont {Hack}, \citenamefont {Kane},\ and\ \citenamefont
  {Greenzweig}}]{Livengood_2011a}%
  \BibitemOpen
  \bibfield  {author} {\bibinfo {author} {\bibfnamefont {R.}~\bibnamefont
  {Livengood}}, \bibinfo {author} {\bibfnamefont {S.}~\bibnamefont {Tan}},
  \bibinfo {author} {\bibfnamefont {P.}~\bibnamefont {Hack}}, \bibinfo {author}
  {\bibfnamefont {M.}~\bibnamefont {Kane}},\ and\ \bibinfo {author}
  {\bibfnamefont {Y.}~\bibnamefont {Greenzweig}},\ }\bibfield  {title}
  {\enquote {\bibinfo {title} {Focused ion beam circuit edit--a look into the
  past, present, and future},}\ }\href
  {https://doi.org/10.1017/s1431927611004235} {\bibfield  {journal} {\bibinfo
  {journal} {Microsc. Microanal.}\ }\textbf {\bibinfo {volume} {17}},\ \bibinfo
  {pages} {672--673} (\bibinfo {year} {2011}{\natexlab{b}})}\BibitemShut
  {NoStop}%
\bibitem [{\citenamefont {Tan}\ \emph {et~al.}(2012{\natexlab{a}})\citenamefont
  {Tan}, \citenamefont {Livengood}, \citenamefont {Greenzweig}, \citenamefont
  {Drezner},\ and\ \citenamefont {Shima}}]{Tan_2012}%
  \BibitemOpen
  \bibfield  {author} {\bibinfo {author} {\bibfnamefont {S.}~\bibnamefont
  {Tan}}, \bibinfo {author} {\bibfnamefont {R.}~\bibnamefont {Livengood}},
  \bibinfo {author} {\bibfnamefont {Y.}~\bibnamefont {Greenzweig}}, \bibinfo
  {author} {\bibfnamefont {Y.}~\bibnamefont {Drezner}},\ and\ \bibinfo {author}
  {\bibfnamefont {D.}~\bibnamefont {Shima}},\ }\bibfield  {title} {\enquote
  {\bibinfo {title} {Probe current distribution characterization technique for
  focused ion beam},}\ }\href {https://doi.org/10.1116/1.4766882} {\bibfield
  {journal} {\bibinfo  {journal} {J. Vac. Sci. Technol. B}\ }\textbf {\bibinfo
  {volume} {30}},\ \bibinfo {pages} {06F606} (\bibinfo {year}
  {2012}{\natexlab{a}})}\BibitemShut {NoStop}%
\bibitem [{\citenamefont {Tan}\ \emph {et~al.}(2012{\natexlab{b}})\citenamefont
  {Tan}, \citenamefont {Livengood}, \citenamefont {Greenzweig}, \citenamefont
  {Drezner}, \citenamefont {Hallstein},\ and\ \citenamefont
  {Scheffler}}]{Tan_2012a}%
  \BibitemOpen
  \bibfield  {author} {\bibinfo {author} {\bibfnamefont {S.}~\bibnamefont
  {Tan}}, \bibinfo {author} {\bibfnamefont {R.~H.}\ \bibnamefont {Livengood}},
  \bibinfo {author} {\bibfnamefont {Y.}~\bibnamefont {Greenzweig}}, \bibinfo
  {author} {\bibfnamefont {Y.}~\bibnamefont {Drezner}}, \bibinfo {author}
  {\bibfnamefont {R.}~\bibnamefont {Hallstein}},\ and\ \bibinfo {author}
  {\bibfnamefont {C.}~\bibnamefont {Scheffler}},\ }\bibfield  {title} {\enquote
  {\bibinfo {title} {Characterization of ion beam current distribution
  influences on nanomachining},}\ }in\ \href@noop {} {\emph {\bibinfo
  {booktitle} {ISTFA 2012}}}\ (\bibinfo {organization} {ASM International},\
  \bibinfo {year} {2012})\ pp.\ \bibinfo {pages} {436--439}\BibitemShut
  {NoStop}%
\bibitem [{\citenamefont {Tan}\ \emph {et~al.}(2016)\citenamefont {Tan},
  \citenamefont {Hallstein}, \citenamefont {Livengood},\ and\ \citenamefont
  {Ali}}]{Tan_2016a}%
  \BibitemOpen
  \bibfield  {author} {\bibinfo {author} {\bibfnamefont {S.}~\bibnamefont
  {Tan}}, \bibinfo {author} {\bibfnamefont {R.}~\bibnamefont {Hallstein}},
  \bibinfo {author} {\bibfnamefont {R.~H.}\ \bibnamefont {Livengood}},\ and\
  \bibinfo {author} {\bibfnamefont {W.}~\bibnamefont {Ali}},\ }\bibfield
  {title} {\enquote {\bibinfo {title} {{GFIS} in semiconductor applications},}\
  }\href {https://doi.org/10.1017/s143192761600163x} {\bibfield  {journal}
  {\bibinfo  {journal} {Microsc. Microanal.}\ }\textbf {\bibinfo {volume}
  {22}},\ \bibinfo {pages} {156--157} (\bibinfo {year} {2016})}\BibitemShut
  {NoStop}%
\bibitem [{\citenamefont {Livengood}\ \emph {et~al.}(2009)\citenamefont
  {Livengood}, \citenamefont {Tan}, \citenamefont {Greenzweig}, \citenamefont
  {Notte},\ and\ \citenamefont {McVey}}]{Livengood2009}%
  \BibitemOpen
  \bibfield  {author} {\bibinfo {author} {\bibfnamefont {R.~H.}\ \bibnamefont
  {Livengood}}, \bibinfo {author} {\bibfnamefont {S.}~\bibnamefont {Tan}},
  \bibinfo {author} {\bibfnamefont {Y.}~\bibnamefont {Greenzweig}}, \bibinfo
  {author} {\bibfnamefont {J.~A.}\ \bibnamefont {Notte}},\ and\ \bibinfo
  {author} {\bibfnamefont {S.}~\bibnamefont {McVey}},\ }\bibfield  {title}
  {\enquote {\bibinfo {title} {{Subsurface damage from helium ions as a
  function of dose, beam energy, and dose rate}},}\ }\href
  {https://doi.org/10.1116/1.3237101} {\bibfield  {journal} {\bibinfo
  {journal} {J. Vac. Sci. Technol. B}\ }\textbf {\bibinfo {volume} {27}},\
  \bibinfo {pages} {3244} (\bibinfo {year} {2009})}\BibitemShut {NoStop}%
\bibitem [{\citenamefont {Basov}, \citenamefont {Averitt},\ and\ \citenamefont
  {Hsieh}(2017)}]{Basov_2017}%
  \BibitemOpen
  \bibfield  {author} {\bibinfo {author} {\bibfnamefont {D.~N.}\ \bibnamefont
  {Basov}}, \bibinfo {author} {\bibfnamefont {R.~D.}\ \bibnamefont {Averitt}},\
  and\ \bibinfo {author} {\bibfnamefont {D.}~\bibnamefont {Hsieh}},\ }\bibfield
   {title} {\enquote {\bibinfo {title} {Towards properties on demand in quantum
  materials},}\ }\href {https://doi.org/10.1038/nmat5017} {\bibfield  {journal}
  {\bibinfo  {journal} {Nat. Mater.}\ }\textbf {\bibinfo {volume} {16}},\
  \bibinfo {pages} {1077--1088} (\bibinfo {year} {2017})}\BibitemShut {NoStop}%
\bibitem [{\citenamefont {Giustino}\ \emph {et~al.}(2020)\citenamefont
  {Giustino}, \citenamefont {Lee}, \citenamefont {Trier}, \citenamefont
  {Bibes}, \citenamefont {Winter}, \citenamefont {Valent{\'{\i}}},
  \citenamefont {Son}, \citenamefont {Taillefer}, \citenamefont {Heil},
  \citenamefont {Figueroa}, \citenamefont {Pla{\c{c}}ais}, \citenamefont {Wu},
  \citenamefont {Yazyev}, \citenamefont {Bakkers}, \citenamefont {Nyg{\aa}rd},
  \citenamefont {Forn-D{\'{\i}}az}, \citenamefont {Franceschi}, \citenamefont
  {McIver}, \citenamefont {Torres}, \citenamefont {Low}, \citenamefont {Kumar},
  \citenamefont {Galceran}, \citenamefont {Valenzuela}, \citenamefont
  {Costache}, \citenamefont {Manchon}, \citenamefont {Kim}, \citenamefont
  {Schleder}, \citenamefont {Fazzio},\ and\ \citenamefont
  {Roche}}]{Giustino_2020}%
  \BibitemOpen
  \bibfield  {author} {\bibinfo {author} {\bibfnamefont {F.}~\bibnamefont
  {Giustino}}, \bibinfo {author} {\bibfnamefont {J.~H.}\ \bibnamefont {Lee}},
  \bibinfo {author} {\bibfnamefont {F.}~\bibnamefont {Trier}}, \bibinfo
  {author} {\bibfnamefont {M.}~\bibnamefont {Bibes}}, \bibinfo {author}
  {\bibfnamefont {S.~M.}\ \bibnamefont {Winter}}, \bibinfo {author}
  {\bibfnamefont {R.}~\bibnamefont {Valent{\'{\i}}}}, \bibinfo {author}
  {\bibfnamefont {Y.-W.}\ \bibnamefont {Son}}, \bibinfo {author} {\bibfnamefont
  {L.}~\bibnamefont {Taillefer}}, \bibinfo {author} {\bibfnamefont
  {C.}~\bibnamefont {Heil}}, \bibinfo {author} {\bibfnamefont {A.~I.}\
  \bibnamefont {Figueroa}}, \bibinfo {author} {\bibfnamefont {B.}~\bibnamefont
  {Pla{\c{c}}ais}}, \bibinfo {author} {\bibfnamefont {Q.}~\bibnamefont {Wu}},
  \bibinfo {author} {\bibfnamefont {O.~V.}\ \bibnamefont {Yazyev}}, \bibinfo
  {author} {\bibfnamefont {E.~P. A.~M.}\ \bibnamefont {Bakkers}}, \bibinfo
  {author} {\bibfnamefont {J.}~\bibnamefont {Nyg{\aa}rd}}, \bibinfo {author}
  {\bibfnamefont {P.}~\bibnamefont {Forn-D{\'{\i}}az}}, \bibinfo {author}
  {\bibfnamefont {S.~D.}\ \bibnamefont {Franceschi}}, \bibinfo {author}
  {\bibfnamefont {J.~W.}\ \bibnamefont {McIver}}, \bibinfo {author}
  {\bibfnamefont {L.~E. F.~F.}\ \bibnamefont {Torres}}, \bibinfo {author}
  {\bibfnamefont {T.}~\bibnamefont {Low}}, \bibinfo {author} {\bibfnamefont
  {A.}~\bibnamefont {Kumar}}, \bibinfo {author} {\bibfnamefont
  {R.}~\bibnamefont {Galceran}}, \bibinfo {author} {\bibfnamefont {S.~O.}\
  \bibnamefont {Valenzuela}}, \bibinfo {author} {\bibfnamefont {M.~V.}\
  \bibnamefont {Costache}}, \bibinfo {author} {\bibfnamefont {A.}~\bibnamefont
  {Manchon}}, \bibinfo {author} {\bibfnamefont {E.-A.}\ \bibnamefont {Kim}},
  \bibinfo {author} {\bibfnamefont {G.~R.}\ \bibnamefont {Schleder}}, \bibinfo
  {author} {\bibfnamefont {A.}~\bibnamefont {Fazzio}},\ and\ \bibinfo {author}
  {\bibfnamefont {S.}~\bibnamefont {Roche}},\ }\bibfield  {title} {\enquote
  {\bibinfo {title} {The 2021 quantum materials roadmap},}\ }\href
  {https://doi.org/10.1088/2515-7639/abb74e} {\bibfield  {journal} {\bibinfo
  {journal} {J. Phys. Mater.}\ }\textbf {\bibinfo {volume} {3}},\ \bibinfo
  {pages} {042006} (\bibinfo {year} {2020})}\BibitemShut {NoStop}%
\bibitem [{\citenamefont {Moll}(2018)}]{Moll_2018}%
  \BibitemOpen
  \bibfield  {author} {\bibinfo {author} {\bibfnamefont {P.~J.~W.}\
  \bibnamefont {Moll}},\ }\bibfield  {title} {\enquote {\bibinfo {title}
  {Focused ion beam microstructuring of quantum matter},}\ }\href
  {https://doi.org/10.1146/annurev-conmatphys-033117-054021} {\bibfield
  {journal} {\bibinfo  {journal} {Annu. Rev. Conden. Ma. P.}\ }\textbf
  {\bibinfo {volume} {9}},\ \bibinfo {pages} {147--162} (\bibinfo {year}
  {2018})}\BibitemShut {NoStop}%
\bibitem [{\citenamefont {Granata}\ and\ \citenamefont
  {Vettoliere}(2016)}]{Granata2016}%
  \BibitemOpen
  \bibfield  {author} {\bibinfo {author} {\bibfnamefont {C.}~\bibnamefont
  {Granata}}\ and\ \bibinfo {author} {\bibfnamefont {A.}~\bibnamefont
  {Vettoliere}},\ }\bibfield  {title} {\enquote {\bibinfo {title} {Nano
  superconducting quantum interference device: a powerful tool for nanoscale
  investigations},}\ }\href {https://doi.org/10.1016/j.physrep.2015.12.001}
  {\bibfield  {journal} {\bibinfo  {journal} {Phys. Rep.}\ }\textbf {\bibinfo
  {volume} {614}},\ \bibinfo {pages} {1--69} (\bibinfo {year}
  {2016})}\BibitemShut {NoStop}%
\bibitem [{\citenamefont {Mart{\'{\i}}nez-P{\'{e}}rez}\ and\ \citenamefont
  {Koelle}(2017)}]{MartinezPerez2017}%
  \BibitemOpen
  \bibfield  {author} {\bibinfo {author} {\bibfnamefont {M.~J.}\ \bibnamefont
  {Mart{\'{\i}}nez-P{\'{e}}rez}}\ and\ \bibinfo {author} {\bibfnamefont
  {D.}~\bibnamefont {Koelle}},\ }\bibfield  {title} {\enquote {\bibinfo {title}
  {{NanoSQUIDs}: basics \& recent advances},}\ }\href
  {https://doi.org/10.1515/psr-2017-5001} {\bibfield  {journal} {\bibinfo
  {journal} {Phys. Sci. Rev.}\ }\textbf {\bibinfo {volume} {2}},\ \bibinfo
  {pages} {20175001} (\bibinfo {year} {2017})}\BibitemShut {NoStop}%
\bibitem [{\citenamefont {Noah}\ \emph {et~al.}(2022)\citenamefont {Noah},
  \citenamefont {Toric}, \citenamefont {Feld}, \citenamefont {Zissman},
  \citenamefont {Gutfreund}, \citenamefont {Tsruya}, \citenamefont {Devidas},
  \citenamefont {Alpern}, \citenamefont {Vakahi}, \citenamefont {Steinberg},
  \citenamefont {Huber}, \citenamefont {Analytis}, \citenamefont {Gazit},
  \citenamefont {Lachman},\ and\ \citenamefont {Anahory}}]{Noah_2022}%
  \BibitemOpen
  \bibfield  {author} {\bibinfo {author} {\bibfnamefont {A.}~\bibnamefont
  {Noah}}, \bibinfo {author} {\bibfnamefont {F.}~\bibnamefont {Toric}},
  \bibinfo {author} {\bibfnamefont {T.~D.}\ \bibnamefont {Feld}}, \bibinfo
  {author} {\bibfnamefont {G.}~\bibnamefont {Zissman}}, \bibinfo {author}
  {\bibfnamefont {A.}~\bibnamefont {Gutfreund}}, \bibinfo {author}
  {\bibfnamefont {D.}~\bibnamefont {Tsruya}}, \bibinfo {author} {\bibfnamefont
  {T.~R.}\ \bibnamefont {Devidas}}, \bibinfo {author} {\bibfnamefont
  {H.}~\bibnamefont {Alpern}}, \bibinfo {author} {\bibfnamefont
  {A.}~\bibnamefont {Vakahi}}, \bibinfo {author} {\bibfnamefont
  {H.}~\bibnamefont {Steinberg}}, \bibinfo {author} {\bibfnamefont {M.~E.}\
  \bibnamefont {Huber}}, \bibinfo {author} {\bibfnamefont {J.~G.}\ \bibnamefont
  {Analytis}}, \bibinfo {author} {\bibfnamefont {S.}~\bibnamefont {Gazit}},
  \bibinfo {author} {\bibfnamefont {E.}~\bibnamefont {Lachman}},\ and\ \bibinfo
  {author} {\bibfnamefont {Y.}~\bibnamefont {Anahory}},\ }\bibfield  {title}
  {\enquote {\bibinfo {title} {Tunable exchange bias in the magnetic {W}eyl
  semimetal {Co$_3$Sn$_2$S$_2$}},}\ }\href
  {https://doi.org/10.1103/physrevb.105.144423} {\bibfield  {journal} {\bibinfo
   {journal} {Phys. Rev. B}\ }\textbf {\bibinfo {volume} {105}},\ \bibinfo
  {pages} {144423} (\bibinfo {year} {2022})}\BibitemShut {NoStop}%
\bibitem [{\citenamefont {Nair}\ \emph {et~al.}(2019)\citenamefont {Nair},
  \citenamefont {Maniv}, \citenamefont {John}, \citenamefont {Doyle},
  \citenamefont {Orenstein},\ and\ \citenamefont {Analytis}}]{Nair_2019}%
  \BibitemOpen
  \bibfield  {author} {\bibinfo {author} {\bibfnamefont {N.~L.}\ \bibnamefont
  {Nair}}, \bibinfo {author} {\bibfnamefont {E.}~\bibnamefont {Maniv}},
  \bibinfo {author} {\bibfnamefont {C.}~\bibnamefont {John}}, \bibinfo {author}
  {\bibfnamefont {S.}~\bibnamefont {Doyle}}, \bibinfo {author} {\bibfnamefont
  {J.}~\bibnamefont {Orenstein}},\ and\ \bibinfo {author} {\bibfnamefont
  {J.~G.}\ \bibnamefont {Analytis}},\ }\bibfield  {title} {\enquote {\bibinfo
  {title} {Electrical switching in a magnetically intercalated transition metal
  dichalcogenide},}\ }\href {https://doi.org/10.1038/s41563-019-0518-x}
  {\bibfield  {journal} {\bibinfo  {journal} {Nat. Mater.}\ }\textbf {\bibinfo
  {volume} {19}},\ \bibinfo {pages} {153--157} (\bibinfo {year}
  {2019})}\BibitemShut {NoStop}%
\bibitem [{\citenamefont {Nair}\ \emph {et~al.}(2020)\citenamefont {Nair},
  \citenamefont {Maniv}, \citenamefont {John}, \citenamefont {Doyle},
  \citenamefont {Orenstein},\ and\ \citenamefont {Analytis}}]{Nair_2020}%
  \BibitemOpen
  \bibfield  {author} {\bibinfo {author} {\bibfnamefont {N.~L.}\ \bibnamefont
  {Nair}}, \bibinfo {author} {\bibfnamefont {E.}~\bibnamefont {Maniv}},
  \bibinfo {author} {\bibfnamefont {C.}~\bibnamefont {John}}, \bibinfo {author}
  {\bibfnamefont {S.}~\bibnamefont {Doyle}}, \bibinfo {author} {\bibfnamefont
  {J.}~\bibnamefont {Orenstein}},\ and\ \bibinfo {author} {\bibfnamefont
  {J.~G.}\ \bibnamefont {Analytis}},\ }\bibfield  {title} {\enquote {\bibinfo
  {title} {Author correction: electrical switching in a magnetically
  intercalated transition metal dichalcogenide},}\ }\href
  {https://doi.org/10.1038/s41563-020-0776-7} {\bibfield  {journal} {\bibinfo
  {journal} {Nat. Mater.}\ }\textbf {\bibinfo {volume} {19}},\ \bibinfo {pages}
  {1036--1036} (\bibinfo {year} {2020})}\BibitemShut {NoStop}%
\bibitem [{\citenamefont {Ronning}\ \emph {et~al.}(2017)\citenamefont
  {Ronning}, \citenamefont {Helm}, \citenamefont {Shirer}, \citenamefont
  {Bachmann}, \citenamefont {Balicas}, \citenamefont {Chan}, \citenamefont
  {Ramshaw}, \citenamefont {McDonald}, \citenamefont {Balakirev}, \citenamefont
  {Jaime}, \citenamefont {Bauer},\ and\ \citenamefont {Moll}}]{Ronning_2017}%
  \BibitemOpen
  \bibfield  {author} {\bibinfo {author} {\bibfnamefont {F.}~\bibnamefont
  {Ronning}}, \bibinfo {author} {\bibfnamefont {T.}~\bibnamefont {Helm}},
  \bibinfo {author} {\bibfnamefont {K.~R.}\ \bibnamefont {Shirer}}, \bibinfo
  {author} {\bibfnamefont {M.~D.}\ \bibnamefont {Bachmann}}, \bibinfo {author}
  {\bibfnamefont {L.}~\bibnamefont {Balicas}}, \bibinfo {author} {\bibfnamefont
  {M.~K.}\ \bibnamefont {Chan}}, \bibinfo {author} {\bibfnamefont {B.~J.}\
  \bibnamefont {Ramshaw}}, \bibinfo {author} {\bibfnamefont {R.~D.}\
  \bibnamefont {McDonald}}, \bibinfo {author} {\bibfnamefont {F.~F.}\
  \bibnamefont {Balakirev}}, \bibinfo {author} {\bibfnamefont {M.}~\bibnamefont
  {Jaime}}, \bibinfo {author} {\bibfnamefont {E.~D.}\ \bibnamefont {Bauer}},\
  and\ \bibinfo {author} {\bibfnamefont {P.~J.~W.}\ \bibnamefont {Moll}},\
  }\bibfield  {title} {\enquote {\bibinfo {title} {Electronic in-plane symmetry
  breaking at field-tuned quantum criticality in {CeRhIn}5},}\ }\href
  {https://doi.org/10.1038/nature23315} {\bibfield  {journal} {\bibinfo
  {journal} {Nature}\ }\textbf {\bibinfo {volume} {548}},\ \bibinfo {pages}
  {313--317} (\bibinfo {year} {2017})}\BibitemShut {NoStop}%
\bibitem [{\citenamefont {Novotny}\ and\ \citenamefont
  {Hecht}(2012)}]{Novotny_2012}%
  \BibitemOpen
  \bibfield  {author} {\bibinfo {author} {\bibfnamefont {L.}~\bibnamefont
  {Novotny}}\ and\ \bibinfo {author} {\bibfnamefont {B.}~\bibnamefont
  {Hecht}},\ }\href {https://doi.org/10.1017/cbo9780511794193} {\emph {\bibinfo
  {title} {Principles of nano-optics}}}\ (\bibinfo  {publisher} {Cambridge
  University Press},\ \bibinfo {year} {2012})\BibitemShut {NoStop}%
\bibitem [{\citenamefont {Garcia-Vidal}\ \emph {et~al.}(2010)\citenamefont
  {Garcia-Vidal}, \citenamefont {Martin-Moreno}, \citenamefont {Ebbesen},\ and\
  \citenamefont {Kuipers}}]{GarciaVidal_2010}%
  \BibitemOpen
  \bibfield  {author} {\bibinfo {author} {\bibfnamefont {F.~J.}\ \bibnamefont
  {Garcia-Vidal}}, \bibinfo {author} {\bibfnamefont {L.}~\bibnamefont
  {Martin-Moreno}}, \bibinfo {author} {\bibfnamefont {T.~W.}\ \bibnamefont
  {Ebbesen}},\ and\ \bibinfo {author} {\bibfnamefont {L.}~\bibnamefont
  {Kuipers}},\ }\bibfield  {title} {\enquote {\bibinfo {title} {Light passing
  through subwavelength apertures},}\ }\href
  {https://doi.org/10.1103/revmodphys.82.729} {\bibfield  {journal} {\bibinfo
  {journal} {Rev. Mod. Phys.}\ }\textbf {\bibinfo {volume} {82}},\ \bibinfo
  {pages} {729--787} (\bibinfo {year} {2010})}\BibitemShut {NoStop}%
\bibitem [{\citenamefont {Novotny}\ and\ \citenamefont {van
  Hulst}(2011)}]{Novotny_2011}%
  \BibitemOpen
  \bibfield  {author} {\bibinfo {author} {\bibfnamefont {L.}~\bibnamefont
  {Novotny}}\ and\ \bibinfo {author} {\bibfnamefont {N.}~\bibnamefont {van
  Hulst}},\ }\bibfield  {title} {\enquote {\bibinfo {title} {Antennas for
  light},}\ }\href {https://doi.org/10.1038/nphoton.2010.237} {\bibfield
  {journal} {\bibinfo  {journal} {Nat. Photonics}\ }\textbf {\bibinfo {volume}
  {5}},\ \bibinfo {pages} {83--90} (\bibinfo {year} {2011})}\BibitemShut
  {NoStop}%
\bibitem [{\citenamefont {Kneipp}\ \emph {et~al.}(1997)\citenamefont {Kneipp},
  \citenamefont {Wang}, \citenamefont {Kneipp}, \citenamefont {Perelman},
  \citenamefont {Itzkan}, \citenamefont {Dasari},\ and\ \citenamefont
  {Feld}}]{Kneipp_1997}%
  \BibitemOpen
  \bibfield  {author} {\bibinfo {author} {\bibfnamefont {K.}~\bibnamefont
  {Kneipp}}, \bibinfo {author} {\bibfnamefont {Y.}~\bibnamefont {Wang}},
  \bibinfo {author} {\bibfnamefont {H.}~\bibnamefont {Kneipp}}, \bibinfo
  {author} {\bibfnamefont {L.~T.}\ \bibnamefont {Perelman}}, \bibinfo {author}
  {\bibfnamefont {I.}~\bibnamefont {Itzkan}}, \bibinfo {author} {\bibfnamefont
  {R.~R.}\ \bibnamefont {Dasari}},\ and\ \bibinfo {author} {\bibfnamefont
  {M.~S.}\ \bibnamefont {Feld}},\ }\bibfield  {title} {\enquote {\bibinfo
  {title} {Single molecule detection using surface-enhanced raman scattering
  ({SERS})},}\ }\href {https://doi.org/10.1103/physrevlett.78.1667} {\bibfield
  {journal} {\bibinfo  {journal} {Phys. Rev. Lett.}\ }\textbf {\bibinfo
  {volume} {78}},\ \bibinfo {pages} {1667--1670} (\bibinfo {year}
  {1997})}\BibitemShut {NoStop}%
\bibitem [{\citenamefont {Punj}\ \emph {et~al.}(2013)\citenamefont {Punj},
  \citenamefont {Mivelle}, \citenamefont {Moparthi}, \citenamefont {van
  Zanten}, \citenamefont {Rigneault}, \citenamefont {van Hulst}, \citenamefont
  {Garc{\'{\i}}a-Paraj{\'{o}}},\ and\ \citenamefont {Wenger}}]{Punj_2013}%
  \BibitemOpen
  \bibfield  {author} {\bibinfo {author} {\bibfnamefont {D.}~\bibnamefont
  {Punj}}, \bibinfo {author} {\bibfnamefont {M.}~\bibnamefont {Mivelle}},
  \bibinfo {author} {\bibfnamefont {S.~B.}\ \bibnamefont {Moparthi}}, \bibinfo
  {author} {\bibfnamefont {T.~S.}\ \bibnamefont {van Zanten}}, \bibinfo
  {author} {\bibfnamefont {H.}~\bibnamefont {Rigneault}}, \bibinfo {author}
  {\bibfnamefont {N.~F.}\ \bibnamefont {van Hulst}}, \bibinfo {author}
  {\bibfnamefont {M.~F.}\ \bibnamefont {Garc{\'{\i}}a-Paraj{\'{o}}}},\ and\
  \bibinfo {author} {\bibfnamefont {J.}~\bibnamefont {Wenger}},\ }\bibfield
  {title} {\enquote {\bibinfo {title} {A plasmonic `antenna-in-box' platform
  for enhanced single-molecule analysis at micromolar concentrations},}\ }\href
  {https://doi.org/10.1038/nnano.2013.98} {\bibfield  {journal} {\bibinfo
  {journal} {Nat. Nanotechnol.}\ }\textbf {\bibinfo {volume} {8}},\ \bibinfo
  {pages} {512--516} (\bibinfo {year} {2013})}\BibitemShut {NoStop}%
\bibitem [{\citenamefont {Ghenuche}\ \emph {et~al.}(2015)\citenamefont
  {Ghenuche}, \citenamefont {Mivelle}, \citenamefont {de~Torres}, \citenamefont
  {Moparthi}, \citenamefont {Rigneault}, \citenamefont {Hulst}, \citenamefont
  {Garc{\'{\i}}a-Paraj{\'{o}}},\ and\ \citenamefont {Wenger}}]{Ghenuche_2015}%
  \BibitemOpen
  \bibfield  {author} {\bibinfo {author} {\bibfnamefont {P.}~\bibnamefont
  {Ghenuche}}, \bibinfo {author} {\bibfnamefont {M.}~\bibnamefont {Mivelle}},
  \bibinfo {author} {\bibfnamefont {J.}~\bibnamefont {de~Torres}}, \bibinfo
  {author} {\bibfnamefont {S.~B.}\ \bibnamefont {Moparthi}}, \bibinfo {author}
  {\bibfnamefont {H.}~\bibnamefont {Rigneault}}, \bibinfo {author}
  {\bibfnamefont {N.~F.~V.}\ \bibnamefont {Hulst}}, \bibinfo {author}
  {\bibfnamefont {M.~F.}\ \bibnamefont {Garc{\'{\i}}a-Paraj{\'{o}}}},\ and\
  \bibinfo {author} {\bibfnamefont {J.}~\bibnamefont {Wenger}},\ }\bibfield
  {title} {\enquote {\bibinfo {title} {Matching nanoantenna field confinement
  to {FRET} distances enhances {F\"{o}}rster energy transfer rates},}\ }\href
  {https://doi.org/10.1021/acs.nanolett.5b02535} {\bibfield  {journal}
  {\bibinfo  {journal} {Nano Lett.}\ }\textbf {\bibinfo {volume} {15}},\
  \bibinfo {pages} {6193--6201} (\bibinfo {year} {2015})}\BibitemShut {NoStop}%
\bibitem [{\citenamefont {Huang}\ \emph {et~al.}(2010)\citenamefont {Huang},
  \citenamefont {Callegari}, \citenamefont {Geisler}, \citenamefont
  {Br{\"{u}}ning}, \citenamefont {Kern}, \citenamefont {Prangsma},
  \citenamefont {Wu}, \citenamefont {Feichtner}, \citenamefont {Ziegler},
  \citenamefont {Weinmann}, \citenamefont {Kamp}, \citenamefont {Forchel},
  \citenamefont {Biagioni}, \citenamefont {Sennhauser},\ and\ \citenamefont
  {Hecht}}]{Huang_2010}%
  \BibitemOpen
  \bibfield  {author} {\bibinfo {author} {\bibfnamefont {J.-S.}\ \bibnamefont
  {Huang}}, \bibinfo {author} {\bibfnamefont {V.}~\bibnamefont {Callegari}},
  \bibinfo {author} {\bibfnamefont {P.}~\bibnamefont {Geisler}}, \bibinfo
  {author} {\bibfnamefont {C.}~\bibnamefont {Br{\"{u}}ning}}, \bibinfo {author}
  {\bibfnamefont {J.}~\bibnamefont {Kern}}, \bibinfo {author} {\bibfnamefont
  {J.~C.}\ \bibnamefont {Prangsma}}, \bibinfo {author} {\bibfnamefont
  {X.}~\bibnamefont {Wu}}, \bibinfo {author} {\bibfnamefont {T.}~\bibnamefont
  {Feichtner}}, \bibinfo {author} {\bibfnamefont {J.}~\bibnamefont {Ziegler}},
  \bibinfo {author} {\bibfnamefont {P.}~\bibnamefont {Weinmann}}, \bibinfo
  {author} {\bibfnamefont {M.}~\bibnamefont {Kamp}}, \bibinfo {author}
  {\bibfnamefont {A.}~\bibnamefont {Forchel}}, \bibinfo {author} {\bibfnamefont
  {P.}~\bibnamefont {Biagioni}}, \bibinfo {author} {\bibfnamefont
  {U.}~\bibnamefont {Sennhauser}},\ and\ \bibinfo {author} {\bibfnamefont
  {B.}~\bibnamefont {Hecht}},\ }\bibfield  {title} {\enquote {\bibinfo {title}
  {Atomically flat single-crystalline gold nanostructures for plasmonic
  nanocircuitry},}\ }\href {https://doi.org/10.1038/ncomms1143} {\bibfield
  {journal} {\bibinfo  {journal} {Nat. Commun.}\ }\textbf {\bibinfo {volume}
  {1}} (\bibinfo {year} {2010}),\ 10.1038/ncomms1143}\BibitemShut {NoStop}%
\bibitem [{\citenamefont {Kern}\ \emph {et~al.}(2015)\citenamefont {Kern},
  \citenamefont {Kullock}, \citenamefont {Prangsma}, \citenamefont {Emmerling},
  \citenamefont {Kamp},\ and\ \citenamefont {Hecht}}]{Kern_2015}%
  \BibitemOpen
  \bibfield  {author} {\bibinfo {author} {\bibfnamefont {J.}~\bibnamefont
  {Kern}}, \bibinfo {author} {\bibfnamefont {R.}~\bibnamefont {Kullock}},
  \bibinfo {author} {\bibfnamefont {J.}~\bibnamefont {Prangsma}}, \bibinfo
  {author} {\bibfnamefont {M.}~\bibnamefont {Emmerling}}, \bibinfo {author}
  {\bibfnamefont {M.}~\bibnamefont {Kamp}},\ and\ \bibinfo {author}
  {\bibfnamefont {B.}~\bibnamefont {Hecht}},\ }\bibfield  {title} {\enquote
  {\bibinfo {title} {Electrically driven optical antennas},}\ }\href
  {https://doi.org/10.1038/nphoton.2015.141} {\bibfield  {journal} {\bibinfo
  {journal} {Nat. Photonics}\ }\textbf {\bibinfo {volume} {9}},\ \bibinfo
  {pages} {582--586} (\bibinfo {year} {2015})}\BibitemShut {NoStop}%
\bibitem [{\citenamefont {Sch{\"{o}}rner}, \citenamefont {Adhikari},\ and\
  \citenamefont {Lippitz}(2019)}]{Schoerner_2019}%
  \BibitemOpen
  \bibfield  {author} {\bibinfo {author} {\bibfnamefont {C.}~\bibnamefont
  {Sch{\"{o}}rner}}, \bibinfo {author} {\bibfnamefont {S.}~\bibnamefont
  {Adhikari}},\ and\ \bibinfo {author} {\bibfnamefont {M.}~\bibnamefont
  {Lippitz}},\ }\bibfield  {title} {\enquote {\bibinfo {title} {A
  single-crystalline silver plasmonic circuit for visible quantum emitters},}\
  }\href {https://doi.org/10.1021/acs.nanolett.9b00773} {\bibfield  {journal}
  {\bibinfo  {journal} {Nano Lett.}\ }\textbf {\bibinfo {volume} {19}},\
  \bibinfo {pages} {3238--3243} (\bibinfo {year} {2019})}\BibitemShut {NoStop}%
\bibitem [{\citenamefont {Ochs}\ \emph {et~al.}(2021)\citenamefont {Ochs},
  \citenamefont {Zurak}, \citenamefont {Krauss}, \citenamefont {Meier},
  \citenamefont {Emmerling}, \citenamefont {Kullock},\ and\ \citenamefont
  {Hecht}}]{Ochs_2021a}%
  \BibitemOpen
  \bibfield  {author} {\bibinfo {author} {\bibfnamefont {M.}~\bibnamefont
  {Ochs}}, \bibinfo {author} {\bibfnamefont {L.}~\bibnamefont {Zurak}},
  \bibinfo {author} {\bibfnamefont {E.}~\bibnamefont {Krauss}}, \bibinfo
  {author} {\bibfnamefont {J.}~\bibnamefont {Meier}}, \bibinfo {author}
  {\bibfnamefont {M.}~\bibnamefont {Emmerling}}, \bibinfo {author}
  {\bibfnamefont {R.}~\bibnamefont {Kullock}},\ and\ \bibinfo {author}
  {\bibfnamefont {B.}~\bibnamefont {Hecht}},\ }\bibfield  {title} {\enquote
  {\bibinfo {title} {Nanoscale electrical excitation of distinct modes in
  plasmonic waveguides},}\ }\href
  {https://doi.org/10.1021/acs.nanolett.1c00182} {\bibfield  {journal}
  {\bibinfo  {journal} {Nano Lett.}\ }\textbf {\bibinfo {volume} {21}},\
  \bibinfo {pages} {4225--4230} (\bibinfo {year} {2021})}\BibitemShut {NoStop}%
\bibitem [{\citenamefont {Hor{\'{a}}k}\ \emph {et~al.}(2018)\citenamefont
  {Hor{\'{a}}k}, \citenamefont {Bukvi{\v{s}}ov{\'{a}}}, \citenamefont
  {{\v{S}}varc}, \citenamefont {Jaskowiec}, \citenamefont
  {K{\v{r}}{\'{a}}pek},\ and\ \citenamefont {{\v{S}}ikola}}]{Horak_2018}%
  \BibitemOpen
  \bibfield  {author} {\bibinfo {author} {\bibfnamefont {M.}~\bibnamefont
  {Hor{\'{a}}k}}, \bibinfo {author} {\bibfnamefont {K.}~\bibnamefont
  {Bukvi{\v{s}}ov{\'{a}}}}, \bibinfo {author} {\bibfnamefont {V.}~\bibnamefont
  {{\v{S}}varc}}, \bibinfo {author} {\bibfnamefont {J.}~\bibnamefont
  {Jaskowiec}}, \bibinfo {author} {\bibfnamefont {V.}~\bibnamefont
  {K{\v{r}}{\'{a}}pek}},\ and\ \bibinfo {author} {\bibfnamefont
  {T.}~\bibnamefont {{\v{S}}ikola}},\ }\bibfield  {title} {\enquote {\bibinfo
  {title} {Comparative study of plasmonic antennas fabricated by electron beam
  and focused ion beam lithography},}\ }\href
  {https://doi.org/10.1038/s41598-018-28037-1} {\bibfield  {journal} {\bibinfo
  {journal} {Sci. Rep.}\ }\textbf {\bibinfo {volume} {8}} (\bibinfo {year}
  {2018}),\ 10.1038/s41598-018-28037-1}\BibitemShut {NoStop}%
\bibitem [{\citenamefont {Chen}, \citenamefont {Yang},\ and\ \citenamefont
  {Gao}(2019)}]{Chen_2019}%
  \BibitemOpen
  \bibfield  {author} {\bibinfo {author} {\bibfnamefont {Y.}~\bibnamefont
  {Chen}}, \bibinfo {author} {\bibfnamefont {X.}~\bibnamefont {Yang}},\ and\
  \bibinfo {author} {\bibfnamefont {J.}~\bibnamefont {Gao}},\ }\bibfield
  {title} {\enquote {\bibinfo {title} {3{D} {J}anus plasmonic helical
  nanoapertures for polarization-encrypted data storage},}\ }\href
  {https://doi.org/10.1038/s41377-019-0156-8} {\bibfield  {journal} {\bibinfo
  {journal} {Light Sci. Appl.}\ }\textbf {\bibinfo {volume} {8}} (\bibinfo
  {year} {2019}),\ 10.1038/s41377-019-0156-8}\BibitemShut {NoStop}%
\bibitem [{\citenamefont {Xia}\ \emph {et~al.}(2006)\citenamefont {Xia},
  \citenamefont {Wu}, \citenamefont {Xu}, \citenamefont {Hao},\ and\
  \citenamefont {Wang}}]{Xia_2006}%
  \BibitemOpen
  \bibfield  {author} {\bibinfo {author} {\bibfnamefont {L.}~\bibnamefont
  {Xia}}, \bibinfo {author} {\bibfnamefont {W.}~\bibnamefont {Wu}}, \bibinfo
  {author} {\bibfnamefont {J.}~\bibnamefont {Xu}}, \bibinfo {author}
  {\bibfnamefont {Y.}~\bibnamefont {Hao}},\ and\ \bibinfo {author}
  {\bibfnamefont {Y.}~\bibnamefont {Wang}},\ }\bibfield  {title} {\enquote
  {\bibinfo {title} {{3D} nanohelix fabrication and 3d nanometer assembly by
  focused ion beam stress-introducing technique},}\ }in\ \href
  {https://doi.org/10.1109/memsys.2006.1627750} {\emph {\bibinfo {booktitle}
  {19th {IEEE} International Conference on Micro Electro Mechanical Systems}}}\
  (\bibinfo  {publisher} {{IEEE}},\ \bibinfo {year} {2006})\BibitemShut
  {NoStop}%
\bibitem [{\citenamefont {Chen}\ \emph {et~al.}(2016)\citenamefont {Chen},
  \citenamefont {Bi}, \citenamefont {Wang}, \citenamefont {Zheng},
  \citenamefont {Liu}, \citenamefont {Han}, \citenamefont {Yang}, \citenamefont
  {Chang}, \citenamefont {Zhang},\ and\ \citenamefont {Duan}}]{Chen_2016a}%
  \BibitemOpen
  \bibfield  {author} {\bibinfo {author} {\bibfnamefont {Y.}~\bibnamefont
  {Chen}}, \bibinfo {author} {\bibfnamefont {K.}~\bibnamefont {Bi}}, \bibinfo
  {author} {\bibfnamefont {Q.}~\bibnamefont {Wang}}, \bibinfo {author}
  {\bibfnamefont {M.}~\bibnamefont {Zheng}}, \bibinfo {author} {\bibfnamefont
  {Q.}~\bibnamefont {Liu}}, \bibinfo {author} {\bibfnamefont {Y.}~\bibnamefont
  {Han}}, \bibinfo {author} {\bibfnamefont {J.}~\bibnamefont {Yang}}, \bibinfo
  {author} {\bibfnamefont {S.}~\bibnamefont {Chang}}, \bibinfo {author}
  {\bibfnamefont {G.}~\bibnamefont {Zhang}},\ and\ \bibinfo {author}
  {\bibfnamefont {H.}~\bibnamefont {Duan}},\ }\bibfield  {title} {\enquote
  {\bibinfo {title} {Rapid focused ion beam milling based fabrication of
  plasmonic nanoparticles and assemblies via ``sketch and peel'' strategy},}\
  }\href {https://doi.org/10.1021/acsnano.6b06290} {\bibfield  {journal}
  {\bibinfo  {journal} {{ACS} Nano}\ }\textbf {\bibinfo {volume} {10}},\
  \bibinfo {pages} {11228--11236} (\bibinfo {year} {2016})}\BibitemShut
  {NoStop}%
\bibitem [{\citenamefont {Laible}\ \emph {et~al.}(2019)\citenamefont {Laible},
  \citenamefont {Dreser}, \citenamefont {Kern},\ and\ \citenamefont
  {Fleischer}}]{Laible_2019}%
  \BibitemOpen
  \bibfield  {author} {\bibinfo {author} {\bibfnamefont {F.}~\bibnamefont
  {Laible}}, \bibinfo {author} {\bibfnamefont {C.}~\bibnamefont {Dreser}},
  \bibinfo {author} {\bibfnamefont {D.~P.}\ \bibnamefont {Kern}},\ and\
  \bibinfo {author} {\bibfnamefont {M.}~\bibnamefont {Fleischer}},\ }\bibfield
  {title} {\enquote {\bibinfo {title} {Time-effective strategies for the
  fabrication of poly- and single-crystalline gold nano-structures by focused
  helium ion beam milling},}\ }\href {https://doi.org/10.1088/1361-6528/ab0506}
  {\bibfield  {journal} {\bibinfo  {journal} {Nanotechnology}\ }\textbf
  {\bibinfo {volume} {30}},\ \bibinfo {pages} {235302} (\bibinfo {year}
  {2019})}\BibitemShut {NoStop}%
\bibitem [{\citenamefont {Frey}\ \emph {et~al.}(2002)\citenamefont {Frey},
  \citenamefont {Keilmann}, \citenamefont {Kriele},\ and\ \citenamefont
  {Guckenberger}}]{Frey_2002}%
  \BibitemOpen
  \bibfield  {author} {\bibinfo {author} {\bibfnamefont {H.~G.}\ \bibnamefont
  {Frey}}, \bibinfo {author} {\bibfnamefont {F.}~\bibnamefont {Keilmann}},
  \bibinfo {author} {\bibfnamefont {A.}~\bibnamefont {Kriele}},\ and\ \bibinfo
  {author} {\bibfnamefont {R.}~\bibnamefont {Guckenberger}},\ }\bibfield
  {title} {\enquote {\bibinfo {title} {Enhancing the resolution of scanning
  near-field optical microscopy by a metal tip grown on an aperture probe},}\
  }\href {https://doi.org/10.1063/1.1530736} {\bibfield  {journal} {\bibinfo
  {journal} {Appl. Phys. Lett.}\ }\textbf {\bibinfo {volume} {81}},\ \bibinfo
  {pages} {5030--5032} (\bibinfo {year} {2002})}\BibitemShut {NoStop}%
\bibitem [{\citenamefont {Frey}\ \emph {et~al.}(2004)\citenamefont {Frey},
  \citenamefont {Witt}, \citenamefont {Felderer},\ and\ \citenamefont
  {Guckenberger}}]{Frey_2004}%
  \BibitemOpen
  \bibfield  {author} {\bibinfo {author} {\bibfnamefont {H.~G.}\ \bibnamefont
  {Frey}}, \bibinfo {author} {\bibfnamefont {S.}~\bibnamefont {Witt}}, \bibinfo
  {author} {\bibfnamefont {K.}~\bibnamefont {Felderer}},\ and\ \bibinfo
  {author} {\bibfnamefont {R.}~\bibnamefont {Guckenberger}},\ }\bibfield
  {title} {\enquote {\bibinfo {title} {High-resolution imaging of single
  fluorescent molecules with the optical near-field of a metal tip},}\ }\href
  {https://doi.org/10.1103/physrevlett.93.200801} {\bibfield  {journal}
  {\bibinfo  {journal} {Phys. Rev. Lett.}\ }\textbf {\bibinfo {volume} {93}},\
  \bibinfo {pages} {200801} (\bibinfo {year} {2004})}\BibitemShut {NoStop}%
\bibitem [{\citenamefont {Bernas}(2010)}]{Bernas2010}%
  \BibitemOpen
  \bibinfo {editor} {\bibfnamefont {H.}~\bibnamefont {Bernas}},\ ed.,\ \href
  {https://doi.org/10.1007/978-3-540-88789-8} {\emph {\bibinfo {title}
  {Materials science with ion beams}}}\ (\bibinfo  {publisher} {Springer Berlin
  Heidelberg},\ \bibinfo {year} {2010})\BibitemShut {NoStop}%
\bibitem [{\citenamefont {Zhang}\ and\ \citenamefont
  {Weber}(2020)}]{Zhang2020}%
  \BibitemOpen
  \bibfield  {author} {\bibinfo {author} {\bibfnamefont {Y.}~\bibnamefont
  {Zhang}}\ and\ \bibinfo {author} {\bibfnamefont {W.~J.}\ \bibnamefont
  {Weber}},\ }\bibfield  {title} {\enquote {\bibinfo {title} {Ion irradiation
  and modification: the role of coupled electronic and nuclear energy
  dissipation and subsequent nonequilibrium processes in materials},}\ }\href
  {https://doi.org/10.1063/5.0027462} {\bibfield  {journal} {\bibinfo
  {journal} {Appl. Phys. Rev.}\ }\textbf {\bibinfo {volume} {7}},\ \bibinfo
  {pages} {041307} (\bibinfo {year} {2020})}\BibitemShut {NoStop}%
\bibitem [{\citenamefont {Kinchin}\ and\ \citenamefont
  {Pease}(1955)}]{Kinchin1955}%
  \BibitemOpen
  \bibfield  {author} {\bibinfo {author} {\bibfnamefont {G.~H.}\ \bibnamefont
  {Kinchin}}\ and\ \bibinfo {author} {\bibfnamefont {R.~S.}\ \bibnamefont
  {Pease}},\ }\bibfield  {title} {{\selectlanguage {english}\enquote {\bibinfo
  {title} {The displacement of atoms in solids by radiation},}\ }}\href
  {https://doi.org/10.1088/0034-4885/18/1/301} {\bibfield  {journal} {\bibinfo
  {journal} {Rep. Prog. Phys.}\ }\textbf {\bibinfo {volume} {18}},\ \bibinfo
  {pages} {1} (\bibinfo {year} {1955})}\BibitemShut {NoStop}%
\bibitem [{\citenamefont {Morimoto}\ \emph {et~al.}(1996)\citenamefont
  {Morimoto}, \citenamefont {Weeks}, \citenamefont {Barnes}, \citenamefont
  {Tolk},\ and\ \citenamefont {Zuhr}}]{Morimoto1996}%
  \BibitemOpen
  \bibfield  {author} {\bibinfo {author} {\bibfnamefont {Y.}~\bibnamefont
  {Morimoto}}, \bibinfo {author} {\bibfnamefont {R.~A.}\ \bibnamefont {Weeks}},
  \bibinfo {author} {\bibfnamefont {A.~V.}\ \bibnamefont {Barnes}}, \bibinfo
  {author} {\bibfnamefont {N.~H.}\ \bibnamefont {Tolk}},\ and\ \bibinfo
  {author} {\bibfnamefont {R.~A.}\ \bibnamefont {Zuhr}},\ }\bibfield  {title}
  {\enquote {\bibinfo {title} {The effect of ion implantation on luminescence
  of a silica},}\ }\href {https://doi.org/10.1016/0022-3093(95)00559-5}
  {\bibfield  {journal} {\bibinfo  {journal} {J. Non-Cryst. Solids}\ }\textbf
  {\bibinfo {volume} {196}},\ \bibinfo {pages} {106--112} (\bibinfo {year}
  {1996})}\BibitemShut {NoStop}%
\bibitem [{\citenamefont {Chen}\ and\ \citenamefont
  {Schrempel}(2016)}]{Chen_2016}%
  \BibitemOpen
  \bibfield  {author} {\bibinfo {author} {\bibfnamefont {F.}~\bibnamefont
  {Chen}}\ and\ \bibinfo {author} {\bibfnamefont {F.}~\bibnamefont
  {Schrempel}},\ }\bibfield  {title} {\enquote {\bibinfo {title} {Modification
  of structure and properties of optical crystals},}\ }in\ \href
  {https://doi.org/10.1007/978-3-319-33561-2_13} {\emph {\bibinfo {booktitle}
  {Ion Beam Modification of Solids: Ion-Solid Interaction and Radiation
  Damage}}},\ \bibinfo {editor} {edited by\ \bibinfo {editor} {\bibfnamefont
  {W.}~\bibnamefont {Wesch}}\ and\ \bibinfo {editor} {\bibfnamefont
  {E.}~\bibnamefont {Wendler}}}\ (\bibinfo  {publisher} {Springer International
  Publishing},\ \bibinfo {address} {Cham},\ \bibinfo {year} {2016})\ pp.\
  \bibinfo {pages} {501--528}\BibitemShut {NoStop}%
\bibitem [{\citenamefont {Ziegler}, \citenamefont {Ziegler},\ and\
  \citenamefont {Biersack}(2010)}]{Ziegler2010}%
  \BibitemOpen
  \bibfield  {author} {\bibinfo {author} {\bibfnamefont {J.~F.}\ \bibnamefont
  {Ziegler}}, \bibinfo {author} {\bibnamefont {Ziegler}},\ and\ \bibinfo
  {author} {\bibfnamefont {J.~P.}\ \bibnamefont {Biersack}},\ }\bibfield
  {title} {\enquote {\bibinfo {title} {{SRIM} -- the stopping and range of ions
  in matter (2010)},}\ }\href {https://doi.org/10.1016/j.nimb.2010.02.091}
  {\bibfield  {journal} {\bibinfo  {journal} {Nucl. Instrum. Meth. Phys. Res.
  B}\ }\textbf {\bibinfo {volume} {268}},\ \bibinfo {pages} {1818--1823}
  (\bibinfo {year} {2010})}\BibitemShut {NoStop}%
\bibitem [{\citenamefont {Nordlund}\ \emph {et~al.}(1998)\citenamefont
  {Nordlund}, \citenamefont {Ghaly}, \citenamefont {Averback}, \citenamefont
  {Caturla}, \citenamefont {Rubia},\ and\ \citenamefont
  {Tarus}}]{nordlund_defect_1998}%
  \BibitemOpen
  \bibfield  {author} {\bibinfo {author} {\bibfnamefont {K.}~\bibnamefont
  {Nordlund}}, \bibinfo {author} {\bibfnamefont {M.}~\bibnamefont {Ghaly}},
  \bibinfo {author} {\bibfnamefont {R.~S.}\ \bibnamefont {Averback}}, \bibinfo
  {author} {\bibfnamefont {M.}~\bibnamefont {Caturla}}, \bibinfo {author}
  {\bibfnamefont {T.~D. d.~l.}\ \bibnamefont {Rubia}},\ and\ \bibinfo {author}
  {\bibfnamefont {J.}~\bibnamefont {Tarus}},\ }\bibfield  {title} {\enquote
  {\bibinfo {title} {Defect production in collision cascades in elemental
  semiconductors and fcc metals},}\ }\href
  {https://doi.org/10.1103/physrevb.57.7556} {\bibfield  {journal} {\bibinfo
  {journal} {Phys. Rev. B}\ }\textbf {\bibinfo {volume} {57}},\ \bibinfo
  {pages} {7556--7570} (\bibinfo {year} {1998})}\BibitemShut {NoStop}%
\bibitem [{\citenamefont {Pelaz}, \citenamefont {Marqu{\'{e}}s},\ and\
  \citenamefont {Barbolla}(2004)}]{pelaz_ion-beam-induced_2004}%
  \BibitemOpen
  \bibfield  {author} {\bibinfo {author} {\bibfnamefont {L.}~\bibnamefont
  {Pelaz}}, \bibinfo {author} {\bibfnamefont {L.~A.}\ \bibnamefont
  {Marqu{\'{e}}s}},\ and\ \bibinfo {author} {\bibfnamefont {J.}~\bibnamefont
  {Barbolla}},\ }\bibfield  {title} {\enquote {\bibinfo {title}
  {Ion-beam-induced amorphization and recrystallization of silicon},}\ }\href
  {https://doi.org/10.1063/1.1808484} {\bibfield  {journal} {\bibinfo
  {journal} {J. Appl. Phys.}\ }\textbf {\bibinfo {volume} {96}},\ \bibinfo
  {pages} {5947--5976} (\bibinfo {year} {2004})}\BibitemShut {NoStop}%
\bibitem [{\citenamefont {Gierak}(2009)}]{Gierak_2009}%
  \BibitemOpen
  \bibfield  {author} {\bibinfo {author} {\bibfnamefont {J.}~\bibnamefont
  {Gierak}},\ }\bibfield  {title} {{\selectlanguage {english}\enquote {\bibinfo
  {title} {Focused ion beam technology and ultimate applications},}\ }}\href
  {https://doi.org/10.1088/0268-1242/24/4/043001} {\bibfield  {journal}
  {\bibinfo  {journal} {Semicond. Sci. Tech.}\ }\textbf {\bibinfo {volume}
  {24}},\ \bibinfo {pages} {043001} (\bibinfo {year} {2009})}\BibitemShut
  {NoStop}%
\bibitem [{\citenamefont {Or{\'{u}}s}, \citenamefont {C{\'{o}}rdoba},\ and\
  \citenamefont {De~Teresa}(2020)}]{Orus_2020}%
  \BibitemOpen
  \bibfield  {author} {\bibinfo {author} {\bibfnamefont {P.}~\bibnamefont
  {Or{\'{u}}s}}, \bibinfo {author} {\bibfnamefont {R.}~\bibnamefont
  {C{\'{o}}rdoba}},\ and\ \bibinfo {author} {\bibfnamefont {J.~M.}\
  \bibnamefont {De~Teresa}},\ }\bibfield  {title} {\enquote {\bibinfo {title}
  {Focused ion beam induced processing},}\ }in\ \href
  {https://doi.org/10.1088/978-0-7503-2608-7ch5} {\emph {\bibinfo {booktitle}
  {Nanofabrication}}}\ (\bibinfo  {publisher} {{IOP} Publishing},\ \bibinfo
  {year} {2020})\ pp.\ \bibinfo {pages} {5--1 to 5--58}\BibitemShut {NoStop}%
\bibitem [{\citenamefont {Nakaharai}\ \emph {et~al.}(2013)\citenamefont
  {Nakaharai}, \citenamefont {Iijima}, \citenamefont {Ogawa}, \citenamefont
  {Suzuki}, \citenamefont {Li}, \citenamefont {Tsukagoshi}, \citenamefont
  {Sato},\ and\ \citenamefont {Yokoyama}}]{Nakaharai2013}%
  \BibitemOpen
  \bibfield  {author} {\bibinfo {author} {\bibfnamefont {S.}~\bibnamefont
  {Nakaharai}}, \bibinfo {author} {\bibfnamefont {T.}~\bibnamefont {Iijima}},
  \bibinfo {author} {\bibfnamefont {S.}~\bibnamefont {Ogawa}}, \bibinfo
  {author} {\bibfnamefont {S.}~\bibnamefont {Suzuki}}, \bibinfo {author}
  {\bibfnamefont {S.-L.}\ \bibnamefont {Li}}, \bibinfo {author} {\bibfnamefont
  {K.}~\bibnamefont {Tsukagoshi}}, \bibinfo {author} {\bibfnamefont
  {S.}~\bibnamefont {Sato}},\ and\ \bibinfo {author} {\bibfnamefont
  {N.}~\bibnamefont {Yokoyama}},\ }\bibfield  {title} {\enquote {\bibinfo
  {title} {Conduction tuning of graphene based on defect-induced
  localization},}\ }\href {https://doi.org/10.1021/nn401992q} {\bibfield
  {journal} {\bibinfo  {journal} {ACS Nano}\ }\textbf {\bibinfo {volume} {7}},\
  \bibinfo {pages} {5694--5700} (\bibinfo {year} {2013})}\BibitemShut {NoStop}%
\bibitem [{\citenamefont {Moktadir}, \citenamefont {Hang},\ and\ \citenamefont
  {Mizuta}(2015)}]{Moktadir2015}%
  \BibitemOpen
  \bibfield  {author} {\bibinfo {author} {\bibfnamefont {Z.}~\bibnamefont
  {Moktadir}}, \bibinfo {author} {\bibfnamefont {S.}~\bibnamefont {Hang}},\
  and\ \bibinfo {author} {\bibfnamefont {H.}~\bibnamefont {Mizuta}},\
  }\bibfield  {title} {\enquote {\bibinfo {title} {Defect-induced fermi level
  pinning and suppression of ambipolar behaviour in graphene},}\ }\href
  {https://doi.org/10.1016/j.carbon.2015.05.049} {\bibfield  {journal}
  {\bibinfo  {journal} {Carbon}\ }\textbf {\bibinfo {volume} {93}},\ \bibinfo
  {pages} {325--334} (\bibinfo {year} {2015})}\BibitemShut {NoStop}%
\bibitem [{\citenamefont {Araujo}\ \emph {et~al.}(2015)\citenamefont {Araujo},
  \citenamefont {Brant}, \citenamefont {Archanjo}, \citenamefont
  {Medeiros-Ribeiro}, \citenamefont {Plentz},\ and\ \citenamefont
  {Alves}}]{Araujo2015}%
  \BibitemOpen
  \bibfield  {author} {\bibinfo {author} {\bibfnamefont {E.~N.~D.}\
  \bibnamefont {Araujo}}, \bibinfo {author} {\bibfnamefont {J.~C.}\
  \bibnamefont {Brant}}, \bibinfo {author} {\bibfnamefont {B.~S.}\ \bibnamefont
  {Archanjo}}, \bibinfo {author} {\bibfnamefont {G.}~\bibnamefont
  {Medeiros-Ribeiro}}, \bibinfo {author} {\bibfnamefont {F.}~\bibnamefont
  {Plentz}},\ and\ \bibinfo {author} {\bibfnamefont {E.~S.}\ \bibnamefont
  {Alves}},\ }\bibfield  {title} {\enquote {\bibinfo {title} {Patterning
  graphene with a helium ion microscope: observation of metal-insulator
  transition induced by disorder},}\ }\href
  {https://doi.org/10.1103/PhysRevB.91.245414} {\bibfield  {journal} {\bibinfo
  {journal} {Phys. Rev. B}\ }\textbf {\bibinfo {volume} {91}},\ \bibinfo
  {pages} {245414} (\bibinfo {year} {2015})}\BibitemShut {NoStop}%
\bibitem [{\citenamefont {Zhou}\ \emph {et~al.}(2016)\citenamefont {Zhou},
  \citenamefont {Maguire}, \citenamefont {Jadwiszczak}, \citenamefont
  {Muruganathan}, \citenamefont {Mizuta},\ and\ \citenamefont
  {Zhang}}]{Zhou2016}%
  \BibitemOpen
  \bibfield  {author} {\bibinfo {author} {\bibfnamefont {Y.}~\bibnamefont
  {Zhou}}, \bibinfo {author} {\bibfnamefont {P.}~\bibnamefont {Maguire}},
  \bibinfo {author} {\bibfnamefont {J.}~\bibnamefont {Jadwiszczak}}, \bibinfo
  {author} {\bibfnamefont {M.}~\bibnamefont {Muruganathan}}, \bibinfo {author}
  {\bibfnamefont {H.}~\bibnamefont {Mizuta}},\ and\ \bibinfo {author}
  {\bibfnamefont {H.}~\bibnamefont {Zhang}},\ }\bibfield  {title} {\enquote
  {\bibinfo {title} {Precise milling of nano-gap chains in graphene with a
  focused helium ion beam},}\ }\href
  {https://doi.org/10.1088/0957-4484/27/32/325302} {\bibfield  {journal}
  {\bibinfo  {journal} {Nanotechnology}\ }\textbf {\bibinfo {volume} {27}},\
  \bibinfo {pages} {325302} (\bibinfo {year} {2016})}\BibitemShut {NoStop}%
\bibitem [{\citenamefont {Naitou}\ and\ \citenamefont
  {Ogawa}(2017)}]{Naitou2017}%
  \BibitemOpen
  \bibfield  {author} {\bibinfo {author} {\bibfnamefont {Y.}~\bibnamefont
  {Naitou}}\ and\ \bibinfo {author} {\bibfnamefont {S.}~\bibnamefont {Ogawa}},\
  }\bibfield  {title} {\enquote {\bibinfo {title} {Conductivity change of
  defective graphene by helium ion beams},}\ }\href
  {https://doi.org/10.1063/1.4979983} {\bibfield  {journal} {\bibinfo
  {journal} {AIP Adv.}\ }\textbf {\bibinfo {volume} {7}},\ \bibinfo {pages}
  {045203} (\bibinfo {year} {2017})}\BibitemShut {NoStop}%
\bibitem [{\citenamefont {Nanda}\ \emph {et~al.}(2017)\citenamefont {Nanda},
  \citenamefont {Hlawacek}, \citenamefont {Goswami}, \citenamefont {Watanabe},
  \citenamefont {Taniguchi},\ and\ \citenamefont {Alkemade}}]{Nanda2017}%
  \BibitemOpen
  \bibfield  {author} {\bibinfo {author} {\bibfnamefont {G.}~\bibnamefont
  {Nanda}}, \bibinfo {author} {\bibfnamefont {G.}~\bibnamefont {Hlawacek}},
  \bibinfo {author} {\bibfnamefont {S.}~\bibnamefont {Goswami}}, \bibinfo
  {author} {\bibfnamefont {K.}~\bibnamefont {Watanabe}}, \bibinfo {author}
  {\bibfnamefont {T.}~\bibnamefont {Taniguchi}},\ and\ \bibinfo {author}
  {\bibfnamefont {P.~F.~A.}\ \bibnamefont {Alkemade}},\ }\bibfield  {title}
  {\enquote {\bibinfo {title} {{Electronic transport in helium-ion-beam etched
  encapsulated graphene nanoribbons}},}\ }\href
  {https://doi.org/10.1016/j.carbon.2017.04.062} {\bibfield  {journal}
  {\bibinfo  {journal} {Carbon}\ }\textbf {\bibinfo {volume} {119}},\ \bibinfo
  {pages} {419--425} (\bibinfo {year} {2017})}\BibitemShut {NoStop}%
\bibitem [{\citenamefont {Fox}\ \emph {et~al.}(2015)\citenamefont {Fox},
  \citenamefont {Zhou}, \citenamefont {Maguire}, \citenamefont {O'Neill},
  \citenamefont {{\'O}'Coile{\'a}in}, \citenamefont {Gatensby}, \citenamefont
  {Glushenkov}, \citenamefont {Tao}, \citenamefont {Duesberg}, \citenamefont
  {Shvets}, \citenamefont {Abid}, \citenamefont {Abid}, \citenamefont {Wu},
  \citenamefont {Chen}, \citenamefont {Coleman}, \citenamefont {Donegan},\ and\
  \citenamefont {Zhang}}]{Fox-2015}%
  \BibitemOpen
  \bibfield  {author} {\bibinfo {author} {\bibfnamefont {D.~S.}\ \bibnamefont
  {Fox}}, \bibinfo {author} {\bibfnamefont {Y.}~\bibnamefont {Zhou}}, \bibinfo
  {author} {\bibfnamefont {P.}~\bibnamefont {Maguire}}, \bibinfo {author}
  {\bibfnamefont {A.}~\bibnamefont {O'Neill}}, \bibinfo {author} {\bibfnamefont
  {C.}~\bibnamefont {{\'O}'Coile{\'a}in}}, \bibinfo {author} {\bibfnamefont
  {R.}~\bibnamefont {Gatensby}}, \bibinfo {author} {\bibfnamefont {A.~M.}\
  \bibnamefont {Glushenkov}}, \bibinfo {author} {\bibfnamefont
  {T.}~\bibnamefont {Tao}}, \bibinfo {author} {\bibfnamefont {G.~S.}\
  \bibnamefont {Duesberg}}, \bibinfo {author} {\bibfnamefont {I.~V.}\
  \bibnamefont {Shvets}}, \bibinfo {author} {\bibfnamefont {M.}~\bibnamefont
  {Abid}}, \bibinfo {author} {\bibfnamefont {M.}~\bibnamefont {Abid}}, \bibinfo
  {author} {\bibfnamefont {H.-C.}\ \bibnamefont {Wu}}, \bibinfo {author}
  {\bibfnamefont {Y.}~\bibnamefont {Chen}}, \bibinfo {author} {\bibfnamefont
  {J.~N.}\ \bibnamefont {Coleman}}, \bibinfo {author} {\bibfnamefont {J.~F.}\
  \bibnamefont {Donegan}},\ and\ \bibinfo {author} {\bibfnamefont
  {H.}~\bibnamefont {Zhang}},\ }\bibfield  {title} {{\selectlanguage
  {english}\enquote {\bibinfo {title} {Nanopatterning and electrical tuning of
  {MoS2} layers with a subnanometer helium ion beam},}\ }}\href
  {https://doi.org/10.1021/acs.nanolett.5b01673} {\bibfield  {journal}
  {\bibinfo  {journal} {Nano Lett.}\ }\textbf {\bibinfo {volume} {15}},\
  \bibinfo {pages} {5307--5313} (\bibinfo {year} {2015})}\BibitemShut {NoStop}%
\bibitem [{\citenamefont {Stanford}\ \emph
  {et~al.}(2017{\natexlab{a}})\citenamefont {Stanford}, \citenamefont
  {Pudasaini}, \citenamefont {Gallmeier}, \citenamefont {Cross}, \citenamefont
  {Liang}, \citenamefont {Oyedele}, \citenamefont {Duscher}, \citenamefont
  {Mahjouri-Samani}, \citenamefont {Wang}, \citenamefont {Xiao}, \citenamefont
  {Geohegan}, \citenamefont {Belianinov}, \citenamefont {Sumpter},\ and\
  \citenamefont {Rack}}]{Stanford2017}%
  \BibitemOpen
  \bibfield  {author} {\bibinfo {author} {\bibfnamefont {M.~G.}\ \bibnamefont
  {Stanford}}, \bibinfo {author} {\bibfnamefont {P.~R.}\ \bibnamefont
  {Pudasaini}}, \bibinfo {author} {\bibfnamefont {E.~T.}\ \bibnamefont
  {Gallmeier}}, \bibinfo {author} {\bibfnamefont {N.}~\bibnamefont {Cross}},
  \bibinfo {author} {\bibfnamefont {L.}~\bibnamefont {Liang}}, \bibinfo
  {author} {\bibfnamefont {A.}~\bibnamefont {Oyedele}}, \bibinfo {author}
  {\bibfnamefont {G.}~\bibnamefont {Duscher}}, \bibinfo {author} {\bibfnamefont
  {M.}~\bibnamefont {Mahjouri-Samani}}, \bibinfo {author} {\bibfnamefont
  {K.}~\bibnamefont {Wang}}, \bibinfo {author} {\bibfnamefont {K.}~\bibnamefont
  {Xiao}}, \bibinfo {author} {\bibfnamefont {D.~B.}\ \bibnamefont {Geohegan}},
  \bibinfo {author} {\bibfnamefont {A.}~\bibnamefont {Belianinov}}, \bibinfo
  {author} {\bibfnamefont {B.~G.}\ \bibnamefont {Sumpter}},\ and\ \bibinfo
  {author} {\bibfnamefont {P.~D.}\ \bibnamefont {Rack}},\ }\bibfield  {title}
  {\enquote {\bibinfo {title} {High conduction hopping behavior induced in
  transition metal dichalcogenides by percolating defect networks: toward
  atomically thin circuits},}\ }\href {https://doi.org/10.1002/adfm.201702829}
  {\bibfield  {journal} {\bibinfo  {journal} {Adv. Funct. Mater.}\ }\textbf
  {\bibinfo {volume} {27}},\ \bibinfo {pages} {1702829} (\bibinfo {year}
  {2017}{\natexlab{a}})}\BibitemShut {NoStop}%
\bibitem [{\citenamefont {Jadwiszczak}\ \emph {et~al.}(2019)\citenamefont
  {Jadwiszczak}, \citenamefont {Keane}, \citenamefont {Maguire}, \citenamefont
  {Cullen}, \citenamefont {Zhou}, \citenamefont {Song}, \citenamefont
  {Downing}, \citenamefont {Fox}, \citenamefont {McEvoy}, \citenamefont {Zhu},
  \citenamefont {Xu}, \citenamefont {Duesberg}, \citenamefont {Liao},
  \citenamefont {Boland},\ and\ \citenamefont {Zhang}}]{Jadwiszczak-2019}%
  \BibitemOpen
  \bibfield  {author} {\bibinfo {author} {\bibfnamefont {J.}~\bibnamefont
  {Jadwiszczak}}, \bibinfo {author} {\bibfnamefont {D.}~\bibnamefont {Keane}},
  \bibinfo {author} {\bibfnamefont {P.}~\bibnamefont {Maguire}}, \bibinfo
  {author} {\bibfnamefont {C.~P.}\ \bibnamefont {Cullen}}, \bibinfo {author}
  {\bibfnamefont {Y.}~\bibnamefont {Zhou}}, \bibinfo {author} {\bibfnamefont
  {H.}~\bibnamefont {Song}}, \bibinfo {author} {\bibfnamefont {C.}~\bibnamefont
  {Downing}}, \bibinfo {author} {\bibfnamefont {D.}~\bibnamefont {Fox}},
  \bibinfo {author} {\bibfnamefont {N.}~\bibnamefont {McEvoy}}, \bibinfo
  {author} {\bibfnamefont {R.}~\bibnamefont {Zhu}}, \bibinfo {author}
  {\bibfnamefont {J.}~\bibnamefont {Xu}}, \bibinfo {author} {\bibfnamefont
  {G.~S.}\ \bibnamefont {Duesberg}}, \bibinfo {author} {\bibfnamefont {Z.-M.}\
  \bibnamefont {Liao}}, \bibinfo {author} {\bibfnamefont {J.~J.}\ \bibnamefont
  {Boland}},\ and\ \bibinfo {author} {\bibfnamefont {H.}~\bibnamefont
  {Zhang}},\ }\bibfield  {title} {\enquote {\bibinfo {title} {{MoS2}
  memtransistors fabricated by localized helium ion beam irradiation},}\ }\href
  {https://doi.org/10.1021/acsnano.9b07421} {\bibfield  {journal} {\bibinfo
  {journal} {ACS Nano}\ }\textbf {\bibinfo {volume} {13}},\ \bibinfo {pages}
  {14262--14273} (\bibinfo {year} {2019})}\BibitemShut {NoStop}%
\bibitem [{\citenamefont {Parida}\ \emph {et~al.}(2022)\citenamefont {Parida},
  \citenamefont {Wang}, \citenamefont {Zhao}, \citenamefont {Htoon},
  \citenamefont {Kucinski}, \citenamefont {Chubarov}, \citenamefont
  {Choudhury}, \citenamefont {Redwing}, \citenamefont {Dongare},\ and\
  \citenamefont {Pettes}}]{Parida_2022}%
  \BibitemOpen
  \bibfield  {author} {\bibinfo {author} {\bibfnamefont {S.}~\bibnamefont
  {Parida}}, \bibinfo {author} {\bibfnamefont {Y.}~\bibnamefont {Wang}},
  \bibinfo {author} {\bibfnamefont {H.}~\bibnamefont {Zhao}}, \bibinfo {author}
  {\bibfnamefont {H.}~\bibnamefont {Htoon}}, \bibinfo {author} {\bibfnamefont
  {T.~M.}\ \bibnamefont {Kucinski}}, \bibinfo {author} {\bibfnamefont
  {M.}~\bibnamefont {Chubarov}}, \bibinfo {author} {\bibfnamefont
  {T.}~\bibnamefont {Choudhury}}, \bibinfo {author} {\bibfnamefont {J.~M.}\
  \bibnamefont {Redwing}}, \bibinfo {author} {\bibfnamefont {A.}~\bibnamefont
  {Dongare}},\ and\ \bibinfo {author} {\bibfnamefont {M.~T.}\ \bibnamefont
  {Pettes}},\ }\bibfield  {title} {\enquote {\bibinfo {title} {Tuning of the
  electronic and vibrational properties of epitaxial {MoS}$_2$ through he-ion
  beam modification},}\ }\href {https://doi.org/10.1088/1361-6528/aca3af}
  {\bibfield  {journal} {\bibinfo  {journal} {Nanotechnology}\ }\textbf
  {\bibinfo {volume} {34}},\ \bibinfo {pages} {085702} (\bibinfo {year}
  {2022})}\BibitemShut {NoStop}%
\bibitem [{\citenamefont {Stanford}\ \emph
  {et~al.}(2016{\natexlab{a}})\citenamefont {Stanford}, \citenamefont
  {Pudasaini}, \citenamefont {Belianinov}, \citenamefont {Cross}, \citenamefont
  {Noh}, \citenamefont {Koehler}, \citenamefont {Mandrus}, \citenamefont
  {Duscher}, \citenamefont {Rondinone}, \citenamefont {Ivanov}, \citenamefont
  {Ward},\ and\ \citenamefont {Rack}}]{Stanford-2016}%
  \BibitemOpen
  \bibfield  {author} {\bibinfo {author} {\bibfnamefont {M.~G.}\ \bibnamefont
  {Stanford}}, \bibinfo {author} {\bibfnamefont {P.~R.}\ \bibnamefont
  {Pudasaini}}, \bibinfo {author} {\bibfnamefont {A.}~\bibnamefont
  {Belianinov}}, \bibinfo {author} {\bibfnamefont {N.}~\bibnamefont {Cross}},
  \bibinfo {author} {\bibfnamefont {J.~H.}\ \bibnamefont {Noh}}, \bibinfo
  {author} {\bibfnamefont {M.~R.}\ \bibnamefont {Koehler}}, \bibinfo {author}
  {\bibfnamefont {D.~G.}\ \bibnamefont {Mandrus}}, \bibinfo {author}
  {\bibfnamefont {G.}~\bibnamefont {Duscher}}, \bibinfo {author} {\bibfnamefont
  {A.~J.}\ \bibnamefont {Rondinone}}, \bibinfo {author} {\bibfnamefont {I.~N.}\
  \bibnamefont {Ivanov}}, \bibinfo {author} {\bibfnamefont {T.~Z.}\
  \bibnamefont {Ward}},\ and\ \bibinfo {author} {\bibfnamefont {P.~D.}\
  \bibnamefont {Rack}},\ }\bibfield  {title} {\enquote {\bibinfo {title}
  {Focused helium-ion beam irradiation effects on electrical transport
  properties of few-layer {WSe2}: enabling nanoscale direct write
  homo-junctions},}\ }\href {https://doi.org/10.1038/srep27276} {\bibfield
  {journal} {\bibinfo  {journal} {Sci. Rep.}\ }\textbf {\bibinfo {volume}
  {6}},\ \bibinfo {pages} {27276} (\bibinfo {year}
  {2016}{\natexlab{a}})}\BibitemShut {NoStop}%
\bibitem [{\citenamefont {Stanford}\ \emph
  {et~al.}(2017{\natexlab{b}})\citenamefont {Stanford}, \citenamefont {Noh},
  \citenamefont {Mahady}, \citenamefont {Ievlev}, \citenamefont {Maksymovych},
  \citenamefont {Ovchinnikova},\ and\ \citenamefont {Rack}}]{Stanford2017b}%
  \BibitemOpen
  \bibfield  {author} {\bibinfo {author} {\bibfnamefont {M.~G.}\ \bibnamefont
  {Stanford}}, \bibinfo {author} {\bibfnamefont {J.~H.}\ \bibnamefont {Noh}},
  \bibinfo {author} {\bibfnamefont {K.}~\bibnamefont {Mahady}}, \bibinfo
  {author} {\bibfnamefont {A.~V.}\ \bibnamefont {Ievlev}}, \bibinfo {author}
  {\bibfnamefont {P.}~\bibnamefont {Maksymovych}}, \bibinfo {author}
  {\bibfnamefont {O.~S.}\ \bibnamefont {Ovchinnikova}},\ and\ \bibinfo {author}
  {\bibfnamefont {P.~D.}\ \bibnamefont {Rack}},\ }\bibfield  {title} {\enquote
  {\bibinfo {title} {{Room-Temperature} activation of {InGaZnO} {Thin-Film}
  transistors via he+ irradiation},}\ }\href
  {https://doi.org/10.1021/acsami.7b10449} {\bibfield  {journal} {\bibinfo
  {journal} {ACS Appl. Mater. Interfaces}\ }\textbf {\bibinfo {volume} {9}},\
  \bibinfo {pages} {35125--35132} (\bibinfo {year}
  {2017}{\natexlab{b}})}\BibitemShut {NoStop}%
\bibitem [{\citenamefont {Hirayama}\ \emph {et~al.}(1988)\citenamefont
  {Hirayama}, \citenamefont {Tarucha}, \citenamefont {Suzuki},\ and\
  \citenamefont {Okamoto}}]{Hirayama_1988a}%
  \BibitemOpen
  \bibfield  {author} {\bibinfo {author} {\bibfnamefont {Y.}~\bibnamefont
  {Hirayama}}, \bibinfo {author} {\bibfnamefont {S.}~\bibnamefont {Tarucha}},
  \bibinfo {author} {\bibfnamefont {Y.}~\bibnamefont {Suzuki}},\ and\ \bibinfo
  {author} {\bibfnamefont {H.}~\bibnamefont {Okamoto}},\ }\bibfield  {title}
  {\enquote {\bibinfo {title} {Fabrication of a {GaAs} quantum-well-wire
  structure by ga focused-ion-beam implantation and its optical properties},}\
  }\href {https://doi.org/10.1103/physrevb.37.2774} {\bibfield  {journal}
  {\bibinfo  {journal} {Phys. Rev. B}\ }\textbf {\bibinfo {volume} {37}},\
  \bibinfo {pages} {2774--2777} (\bibinfo {year} {1988})}\BibitemShut {NoStop}%
\bibitem [{\citenamefont {Itoh}, \citenamefont {Saku},\ and\ \citenamefont
  {Tarucha}(1992)}]{Itoh_1992}%
  \BibitemOpen
  \bibfield  {author} {\bibinfo {author} {\bibfnamefont {M.}~\bibnamefont
  {Itoh}}, \bibinfo {author} {\bibfnamefont {T.~S.~T.}\ \bibnamefont {Saku}},\
  and\ \bibinfo {author} {\bibfnamefont {S.~T.~S.}\ \bibnamefont {Tarucha}},\
  }\bibfield  {title} {{\selectlanguage {english}\enquote {\bibinfo {title}
  {{High-Mobility} quantum wires fabricated by ga focused ion beam shallow
  implantation},}\ }}\href {https://doi.org/10.1143/JJAP.31.4487} {\bibfield
  {journal} {\bibinfo  {journal} {Jpn. J. Appl. Phys.}\ }\textbf {\bibinfo
  {volume} {31}},\ \bibinfo {pages} {4487} (\bibinfo {year}
  {1992})}\BibitemShut {NoStop}%
\bibitem [{\citenamefont {Wieck}\ and\ \citenamefont
  {Ploog}(1990)}]{Wieck_1990}%
  \BibitemOpen
  \bibfield  {author} {\bibinfo {author} {\bibfnamefont {A.~D.}\ \bibnamefont
  {Wieck}}\ and\ \bibinfo {author} {\bibfnamefont {K.}~\bibnamefont {Ploog}},\
  }\bibfield  {title} {\enquote {\bibinfo {title} {In‐plane‐gated quantum
  wire transistor fabricated with directly written focused ion beams},}\ }\href
  {https://doi.org/10.1063/1.102628} {\bibfield  {journal} {\bibinfo  {journal}
  {Appl. Phys. Lett.}\ }\textbf {\bibinfo {volume} {56}},\ \bibinfo {pages}
  {928--930} (\bibinfo {year} {1990})}\BibitemShut {NoStop}%
\bibitem [{\citenamefont {Nakata}\ \emph {et~al.}(1990)\citenamefont {Nakata},
  \citenamefont {Yamada}, \citenamefont {Hirayama}, \citenamefont {Saku},\ and\
  \citenamefont {Horikoshi}}]{Nakata1990}%
  \BibitemOpen
  \bibfield  {author} {\bibinfo {author} {\bibfnamefont {S.}~\bibnamefont
  {Nakata}}, \bibinfo {author} {\bibfnamefont {S.}~\bibnamefont {Yamada}},
  \bibinfo {author} {\bibfnamefont {Y.}~\bibnamefont {Hirayama}}, \bibinfo
  {author} {\bibfnamefont {T.}~\bibnamefont {Saku}},\ and\ \bibinfo {author}
  {\bibfnamefont {Y.}~\bibnamefont {Horikoshi}},\ }\bibfield  {title} {\enquote
  {\bibinfo {title} {Fabrication of quantum wires by {Ga} focused-ion-beam
  implantation and their transport properties},}\ }\href
  {https://doi.org/10.1143/jjap.29.48} {\bibfield  {journal} {\bibinfo
  {journal} {Jpn. J. Appl. Phys.}\ }\textbf {\bibinfo {volume} {29}},\ \bibinfo
  {pages} {48} (\bibinfo {year} {1990})}\BibitemShut {NoStop}%
\bibitem [{\citenamefont {Hirayama}, \citenamefont {Saku},\ and\ \citenamefont
  {Horikoshi}(1989)}]{Hirayama_1989}%
  \BibitemOpen
  \bibfield  {author} {\bibinfo {author} {\bibfnamefont {Y.}~\bibnamefont
  {Hirayama}}, \bibinfo {author} {\bibfnamefont {T.}~\bibnamefont {Saku}},\
  and\ \bibinfo {author} {\bibfnamefont {Y.}~\bibnamefont {Horikoshi}},\
  }\bibfield  {title} {{\selectlanguage {english}\enquote {\bibinfo {title}
  {Electronic transport through very short and narrow channels constricted in
  {GaAs} by highly resistive ga-implanted regions},}\ }}\href
  {https://doi.org/10.1103/physrevb.39.5535} {\bibfield  {journal} {\bibinfo
  {journal} {Phys. Rev. B}\ }\textbf {\bibinfo {volume} {39}},\ \bibinfo
  {pages} {5535--5537} (\bibinfo {year} {1989})}\BibitemShut {NoStop}%
\bibitem [{\citenamefont {Dhara}\ \emph {et~al.}(2010)\citenamefont {Dhara},
  \citenamefont {Lu}, \citenamefont {Wu}, \citenamefont {Hsu}, \citenamefont
  {Tu}, \citenamefont {Chen}, \citenamefont {Wang}, \citenamefont {Chen},\ and\
  \citenamefont {Raj}}]{Dhara_2010}%
  \BibitemOpen
  \bibfield  {author} {\bibinfo {author} {\bibfnamefont {S.}~\bibnamefont
  {Dhara}}, \bibinfo {author} {\bibfnamefont {C.~Y.}\ \bibnamefont {Lu}},
  \bibinfo {author} {\bibfnamefont {C.~T.}\ \bibnamefont {Wu}}, \bibinfo
  {author} {\bibfnamefont {C.~W.}\ \bibnamefont {Hsu}}, \bibinfo {author}
  {\bibfnamefont {W.~S.}\ \bibnamefont {Tu}}, \bibinfo {author} {\bibfnamefont
  {K.~H.}\ \bibnamefont {Chen}}, \bibinfo {author} {\bibfnamefont {Y.~L.}\
  \bibnamefont {Wang}}, \bibinfo {author} {\bibfnamefont {L.~C.}\ \bibnamefont
  {Chen}},\ and\ \bibinfo {author} {\bibfnamefont {B.}~\bibnamefont {Raj}},\
  }\bibfield  {title} {\enquote {\bibinfo {title} {Focused ion beam induced
  nanojunction and defect doping as a building block for nanoscale electronics
  in {GaN} nanowires},}\ }\href {https://doi.org/10.1021/jp1020998} {\bibfield
  {journal} {\bibinfo  {journal} {J. Phys. Chem. C}\ }\textbf {\bibinfo
  {volume} {114}},\ \bibinfo {pages} {15260--15265} (\bibinfo {year}
  {2010})}\BibitemShut {NoStop}%
\bibitem [{\citenamefont {Liu}\ \emph {et~al.}(2021{\natexlab{b}})\citenamefont
  {Liu}, \citenamefont {Qu}, \citenamefont {Liu}, \citenamefont {Yin},
  \citenamefont {Liu}, \citenamefont {Tan},\ and\ \citenamefont
  {Chen}}]{Liu2021}%
  \BibitemOpen
  \bibfield  {author} {\bibinfo {author} {\bibfnamefont {Y.}~\bibnamefont
  {Liu}}, \bibinfo {author} {\bibfnamefont {Y.}~\bibnamefont {Qu}}, \bibinfo
  {author} {\bibfnamefont {Y.}~\bibnamefont {Liu}}, \bibinfo {author}
  {\bibfnamefont {H.}~\bibnamefont {Yin}}, \bibinfo {author} {\bibfnamefont
  {J.}~\bibnamefont {Liu}}, \bibinfo {author} {\bibfnamefont {Y.}~\bibnamefont
  {Tan}},\ and\ \bibinfo {author} {\bibfnamefont {F.}~\bibnamefont {Chen}},\
  }\bibfield  {title} {\enquote {\bibinfo {title} {Direct-{{Writing}} of {{2D
  Diodes}} by {{Focused Ion Beams}}},}\ }\href
  {https://doi.org/10.1002/adfm.202102708} {\bibfield  {journal} {\bibinfo
  {journal} {Adv. Funct. Mater.}\ }\textbf {\bibinfo {volume} {31}},\ \bibinfo
  {pages} {2102708} (\bibinfo {year} {2021}{\natexlab{b}})}\BibitemShut
  {NoStop}%
\bibitem [{\citenamefont {Mykk{\"a}nen}\ \emph {et~al.}(2020)\citenamefont
  {Mykk{\"a}nen}, \citenamefont {Bera}, \citenamefont {Lehtinen}, \citenamefont
  {Ronzani}, \citenamefont {Kohop{\"a}{\"a}}, \citenamefont
  {H{\"o}nigl-Decrinis}, \citenamefont {Shaikhaidarov}, \citenamefont
  {de~Graaf}, \citenamefont {Govenius},\ and\ \citenamefont
  {Prunnila}}]{Mykkanen2020}%
  \BibitemOpen
  \bibfield  {author} {\bibinfo {author} {\bibfnamefont {E.}~\bibnamefont
  {Mykk{\"a}nen}}, \bibinfo {author} {\bibfnamefont {A.}~\bibnamefont {Bera}},
  \bibinfo {author} {\bibfnamefont {J.~S.}\ \bibnamefont {Lehtinen}}, \bibinfo
  {author} {\bibfnamefont {A.}~\bibnamefont {Ronzani}}, \bibinfo {author}
  {\bibfnamefont {K.}~\bibnamefont {Kohop{\"a}{\"a}}}, \bibinfo {author}
  {\bibfnamefont {T.}~\bibnamefont {H{\"o}nigl-Decrinis}}, \bibinfo {author}
  {\bibfnamefont {R.}~\bibnamefont {Shaikhaidarov}}, \bibinfo {author}
  {\bibfnamefont {S.~E.}\ \bibnamefont {de~Graaf}}, \bibinfo {author}
  {\bibfnamefont {J.}~\bibnamefont {Govenius}},\ and\ \bibinfo {author}
  {\bibfnamefont {M.}~\bibnamefont {Prunnila}},\ }\bibfield  {title} {\enquote
  {\bibinfo {title} {Enhancement of {{Superconductivity}} by {{Amorphizing
  Molybdenum Silicide Films Using}} a {{Focused Ion Beam}}},}\ }\href
  {https://doi.org/10.3390/nano10050950} {\bibfield  {journal} {\bibinfo
  {journal} {Nanomaterials}\ }\textbf {\bibinfo {volume} {10}},\ \bibinfo
  {pages} {950} (\bibinfo {year} {2020})}\BibitemShut {NoStop}%
\bibitem [{\citenamefont {Sosa}\ \emph {et~al.}(2010)\citenamefont {Sosa},
  \citenamefont {Chen}, \citenamefont {Liu}, \citenamefont {Xie}, \citenamefont
  {Marks},\ and\ \citenamefont {Hersam}}]{Sosa2010}%
  \BibitemOpen
  \bibfield  {author} {\bibinfo {author} {\bibfnamefont {N.~E.}\ \bibnamefont
  {Sosa}}, \bibinfo {author} {\bibfnamefont {C.}~\bibnamefont {Chen}}, \bibinfo
  {author} {\bibfnamefont {J.}~\bibnamefont {Liu}}, \bibinfo {author}
  {\bibfnamefont {S.}~\bibnamefont {Xie}}, \bibinfo {author} {\bibfnamefont
  {T.~J.}\ \bibnamefont {Marks}},\ and\ \bibinfo {author} {\bibfnamefont
  {M.~C.}\ \bibnamefont {Hersam}},\ }\bibfield  {title} {{\selectlanguage
  {english}\enquote {\bibinfo {title} {Nanoscale structure, composition, and
  charge transport analysis of transparent conducting oxide nanowires written
  by focused ion beam implantation},}\ }}\href
  {https://doi.org/10.1021/ja9092242} {\bibfield  {journal} {\bibinfo
  {journal} {J. Am. Chem. Soc.}\ }\textbf {\bibinfo {volume} {132}},\ \bibinfo
  {pages} {7347--7354} (\bibinfo {year} {2010})}\BibitemShut {NoStop}%
\bibitem [{\citenamefont {Bianconi}, \citenamefont {Park},\ and\ \citenamefont
  {Mohseni}(2019)}]{Bianconi2019}%
  \BibitemOpen
  \bibfield  {author} {\bibinfo {author} {\bibfnamefont {S.}~\bibnamefont
  {Bianconi}}, \bibinfo {author} {\bibfnamefont {M.-S.}\ \bibnamefont {Park}},\
  and\ \bibinfo {author} {\bibfnamefont {H.}~\bibnamefont {Mohseni}},\
  }\bibfield  {title} {\enquote {\bibinfo {title} {Giant conductivity
  modulation of aluminum oxide using focused ion beam},}\ }\href
  {https://doi.org/10.1021/acsaelm.9b00185} {\bibfield  {journal} {\bibinfo
  {journal} {ACS Appl. Electron. Mater.}\ }\textbf {\bibinfo {volume} {1}},\
  \bibinfo {pages} {1208--1214} (\bibinfo {year} {2019})}\BibitemShut {NoStop}%
\bibitem [{\citenamefont {Mei}\ \emph {et~al.}(2022)\citenamefont {Mei},
  \citenamefont {Koch}, \citenamefont {Wan}, \citenamefont {Rensberg},
  \citenamefont {Zhang}, \citenamefont {Salman}, \citenamefont {Hafermann},
  \citenamefont {Schaal}, \citenamefont {Xiao}, \citenamefont {Wambold},
  \citenamefont {Ramanathan}, \citenamefont {Ronning},\ and\ \citenamefont
  {Kats}}]{Mei_2022}%
  \BibitemOpen
  \bibfield  {author} {\bibinfo {author} {\bibfnamefont {H.}~\bibnamefont
  {Mei}}, \bibinfo {author} {\bibfnamefont {A.}~\bibnamefont {Koch}}, \bibinfo
  {author} {\bibfnamefont {C.}~\bibnamefont {Wan}}, \bibinfo {author}
  {\bibfnamefont {J.}~\bibnamefont {Rensberg}}, \bibinfo {author}
  {\bibfnamefont {Z.}~\bibnamefont {Zhang}}, \bibinfo {author} {\bibfnamefont
  {J.}~\bibnamefont {Salman}}, \bibinfo {author} {\bibfnamefont
  {M.}~\bibnamefont {Hafermann}}, \bibinfo {author} {\bibfnamefont
  {M.}~\bibnamefont {Schaal}}, \bibinfo {author} {\bibfnamefont
  {Y.}~\bibnamefont {Xiao}}, \bibinfo {author} {\bibfnamefont {R.}~\bibnamefont
  {Wambold}}, \bibinfo {author} {\bibfnamefont {S.}~\bibnamefont {Ramanathan}},
  \bibinfo {author} {\bibfnamefont {C.}~\bibnamefont {Ronning}},\ and\ \bibinfo
  {author} {\bibfnamefont {M.~A.}\ \bibnamefont {Kats}},\ }\bibfield  {title}
  {\enquote {\bibinfo {title} {Tuning carrier density and phase transitions in
  oxide semiconductors using focused ion beams},}\ }\href
  {https://doi.org/10.1515/nanoph-2022-0050} {\bibfield  {journal} {\bibinfo
  {journal} {Nanophotonics}\ }\textbf {\bibinfo {volume} {11}},\ \bibinfo
  {pages} {3923--3932} (\bibinfo {year} {2022})}\BibitemShut {NoStop}%
\bibitem [{\citenamefont {Hiramoto}\ \emph {et~al.}(1989)\citenamefont
  {Hiramoto}, \citenamefont {Hirakawa}, \citenamefont {Iye},\ and\
  \citenamefont {Ikoma}}]{Hiramoto1989}%
  \BibitemOpen
  \bibfield  {author} {\bibinfo {author} {\bibfnamefont {T.}~\bibnamefont
  {Hiramoto}}, \bibinfo {author} {\bibfnamefont {K.}~\bibnamefont {Hirakawa}},
  \bibinfo {author} {\bibfnamefont {Y.}~\bibnamefont {Iye}},\ and\ \bibinfo
  {author} {\bibfnamefont {T.}~\bibnamefont {Ikoma}},\ }\bibfield  {title}
  {\enquote {\bibinfo {title} {Phase coherence length of electron waves in
  narrow {AlGaAs}/{GaAs} quantum wires fabricated by focused ion beam
  implantation},}\ }\href {https://doi.org/10.1063/1.101177} {\bibfield
  {journal} {\bibinfo  {journal} {Appl. Phys. Lett.}\ }\textbf {\bibinfo
  {volume} {54}},\ \bibinfo {pages} {2103--2105} (\bibinfo {year}
  {1989})}\BibitemShut {NoStop}%
\bibitem [{\citenamefont {Cybart}\ \emph {et~al.}(2015)\citenamefont {Cybart},
  \citenamefont {Cho}, \citenamefont {Wong}, \citenamefont {Wehlin},
  \citenamefont {Ma}, \citenamefont {Huynh},\ and\ \citenamefont
  {Dynes}}]{Cybart2015}%
  \BibitemOpen
  \bibfield  {author} {\bibinfo {author} {\bibfnamefont {S.~A.}\ \bibnamefont
  {Cybart}}, \bibinfo {author} {\bibfnamefont {E.~Y.}\ \bibnamefont {Cho}},
  \bibinfo {author} {\bibfnamefont {T.~J.}\ \bibnamefont {Wong}}, \bibinfo
  {author} {\bibfnamefont {B.~H.}\ \bibnamefont {Wehlin}}, \bibinfo {author}
  {\bibfnamefont {M.~K.}\ \bibnamefont {Ma}}, \bibinfo {author} {\bibfnamefont
  {C.}~\bibnamefont {Huynh}},\ and\ \bibinfo {author} {\bibfnamefont {R.~C.}\
  \bibnamefont {Dynes}},\ }\bibfield  {title} {{\selectlanguage
  {english}\enquote {\bibinfo {title} {Nano josephson superconducting tunnel
  junctions in {YBa2Cu3O(7-$\delta$}) directly patterned with a focused helium
  ion beam},}\ }}\href {https://doi.org/10.1038/nnano.2015.76} {\bibfield
  {journal} {\bibinfo  {journal} {Nat. Nanotechnol.}\ }\textbf {\bibinfo
  {volume} {10}},\ \bibinfo {pages} {598--602} (\bibinfo {year}
  {2015})}\BibitemShut {NoStop}%
\bibitem [{\citenamefont {Cho}\ \emph {et~al.}(2018)\citenamefont {Cho},
  \citenamefont {Zhou}, \citenamefont {Cho},\ and\ \citenamefont
  {Cybart}}]{Cho2018}%
  \BibitemOpen
  \bibfield  {author} {\bibinfo {author} {\bibfnamefont {E.~Y.}\ \bibnamefont
  {Cho}}, \bibinfo {author} {\bibfnamefont {Y.~W.}\ \bibnamefont {Zhou}},
  \bibinfo {author} {\bibfnamefont {J.~Y.}\ \bibnamefont {Cho}},\ and\ \bibinfo
  {author} {\bibfnamefont {S.~A.}\ \bibnamefont {Cybart}},\ }\bibfield  {title}
  {\enquote {\bibinfo {title} {Superconducting nano josephson junctions
  patterned with a focused helium ion beam},}\ }\href
  {https://doi.org/10.1063/1.5042105} {\bibfield  {journal} {\bibinfo
  {journal} {Appl. Phys. Lett.}\ }\textbf {\bibinfo {volume} {113}},\ \bibinfo
  {pages} {022604} (\bibinfo {year} {2018})}\BibitemShut {NoStop}%
\bibitem [{\citenamefont {M{\"u}ller}\ \emph {et~al.}(2019)\citenamefont
  {M{\"u}ller}, \citenamefont {Karrer}, \citenamefont {Limberger},
  \citenamefont {Becker}, \citenamefont {Schr{\"o}ppel}, \citenamefont
  {Burkhardt}, \citenamefont {Kleiner}, \citenamefont {Goldobin},\ and\
  \citenamefont {Koelle}}]{Muller2019}%
  \BibitemOpen
  \bibfield  {author} {\bibinfo {author} {\bibfnamefont {B.}~\bibnamefont
  {M{\"u}ller}}, \bibinfo {author} {\bibfnamefont {M.}~\bibnamefont {Karrer}},
  \bibinfo {author} {\bibfnamefont {F.}~\bibnamefont {Limberger}}, \bibinfo
  {author} {\bibfnamefont {M.}~\bibnamefont {Becker}}, \bibinfo {author}
  {\bibfnamefont {B.}~\bibnamefont {Schr{\"o}ppel}}, \bibinfo {author}
  {\bibfnamefont {C.~J.}\ \bibnamefont {Burkhardt}}, \bibinfo {author}
  {\bibfnamefont {R.}~\bibnamefont {Kleiner}}, \bibinfo {author} {\bibfnamefont
  {E.}~\bibnamefont {Goldobin}},\ and\ \bibinfo {author} {\bibfnamefont
  {D.}~\bibnamefont {Koelle}},\ }\bibfield  {title} {\enquote {\bibinfo {title}
  {Josephson junctions and {SQUIDs} created by focused {Helium-Ion-Beam}
  irradiation of {YBa$_{2}$Cu$_{3}$O$_{7}$}},}\ }\href
  {https://doi.org/10.1103/PhysRevApplied.11.044082} {\bibfield  {journal}
  {\bibinfo  {journal} {Phys. Rev. Applied}\ }\textbf {\bibinfo {volume}
  {11}},\ \bibinfo {pages} {044082} (\bibinfo {year} {2019})}\BibitemShut
  {NoStop}%
\bibitem [{\citenamefont {Cou{\"e}do}\ \emph {et~al.}(2020)\citenamefont
  {Cou{\"e}do}, \citenamefont {Amari}, \citenamefont {Feuillet-Palma},
  \citenamefont {Ulysse}, \citenamefont {Srivastava}, \citenamefont {Singh},
  \citenamefont {Bergeal},\ and\ \citenamefont {Lesueur}}]{Couedo2020}%
  \BibitemOpen
  \bibfield  {author} {\bibinfo {author} {\bibfnamefont {F.}~\bibnamefont
  {Cou{\"e}do}}, \bibinfo {author} {\bibfnamefont {P.}~\bibnamefont {Amari}},
  \bibinfo {author} {\bibfnamefont {C.}~\bibnamefont {Feuillet-Palma}},
  \bibinfo {author} {\bibfnamefont {C.}~\bibnamefont {Ulysse}}, \bibinfo
  {author} {\bibfnamefont {Y.~K.}\ \bibnamefont {Srivastava}}, \bibinfo
  {author} {\bibfnamefont {R.}~\bibnamefont {Singh}}, \bibinfo {author}
  {\bibfnamefont {N.}~\bibnamefont {Bergeal}},\ and\ \bibinfo {author}
  {\bibfnamefont {J.}~\bibnamefont {Lesueur}},\ }\bibfield  {title} {\enquote
  {\bibinfo {title} {Dynamic properties of high-tc superconducting
  nano-junctions made with a focused helium ion beam},}\ }\href
  {https://doi.org/10.1038/s41598-020-66882-1} {\bibfield  {journal} {\bibinfo
  {journal} {Sci. Rep.}\ }\textbf {\bibinfo {volume} {10}},\ \bibinfo {pages}
  {10256} (\bibinfo {year} {2020})}\BibitemShut {NoStop}%
\bibitem [{\citenamefont {LeFebvre}\ \emph {et~al.}(2022)\citenamefont
  {LeFebvre}, \citenamefont {Cho}, \citenamefont {Li}, \citenamefont {Cai},\
  and\ \citenamefont {Cybart}}]{LeFebvre_2022}%
  \BibitemOpen
  \bibfield  {author} {\bibinfo {author} {\bibfnamefont {J.~C.}\ \bibnamefont
  {LeFebvre}}, \bibinfo {author} {\bibfnamefont {E.}~\bibnamefont {Cho}},
  \bibinfo {author} {\bibfnamefont {H.}~\bibnamefont {Li}}, \bibinfo {author}
  {\bibfnamefont {H.}~\bibnamefont {Cai}},\ and\ \bibinfo {author}
  {\bibfnamefont {S.~A.}\ \bibnamefont {Cybart}},\ }\bibfield  {title}
  {\enquote {\bibinfo {title} {Flux focused series arrays of long josephson
  junctions for high-dynamic range magnetic field sensing},}\ }\href
  {https://doi.org/10.1063/5.0087611} {\bibfield  {journal} {\bibinfo
  {journal} {J. Appl. Phys.}\ }\textbf {\bibinfo {volume} {131}},\ \bibinfo
  {pages} {163902} (\bibinfo {year} {2022})}\BibitemShut {NoStop}%
\bibitem [{\citenamefont {Chen}\ \emph {et~al.}(2022)\citenamefont {Chen},
  \citenamefont {Li}, \citenamefont {Zhu}, \citenamefont {Xu}, \citenamefont
  {Xu}, \citenamefont {Yin}, \citenamefont {Cai}, \citenamefont {Wang},
  \citenamefont {Lu}, \citenamefont {Zhang},\ and\ \citenamefont
  {Ma}}]{Chen2022}%
  \BibitemOpen
  \bibfield  {author} {\bibinfo {author} {\bibfnamefont {Z.}~\bibnamefont
  {Chen}}, \bibinfo {author} {\bibfnamefont {Y.}~\bibnamefont {Li}}, \bibinfo
  {author} {\bibfnamefont {R.}~\bibnamefont {Zhu}}, \bibinfo {author}
  {\bibfnamefont {J.}~\bibnamefont {Xu}}, \bibinfo {author} {\bibfnamefont
  {T.}~\bibnamefont {Xu}}, \bibinfo {author} {\bibfnamefont {D.}~\bibnamefont
  {Yin}}, \bibinfo {author} {\bibfnamefont {X.}~\bibnamefont {Cai}}, \bibinfo
  {author} {\bibfnamefont {Y.}~\bibnamefont {Wang}}, \bibinfo {author}
  {\bibfnamefont {J.}~\bibnamefont {Lu}}, \bibinfo {author} {\bibfnamefont
  {Y.}~\bibnamefont {Zhang}},\ and\ \bibinfo {author} {\bibfnamefont
  {P.}~\bibnamefont {Ma}},\ }\bibfield  {title} {\enquote {\bibinfo {title}
  {High-temperature superconducting {YBa}$_{2}${Cu}$_{3}${O}$_{7-\updelta}$
  {J}osephson junction fabricated with a focused helium ion beam},}\ }\href
  {https://doi.org/10.1088/0256-307x/39/7/077402} {\bibfield  {journal}
  {\bibinfo  {journal} {Chinese Phys. Lett.}\ }\textbf {\bibinfo {volume}
  {39}},\ \bibinfo {pages} {077402} (\bibinfo {year} {2022})}\BibitemShut
  {NoStop}%
\bibitem [{\citenamefont {Karrer}\ \emph {et~al.}(2023)\citenamefont {Karrer},
  \citenamefont {Wurster}, \citenamefont {Linek}, \citenamefont {Meichsner},
  \citenamefont {Kleiner}, \citenamefont {Goldobin},\ and\ \citenamefont
  {Koelle}}]{Karrer2023}%
  \BibitemOpen
  \bibfield  {author} {\bibinfo {author} {\bibfnamefont {M.}~\bibnamefont
  {Karrer}}, \bibinfo {author} {\bibfnamefont {K.}~\bibnamefont {Wurster}},
  \bibinfo {author} {\bibfnamefont {J.}~\bibnamefont {Linek}}, \bibinfo
  {author} {\bibfnamefont {M.}~\bibnamefont {Meichsner}}, \bibinfo {author}
  {\bibfnamefont {R.}~\bibnamefont {Kleiner}}, \bibinfo {author} {\bibfnamefont
  {E.}~\bibnamefont {Goldobin}},\ and\ \bibinfo {author} {\bibfnamefont
  {D.}~\bibnamefont {Koelle}},\ }\href
  {https://doi.org/10.48550/ARXIV.2304.02749} {\enquote {\bibinfo {title}
  {Temporal evolution of electric transport properties of {YBCO} josephson
  junctions produced by focused helium ion beam irradiation},}\ } (\bibinfo
  {year} {2023}),\ \bibinfo {note} {preprint},\ \Eprint
  {https://arxiv.org/abs/2304.02749} {arXiv:2304.02749 [cond-mat.supr-con]}
  \BibitemShut {NoStop}%
\bibitem [{\citenamefont {Kasaei}\ \emph {et~al.}(2018)\citenamefont {Kasaei},
  \citenamefont {Melbourne}, \citenamefont {Manichev}, \citenamefont {Feldman},
  \citenamefont {Gustafsson}, \citenamefont {Chen}, \citenamefont {Xi},\ and\
  \citenamefont {Davidson}}]{Kasaei2018}%
  \BibitemOpen
  \bibfield  {author} {\bibinfo {author} {\bibfnamefont {L.}~\bibnamefont
  {Kasaei}}, \bibinfo {author} {\bibfnamefont {T.}~\bibnamefont {Melbourne}},
  \bibinfo {author} {\bibfnamefont {V.}~\bibnamefont {Manichev}}, \bibinfo
  {author} {\bibfnamefont {L.~C.}\ \bibnamefont {Feldman}}, \bibinfo {author}
  {\bibfnamefont {T.}~\bibnamefont {Gustafsson}}, \bibinfo {author}
  {\bibfnamefont {K.}~\bibnamefont {Chen}}, \bibinfo {author} {\bibfnamefont
  {X.~X.}\ \bibnamefont {Xi}},\ and\ \bibinfo {author} {\bibfnamefont {B.~A.}\
  \bibnamefont {Davidson}},\ }\bibfield  {title} {\enquote {\bibinfo {title}
  {{MgB2} josephson junctions produced by focused helium ion beam
  irradiation},}\ }\href {https://doi.org/10.1063/1.5030751} {\bibfield
  {journal} {\bibinfo  {journal} {AIP Adv.}\ }\textbf {\bibinfo {volume} {8}},\
  \bibinfo {pages} {075020} (\bibinfo {year} {2018})}\BibitemShut {NoStop}%
\bibitem [{\citenamefont {Seifert}\ \emph {et~al.}(2021)\citenamefont
  {Seifert}, \citenamefont {Retamal}, \citenamefont {Merino}, \citenamefont
  {Sheinfux}, \citenamefont {Moore}, \citenamefont {Aamir}, \citenamefont
  {Taniguchi}, \citenamefont {Watanabe}, \citenamefont {Kadowaki},
  \citenamefont {Artiglia}, \citenamefont {Romagnoli},\ and\ \citenamefont
  {Efetov}}]{Seifert2021}%
  \BibitemOpen
  \bibfield  {author} {\bibinfo {author} {\bibfnamefont {P.}~\bibnamefont
  {Seifert}}, \bibinfo {author} {\bibfnamefont {J.~R.~D.}\ \bibnamefont
  {Retamal}}, \bibinfo {author} {\bibfnamefont {R.~L.}\ \bibnamefont {Merino}},
  \bibinfo {author} {\bibfnamefont {H.~H.}\ \bibnamefont {Sheinfux}}, \bibinfo
  {author} {\bibfnamefont {J.~N.}\ \bibnamefont {Moore}}, \bibinfo {author}
  {\bibfnamefont {M.~A.}\ \bibnamefont {Aamir}}, \bibinfo {author}
  {\bibfnamefont {T.}~\bibnamefont {Taniguchi}}, \bibinfo {author}
  {\bibfnamefont {K.}~\bibnamefont {Watanabe}}, \bibinfo {author}
  {\bibfnamefont {K.}~\bibnamefont {Kadowaki}}, \bibinfo {author}
  {\bibfnamefont {M.}~\bibnamefont {Artiglia}}, \bibinfo {author}
  {\bibfnamefont {M.}~\bibnamefont {Romagnoli}},\ and\ \bibinfo {author}
  {\bibfnamefont {D.~K.}\ \bibnamefont {Efetov}},\ }\bibfield  {title}
  {\enquote {\bibinfo {title} {A high-t$_{c}$ van der waals superconductor
  based photodetector with ultra-high responsivity and nanosecond relaxation
  time},}\ }\href {https://doi.org/10.1088/2053-1583/ac072f} {\bibfield
  {journal} {\bibinfo  {journal} {2D Mater.}\ }\textbf {\bibinfo {volume}
  {8}},\ \bibinfo {pages} {035053} (\bibinfo {year} {2021})}\BibitemShut
  {NoStop}%
\bibitem [{\citenamefont {Merino}\ \emph {et~al.}(2023)\citenamefont {Merino},
  \citenamefont {Seifert}, \citenamefont {Retamal}, \citenamefont {Mech},
  \citenamefont {Taniguchi}, \citenamefont {Watanabe}, \citenamefont
  {Kadowaki}, \citenamefont {Hadfield},\ and\ \citenamefont
  {Efetov}}]{Merino2023}%
  \BibitemOpen
  \bibfield  {author} {\bibinfo {author} {\bibfnamefont {R.~L.}\ \bibnamefont
  {Merino}}, \bibinfo {author} {\bibfnamefont {P.}~\bibnamefont {Seifert}},
  \bibinfo {author} {\bibfnamefont {J.~D.}\ \bibnamefont {Retamal}}, \bibinfo
  {author} {\bibfnamefont {R.~K.}\ \bibnamefont {Mech}}, \bibinfo {author}
  {\bibfnamefont {T.}~\bibnamefont {Taniguchi}}, \bibinfo {author}
  {\bibfnamefont {K.}~\bibnamefont {Watanabe}}, \bibinfo {author}
  {\bibfnamefont {K.}~\bibnamefont {Kadowaki}}, \bibinfo {author}
  {\bibfnamefont {R.~H.}\ \bibnamefont {Hadfield}},\ and\ \bibinfo {author}
  {\bibfnamefont {D.~K.}\ \bibnamefont {Efetov}},\ }\bibfield  {title}
  {\enquote {\bibinfo {title} {Two-dimensional cuprate nanodetector with single
  telecom photon sensitivity at {T} = 20 {K}},}\ }\href
  {https://doi.org/10.1088/2053-1583/acb4a8} {\bibfield  {journal} {\bibinfo
  {journal} {2D Mater.}\ }\textbf {\bibinfo {volume} {10}},\ \bibinfo {pages}
  {021001} (\bibinfo {year} {2023})}\BibitemShut {NoStop}%
\bibitem [{\citenamefont {Maehashi}\ \emph {et~al.}(2007)\citenamefont
  {Maehashi}, \citenamefont {Ozaki}, \citenamefont {Ohno}, \citenamefont
  {Inoue}, \citenamefont {Matsumoto}, \citenamefont {Seki},\ and\ \citenamefont
  {Tagawa}}]{Maehashi_2007}%
  \BibitemOpen
  \bibfield  {author} {\bibinfo {author} {\bibfnamefont {K.}~\bibnamefont
  {Maehashi}}, \bibinfo {author} {\bibfnamefont {H.}~\bibnamefont {Ozaki}},
  \bibinfo {author} {\bibfnamefont {Y.}~\bibnamefont {Ohno}}, \bibinfo {author}
  {\bibfnamefont {K.}~\bibnamefont {Inoue}}, \bibinfo {author} {\bibfnamefont
  {K.}~\bibnamefont {Matsumoto}}, \bibinfo {author} {\bibfnamefont
  {S.}~\bibnamefont {Seki}},\ and\ \bibinfo {author} {\bibfnamefont
  {S.}~\bibnamefont {Tagawa}},\ }\bibfield  {title} {\enquote {\bibinfo {title}
  {Formation of single quantum dot in single-walled carbon nanotube channel
  using focused-ion-beam technique},}\ }\href
  {https://doi.org/10.1063/1.2430680} {\bibfield  {journal} {\bibinfo
  {journal} {Appl. Phys. Lett.}\ }\textbf {\bibinfo {volume} {90}},\ \bibinfo
  {pages} {023103} (\bibinfo {year} {2007})}\BibitemShut {NoStop}%
\bibitem [{\citenamefont {Kim}\ \emph {et~al.}(2022)\citenamefont {Kim},
  \citenamefont {Quang}, \citenamefont {Nguyen}, \citenamefont {Kim},
  \citenamefont {Lee}, \citenamefont {Lee}, \citenamefont {Cho}, \citenamefont
  {Seong}, \citenamefont {Kim},\ and\ \citenamefont {Chang}}]{Kim_2022}%
  \BibitemOpen
  \bibfield  {author} {\bibinfo {author} {\bibfnamefont {H.~J.}\ \bibnamefont
  {Kim}}, \bibinfo {author} {\bibfnamefont {N.~V.}\ \bibnamefont {Quang}},
  \bibinfo {author} {\bibfnamefont {T.~H.}\ \bibnamefont {Nguyen}}, \bibinfo
  {author} {\bibfnamefont {S.}~\bibnamefont {Kim}}, \bibinfo {author}
  {\bibfnamefont {Y.}~\bibnamefont {Lee}}, \bibinfo {author} {\bibfnamefont
  {I.~H.}\ \bibnamefont {Lee}}, \bibinfo {author} {\bibfnamefont
  {S.}~\bibnamefont {Cho}}, \bibinfo {author} {\bibfnamefont {M.-J.}\
  \bibnamefont {Seong}}, \bibinfo {author} {\bibfnamefont {K.}~\bibnamefont
  {Kim}},\ and\ \bibinfo {author} {\bibfnamefont {Y.~J.}\ \bibnamefont
  {Chang}},\ }\bibfield  {title} {\enquote {\bibinfo {title} {Tuning of
  thermoelectric properties of {MoSe}$_2$ thin films under helium ion
  irradiation},}\ }\href {https://doi.org/10.1186/s11671-022-03665-9}
  {\bibfield  {journal} {\bibinfo  {journal} {Nanoscale Res. Lett.}\ }\textbf
  {\bibinfo {volume} {17}} (\bibinfo {year} {2022}),\
  10.1186/s11671-022-03665-9}\BibitemShut {NoStop}%
\bibitem [{\citenamefont {Aichner}\ \emph {et~al.}(2019)\citenamefont
  {Aichner}, \citenamefont {M{\"u}ller}, \citenamefont {Karrer}, \citenamefont
  {Misko}, \citenamefont {Limberger}, \citenamefont {Mletschnig}, \citenamefont
  {Dosmailov}, \citenamefont {Pedarnig}, \citenamefont {Nori}, \citenamefont
  {Kleiner}, \citenamefont {Koelle},\ and\ \citenamefont
  {Lang}}]{Aichner_2019}%
  \BibitemOpen
  \bibfield  {author} {\bibinfo {author} {\bibfnamefont {B.}~\bibnamefont
  {Aichner}}, \bibinfo {author} {\bibfnamefont {B.}~\bibnamefont {M{\"u}ller}},
  \bibinfo {author} {\bibfnamefont {M.}~\bibnamefont {Karrer}}, \bibinfo
  {author} {\bibfnamefont {V.~R.}\ \bibnamefont {Misko}}, \bibinfo {author}
  {\bibfnamefont {F.}~\bibnamefont {Limberger}}, \bibinfo {author}
  {\bibfnamefont {K.~L.}\ \bibnamefont {Mletschnig}}, \bibinfo {author}
  {\bibfnamefont {M.}~\bibnamefont {Dosmailov}}, \bibinfo {author}
  {\bibfnamefont {J.~D.}\ \bibnamefont {Pedarnig}}, \bibinfo {author}
  {\bibfnamefont {F.}~\bibnamefont {Nori}}, \bibinfo {author} {\bibfnamefont
  {R.}~\bibnamefont {Kleiner}}, \bibinfo {author} {\bibfnamefont
  {D.}~\bibnamefont {Koelle}},\ and\ \bibinfo {author} {\bibfnamefont
  {W.}~\bibnamefont {Lang}},\ }\bibfield  {title} {\enquote {\bibinfo {title}
  {Ultradense tailored vortex pinning arrays in superconducting {YBa2Cu3O7-d }
  thin films created by focused {He} ion beam irradiation for fluxonics
  applications},}\ }\href {https://doi.org/10.1021/acsanm.9b01006} {\bibfield
  {journal} {\bibinfo  {journal} {ACS Appl. Nano Mater.}\ }\textbf {\bibinfo
  {volume} {2}},\ \bibinfo {pages} {5108--5115} (\bibinfo {year}
  {2019})}\BibitemShut {NoStop}%
\bibitem [{\citenamefont {Backmeister}\ \emph {et~al.}(2022)\citenamefont
  {Backmeister}, \citenamefont {Aichner}, \citenamefont {Karrer}, \citenamefont
  {Wurster}, \citenamefont {Kleiner}, \citenamefont {Goldobin}, \citenamefont
  {Koelle},\ and\ \citenamefont {Lang}}]{Backmeister2022}%
  \BibitemOpen
  \bibfield  {author} {\bibinfo {author} {\bibfnamefont {L.}~\bibnamefont
  {Backmeister}}, \bibinfo {author} {\bibfnamefont {B.}~\bibnamefont
  {Aichner}}, \bibinfo {author} {\bibfnamefont {M.}~\bibnamefont {Karrer}},
  \bibinfo {author} {\bibfnamefont {K.}~\bibnamefont {Wurster}}, \bibinfo
  {author} {\bibfnamefont {R.}~\bibnamefont {Kleiner}}, \bibinfo {author}
  {\bibfnamefont {E.}~\bibnamefont {Goldobin}}, \bibinfo {author}
  {\bibfnamefont {D.}~\bibnamefont {Koelle}},\ and\ \bibinfo {author}
  {\bibfnamefont {W.}~\bibnamefont {Lang}},\ }\bibfield  {title} {\enquote
  {\bibinfo {title} {Ordered bose glass of vortices in superconducting
  {YBa}$_{2}${Cu}$_{3}${O}$_{7-\updelta}$ thin films with a periodic pin
  lattice created by focused helium ion irradiation},}\ }\href
  {https://doi.org/10.3390/nano12193491} {\bibfield  {journal} {\bibinfo
  {journal} {Nanomaterials}\ }\textbf {\bibinfo {volume} {12}},\ \bibinfo
  {pages} {3491} (\bibinfo {year} {2022})}\BibitemShut {NoStop}%
\bibitem [{\citenamefont {Saremi}\ \emph {et~al.}(2018)\citenamefont {Saremi},
  \citenamefont {Xu}, \citenamefont {Allen}, \citenamefont {Maher},
  \citenamefont {Agar}, \citenamefont {Gao}, \citenamefont {Hosemann},\ and\
  \citenamefont {Martin}}]{Saremi2018}%
  \BibitemOpen
  \bibfield  {author} {\bibinfo {author} {\bibfnamefont {S.}~\bibnamefont
  {Saremi}}, \bibinfo {author} {\bibfnamefont {R.}~\bibnamefont {Xu}}, \bibinfo
  {author} {\bibfnamefont {F.~I.}\ \bibnamefont {Allen}}, \bibinfo {author}
  {\bibfnamefont {J.}~\bibnamefont {Maher}}, \bibinfo {author} {\bibfnamefont
  {J.~C.}\ \bibnamefont {Agar}}, \bibinfo {author} {\bibfnamefont
  {R.}~\bibnamefont {Gao}}, \bibinfo {author} {\bibfnamefont {P.}~\bibnamefont
  {Hosemann}},\ and\ \bibinfo {author} {\bibfnamefont {L.~W.}\ \bibnamefont
  {Martin}},\ }\bibfield  {title} {\enquote {\bibinfo {title} {Local control of
  defects and switching properties in ferroelectric thin films},}\ }\href
  {https://doi.org/10.1103/PhysRevMaterials.2.084414} {\bibfield  {journal}
  {\bibinfo  {journal} {Phys. Rev. Materials}\ }\textbf {\bibinfo {volume}
  {2}},\ \bibinfo {pages} {084414} (\bibinfo {year} {2018})}\BibitemShut
  {NoStop}%
\bibitem [{\citenamefont {Seol}\ \emph {et~al.}(2021)\citenamefont {Seol},
  \citenamefont {Kim}, \citenamefont {Jang}, \citenamefont {Jin}, \citenamefont
  {Kang}, \citenamefont {Kim}, \citenamefont {Won}, \citenamefont {Lee},
  \citenamefont {Kim}, \citenamefont {Lee}, \citenamefont {Yang}, \citenamefont
  {Jeong}, \citenamefont {Belianinov}, \citenamefont {Tselev}, \citenamefont
  {Somnath}, \citenamefont {Smith}, \citenamefont {Ovchinnikova}, \citenamefont
  {Balke},\ and\ \citenamefont {Kim}}]{Seol_2021}%
  \BibitemOpen
  \bibfield  {author} {\bibinfo {author} {\bibfnamefont {D.}~\bibnamefont
  {Seol}}, \bibinfo {author} {\bibfnamefont {S.}~\bibnamefont {Kim}}, \bibinfo
  {author} {\bibfnamefont {W.-S.}\ \bibnamefont {Jang}}, \bibinfo {author}
  {\bibfnamefont {Y.}~\bibnamefont {Jin}}, \bibinfo {author} {\bibfnamefont
  {S.}~\bibnamefont {Kang}}, \bibinfo {author} {\bibfnamefont {S.}~\bibnamefont
  {Kim}}, \bibinfo {author} {\bibfnamefont {D.}~\bibnamefont {Won}}, \bibinfo
  {author} {\bibfnamefont {C.}~\bibnamefont {Lee}}, \bibinfo {author}
  {\bibfnamefont {Y.-M.}\ \bibnamefont {Kim}}, \bibinfo {author} {\bibfnamefont
  {J.}~\bibnamefont {Lee}}, \bibinfo {author} {\bibfnamefont {H.}~\bibnamefont
  {Yang}}, \bibinfo {author} {\bibfnamefont {M.~S.}\ \bibnamefont {Jeong}},
  \bibinfo {author} {\bibfnamefont {A.}~\bibnamefont {Belianinov}}, \bibinfo
  {author} {\bibfnamefont {A.}~\bibnamefont {Tselev}}, \bibinfo {author}
  {\bibfnamefont {S.}~\bibnamefont {Somnath}}, \bibinfo {author} {\bibfnamefont
  {C.~R.}\ \bibnamefont {Smith}}, \bibinfo {author} {\bibfnamefont {O.~S.}\
  \bibnamefont {Ovchinnikova}}, \bibinfo {author} {\bibfnamefont
  {N.}~\bibnamefont {Balke}},\ and\ \bibinfo {author} {\bibfnamefont
  {Y.}~\bibnamefont {Kim}},\ }\bibfield  {title} {\enquote {\bibinfo {title}
  {Selective patterning of out-of-plane piezoelectricity in {MoTe2} via focused
  ion beam},}\ }\href {https://doi.org/10.1016/j.nanoen.2020.105451} {\bibfield
   {journal} {\bibinfo  {journal} {Nano Energy}\ }\textbf {\bibinfo {volume}
  {79}},\ \bibinfo {pages} {105451} (\bibinfo {year} {2021})}\BibitemShut
  {NoStop}%
\bibitem [{\citenamefont {Kang}\ \emph {et~al.}(2022)\citenamefont {Kang},
  \citenamefont {Jang}, \citenamefont {Morozovska}, \citenamefont {Kwon},
  \citenamefont {Jin}, \citenamefont {Kim}, \citenamefont {Bae}, \citenamefont
  {Wang}, \citenamefont {Yang}, \citenamefont {Belianinov}, \citenamefont
  {Randolph}, \citenamefont {Eliseev}, \citenamefont {Collins}, \citenamefont
  {Park}, \citenamefont {Jo}, \citenamefont {Jung}, \citenamefont {Go},
  \citenamefont {Cho}, \citenamefont {Choi}, \citenamefont {Jang},
  \citenamefont {Kim}, \citenamefont {Jeong}, \citenamefont {Lee},
  \citenamefont {Ovchinnikova}, \citenamefont {Heo}, \citenamefont {Kalinin},
  \citenamefont {Kim},\ and\ \citenamefont {Kim}}]{Kang_2022}%
  \BibitemOpen
  \bibfield  {author} {\bibinfo {author} {\bibfnamefont {S.}~\bibnamefont
  {Kang}}, \bibinfo {author} {\bibfnamefont {W.-S.}\ \bibnamefont {Jang}},
  \bibinfo {author} {\bibfnamefont {A.~N.}\ \bibnamefont {Morozovska}},
  \bibinfo {author} {\bibfnamefont {O.}~\bibnamefont {Kwon}}, \bibinfo {author}
  {\bibfnamefont {Y.}~\bibnamefont {Jin}}, \bibinfo {author} {\bibfnamefont
  {Y.-H.}\ \bibnamefont {Kim}}, \bibinfo {author} {\bibfnamefont
  {H.}~\bibnamefont {Bae}}, \bibinfo {author} {\bibfnamefont {C.}~\bibnamefont
  {Wang}}, \bibinfo {author} {\bibfnamefont {S.-H.}\ \bibnamefont {Yang}},
  \bibinfo {author} {\bibfnamefont {A.}~\bibnamefont {Belianinov}}, \bibinfo
  {author} {\bibfnamefont {S.}~\bibnamefont {Randolph}}, \bibinfo {author}
  {\bibfnamefont {E.~A.}\ \bibnamefont {Eliseev}}, \bibinfo {author}
  {\bibfnamefont {L.}~\bibnamefont {Collins}}, \bibinfo {author} {\bibfnamefont
  {Y.}~\bibnamefont {Park}}, \bibinfo {author} {\bibfnamefont {S.}~\bibnamefont
  {Jo}}, \bibinfo {author} {\bibfnamefont {M.-H.}\ \bibnamefont {Jung}},
  \bibinfo {author} {\bibfnamefont {K.-J.}\ \bibnamefont {Go}}, \bibinfo
  {author} {\bibfnamefont {H.~W.}\ \bibnamefont {Cho}}, \bibinfo {author}
  {\bibfnamefont {S.-Y.}\ \bibnamefont {Choi}}, \bibinfo {author}
  {\bibfnamefont {J.~H.}\ \bibnamefont {Jang}}, \bibinfo {author}
  {\bibfnamefont {S.}~\bibnamefont {Kim}}, \bibinfo {author} {\bibfnamefont
  {H.~Y.}\ \bibnamefont {Jeong}}, \bibinfo {author} {\bibfnamefont
  {J.}~\bibnamefont {Lee}}, \bibinfo {author} {\bibfnamefont {O.~S.}\
  \bibnamefont {Ovchinnikova}}, \bibinfo {author} {\bibfnamefont
  {J.}~\bibnamefont {Heo}}, \bibinfo {author} {\bibfnamefont {S.~V.}\
  \bibnamefont {Kalinin}}, \bibinfo {author} {\bibfnamefont {Y.-M.}\
  \bibnamefont {Kim}},\ and\ \bibinfo {author} {\bibfnamefont {Y.}~\bibnamefont
  {Kim}},\ }\bibfield  {title} {\enquote {\bibinfo {title} {Highly enhanced
  ferroelectricity in {HfO}$_2$-based ferroelectric thin film by light ion
  bombardment},}\ }\href {https://doi.org/10.1126/science.abk3195} {\bibfield
  {journal} {\bibinfo  {journal} {Science}\ }\textbf {\bibinfo {volume}
  {376}},\ \bibinfo {pages} {731--738} (\bibinfo {year} {2022})}\BibitemShut
  {NoStop}%
\bibitem [{\citenamefont {Belianinov}\ \emph
  {et~al.}(2017{\natexlab{a}})\citenamefont {Belianinov}, \citenamefont
  {Burch}, \citenamefont {Hysmith}, \citenamefont {Ievlev}, \citenamefont
  {Iberi}, \citenamefont {Susner}, \citenamefont {McGuire}, \citenamefont
  {Maksymovych}, \citenamefont {Chyasnavichyus}, \citenamefont {Jesse},\ and\
  \citenamefont {Ovchinnikova}}]{Belianinov_2017}%
  \BibitemOpen
  \bibfield  {author} {\bibinfo {author} {\bibfnamefont {A.}~\bibnamefont
  {Belianinov}}, \bibinfo {author} {\bibfnamefont {M.~J.}\ \bibnamefont
  {Burch}}, \bibinfo {author} {\bibfnamefont {H.~E.}\ \bibnamefont {Hysmith}},
  \bibinfo {author} {\bibfnamefont {A.~V.}\ \bibnamefont {Ievlev}}, \bibinfo
  {author} {\bibfnamefont {V.}~\bibnamefont {Iberi}}, \bibinfo {author}
  {\bibfnamefont {M.~A.}\ \bibnamefont {Susner}}, \bibinfo {author}
  {\bibfnamefont {M.~A.}\ \bibnamefont {McGuire}}, \bibinfo {author}
  {\bibfnamefont {P.}~\bibnamefont {Maksymovych}}, \bibinfo {author}
  {\bibfnamefont {M.}~\bibnamefont {Chyasnavichyus}}, \bibinfo {author}
  {\bibfnamefont {S.}~\bibnamefont {Jesse}},\ and\ \bibinfo {author}
  {\bibfnamefont {O.~S.}\ \bibnamefont {Ovchinnikova}},\ }\bibfield  {title}
  {\enquote {\bibinfo {title} {Chemical changes in layered ferroelectric
  semiconductors induced by helium ion beam},}\ }\href
  {https://doi.org/10.1038/s41598-017-16949-3} {\bibfield  {journal} {\bibinfo
  {journal} {Sci. Rep.}\ }\textbf {\bibinfo {volume} {7}} (\bibinfo {year}
  {2017}{\natexlab{a}}),\ 10.1038/s41598-017-16949-3}\BibitemShut {NoStop}%
\bibitem [{\citenamefont {Guberna}\ \emph {et~al.}(2021)\citenamefont
  {Guberna}, \citenamefont {Chouprik}, \citenamefont {Kirtaev}, \citenamefont
  {Zarubin}, \citenamefont {Margolin}, \citenamefont {Spiridonov},\ and\
  \citenamefont {Negrov}}]{Guberna_2021}%
  \BibitemOpen
  \bibfield  {author} {\bibinfo {author} {\bibfnamefont {E.}~\bibnamefont
  {Guberna}}, \bibinfo {author} {\bibfnamefont {A.}~\bibnamefont {Chouprik}},
  \bibinfo {author} {\bibfnamefont {R.}~\bibnamefont {Kirtaev}}, \bibinfo
  {author} {\bibfnamefont {S.}~\bibnamefont {Zarubin}}, \bibinfo {author}
  {\bibfnamefont {I.}~\bibnamefont {Margolin}}, \bibinfo {author}
  {\bibfnamefont {M.}~\bibnamefont {Spiridonov}},\ and\ \bibinfo {author}
  {\bibfnamefont {D.}~\bibnamefont {Negrov}},\ }\bibfield  {title} {\enquote
  {\bibinfo {title} {Local ga ion implantation as a source of diverse
  ferroelectric properties of hafnium oxide},}\ }\href
  {https://doi.org/10.1002/pssr.202100485} {\bibfield  {journal} {\bibinfo
  {journal} {Phys. Status Solidi RRL}\ }\textbf {\bibinfo {volume} {16}},\
  \bibinfo {pages} {2100485} (\bibinfo {year} {2021})}\BibitemShut {NoStop}%
\bibitem [{\citenamefont {Stanishevsky}\ \emph {et~al.}(2002)\citenamefont
  {Stanishevsky}, \citenamefont {Nagaraj}, \citenamefont {Melngailis},
  \citenamefont {Ramesh}, \citenamefont {Khriachtchev},\ and\ \citenamefont
  {McDaniel}}]{Stanishevsky_2002}%
  \BibitemOpen
  \bibfield  {author} {\bibinfo {author} {\bibfnamefont {A.}~\bibnamefont
  {Stanishevsky}}, \bibinfo {author} {\bibfnamefont {B.}~\bibnamefont
  {Nagaraj}}, \bibinfo {author} {\bibfnamefont {J.}~\bibnamefont {Melngailis}},
  \bibinfo {author} {\bibfnamefont {R.}~\bibnamefont {Ramesh}}, \bibinfo
  {author} {\bibfnamefont {L.}~\bibnamefont {Khriachtchev}},\ and\ \bibinfo
  {author} {\bibfnamefont {E.}~\bibnamefont {McDaniel}},\ }\bibfield  {title}
  {\enquote {\bibinfo {title} {Radiation damage and its recovery in focused ion
  beam fabricated ferroelectric capacitors},}\ }\href
  {https://doi.org/10.1063/1.1489069} {\bibfield  {journal} {\bibinfo
  {journal} {J. Appl. Phys.}\ }\textbf {\bibinfo {volume} {92}},\ \bibinfo
  {pages} {3275--3278} (\bibinfo {year} {2002})}\BibitemShut {NoStop}%
\bibitem [{\citenamefont {Ganpule}\ \emph {et~al.}(1999)\citenamefont
  {Ganpule}, \citenamefont {Stanishevsky}, \citenamefont {Su}, \citenamefont
  {Aggarwal}, \citenamefont {Melngailis}, \citenamefont {Williams},\ and\
  \citenamefont {Ramesh}}]{Ganpule_1999}%
  \BibitemOpen
  \bibfield  {author} {\bibinfo {author} {\bibfnamefont {C.~S.}\ \bibnamefont
  {Ganpule}}, \bibinfo {author} {\bibfnamefont {A.}~\bibnamefont
  {Stanishevsky}}, \bibinfo {author} {\bibfnamefont {Q.}~\bibnamefont {Su}},
  \bibinfo {author} {\bibfnamefont {S.}~\bibnamefont {Aggarwal}}, \bibinfo
  {author} {\bibfnamefont {J.}~\bibnamefont {Melngailis}}, \bibinfo {author}
  {\bibfnamefont {E.}~\bibnamefont {Williams}},\ and\ \bibinfo {author}
  {\bibfnamefont {R.}~\bibnamefont {Ramesh}},\ }\bibfield  {title} {\enquote
  {\bibinfo {title} {Scaling of ferroelectric properties in thin films},}\
  }\href {https://doi.org/10.1063/1.124391} {\bibfield  {journal} {\bibinfo
  {journal} {Appl. Phys. Lett.}\ }\textbf {\bibinfo {volume} {75}},\ \bibinfo
  {pages} {409--411} (\bibinfo {year} {1999})}\BibitemShut {NoStop}%
\bibitem [{\citenamefont {Franken}\ \emph {et~al.}(2011)\citenamefont
  {Franken}, \citenamefont {Hoeijmakers}, \citenamefont {Lavrijsen},
  \citenamefont {Kohlhepp}, \citenamefont {Swagten}, \citenamefont {Koopmans},
  \citenamefont {van Veldhoven},\ and\ \citenamefont {Maas}}]{Franken2011}%
  \BibitemOpen
  \bibfield  {author} {\bibinfo {author} {\bibfnamefont {J.~H.}\ \bibnamefont
  {Franken}}, \bibinfo {author} {\bibfnamefont {M.}~\bibnamefont
  {Hoeijmakers}}, \bibinfo {author} {\bibfnamefont {R.}~\bibnamefont
  {Lavrijsen}}, \bibinfo {author} {\bibfnamefont {J.~T.}\ \bibnamefont
  {Kohlhepp}}, \bibinfo {author} {\bibfnamefont {H.~J.~M.}\ \bibnamefont
  {Swagten}}, \bibinfo {author} {\bibfnamefont {B.}~\bibnamefont {Koopmans}},
  \bibinfo {author} {\bibfnamefont {E.}~\bibnamefont {van Veldhoven}},\ and\
  \bibinfo {author} {\bibfnamefont {D.~J.}\ \bibnamefont {Maas}},\ }\bibfield
  {title} {\enquote {\bibinfo {title} {Precise control of domain wall injection
  and pinning using helium and gallium focused ion beams},}\ }\href
  {https://doi.org/10.1063/1.3549589} {\bibfield  {journal} {\bibinfo
  {journal} {J. Appl. Phys.}\ }\textbf {\bibinfo {volume} {109}},\ \bibinfo
  {pages} {07D504} (\bibinfo {year} {2011})}\BibitemShut {NoStop}%
\bibitem [{\citenamefont {Huckfeldt}\ \emph {et~al.}(2017)\citenamefont
  {Huckfeldt}, \citenamefont {Gaul}, \citenamefont {David~M{\"u}glich},
  \citenamefont {Holzinger}, \citenamefont {Nissen}, \citenamefont {Albrecht},
  \citenamefont {Emmrich}, \citenamefont {Beyer}, \citenamefont
  {G{\"o}lzh{\"a}user},\ and\ \citenamefont {Ehresmann}}]{Huckfeldt2017}%
  \BibitemOpen
  \bibfield  {author} {\bibinfo {author} {\bibfnamefont {H.}~\bibnamefont
  {Huckfeldt}}, \bibinfo {author} {\bibfnamefont {A.}~\bibnamefont {Gaul}},
  \bibinfo {author} {\bibfnamefont {N.}~\bibnamefont {David~M{\"u}glich}},
  \bibinfo {author} {\bibfnamefont {D.}~\bibnamefont {Holzinger}}, \bibinfo
  {author} {\bibfnamefont {D.}~\bibnamefont {Nissen}}, \bibinfo {author}
  {\bibfnamefont {M.}~\bibnamefont {Albrecht}}, \bibinfo {author}
  {\bibfnamefont {D.}~\bibnamefont {Emmrich}}, \bibinfo {author} {\bibfnamefont
  {A.}~\bibnamefont {Beyer}}, \bibinfo {author} {\bibfnamefont
  {A.}~\bibnamefont {G{\"o}lzh{\"a}user}},\ and\ \bibinfo {author}
  {\bibfnamefont {A.}~\bibnamefont {Ehresmann}},\ }\bibfield  {title} {\enquote
  {\bibinfo {title} {Modification of the saturation magnetization of exchange
  bias thin film systems upon light-ion bombardment},}\ }\href
  {https://doi.org/10.1088/1361-648x/aa5ad5} {\bibfield  {journal} {\bibinfo
  {journal} {J. Phys.: Condens. Matter}\ }\textbf {\bibinfo {volume} {29}},\
  \bibinfo {pages} {125801} (\bibinfo {year} {2017})}\BibitemShut {NoStop}%
\bibitem [{\citenamefont {Dunne}\ \emph {et~al.}(2020)\citenamefont {Dunne},
  \citenamefont {Fowley}, \citenamefont {Hlawacek}, \citenamefont {Kurian},
  \citenamefont {Atcheson}, \citenamefont {Colis}, \citenamefont {Teichert},
  \citenamefont {Kundys}, \citenamefont {Venkatesan}, \citenamefont {Lindner},
  \citenamefont {Deac}, \citenamefont {Hermans}, \citenamefont {Coey},\ and\
  \citenamefont {Doudin}}]{Dunne2020}%
  \BibitemOpen
  \bibfield  {author} {\bibinfo {author} {\bibfnamefont {P.}~\bibnamefont
  {Dunne}}, \bibinfo {author} {\bibfnamefont {C.}~\bibnamefont {Fowley}},
  \bibinfo {author} {\bibfnamefont {G.}~\bibnamefont {Hlawacek}}, \bibinfo
  {author} {\bibfnamefont {J.}~\bibnamefont {Kurian}}, \bibinfo {author}
  {\bibfnamefont {G.}~\bibnamefont {Atcheson}}, \bibinfo {author}
  {\bibfnamefont {S.}~\bibnamefont {Colis}}, \bibinfo {author} {\bibfnamefont
  {N.}~\bibnamefont {Teichert}}, \bibinfo {author} {\bibfnamefont
  {B.}~\bibnamefont {Kundys}}, \bibinfo {author} {\bibfnamefont
  {M.}~\bibnamefont {Venkatesan}}, \bibinfo {author} {\bibfnamefont
  {J.}~\bibnamefont {Lindner}}, \bibinfo {author} {\bibfnamefont {A.~M.}\
  \bibnamefont {Deac}}, \bibinfo {author} {\bibfnamefont {T.~M.}\ \bibnamefont
  {Hermans}}, \bibinfo {author} {\bibfnamefont {J.~M.~D.}\ \bibnamefont
  {Coey}},\ and\ \bibinfo {author} {\bibfnamefont {B.}~\bibnamefont {Doudin}},\
  }\bibfield  {title} {{\selectlanguage {english}\enquote {\bibinfo {title}
  {Helium ion microscopy for reduced spin orbit torque switching currents},}\
  }}\href {https://doi.org/10.1021/acs.nanolett.0c02060} {\bibfield  {journal}
  {\bibinfo  {journal} {Nano Lett.}\ }\textbf {\bibinfo {volume} {20}},\
  \bibinfo {pages} {7036--7042} (\bibinfo {year} {2020})},\ \Eprint
  {https://arxiv.org/abs/2005.07626} {arXiv:2005.07626 [cond-mat.mes-hall]}
  \BibitemShut {NoStop}%
\bibitem [{\citenamefont {Kurian}\ \emph {et~al.}(2023)\citenamefont {Kurian},
  \citenamefont {Joseph}, \citenamefont {Cherifi-Hertel}, \citenamefont
  {Fowley}, \citenamefont {Hlawacek}, \citenamefont {Dunne}, \citenamefont
  {Romeo}, \citenamefont {Atcheson}, \citenamefont {Coey},\ and\ \citenamefont
  {Doudin}}]{Kurian_2023}%
  \BibitemOpen
  \bibfield  {author} {\bibinfo {author} {\bibfnamefont {J.}~\bibnamefont
  {Kurian}}, \bibinfo {author} {\bibfnamefont {A.}~\bibnamefont {Joseph}},
  \bibinfo {author} {\bibfnamefont {S.}~\bibnamefont {Cherifi-Hertel}},
  \bibinfo {author} {\bibfnamefont {C.}~\bibnamefont {Fowley}}, \bibinfo
  {author} {\bibfnamefont {G.}~\bibnamefont {Hlawacek}}, \bibinfo {author}
  {\bibfnamefont {P.}~\bibnamefont {Dunne}}, \bibinfo {author} {\bibfnamefont
  {M.}~\bibnamefont {Romeo}}, \bibinfo {author} {\bibfnamefont
  {G.}~\bibnamefont {Atcheson}}, \bibinfo {author} {\bibfnamefont {J.~M.~D.}\
  \bibnamefont {Coey}},\ and\ \bibinfo {author} {\bibfnamefont
  {B.}~\bibnamefont {Doudin}},\ }\bibfield  {title} {\enquote {\bibinfo {title}
  {Deterministic multi-level spin orbit torque switching using {He}$^{+}$
  microscopy patterning},}\ }\href {https://doi.org/10.1063/5.0131188}
  {\bibfield  {journal} {\bibinfo  {journal} {Appl. Phys. Lett.}\ }\textbf
  {\bibinfo {volume} {122}},\ \bibinfo {pages} {032402} (\bibinfo {year}
  {2023})},\ \Eprint {https://arxiv.org/abs/2301.02188} {arxiv:2301.02188}
  \BibitemShut {NoStop}%
\bibitem [{\citenamefont {Samad}\ \emph {et~al.}(2021)\citenamefont {Samad},
  \citenamefont {Hlawacek}, \citenamefont {Arekapudi}, \citenamefont {Xu},
  \citenamefont {Koch}, \citenamefont {Lenz},\ and\ \citenamefont
  {Hellwig}}]{Samad_2021}%
  \BibitemOpen
  \bibfield  {author} {\bibinfo {author} {\bibfnamefont {F.}~\bibnamefont
  {Samad}}, \bibinfo {author} {\bibfnamefont {G.}~\bibnamefont {Hlawacek}},
  \bibinfo {author} {\bibfnamefont {S.~S. P.~K.}\ \bibnamefont {Arekapudi}},
  \bibinfo {author} {\bibfnamefont {X.}~\bibnamefont {Xu}}, \bibinfo {author}
  {\bibfnamefont {L.}~\bibnamefont {Koch}}, \bibinfo {author} {\bibfnamefont
  {M.}~\bibnamefont {Lenz}},\ and\ \bibinfo {author} {\bibfnamefont
  {O.}~\bibnamefont {Hellwig}},\ }\bibfield  {title} {\enquote {\bibinfo
  {title} {Controlled and deterministic creation of synthetic antiferromagnetic
  domains by focused ion beam irradiation},}\ }\href
  {https://doi.org/10.1063/5.0049926} {\bibfield  {journal} {\bibinfo
  {journal} {Appl. Phys. Lett.}\ }\textbf {\bibinfo {volume} {119}},\ \bibinfo
  {pages} {022409} (\bibinfo {year} {2021})}\BibitemShut {NoStop}%
\bibitem [{\citenamefont {Juge}\ \emph {et~al.}(2021)\citenamefont {Juge},
  \citenamefont {Bairagi}, \citenamefont {Rana}, \citenamefont {Vogel},
  \citenamefont {Sall}, \citenamefont {Mailly}, \citenamefont {Pham},
  \citenamefont {Zhang}, \citenamefont {Sisodia}, \citenamefont {Foerster},
  \citenamefont {Aballe}, \citenamefont {Belmeguenai}, \citenamefont
  {Roussign{\'e}}, \citenamefont {Auffret}, \citenamefont {Buda-Prejbeanu},
  \citenamefont {Gaudin}, \citenamefont {Ravelosona},\ and\ \citenamefont
  {Boulle}}]{Juge_2021}%
  \BibitemOpen
  \bibfield  {author} {\bibinfo {author} {\bibfnamefont {R.}~\bibnamefont
  {Juge}}, \bibinfo {author} {\bibfnamefont {K.}~\bibnamefont {Bairagi}},
  \bibinfo {author} {\bibfnamefont {K.~G.}\ \bibnamefont {Rana}}, \bibinfo
  {author} {\bibfnamefont {J.}~\bibnamefont {Vogel}}, \bibinfo {author}
  {\bibfnamefont {M.}~\bibnamefont {Sall}}, \bibinfo {author} {\bibfnamefont
  {D.}~\bibnamefont {Mailly}}, \bibinfo {author} {\bibfnamefont {V.~T.}\
  \bibnamefont {Pham}}, \bibinfo {author} {\bibfnamefont {Q.}~\bibnamefont
  {Zhang}}, \bibinfo {author} {\bibfnamefont {N.}~\bibnamefont {Sisodia}},
  \bibinfo {author} {\bibfnamefont {M.}~\bibnamefont {Foerster}}, \bibinfo
  {author} {\bibfnamefont {L.}~\bibnamefont {Aballe}}, \bibinfo {author}
  {\bibfnamefont {M.}~\bibnamefont {Belmeguenai}}, \bibinfo {author}
  {\bibfnamefont {Y.}~\bibnamefont {Roussign{\'e}}}, \bibinfo {author}
  {\bibfnamefont {S.}~\bibnamefont {Auffret}}, \bibinfo {author} {\bibfnamefont
  {L.~D.}\ \bibnamefont {Buda-Prejbeanu}}, \bibinfo {author} {\bibfnamefont
  {G.}~\bibnamefont {Gaudin}}, \bibinfo {author} {\bibfnamefont
  {D.}~\bibnamefont {Ravelosona}},\ and\ \bibinfo {author} {\bibfnamefont
  {O.}~\bibnamefont {Boulle}},\ }\bibfield  {title} {{\selectlanguage
  {english}\enquote {\bibinfo {title} {Helium ions put magnetic skyrmions on
  the track},}\ }}\href {https://doi.org/10.1021/acs.nanolett.1c00136}
  {\bibfield  {journal} {\bibinfo  {journal} {Nano Lett.}\ }\textbf {\bibinfo
  {volume} {21}},\ \bibinfo {pages} {2989--2996} (\bibinfo {year}
  {2021})}\BibitemShut {NoStop}%
\bibitem [{\citenamefont {Kern}\ \emph {et~al.}(2022)\citenamefont {Kern},
  \citenamefont {Pfau}, \citenamefont {Deinhart}, \citenamefont {Schneider},
  \citenamefont {Klose}, \citenamefont {Gerlinger}, \citenamefont {Wittrock},
  \citenamefont {Engel}, \citenamefont {Will}, \citenamefont {G{\"u}nther},
  \citenamefont {Liefferink}, \citenamefont {Mentink}, \citenamefont {Wintz},
  \citenamefont {Weigand}, \citenamefont {Huang}, \citenamefont {Battistelli},
  \citenamefont {Metternich}, \citenamefont {B{\"u}ttner}, \citenamefont
  {H{\"o}flich},\ and\ \citenamefont {Eisebitt}}]{Kern_2022}%
  \BibitemOpen
  \bibfield  {author} {\bibinfo {author} {\bibfnamefont {L.-M.}\ \bibnamefont
  {Kern}}, \bibinfo {author} {\bibfnamefont {B.}~\bibnamefont {Pfau}}, \bibinfo
  {author} {\bibfnamefont {V.}~\bibnamefont {Deinhart}}, \bibinfo {author}
  {\bibfnamefont {M.}~\bibnamefont {Schneider}}, \bibinfo {author}
  {\bibfnamefont {C.}~\bibnamefont {Klose}}, \bibinfo {author} {\bibfnamefont
  {K.}~\bibnamefont {Gerlinger}}, \bibinfo {author} {\bibfnamefont
  {S.}~\bibnamefont {Wittrock}}, \bibinfo {author} {\bibfnamefont
  {D.}~\bibnamefont {Engel}}, \bibinfo {author} {\bibfnamefont
  {I.}~\bibnamefont {Will}}, \bibinfo {author} {\bibfnamefont {C.~M.}\
  \bibnamefont {G{\"u}nther}}, \bibinfo {author} {\bibfnamefont
  {R.}~\bibnamefont {Liefferink}}, \bibinfo {author} {\bibfnamefont {J.~H.}\
  \bibnamefont {Mentink}}, \bibinfo {author} {\bibfnamefont {S.}~\bibnamefont
  {Wintz}}, \bibinfo {author} {\bibfnamefont {M.}~\bibnamefont {Weigand}},
  \bibinfo {author} {\bibfnamefont {M.-J.}\ \bibnamefont {Huang}}, \bibinfo
  {author} {\bibfnamefont {R.}~\bibnamefont {Battistelli}}, \bibinfo {author}
  {\bibfnamefont {D.}~\bibnamefont {Metternich}}, \bibinfo {author}
  {\bibfnamefont {F.}~\bibnamefont {B{\"u}ttner}}, \bibinfo {author}
  {\bibfnamefont {K.}~\bibnamefont {H{\"o}flich}},\ and\ \bibinfo {author}
  {\bibfnamefont {S.}~\bibnamefont {Eisebitt}},\ }\href@noop {} {\enquote
  {\bibinfo {title} {Deterministic generation and guided motion of magnetic
  skyrmions by focused {He$^+$-Ion} irradiation},}\ } (\bibinfo {year}
  {2022}),\ \bibinfo {note} {preprint},\ \Eprint
  {https://arxiv.org/abs/2202.12057} {arXiv:2202.12057 [cond-mat.mtrl-sci]}
  \BibitemShut {NoStop}%
\bibitem [{\citenamefont {Ishii}, \citenamefont {Nonaka},\ and\ \citenamefont
  {Hatakeyama}(1989)}]{Ishii_1989}%
  \BibitemOpen
  \bibfield  {author} {\bibinfo {author} {\bibfnamefont {O.}~\bibnamefont
  {Ishii}}, \bibinfo {author} {\bibfnamefont {K.}~\bibnamefont {Nonaka}},\ and\
  \bibinfo {author} {\bibfnamefont {I.}~\bibnamefont {Hatakeyama}},\ }\bibfield
   {title} {\enquote {\bibinfo {title} {Magnetization reversal in focussed
  ion-beam implanted {YIG} patterns},}\ }\href
  {https://doi.org/10.1109/tjmj.1989.4563991} {\bibfield  {journal} {\bibinfo
  {journal} {IEEE Transl. J. Magn. Jpn.}\ }\textbf {\bibinfo {volume} {4}},\
  \bibinfo {pages} {197--203} (\bibinfo {year} {1989})}\BibitemShut {NoStop}%
\bibitem [{\citenamefont {Aign}\ \emph {et~al.}(1998)\citenamefont {Aign},
  \citenamefont {Meyer}, \citenamefont {Lemerle}, \citenamefont {Jamet},
  \citenamefont {Ferr{\'{e}}}, \citenamefont {Mathet}, \citenamefont
  {Chappert}, \citenamefont {Gierak}, \citenamefont {Vieu}, \citenamefont
  {Rousseaux}, \citenamefont {Launois},\ and\ \citenamefont
  {Bernas}}]{Aign_1998}%
  \BibitemOpen
  \bibfield  {author} {\bibinfo {author} {\bibfnamefont {T.}~\bibnamefont
  {Aign}}, \bibinfo {author} {\bibfnamefont {P.}~\bibnamefont {Meyer}},
  \bibinfo {author} {\bibfnamefont {S.}~\bibnamefont {Lemerle}}, \bibinfo
  {author} {\bibfnamefont {J.~P.}\ \bibnamefont {Jamet}}, \bibinfo {author}
  {\bibfnamefont {J.}~\bibnamefont {Ferr{\'{e}}}}, \bibinfo {author}
  {\bibfnamefont {V.}~\bibnamefont {Mathet}}, \bibinfo {author} {\bibfnamefont
  {C.}~\bibnamefont {Chappert}}, \bibinfo {author} {\bibfnamefont
  {J.}~\bibnamefont {Gierak}}, \bibinfo {author} {\bibfnamefont
  {C.}~\bibnamefont {Vieu}}, \bibinfo {author} {\bibfnamefont {F.}~\bibnamefont
  {Rousseaux}}, \bibinfo {author} {\bibfnamefont {H.}~\bibnamefont {Launois}},\
  and\ \bibinfo {author} {\bibfnamefont {H.}~\bibnamefont {Bernas}},\
  }\bibfield  {title} {\enquote {\bibinfo {title} {Magnetization reversal in
  arrays of perpendicularly magnetized ultrathin dots coupled by dipolar
  interaction},}\ }\href {https://doi.org/10.1103/physrevlett.81.5656}
  {\bibfield  {journal} {\bibinfo  {journal} {Phys. Rev. Lett.}\ }\textbf
  {\bibinfo {volume} {81}},\ \bibinfo {pages} {5656--5659} (\bibinfo {year}
  {1998})}\BibitemShut {NoStop}%
\bibitem [{\citenamefont {Flaj{\v s}man}\ \emph {et~al.}(2020)\citenamefont
  {Flaj{\v s}man}, \citenamefont {Wagner}, \citenamefont {Va{\v n}atka},
  \citenamefont {Gloss}, \citenamefont {K{\v r}i{\v z}{\'a}kov{\'a}},
  \citenamefont {Schmid}, \citenamefont {Schultheiss},\ and\ \citenamefont
  {Urb{\'a}nek}}]{Flajsman_2020}%
  \BibitemOpen
  \bibfield  {author} {\bibinfo {author} {\bibfnamefont {L.}~\bibnamefont
  {Flaj{\v s}man}}, \bibinfo {author} {\bibfnamefont {K.}~\bibnamefont
  {Wagner}}, \bibinfo {author} {\bibfnamefont {M.}~\bibnamefont {Va{\v
  n}atka}}, \bibinfo {author} {\bibfnamefont {J.}~\bibnamefont {Gloss}},
  \bibinfo {author} {\bibfnamefont {V.}~\bibnamefont {K{\v r}i{\v
  z}{\'a}kov{\'a}}}, \bibinfo {author} {\bibfnamefont {M.}~\bibnamefont
  {Schmid}}, \bibinfo {author} {\bibfnamefont {H.}~\bibnamefont
  {Schultheiss}},\ and\ \bibinfo {author} {\bibfnamefont {M.}~\bibnamefont
  {Urb{\'a}nek}},\ }\bibfield  {title} {\enquote {\bibinfo {title} {Zero-field
  propagation of spin waves in waveguides prepared by focused ion beam direct
  writing},}\ }\href {https://doi.org/10.1103/PhysRevB.101.014436} {\bibfield
  {journal} {\bibinfo  {journal} {Phys. Rev. B}\ }\textbf {\bibinfo {volume}
  {101}},\ \bibinfo {pages} {014436} (\bibinfo {year} {2020})}\BibitemShut
  {NoStop}%
\bibitem [{\citenamefont {de~Jong}\ \emph {et~al.}(2022)\citenamefont
  {de~Jong}, \citenamefont {Meijer}, \citenamefont {Lucassen}, \citenamefont
  {van Liempt}, \citenamefont {Swagten}, \citenamefont {Koopmans},\ and\
  \citenamefont {Lavrijsen}}]{Jong_2022}%
  \BibitemOpen
  \bibfield  {author} {\bibinfo {author} {\bibfnamefont {M.~C.~H.}\
  \bibnamefont {de~Jong}}, \bibinfo {author} {\bibfnamefont {M.~J.}\
  \bibnamefont {Meijer}}, \bibinfo {author} {\bibfnamefont {J.}~\bibnamefont
  {Lucassen}}, \bibinfo {author} {\bibfnamefont {J.}~\bibnamefont {van
  Liempt}}, \bibinfo {author} {\bibfnamefont {H.~J.~M.}\ \bibnamefont
  {Swagten}}, \bibinfo {author} {\bibfnamefont {B.}~\bibnamefont {Koopmans}},\
  and\ \bibinfo {author} {\bibfnamefont {R.}~\bibnamefont {Lavrijsen}},\
  }\bibfield  {title} {\enquote {\bibinfo {title} {Local control of magnetic
  interface effects in chiral {Ir|Co|Pt} multilayers using {G}a$^+$ ion
  irradiation},}\ }\href {https://doi.org/10.1103/physrevb.105.064429}
  {\bibfield  {journal} {\bibinfo  {journal} {Phys. Rev. B}\ }\textbf {\bibinfo
  {volume} {105}},\ \bibinfo {pages} {064429} (\bibinfo {year}
  {2022})}\BibitemShut {NoStop}%
\bibitem [{\citenamefont {Shibata}\ \emph {et~al.}(2015)\citenamefont
  {Shibata}, \citenamefont {Iwasaki}, \citenamefont {Kanazawa}, \citenamefont
  {Aizawa}, \citenamefont {Tanigaki}, \citenamefont {Shirai}, \citenamefont
  {Nakajima}, \citenamefont {Kubota}, \citenamefont {Kawasaki}, \citenamefont
  {Park}, \citenamefont {Shindo}, \citenamefont {Nagaosa},\ and\ \citenamefont
  {Tokura}}]{Shibata_2015}%
  \BibitemOpen
  \bibfield  {author} {\bibinfo {author} {\bibfnamefont {K.}~\bibnamefont
  {Shibata}}, \bibinfo {author} {\bibfnamefont {J.}~\bibnamefont {Iwasaki}},
  \bibinfo {author} {\bibfnamefont {N.}~\bibnamefont {Kanazawa}}, \bibinfo
  {author} {\bibfnamefont {S.}~\bibnamefont {Aizawa}}, \bibinfo {author}
  {\bibfnamefont {T.}~\bibnamefont {Tanigaki}}, \bibinfo {author}
  {\bibfnamefont {M.}~\bibnamefont {Shirai}}, \bibinfo {author} {\bibfnamefont
  {T.}~\bibnamefont {Nakajima}}, \bibinfo {author} {\bibfnamefont
  {M.}~\bibnamefont {Kubota}}, \bibinfo {author} {\bibfnamefont
  {M.}~\bibnamefont {Kawasaki}}, \bibinfo {author} {\bibfnamefont {H.~S.}\
  \bibnamefont {Park}}, \bibinfo {author} {\bibfnamefont {D.}~\bibnamefont
  {Shindo}}, \bibinfo {author} {\bibfnamefont {N.}~\bibnamefont {Nagaosa}},\
  and\ \bibinfo {author} {\bibfnamefont {Y.}~\bibnamefont {Tokura}},\
  }\bibfield  {title} {\enquote {\bibinfo {title} {Large anisotropic
  deformation of skyrmions in strained crystal},}\ }\href
  {https://doi.org/10.1038/nnano.2015.113} {\bibfield  {journal} {\bibinfo
  {journal} {Nat. Nanotechnol.}\ }\textbf {\bibinfo {volume} {10}},\ \bibinfo
  {pages} {589--592} (\bibinfo {year} {2015})}\BibitemShut {NoStop}%
\bibitem [{\citenamefont {Sapozhnikov}\ \emph {et~al.}(2019)\citenamefont
  {Sapozhnikov}, \citenamefont {Petrov}, \citenamefont {Gusev}, \citenamefont
  {Temiryazev}, \citenamefont {Ermolaeva}, \citenamefont {Mironov},\ and\
  \citenamefont {Udalov}}]{Sapozhnikov2019}%
  \BibitemOpen
  \bibfield  {author} {\bibinfo {author} {\bibfnamefont {M.~V.}\ \bibnamefont
  {Sapozhnikov}}, \bibinfo {author} {\bibfnamefont {Y.~V.}\ \bibnamefont
  {Petrov}}, \bibinfo {author} {\bibfnamefont {N.~S.}\ \bibnamefont {Gusev}},
  \bibinfo {author} {\bibfnamefont {A.~G.}\ \bibnamefont {Temiryazev}},
  \bibinfo {author} {\bibfnamefont {O.~L.}\ \bibnamefont {Ermolaeva}}, \bibinfo
  {author} {\bibfnamefont {V.~L.}\ \bibnamefont {Mironov}},\ and\ \bibinfo
  {author} {\bibfnamefont {O.~G.}\ \bibnamefont {Udalov}},\ }\bibfield  {title}
  {\enquote {\bibinfo {title} {Artificial dense lattices of magnetic
  skyrmions},}\ }\href {https://doi.org/10.3390/ma13010099} {\bibfield
  {journal} {\bibinfo  {journal} {Materials}\ }\textbf {\bibinfo {volume}
  {13}},\ \bibinfo {pages} {99} (\bibinfo {year} {2019})}\BibitemShut {NoStop}%
\bibitem [{\citenamefont {Fallon}\ \emph {et~al.}(2020)\citenamefont {Fallon},
  \citenamefont {Hughes}, \citenamefont {Zeissler}, \citenamefont {Legrand},
  \citenamefont {Ajejas}, \citenamefont {Maccariello}, \citenamefont
  {McFadzean}, \citenamefont {Smith}, \citenamefont {McGrouther}, \citenamefont
  {Collin}, \citenamefont {Reyren}, \citenamefont {Cros}, \citenamefont
  {Marrows},\ and\ \citenamefont {McVitie}}]{Fallon2020}%
  \BibitemOpen
  \bibfield  {author} {\bibinfo {author} {\bibfnamefont {K.}~\bibnamefont
  {Fallon}}, \bibinfo {author} {\bibfnamefont {S.}~\bibnamefont {Hughes}},
  \bibinfo {author} {\bibfnamefont {K.}~\bibnamefont {Zeissler}}, \bibinfo
  {author} {\bibfnamefont {W.}~\bibnamefont {Legrand}}, \bibinfo {author}
  {\bibfnamefont {F.}~\bibnamefont {Ajejas}}, \bibinfo {author} {\bibfnamefont
  {D.}~\bibnamefont {Maccariello}}, \bibinfo {author} {\bibfnamefont
  {S.}~\bibnamefont {McFadzean}}, \bibinfo {author} {\bibfnamefont
  {W.}~\bibnamefont {Smith}}, \bibinfo {author} {\bibfnamefont
  {D.}~\bibnamefont {McGrouther}}, \bibinfo {author} {\bibfnamefont
  {S.}~\bibnamefont {Collin}}, \bibinfo {author} {\bibfnamefont
  {N.}~\bibnamefont {Reyren}}, \bibinfo {author} {\bibfnamefont
  {V.}~\bibnamefont {Cros}}, \bibinfo {author} {\bibfnamefont {C.~H.}\
  \bibnamefont {Marrows}},\ and\ \bibinfo {author} {\bibfnamefont
  {S.}~\bibnamefont {McVitie}},\ }\bibfield  {title} {{\selectlanguage
  {english}\enquote {\bibinfo {title} {Controlled individual skyrmion
  nucleation at artificial defects formed by ion irradiation},}\ }}\href
  {https://doi.org/10.1002/smll.201907450} {\bibfield  {journal} {\bibinfo
  {journal} {Small}\ }\textbf {\bibinfo {volume} {16}},\ \bibinfo {pages}
  {e1907450} (\bibinfo {year} {2020})}\BibitemShut {NoStop}%
\bibitem [{\citenamefont {Hu}\ \emph {et~al.}(2022)\citenamefont {Hu},
  \citenamefont {Zhang}, \citenamefont {Zhu}, \citenamefont {Song},
  \citenamefont {Huang}, \citenamefont {Liu}, \citenamefont {Meng},
  \citenamefont {Deng}, \citenamefont {Zhu}, \citenamefont {Guan},
  \citenamefont {Yang}, \citenamefont {Si}, \citenamefont {Zhang},\ and\
  \citenamefont {Peng}}]{Hu_2022}%
  \BibitemOpen
  \bibfield  {author} {\bibinfo {author} {\bibfnamefont {Y.}~\bibnamefont
  {Hu}}, \bibinfo {author} {\bibfnamefont {S.}~\bibnamefont {Zhang}}, \bibinfo
  {author} {\bibfnamefont {Y.}~\bibnamefont {Zhu}}, \bibinfo {author}
  {\bibfnamefont {C.}~\bibnamefont {Song}}, \bibinfo {author} {\bibfnamefont
  {J.}~\bibnamefont {Huang}}, \bibinfo {author} {\bibfnamefont
  {C.}~\bibnamefont {Liu}}, \bibinfo {author} {\bibfnamefont {X.}~\bibnamefont
  {Meng}}, \bibinfo {author} {\bibfnamefont {X.}~\bibnamefont {Deng}}, \bibinfo
  {author} {\bibfnamefont {L.}~\bibnamefont {Zhu}}, \bibinfo {author}
  {\bibfnamefont {C.}~\bibnamefont {Guan}}, \bibinfo {author} {\bibfnamefont
  {H.}~\bibnamefont {Yang}}, \bibinfo {author} {\bibfnamefont {M.}~\bibnamefont
  {Si}}, \bibinfo {author} {\bibfnamefont {J.}~\bibnamefont {Zhang}},\ and\
  \bibinfo {author} {\bibfnamefont {Y.}~\bibnamefont {Peng}},\ }\bibfield
  {title} {\enquote {\bibinfo {title} {Precise tuning of skyrmion density in a
  controllable manner by ion irradiation},}\ }\href
  {https://doi.org/10.1021/acsami.2c07268} {\bibfield  {journal} {\bibinfo
  {journal} {ACS Appl. Mater. Interfaces}\ }\textbf {\bibinfo {volume} {14}},\
  \bibinfo {pages} {34011--34019} (\bibinfo {year} {2022})}\BibitemShut
  {NoStop}%
\bibitem [{\citenamefont {Cress}\ \emph {et~al.}(2021)\citenamefont {Cress},
  \citenamefont {Wickramaratne}, \citenamefont {Rosenberger}, \citenamefont
  {Hennighausen}, \citenamefont {Callahan}, \citenamefont {LaGasse},
  \citenamefont {Bernstein}, \citenamefont {van~'t Erve}, \citenamefont
  {Jonker}, \citenamefont {Qadri}, \citenamefont {Prestigiacomo}, \citenamefont
  {Currie}, \citenamefont {Mazin},\ and\ \citenamefont {Bennett}}]{Cress2021}%
  \BibitemOpen
  \bibfield  {author} {\bibinfo {author} {\bibfnamefont {C.~D.}\ \bibnamefont
  {Cress}}, \bibinfo {author} {\bibfnamefont {D.}~\bibnamefont
  {Wickramaratne}}, \bibinfo {author} {\bibfnamefont {M.~R.}\ \bibnamefont
  {Rosenberger}}, \bibinfo {author} {\bibfnamefont {Z.}~\bibnamefont
  {Hennighausen}}, \bibinfo {author} {\bibfnamefont {P.~G.}\ \bibnamefont
  {Callahan}}, \bibinfo {author} {\bibfnamefont {S.~W.}\ \bibnamefont
  {LaGasse}}, \bibinfo {author} {\bibfnamefont {N.}~\bibnamefont {Bernstein}},
  \bibinfo {author} {\bibfnamefont {O.~M.}\ \bibnamefont {van~'t Erve}},
  \bibinfo {author} {\bibfnamefont {B.~T.}\ \bibnamefont {Jonker}}, \bibinfo
  {author} {\bibfnamefont {S.~B.}\ \bibnamefont {Qadri}}, \bibinfo {author}
  {\bibfnamefont {J.~C.}\ \bibnamefont {Prestigiacomo}}, \bibinfo {author}
  {\bibfnamefont {M.}~\bibnamefont {Currie}}, \bibinfo {author} {\bibfnamefont
  {I.~I.}\ \bibnamefont {Mazin}},\ and\ \bibinfo {author} {\bibfnamefont
  {S.~P.}\ \bibnamefont {Bennett}},\ }\bibfield  {title} {{\selectlanguage
  {english}\enquote {\bibinfo {title} {{Direct-Write} of nanoscale domains with
  tunable metamagnetic order in {FeRh} thin films},}\ }}\href
  {https://doi.org/10.1021/acsami.0c13565} {\bibfield  {journal} {\bibinfo
  {journal} {ACS Appl. Mater. Interfaces}\ }\textbf {\bibinfo {volume} {13}},\
  \bibinfo {pages} {836--847} (\bibinfo {year} {2021})}\BibitemShut {NoStop}%
\bibitem [{\citenamefont {R{\"o}der}\ \emph {et~al.}(2015)\citenamefont
  {R{\"o}der}, \citenamefont {Hlawacek}, \citenamefont {Wintz}, \citenamefont
  {H{\"u}bner}, \citenamefont {Bischoff}, \citenamefont {Lichte}, \citenamefont
  {Potzger}, \citenamefont {Lindner}, \citenamefont {Fassbender},\ and\
  \citenamefont {Bali}}]{Roder2015}%
  \BibitemOpen
  \bibfield  {author} {\bibinfo {author} {\bibfnamefont {F.}~\bibnamefont
  {R{\"o}der}}, \bibinfo {author} {\bibfnamefont {G.}~\bibnamefont {Hlawacek}},
  \bibinfo {author} {\bibfnamefont {S.}~\bibnamefont {Wintz}}, \bibinfo
  {author} {\bibfnamefont {R.}~\bibnamefont {H{\"u}bner}}, \bibinfo {author}
  {\bibfnamefont {L.}~\bibnamefont {Bischoff}}, \bibinfo {author}
  {\bibfnamefont {H.}~\bibnamefont {Lichte}}, \bibinfo {author} {\bibfnamefont
  {K.}~\bibnamefont {Potzger}}, \bibinfo {author} {\bibfnamefont
  {J.}~\bibnamefont {Lindner}}, \bibinfo {author} {\bibfnamefont
  {J.}~\bibnamefont {Fassbender}},\ and\ \bibinfo {author} {\bibfnamefont
  {R.}~\bibnamefont {Bali}},\ }\bibfield  {title} {\enquote {\bibinfo {title}
  {Direct depth- and lateral- imaging of nanoscale magnets generated by ion
  impact},}\ }\href {https://doi.org/10.1038/srep16786} {\bibfield  {journal}
  {\bibinfo  {journal} {Sci. Rep.}\ }\textbf {\bibinfo {volume} {5}},\ \bibinfo
  {pages} {16786} (\bibinfo {year} {2015})}\BibitemShut {NoStop}%
\bibitem [{\citenamefont {Men{\'{e}}ndez}\ \emph {et~al.}(2008)\citenamefont
  {Men{\'{e}}ndez}, \citenamefont {Liedke}, \citenamefont {Fassbender},
  \citenamefont {Gemming}, \citenamefont {Weber}, \citenamefont {Heyderman},
  \citenamefont {Rao}, \citenamefont {Deevi}, \citenamefont {Suri{\~{n}}ach},
  \citenamefont {Bar{\'{o}}}, \citenamefont {Sort},\ and\ \citenamefont
  {Nogu{\'{e}}s}}]{Menendez_2008}%
  \BibitemOpen
  \bibfield  {author} {\bibinfo {author} {\bibfnamefont {E.}~\bibnamefont
  {Men{\'{e}}ndez}}, \bibinfo {author} {\bibfnamefont {M.~O.}\ \bibnamefont
  {Liedke}}, \bibinfo {author} {\bibfnamefont {J.}~\bibnamefont {Fassbender}},
  \bibinfo {author} {\bibfnamefont {T.}~\bibnamefont {Gemming}}, \bibinfo
  {author} {\bibfnamefont {A.}~\bibnamefont {Weber}}, \bibinfo {author}
  {\bibfnamefont {L.~J.}\ \bibnamefont {Heyderman}}, \bibinfo {author}
  {\bibfnamefont {K.~V.}\ \bibnamefont {Rao}}, \bibinfo {author} {\bibfnamefont
  {S.~C.}\ \bibnamefont {Deevi}}, \bibinfo {author} {\bibfnamefont
  {S.}~\bibnamefont {Suri{\~{n}}ach}}, \bibinfo {author} {\bibfnamefont
  {M.~D.}\ \bibnamefont {Bar{\'{o}}}}, \bibinfo {author} {\bibfnamefont
  {J.}~\bibnamefont {Sort}},\ and\ \bibinfo {author} {\bibfnamefont
  {J.}~\bibnamefont {Nogu{\'{e}}s}},\ }\bibfield  {title} {\enquote {\bibinfo
  {title} {Direct magnetic patterning due to the generation of ferromagnetism
  by selective ion irradiation of paramagnetic {FeAl} alloys},}\ }\href
  {https://doi.org/10.1002/smll.200800783} {\bibfield  {journal} {\bibinfo
  {journal} {Small}\ }\textbf {\bibinfo {volume} {5}},\ \bibinfo {pages}
  {229--234} (\bibinfo {year} {2008})}\BibitemShut {NoStop}%
\bibitem [{\citenamefont {Urb{\'{a}}nek}\ \emph {et~al.}(2018)\citenamefont
  {Urb{\'{a}}nek}, \citenamefont {Flaj{\v{s}}man}, \citenamefont
  {K{\v{r}}i{\v{z}}{\'{a}}kov{\'{a}}}, \citenamefont {Gloss}, \citenamefont
  {Hork{\'{y}}}, \citenamefont {Schmid},\ and\ \citenamefont
  {Varga}}]{Urbanek_2018}%
  \BibitemOpen
  \bibfield  {author} {\bibinfo {author} {\bibfnamefont {M.}~\bibnamefont
  {Urb{\'{a}}nek}}, \bibinfo {author} {\bibfnamefont {L.}~\bibnamefont
  {Flaj{\v{s}}man}}, \bibinfo {author} {\bibfnamefont {V.}~\bibnamefont
  {K{\v{r}}i{\v{z}}{\'{a}}kov{\'{a}}}}, \bibinfo {author} {\bibfnamefont
  {J.}~\bibnamefont {Gloss}}, \bibinfo {author} {\bibfnamefont
  {M.}~\bibnamefont {Hork{\'{y}}}}, \bibinfo {author} {\bibfnamefont
  {M.}~\bibnamefont {Schmid}},\ and\ \bibinfo {author} {\bibfnamefont
  {P.}~\bibnamefont {Varga}},\ }\bibfield  {title} {\enquote {\bibinfo {title}
  {Research update: focused ion beam direct writing of magnetic patterns with
  controlled structural and magnetic properties},}\ }\href
  {https://doi.org/10.1063/1.5029367} {\bibfield  {journal} {\bibinfo
  {journal} {{APL} Materials}\ }\textbf {\bibinfo {volume} {6}},\ \bibinfo
  {pages} {060701} (\bibinfo {year} {2018})}\BibitemShut {NoStop}%
\bibitem [{\citenamefont {Kiechle}\ \emph {et~al.}(2023)\citenamefont
  {Kiechle}, \citenamefont {Papp}, \citenamefont {Mendisch}, \citenamefont
  {Ahrens}, \citenamefont {Golibrzuch}, \citenamefont {Bernstein},
  \citenamefont {Porod}, \citenamefont {Csaba},\ and\ \citenamefont
  {Becherer}}]{Kiechle_2023}%
  \BibitemOpen
  \bibfield  {author} {\bibinfo {author} {\bibfnamefont {M.}~\bibnamefont
  {Kiechle}}, \bibinfo {author} {\bibfnamefont {A.}~\bibnamefont {Papp}},
  \bibinfo {author} {\bibfnamefont {S.}~\bibnamefont {Mendisch}}, \bibinfo
  {author} {\bibfnamefont {V.}~\bibnamefont {Ahrens}}, \bibinfo {author}
  {\bibfnamefont {M.}~\bibnamefont {Golibrzuch}}, \bibinfo {author}
  {\bibfnamefont {G.~H.}\ \bibnamefont {Bernstein}}, \bibinfo {author}
  {\bibfnamefont {W.}~\bibnamefont {Porod}}, \bibinfo {author} {\bibfnamefont
  {G.}~\bibnamefont {Csaba}},\ and\ \bibinfo {author} {\bibfnamefont
  {M.}~\bibnamefont {Becherer}},\ }\bibfield  {title} {\enquote {\bibinfo
  {title} {Spin-wave optics in {YIG} realized by ion-beam irradiation},}\
  }\href {https://doi.org/10.1002/smll.202207293} {\bibfield  {journal}
  {\bibinfo  {journal} {Small}\ ,\ \bibinfo {pages} {2207293}} (\bibinfo {year}
  {2023})}\BibitemShut {NoStop}%
\bibitem [{\citenamefont {Sawafta}\ \emph {et~al.}(2014)\citenamefont
  {Sawafta}, \citenamefont {Clancy}, \citenamefont {Carlsen}, \citenamefont
  {Huber},\ and\ \citenamefont {Hall}}]{Sawafta2014}%
  \BibitemOpen
  \bibfield  {author} {\bibinfo {author} {\bibfnamefont {F.}~\bibnamefont
  {Sawafta}}, \bibinfo {author} {\bibfnamefont {B.}~\bibnamefont {Clancy}},
  \bibinfo {author} {\bibfnamefont {A.~T.}\ \bibnamefont {Carlsen}}, \bibinfo
  {author} {\bibfnamefont {M.}~\bibnamefont {Huber}},\ and\ \bibinfo {author}
  {\bibfnamefont {A.~R.}\ \bibnamefont {Hall}},\ }\bibfield  {title} {\enquote
  {\bibinfo {title} {Solid-state nanopores and nanopore arrays optimized for
  optical detection},}\ }\href {https://doi.org/10.1039/c4nr00305e} {\bibfield
  {journal} {\bibinfo  {journal} {Nanoscale}\ }\textbf {\bibinfo {volume}
  {6}},\ \bibinfo {pages} {6991--6996} (\bibinfo {year} {2014})}\BibitemShut
  {NoStop}%
\bibitem [{\citenamefont {Iberi}\ \emph {et~al.}(2016)\citenamefont {Iberi},
  \citenamefont {Liang}, \citenamefont {Ievlev}, \citenamefont {Stanford},
  \citenamefont {Lin}, \citenamefont {Li}, \citenamefont {Mahjouri-Samani},
  \citenamefont {Jesse}, \citenamefont {Sumpter}, \citenamefont {Kalinin},
  \citenamefont {Joy}, \citenamefont {Xiao}, \citenamefont {Belianinov},\ and\
  \citenamefont {Ovchinnikova}}]{Iberi2016a}%
  \BibitemOpen
  \bibfield  {author} {\bibinfo {author} {\bibfnamefont {V.}~\bibnamefont
  {Iberi}}, \bibinfo {author} {\bibfnamefont {L.}~\bibnamefont {Liang}},
  \bibinfo {author} {\bibfnamefont {A.~V.}\ \bibnamefont {Ievlev}}, \bibinfo
  {author} {\bibfnamefont {M.~G.}\ \bibnamefont {Stanford}}, \bibinfo {author}
  {\bibfnamefont {M.-W.}\ \bibnamefont {Lin}}, \bibinfo {author} {\bibfnamefont
  {X.}~\bibnamefont {Li}}, \bibinfo {author} {\bibfnamefont {M.}~\bibnamefont
  {Mahjouri-Samani}}, \bibinfo {author} {\bibfnamefont {S.}~\bibnamefont
  {Jesse}}, \bibinfo {author} {\bibfnamefont {B.~G.}\ \bibnamefont {Sumpter}},
  \bibinfo {author} {\bibfnamefont {S.~V.}\ \bibnamefont {Kalinin}}, \bibinfo
  {author} {\bibfnamefont {D.~C.}\ \bibnamefont {Joy}}, \bibinfo {author}
  {\bibfnamefont {K.}~\bibnamefont {Xiao}}, \bibinfo {author} {\bibfnamefont
  {A.}~\bibnamefont {Belianinov}},\ and\ \bibinfo {author} {\bibfnamefont
  {O.~S.}\ \bibnamefont {Ovchinnikova}},\ }\bibfield  {title} {\enquote
  {\bibinfo {title} {Nanoforging single layer {MoSe2} through defect
  engineering with focused helium ion beams},}\ }\href
  {https://doi.org/10.1038/srep30481} {\bibfield  {journal} {\bibinfo
  {journal} {Sci. Rep.}\ }\textbf {\bibinfo {volume} {6}},\ \bibinfo {pages}
  {30481} (\bibinfo {year} {2016})}\BibitemShut {NoStop}%
\bibitem [{\citenamefont {Kapitonov}\ \emph {et~al.}(2016)\citenamefont
  {Kapitonov}, \citenamefont {Shapochkin}, \citenamefont {Beliaev},
  \citenamefont {Petrov}, \citenamefont {Efimov}, \citenamefont {Eliseev},
  \citenamefont {Lovtcius}, \citenamefont {Petrov},\ and\ \citenamefont
  {Ovsyankin}}]{Kapitonov_2016}%
  \BibitemOpen
  \bibfield  {author} {\bibinfo {author} {\bibfnamefont {Y.~V.}\ \bibnamefont
  {Kapitonov}}, \bibinfo {author} {\bibfnamefont {P.~Y.}\ \bibnamefont
  {Shapochkin}}, \bibinfo {author} {\bibfnamefont {L.~Y.}\ \bibnamefont
  {Beliaev}}, \bibinfo {author} {\bibfnamefont {Y.~V.}\ \bibnamefont {Petrov}},
  \bibinfo {author} {\bibfnamefont {Y.~P.}\ \bibnamefont {Efimov}}, \bibinfo
  {author} {\bibfnamefont {S.~A.}\ \bibnamefont {Eliseev}}, \bibinfo {author}
  {\bibfnamefont {V.~A.}\ \bibnamefont {Lovtcius}}, \bibinfo {author}
  {\bibfnamefont {V.~V.}\ \bibnamefont {Petrov}},\ and\ \bibinfo {author}
  {\bibfnamefont {V.~V.}\ \bibnamefont {Ovsyankin}},\ }\bibfield  {title}
  {{\selectlanguage {english}\enquote {\bibinfo {title} {Ion-beam-assisted
  spatial modulation of inhomogeneous broadening of a quantum well resonance:
  excitonic diffraction grating},}\ }}\href
  {https://doi.org/10.1364/OL.41.000104} {\bibfield  {journal} {\bibinfo
  {journal} {Opt. Lett.}\ }\textbf {\bibinfo {volume} {41}},\ \bibinfo {pages}
  {104--106} (\bibinfo {year} {2016})}\BibitemShut {NoStop}%
\bibitem [{\citenamefont {Grosso}\ \emph {et~al.}(2017)\citenamefont {Grosso},
  \citenamefont {Moon}, \citenamefont {Lienhard}, \citenamefont {Ali},
  \citenamefont {Efetov}, \citenamefont {Furchi}, \citenamefont
  {Jarillo-Herrero}, \citenamefont {Ford}, \citenamefont {Aharonovich},\ and\
  \citenamefont {Englund}}]{Grosso_2017}%
  \BibitemOpen
  \bibfield  {author} {\bibinfo {author} {\bibfnamefont {G.}~\bibnamefont
  {Grosso}}, \bibinfo {author} {\bibfnamefont {H.}~\bibnamefont {Moon}},
  \bibinfo {author} {\bibfnamefont {B.}~\bibnamefont {Lienhard}}, \bibinfo
  {author} {\bibfnamefont {S.}~\bibnamefont {Ali}}, \bibinfo {author}
  {\bibfnamefont {D.~K.}\ \bibnamefont {Efetov}}, \bibinfo {author}
  {\bibfnamefont {M.~M.}\ \bibnamefont {Furchi}}, \bibinfo {author}
  {\bibfnamefont {P.}~\bibnamefont {Jarillo-Herrero}}, \bibinfo {author}
  {\bibfnamefont {M.~J.}\ \bibnamefont {Ford}}, \bibinfo {author}
  {\bibfnamefont {I.}~\bibnamefont {Aharonovich}},\ and\ \bibinfo {author}
  {\bibfnamefont {D.}~\bibnamefont {Englund}},\ }\bibfield  {title} {\enquote
  {\bibinfo {title} {Tunable and high-purity room temperature single-photon
  emission from atomic defects in hexagonal boron nitride},}\ }\href
  {https://doi.org/10.1038/s41467-017-00810-2} {\bibfield  {journal} {\bibinfo
  {journal} {Nat. Commun.}\ }\textbf {\bibinfo {volume} {8}} (\bibinfo {year}
  {2017}),\ 10.1038/s41467-017-00810-2}\BibitemShut {NoStop}%
\bibitem [{\citenamefont {Reithmaier}(1998)}]{Reithmaier_1998}%
  \BibitemOpen
  \bibfield  {author} {\bibinfo {author} {\bibfnamefont {J.~P.}\ \bibnamefont
  {Reithmaier}},\ }\bibfield  {title} {\enquote {\bibinfo {title} {Focused
  ion-beam implantation induced thermal quantum-well intermixing for monolithic
  optoelectronic device integration},}\ }\href
  {https://doi.org/10.1109/2944.720469} {\bibfield  {journal} {\bibinfo
  {journal} {IEEE J. Sel. Top. Quantum Electron.}\ }\textbf {\bibinfo {volume}
  {4}},\ \bibinfo {pages} {595--605} (\bibinfo {year} {1998})}\BibitemShut
  {NoStop}%
\bibitem [{\citenamefont {Jiang}\ \emph {et~al.}(2018)\citenamefont {Jiang},
  \citenamefont {Cai}, \citenamefont {Luo}, \citenamefont {Xiang},
  \citenamefont {Zhang}, \citenamefont {Ren}, \citenamefont {Zhang},\ and\
  \citenamefont {Xu}}]{Jiang2018}%
  \BibitemOpen
  \bibfield  {author} {\bibinfo {author} {\bibfnamefont {X.}~\bibnamefont
  {Jiang}}, \bibinfo {author} {\bibfnamefont {W.}~\bibnamefont {Cai}}, \bibinfo
  {author} {\bibfnamefont {W.}~\bibnamefont {Luo}}, \bibinfo {author}
  {\bibfnamefont {Y.}~\bibnamefont {Xiang}}, \bibinfo {author} {\bibfnamefont
  {N.}~\bibnamefont {Zhang}}, \bibinfo {author} {\bibfnamefont
  {M.}~\bibnamefont {Ren}}, \bibinfo {author} {\bibfnamefont {X.}~\bibnamefont
  {Zhang}},\ and\ \bibinfo {author} {\bibfnamefont {J.}~\bibnamefont {Xu}},\
  }\bibfield  {title} {\enquote {\bibinfo {title} {Near-field imaging of
  graphene triangles patterned by helium ion lithography},}\ }\href
  {https://doi.org/10.1088/1361-6528/aad0b4} {\bibfield  {journal} {\bibinfo
  {journal} {Nanotechnology}\ }\textbf {\bibinfo {volume} {29}},\ \bibinfo
  {pages} {385205} (\bibinfo {year} {2018})}\BibitemShut {NoStop}%
\bibitem [{\citenamefont {Takashima}\ \emph {et~al.}(2019)\citenamefont
  {Takashima}, \citenamefont {Fukuda}, \citenamefont {Maruya}, \citenamefont
  {Tashima}, \citenamefont {Schell},\ and\ \citenamefont
  {Takeuchi}}]{Takashima_2019a}%
  \BibitemOpen
  \bibfield  {author} {\bibinfo {author} {\bibfnamefont {H.}~\bibnamefont
  {Takashima}}, \bibinfo {author} {\bibfnamefont {A.}~\bibnamefont {Fukuda}},
  \bibinfo {author} {\bibfnamefont {H.}~\bibnamefont {Maruya}}, \bibinfo
  {author} {\bibfnamefont {T.}~\bibnamefont {Tashima}}, \bibinfo {author}
  {\bibfnamefont {A.~W.}\ \bibnamefont {Schell}},\ and\ \bibinfo {author}
  {\bibfnamefont {S.}~\bibnamefont {Takeuchi}},\ }\bibfield  {title} {\enquote
  {\bibinfo {title} {Fabrication of a nanofiber {B}ragg cavity with high
  quality factor using a focused helium ion beam},}\ }\href
  {https://doi.org/10.1364/oe.27.006792} {\bibfield  {journal} {\bibinfo
  {journal} {Opt. Express}\ }\textbf {\bibinfo {volume} {27}},\ \bibinfo
  {pages} {6792} (\bibinfo {year} {2019})}\BibitemShut {NoStop}%
\bibitem [{\citenamefont {McGehee}\ \emph {et~al.}(2017)\citenamefont
  {McGehee}, \citenamefont {Michels}, \citenamefont {Aksyuk},\ and\
  \citenamefont {McClelland}}]{McGehee_2017a}%
  \BibitemOpen
  \bibfield  {author} {\bibinfo {author} {\bibfnamefont {W.~R.}\ \bibnamefont
  {McGehee}}, \bibinfo {author} {\bibfnamefont {T.}~\bibnamefont {Michels}},
  \bibinfo {author} {\bibfnamefont {V.}~\bibnamefont {Aksyuk}},\ and\ \bibinfo
  {author} {\bibfnamefont {J.~J.}\ \bibnamefont {McClelland}},\ }\bibfield
  {title} {\enquote {\bibinfo {title} {Two-dimensional imaging and modification
  of nanophotonic resonator modes using a focused ion beam},}\ }\href
  {https://doi.org/10.1364/optica.4.001444} {\bibfield  {journal} {\bibinfo
  {journal} {Optica}\ }\textbf {\bibinfo {volume} {4}},\ \bibinfo {pages}
  {1444} (\bibinfo {year} {2017})}\BibitemShut {NoStop}%
\bibitem [{\citenamefont {Lohmeyer}\ \emph {et~al.}(2008)\citenamefont
  {Lohmeyer}, \citenamefont {Kalden}, \citenamefont {Sebald}, \citenamefont
  {Kruse}, \citenamefont {Hommel},\ and\ \citenamefont
  {Gutowski}}]{Lohmeyer_2008}%
  \BibitemOpen
  \bibfield  {author} {\bibinfo {author} {\bibfnamefont {H.}~\bibnamefont
  {Lohmeyer}}, \bibinfo {author} {\bibfnamefont {J.}~\bibnamefont {Kalden}},
  \bibinfo {author} {\bibfnamefont {K.}~\bibnamefont {Sebald}}, \bibinfo
  {author} {\bibfnamefont {C.}~\bibnamefont {Kruse}}, \bibinfo {author}
  {\bibfnamefont {D.}~\bibnamefont {Hommel}},\ and\ \bibinfo {author}
  {\bibfnamefont {J.}~\bibnamefont {Gutowski}},\ }\bibfield  {title} {\enquote
  {\bibinfo {title} {Fine tuning of quantum-dot pillar microcavities by focused
  ion beam milling},}\ }\href {https://doi.org/10.1063/1.2827574} {\bibfield
  {journal} {\bibinfo  {journal} {Appl. Phys. Lett.}\ }\textbf {\bibinfo
  {volume} {92}},\ \bibinfo {pages} {011116} (\bibinfo {year}
  {2008})}\BibitemShut {NoStop}%
\bibitem [{\citenamefont {Chezganov}\ \emph {et~al.}(2023)\citenamefont
  {Chezganov}, \citenamefont {Vlasov}, \citenamefont {Gimadeeva}, \citenamefont
  {Neradovskiy}, \citenamefont {Akhmatkhanov}, \citenamefont {Chuvakova},
  \citenamefont {Alikin}, \citenamefont {Tronche}, \citenamefont {Doutre},
  \citenamefont {Baldi},\ and\ \citenamefont {Shur}}]{Chezganov_2023}%
  \BibitemOpen
  \bibfield  {author} {\bibinfo {author} {\bibfnamefont {D.~S.}\ \bibnamefont
  {Chezganov}}, \bibinfo {author} {\bibfnamefont {E.~O.}\ \bibnamefont
  {Vlasov}}, \bibinfo {author} {\bibfnamefont {L.~V.}\ \bibnamefont
  {Gimadeeva}}, \bibinfo {author} {\bibfnamefont {M.~M.}\ \bibnamefont
  {Neradovskiy}}, \bibinfo {author} {\bibfnamefont {A.~R.}\ \bibnamefont
  {Akhmatkhanov}}, \bibinfo {author} {\bibfnamefont {M.~A.}\ \bibnamefont
  {Chuvakova}}, \bibinfo {author} {\bibfnamefont {D.~O.}\ \bibnamefont
  {Alikin}}, \bibinfo {author} {\bibfnamefont {H.}~\bibnamefont {Tronche}},
  \bibinfo {author} {\bibfnamefont {F.}~\bibnamefont {Doutre}}, \bibinfo
  {author} {\bibfnamefont {P.}~\bibnamefont {Baldi}},\ and\ \bibinfo {author}
  {\bibfnamefont {V.~Y.}\ \bibnamefont {Shur}},\ }\bibfield  {title} {\enquote
  {\bibinfo {title} {Short-period domain patterning by ion beam irradiation in
  lithium niobate waveguides produced by soft proton exchange},}\ }\href
  {https://doi.org/10.1016/j.optlastec.2022.108813} {\bibfield  {journal}
  {\bibinfo  {journal} {Optics \& Laser Technology}\ }\textbf {\bibinfo
  {volume} {158}},\ \bibinfo {pages} {108813} (\bibinfo {year}
  {2023})}\BibitemShut {NoStop}%
\bibitem [{\citenamefont {Huang}\ \emph {et~al.}(2013)\citenamefont {Huang},
  \citenamefont {Li}, \citenamefont {Santori}, \citenamefont {Acosta},
  \citenamefont {Faraon}, \citenamefont {Ishikawa}, \citenamefont {Wu},
  \citenamefont {Winston}, \citenamefont {Williams},\ and\ \citenamefont
  {Beausoleil}}]{Huang2013}%
  \BibitemOpen
  \bibfield  {author} {\bibinfo {author} {\bibfnamefont {Z.}~\bibnamefont
  {Huang}}, \bibinfo {author} {\bibfnamefont {W.-D.}\ \bibnamefont {Li}},
  \bibinfo {author} {\bibfnamefont {C.}~\bibnamefont {Santori}}, \bibinfo
  {author} {\bibfnamefont {V.~M.}\ \bibnamefont {Acosta}}, \bibinfo {author}
  {\bibfnamefont {A.}~\bibnamefont {Faraon}}, \bibinfo {author} {\bibfnamefont
  {T.}~\bibnamefont {Ishikawa}}, \bibinfo {author} {\bibfnamefont
  {W.}~\bibnamefont {Wu}}, \bibinfo {author} {\bibfnamefont {D.}~\bibnamefont
  {Winston}}, \bibinfo {author} {\bibfnamefont {R.~S.}\ \bibnamefont
  {Williams}},\ and\ \bibinfo {author} {\bibfnamefont {R.~G.}\ \bibnamefont
  {Beausoleil}},\ }\bibfield  {title} {\enquote {\bibinfo {title} {Diamond
  nitrogen-vacancy centers created by scanning focused helium ion beam and
  annealing},}\ }\href {https://doi.org/10.1063/1.4819339} {\bibfield
  {journal} {\bibinfo  {journal} {Appl. Phys. Lett.}\ }\textbf {\bibinfo
  {volume} {103}},\ \bibinfo {pages} {081906} (\bibinfo {year}
  {2013})}\BibitemShut {NoStop}%
\bibitem [{\citenamefont {McCloskey}\ \emph {et~al.}(2014)\citenamefont
  {McCloskey}, \citenamefont {Fox}, \citenamefont {O'Hara}, \citenamefont
  {Usov}, \citenamefont {Scanlan}, \citenamefont {McEvoy}, \citenamefont
  {Duesberg}, \citenamefont {Cross}, \citenamefont {Zhang},\ and\ \citenamefont
  {Donegan}}]{McCloskey2014}%
  \BibitemOpen
  \bibfield  {author} {\bibinfo {author} {\bibfnamefont {D.}~\bibnamefont
  {McCloskey}}, \bibinfo {author} {\bibfnamefont {D.}~\bibnamefont {Fox}},
  \bibinfo {author} {\bibfnamefont {N.}~\bibnamefont {O'Hara}}, \bibinfo
  {author} {\bibfnamefont {V.}~\bibnamefont {Usov}}, \bibinfo {author}
  {\bibfnamefont {D.}~\bibnamefont {Scanlan}}, \bibinfo {author} {\bibfnamefont
  {N.}~\bibnamefont {McEvoy}}, \bibinfo {author} {\bibfnamefont {G.~S.}\
  \bibnamefont {Duesberg}}, \bibinfo {author} {\bibfnamefont {G.~L.~W.}\
  \bibnamefont {Cross}}, \bibinfo {author} {\bibfnamefont {H.~Z.}\ \bibnamefont
  {Zhang}},\ and\ \bibinfo {author} {\bibfnamefont {J.~F.}\ \bibnamefont
  {Donegan}},\ }\bibfield  {title} {\enquote {\bibinfo {title} {Helium ion
  microscope generated nitrogen-vacancy centres in type ib diamond},}\ }\href
  {https://doi.org/10.1063/1.4862331} {\bibfield  {journal} {\bibinfo
  {journal} {Appl. Phys. Lett.}\ }\textbf {\bibinfo {volume} {104}},\ \bibinfo
  {pages} {031109} (\bibinfo {year} {2014})}\BibitemShut {NoStop}%
\bibitem [{\citenamefont {Song}\ \emph {et~al.}(2020)\citenamefont {Song},
  \citenamefont {Xu}, \citenamefont {Li}, \citenamefont {Wang}, \citenamefont
  {Fan}, \citenamefont {Rommel}, \citenamefont {Liu}, \citenamefont {Astakhov},
  \citenamefont {Hlawacek}, \citenamefont {Li}, \citenamefont {Xu},\ and\
  \citenamefont {Fang}}]{Song_2020}%
  \BibitemOpen
  \bibfield  {author} {\bibinfo {author} {\bibfnamefont {Y.}~\bibnamefont
  {Song}}, \bibinfo {author} {\bibfnamefont {Z.}~\bibnamefont {Xu}}, \bibinfo
  {author} {\bibfnamefont {R.}~\bibnamefont {Li}}, \bibinfo {author}
  {\bibfnamefont {H.}~\bibnamefont {Wang}}, \bibinfo {author} {\bibfnamefont
  {Y.}~\bibnamefont {Fan}}, \bibinfo {author} {\bibfnamefont {M.}~\bibnamefont
  {Rommel}}, \bibinfo {author} {\bibfnamefont {J.}~\bibnamefont {Liu}},
  \bibinfo {author} {\bibfnamefont {G.~V.}\ \bibnamefont {Astakhov}}, \bibinfo
  {author} {\bibfnamefont {G.}~\bibnamefont {Hlawacek}}, \bibinfo {author}
  {\bibfnamefont {B.}~\bibnamefont {Li}}, \bibinfo {author} {\bibfnamefont
  {J.}~\bibnamefont {Xu}},\ and\ \bibinfo {author} {\bibfnamefont
  {F.}~\bibnamefont {Fang}},\ }\bibfield  {title} {\enquote {\bibinfo {title}
  {Photoluminescence and {R}aman spectroscopy study on color centers of helium
  {Ion-Implanted} {4H--SiC}},}\ }\href
  {https://doi.org/10.1007/s41871-020-00061-8} {\bibfield  {journal} {\bibinfo
  {journal} {Nanomanufacturing and Metrology}\ }\textbf {\bibinfo {volume}
  {3}},\ \bibinfo {pages} {205--217} (\bibinfo {year} {2020})}\BibitemShut
  {NoStop}%
\bibitem [{\citenamefont {He}\ \emph {et~al.}(2022)\citenamefont {He},
  \citenamefont {Li}, \citenamefont {Wen}, \citenamefont {Zhou}, \citenamefont
  {Lin}, \citenamefont {Hao}, \citenamefont {Xu}, \citenamefont {Li},\ and\
  \citenamefont {Guo}}]{He_2022}%
  \BibitemOpen
  \bibfield  {author} {\bibinfo {author} {\bibfnamefont {Z.-X.}\ \bibnamefont
  {He}}, \bibinfo {author} {\bibfnamefont {Q.}~\bibnamefont {Li}}, \bibinfo
  {author} {\bibfnamefont {X.-L.}\ \bibnamefont {Wen}}, \bibinfo {author}
  {\bibfnamefont {J.-Y.}\ \bibnamefont {Zhou}}, \bibinfo {author}
  {\bibfnamefont {W.-X.}\ \bibnamefont {Lin}}, \bibinfo {author} {\bibfnamefont
  {Z.-H.}\ \bibnamefont {Hao}}, \bibinfo {author} {\bibfnamefont {J.-S.}\
  \bibnamefont {Xu}}, \bibinfo {author} {\bibfnamefont {C.-F.}\ \bibnamefont
  {Li}},\ and\ \bibinfo {author} {\bibfnamefont {G.-C.}\ \bibnamefont {Guo}},\
  }\bibfield  {title} {\enquote {\bibinfo {title} {Maskless generation of
  single silicon vacancy arrays in silicon carbide by a focused he$^+$ ion
  beam},}\ }\href {https://doi.org/10.1021/acsphotonics.2c01209} {\bibfield
  {journal} {\bibinfo  {journal} {{ACS} Photonics}\ } (\bibinfo {year}
  {2022}),\ 10.1021/acsphotonics.2c01209}\BibitemShut {NoStop}%
\bibitem [{\citenamefont {Klein}\ \emph {et~al.}(2017)\citenamefont {Klein},
  \citenamefont {Kuc}, \citenamefont {Nolinder}, \citenamefont {Altzschner},
  \citenamefont {Wierzbowski}, \citenamefont {Sigger}, \citenamefont {Kreupl},
  \citenamefont {Finley}, \citenamefont {Wurstbauer}, \citenamefont
  {Holleitner},\ and\ \citenamefont {Kaniber}}]{Klein2017}%
  \BibitemOpen
  \bibfield  {author} {\bibinfo {author} {\bibfnamefont {J.}~\bibnamefont
  {Klein}}, \bibinfo {author} {\bibfnamefont {A.}~\bibnamefont {Kuc}}, \bibinfo
  {author} {\bibfnamefont {A.}~\bibnamefont {Nolinder}}, \bibinfo {author}
  {\bibfnamefont {M.}~\bibnamefont {Altzschner}}, \bibinfo {author}
  {\bibfnamefont {J.}~\bibnamefont {Wierzbowski}}, \bibinfo {author}
  {\bibfnamefont {F.}~\bibnamefont {Sigger}}, \bibinfo {author} {\bibfnamefont
  {F.}~\bibnamefont {Kreupl}}, \bibinfo {author} {\bibfnamefont {J.~J.}\
  \bibnamefont {Finley}}, \bibinfo {author} {\bibfnamefont {U.}~\bibnamefont
  {Wurstbauer}}, \bibinfo {author} {\bibfnamefont {A.~W.}\ \bibnamefont
  {Holleitner}},\ and\ \bibinfo {author} {\bibfnamefont {M.}~\bibnamefont
  {Kaniber}},\ }\bibfield  {title} {{\selectlanguage {english}\enquote
  {\bibinfo {title} {Robust valley polarization of helium ion modified
  atomically thin {MoS2}},}\ }}\href {https://doi.org/10.1088/2053-1583/aa9642}
  {\bibfield  {journal} {\bibinfo  {journal} {2D Mater.}\ }\textbf {\bibinfo
  {volume} {5}},\ \bibinfo {pages} {011007} (\bibinfo {year}
  {2017})}\BibitemShut {NoStop}%
\bibitem [{\citenamefont {Klein}\ \emph {et~al.}(2019)\citenamefont {Klein},
  \citenamefont {Lorke}, \citenamefont {Florian}, \citenamefont {Sigger},
  \citenamefont {Sigl}, \citenamefont {Rey}, \citenamefont {Wierzbowski},
  \citenamefont {Cerne}, \citenamefont {M{\"u}ller}, \citenamefont
  {Mitterreiter}, \citenamefont {Zimmermann}, \citenamefont {Taniguchi},
  \citenamefont {Watanabe}, \citenamefont {Wurstbauer}, \citenamefont
  {Kaniber}, \citenamefont {Knap}, \citenamefont {Schmidt}, \citenamefont
  {Finley},\ and\ \citenamefont {Holleitner}}]{Klein-2019}%
  \BibitemOpen
  \bibfield  {author} {\bibinfo {author} {\bibfnamefont {J.}~\bibnamefont
  {Klein}}, \bibinfo {author} {\bibfnamefont {M.}~\bibnamefont {Lorke}},
  \bibinfo {author} {\bibfnamefont {M.}~\bibnamefont {Florian}}, \bibinfo
  {author} {\bibfnamefont {F.}~\bibnamefont {Sigger}}, \bibinfo {author}
  {\bibfnamefont {L.}~\bibnamefont {Sigl}}, \bibinfo {author} {\bibfnamefont
  {S.}~\bibnamefont {Rey}}, \bibinfo {author} {\bibfnamefont {J.}~\bibnamefont
  {Wierzbowski}}, \bibinfo {author} {\bibfnamefont {J.}~\bibnamefont {Cerne}},
  \bibinfo {author} {\bibfnamefont {K.}~\bibnamefont {M{\"u}ller}}, \bibinfo
  {author} {\bibfnamefont {E.}~\bibnamefont {Mitterreiter}}, \bibinfo {author}
  {\bibfnamefont {P.}~\bibnamefont {Zimmermann}}, \bibinfo {author}
  {\bibfnamefont {T.}~\bibnamefont {Taniguchi}}, \bibinfo {author}
  {\bibfnamefont {K.}~\bibnamefont {Watanabe}}, \bibinfo {author}
  {\bibfnamefont {U.}~\bibnamefont {Wurstbauer}}, \bibinfo {author}
  {\bibfnamefont {M.}~\bibnamefont {Kaniber}}, \bibinfo {author} {\bibfnamefont
  {M.}~\bibnamefont {Knap}}, \bibinfo {author} {\bibfnamefont {R.}~\bibnamefont
  {Schmidt}}, \bibinfo {author} {\bibfnamefont {J.~J.}\ \bibnamefont
  {Finley}},\ and\ \bibinfo {author} {\bibfnamefont {A.~W.}\ \bibnamefont
  {Holleitner}},\ }\bibfield  {title} {\enquote {\bibinfo {title}
  {Site-selectively generated photon emitters in monolayer {MoS2} via local
  helium ion irradiation},}\ }\href
  {https://doi.org/10.1038/s41467-019-10632-z} {\bibfield  {journal} {\bibinfo
  {journal} {Nat. Commun.}\ }\textbf {\bibinfo {volume} {10}},\ \bibinfo
  {pages} {2755} (\bibinfo {year} {2019})}\BibitemShut {NoStop}%
\bibitem [{\citenamefont {Pavunny}\ \emph {et~al.}(2021)\citenamefont
  {Pavunny}, \citenamefont {Yeats}, \citenamefont {Banks}, \citenamefont
  {Bielejec}, \citenamefont {Myers-Ward}, \citenamefont {DeJarld},
  \citenamefont {Bracker}, \citenamefont {Gaskill},\ and\ \citenamefont
  {Carter}}]{Pavunny_2021}%
  \BibitemOpen
  \bibfield  {author} {\bibinfo {author} {\bibfnamefont {S.~P.}\ \bibnamefont
  {Pavunny}}, \bibinfo {author} {\bibfnamefont {A.~L.}\ \bibnamefont {Yeats}},
  \bibinfo {author} {\bibfnamefont {H.~B.}\ \bibnamefont {Banks}}, \bibinfo
  {author} {\bibfnamefont {E.}~\bibnamefont {Bielejec}}, \bibinfo {author}
  {\bibfnamefont {R.~L.}\ \bibnamefont {Myers-Ward}}, \bibinfo {author}
  {\bibfnamefont {M.~T.}\ \bibnamefont {DeJarld}}, \bibinfo {author}
  {\bibfnamefont {A.~S.}\ \bibnamefont {Bracker}}, \bibinfo {author}
  {\bibfnamefont {D.~K.}\ \bibnamefont {Gaskill}},\ and\ \bibinfo {author}
  {\bibfnamefont {S.~G.}\ \bibnamefont {Carter}},\ }\bibfield  {title}
  {\enquote {\bibinfo {title} {Arrays of si vacancies in 4h-{SiC} produced by
  focused li ion beam implantation},}\ }\href
  {https://doi.org/10.1038/s41598-021-82832-x} {\bibfield  {journal} {\bibinfo
  {journal} {Sci. Rep.}\ }\textbf {\bibinfo {volume} {11}} (\bibinfo {year}
  {2021}),\ 10.1038/s41598-021-82832-x}\BibitemShut {NoStop}%
\bibitem [{\citenamefont {Tamura}\ \emph {et~al.}(2014)\citenamefont {Tamura},
  \citenamefont {Koike}, \citenamefont {Komatsubara}, \citenamefont {Teraji},
  \citenamefont {Onoda}, \citenamefont {McGuinness}, \citenamefont {Rogers},
  \citenamefont {Naydenov}, \citenamefont {Wu}, \citenamefont {Yan},
  \citenamefont {Jelezko}, \citenamefont {Ohshima}, \citenamefont {Isoya},
  \citenamefont {Shinada},\ and\ \citenamefont {Tanii}}]{Tamura_2014}%
  \BibitemOpen
  \bibfield  {author} {\bibinfo {author} {\bibfnamefont {S.}~\bibnamefont
  {Tamura}}, \bibinfo {author} {\bibfnamefont {G.}~\bibnamefont {Koike}},
  \bibinfo {author} {\bibfnamefont {A.}~\bibnamefont {Komatsubara}}, \bibinfo
  {author} {\bibfnamefont {T.}~\bibnamefont {Teraji}}, \bibinfo {author}
  {\bibfnamefont {S.}~\bibnamefont {Onoda}}, \bibinfo {author} {\bibfnamefont
  {L.~P.}\ \bibnamefont {McGuinness}}, \bibinfo {author} {\bibfnamefont
  {L.}~\bibnamefont {Rogers}}, \bibinfo {author} {\bibfnamefont
  {B.}~\bibnamefont {Naydenov}}, \bibinfo {author} {\bibfnamefont
  {E.}~\bibnamefont {Wu}}, \bibinfo {author} {\bibfnamefont {L.}~\bibnamefont
  {Yan}}, \bibinfo {author} {\bibfnamefont {F.}~\bibnamefont {Jelezko}},
  \bibinfo {author} {\bibfnamefont {T.}~\bibnamefont {Ohshima}}, \bibinfo
  {author} {\bibfnamefont {J.}~\bibnamefont {Isoya}}, \bibinfo {author}
  {\bibfnamefont {T.}~\bibnamefont {Shinada}},\ and\ \bibinfo {author}
  {\bibfnamefont {T.}~\bibnamefont {Tanii}},\ }\bibfield  {title} {\enquote
  {\bibinfo {title} {Array of bright silicon-vacancy centers in diamond
  fabricated by low-energy focused ion beam implantation},}\ }\href
  {https://doi.org/10.7567/apex.7.115201} {\bibfield  {journal} {\bibinfo
  {journal} {Appl. Phys. Express}\ }\textbf {\bibinfo {volume} {7}},\ \bibinfo
  {pages} {115201} (\bibinfo {year} {2014})}\BibitemShut {NoStop}%
\bibitem [{\citenamefont {Schr{\"{o}}der}\ \emph {et~al.}(2017)\citenamefont
  {Schr{\"{o}}der}, \citenamefont {Trusheim}, \citenamefont {Walsh},
  \citenamefont {Li}, \citenamefont {Zheng}, \citenamefont {Schukraft},
  \citenamefont {Sipahigil}, \citenamefont {Evans}, \citenamefont {Sukachev},
  \citenamefont {Nguyen}, \citenamefont {Pacheco}, \citenamefont {Camacho},
  \citenamefont {Bielejec}, \citenamefont {Lukin},\ and\ \citenamefont
  {Englund}}]{Schroeder_2017}%
  \BibitemOpen
  \bibfield  {author} {\bibinfo {author} {\bibfnamefont {T.}~\bibnamefont
  {Schr{\"{o}}der}}, \bibinfo {author} {\bibfnamefont {M.~E.}\ \bibnamefont
  {Trusheim}}, \bibinfo {author} {\bibfnamefont {M.}~\bibnamefont {Walsh}},
  \bibinfo {author} {\bibfnamefont {L.}~\bibnamefont {Li}}, \bibinfo {author}
  {\bibfnamefont {J.}~\bibnamefont {Zheng}}, \bibinfo {author} {\bibfnamefont
  {M.}~\bibnamefont {Schukraft}}, \bibinfo {author} {\bibfnamefont
  {A.}~\bibnamefont {Sipahigil}}, \bibinfo {author} {\bibfnamefont {R.~E.}\
  \bibnamefont {Evans}}, \bibinfo {author} {\bibfnamefont {D.~D.}\ \bibnamefont
  {Sukachev}}, \bibinfo {author} {\bibfnamefont {C.~T.}\ \bibnamefont
  {Nguyen}}, \bibinfo {author} {\bibfnamefont {J.~L.}\ \bibnamefont {Pacheco}},
  \bibinfo {author} {\bibfnamefont {R.~M.}\ \bibnamefont {Camacho}}, \bibinfo
  {author} {\bibfnamefont {E.~S.}\ \bibnamefont {Bielejec}}, \bibinfo {author}
  {\bibfnamefont {M.~D.}\ \bibnamefont {Lukin}},\ and\ \bibinfo {author}
  {\bibfnamefont {D.}~\bibnamefont {Englund}},\ }\bibfield  {title} {\enquote
  {\bibinfo {title} {Scalable focused ion beam creation of nearly
  lifetime-limited single quantum emitters in diamond nanostructures},}\ }\href
  {https://doi.org/10.1038/ncomms15376} {\bibfield  {journal} {\bibinfo
  {journal} {Nat. Commun.}\ }\textbf {\bibinfo {volume} {8}} (\bibinfo {year}
  {2017}),\ 10.1038/ncomms15376}\BibitemShut {NoStop}%
\bibitem [{\citenamefont {Hollenbach}\ \emph {et~al.}(2022)\citenamefont
  {Hollenbach}, \citenamefont {Klingner}, \citenamefont {Jagtap}, \citenamefont
  {Bischoff}, \citenamefont {Fowley}, \citenamefont {Kentsch}, \citenamefont
  {Hlawacek}, \citenamefont {Erbe}, \citenamefont {Abrosimov}, \citenamefont
  {Helm}, \citenamefont {Berenc{\'{e}}n},\ and\ \citenamefont
  {Astakhov}}]{Hollenbach_2022}%
  \BibitemOpen
  \bibfield  {author} {\bibinfo {author} {\bibfnamefont {M.}~\bibnamefont
  {Hollenbach}}, \bibinfo {author} {\bibfnamefont {N.}~\bibnamefont
  {Klingner}}, \bibinfo {author} {\bibfnamefont {N.~S.}\ \bibnamefont
  {Jagtap}}, \bibinfo {author} {\bibfnamefont {L.}~\bibnamefont {Bischoff}},
  \bibinfo {author} {\bibfnamefont {C.}~\bibnamefont {Fowley}}, \bibinfo
  {author} {\bibfnamefont {U.}~\bibnamefont {Kentsch}}, \bibinfo {author}
  {\bibfnamefont {G.}~\bibnamefont {Hlawacek}}, \bibinfo {author}
  {\bibfnamefont {A.}~\bibnamefont {Erbe}}, \bibinfo {author} {\bibfnamefont
  {N.~V.}\ \bibnamefont {Abrosimov}}, \bibinfo {author} {\bibfnamefont
  {M.}~\bibnamefont {Helm}}, \bibinfo {author} {\bibfnamefont {Y.}~\bibnamefont
  {Berenc{\'{e}}n}},\ and\ \bibinfo {author} {\bibfnamefont {G.~V.}\
  \bibnamefont {Astakhov}},\ }\bibfield  {title} {\enquote {\bibinfo {title}
  {Wafer-scale nanofabrication of telecom single-photon emitters in silicon},}\
  }\href {https://doi.org/10.1038/s41467-022-35051-5} {\bibfield  {journal}
  {\bibinfo  {journal} {Nat. Commun.}\ }\textbf {\bibinfo {volume} {13}}
  (\bibinfo {year} {2022}),\ 10.1038/s41467-022-35051-5},\ \Eprint
  {https://arxiv.org/abs/2204.13173} {arXiv:2204.13173 [quant-ph]} \BibitemShut
  {NoStop}%
\bibitem [{\citenamefont {Acharya}\ \emph {et~al.}(2023)\citenamefont
  {Acharya}, \citenamefont {Coke}, \citenamefont {Adshead}, \citenamefont {Li},
  \citenamefont {Achinuq}, \citenamefont {Cai}, \citenamefont {Gholizadeh},
  \citenamefont {Jacobs}, \citenamefont {Boland}, \citenamefont {Haigh},
  \citenamefont {Moore}, \citenamefont {Jamieson},\ and\ \citenamefont
  {Curry}}]{Acharya2023}%
  \BibitemOpen
  \bibfield  {author} {\bibinfo {author} {\bibfnamefont {R.}~\bibnamefont
  {Acharya}}, \bibinfo {author} {\bibfnamefont {M.}~\bibnamefont {Coke}},
  \bibinfo {author} {\bibfnamefont {M.}~\bibnamefont {Adshead}}, \bibinfo
  {author} {\bibfnamefont {K.}~\bibnamefont {Li}}, \bibinfo {author}
  {\bibfnamefont {B.}~\bibnamefont {Achinuq}}, \bibinfo {author} {\bibfnamefont
  {R.}~\bibnamefont {Cai}}, \bibinfo {author} {\bibfnamefont {A.~B.}\
  \bibnamefont {Gholizadeh}}, \bibinfo {author} {\bibfnamefont
  {J.}~\bibnamefont {Jacobs}}, \bibinfo {author} {\bibfnamefont {J.~L.}\
  \bibnamefont {Boland}}, \bibinfo {author} {\bibfnamefont {S.~J.}\
  \bibnamefont {Haigh}}, \bibinfo {author} {\bibfnamefont {K.~L.}\ \bibnamefont
  {Moore}}, \bibinfo {author} {\bibfnamefont {D.~N.}\ \bibnamefont
  {Jamieson}},\ and\ \bibinfo {author} {\bibfnamefont {R.~J.}\ \bibnamefont
  {Curry}},\ }\href {https://doi.org/10.48550/ARXIV.2308.12471} {\enquote
  {\bibinfo {title} {Highly ${ }^{28} \mathrm{Si}$ enriched silicon by
  localised focused ion beam implantation},}\ } (\bibinfo {year}
  {2023})\BibitemShut {NoStop}%
\bibitem [{\citenamefont {R{\"{a}}cke}\ \emph {et~al.}(2021)\citenamefont
  {R{\"{a}}cke}, \citenamefont {Pietzonka}, \citenamefont {Meijer},
  \citenamefont {Spemann},\ and\ \citenamefont {Wunderlich}}]{Raecke_2021}%
  \BibitemOpen
  \bibfield  {author} {\bibinfo {author} {\bibfnamefont {P.}~\bibnamefont
  {R{\"{a}}cke}}, \bibinfo {author} {\bibfnamefont {L.}~\bibnamefont
  {Pietzonka}}, \bibinfo {author} {\bibfnamefont {J.}~\bibnamefont {Meijer}},
  \bibinfo {author} {\bibfnamefont {D.}~\bibnamefont {Spemann}},\ and\ \bibinfo
  {author} {\bibfnamefont {R.}~\bibnamefont {Wunderlich}},\ }\bibfield  {title}
  {\enquote {\bibinfo {title} {Vacancy diffusion and nitrogen-vacancy center
  formation near the diamond surface},}\ }\href
  {https://doi.org/10.1063/5.0046031} {\bibfield  {journal} {\bibinfo
  {journal} {Appl. Phys. Lett.}\ }\textbf {\bibinfo {volume} {118}},\ \bibinfo
  {pages} {204003} (\bibinfo {year} {2021})}\BibitemShut {NoStop}%
\bibitem [{\citenamefont {Ziegler}\ \emph {et~al.}(2019)\citenamefont
  {Ziegler}, \citenamefont {Klaiss}, \citenamefont {Blaikie}, \citenamefont
  {Miller}, \citenamefont {Horowitz},\ and\ \citenamefont
  {Alem{\'{a}}n}}]{Ziegler_2019}%
  \BibitemOpen
  \bibfield  {author} {\bibinfo {author} {\bibfnamefont {J.}~\bibnamefont
  {Ziegler}}, \bibinfo {author} {\bibfnamefont {R.}~\bibnamefont {Klaiss}},
  \bibinfo {author} {\bibfnamefont {A.}~\bibnamefont {Blaikie}}, \bibinfo
  {author} {\bibfnamefont {D.}~\bibnamefont {Miller}}, \bibinfo {author}
  {\bibfnamefont {V.~R.}\ \bibnamefont {Horowitz}},\ and\ \bibinfo {author}
  {\bibfnamefont {B.~J.}\ \bibnamefont {Alem{\'{a}}n}},\ }\bibfield  {title}
  {\enquote {\bibinfo {title} {Deterministic quantum emitter formation in
  hexagonal boron nitride via controlled edge creation},}\ }\href
  {https://doi.org/10.1021/acs.nanolett.9b00357} {\bibfield  {journal}
  {\bibinfo  {journal} {Nano Lett.}\ }\textbf {\bibinfo {volume} {19}},\
  \bibinfo {pages} {2121--2127} (\bibinfo {year} {2019})}\BibitemShut {NoStop}%
\bibitem [{\citenamefont {Klaiss}\ \emph {et~al.}(2022)\citenamefont {Klaiss},
  \citenamefont {Ziegler}, \citenamefont {Miller}, \citenamefont {Zappitelli},
  \citenamefont {Watanabe}, \citenamefont {Taniguchi},\ and\ \citenamefont
  {Alem{\'{a}}n}}]{Klaiss_2022}%
  \BibitemOpen
  \bibfield  {author} {\bibinfo {author} {\bibfnamefont {R.}~\bibnamefont
  {Klaiss}}, \bibinfo {author} {\bibfnamefont {J.}~\bibnamefont {Ziegler}},
  \bibinfo {author} {\bibfnamefont {D.}~\bibnamefont {Miller}}, \bibinfo
  {author} {\bibfnamefont {K.}~\bibnamefont {Zappitelli}}, \bibinfo {author}
  {\bibfnamefont {K.}~\bibnamefont {Watanabe}}, \bibinfo {author}
  {\bibfnamefont {T.}~\bibnamefont {Taniguchi}},\ and\ \bibinfo {author}
  {\bibfnamefont {B.}~\bibnamefont {Alem{\'{a}}n}},\ }\bibfield  {title}
  {\enquote {\bibinfo {title} {Uncovering the morphological effects of
  high-energy ga$^+$ focused ion beam milling on {hBN} single-photon emitter
  fabrication},}\ }\href {https://doi.org/10.1063/5.0097581} {\bibfield
  {journal} {\bibinfo  {journal} {J. Chem. Phys.}\ }\textbf {\bibinfo {volume}
  {157}},\ \bibinfo {pages} {074703} (\bibinfo {year} {2022})}\BibitemShut
  {NoStop}%
\bibitem [{\citenamefont {Qian}\ \emph {et~al.}(2020)\citenamefont {Qian},
  \citenamefont {Peng}, \citenamefont {Perea-Lopez}, \citenamefont {Fujisawa},
  \citenamefont {Zhang}, \citenamefont {Zhang}, \citenamefont {Choudhury},
  \citenamefont {Redwing}, \citenamefont {Terrones}, \citenamefont {Ma},\ and\
  \citenamefont {Huang}}]{Qian_2020a}%
  \BibitemOpen
  \bibfield  {author} {\bibinfo {author} {\bibfnamefont {Q.}~\bibnamefont
  {Qian}}, \bibinfo {author} {\bibfnamefont {L.}~\bibnamefont {Peng}}, \bibinfo
  {author} {\bibfnamefont {N.}~\bibnamefont {Perea-Lopez}}, \bibinfo {author}
  {\bibfnamefont {K.}~\bibnamefont {Fujisawa}}, \bibinfo {author}
  {\bibfnamefont {K.}~\bibnamefont {Zhang}}, \bibinfo {author} {\bibfnamefont
  {X.}~\bibnamefont {Zhang}}, \bibinfo {author} {\bibfnamefont {T.~H.}\
  \bibnamefont {Choudhury}}, \bibinfo {author} {\bibfnamefont {J.~M.}\
  \bibnamefont {Redwing}}, \bibinfo {author} {\bibfnamefont {M.}~\bibnamefont
  {Terrones}}, \bibinfo {author} {\bibfnamefont {X.}~\bibnamefont {Ma}},\ and\
  \bibinfo {author} {\bibfnamefont {S.}~\bibnamefont {Huang}},\ }\bibfield
  {title} {\enquote {\bibinfo {title} {Defect creation in {WSe}2 with a
  microsecond photoluminescence lifetime by focused ion beam irradiation},}\
  }\href {https://doi.org/10.1039/c9nr08390a} {\bibfield  {journal} {\bibinfo
  {journal} {Nanoscale}\ }\textbf {\bibinfo {volume} {12}},\ \bibinfo {pages}
  {2047--2056} (\bibinfo {year} {2020})}\BibitemShut {NoStop}%
\bibitem [{\citenamefont {Thiruraman}, \citenamefont {Das},\ and\ \citenamefont
  {Drndi{\'{c}}}(2019)}]{Thiruraman_2019a}%
  \BibitemOpen
  \bibfield  {author} {\bibinfo {author} {\bibfnamefont {J.~P.}\ \bibnamefont
  {Thiruraman}}, \bibinfo {author} {\bibfnamefont {P.~M.}\ \bibnamefont
  {Das}},\ and\ \bibinfo {author} {\bibfnamefont {M.}~\bibnamefont
  {Drndi{\'{c}}}},\ }\bibfield  {title} {\enquote {\bibinfo {title}
  {Irradiation of transition metal dichalcogenides using a focused ion beam:
  controlled single-atom defect creation},}\ }\href
  {https://doi.org/10.1002/adfm.201904668} {\bibfield  {journal} {\bibinfo
  {journal} {Adv. Funct. Mater.}\ }\textbf {\bibinfo {volume} {29}},\ \bibinfo
  {pages} {1904668} (\bibinfo {year} {2019})}\BibitemShut {NoStop}%
\bibitem [{\citenamefont {Zhou}\ \emph {et~al.}(2018)\citenamefont {Zhou},
  \citenamefont {Mu}, \citenamefont {Adamo}, \citenamefont {Bauerdick},
  \citenamefont {Rudzinski}, \citenamefont {Aharonovich},\ and\ \citenamefont
  {Gao}}]{Zhou_2018a}%
  \BibitemOpen
  \bibfield  {author} {\bibinfo {author} {\bibfnamefont {Y.}~\bibnamefont
  {Zhou}}, \bibinfo {author} {\bibfnamefont {Z.}~\bibnamefont {Mu}}, \bibinfo
  {author} {\bibfnamefont {G.}~\bibnamefont {Adamo}}, \bibinfo {author}
  {\bibfnamefont {S.}~\bibnamefont {Bauerdick}}, \bibinfo {author}
  {\bibfnamefont {A.}~\bibnamefont {Rudzinski}}, \bibinfo {author}
  {\bibfnamefont {I.}~\bibnamefont {Aharonovich}},\ and\ \bibinfo {author}
  {\bibfnamefont {W.-b.}\ \bibnamefont {Gao}},\ }\bibfield  {title} {\enquote
  {\bibinfo {title} {Direct writing of single germanium vacancy center arrays
  in diamond},}\ }\href {https://doi.org/10.1088/1367-2630/aaf2ac} {\bibfield
  {journal} {\bibinfo  {journal} {New J. Phys.}\ }\textbf {\bibinfo {volume}
  {20}},\ \bibinfo {pages} {125004} (\bibinfo {year} {2018})}\BibitemShut
  {NoStop}%
\bibitem [{\citenamefont {Pacheco}\ \emph {et~al.}(2017)\citenamefont
  {Pacheco}, \citenamefont {Singh}, \citenamefont {Perry}, \citenamefont
  {Wendt}, \citenamefont {Eyck}, \citenamefont {Manginell}, \citenamefont
  {Pluym}, \citenamefont {Luhman}, \citenamefont {Lilly}, \citenamefont
  {Carroll},\ and\ \citenamefont {Bielejec}}]{Pacheco_2017a}%
  \BibitemOpen
  \bibfield  {author} {\bibinfo {author} {\bibfnamefont {J.~L.}\ \bibnamefont
  {Pacheco}}, \bibinfo {author} {\bibfnamefont {M.}~\bibnamefont {Singh}},
  \bibinfo {author} {\bibfnamefont {D.~L.}\ \bibnamefont {Perry}}, \bibinfo
  {author} {\bibfnamefont {J.~R.}\ \bibnamefont {Wendt}}, \bibinfo {author}
  {\bibfnamefont {G.~T.}\ \bibnamefont {Eyck}}, \bibinfo {author}
  {\bibfnamefont {R.~P.}\ \bibnamefont {Manginell}}, \bibinfo {author}
  {\bibfnamefont {T.}~\bibnamefont {Pluym}}, \bibinfo {author} {\bibfnamefont
  {D.~R.}\ \bibnamefont {Luhman}}, \bibinfo {author} {\bibfnamefont {M.~P.}\
  \bibnamefont {Lilly}}, \bibinfo {author} {\bibfnamefont {M.~S.}\ \bibnamefont
  {Carroll}},\ and\ \bibinfo {author} {\bibfnamefont {E.}~\bibnamefont
  {Bielejec}},\ }\bibfield  {title} {\enquote {\bibinfo {title} {Ion
  implantation for deterministic single atom devices},}\ }\href
  {https://doi.org/10.1063/1.5001520} {\bibfield  {journal} {\bibinfo
  {journal} {Rev. Sci. Instrum.}\ }\textbf {\bibinfo {volume} {88}},\ \bibinfo
  {pages} {123301} (\bibinfo {year} {2017})}\BibitemShut {NoStop}%
\bibitem [{\citenamefont {Fr{\"{o}}ch}\ \emph {et~al.}(2020)\citenamefont
  {Fr{\"{o}}ch}, \citenamefont {Bahm}, \citenamefont {Kianinia}, \citenamefont
  {Mu}, \citenamefont {Bhatia}, \citenamefont {Kim}, \citenamefont {Cairney},
  \citenamefont {Gao}, \citenamefont {Bradac}, \citenamefont {Aharonovich},\
  and\ \citenamefont {Toth}}]{Froech_2020}%
  \BibitemOpen
  \bibfield  {author} {\bibinfo {author} {\bibfnamefont {J.~E.}\ \bibnamefont
  {Fr{\"{o}}ch}}, \bibinfo {author} {\bibfnamefont {A.}~\bibnamefont {Bahm}},
  \bibinfo {author} {\bibfnamefont {M.}~\bibnamefont {Kianinia}}, \bibinfo
  {author} {\bibfnamefont {Z.}~\bibnamefont {Mu}}, \bibinfo {author}
  {\bibfnamefont {V.}~\bibnamefont {Bhatia}}, \bibinfo {author} {\bibfnamefont
  {S.}~\bibnamefont {Kim}}, \bibinfo {author} {\bibfnamefont {J.~M.}\
  \bibnamefont {Cairney}}, \bibinfo {author} {\bibfnamefont {W.}~\bibnamefont
  {Gao}}, \bibinfo {author} {\bibfnamefont {C.}~\bibnamefont {Bradac}},
  \bibinfo {author} {\bibfnamefont {I.}~\bibnamefont {Aharonovich}},\ and\
  \bibinfo {author} {\bibfnamefont {M.}~\bibnamefont {Toth}},\ }\bibfield
  {title} {\enquote {\bibinfo {title} {Versatile direct-writing of dopants in a
  solid state host through recoil implantation},}\ }\href
  {https://doi.org/10.1038/s41467-020-18749-2} {\bibfield  {journal} {\bibinfo
  {journal} {Nat. Commun.}\ }\textbf {\bibinfo {volume} {11}} (\bibinfo {year}
  {2020}),\ 10.1038/s41467-020-18749-2}\BibitemShut {NoStop}%
\bibitem [{\citenamefont {Kianinia}\ \emph {et~al.}(2020)\citenamefont
  {Kianinia}, \citenamefont {White}, \citenamefont {Fr{\"{o}}ch}, \citenamefont
  {Bradac},\ and\ \citenamefont {Aharonovich}}]{Kianinia_2020}%
  \BibitemOpen
  \bibfield  {author} {\bibinfo {author} {\bibfnamefont {M.}~\bibnamefont
  {Kianinia}}, \bibinfo {author} {\bibfnamefont {S.}~\bibnamefont {White}},
  \bibinfo {author} {\bibfnamefont {J.~E.}\ \bibnamefont {Fr{\"{o}}ch}},
  \bibinfo {author} {\bibfnamefont {C.}~\bibnamefont {Bradac}},\ and\ \bibinfo
  {author} {\bibfnamefont {I.}~\bibnamefont {Aharonovich}},\ }\bibfield
  {title} {\enquote {\bibinfo {title} {Generation of spin defects in hexagonal
  boron nitride},}\ }\href {https://doi.org/10.1021/acsphotonics.0c00614}
  {\bibfield  {journal} {\bibinfo  {journal} {{ACS} Photonics}\ }\textbf
  {\bibinfo {volume} {7}},\ \bibinfo {pages} {2147--2152} (\bibinfo {year}
  {2020})}\BibitemShut {NoStop}%
\bibitem [{\citenamefont {Glushkov}\ \emph {et~al.}(2022)\citenamefont
  {Glushkov}, \citenamefont {Macha}, \citenamefont {R{\"{a}}th}, \citenamefont
  {Navikas}, \citenamefont {Ronceray}, \citenamefont {Cheon}, \citenamefont
  {Ahmed}, \citenamefont {Avsar}, \citenamefont {Watanabe}, \citenamefont
  {Taniguchi}, \citenamefont {Shorubalko}, \citenamefont {Kis}, \citenamefont
  {Fantner},\ and\ \citenamefont {Radenovic}}]{Glushkov_2022}%
  \BibitemOpen
  \bibfield  {author} {\bibinfo {author} {\bibfnamefont {E.}~\bibnamefont
  {Glushkov}}, \bibinfo {author} {\bibfnamefont {M.}~\bibnamefont {Macha}},
  \bibinfo {author} {\bibfnamefont {E.}~\bibnamefont {R{\"{a}}th}}, \bibinfo
  {author} {\bibfnamefont {V.}~\bibnamefont {Navikas}}, \bibinfo {author}
  {\bibfnamefont {N.}~\bibnamefont {Ronceray}}, \bibinfo {author}
  {\bibfnamefont {C.~Y.}\ \bibnamefont {Cheon}}, \bibinfo {author}
  {\bibfnamefont {A.}~\bibnamefont {Ahmed}}, \bibinfo {author} {\bibfnamefont
  {A.}~\bibnamefont {Avsar}}, \bibinfo {author} {\bibfnamefont
  {K.}~\bibnamefont {Watanabe}}, \bibinfo {author} {\bibfnamefont
  {T.}~\bibnamefont {Taniguchi}}, \bibinfo {author} {\bibfnamefont
  {I.}~\bibnamefont {Shorubalko}}, \bibinfo {author} {\bibfnamefont
  {A.}~\bibnamefont {Kis}}, \bibinfo {author} {\bibfnamefont {G.}~\bibnamefont
  {Fantner}},\ and\ \bibinfo {author} {\bibfnamefont {A.}~\bibnamefont
  {Radenovic}},\ }\bibfield  {title} {\enquote {\bibinfo {title} {Engineering
  optically active defects in hexagonal boron nitride using focused ion beam
  and water},}\ }\href {https://doi.org/10.1021/acsnano.1c07086} {\bibfield
  {journal} {\bibinfo  {journal} {{ACS} Nano}\ }\textbf {\bibinfo {volume}
  {16}},\ \bibinfo {pages} {3695--3703} (\bibinfo {year} {2022})}\BibitemShut
  {NoStop}%
\bibitem [{\citenamefont {Cassidy}\ \emph {et~al.}(2020)\citenamefont
  {Cassidy}, \citenamefont {Blenkinsopp}, \citenamefont {Brown}, \citenamefont
  {Curry}, \citenamefont {Murdin}, \citenamefont {Webb},\ and\ \citenamefont
  {Cox}}]{Cassidy_2020}%
  \BibitemOpen
  \bibfield  {author} {\bibinfo {author} {\bibfnamefont {N.}~\bibnamefont
  {Cassidy}}, \bibinfo {author} {\bibfnamefont {P.}~\bibnamefont
  {Blenkinsopp}}, \bibinfo {author} {\bibfnamefont {I.}~\bibnamefont {Brown}},
  \bibinfo {author} {\bibfnamefont {R.~J.}\ \bibnamefont {Curry}}, \bibinfo
  {author} {\bibfnamefont {B.~N.}\ \bibnamefont {Murdin}}, \bibinfo {author}
  {\bibfnamefont {R.}~\bibnamefont {Webb}},\ and\ \bibinfo {author}
  {\bibfnamefont {D.}~\bibnamefont {Cox}},\ }\bibfield  {title} {\enquote
  {\bibinfo {title} {Single ion implantation of bismuth},}\ }\href
  {https://doi.org/10.1002/pssa.202000237} {\bibfield  {journal} {\bibinfo
  {journal} {Phys. Status Solidi A}\ }\textbf {\bibinfo {volume} {218}},\
  \bibinfo {pages} {2000237} (\bibinfo {year} {2020})}\BibitemShut {NoStop}%
\bibitem [{\citenamefont {Petrov}\ \emph {et~al.}(2015)\citenamefont {Petrov},
  \citenamefont {Sharov}, \citenamefont {Baraban},\ and\ \citenamefont
  {Vyvenko}}]{Petrov2015}%
  \BibitemOpen
  \bibfield  {author} {\bibinfo {author} {\bibfnamefont {Y.~V.}\ \bibnamefont
  {Petrov}}, \bibinfo {author} {\bibfnamefont {T.~V.}\ \bibnamefont {Sharov}},
  \bibinfo {author} {\bibfnamefont {A.~P.}\ \bibnamefont {Baraban}},\ and\
  \bibinfo {author} {\bibfnamefont {O.~F.}\ \bibnamefont {Vyvenko}},\
  }\bibfield  {title} {\enquote {\bibinfo {title} {Effect of helium ion beam
  treatment on the etching rate of silicon nitride},}\ }\href
  {https://doi.org/10.1016/j.nimb.2015.02.054} {\bibfield  {journal} {\bibinfo
  {journal} {Nucl. Instrum. Meth. Phys. Res. B}\ }\textbf {\bibinfo {volume}
  {349}},\ \bibinfo {pages} {90--95} (\bibinfo {year} {2015})}\BibitemShut
  {NoStop}%
\bibitem [{\citenamefont {Petrov}, \citenamefont {Grigoryev},\ and\
  \citenamefont {Baraban}(2020)}]{Petrov2020}%
  \BibitemOpen
  \bibfield  {author} {\bibinfo {author} {\bibfnamefont {Y.~V.}\ \bibnamefont
  {Petrov}}, \bibinfo {author} {\bibfnamefont {E.~A.}\ \bibnamefont
  {Grigoryev}},\ and\ \bibinfo {author} {\bibfnamefont {A.~P.}\ \bibnamefont
  {Baraban}},\ }\bibfield  {title} {\enquote {\bibinfo {title} {Helium focused
  ion beam irradiation with subsequent chemical etching for the fabrication of
  nanostructures},}\ }\href {https://doi.org/10.1088/1361-6528/ab6fe3}
  {\bibfield  {journal} {\bibinfo  {journal} {Nanotechnology}\ }\textbf
  {\bibinfo {volume} {31}},\ \bibinfo {pages} {215301} (\bibinfo {year}
  {2020})}\BibitemShut {NoStop}%
\bibitem [{\citenamefont {Maguire}\ \emph {et~al.}(2019)\citenamefont
  {Maguire}, \citenamefont {Jadwiszczak}, \citenamefont {O'Brien},
  \citenamefont {Keane}, \citenamefont {Duesberg}, \citenamefont {McEvoy},\
  and\ \citenamefont {Zhang}}]{Maguire2019}%
  \BibitemOpen
  \bibfield  {author} {\bibinfo {author} {\bibfnamefont {P.}~\bibnamefont
  {Maguire}}, \bibinfo {author} {\bibfnamefont {J.}~\bibnamefont
  {Jadwiszczak}}, \bibinfo {author} {\bibfnamefont {M.}~\bibnamefont
  {O'Brien}}, \bibinfo {author} {\bibfnamefont {D.}~\bibnamefont {Keane}},
  \bibinfo {author} {\bibfnamefont {G.~S.}\ \bibnamefont {Duesberg}}, \bibinfo
  {author} {\bibfnamefont {N.}~\bibnamefont {McEvoy}},\ and\ \bibinfo {author}
  {\bibfnamefont {H.}~\bibnamefont {Zhang}},\ }\bibfield  {title} {\enquote
  {\bibinfo {title} {Defect-moderated oxidative etching of {MoS2}},}\ }\href
  {https://doi.org/10.1063/1.5115036} {\bibfield  {journal} {\bibinfo
  {journal} {J. Appl. Phys.}\ }\textbf {\bibinfo {volume} {126}},\ \bibinfo
  {pages} {164301} (\bibinfo {year} {2019})}\BibitemShut {NoStop}%
\bibitem [{\citenamefont {Mitterreiter}\ \emph {et~al.}(2019)\citenamefont
  {Mitterreiter}, \citenamefont {Liang}, \citenamefont {Golibrzuch},
  \citenamefont {McLaughlin}, \citenamefont {Csoklich}, \citenamefont {Bartl},
  \citenamefont {Holleitner}, \citenamefont {Wurstbauer},\ and\ \citenamefont
  {Bandarenka}}]{Mitterreiter2019}%
  \BibitemOpen
  \bibfield  {author} {\bibinfo {author} {\bibfnamefont {E.}~\bibnamefont
  {Mitterreiter}}, \bibinfo {author} {\bibfnamefont {Y.}~\bibnamefont {Liang}},
  \bibinfo {author} {\bibfnamefont {M.}~\bibnamefont {Golibrzuch}}, \bibinfo
  {author} {\bibfnamefont {D.}~\bibnamefont {McLaughlin}}, \bibinfo {author}
  {\bibfnamefont {C.}~\bibnamefont {Csoklich}}, \bibinfo {author}
  {\bibfnamefont {J.~D.}\ \bibnamefont {Bartl}}, \bibinfo {author}
  {\bibfnamefont {A.}~\bibnamefont {Holleitner}}, \bibinfo {author}
  {\bibfnamefont {U.}~\bibnamefont {Wurstbauer}},\ and\ \bibinfo {author}
  {\bibfnamefont {A.~S.}\ \bibnamefont {Bandarenka}},\ }\bibfield  {title}
  {\enquote {\bibinfo {title} {In-situ visualization of hydrogen evolution
  sites on helium ion treated molybdenum dichalcogenides under reaction
  conditions},}\ }\href {https://doi.org/10.1038/s41699-019-0107-5} {\bibfield
  {journal} {\bibinfo  {journal} {npj 2D Materials and Applications}\ }\textbf
  {\bibinfo {volume} {3}},\ \bibinfo {pages} {25} (\bibinfo {year}
  {2019})}\BibitemShut {NoStop}%
\bibitem [{\citenamefont {Heilmann}\ \emph {et~al.}(2021)\citenamefont
  {Heilmann}, \citenamefont {Deinhart}, \citenamefont {Tahraoui}, \citenamefont
  {H{\"o}flich},\ and\ \citenamefont {Lopes}}]{Heilmann2021}%
  \BibitemOpen
  \bibfield  {author} {\bibinfo {author} {\bibfnamefont {M.}~\bibnamefont
  {Heilmann}}, \bibinfo {author} {\bibfnamefont {V.}~\bibnamefont {Deinhart}},
  \bibinfo {author} {\bibfnamefont {A.}~\bibnamefont {Tahraoui}}, \bibinfo
  {author} {\bibfnamefont {K.}~\bibnamefont {H{\"o}flich}},\ and\ \bibinfo
  {author} {\bibfnamefont {J.~M.~J.}\ \bibnamefont {Lopes}},\ }\bibfield
  {title} {{\selectlanguage {english}\enquote {\bibinfo {title} {Spatially
  controlled epitaxial growth of {2D} heterostructures via defect engineering
  using a focused he ion beam},}\ }}\href
  {https://doi.org/10.1038/s41699-021-00250-z} {\bibfield  {journal} {\bibinfo
  {journal} {npj 2D Materials and Applications}\ }\textbf {\bibinfo {volume}
  {5}},\ \bibinfo {pages} {1--7} (\bibinfo {year} {2021})}\BibitemShut
  {NoStop}%
\bibitem [{\citenamefont {Akhmadaliev}, \citenamefont {Schmidt},\ and\
  \citenamefont {Bischoff}(2006)}]{Akhmadaliev_2006}%
  \BibitemOpen
  \bibfield  {author} {\bibinfo {author} {\bibfnamefont {C.}~\bibnamefont
  {Akhmadaliev}}, \bibinfo {author} {\bibfnamefont {B.}~\bibnamefont
  {Schmidt}},\ and\ \bibinfo {author} {\bibfnamefont {L.}~\bibnamefont
  {Bischoff}},\ }\bibfield  {title} {\enquote {\bibinfo {title} {Defect induced
  formation of {CoSi}$_2$ nanowires by focused ion beam synthesis},}\ }\href
  {https://doi.org/10.1063/1.2400068} {\bibfield  {journal} {\bibinfo
  {journal} {Appl. Phys. Lett.}\ }\textbf {\bibinfo {volume} {89}},\ \bibinfo
  {pages} {223129} (\bibinfo {year} {2006})}\BibitemShut {NoStop}%
\bibitem [{\citenamefont {Lee}\ \emph {et~al.}(2011)\citenamefont {Lee},
  \citenamefont {Saucer}, \citenamefont {Martin}, \citenamefont {Tien},
  \citenamefont {Millunchick},\ and\ \citenamefont {Sih}}]{Lee_2011a}%
  \BibitemOpen
  \bibfield  {author} {\bibinfo {author} {\bibfnamefont {J.}~\bibnamefont
  {Lee}}, \bibinfo {author} {\bibfnamefont {T.~W.}\ \bibnamefont {Saucer}},
  \bibinfo {author} {\bibfnamefont {A.~J.}\ \bibnamefont {Martin}}, \bibinfo
  {author} {\bibfnamefont {D.}~\bibnamefont {Tien}}, \bibinfo {author}
  {\bibfnamefont {J.~M.}\ \bibnamefont {Millunchick}},\ and\ \bibinfo {author}
  {\bibfnamefont {V.}~\bibnamefont {Sih}},\ }\bibfield  {title} {\enquote
  {\bibinfo {title} {Photoluminescence imaging of focused ion beam induced
  individual quantum dots},}\ }\href {https://doi.org/10.1021/nl1038902}
  {\bibfield  {journal} {\bibinfo  {journal} {Nano Lett.}\ }\textbf {\bibinfo
  {volume} {11}},\ \bibinfo {pages} {1040--1043} (\bibinfo {year}
  {2011})}\BibitemShut {NoStop}%
\bibitem [{\citenamefont {Portavoce}\ \emph {et~al.}(2007)\citenamefont
  {Portavoce}, \citenamefont {Hull}, \citenamefont {Reuter},\ and\
  \citenamefont {Ross}}]{Portavoce_2007a}%
  \BibitemOpen
  \bibfield  {author} {\bibinfo {author} {\bibfnamefont {A.}~\bibnamefont
  {Portavoce}}, \bibinfo {author} {\bibfnamefont {R.}~\bibnamefont {Hull}},
  \bibinfo {author} {\bibfnamefont {M.~C.}\ \bibnamefont {Reuter}},\ and\
  \bibinfo {author} {\bibfnamefont {F.~M.}\ \bibnamefont {Ross}},\ }\bibfield
  {title} {\enquote {\bibinfo {title} {Nanometer-scale control of single
  quantum dot nucleation through focused ion-beam implantation},}\ }\href
  {https://doi.org/10.1103/physrevb.76.235301} {\bibfield  {journal} {\bibinfo
  {journal} {Phys. Rev. B}\ }\textbf {\bibinfo {volume} {76}},\ \bibinfo
  {pages} {235301} (\bibinfo {year} {2007})}\BibitemShut {NoStop}%
\bibitem [{\citenamefont {Gherasimova}\ \emph {et~al.}(2008)\citenamefont
  {Gherasimova}, \citenamefont {Hull}, \citenamefont {Reuter},\ and\
  \citenamefont {Ross}}]{Gherasimova_2008a}%
  \BibitemOpen
  \bibfield  {author} {\bibinfo {author} {\bibfnamefont {M.}~\bibnamefont
  {Gherasimova}}, \bibinfo {author} {\bibfnamefont {R.}~\bibnamefont {Hull}},
  \bibinfo {author} {\bibfnamefont {M.~C.}\ \bibnamefont {Reuter}},\ and\
  \bibinfo {author} {\bibfnamefont {F.~M.}\ \bibnamefont {Ross}},\ }\bibfield
  {title} {\enquote {\bibinfo {title} {Pattern level assembly of ge quantum
  dots on si with focused ion beam templating},}\ }\href
  {https://doi.org/10.1063/1.2957676} {\bibfield  {journal} {\bibinfo
  {journal} {Appl. Phys. Lett.}\ }\textbf {\bibinfo {volume} {93}},\ \bibinfo
  {pages} {023106} (\bibinfo {year} {2008})}\BibitemShut {NoStop}%
\bibitem [{\citenamefont {Li}\ \emph {et~al.}(2016)\citenamefont {Li},
  \citenamefont {Daukiya}, \citenamefont {Haldar}, \citenamefont {Lindblad},
  \citenamefont {Sanyal}, \citenamefont {Eriksson}, \citenamefont {Aubel},
  \citenamefont {Hajjar-Garreau}, \citenamefont {Simon},\ and\ \citenamefont
  {Leifer}}]{Li-2016}%
  \BibitemOpen
  \bibfield  {author} {\bibinfo {author} {\bibfnamefont {H.}~\bibnamefont
  {Li}}, \bibinfo {author} {\bibfnamefont {L.}~\bibnamefont {Daukiya}},
  \bibinfo {author} {\bibfnamefont {S.}~\bibnamefont {Haldar}}, \bibinfo
  {author} {\bibfnamefont {A.}~\bibnamefont {Lindblad}}, \bibinfo {author}
  {\bibfnamefont {B.}~\bibnamefont {Sanyal}}, \bibinfo {author} {\bibfnamefont
  {O.}~\bibnamefont {Eriksson}}, \bibinfo {author} {\bibfnamefont
  {D.}~\bibnamefont {Aubel}}, \bibinfo {author} {\bibfnamefont
  {S.}~\bibnamefont {Hajjar-Garreau}}, \bibinfo {author} {\bibfnamefont
  {L.}~\bibnamefont {Simon}},\ and\ \bibinfo {author} {\bibfnamefont
  {K.}~\bibnamefont {Leifer}},\ }\bibfield  {title} {{\selectlanguage
  {english}\enquote {\bibinfo {title} {Site-selective local fluorination of
  graphene induced by focused ion beam irradiation},}\ }}\href
  {https://doi.org/10.1038/srep19719} {\bibfield  {journal} {\bibinfo
  {journal} {Sci. Rep.}\ }\textbf {\bibinfo {volume} {6}},\ \bibinfo {pages}
  {19719} (\bibinfo {year} {2016})}\BibitemShut {NoStop}%
\bibitem [{\citenamefont {Yamada}\ \emph {et~al.}(2015)\citenamefont {Yamada},
  \citenamefont {Takahashi}, \citenamefont {Ikuta}, \citenamefont {Nishiyama},
  \citenamefont {Takata}, \citenamefont {Ma},\ and\ \citenamefont
  {Takahara}}]{Yamada_2015a}%
  \BibitemOpen
  \bibfield  {author} {\bibinfo {author} {\bibfnamefont {Y.}~\bibnamefont
  {Yamada}}, \bibinfo {author} {\bibfnamefont {K.}~\bibnamefont {Takahashi}},
  \bibinfo {author} {\bibfnamefont {T.}~\bibnamefont {Ikuta}}, \bibinfo
  {author} {\bibfnamefont {T.}~\bibnamefont {Nishiyama}}, \bibinfo {author}
  {\bibfnamefont {Y.}~\bibnamefont {Takata}}, \bibinfo {author} {\bibfnamefont
  {W.}~\bibnamefont {Ma}},\ and\ \bibinfo {author} {\bibfnamefont
  {A.}~\bibnamefont {Takahara}},\ }\bibfield  {title} {\enquote {\bibinfo
  {title} {Tuning surface wettability at the submicron-scale: effect of focused
  ion beam irradiation on a self-assembled monolayer},}\ }\href
  {https://doi.org/10.1021/acs.jpcc.5b09019} {\bibfield  {journal} {\bibinfo
  {journal} {J. Phys. Chem. C}\ }\textbf {\bibinfo {volume} {120}},\ \bibinfo
  {pages} {274--280} (\bibinfo {year} {2015})}\BibitemShut {NoStop}%
\bibitem [{\citenamefont {Zhao}\ \emph {et~al.}(2017)\citenamefont {Zhao},
  \citenamefont {Liu}, \citenamefont {Chen}, \citenamefont {Zhu}, \citenamefont
  {Belianinov}, \citenamefont {Ovchinnikova}, \citenamefont {Unocic},
  \citenamefont {Burch}, \citenamefont {Kim}, \citenamefont {Hao},
  \citenamefont {Pickard}, \citenamefont {Li},\ and\ \citenamefont
  {Thong}}]{Zhao2017}%
  \BibitemOpen
  \bibfield  {author} {\bibinfo {author} {\bibfnamefont {Y.}~\bibnamefont
  {Zhao}}, \bibinfo {author} {\bibfnamefont {D.}~\bibnamefont {Liu}}, \bibinfo
  {author} {\bibfnamefont {J.}~\bibnamefont {Chen}}, \bibinfo {author}
  {\bibfnamefont {L.}~\bibnamefont {Zhu}}, \bibinfo {author} {\bibfnamefont
  {A.}~\bibnamefont {Belianinov}}, \bibinfo {author} {\bibfnamefont {O.~S.}\
  \bibnamefont {Ovchinnikova}}, \bibinfo {author} {\bibfnamefont {R.~R.}\
  \bibnamefont {Unocic}}, \bibinfo {author} {\bibfnamefont {M.~J.}\
  \bibnamefont {Burch}}, \bibinfo {author} {\bibfnamefont {S.}~\bibnamefont
  {Kim}}, \bibinfo {author} {\bibfnamefont {H.}~\bibnamefont {Hao}}, \bibinfo
  {author} {\bibfnamefont {D.~S.}\ \bibnamefont {Pickard}}, \bibinfo {author}
  {\bibfnamefont {B.}~\bibnamefont {Li}},\ and\ \bibinfo {author}
  {\bibfnamefont {J.~T.~L.}\ \bibnamefont {Thong}},\ }\bibfield  {title}
  {\enquote {\bibinfo {title} {Engineering the thermal conductivity along an
  individual silicon nanowire by selective helium ion irradiation},}\ }\href
  {https://doi.org/10.1038/ncomms15919} {\bibfield  {journal} {\bibinfo
  {journal} {Nat. Commun.}\ }\textbf {\bibinfo {volume} {8}},\ \bibinfo {pages}
  {15919} (\bibinfo {year} {2017})}\BibitemShut {NoStop}%
\bibitem [{\citenamefont {Choe}\ \emph
  {et~al.}(2019{\natexlab{a}})\citenamefont {Choe}, \citenamefont {Prabhakar},
  \citenamefont {Wehmeyer}, \citenamefont {Allen}, \citenamefont {Lee},
  \citenamefont {Jin}, \citenamefont {Li}, \citenamefont {Yang}, \citenamefont
  {Qiu}, \citenamefont {Dames}, \citenamefont {Scott}, \citenamefont {Minor},
  \citenamefont {Bahk},\ and\ \citenamefont {Wu}}]{Choe2019}%
  \BibitemOpen
  \bibfield  {author} {\bibinfo {author} {\bibfnamefont {H.~S.}\ \bibnamefont
  {Choe}}, \bibinfo {author} {\bibfnamefont {R.}~\bibnamefont {Prabhakar}},
  \bibinfo {author} {\bibfnamefont {G.}~\bibnamefont {Wehmeyer}}, \bibinfo
  {author} {\bibfnamefont {F.~I.}\ \bibnamefont {Allen}}, \bibinfo {author}
  {\bibfnamefont {W.}~\bibnamefont {Lee}}, \bibinfo {author} {\bibfnamefont
  {L.}~\bibnamefont {Jin}}, \bibinfo {author} {\bibfnamefont {Y.}~\bibnamefont
  {Li}}, \bibinfo {author} {\bibfnamefont {P.}~\bibnamefont {Yang}}, \bibinfo
  {author} {\bibfnamefont {C.-W.}\ \bibnamefont {Qiu}}, \bibinfo {author}
  {\bibfnamefont {C.}~\bibnamefont {Dames}}, \bibinfo {author} {\bibfnamefont
  {M.}~\bibnamefont {Scott}}, \bibinfo {author} {\bibfnamefont
  {A.}~\bibnamefont {Minor}}, \bibinfo {author} {\bibfnamefont {J.-H.}\
  \bibnamefont {Bahk}},\ and\ \bibinfo {author} {\bibfnamefont
  {J.}~\bibnamefont {Wu}},\ }\bibfield  {title} {{\selectlanguage
  {english}\enquote {\bibinfo {title} {Ion write microthermotics: programing
  thermal metamaterials at the microscale},}\ }}\href
  {https://doi.org/10.1021/acs.nanolett.9b00984} {\bibfield  {journal}
  {\bibinfo  {journal} {Nano Lett.}\ }\textbf {\bibinfo {volume} {19}},\
  \bibinfo {pages} {3830--3837} (\bibinfo {year}
  {2019}{\natexlab{a}})}\BibitemShut {NoStop}%
\bibitem [{\citenamefont {Jin}\ \emph {et~al.}(2020)\citenamefont {Jin},
  \citenamefont {Zeltmann}, \citenamefont {Choe}, \citenamefont {Liu},
  \citenamefont {Allen}, \citenamefont {Minor},\ and\ \citenamefont
  {Wu}}]{Jin2020}%
  \BibitemOpen
  \bibfield  {author} {\bibinfo {author} {\bibfnamefont {L.}~\bibnamefont
  {Jin}}, \bibinfo {author} {\bibfnamefont {S.~E.}\ \bibnamefont {Zeltmann}},
  \bibinfo {author} {\bibfnamefont {H.~S.}\ \bibnamefont {Choe}}, \bibinfo
  {author} {\bibfnamefont {H.}~\bibnamefont {Liu}}, \bibinfo {author}
  {\bibfnamefont {F.~I.}\ \bibnamefont {Allen}}, \bibinfo {author}
  {\bibfnamefont {A.~M.}\ \bibnamefont {Minor}},\ and\ \bibinfo {author}
  {\bibfnamefont {J.}~\bibnamefont {Wu}},\ }\bibfield  {title} {\enquote
  {\bibinfo {title} {Disorder recovers the {Wiedemann-Franz} law in the
  metallic phase of {VO$_{2}$}},}\ }\href
  {https://doi.org/10.1103/PhysRevB.102.041120} {\bibfield  {journal} {\bibinfo
   {journal} {Phys. Rev. B}\ }\textbf {\bibinfo {volume} {102}},\ \bibinfo
  {pages} {041120} (\bibinfo {year} {2020})}\BibitemShut {NoStop}%
\bibitem [{\citenamefont {Choe}\ \emph
  {et~al.}(2019{\natexlab{b}})\citenamefont {Choe}, \citenamefont {Li},
  \citenamefont {Zheng}, \citenamefont {Lee}, \citenamefont {Suh},
  \citenamefont {Allen}, \citenamefont {Liu}, \citenamefont {Choi},
  \citenamefont {Walukiewicz}, \citenamefont {Zheng},\ and\ \citenamefont
  {Wu}}]{Choe2019b}%
  \BibitemOpen
  \bibfield  {author} {\bibinfo {author} {\bibfnamefont {H.~S.}\ \bibnamefont
  {Choe}}, \bibinfo {author} {\bibfnamefont {J.}~\bibnamefont {Li}}, \bibinfo
  {author} {\bibfnamefont {W.}~\bibnamefont {Zheng}}, \bibinfo {author}
  {\bibfnamefont {J.}~\bibnamefont {Lee}}, \bibinfo {author} {\bibfnamefont
  {J.}~\bibnamefont {Suh}}, \bibinfo {author} {\bibfnamefont {F.~I.}\
  \bibnamefont {Allen}}, \bibinfo {author} {\bibfnamefont {H.}~\bibnamefont
  {Liu}}, \bibinfo {author} {\bibfnamefont {H.-J.}\ \bibnamefont {Choi}},
  \bibinfo {author} {\bibfnamefont {W.}~\bibnamefont {Walukiewicz}}, \bibinfo
  {author} {\bibfnamefont {H.}~\bibnamefont {Zheng}},\ and\ \bibinfo {author}
  {\bibfnamefont {J.}~\bibnamefont {Wu}},\ }\bibfield  {title} {\enquote
  {\bibinfo {title} {Anomalously high electronic thermal conductivity and
  lorenz ratio in {Bi2Te3} nanoribbons far from the bipolar condition},}\
  }\href {https://doi.org/10.1063/1.5092221} {\bibfield  {journal} {\bibinfo
  {journal} {Appl. Phys. Lett.}\ }\textbf {\bibinfo {volume} {114}},\ \bibinfo
  {pages} {152101} (\bibinfo {year} {2019}{\natexlab{b}})}\BibitemShut
  {NoStop}%
\bibitem [{\citenamefont {Liu}\ and\ \citenamefont {Fu}(2015)}]{Liu2015}%
  \BibitemOpen
  \bibfield  {author} {\bibinfo {author} {\bibfnamefont {B.}~\bibnamefont
  {Liu}}\ and\ \bibinfo {author} {\bibfnamefont {J.}~\bibnamefont {Fu}},\
  }\bibfield  {title} {\enquote {\bibinfo {title} {Modulating surface stiffness
  of polydimethylsiloxane ({{PDMS}}) with kiloelectronvolt ion patterning},}\
  }\href {https://doi.org/10.1088/0960-1317/25/6/065006} {\bibfield  {journal}
  {\bibinfo  {journal} {J. Micromech. Microeng.}\ }\textbf {\bibinfo {volume}
  {25}},\ \bibinfo {pages} {065006} (\bibinfo {year} {2015})}\BibitemShut
  {NoStop}%
\bibitem [{\citenamefont {Kim}\ \emph {et~al.}(2014)\citenamefont {Kim},
  \citenamefont {Abuelfilat}, \citenamefont {Hoo}, \citenamefont {Al-Abboodi},
  \citenamefont {Liu}, \citenamefont {Ng}, \citenamefont {Chan},\ and\
  \citenamefont {Fu}}]{Kim2014}%
  \BibitemOpen
  \bibfield  {author} {\bibinfo {author} {\bibfnamefont {Y.}~\bibnamefont
  {Kim}}, \bibinfo {author} {\bibfnamefont {A.~Y.}\ \bibnamefont {Abuelfilat}},
  \bibinfo {author} {\bibfnamefont {S.~P.}\ \bibnamefont {Hoo}}, \bibinfo
  {author} {\bibfnamefont {A.}~\bibnamefont {Al-Abboodi}}, \bibinfo {author}
  {\bibfnamefont {B.}~\bibnamefont {Liu}}, \bibinfo {author} {\bibfnamefont
  {T.}~\bibnamefont {Ng}}, \bibinfo {author} {\bibfnamefont {P.}~\bibnamefont
  {Chan}},\ and\ \bibinfo {author} {\bibfnamefont {J.}~\bibnamefont {Fu}},\
  }\bibfield  {title} {\enquote {\bibinfo {title} {Tuning the surface
  properties of hydrogel at the nanoscale with focused ion irradiation},}\
  }\href {https://doi.org/10.1039/C4SM01061B} {\bibfield  {journal} {\bibinfo
  {journal} {Soft Matter}\ }\textbf {\bibinfo {volume} {10}},\ \bibinfo {pages}
  {8448--8456} (\bibinfo {year} {2014})}\BibitemShut {NoStop}%
\bibitem [{\citenamefont {Wang}\ \emph {et~al.}(2016)\citenamefont {Wang},
  \citenamefont {Allen}, \citenamefont {Shan},\ and\ \citenamefont
  {Hosemann}}]{Wang_2016}%
  \BibitemOpen
  \bibfield  {author} {\bibinfo {author} {\bibfnamefont {Z.-J.}\ \bibnamefont
  {Wang}}, \bibinfo {author} {\bibfnamefont {F.~I.}\ \bibnamefont {Allen}},
  \bibinfo {author} {\bibfnamefont {Z.-W.}\ \bibnamefont {Shan}},\ and\
  \bibinfo {author} {\bibfnamefont {P.}~\bibnamefont {Hosemann}},\ }\bibfield
  {title} {\enquote {\bibinfo {title} {Mechanical behavior of copper containing
  a gas-bubble superlattice},}\ }\href
  {https://doi.org/10.1016/j.actamat.2016.08.085} {\bibfield  {journal}
  {\bibinfo  {journal} {Acta Mater.}\ }\textbf {\bibinfo {volume} {121}},\
  \bibinfo {pages} {78--84} (\bibinfo {year} {2016})}\BibitemShut {NoStop}%
\bibitem [{\citenamefont {Wang}\ \emph
  {et~al.}(2017{\natexlab{a}})\citenamefont {Wang}, \citenamefont {Tian},
  \citenamefont {Liu}, \citenamefont {Qin}, \citenamefont {Zheng},
  \citenamefont {Wang}, \citenamefont {Ma},\ and\ \citenamefont
  {Shan}}]{Wang_2017b}%
  \BibitemOpen
  \bibfield  {author} {\bibinfo {author} {\bibfnamefont {Y.-C.}\ \bibnamefont
  {Wang}}, \bibinfo {author} {\bibfnamefont {L.}~\bibnamefont {Tian}}, \bibinfo
  {author} {\bibfnamefont {F.}~\bibnamefont {Liu}}, \bibinfo {author}
  {\bibfnamefont {Y.-B.}\ \bibnamefont {Qin}}, \bibinfo {author} {\bibfnamefont
  {G.}~\bibnamefont {Zheng}}, \bibinfo {author} {\bibfnamefont {J.-T.}\
  \bibnamefont {Wang}}, \bibinfo {author} {\bibfnamefont {E.}~\bibnamefont
  {Ma}},\ and\ \bibinfo {author} {\bibfnamefont {Z.-W.}\ \bibnamefont {Shan}},\
  }\bibfield  {title} {{\selectlanguage {english}\enquote {\bibinfo {title}
  {Helium ion microscope fabrication causing changes in the structure and
  mechanical behavior of silicon micropillars},}\ }}\href
  {https://doi.org/10.1002/smll.201601753} {\bibfield  {journal} {\bibinfo
  {journal} {Small}\ }\textbf {\bibinfo {volume} {13}} (\bibinfo {year}
  {2017}{\natexlab{a}}),\ 10.1002/smll.201601753}\BibitemShut {NoStop}%
\bibitem [{\citenamefont {Chen}\ \emph
  {et~al.}(2020{\natexlab{b}})\citenamefont {Chen}, \citenamefont {Best},
  \citenamefont {Shorubalko}, \citenamefont {Michler}, \citenamefont
  {Spolenak},\ and\ \citenamefont {Wheeler}}]{Chen_2020}%
  \BibitemOpen
  \bibfield  {author} {\bibinfo {author} {\bibfnamefont {M.}~\bibnamefont
  {Chen}}, \bibinfo {author} {\bibfnamefont {J.~P.}\ \bibnamefont {Best}},
  \bibinfo {author} {\bibfnamefont {I.}~\bibnamefont {Shorubalko}}, \bibinfo
  {author} {\bibfnamefont {J.}~\bibnamefont {Michler}}, \bibinfo {author}
  {\bibfnamefont {R.}~\bibnamefont {Spolenak}},\ and\ \bibinfo {author}
  {\bibfnamefont {J.~M.}\ \bibnamefont {Wheeler}},\ }\bibfield  {title}
  {\enquote {\bibinfo {title} {Influence of helium ion irradiation on the
  structure and strength of diamond},}\ }\href
  {https://doi.org/10.1016/j.carbon.2019.10.078} {\bibfield  {journal}
  {\bibinfo  {journal} {Carbon}\ }\textbf {\bibinfo {volume} {158}},\ \bibinfo
  {pages} {337--345} (\bibinfo {year} {2020}{\natexlab{b}})}\BibitemShut
  {NoStop}%
\bibitem [{\citenamefont {Bergner}, \citenamefont {Hlawacek},\ and\
  \citenamefont {Heintze}(2018)}]{Bergner_2018}%
  \BibitemOpen
  \bibfield  {author} {\bibinfo {author} {\bibfnamefont {F.}~\bibnamefont
  {Bergner}}, \bibinfo {author} {\bibfnamefont {G.}~\bibnamefont {Hlawacek}},\
  and\ \bibinfo {author} {\bibfnamefont {C.}~\bibnamefont {Heintze}},\
  }\bibfield  {title} {\enquote {\bibinfo {title} {Helium-ion microscopy,
  helium-ion irradiation and nanoindentation of eurofer 97 and {ODS}
  eurofer},}\ }\href {https://doi.org/10.1016/j.jnucmat.2017.07.054} {\bibfield
   {journal} {\bibinfo  {journal} {J. Nucl. Mater.}\ }\textbf {\bibinfo
  {volume} {505}},\ \bibinfo {pages} {267--275} (\bibinfo {year}
  {2018})}\BibitemShut {NoStop}%
\bibitem [{\citenamefont {Allen}, \citenamefont {Hosemann},\ and\ \citenamefont
  {Balooch}(2020)}]{Allen_2020}%
  \BibitemOpen
  \bibfield  {author} {\bibinfo {author} {\bibfnamefont {F.~I.}\ \bibnamefont
  {Allen}}, \bibinfo {author} {\bibfnamefont {P.}~\bibnamefont {Hosemann}},\
  and\ \bibinfo {author} {\bibfnamefont {M.}~\bibnamefont {Balooch}},\
  }\bibfield  {title} {\enquote {\bibinfo {title} {Key mechanistic features of
  swelling and blistering of helium-ion-irradiated tungsten},}\ }\href
  {https://doi.org/10.1016/j.scriptamat.2019.11.039} {\bibfield  {journal}
  {\bibinfo  {journal} {Scripta Mater.}\ }\textbf {\bibinfo {volume} {178}},\
  \bibinfo {pages} {256--260} (\bibinfo {year} {2020})}\BibitemShut {NoStop}%
\bibitem [{\citenamefont {Wurmshuber}\ \emph {et~al.}(2022)\citenamefont
  {Wurmshuber}, \citenamefont {Balooch}, \citenamefont {Huang}, \citenamefont
  {Hosemann},\ and\ \citenamefont {Kiener}}]{Wurmshuber_2022}%
  \BibitemOpen
  \bibfield  {author} {\bibinfo {author} {\bibfnamefont {M.}~\bibnamefont
  {Wurmshuber}}, \bibinfo {author} {\bibfnamefont {M.}~\bibnamefont {Balooch}},
  \bibinfo {author} {\bibfnamefont {X.}~\bibnamefont {Huang}}, \bibinfo
  {author} {\bibfnamefont {P.}~\bibnamefont {Hosemann}},\ and\ \bibinfo
  {author} {\bibfnamefont {D.}~\bibnamefont {Kiener}},\ }\bibfield  {title}
  {\enquote {\bibinfo {title} {Helium-induced swelling and mechanical property
  degradation in ultrafine-grained w and w-cu nanocomposites for fusion
  applications},}\ }\href {https://doi.org/10.1016/j.scriptamat.2022.114641}
  {\bibfield  {journal} {\bibinfo  {journal} {Scripta Mater.}\ }\textbf
  {\bibinfo {volume} {213}},\ \bibinfo {pages} {114641} (\bibinfo {year}
  {2022})}\BibitemShut {NoStop}%
\bibitem [{\citenamefont {Fassbender}, \citenamefont {Ravelosona},\ and\
  \citenamefont {Samson}(2004)}]{Fassbender_2004}%
  \BibitemOpen
  \bibfield  {author} {\bibinfo {author} {\bibfnamefont {J.}~\bibnamefont
  {Fassbender}}, \bibinfo {author} {\bibfnamefont {D.}~\bibnamefont
  {Ravelosona}},\ and\ \bibinfo {author} {\bibfnamefont {Y.}~\bibnamefont
  {Samson}},\ }\bibfield  {title} {\enquote {\bibinfo {title} {Tailoring
  magnetism by light-ion irradiation},}\ }\href
  {https://doi.org/10.1088/0022-3727/37/16/r01} {\bibfield  {journal} {\bibinfo
   {journal} {J. Phys. D: Appl. Phys.}\ }\textbf {\bibinfo {volume} {37}},\
  \bibinfo {pages} {R179--R196} (\bibinfo {year} {2004})}\BibitemShut {NoStop}%
\bibitem [{\citenamefont {Pan}\ \emph {et~al.}(1999)\citenamefont {Pan},
  \citenamefont {Takahashi}, \citenamefont {Walker}, \citenamefont {Mapps},
  \citenamefont {Yamakawa}, \citenamefont {Honda},\ and\ \citenamefont
  {Ouchi}}]{Pan_1999}%
  \BibitemOpen
  \bibfield  {author} {\bibinfo {author} {\bibfnamefont {G.}~\bibnamefont
  {Pan}}, \bibinfo {author} {\bibfnamefont {S.}~\bibnamefont {Takahashi}},
  \bibinfo {author} {\bibfnamefont {J.~F.}\ \bibnamefont {Walker}}, \bibinfo
  {author} {\bibfnamefont {D.~J.}\ \bibnamefont {Mapps}}, \bibinfo {author}
  {\bibfnamefont {K.}~\bibnamefont {Yamakawa}}, \bibinfo {author}
  {\bibfnamefont {N.}~\bibnamefont {Honda}},\ and\ \bibinfo {author}
  {\bibfnamefont {K.}~\bibnamefont {Ouchi}},\ }\bibfield  {title} {\enquote
  {\bibinfo {title} {A new fabrication method for submicron trackwidth thin
  film heads},}\ }\href {https://doi.org/10.1016/s0304-8853(99)00421-7}
  {\bibfield  {journal} {\bibinfo  {journal} {J. Magn. Magn. Mater.}\ }\textbf
  {\bibinfo {volume} {202}},\ \bibinfo {pages} {583--592} (\bibinfo {year}
  {1999})}\BibitemShut {NoStop}%
\bibitem [{\citenamefont {Nishimura}\ \emph {et~al.}(2002)\citenamefont
  {Nishimura}, \citenamefont {Hirai}, \citenamefont {Koganei}, \citenamefont
  {Ikeda}, \citenamefont {Okano}, \citenamefont {Sekiguchi},\ and\
  \citenamefont {Osada}}]{Nishimura_2002}%
  \BibitemOpen
  \bibfield  {author} {\bibinfo {author} {\bibfnamefont {N.}~\bibnamefont
  {Nishimura}}, \bibinfo {author} {\bibfnamefont {T.}~\bibnamefont {Hirai}},
  \bibinfo {author} {\bibfnamefont {A.}~\bibnamefont {Koganei}}, \bibinfo
  {author} {\bibfnamefont {T.}~\bibnamefont {Ikeda}}, \bibinfo {author}
  {\bibfnamefont {K.}~\bibnamefont {Okano}}, \bibinfo {author} {\bibfnamefont
  {Y.}~\bibnamefont {Sekiguchi}},\ and\ \bibinfo {author} {\bibfnamefont
  {Y.}~\bibnamefont {Osada}},\ }\bibfield  {title} {\enquote {\bibinfo {title}
  {Magnetic tunnel junction device with perpendicular magnetization films for
  high-density magnetic random access memory},}\ }\href
  {https://doi.org/10.1063/1.1459605} {\bibfield  {journal} {\bibinfo
  {journal} {J. Appl. Phys.}\ }\textbf {\bibinfo {volume} {91}},\ \bibinfo
  {pages} {5246--5249} (\bibinfo {year} {2002})}\BibitemShut {NoStop}%
\bibitem [{\citenamefont {Albrecht}\ \emph {et~al.}(2003)\citenamefont
  {Albrecht}, \citenamefont {Ganesan}, \citenamefont {Rettner}, \citenamefont
  {Moser}, \citenamefont {Best}, \citenamefont {White},\ and\ \citenamefont
  {Terris}}]{Albrecht_2003}%
  \BibitemOpen
  \bibfield  {author} {\bibinfo {author} {\bibfnamefont {M.}~\bibnamefont
  {Albrecht}}, \bibinfo {author} {\bibfnamefont {S.}~\bibnamefont {Ganesan}},
  \bibinfo {author} {\bibfnamefont {C.~T.}\ \bibnamefont {Rettner}}, \bibinfo
  {author} {\bibfnamefont {A.}~\bibnamefont {Moser}}, \bibinfo {author}
  {\bibfnamefont {M.~E.}\ \bibnamefont {Best}}, \bibinfo {author}
  {\bibfnamefont {R.~L.}\ \bibnamefont {White}},\ and\ \bibinfo {author}
  {\bibfnamefont {B.~D.}\ \bibnamefont {Terris}},\ }\bibfield  {title}
  {\enquote {\bibinfo {title} {Patterned perpendicular and longitudinal media:
  a magnetic recording study},}\ }\href
  {https://doi.org/10.1109/tmag.2003.816285} {\bibfield  {journal} {\bibinfo
  {journal} {IEEE Trans. Magn.}\ }\textbf {\bibinfo {volume} {39}},\ \bibinfo
  {pages} {2323--2325} (\bibinfo {year} {2003})}\BibitemShut {NoStop}%
\bibitem [{\citenamefont {Serrano-Ram{\'{o}}n}\ \emph
  {et~al.}(2013)\citenamefont {Serrano-Ram{\'{o}}n}, \citenamefont
  {Fern{\'{a}}ndez-Pacheco}, \citenamefont {Ibarra}, \citenamefont {Petit},
  \citenamefont {Cowburn}, \citenamefont {Tyliszczak},\ and\ \citenamefont
  {De~Teresa}}]{SerranoRamon_2013}%
  \BibitemOpen
  \bibfield  {author} {\bibinfo {author} {\bibfnamefont {L.}~\bibnamefont
  {Serrano-Ram{\'{o}}n}}, \bibinfo {author} {\bibfnamefont {A.}~\bibnamefont
  {Fern{\'{a}}ndez-Pacheco}}, \bibinfo {author} {\bibfnamefont {M.~R.}\
  \bibnamefont {Ibarra}}, \bibinfo {author} {\bibfnamefont {D.}~\bibnamefont
  {Petit}}, \bibinfo {author} {\bibfnamefont {R.~P.}\ \bibnamefont {Cowburn}},
  \bibinfo {author} {\bibfnamefont {T.}~\bibnamefont {Tyliszczak}},\ and\
  \bibinfo {author} {\bibfnamefont {J.~M.}\ \bibnamefont {De~Teresa}},\
  }\bibfield  {title} {\enquote {\bibinfo {title} {Modification of domain-wall
  propagation in co nanowires via {G}a$^+$ irradiation},}\ }\href
  {https://doi.org/10.1140/epjb/e2013-30926-5} {\bibfield  {journal} {\bibinfo
  {journal} {Eur. Phys. J. B}\ }\textbf {\bibinfo {volume} {86}},\ \bibinfo
  {pages} {97} (\bibinfo {year} {2013})}\BibitemShut {NoStop}%
\bibitem [{\citenamefont {Nguyen}\ \emph {et~al.}(2013)\citenamefont {Nguyen},
  \citenamefont {Balocchi}, \citenamefont {Lagarde}, \citenamefont {Zhang},
  \citenamefont {Carr{\`{e}}re}, \citenamefont {Mazzucato}, \citenamefont
  {Barate}, \citenamefont {Galopin}, \citenamefont {Gierak}, \citenamefont
  {Bourhis}, \citenamefont {Harmand}, \citenamefont {Amand},\ and\
  \citenamefont {Marie}}]{Nguyen_2013}%
  \BibitemOpen
  \bibfield  {author} {\bibinfo {author} {\bibfnamefont {C.~T.}\ \bibnamefont
  {Nguyen}}, \bibinfo {author} {\bibfnamefont {A.}~\bibnamefont {Balocchi}},
  \bibinfo {author} {\bibfnamefont {D.}~\bibnamefont {Lagarde}}, \bibinfo
  {author} {\bibfnamefont {T.~T.}\ \bibnamefont {Zhang}}, \bibinfo {author}
  {\bibfnamefont {H.}~\bibnamefont {Carr{\`{e}}re}}, \bibinfo {author}
  {\bibfnamefont {S.}~\bibnamefont {Mazzucato}}, \bibinfo {author}
  {\bibfnamefont {P.}~\bibnamefont {Barate}}, \bibinfo {author} {\bibfnamefont
  {E.}~\bibnamefont {Galopin}}, \bibinfo {author} {\bibfnamefont
  {J.}~\bibnamefont {Gierak}}, \bibinfo {author} {\bibfnamefont
  {E.}~\bibnamefont {Bourhis}}, \bibinfo {author} {\bibfnamefont {J.~C.}\
  \bibnamefont {Harmand}}, \bibinfo {author} {\bibfnamefont {T.}~\bibnamefont
  {Amand}},\ and\ \bibinfo {author} {\bibfnamefont {X.}~\bibnamefont {Marie}},\
  }\bibfield  {title} {\enquote {\bibinfo {title} {Fabrication of an {InGaAs}
  spin filter by implantation of paramagnetic centers},}\ }\href
  {https://doi.org/10.1063/1.4816970} {\bibfield  {journal} {\bibinfo
  {journal} {Appl. Phys. Lett.}\ }\textbf {\bibinfo {volume} {103}},\ \bibinfo
  {pages} {052403} (\bibinfo {year} {2013})}\BibitemShut {NoStop}%
\bibitem [{\citenamefont {Devolder}\ \emph {et~al.}(1999)\citenamefont
  {Devolder}, \citenamefont {Vieu}, \citenamefont {Bernas}, \citenamefont
  {Ferri{\'{e}}}, \citenamefont {Chappert}, \citenamefont {Gierak},
  \citenamefont {Jamet}, \citenamefont {Aign}, \citenamefont {Meyer},
  \citenamefont {Chen}, \citenamefont {Rousseaux}, \citenamefont {Mathet},
  \citenamefont {Launois},\ and\ \citenamefont {Kaitasov}}]{Devolder_1999}%
  \BibitemOpen
  \bibfield  {author} {\bibinfo {author} {\bibfnamefont {T.}~\bibnamefont
  {Devolder}}, \bibinfo {author} {\bibfnamefont {C.}~\bibnamefont {Vieu}},
  \bibinfo {author} {\bibfnamefont {H.}~\bibnamefont {Bernas}}, \bibinfo
  {author} {\bibfnamefont {J.}~\bibnamefont {Ferri{\'{e}}}}, \bibinfo {author}
  {\bibfnamefont {C.}~\bibnamefont {Chappert}}, \bibinfo {author}
  {\bibfnamefont {J.}~\bibnamefont {Gierak}}, \bibinfo {author} {\bibfnamefont
  {J.-P.}\ \bibnamefont {Jamet}}, \bibinfo {author} {\bibfnamefont
  {T.}~\bibnamefont {Aign}}, \bibinfo {author} {\bibfnamefont {P.}~\bibnamefont
  {Meyer}}, \bibinfo {author} {\bibfnamefont {Y.}~\bibnamefont {Chen}},
  \bibinfo {author} {\bibfnamefont {F.}~\bibnamefont {Rousseaux}}, \bibinfo
  {author} {\bibfnamefont {V.}~\bibnamefont {Mathet}}, \bibinfo {author}
  {\bibfnamefont {H.}~\bibnamefont {Launois}},\ and\ \bibinfo {author}
  {\bibfnamefont {O.}~\bibnamefont {Kaitasov}},\ }\bibfield  {title} {\enquote
  {\bibinfo {title} {Ion beam-induced magnetic patterning at the sub-0.1
  \textmu{}m level},}\ }\href {https://doi.org/10.1016/s1287-4620(99)80155-0}
  {\bibfield  {journal} {\bibinfo  {journal} {Comptes Rendus de
  l'Acad{\'{e}}mie des Sciences - Series {IIB} - Mechanics-Physics-Astronomy}\
  }\textbf {\bibinfo {volume} {327}},\ \bibinfo {pages} {915--923} (\bibinfo
  {year} {1999})}\BibitemShut {NoStop}%
\bibitem [{\citenamefont {Warin}\ \emph {et~al.}(2001)\citenamefont {Warin},
  \citenamefont {Hyndman}, \citenamefont {Glerak}, \citenamefont {Chapman},
  \citenamefont {Ferr{\'{e}}}, \citenamefont {Jamet}, \citenamefont {Mathet},\
  and\ \citenamefont {Chappert}}]{Warin_2001}%
  \BibitemOpen
  \bibfield  {author} {\bibinfo {author} {\bibfnamefont {P.}~\bibnamefont
  {Warin}}, \bibinfo {author} {\bibfnamefont {R.}~\bibnamefont {Hyndman}},
  \bibinfo {author} {\bibfnamefont {J.}~\bibnamefont {Glerak}}, \bibinfo
  {author} {\bibfnamefont {J.~N.}\ \bibnamefont {Chapman}}, \bibinfo {author}
  {\bibfnamefont {J.}~\bibnamefont {Ferr{\'{e}}}}, \bibinfo {author}
  {\bibfnamefont {J.~P.}\ \bibnamefont {Jamet}}, \bibinfo {author}
  {\bibfnamefont {V.}~\bibnamefont {Mathet}},\ and\ \bibinfo {author}
  {\bibfnamefont {C.}~\bibnamefont {Chappert}},\ }\bibfield  {title} {\enquote
  {\bibinfo {title} {Modification of {C}o/{P}t multilayers by gallium
  irradiation---part 2: the effect of patterning using a highly focused ion
  beam},}\ }\href {https://doi.org/10.1063/1.1401794} {\bibfield  {journal}
  {\bibinfo  {journal} {J. Appl. Phys.}\ }\textbf {\bibinfo {volume} {90}},\
  \bibinfo {pages} {3850--3855} (\bibinfo {year} {2001})}\BibitemShut {NoStop}%
\bibitem [{\citenamefont {Gierak}\ \emph {et~al.}(2005)\citenamefont {Gierak},
  \citenamefont {Mailly}, \citenamefont {Hawkes}, \citenamefont {Jede},
  \citenamefont {Bruchhaus}, \citenamefont {Bardotti}, \citenamefont
  {Pr{\'{e}}vel}, \citenamefont {M{\'{e}}linon}, \citenamefont {Perez},
  \citenamefont {Hyndman}, \citenamefont {Jamet}, \citenamefont {Ferr{\'{e}}},
  \citenamefont {Mougin}, \citenamefont {Chappert}, \citenamefont {Mathet},
  \citenamefont {Warin},\ and\ \citenamefont {Chapman}}]{Gierak_2005}%
  \BibitemOpen
  \bibfield  {author} {\bibinfo {author} {\bibfnamefont {J.}~\bibnamefont
  {Gierak}}, \bibinfo {author} {\bibfnamefont {D.}~\bibnamefont {Mailly}},
  \bibinfo {author} {\bibfnamefont {P.}~\bibnamefont {Hawkes}}, \bibinfo
  {author} {\bibfnamefont {R.}~\bibnamefont {Jede}}, \bibinfo {author}
  {\bibfnamefont {L.}~\bibnamefont {Bruchhaus}}, \bibinfo {author}
  {\bibfnamefont {L.}~\bibnamefont {Bardotti}}, \bibinfo {author}
  {\bibfnamefont {B.}~\bibnamefont {Pr{\'{e}}vel}}, \bibinfo {author}
  {\bibfnamefont {P.}~\bibnamefont {M{\'{e}}linon}}, \bibinfo {author}
  {\bibfnamefont {A.}~\bibnamefont {Perez}}, \bibinfo {author} {\bibfnamefont
  {R.}~\bibnamefont {Hyndman}}, \bibinfo {author} {\bibfnamefont {J.-P.}\
  \bibnamefont {Jamet}}, \bibinfo {author} {\bibfnamefont {J.}~\bibnamefont
  {Ferr{\'{e}}}}, \bibinfo {author} {\bibfnamefont {A.}~\bibnamefont {Mougin}},
  \bibinfo {author} {\bibfnamefont {C.}~\bibnamefont {Chappert}}, \bibinfo
  {author} {\bibfnamefont {V.}~\bibnamefont {Mathet}}, \bibinfo {author}
  {\bibfnamefont {P.}~\bibnamefont {Warin}},\ and\ \bibinfo {author}
  {\bibfnamefont {J.}~\bibnamefont {Chapman}},\ }\bibfield  {title} {\enquote
  {\bibinfo {title} {Exploration of the ultimate patterning potential
  achievable with high resolution focused ion beams},}\ }\href
  {https://doi.org/10.1007/s00339-004-2551-z} {\bibfield  {journal} {\bibinfo
  {journal} {Appl. Phys. A}\ }\textbf {\bibinfo {volume} {80}},\ \bibinfo
  {pages} {187--194} (\bibinfo {year} {2005})}\BibitemShut {NoStop}%
\bibitem [{\citenamefont {Townsend}(1987)}]{Townsend_1987}%
  \BibitemOpen
  \bibfield  {author} {\bibinfo {author} {\bibfnamefont {P.~D.}\ \bibnamefont
  {Townsend}},\ }\bibfield  {title} {\enquote {\bibinfo {title} {Optical
  effects of ion implantation},}\ }\href
  {https://doi.org/10.1088/0034-4885/50/5/001} {\bibfield  {journal} {\bibinfo
  {journal} {Rep. Prog. Phys.}\ }\textbf {\bibinfo {volume} {50}},\ \bibinfo
  {pages} {501--558} (\bibinfo {year} {1987})}\BibitemShut {NoStop}%
\bibitem [{\citenamefont {Lagomarsino}\ \emph {et~al.}(2021)\citenamefont
  {Lagomarsino}, \citenamefont {Flatae}, \citenamefont {Kambalathmana},
  \citenamefont {Sledz}, \citenamefont {Hunold}, \citenamefont {Soltani},
  \citenamefont {Reuschel}, \citenamefont {Sciortino}, \citenamefont {Gelli},
  \citenamefont {Massi}, \citenamefont {Czelusniak}, \citenamefont {Giuntini},\
  and\ \citenamefont {Agio}}]{Lagomarsino_2021}%
  \BibitemOpen
  \bibfield  {author} {\bibinfo {author} {\bibfnamefont {S.}~\bibnamefont
  {Lagomarsino}}, \bibinfo {author} {\bibfnamefont {A.~M.}\ \bibnamefont
  {Flatae}}, \bibinfo {author} {\bibfnamefont {H.}~\bibnamefont
  {Kambalathmana}}, \bibinfo {author} {\bibfnamefont {F.}~\bibnamefont
  {Sledz}}, \bibinfo {author} {\bibfnamefont {L.}~\bibnamefont {Hunold}},
  \bibinfo {author} {\bibfnamefont {N.}~\bibnamefont {Soltani}}, \bibinfo
  {author} {\bibfnamefont {P.}~\bibnamefont {Reuschel}}, \bibinfo {author}
  {\bibfnamefont {S.}~\bibnamefont {Sciortino}}, \bibinfo {author}
  {\bibfnamefont {N.}~\bibnamefont {Gelli}}, \bibinfo {author} {\bibfnamefont
  {M.}~\bibnamefont {Massi}}, \bibinfo {author} {\bibfnamefont
  {C.}~\bibnamefont {Czelusniak}}, \bibinfo {author} {\bibfnamefont
  {L.}~\bibnamefont {Giuntini}},\ and\ \bibinfo {author} {\bibfnamefont
  {M.}~\bibnamefont {Agio}},\ }\bibfield  {title} {\enquote {\bibinfo {title}
  {Creation of silicon-vacancy color centers in diamond by ion implantation},}\
  }\href {https://doi.org/10.3389/fphy.2020.601362} {\bibfield  {journal}
  {\bibinfo  {journal} {Aip. Conf. Proc.}\ }\textbf {\bibinfo {volume} {8}}
  (\bibinfo {year} {2021}),\ 10.3389/fphy.2020.601362}\BibitemShut {NoStop}%
\bibitem [{\citenamefont {Wang}\ \emph
  {et~al.}(2017{\natexlab{b}})\citenamefont {Wang}, \citenamefont {Zhang},
  \citenamefont {Zhou}, \citenamefont {Li}, \citenamefont {Wang}, \citenamefont
  {Peddibhotla}, \citenamefont {Liu}, \citenamefont {Bauerdick}, \citenamefont
  {Rudzinski}, \citenamefont {Liu},\ and\ \citenamefont {Gao}}]{Wang_2017a}%
  \BibitemOpen
  \bibfield  {author} {\bibinfo {author} {\bibfnamefont {J.}~\bibnamefont
  {Wang}}, \bibinfo {author} {\bibfnamefont {X.}~\bibnamefont {Zhang}},
  \bibinfo {author} {\bibfnamefont {Y.}~\bibnamefont {Zhou}}, \bibinfo {author}
  {\bibfnamefont {K.}~\bibnamefont {Li}}, \bibinfo {author} {\bibfnamefont
  {Z.}~\bibnamefont {Wang}}, \bibinfo {author} {\bibfnamefont {P.}~\bibnamefont
  {Peddibhotla}}, \bibinfo {author} {\bibfnamefont {F.}~\bibnamefont {Liu}},
  \bibinfo {author} {\bibfnamefont {S.}~\bibnamefont {Bauerdick}}, \bibinfo
  {author} {\bibfnamefont {A.}~\bibnamefont {Rudzinski}}, \bibinfo {author}
  {\bibfnamefont {Z.}~\bibnamefont {Liu}},\ and\ \bibinfo {author}
  {\bibfnamefont {W.}~\bibnamefont {Gao}},\ }\bibfield  {title} {\enquote
  {\bibinfo {title} {Scalable fabrication of single silicon vacancy defect
  arrays in silicon carbide using focused ion beam},}\ }\href
  {https://doi.org/10.1021/acsphotonics.7b00230} {\bibfield  {journal}
  {\bibinfo  {journal} {{ACS} Photonics}\ }\textbf {\bibinfo {volume} {4}},\
  \bibinfo {pages} {1054--1059} (\bibinfo {year}
  {2017}{\natexlab{b}})}\BibitemShut {NoStop}%
\bibitem [{\citenamefont {Wang}\ \emph
  {et~al.}(2019{\natexlab{a}})\citenamefont {Wang}, \citenamefont {Li},
  \citenamefont {Yan}, \citenamefont {Liu}, \citenamefont {Guo}, \citenamefont
  {Zhang}, \citenamefont {Zhou}, \citenamefont {Guo}, \citenamefont {Lin},
  \citenamefont {Cui}, \citenamefont {Xu}, \citenamefont {Xu}, \citenamefont
  {Li},\ and\ \citenamefont {Guo}}]{Wang_2019b}%
  \BibitemOpen
  \bibfield  {author} {\bibinfo {author} {\bibfnamefont {J.-F.}\ \bibnamefont
  {Wang}}, \bibinfo {author} {\bibfnamefont {Q.}~\bibnamefont {Li}}, \bibinfo
  {author} {\bibfnamefont {F.-F.}\ \bibnamefont {Yan}}, \bibinfo {author}
  {\bibfnamefont {H.}~\bibnamefont {Liu}}, \bibinfo {author} {\bibfnamefont
  {G.-P.}\ \bibnamefont {Guo}}, \bibinfo {author} {\bibfnamefont {W.-P.}\
  \bibnamefont {Zhang}}, \bibinfo {author} {\bibfnamefont {X.}~\bibnamefont
  {Zhou}}, \bibinfo {author} {\bibfnamefont {L.-P.}\ \bibnamefont {Guo}},
  \bibinfo {author} {\bibfnamefont {Z.-H.}\ \bibnamefont {Lin}}, \bibinfo
  {author} {\bibfnamefont {J.-M.}\ \bibnamefont {Cui}}, \bibinfo {author}
  {\bibfnamefont {X.-Y.}\ \bibnamefont {Xu}}, \bibinfo {author} {\bibfnamefont
  {J.-S.}\ \bibnamefont {Xu}}, \bibinfo {author} {\bibfnamefont {C.-F.}\
  \bibnamefont {Li}},\ and\ \bibinfo {author} {\bibfnamefont {G.-C.}\
  \bibnamefont {Guo}},\ }\bibfield  {title} {\enquote {\bibinfo {title}
  {On-demand generation of single silicon vacancy defects in silicon
  carbide},}\ }\href {https://doi.org/10.1021/acsphotonics.9b00451} {\bibfield
  {journal} {\bibinfo  {journal} {{ACS} Photonics}\ }\textbf {\bibinfo {volume}
  {6}},\ \bibinfo {pages} {1736--1743} (\bibinfo {year}
  {2019}{\natexlab{a}})}\BibitemShut {NoStop}%
\bibitem [{\citenamefont {Adshead}\ \emph {et~al.}(2023)\citenamefont
  {Adshead}, \citenamefont {Coke}, \citenamefont {Aresta}, \citenamefont
  {Bellew}, \citenamefont {Lagator}, \citenamefont {Li}, \citenamefont {Cui},
  \citenamefont {Cai}, \citenamefont {Almutawa}, \citenamefont {Haigh},
  \citenamefont {Moore}, \citenamefont {Lockyer}, \citenamefont {Gourlay},\
  and\ \citenamefont {Curry}}]{Adshead2023}%
  \BibitemOpen
  \bibfield  {author} {\bibinfo {author} {\bibfnamefont {M.}~\bibnamefont
  {Adshead}}, \bibinfo {author} {\bibfnamefont {M.}~\bibnamefont {Coke}},
  \bibinfo {author} {\bibfnamefont {G.}~\bibnamefont {Aresta}}, \bibinfo
  {author} {\bibfnamefont {A.}~\bibnamefont {Bellew}}, \bibinfo {author}
  {\bibfnamefont {M.}~\bibnamefont {Lagator}}, \bibinfo {author} {\bibfnamefont
  {K.}~\bibnamefont {Li}}, \bibinfo {author} {\bibfnamefont {Y.}~\bibnamefont
  {Cui}}, \bibinfo {author} {\bibfnamefont {R.}~\bibnamefont {Cai}}, \bibinfo
  {author} {\bibfnamefont {A.}~\bibnamefont {Almutawa}}, \bibinfo {author}
  {\bibfnamefont {S.~J.}\ \bibnamefont {Haigh}}, \bibinfo {author}
  {\bibfnamefont {K.}~\bibnamefont {Moore}}, \bibinfo {author} {\bibfnamefont
  {N.}~\bibnamefont {Lockyer}}, \bibinfo {author} {\bibfnamefont {C.~M.}\
  \bibnamefont {Gourlay}},\ and\ \bibinfo {author} {\bibfnamefont {R.~J.}\
  \bibnamefont {Curry}},\ }\bibfield  {title} {\enquote {\bibinfo {title} {A
  high-resolution versatile focused ion implantation platform for nanoscale
  engineering},}\ }\href {https://doi.org/10.1002/adem.202300889} {\bibfield
  {journal} {\bibinfo  {journal} {Adv. Eng. Mater.}\ } (\bibinfo {year}
  {2023}),\ 10.1002/adem.202300889}\BibitemShut {NoStop}%
\bibitem [{\citenamefont {Kianinia}\ \emph {et~al.}(2022)\citenamefont
  {Kianinia}, \citenamefont {Xu}, \citenamefont {Toth},\ and\ \citenamefont
  {Aharonovich}}]{Kianinia_2022}%
  \BibitemOpen
  \bibfield  {author} {\bibinfo {author} {\bibfnamefont {M.}~\bibnamefont
  {Kianinia}}, \bibinfo {author} {\bibfnamefont {Z.-Q.}\ \bibnamefont {Xu}},
  \bibinfo {author} {\bibfnamefont {M.}~\bibnamefont {Toth}},\ and\ \bibinfo
  {author} {\bibfnamefont {I.}~\bibnamefont {Aharonovich}},\ }\bibfield
  {title} {\enquote {\bibinfo {title} {Quantum emitters in 2{D} materials:
  emitter engineering, photophysics, and integration in photonic
  nanostructures},}\ }\href {https://doi.org/10.1063/5.0072091} {\bibfield
  {journal} {\bibinfo  {journal} {Appl. Phys. Rev.}\ }\textbf {\bibinfo
  {volume} {9}},\ \bibinfo {pages} {011306} (\bibinfo {year}
  {2022})}\BibitemShut {NoStop}%
\bibitem [{\citenamefont {Ma}\ \emph {et~al.}(2017{\natexlab{b}})\citenamefont
  {Ma}, \citenamefont {Tan}, \citenamefont {Ghorbani-Asl}, \citenamefont
  {Boettger}, \citenamefont {Kretschmer}, \citenamefont {Zhou}, \citenamefont
  {Huang}, \citenamefont {Krasheninnikov},\ and\ \citenamefont
  {Chen}}]{Ma2017aNS}%
  \BibitemOpen
  \bibfield  {author} {\bibinfo {author} {\bibfnamefont {L.}~\bibnamefont
  {Ma}}, \bibinfo {author} {\bibfnamefont {Y.}~\bibnamefont {Tan}}, \bibinfo
  {author} {\bibfnamefont {M.}~\bibnamefont {Ghorbani-Asl}}, \bibinfo {author}
  {\bibfnamefont {R.}~\bibnamefont {Boettger}}, \bibinfo {author}
  {\bibfnamefont {S.}~\bibnamefont {Kretschmer}}, \bibinfo {author}
  {\bibfnamefont {S.}~\bibnamefont {Zhou}}, \bibinfo {author} {\bibfnamefont
  {Z.}~\bibnamefont {Huang}}, \bibinfo {author} {\bibfnamefont {A.~V.}\
  \bibnamefont {Krasheninnikov}},\ and\ \bibinfo {author} {\bibfnamefont
  {F.}~\bibnamefont {Chen}},\ }\bibfield  {title} {\enquote {\bibinfo {title}
  {Tailoring the optical properties of atomically-thin {WS}$_{2}$ via ion
  irradiation},}\ }\href {https://doi.org/10.1039/C7NR02025B} {\bibfield
  {journal} {\bibinfo  {journal} {Nanoscale}\ }\textbf {\bibinfo {volume}
  {9}},\ \bibinfo {pages} {11027--11034} (\bibinfo {year}
  {2017}{\natexlab{b}})}\BibitemShut {NoStop}%
\bibitem [{\citenamefont {Choi}\ \emph {et~al.}(2016)\citenamefont {Choi},
  \citenamefont {Tran}, \citenamefont {Elbadawi}, \citenamefont {Lobo},
  \citenamefont {Wang}, \citenamefont {Juodkazis}, \citenamefont {Seniutinas},
  \citenamefont {Toth},\ and\ \citenamefont {Aharonovich}}]{Choi2016acs}%
  \BibitemOpen
  \bibfield  {author} {\bibinfo {author} {\bibfnamefont {S.}~\bibnamefont
  {Choi}}, \bibinfo {author} {\bibfnamefont {T.~T.}\ \bibnamefont {Tran}},
  \bibinfo {author} {\bibfnamefont {C.}~\bibnamefont {Elbadawi}}, \bibinfo
  {author} {\bibfnamefont {C.}~\bibnamefont {Lobo}}, \bibinfo {author}
  {\bibfnamefont {X.}~\bibnamefont {Wang}}, \bibinfo {author} {\bibfnamefont
  {S.}~\bibnamefont {Juodkazis}}, \bibinfo {author} {\bibfnamefont
  {G.}~\bibnamefont {Seniutinas}}, \bibinfo {author} {\bibfnamefont
  {M.}~\bibnamefont {Toth}},\ and\ \bibinfo {author} {\bibfnamefont
  {I.}~\bibnamefont {Aharonovich}},\ }\bibfield  {title} {\enquote {\bibinfo
  {title} {{Engineering and Localization of Quantum Emitters in Large Hexagonal
  Boron Nitride Layers}},}\ }\href {https://doi.org/10.1021/acsami.6b09875}
  {\bibfield  {journal} {\bibinfo  {journal} {ACS Appl. Mater. Interfaces}\
  }\textbf {\bibinfo {volume} {8}},\ \bibinfo {pages} {29642--29648} (\bibinfo
  {year} {2016})}\BibitemShut {NoStop}%
\bibitem [{\citenamefont {Fischer}\ \emph {et~al.}(2021)\citenamefont
  {Fischer}, \citenamefont {Caridad}, \citenamefont {Sajid}, \citenamefont
  {Ghaderzadeh}, \citenamefont {Ghorbani-Asl}, \citenamefont {Gammelgaard},
  \citenamefont {B{\o}ggild}, \citenamefont {Thygesen}, \citenamefont
  {Krasheninnikov}, \citenamefont {Xiao}, \citenamefont {Wubs},\ and\
  \citenamefont {Stenger}}]{Fischer2021}%
  \BibitemOpen
  \bibfield  {author} {\bibinfo {author} {\bibfnamefont {M.}~\bibnamefont
  {Fischer}}, \bibinfo {author} {\bibfnamefont {J.~M.}\ \bibnamefont
  {Caridad}}, \bibinfo {author} {\bibfnamefont {A.}~\bibnamefont {Sajid}},
  \bibinfo {author} {\bibfnamefont {S.}~\bibnamefont {Ghaderzadeh}}, \bibinfo
  {author} {\bibfnamefont {M.}~\bibnamefont {Ghorbani-Asl}}, \bibinfo {author}
  {\bibfnamefont {L.}~\bibnamefont {Gammelgaard}}, \bibinfo {author}
  {\bibfnamefont {P.}~\bibnamefont {B{\o}ggild}}, \bibinfo {author}
  {\bibfnamefont {K.~S.}\ \bibnamefont {Thygesen}}, \bibinfo {author}
  {\bibfnamefont {A.~V.}\ \bibnamefont {Krasheninnikov}}, \bibinfo {author}
  {\bibfnamefont {S.}~\bibnamefont {Xiao}}, \bibinfo {author} {\bibfnamefont
  {M.}~\bibnamefont {Wubs}},\ and\ \bibinfo {author} {\bibfnamefont
  {N.}~\bibnamefont {Stenger}},\ }\bibfield  {title} {\enquote {\bibinfo
  {title} {{Controlled generation of luminescent centers in hexagonal boron
  nitride by irradiation engineering}},}\ }\href
  {https://doi.org/10.1126/sciadv.abe7138} {\bibfield  {journal} {\bibinfo
  {journal} {Sci. Adv.}\ }\textbf {\bibinfo {volume} {7}},\ \bibinfo {pages}
  {eabe7138} (\bibinfo {year} {2021})}\BibitemShut {NoStop}%
\bibitem [{\citenamefont {Krasheninnikov}(2020)}]{Krasheninnikov2020}%
  \BibitemOpen
  \bibfield  {author} {\bibinfo {author} {\bibfnamefont {A.~V.}\ \bibnamefont
  {Krasheninnikov}},\ }\bibfield  {title} {\enquote {\bibinfo {title} {{Are
  two-dimensional materials radiation tolerant?}}}\ }\href
  {https://doi.org/10.1039/d0nh00465k} {\bibfield  {journal} {\bibinfo
  {journal} {Nanoscale Horiz.}\ }\textbf {\bibinfo {volume} {5}},\ \bibinfo
  {pages} {1447--1452} (\bibinfo {year} {2020})}\BibitemShut {NoStop}%
\bibitem [{\citenamefont {Wu}\ \emph {et~al.}(2018)\citenamefont {Wu},
  \citenamefont {Lin}, \citenamefont {Chen}, \citenamefont {Lin}, \citenamefont
  {Shih},\ and\ \citenamefont {Pao}}]{Wu-2018-1}%
  \BibitemOpen
  \bibfield  {author} {\bibinfo {author} {\bibfnamefont {C.~L.}\ \bibnamefont
  {Wu}}, \bibinfo {author} {\bibfnamefont {H.~T.}\ \bibnamefont {Lin}},
  \bibinfo {author} {\bibfnamefont {H.~A.}\ \bibnamefont {Chen}}, \bibinfo
  {author} {\bibfnamefont {S.~Y.}\ \bibnamefont {Lin}}, \bibinfo {author}
  {\bibfnamefont {M.~H.}\ \bibnamefont {Shih}},\ and\ \bibinfo {author}
  {\bibfnamefont {C.~W.}\ \bibnamefont {Pao}},\ }\bibfield  {title} {\enquote
  {\bibinfo {title} {{Defect formation and modulation during patterning
  supported graphene sheets using focused ion beams}},}\ }\href
  {https://doi.org/10.1016/j.mtcomm.2018.08.006} {\bibfield  {journal}
  {\bibinfo  {journal} {Mater. Today Commun.}\ }\textbf {\bibinfo {volume}
  {17}},\ \bibinfo {pages} {60--68} (\bibinfo {year} {2018})}\BibitemShut
  {NoStop}%
\bibitem [{\citenamefont {Chen}\ \emph
  {et~al.}(2018{\natexlab{b}})\citenamefont {Chen}, \citenamefont {Niu},
  \citenamefont {Wang}, \citenamefont {Tian}, \citenamefont {Kecskes},
  \citenamefont {Zhu},\ and\ \citenamefont {Wei}}]{Chen_2018}%
  \BibitemOpen
  \bibfield  {author} {\bibinfo {author} {\bibfnamefont {Z.}~\bibnamefont
  {Chen}}, \bibinfo {author} {\bibfnamefont {L.-L.}\ \bibnamefont {Niu}},
  \bibinfo {author} {\bibfnamefont {Z.}~\bibnamefont {Wang}}, \bibinfo {author}
  {\bibfnamefont {L.}~\bibnamefont {Tian}}, \bibinfo {author} {\bibfnamefont
  {L.}~\bibnamefont {Kecskes}}, \bibinfo {author} {\bibfnamefont
  {K.}~\bibnamefont {Zhu}},\ and\ \bibinfo {author} {\bibfnamefont
  {Q.}~\bibnamefont {Wei}},\ }\bibfield  {title} {\enquote {\bibinfo {title} {A
  comparative study on the in situ helium irradiation behavior of tungsten:
  coarse grain vs. nanocrystalline grain},}\ }\href
  {https://doi.org/10.1016/j.actamat.2018.01.015} {\bibfield  {journal}
  {\bibinfo  {journal} {Acta Mater.}\ }\textbf {\bibinfo {volume} {147}},\
  \bibinfo {pages} {100--112} (\bibinfo {year}
  {2018}{\natexlab{b}})}\BibitemShut {NoStop}%
\bibitem [{\citenamefont {Balooch}\ \emph {et~al.}(2022)\citenamefont
  {Balooch}, \citenamefont {Allen}, \citenamefont {Popovic},\ and\
  \citenamefont {Hosemann}}]{Balooch_2022}%
  \BibitemOpen
  \bibfield  {author} {\bibinfo {author} {\bibfnamefont {M.}~\bibnamefont
  {Balooch}}, \bibinfo {author} {\bibfnamefont {F.~I.}\ \bibnamefont {Allen}},
  \bibinfo {author} {\bibfnamefont {M.~P.}\ \bibnamefont {Popovic}},\ and\
  \bibinfo {author} {\bibfnamefont {P.}~\bibnamefont {Hosemann}},\ }\bibfield
  {title} {\enquote {\bibinfo {title} {Mechanical and structural
  transformations of tungsten implanted with helium ions},}\ }\href
  {https://doi.org/10.1016/j.jnucmat.2021.153436} {\bibfield  {journal}
  {\bibinfo  {journal} {J. Nucl. Mater.}\ }\textbf {\bibinfo {volume} {559}},\
  \bibinfo {pages} {153436} (\bibinfo {year} {2022})}\BibitemShut {NoStop}%
\bibitem [{\citenamefont {Fox}\ \emph {et~al.}(2012)\citenamefont {Fox},
  \citenamefont {Chen}, \citenamefont {Faulkner},\ and\ \citenamefont
  {Zhang}}]{Fox2012}%
  \BibitemOpen
  \bibfield  {author} {\bibinfo {author} {\bibfnamefont {D.}~\bibnamefont
  {Fox}}, \bibinfo {author} {\bibfnamefont {Y.}~\bibnamefont {Chen}}, \bibinfo
  {author} {\bibfnamefont {C.~C.}\ \bibnamefont {Faulkner}},\ and\ \bibinfo
  {author} {\bibfnamefont {H.}~\bibnamefont {Zhang}},\ }\bibfield  {title}
  {\enquote {\bibinfo {title} {Nano-structuring, surface and bulk modification
  with a focused helium ion beam},}\ }\href
  {https://doi.org/10.3762/bjnano.3.67} {\bibfield  {journal} {\bibinfo
  {journal} {Beilstein J. Nanotechnol.}\ }\textbf {\bibinfo {volume} {3}},\
  \bibinfo {pages} {579--585} (\bibinfo {year} {2012})}\BibitemShut {NoStop}%
\bibitem [{\citenamefont {Zhang}\ \emph {et~al.}(2015)\citenamefont {Zhang},
  \citenamefont {Heinig}, \citenamefont {Bazargan}, \citenamefont {Abd-Ellah},
  \citenamefont {Moghimi},\ and\ \citenamefont {Leung}}]{Zhang_2015}%
  \BibitemOpen
  \bibfield  {author} {\bibinfo {author} {\bibfnamefont {L.}~\bibnamefont
  {Zhang}}, \bibinfo {author} {\bibfnamefont {N.~F.}\ \bibnamefont {Heinig}},
  \bibinfo {author} {\bibfnamefont {S.}~\bibnamefont {Bazargan}}, \bibinfo
  {author} {\bibfnamefont {M.}~\bibnamefont {Abd-Ellah}}, \bibinfo {author}
  {\bibfnamefont {N.}~\bibnamefont {Moghimi}},\ and\ \bibinfo {author}
  {\bibfnamefont {K.~T.}\ \bibnamefont {Leung}},\ }\bibfield  {title}
  {{\selectlanguage {english}\enquote {\bibinfo {title} {Direct-write
  three-dimensional nanofabrication of nanopyramids and nanocones on {S}i by
  nanotumefaction using a helium ion microscope},}\ }}\href
  {https://doi.org/10.1088/0957-4484/26/25/255303} {\bibfield  {journal}
  {\bibinfo  {journal} {Nanotechnology}\ }\textbf {\bibinfo {volume} {26}},\
  \bibinfo {pages} {255303} (\bibinfo {year} {2015})}\BibitemShut {NoStop}%
\bibitem [{\citenamefont {Kim}\ \emph {et~al.}(2019{\natexlab{b}})\citenamefont
  {Kim}, \citenamefont {Hobbs}, \citenamefont {Agarwal}, \citenamefont {Yang},
  \citenamefont {Manfrinato}, \citenamefont {Short}, \citenamefont {Li},\ and\
  \citenamefont {Berggren}}]{Kim_2020b}%
  \BibitemOpen
  \bibfield  {author} {\bibinfo {author} {\bibfnamefont {C.-S.}\ \bibnamefont
  {Kim}}, \bibinfo {author} {\bibfnamefont {R.~G.}\ \bibnamefont {Hobbs}},
  \bibinfo {author} {\bibfnamefont {A.}~\bibnamefont {Agarwal}}, \bibinfo
  {author} {\bibfnamefont {Y.}~\bibnamefont {Yang}}, \bibinfo {author}
  {\bibfnamefont {V.~R.}\ \bibnamefont {Manfrinato}}, \bibinfo {author}
  {\bibfnamefont {M.~P.}\ \bibnamefont {Short}}, \bibinfo {author}
  {\bibfnamefont {J.}~\bibnamefont {Li}},\ and\ \bibinfo {author}
  {\bibfnamefont {K.~K.}\ \bibnamefont {Berggren}},\ }\bibfield  {title}
  {{\selectlanguage {english}\enquote {\bibinfo {title}
  {Focused-helium-ion-beam blow forming of nanostructures: radiation damage and
  nanofabrication},}\ }}\href {https://doi.org/10.1088/1361-6528/ab4a65}
  {\bibfield  {journal} {\bibinfo  {journal} {Nanotechnology}\ }\textbf
  {\bibinfo {volume} {31}},\ \bibinfo {pages} {045302} (\bibinfo {year}
  {2019}{\natexlab{b}})}\BibitemShut {NoStop}%
\bibitem [{\citenamefont {Wen}, \citenamefont {Mao},\ and\ \citenamefont
  {Hu}(2022)}]{Wen_2022}%
  \BibitemOpen
  \bibfield  {author} {\bibinfo {author} {\bibfnamefont {X.}~\bibnamefont
  {Wen}}, \bibinfo {author} {\bibfnamefont {R.}~\bibnamefont {Mao}},\ and\
  \bibinfo {author} {\bibfnamefont {H.}~\bibnamefont {Hu}},\ }\bibfield
  {title} {{\selectlanguage {english}\enquote {\bibinfo {title} {3‐d
  nanofabrication of silicon and nanostructure fine‐tuning via helium ion
  implantation},}\ }}\href {https://doi.org/10.1002/admi.202101643} {\bibfield
  {journal} {\bibinfo  {journal} {Adv. Mater. Interfaces}\ ,\ \bibinfo {pages}
  {2101643}} (\bibinfo {year} {2022})}\BibitemShut {NoStop}%
\bibitem [{\citenamefont {Groot-Berning}\ \emph {et~al.}(2021)\citenamefont
  {Groot-Berning}, \citenamefont {Jacob}, \citenamefont {Osterkamp},
  \citenamefont {Jelezko},\ and\ \citenamefont
  {Schmidt-Kaler}}]{GrootBerning_2021}%
  \BibitemOpen
  \bibfield  {author} {\bibinfo {author} {\bibfnamefont {K.}~\bibnamefont
  {Groot-Berning}}, \bibinfo {author} {\bibfnamefont {G.}~\bibnamefont
  {Jacob}}, \bibinfo {author} {\bibfnamefont {C.}~\bibnamefont {Osterkamp}},
  \bibinfo {author} {\bibfnamefont {F.}~\bibnamefont {Jelezko}},\ and\ \bibinfo
  {author} {\bibfnamefont {F.}~\bibnamefont {Schmidt-Kaler}},\ }\bibfield
  {title} {\enquote {\bibinfo {title} {Fabrication of $^{15}$nv$^-$ centers in
  diamond using a deterministic single ion implanter},}\ }\href
  {https://doi.org/10.1088/1367-2630/ac0753} {\bibfield  {journal} {\bibinfo
  {journal} {New J. Phys.}\ }\textbf {\bibinfo {volume} {23}},\ \bibinfo
  {pages} {063067} (\bibinfo {year} {2021})}\BibitemShut {NoStop}%
\bibitem [{\citenamefont {Groot-Berning}\ \emph {et~al.}(2019)\citenamefont
  {Groot-Berning}, \citenamefont {Kornher}, \citenamefont {Jacob},
  \citenamefont {Stopp}, \citenamefont {Dawkins}, \citenamefont {Kolesov},
  \citenamefont {Wrachtrup}, \citenamefont {Singer},\ and\ \citenamefont
  {Schmidt-Kaler}}]{GrootBerning_2019}%
  \BibitemOpen
  \bibfield  {author} {\bibinfo {author} {\bibfnamefont {K.}~\bibnamefont
  {Groot-Berning}}, \bibinfo {author} {\bibfnamefont {T.}~\bibnamefont
  {Kornher}}, \bibinfo {author} {\bibfnamefont {G.}~\bibnamefont {Jacob}},
  \bibinfo {author} {\bibfnamefont {F.}~\bibnamefont {Stopp}}, \bibinfo
  {author} {\bibfnamefont {S.~T.}\ \bibnamefont {Dawkins}}, \bibinfo {author}
  {\bibfnamefont {R.}~\bibnamefont {Kolesov}}, \bibinfo {author} {\bibfnamefont
  {J.}~\bibnamefont {Wrachtrup}}, \bibinfo {author} {\bibfnamefont
  {K.}~\bibnamefont {Singer}},\ and\ \bibinfo {author} {\bibfnamefont
  {F.}~\bibnamefont {Schmidt-Kaler}},\ }\bibfield  {title} {\enquote {\bibinfo
  {title} {Deterministic single-ion implantation of rare-earth ions for
  nanometer-resolution color-center generation},}\ }\href
  {https://doi.org/10.1103/physrevlett.123.106802} {\bibfield  {journal}
  {\bibinfo  {journal} {Phys. Rev. Lett.}\ }\textbf {\bibinfo {volume} {123}},\
  \bibinfo {pages} {106802} (\bibinfo {year} {2019})}\BibitemShut {NoStop}%
\bibitem [{\citenamefont {Collins}\ \emph {et~al.}(2023)\citenamefont
  {Collins}, \citenamefont {Jakob}, \citenamefont {Robson}, \citenamefont
  {Lim}, \citenamefont {R{\"{a}}cke}, \citenamefont {Johnson}, \citenamefont
  {Liu}, \citenamefont {Lu}, \citenamefont {Spemann}, \citenamefont
  {McCallum},\ and\ \citenamefont {Jamieson}}]{Collins2023}%
  \BibitemOpen
  \bibfield  {author} {\bibinfo {author} {\bibfnamefont {N.~F.~L.}\
  \bibnamefont {Collins}}, \bibinfo {author} {\bibfnamefont {A.~M.}\
  \bibnamefont {Jakob}}, \bibinfo {author} {\bibfnamefont {S.~G.}\ \bibnamefont
  {Robson}}, \bibinfo {author} {\bibfnamefont {S.~Q.}\ \bibnamefont {Lim}},
  \bibinfo {author} {\bibfnamefont {P.}~\bibnamefont {R{\"{a}}cke}}, \bibinfo
  {author} {\bibfnamefont {B.~C.}\ \bibnamefont {Johnson}}, \bibinfo {author}
  {\bibfnamefont {B.}~\bibnamefont {Liu}}, \bibinfo {author} {\bibfnamefont
  {Y.}~\bibnamefont {Lu}}, \bibinfo {author} {\bibfnamefont {D.}~\bibnamefont
  {Spemann}}, \bibinfo {author} {\bibfnamefont {J.~C.}\ \bibnamefont
  {McCallum}},\ and\ \bibinfo {author} {\bibfnamefont {D.~N.}\ \bibnamefont
  {Jamieson}},\ }\href {https://doi.org/10.48550/ARXIV.2306.07496} {\enquote
  {\bibinfo {title} {Graphene-enhanced single ion detectors for deterministic
  near-surface dopant implantation in diamond},}\ } (\bibinfo {year} {2023}),\
  \bibinfo {note} {preprint},\ \Eprint {https://arxiv.org/abs/2306.07496}
  {arXiv:2306.07496 [cond-mat.mes-hall]} \BibitemShut {NoStop}%
\bibitem [{\citenamefont {Rangelow}, \citenamefont {Meijer},\ and\
  \citenamefont {Schenkel}(2006)}]{Rangelow_2006}%
  \BibitemOpen
  \bibfield  {author} {\bibinfo {author} {\bibfnamefont {I.~W.}\ \bibnamefont
  {Rangelow}}, \bibinfo {author} {\bibfnamefont {J.}~\bibnamefont {Meijer}},\
  and\ \bibinfo {author} {\bibfnamefont {T.}~\bibnamefont {Schenkel}},\ }\href
  {https://patents.google.com/patent/DE10347969A1} {{\selectlanguage
  {english}\enquote {\bibinfo {title} {Device for accurately positioning
  individual particles on substrate surface, e.g. for quantum computer, has
  aperture at tip of end portion of cantilever beam},}\ }}\bibinfo {type}
  {patent}\ \bibinfo {number} {{DE}10347969A1}\ (\bibinfo  {institution}
  {Universität Kassel},\ \bibinfo {year} {2006})\BibitemShut {NoStop}%
\bibitem [{\citenamefont {Meijer}\ \emph {et~al.}(2006)\citenamefont {Meijer},
  \citenamefont {Vogel}, \citenamefont {Burchard}, \citenamefont {Rangelow},
  \citenamefont {Bischoff}, \citenamefont {Wrachtrup}, \citenamefont {Domhan},
  \citenamefont {Jelezko}, \citenamefont {Schnitzler}, \citenamefont {Schulz},
  \citenamefont {Singer},\ and\ \citenamefont {Schmidt-Kaler}}]{Meijer_2006}%
  \BibitemOpen
  \bibfield  {author} {\bibinfo {author} {\bibfnamefont {J.}~\bibnamefont
  {Meijer}}, \bibinfo {author} {\bibfnamefont {T.}~\bibnamefont {Vogel}},
  \bibinfo {author} {\bibfnamefont {B.}~\bibnamefont {Burchard}}, \bibinfo
  {author} {\bibfnamefont {I.~W.}\ \bibnamefont {Rangelow}}, \bibinfo {author}
  {\bibfnamefont {L.}~\bibnamefont {Bischoff}}, \bibinfo {author}
  {\bibfnamefont {J.}~\bibnamefont {Wrachtrup}}, \bibinfo {author}
  {\bibfnamefont {M.}~\bibnamefont {Domhan}}, \bibinfo {author} {\bibfnamefont
  {F.}~\bibnamefont {Jelezko}}, \bibinfo {author} {\bibfnamefont
  {W.}~\bibnamefont {Schnitzler}}, \bibinfo {author} {\bibfnamefont {S.~A.}\
  \bibnamefont {Schulz}}, \bibinfo {author} {\bibfnamefont {K.}~\bibnamefont
  {Singer}},\ and\ \bibinfo {author} {\bibfnamefont {F.}~\bibnamefont
  {Schmidt-Kaler}},\ }\bibfield  {title} {\enquote {\bibinfo {title} {Concept
  of deterministic single ion doping with sub-nm spatial resolution},}\ }\href
  {https://doi.org/10.1007/s00339-006-3497-0} {\bibfield  {journal} {\bibinfo
  {journal} {Appl. Phys. A}\ }\textbf {\bibinfo {volume} {83}},\ \bibinfo
  {pages} {321--327} (\bibinfo {year} {2006})}\BibitemShut {NoStop}%
\bibitem [{\citenamefont {Meijer}\ \emph {et~al.}(2008)\citenamefont {Meijer},
  \citenamefont {Pezzagna}, \citenamefont {Vogel}, \citenamefont {Burchard},
  \citenamefont {Bukow}, \citenamefont {Rangelow}, \citenamefont {Sarov},
  \citenamefont {Wiggers}, \citenamefont {Pl{\"{u}}mel}, \citenamefont
  {Jelezko}, \citenamefont {Wrachtrup}, \citenamefont {Schmidt-Kaler},
  \citenamefont {Schnitzler},\ and\ \citenamefont {Singer}}]{Meijer_2008}%
  \BibitemOpen
  \bibfield  {author} {\bibinfo {author} {\bibfnamefont {J.}~\bibnamefont
  {Meijer}}, \bibinfo {author} {\bibfnamefont {S.}~\bibnamefont {Pezzagna}},
  \bibinfo {author} {\bibfnamefont {T.}~\bibnamefont {Vogel}}, \bibinfo
  {author} {\bibfnamefont {B.}~\bibnamefont {Burchard}}, \bibinfo {author}
  {\bibfnamefont {H.~H.}\ \bibnamefont {Bukow}}, \bibinfo {author}
  {\bibfnamefont {I.~W.}\ \bibnamefont {Rangelow}}, \bibinfo {author}
  {\bibfnamefont {Y.}~\bibnamefont {Sarov}}, \bibinfo {author} {\bibfnamefont
  {H.}~\bibnamefont {Wiggers}}, \bibinfo {author} {\bibfnamefont
  {I.}~\bibnamefont {Pl{\"{u}}mel}}, \bibinfo {author} {\bibfnamefont
  {F.}~\bibnamefont {Jelezko}}, \bibinfo {author} {\bibfnamefont
  {J.}~\bibnamefont {Wrachtrup}}, \bibinfo {author} {\bibfnamefont
  {F.}~\bibnamefont {Schmidt-Kaler}}, \bibinfo {author} {\bibfnamefont
  {W.}~\bibnamefont {Schnitzler}},\ and\ \bibinfo {author} {\bibfnamefont
  {K.}~\bibnamefont {Singer}},\ }\bibfield  {title} {\enquote {\bibinfo {title}
  {Towards the implanting of ions and positioning of nanoparticles with nm
  spatial resolution},}\ }\href {https://doi.org/10.1007/s00339-008-4515-1}
  {\bibfield  {journal} {\bibinfo  {journal} {Appl. Phys. A}\ }\textbf
  {\bibinfo {volume} {91}},\ \bibinfo {pages} {567--571} (\bibinfo {year}
  {2008})}\BibitemShut {NoStop}%
\bibitem [{\citenamefont {Murdin}\ \emph {et~al.}(2021)\citenamefont {Murdin},
  \citenamefont {Cassidy}, \citenamefont {Cox}, \citenamefont {Webb},\ and\
  \citenamefont {Curry}}]{Murdin_2021}%
  \BibitemOpen
  \bibfield  {author} {\bibinfo {author} {\bibfnamefont {B.~N.}\ \bibnamefont
  {Murdin}}, \bibinfo {author} {\bibfnamefont {N.}~\bibnamefont {Cassidy}},
  \bibinfo {author} {\bibfnamefont {D.}~\bibnamefont {Cox}}, \bibinfo {author}
  {\bibfnamefont {R.}~\bibnamefont {Webb}},\ and\ \bibinfo {author}
  {\bibfnamefont {R.~J.}\ \bibnamefont {Curry}},\ }\bibfield  {title} {\enquote
  {\bibinfo {title} {Error rates in deterministic ion implantation for qubit
  arrays},}\ }\href {https://doi.org/10.1002/pssb.202000615} {\bibfield
  {journal} {\bibinfo  {journal} {Phys. Status Solidi B}\ }\textbf {\bibinfo
  {volume} {258}},\ \bibinfo {pages} {2000615} (\bibinfo {year}
  {2021})}\BibitemShut {NoStop}%
\bibitem [{\citenamefont {Budnik}\ \emph {et~al.}(2022)\citenamefont {Budnik},
  \citenamefont {Scott}, \citenamefont {Jiao}, \citenamefont {Maazouz},
  \citenamefont {Gledhill}, \citenamefont {Fu}, \citenamefont {Tan},\ and\
  \citenamefont {Toth}}]{Budnik_2022}%
  \BibitemOpen
  \bibfield  {author} {\bibinfo {author} {\bibfnamefont {G.}~\bibnamefont
  {Budnik}}, \bibinfo {author} {\bibfnamefont {J.~A.}\ \bibnamefont {Scott}},
  \bibinfo {author} {\bibfnamefont {C.}~\bibnamefont {Jiao}}, \bibinfo {author}
  {\bibfnamefont {M.}~\bibnamefont {Maazouz}}, \bibinfo {author} {\bibfnamefont
  {G.}~\bibnamefont {Gledhill}}, \bibinfo {author} {\bibfnamefont
  {L.}~\bibnamefont {Fu}}, \bibinfo {author} {\bibfnamefont {H.~H.}\
  \bibnamefont {Tan}},\ and\ \bibinfo {author} {\bibfnamefont {M.}~\bibnamefont
  {Toth}},\ }\bibfield  {title} {\enquote {\bibinfo {title} {Nanoscale 3{D}
  tomography by in-flight fluorescence spectroscopy of atoms sputtered by a
  focused ion beam},}\ }\href {https://doi.org/10.1021/acs.nanolett.2c03101}
  {\bibfield  {journal} {\bibinfo  {journal} {Nano Lett.}\ }\textbf {\bibinfo
  {volume} {22}},\ \bibinfo {pages} {8287--8293} (\bibinfo {year}
  {2022})}\BibitemShut {NoStop}%
\bibitem [{\citenamefont {Chandrasekaran}\ \emph {et~al.}(2023)\citenamefont
  {Chandrasekaran}, \citenamefont {Titze}, \citenamefont {Flores},
  \citenamefont {Campbell}, \citenamefont {Henshaw}, \citenamefont {Jones},
  \citenamefont {Bielejec},\ and\ \citenamefont {Htoon}}]{Chandrasekaran_2023}%
  \BibitemOpen
  \bibfield  {author} {\bibinfo {author} {\bibfnamefont {V.}~\bibnamefont
  {Chandrasekaran}}, \bibinfo {author} {\bibfnamefont {M.}~\bibnamefont
  {Titze}}, \bibinfo {author} {\bibfnamefont {A.~R.}\ \bibnamefont {Flores}},
  \bibinfo {author} {\bibfnamefont {D.}~\bibnamefont {Campbell}}, \bibinfo
  {author} {\bibfnamefont {J.}~\bibnamefont {Henshaw}}, \bibinfo {author}
  {\bibfnamefont {A.~C.}\ \bibnamefont {Jones}}, \bibinfo {author}
  {\bibfnamefont {E.~S.}\ \bibnamefont {Bielejec}},\ and\ \bibinfo {author}
  {\bibfnamefont {H.}~\bibnamefont {Htoon}},\ }\bibfield  {title} {\enquote
  {\bibinfo {title} {High-yield deterministic focused ion beam implantation of
  quantum defects enabled by in situ photoluminescence feedback},}\ }\href
  {https://doi.org/10.1002/advs.202300190} {\bibfield  {journal} {\bibinfo
  {journal} {Adv. Sci.}\ }\textbf {\bibinfo {volume} {n/a}},\ \bibinfo {pages}
  {2300190} (\bibinfo {year} {2023})}\BibitemShut {NoStop}%
\bibitem [{\citenamefont {Knoll}\ and\ \citenamefont
  {Ruska}(1932)}]{Knoll_1932}%
  \BibitemOpen
  \bibfield  {author} {\bibinfo {author} {\bibfnamefont {M.}~\bibnamefont
  {Knoll}}\ and\ \bibinfo {author} {\bibfnamefont {E.}~\bibnamefont {Ruska}},\
  }\bibfield  {title} {\enquote {\bibinfo {title} {{Das
  Elektronenmikroskop}},}\ }\href {https://doi.org/10.1007/BF01342199}
  {\bibfield  {journal} {\bibinfo  {journal} {Zeitschrift f{\"{u}}r Phys.}\
  }\textbf {\bibinfo {volume} {78}},\ \bibinfo {pages} {318--339} (\bibinfo
  {year} {1932})}\BibitemShut {NoStop}%
\bibitem [{\citenamefont {Krause}(1937)}]{Krause_1937}%
  \BibitemOpen
  \bibfield  {author} {\bibinfo {author} {\bibfnamefont {F.}~\bibnamefont
  {Krause}},\ }\bibfield  {title} {\enquote {\bibinfo {title} {Das magnetische
  elektronenmikroskop und seine anwendung in der biologie},}\ }\href
  {https://doi.org/10.1007/bf01734636} {\bibfield  {journal} {\bibinfo
  {journal} {Die Naturwissenschaften}\ }\textbf {\bibinfo {volume} {25}},\
  \bibinfo {pages} {817--825} (\bibinfo {year} {1937})}\BibitemShut {NoStop}%
\bibitem [{\citenamefont {Br{\"{u}}che}\ and\ \citenamefont
  {Haagen}(1939)}]{Brueche_1939}%
  \BibitemOpen
  \bibfield  {author} {\bibinfo {author} {\bibfnamefont {E.}~\bibnamefont
  {Br{\"{u}}che}}\ and\ \bibinfo {author} {\bibfnamefont {E.}~\bibnamefont
  {Haagen}},\ }\bibfield  {title} {\enquote {\bibinfo {title} {Ein neues,
  einfaches {\"{u}}bermikroskop und seine {A}nwendung in der
  {B}akteriologie},}\ }\href {https://doi.org/10.1007/bf01806606} {\bibfield
  {journal} {\bibinfo  {journal} {Die Naturwissenschaften}\ }\textbf {\bibinfo
  {volume} {27}},\ \bibinfo {pages} {809--811} (\bibinfo {year}
  {1939})}\BibitemShut {NoStop}%
\bibitem [{\citenamefont {Ruska}, \citenamefont {v.~Borries},\ and\
  \citenamefont {Ruska}(1939)}]{Ruska_1939}%
  \BibitemOpen
  \bibfield  {author} {\bibinfo {author} {\bibfnamefont {H.}~\bibnamefont
  {Ruska}}, \bibinfo {author} {\bibfnamefont {B.}~\bibnamefont {v.~Borries}},\
  and\ \bibinfo {author} {\bibfnamefont {E.}~\bibnamefont {Ruska}},\ }\bibfield
   {title} {\enquote {\bibinfo {title} {{Die Bedeutung der Übermikroskopie
  für die Virusforschung}},}\ }\href {https://doi.org/10.1007/bf01243399}
  {\bibfield  {journal} {\bibinfo  {journal} {Archiv für die gesamte
  Virusforschung}\ }\textbf {\bibinfo {volume} {1}},\ \bibinfo {pages}
  {155--169} (\bibinfo {year} {1939})}\BibitemShut {NoStop}%
\bibitem [{\citenamefont {Ballerini}\ \emph {et~al.}(2001)\citenamefont
  {Ballerini}, \citenamefont {Milani}, \citenamefont {Batani},\ and\
  \citenamefont {Squadrini}}]{Ballerini_2001}%
  \BibitemOpen
  \bibfield  {author} {\bibinfo {author} {\bibfnamefont {M.}~\bibnamefont
  {Ballerini}}, \bibinfo {author} {\bibfnamefont {M.}~\bibnamefont {Milani}},
  \bibinfo {author} {\bibfnamefont {D.}~\bibnamefont {Batani}},\ and\ \bibinfo
  {author} {\bibfnamefont {F.}~\bibnamefont {Squadrini}},\ }\bibfield  {title}
  {\enquote {\bibinfo {title} {Focused ion beam techniques for the analysis of
  biological samples: a revolution in ultramicroscopy?}}\ }in\ \href
  {https://doi.org/10.1117/12.424523} {\emph {\bibinfo {booktitle} {{SPIE}
  Proceedings}}},\ \bibinfo {editor} {edited by\ \bibinfo {editor}
  {\bibfnamefont {J.-A.}\ \bibnamefont {Conchello}}, \bibinfo {editor}
  {\bibfnamefont {C.~J.}\ \bibnamefont {Cogswell}},\ and\ \bibinfo {editor}
  {\bibfnamefont {T.}~\bibnamefont {Wilson}}}\ (\bibinfo  {publisher}
  {{SPIE}},\ \bibinfo {year} {2001})\BibitemShut {NoStop}%
\bibitem [{\citenamefont {Gierak}, \citenamefont {Septier},\ and\ \citenamefont
  {Vieu}(1999)}]{Gierak_1999}%
  \BibitemOpen
  \bibfield  {author} {\bibinfo {author} {\bibfnamefont {J.}~\bibnamefont
  {Gierak}}, \bibinfo {author} {\bibfnamefont {A.}~\bibnamefont {Septier}},\
  and\ \bibinfo {author} {\bibfnamefont {C.}~\bibnamefont {Vieu}},\ }\bibfield
  {title} {\enquote {\bibinfo {title} {Design and realization of a very
  high-resolution {FIB} nanofabrication instrument},}\ }\href
  {https://doi.org/10.1016/S0168-9002(98)01541-1} {\bibfield  {journal}
  {\bibinfo  {journal} {Nucl. Instrum. Meth. Phys. Res. A}\ }\textbf {\bibinfo
  {volume} {427}},\ \bibinfo {pages} {91--98} (\bibinfo {year}
  {1999})}\BibitemShut {NoStop}%
\bibitem [{\citenamefont {Hekking}\ \emph {et~al.}(2009)\citenamefont
  {Hekking}, \citenamefont {Lebbink}, \citenamefont {Winter}, \citenamefont
  {Schneijdenberg}, \citenamefont {Brand}, \citenamefont {Humbel},
  \citenamefont {Verkleij},\ and\ \citenamefont {Post}}]{HEKKING_2009}%
  \BibitemOpen
  \bibfield  {author} {\bibinfo {author} {\bibfnamefont {L.~H.~P.}\
  \bibnamefont {Hekking}}, \bibinfo {author} {\bibfnamefont {M.~N.}\
  \bibnamefont {Lebbink}}, \bibinfo {author} {\bibfnamefont {D.~A. M.~D.}\
  \bibnamefont {Winter}}, \bibinfo {author} {\bibfnamefont {C.~T. W.~M.}\
  \bibnamefont {Schneijdenberg}}, \bibinfo {author} {\bibfnamefont {C.~M.}\
  \bibnamefont {Brand}}, \bibinfo {author} {\bibfnamefont {B.~M.}\ \bibnamefont
  {Humbel}}, \bibinfo {author} {\bibfnamefont {A.~J.}\ \bibnamefont
  {Verkleij}},\ and\ \bibinfo {author} {\bibfnamefont {J.~A.}\ \bibnamefont
  {Post}},\ }\bibfield  {title} {\enquote {\bibinfo {title} {Focused ion
  beam-scanning electron microscope: exploring large volumes of atherosclerotic
  tissue},}\ }\href {https://doi.org/10.1111/j.1365-2818.2009.03274.x}
  {\bibfield  {journal} {\bibinfo  {journal} {J. Microsc.}\ }\textbf {\bibinfo
  {volume} {235}},\ \bibinfo {pages} {336--347} (\bibinfo {year}
  {2009})}\BibitemShut {NoStop}%
\bibitem [{\citenamefont {Inkson}, \citenamefont {Mulvihill},\ and\
  \citenamefont {M{\"{o}}bus}(2001)}]{Inkson_2001}%
  \BibitemOpen
  \bibfield  {author} {\bibinfo {author} {\bibfnamefont {B.~J.}\ \bibnamefont
  {Inkson}}, \bibinfo {author} {\bibfnamefont {M.}~\bibnamefont {Mulvihill}},\
  and\ \bibinfo {author} {\bibfnamefont {G.}~\bibnamefont {M{\"{o}}bus}},\
  }\bibfield  {title} {\enquote {\bibinfo {title} {3{D} determination of grain
  shape in a {FeAl}-based nanocomposite by 3{D} {FIB} tomography},}\ }\href
  {https://doi.org/10.1016/s1359-6462(01)01090-9} {\bibfield  {journal}
  {\bibinfo  {journal} {Scripta Mater.}\ }\textbf {\bibinfo {volume} {45}},\
  \bibinfo {pages} {753--758} (\bibinfo {year} {2001})}\BibitemShut {NoStop}%
\bibitem [{\citenamefont {Lasagni}\ \emph {et~al.}(2008)\citenamefont
  {Lasagni}, \citenamefont {Lasagni}, \citenamefont {Engstler}, \citenamefont
  {Degischer},\ and\ \citenamefont {M{\"{u}}cklich}}]{Lasagni_2008}%
  \BibitemOpen
  \bibfield  {author} {\bibinfo {author} {\bibfnamefont {F.}~\bibnamefont
  {Lasagni}}, \bibinfo {author} {\bibfnamefont {A.}~\bibnamefont {Lasagni}},
  \bibinfo {author} {\bibfnamefont {M.}~\bibnamefont {Engstler}}, \bibinfo
  {author} {\bibfnamefont {H.}~\bibnamefont {Degischer}},\ and\ \bibinfo
  {author} {\bibfnamefont {F.}~\bibnamefont {M{\"{u}}cklich}},\ }\bibfield
  {title} {\enquote {\bibinfo {title} {Nano-characterization of cast structures
  by {FIB}-tomography},}\ }\href {https://doi.org/10.1002/adem.200700249}
  {\bibfield  {journal} {\bibinfo  {journal} {Adv. Eng. Mater.}\ }\textbf
  {\bibinfo {volume} {10}},\ \bibinfo {pages} {62--66} (\bibinfo {year}
  {2008})}\BibitemShut {NoStop}%
\bibitem [{\citenamefont {Ostadi}\ \emph {et~al.}(2010)\citenamefont {Ostadi},
  \citenamefont {Rama}, \citenamefont {Liu}, \citenamefont {Chen},
  \citenamefont {Zhang},\ and\ \citenamefont {Jiang}}]{Ostadi_2010}%
  \BibitemOpen
  \bibfield  {author} {\bibinfo {author} {\bibfnamefont {H.}~\bibnamefont
  {Ostadi}}, \bibinfo {author} {\bibfnamefont {P.}~\bibnamefont {Rama}},
  \bibinfo {author} {\bibfnamefont {Y.}~\bibnamefont {Liu}}, \bibinfo {author}
  {\bibfnamefont {R.}~\bibnamefont {Chen}}, \bibinfo {author} {\bibfnamefont
  {X.~X.}\ \bibnamefont {Zhang}},\ and\ \bibinfo {author} {\bibfnamefont
  {K.}~\bibnamefont {Jiang}},\ }\bibfield  {title} {\enquote {\bibinfo {title}
  {3{D} reconstruction of a gas diffusion layer and a microporous layer},}\
  }\href {https://doi.org/10.1016/j.memsci.2010.01.031} {\bibfield  {journal}
  {\bibinfo  {journal} {J. Membrane Sci.}\ }\textbf {\bibinfo {volume} {351}},\
  \bibinfo {pages} {69--74} (\bibinfo {year} {2010})}\BibitemShut {NoStop}%
\bibitem [{\citenamefont {Winter}\ \emph {et~al.}(2009)\citenamefont {Winter},
  \citenamefont {Schneijdenberg}, \citenamefont {Lebbink}, \citenamefont
  {Lich}, \citenamefont {Verkleij}, \citenamefont {Drury},\ and\ \citenamefont
  {Humbel}}]{WINTER_2009}%
  \BibitemOpen
  \bibfield  {author} {\bibinfo {author} {\bibfnamefont {D.~A. M.~D.}\
  \bibnamefont {Winter}}, \bibinfo {author} {\bibfnamefont {C.~T. W.~M.}\
  \bibnamefont {Schneijdenberg}}, \bibinfo {author} {\bibfnamefont {M.~M.}\
  \bibnamefont {Lebbink}}, \bibinfo {author} {\bibfnamefont {B.}~\bibnamefont
  {Lich}}, \bibinfo {author} {\bibfnamefont {A.~J.}\ \bibnamefont {Verkleij}},
  \bibinfo {author} {\bibfnamefont {M.~R.}\ \bibnamefont {Drury}},\ and\
  \bibinfo {author} {\bibfnamefont {B.~M.}\ \bibnamefont {Humbel}},\ }\bibfield
   {title} {\enquote {\bibinfo {title} {Tomography of insulating biological and
  geological materials using focused ion beam ({FIB}) sectioning and low-{kV}
  {BSE} imaging},}\ }\href {https://doi.org/10.1111/j.1365-2818.2009.03139.x}
  {\bibfield  {journal} {\bibinfo  {journal} {J. Microsc.}\ }\textbf {\bibinfo
  {volume} {233}},\ \bibinfo {pages} {372--383} (\bibinfo {year}
  {2009})}\BibitemShut {NoStop}%
\bibitem [{\citenamefont {Haridy}\ \emph {et~al.}(2021)\citenamefont {Haridy},
  \citenamefont {Osenberg}, \citenamefont {Hilger}, \citenamefont {Manke},
  \citenamefont {Davesne},\ and\ \citenamefont {Witzmann}}]{M_Yara_2021}%
  \BibitemOpen
  \bibfield  {author} {\bibinfo {author} {\bibfnamefont {Y.}~\bibnamefont
  {Haridy}}, \bibinfo {author} {\bibfnamefont {M.}~\bibnamefont {Osenberg}},
  \bibinfo {author} {\bibfnamefont {A.}~\bibnamefont {Hilger}}, \bibinfo
  {author} {\bibfnamefont {I.}~\bibnamefont {Manke}}, \bibinfo {author}
  {\bibfnamefont {D.}~\bibnamefont {Davesne}},\ and\ \bibinfo {author}
  {\bibfnamefont {F.}~\bibnamefont {Witzmann}},\ }\bibfield  {title} {\enquote
  {\bibinfo {title} {Bone metabolism and evolutionary origin of osteocytes:
  novel application of {FIB-SEM} tomography},}\ }\href
  {https://doi.org/10.1126/sciadv.abb9113} {\bibfield  {journal} {\bibinfo
  {journal} {Sci. Adv.}\ }\textbf {\bibinfo {volume} {7}},\ \bibinfo {pages}
  {eabb9113} (\bibinfo {year} {2021})}\BibitemShut {NoStop}%
\bibitem [{\citenamefont {Pham}, \citenamefont {Xu},\ and\ \citenamefont
  {Prince}(2000)}]{Pham_2000}%
  \BibitemOpen
  \bibfield  {author} {\bibinfo {author} {\bibfnamefont {D.~L.}\ \bibnamefont
  {Pham}}, \bibinfo {author} {\bibfnamefont {C.}~\bibnamefont {Xu}},\ and\
  \bibinfo {author} {\bibfnamefont {J.~L.}\ \bibnamefont {Prince}},\ }\bibfield
   {title} {\enquote {\bibinfo {title} {Current methods in medical image
  segmentation},}\ }\href {https://doi.org/10.1146/annurev.bioeng.2.1.315}
  {\bibfield  {journal} {\bibinfo  {journal} {Annu. Rev. Biomed. Eng.}\
  }\textbf {\bibinfo {volume} {2}},\ \bibinfo {pages} {315--337} (\bibinfo
  {year} {2000})}\BibitemShut {NoStop}%
\bibitem [{\citenamefont {Heymann}\ \emph {et~al.}(2006)\citenamefont
  {Heymann}, \citenamefont {Hayles}, \citenamefont {Gestmann}, \citenamefont
  {Giannuzzi}, \citenamefont {Lich},\ and\ \citenamefont
  {Subramaniam}}]{Heymann_2006}%
  \BibitemOpen
  \bibfield  {author} {\bibinfo {author} {\bibfnamefont {J.~A.~W.}\
  \bibnamefont {Heymann}}, \bibinfo {author} {\bibfnamefont {M.}~\bibnamefont
  {Hayles}}, \bibinfo {author} {\bibfnamefont {I.}~\bibnamefont {Gestmann}},
  \bibinfo {author} {\bibfnamefont {L.~A.}\ \bibnamefont {Giannuzzi}}, \bibinfo
  {author} {\bibfnamefont {B.}~\bibnamefont {Lich}},\ and\ \bibinfo {author}
  {\bibfnamefont {S.}~\bibnamefont {Subramaniam}},\ }\bibfield  {title}
  {\enquote {\bibinfo {title} {Site-specific 3{D} imaging of cells and tissues
  with a dual beam microscope},}\ }\href
  {https://doi.org/10.1016/j.jsb.2006.03.006} {\bibfield  {journal} {\bibinfo
  {journal} {J. Struct. Biol.}\ }\textbf {\bibinfo {volume} {155}},\ \bibinfo
  {pages} {63--73} (\bibinfo {year} {2006})}\BibitemShut {NoStop}%
\bibitem [{\citenamefont {Hayles}\ \emph {et~al.}(2007)\citenamefont {Hayles},
  \citenamefont {Stokes}, \citenamefont {Phifer},\ and\ \citenamefont
  {Findlay}}]{Hayles_2007}%
  \BibitemOpen
  \bibfield  {author} {\bibinfo {author} {\bibfnamefont {M.~F.}\ \bibnamefont
  {Hayles}}, \bibinfo {author} {\bibfnamefont {D.~J.}\ \bibnamefont {Stokes}},
  \bibinfo {author} {\bibfnamefont {D.}~\bibnamefont {Phifer}},\ and\ \bibinfo
  {author} {\bibfnamefont {K.~C.}\ \bibnamefont {Findlay}},\ }\bibfield
  {title} {\enquote {\bibinfo {title} {A technique for improved focused ion
  beam milling of cryo-prepared life science specimens},}\ }\href
  {https://doi.org/10.1111/j.1365-2818.2007.01775.x} {\bibfield  {journal}
  {\bibinfo  {journal} {J. Microsc.}\ }\textbf {\bibinfo {volume} {226}},\
  \bibinfo {pages} {263--269} (\bibinfo {year} {2007})}\BibitemShut {NoStop}%
\bibitem [{\citenamefont {Rigort}\ and\ \citenamefont
  {Plitzko}(2015)}]{Rigort_2015}%
  \BibitemOpen
  \bibfield  {author} {\bibinfo {author} {\bibfnamefont {A.}~\bibnamefont
  {Rigort}}\ and\ \bibinfo {author} {\bibfnamefont {J.~M.}\ \bibnamefont
  {Plitzko}},\ }\bibfield  {title} {{\selectlanguage {en}\enquote {\bibinfo
  {title} {Cryo-focused-ion-beam applications in structural biology},}\ }}\href
  {https://doi.org/10.1016/j.abb.2015.02.009} {\bibfield  {journal} {\bibinfo
  {journal} {Arch. Biochem. Biophys.}\ }\textbf {\bibinfo {volume} {581}},\
  \bibinfo {pages} {122--130} (\bibinfo {year} {2015})}\BibitemShut {NoStop}%
\bibitem [{\citenamefont {Schaffer}\ \emph {et~al.}(2019)\citenamefont
  {Schaffer}, \citenamefont {Pfeffer}, \citenamefont {Mahamid}, \citenamefont
  {Kleindiek}, \citenamefont {Laugks}, \citenamefont {Albert}, \citenamefont
  {Engel}, \citenamefont {Rummel}, \citenamefont {Smith}, \citenamefont
  {Baumeister},\ and\ \citenamefont {Plitzko}}]{Schaffer_2019}%
  \BibitemOpen
  \bibfield  {author} {\bibinfo {author} {\bibfnamefont {M.}~\bibnamefont
  {Schaffer}}, \bibinfo {author} {\bibfnamefont {S.}~\bibnamefont {Pfeffer}},
  \bibinfo {author} {\bibfnamefont {J.}~\bibnamefont {Mahamid}}, \bibinfo
  {author} {\bibfnamefont {S.}~\bibnamefont {Kleindiek}}, \bibinfo {author}
  {\bibfnamefont {T.}~\bibnamefont {Laugks}}, \bibinfo {author} {\bibfnamefont
  {S.}~\bibnamefont {Albert}}, \bibinfo {author} {\bibfnamefont {B.~D.}\
  \bibnamefont {Engel}}, \bibinfo {author} {\bibfnamefont {A.}~\bibnamefont
  {Rummel}}, \bibinfo {author} {\bibfnamefont {A.~J.}\ \bibnamefont {Smith}},
  \bibinfo {author} {\bibfnamefont {W.}~\bibnamefont {Baumeister}},\ and\
  \bibinfo {author} {\bibfnamefont {J.~M.}\ \bibnamefont {Plitzko}},\
  }\bibfield  {title} {{\selectlanguage {en}\enquote {\bibinfo {title} {A
  {cryo-FIB} lift-out technique enables molecular-resolution {cryo-ET} within
  native caenorhabditis elegans tissue},}\ }}\href
  {https://doi.org/10.1038/s41592-019-0497-5} {\bibfield  {journal} {\bibinfo
  {journal} {Nat. Methods}\ }\textbf {\bibinfo {volume} {16}},\ \bibinfo
  {pages} {757--762} (\bibinfo {year} {2019})}\BibitemShut {NoStop}%
\bibitem [{\citenamefont {Spehner}\ \emph {et~al.}(2020)\citenamefont
  {Spehner}, \citenamefont {Steyer}, \citenamefont {Bertinetti}, \citenamefont
  {Orlov}, \citenamefont {Benoit}, \citenamefont {Pernet-Gallay}, \citenamefont
  {Schertel},\ and\ \citenamefont {Schultz}}]{Spehner_2020}%
  \BibitemOpen
  \bibfield  {author} {\bibinfo {author} {\bibfnamefont {D.}~\bibnamefont
  {Spehner}}, \bibinfo {author} {\bibfnamefont {A.~M.}\ \bibnamefont {Steyer}},
  \bibinfo {author} {\bibfnamefont {L.}~\bibnamefont {Bertinetti}}, \bibinfo
  {author} {\bibfnamefont {I.}~\bibnamefont {Orlov}}, \bibinfo {author}
  {\bibfnamefont {L.}~\bibnamefont {Benoit}}, \bibinfo {author} {\bibfnamefont
  {K.}~\bibnamefont {Pernet-Gallay}}, \bibinfo {author} {\bibfnamefont
  {A.}~\bibnamefont {Schertel}},\ and\ \bibinfo {author} {\bibfnamefont
  {P.}~\bibnamefont {Schultz}},\ }\bibfield  {title} {\enquote {\bibinfo
  {title} {Cryo-{FIB-SEM} as a promising tool for localizing proteins in
  3{D}},}\ }\href {https://doi.org/10.1016/j.jsb.2020.107528} {\bibfield
  {journal} {\bibinfo  {journal} {J. Struct. Biol.}\ }\textbf {\bibinfo
  {volume} {211}},\ \bibinfo {pages} {107528} (\bibinfo {year}
  {2020})}\BibitemShut {NoStop}%
\bibitem [{\citenamefont {Lim}\ and\ \citenamefont {Moon}(2020)}]{Lim_2020}%
  \BibitemOpen
  \bibfield  {author} {\bibinfo {author} {\bibfnamefont {H.}~\bibnamefont
  {Lim}}\ and\ \bibinfo {author} {\bibfnamefont {D.}~\bibnamefont {Moon}},\
  }\bibfield  {title} {\enquote {\bibinfo {title} {Microscopic and nanoscopic
  protein imaging by sims and helium ion microscopy},}\ }\href
  {https://doi.org/10.1116/6.0000220} {\bibfield  {journal} {\bibinfo
  {journal} {Biointerphases}\ }\textbf {\bibinfo {volume} {15}},\ \bibinfo
  {pages} {038501} (\bibinfo {year} {2020})}\BibitemShut {NoStop}%
\bibitem [{\citenamefont {Arlinghaus}(2008)}]{Arlinghaus_2008}%
  \BibitemOpen
  \bibfield  {author} {\bibinfo {author} {\bibfnamefont {H.~F.}\ \bibnamefont
  {Arlinghaus}},\ }\bibfield  {title} {\enquote {\bibinfo {title}
  {Possibilities and limitations of high-resolution mass spectrometry in life
  sciences},}\ }\href {https://doi.org/10.1016/j.apsusc.2008.05.047} {\bibfield
   {journal} {\bibinfo  {journal} {Appl. Surf. Sci.}\ }\textbf {\bibinfo
  {volume} {255}},\ \bibinfo {pages} {1058--1063} (\bibinfo {year} {2008})},\
  \bibinfo {note} {proceedings of the Sixteenth International Conference on
  Secondary Ion Mass Spectrometry, SIMS XVI}\BibitemShut {NoStop}%
\bibitem [{\citenamefont {Passarelli}\ \emph {et~al.}(2017)\citenamefont
  {Passarelli}, \citenamefont {Pirkl}, \citenamefont {Moellers}, \citenamefont
  {Grinfeld}, \citenamefont {Kollmer}, \citenamefont {Havelund}, \citenamefont
  {Newman}, \citenamefont {Marshall}, \citenamefont {Arlinghaus}, \citenamefont
  {Alexander}, \citenamefont {West}, \citenamefont {Horning}, \citenamefont
  {Niehuis}, \citenamefont {Makarov}, \citenamefont {Dollery},\ and\
  \citenamefont {Gilmore}}]{Passarelli_2017a}%
  \BibitemOpen
  \bibfield  {author} {\bibinfo {author} {\bibfnamefont {M.~K.}\ \bibnamefont
  {Passarelli}}, \bibinfo {author} {\bibfnamefont {A.}~\bibnamefont {Pirkl}},
  \bibinfo {author} {\bibfnamefont {R.}~\bibnamefont {Moellers}}, \bibinfo
  {author} {\bibfnamefont {D.}~\bibnamefont {Grinfeld}}, \bibinfo {author}
  {\bibfnamefont {F.}~\bibnamefont {Kollmer}}, \bibinfo {author} {\bibfnamefont
  {R.}~\bibnamefont {Havelund}}, \bibinfo {author} {\bibfnamefont {C.~F.}\
  \bibnamefont {Newman}}, \bibinfo {author} {\bibfnamefont {P.~S.}\
  \bibnamefont {Marshall}}, \bibinfo {author} {\bibfnamefont {H.}~\bibnamefont
  {Arlinghaus}}, \bibinfo {author} {\bibfnamefont {M.~R.}\ \bibnamefont
  {Alexander}}, \bibinfo {author} {\bibfnamefont {A.}~\bibnamefont {West}},
  \bibinfo {author} {\bibfnamefont {S.}~\bibnamefont {Horning}}, \bibinfo
  {author} {\bibfnamefont {E.}~\bibnamefont {Niehuis}}, \bibinfo {author}
  {\bibfnamefont {A.}~\bibnamefont {Makarov}}, \bibinfo {author} {\bibfnamefont
  {C.~T.}\ \bibnamefont {Dollery}},\ and\ \bibinfo {author} {\bibfnamefont
  {I.~S.}\ \bibnamefont {Gilmore}},\ }\bibfield  {title} {\enquote {\bibinfo
  {title} {The {3D OrbiSIMS}---label-free metabolic imaging with subcellular
  lateral resolution and high mass-resolving power},}\ }\href
  {https://doi.org/10.1038/nmeth.4504} {\bibfield  {journal} {\bibinfo
  {journal} {Nat. Methods}\ }\textbf {\bibinfo {volume} {14}},\ \bibinfo
  {pages} {1175--1183} (\bibinfo {year} {2017})}\BibitemShut {NoStop}%
\bibitem [{\citenamefont {Gilmore}, \citenamefont {Heiles},\ and\ \citenamefont
  {Pieterse}(2019)}]{Gilmore_2019}%
  \BibitemOpen
  \bibfield  {author} {\bibinfo {author} {\bibfnamefont {I.~S.}\ \bibnamefont
  {Gilmore}}, \bibinfo {author} {\bibfnamefont {S.}~\bibnamefont {Heiles}},\
  and\ \bibinfo {author} {\bibfnamefont {C.~L.}\ \bibnamefont {Pieterse}},\
  }\bibfield  {title} {\enquote {\bibinfo {title} {Metabolic imaging at the
  single-cell scale: recent advances in mass spectrometry imaging},}\ }\href
  {https://doi.org/10.1146/annurev-anchem-061318-115516} {\bibfield  {journal}
  {\bibinfo  {journal} {Annu. Rev. Anal. Chem.}\ }\textbf {\bibinfo {volume}
  {12}},\ \bibinfo {pages} {201--224} (\bibinfo {year} {2019})},\ \bibinfo
  {note} {pMID: 30848927}\BibitemShut {NoStop}%
\bibitem [{\citenamefont {Lepp{\"{a}}nen}\ \emph {et~al.}(2017)\citenamefont
  {Lepp{\"{a}}nen}, \citenamefont {Sundberg}, \citenamefont {Laanto},
  \citenamefont {de~Freitas~Almeida}, \citenamefont {Papponen},\ and\
  \citenamefont {Maasilta}}]{Leppaenen_2017}%
  \BibitemOpen
  \bibfield  {author} {\bibinfo {author} {\bibfnamefont {M.}~\bibnamefont
  {Lepp{\"{a}}nen}}, \bibinfo {author} {\bibfnamefont {L.-R.}\ \bibnamefont
  {Sundberg}}, \bibinfo {author} {\bibfnamefont {E.}~\bibnamefont {Laanto}},
  \bibinfo {author} {\bibfnamefont {G.~M.}\ \bibnamefont {de~Freitas~Almeida}},
  \bibinfo {author} {\bibfnamefont {P.}~\bibnamefont {Papponen}},\ and\
  \bibinfo {author} {\bibfnamefont {I.~J.}\ \bibnamefont {Maasilta}},\
  }\bibfield  {title} {\enquote {\bibinfo {title} {Imaging bacterial colonies
  and phage-bacterium interaction at sub-nanometer resolution using helium-ion
  microscopy},}\ }\href {https://doi.org/10.1002/adbi.201700070} {\bibfield
  {journal} {\bibinfo  {journal} {Advanced Biosystems}\ }\textbf {\bibinfo
  {volume} {1}},\ \bibinfo {pages} {1700070} (\bibinfo {year}
  {2017})}\BibitemShut {NoStop}%
\bibitem [{\citenamefont {Sharma}\ \emph {et~al.}(2018)\citenamefont {Sharma},
  \citenamefont {Schmidt}, \citenamefont {Kiesel}, \citenamefont {Mahato},
  \citenamefont {Cralle}, \citenamefont {Singh}, \citenamefont {Richnow},
  \citenamefont {Gilbert}, \citenamefont {Arnold},\ and\ \citenamefont
  {Lal}}]{Sharma_2018a}%
  \BibitemOpen
  \bibfield  {author} {\bibinfo {author} {\bibfnamefont {A.}~\bibnamefont
  {Sharma}}, \bibinfo {author} {\bibfnamefont {M.}~\bibnamefont {Schmidt}},
  \bibinfo {author} {\bibfnamefont {B.}~\bibnamefont {Kiesel}}, \bibinfo
  {author} {\bibfnamefont {N.~K.}\ \bibnamefont {Mahato}}, \bibinfo {author}
  {\bibfnamefont {L.}~\bibnamefont {Cralle}}, \bibinfo {author} {\bibfnamefont
  {Y.}~\bibnamefont {Singh}}, \bibinfo {author} {\bibfnamefont {H.~H.}\
  \bibnamefont {Richnow}}, \bibinfo {author} {\bibfnamefont {J.~A.}\
  \bibnamefont {Gilbert}}, \bibinfo {author} {\bibfnamefont {W.}~\bibnamefont
  {Arnold}},\ and\ \bibinfo {author} {\bibfnamefont {R.}~\bibnamefont {Lal}},\
  }\bibfield  {title} {\enquote {\bibinfo {title} {Bacterial and archaeal
  viruses of himalayan hot springs at manikaran modulate host genomes},}\
  }\href {https://doi.org/10.3389/fmicb.2018.03095} {\bibfield  {journal}
  {\bibinfo  {journal} {Front. Microbiol.}\ }\textbf {\bibinfo {volume} {9}}
  (\bibinfo {year} {2018}),\ 10.3389/fmicb.2018.03095}\BibitemShut {NoStop}%
\bibitem [{\citenamefont {You}\ \emph {et~al.}(2021)\citenamefont {You},
  \citenamefont {Kallies}, \citenamefont {K{\"{u}}hn}, \citenamefont {Schmidt},
  \citenamefont {Harms}, \citenamefont {Chatzinotas},\ and\ \citenamefont
  {Wick}}]{You_2021a}%
  \BibitemOpen
  \bibfield  {author} {\bibinfo {author} {\bibfnamefont {X.}~\bibnamefont
  {You}}, \bibinfo {author} {\bibfnamefont {R.}~\bibnamefont {Kallies}},
  \bibinfo {author} {\bibfnamefont {I.}~\bibnamefont {K{\"{u}}hn}}, \bibinfo
  {author} {\bibfnamefont {M.}~\bibnamefont {Schmidt}}, \bibinfo {author}
  {\bibfnamefont {H.}~\bibnamefont {Harms}}, \bibinfo {author} {\bibfnamefont
  {A.}~\bibnamefont {Chatzinotas}},\ and\ \bibinfo {author} {\bibfnamefont
  {L.~Y.}\ \bibnamefont {Wick}},\ }\bibfield  {title} {\enquote {\bibinfo
  {title} {Phage co-transport with hyphal-riding bacteria fuels bacterial
  invasion in a water-unsaturated microbial model system},}\ }\href
  {https://doi.org/10.1038/s41396-021-01155-x} {\bibfield  {journal} {\bibinfo
  {journal} {The {ISME} Journal}\ } (\bibinfo {year} {2021}),\
  10.1038/s41396-021-01155-x}\BibitemShut {NoStop}%
\bibitem [{\citenamefont {Frese}\ \emph {et~al.}(2021)\citenamefont {Frese},
  \citenamefont {Schmerer}, \citenamefont {Wortmann}, \citenamefont
  {Sch{\"{u}}rmann}, \citenamefont {K{\"{o}}nig}, \citenamefont {Westphal},
  \citenamefont {Weber}, \citenamefont {Sudhoff},\ and\ \citenamefont
  {G{\"{o}}lzh{\"{a}}user}}]{Frese2021}%
  \BibitemOpen
  \bibfield  {author} {\bibinfo {author} {\bibfnamefont {N.}~\bibnamefont
  {Frese}}, \bibinfo {author} {\bibfnamefont {P.}~\bibnamefont {Schmerer}},
  \bibinfo {author} {\bibfnamefont {M.}~\bibnamefont {Wortmann}}, \bibinfo
  {author} {\bibfnamefont {M.}~\bibnamefont {Sch{\"{u}}rmann}}, \bibinfo
  {author} {\bibfnamefont {M.}~\bibnamefont {K{\"{o}}nig}}, \bibinfo {author}
  {\bibfnamefont {M.}~\bibnamefont {Westphal}}, \bibinfo {author}
  {\bibfnamefont {F.}~\bibnamefont {Weber}}, \bibinfo {author} {\bibfnamefont
  {H.}~\bibnamefont {Sudhoff}},\ and\ \bibinfo {author} {\bibfnamefont
  {A.}~\bibnamefont {G{\"{o}}lzh{\"{a}}user}},\ }\bibfield  {title} {\enquote
  {\bibinfo {title} {Imaging of sars-cov-2 infected vero e6 cells by helium ion
  microscopy},}\ }\href {https://doi.org/10.3762/bjnano.12.13} {\bibfield
  {journal} {\bibinfo  {journal} {Beilstein J. Nanotechnol.}\ }\textbf
  {\bibinfo {volume} {12}},\ \bibinfo {pages} {172--179} (\bibinfo {year}
  {2021})}\BibitemShut {NoStop}%
\bibitem [{\citenamefont {Merolli}\ \emph {et~al.}(2022)\citenamefont
  {Merolli}, \citenamefont {Kasaei}, \citenamefont {Ramasamy}, \citenamefont
  {Kolloli}, \citenamefont {Kumar}, \citenamefont {Subbian},\ and\
  \citenamefont {Feldman}}]{Merolli_2022}%
  \BibitemOpen
  \bibfield  {author} {\bibinfo {author} {\bibfnamefont {A.}~\bibnamefont
  {Merolli}}, \bibinfo {author} {\bibfnamefont {L.}~\bibnamefont {Kasaei}},
  \bibinfo {author} {\bibfnamefont {S.}~\bibnamefont {Ramasamy}}, \bibinfo
  {author} {\bibfnamefont {A.}~\bibnamefont {Kolloli}}, \bibinfo {author}
  {\bibfnamefont {R.}~\bibnamefont {Kumar}}, \bibinfo {author} {\bibfnamefont
  {S.}~\bibnamefont {Subbian}},\ and\ \bibinfo {author} {\bibfnamefont {L.~C.}\
  \bibnamefont {Feldman}},\ }\href {https://doi.org/10.1038/s41598-022-07867-0}
  {\bibfield  {journal} {\bibinfo  {journal} {Sci. Rep.}\ }\textbf {\bibinfo
  {volume} {12}},\ \bibinfo {pages} {3794} (\bibinfo {year}
  {2022})}\BibitemShut {NoStop}%
\bibitem [{\citenamefont {Baena}\ \emph {et~al.}(2021)\citenamefont {Baena},
  \citenamefont {Conrad}, \citenamefont {Friday}, \citenamefont {Fitzgerald},
  \citenamefont {Kim}, \citenamefont {Bernbaum}, \citenamefont {Berensmann},
  \citenamefont {Harned}, \citenamefont {Nagashima},\ and\ \citenamefont
  {Narayan}}]{Baena_2021}%
  \BibitemOpen
  \bibfield  {author} {\bibinfo {author} {\bibfnamefont {V.}~\bibnamefont
  {Baena}}, \bibinfo {author} {\bibfnamefont {R.}~\bibnamefont {Conrad}},
  \bibinfo {author} {\bibfnamefont {P.}~\bibnamefont {Friday}}, \bibinfo
  {author} {\bibfnamefont {E.}~\bibnamefont {Fitzgerald}}, \bibinfo {author}
  {\bibfnamefont {T.}~\bibnamefont {Kim}}, \bibinfo {author} {\bibfnamefont
  {J.}~\bibnamefont {Bernbaum}}, \bibinfo {author} {\bibfnamefont
  {H.}~\bibnamefont {Berensmann}}, \bibinfo {author} {\bibfnamefont
  {A.}~\bibnamefont {Harned}}, \bibinfo {author} {\bibfnamefont
  {K.}~\bibnamefont {Nagashima}},\ and\ \bibinfo {author} {\bibfnamefont
  {K.}~\bibnamefont {Narayan}},\ }\bibfield  {title} {\enquote {\bibinfo
  {title} {{FIB-SEM} as a volume electron microscopy approach to study cellular
  architectures in {SARS-CoV-2} and other viral infections: a practical primer
  for a virologist},}\ }\href {https://doi.org/10.3390/v13040611} {\bibfield
  {journal} {\bibinfo  {journal} {Viruses}\ }\textbf {\bibinfo {volume} {13}}
  (\bibinfo {year} {2021}),\ 10.3390/v13040611}\BibitemShut {NoStop}%
\bibitem [{\citenamefont {Guillermier}\ \emph {et~al.}(2019)\citenamefont
  {Guillermier}, \citenamefont {Medina~Cruz}, \citenamefont {Audinot},\ and\
  \citenamefont {Wirtz}}]{Guillermier_2019}%
  \BibitemOpen
  \bibfield  {author} {\bibinfo {author} {\bibfnamefont {C.}~\bibnamefont
  {Guillermier}}, \bibinfo {author} {\bibfnamefont {D.}~\bibnamefont
  {Medina~Cruz}}, \bibinfo {author} {\bibfnamefont {J.-N.}\ \bibnamefont
  {Audinot}},\ and\ \bibinfo {author} {\bibfnamefont {T.}~\bibnamefont
  {Wirtz}},\ }\bibfield  {title} {\enquote {\bibinfo {title} {{Characterization
  of Biogenic Nanoparticles Via In-Situ Correlative Secondary Electron Helium
  Microscopy and Secondary Ion Mass Spectrometry}},}\ }\href
  {https://doi.org/10.1017/s1431927619006044} {\bibfield  {journal} {\bibinfo
  {journal} {Microsc. Microanal.}\ }\textbf {\bibinfo {volume} {25}},\ \bibinfo
  {pages} {1062--1063} (\bibinfo {year} {2019})}\BibitemShut {NoStop}%
\bibitem [{\citenamefont {Benettoni}\ \emph {et~al.}(2019)\citenamefont
  {Benettoni}, \citenamefont {Stryhanyuk}, \citenamefont {Wagner},
  \citenamefont {Kollmer}, \citenamefont {Moreno-Osorio}, \citenamefont
  {Schmidt},\ and\ \citenamefont {Richnow}}]{Benettoni_2019}%
  \BibitemOpen
  \bibfield  {author} {\bibinfo {author} {\bibfnamefont {P.}~\bibnamefont
  {Benettoni}}, \bibinfo {author} {\bibfnamefont {H.}~\bibnamefont
  {Stryhanyuk}}, \bibinfo {author} {\bibfnamefont {S.}~\bibnamefont {Wagner}},
  \bibinfo {author} {\bibfnamefont {F.}~\bibnamefont {Kollmer}}, \bibinfo
  {author} {\bibfnamefont {J.~H.}\ \bibnamefont {Moreno-Osorio}}, \bibinfo
  {author} {\bibfnamefont {T.}~\bibnamefont {Schmidt}, \bibfnamefont
  {M.~Reemtsma}},\ and\ \bibinfo {author} {\bibfnamefont {H.-H.}\ \bibnamefont
  {Richnow}},\ }\href {https://doi.org/10.1039/c8ja00439k} {\bibfield
  {journal} {\bibinfo  {journal} {J. Anal. At. Spectrom.}\ ,\ \bibinfo {pages}
  {1098}} (\bibinfo {year} {2019})}\BibitemShut {NoStop}%
\bibitem [{\citenamefont {Joens}\ \emph {et~al.}(2013)\citenamefont {Joens},
  \citenamefont {Huynh}, \citenamefont {Kasuboski}, \citenamefont {Ferranti},
  \citenamefont {Sigal}, \citenamefont {Zeitvogel}, \citenamefont {Obst},
  \citenamefont {Burkhardt}, \citenamefont {Curran}, \citenamefont {Chalasani},
  \citenamefont {Stern}, \citenamefont {Goetze},\ and\ \citenamefont
  {Fitzpatrick}}]{Joens_2013}%
  \BibitemOpen
  \bibfield  {author} {\bibinfo {author} {\bibfnamefont {M.~S.}\ \bibnamefont
  {Joens}}, \bibinfo {author} {\bibfnamefont {C.}~\bibnamefont {Huynh}},
  \bibinfo {author} {\bibfnamefont {J.~M.}\ \bibnamefont {Kasuboski}}, \bibinfo
  {author} {\bibfnamefont {D.~C.}\ \bibnamefont {Ferranti}}, \bibinfo {author}
  {\bibfnamefont {Y.~J.}\ \bibnamefont {Sigal}}, \bibinfo {author}
  {\bibfnamefont {F.}~\bibnamefont {Zeitvogel}}, \bibinfo {author}
  {\bibfnamefont {M.}~\bibnamefont {Obst}}, \bibinfo {author} {\bibfnamefont
  {C.~J.}\ \bibnamefont {Burkhardt}}, \bibinfo {author} {\bibfnamefont {K.~P.}\
  \bibnamefont {Curran}}, \bibinfo {author} {\bibfnamefont {S.~H.}\
  \bibnamefont {Chalasani}}, \bibinfo {author} {\bibfnamefont {L.~A.}\
  \bibnamefont {Stern}}, \bibinfo {author} {\bibfnamefont {B.}~\bibnamefont
  {Goetze}},\ and\ \bibinfo {author} {\bibfnamefont {J.~A.~J.}\ \bibnamefont
  {Fitzpatrick}},\ }\bibfield  {title} {\enquote {\bibinfo {title} {{Helium Ion
  Microscopy (HIM) for the imaging of biological samples at sub--nanometer
  resolution.}}}\ }\href {https://doi.org/10.1038/srep03514} {\bibfield
  {journal} {\bibinfo  {journal} {Sci. Rep.}\ }\textbf {\bibinfo {volume}
  {3}},\ \bibinfo {pages} {3514} (\bibinfo {year} {2013})}\BibitemShut
  {NoStop}%
\bibitem [{\citenamefont {Byrne}\ \emph {et~al.}(2018)\citenamefont {Byrne},
  \citenamefont {Schmidt}, \citenamefont {Gauger}, \citenamefont {Bryce},\ and\
  \citenamefont {Kappler}}]{Byrne_2018}%
  \BibitemOpen
  \bibfield  {author} {\bibinfo {author} {\bibfnamefont {J.~M.}\ \bibnamefont
  {Byrne}}, \bibinfo {author} {\bibfnamefont {M.}~\bibnamefont {Schmidt}},
  \bibinfo {author} {\bibfnamefont {T.}~\bibnamefont {Gauger}}, \bibinfo
  {author} {\bibfnamefont {C.}~\bibnamefont {Bryce}},\ and\ \bibinfo {author}
  {\bibfnamefont {A.}~\bibnamefont {Kappler}},\ }\bibfield  {title} {\enquote
  {\bibinfo {title} {Imaging organic--mineral aggregates formed by
  {F}e({II})-oxidizing bacteria using helium ion microscopy},}\ }\href
  {https://doi.org/10.1021/acs.estlett.8b00077} {\bibfield  {journal} {\bibinfo
   {journal} {Environ. Sci. Technol. Lett.}\ }\textbf {\bibinfo {volume} {5}},\
  \bibinfo {pages} {209--213} (\bibinfo {year} {2018})}\BibitemShut {NoStop}%
\bibitem [{\citenamefont {Chen}\ \emph {et~al.}(2019)\citenamefont {Chen},
  \citenamefont {Musat}, \citenamefont {Lechtenfeld}, \citenamefont {Paschke},
  \citenamefont {Schmidt}, \citenamefont {Said}, \citenamefont {Popp},
  \citenamefont {Calabrese}, \citenamefont {Stryhanyuk}, \citenamefont
  {Jaekel}, \citenamefont {Zhu}, \citenamefont {Joye}, \citenamefont {Richnow},
  \citenamefont {Widdel},\ and\ \citenamefont {Musat}}]{Chen_2019a}%
  \BibitemOpen
  \bibfield  {author} {\bibinfo {author} {\bibfnamefont {S.-C.}\ \bibnamefont
  {Chen}}, \bibinfo {author} {\bibfnamefont {N.}~\bibnamefont {Musat}},
  \bibinfo {author} {\bibfnamefont {O.~J.}\ \bibnamefont {Lechtenfeld}},
  \bibinfo {author} {\bibfnamefont {H.}~\bibnamefont {Paschke}}, \bibinfo
  {author} {\bibfnamefont {M.}~\bibnamefont {Schmidt}}, \bibinfo {author}
  {\bibfnamefont {N.}~\bibnamefont {Said}}, \bibinfo {author} {\bibfnamefont
  {D.}~\bibnamefont {Popp}}, \bibinfo {author} {\bibfnamefont {F.}~\bibnamefont
  {Calabrese}}, \bibinfo {author} {\bibfnamefont {H.}~\bibnamefont
  {Stryhanyuk}}, \bibinfo {author} {\bibfnamefont {U.}~\bibnamefont {Jaekel}},
  \bibinfo {author} {\bibfnamefont {Y.-G.}\ \bibnamefont {Zhu}}, \bibinfo
  {author} {\bibfnamefont {S.~B.}\ \bibnamefont {Joye}}, \bibinfo {author}
  {\bibfnamefont {H.-H.}\ \bibnamefont {Richnow}}, \bibinfo {author}
  {\bibfnamefont {F.}~\bibnamefont {Widdel}},\ and\ \bibinfo {author}
  {\bibfnamefont {F.}~\bibnamefont {Musat}},\ }\bibfield  {title} {\enquote
  {\bibinfo {title} {Anaerobic oxidation of ethane by archaea from a marine
  hydrocarbon seep},}\ }\href {https://doi.org/10.1038/s41586-019-1063-0}
  {\bibfield  {journal} {\bibinfo  {journal} {Nature}\ }\textbf {\bibinfo
  {volume} {568}},\ \bibinfo {pages} {108--111} (\bibinfo {year}
  {2019})}\BibitemShut {NoStop}%
\bibitem [{\citenamefont {Said}, \citenamefont {Chatzinotas},\ and\
  \citenamefont {Schmidt}(2018)}]{Said2018}%
  \BibitemOpen
  \bibfield  {author} {\bibinfo {author} {\bibfnamefont {N.}~\bibnamefont
  {Said}}, \bibinfo {author} {\bibfnamefont {A.}~\bibnamefont {Chatzinotas}},\
  and\ \bibinfo {author} {\bibfnamefont {M.}~\bibnamefont {Schmidt}},\
  }\bibfield  {title} {\enquote {\bibinfo {title} {Have an ion on it: the
  life-cycle of bdellovibrio bacteriovorus viewed by helium-ion microscopy},}\
  }\href {https://doi.org/10.1002/adbi.201800250} {\bibfield  {journal}
  {\bibinfo  {journal} {Advanced Biosystems}\ ,\ \bibinfo {pages} {1800250}}
  (\bibinfo {year} {2018})}\BibitemShut {NoStop}%
\bibitem [{\citenamefont {Belianinov}\ \emph
  {et~al.}(2017{\natexlab{b}})\citenamefont {Belianinov}, \citenamefont
  {Halsted}, \citenamefont {Burch}, \citenamefont {Songkil},\ and\
  \citenamefont {Retterer}}]{Belianinov_2017a}%
  \BibitemOpen
  \bibfield  {author} {\bibinfo {author} {\bibfnamefont {A.}~\bibnamefont
  {Belianinov}}, \bibinfo {author} {\bibfnamefont {M.~C.}\ \bibnamefont
  {Halsted}}, \bibinfo {author} {\bibfnamefont {M.~J.}\ \bibnamefont {Burch}},
  \bibinfo {author} {\bibfnamefont {K.}~\bibnamefont {Songkil}},\ and\ \bibinfo
  {author} {\bibfnamefont {S.~T.}\ \bibnamefont {Retterer}},\ }\bibfield
  {title} {\enquote {\bibinfo {title} {Biofilm structure of geobacter
  sulfurreducens by helium ion microscopy},}\ }\href
  {https://doi.org/10.1017/s1431927617006420} {\bibfield  {journal} {\bibinfo
  {journal} {Microsc. Microanal.}\ }\textbf {\bibinfo {volume} {23}},\ \bibinfo
  {pages} {1152--1153} (\bibinfo {year} {2017}{\natexlab{b}})}\BibitemShut
  {NoStop}%
\bibitem [{\citenamefont {Moreno~Osorio}\ \emph {et~al.}(2019)\citenamefont
  {Moreno~Osorio}, \citenamefont {Benettoni}, \citenamefont {Schmidt},
  \citenamefont {Stryhanyuk}, \citenamefont {Schmitt-Jansen}, \citenamefont
  {Pinto}, \citenamefont {Pollio}, \citenamefont {Frunzo}, \citenamefont
  {Lens}, \citenamefont {Richnow}, \citenamefont {Esposito},\ and\
  \citenamefont {Musat}}]{MorenoOsorio_2019}%
  \BibitemOpen
  \bibfield  {author} {\bibinfo {author} {\bibfnamefont {J.~H.}\ \bibnamefont
  {Moreno~Osorio}}, \bibinfo {author} {\bibfnamefont {P.}~\bibnamefont
  {Benettoni}}, \bibinfo {author} {\bibfnamefont {M.}~\bibnamefont {Schmidt}},
  \bibinfo {author} {\bibfnamefont {H.}~\bibnamefont {Stryhanyuk}}, \bibinfo
  {author} {\bibfnamefont {M.}~\bibnamefont {Schmitt-Jansen}}, \bibinfo
  {author} {\bibfnamefont {G.}~\bibnamefont {Pinto}}, \bibinfo {author}
  {\bibfnamefont {A.}~\bibnamefont {Pollio}}, \bibinfo {author} {\bibfnamefont
  {L.}~\bibnamefont {Frunzo}}, \bibinfo {author} {\bibfnamefont {P.~N.~L.}\
  \bibnamefont {Lens}}, \bibinfo {author} {\bibfnamefont {H.~H.}\ \bibnamefont
  {Richnow}}, \bibinfo {author} {\bibfnamefont {G.}~\bibnamefont {Esposito}},\
  and\ \bibinfo {author} {\bibfnamefont {N.}~\bibnamefont {Musat}},\ }\bibfield
   {title} {\enquote {\bibinfo {title} {Investigation of architecture
  development and phosphate distribution in chlorella biofilm by complementary
  microscopy techniques},}\ }\href {https://doi.org/10.1093/femsec/fiz029}
  {\bibfield  {journal} {\bibinfo  {journal} {{FEMS} Microbiology Ecology}\
  }\textbf {\bibinfo {volume} {95}} (\bibinfo {year} {2019}),\
  10.1093/femsec/fiz029}\BibitemShut {NoStop}%
\bibitem [{\citenamefont {Sch{\"{u}}rmann}\ \emph {et~al.}(2015)\citenamefont
  {Sch{\"{u}}rmann}, \citenamefont {Frese}, \citenamefont {Beyer},
  \citenamefont {Heimann}, \citenamefont {Widera}, \citenamefont
  {M{\"{o}}nkem{\"{o}}ller}, \citenamefont {Huser}, \citenamefont
  {Kaltschmidt}, \citenamefont {Kaltschmidt},\ and\ \citenamefont
  {G{\"{o}}lzh{\"{a}}user}}]{Schuermann_2015}%
  \BibitemOpen
  \bibfield  {author} {\bibinfo {author} {\bibfnamefont {M.}~\bibnamefont
  {Sch{\"{u}}rmann}}, \bibinfo {author} {\bibfnamefont {N.}~\bibnamefont
  {Frese}}, \bibinfo {author} {\bibfnamefont {A.}~\bibnamefont {Beyer}},
  \bibinfo {author} {\bibfnamefont {P.}~\bibnamefont {Heimann}}, \bibinfo
  {author} {\bibfnamefont {D.}~\bibnamefont {Widera}}, \bibinfo {author}
  {\bibfnamefont {V.}~\bibnamefont {M{\"{o}}nkem{\"{o}}ller}}, \bibinfo
  {author} {\bibfnamefont {T.}~\bibnamefont {Huser}}, \bibinfo {author}
  {\bibfnamefont {B.}~\bibnamefont {Kaltschmidt}}, \bibinfo {author}
  {\bibfnamefont {C.}~\bibnamefont {Kaltschmidt}},\ and\ \bibinfo {author}
  {\bibfnamefont {A.}~\bibnamefont {G{\"{o}}lzh{\"{a}}user}},\ }\bibfield
  {title} {\enquote {\bibinfo {title} {Helium ion microscopy visualizes lipid
  nanodomains in mammalian cells},}\ }\href
  {https://doi.org/10.1002/smll.201501540} {\bibfield  {journal} {\bibinfo
  {journal} {Small}\ }\textbf {\bibinfo {volume} {11}},\ \bibinfo {pages}
  {5781--5789} (\bibinfo {year} {2015})}\BibitemShut {NoStop}%
\bibitem [{\citenamefont {Paunescu}, \citenamefont {Breton},\ and\
  \citenamefont {Brown}(2014)}]{Paunescu_2014}%
  \BibitemOpen
  \bibfield  {author} {\bibinfo {author} {\bibfnamefont {T.}~\bibnamefont
  {Paunescu}}, \bibinfo {author} {\bibfnamefont {S.}~\bibnamefont {Breton}},\
  and\ \bibinfo {author} {\bibfnamefont {D.}~\bibnamefont {Brown}},\ }\bibfield
   {title} {\enquote {\bibinfo {title} {Himaging of the kidney: high resolution
  helium ion microscopy},}\ }\href {https://doi.org/10.36866/pn.95.32}
  {\bibfield  {journal} {\bibinfo  {journal} {Physiology News}\ ,\ \bibinfo
  {pages} {32--35}} (\bibinfo {year} {2014})}\BibitemShut {NoStop}%
\bibitem [{\citenamefont {Decelle}\ \emph {et~al.}(2019)\citenamefont
  {Decelle}, \citenamefont {Stryhanyuk}, \citenamefont {Gallet}, \citenamefont
  {Veronesi}, \citenamefont {Schmidt}, \citenamefont {Balzano}, \citenamefont
  {Marro}, \citenamefont {Uwizeye}, \citenamefont {Jouneau}, \citenamefont
  {Lupette}, \citenamefont {Jouhet}, \citenamefont {Mar{\'{e}}chal},
  \citenamefont {Schwab}, \citenamefont {Schieber}, \citenamefont {Tucoulou},
  \citenamefont {Richnow}, \citenamefont {Finazzi},\ and\ \citenamefont
  {Musat}}]{Decelle_2019}%
  \BibitemOpen
  \bibfield  {author} {\bibinfo {author} {\bibfnamefont {J.}~\bibnamefont
  {Decelle}}, \bibinfo {author} {\bibfnamefont {H.}~\bibnamefont {Stryhanyuk}},
  \bibinfo {author} {\bibfnamefont {B.}~\bibnamefont {Gallet}}, \bibinfo
  {author} {\bibfnamefont {G.}~\bibnamefont {Veronesi}}, \bibinfo {author}
  {\bibfnamefont {M.}~\bibnamefont {Schmidt}}, \bibinfo {author} {\bibfnamefont
  {S.}~\bibnamefont {Balzano}}, \bibinfo {author} {\bibfnamefont
  {S.}~\bibnamefont {Marro}}, \bibinfo {author} {\bibfnamefont
  {C.}~\bibnamefont {Uwizeye}}, \bibinfo {author} {\bibfnamefont {P.-H.}\
  \bibnamefont {Jouneau}}, \bibinfo {author} {\bibfnamefont {J.}~\bibnamefont
  {Lupette}}, \bibinfo {author} {\bibfnamefont {J.}~\bibnamefont {Jouhet}},
  \bibinfo {author} {\bibfnamefont {E.}~\bibnamefont {Mar{\'{e}}chal}},
  \bibinfo {author} {\bibfnamefont {Y.}~\bibnamefont {Schwab}}, \bibinfo
  {author} {\bibfnamefont {N.~L.}\ \bibnamefont {Schieber}}, \bibinfo {author}
  {\bibfnamefont {R.}~\bibnamefont {Tucoulou}}, \bibinfo {author}
  {\bibfnamefont {H.}~\bibnamefont {Richnow}}, \bibinfo {author} {\bibfnamefont
  {G.}~\bibnamefont {Finazzi}},\ and\ \bibinfo {author} {\bibfnamefont
  {N.}~\bibnamefont {Musat}},\ }\bibfield  {title} {\enquote {\bibinfo {title}
  {Algal remodeling in a ubiquitous planktonic photosymbiosis},}\ }\href
  {https://doi.org/10.1016/j.cub.2019.01.073} {\bibfield  {journal} {\bibinfo
  {journal} {Curr. Biol.}\ }\textbf {\bibinfo {volume} {29}},\ \bibinfo {pages}
  {968--978.e4} (\bibinfo {year} {2019})}\BibitemShut {NoStop}%
\bibitem [{\citenamefont {Drobne}\ \emph {et~al.}(2005)\citenamefont {Drobne},
  \citenamefont {Milani}, \citenamefont {Zrimec}, \citenamefont {Leser},\ and\
  \citenamefont {Berden~Zrimec}}]{Drobne_2005}%
  \BibitemOpen
  \bibfield  {author} {\bibinfo {author} {\bibfnamefont {D.}~\bibnamefont
  {Drobne}}, \bibinfo {author} {\bibfnamefont {M.}~\bibnamefont {Milani}},
  \bibinfo {author} {\bibfnamefont {A.}~\bibnamefont {Zrimec}}, \bibinfo
  {author} {\bibfnamefont {V.}~\bibnamefont {Leser}},\ and\ \bibinfo {author}
  {\bibfnamefont {M.}~\bibnamefont {Berden~Zrimec}},\ }\bibfield  {title}
  {\enquote {\bibinfo {title} {Electron and ion imaging of gland cells using
  the {FIB}/{SEM} system},}\ }\href
  {https://doi.org/10.1111/j.1365-2818.2005.01490.x} {\bibfield  {journal}
  {\bibinfo  {journal} {J. Microsc.}\ }\textbf {\bibinfo {volume} {219}},\
  \bibinfo {pages} {29--35} (\bibinfo {year} {2005})}\BibitemShut {NoStop}%
\bibitem [{\citenamefont {Kobler}\ \emph {et~al.}(2014)\citenamefont {Kobler},
  \citenamefont {Saber}, \citenamefont {Jacobsen}, \citenamefont {Wallin},
  \citenamefont {Vogel}, \citenamefont {Qvortrup},\ and\ \citenamefont
  {Molhave}}]{Kobler_2014}%
  \BibitemOpen
  \bibfield  {author} {\bibinfo {author} {\bibfnamefont {C.}~\bibnamefont
  {Kobler}}, \bibinfo {author} {\bibfnamefont {A.~T.}\ \bibnamefont {Saber}},
  \bibinfo {author} {\bibfnamefont {N.~R.}\ \bibnamefont {Jacobsen}}, \bibinfo
  {author} {\bibfnamefont {H.}~\bibnamefont {Wallin}}, \bibinfo {author}
  {\bibfnamefont {U.}~\bibnamefont {Vogel}}, \bibinfo {author} {\bibfnamefont
  {K.}~\bibnamefont {Qvortrup}},\ and\ \bibinfo {author} {\bibfnamefont
  {K.}~\bibnamefont {Molhave}},\ }\href
  {https://doi.org/10.1007/s00216-013-7566-x} {\bibfield  {journal} {\bibinfo
  {journal} {Anal. Bioanal. Chem.}\ ,\ \bibinfo {pages} {3863--3873}} (\bibinfo
  {year} {2014})}\BibitemShut {NoStop}%
\bibitem [{\citenamefont {Binkley}\ \emph {et~al.}(2020)\citenamefont
  {Binkley}, \citenamefont {Deering}, \citenamefont {Yuan}, \citenamefont
  {Gourrier},\ and\ \citenamefont {Grandfield}}]{Binkley_2020}%
  \BibitemOpen
  \bibfield  {author} {\bibinfo {author} {\bibfnamefont {D.~M.}\ \bibnamefont
  {Binkley}}, \bibinfo {author} {\bibfnamefont {J.}~\bibnamefont {Deering}},
  \bibinfo {author} {\bibfnamefont {H.}~\bibnamefont {Yuan}}, \bibinfo {author}
  {\bibfnamefont {A.}~\bibnamefont {Gourrier}},\ and\ \bibinfo {author}
  {\bibfnamefont {K.}~\bibnamefont {Grandfield}},\ }\bibfield  {title}
  {\enquote {\bibinfo {title} {Ellipsoidal mesoscale mineralization pattern in
  human cortical bone revealed in 3{D} by plasma focused ion beam serial
  sectioning},}\ }\href {https://doi.org/10.1016/j.jsb.2020.107615} {\bibfield
  {journal} {\bibinfo  {journal} {J. Struct. Biol.}\ }\textbf {\bibinfo
  {volume} {212}},\ \bibinfo {pages} {107615} (\bibinfo {year}
  {2020})}\BibitemShut {NoStop}%
\bibitem [{\citenamefont {Morrison}, \citenamefont {Misra},\ and\ \citenamefont
  {Williams}(2016)}]{Morrisson_2016}%
  \BibitemOpen
  \bibfield  {author} {\bibinfo {author} {\bibfnamefont {K.~D.}\ \bibnamefont
  {Morrison}}, \bibinfo {author} {\bibfnamefont {R.}~\bibnamefont {Misra}},\
  and\ \bibinfo {author} {\bibfnamefont {L.~B.}\ \bibnamefont {Williams}},\
  }\bibfield  {title} {\enquote {\bibinfo {title} {Unearthing the antibacterial
  mechanism of medicinal clay: a geochemical approach to combating antibiotic
  resistance},}\ }\href {https://doi.org/10.1038/srep19043} {\bibfield
  {journal} {\bibinfo  {journal} {Sci. Rep.}\ }\textbf {\bibinfo {volume}
  {6}},\ \bibinfo {pages} {19043} (\bibinfo {year} {2016})}\BibitemShut
  {NoStop}%
\bibitem [{\citenamefont {Forslind}, \citenamefont {Malmqvist},\ and\
  \citenamefont {Pallon}(1991)}]{Forslind_1991}%
  \BibitemOpen
  \bibfield  {author} {\bibinfo {author} {\bibfnamefont {B.}~\bibnamefont
  {Forslind}}, \bibinfo {author} {\bibfnamefont {K.~G.}\ \bibnamefont
  {Malmqvist}},\ and\ \bibinfo {author} {\bibfnamefont {J.}~\bibnamefont
  {Pallon}},\ }\bibfield  {title} {\enquote {\bibinfo {title} {Proton induced
  {X}-ray emission analysis of biological specimens--past and future.}}\
  }\href@noop {} {\bibfield  {journal} {\bibinfo  {journal} {Scanning
  microscopy}\ }\textbf {\bibinfo {volume} {5}},\ \bibinfo {pages} {877--884}
  (\bibinfo {year} {1991})}\BibitemShut {NoStop}%
\bibitem [{\citenamefont {Basnakova}\ \emph {et~al.}(1998)\citenamefont
  {Basnakova}, \citenamefont {Spencer}, \citenamefont {Palsgard}, \citenamefont
  {Grime},\ and\ \citenamefont {Macaskie}}]{Basnakova_1998}%
  \BibitemOpen
  \bibfield  {author} {\bibinfo {author} {\bibfnamefont {G.}~\bibnamefont
  {Basnakova}}, \bibinfo {author} {\bibfnamefont {A.~J.}\ \bibnamefont
  {Spencer}}, \bibinfo {author} {\bibfnamefont {E.}~\bibnamefont {Palsgard}},
  \bibinfo {author} {\bibfnamefont {G.~W.}\ \bibnamefont {Grime}},\ and\
  \bibinfo {author} {\bibfnamefont {L.~E.}\ \bibnamefont {Macaskie}},\
  }\bibfield  {title} {\enquote {\bibinfo {title} {Identification of the nickel
  uranyl phosphate deposits on citrobacter sp. cells by electron microscopy
  with electron probe x-ray microanalysis and by proton-induced x-ray emission
  analysis},}\ }\href {https://doi.org/10.1021/es9705553} {\bibfield  {journal}
  {\bibinfo  {journal} {Environ. Sci. Technol.}\ }\textbf {\bibinfo {volume}
  {32}},\ \bibinfo {pages} {760--765} (\bibinfo {year} {1998})}\BibitemShut
  {NoStop}%
\bibitem [{\citenamefont {Musat}\ \emph {et~al.}(2012)\citenamefont {Musat},
  \citenamefont {Foster}, \citenamefont {Vagner}, \citenamefont {Adam},\ and\
  \citenamefont {Kuypers}}]{Musat_2012}%
  \BibitemOpen
  \bibfield  {author} {\bibinfo {author} {\bibfnamefont {N.}~\bibnamefont
  {Musat}}, \bibinfo {author} {\bibfnamefont {R.}~\bibnamefont {Foster}},
  \bibinfo {author} {\bibfnamefont {T.}~\bibnamefont {Vagner}}, \bibinfo
  {author} {\bibfnamefont {B.}~\bibnamefont {Adam}},\ and\ \bibinfo {author}
  {\bibfnamefont {M.~M.~M.}\ \bibnamefont {Kuypers}},\ }\bibfield  {title}
  {\enquote {\bibinfo {title} {{Detecting metabolic activities in single cells,
  with emphasis on nanoSIMS}},}\ }\href
  {https://doi.org/10.1111/j.1574-6976.2011.00303.x} {\bibfield  {journal}
  {\bibinfo  {journal} {FEMS Microbiol. Rev.}\ }\textbf {\bibinfo {volume}
  {36}},\ \bibinfo {pages} {486--511} (\bibinfo {year} {2012})}\BibitemShut
  {NoStop}%
\bibitem [{\citenamefont {McNichol}\ \emph {et~al.}(2018)\citenamefont
  {McNichol}, \citenamefont {Stryhanyuk}, \citenamefont {Sylva}, \citenamefont
  {Thomas}, \citenamefont {Musat}, \citenamefont {Seewald},\ and\ \citenamefont
  {Sievert}}]{McNichol_2018}%
  \BibitemOpen
  \bibfield  {author} {\bibinfo {author} {\bibfnamefont {J.}~\bibnamefont
  {McNichol}}, \bibinfo {author} {\bibfnamefont {H.}~\bibnamefont
  {Stryhanyuk}}, \bibinfo {author} {\bibfnamefont {S.~P.}\ \bibnamefont
  {Sylva}}, \bibinfo {author} {\bibfnamefont {F.}~\bibnamefont {Thomas}},
  \bibinfo {author} {\bibfnamefont {N.}~\bibnamefont {Musat}}, \bibinfo
  {author} {\bibfnamefont {J.~S.}\ \bibnamefont {Seewald}},\ and\ \bibinfo
  {author} {\bibfnamefont {S.~M.}\ \bibnamefont {Sievert}},\ }\bibfield
  {title} {\enquote {\bibinfo {title} {Primary productivity below the seafloor
  at deep-sea hot springs},}\ }\href {https://doi.org/10.1073/pnas.1804351115}
  {\bibfield  {journal} {\bibinfo  {journal} {Proc. Natl. Acad. Sci.}\ }\textbf
  {\bibinfo {volume} {115}},\ \bibinfo {pages} {6756--6761} (\bibinfo {year}
  {2018})}\BibitemShut {NoStop}%
\bibitem [{\citenamefont {Stryhanyuk}\ \emph {et~al.}(2018)\citenamefont
  {Stryhanyuk}, \citenamefont {Calabrese}, \citenamefont {K{\"{u}}mmel},
  \citenamefont {Musat}, \citenamefont {Richnow},\ and\ \citenamefont
  {Musat}}]{Stryhanyuk_2018}%
  \BibitemOpen
  \bibfield  {author} {\bibinfo {author} {\bibfnamefont {H.}~\bibnamefont
  {Stryhanyuk}}, \bibinfo {author} {\bibfnamefont {F.}~\bibnamefont
  {Calabrese}}, \bibinfo {author} {\bibfnamefont {S.}~\bibnamefont
  {K{\"{u}}mmel}}, \bibinfo {author} {\bibfnamefont {F.}~\bibnamefont {Musat}},
  \bibinfo {author} {\bibfnamefont {H.-H.}\ \bibnamefont {Richnow}},\ and\
  \bibinfo {author} {\bibfnamefont {N.}~\bibnamefont {Musat}},\ }\bibfield
  {title} {\enquote {\bibinfo {title} {Calculation of single cell assimilation
  rates from sip-nanosims-derived isotope ratios: a comprehensive approach},}\
  }\href {https://doi.org/10.3389/fmicb.2018.02342} {\bibfield  {journal}
  {\bibinfo  {journal} {Front. Microbiol.}\ }\textbf {\bibinfo {volume} {9}},\
  \bibinfo {pages} {2342} (\bibinfo {year} {2018})}\BibitemShut {NoStop}%
\bibitem [{\citenamefont {Calabrese}\ \emph {et~al.}(2019)\citenamefont
  {Calabrese}, \citenamefont {Voloshynovska}, \citenamefont {Musat},
  \citenamefont {Thullner}, \citenamefont {Schl{\"{o}}mann}, \citenamefont
  {Richnow}, \citenamefont {Lambrecht}, \citenamefont {M{\"{u}}ller},
  \citenamefont {Wick}, \citenamefont {Musat},\ and\ \citenamefont
  {Stryhanyuk}}]{Calabrese_2019}%
  \BibitemOpen
  \bibfield  {author} {\bibinfo {author} {\bibfnamefont {F.}~\bibnamefont
  {Calabrese}}, \bibinfo {author} {\bibfnamefont {I.}~\bibnamefont
  {Voloshynovska}}, \bibinfo {author} {\bibfnamefont {F.}~\bibnamefont
  {Musat}}, \bibinfo {author} {\bibfnamefont {M.}~\bibnamefont {Thullner}},
  \bibinfo {author} {\bibfnamefont {M.}~\bibnamefont {Schl{\"{o}}mann}},
  \bibinfo {author} {\bibfnamefont {H.~H.}\ \bibnamefont {Richnow}}, \bibinfo
  {author} {\bibfnamefont {J.}~\bibnamefont {Lambrecht}}, \bibinfo {author}
  {\bibfnamefont {S.}~\bibnamefont {M{\"{u}}ller}}, \bibinfo {author}
  {\bibfnamefont {L.~Y.}\ \bibnamefont {Wick}}, \bibinfo {author}
  {\bibfnamefont {N.}~\bibnamefont {Musat}},\ and\ \bibinfo {author}
  {\bibfnamefont {H.}~\bibnamefont {Stryhanyuk}},\ }\bibfield  {title}
  {\enquote {\bibinfo {title} {Quantitation and comparison of phenotypic
  heterogeneity among single cells of monoclonal microbial populations},}\
  }\href {https://doi.org/10.3389/fmicb.2019.02814} {\bibfield  {journal}
  {\bibinfo  {journal} {Front. Microbiol.}\ }\textbf {\bibinfo {volume} {10}}
  (\bibinfo {year} {2019}),\ 10.3389/fmicb.2019.02814}\BibitemShut {NoStop}%
\bibitem [{\citenamefont {Leefmann}\ \emph {et~al.}(2013)\citenamefont
  {Leefmann}, \citenamefont {Heim}, \citenamefont {Kryvenda}, \citenamefont
  {Siljestr{\"{o}}m}, \citenamefont {Sj{\"{o}}vall},\ and\ \citenamefont
  {Thiel}}]{Leefmann_2013}%
  \BibitemOpen
  \bibfield  {author} {\bibinfo {author} {\bibfnamefont {T.}~\bibnamefont
  {Leefmann}}, \bibinfo {author} {\bibfnamefont {C.}~\bibnamefont {Heim}},
  \bibinfo {author} {\bibfnamefont {A.}~\bibnamefont {Kryvenda}}, \bibinfo
  {author} {\bibfnamefont {S.}~\bibnamefont {Siljestr{\"{o}}m}}, \bibinfo
  {author} {\bibfnamefont {P.}~\bibnamefont {Sj{\"{o}}vall}},\ and\ \bibinfo
  {author} {\bibfnamefont {V.}~\bibnamefont {Thiel}},\ }\bibfield  {title}
  {\enquote {\bibinfo {title} {Biomarker imaging of single diatom cells in a
  microbial mat using time-of-flight secondary ion mass spectrometry
  ({ToF}-{SIMS})},}\ }\href {https://doi.org/10.1016/j.orggeochem.2013.01.005}
  {\bibfield  {journal} {\bibinfo  {journal} {Org. Geochem.}\ }\textbf
  {\bibinfo {volume} {57}},\ \bibinfo {pages} {23--33} (\bibinfo {year}
  {2013})}\BibitemShut {NoStop}%
\bibitem [{\citenamefont {Moreno}\ \emph {et~al.}(2018)\citenamefont {Moreno},
  \citenamefont {Mouton}, \citenamefont {Chevalier}, \citenamefont {Barnes},
  \citenamefont {Bassani},\ and\ \citenamefont {Gautier}}]{Moreno_2018}%
  \BibitemOpen
  \bibfield  {author} {\bibinfo {author} {\bibfnamefont {M.~A.}\ \bibnamefont
  {Moreno}}, \bibinfo {author} {\bibfnamefont {I.}~\bibnamefont {Mouton}},
  \bibinfo {author} {\bibfnamefont {N.}~\bibnamefont {Chevalier}}, \bibinfo
  {author} {\bibfnamefont {J.-P.}\ \bibnamefont {Barnes}}, \bibinfo {author}
  {\bibfnamefont {F.}~\bibnamefont {Bassani}},\ and\ \bibinfo {author}
  {\bibfnamefont {B.}~\bibnamefont {Gautier}},\ }\bibfield  {title} {\enquote
  {\bibinfo {title} {Combined {ToF}-{SIMS} and {AFM} protocol for accurate 3{D}
  chemical analysis and data visualization},}\ }\href
  {https://doi.org/10.1116/1.5019464} {\bibfield  {journal} {\bibinfo
  {journal} {J. Vac. Sci. Technol. B}\ }\textbf {\bibinfo {volume} {36}},\
  \bibinfo {pages} {03F122} (\bibinfo {year} {2018})}\BibitemShut {NoStop}%
\bibitem [{\citenamefont {Laufer}\ \emph {et~al.}(2017)\citenamefont {Laufer},
  \citenamefont {Nordhoff}, \citenamefont {Halama}, \citenamefont {Martinez},
  \citenamefont {Obst}, \citenamefont {Nowak}, \citenamefont {Stryhanyuk},
  \citenamefont {Richnow},\ and\ \citenamefont {Kappler}}]{Laufer_2017}%
  \BibitemOpen
  \bibfield  {author} {\bibinfo {author} {\bibfnamefont {K.}~\bibnamefont
  {Laufer}}, \bibinfo {author} {\bibfnamefont {M.}~\bibnamefont {Nordhoff}},
  \bibinfo {author} {\bibfnamefont {M.}~\bibnamefont {Halama}}, \bibinfo
  {author} {\bibfnamefont {R.~E.}\ \bibnamefont {Martinez}}, \bibinfo {author}
  {\bibfnamefont {M.}~\bibnamefont {Obst}}, \bibinfo {author} {\bibfnamefont
  {M.}~\bibnamefont {Nowak}}, \bibinfo {author} {\bibfnamefont
  {H.}~\bibnamefont {Stryhanyuk}}, \bibinfo {author} {\bibfnamefont {H.~H.}\
  \bibnamefont {Richnow}},\ and\ \bibinfo {author} {\bibfnamefont
  {A.}~\bibnamefont {Kappler}},\ }\bibfield  {title} {\enquote {\bibinfo
  {title} {Microaerophilic fe(ii)-oxidizing zetaproteobacteria isolated from
  low-fe marine coastal sediments: physiology and composition of their twisted
  stalks},}\ }\href {https://doi.org/10.1128/AEM.03118-16} {\bibfield
  {journal} {\bibinfo  {journal} {Appl. Environ. Microbiol.}\ }\textbf
  {\bibinfo {volume} {83}},\ \bibinfo {pages} {e03118--16} (\bibinfo {year}
  {2017})}\BibitemShut {NoStop}%
\bibitem [{\citenamefont {Tamisier}\ \emph {et~al.}(2022)\citenamefont
  {Tamisier}, \citenamefont {Schmidt}, \citenamefont {Vogt}, \citenamefont
  {K{\"{u}}mmel}, \citenamefont {Stryhanyuk}, \citenamefont {Musat},
  \citenamefont {Richnow},\ and\ \citenamefont {Musat}}]{Tamisier_2022}%
  \BibitemOpen
  \bibfield  {author} {\bibinfo {author} {\bibfnamefont {M.}~\bibnamefont
  {Tamisier}}, \bibinfo {author} {\bibfnamefont {M.}~\bibnamefont {Schmidt}},
  \bibinfo {author} {\bibfnamefont {C.}~\bibnamefont {Vogt}}, \bibinfo {author}
  {\bibfnamefont {S.}~\bibnamefont {K{\"{u}}mmel}}, \bibinfo {author}
  {\bibfnamefont {H.}~\bibnamefont {Stryhanyuk}}, \bibinfo {author}
  {\bibfnamefont {N.}~\bibnamefont {Musat}}, \bibinfo {author} {\bibfnamefont
  {H.-H.}\ \bibnamefont {Richnow}},\ and\ \bibinfo {author} {\bibfnamefont
  {F.}~\bibnamefont {Musat}},\ }\bibfield  {title} {\enquote {\bibinfo {title}
  {Iron corrosion by methanogenic archaea characterized by stable isotope
  effects and crust mineralogy},}\ }\href
  {https://doi.org/10.1111/1462-2920.15658} {\bibfield  {journal} {\bibinfo
  {journal} {Environ. Microbiol.}\ }\textbf {\bibinfo {volume} {24}},\ \bibinfo
  {pages} {583--595} (\bibinfo {year} {2022})}\BibitemShut {NoStop}%
\bibitem [{\citenamefont {Bandara}\ \emph {et~al.}(2020)\citenamefont
  {Bandara}, \citenamefont {Ballerin}, \citenamefont {Lepp{\"{a}}nen},
  \citenamefont {Tesfamichael}, \citenamefont {Ostrikov},\ and\ \citenamefont
  {Whitchurch}}]{Bandara_2020a}%
  \BibitemOpen
  \bibfield  {author} {\bibinfo {author} {\bibfnamefont {C.~D.}\ \bibnamefont
  {Bandara}}, \bibinfo {author} {\bibfnamefont {G.}~\bibnamefont {Ballerin}},
  \bibinfo {author} {\bibfnamefont {M.}~\bibnamefont {Lepp{\"{a}}nen}},
  \bibinfo {author} {\bibfnamefont {T.}~\bibnamefont {Tesfamichael}}, \bibinfo
  {author} {\bibfnamefont {K.~K.}\ \bibnamefont {Ostrikov}},\ and\ \bibinfo
  {author} {\bibfnamefont {C.~B.}\ \bibnamefont {Whitchurch}},\ }\bibfield
  {title} {\enquote {\bibinfo {title} {Resolving bio{\textendash}nano
  interactions of {E}. {C}oli bacteria{\textendash}dragonfly wing interface
  with helium ion and 3{D}-structured illumination microscopy to understand
  bacterial death on nanotopography},}\ }\href
  {https://doi.org/10.1021/acsbiomaterials.9b01973} {\bibfield  {journal}
  {\bibinfo  {journal} {ACS Biomater. Sci. Eng.}\ }\textbf {\bibinfo {volume}
  {6}},\ \bibinfo {pages} {3925--3932} (\bibinfo {year} {2020})}\BibitemShut
  {NoStop}%
\bibitem [{\citenamefont {Hlawacek}\ \emph {et~al.}(2013)\citenamefont
  {Hlawacek}, \citenamefont {Ahmad}, \citenamefont {Smithers},\ and\
  \citenamefont {Kooij}}]{Hlawacek_2013}%
  \BibitemOpen
  \bibfield  {author} {\bibinfo {author} {\bibfnamefont {G.}~\bibnamefont
  {Hlawacek}}, \bibinfo {author} {\bibfnamefont {I.}~\bibnamefont {Ahmad}},
  \bibinfo {author} {\bibfnamefont {M.~A.}\ \bibnamefont {Smithers}},\ and\
  \bibinfo {author} {\bibfnamefont {E.~S.}\ \bibnamefont {Kooij}},\ }\bibfield
  {title} {{\selectlanguage {english}\enquote {\bibinfo {title} {To see or not
  to see: imaging surfactant coated nano-particles using {HIM} and {SEM}},}\
  }}\href {https://doi.org/10.1016/j.ultramic.2013.07.010} {\bibfield
  {journal} {\bibinfo  {journal} {Ultramicroscopy}\ }\textbf {\bibinfo {volume}
  {135}},\ \bibinfo {pages} {89--94} (\bibinfo {year} {2013})}\BibitemShut
  {NoStop}%
\bibitem [{\citenamefont {Golding}\ \emph {et~al.}(2016)\citenamefont
  {Golding}, \citenamefont {Lamboo}, \citenamefont {Beniac},\ and\
  \citenamefont {Booth}}]{Golding_2016a}%
  \BibitemOpen
  \bibfield  {author} {\bibinfo {author} {\bibfnamefont {C.~G.}\ \bibnamefont
  {Golding}}, \bibinfo {author} {\bibfnamefont {L.~L.}\ \bibnamefont {Lamboo}},
  \bibinfo {author} {\bibfnamefont {D.~R.}\ \bibnamefont {Beniac}},\ and\
  \bibinfo {author} {\bibfnamefont {T.~F.}\ \bibnamefont {Booth}},\ }\bibfield
  {title} {\enquote {\bibinfo {title} {The scanning electron microscope in
  microbiology and diagnosis of infectious disease},}\ }\href
  {https://doi.org/10.1038/srep26516} {\bibfield  {journal} {\bibinfo
  {journal} {Sci. Rep.}\ }\textbf {\bibinfo {volume} {6}} (\bibinfo {year}
  {2016}),\ 10.1038/srep26516}\BibitemShut {NoStop}%
\bibitem [{\citenamefont {Schmidt}, \citenamefont {Byrne},\ and\ \citenamefont
  {Maasilta}(2021)}]{Schmidt_2021a}%
  \BibitemOpen
  \bibfield  {author} {\bibinfo {author} {\bibfnamefont {M.}~\bibnamefont
  {Schmidt}}, \bibinfo {author} {\bibfnamefont {J.~M.}\ \bibnamefont {Byrne}},\
  and\ \bibinfo {author} {\bibfnamefont {I.~J.}\ \bibnamefont {Maasilta}},\
  }\bibfield  {title} {\enquote {\bibinfo {title} {Bio-imaging with the
  helium-ion microscope: a review},}\ }\href
  {https://doi.org/10.3762/bjnano.12.1} {\bibfield  {journal} {\bibinfo
  {journal} {Beilstein J. Nanotechnol.}\ }\textbf {\bibinfo {volume} {12}},\
  \bibinfo {pages} {1--23} (\bibinfo {year} {2021})}\BibitemShut {NoStop}%
\bibitem [{\citenamefont {Vinner}\ \emph {et~al.}(2019)\citenamefont {Vinner},
  \citenamefont {Rezaie-Yazdi}, \citenamefont {Leppanen}, \citenamefont
  {Stapley}, \citenamefont {Leaper},\ and\ \citenamefont
  {Malik}}]{Vinner_2019a}%
  \BibitemOpen
  \bibfield  {author} {\bibinfo {author} {\bibfnamefont {G.}~\bibnamefont
  {Vinner}}, \bibinfo {author} {\bibfnamefont {Z.}~\bibnamefont
  {Rezaie-Yazdi}}, \bibinfo {author} {\bibfnamefont {M.}~\bibnamefont
  {Leppanen}}, \bibinfo {author} {\bibfnamefont {A.}~\bibnamefont {Stapley}},
  \bibinfo {author} {\bibfnamefont {M.}~\bibnamefont {Leaper}},\ and\ \bibinfo
  {author} {\bibfnamefont {D.}~\bibnamefont {Malik}},\ }\bibfield  {title}
  {\enquote {\bibinfo {title} {Microencapsulation of salmonella-specific
  bacteriophage felix o1 using spray-drying in a {pH}-responsive formulation
  and direct compression tableting of powders into a solid oral dosage form},}\
  }\href {https://doi.org/10.3390/ph12010043} {\bibfield  {journal} {\bibinfo
  {journal} {Pharmaceuticals}\ }\textbf {\bibinfo {volume} {12}},\ \bibinfo
  {pages} {43} (\bibinfo {year} {2019})}\BibitemShut {NoStop}%
\bibitem [{\citenamefont {Almeida}\ \emph {et~al.}(2018)\citenamefont
  {Almeida}, \citenamefont {Lepp{\"{a}}nen}, \citenamefont {Maasilta},\ and\
  \citenamefont {Sundberg}}]{Almeida_2018a}%
  \BibitemOpen
  \bibfield  {author} {\bibinfo {author} {\bibfnamefont {G.~M.~F.}\
  \bibnamefont {Almeida}}, \bibinfo {author} {\bibfnamefont {M.}~\bibnamefont
  {Lepp{\"{a}}nen}}, \bibinfo {author} {\bibfnamefont {I.~J.}\ \bibnamefont
  {Maasilta}},\ and\ \bibinfo {author} {\bibfnamefont {L.-R.}\ \bibnamefont
  {Sundberg}},\ }\bibfield  {title} {\enquote {\bibinfo {title} {Bacteriophage
  imaging: past, present and future},}\ }\href
  {https://doi.org/10.1016/j.resmic.2018.05.006} {\bibfield  {journal}
  {\bibinfo  {journal} {Res. Microbiol.}\ }\textbf {\bibinfo {volume} {169}},\
  \bibinfo {pages} {488--494} (\bibinfo {year} {2018})}\BibitemShut {NoStop}%
\bibitem [{\citenamefont {Barreto-Vieira}\ \emph {et~al.}(2022)\citenamefont
  {Barreto-Vieira}, \citenamefont {da~Silva}, \citenamefont {de~Almeida},
  \citenamefont {da~Costa~Rasinhas}, \citenamefont {Monteiro}, \citenamefont
  {Miranda}, \citenamefont {Motta}, \citenamefont {Siqueira}, \citenamefont
  {Girard-Dias}, \citenamefont {Archanjo}, \citenamefont {Bozza}, \citenamefont
  {Souza}, \citenamefont {Dias}, \citenamefont {Soares},\ and\ \citenamefont
  {Barth}}]{BarretoVieira_2022}%
  \BibitemOpen
  \bibfield  {author} {\bibinfo {author} {\bibfnamefont {D.~F.}\ \bibnamefont
  {Barreto-Vieira}}, \bibinfo {author} {\bibfnamefont {M.~A.~N.}\ \bibnamefont
  {da~Silva}}, \bibinfo {author} {\bibfnamefont {A.~L.~T.}\ \bibnamefont
  {de~Almeida}}, \bibinfo {author} {\bibfnamefont {A.}~\bibnamefont
  {da~Costa~Rasinhas}}, \bibinfo {author} {\bibfnamefont {M.~E.}\ \bibnamefont
  {Monteiro}}, \bibinfo {author} {\bibfnamefont {M.~D.}\ \bibnamefont
  {Miranda}}, \bibinfo {author} {\bibfnamefont {F.~C.}\ \bibnamefont {Motta}},
  \bibinfo {author} {\bibfnamefont {M.~M.}\ \bibnamefont {Siqueira}}, \bibinfo
  {author} {\bibfnamefont {W.}~\bibnamefont {Girard-Dias}}, \bibinfo {author}
  {\bibfnamefont {B.~S.}\ \bibnamefont {Archanjo}}, \bibinfo {author}
  {\bibfnamefont {P.~T.}\ \bibnamefont {Bozza}}, \bibinfo {author}
  {\bibfnamefont {T.~M.~L.}\ \bibnamefont {Souza}}, \bibinfo {author}
  {\bibfnamefont {S.~S.~G.}\ \bibnamefont {Dias}}, \bibinfo {author}
  {\bibfnamefont {V.~C.}\ \bibnamefont {Soares}},\ and\ \bibinfo {author}
  {\bibfnamefont {O.~M.}\ \bibnamefont {Barth}},\ }\bibfield  {title} {\enquote
  {\bibinfo {title} {{SARS}-{CoV}-2: ultrastructural characterization of
  morphogenesis in an in vitro system},}\ }\href
  {https://doi.org/10.3390/v14020201} {\bibfield  {journal} {\bibinfo
  {journal} {Viruses}\ }\textbf {\bibinfo {volume} {14}},\ \bibinfo {pages}
  {201} (\bibinfo {year} {2022})}\BibitemShut {NoStop}%
\bibitem [{\citenamefont {LeTourneau}\ \emph {et~al.}(2015)\citenamefont
  {LeTourneau}, \citenamefont {Marshall}, \citenamefont {Thomashow},\ and\
  \citenamefont {Harsh}}]{LeTourneau_2015a}%
  \BibitemOpen
  \bibfield  {author} {\bibinfo {author} {\bibfnamefont {M.~K.}\ \bibnamefont
  {LeTourneau}}, \bibinfo {author} {\bibfnamefont {M.~M.}\ \bibnamefont
  {Marshall}}, \bibinfo {author} {\bibfnamefont {L.~S.}\ \bibnamefont
  {Thomashow}},\ and\ \bibinfo {author} {\bibfnamefont {J.~B.}\ \bibnamefont
  {Harsh}},\ }\bibfield  {title} {\enquote {\bibinfo {title} {Impact of
  phenazine-1-carboxylic acid upon biofilm development in the rhizosphere of
  dryland and irrigated wheat},}\ }\href
  {https://doi.org/10.1017/s1431927615004353} {\bibfield  {journal} {\bibinfo
  {journal} {Microsc. Microanal.}\ }\textbf {\bibinfo {volume} {21}},\ \bibinfo
  {pages} {711--712} (\bibinfo {year} {2015})}\BibitemShut {NoStop}%
\bibitem [{\citenamefont {Rice}\ \emph {et~al.}(2013)\citenamefont {Rice},
  \citenamefont {Hoek}, \citenamefont {P{\u{a}}unescu}, \citenamefont {Huynh},
  \citenamefont {Goetze}, \citenamefont {Singh}, \citenamefont {Scipioni},
  \citenamefont {Stern},\ and\ \citenamefont {Brown}}]{Rice_2013a}%
  \BibitemOpen
  \bibfield  {author} {\bibinfo {author} {\bibfnamefont {W.~L.}\ \bibnamefont
  {Rice}}, \bibinfo {author} {\bibfnamefont {A.~N.~V.}\ \bibnamefont {Hoek}},
  \bibinfo {author} {\bibfnamefont {T.~G.}\ \bibnamefont {P{\u{a}}unescu}},
  \bibinfo {author} {\bibfnamefont {C.}~\bibnamefont {Huynh}}, \bibinfo
  {author} {\bibfnamefont {B.}~\bibnamefont {Goetze}}, \bibinfo {author}
  {\bibfnamefont {B.}~\bibnamefont {Singh}}, \bibinfo {author} {\bibfnamefont
  {L.}~\bibnamefont {Scipioni}}, \bibinfo {author} {\bibfnamefont {L.~A.}\
  \bibnamefont {Stern}},\ and\ \bibinfo {author} {\bibfnamefont
  {D.}~\bibnamefont {Brown}},\ }\bibfield  {title} {\enquote {\bibinfo {title}
  {High resolution helium ion scanning microscopy of the rat kidney},}\ }\href
  {https://doi.org/10.1371/journal.pone.0057051} {\bibfield  {journal}
  {\bibinfo  {journal} {{PLoS} {ONE}}\ }\textbf {\bibinfo {volume} {8}},\
  \bibinfo {pages} {e57051} (\bibinfo {year} {2013})}\BibitemShut {NoStop}%
\bibitem [{\citenamefont {Chen}\ \emph {et~al.}(2011)\citenamefont {Chen},
  \citenamefont {Udalagama}, \citenamefont {Chen}, \citenamefont {Bettiol},
  \citenamefont {Pickard}, \citenamefont {Venkatesan},\ and\ \citenamefont
  {Watt}}]{Chen_2011a}%
  \BibitemOpen
  \bibfield  {author} {\bibinfo {author} {\bibfnamefont {X.}~\bibnamefont
  {Chen}}, \bibinfo {author} {\bibfnamefont {C.~N.~B.}\ \bibnamefont
  {Udalagama}}, \bibinfo {author} {\bibfnamefont {C.-B.}\ \bibnamefont {Chen}},
  \bibinfo {author} {\bibfnamefont {A.~A.}\ \bibnamefont {Bettiol}}, \bibinfo
  {author} {\bibfnamefont {D.~S.}\ \bibnamefont {Pickard}}, \bibinfo {author}
  {\bibfnamefont {T.}~\bibnamefont {Venkatesan}},\ and\ \bibinfo {author}
  {\bibfnamefont {F.}~\bibnamefont {Watt}},\ }\bibfield  {title} {\enquote
  {\bibinfo {title} {Whole-cell imaging at nanometer resolutions using fast and
  slow focused helium ions},}\ }\href
  {https://doi.org/10.1016/j.bpj.2011.08.028} {\bibfield  {journal} {\bibinfo
  {journal} {Biophys. J.}\ }\textbf {\bibinfo {volume} {101}},\ \bibinfo
  {pages} {1788--1793} (\bibinfo {year} {2011})}\BibitemShut {NoStop}%
\bibitem [{\citenamefont {Sch{\"{u}}rmann}\ \emph {et~al.}(2018)\citenamefont
  {Sch{\"{u}}rmann}, \citenamefont {Shepheard}, \citenamefont {Frese},
  \citenamefont {Geishendorf}, \citenamefont {Sudhoff}, \citenamefont
  {G{\"{o}}lzh{\"{a}}user}, \citenamefont {R{\"{u}}ckert}, \citenamefont
  {Kaltschmidt}, \citenamefont {Kaltschmidt},\ and\ \citenamefont
  {Thomas}}]{Schuermann_2018}%
  \BibitemOpen
  \bibfield  {author} {\bibinfo {author} {\bibfnamefont {M.}~\bibnamefont
  {Sch{\"{u}}rmann}}, \bibinfo {author} {\bibfnamefont {N.}~\bibnamefont
  {Shepheard}}, \bibinfo {author} {\bibfnamefont {N.}~\bibnamefont {Frese}},
  \bibinfo {author} {\bibfnamefont {K.}~\bibnamefont {Geishendorf}}, \bibinfo
  {author} {\bibfnamefont {H.}~\bibnamefont {Sudhoff}}, \bibinfo {author}
  {\bibfnamefont {A.}~\bibnamefont {G{\"{o}}lzh{\"{a}}user}}, \bibinfo {author}
  {\bibfnamefont {U.}~\bibnamefont {R{\"{u}}ckert}}, \bibinfo {author}
  {\bibfnamefont {C.}~\bibnamefont {Kaltschmidt}}, \bibinfo {author}
  {\bibfnamefont {B.}~\bibnamefont {Kaltschmidt}},\ and\ \bibinfo {author}
  {\bibfnamefont {A.}~\bibnamefont {Thomas}},\ }\bibfield  {title} {\enquote
  {\bibinfo {title} {Technical feasibility study for production of tailored
  multielectrode arrays and patterning of arranged neuronal networks},}\ }\href
  {https://doi.org/10.1371/journal.pone.0192647} {\bibfield  {journal}
  {\bibinfo  {journal} {{PLOS} {ONE}}\ }\textbf {\bibinfo {volume} {13}},\
  \bibinfo {pages} {e0192647} (\bibinfo {year} {2018})}\BibitemShut {NoStop}%
\bibitem [{\citenamefont {Sato}, \citenamefont {Sato},\ and\ \citenamefont
  {Ogawa}(2018)}]{Sato_2018a}%
  \BibitemOpen
  \bibfield  {author} {\bibinfo {author} {\bibfnamefont {C.}~\bibnamefont
  {Sato}}, \bibinfo {author} {\bibfnamefont {M.}~\bibnamefont {Sato}},\ and\
  \bibinfo {author} {\bibfnamefont {S.}~\bibnamefont {Ogawa}},\ }\bibfield
  {title} {\enquote {\bibinfo {title} {Imaging of immunogold labeling in cells
  and tissues by helium ion microscopy},}\ }\href
  {https://doi.org/10.3892/ijmm.2018.3604} {\bibfield  {journal} {\bibinfo
  {journal} {Int. J. Mol. Med.}\ } (\bibinfo {year} {2018}),\
  10.3892/ijmm.2018.3604}\BibitemShut {NoStop}%
\bibitem [{\citenamefont {Takei}\ \emph {et~al.}(2021)\citenamefont {Takei},
  \citenamefont {Nagata}, \citenamefont {Frese}, \citenamefont
  {G{\"{o}}lzh{\"{a}}user},\ and\ \citenamefont {Okamoto}}]{Takei_2021a}%
  \BibitemOpen
  \bibfield  {author} {\bibinfo {author} {\bibfnamefont {H.}~\bibnamefont
  {Takei}}, \bibinfo {author} {\bibfnamefont {K.}~\bibnamefont {Nagata}},
  \bibinfo {author} {\bibfnamefont {N.}~\bibnamefont {Frese}}, \bibinfo
  {author} {\bibfnamefont {A.}~\bibnamefont {G{\"{o}}lzh{\"{a}}user}},\ and\
  \bibinfo {author} {\bibfnamefont {T.}~\bibnamefont {Okamoto}},\ }\bibfield
  {title} {\enquote {\bibinfo {title} {Surface-enhanced raman spectroscopy for
  molecule characterization: {HIM} investigation into sources of {SERS}
  activity of silver-coated butterfly scales},}\ }\href
  {https://doi.org/10.3390/nano11071741} {\bibfield  {journal} {\bibinfo
  {journal} {Nanomaterials}\ }\textbf {\bibinfo {volume} {11}},\ \bibinfo
  {pages} {1741} (\bibinfo {year} {2021})}\BibitemShut {NoStop}%
\bibitem [{\citenamefont {Bandara}\ \emph {et~al.}(2021)\citenamefont
  {Bandara}, \citenamefont {Schmidt}, \citenamefont {Davoudpour}, \citenamefont
  {Stryhanyuk}, \citenamefont {Richnow},\ and\ \citenamefont
  {Musat}}]{Bandara_2021a}%
  \BibitemOpen
  \bibfield  {author} {\bibinfo {author} {\bibfnamefont {C.~D.}\ \bibnamefont
  {Bandara}}, \bibinfo {author} {\bibfnamefont {M.}~\bibnamefont {Schmidt}},
  \bibinfo {author} {\bibfnamefont {Y.}~\bibnamefont {Davoudpour}}, \bibinfo
  {author} {\bibfnamefont {H.}~\bibnamefont {Stryhanyuk}}, \bibinfo {author}
  {\bibfnamefont {H.~H.}\ \bibnamefont {Richnow}},\ and\ \bibinfo {author}
  {\bibfnamefont {N.}~\bibnamefont {Musat}},\ }\bibfield  {title} {\enquote
  {\bibinfo {title} {Microbial identification, high-resolution microscopy and
  spectrometry of the rhizosphere in its native spatial context},}\ }\href
  {https://doi.org/10.3389/fpls.2021.668929} {\bibfield  {journal} {\bibinfo
  {journal} {Front. Plant Sci.}\ }\textbf {\bibinfo {volume} {12}} (\bibinfo
  {year} {2021}),\ 10.3389/fpls.2021.668929}\BibitemShut {NoStop}%
\bibitem [{\citenamefont {Denk}\ and\ \citenamefont
  {Horstmann}(2004)}]{Denk_2004}%
  \BibitemOpen
  \bibfield  {author} {\bibinfo {author} {\bibfnamefont {W.}~\bibnamefont
  {Denk}}\ and\ \bibinfo {author} {\bibfnamefont {H.}~\bibnamefont
  {Horstmann}},\ }\bibfield  {title} {\enquote {\bibinfo {title} {Serial
  block-face scanning electron microscopy to reconstruct three-dimensional
  tissue nanostructure},}\ }\href
  {https://doi.org/10.1371/journal.pbio.0020329} {\bibfield  {journal}
  {\bibinfo  {journal} {PLOS Biol.}\ }\textbf {\bibinfo {volume} {2}},\
  \bibinfo {pages} {null} (\bibinfo {year} {2004})}\BibitemShut {NoStop}%
\bibitem [{\citenamefont {Harris}\ \emph {et~al.}(2006)\citenamefont {Harris},
  \citenamefont {Perry}, \citenamefont {Bourne}, \citenamefont {Feinberg},
  \citenamefont {Ostroff},\ and\ \citenamefont {Hurlburt}}]{Harris_2006}%
  \BibitemOpen
  \bibfield  {author} {\bibinfo {author} {\bibfnamefont {K.~M.}\ \bibnamefont
  {Harris}}, \bibinfo {author} {\bibfnamefont {E.}~\bibnamefont {Perry}},
  \bibinfo {author} {\bibfnamefont {J.}~\bibnamefont {Bourne}}, \bibinfo
  {author} {\bibfnamefont {M.}~\bibnamefont {Feinberg}}, \bibinfo {author}
  {\bibfnamefont {L.}~\bibnamefont {Ostroff}},\ and\ \bibinfo {author}
  {\bibfnamefont {J.}~\bibnamefont {Hurlburt}},\ }\bibfield  {title} {\enquote
  {\bibinfo {title} {Uniform serial sectioning for transmission electron
  microscopy},}\ }\href {https://doi.org/10.1523/JNEUROSCI.3994-06.2006}
  {\bibfield  {journal} {\bibinfo  {journal} {J. Neurosci.}\ }\textbf {\bibinfo
  {volume} {26}},\ \bibinfo {pages} {12101--12103} (\bibinfo {year}
  {2006})}\BibitemShut {NoStop}%
\bibitem [{\citenamefont {Narayan}\ and\ \citenamefont
  {Subramaniam}(2015)}]{Narayan_2015}%
  \BibitemOpen
  \bibfield  {author} {\bibinfo {author} {\bibfnamefont {K.}~\bibnamefont
  {Narayan}}\ and\ \bibinfo {author} {\bibfnamefont {S.}~\bibnamefont
  {Subramaniam}},\ }\bibfield  {title} {\enquote {\bibinfo {title} {{Focused
  ion beams in biology}},}\ }\href {https://doi.org/10.1038/nmeth.3623}
  {\bibfield  {journal} {\bibinfo  {journal} {Nat. Methods}\ }\textbf {\bibinfo
  {volume} {12}},\ \bibinfo {pages} {1021--1031} (\bibinfo {year}
  {2015})}\BibitemShut {NoStop}%
\bibitem [{\citenamefont {Gorelick}\ \emph {et~al.}(2018)\citenamefont
  {Gorelick}, \citenamefont {Korneev}, \citenamefont {Handley}, \citenamefont
  {Gervinskas}, \citenamefont {Oorschot}, \citenamefont {Kaluza}, \citenamefont
  {Law}, \citenamefont {O'Bryan}, \citenamefont {Pocock}, \citenamefont
  {Whisstock},\ and\ \citenamefont {de~Marco}}]{Sergey_2018}%
  \BibitemOpen
  \bibfield  {author} {\bibinfo {author} {\bibfnamefont {S.}~\bibnamefont
  {Gorelick}}, \bibinfo {author} {\bibfnamefont {D.}~\bibnamefont {Korneev}},
  \bibinfo {author} {\bibfnamefont {A.}~\bibnamefont {Handley}}, \bibinfo
  {author} {\bibfnamefont {G.}~\bibnamefont {Gervinskas}}, \bibinfo {author}
  {\bibfnamefont {V.}~\bibnamefont {Oorschot}}, \bibinfo {author}
  {\bibfnamefont {O.~L.}\ \bibnamefont {Kaluza}}, \bibinfo {author}
  {\bibfnamefont {R.~H.~P.}\ \bibnamefont {Law}}, \bibinfo {author}
  {\bibfnamefont {M.}~\bibnamefont {O'Bryan}}, \bibinfo {author} {\bibfnamefont
  {R.}~\bibnamefont {Pocock}}, \bibinfo {author} {\bibfnamefont {J.~C.}\
  \bibnamefont {Whisstock}},\ and\ \bibinfo {author} {\bibfnamefont
  {A.}~\bibnamefont {de~Marco}},\ }\href {https://doi.org/10.1101/457820}
  {\enquote {\bibinfo {title} {Oxygen plasma focused ion beam scanning electron
  microscopy for biological samples},}\ } (\bibinfo {year} {2018}),\ \bibinfo
  {note} {preprint},\ \Eprint {https://arxiv.org/abs/457820} {bioRxiv:457820}
  \BibitemShut {NoStop}%
\bibitem [{\citenamefont {Hartnell}\ \emph {et~al.}(2016)\citenamefont
  {Hartnell}, \citenamefont {Earl}, \citenamefont {Bliss}, \citenamefont
  {Moran},\ and\ \citenamefont {Subramaniam}}]{Hartnell_2016}%
  \BibitemOpen
  \bibfield  {author} {\bibinfo {author} {\bibfnamefont {L.~M.}\ \bibnamefont
  {Hartnell}}, \bibinfo {author} {\bibfnamefont {L.~A.}\ \bibnamefont {Earl}},
  \bibinfo {author} {\bibfnamefont {D.}~\bibnamefont {Bliss}}, \bibinfo
  {author} {\bibfnamefont {A.}~\bibnamefont {Moran}},\ and\ \bibinfo {author}
  {\bibfnamefont {S.}~\bibnamefont {Subramaniam}},\ }\bibfield  {title}
  {\enquote {\bibinfo {title} {Imaging cellular architecture with 3{D}
  {SEM}},}\ }in\ \href {https://doi.org/10.1016/b978-0-12-394447-4.20085-0}
  {\emph {\bibinfo {booktitle} {Encyclopedia of Cell Biology}}}\ (\bibinfo
  {publisher} {Elsevier},\ \bibinfo {year} {2016})\ pp.\ \bibinfo {pages}
  {44--50}\BibitemShut {NoStop}%
\bibitem [{\citenamefont {Leser}\ \emph {et~al.}(2009)\citenamefont {Leser},
  \citenamefont {Drobne}, \citenamefont {Pipan}, \citenamefont {Milani},\ and\
  \citenamefont {Tatti}}]{Leser_2009}%
  \BibitemOpen
  \bibfield  {author} {\bibinfo {author} {\bibfnamefont {V.}~\bibnamefont
  {Leser}}, \bibinfo {author} {\bibfnamefont {D.}~\bibnamefont {Drobne}},
  \bibinfo {author} {\bibfnamefont {Z.}~\bibnamefont {Pipan}}, \bibinfo
  {author} {\bibfnamefont {M.}~\bibnamefont {Milani}},\ and\ \bibinfo {author}
  {\bibfnamefont {F.}~\bibnamefont {Tatti}},\ }\bibfield  {title} {\enquote
  {\bibinfo {title} {Comparison of different preparation methods of biological
  samples for {FIB} milling and {SEM} investigation},}\ }\href
  {https://doi.org/10.1111/j.1365-2818.2009.03121.x} {\bibfield  {journal}
  {\bibinfo  {journal} {J. Microsc.}\ }\textbf {\bibinfo {volume} {233}},\
  \bibinfo {pages} {309--319} (\bibinfo {year} {2009})}\BibitemShut {NoStop}%
\bibitem [{\citenamefont {Knott}, \citenamefont {Rosset},\ and\ \citenamefont
  {Cantoni}(2011)}]{Knott_2011}%
  \BibitemOpen
  \bibfield  {author} {\bibinfo {author} {\bibfnamefont {G.}~\bibnamefont
  {Knott}}, \bibinfo {author} {\bibfnamefont {S.}~\bibnamefont {Rosset}},\ and\
  \bibinfo {author} {\bibfnamefont {M.}~\bibnamefont {Cantoni}},\ }\bibfield
  {title} {\enquote {\bibinfo {title} {Focussed ion beam milling and scanning
  electron microscopy of brain tissue},}\ }\href
  {https://doi.org/10.3791/2588(2011)} {\bibfield  {journal} {\bibinfo
  {journal} {J. Vis. Exp.}\ }\textbf {\bibinfo {volume} {53}},\ \bibinfo
  {pages} {e2588} (\bibinfo {year} {2011})}\BibitemShut {NoStop}%
\bibitem [{\citenamefont {Kizilyaprak}\ \emph {et~al.}(2014)\citenamefont
  {Kizilyaprak}, \citenamefont {Bittermann}, \citenamefont {Daraspe},\ and\
  \citenamefont {Humbel}}]{Kizilyaprak_2014}%
  \BibitemOpen
  \bibfield  {author} {\bibinfo {author} {\bibfnamefont {C.}~\bibnamefont
  {Kizilyaprak}}, \bibinfo {author} {\bibfnamefont {A.~G.}\ \bibnamefont
  {Bittermann}}, \bibinfo {author} {\bibfnamefont {J.}~\bibnamefont
  {Daraspe}},\ and\ \bibinfo {author} {\bibfnamefont {B.~M.}\ \bibnamefont
  {Humbel}},\ }\bibfield  {title} {\enquote {\bibinfo {title} {{FIB-SEM}
  tomography in biology},}\ }in\ \href
  {https://doi.org/10.1007/978-1-62703-776-1_24} {\emph {\bibinfo {booktitle}
  {Electron Microscopy: Methods and Protocols, Methods in Molecular
  Biology}}},\ Vol.\ \bibinfo {volume} {1117},\ \bibinfo {editor} {edited by\
  \bibinfo {editor} {\bibfnamefont {J.}~\bibnamefont {Kuo}}}\ (\bibinfo
  {publisher} {Springer Science+Business Media New York},\ \bibinfo {year}
  {2014})\ Chap.~\bibinfo {chapter} {24}\BibitemShut {NoStop}%
\bibitem [{\citenamefont {Drobne}(2013)}]{Drobne_2013}%
  \BibitemOpen
  \bibfield  {author} {\bibinfo {author} {\bibfnamefont {D.}~\bibnamefont
  {Drobne}},\ }\bibfield  {title} {\enquote {\bibinfo {title} {3{D} imaging of
  cells and tissues by focused ion beam/scanning electron microscopy
  ({FIB/SEM})},}\ }in\ \href {https://doi.org/10.1007/978-1-62703-137-0_16}
  {\emph {\bibinfo {booktitle} {Nanoimaging: Methods and Protocols, Methods in
  Molecular Biology}}},\ Vol.\ \bibinfo {volume} {950},\ \bibinfo {editor}
  {edited by\ \bibinfo {editor} {\bibfnamefont {A.~A.}\ \bibnamefont {Sousa}}\
  and\ \bibinfo {editor} {\bibfnamefont {M.~J.}\ \bibnamefont {Kruhlak}}}\
  (\bibinfo  {publisher} {Springer Science+Business Media},\ \bibinfo {year}
  {2013})\ Chap.~\bibinfo {chapter} {16}\BibitemShut {NoStop}%
\bibitem [{\citenamefont {Schmid}\ \emph {et~al.}(2014)\citenamefont {Schmid},
  \citenamefont {Zeitvogel}, \citenamefont {Hao}, \citenamefont {Ingino},
  \citenamefont {Floetenmeyer}, \citenamefont {Stierhof}, \citenamefont
  {Schroeppel}, \citenamefont {Burkhardt}, \citenamefont {Kappler},\ and\
  \citenamefont {Obst}}]{Schmid_2014}%
  \BibitemOpen
  \bibfield  {author} {\bibinfo {author} {\bibfnamefont {G.}~\bibnamefont
  {Schmid}}, \bibinfo {author} {\bibfnamefont {F.}~\bibnamefont {Zeitvogel}},
  \bibinfo {author} {\bibfnamefont {L.}~\bibnamefont {Hao}}, \bibinfo {author}
  {\bibfnamefont {P.}~\bibnamefont {Ingino}}, \bibinfo {author} {\bibfnamefont
  {M.}~\bibnamefont {Floetenmeyer}}, \bibinfo {author} {\bibfnamefont {Y.-D.}\
  \bibnamefont {Stierhof}}, \bibinfo {author} {\bibfnamefont {B.}~\bibnamefont
  {Schroeppel}}, \bibinfo {author} {\bibfnamefont {C.~J.}\ \bibnamefont
  {Burkhardt}}, \bibinfo {author} {\bibfnamefont {A.}~\bibnamefont {Kappler}},\
  and\ \bibinfo {author} {\bibfnamefont {M.}~\bibnamefont {Obst}},\ }\bibfield
  {title} {\enquote {\bibinfo {title} {3-d analysis of bacterial
  cell-(iron)mineral aggregates formed during fe(ii) oxidation by the
  nitrate-reducing acidovorax sp. strain bofen1 using complementary microscopy
  tomography approaches},}\ }\href {https://doi.org/10.1111/gbi.12088}
  {\bibfield  {journal} {\bibinfo  {journal} {Geobiology}\ }\textbf {\bibinfo
  {volume} {12}},\ \bibinfo {pages} {340--361} (\bibinfo {year}
  {2014})}\BibitemShut {NoStop}%
\bibitem [{\citenamefont {Marko}\ \emph {et~al.}(2006)\citenamefont {Marko},
  \citenamefont {Hsieh}, \citenamefont {Moberlychan}, \citenamefont
  {Mannella},\ and\ \citenamefont {Frank}}]{Marko_2006}%
  \BibitemOpen
  \bibfield  {author} {\bibinfo {author} {\bibfnamefont {M.}~\bibnamefont
  {Marko}}, \bibinfo {author} {\bibfnamefont {C.}~\bibnamefont {Hsieh}},
  \bibinfo {author} {\bibfnamefont {W.}~\bibnamefont {Moberlychan}}, \bibinfo
  {author} {\bibfnamefont {C.~A.}\ \bibnamefont {Mannella}},\ and\ \bibinfo
  {author} {\bibfnamefont {J.}~\bibnamefont {Frank}},\ }\bibfield  {title}
  {\enquote {\bibinfo {title} {Focused ion beam milling of vitreous water:
  prospects for an alternative to cryo-ultramicrotomy of frozen-hydrated
  biological samples},}\ }\href
  {https://doi.org/10.1111/j.1365-2818.2006.01567.x} {\bibfield  {journal}
  {\bibinfo  {journal} {J. Microsc.}\ }\textbf {\bibinfo {volume} {222}},\
  \bibinfo {pages} {42--47} (\bibinfo {year} {2006})}\BibitemShut {NoStop}%
\bibitem [{\citenamefont {Hayles}\ and\ \citenamefont
  {De~Winter}(2021)}]{Hayles_2021}%
  \BibitemOpen
  \bibfield  {author} {\bibinfo {author} {\bibfnamefont {M.~F.}\ \bibnamefont
  {Hayles}}\ and\ \bibinfo {author} {\bibfnamefont {D.~A.~M.}\ \bibnamefont
  {De~Winter}},\ }\bibfield  {title} {\enquote {\bibinfo {title} {An
  introduction to cryo-{FIB-SEM} cross-sectioning of frozen hydrated life
  cience samples},}\ }\href {https://doi.org/10.1111/jmi.12951} {\bibfield
  {journal} {\bibinfo  {journal} {J. Microsc.}\ }\textbf {\bibinfo {volume}
  {281}},\ \bibinfo {pages} {138--156} (\bibinfo {year} {2021})}\BibitemShut
  {NoStop}%
\bibitem [{\citenamefont {Lucas}\ \emph {et~al.}(2014)\citenamefont {Lucas},
  \citenamefont {Guenthert}, \citenamefont {Gasser}, \citenamefont {Lucas},\
  and\ \citenamefont {Wepf}}]{Lucas_2014}%
  \BibitemOpen
  \bibfield  {author} {\bibinfo {author} {\bibfnamefont {M.}~\bibnamefont
  {Lucas}}, \bibinfo {author} {\bibfnamefont {M.}~\bibnamefont {Guenthert}},
  \bibinfo {author} {\bibfnamefont {P.}~\bibnamefont {Gasser}}, \bibinfo
  {author} {\bibfnamefont {F.}~\bibnamefont {Lucas}},\ and\ \bibinfo {author}
  {\bibfnamefont {R.}~\bibnamefont {Wepf}},\ }\bibfield  {title} {\enquote
  {\bibinfo {title} {Correlative 3{D} imaging: {CLSM} and {FIB-SEM} tomography
  using high-pressure frozen, freeze-substituted biological samples},}\ }in\
  \href {https://doi.org/10.1007/978-1-62703-776-1_26} {\emph {\bibinfo
  {booktitle} {Electron Microscopy: Methods and Protocols, Methods in Molecular
  Biology}}},\ Vol.\ \bibinfo {volume} {1117},\ \bibinfo {editor} {edited by\
  \bibinfo {editor} {\bibfnamefont {J.}~\bibnamefont {Kuo}}}\ (\bibinfo
  {publisher} {Springer Science+Business Media New York},\ \bibinfo {year}
  {2014})\ Chap.~\bibinfo {chapter} {26}\BibitemShut {NoStop}%
\bibitem [{\citenamefont {Gorelick}\ \emph {et~al.}(2019)\citenamefont
  {Gorelick}, \citenamefont {Buckley}, \citenamefont {Gervinskas},
  \citenamefont {Johnson}, \citenamefont {Handley}, \citenamefont {Caggiano},
  \citenamefont {Whisstock}, \citenamefont {Pocock},\ and\ \citenamefont
  {de~Marco}}]{Gorelick_2019}%
  \BibitemOpen
  \bibfield  {author} {\bibinfo {author} {\bibfnamefont {S.}~\bibnamefont
  {Gorelick}}, \bibinfo {author} {\bibfnamefont {G.}~\bibnamefont {Buckley}},
  \bibinfo {author} {\bibfnamefont {G.}~\bibnamefont {Gervinskas}}, \bibinfo
  {author} {\bibfnamefont {T.~K.}\ \bibnamefont {Johnson}}, \bibinfo {author}
  {\bibfnamefont {A.}~\bibnamefont {Handley}}, \bibinfo {author} {\bibfnamefont
  {M.~P.}\ \bibnamefont {Caggiano}}, \bibinfo {author} {\bibfnamefont {J.~C.}\
  \bibnamefont {Whisstock}}, \bibinfo {author} {\bibfnamefont {R.}~\bibnamefont
  {Pocock}},\ and\ \bibinfo {author} {\bibfnamefont {A.}~\bibnamefont
  {de~Marco}},\ }\bibfield  {title} {\enquote {\bibinfo {title} {{PIE}-scope,
  integrated cryo-correlative light and {FIB}/{SEM} microscopy},}\ }\href
  {https://doi.org/10.7554/elife.45919} {\bibfield  {journal} {\bibinfo
  {journal} {{eLife}}\ }\textbf {\bibinfo {volume} {8}} (\bibinfo {year}
  {2019}),\ 10.7554/elife.45919}\BibitemShut {NoStop}%
\bibitem [{\citenamefont {Wang}\ \emph
  {et~al.}(2021{\natexlab{a}})\citenamefont {Wang}, \citenamefont {Randolph},
  \citenamefont {Wu}, \citenamefont {Botman}, \citenamefont {Schardt},
  \citenamefont {Bouchet-Marquis}, \citenamefont {Nan}, \citenamefont {Rue},\
  and\ \citenamefont {Straw}}]{Wang_2021b}%
  \BibitemOpen
  \bibfield  {author} {\bibinfo {author} {\bibfnamefont {J.}~\bibnamefont
  {Wang}}, \bibinfo {author} {\bibfnamefont {S.}~\bibnamefont {Randolph}},
  \bibinfo {author} {\bibfnamefont {Q.}~\bibnamefont {Wu}}, \bibinfo {author}
  {\bibfnamefont {A.}~\bibnamefont {Botman}}, \bibinfo {author} {\bibfnamefont
  {J.}~\bibnamefont {Schardt}}, \bibinfo {author} {\bibfnamefont
  {C.}~\bibnamefont {Bouchet-Marquis}}, \bibinfo {author} {\bibfnamefont
  {X.}~\bibnamefont {Nan}}, \bibinfo {author} {\bibfnamefont {C.}~\bibnamefont
  {Rue}},\ and\ \bibinfo {author} {\bibfnamefont {M.}~\bibnamefont {Straw}},\
  }\bibfield  {title} {\enquote {\bibinfo {title} {Reactive oxygen {FIB} spin
  milling enables correlative workflow for {3D} super-resolution light
  microscopy and serial {FIB/SEM} of cultured cells},}\ }\href
  {https://doi.org/10.1038/s41598-021-92608-y} {\bibfield  {journal} {\bibinfo
  {journal} {Sci. Rep.}\ }\textbf {\bibinfo {volume} {11}},\ \bibinfo {pages}
  {13162} (\bibinfo {year} {2021}{\natexlab{a}})}\BibitemShut {NoStop}%
\bibitem [{\citenamefont {Xu}\ \emph {et~al.}(2017{\natexlab{b}})\citenamefont
  {Xu}, \citenamefont {Hayworth}, \citenamefont {Lu}, \citenamefont {Grob},
  \citenamefont {Hassan}, \citenamefont {Garc{\'{i}}a-Cerd{\'{a}}n},
  \citenamefont {Niyogi}, \citenamefont {Nogales}, \citenamefont {Weinberg},\
  and\ \citenamefont {Hess}}]{Xu_2017a}%
  \BibitemOpen
  \bibfield  {author} {\bibinfo {author} {\bibfnamefont {C.~S.}\ \bibnamefont
  {Xu}}, \bibinfo {author} {\bibfnamefont {K.~J.}\ \bibnamefont {Hayworth}},
  \bibinfo {author} {\bibfnamefont {Z.}~\bibnamefont {Lu}}, \bibinfo {author}
  {\bibfnamefont {P.}~\bibnamefont {Grob}}, \bibinfo {author} {\bibfnamefont
  {A.~M.}\ \bibnamefont {Hassan}}, \bibinfo {author} {\bibfnamefont {J.~G.}\
  \bibnamefont {Garc{\'{i}}a-Cerd{\'{a}}n}}, \bibinfo {author} {\bibfnamefont
  {K.~K.}\ \bibnamefont {Niyogi}}, \bibinfo {author} {\bibfnamefont
  {E.}~\bibnamefont {Nogales}}, \bibinfo {author} {\bibfnamefont {R.~J.}\
  \bibnamefont {Weinberg}},\ and\ \bibinfo {author} {\bibfnamefont {H.~F.}\
  \bibnamefont {Hess}},\ }\bibfield  {title} {\enquote {\bibinfo {title}
  {Enhanced {FIB-SEM} systems for large-volume 3{D} imaging},}\ }\href
  {https://doi.org/10.7554/eLife.25916} {\bibfield  {journal} {\bibinfo
  {journal} {eLife}\ }\textbf {\bibinfo {volume} {6}},\ \bibinfo {pages}
  {e25916} (\bibinfo {year} {2017}{\natexlab{b}})}\BibitemShut {NoStop}%
\bibitem [{\citenamefont {Cortese}\ \emph {et~al.}(2020)\citenamefont
  {Cortese}, \citenamefont {Lee}, \citenamefont {Cerikan}, \citenamefont
  {Neufeldt}, \citenamefont {Oorschot}, \citenamefont {K{\"{o}}hrer},
  \citenamefont {Hennies}, \citenamefont {Schieber}, \citenamefont {Ronchi},
  \citenamefont {Mizzon}, \citenamefont {Romero-Brey}, \citenamefont
  {Santarella-Mellwig}, \citenamefont {Schorb}, \citenamefont {Boermel},
  \citenamefont {Mocaer}, \citenamefont {Beckwith}, \citenamefont {Templin},
  \citenamefont {Gross}, \citenamefont {Pape}, \citenamefont {Tischer},
  \citenamefont {Frankish}, \citenamefont {Horvat}, \citenamefont {Laketa},
  \citenamefont {Stanifer}, \citenamefont {Boulant}, \citenamefont {Ruggieri},
  \citenamefont {Chatel-Chaix}, \citenamefont {Schwab},\ and\ \citenamefont
  {Bartenschlager}}]{Cortese_2020}%
  \BibitemOpen
  \bibfield  {author} {\bibinfo {author} {\bibfnamefont {M.}~\bibnamefont
  {Cortese}}, \bibinfo {author} {\bibfnamefont {J.-Y.}\ \bibnamefont {Lee}},
  \bibinfo {author} {\bibfnamefont {B.}~\bibnamefont {Cerikan}}, \bibinfo
  {author} {\bibfnamefont {C.~J.}\ \bibnamefont {Neufeldt}}, \bibinfo {author}
  {\bibfnamefont {V.~M.~J.}\ \bibnamefont {Oorschot}}, \bibinfo {author}
  {\bibfnamefont {S.}~\bibnamefont {K{\"{o}}hrer}}, \bibinfo {author}
  {\bibfnamefont {J.}~\bibnamefont {Hennies}}, \bibinfo {author} {\bibfnamefont
  {N.~L.}\ \bibnamefont {Schieber}}, \bibinfo {author} {\bibfnamefont
  {P.}~\bibnamefont {Ronchi}}, \bibinfo {author} {\bibfnamefont
  {G.}~\bibnamefont {Mizzon}}, \bibinfo {author} {\bibfnamefont
  {I.}~\bibnamefont {Romero-Brey}}, \bibinfo {author} {\bibfnamefont
  {R.}~\bibnamefont {Santarella-Mellwig}}, \bibinfo {author} {\bibfnamefont
  {M.}~\bibnamefont {Schorb}}, \bibinfo {author} {\bibfnamefont
  {M.}~\bibnamefont {Boermel}}, \bibinfo {author} {\bibfnamefont
  {K.}~\bibnamefont {Mocaer}}, \bibinfo {author} {\bibfnamefont {M.~S.}\
  \bibnamefont {Beckwith}}, \bibinfo {author} {\bibfnamefont {R.~M.}\
  \bibnamefont {Templin}}, \bibinfo {author} {\bibfnamefont {V.}~\bibnamefont
  {Gross}}, \bibinfo {author} {\bibfnamefont {C.}~\bibnamefont {Pape}},
  \bibinfo {author} {\bibfnamefont {C.}~\bibnamefont {Tischer}}, \bibinfo
  {author} {\bibfnamefont {J.}~\bibnamefont {Frankish}}, \bibinfo {author}
  {\bibfnamefont {N.~K.}\ \bibnamefont {Horvat}}, \bibinfo {author}
  {\bibfnamefont {V.}~\bibnamefont {Laketa}}, \bibinfo {author} {\bibfnamefont
  {M.}~\bibnamefont {Stanifer}}, \bibinfo {author} {\bibfnamefont
  {S.}~\bibnamefont {Boulant}}, \bibinfo {author} {\bibfnamefont
  {A.}~\bibnamefont {Ruggieri}}, \bibinfo {author} {\bibfnamefont
  {L.}~\bibnamefont {Chatel-Chaix}}, \bibinfo {author} {\bibfnamefont
  {Y.}~\bibnamefont {Schwab}},\ and\ \bibinfo {author} {\bibfnamefont
  {R.}~\bibnamefont {Bartenschlager}},\ }\bibfield  {title} {\enquote {\bibinfo
  {title} {Integrative imaging reveals sars-cov-2-induced reshaping of
  subcellular morphologies},}\ }\href
  {https://doi.org/10.1016/j.chom.2020.11.003} {\bibfield  {journal} {\bibinfo
  {journal} {Cell Host Microbe}\ }\textbf {\bibinfo {volume} {28}},\ \bibinfo
  {pages} {853--866.e5} (\bibinfo {year} {2020})}\BibitemShut {NoStop}%
\bibitem [{\citenamefont {Jansen}\ \emph {et~al.}(2022)\citenamefont {Jansen},
  \citenamefont {Reimer}, \citenamefont {Nagai}, \citenamefont {Varghese},
  \citenamefont {Overheul}, \citenamefont {de~Beer}, \citenamefont {Roverts},
  \citenamefont {Daviran}, \citenamefont {Fermin}, \citenamefont {Willemsen},
  \citenamefont {Beukenboom}, \citenamefont {Djudjaj}, \citenamefont {von
  Stillfried}, \citenamefont {van Eijk}, \citenamefont {Mastik}, \citenamefont
  {Bulthuis}, \citenamefont {den Dunnen}, \citenamefont {van Goor},
  \citenamefont {Hillebrands}, \citenamefont {Triana}, \citenamefont
  {Alexandrov}, \citenamefont {Timm}, \citenamefont {van~den Berge},
  \citenamefont {van~den Broek}, \citenamefont {Nlandu}, \citenamefont
  {Heijnert}, \citenamefont {Bindels}, \citenamefont {Hoogenboezem},
  \citenamefont {Mooren}, \citenamefont {Kuppe}, \citenamefont {Miesen},
  \citenamefont {Gr{\"{u}}nberg}, \citenamefont {Ijzermans}, \citenamefont
  {Steenbergen}, \citenamefont {Czogalla}, \citenamefont {Schreuder},
  \citenamefont {Sommerdijk}, \citenamefont {Akiva}, \citenamefont {Boor},
  \citenamefont {Puelles}, \citenamefont {Floege}, \citenamefont {Huber},
  \citenamefont {Achdout}, \citenamefont {Aimon}, \citenamefont {Bar-David},
  \citenamefont {Barr}, \citenamefont {Ben-Shmuel}, \citenamefont {Bennett},
  \citenamefont {Boby}, \citenamefont {Borden}, \citenamefont {Bowman},
  \citenamefont {Brun}, \citenamefont {BVNBS}, \citenamefont {Calmiano},
  \citenamefont {Carbery}, \citenamefont {Cattermole}, \citenamefont
  {Chernychenko}, \citenamefont {Choder}, \citenamefont {Clyde}, \citenamefont
  {Coffland}, \citenamefont {Cohen}, \citenamefont {Cole}, \citenamefont
  {Contini}, \citenamefont {Cox}, \citenamefont {Cvitkovic}, \citenamefont
  {Dias}, \citenamefont {Donckers}, \citenamefont {Dotson}, \citenamefont
  {Douangamath}, \citenamefont {Duberstein}, \citenamefont {Dudgeon},
  \citenamefont {Dunnett}, \citenamefont {Eastman}, \citenamefont {Erez},
  \citenamefont {Eyermann}, \citenamefont {Fairhead}, \citenamefont {Fate},
  \citenamefont {Fearon}, \citenamefont {Federov}, \citenamefont {Ferla},
  \citenamefont {Fernandes}, \citenamefont {Ferrins}, \citenamefont {Foster},
  \citenamefont {Foster}, \citenamefont {Gabizon}, \citenamefont
  {Garcia-Sastre}, \citenamefont {Gawriljuk}, \citenamefont {Gehrtz},
  \citenamefont {Gileadi}, \citenamefont {Giroud}, \citenamefont {Glass},
  \citenamefont {Glen}, \citenamefont {glinert}, \citenamefont {Godoy},
  \citenamefont {Gorichko}, \citenamefont {Gorrie-Stone}, \citenamefont
  {Griffen}, \citenamefont {Hart}, \citenamefont {Heer}, \citenamefont {Henry},
  \citenamefont {Hill}, \citenamefont {Horrell}, \citenamefont {Hurley},
  \citenamefont {Israely}, \citenamefont {Jajack}, \citenamefont {Jnoff},
  \citenamefont {Jochmans}, \citenamefont {John}, \citenamefont {De~Jonghe},
  \citenamefont {Kantsadi}, \citenamefont {Kenny}, \citenamefont {Kiappes},
  \citenamefont {Koekemoer}, \citenamefont {Kovar}, \citenamefont {Krojer},
  \citenamefont {Lee}, \citenamefont {Lefker}, \citenamefont {Levy},
  \citenamefont {London}, \citenamefont {Lukacik}, \citenamefont {Macdonald},
  \citenamefont {Maclean}, \citenamefont {Malla}, \citenamefont {Matviiuk},
  \citenamefont {McCorkindale}, \citenamefont {McGovern}, \citenamefont
  {Melamed}, \citenamefont {Michurin}, \citenamefont {Mikolajek}, \citenamefont
  {Milne}, \citenamefont {Morris}, \citenamefont {Morris}, \citenamefont
  {Morwitzer}, \citenamefont {Moustakas}, \citenamefont {Nakamura},
  \citenamefont {Neto}, \citenamefont {Neyts}, \citenamefont {Nguyen},
  \citenamefont {Noske}, \citenamefont {Oleinikovas}, \citenamefont {Oliva},
  \citenamefont {Overheul}, \citenamefont {Owen}, \citenamefont {Psenak},
  \citenamefont {Pai}, \citenamefont {Pan}, \citenamefont {Paran},
  \citenamefont {Perry}, \citenamefont {Pingle}, \citenamefont {Pinjari},
  \citenamefont {Politi}, \citenamefont {Powell}, \citenamefont {Puni},
  \citenamefont {Rangel}, \citenamefont {Reddi}, \citenamefont {Reid},
  \citenamefont {Resnick}, \citenamefont {Ripka}, \citenamefont {Robinson},
  \citenamefont {Robinson}, \citenamefont {Rodriguez-Guerra}, \citenamefont
  {Rosales}, \citenamefont {Rufa}, \citenamefont {Schofield}, \citenamefont
  {Shafeev}, \citenamefont {Shaikh}, \citenamefont {Shi}, \citenamefont
  {Shurrush}, \citenamefont {Sing}, \citenamefont {Sittner}, \citenamefont
  {Skyner}, \citenamefont {Smalley}, \citenamefont {Smilova}, \citenamefont
  {Solmesky}, \citenamefont {Spencer}, \citenamefont {Strain-Damarell},
  \citenamefont {Swamy}, \citenamefont {Tamir}, \citenamefont {Tennant},
  \citenamefont {Thompson}, \citenamefont {Thompson}, \citenamefont {Thompson},
  \citenamefont {Tomasia}, \citenamefont {Tumber}, \citenamefont {Vakonakis},
  \citenamefont {van Rij}, \citenamefont {van Geel}, \citenamefont {Varghese},
  \citenamefont {Vaschetto}, \citenamefont {Vitner}, \citenamefont {Voelz},
  \citenamefont {Volkamer}, \citenamefont {von Delft}, \citenamefont {von
  Delft}, \citenamefont {Walsh}, \citenamefont {Ward}, \citenamefont
  {Weatherall}, \citenamefont {Weiss}, \citenamefont {White}, \citenamefont
  {Wild}, \citenamefont {Wittmann}, \citenamefont {Wright}, \citenamefont
  {Yahalom-Ronen}, \citenamefont {Zaidmann}, \citenamefont {Zidane},
  \citenamefont {Zitzmann}, \citenamefont {van Rij}, \citenamefont {Costa},
  \citenamefont {Schneider}, \citenamefont {Smeets},\ and\ \citenamefont
  {Kramann}}]{Jansen_2022}%
  \BibitemOpen
  \bibfield  {author} {\bibinfo {author} {\bibfnamefont {J.}~\bibnamefont
  {Jansen}}, \bibinfo {author} {\bibfnamefont {K.~C.}\ \bibnamefont {Reimer}},
  \bibinfo {author} {\bibfnamefont {J.~S.}\ \bibnamefont {Nagai}}, \bibinfo
  {author} {\bibfnamefont {F.~S.}\ \bibnamefont {Varghese}}, \bibinfo {author}
  {\bibfnamefont {G.~J.}\ \bibnamefont {Overheul}}, \bibinfo {author}
  {\bibfnamefont {M.}~\bibnamefont {de~Beer}}, \bibinfo {author} {\bibfnamefont
  {R.}~\bibnamefont {Roverts}}, \bibinfo {author} {\bibfnamefont
  {D.}~\bibnamefont {Daviran}}, \bibinfo {author} {\bibfnamefont {L.~A.~S.}\
  \bibnamefont {Fermin}}, \bibinfo {author} {\bibfnamefont {B.}~\bibnamefont
  {Willemsen}}, \bibinfo {author} {\bibfnamefont {M.}~\bibnamefont
  {Beukenboom}}, \bibinfo {author} {\bibfnamefont {S.}~\bibnamefont {Djudjaj}},
  \bibinfo {author} {\bibfnamefont {S.}~\bibnamefont {von Stillfried}},
  \bibinfo {author} {\bibfnamefont {L.~E.}\ \bibnamefont {van Eijk}}, \bibinfo
  {author} {\bibfnamefont {M.}~\bibnamefont {Mastik}}, \bibinfo {author}
  {\bibfnamefont {M.}~\bibnamefont {Bulthuis}}, \bibinfo {author}
  {\bibfnamefont {W.}~\bibnamefont {den Dunnen}}, \bibinfo {author}
  {\bibfnamefont {H.}~\bibnamefont {van Goor}}, \bibinfo {author}
  {\bibfnamefont {J.-L.}\ \bibnamefont {Hillebrands}}, \bibinfo {author}
  {\bibfnamefont {S.~H.}\ \bibnamefont {Triana}}, \bibinfo {author}
  {\bibfnamefont {T.}~\bibnamefont {Alexandrov}}, \bibinfo {author}
  {\bibfnamefont {M.-C.}\ \bibnamefont {Timm}}, \bibinfo {author}
  {\bibfnamefont {B.~T.}\ \bibnamefont {van~den Berge}}, \bibinfo {author}
  {\bibfnamefont {M.}~\bibnamefont {van~den Broek}}, \bibinfo {author}
  {\bibfnamefont {Q.}~\bibnamefont {Nlandu}}, \bibinfo {author} {\bibfnamefont
  {J.}~\bibnamefont {Heijnert}}, \bibinfo {author} {\bibfnamefont {E.~M.~J.}\
  \bibnamefont {Bindels}}, \bibinfo {author} {\bibfnamefont {R.~M.}\
  \bibnamefont {Hoogenboezem}}, \bibinfo {author} {\bibfnamefont
  {F.}~\bibnamefont {Mooren}}, \bibinfo {author} {\bibfnamefont
  {C.}~\bibnamefont {Kuppe}}, \bibinfo {author} {\bibfnamefont
  {P.}~\bibnamefont {Miesen}}, \bibinfo {author} {\bibfnamefont
  {K.}~\bibnamefont {Gr{\"{u}}nberg}}, \bibinfo {author} {\bibfnamefont
  {T.}~\bibnamefont {Ijzermans}}, \bibinfo {author} {\bibfnamefont {E.~J.}\
  \bibnamefont {Steenbergen}}, \bibinfo {author} {\bibfnamefont
  {J.}~\bibnamefont {Czogalla}}, \bibinfo {author} {\bibfnamefont {M.~F.}\
  \bibnamefont {Schreuder}}, \bibinfo {author} {\bibfnamefont {N.}~\bibnamefont
  {Sommerdijk}}, \bibinfo {author} {\bibfnamefont {A.}~\bibnamefont {Akiva}},
  \bibinfo {author} {\bibfnamefont {P.}~\bibnamefont {Boor}}, \bibinfo {author}
  {\bibfnamefont {V.~G.}\ \bibnamefont {Puelles}}, \bibinfo {author}
  {\bibfnamefont {J.}~\bibnamefont {Floege}}, \bibinfo {author} {\bibfnamefont
  {T.~B.}\ \bibnamefont {Huber}}, \bibinfo {author} {\bibfnamefont
  {H.}~\bibnamefont {Achdout}}, \bibinfo {author} {\bibfnamefont
  {A.}~\bibnamefont {Aimon}}, \bibinfo {author} {\bibfnamefont
  {E.}~\bibnamefont {Bar-David}}, \bibinfo {author} {\bibfnamefont
  {H.}~\bibnamefont {Barr}}, \bibinfo {author} {\bibfnamefont {A.}~\bibnamefont
  {Ben-Shmuel}}, \bibinfo {author} {\bibfnamefont {J.}~\bibnamefont {Bennett}},
  \bibinfo {author} {\bibfnamefont {M.~L.}\ \bibnamefont {Boby}}, \bibinfo
  {author} {\bibfnamefont {B.}~\bibnamefont {Borden}}, \bibinfo {author}
  {\bibfnamefont {G.~R.}\ \bibnamefont {Bowman}}, \bibinfo {author}
  {\bibfnamefont {J.}~\bibnamefont {Brun}}, \bibinfo {author} {\bibfnamefont
  {S.}~\bibnamefont {BVNBS}}, \bibinfo {author} {\bibfnamefont
  {M.}~\bibnamefont {Calmiano}}, \bibinfo {author} {\bibfnamefont
  {A.}~\bibnamefont {Carbery}}, \bibinfo {author} {\bibfnamefont
  {E.}~\bibnamefont {Cattermole}}, \bibinfo {author} {\bibfnamefont
  {E.}~\bibnamefont {Chernychenko}}, \bibinfo {author} {\bibfnamefont {J.~D.}\
  \bibnamefont {Choder}}, \bibinfo {author} {\bibfnamefont {A.}~\bibnamefont
  {Clyde}}, \bibinfo {author} {\bibfnamefont {J.~E.}\ \bibnamefont {Coffland}},
  \bibinfo {author} {\bibfnamefont {G.}~\bibnamefont {Cohen}}, \bibinfo
  {author} {\bibfnamefont {J.}~\bibnamefont {Cole}}, \bibinfo {author}
  {\bibfnamefont {A.}~\bibnamefont {Contini}}, \bibinfo {author} {\bibfnamefont
  {L.}~\bibnamefont {Cox}}, \bibinfo {author} {\bibfnamefont {M.}~\bibnamefont
  {Cvitkovic}}, \bibinfo {author} {\bibfnamefont {A.}~\bibnamefont {Dias}},
  \bibinfo {author} {\bibfnamefont {K.}~\bibnamefont {Donckers}}, \bibinfo
  {author} {\bibfnamefont {D.~L.}\ \bibnamefont {Dotson}}, \bibinfo {author}
  {\bibfnamefont {A.}~\bibnamefont {Douangamath}}, \bibinfo {author}
  {\bibfnamefont {S.}~\bibnamefont {Duberstein}}, \bibinfo {author}
  {\bibfnamefont {T.}~\bibnamefont {Dudgeon}}, \bibinfo {author} {\bibfnamefont
  {L.}~\bibnamefont {Dunnett}}, \bibinfo {author} {\bibfnamefont {P.~K.}\
  \bibnamefont {Eastman}}, \bibinfo {author} {\bibfnamefont {N.}~\bibnamefont
  {Erez}}, \bibinfo {author} {\bibfnamefont {C.~J.}\ \bibnamefont {Eyermann}},
  \bibinfo {author} {\bibfnamefont {M.}~\bibnamefont {Fairhead}}, \bibinfo
  {author} {\bibfnamefont {G.}~\bibnamefont {Fate}}, \bibinfo {author}
  {\bibfnamefont {D.}~\bibnamefont {Fearon}}, \bibinfo {author} {\bibfnamefont
  {O.}~\bibnamefont {Federov}}, \bibinfo {author} {\bibfnamefont
  {M.}~\bibnamefont {Ferla}}, \bibinfo {author} {\bibfnamefont {R.~S.}\
  \bibnamefont {Fernandes}}, \bibinfo {author} {\bibfnamefont {L.}~\bibnamefont
  {Ferrins}}, \bibinfo {author} {\bibfnamefont {R.}~\bibnamefont {Foster}},
  \bibinfo {author} {\bibfnamefont {H.}~\bibnamefont {Foster}}, \bibinfo
  {author} {\bibfnamefont {R.}~\bibnamefont {Gabizon}}, \bibinfo {author}
  {\bibfnamefont {A.}~\bibnamefont {Garcia-Sastre}}, \bibinfo {author}
  {\bibfnamefont {V.~O.}\ \bibnamefont {Gawriljuk}}, \bibinfo {author}
  {\bibfnamefont {P.}~\bibnamefont {Gehrtz}}, \bibinfo {author} {\bibfnamefont
  {C.}~\bibnamefont {Gileadi}}, \bibinfo {author} {\bibfnamefont
  {C.}~\bibnamefont {Giroud}}, \bibinfo {author} {\bibfnamefont {W.~G.}\
  \bibnamefont {Glass}}, \bibinfo {author} {\bibfnamefont {R.}~\bibnamefont
  {Glen}}, \bibinfo {author} {\bibfnamefont {I.}~\bibnamefont {glinert}},
  \bibinfo {author} {\bibfnamefont {A.~S.}\ \bibnamefont {Godoy}}, \bibinfo
  {author} {\bibfnamefont {M.}~\bibnamefont {Gorichko}}, \bibinfo {author}
  {\bibfnamefont {T.}~\bibnamefont {Gorrie-Stone}}, \bibinfo {author}
  {\bibfnamefont {E.~J.}\ \bibnamefont {Griffen}}, \bibinfo {author}
  {\bibfnamefont {S.~H.}\ \bibnamefont {Hart}}, \bibinfo {author}
  {\bibfnamefont {J.}~\bibnamefont {Heer}}, \bibinfo {author} {\bibfnamefont
  {M.}~\bibnamefont {Henry}}, \bibinfo {author} {\bibfnamefont
  {M.}~\bibnamefont {Hill}}, \bibinfo {author} {\bibfnamefont {S.}~\bibnamefont
  {Horrell}}, \bibinfo {author} {\bibfnamefont {M.~F.~D.}\ \bibnamefont
  {Hurley}}, \bibinfo {author} {\bibfnamefont {T.}~\bibnamefont {Israely}},
  \bibinfo {author} {\bibfnamefont {A.}~\bibnamefont {Jajack}}, \bibinfo
  {author} {\bibfnamefont {E.}~\bibnamefont {Jnoff}}, \bibinfo {author}
  {\bibfnamefont {D.}~\bibnamefont {Jochmans}}, \bibinfo {author}
  {\bibfnamefont {T.}~\bibnamefont {John}}, \bibinfo {author} {\bibfnamefont
  {S.}~\bibnamefont {De~Jonghe}}, \bibinfo {author} {\bibfnamefont {A.~L.}\
  \bibnamefont {Kantsadi}}, \bibinfo {author} {\bibfnamefont {P.~W.}\
  \bibnamefont {Kenny}}, \bibinfo {author} {\bibfnamefont {J.~L.}\ \bibnamefont
  {Kiappes}}, \bibinfo {author} {\bibfnamefont {L.}~\bibnamefont {Koekemoer}},
  \bibinfo {author} {\bibfnamefont {B.}~\bibnamefont {Kovar}}, \bibinfo
  {author} {\bibfnamefont {T.}~\bibnamefont {Krojer}}, \bibinfo {author}
  {\bibfnamefont {A.~A.}\ \bibnamefont {Lee}}, \bibinfo {author} {\bibfnamefont
  {B.~A.}\ \bibnamefont {Lefker}}, \bibinfo {author} {\bibfnamefont
  {H.}~\bibnamefont {Levy}}, \bibinfo {author} {\bibfnamefont {N.}~\bibnamefont
  {London}}, \bibinfo {author} {\bibfnamefont {P.}~\bibnamefont {Lukacik}},
  \bibinfo {author} {\bibfnamefont {H.~B.}\ \bibnamefont {Macdonald}}, \bibinfo
  {author} {\bibfnamefont {B.}~\bibnamefont {Maclean}}, \bibinfo {author}
  {\bibfnamefont {T.~R.}\ \bibnamefont {Malla}}, \bibinfo {author}
  {\bibfnamefont {T.}~\bibnamefont {Matviiuk}}, \bibinfo {author}
  {\bibfnamefont {W.}~\bibnamefont {McCorkindale}}, \bibinfo {author}
  {\bibfnamefont {B.~L.}\ \bibnamefont {McGovern}}, \bibinfo {author}
  {\bibfnamefont {S.}~\bibnamefont {Melamed}}, \bibinfo {author} {\bibfnamefont
  {O.}~\bibnamefont {Michurin}}, \bibinfo {author} {\bibfnamefont
  {H.}~\bibnamefont {Mikolajek}}, \bibinfo {author} {\bibfnamefont {B.~F.}\
  \bibnamefont {Milne}}, \bibinfo {author} {\bibfnamefont {A.}~\bibnamefont
  {Morris}}, \bibinfo {author} {\bibfnamefont {G.~M.}\ \bibnamefont {Morris}},
  \bibinfo {author} {\bibfnamefont {M.~J.}\ \bibnamefont {Morwitzer}}, \bibinfo
  {author} {\bibfnamefont {D.}~\bibnamefont {Moustakas}}, \bibinfo {author}
  {\bibfnamefont {A.~M.}\ \bibnamefont {Nakamura}}, \bibinfo {author}
  {\bibfnamefont {J.~B.}\ \bibnamefont {Neto}}, \bibinfo {author}
  {\bibfnamefont {J.}~\bibnamefont {Neyts}}, \bibinfo {author} {\bibfnamefont
  {L.}~\bibnamefont {Nguyen}}, \bibinfo {author} {\bibfnamefont {G.~D.}\
  \bibnamefont {Noske}}, \bibinfo {author} {\bibfnamefont {V.}~\bibnamefont
  {Oleinikovas}}, \bibinfo {author} {\bibfnamefont {G.}~\bibnamefont {Oliva}},
  \bibinfo {author} {\bibfnamefont {G.~J.}\ \bibnamefont {Overheul}}, \bibinfo
  {author} {\bibfnamefont {D.}~\bibnamefont {Owen}}, \bibinfo {author}
  {\bibfnamefont {V.}~\bibnamefont {Psenak}}, \bibinfo {author} {\bibfnamefont
  {R.}~\bibnamefont {Pai}}, \bibinfo {author} {\bibfnamefont {J.}~\bibnamefont
  {Pan}}, \bibinfo {author} {\bibfnamefont {N.}~\bibnamefont {Paran}}, \bibinfo
  {author} {\bibfnamefont {B.}~\bibnamefont {Perry}}, \bibinfo {author}
  {\bibfnamefont {M.}~\bibnamefont {Pingle}}, \bibinfo {author} {\bibfnamefont
  {J.}~\bibnamefont {Pinjari}}, \bibinfo {author} {\bibfnamefont
  {B.}~\bibnamefont {Politi}}, \bibinfo {author} {\bibfnamefont
  {A.}~\bibnamefont {Powell}}, \bibinfo {author} {\bibfnamefont
  {R.}~\bibnamefont {Puni}}, \bibinfo {author} {\bibfnamefont {V.~L.}\
  \bibnamefont {Rangel}}, \bibinfo {author} {\bibfnamefont {R.~N.}\
  \bibnamefont {Reddi}}, \bibinfo {author} {\bibfnamefont {S.~P.}\ \bibnamefont
  {Reid}}, \bibinfo {author} {\bibfnamefont {E.}~\bibnamefont {Resnick}},
  \bibinfo {author} {\bibfnamefont {E.~G.}\ \bibnamefont {Ripka}}, \bibinfo
  {author} {\bibfnamefont {M.~C.}\ \bibnamefont {Robinson}}, \bibinfo {author}
  {\bibfnamefont {R.~P.}\ \bibnamefont {Robinson}}, \bibinfo {author}
  {\bibfnamefont {J.}~\bibnamefont {Rodriguez-Guerra}}, \bibinfo {author}
  {\bibfnamefont {R.}~\bibnamefont {Rosales}}, \bibinfo {author} {\bibfnamefont
  {D.}~\bibnamefont {Rufa}}, \bibinfo {author} {\bibfnamefont {C.}~\bibnamefont
  {Schofield}}, \bibinfo {author} {\bibfnamefont {M.}~\bibnamefont {Shafeev}},
  \bibinfo {author} {\bibfnamefont {A.}~\bibnamefont {Shaikh}}, \bibinfo
  {author} {\bibfnamefont {J.}~\bibnamefont {Shi}}, \bibinfo {author}
  {\bibfnamefont {K.}~\bibnamefont {Shurrush}}, \bibinfo {author}
  {\bibfnamefont {S.}~\bibnamefont {Sing}}, \bibinfo {author} {\bibfnamefont
  {A.}~\bibnamefont {Sittner}}, \bibinfo {author} {\bibfnamefont
  {R.}~\bibnamefont {Skyner}}, \bibinfo {author} {\bibfnamefont
  {A.}~\bibnamefont {Smalley}}, \bibinfo {author} {\bibfnamefont {M.~D.}\
  \bibnamefont {Smilova}}, \bibinfo {author} {\bibfnamefont {L.~J.}\
  \bibnamefont {Solmesky}}, \bibinfo {author} {\bibfnamefont {J.}~\bibnamefont
  {Spencer}}, \bibinfo {author} {\bibfnamefont {C.}~\bibnamefont
  {Strain-Damarell}}, \bibinfo {author} {\bibfnamefont {V.}~\bibnamefont
  {Swamy}}, \bibinfo {author} {\bibfnamefont {H.}~\bibnamefont {Tamir}},
  \bibinfo {author} {\bibfnamefont {R.}~\bibnamefont {Tennant}}, \bibinfo
  {author} {\bibfnamefont {W.}~\bibnamefont {Thompson}}, \bibinfo {author}
  {\bibfnamefont {A.}~\bibnamefont {Thompson}}, \bibinfo {author}
  {\bibfnamefont {W.}~\bibnamefont {Thompson}}, \bibinfo {author}
  {\bibfnamefont {S.}~\bibnamefont {Tomasia}}, \bibinfo {author} {\bibfnamefont
  {A.}~\bibnamefont {Tumber}}, \bibinfo {author} {\bibfnamefont
  {I.}~\bibnamefont {Vakonakis}}, \bibinfo {author} {\bibfnamefont {R.~P.}\
  \bibnamefont {van Rij}}, \bibinfo {author} {\bibfnamefont {L.}~\bibnamefont
  {van Geel}}, \bibinfo {author} {\bibfnamefont {F.~S.}\ \bibnamefont
  {Varghese}}, \bibinfo {author} {\bibfnamefont {M.}~\bibnamefont {Vaschetto}},
  \bibinfo {author} {\bibfnamefont {E.~B.}\ \bibnamefont {Vitner}}, \bibinfo
  {author} {\bibfnamefont {V.}~\bibnamefont {Voelz}}, \bibinfo {author}
  {\bibfnamefont {A.}~\bibnamefont {Volkamer}}, \bibinfo {author}
  {\bibfnamefont {F.}~\bibnamefont {von Delft}}, \bibinfo {author}
  {\bibfnamefont {A.}~\bibnamefont {von Delft}}, \bibinfo {author}
  {\bibfnamefont {M.}~\bibnamefont {Walsh}}, \bibinfo {author} {\bibfnamefont
  {W.}~\bibnamefont {Ward}}, \bibinfo {author} {\bibfnamefont {C.}~\bibnamefont
  {Weatherall}}, \bibinfo {author} {\bibfnamefont {S.}~\bibnamefont {Weiss}},
  \bibinfo {author} {\bibfnamefont {K.~M.}\ \bibnamefont {White}}, \bibinfo
  {author} {\bibfnamefont {C.~F.}\ \bibnamefont {Wild}}, \bibinfo {author}
  {\bibfnamefont {M.}~\bibnamefont {Wittmann}}, \bibinfo {author}
  {\bibfnamefont {N.}~\bibnamefont {Wright}}, \bibinfo {author} {\bibfnamefont
  {Y.}~\bibnamefont {Yahalom-Ronen}}, \bibinfo {author} {\bibfnamefont
  {D.}~\bibnamefont {Zaidmann}}, \bibinfo {author} {\bibfnamefont
  {H.}~\bibnamefont {Zidane}}, \bibinfo {author} {\bibfnamefont
  {N.}~\bibnamefont {Zitzmann}}, \bibinfo {author} {\bibfnamefont {R.~P.}\
  \bibnamefont {van Rij}}, \bibinfo {author} {\bibfnamefont {I.~G.}\
  \bibnamefont {Costa}}, \bibinfo {author} {\bibfnamefont {R.~K.}\ \bibnamefont
  {Schneider}}, \bibinfo {author} {\bibfnamefont {B.}~\bibnamefont {Smeets}},\
  and\ \bibinfo {author} {\bibfnamefont {R.}~\bibnamefont {Kramann}},\
  }\bibfield  {title} {\enquote {\bibinfo {title} {{SARS}-{CoV}-2 infects the
  human kidney and drives fibrosis in kidney organoids},}\ }\href
  {https://doi.org/10.1016/j.stem.2021.12.010} {\bibfield  {journal} {\bibinfo
  {journal} {Cell Stem Cell}\ }\textbf {\bibinfo {volume} {29}},\ \bibinfo
  {pages} {217--231.e8} (\bibinfo {year} {2022})}\BibitemShut {NoStop}%
\bibitem [{\citenamefont {Wilson}\ \emph {et~al.}(2006)\citenamefont {Wilson},
  \citenamefont {Kobsiriphat}, \citenamefont {Mendoza}, \citenamefont {Chen},
  \citenamefont {Hiller}, \citenamefont {Miller}, \citenamefont {Thornton},
  \citenamefont {Voorhees}, \citenamefont {Adler},\ and\ \citenamefont
  {Barnett}}]{Wilson_2006}%
  \BibitemOpen
  \bibfield  {author} {\bibinfo {author} {\bibfnamefont {J.~R.}\ \bibnamefont
  {Wilson}}, \bibinfo {author} {\bibfnamefont {W.}~\bibnamefont {Kobsiriphat}},
  \bibinfo {author} {\bibfnamefont {R.}~\bibnamefont {Mendoza}}, \bibinfo
  {author} {\bibfnamefont {H.-Y.}\ \bibnamefont {Chen}}, \bibinfo {author}
  {\bibfnamefont {J.~M.}\ \bibnamefont {Hiller}}, \bibinfo {author}
  {\bibfnamefont {D.~J.}\ \bibnamefont {Miller}}, \bibinfo {author}
  {\bibfnamefont {K.}~\bibnamefont {Thornton}}, \bibinfo {author}
  {\bibfnamefont {P.~W.}\ \bibnamefont {Voorhees}}, \bibinfo {author}
  {\bibfnamefont {S.~B.}\ \bibnamefont {Adler}},\ and\ \bibinfo {author}
  {\bibfnamefont {S.~A.}\ \bibnamefont {Barnett}},\ }\bibfield  {title}
  {\enquote {\bibinfo {title} {Three-dimensional reconstruction of a
  solid-oxide fuel-cell anode},}\ }\href {https://doi.org/10.1038/nmat1668}
  {\bibfield  {journal} {\bibinfo  {journal} {Nat. Mater.}\ }\textbf {\bibinfo
  {volume} {5}},\ \bibinfo {pages} {541--544} (\bibinfo {year}
  {2006})}\BibitemShut {NoStop}%
\bibitem [{\citenamefont {Zils}\ \emph {et~al.}(2010)\citenamefont {Zils},
  \citenamefont {Timpel}, \citenamefont {Arlt}, \citenamefont {Wolz},
  \citenamefont {Manke},\ and\ \citenamefont {Roth}}]{Zils_2010}%
  \BibitemOpen
  \bibfield  {author} {\bibinfo {author} {\bibfnamefont {S.}~\bibnamefont
  {Zils}}, \bibinfo {author} {\bibfnamefont {M.}~\bibnamefont {Timpel}},
  \bibinfo {author} {\bibfnamefont {T.}~\bibnamefont {Arlt}}, \bibinfo {author}
  {\bibfnamefont {A.}~\bibnamefont {Wolz}}, \bibinfo {author} {\bibfnamefont
  {I.}~\bibnamefont {Manke}},\ and\ \bibinfo {author} {\bibfnamefont
  {C.}~\bibnamefont {Roth}},\ }\bibfield  {title} {\enquote {\bibinfo {title}
  {3{D} visualisation of {PEMFC} electrode structures using {FIB}
  nanotomography},}\ }\href {https://doi.org/10.1002/fuce.201000133} {\bibfield
   {journal} {\bibinfo  {journal} {Fuel Cells}\ }\textbf {\bibinfo {volume}
  {10}},\ \bibinfo {pages} {966--972} (\bibinfo {year} {2010})}\BibitemShut
  {NoStop}%
\bibitem [{\citenamefont {Schulenburg}\ \emph {et~al.}(2011)\citenamefont
  {Schulenburg}, \citenamefont {Schwanitz}, \citenamefont {Linse},
  \citenamefont {Scherer}, \citenamefont {Wokaun}, \citenamefont {Krbanjevic},
  \citenamefont {Grothausmann},\ and\ \citenamefont
  {Manke}}]{Schulenburg_2011}%
  \BibitemOpen
  \bibfield  {author} {\bibinfo {author} {\bibfnamefont {H.}~\bibnamefont
  {Schulenburg}}, \bibinfo {author} {\bibfnamefont {B.}~\bibnamefont
  {Schwanitz}}, \bibinfo {author} {\bibfnamefont {N.}~\bibnamefont {Linse}},
  \bibinfo {author} {\bibfnamefont {G.~G.}\ \bibnamefont {Scherer}}, \bibinfo
  {author} {\bibfnamefont {A.}~\bibnamefont {Wokaun}}, \bibinfo {author}
  {\bibfnamefont {J.}~\bibnamefont {Krbanjevic}}, \bibinfo {author}
  {\bibfnamefont {R.}~\bibnamefont {Grothausmann}},\ and\ \bibinfo {author}
  {\bibfnamefont {I.}~\bibnamefont {Manke}},\ }\bibfield  {title} {\enquote
  {\bibinfo {title} {3{D} imaging of catalyst support corrosion in polymer
  electrolyte fuel cells},}\ }\href {https://doi.org/10.1021/jp203016u}
  {\bibfield  {journal} {\bibinfo  {journal} {J. Phys. Chem. C}\ }\textbf
  {\bibinfo {volume} {115}},\ \bibinfo {pages} {14236--14243} (\bibinfo {year}
  {2011})}\BibitemShut {NoStop}%
\bibitem [{\citenamefont {Moroni}\ \emph {et~al.}(2016)\citenamefont {Moroni},
  \citenamefont {B{\"{o}}rner}, \citenamefont {Zielke}, \citenamefont
  {Schroeder}, \citenamefont {Nowak}, \citenamefont {Winter}, \citenamefont
  {Manke}, \citenamefont {Zengerle},\ and\ \citenamefont
  {Thiele}}]{Moroni_2016}%
  \BibitemOpen
  \bibfield  {author} {\bibinfo {author} {\bibfnamefont {R.}~\bibnamefont
  {Moroni}}, \bibinfo {author} {\bibfnamefont {M.}~\bibnamefont
  {B{\"{o}}rner}}, \bibinfo {author} {\bibfnamefont {L.}~\bibnamefont
  {Zielke}}, \bibinfo {author} {\bibfnamefont {M.}~\bibnamefont {Schroeder}},
  \bibinfo {author} {\bibfnamefont {S.}~\bibnamefont {Nowak}}, \bibinfo
  {author} {\bibfnamefont {M.}~\bibnamefont {Winter}}, \bibinfo {author}
  {\bibfnamefont {I.}~\bibnamefont {Manke}}, \bibinfo {author} {\bibfnamefont
  {R.}~\bibnamefont {Zengerle}},\ and\ \bibinfo {author} {\bibfnamefont
  {S.}~\bibnamefont {Thiele}},\ }\bibfield  {title} {\enquote {\bibinfo {title}
  {Multi-scale correlative tomography of a li-ion battery composite cathode},}\
  }\href {https://doi.org/10.1038/srep30109} {\bibfield  {journal} {\bibinfo
  {journal} {Sci. Rep.}\ }\textbf {\bibinfo {volume} {6}} (\bibinfo {year}
  {2016}),\ 10.1038/srep30109}\BibitemShut {NoStop}%
\bibitem [{\citenamefont {Zielke}\ \emph {et~al.}(2014)\citenamefont {Zielke},
  \citenamefont {Hutzenlaub}, \citenamefont {Wheeler}, \citenamefont {Chao},
  \citenamefont {Manke}, \citenamefont {Hilger}, \citenamefont {Paust},
  \citenamefont {Zengerle},\ and\ \citenamefont {Thiele}}]{Zielke_2014}%
  \BibitemOpen
  \bibfield  {author} {\bibinfo {author} {\bibfnamefont {L.}~\bibnamefont
  {Zielke}}, \bibinfo {author} {\bibfnamefont {T.}~\bibnamefont {Hutzenlaub}},
  \bibinfo {author} {\bibfnamefont {D.~R.}\ \bibnamefont {Wheeler}}, \bibinfo
  {author} {\bibfnamefont {C.-W.}\ \bibnamefont {Chao}}, \bibinfo {author}
  {\bibfnamefont {I.}~\bibnamefont {Manke}}, \bibinfo {author} {\bibfnamefont
  {A.}~\bibnamefont {Hilger}}, \bibinfo {author} {\bibfnamefont
  {N.}~\bibnamefont {Paust}}, \bibinfo {author} {\bibfnamefont
  {R.}~\bibnamefont {Zengerle}},\ and\ \bibinfo {author} {\bibfnamefont
  {S.}~\bibnamefont {Thiele}},\ }\bibfield  {title} {\enquote {\bibinfo {title}
  {Three-phase multiscale modeling of a {LiCoO}$_2$ cathode: combining the
  advantages of {FIB}-{SEM} imaging and {X}-ray tomography},}\ }\href
  {https://doi.org/10.1002/aenm.201401612} {\bibfield  {journal} {\bibinfo
  {journal} {Adv. Energy Mater.}\ }\textbf {\bibinfo {volume} {5}},\ \bibinfo
  {pages} {1401612} (\bibinfo {year} {2014})}\BibitemShut {NoStop}%
\bibitem [{\citenamefont {Paulisch}\ \emph {et~al.}(2021)\citenamefont
  {Paulisch}, \citenamefont {Gebhard}, \citenamefont {Franzen}, \citenamefont
  {Hilger}, \citenamefont {Osenberg}, \citenamefont {Marathe}, \citenamefont
  {Rau}, \citenamefont {Ellendorff}, \citenamefont {Turek}, \citenamefont
  {Roth},\ and\ \citenamefont {Manke}}]{Paulisch_2021}%
  \BibitemOpen
  \bibfield  {author} {\bibinfo {author} {\bibfnamefont {M.~C.}\ \bibnamefont
  {Paulisch}}, \bibinfo {author} {\bibfnamefont {M.}~\bibnamefont {Gebhard}},
  \bibinfo {author} {\bibfnamefont {D.}~\bibnamefont {Franzen}}, \bibinfo
  {author} {\bibfnamefont {A.}~\bibnamefont {Hilger}}, \bibinfo {author}
  {\bibfnamefont {M.}~\bibnamefont {Osenberg}}, \bibinfo {author}
  {\bibfnamefont {S.}~\bibnamefont {Marathe}}, \bibinfo {author} {\bibfnamefont
  {C.}~\bibnamefont {Rau}}, \bibinfo {author} {\bibfnamefont {B.}~\bibnamefont
  {Ellendorff}}, \bibinfo {author} {\bibfnamefont {T.}~\bibnamefont {Turek}},
  \bibinfo {author} {\bibfnamefont {C.}~\bibnamefont {Roth}},\ and\ \bibinfo
  {author} {\bibfnamefont {I.}~\bibnamefont {Manke}},\ }\bibfield  {title}
  {\enquote {\bibinfo {title} {Operando synchrotron imaging of electrolyte
  distribution in silver-based gas diffusion electrodes during oxygen reduction
  reaction in highly alkaline media},}\ }\href
  {https://doi.org/10.1021/acsaem.1c01524} {\bibfield  {journal} {\bibinfo
  {journal} {ACS Appl. Energy Mater.}\ }\textbf {\bibinfo {volume} {4}},\
  \bibinfo {pages} {7497--7503} (\bibinfo {year} {2021})}\BibitemShut {NoStop}%
\bibitem [{\citenamefont {Neumann}\ \emph {et~al.}(2019)\citenamefont
  {Neumann}, \citenamefont {Osenberg}, \citenamefont {Hilger}, \citenamefont
  {Franzen}, \citenamefont {Turek}, \citenamefont {Manke},\ and\ \citenamefont
  {Schmidt}}]{Neumann_2019}%
  \BibitemOpen
  \bibfield  {author} {\bibinfo {author} {\bibfnamefont {M.}~\bibnamefont
  {Neumann}}, \bibinfo {author} {\bibfnamefont {M.}~\bibnamefont {Osenberg}},
  \bibinfo {author} {\bibfnamefont {A.}~\bibnamefont {Hilger}}, \bibinfo
  {author} {\bibfnamefont {D.}~\bibnamefont {Franzen}}, \bibinfo {author}
  {\bibfnamefont {T.}~\bibnamefont {Turek}}, \bibinfo {author} {\bibfnamefont
  {I.}~\bibnamefont {Manke}},\ and\ \bibinfo {author} {\bibfnamefont
  {V.}~\bibnamefont {Schmidt}},\ }\bibfield  {title} {\enquote {\bibinfo
  {title} {On a pluri-gaussian model for three-phase microstructures, with
  applications to 3{D} image data of gas-diffusion electrodes},}\ }\href
  {https://doi.org/10.1016/j.commatsci.2018.09.033} {\bibfield  {journal}
  {\bibinfo  {journal} {Computational Materials Science}\ }\textbf {\bibinfo
  {volume} {156}},\ \bibinfo {pages} {325--331} (\bibinfo {year}
  {2019})}\BibitemShut {NoStop}%
\bibitem [{\citenamefont {Vierrath}\ \emph {et~al.}(2015)\citenamefont
  {Vierrath}, \citenamefont {Zielke}, \citenamefont {Moroni}, \citenamefont
  {Mondon}, \citenamefont {Wheeler}, \citenamefont {Zengerle},\ and\
  \citenamefont {Thiele}}]{Vierrath_2015}%
  \BibitemOpen
  \bibfield  {author} {\bibinfo {author} {\bibfnamefont {S.}~\bibnamefont
  {Vierrath}}, \bibinfo {author} {\bibfnamefont {L.}~\bibnamefont {Zielke}},
  \bibinfo {author} {\bibfnamefont {R.}~\bibnamefont {Moroni}}, \bibinfo
  {author} {\bibfnamefont {A.}~\bibnamefont {Mondon}}, \bibinfo {author}
  {\bibfnamefont {D.~R.}\ \bibnamefont {Wheeler}}, \bibinfo {author}
  {\bibfnamefont {R.}~\bibnamefont {Zengerle}},\ and\ \bibinfo {author}
  {\bibfnamefont {S.}~\bibnamefont {Thiele}},\ }\bibfield  {title} {\enquote
  {\bibinfo {title} {Morphology of nanoporous carbon-binder domains in {L}i-ion
  batteries{\textemdash}a {FIB}-{SEM} study},}\ }\href
  {https://doi.org/10.1016/j.elecom.2015.09.010} {\bibfield  {journal}
  {\bibinfo  {journal} {Electrochem. Commun.}\ }\textbf {\bibinfo {volume}
  {60}},\ \bibinfo {pages} {176--179} (\bibinfo {year} {2015})}\BibitemShut
  {NoStop}%
\bibitem [{\citenamefont {Etiemble}\ \emph {et~al.}(2016)\citenamefont
  {Etiemble}, \citenamefont {Tranchot}, \citenamefont {Douillard},
  \citenamefont {Idrissi}, \citenamefont {Maire},\ and\ \citenamefont
  {Rou{\'{e}}}}]{Etiemble_2016}%
  \BibitemOpen
  \bibfield  {author} {\bibinfo {author} {\bibfnamefont {A.}~\bibnamefont
  {Etiemble}}, \bibinfo {author} {\bibfnamefont {A.}~\bibnamefont {Tranchot}},
  \bibinfo {author} {\bibfnamefont {T.}~\bibnamefont {Douillard}}, \bibinfo
  {author} {\bibfnamefont {H.}~\bibnamefont {Idrissi}}, \bibinfo {author}
  {\bibfnamefont {E.}~\bibnamefont {Maire}},\ and\ \bibinfo {author}
  {\bibfnamefont {L.}~\bibnamefont {Rou{\'{e}}}},\ }\bibfield  {title}
  {\enquote {\bibinfo {title} {Evolution of the 3d microstructure of a
  {S}i-based electrode for {L}i-ion batteries investigated by {FIB}/{SEM}
  tomography},}\ }\href {https://doi.org/10.1149/2.0421608jes} {\bibfield
  {journal} {\bibinfo  {journal} {J. Electrochem. Soc.}\ }\textbf {\bibinfo
  {volume} {163}},\ \bibinfo {pages} {A1550--A1559} (\bibinfo {year}
  {2016})}\BibitemShut {NoStop}%
\bibitem [{\citenamefont {Billaud}\ \emph {et~al.}(2016)\citenamefont
  {Billaud}, \citenamefont {Bouville}, \citenamefont {Magrini}, \citenamefont
  {Villevieille},\ and\ \citenamefont {Studart}}]{Billaud_2016}%
  \BibitemOpen
  \bibfield  {author} {\bibinfo {author} {\bibfnamefont {J.}~\bibnamefont
  {Billaud}}, \bibinfo {author} {\bibfnamefont {F.}~\bibnamefont {Bouville}},
  \bibinfo {author} {\bibfnamefont {T.}~\bibnamefont {Magrini}}, \bibinfo
  {author} {\bibfnamefont {C.}~\bibnamefont {Villevieille}},\ and\ \bibinfo
  {author} {\bibfnamefont {A.~R.}\ \bibnamefont {Studart}},\ }\bibfield
  {title} {\enquote {\bibinfo {title} {Magnetically aligned graphite electrodes
  for high-rate performance li-ion batteries},}\ }\href
  {https://doi.org/10.1038/nenergy.2016.97} {\bibfield  {journal} {\bibinfo
  {journal} {Nat. Energy}\ }\textbf {\bibinfo {volume} {1}} (\bibinfo {year}
  {2016}),\ 10.1038/nenergy.2016.97}\BibitemShut {NoStop}%
\bibitem [{\citenamefont {Liu}\ \emph {et~al.}(2016)\citenamefont {Liu},
  \citenamefont {Chen-Wiegart}, \citenamefont {Wang}, \citenamefont {Barnett},\
  and\ \citenamefont {Faber}}]{Liu_2016}%
  \BibitemOpen
  \bibfield  {author} {\bibinfo {author} {\bibfnamefont {Z.}~\bibnamefont
  {Liu}}, \bibinfo {author} {\bibfnamefont {Y.-c.~K.}\ \bibnamefont
  {Chen-Wiegart}}, \bibinfo {author} {\bibfnamefont {J.}~\bibnamefont {Wang}},
  \bibinfo {author} {\bibfnamefont {S.~A.}\ \bibnamefont {Barnett}},\ and\
  \bibinfo {author} {\bibfnamefont {K.~T.}\ \bibnamefont {Faber}},\ }\bibfield
  {title} {\enquote {\bibinfo {title} {Three-phase 3d{D} reconstruction of a
  {LiCoO}$_2$ cathode via {FIB}-{SEM} tomography},}\ }\href
  {https://doi.org/10.1017/s1431927615015640} {\bibfield  {journal} {\bibinfo
  {journal} {Microsc. Microanal.}\ }\textbf {\bibinfo {volume} {22}},\ \bibinfo
  {pages} {140--148} (\bibinfo {year} {2016})}\BibitemShut {NoStop}%
\bibitem [{\citenamefont {Shearing}\ \emph {et~al.}(2012)\citenamefont
  {Shearing}, \citenamefont {Brandon}, \citenamefont {Gelb}, \citenamefont
  {Bradley}, \citenamefont {Withers}, \citenamefont {Marquis}, \citenamefont
  {Cooper},\ and\ \citenamefont {Harris}}]{Shearing_2012}%
  \BibitemOpen
  \bibfield  {author} {\bibinfo {author} {\bibfnamefont {P.~R.}\ \bibnamefont
  {Shearing}}, \bibinfo {author} {\bibfnamefont {N.~P.}\ \bibnamefont
  {Brandon}}, \bibinfo {author} {\bibfnamefont {J.}~\bibnamefont {Gelb}},
  \bibinfo {author} {\bibfnamefont {R.}~\bibnamefont {Bradley}}, \bibinfo
  {author} {\bibfnamefont {P.~J.}\ \bibnamefont {Withers}}, \bibinfo {author}
  {\bibfnamefont {A.~J.}\ \bibnamefont {Marquis}}, \bibinfo {author}
  {\bibfnamefont {S.}~\bibnamefont {Cooper}},\ and\ \bibinfo {author}
  {\bibfnamefont {S.~J.}\ \bibnamefont {Harris}},\ }\bibfield  {title}
  {\enquote {\bibinfo {title} {Multi length scale microstructural
  investigations of a commercially available li-ion battery electrode},}\
  }\href {https://doi.org/10.1149/2.053207jes} {\bibfield  {journal} {\bibinfo
  {journal} {J. Electrochem. Soc.}\ }\textbf {\bibinfo {volume} {159}},\
  \bibinfo {pages} {A1023--A1027} (\bibinfo {year} {2012})}\BibitemShut
  {NoStop}%
\bibitem [{\citenamefont {Liu}\ \emph {et~al.}(2015)\citenamefont {Liu},
  \citenamefont {Wang}, \citenamefont {Singh}, \citenamefont {Wagemaker},
  \citenamefont {Faber},\ and\ \citenamefont {Barnett}}]{Liu_2015}%
  \BibitemOpen
  \bibfield  {author} {\bibinfo {author} {\bibfnamefont {Z.}~\bibnamefont
  {Liu}}, \bibinfo {author} {\bibfnamefont {H.}~\bibnamefont {Wang}}, \bibinfo
  {author} {\bibfnamefont {D.~P.}\ \bibnamefont {Singh}}, \bibinfo {author}
  {\bibfnamefont {M.}~\bibnamefont {Wagemaker}}, \bibinfo {author}
  {\bibfnamefont {K.~T.}\ \bibnamefont {Faber}},\ and\ \bibinfo {author}
  {\bibfnamefont {S.~A.}\ \bibnamefont {Barnett}},\ }\bibfield  {title}
  {\enquote {\bibinfo {title} {Three-phase 3{D} reconstruction of li-ion
  batteries electrodes via {FIB}-{SEM} tomography},}\ }\href
  {https://doi.org/10.1149/ma2015-01/2/548} {\bibfield  {journal} {\bibinfo
  {journal} {{ECS} Meeting Abstracts}\ }\textbf {\bibinfo {volume}
  {{MA}2015-01}},\ \bibinfo {pages} {548--548} (\bibinfo {year}
  {2015})}\BibitemShut {NoStop}%
\bibitem [{\citenamefont {Lagadec}, \citenamefont {Zahn},\ and\ \citenamefont
  {Wood}(2018)}]{Lagadec_2018}%
  \BibitemOpen
  \bibfield  {author} {\bibinfo {author} {\bibfnamefont {M.~F.}\ \bibnamefont
  {Lagadec}}, \bibinfo {author} {\bibfnamefont {R.}~\bibnamefont {Zahn}},\ and\
  \bibinfo {author} {\bibfnamefont {V.}~\bibnamefont {Wood}},\ }\bibfield
  {title} {\enquote {\bibinfo {title} {Characterization and performance
  evaluation of lithium-ion battery separators},}\ }\href
  {https://doi.org/10.1038/s41560-018-0295-9} {\bibfield  {journal} {\bibinfo
  {journal} {Nat. Energy}\ }\textbf {\bibinfo {volume} {4}},\ \bibinfo {pages}
  {16--25} (\bibinfo {year} {2018})}\BibitemShut {NoStop}%
\bibitem [{\citenamefont {Shi}\ \emph {et~al.}(2020)\citenamefont {Shi},
  \citenamefont {Zhang}, \citenamefont {Tu}, \citenamefont {Wang},
  \citenamefont {Scott},\ and\ \citenamefont {Ceder}}]{Shi_2020}%
  \BibitemOpen
  \bibfield  {author} {\bibinfo {author} {\bibfnamefont {T.}~\bibnamefont
  {Shi}}, \bibinfo {author} {\bibfnamefont {Y.-Q.}\ \bibnamefont {Zhang}},
  \bibinfo {author} {\bibfnamefont {Q.}~\bibnamefont {Tu}}, \bibinfo {author}
  {\bibfnamefont {Y.}~\bibnamefont {Wang}}, \bibinfo {author} {\bibfnamefont
  {M.~C.}\ \bibnamefont {Scott}},\ and\ \bibinfo {author} {\bibfnamefont
  {G.}~\bibnamefont {Ceder}},\ }\bibfield  {title} {\enquote {\bibinfo {title}
  {Characterization of mechanical degradation in an all-solid-state battery
  cathode},}\ }\href {https://doi.org/10.1039/d0ta06985j} {\bibfield  {journal}
  {\bibinfo  {journal} {J. Mater. Chem.A}\ }\textbf {\bibinfo {volume} {8}},\
  \bibinfo {pages} {17399--17404} (\bibinfo {year} {2020})}\BibitemShut
  {NoStop}%
\bibitem [{\citenamefont {Houx}\ \emph {et~al.}(2020)\citenamefont {Houx},
  \citenamefont {Osenberg}, \citenamefont {Neumann}, \citenamefont {Binder},
  \citenamefont {Schmidt}, \citenamefont {Manke}, \citenamefont {Carraro},\
  and\ \citenamefont {Kramer}}]{Houx_2020}%
  \BibitemOpen
  \bibfield  {author} {\bibinfo {author} {\bibfnamefont {J.~L.}\ \bibnamefont
  {Houx}}, \bibinfo {author} {\bibfnamefont {M.}~\bibnamefont {Osenberg}},
  \bibinfo {author} {\bibfnamefont {M.}~\bibnamefont {Neumann}}, \bibinfo
  {author} {\bibfnamefont {J.~R.}\ \bibnamefont {Binder}}, \bibinfo {author}
  {\bibfnamefont {V.}~\bibnamefont {Schmidt}}, \bibinfo {author} {\bibfnamefont
  {I.}~\bibnamefont {Manke}}, \bibinfo {author} {\bibfnamefont
  {T.}~\bibnamefont {Carraro}},\ and\ \bibinfo {author} {\bibfnamefont
  {D.}~\bibnamefont {Kramer}},\ }\bibfield  {title} {\enquote {\bibinfo {title}
  {Effect of tomography resolution on calculation of microstructural properties
  for lithium ion porous electrodes},}\ }\href
  {https://doi.org/10.1149/09707.0255ecst} {\bibfield  {journal} {\bibinfo
  {journal} {{ECS} Transactions}\ }\textbf {\bibinfo {volume} {97}},\ \bibinfo
  {pages} {255--266} (\bibinfo {year} {2020})}\BibitemShut {NoStop}%
\bibitem [{\citenamefont {Kroll}\ \emph {et~al.}(2021)\citenamefont {Kroll},
  \citenamefont {Karstens}, \citenamefont {Cronau}, \citenamefont
  {H{\"{o}}ltzel}, \citenamefont {Schlabach}, \citenamefont {Nobel},
  \citenamefont {Redenbach}, \citenamefont {Roling},\ and\ \citenamefont
  {Tallarek}}]{Kroll_2021}%
  \BibitemOpen
  \bibfield  {author} {\bibinfo {author} {\bibfnamefont {M.}~\bibnamefont
  {Kroll}}, \bibinfo {author} {\bibfnamefont {S.~L.}\ \bibnamefont {Karstens}},
  \bibinfo {author} {\bibfnamefont {M.}~\bibnamefont {Cronau}}, \bibinfo
  {author} {\bibfnamefont {A.}~\bibnamefont {H{\"{o}}ltzel}}, \bibinfo {author}
  {\bibfnamefont {S.}~\bibnamefont {Schlabach}}, \bibinfo {author}
  {\bibfnamefont {N.}~\bibnamefont {Nobel}}, \bibinfo {author} {\bibfnamefont
  {C.}~\bibnamefont {Redenbach}}, \bibinfo {author} {\bibfnamefont
  {B.}~\bibnamefont {Roling}},\ and\ \bibinfo {author} {\bibfnamefont
  {U.}~\bibnamefont {Tallarek}},\ }\bibfield  {title} {\enquote {\bibinfo
  {title} {Three-phase reconstruction reveals how the microscopic structure of
  the carbon-binder domain affects ion transport in lithium-ion batteries},}\
  }\href {https://doi.org/10.1002/batt.202100057} {\bibfield  {journal}
  {\bibinfo  {journal} {Batteries \& Supercaps}\ }\textbf {\bibinfo {volume}
  {4}},\ \bibinfo {pages} {1363--1373} (\bibinfo {year} {2021})}\BibitemShut
  {NoStop}%
\bibitem [{\citenamefont {Ghadban}\ \emph {et~al.}(2023)\citenamefont
  {Ghadban}, \citenamefont {Sabharwal}, \citenamefont {Rea}, \citenamefont
  {Li}, \citenamefont {Goode}, \citenamefont {Smith}, \citenamefont {Murphy},\
  and\ \citenamefont {Secanell}}]{Ghadban_2023}%
  \BibitemOpen
  \bibfield  {author} {\bibinfo {author} {\bibfnamefont {M.}~\bibnamefont
  {Ghadban}}, \bibinfo {author} {\bibfnamefont {M.}~\bibnamefont {Sabharwal}},
  \bibinfo {author} {\bibfnamefont {C.}~\bibnamefont {Rea}}, \bibinfo {author}
  {\bibfnamefont {X.}~\bibnamefont {Li}}, \bibinfo {author} {\bibfnamefont
  {A.~E.}\ \bibnamefont {Goode}}, \bibinfo {author} {\bibfnamefont
  {M.}~\bibnamefont {Smith}}, \bibinfo {author} {\bibfnamefont
  {C.}~\bibnamefont {Murphy}},\ and\ \bibinfo {author} {\bibfnamefont
  {M.}~\bibnamefont {Secanell}},\ }\bibfield  {title} {\enquote {\bibinfo
  {title} {3{D} microscale modeling of {NMC} cathodes using multi-resolution
  {FIB}-{SEM} tomography},}\ }\href
  {https://doi.org/10.1016/j.jpowsour.2023.232745} {\bibfield  {journal}
  {\bibinfo  {journal} {J. Power Sources}\ }\textbf {\bibinfo {volume} {562}},\
  \bibinfo {pages} {232745} (\bibinfo {year} {2023})}\BibitemShut {NoStop}%
\bibitem [{\citenamefont {Star}\ and\ \citenamefont
  {Fuller}(2017)}]{Star_2017}%
  \BibitemOpen
  \bibfield  {author} {\bibinfo {author} {\bibfnamefont {A.~G.}\ \bibnamefont
  {Star}}\ and\ \bibinfo {author} {\bibfnamefont {T.~F.}\ \bibnamefont
  {Fuller}},\ }\bibfield  {title} {\enquote {\bibinfo {title} {{FIB}-{SEM}
  tomography connects microstructure to corrosion-induced performance loss in
  {PEMFC} cathodes},}\ }\href {https://doi.org/10.1149/2.0321709jes} {\bibfield
   {journal} {\bibinfo  {journal} {J. Electrochem. Soc.}\ }\textbf {\bibinfo
  {volume} {164}},\ \bibinfo {pages} {F901--F907} (\bibinfo {year}
  {2017})}\BibitemShut {NoStop}%
\bibitem [{\citenamefont {Bevilacqua}\ \emph {et~al.}(2021)\citenamefont
  {Bevilacqua}, \citenamefont {Asset}, \citenamefont {Schmid}, \citenamefont
  {Mark{\"{o}}tter}, \citenamefont {Manke}, \citenamefont {Atanassov},\ and\
  \citenamefont {Zeis}}]{Bevilacqua_2021}%
  \BibitemOpen
  \bibfield  {author} {\bibinfo {author} {\bibfnamefont {N.}~\bibnamefont
  {Bevilacqua}}, \bibinfo {author} {\bibfnamefont {T.}~\bibnamefont {Asset}},
  \bibinfo {author} {\bibfnamefont {M.~A.}\ \bibnamefont {Schmid}}, \bibinfo
  {author} {\bibfnamefont {H.}~\bibnamefont {Mark{\"{o}}tter}}, \bibinfo
  {author} {\bibfnamefont {I.}~\bibnamefont {Manke}}, \bibinfo {author}
  {\bibfnamefont {P.}~\bibnamefont {Atanassov}},\ and\ \bibinfo {author}
  {\bibfnamefont {R.}~\bibnamefont {Zeis}},\ }\bibfield  {title} {\enquote
  {\bibinfo {title} {Impact of catalyst layer morphology on the operation of
  high temperature {PEM} fuel cells},}\ }\href
  {https://doi.org/10.1016/j.powera.2020.100042} {\bibfield  {journal}
  {\bibinfo  {journal} {Journal of Power Sources Advances}\ }\textbf {\bibinfo
  {volume} {7}},\ \bibinfo {pages} {100042} (\bibinfo {year}
  {2021})}\BibitemShut {NoStop}%
\bibitem [{\citenamefont {Netzeband}\ \emph {et~al.}(2016)\citenamefont
  {Netzeband}, \citenamefont {Arlt}, \citenamefont {Wippermann}, \citenamefont
  {Lehnert},\ and\ \citenamefont {Manke}}]{Netzeband_2016}%
  \BibitemOpen
  \bibfield  {author} {\bibinfo {author} {\bibfnamefont {C.}~\bibnamefont
  {Netzeband}}, \bibinfo {author} {\bibfnamefont {T.}~\bibnamefont {Arlt}},
  \bibinfo {author} {\bibfnamefont {K.}~\bibnamefont {Wippermann}}, \bibinfo
  {author} {\bibfnamefont {W.}~\bibnamefont {Lehnert}},\ and\ \bibinfo {author}
  {\bibfnamefont {I.}~\bibnamefont {Manke}},\ }\bibfield  {title} {\enquote
  {\bibinfo {title} {Three-dimensional multiscale analysis of degradation of
  nano- and micro-structure in direct methanol fuel cell electrodes after
  methanol starvation},}\ }\href
  {https://doi.org/10.1016/j.jpowsour.2016.07.094} {\bibfield  {journal}
  {\bibinfo  {journal} {J. Power Sources}\ }\textbf {\bibinfo {volume} {327}},\
  \bibinfo {pages} {481--487} (\bibinfo {year} {2016})}\BibitemShut {NoStop}%
\bibitem [{\citenamefont {Yakal-Kremski}\ \emph {et~al.}(2012)\citenamefont
  {Yakal-Kremski}, \citenamefont {Hughes}, \citenamefont {Call},\ and\
  \citenamefont {Barnett}}]{YakalKremski_2012}%
  \BibitemOpen
  \bibfield  {author} {\bibinfo {author} {\bibfnamefont {K.}~\bibnamefont
  {Yakal-Kremski}}, \bibinfo {author} {\bibfnamefont {G.~A.}\ \bibnamefont
  {Hughes}}, \bibinfo {author} {\bibfnamefont {A.~V.}\ \bibnamefont {Call}},\
  and\ \bibinfo {author} {\bibfnamefont {S.~A.}\ \bibnamefont {Barnett}},\
  }\bibfield  {title} {\enquote {\bibinfo {title} {Effect of current switching
  on {LSM}-{YSZ} composite electrode durability},}\ }\href
  {https://doi.org/10.1149/1.3702420} {\bibfield  {journal} {\bibinfo
  {journal} {{ECS} Transactions}\ }\textbf {\bibinfo {volume} {41}},\ \bibinfo
  {pages} {129--136} (\bibinfo {year} {2012})}\BibitemShut {NoStop}%
\bibitem [{\citenamefont {Etzold}\ \emph {et~al.}(2021)\citenamefont {Etzold},
  \citenamefont {Krewer}, \citenamefont {Thiele}, \citenamefont {Dreizler},
  \citenamefont {Klemm},\ and\ \citenamefont {Turek}}]{Etzold_2021}%
  \BibitemOpen
  \bibfield  {author} {\bibinfo {author} {\bibfnamefont {B.~J.~M.}\
  \bibnamefont {Etzold}}, \bibinfo {author} {\bibfnamefont {U.}~\bibnamefont
  {Krewer}}, \bibinfo {author} {\bibfnamefont {S.}~\bibnamefont {Thiele}},
  \bibinfo {author} {\bibfnamefont {A.}~\bibnamefont {Dreizler}}, \bibinfo
  {author} {\bibfnamefont {E.}~\bibnamefont {Klemm}},\ and\ \bibinfo {author}
  {\bibfnamefont {T.}~\bibnamefont {Turek}},\ }\bibfield  {title} {\enquote
  {\bibinfo {title} {Understanding the activity transport nexus in water and
  {CO}$_2$ electrolysis: state of the art, challenges and perspectives},}\
  }\href {https://doi.org/10.1016/j.cej.2021.130501} {\bibfield  {journal}
  {\bibinfo  {journal} {Chem. Eng. J.}\ }\textbf {\bibinfo {volume} {424}},\
  \bibinfo {pages} {130501} (\bibinfo {year} {2021})}\BibitemShut {NoStop}%
\bibitem [{\citenamefont {Wang}\ \emph
  {et~al.}(2021{\natexlab{b}})\citenamefont {Wang}, \citenamefont {Lin},
  \citenamefont {Zhang}, \citenamefont {Xiao}, \citenamefont {Guan},
  \citenamefont {Yang}, \citenamefont {Lv}, \citenamefont {Liu},\ and\
  \citenamefont {Wang}}]{Wang_2021}%
  \BibitemOpen
  \bibfield  {author} {\bibinfo {author} {\bibfnamefont {Y.}~\bibnamefont
  {Wang}}, \bibinfo {author} {\bibfnamefont {X.}~\bibnamefont {Lin}}, \bibinfo
  {author} {\bibfnamefont {L.}~\bibnamefont {Zhang}}, \bibinfo {author}
  {\bibfnamefont {G.}~\bibnamefont {Xiao}}, \bibinfo {author} {\bibfnamefont
  {C.}~\bibnamefont {Guan}}, \bibinfo {author} {\bibfnamefont {J.}~\bibnamefont
  {Yang}}, \bibinfo {author} {\bibfnamefont {X.}~\bibnamefont {Lv}}, \bibinfo
  {author} {\bibfnamefont {D.}~\bibnamefont {Liu}},\ and\ \bibinfo {author}
  {\bibfnamefont {J.-Q.}\ \bibnamefont {Wang}},\ }\bibfield  {title} {\enquote
  {\bibinfo {title} {Three-dimensional microstructural characterization of
  solid oxide electrolysis cell with ce0.8gd0.2o2-infiltrated ni/{YSZ}
  electrode using focused ion beam-scanning electron microscopy},}\ }\href
  {https://doi.org/10.1007/s10008-021-04926-w} {\bibfield  {journal} {\bibinfo
  {journal} {J. Solid State Electr.}\ }\textbf {\bibinfo {volume} {25}},\
  \bibinfo {pages} {1633--1644} (\bibinfo {year}
  {2021}{\natexlab{b}})}\BibitemShut {NoStop}%
\bibitem [{\citenamefont {Hegge}\ \emph {et~al.}(2018)\citenamefont {Hegge},
  \citenamefont {Moroni}, \citenamefont {Trinke}, \citenamefont {Bensmann},
  \citenamefont {Hanke-Rauschenbach}, \citenamefont {Thiele},\ and\
  \citenamefont {Vierrath}}]{Hegge_2018}%
  \BibitemOpen
  \bibfield  {author} {\bibinfo {author} {\bibfnamefont {F.}~\bibnamefont
  {Hegge}}, \bibinfo {author} {\bibfnamefont {R.}~\bibnamefont {Moroni}},
  \bibinfo {author} {\bibfnamefont {P.}~\bibnamefont {Trinke}}, \bibinfo
  {author} {\bibfnamefont {B.}~\bibnamefont {Bensmann}}, \bibinfo {author}
  {\bibfnamefont {R.}~\bibnamefont {Hanke-Rauschenbach}}, \bibinfo {author}
  {\bibfnamefont {S.}~\bibnamefont {Thiele}},\ and\ \bibinfo {author}
  {\bibfnamefont {S.}~\bibnamefont {Vierrath}},\ }\bibfield  {title} {\enquote
  {\bibinfo {title} {Three-dimensional microstructure analysis of a polymer
  electrolyte membrane water electrolyzer anode},}\ }\href
  {https://doi.org/10.1016/j.jpowsour.2018.04.089} {\bibfield  {journal}
  {\bibinfo  {journal} {J. Power Sources}\ }\textbf {\bibinfo {volume} {393}},\
  \bibinfo {pages} {62--66} (\bibinfo {year} {2018})}\BibitemShut {NoStop}%
\bibitem [{\citenamefont {McLaughlin}\ and\ \citenamefont
  {Thiele}(2020)}]{McLaughlin_2020a}%
  \BibitemOpen
  \bibfield  {author} {\bibinfo {author} {\bibfnamefont {D.}~\bibnamefont
  {McLaughlin}}\ and\ \bibinfo {author} {\bibfnamefont {S.}~\bibnamefont
  {Thiele}},\ }\bibfield  {title} {\enquote {\bibinfo {title} {{FIB}/{SEM}
  nanotomography of a {CO}$_2$ electrolyzer cathode catalyst layer},}\ }\href
  {https://doi.org/10.1149/ma2020-01482705mtgabs} {\bibfield  {journal}
  {\bibinfo  {journal} {{ECS} Meeting Abstracts}\ }\textbf {\bibinfo {volume}
  {{MA}2020-01}},\ \bibinfo {pages} {2705--2705} (\bibinfo {year}
  {2020})}\BibitemShut {NoStop}%
\bibitem [{\citenamefont {Gebhard}\ \emph {et~al.}(2020)\citenamefont
  {Gebhard}, \citenamefont {Tichter}, \citenamefont {Franzen}, \citenamefont
  {Paulisch}, \citenamefont {Schutjajew}, \citenamefont {Turek}, \citenamefont
  {Manke},\ and\ \citenamefont {Roth}}]{Gebhard_2020}%
  \BibitemOpen
  \bibfield  {author} {\bibinfo {author} {\bibfnamefont {M.}~\bibnamefont
  {Gebhard}}, \bibinfo {author} {\bibfnamefont {T.}~\bibnamefont {Tichter}},
  \bibinfo {author} {\bibfnamefont {D.}~\bibnamefont {Franzen}}, \bibinfo
  {author} {\bibfnamefont {M.~C.}\ \bibnamefont {Paulisch}}, \bibinfo {author}
  {\bibfnamefont {K.}~\bibnamefont {Schutjajew}}, \bibinfo {author}
  {\bibfnamefont {T.}~\bibnamefont {Turek}}, \bibinfo {author} {\bibfnamefont
  {I.}~\bibnamefont {Manke}},\ and\ \bibinfo {author} {\bibfnamefont
  {C.}~\bibnamefont {Roth}},\ }\bibfield  {title} {\enquote {\bibinfo {title}
  {Improvement of oxygen-depolarized cathodes in highly alkaline media by
  electrospinning of poly(vinylidene fluoride) barrier layers},}\ }\href
  {https://doi.org/10.1002/celc.201902115} {\bibfield  {journal} {\bibinfo
  {journal} {{ChemElectroChem}}\ }\textbf {\bibinfo {volume} {7}},\ \bibinfo
  {pages} {830--837} (\bibinfo {year} {2020})}\BibitemShut {NoStop}%
\bibitem [{\citenamefont {Grothausmann}\ \emph {et~al.}(2012)\citenamefont
  {Grothausmann}, \citenamefont {Fiechter}, \citenamefont {Beare},
  \citenamefont {Lehmann}, \citenamefont {Kropf}, \citenamefont {Kumar},
  \citenamefont {Manke},\ and\ \citenamefont {Banhart}}]{Grothausmann_2012}%
  \BibitemOpen
  \bibfield  {author} {\bibinfo {author} {\bibfnamefont {R.}~\bibnamefont
  {Grothausmann}}, \bibinfo {author} {\bibfnamefont {S.}~\bibnamefont
  {Fiechter}}, \bibinfo {author} {\bibfnamefont {R.}~\bibnamefont {Beare}},
  \bibinfo {author} {\bibfnamefont {G.}~\bibnamefont {Lehmann}}, \bibinfo
  {author} {\bibfnamefont {H.}~\bibnamefont {Kropf}}, \bibinfo {author}
  {\bibfnamefont {G.~S.~V.}\ \bibnamefont {Kumar}}, \bibinfo {author}
  {\bibfnamefont {I.}~\bibnamefont {Manke}},\ and\ \bibinfo {author}
  {\bibfnamefont {J.}~\bibnamefont {Banhart}},\ }\bibfield  {title} {\enquote
  {\bibinfo {title} {Automated quantitative 3{D} analysis of faceting of
  particles in tomographic datasets},}\ }\href
  {https://doi.org/10.1016/j.ultramic.2012.07.024} {\bibfield  {journal}
  {\bibinfo  {journal} {Ultramicroscopy}\ }\textbf {\bibinfo {volume} {122}},\
  \bibinfo {pages} {65--75} (\bibinfo {year} {2012})}\BibitemShut {NoStop}%
\bibitem [{\citenamefont {Franzen}\ \emph {et~al.}(2019)\citenamefont
  {Franzen}, \citenamefont {Ellendorff}, \citenamefont {Paulisch},
  \citenamefont {Hilger}, \citenamefont {Osenberg}, \citenamefont {Manke},\
  and\ \citenamefont {Turek}}]{Franzen_2019}%
  \BibitemOpen
  \bibfield  {author} {\bibinfo {author} {\bibfnamefont {D.}~\bibnamefont
  {Franzen}}, \bibinfo {author} {\bibfnamefont {B.}~\bibnamefont {Ellendorff}},
  \bibinfo {author} {\bibfnamefont {M.~C.}\ \bibnamefont {Paulisch}}, \bibinfo
  {author} {\bibfnamefont {A.}~\bibnamefont {Hilger}}, \bibinfo {author}
  {\bibfnamefont {M.}~\bibnamefont {Osenberg}}, \bibinfo {author}
  {\bibfnamefont {I.}~\bibnamefont {Manke}},\ and\ \bibinfo {author}
  {\bibfnamefont {T.}~\bibnamefont {Turek}},\ }\bibfield  {title} {\enquote
  {\bibinfo {title} {Influence of binder content in silver-based gas diffusion
  electrodes on pore system and electrochemical performance},}\ }\href
  {https://doi.org/10.1007/s10800-019-01311-4} {\bibfield  {journal} {\bibinfo
  {journal} {J. Appl. Electrochem.}\ }\textbf {\bibinfo {volume} {49}},\
  \bibinfo {pages} {705--713} (\bibinfo {year} {2019})}\BibitemShut {NoStop}%
\bibitem [{\citenamefont {McLaughlin}\ \emph {et~al.}(2020)\citenamefont
  {McLaughlin}, \citenamefont {Bierling}, \citenamefont {Moroni}, \citenamefont
  {Vogl}, \citenamefont {Schmid},\ and\ \citenamefont
  {Thiele}}]{McLaughlin_2020}%
  \BibitemOpen
  \bibfield  {author} {\bibinfo {author} {\bibfnamefont {D.}~\bibnamefont
  {McLaughlin}}, \bibinfo {author} {\bibfnamefont {M.}~\bibnamefont
  {Bierling}}, \bibinfo {author} {\bibfnamefont {R.}~\bibnamefont {Moroni}},
  \bibinfo {author} {\bibfnamefont {C.}~\bibnamefont {Vogl}}, \bibinfo {author}
  {\bibfnamefont {G.}~\bibnamefont {Schmid}},\ and\ \bibinfo {author}
  {\bibfnamefont {S.}~\bibnamefont {Thiele}},\ }\bibfield  {title} {\enquote
  {\bibinfo {title} {Tomographic reconstruction and analysis of a silver
  {CO}$_2$ reduction cathode},}\ }\href
  {https://doi.org/10.1002/aenm.202000488} {\bibfield  {journal} {\bibinfo
  {journal} {Adv. Energy Mater.}\ }\textbf {\bibinfo {volume} {10}},\ \bibinfo
  {pages} {2000488} (\bibinfo {year} {2020})}\BibitemShut {NoStop}%
\bibitem [{\citenamefont {Fang}\ \emph {et~al.}(2019)\citenamefont {Fang},
  \citenamefont {Li}, \citenamefont {Zhang}, \citenamefont {Zhang},
  \citenamefont {Yang}, \citenamefont {Lee}, \citenamefont {Lee}, \citenamefont
  {Alvarado}, \citenamefont {Schroeder}, \citenamefont {Yang}, \citenamefont
  {Lu}, \citenamefont {Williams}, \citenamefont {Ceja}, \citenamefont {Yang},
  \citenamefont {Cai}, \citenamefont {Gu}, \citenamefont {Xu}, \citenamefont
  {Wang},\ and\ \citenamefont {Meng}}]{Fang_2019}%
  \BibitemOpen
  \bibfield  {author} {\bibinfo {author} {\bibfnamefont {C.}~\bibnamefont
  {Fang}}, \bibinfo {author} {\bibfnamefont {J.}~\bibnamefont {Li}}, \bibinfo
  {author} {\bibfnamefont {M.}~\bibnamefont {Zhang}}, \bibinfo {author}
  {\bibfnamefont {Y.}~\bibnamefont {Zhang}}, \bibinfo {author} {\bibfnamefont
  {F.}~\bibnamefont {Yang}}, \bibinfo {author} {\bibfnamefont {J.~Z.}\
  \bibnamefont {Lee}}, \bibinfo {author} {\bibfnamefont {M.-H.}\ \bibnamefont
  {Lee}}, \bibinfo {author} {\bibfnamefont {J.}~\bibnamefont {Alvarado}},
  \bibinfo {author} {\bibfnamefont {M.~A.}\ \bibnamefont {Schroeder}}, \bibinfo
  {author} {\bibfnamefont {Y.}~\bibnamefont {Yang}}, \bibinfo {author}
  {\bibfnamefont {B.}~\bibnamefont {Lu}}, \bibinfo {author} {\bibfnamefont
  {N.}~\bibnamefont {Williams}}, \bibinfo {author} {\bibfnamefont
  {M.}~\bibnamefont {Ceja}}, \bibinfo {author} {\bibfnamefont {L.}~\bibnamefont
  {Yang}}, \bibinfo {author} {\bibfnamefont {M.}~\bibnamefont {Cai}}, \bibinfo
  {author} {\bibfnamefont {J.}~\bibnamefont {Gu}}, \bibinfo {author}
  {\bibfnamefont {K.}~\bibnamefont {Xu}}, \bibinfo {author} {\bibfnamefont
  {X.}~\bibnamefont {Wang}},\ and\ \bibinfo {author} {\bibfnamefont {Y.~S.}\
  \bibnamefont {Meng}},\ }\bibfield  {title} {{\selectlanguage {en}\enquote
  {\bibinfo {title} {Quantifying inactive lithium in lithium metal
  batteries},}\ }}\href {https://doi.org/10.1038/s41586-019-1481-z} {\bibfield
  {journal} {\bibinfo  {journal} {Nature}\ }\textbf {\bibinfo {volume} {572}},\
  \bibinfo {pages} {511--515} (\bibinfo {year} {2019})}\BibitemShut {NoStop}%
\bibitem [{\citenamefont {Lee}\ \emph {et~al.}(2019)\citenamefont {Lee},
  \citenamefont {Wynn}, \citenamefont {Schroeder}, \citenamefont {Alvarado},
  \citenamefont {Wang}, \citenamefont {Xu},\ and\ \citenamefont
  {Meng}}]{Lee_2019}%
  \BibitemOpen
  \bibfield  {author} {\bibinfo {author} {\bibfnamefont {J.~Z.}\ \bibnamefont
  {Lee}}, \bibinfo {author} {\bibfnamefont {T.~A.}\ \bibnamefont {Wynn}},
  \bibinfo {author} {\bibfnamefont {M.~A.}\ \bibnamefont {Schroeder}}, \bibinfo
  {author} {\bibfnamefont {J.}~\bibnamefont {Alvarado}}, \bibinfo {author}
  {\bibfnamefont {X.}~\bibnamefont {Wang}}, \bibinfo {author} {\bibfnamefont
  {K.}~\bibnamefont {Xu}},\ and\ \bibinfo {author} {\bibfnamefont {Y.~S.}\
  \bibnamefont {Meng}},\ }\bibfield  {title} {\enquote {\bibinfo {title}
  {Cryogenic focused ion beam characterization of lithium metal anodes},}\
  }\href {https://doi.org/10.1021/acsenergylett.8b02381} {\bibfield  {journal}
  {\bibinfo  {journal} {ACS Energy Lett.}\ }\textbf {\bibinfo {volume} {4}},\
  \bibinfo {pages} {489--493} (\bibinfo {year} {2019})}\BibitemShut {NoStop}%
\bibitem [{\citenamefont {Cui}, \citenamefont {Zheng},\ and\ \citenamefont
  {He}(2020)}]{Cui_2020}%
  \BibitemOpen
  \bibfield  {author} {\bibinfo {author} {\bibfnamefont {J.}~\bibnamefont
  {Cui}}, \bibinfo {author} {\bibfnamefont {H.}~\bibnamefont {Zheng}},\ and\
  \bibinfo {author} {\bibfnamefont {K.}~\bibnamefont {He}},\ }\bibfield
  {title} {\enquote {\bibinfo {title} {Air-protective cryo-{FIB} tomography of
  sensitive materials for energy applications},}\ }\href
  {https://doi.org/10.1017/s1431927620019522} {\bibfield  {journal} {\bibinfo
  {journal} {Microsc. Microanal.}\ }\textbf {\bibinfo {volume} {26}},\ \bibinfo
  {pages} {1828--1829} (\bibinfo {year} {2020})}\BibitemShut {NoStop}%
\bibitem [{\citenamefont {Ren}\ \emph {et~al.}(2020)\citenamefont {Ren},
  \citenamefont {Zhang}, \citenamefont {Xu}, \citenamefont {Huang},\ and\
  \citenamefont {Zhang}}]{Ren_2020}%
  \BibitemOpen
  \bibfield  {author} {\bibinfo {author} {\bibfnamefont {X.-C.}\ \bibnamefont
  {Ren}}, \bibinfo {author} {\bibfnamefont {X.-Q.}\ \bibnamefont {Zhang}},
  \bibinfo {author} {\bibfnamefont {R.}~\bibnamefont {Xu}}, \bibinfo {author}
  {\bibfnamefont {J.-Q.}\ \bibnamefont {Huang}},\ and\ \bibinfo {author}
  {\bibfnamefont {Q.}~\bibnamefont {Zhang}},\ }\bibfield  {title}
  {{\selectlanguage {en}\enquote {\bibinfo {title} {Analyzing energy materials
  by cryogenic electron microscopy},}\ }}\href
  {https://doi.org/10.1002/adma.201908293} {\bibfield  {journal} {\bibinfo
  {journal} {Adv. Mater.}\ }\textbf {\bibinfo {volume} {32}},\ \bibinfo {pages}
  {e1908293} (\bibinfo {year} {2020})}\BibitemShut {NoStop}%
\bibitem [{\citenamefont {Dong}\ \emph {et~al.}(2021)\citenamefont {Dong},
  \citenamefont {Xu}, \citenamefont {Tan}, \citenamefont {Osenberg},
  \citenamefont {Sun}, \citenamefont {Kochovski}, \citenamefont {Pham},
  \citenamefont {Mei}, \citenamefont {Hilger}, \citenamefont {Ryan},
  \citenamefont {Lu}, \citenamefont {Banhart},\ and\ \citenamefont
  {Manke}}]{Dong_2021}%
  \BibitemOpen
  \bibfield  {author} {\bibinfo {author} {\bibfnamefont {K.}~\bibnamefont
  {Dong}}, \bibinfo {author} {\bibfnamefont {Y.}~\bibnamefont {Xu}}, \bibinfo
  {author} {\bibfnamefont {J.}~\bibnamefont {Tan}}, \bibinfo {author}
  {\bibfnamefont {M.}~\bibnamefont {Osenberg}}, \bibinfo {author}
  {\bibfnamefont {F.}~\bibnamefont {Sun}}, \bibinfo {author} {\bibfnamefont
  {Z.}~\bibnamefont {Kochovski}}, \bibinfo {author} {\bibfnamefont {D.~T.}\
  \bibnamefont {Pham}}, \bibinfo {author} {\bibfnamefont {S.}~\bibnamefont
  {Mei}}, \bibinfo {author} {\bibfnamefont {A.}~\bibnamefont {Hilger}},
  \bibinfo {author} {\bibfnamefont {E.}~\bibnamefont {Ryan}}, \bibinfo {author}
  {\bibfnamefont {Y.}~\bibnamefont {Lu}}, \bibinfo {author} {\bibfnamefont
  {J.}~\bibnamefont {Banhart}},\ and\ \bibinfo {author} {\bibfnamefont
  {I.}~\bibnamefont {Manke}},\ }\bibfield  {title} {\enquote {\bibinfo {title}
  {Unravelling the mechanism of lithium nucleation and growth and the
  interaction with the solid electrolyte interface},}\ }\href
  {https://doi.org/10.1021/acsenergylett.1c00551} {\bibfield  {journal}
  {\bibinfo  {journal} {ACS Energy Lett.}\ }\textbf {\bibinfo {volume} {6}},\
  \bibinfo {pages} {1719--1728} (\bibinfo {year} {2021})}\BibitemShut {NoStop}%
\bibitem [{\citenamefont {Xu}\ \emph {et~al.}(2022)\citenamefont {Xu},
  \citenamefont {Dong}, \citenamefont {Jie}, \citenamefont {Adelhelm},
  \citenamefont {Chen}, \citenamefont {Xu}, \citenamefont {Yu}, \citenamefont
  {Kim}, \citenamefont {Kochovski}, \citenamefont {Yu}, \citenamefont {Li},
  \citenamefont {LeBeau}, \citenamefont {Shao-Horn}, \citenamefont {Cao},
  \citenamefont {Jiao}, \citenamefont {Cheng}, \citenamefont {Manke},\ and\
  \citenamefont {Lu}}]{Xu_2022}%
  \BibitemOpen
  \bibfield  {author} {\bibinfo {author} {\bibfnamefont {Y.}~\bibnamefont
  {Xu}}, \bibinfo {author} {\bibfnamefont {K.}~\bibnamefont {Dong}}, \bibinfo
  {author} {\bibfnamefont {Y.}~\bibnamefont {Jie}}, \bibinfo {author}
  {\bibfnamefont {P.}~\bibnamefont {Adelhelm}}, \bibinfo {author}
  {\bibfnamefont {Y.}~\bibnamefont {Chen}}, \bibinfo {author} {\bibfnamefont
  {L.}~\bibnamefont {Xu}}, \bibinfo {author} {\bibfnamefont {P.}~\bibnamefont
  {Yu}}, \bibinfo {author} {\bibfnamefont {J.}~\bibnamefont {Kim}}, \bibinfo
  {author} {\bibfnamefont {Z.}~\bibnamefont {Kochovski}}, \bibinfo {author}
  {\bibfnamefont {Z.}~\bibnamefont {Yu}}, \bibinfo {author} {\bibfnamefont
  {W.}~\bibnamefont {Li}}, \bibinfo {author} {\bibfnamefont {J.}~\bibnamefont
  {LeBeau}}, \bibinfo {author} {\bibfnamefont {Y.}~\bibnamefont {Shao-Horn}},
  \bibinfo {author} {\bibfnamefont {R.}~\bibnamefont {Cao}}, \bibinfo {author}
  {\bibfnamefont {S.}~\bibnamefont {Jiao}}, \bibinfo {author} {\bibfnamefont
  {T.}~\bibnamefont {Cheng}}, \bibinfo {author} {\bibfnamefont
  {I.}~\bibnamefont {Manke}},\ and\ \bibinfo {author} {\bibfnamefont
  {Y.}~\bibnamefont {Lu}},\ }\bibfield  {title} {\enquote {\bibinfo {title}
  {Promoting mechanistic understanding of lithium deposition and
  solid-electrolyte interphase ({SEI}) formation using advanced
  characterization and simulation methods: recent progress, limitations, and
  future perspectives},}\ }\href {https://doi.org/10.1002/aenm.202200398}
  {\bibfield  {journal} {\bibinfo  {journal} {Adv. Energy Mater.}\ }\textbf
  {\bibinfo {volume} {12}},\ \bibinfo {pages} {2200398} (\bibinfo {year}
  {2022})}\BibitemShut {NoStop}%
\bibitem [{\citenamefont {Berg}\ \emph {et~al.}(2019)\citenamefont {Berg},
  \citenamefont {Kutra}, \citenamefont {Kroeger}, \citenamefont {Straehle},
  \citenamefont {Kausler}, \citenamefont {Haubold}, \citenamefont {Schiegg},
  \citenamefont {Ales}, \citenamefont {Beier}, \citenamefont {Rudy},
  \citenamefont {Eren}, \citenamefont {Cervantes}, \citenamefont {Xu},
  \citenamefont {Beuttenmueller}, \citenamefont {Wolny}, \citenamefont {Zhang},
  \citenamefont {Koethe}, \citenamefont {Hamprecht},\ and\ \citenamefont
  {Kreshuk}}]{berg2019}%
  \BibitemOpen
  \bibfield  {author} {\bibinfo {author} {\bibfnamefont {S.}~\bibnamefont
  {Berg}}, \bibinfo {author} {\bibfnamefont {D.}~\bibnamefont {Kutra}},
  \bibinfo {author} {\bibfnamefont {T.}~\bibnamefont {Kroeger}}, \bibinfo
  {author} {\bibfnamefont {C.~N.}\ \bibnamefont {Straehle}}, \bibinfo {author}
  {\bibfnamefont {B.~X.}\ \bibnamefont {Kausler}}, \bibinfo {author}
  {\bibfnamefont {C.}~\bibnamefont {Haubold}}, \bibinfo {author} {\bibfnamefont
  {M.}~\bibnamefont {Schiegg}}, \bibinfo {author} {\bibfnamefont
  {J.}~\bibnamefont {Ales}}, \bibinfo {author} {\bibfnamefont {T.}~\bibnamefont
  {Beier}}, \bibinfo {author} {\bibfnamefont {M.}~\bibnamefont {Rudy}},
  \bibinfo {author} {\bibfnamefont {K.}~\bibnamefont {Eren}}, \bibinfo {author}
  {\bibfnamefont {J.~I.}\ \bibnamefont {Cervantes}}, \bibinfo {author}
  {\bibfnamefont {B.}~\bibnamefont {Xu}}, \bibinfo {author} {\bibfnamefont
  {F.}~\bibnamefont {Beuttenmueller}}, \bibinfo {author} {\bibfnamefont
  {A.}~\bibnamefont {Wolny}}, \bibinfo {author} {\bibfnamefont
  {C.}~\bibnamefont {Zhang}}, \bibinfo {author} {\bibfnamefont
  {U.}~\bibnamefont {Koethe}}, \bibinfo {author} {\bibfnamefont {F.~A.}\
  \bibnamefont {Hamprecht}},\ and\ \bibinfo {author} {\bibfnamefont
  {A.}~\bibnamefont {Kreshuk}},\ }\bibfield  {title} {\enquote {\bibinfo
  {title} {Ilastik: interactive machine learning for (bio)image analysis},}\
  }\href {https://doi.org/10.1038/s41592-019-0582-9} {\bibfield  {journal}
  {\bibinfo  {journal} {Nat. Methods}\ } (\bibinfo {year} {2019}),\
  10.1038/s41592-019-0582-9}\BibitemShut {NoStop}%
\bibitem [{\citenamefont {Arganda-Carreras}\ \emph {et~al.}(2017)\citenamefont
  {Arganda-Carreras}, \citenamefont {Kaynig}, \citenamefont {Rueden},
  \citenamefont {Eliceiri}, \citenamefont {Schindelin}, \citenamefont
  {Cardona},\ and\ \citenamefont {Sebastian~Seung}}]{M_ArgandaC_2017}%
  \BibitemOpen
  \bibfield  {author} {\bibinfo {author} {\bibfnamefont {I.}~\bibnamefont
  {Arganda-Carreras}}, \bibinfo {author} {\bibfnamefont {V.}~\bibnamefont
  {Kaynig}}, \bibinfo {author} {\bibfnamefont {C.}~\bibnamefont {Rueden}},
  \bibinfo {author} {\bibfnamefont {K.~W.}\ \bibnamefont {Eliceiri}}, \bibinfo
  {author} {\bibfnamefont {J.}~\bibnamefont {Schindelin}}, \bibinfo {author}
  {\bibfnamefont {A.}~\bibnamefont {Cardona}},\ and\ \bibinfo {author}
  {\bibfnamefont {H.}~\bibnamefont {Sebastian~Seung}},\ }\bibfield  {title}
  {\enquote {\bibinfo {title} {{Trainable Weka Segmentation: a machine learning
  tool for microscopy pixel classification}},}\ }\href
  {https://doi.org/10.1093/bioinformatics/btx180} {\bibfield  {journal}
  {\bibinfo  {journal} {Bioinformatics}\ }\textbf {\bibinfo {volume} {33}},\
  \bibinfo {pages} {2424--2426} (\bibinfo {year} {2017})}\BibitemShut {NoStop}%
\bibitem [{\citenamefont {Ender}\ \emph {et~al.}(2011)\citenamefont {Ender},
  \citenamefont {Joos}, \citenamefont {Carraro},\ and\ \citenamefont
  {Ivers-Tiff{\'{e}}e}}]{Ender_2011}%
  \BibitemOpen
  \bibfield  {author} {\bibinfo {author} {\bibfnamefont {M.}~\bibnamefont
  {Ender}}, \bibinfo {author} {\bibfnamefont {J.}~\bibnamefont {Joos}},
  \bibinfo {author} {\bibfnamefont {T.}~\bibnamefont {Carraro}},\ and\ \bibinfo
  {author} {\bibfnamefont {E.}~\bibnamefont {Ivers-Tiff{\'{e}}e}},\ }\bibfield
  {title} {\enquote {\bibinfo {title} {Three-dimensional reconstruction of a
  composite cathode for lithium-ion cells},}\ }\href
  {https://doi.org/10.1016/j.elecom.2010.12.004} {\bibfield  {journal}
  {\bibinfo  {journal} {Electrochem. Commun.}\ }\textbf {\bibinfo {volume}
  {13}},\ \bibinfo {pages} {166--168} (\bibinfo {year} {2011})}\BibitemShut
  {NoStop}%
\bibitem [{\citenamefont {Moroni}\ and\ \citenamefont
  {Thiele}(2020)}]{Moroni_2020}%
  \BibitemOpen
  \bibfield  {author} {\bibinfo {author} {\bibfnamefont {R.}~\bibnamefont
  {Moroni}}\ and\ \bibinfo {author} {\bibfnamefont {S.}~\bibnamefont
  {Thiele}},\ }\bibfield  {title} {\enquote {\bibinfo {title} {{FIB}/{SEM}
  tomography segmentation by optical flow estimation},}\ }\href
  {https://doi.org/10.1016/j.ultramic.2020.113090} {\bibfield  {journal}
  {\bibinfo  {journal} {Ultramicroscopy}\ }\textbf {\bibinfo {volume} {219}},\
  \bibinfo {pages} {113090} (\bibinfo {year} {2020})}\BibitemShut {NoStop}%
\bibitem [{\citenamefont {Salzer}\ \emph {et~al.}(2012)\citenamefont {Salzer},
  \citenamefont {Spettl}, \citenamefont {Stenzel}, \citenamefont {Sm{\aa}tt},
  \citenamefont {Lind{\'{e}}n}, \citenamefont {Manke},\ and\ \citenamefont
  {Schmidt}}]{Salzer_2012}%
  \BibitemOpen
  \bibfield  {author} {\bibinfo {author} {\bibfnamefont {M.}~\bibnamefont
  {Salzer}}, \bibinfo {author} {\bibfnamefont {A.}~\bibnamefont {Spettl}},
  \bibinfo {author} {\bibfnamefont {O.}~\bibnamefont {Stenzel}}, \bibinfo
  {author} {\bibfnamefont {J.-H.}\ \bibnamefont {Sm{\aa}tt}}, \bibinfo {author}
  {\bibfnamefont {M.}~\bibnamefont {Lind{\'{e}}n}}, \bibinfo {author}
  {\bibfnamefont {I.}~\bibnamefont {Manke}},\ and\ \bibinfo {author}
  {\bibfnamefont {V.}~\bibnamefont {Schmidt}},\ }\bibfield  {title} {\enquote
  {\bibinfo {title} {A two-stage approach to the segmentation of {FIB}-{SEM}
  images of highly porous materials},}\ }\href
  {https://doi.org/10.1016/j.matchar.2012.04.003} {\bibfield  {journal}
  {\bibinfo  {journal} {Mater. Charact.}\ }\textbf {\bibinfo {volume} {69}},\
  \bibinfo {pages} {115--126} (\bibinfo {year} {2012})}\BibitemShut {NoStop}%
\bibitem [{\citenamefont {Salzer}\ \emph
  {et~al.}(2014{\natexlab{a}})\citenamefont {Salzer}, \citenamefont {Thiele},
  \citenamefont {Zengerle},\ and\ \citenamefont {Schmidt}}]{Salzer_2014}%
  \BibitemOpen
  \bibfield  {author} {\bibinfo {author} {\bibfnamefont {M.}~\bibnamefont
  {Salzer}}, \bibinfo {author} {\bibfnamefont {S.}~\bibnamefont {Thiele}},
  \bibinfo {author} {\bibfnamefont {R.}~\bibnamefont {Zengerle}},\ and\
  \bibinfo {author} {\bibfnamefont {V.}~\bibnamefont {Schmidt}},\ }\bibfield
  {title} {\enquote {\bibinfo {title} {On the importance of {FIB}-{SEM}
  specific segmentation algorithms for porous media},}\ }\href
  {https://doi.org/10.1016/j.matchar.2014.05.014} {\bibfield  {journal}
  {\bibinfo  {journal} {Mater. Charact.}\ }\textbf {\bibinfo {volume} {95}},\
  \bibinfo {pages} {36--43} (\bibinfo {year} {2014}{\natexlab{a}})}\BibitemShut
  {NoStop}%
\bibitem [{\citenamefont {Salzer}\ \emph
  {et~al.}(2014{\natexlab{b}})\citenamefont {Salzer}, \citenamefont {Prill},
  \citenamefont {Spettl}, \citenamefont {Jeulin}, \citenamefont {Schladitz},\
  and\ \citenamefont {Schmidt}}]{Salzer_2014a}%
  \BibitemOpen
  \bibfield  {author} {\bibinfo {author} {\bibfnamefont {M.}~\bibnamefont
  {Salzer}}, \bibinfo {author} {\bibfnamefont {T.}~\bibnamefont {Prill}},
  \bibinfo {author} {\bibfnamefont {A.}~\bibnamefont {Spettl}}, \bibinfo
  {author} {\bibfnamefont {D.}~\bibnamefont {Jeulin}}, \bibinfo {author}
  {\bibfnamefont {K.}~\bibnamefont {Schladitz}},\ and\ \bibinfo {author}
  {\bibfnamefont {V.}~\bibnamefont {Schmidt}},\ }\bibfield  {title} {\enquote
  {\bibinfo {title} {Quantitative comparison of segmentation algorithms for
  {FIB}-{SEM} images of porous media},}\ }\href
  {https://doi.org/10.1111/jmi.12182} {\bibfield  {journal} {\bibinfo
  {journal} {J. Microsc.}\ }\textbf {\bibinfo {volume} {257}},\ \bibinfo
  {pages} {23--30} (\bibinfo {year} {2014}{\natexlab{b}})}\BibitemShut
  {NoStop}%
\bibitem [{\citenamefont {Furat}\ \emph {et~al.}(2019)\citenamefont {Furat},
  \citenamefont {Wang}, \citenamefont {Neumann}, \citenamefont {Petrich},
  \citenamefont {Weber}, \citenamefont {Krill},\ and\ \citenamefont
  {Schmidt}}]{Furat_2019}%
  \BibitemOpen
  \bibfield  {author} {\bibinfo {author} {\bibfnamefont {O.}~\bibnamefont
  {Furat}}, \bibinfo {author} {\bibfnamefont {M.}~\bibnamefont {Wang}},
  \bibinfo {author} {\bibfnamefont {M.}~\bibnamefont {Neumann}}, \bibinfo
  {author} {\bibfnamefont {L.}~\bibnamefont {Petrich}}, \bibinfo {author}
  {\bibfnamefont {M.}~\bibnamefont {Weber}}, \bibinfo {author} {\bibfnamefont
  {C.~E.}\ \bibnamefont {Krill}},\ and\ \bibinfo {author} {\bibfnamefont
  {V.}~\bibnamefont {Schmidt}},\ }\bibfield  {title} {\enquote {\bibinfo
  {title} {Machine learning techniques for the segmentation of tomographic
  image data of functional materials},}\ }\href
  {https://doi.org/10.3389/fmats.2019.00145} {\bibfield  {journal} {\bibinfo
  {journal} {Front. Mater.}\ }\textbf {\bibinfo {volume} {6}} (\bibinfo {year}
  {2019}),\ 10.3389/fmats.2019.00145}\BibitemShut {NoStop}%
\bibitem [{\citenamefont {Arlt}\ \emph {et~al.}(2020)\citenamefont {Arlt},
  \citenamefont {Liebert}, \citenamefont {Paulisch}, \citenamefont
  {L{\"{u}}deking}, \citenamefont {Bergbreiter}, \citenamefont
  {J{\"{o}}rissen},\ and\ \citenamefont {Manke}}]{Arlt_2020}%
  \BibitemOpen
  \bibfield  {author} {\bibinfo {author} {\bibfnamefont {T.}~\bibnamefont
  {Arlt}}, \bibinfo {author} {\bibfnamefont {M.}~\bibnamefont {Liebert}},
  \bibinfo {author} {\bibfnamefont {M.}~\bibnamefont {Paulisch}}, \bibinfo
  {author} {\bibfnamefont {I.}~\bibnamefont {L{\"{u}}deking}}, \bibinfo
  {author} {\bibfnamefont {C.}~\bibnamefont {Bergbreiter}}, \bibinfo {author}
  {\bibfnamefont {L.}~\bibnamefont {J{\"{o}}rissen}},\ and\ \bibinfo {author}
  {\bibfnamefont {I.}~\bibnamefont {Manke}},\ }\bibfield  {title} {\enquote
  {\bibinfo {title} {Multi-scale analysis and phase segmentation of {FIB} and
  {X}-ray tomographic data of electrolyzer electrodes using machine learning
  algorithms},}\ }\href {https://doi.org/10.1149/09707.0639ecst} {\bibfield
  {journal} {\bibinfo  {journal} {{ECS} Transactions}\ }\textbf {\bibinfo
  {volume} {97}},\ \bibinfo {pages} {639--649} (\bibinfo {year}
  {2020})}\BibitemShut {NoStop}%
\bibitem [{\citenamefont {Bailey}\ \emph {et~al.}(2023)\citenamefont {Bailey},
  \citenamefont {Wade}, \citenamefont {Boyce}, \citenamefont {Zhang},
  \citenamefont {Brett},\ and\ \citenamefont {Shearing}}]{Bailey_2023}%
  \BibitemOpen
  \bibfield  {author} {\bibinfo {author} {\bibfnamefont {J.~J.}\ \bibnamefont
  {Bailey}}, \bibinfo {author} {\bibfnamefont {A.}~\bibnamefont {Wade}},
  \bibinfo {author} {\bibfnamefont {A.~M.}\ \bibnamefont {Boyce}}, \bibinfo
  {author} {\bibfnamefont {Y.~S.}\ \bibnamefont {Zhang}}, \bibinfo {author}
  {\bibfnamefont {D.~J.~L.}\ \bibnamefont {Brett}},\ and\ \bibinfo {author}
  {\bibfnamefont {P.~R.}\ \bibnamefont {Shearing}},\ }\bibfield  {title}
  {\enquote {\bibinfo {title} {Quantitative assessment of machine-learning
  segmentation of battery electrode materials for active material
  quantification},}\ }\href {https://doi.org/10.1016/j.jpowsour.2022.232503}
  {\bibfield  {journal} {\bibinfo  {journal} {J. Power Sources}\ }\textbf
  {\bibinfo {volume} {557}},\ \bibinfo {pages} {232503} (\bibinfo {year}
  {2023})}\BibitemShut {NoStop}%
\bibitem [{\citenamefont {Osenberg}\ \emph {et~al.}(2022)\citenamefont
  {Osenberg}, \citenamefont {Hilger}, \citenamefont {Neumann}, \citenamefont
  {Wagner}, \citenamefont {Bohn}, \citenamefont {Binder}, \citenamefont
  {Schmidt}, \citenamefont {Banhart},\ and\ \citenamefont
  {Manke}}]{M_Osenberg_2023}%
  \BibitemOpen
  \bibfield  {author} {\bibinfo {author} {\bibfnamefont {M.}~\bibnamefont
  {Osenberg}}, \bibinfo {author} {\bibfnamefont {A.}~\bibnamefont {Hilger}},
  \bibinfo {author} {\bibfnamefont {M.}~\bibnamefont {Neumann}}, \bibinfo
  {author} {\bibfnamefont {A.}~\bibnamefont {Wagner}}, \bibinfo {author}
  {\bibfnamefont {N.}~\bibnamefont {Bohn}}, \bibinfo {author} {\bibfnamefont
  {J.~R.}\ \bibnamefont {Binder}}, \bibinfo {author} {\bibfnamefont
  {V.}~\bibnamefont {Schmidt}}, \bibinfo {author} {\bibfnamefont
  {J.}~\bibnamefont {Banhart}},\ and\ \bibinfo {author} {\bibfnamefont
  {I.}~\bibnamefont {Manke}},\ }\href
  {https://doi.org/10.48550/ARXIV.2207.14114} {\enquote {\bibinfo {title}
  {Classification of {FIB/SEM}-tomography images for highly porous multiphase
  materials using random forest classifiers},}\ } (\bibinfo {year} {2022}),\
  \bibinfo {note} {preprint},\ \Eprint {https://arxiv.org/abs/2207.14114}
  {arXiv:2207.14114 [cond-mat.mtrl-sci]} \BibitemShut {NoStop}%
\bibitem [{\citenamefont {Orthacker}\ \emph {et~al.}(2014)\citenamefont
  {Orthacker}, \citenamefont {Schmied}, \citenamefont {Chernev}, \citenamefont
  {Fr{\"{o}}ch}, \citenamefont {Winkler}, \citenamefont {Hobisch},
  \citenamefont {Trimmel},\ and\ \citenamefont {Plank}}]{Orthacker2014}%
  \BibitemOpen
  \bibfield  {author} {\bibinfo {author} {\bibfnamefont {A.}~\bibnamefont
  {Orthacker}}, \bibinfo {author} {\bibfnamefont {R.}~\bibnamefont {Schmied}},
  \bibinfo {author} {\bibfnamefont {B.}~\bibnamefont {Chernev}}, \bibinfo
  {author} {\bibfnamefont {J.~E.}\ \bibnamefont {Fr{\"{o}}ch}}, \bibinfo
  {author} {\bibfnamefont {R.}~\bibnamefont {Winkler}}, \bibinfo {author}
  {\bibfnamefont {J.}~\bibnamefont {Hobisch}}, \bibinfo {author} {\bibfnamefont
  {G.}~\bibnamefont {Trimmel}},\ and\ \bibinfo {author} {\bibfnamefont
  {H.}~\bibnamefont {Plank}},\ }\bibfield  {title} {\enquote {\bibinfo {title}
  {Chemical degradation and morphological instabilities during focused ion beam
  prototyping of polymers},}\ }\href {https://doi.org/10.1039/c3cp54037e}
  {\bibfield  {journal} {\bibinfo  {journal} {Phys. Chem. Chem. Phys.}\
  }\textbf {\bibinfo {volume} {16}},\ \bibinfo {pages} {1658--1666} (\bibinfo
  {year} {2014})}\BibitemShut {NoStop}%
\bibitem [{\citenamefont {Scheffler}\ \emph {et~al.}(2022)\citenamefont
  {Scheffler}, \citenamefont {Aeschlimann}, \citenamefont {Albrecht},
  \citenamefont {Bereau}, \citenamefont {Bungartz}, \citenamefont {Felser},
  \citenamefont {Greiner}, \citenamefont {Gro{\ss}}, \citenamefont {Koch},
  \citenamefont {Kremer}, \citenamefont {Nagel}, \citenamefont {Scheidgen},
  \citenamefont {W{\"{o}}ll},\ and\ \citenamefont {Draxl}}]{Scheffler2022}%
  \BibitemOpen
  \bibfield  {author} {\bibinfo {author} {\bibfnamefont {M.}~\bibnamefont
  {Scheffler}}, \bibinfo {author} {\bibfnamefont {M.}~\bibnamefont
  {Aeschlimann}}, \bibinfo {author} {\bibfnamefont {M.}~\bibnamefont
  {Albrecht}}, \bibinfo {author} {\bibfnamefont {T.}~\bibnamefont {Bereau}},
  \bibinfo {author} {\bibfnamefont {H.-J.}\ \bibnamefont {Bungartz}}, \bibinfo
  {author} {\bibfnamefont {C.}~\bibnamefont {Felser}}, \bibinfo {author}
  {\bibfnamefont {M.}~\bibnamefont {Greiner}}, \bibinfo {author} {\bibfnamefont
  {A.}~\bibnamefont {Gro{\ss}}}, \bibinfo {author} {\bibfnamefont {C.~T.}\
  \bibnamefont {Koch}}, \bibinfo {author} {\bibfnamefont {K.}~\bibnamefont
  {Kremer}}, \bibinfo {author} {\bibfnamefont {W.~E.}\ \bibnamefont {Nagel}},
  \bibinfo {author} {\bibfnamefont {M.}~\bibnamefont {Scheidgen}}, \bibinfo
  {author} {\bibfnamefont {C.}~\bibnamefont {W{\"{o}}ll}},\ and\ \bibinfo
  {author} {\bibfnamefont {C.}~\bibnamefont {Draxl}},\ }\bibfield  {title}
  {\enquote {\bibinfo {title} {{FAIR} data enabling new horizons for materials
  research},}\ }\href {https://doi.org/10.1038/s41586-022-04501-x} {\bibfield
  {journal} {\bibinfo  {journal} {Nature}\ }\textbf {\bibinfo {volume} {604}},\
  \bibinfo {pages} {635--642} (\bibinfo {year} {2022})}\BibitemShut {NoStop}%
\bibitem [{\citenamefont {Ronneberger}, \citenamefont {Fischer},\ and\
  \citenamefont {Brox}(2015)}]{ronneberger2015u}%
  \BibitemOpen
  \bibfield  {author} {\bibinfo {author} {\bibfnamefont {O.}~\bibnamefont
  {Ronneberger}}, \bibinfo {author} {\bibfnamefont {P.}~\bibnamefont
  {Fischer}},\ and\ \bibinfo {author} {\bibfnamefont {T.}~\bibnamefont
  {Brox}},\ }\bibfield  {title} {\enquote {\bibinfo {title} {U-net:
  convolutional networks for biomedical image segmentation},}\ }in\ \href@noop
  {} {\emph {\bibinfo {booktitle} {Medical Image Computing and
  Computer-Assisted Intervention--MICCAI 2015: 18th International Conference,
  Munich, Germany, October 5-9, 2015, Proceedings, Part III 18}}}\ (\bibinfo
  {organization} {Springer},\ \bibinfo {year} {2015})\ pp.\ \bibinfo {pages}
  {234--241}\BibitemShut {NoStop}%
\bibitem [{\citenamefont {Scheffer}\ \emph {et~al.}(2020)\citenamefont
  {Scheffer}, \citenamefont {Xu}, \citenamefont {Januszewski}, \citenamefont
  {Lu}, \citenamefont {Takemura}, \citenamefont {Hayworth}, \citenamefont
  {Huang}, \citenamefont {Shinomiya}, \citenamefont {Maitlin-Shepard},
  \citenamefont {Berg}, \citenamefont {Clements}, \citenamefont {Hubbard},
  \citenamefont {Katz}, \citenamefont {Umayam}, \citenamefont {Zhao},
  \citenamefont {Ackerman}, \citenamefont {Blakely}, \citenamefont {Bogovic},
  \citenamefont {Dolafi}, \citenamefont {Kainmueller}, \citenamefont {Kawase},
  \citenamefont {Khairy}, \citenamefont {Leavitt}, \citenamefont {Li},
  \citenamefont {Lindsey}, \citenamefont {Neubarth}, \citenamefont {Olbris},
  \citenamefont {Otsuna}, \citenamefont {Trautman}, \citenamefont {Ito},
  \citenamefont {Bates}, \citenamefont {Goldammer}, \citenamefont {Wolff},
  \citenamefont {Svirskas}, \citenamefont {Schlegel}, \citenamefont {Neace},
  \citenamefont {Knecht}, \citenamefont {Alvarado}, \citenamefont {Bailey},
  \citenamefont {Ballinger}, \citenamefont {Borycz}, \citenamefont {Canino},
  \citenamefont {Cheatham}, \citenamefont {Cook}, \citenamefont {Dreher},
  \citenamefont {Duclos}, \citenamefont {Eubanks}, \citenamefont {Fairbanks},
  \citenamefont {Finley}, \citenamefont {Forknall}, \citenamefont {Francis},
  \citenamefont {Hopkins}, \citenamefont {Joyce}, \citenamefont {Kim},
  \citenamefont {Kirk}, \citenamefont {Kovalyak}, \citenamefont {Lauchie},
  \citenamefont {Lohff}, \citenamefont {Maldonado}, \citenamefont {Manley},
  \citenamefont {McLin}, \citenamefont {Mooney}, \citenamefont {Ndama},
  \citenamefont {Ogundeyi}, \citenamefont {Okeoma}, \citenamefont {Ordish},
  \citenamefont {Padilla}, \citenamefont {Patrick}, \citenamefont {Paterson},
  \citenamefont {Phillips}, \citenamefont {Phillips}, \citenamefont {Rampally},
  \citenamefont {Ribeiro}, \citenamefont {Robertson}, \citenamefont {Rymer},
  \citenamefont {Ryan}, \citenamefont {Sammons}, \citenamefont {Scott},
  \citenamefont {Scott}, \citenamefont {Shinomiya}, \citenamefont {Smith},
  \citenamefont {Smith}, \citenamefont {Smith}, \citenamefont {Sobeski},
  \citenamefont {Suleiman}, \citenamefont {Swift}, \citenamefont {Takemura},
  \citenamefont {Talebi}, \citenamefont {Tarnogorska}, \citenamefont {Tenshaw},
  \citenamefont {Tokhi}, \citenamefont {Walsh}, \citenamefont {Yang},
  \citenamefont {Horne}, \citenamefont {Li}, \citenamefont {Parekh},
  \citenamefont {Rivlin}, \citenamefont {Jayaraman}, \citenamefont {Costa},
  \citenamefont {Jefferis}, \citenamefont {Ito}, \citenamefont {Saalfeld},
  \citenamefont {George}, \citenamefont {Meinertzhagen}, \citenamefont {Rubin},
  \citenamefont {Hess}, \citenamefont {Jain},\ and\ \citenamefont
  {Plaza}}]{Scheffer_2020}%
  \BibitemOpen
  \bibfield  {author} {\bibinfo {author} {\bibfnamefont {L.~K.}\ \bibnamefont
  {Scheffer}}, \bibinfo {author} {\bibfnamefont {C.~S.}\ \bibnamefont {Xu}},
  \bibinfo {author} {\bibfnamefont {M.}~\bibnamefont {Januszewski}}, \bibinfo
  {author} {\bibfnamefont {Z.}~\bibnamefont {Lu}}, \bibinfo {author}
  {\bibfnamefont {S.-y.}\ \bibnamefont {Takemura}}, \bibinfo {author}
  {\bibfnamefont {K.~J.}\ \bibnamefont {Hayworth}}, \bibinfo {author}
  {\bibfnamefont {G.~B.}\ \bibnamefont {Huang}}, \bibinfo {author}
  {\bibfnamefont {K.}~\bibnamefont {Shinomiya}}, \bibinfo {author}
  {\bibfnamefont {J.}~\bibnamefont {Maitlin-Shepard}}, \bibinfo {author}
  {\bibfnamefont {S.}~\bibnamefont {Berg}}, \bibinfo {author} {\bibfnamefont
  {J.}~\bibnamefont {Clements}}, \bibinfo {author} {\bibfnamefont {P.~M.}\
  \bibnamefont {Hubbard}}, \bibinfo {author} {\bibfnamefont {W.~T.}\
  \bibnamefont {Katz}}, \bibinfo {author} {\bibfnamefont {L.}~\bibnamefont
  {Umayam}}, \bibinfo {author} {\bibfnamefont {T.}~\bibnamefont {Zhao}},
  \bibinfo {author} {\bibfnamefont {D.}~\bibnamefont {Ackerman}}, \bibinfo
  {author} {\bibfnamefont {T.}~\bibnamefont {Blakely}}, \bibinfo {author}
  {\bibfnamefont {J.}~\bibnamefont {Bogovic}}, \bibinfo {author} {\bibfnamefont
  {T.}~\bibnamefont {Dolafi}}, \bibinfo {author} {\bibfnamefont
  {D.}~\bibnamefont {Kainmueller}}, \bibinfo {author} {\bibfnamefont
  {T.}~\bibnamefont {Kawase}}, \bibinfo {author} {\bibfnamefont {K.~A.}\
  \bibnamefont {Khairy}}, \bibinfo {author} {\bibfnamefont {L.}~\bibnamefont
  {Leavitt}}, \bibinfo {author} {\bibfnamefont {P.~H.}\ \bibnamefont {Li}},
  \bibinfo {author} {\bibfnamefont {L.}~\bibnamefont {Lindsey}}, \bibinfo
  {author} {\bibfnamefont {N.}~\bibnamefont {Neubarth}}, \bibinfo {author}
  {\bibfnamefont {D.~J.}\ \bibnamefont {Olbris}}, \bibinfo {author}
  {\bibfnamefont {H.}~\bibnamefont {Otsuna}}, \bibinfo {author} {\bibfnamefont
  {E.~T.}\ \bibnamefont {Trautman}}, \bibinfo {author} {\bibfnamefont
  {M.}~\bibnamefont {Ito}}, \bibinfo {author} {\bibfnamefont {A.~S.}\
  \bibnamefont {Bates}}, \bibinfo {author} {\bibfnamefont {J.}~\bibnamefont
  {Goldammer}}, \bibinfo {author} {\bibfnamefont {T.}~\bibnamefont {Wolff}},
  \bibinfo {author} {\bibfnamefont {R.}~\bibnamefont {Svirskas}}, \bibinfo
  {author} {\bibfnamefont {P.}~\bibnamefont {Schlegel}}, \bibinfo {author}
  {\bibfnamefont {E.}~\bibnamefont {Neace}}, \bibinfo {author} {\bibfnamefont
  {C.~J.}\ \bibnamefont {Knecht}}, \bibinfo {author} {\bibfnamefont {C.~X.}\
  \bibnamefont {Alvarado}}, \bibinfo {author} {\bibfnamefont {D.~A.}\
  \bibnamefont {Bailey}}, \bibinfo {author} {\bibfnamefont {S.}~\bibnamefont
  {Ballinger}}, \bibinfo {author} {\bibfnamefont {J.~A.}\ \bibnamefont
  {Borycz}}, \bibinfo {author} {\bibfnamefont {B.~S.}\ \bibnamefont {Canino}},
  \bibinfo {author} {\bibfnamefont {N.}~\bibnamefont {Cheatham}}, \bibinfo
  {author} {\bibfnamefont {M.}~\bibnamefont {Cook}}, \bibinfo {author}
  {\bibfnamefont {M.}~\bibnamefont {Dreher}}, \bibinfo {author} {\bibfnamefont
  {O.}~\bibnamefont {Duclos}}, \bibinfo {author} {\bibfnamefont
  {B.}~\bibnamefont {Eubanks}}, \bibinfo {author} {\bibfnamefont
  {K.}~\bibnamefont {Fairbanks}}, \bibinfo {author} {\bibfnamefont
  {S.}~\bibnamefont {Finley}}, \bibinfo {author} {\bibfnamefont
  {N.}~\bibnamefont {Forknall}}, \bibinfo {author} {\bibfnamefont
  {A.}~\bibnamefont {Francis}}, \bibinfo {author} {\bibfnamefont {G.~P.}\
  \bibnamefont {Hopkins}}, \bibinfo {author} {\bibfnamefont {E.~M.}\
  \bibnamefont {Joyce}}, \bibinfo {author} {\bibfnamefont {S.}~\bibnamefont
  {Kim}}, \bibinfo {author} {\bibfnamefont {N.~A.}\ \bibnamefont {Kirk}},
  \bibinfo {author} {\bibfnamefont {J.}~\bibnamefont {Kovalyak}}, \bibinfo
  {author} {\bibfnamefont {S.~A.}\ \bibnamefont {Lauchie}}, \bibinfo {author}
  {\bibfnamefont {A.}~\bibnamefont {Lohff}}, \bibinfo {author} {\bibfnamefont
  {C.}~\bibnamefont {Maldonado}}, \bibinfo {author} {\bibfnamefont {E.~A.}\
  \bibnamefont {Manley}}, \bibinfo {author} {\bibfnamefont {S.}~\bibnamefont
  {McLin}}, \bibinfo {author} {\bibfnamefont {C.}~\bibnamefont {Mooney}},
  \bibinfo {author} {\bibfnamefont {M.}~\bibnamefont {Ndama}}, \bibinfo
  {author} {\bibfnamefont {O.}~\bibnamefont {Ogundeyi}}, \bibinfo {author}
  {\bibfnamefont {N.}~\bibnamefont {Okeoma}}, \bibinfo {author} {\bibfnamefont
  {C.}~\bibnamefont {Ordish}}, \bibinfo {author} {\bibfnamefont
  {N.}~\bibnamefont {Padilla}}, \bibinfo {author} {\bibfnamefont {C.~M.}\
  \bibnamefont {Patrick}}, \bibinfo {author} {\bibfnamefont {T.}~\bibnamefont
  {Paterson}}, \bibinfo {author} {\bibfnamefont {E.~E.}\ \bibnamefont
  {Phillips}}, \bibinfo {author} {\bibfnamefont {E.~M.}\ \bibnamefont
  {Phillips}}, \bibinfo {author} {\bibfnamefont {N.}~\bibnamefont {Rampally}},
  \bibinfo {author} {\bibfnamefont {C.}~\bibnamefont {Ribeiro}}, \bibinfo
  {author} {\bibfnamefont {M.~K.}\ \bibnamefont {Robertson}}, \bibinfo {author}
  {\bibfnamefont {J.~T.}\ \bibnamefont {Rymer}}, \bibinfo {author}
  {\bibfnamefont {S.~M.}\ \bibnamefont {Ryan}}, \bibinfo {author}
  {\bibfnamefont {M.}~\bibnamefont {Sammons}}, \bibinfo {author} {\bibfnamefont
  {A.~K.}\ \bibnamefont {Scott}}, \bibinfo {author} {\bibfnamefont {A.~L.}\
  \bibnamefont {Scott}}, \bibinfo {author} {\bibfnamefont {A.}~\bibnamefont
  {Shinomiya}}, \bibinfo {author} {\bibfnamefont {C.}~\bibnamefont {Smith}},
  \bibinfo {author} {\bibfnamefont {K.}~\bibnamefont {Smith}}, \bibinfo
  {author} {\bibfnamefont {N.~L.}\ \bibnamefont {Smith}}, \bibinfo {author}
  {\bibfnamefont {M.~A.}\ \bibnamefont {Sobeski}}, \bibinfo {author}
  {\bibfnamefont {A.}~\bibnamefont {Suleiman}}, \bibinfo {author}
  {\bibfnamefont {J.}~\bibnamefont {Swift}}, \bibinfo {author} {\bibfnamefont
  {S.}~\bibnamefont {Takemura}}, \bibinfo {author} {\bibfnamefont
  {I.}~\bibnamefont {Talebi}}, \bibinfo {author} {\bibfnamefont
  {D.}~\bibnamefont {Tarnogorska}}, \bibinfo {author} {\bibfnamefont
  {E.}~\bibnamefont {Tenshaw}}, \bibinfo {author} {\bibfnamefont
  {T.}~\bibnamefont {Tokhi}}, \bibinfo {author} {\bibfnamefont {J.~J.}\
  \bibnamefont {Walsh}}, \bibinfo {author} {\bibfnamefont {T.}~\bibnamefont
  {Yang}}, \bibinfo {author} {\bibfnamefont {J.~A.}\ \bibnamefont {Horne}},
  \bibinfo {author} {\bibfnamefont {F.}~\bibnamefont {Li}}, \bibinfo {author}
  {\bibfnamefont {R.}~\bibnamefont {Parekh}}, \bibinfo {author} {\bibfnamefont
  {P.~K.}\ \bibnamefont {Rivlin}}, \bibinfo {author} {\bibfnamefont
  {V.}~\bibnamefont {Jayaraman}}, \bibinfo {author} {\bibfnamefont
  {M.}~\bibnamefont {Costa}}, \bibinfo {author} {\bibfnamefont {G.~S. X.~E.}\
  \bibnamefont {Jefferis}}, \bibinfo {author} {\bibfnamefont {K.}~\bibnamefont
  {Ito}}, \bibinfo {author} {\bibfnamefont {S.}~\bibnamefont {Saalfeld}},
  \bibinfo {author} {\bibfnamefont {R.}~\bibnamefont {George}}, \bibinfo
  {author} {\bibfnamefont {I.~A.}\ \bibnamefont {Meinertzhagen}}, \bibinfo
  {author} {\bibfnamefont {G.~M.}\ \bibnamefont {Rubin}}, \bibinfo {author}
  {\bibfnamefont {H.~F.}\ \bibnamefont {Hess}}, \bibinfo {author}
  {\bibfnamefont {V.}~\bibnamefont {Jain}},\ and\ \bibinfo {author}
  {\bibfnamefont {S.~M.}\ \bibnamefont {Plaza}},\ }\bibfield  {title} {\enquote
  {\bibinfo {title} {A connectome and analysis of the adult drosophila central
  brain},}\ }\href {https://doi.org/10.7554/elife.57443} {\bibfield  {journal}
  {\bibinfo  {journal} {{eLife}}\ }\textbf {\bibinfo {volume} {9}} (\bibinfo
  {year} {2020}),\ 10.7554/elife.57443}\BibitemShut {NoStop}%
\bibitem [{\citenamefont {Court}\ \emph {et~al.}(2023)\citenamefont {Court},
  \citenamefont {Costa}, \citenamefont {Pilgrim}, \citenamefont {Millburn},
  \citenamefont {Holmes}, \citenamefont {McLachlan}, \citenamefont {Larkin},
  \citenamefont {Matentzoglu}, \citenamefont {Kir}, \citenamefont {Parkinson},
  \citenamefont {Brown}, \citenamefont {O'Kane}, \citenamefont {Armstrong},
  \citenamefont {Jefferis},\ and\ \citenamefont
  {Osumi-Sutherland}}]{Court_2023}%
  \BibitemOpen
  \bibfield  {author} {\bibinfo {author} {\bibfnamefont {R.}~\bibnamefont
  {Court}}, \bibinfo {author} {\bibfnamefont {M.}~\bibnamefont {Costa}},
  \bibinfo {author} {\bibfnamefont {C.}~\bibnamefont {Pilgrim}}, \bibinfo
  {author} {\bibfnamefont {G.}~\bibnamefont {Millburn}}, \bibinfo {author}
  {\bibfnamefont {A.}~\bibnamefont {Holmes}}, \bibinfo {author} {\bibfnamefont
  {A.}~\bibnamefont {McLachlan}}, \bibinfo {author} {\bibfnamefont
  {A.}~\bibnamefont {Larkin}}, \bibinfo {author} {\bibfnamefont
  {N.}~\bibnamefont {Matentzoglu}}, \bibinfo {author} {\bibfnamefont
  {H.}~\bibnamefont {Kir}}, \bibinfo {author} {\bibfnamefont {H.}~\bibnamefont
  {Parkinson}}, \bibinfo {author} {\bibfnamefont {N.~H.}\ \bibnamefont
  {Brown}}, \bibinfo {author} {\bibfnamefont {C.~J.}\ \bibnamefont {O'Kane}},
  \bibinfo {author} {\bibfnamefont {J.~D.}\ \bibnamefont {Armstrong}}, \bibinfo
  {author} {\bibfnamefont {G.~S. X.~E.}\ \bibnamefont {Jefferis}},\ and\
  \bibinfo {author} {\bibfnamefont {D.}~\bibnamefont {Osumi-Sutherland}},\
  }\bibfield  {title} {\enquote {\bibinfo {title} {Virtual fly
  brain{\textemdash}an interactive atlas of the drosophila nervous system},}\
  }\href {https://doi.org/10.3389/fphys.2023.1076533} {\bibfield  {journal}
  {\bibinfo  {journal} {Front. Physiol.}\ }\textbf {\bibinfo {volume} {14}}
  (\bibinfo {year} {2023}),\ 10.3389/fphys.2023.1076533}\BibitemShut {NoStop}%
\bibitem [{\citenamefont {Scheffer}\ and\ \citenamefont
  {Meinertzhagen}(2019)}]{Scheffer_2019}%
  \BibitemOpen
  \bibfield  {author} {\bibinfo {author} {\bibfnamefont {L.~K.}\ \bibnamefont
  {Scheffer}}\ and\ \bibinfo {author} {\bibfnamefont {I.~A.}\ \bibnamefont
  {Meinertzhagen}},\ }\bibfield  {title} {\enquote {\bibinfo {title} {The fly
  brain atlas},}\ }\href
  {https://doi.org/10.1146/annurev-cellbio-100818-125444} {\bibfield  {journal}
  {\bibinfo  {journal} {Annu. Rev. Cell Dev. Biol.}\ }\textbf {\bibinfo
  {volume} {35}},\ \bibinfo {pages} {637--653} (\bibinfo {year}
  {2019})}\BibitemShut {NoStop}%
\bibitem [{\citenamefont {Ag{\"{u}}i-Gonzalez}, \citenamefont {J{\"{a}}hne},\
  and\ \citenamefont {Phan}(2019)}]{AgueiGonzalez_2019}%
  \BibitemOpen
  \bibfield  {author} {\bibinfo {author} {\bibfnamefont {P.}~\bibnamefont
  {Ag{\"{u}}i-Gonzalez}}, \bibinfo {author} {\bibfnamefont {S.}~\bibnamefont
  {J{\"{a}}hne}},\ and\ \bibinfo {author} {\bibfnamefont {N.~T.~N.}\
  \bibnamefont {Phan}},\ }\bibfield  {title} {\enquote {\bibinfo {title}
  {{SIMS} imaging in neurobiology and cell biology},}\ }\href
  {https://doi.org/10.1039/c9ja00118b} {\bibfield  {journal} {\bibinfo
  {journal} {J. Anal. Atom. Spectrom.}\ }\textbf {\bibinfo {volume} {34}},\
  \bibinfo {pages} {1355--1368} (\bibinfo {year} {2019})}\BibitemShut {NoStop}%
\bibitem [{\citenamefont {McPhail}(2006{\natexlab{a}})}]{McPhail_2006a}%
  \BibitemOpen
  \bibfield  {author} {\bibinfo {author} {\bibfnamefont {D.~S.}\ \bibnamefont
  {McPhail}},\ }\bibfield  {title} {\enquote {\bibinfo {title} {Applications of
  secondary ion mass spectrometry ({SIMS}) in materials science},}\ }\href
  {https://doi.org/10.1007/s10853-006-6568-x} {\bibfield  {journal} {\bibinfo
  {journal} {J. Mater. Sci.}\ }\textbf {\bibinfo {volume} {41}},\ \bibinfo
  {pages} {873--903} (\bibinfo {year} {2006}{\natexlab{a}})}\BibitemShut
  {NoStop}%
\bibitem [{\citenamefont {Wirtz}, \citenamefont {Dowsett},\ and\ \citenamefont
  {Philipp}(2016)}]{Wirtz_2016}%
  \BibitemOpen
  \bibfield  {author} {\bibinfo {author} {\bibfnamefont {T.}~\bibnamefont
  {Wirtz}}, \bibinfo {author} {\bibfnamefont {D.}~\bibnamefont {Dowsett}},\
  and\ \bibinfo {author} {\bibfnamefont {P.}~\bibnamefont {Philipp}},\
  }\bibfield  {title} {\enquote {\bibinfo {title} {{SIMS on the Helium Ion
  Microscope: A Powerful Tool for High-Resolution High-Sensitivity
  Nano-Analytics}},}\ }in\  \cite{Hlawacek_2016},\ Chap.~\bibinfo {chapter}
  {13}, pp.\ \bibinfo {pages} {297--323}\BibitemShut {NoStop}%
\bibitem [{\citenamefont {Steele}\ and\ \citenamefont
  {Knuffman}(2023)}]{NanoK}%
  \BibitemOpen
  \bibfield  {author} {\bibinfo {author} {\bibfnamefont {A.}~\bibnamefont
  {Steele}}\ and\ \bibinfo {author} {\bibfnamefont {B.}~\bibnamefont
  {Knuffman}},\ }\href {https://www.zerok.com/products.html#SIMSZERO_GOTO}
  {\enquote {\bibinfo {title} {{SIMS:ZERO}},}\ }\bibinfo {howpublished}
  {Online} (\bibinfo {year} {2023})\BibitemShut {NoStop}%
\bibitem [{\citenamefont {Stevie}\ \emph {et~al.}(1999)\citenamefont {Stevie},
  \citenamefont {Downey}, \citenamefont {Brown}, \citenamefont {Shofner},
  \citenamefont {Decker}, \citenamefont {Dingle},\ and\ \citenamefont
  {Christman}}]{Stevie_1999}%
  \BibitemOpen
  \bibfield  {author} {\bibinfo {author} {\bibfnamefont {F.~A.}\ \bibnamefont
  {Stevie}}, \bibinfo {author} {\bibfnamefont {S.~W.}\ \bibnamefont {Downey}},
  \bibinfo {author} {\bibfnamefont {S.~R.}\ \bibnamefont {Brown}}, \bibinfo
  {author} {\bibfnamefont {T.~L.}\ \bibnamefont {Shofner}}, \bibinfo {author}
  {\bibfnamefont {M.~A.}\ \bibnamefont {Decker}}, \bibinfo {author}
  {\bibfnamefont {T.}~\bibnamefont {Dingle}},\ and\ \bibinfo {author}
  {\bibfnamefont {L.}~\bibnamefont {Christman}},\ }\bibfield  {title} {\enquote
  {\bibinfo {title} {Nanoscale elemental imaging of semiconductor materials
  using focused ion beam secondary ion mass spectrometry},}\ }\href
  {https://doi.org/10.1116/1.591115} {\bibfield  {journal} {\bibinfo  {journal}
  {J. Vac. Sci. Technol. B}\ }\textbf {\bibinfo {volume} {17}},\ \bibinfo
  {pages} {2476} (\bibinfo {year} {1999})}\BibitemShut {NoStop}%
\bibitem [{\citenamefont {Gratia}\ \emph {et~al.}(2016)\citenamefont {Gratia},
  \citenamefont {Grancini}, \citenamefont {Audinot}, \citenamefont
  {Jeanbourquin}, \citenamefont {Mosconi}, \citenamefont {Zimmermann},
  \citenamefont {Dowsett}, \citenamefont {Lee}, \citenamefont {Gr{\"{a}}tzel},
  \citenamefont {Angelis}, \citenamefont {Sivula}, \citenamefont {Wirtz},\ and\
  \citenamefont {Nazeeruddin}}]{Gratia_2016}%
  \BibitemOpen
  \bibfield  {author} {\bibinfo {author} {\bibfnamefont {P.}~\bibnamefont
  {Gratia}}, \bibinfo {author} {\bibfnamefont {G.}~\bibnamefont {Grancini}},
  \bibinfo {author} {\bibfnamefont {J.-N.}\ \bibnamefont {Audinot}}, \bibinfo
  {author} {\bibfnamefont {X.}~\bibnamefont {Jeanbourquin}}, \bibinfo {author}
  {\bibfnamefont {E.}~\bibnamefont {Mosconi}}, \bibinfo {author} {\bibfnamefont
  {I.}~\bibnamefont {Zimmermann}}, \bibinfo {author} {\bibfnamefont
  {D.}~\bibnamefont {Dowsett}}, \bibinfo {author} {\bibfnamefont
  {Y.}~\bibnamefont {Lee}}, \bibinfo {author} {\bibfnamefont {M.}~\bibnamefont
  {Gr{\"{a}}tzel}}, \bibinfo {author} {\bibfnamefont {F.~D.}\ \bibnamefont
  {Angelis}}, \bibinfo {author} {\bibfnamefont {K.}~\bibnamefont {Sivula}},
  \bibinfo {author} {\bibfnamefont {T.}~\bibnamefont {Wirtz}},\ and\ \bibinfo
  {author} {\bibfnamefont {M.~K.}\ \bibnamefont {Nazeeruddin}},\ }\bibfield
  {title} {\enquote {\bibinfo {title} {Intrinsic halide segregation at
  nanometer scale determines the high efficiency of mixed cation/mixed halide
  perovskite solar cells},}\ }\href {https://doi.org/10.1021/jacs.6b10049}
  {\bibfield  {journal} {\bibinfo  {journal} {J. Am. Chem. Soc.}\ }\textbf
  {\bibinfo {volume} {138}},\ \bibinfo {pages} {15821--15824} (\bibinfo {year}
  {2016})}\BibitemShut {NoStop}%
\bibitem [{\citenamefont {Usiobo}\ \emph {et~al.}(2020)\citenamefont {Usiobo},
  \citenamefont {Kanda}, \citenamefont {Gratia}, \citenamefont {Zimmermann},
  \citenamefont {Wirtz}, \citenamefont {Nazeeruddin},\ and\ \citenamefont
  {Audinot}}]{Usiobo_2020}%
  \BibitemOpen
  \bibfield  {author} {\bibinfo {author} {\bibfnamefont {O.~J.}\ \bibnamefont
  {Usiobo}}, \bibinfo {author} {\bibfnamefont {H.}~\bibnamefont {Kanda}},
  \bibinfo {author} {\bibfnamefont {P.}~\bibnamefont {Gratia}}, \bibinfo
  {author} {\bibfnamefont {I.}~\bibnamefont {Zimmermann}}, \bibinfo {author}
  {\bibfnamefont {T.}~\bibnamefont {Wirtz}}, \bibinfo {author} {\bibfnamefont
  {M.~K.}\ \bibnamefont {Nazeeruddin}},\ and\ \bibinfo {author} {\bibfnamefont
  {J.-N.}\ \bibnamefont {Audinot}},\ }\bibfield  {title} {\enquote {\bibinfo
  {title} {Nanoscale mass-spectrometry imaging of grain boundaries in
  perovskite semiconductors},}\ }\href
  {https://doi.org/10.1021/acs.jpcc.0c07464} {\bibfield  {journal} {\bibinfo
  {journal} {J. Phys. Chem. C}\ }\textbf {\bibinfo {volume} {124}},\ \bibinfo
  {pages} {23230--23236} (\bibinfo {year} {2020})}\BibitemShut {NoStop}%
\bibitem [{\citenamefont {Gratia}\ \emph {et~al.}(2017)\citenamefont {Gratia},
  \citenamefont {Zimmermann}, \citenamefont {Schouwink}, \citenamefont {Yum},
  \citenamefont {Audinot}, \citenamefont {Sivula}, \citenamefont {Wirtz},\ and\
  \citenamefont {Nazeeruddin}}]{Gratia_2017}%
  \BibitemOpen
  \bibfield  {author} {\bibinfo {author} {\bibfnamefont {P.}~\bibnamefont
  {Gratia}}, \bibinfo {author} {\bibfnamefont {I.}~\bibnamefont {Zimmermann}},
  \bibinfo {author} {\bibfnamefont {P.}~\bibnamefont {Schouwink}}, \bibinfo
  {author} {\bibfnamefont {J.-H.}\ \bibnamefont {Yum}}, \bibinfo {author}
  {\bibfnamefont {J.-N.}\ \bibnamefont {Audinot}}, \bibinfo {author}
  {\bibfnamefont {K.}~\bibnamefont {Sivula}}, \bibinfo {author} {\bibfnamefont
  {T.}~\bibnamefont {Wirtz}},\ and\ \bibinfo {author} {\bibfnamefont {M.~K.}\
  \bibnamefont {Nazeeruddin}},\ }\bibfield  {title} {\enquote {\bibinfo {title}
  {{The Many Faces of Mixed Ion Perovskites: Unraveling and Understanding the
  Crystallization Process}},}\ }\href
  {https://doi.org/10.1021/acsenergylett.7b00981} {\bibfield  {journal}
  {\bibinfo  {journal} {ACS Energy Lett.}\ }\textbf {\bibinfo {volume} {2}},\
  \bibinfo {pages} {2686--2693} (\bibinfo {year} {2017})}\BibitemShut {NoStop}%
\bibitem [{\citenamefont {Liu}\ \emph {et~al.}(2021{\natexlab{c}})\citenamefont
  {Liu}, \citenamefont {Yang}, \citenamefont {Rakstys}, \citenamefont {Mahata},
  \citenamefont {Franckevicius}, \citenamefont {Mosconi}, \citenamefont
  {Skackauskaite}, \citenamefont {Ding}, \citenamefont {Brooks}, \citenamefont
  {Usiobo}, \citenamefont {Audinot}, \citenamefont {Kanda}, \citenamefont
  {Driukas}, \citenamefont {Kavaliauskaite}, \citenamefont {Gulbinas},
  \citenamefont {Dessimoz}, \citenamefont {Getautis}, \citenamefont {Angelis},
  \citenamefont {Ding}, \citenamefont {Dai}, \citenamefont {Dyson},\ and\
  \citenamefont {Nazeeruddin}}]{Liu_2021}%
  \BibitemOpen
  \bibfield  {author} {\bibinfo {author} {\bibfnamefont {C.}~\bibnamefont
  {Liu}}, \bibinfo {author} {\bibfnamefont {Y.}~\bibnamefont {Yang}}, \bibinfo
  {author} {\bibfnamefont {K.}~\bibnamefont {Rakstys}}, \bibinfo {author}
  {\bibfnamefont {A.}~\bibnamefont {Mahata}}, \bibinfo {author} {\bibfnamefont
  {M.}~\bibnamefont {Franckevicius}}, \bibinfo {author} {\bibfnamefont
  {E.}~\bibnamefont {Mosconi}}, \bibinfo {author} {\bibfnamefont
  {R.}~\bibnamefont {Skackauskaite}}, \bibinfo {author} {\bibfnamefont
  {B.}~\bibnamefont {Ding}}, \bibinfo {author} {\bibfnamefont {K.~G.}\
  \bibnamefont {Brooks}}, \bibinfo {author} {\bibfnamefont {O.~J.}\
  \bibnamefont {Usiobo}}, \bibinfo {author} {\bibfnamefont {J.-N.}\
  \bibnamefont {Audinot}}, \bibinfo {author} {\bibfnamefont {H.}~\bibnamefont
  {Kanda}}, \bibinfo {author} {\bibfnamefont {S.}~\bibnamefont {Driukas}},
  \bibinfo {author} {\bibfnamefont {G.}~\bibnamefont {Kavaliauskaite}},
  \bibinfo {author} {\bibfnamefont {V.}~\bibnamefont {Gulbinas}}, \bibinfo
  {author} {\bibfnamefont {M.}~\bibnamefont {Dessimoz}}, \bibinfo {author}
  {\bibfnamefont {V.}~\bibnamefont {Getautis}}, \bibinfo {author}
  {\bibfnamefont {F.~D.}\ \bibnamefont {Angelis}}, \bibinfo {author}
  {\bibfnamefont {Y.}~\bibnamefont {Ding}}, \bibinfo {author} {\bibfnamefont
  {S.}~\bibnamefont {Dai}}, \bibinfo {author} {\bibfnamefont {P.~J.}\
  \bibnamefont {Dyson}},\ and\ \bibinfo {author} {\bibfnamefont {M.~K.}\
  \bibnamefont {Nazeeruddin}},\ }\bibfield  {title} {\enquote {\bibinfo {title}
  {Tuning structural isomers of phenylenediammonium to afford efficient and
  stable perovskite solar cells and modules},}\ }\href
  {https://doi.org/10.1038/s41467-021-26754-2} {\bibfield  {journal} {\bibinfo
  {journal} {Nat. Commun.}\ }\textbf {\bibinfo {volume} {12}} (\bibinfo {year}
  {2021}{\natexlab{c}}),\ 10.1038/s41467-021-26754-2}\BibitemShut {NoStop}%
\bibitem [{\citenamefont {Zimmermann}\ \emph {et~al.}(2019)\citenamefont
  {Zimmermann}, \citenamefont {Gratia}, \citenamefont {Martineau},
  \citenamefont {Grancini}, \citenamefont {Audinot}, \citenamefont {Wirtz},\
  and\ \citenamefont {Nazeeruddin}}]{Zimmermann_2019}%
  \BibitemOpen
  \bibfield  {author} {\bibinfo {author} {\bibfnamefont {I.}~\bibnamefont
  {Zimmermann}}, \bibinfo {author} {\bibfnamefont {P.}~\bibnamefont {Gratia}},
  \bibinfo {author} {\bibfnamefont {D.}~\bibnamefont {Martineau}}, \bibinfo
  {author} {\bibfnamefont {G.}~\bibnamefont {Grancini}}, \bibinfo {author}
  {\bibfnamefont {J.-N.}\ \bibnamefont {Audinot}}, \bibinfo {author}
  {\bibfnamefont {T.}~\bibnamefont {Wirtz}},\ and\ \bibinfo {author}
  {\bibfnamefont {M.~K.}\ \bibnamefont {Nazeeruddin}},\ }\bibfield  {title}
  {\enquote {\bibinfo {title} {Improved efficiency and reduced hysteresis in
  ultra-stable fully printable mesoscopic perovskite solar cells through
  incorporation of {CuSCN} into the perovskite layer},}\ }\href
  {https://doi.org/10.1039/c9ta00669a} {\bibfield  {journal} {\bibinfo
  {journal} {J. Mater. Chem.A}\ }\textbf {\bibinfo {volume} {7}},\ \bibinfo
  {pages} {8073--8077} (\bibinfo {year} {2019})}\BibitemShut {NoStop}%
\bibitem [{\citenamefont {Sarbada}\ \emph {et~al.}(2020)\citenamefont
  {Sarbada}, \citenamefont {Yedra}, \citenamefont {Pshenova}, \citenamefont
  {Kercher}, \citenamefont {Marschilok}, \citenamefont {Takeuchi},
  \citenamefont {Takeuchi}, \citenamefont {Dudney}, \citenamefont {Wirtz},
  \citenamefont {Eswara},\ and\ \citenamefont {Hull}}]{Sarbada_2020}%
  \BibitemOpen
  \bibfield  {author} {\bibinfo {author} {\bibfnamefont {V.}~\bibnamefont
  {Sarbada}}, \bibinfo {author} {\bibfnamefont {L.}~\bibnamefont {Yedra}},
  \bibinfo {author} {\bibfnamefont {A.}~\bibnamefont {Pshenova}}, \bibinfo
  {author} {\bibfnamefont {A.}~\bibnamefont {Kercher}}, \bibinfo {author}
  {\bibfnamefont {A.}~\bibnamefont {Marschilok}}, \bibinfo {author}
  {\bibfnamefont {K.~J.}\ \bibnamefont {Takeuchi}}, \bibinfo {author}
  {\bibfnamefont {E.}~\bibnamefont {Takeuchi}}, \bibinfo {author}
  {\bibfnamefont {N.}~\bibnamefont {Dudney}}, \bibinfo {author} {\bibfnamefont
  {T.}~\bibnamefont {Wirtz}}, \bibinfo {author} {\bibfnamefont
  {S.}~\bibnamefont {Eswara}},\ and\ \bibinfo {author} {\bibfnamefont
  {R.}~\bibnamefont {Hull}},\ }\bibfield  {title} {\enquote {\bibinfo {title}
  {Correlative electron and ion beam analysis of the electrochemical
  performances of {LiV}3o8 cathode films as a function of microstructures},}\
  }\href {https://doi.org/10.1016/j.jpowsour.2020.228177} {\bibfield  {journal}
  {\bibinfo  {journal} {J. Power Sources}\ }\textbf {\bibinfo {volume} {463}},\
  \bibinfo {pages} {228177} (\bibinfo {year} {2020})}\BibitemShut {NoStop}%
\bibitem [{\citenamefont {Wheatcroft}\ \emph {et~al.}(2020)\citenamefont
  {Wheatcroft}, \citenamefont {Klingner}, \citenamefont {Heller}, \citenamefont
  {Hlawacek}, \citenamefont {{{\"{O}}}zkaya}, \citenamefont {Cookson},\ and\
  \citenamefont {Inkson}}]{Wheatcroft_2020}%
  \BibitemOpen
  \bibfield  {author} {\bibinfo {author} {\bibfnamefont {L.}~\bibnamefont
  {Wheatcroft}}, \bibinfo {author} {\bibfnamefont {N.}~\bibnamefont
  {Klingner}}, \bibinfo {author} {\bibfnamefont {R.}~\bibnamefont {Heller}},
  \bibinfo {author} {\bibfnamefont {G.}~\bibnamefont {Hlawacek}}, \bibinfo
  {author} {\bibfnamefont {D.}~\bibnamefont {{{\"{O}}}zkaya}}, \bibinfo
  {author} {\bibfnamefont {J.}~\bibnamefont {Cookson}},\ and\ \bibinfo {author}
  {\bibfnamefont {B.~J.}\ \bibnamefont {Inkson}},\ }\bibfield  {title}
  {\enquote {\bibinfo {title} {Visualization and chemical characterization of
  the cathode electrolyte interphase using he-ion microscopy and in situ
  time-of-flight secondary ion mass spectrometry},}\ }\href
  {https://doi.org/10.1021/acsaem.0c01333} {\bibfield  {journal} {\bibinfo
  {journal} {ACS Appl. Energy Mater.}\ }\textbf {\bibinfo {volume} {3}},\
  \bibinfo {pages} {8822--8832} (\bibinfo {year} {2020})}\BibitemShut {NoStop}%
\bibitem [{\citenamefont {Taeye}, \citenamefont {Mees},\ and\ \citenamefont
  {Vereecken}(2021)}]{Taeye_2021}%
  \BibitemOpen
  \bibfield  {author} {\bibinfo {author} {\bibfnamefont {L.~L.~D.}\
  \bibnamefont {Taeye}}, \bibinfo {author} {\bibfnamefont {M.~J.}\ \bibnamefont
  {Mees}},\ and\ \bibinfo {author} {\bibfnamefont {P.~M.}\ \bibnamefont
  {Vereecken}},\ }\bibfield  {title} {\enquote {\bibinfo {title} {Surpassing
  the 1 li/ti capacity limit in chlorine modified {TiO}2-{yCl}2y},}\ }\href
  {https://doi.org/10.1016/j.ensm.2020.12.030} {\bibfield  {journal} {\bibinfo
  {journal} {Energy Storage Materials}\ }\textbf {\bibinfo {volume} {36}},\
  \bibinfo {pages} {279--290} (\bibinfo {year} {2021})}\BibitemShut {NoStop}%
\bibitem [{\citenamefont {Sastre}\ \emph {et~al.}(2020)\citenamefont {Sastre},
  \citenamefont {Priebe}, \citenamefont {D{\"{o}}beli}, \citenamefont
  {Michler}, \citenamefont {Tiwari},\ and\ \citenamefont
  {Romanyuk}}]{Sastre_2020a}%
  \BibitemOpen
  \bibfield  {author} {\bibinfo {author} {\bibfnamefont {J.}~\bibnamefont
  {Sastre}}, \bibinfo {author} {\bibfnamefont {A.}~\bibnamefont {Priebe}},
  \bibinfo {author} {\bibfnamefont {M.}~\bibnamefont {D{\"{o}}beli}}, \bibinfo
  {author} {\bibfnamefont {J.}~\bibnamefont {Michler}}, \bibinfo {author}
  {\bibfnamefont {A.~N.}\ \bibnamefont {Tiwari}},\ and\ \bibinfo {author}
  {\bibfnamefont {Y.~E.}\ \bibnamefont {Romanyuk}},\ }\bibfield  {title}
  {\enquote {\bibinfo {title} {Lithium garnet {Li}$_7${La}$_3${Zr}$_
  2${O}$_{12}$ electrolyte for all-solid-state batteries: closing the gap
  between bulk and thin film {L}i-ion conductivities},}\ }\href
  {https://doi.org/10.1002/admi.202000425} {\bibfield  {journal} {\bibinfo
  {journal} {Adv. Mater. Interfaces}\ }\textbf {\bibinfo {volume} {7}},\
  \bibinfo {pages} {2000425} (\bibinfo {year} {2020})}\BibitemShut {NoStop}%
\bibitem [{\citenamefont {Dubey}\ \emph {et~al.}(2021)\citenamefont {Dubey},
  \citenamefont {Sastre}, \citenamefont {Cancellieri}, \citenamefont {Okur},
  \citenamefont {Forster}, \citenamefont {Pompizii}, \citenamefont {Priebe},
  \citenamefont {Romanyuk}, \citenamefont {Jeurgens}, \citenamefont
  {Kovalenko},\ and\ \citenamefont {Kravchyk}}]{Dubey_2021}%
  \BibitemOpen
  \bibfield  {author} {\bibinfo {author} {\bibfnamefont {R.}~\bibnamefont
  {Dubey}}, \bibinfo {author} {\bibfnamefont {J.}~\bibnamefont {Sastre}},
  \bibinfo {author} {\bibfnamefont {C.}~\bibnamefont {Cancellieri}}, \bibinfo
  {author} {\bibfnamefont {F.}~\bibnamefont {Okur}}, \bibinfo {author}
  {\bibfnamefont {A.}~\bibnamefont {Forster}}, \bibinfo {author} {\bibfnamefont
  {L.}~\bibnamefont {Pompizii}}, \bibinfo {author} {\bibfnamefont
  {A.}~\bibnamefont {Priebe}}, \bibinfo {author} {\bibfnamefont {Y.~E.}\
  \bibnamefont {Romanyuk}}, \bibinfo {author} {\bibfnamefont {L.~P.~H.}\
  \bibnamefont {Jeurgens}}, \bibinfo {author} {\bibfnamefont {M.~V.}\
  \bibnamefont {Kovalenko}},\ and\ \bibinfo {author} {\bibfnamefont {K.~V.}\
  \bibnamefont {Kravchyk}},\ }\bibfield  {title} {\enquote {\bibinfo {title}
  {Building a better {L}i-garnet solid electrolyte/metallic {L}i interface with
  antimony},}\ }\href {https://doi.org/10.1002/aenm.202102086} {\bibfield
  {journal} {\bibinfo  {journal} {Adv. Energy Mater.}\ }\textbf {\bibinfo
  {volume} {11}},\ \bibinfo {pages} {2102086} (\bibinfo {year}
  {2021})}\BibitemShut {NoStop}%
\bibitem [{\citenamefont {Nishinomiya}\ \emph {et~al.}(2016)\citenamefont
  {Nishinomiya}, \citenamefont {Toshin}, \citenamefont {Shishido},\ and\
  \citenamefont {Suzuki}}]{Nishinomiya_2016a}%
  \BibitemOpen
  \bibfield  {author} {\bibinfo {author} {\bibfnamefont {S.}~\bibnamefont
  {Nishinomiya}}, \bibinfo {author} {\bibfnamefont {K.}~\bibnamefont {Toshin}},
  \bibinfo {author} {\bibfnamefont {R.}~\bibnamefont {Shishido}},\ and\
  \bibinfo {author} {\bibfnamefont {S.}~\bibnamefont {Suzuki}},\ }\bibfield
  {title} {\enquote {\bibinfo {title} {{TOF}-{SIMS} imaging of
  polyester/melamine resin with bismuth cluster ions},}\ }\href
  {https://doi.org/10.1002/sia.6087} {\bibfield  {journal} {\bibinfo  {journal}
  {Surf. Interface Anal.}\ }\textbf {\bibinfo {volume} {48}},\ \bibinfo {pages}
  {1114--1118} (\bibinfo {year} {2016})}\BibitemShut {NoStop}%
\bibitem [{\citenamefont {Gaderbauer}\ \emph {et~al.}(2020)\citenamefont
  {Gaderbauer}, \citenamefont {Arndt}, \citenamefont {Truglas}, \citenamefont
  {Steck}, \citenamefont {Klingner}, \citenamefont {Stifter}, \citenamefont
  {Faderl},\ and\ \citenamefont {Groiss}}]{Gaderbauer_2020}%
  \BibitemOpen
  \bibfield  {author} {\bibinfo {author} {\bibfnamefont {W.}~\bibnamefont
  {Gaderbauer}}, \bibinfo {author} {\bibfnamefont {M.}~\bibnamefont {Arndt}},
  \bibinfo {author} {\bibfnamefont {T.}~\bibnamefont {Truglas}}, \bibinfo
  {author} {\bibfnamefont {T.}~\bibnamefont {Steck}}, \bibinfo {author}
  {\bibfnamefont {N.}~\bibnamefont {Klingner}}, \bibinfo {author}
  {\bibfnamefont {D.}~\bibnamefont {Stifter}}, \bibinfo {author} {\bibfnamefont
  {J.}~\bibnamefont {Faderl}},\ and\ \bibinfo {author} {\bibfnamefont
  {H.}~\bibnamefont {Groiss}},\ }\bibfield  {title} {\enquote {\bibinfo {title}
  {Effects of alloying elements on surface oxides of hot{\textendash}dip
  galvanized press hardened steel},}\ }\href
  {https://doi.org/10.1016/j.surfcoat.2020.126466} {\bibfield  {journal}
  {\bibinfo  {journal} {Surf. Coat. Technol.}\ }\textbf {\bibinfo {volume}
  {404}},\ \bibinfo {pages} {126466} (\bibinfo {year} {2020})}\BibitemShut
  {NoStop}%
\bibitem [{\citenamefont {Kosari}\ \emph {et~al.}(2020)\citenamefont {Kosari},
  \citenamefont {Visser}, \citenamefont {Tichelaar}, \citenamefont {Eswara},
  \citenamefont {Audinot}, \citenamefont {Wirtz}, \citenamefont {Zandbergen},
  \citenamefont {Terryn},\ and\ \citenamefont {Mol}}]{Kosari_2020}%
  \BibitemOpen
  \bibfield  {author} {\bibinfo {author} {\bibfnamefont {A.}~\bibnamefont
  {Kosari}}, \bibinfo {author} {\bibfnamefont {P.}~\bibnamefont {Visser}},
  \bibinfo {author} {\bibfnamefont {F.}~\bibnamefont {Tichelaar}}, \bibinfo
  {author} {\bibfnamefont {S.}~\bibnamefont {Eswara}}, \bibinfo {author}
  {\bibfnamefont {J.-N.}\ \bibnamefont {Audinot}}, \bibinfo {author}
  {\bibfnamefont {T.}~\bibnamefont {Wirtz}}, \bibinfo {author} {\bibfnamefont
  {H.}~\bibnamefont {Zandbergen}}, \bibinfo {author} {\bibfnamefont
  {H.}~\bibnamefont {Terryn}},\ and\ \bibinfo {author} {\bibfnamefont
  {J.~M.~C.}\ \bibnamefont {Mol}},\ }\bibfield  {title} {\enquote {\bibinfo
  {title} {Cross-sectional characterization of the conversion layer formed on
  {AA}2024-t3 by a lithium-leaching coating},}\ }\href
  {https://doi.org/10.1016/j.apsusc.2020.145665} {\bibfield  {journal}
  {\bibinfo  {journal} {Appl. Surf. Sci.}\ }\textbf {\bibinfo {volume} {512}},\
  \bibinfo {pages} {145665} (\bibinfo {year} {2020})}\BibitemShut {NoStop}%
\bibitem [{\citenamefont {Wieczerzak}\ \emph {et~al.}(2021)\citenamefont
  {Wieczerzak}, \citenamefont {Priebe}, \citenamefont {Utke},\ and\
  \citenamefont {Michler}}]{Wieczerzak_2021a}%
  \BibitemOpen
  \bibfield  {author} {\bibinfo {author} {\bibfnamefont {K.}~\bibnamefont
  {Wieczerzak}}, \bibinfo {author} {\bibfnamefont {A.}~\bibnamefont {Priebe}},
  \bibinfo {author} {\bibfnamefont {I.}~\bibnamefont {Utke}},\ and\ \bibinfo
  {author} {\bibfnamefont {J.}~\bibnamefont {Michler}},\ }\bibfield  {title}
  {\enquote {\bibinfo {title} {Practical aspects of focused ion beam
  time-of-flight secondary ion mass spectrometry analysis enhanced by fluorine
  gas coinjection},}\ }\href {https://doi.org/10.1021/acs.chemmater.1c00052}
  {\bibfield  {journal} {\bibinfo  {journal} {Chem. Mater.}\ }\textbf {\bibinfo
  {volume} {33}},\ \bibinfo {pages} {1581--1593} (\bibinfo {year}
  {2021})}\BibitemShut {NoStop}%
\bibitem [{\citenamefont {Kennedy}\ \emph {et~al.}(2020)\citenamefont
  {Kennedy}, \citenamefont {Zapf}, \citenamefont {Audinot}, \citenamefont
  {Pal}, \citenamefont {Eswara}, \citenamefont {Wirtz}, \citenamefont
  {Ronning},\ and\ \citenamefont {Warburton}}]{Kennedy_2020}%
  \BibitemOpen
  \bibfield  {author} {\bibinfo {author} {\bibfnamefont {O.~W.}\ \bibnamefont
  {Kennedy}}, \bibinfo {author} {\bibfnamefont {M.}~\bibnamefont {Zapf}},
  \bibinfo {author} {\bibfnamefont {J.-N.}\ \bibnamefont {Audinot}}, \bibinfo
  {author} {\bibfnamefont {S.}~\bibnamefont {Pal}}, \bibinfo {author}
  {\bibfnamefont {S.}~\bibnamefont {Eswara}}, \bibinfo {author} {\bibfnamefont
  {T.}~\bibnamefont {Wirtz}}, \bibinfo {author} {\bibfnamefont
  {C.}~\bibnamefont {Ronning}},\ and\ \bibinfo {author} {\bibfnamefont {P.~A.}\
  \bibnamefont {Warburton}},\ }\bibfield  {title} {\enquote {\bibinfo {title}
  {Photoluminescence of {ZnO}/{ZnMgO} heterostructure nanobelts grown by
  {MBE}},}\ }\href {https://doi.org/10.1088/1361-6528/ab60cb} {\bibfield
  {journal} {\bibinfo  {journal} {Nanotechnology}\ }\textbf {\bibinfo {volume}
  {31}},\ \bibinfo {pages} {135604} (\bibinfo {year} {2020})}\BibitemShut
  {NoStop}%
\bibitem [{\citenamefont {Fizesan}\ \emph {et~al.}(2019)\citenamefont
  {Fizesan}, \citenamefont {Cambier}, \citenamefont {Moschini}, \citenamefont
  {Chary}, \citenamefont {Nelissen}, \citenamefont {Ziebel}, \citenamefont
  {Audinot}, \citenamefont {Wirtz}, \citenamefont {Kruszewski}, \citenamefont
  {Pop}, \citenamefont {Kiss}, \citenamefont {Serchi}, \citenamefont {Loghin},\
  and\ \citenamefont {Gutleb}}]{Fizesan_2019}%
  \BibitemOpen
  \bibfield  {author} {\bibinfo {author} {\bibfnamefont {I.}~\bibnamefont
  {Fizesan}}, \bibinfo {author} {\bibfnamefont {S.}~\bibnamefont {Cambier}},
  \bibinfo {author} {\bibfnamefont {E.}~\bibnamefont {Moschini}}, \bibinfo
  {author} {\bibfnamefont {A.}~\bibnamefont {Chary}}, \bibinfo {author}
  {\bibfnamefont {I.}~\bibnamefont {Nelissen}}, \bibinfo {author}
  {\bibfnamefont {J.}~\bibnamefont {Ziebel}}, \bibinfo {author} {\bibfnamefont
  {J.-N.}\ \bibnamefont {Audinot}}, \bibinfo {author} {\bibfnamefont
  {T.}~\bibnamefont {Wirtz}}, \bibinfo {author} {\bibfnamefont
  {M.}~\bibnamefont {Kruszewski}}, \bibinfo {author} {\bibfnamefont
  {A.}~\bibnamefont {Pop}}, \bibinfo {author} {\bibfnamefont {B.}~\bibnamefont
  {Kiss}}, \bibinfo {author} {\bibfnamefont {T.}~\bibnamefont {Serchi}},
  \bibinfo {author} {\bibfnamefont {F.}~\bibnamefont {Loghin}},\ and\ \bibinfo
  {author} {\bibfnamefont {A.~C.}\ \bibnamefont {Gutleb}},\ }\bibfield  {title}
  {\enquote {\bibinfo {title} {In vitro exposure of a 3{D}-tetraculture
  representative for the alveolar barrier at the air-liquid interface to silver
  particles and nanowires},}\ }\href
  {https://doi.org/10.1186/s12989-019-0297-1} {\bibfield  {journal} {\bibinfo
  {journal} {Part. Fibre Toxicol.}\ }\textbf {\bibinfo {volume} {16}} (\bibinfo
  {year} {2019}),\ 10.1186/s12989-019-0297-1}\BibitemShut {NoStop}%
\bibitem [{\citenamefont {Priebe}\ \emph
  {et~al.}(2019{\natexlab{b}})\citenamefont {Priebe}, \citenamefont {Barnes},
  \citenamefont {Edwards}, \citenamefont {Peth{\"{o}}}, \citenamefont
  {Balogh},\ and\ \citenamefont {Michler}}]{Priebe_2019a}%
  \BibitemOpen
  \bibfield  {author} {\bibinfo {author} {\bibfnamefont {A.}~\bibnamefont
  {Priebe}}, \bibinfo {author} {\bibfnamefont {J.-P.}\ \bibnamefont {Barnes}},
  \bibinfo {author} {\bibfnamefont {T.~E.~J.}\ \bibnamefont {Edwards}},
  \bibinfo {author} {\bibfnamefont {L.}~\bibnamefont {Peth{\"{o}}}}, \bibinfo
  {author} {\bibfnamefont {I.}~\bibnamefont {Balogh}},\ and\ \bibinfo {author}
  {\bibfnamefont {J.}~\bibnamefont {Michler}},\ }\bibfield  {title} {\enquote
  {\bibinfo {title} {{3D} imaging of nanoparticles in an inorganic matrix using
  {TOF}-{SIMS} validated with {STEM} and {EDX}},}\ }\href
  {https://doi.org/10.1021/acs.analchem.9b02545} {\bibfield  {journal}
  {\bibinfo  {journal} {Anal. Chem.}\ }\textbf {\bibinfo {volume} {91}},\
  \bibinfo {pages} {11834--11839} (\bibinfo {year}
  {2019}{\natexlab{b}})}\BibitemShut {NoStop}%
\bibitem [{\citenamefont {Priebe}\ \emph {et~al.}(2020)\citenamefont {Priebe},
  \citenamefont {Barnes}, \citenamefont {Edwards}, \citenamefont
  {Husz{\'{a}}r}, \citenamefont {Peth{\"{o}}},\ and\ \citenamefont
  {Michler}}]{Priebe_2020a}%
  \BibitemOpen
  \bibfield  {author} {\bibinfo {author} {\bibfnamefont {A.}~\bibnamefont
  {Priebe}}, \bibinfo {author} {\bibfnamefont {J.-P.}\ \bibnamefont {Barnes}},
  \bibinfo {author} {\bibfnamefont {T.~E.~J.}\ \bibnamefont {Edwards}},
  \bibinfo {author} {\bibfnamefont {E.}~\bibnamefont {Husz{\'{a}}r}}, \bibinfo
  {author} {\bibfnamefont {L.}~\bibnamefont {Peth{\"{o}}}},\ and\ \bibinfo
  {author} {\bibfnamefont {J.}~\bibnamefont {Michler}},\ }\bibfield  {title}
  {\enquote {\bibinfo {title} {Elemental characterization of {A}l nanoparticles
  buried under a {C}u thin film: {TOF}-{SIMS} vs {STEM}/{EDX}},}\ }\href
  {https://doi.org/10.1021/acs.analchem.0c02361} {\bibfield  {journal}
  {\bibinfo  {journal} {Anal. Chem.}\ }\textbf {\bibinfo {volume} {92}},\
  \bibinfo {pages} {12518--12527} (\bibinfo {year} {2020})}\BibitemShut
  {NoStop}%
\bibitem [{\citenamefont {Lunca-Popa}\ \emph {et~al.}(2020)\citenamefont
  {Lunca-Popa}, \citenamefont {Botsoa}, \citenamefont {Bahri}, \citenamefont
  {Cr{\^{e}}pelli{\`{e}}re}, \citenamefont {Desgardin}, \citenamefont
  {Audinot}, \citenamefont {Wirtz}, \citenamefont {Arl}, \citenamefont {Ersen},
  \citenamefont {Barthe},\ and\ \citenamefont {Lenoble}}]{LuncaPopa_2020}%
  \BibitemOpen
  \bibfield  {author} {\bibinfo {author} {\bibfnamefont {P.}~\bibnamefont
  {Lunca-Popa}}, \bibinfo {author} {\bibfnamefont {J.}~\bibnamefont {Botsoa}},
  \bibinfo {author} {\bibfnamefont {M.}~\bibnamefont {Bahri}}, \bibinfo
  {author} {\bibfnamefont {J.}~\bibnamefont {Cr{\^{e}}pelli{\`{e}}re}},
  \bibinfo {author} {\bibfnamefont {P.}~\bibnamefont {Desgardin}}, \bibinfo
  {author} {\bibfnamefont {J.-N.}\ \bibnamefont {Audinot}}, \bibinfo {author}
  {\bibfnamefont {T.}~\bibnamefont {Wirtz}}, \bibinfo {author} {\bibfnamefont
  {D.}~\bibnamefont {Arl}}, \bibinfo {author} {\bibfnamefont {O.}~\bibnamefont
  {Ersen}}, \bibinfo {author} {\bibfnamefont {M.-F.}\ \bibnamefont {Barthe}},\
  and\ \bibinfo {author} {\bibfnamefont {D.}~\bibnamefont {Lenoble}},\
  }\bibfield  {title} {\enquote {\bibinfo {title} {Tuneable interplay between
  atomistic defects morphology and electrical properties of transparent p-type
  highly conductive off-stoichiometric cu-cr-o delafossite thin films},}\
  }\href {https://doi.org/10.1038/s41598-020-58312-z} {\bibfield  {journal}
  {\bibinfo  {journal} {Sci. Rep.}\ }\textbf {\bibinfo {volume} {10}} (\bibinfo
  {year} {2020}),\ 10.1038/s41598-020-58312-z}\BibitemShut {NoStop}%
\bibitem [{\citenamefont {Ooi}\ \emph {et~al.}(2006)\citenamefont {Ooi},
  \citenamefont {Mcphail}, \citenamefont {Chater},\ and\ \citenamefont
  {Shollock}}]{Ooi_2006}%
  \BibitemOpen
  \bibfield  {author} {\bibinfo {author} {\bibfnamefont {T.~N.}\ \bibnamefont
  {Ooi}}, \bibinfo {author} {\bibfnamefont {D.~S.}\ \bibnamefont {Mcphail}},
  \bibinfo {author} {\bibfnamefont {R.~J.}\ \bibnamefont {Chater}},\ and\
  \bibinfo {author} {\bibfnamefont {B.~A.}\ \bibnamefont {Shollock}},\
  }\bibfield  {title} {\enquote {\bibinfo {title} {Isotope exchange studies of
  oxidation mechanisms in nickel-base superalloys using {FIB}-{SIMS}
  techniques},}\ }\href {https://doi.org/10.1016/j.surfcoat.2006.07.243}
  {\bibfield  {journal} {\bibinfo  {journal} {Surf. Coat. Technol.}\ }\textbf
  {\bibinfo {volume} {201}},\ \bibinfo {pages} {3885--3888} (\bibinfo {year}
  {2006})}\BibitemShut {NoStop}%
\bibitem [{\citenamefont {Sedl{\'{a}}{\v{c}}ek}\ \emph
  {et~al.}(2014)\citenamefont {Sedl{\'{a}}{\v{c}}ek}, \citenamefont
  {Kolo{\v{s}}ov{\'{a}}}, \citenamefont {Jiru{\v{s}}e},\ and\ \citenamefont
  {Stevie}}]{Sedlacek_2014}%
  \BibitemOpen
  \bibfield  {author} {\bibinfo {author} {\bibfnamefont {L.}~\bibnamefont
  {Sedl{\'{a}}{\v{c}}ek}}, \bibinfo {author} {\bibfnamefont {J.}~\bibnamefont
  {Kolo{\v{s}}ov{\'{a}}}}, \bibinfo {author} {\bibfnamefont {J.}~\bibnamefont
  {Jiru{\v{s}}e}},\ and\ \bibinfo {author} {\bibfnamefont {F.~A.}\ \bibnamefont
  {Stevie}},\ }\bibfield  {title} {\enquote {\bibinfo {title} {Insulator
  analysis using combined {FIB}-{SEM} instrument with {TOF}-{SIMS}},}\ }\href
  {https://doi.org/10.1017/s1431927614003250} {\bibfield  {journal} {\bibinfo
  {journal} {Microsc. Microanal.}\ }\textbf {\bibinfo {volume} {20}},\ \bibinfo
  {pages} {306--307} (\bibinfo {year} {2014})}\BibitemShut {NoStop}%
\bibitem [{\citenamefont {Lorin{\v{c}}{\'{\i}}k}\ \emph
  {et~al.}(2014)\citenamefont {Lorin{\v{c}}{\'{\i}}k}, \citenamefont
  {Ka{\v{s}}{\'{\i}}k}, \citenamefont {Vani{\v{s}}}, \citenamefont
  {Sedl{\'{a}}{\v{c}}ek},\ and\ \citenamefont {Dluho{\v{s}}}}]{Lorincik_2014}%
  \BibitemOpen
  \bibfield  {author} {\bibinfo {author} {\bibfnamefont {J.}~\bibnamefont
  {Lorin{\v{c}}{\'{\i}}k}}, \bibinfo {author} {\bibfnamefont {I.}~\bibnamefont
  {Ka{\v{s}}{\'{\i}}k}}, \bibinfo {author} {\bibfnamefont {J.}~\bibnamefont
  {Vani{\v{s}}}}, \bibinfo {author} {\bibfnamefont {L.}~\bibnamefont
  {Sedl{\'{a}}{\v{c}}ek}},\ and\ \bibinfo {author} {\bibfnamefont
  {J.}~\bibnamefont {Dluho{\v{s}}}},\ }\bibfield  {title} {\enquote {\bibinfo
  {title} {Imaging of dopant distribution in optical fibers with an orthogonal
  {TOF} {SIMS}},}\ }\href {https://doi.org/10.1002/sia.5536} {\bibfield
  {journal} {\bibinfo  {journal} {Surf. Interface Anal.}\ }\textbf {\bibinfo
  {volume} {46}},\ \bibinfo {pages} {238--240} (\bibinfo {year}
  {2014})}\BibitemShut {NoStop}%
\bibitem [{\citenamefont {Priebe}\ \emph
  {et~al.}(2021{\natexlab{c}})\citenamefont {Priebe}, \citenamefont
  {Peth{\"{o}}}, \citenamefont {Huszar}, \citenamefont {Xie}, \citenamefont
  {Utke},\ and\ \citenamefont {Michler}}]{Priebe_2021c}%
  \BibitemOpen
  \bibfield  {author} {\bibinfo {author} {\bibfnamefont {A.}~\bibnamefont
  {Priebe}}, \bibinfo {author} {\bibfnamefont {L.}~\bibnamefont {Peth{\"{o}}}},
  \bibinfo {author} {\bibfnamefont {E.}~\bibnamefont {Huszar}}, \bibinfo
  {author} {\bibfnamefont {T.}~\bibnamefont {Xie}}, \bibinfo {author}
  {\bibfnamefont {I.}~\bibnamefont {Utke}},\ and\ \bibinfo {author}
  {\bibfnamefont {J.}~\bibnamefont {Michler}},\ }\bibfield  {title} {\enquote
  {\bibinfo {title} {High sensitivity of fluorine gas-assisted
  {FIB}-{TOF}-{SIMS} for chemical characterization of buried sublayers in thin
  films},}\ }\href {https://doi.org/10.1021/acsami.1c01627} {\bibfield
  {journal} {\bibinfo  {journal} {ACS Appl. Mater. Interfaces}\ }\textbf
  {\bibinfo {volume} {13}},\ \bibinfo {pages} {15890--15900} (\bibinfo {year}
  {2021}{\natexlab{c}})}\BibitemShut {NoStop}%
\bibitem [{\citenamefont {Jundt}\ \emph {et~al.}(1974)\citenamefont {Jundt},
  \citenamefont {Purser}, \citenamefont {Kubo},\ and\ \citenamefont
  {Schenk}}]{Jundt_1974}%
  \BibitemOpen
  \bibfield  {author} {\bibinfo {author} {\bibfnamefont {F.~C.}\ \bibnamefont
  {Jundt}}, \bibinfo {author} {\bibfnamefont {K.~H.}\ \bibnamefont {Purser}},
  \bibinfo {author} {\bibfnamefont {H.}~\bibnamefont {Kubo}},\ and\ \bibinfo
  {author} {\bibfnamefont {E.~A.}\ \bibnamefont {Schenk}},\ }\bibfield  {title}
  {\enquote {\bibinfo {title} {Proton-induced x-ray analysis of trace elements
  in tissue sections},}\ }\href {https://doi.org/10.1177/22.1.1} {\bibfield
  {journal} {\bibinfo  {journal} {J Histochem Cytochem}\ }\textbf {\bibinfo
  {volume} {22}},\ \bibinfo {pages} {1--6} (\bibinfo {year}
  {1974})}\BibitemShut {NoStop}%
\bibitem [{\citenamefont {Levi-Setti}, \citenamefont {Wang},\ and\
  \citenamefont {Crow}(1986)}]{LeviSetti_1986}%
  \BibitemOpen
  \bibfield  {author} {\bibinfo {author} {\bibfnamefont {R.}~\bibnamefont
  {Levi-Setti}}, \bibinfo {author} {\bibfnamefont {Y.~L.}\ \bibnamefont
  {Wang}},\ and\ \bibinfo {author} {\bibfnamefont {G.}~\bibnamefont {Crow}},\
  }\bibfield  {title} {\enquote {\bibinfo {title} {Scanning ion microscopy:
  elemental maps at high lateral resolution},}\ }\href
  {https://doi.org/10.1016/0169-4332(86)90067-X} {\bibfield  {journal}
  {\bibinfo  {journal} {Appl. Surf. Sci.}\ }\textbf {\bibinfo {volume} {26}},\
  \bibinfo {pages} {249--264} (\bibinfo {year} {1986})}\BibitemShut {NoStop}%
\bibitem [{\citenamefont {Levi-Setti}(1988)}]{LeviSetti_1988}%
  \BibitemOpen
  \bibfield  {author} {\bibinfo {author} {\bibfnamefont {R.}~\bibnamefont
  {Levi-Setti}},\ }\bibfield  {title} {\enquote {\bibinfo {title} {Structural
  and microanalytical imaging of biological materials by scanning microscopy
  with heavy-ion probes},}\ }\href
  {https://doi.org/10.1146/annurev.bb.17.060188.001545} {\bibfield  {journal}
  {\bibinfo  {journal} {Annual review of biophysics and biophysical chemistry}\
  }\textbf {\bibinfo {volume} {17}},\ \bibinfo {pages} {325--347} (\bibinfo
  {year} {1988})}\BibitemShut {NoStop}%
\bibitem [{\citenamefont {Boxer}, \citenamefont {Kraft},\ and\ \citenamefont
  {Weber}(2009)}]{Boxer_2009}%
  \BibitemOpen
  \bibfield  {author} {\bibinfo {author} {\bibfnamefont {S.~G.}\ \bibnamefont
  {Boxer}}, \bibinfo {author} {\bibfnamefont {M.~L.}\ \bibnamefont {Kraft}},\
  and\ \bibinfo {author} {\bibfnamefont {P.~K.}\ \bibnamefont {Weber}},\
  }\bibfield  {title} {\enquote {\bibinfo {title} {Advances in imaging
  secondary ion mass spectrometry for biological samples},}\ }\href
  {https://doi.org/10.1146/annurev.biophys.050708.133634} {\bibfield  {journal}
  {\bibinfo  {journal} {Annu. Rev. Biophys.}\ }\textbf {\bibinfo {volume}
  {38}},\ \bibinfo {pages} {53--74} (\bibinfo {year} {2009})},\ \bibinfo {note}
  {pMID: 19086820}\BibitemShut {NoStop}%
\bibitem [{\citenamefont {Szakal}\ \emph {et~al.}(2011)\citenamefont {Szakal},
  \citenamefont {Narayan}, \citenamefont {Fu}, \citenamefont {Lefman},\ and\
  \citenamefont {Subramaniam}}]{Szakal_2011}%
  \BibitemOpen
  \bibfield  {author} {\bibinfo {author} {\bibfnamefont {C.}~\bibnamefont
  {Szakal}}, \bibinfo {author} {\bibfnamefont {K.}~\bibnamefont {Narayan}},
  \bibinfo {author} {\bibfnamefont {J.}~\bibnamefont {Fu}}, \bibinfo {author}
  {\bibfnamefont {J.}~\bibnamefont {Lefman}},\ and\ \bibinfo {author}
  {\bibfnamefont {S.}~\bibnamefont {Subramaniam}},\ }\bibfield  {title}
  {\enquote {\bibinfo {title} {Compositional mapping of the surface and
  interior of mammalian cells at submicrometer resolution},}\ }\href
  {https://doi.org/10.1021/ac1030607} {\bibfield  {journal} {\bibinfo
  {journal} {Anal. Chem.}\ }\textbf {\bibinfo {volume} {83}},\ \bibinfo {pages}
  {1207--1213} (\bibinfo {year} {2011})}\BibitemShut {NoStop}%
\bibitem [{\citenamefont {Malmberg}\ and\ \citenamefont
  {Nygren}(2008)}]{Malmberg_2008}%
  \BibitemOpen
  \bibfield  {author} {\bibinfo {author} {\bibfnamefont {P.}~\bibnamefont
  {Malmberg}}\ and\ \bibinfo {author} {\bibfnamefont {H.}~\bibnamefont
  {Nygren}},\ }\bibfield  {title} {\enquote {\bibinfo {title} {Methods for the
  analysis of the composition of bone tissue, with a focus on imaging mass
  spectrometry ({TOF}-{SIMS})},}\ }\href
  {https://doi.org/10.1002/pmic.200800198} {\bibfield  {journal} {\bibinfo
  {journal} {Proteomics}\ }\textbf {\bibinfo {volume} {8}},\ \bibinfo {pages}
  {3755--3762} (\bibinfo {year} {2008})}\BibitemShut {NoStop}%
\bibitem [{\citenamefont {Nu{\~{n}}ez}\ \emph {et~al.}(2018)\citenamefont
  {Nu{\~{n}}ez}, \citenamefont {Renslow}, \citenamefont {Cliff},\ and\
  \citenamefont {Anderton}}]{Nunez_2017}%
  \BibitemOpen
  \bibfield  {author} {\bibinfo {author} {\bibfnamefont {J.}~\bibnamefont
  {Nu{\~{n}}ez}}, \bibinfo {author} {\bibfnamefont {R.}~\bibnamefont
  {Renslow}}, \bibinfo {author} {\bibfnamefont {J.~B.}\ \bibnamefont {Cliff}},\
  and\ \bibinfo {author} {\bibfnamefont {C.~R.}\ \bibnamefont {Anderton}},\
  }\bibfield  {title} {\enquote {\bibinfo {title} {Nanosims for biological
  applications: current practices and analyses},}\ }\href
  {https://doi.org/10.1116/1.4993628} {\bibfield  {journal} {\bibinfo
  {journal} {Biointerphases}\ }\textbf {\bibinfo {volume} {13}},\ \bibinfo
  {pages} {03B301} (\bibinfo {year} {2018})}\BibitemShut {NoStop}%
\bibitem [{\citenamefont {Gillois}\ \emph {et~al.}(2021)\citenamefont
  {Gillois}, \citenamefont {Stoffels}, \citenamefont {Leveque}, \citenamefont
  {Fourquaux}, \citenamefont {Blesson}, \citenamefont {Mils}, \citenamefont
  {Cambier}, \citenamefont {Vignard}, \citenamefont {Terrisse}, \citenamefont
  {Mirey}, \citenamefont {Audinot}, \citenamefont {Theodorou}, \citenamefont
  {Ropers}, \citenamefont {Robert},\ and\ \citenamefont
  {Mercier-Bonin}}]{Gillois_2021}%
  \BibitemOpen
  \bibfield  {author} {\bibinfo {author} {\bibfnamefont {K.}~\bibnamefont
  {Gillois}}, \bibinfo {author} {\bibfnamefont {C.}~\bibnamefont {Stoffels}},
  \bibinfo {author} {\bibfnamefont {M.}~\bibnamefont {Leveque}}, \bibinfo
  {author} {\bibfnamefont {I.}~\bibnamefont {Fourquaux}}, \bibinfo {author}
  {\bibfnamefont {J.}~\bibnamefont {Blesson}}, \bibinfo {author} {\bibfnamefont
  {V.}~\bibnamefont {Mils}}, \bibinfo {author} {\bibfnamefont {S.}~\bibnamefont
  {Cambier}}, \bibinfo {author} {\bibfnamefont {J.}~\bibnamefont {Vignard}},
  \bibinfo {author} {\bibfnamefont {H.}~\bibnamefont {Terrisse}}, \bibinfo
  {author} {\bibfnamefont {G.}~\bibnamefont {Mirey}}, \bibinfo {author}
  {\bibfnamefont {J.-N.}\ \bibnamefont {Audinot}}, \bibinfo {author}
  {\bibfnamefont {V.}~\bibnamefont {Theodorou}}, \bibinfo {author}
  {\bibfnamefont {M.-H.}\ \bibnamefont {Ropers}}, \bibinfo {author}
  {\bibfnamefont {H.}~\bibnamefont {Robert}},\ and\ \bibinfo {author}
  {\bibfnamefont {M.}~\bibnamefont {Mercier-Bonin}},\ }\bibfield  {title}
  {\enquote {\bibinfo {title} {Repeated exposure of caco-2 versus
  caco-2/{HT}29-{MTX} intestinal cell models to (nano)silver in vitro:
  comparison of two commercially available colloidal silver products},}\ }\href
  {https://doi.org/10.1016/j.scitotenv.2020.142324} {\bibfield  {journal}
  {\bibinfo  {journal} {Sci. Total Environ.}\ }\textbf {\bibinfo {volume}
  {754}},\ \bibinfo {pages} {142324} (\bibinfo {year} {2021})}\BibitemShut
  {NoStop}%
\bibitem [{\citenamefont {Ost}\ \emph {et~al.}(2021)\citenamefont {Ost},
  \citenamefont {Wu}, \citenamefont {H{\"{o}}schen}, \citenamefont {Mueller},
  \citenamefont {Wirtz},\ and\ \citenamefont {Audinot}}]{Ost_2021}%
  \BibitemOpen
  \bibfield  {author} {\bibinfo {author} {\bibfnamefont {A.~D.}\ \bibnamefont
  {Ost}}, \bibinfo {author} {\bibfnamefont {T.}~\bibnamefont {Wu}}, \bibinfo
  {author} {\bibfnamefont {C.}~\bibnamefont {H{\"{o}}schen}}, \bibinfo {author}
  {\bibfnamefont {C.~W.}\ \bibnamefont {Mueller}}, \bibinfo {author}
  {\bibfnamefont {T.}~\bibnamefont {Wirtz}},\ and\ \bibinfo {author}
  {\bibfnamefont {J.-N.}\ \bibnamefont {Audinot}},\ }\bibfield  {title}
  {\enquote {\bibinfo {title} {4d surface reconstructions to study microscale
  structures and functions in soil biogeochemistry},}\ }\href
  {https://doi.org/10.1021/acs.est.1c02971} {\bibfield  {journal} {\bibinfo
  {journal} {Environmental Science {\&} Technology}\ }\textbf {\bibinfo
  {volume} {55}},\ \bibinfo {pages} {9384--9393} (\bibinfo {year}
  {2021})}\BibitemShut {NoStop}%
\bibitem [{\citenamefont {Ball}\ \emph {et~al.}(2019)\citenamefont {Ball},
  \citenamefont {Taylor}, \citenamefont {Einsle}, \citenamefont {Audinot},\
  and\ \citenamefont {Harrison}}]{Ball_2019}%
  \BibitemOpen
  \bibfield  {author} {\bibinfo {author} {\bibfnamefont {M.~R.}\ \bibnamefont
  {Ball}}, \bibinfo {author} {\bibfnamefont {R.~J.~M.}\ \bibnamefont {Taylor}},
  \bibinfo {author} {\bibfnamefont {J.~F.}\ \bibnamefont {Einsle}}, \bibinfo
  {author} {\bibfnamefont {J.-N.}\ \bibnamefont {Audinot}},\ and\ \bibinfo
  {author} {\bibfnamefont {R.~J.}\ \bibnamefont {Harrison}},\ }\bibfield
  {title} {\enquote {\bibinfo {title} {Analytics on the {FIB}: {ORION}-{SIMS}
  and the discovery of a unique chondrite-like, precambrian impactor},}\ }\href
  {https://doi.org/10.1017/s143192761900518x} {\bibfield  {journal} {\bibinfo
  {journal} {Microsc. Microanal.}\ }\textbf {\bibinfo {volume} {25}},\ \bibinfo
  {pages} {890--891} (\bibinfo {year} {2019})}\BibitemShut {NoStop}%
\bibitem [{\citenamefont {McPhail}(2006{\natexlab{b}})}]{McPhail_2006}%
  \BibitemOpen
  \bibfield  {author} {\bibinfo {author} {\bibfnamefont {D.~S.}\ \bibnamefont
  {McPhail}},\ }\bibfield  {title} {\enquote {\bibinfo {title} {Some
  applications of {SIMS} in conservation science, archaeometry and
  cosmochemistry},}\ }\href {https://doi.org/10.1016/j.apsusc.2006.02.132}
  {\bibfield  {journal} {\bibinfo  {journal} {Appl. Surf. Sci.}\ }\textbf
  {\bibinfo {volume} {252}},\ \bibinfo {pages} {7107--7112} (\bibinfo {year}
  {2006}{\natexlab{b}})}\BibitemShut {NoStop}%
\bibitem [{\citenamefont {Keune}\ and\ \citenamefont
  {Boon}(2004)}]{Keune_2004}%
  \BibitemOpen
  \bibfield  {author} {\bibinfo {author} {\bibfnamefont {K.}~\bibnamefont
  {Keune}}\ and\ \bibinfo {author} {\bibfnamefont {J.~J.}\ \bibnamefont
  {Boon}},\ }\bibfield  {title} {\enquote {\bibinfo {title} {Imaging secondary
  ion mass spectrometry of a paint cross section taken from an early
  netherlandish painting by rogier van der weyden},}\ }\href
  {https://doi.org/10.1021/ac035201a} {\bibfield  {journal} {\bibinfo
  {journal} {Anal. Chem.}\ }\textbf {\bibinfo {volume} {76}},\ \bibinfo {pages}
  {1374--1385} (\bibinfo {year} {2004})}\BibitemShut {NoStop}%
\bibitem [{\citenamefont {Adriaens}\ and\ \citenamefont
  {Dowsett}(2006)}]{Adriaens_2006}%
  \BibitemOpen
  \bibfield  {author} {\bibinfo {author} {\bibfnamefont {A.}~\bibnamefont
  {Adriaens}}\ and\ \bibinfo {author} {\bibfnamefont {M.~G.}\ \bibnamefont
  {Dowsett}},\ }\bibfield  {title} {\enquote {\bibinfo {title} {Applications of
  {SIMS} to cultural heritage studies},}\ }\href
  {https://doi.org/10.1016/j.apsusc.2006.02.127} {\bibfield  {journal}
  {\bibinfo  {journal} {Appl. Surf. Sci.}\ }\textbf {\bibinfo {volume} {252}},\
  \bibinfo {pages} {7096--7101} (\bibinfo {year} {2006})}\BibitemShut {NoStop}%
\bibitem [{\citenamefont {Voras}\ \emph {et~al.}(2016)\citenamefont {Voras},
  \citenamefont {deGhetaldi}, \citenamefont {Baade}, \citenamefont {Gordon},
  \citenamefont {Gates},\ and\ \citenamefont {Beebe}}]{Voras_2016}%
  \BibitemOpen
  \bibfield  {author} {\bibinfo {author} {\bibfnamefont {Z.~E.}\ \bibnamefont
  {Voras}}, \bibinfo {author} {\bibfnamefont {K.}~\bibnamefont {deGhetaldi}},
  \bibinfo {author} {\bibfnamefont {B.}~\bibnamefont {Baade}}, \bibinfo
  {author} {\bibfnamefont {E.}~\bibnamefont {Gordon}}, \bibinfo {author}
  {\bibfnamefont {G.}~\bibnamefont {Gates}},\ and\ \bibinfo {author}
  {\bibfnamefont {T.~P.}\ \bibnamefont {Beebe}},\ }\bibfield  {title} {\enquote
  {\bibinfo {title} {Comparison of oil and egg tempera paint systems using
  time-of-flight secondary ion mass spectrometry},}\ }\href
  {https://doi.org/10.1179/2047058414y.0000000154} {\bibfield  {journal}
  {\bibinfo  {journal} {Stud. Conserv.}\ }\textbf {\bibinfo {volume} {61}},\
  \bibinfo {pages} {222--235} (\bibinfo {year} {2016})}\BibitemShut {NoStop}%
\bibitem [{\citenamefont {Philipp}, \citenamefont {Rzeznik},\ and\
  \citenamefont {Wirtz}(2016)}]{Philipp_2016}%
  \BibitemOpen
  \bibfield  {author} {\bibinfo {author} {\bibfnamefont {P.}~\bibnamefont
  {Philipp}}, \bibinfo {author} {\bibfnamefont {L.}~\bibnamefont {Rzeznik}},\
  and\ \bibinfo {author} {\bibfnamefont {T.}~\bibnamefont {Wirtz}},\ }\bibfield
   {title} {\enquote {\bibinfo {title} {Numerical investigation of depth
  profiling capabilities of helium and neon ions in ion microscopy},}\ }\href
  {https://doi.org/10.3762/bjnano.7.168} {\bibfield  {journal} {\bibinfo
  {journal} {Beilstein J. Nanotechnol.}\ }\textbf {\bibinfo {volume} {7}},\
  \bibinfo {pages} {1749--1760} (\bibinfo {year} {2016})}\BibitemShut {NoStop}%
\bibitem [{\citenamefont {Kollmer}(2004)}]{Kollmer_2004}%
  \BibitemOpen
  \bibfield  {author} {\bibinfo {author} {\bibfnamefont {F.}~\bibnamefont
  {Kollmer}},\ }\bibfield  {title} {\enquote {\bibinfo {title} {Cluster primary
  ion bombardment of organic materials},}\ }\href
  {https://doi.org/10.1016/j.apsusc.2004.03.101} {\bibfield  {journal}
  {\bibinfo  {journal} {Appl. Surf. Sci.}\ }\textbf {\bibinfo {volume}
  {231-232}},\ \bibinfo {pages} {153--158} (\bibinfo {year}
  {2004})}\BibitemShut {NoStop}%
\bibitem [{\citenamefont {van Gastel}\ \emph {et~al.}(2011)\citenamefont {van
  Gastel}, \citenamefont {Barriss}, \citenamefont {Sanford}, \citenamefont
  {Hlawacek}, \citenamefont {Scipioni}, \citenamefont {Merkle}, \citenamefont
  {Voci}, \citenamefont {Fenner}, \citenamefont {Zandvliet},\ and\
  \citenamefont {Poelsema}}]{Gastel2011}%
  \BibitemOpen
  \bibfield  {author} {\bibinfo {author} {\bibfnamefont {R.}~\bibnamefont {van
  Gastel}}, \bibinfo {author} {\bibfnamefont {L.}~\bibnamefont {Barriss}},
  \bibinfo {author} {\bibfnamefont {C.~A.}\ \bibnamefont {Sanford}}, \bibinfo
  {author} {\bibfnamefont {G.}~\bibnamefont {Hlawacek}}, \bibinfo {author}
  {\bibfnamefont {L.}~\bibnamefont {Scipioni}}, \bibinfo {author}
  {\bibfnamefont {A.}~\bibnamefont {Merkle}}, \bibinfo {author} {\bibfnamefont
  {D.}~\bibnamefont {Voci}}, \bibinfo {author} {\bibfnamefont {C.}~\bibnamefont
  {Fenner}}, \bibinfo {author} {\bibfnamefont {H.~J.~W.}\ \bibnamefont
  {Zandvliet}},\ and\ \bibinfo {author} {\bibfnamefont {B.}~\bibnamefont
  {Poelsema}},\ }\bibfield  {title} {\enquote {\bibinfo {title} {{Design and
  performance of a Near Ultra High Vacuum Helium Ion Microscope}},}\ }\href
  {https://doi.org/10.1017/S1431927611005514} {\bibfield  {journal} {\bibinfo
  {journal} {Microsc. Microanal.}\ }\textbf {\bibinfo {volume} {17}},\ \bibinfo
  {pages} {928--929} (\bibinfo {year} {2011})}\BibitemShut {NoStop}%
\bibitem [{\citenamefont {Wu}\ \emph {et~al.}(2013)\citenamefont {Wu},
  \citenamefont {Stern}, \citenamefont {Chen}, \citenamefont {Huth},
  \citenamefont {Schwalb}, \citenamefont {Winhold}, \citenamefont {Porrati},
  \citenamefont {Gonzalez}, \citenamefont {Timilsina},\ and\ \citenamefont
  {Rack}}]{Wu_2013}%
  \BibitemOpen
  \bibfield  {author} {\bibinfo {author} {\bibfnamefont {H.~M.}\ \bibnamefont
  {Wu}}, \bibinfo {author} {\bibfnamefont {L.~A.}\ \bibnamefont {Stern}},
  \bibinfo {author} {\bibfnamefont {J.~H.}\ \bibnamefont {Chen}}, \bibinfo
  {author} {\bibfnamefont {M.}~\bibnamefont {Huth}}, \bibinfo {author}
  {\bibfnamefont {C.~H.}\ \bibnamefont {Schwalb}}, \bibinfo {author}
  {\bibfnamefont {M.}~\bibnamefont {Winhold}}, \bibinfo {author} {\bibfnamefont
  {F.}~\bibnamefont {Porrati}}, \bibinfo {author} {\bibfnamefont {C.~M.}\
  \bibnamefont {Gonzalez}}, \bibinfo {author} {\bibfnamefont {R.}~\bibnamefont
  {Timilsina}},\ and\ \bibinfo {author} {\bibfnamefont {P.~D.}\ \bibnamefont
  {Rack}},\ }\bibfield  {title} {\enquote {\bibinfo {title} {Synthesis of
  nanowires via helium and neon focused ion beam induced deposition with the
  gas field ion microscope},}\ }\href
  {https://doi.org/10.1088/0957-4484/24/17/175302} {\bibfield  {journal}
  {\bibinfo  {journal} {Nanotechnology}\ }\textbf {\bibinfo {volume} {24}},\
  \bibinfo {pages} {175302} (\bibinfo {year} {2013})}\BibitemShut {NoStop}%
\bibitem [{\citenamefont {Wu}\ \emph {et~al.}(2014)\citenamefont {Wu},
  \citenamefont {Stern}, \citenamefont {Xia}, \citenamefont {Ferranti},
  \citenamefont {Thompson}, \citenamefont {Klein}, \citenamefont {Gonzalez},\
  and\ \citenamefont {Rack}}]{Wu_2014}%
  \BibitemOpen
  \bibfield  {author} {\bibinfo {author} {\bibfnamefont {H.}~\bibnamefont
  {Wu}}, \bibinfo {author} {\bibfnamefont {L.~A.}\ \bibnamefont {Stern}},
  \bibinfo {author} {\bibfnamefont {D.}~\bibnamefont {Xia}}, \bibinfo {author}
  {\bibfnamefont {D.}~\bibnamefont {Ferranti}}, \bibinfo {author}
  {\bibfnamefont {B.}~\bibnamefont {Thompson}}, \bibinfo {author}
  {\bibfnamefont {K.~L.}\ \bibnamefont {Klein}}, \bibinfo {author}
  {\bibfnamefont {C.~M.}\ \bibnamefont {Gonzalez}},\ and\ \bibinfo {author}
  {\bibfnamefont {P.~D.}\ \bibnamefont {Rack}},\ }\bibfield  {title}
  {{\selectlanguage {english}\enquote {\bibinfo {title} {Focused helium ion
  beam deposited low resistivity cobalt metal lines with 10 nm resolution:
  implications for advanced circuit editing},}\ }}\href
  {https://doi.org/10.1007/s10854-013-1522-6} {\bibfield  {journal} {\bibinfo
  {journal} {J. Mater. Sci. Mater. Electron.}\ }\textbf {\bibinfo {volume}
  {25}},\ \bibinfo {pages} {587--595} (\bibinfo {year} {2014})}\BibitemShut
  {NoStop}%
\bibitem [{\citenamefont {Lin}\ \emph {et~al.}(2003)\citenamefont {Lin},
  \citenamefont {Bird}, \citenamefont {Rotkina},\ and\ \citenamefont
  {Bennett}}]{Lin_2003}%
  \BibitemOpen
  \bibfield  {author} {\bibinfo {author} {\bibfnamefont {J.-F.}\ \bibnamefont
  {Lin}}, \bibinfo {author} {\bibfnamefont {J.~P.}\ \bibnamefont {Bird}},
  \bibinfo {author} {\bibfnamefont {L.}~\bibnamefont {Rotkina}},\ and\ \bibinfo
  {author} {\bibfnamefont {P.~A.}\ \bibnamefont {Bennett}},\ }\bibfield
  {title} {\enquote {\bibinfo {title} {Classical and quantum transport in
  focused-ion-beam-deposited pt nanointerconnects},}\ }\href
  {https://doi.org/10.1063/1.1541940} {\bibfield  {journal} {\bibinfo
  {journal} {Appl. Phys. Lett.}\ }\textbf {\bibinfo {volume} {82}},\ \bibinfo
  {pages} {802--804} (\bibinfo {year} {2003})}\BibitemShut {NoStop}%
\bibitem [{\citenamefont {Lin}, \citenamefont {Bird},\ and\ \citenamefont
  {Rotkina}(2003)}]{Lin_2003a}%
  \BibitemOpen
  \bibfield  {author} {\bibinfo {author} {\bibfnamefont {J.-F.}\ \bibnamefont
  {Lin}}, \bibinfo {author} {\bibfnamefont {J.~P.}\ \bibnamefont {Bird}},\ and\
  \bibinfo {author} {\bibfnamefont {L.}~\bibnamefont {Rotkina}},\ }\bibfield
  {title} {\enquote {\bibinfo {title} {Low-temperature decoherence in
  disordered pt nanowires},}\ }\href
  {https://doi.org/10.1016/s1386-9477(03)00317-5} {\bibfield  {journal}
  {\bibinfo  {journal} {Phys. E: Low-Dimens. Syst. Nanostructures}\ }\textbf
  {\bibinfo {volume} {19}},\ \bibinfo {pages} {112--116} (\bibinfo {year}
  {2003})}\BibitemShut {NoStop}%
\bibitem [{\citenamefont {Tao}\ \emph {et~al.}(1990)\citenamefont {Tao},
  \citenamefont {Ro}, \citenamefont {Melngailis}, \citenamefont {Xue},\ and\
  \citenamefont {Kaesz}}]{Tao_1990}%
  \BibitemOpen
  \bibfield  {author} {\bibinfo {author} {\bibfnamefont {T.}~\bibnamefont
  {Tao}}, \bibinfo {author} {\bibfnamefont {J.}~\bibnamefont {Ro}}, \bibinfo
  {author} {\bibfnamefont {J.}~\bibnamefont {Melngailis}}, \bibinfo {author}
  {\bibfnamefont {Z.}~\bibnamefont {Xue}},\ and\ \bibinfo {author}
  {\bibfnamefont {H.~D.}\ \bibnamefont {Kaesz}},\ }\bibfield  {title} {\enquote
  {\bibinfo {title} {Focused ion beam induced deposition of platinum},}\ }\href
  {https://doi.org/10.1116/1.585167} {\bibfield  {journal} {\bibinfo  {journal}
  {J. Vac. Sci. Technol. B}\ }\textbf {\bibinfo {volume} {8}},\ \bibinfo
  {pages} {1826--1829} (\bibinfo {year} {1990})}\BibitemShut {NoStop}%
\bibitem [{\citenamefont {Fern{\'{a}}ndez-Pacheco}\ \emph
  {et~al.}(2009)\citenamefont {Fern{\'{a}}ndez-Pacheco}, \citenamefont
  {De~Teresa}, \citenamefont {C{\'{o}}rdoba},\ and\ \citenamefont
  {Ibarra}}]{FernandezPacheco_2009}%
  \BibitemOpen
  \bibfield  {author} {\bibinfo {author} {\bibfnamefont {A.}~\bibnamefont
  {Fern{\'{a}}ndez-Pacheco}}, \bibinfo {author} {\bibfnamefont {J.~M.}\
  \bibnamefont {De~Teresa}}, \bibinfo {author} {\bibfnamefont {R.}~\bibnamefont
  {C{\'{o}}rdoba}},\ and\ \bibinfo {author} {\bibfnamefont {M.~R.}\
  \bibnamefont {Ibarra}},\ }\bibfield  {title} {\enquote {\bibinfo {title}
  {Metal-insulator transition in pt-c nanowires grown by
  focused-ion-beam-induced deposition},}\ }\href
  {https://doi.org/10.1103/physrevb.79.174204} {\bibfield  {journal} {\bibinfo
  {journal} {Phys. Rev. B}\ }\textbf {\bibinfo {volume} {79}},\ \bibinfo
  {pages} {174204} (\bibinfo {year} {2009})}\BibitemShut {NoStop}%
\bibitem [{\citenamefont {De~Teresa}\ \emph {et~al.}(2009)\citenamefont
  {De~Teresa}, \citenamefont {C{\'{o}}rdoba}, \citenamefont
  {Fern{\'{a}}ndez-Pacheco}, \citenamefont {Montero}, \citenamefont
  {Strichovanec},\ and\ \citenamefont {Ibarra}}]{DeTeresa_2009}%
  \BibitemOpen
  \bibfield  {author} {\bibinfo {author} {\bibfnamefont {J.~M.}\ \bibnamefont
  {De~Teresa}}, \bibinfo {author} {\bibfnamefont {R.}~\bibnamefont
  {C{\'{o}}rdoba}}, \bibinfo {author} {\bibfnamefont {A.}~\bibnamefont
  {Fern{\'{a}}ndez-Pacheco}}, \bibinfo {author} {\bibfnamefont
  {O.}~\bibnamefont {Montero}}, \bibinfo {author} {\bibfnamefont
  {P.}~\bibnamefont {Strichovanec}},\ and\ \bibinfo {author} {\bibfnamefont
  {M.~R.}\ \bibnamefont {Ibarra}},\ }\bibfield  {title} {\enquote {\bibinfo
  {title} {Origin of the difference in the resistivity of as-grown focused-ion-
  and focused-electron-beam-induced pt nanodeposits},}\ }\href
  {https://doi.org/10.1155/2009/936863} {\bibfield  {journal} {\bibinfo
  {journal} {J. Nanomater.}\ }\textbf {\bibinfo {volume} {2009}},\ \bibinfo
  {pages} {936863} (\bibinfo {year} {2009})}\BibitemShut {NoStop}%
\bibitem [{\citenamefont {Blauner}(1989)}]{Blauner_1989}%
  \BibitemOpen
  \bibfield  {author} {\bibinfo {author} {\bibfnamefont {P.~G.}\ \bibnamefont
  {Blauner}},\ }\bibfield  {title} {\enquote {\bibinfo {title} {Focused ion
  beam fabrication of submicron gold structures},}\ }\href
  {https://doi.org/10.1116/1.584803} {\bibfield  {journal} {\bibinfo  {journal}
  {J. Vac. Sci. Technol. B}\ }\textbf {\bibinfo {volume} {7}},\ \bibinfo
  {pages} {609} (\bibinfo {year} {1989})}\BibitemShut {NoStop}%
\bibitem [{\citenamefont {Shedd}\ \emph {et~al.}(1986)\citenamefont {Shedd},
  \citenamefont {Lezec}, \citenamefont {Dubner},\ and\ \citenamefont
  {Melngailis}}]{Shedd_1986}%
  \BibitemOpen
  \bibfield  {author} {\bibinfo {author} {\bibfnamefont {G.~M.}\ \bibnamefont
  {Shedd}}, \bibinfo {author} {\bibfnamefont {H.}~\bibnamefont {Lezec}},
  \bibinfo {author} {\bibfnamefont {A.~D.}\ \bibnamefont {Dubner}},\ and\
  \bibinfo {author} {\bibfnamefont {J.}~\bibnamefont {Melngailis}},\ }\bibfield
   {title} {\enquote {\bibinfo {title} {Focused ion beam induced deposition of
  gold},}\ }\href {https://doi.org/10.1063/1.97287} {\bibfield  {journal}
  {\bibinfo  {journal} {Appl. Phys. Lett.}\ }\textbf {\bibinfo {volume} {49}},\
  \bibinfo {pages} {1584--1586} (\bibinfo {year} {1986})}\BibitemShut {NoStop}%
\bibitem [{\citenamefont {Gazzadi}\ \emph {et~al.}(2011)\citenamefont
  {Gazzadi}, \citenamefont {Mulders}, \citenamefont {Trompenaars},
  \citenamefont {Ghirri}, \citenamefont {Rota}, \citenamefont {Affronte},\ and\
  \citenamefont {Frabboni}}]{Gazzadi_2011}%
  \BibitemOpen
  \bibfield  {author} {\bibinfo {author} {\bibfnamefont {G.~C.}\ \bibnamefont
  {Gazzadi}}, \bibinfo {author} {\bibfnamefont {J.~J.~L.}\ \bibnamefont
  {Mulders}}, \bibinfo {author} {\bibfnamefont {P.}~\bibnamefont
  {Trompenaars}}, \bibinfo {author} {\bibfnamefont {A.}~\bibnamefont {Ghirri}},
  \bibinfo {author} {\bibfnamefont {A.}~\bibnamefont {Rota}}, \bibinfo {author}
  {\bibfnamefont {M.}~\bibnamefont {Affronte}},\ and\ \bibinfo {author}
  {\bibfnamefont {S.}~\bibnamefont {Frabboni}},\ }\bibfield  {title} {\enquote
  {\bibinfo {title} {Characterization of a new cobalt precursor for focused
  beam deposition of magnetic nanostructures},}\ }\href
  {https://doi.org/10.1016/j.mee.2010.12.031} {\bibfield  {journal} {\bibinfo
  {journal} {Microelectron. Eng.}\ }\textbf {\bibinfo {volume} {88}},\ \bibinfo
  {pages} {1955--1958} (\bibinfo {year} {2011})}\BibitemShut {NoStop}%
\bibitem [{\citenamefont {Ratta}, \citenamefont {Melngailis},\ and\
  \citenamefont {Thompson}(1993)}]{DellaRatta_1993}%
  \BibitemOpen
  \bibfield  {author} {\bibinfo {author} {\bibfnamefont {A.~D.~D.}\
  \bibnamefont {Ratta}}, \bibinfo {author} {\bibfnamefont {J.}~\bibnamefont
  {Melngailis}},\ and\ \bibinfo {author} {\bibfnamefont {C.~V.}\ \bibnamefont
  {Thompson}},\ }\bibfield  {title} {{\selectlanguage {english}\enquote
  {\bibinfo {title} {Focused-ion beam induced deposition of copper},}\ }}\href
  {https://doi.org/10.1116/1.586455} {\bibfield  {journal} {\bibinfo  {journal}
  {J. Vac. Sci. Technol. B}\ }\textbf {\bibinfo {volume} {11}},\ \bibinfo
  {pages} {2195--2199} (\bibinfo {year} {1993})}\BibitemShut {NoStop}%
\bibitem [{\citenamefont {Wang}\ \emph {et~al.}(2010)\citenamefont {Wang},
  \citenamefont {Singh}, \citenamefont {Tian}, \citenamefont {Kumar},
  \citenamefont {Liu}, \citenamefont {Shi}, \citenamefont {Jain}, \citenamefont
  {Samarth}, \citenamefont {Mallouk},\ and\ \citenamefont {Chan}}]{Wang_2010}%
  \BibitemOpen
  \bibfield  {author} {\bibinfo {author} {\bibfnamefont {J.}~\bibnamefont
  {Wang}}, \bibinfo {author} {\bibfnamefont {M.}~\bibnamefont {Singh}},
  \bibinfo {author} {\bibfnamefont {M.}~\bibnamefont {Tian}}, \bibinfo {author}
  {\bibfnamefont {N.}~\bibnamefont {Kumar}}, \bibinfo {author} {\bibfnamefont
  {B.}~\bibnamefont {Liu}}, \bibinfo {author} {\bibfnamefont {C.}~\bibnamefont
  {Shi}}, \bibinfo {author} {\bibfnamefont {J.~K.}\ \bibnamefont {Jain}},
  \bibinfo {author} {\bibfnamefont {N.}~\bibnamefont {Samarth}}, \bibinfo
  {author} {\bibfnamefont {T.~E.}\ \bibnamefont {Mallouk}},\ and\ \bibinfo
  {author} {\bibfnamefont {M.~H.~W.}\ \bibnamefont {Chan}},\ }\bibfield
  {title} {\enquote {\bibinfo {title} {Interplay between superconductivity and
  ferromagnetism in crystalline nanowires},}\ }\href
  {https://doi.org/10.1038/nphys1621} {\bibfield  {journal} {\bibinfo
  {journal} {Nat. Phys.}\ }\textbf {\bibinfo {volume} {6}},\ \bibinfo {pages}
  {389 EP--} (\bibinfo {year} {2010})}\BibitemShut {NoStop}%
\bibitem [{\citenamefont {Kompaniiets}\ \emph
  {et~al.}(2014{\natexlab{a}})\citenamefont {Kompaniiets}, \citenamefont
  {Dobrovolskiy}, \citenamefont {Neetzel}, \citenamefont {Begun}, \citenamefont
  {Porrati}, \citenamefont {Ensinger},\ and\ \citenamefont
  {Huth}}]{Kompaniiets_2014}%
  \BibitemOpen
  \bibfield  {author} {\bibinfo {author} {\bibfnamefont {M.}~\bibnamefont
  {Kompaniiets}}, \bibinfo {author} {\bibfnamefont {O.~V.}\ \bibnamefont
  {Dobrovolskiy}}, \bibinfo {author} {\bibfnamefont {C.}~\bibnamefont
  {Neetzel}}, \bibinfo {author} {\bibfnamefont {E.}~\bibnamefont {Begun}},
  \bibinfo {author} {\bibfnamefont {F.}~\bibnamefont {Porrati}}, \bibinfo
  {author} {\bibfnamefont {W.}~\bibnamefont {Ensinger}},\ and\ \bibinfo
  {author} {\bibfnamefont {M.}~\bibnamefont {Huth}},\ }\bibfield  {title}
  {\enquote {\bibinfo {title} {Proximity-induced superconductivity in
  crystalline cu and co nanowires and nanogranular co structures},}\ }\href
  {https://doi.org/10.1063/1.4893549} {\bibfield  {journal} {\bibinfo
  {journal} {J. Appl. Phys.}\ }\textbf {\bibinfo {volume} {116}},\ \bibinfo
  {eid} {073906} (\bibinfo {year} {2014}{\natexlab{a}})}\BibitemShut {NoStop}%
\bibitem [{\citenamefont {Kompaniiets}\ \emph
  {et~al.}(2014{\natexlab{b}})\citenamefont {Kompaniiets}, \citenamefont
  {Dobrovolskiy}, \citenamefont {Neetzel}, \citenamefont {Porrati},
  \citenamefont {Br\"otz}, \citenamefont {Ensinger},\ and\ \citenamefont
  {Huth}}]{Kompaniiets_2014a}%
  \BibitemOpen
  \bibfield  {author} {\bibinfo {author} {\bibfnamefont {M.}~\bibnamefont
  {Kompaniiets}}, \bibinfo {author} {\bibfnamefont {O.~V.}\ \bibnamefont
  {Dobrovolskiy}}, \bibinfo {author} {\bibfnamefont {C.}~\bibnamefont
  {Neetzel}}, \bibinfo {author} {\bibfnamefont {F.}~\bibnamefont {Porrati}},
  \bibinfo {author} {\bibfnamefont {J.}~\bibnamefont {Br\"otz}}, \bibinfo
  {author} {\bibfnamefont {W.}~\bibnamefont {Ensinger}},\ and\ \bibinfo
  {author} {\bibfnamefont {M.}~\bibnamefont {Huth}},\ }\bibfield  {title}
  {\enquote {\bibinfo {title} {Long-range superconducting proximity effect in
  polycrystalline {Co} nanowires},}\ }\href {https://doi.org/10.1063/1.4863980}
  {\bibfield  {journal} {\bibinfo  {journal} {Appl. Phys. Lett.}\ }\textbf
  {\bibinfo {volume} {104}},\ \bibinfo {pages} {052603} (\bibinfo {year}
  {2014}{\natexlab{b}})}\BibitemShut {NoStop}%
\bibitem [{\citenamefont {Dobrovolskiy}\ \emph {et~al.}(2018)\citenamefont
  {Dobrovolskiy}, \citenamefont {Bevz}, \citenamefont {Mikhailov},
  \citenamefont {Yuzephovich}, \citenamefont {Shklovskij}, \citenamefont
  {Vovk}, \citenamefont {Tsindlekht}, \citenamefont {Sachser},\ and\
  \citenamefont {Huth}}]{Dobrovolskiy_2018}%
  \BibitemOpen
  \bibfield  {author} {\bibinfo {author} {\bibfnamefont {O.~V.}\ \bibnamefont
  {Dobrovolskiy}}, \bibinfo {author} {\bibfnamefont {V.~M.}\ \bibnamefont
  {Bevz}}, \bibinfo {author} {\bibfnamefont {M.~Y.}\ \bibnamefont {Mikhailov}},
  \bibinfo {author} {\bibfnamefont {O.~I.}\ \bibnamefont {Yuzephovich}},
  \bibinfo {author} {\bibfnamefont {V.~A.}\ \bibnamefont {Shklovskij}},
  \bibinfo {author} {\bibfnamefont {R.~V.}\ \bibnamefont {Vovk}}, \bibinfo
  {author} {\bibfnamefont {M.~I.}\ \bibnamefont {Tsindlekht}}, \bibinfo
  {author} {\bibfnamefont {R.}~\bibnamefont {Sachser}},\ and\ \bibinfo {author}
  {\bibfnamefont {M.}~\bibnamefont {Huth}},\ }\bibfield  {title} {\enquote
  {\bibinfo {title} {Microwave emission from superconducting vortices in
  {Mo/Si} superlattices},}\ }\href {https://doi.org/10.1038/s41467-018-07256-0}
  {\bibfield  {journal} {\bibinfo  {journal} {Nat. Commun.}\ }\textbf {\bibinfo
  {volume} {9}},\ \bibinfo {pages} {4927} (\bibinfo {year} {2018})}\BibitemShut
  {NoStop}%
\bibitem [{\citenamefont {Komano}, \citenamefont {Ogawa},\ and\ \citenamefont
  {Takigawa}(1989)}]{Komano_1989}%
  \BibitemOpen
  \bibfield  {author} {\bibinfo {author} {\bibfnamefont {H.}~\bibnamefont
  {Komano}}, \bibinfo {author} {\bibfnamefont {Y.}~\bibnamefont {Ogawa}},\ and\
  \bibinfo {author} {\bibfnamefont {T.}~\bibnamefont {Takigawa}},\ }\bibfield
  {title} {\enquote {\bibinfo {title} {Silicon oxide film formation by focused
  ion beam ({FIB})-assisted deposition},}\ }\href
  {https://doi.org/10.1143/jjap.28.2372} {\bibfield  {journal} {\bibinfo
  {journal} {Jpn. J. Appl. Phys.}\ }\textbf {\bibinfo {volume} {28}},\ \bibinfo
  {pages} {2372--2375} (\bibinfo {year} {1989})}\BibitemShut {NoStop}%
\bibitem [{\citenamefont {Young}\ and\ \citenamefont
  {Puretz}(1995)}]{Young_1995}%
  \BibitemOpen
  \bibfield  {author} {\bibinfo {author} {\bibfnamefont {R.~J.}\ \bibnamefont
  {Young}}\ and\ \bibinfo {author} {\bibfnamefont {J.}~\bibnamefont {Puretz}},\
  }\bibfield  {title} {\enquote {\bibinfo {title} {Focused ion beam insulator
  deposition},}\ }\href {https://doi.org/10.1116/1.588026} {\bibfield
  {journal} {\bibinfo  {journal} {J. Vac. Sci. Technol. B}\ }\textbf {\bibinfo
  {volume} {13}},\ \bibinfo {pages} {2576} (\bibinfo {year}
  {1995})}\BibitemShut {NoStop}%
\bibitem [{\citenamefont {Lipp}(1996)}]{Lipp_1996}%
  \BibitemOpen
  \bibfield  {author} {\bibinfo {author} {\bibfnamefont {S.}~\bibnamefont
  {Lipp}},\ }\bibfield  {title} {\enquote {\bibinfo {title} {Tetramethoxysilane
  as a precursor for focused ion beam and electron beam assisted insulator
  ({SiO}$_x$) deposition},}\ }\href {https://doi.org/10.1116/1.588695}
  {\bibfield  {journal} {\bibinfo  {journal} {J. Vac. Sci. Technol. B}\
  }\textbf {\bibinfo {volume} {14}},\ \bibinfo {pages} {3920} (\bibinfo {year}
  {1996})}\BibitemShut {NoStop}%
\bibitem [{\citenamefont {Edinger}(1998)}]{Edinger_1998}%
  \BibitemOpen
  \bibfield  {author} {\bibinfo {author} {\bibfnamefont {K.}~\bibnamefont
  {Edinger}},\ }\bibfield  {title} {\enquote {\bibinfo {title} {Study of
  precursor gases for focused ion beam insulator deposition},}\ }\href
  {https://doi.org/10.1116/1.590497} {\bibfield  {journal} {\bibinfo  {journal}
  {J. Vac. Sci. Technol. B}\ }\textbf {\bibinfo {volume} {16}},\ \bibinfo
  {pages} {3311} (\bibinfo {year} {1998})}\BibitemShut {NoStop}%
\bibitem [{\citenamefont {C{\'{o}}rdoba}\ \emph {et~al.}(2018)\citenamefont
  {C{\'{o}}rdoba}, \citenamefont {Ibarra}, \citenamefont {Mailly},\ and\
  \citenamefont {De~Teresa}}]{Cordoba_2018}%
  \BibitemOpen
  \bibfield  {author} {\bibinfo {author} {\bibfnamefont {R.}~\bibnamefont
  {C{\'{o}}rdoba}}, \bibinfo {author} {\bibfnamefont {A.}~\bibnamefont
  {Ibarra}}, \bibinfo {author} {\bibfnamefont {D.}~\bibnamefont {Mailly}},\
  and\ \bibinfo {author} {\bibfnamefont {J.~M.}\ \bibnamefont {De~Teresa}},\
  }\bibfield  {title} {\enquote {\bibinfo {title} {Vertical growth of
  superconducting crystalline hollow nanowires by he + focused ion beam induced
  deposition},}\ }\href {https://doi.org/10.1021/acs.nanolett.7b05103}
  {\bibfield  {journal} {\bibinfo  {journal} {Nano Lett.}\ }\textbf {\bibinfo
  {volume} {18}},\ \bibinfo {pages} {1379--1386} (\bibinfo {year}
  {2018})}\BibitemShut {NoStop}%
\bibitem [{\citenamefont {C{\'{o}}rdoba}\ \emph
  {et~al.}(2019{\natexlab{a}})\citenamefont {C{\'{o}}rdoba}, \citenamefont
  {Mailly}, \citenamefont {Rezaev}, \citenamefont {Smirnova}, \citenamefont
  {Schmidt}, \citenamefont {Fomin}, \citenamefont {Zeitler}, \citenamefont
  {Guillamon}, \citenamefont {Suderow},\ and\ \citenamefont
  {De~Teresa}}]{Cordoba_2019a}%
  \BibitemOpen
  \bibfield  {author} {\bibinfo {author} {\bibfnamefont {R.}~\bibnamefont
  {C{\'{o}}rdoba}}, \bibinfo {author} {\bibfnamefont {D.}~\bibnamefont
  {Mailly}}, \bibinfo {author} {\bibfnamefont {R.}~\bibnamefont {Rezaev}},
  \bibinfo {author} {\bibfnamefont {E.}~\bibnamefont {Smirnova}}, \bibinfo
  {author} {\bibfnamefont {O.~G.}\ \bibnamefont {Schmidt}}, \bibinfo {author}
  {\bibfnamefont {V.~M.}\ \bibnamefont {Fomin}}, \bibinfo {author}
  {\bibfnamefont {U.}~\bibnamefont {Zeitler}}, \bibinfo {author} {\bibfnamefont
  {I.}~\bibnamefont {Guillamon}}, \bibinfo {author} {\bibfnamefont
  {h.}~\bibnamefont {Suderow}},\ and\ \bibinfo {author} {\bibfnamefont {J.~M.}\
  \bibnamefont {De~Teresa}},\ }\bibfield  {title} {\enquote {\bibinfo {title}
  {Three-dimensional superconducting nano-helices grown by
  he$^+$-focused-ion-beam direct writing},}\ }\href
  {https://doi.org/10.1021/acs.nanolett.9b03153} {\bibfield  {journal}
  {\bibinfo  {journal} {Nano Lett.}\ }\textbf {\bibinfo {volume} {19}},\
  \bibinfo {pages} {8597--8604} (\bibinfo {year}
  {2019}{\natexlab{a}})}\BibitemShut {NoStop}%
\bibitem [{\citenamefont {C{\'{o}}rdoba}\ \emph {et~al.}(2020)\citenamefont
  {C{\'{o}}rdoba}, \citenamefont {Ibarra}, \citenamefont {Mailly},
  \citenamefont {Guillam{\'{o}}n}, \citenamefont {Suderow},\ and\ \citenamefont
  {De~Teresa}}]{Cordoba_2020}%
  \BibitemOpen
  \bibfield  {author} {\bibinfo {author} {\bibfnamefont {R.}~\bibnamefont
  {C{\'{o}}rdoba}}, \bibinfo {author} {\bibfnamefont {A.}~\bibnamefont
  {Ibarra}}, \bibinfo {author} {\bibfnamefont {D.}~\bibnamefont {Mailly}},
  \bibinfo {author} {\bibfnamefont {I.}~\bibnamefont {Guillam{\'{o}}n}},
  \bibinfo {author} {\bibfnamefont {H.}~\bibnamefont {Suderow}},\ and\ \bibinfo
  {author} {\bibfnamefont {J.~M.}\ \bibnamefont {De~Teresa}},\ }\bibfield
  {title} {\enquote {\bibinfo {title} {3{D} superconducting hollow nanowires
  with tailored diameters grown by focused {H}e $^+$ beam direct writing},}\
  }\href {https://doi.org/10.3762/bjnano.11.104} {\bibfield  {journal}
  {\bibinfo  {journal} {Beilstein J. Nanotechnol.}\ }\textbf {\bibinfo {volume}
  {11}},\ \bibinfo {pages} {1198--1206} (\bibinfo {year} {2020})}\BibitemShut
  {NoStop}%
\bibitem [{\citenamefont {Or{\'{u}}s}\ \emph
  {et~al.}(2021{\natexlab{a}})\citenamefont {Or{\'{u}}s}, \citenamefont
  {C{\'{o}}rdoba}, \citenamefont {Hlawacek},\ and\ \citenamefont
  {De~Teresa}}]{Orus_2021a}%
  \BibitemOpen
  \bibfield  {author} {\bibinfo {author} {\bibfnamefont {P.}~\bibnamefont
  {Or{\'{u}}s}}, \bibinfo {author} {\bibfnamefont {R.}~\bibnamefont
  {C{\'{o}}rdoba}}, \bibinfo {author} {\bibfnamefont {G.}~\bibnamefont
  {Hlawacek}},\ and\ \bibinfo {author} {\bibfnamefont {J.~M.}\ \bibnamefont
  {De~Teresa}},\ }\bibfield  {title} {\enquote {\bibinfo {title}
  {Superconducting properties of in-plane {W}-{C} nanowires grown by {H}e$^{+}$
  focused ion beam induced deposition},}\ }\href
  {https://doi.org/10.1088/1361-6528/abc91c} {\bibfield  {journal} {\bibinfo
  {journal} {Nanotechnology}\ }\textbf {\bibinfo {volume} {32}},\ \bibinfo
  {pages} {085301} (\bibinfo {year} {2021}{\natexlab{a}})}\BibitemShut
  {NoStop}%
\bibitem [{\citenamefont {Sadki}, \citenamefont {Ooi},\ and\ \citenamefont
  {Hirata}(2004)}]{Sadki_2004}%
  \BibitemOpen
  \bibfield  {author} {\bibinfo {author} {\bibfnamefont {E.~S.}\ \bibnamefont
  {Sadki}}, \bibinfo {author} {\bibfnamefont {S.}~\bibnamefont {Ooi}},\ and\
  \bibinfo {author} {\bibfnamefont {K.}~\bibnamefont {Hirata}},\ }\bibfield
  {title} {\enquote {\bibinfo {title} {Focused-ion-beam-induced deposition of
  superconducting nanowires},}\ }\href {https://doi.org/10.1063/1.1842367}
  {\bibfield  {journal} {\bibinfo  {journal} {Appl. Phys. Lett.}\ }\textbf
  {\bibinfo {volume} {85}},\ \bibinfo {pages} {6206--6208} (\bibinfo {year}
  {2004})}\BibitemShut {NoStop}%
\bibitem [{\citenamefont {Sadki}, \citenamefont {Ooi},\ and\ \citenamefont
  {Hirata}(2005)}]{Sadki_2005}%
  \BibitemOpen
  \bibfield  {author} {\bibinfo {author} {\bibfnamefont {E.~S.}\ \bibnamefont
  {Sadki}}, \bibinfo {author} {\bibfnamefont {S.}~\bibnamefont {Ooi}},\ and\
  \bibinfo {author} {\bibfnamefont {K.}~\bibnamefont {Hirata}},\ }\bibfield
  {title} {\enquote {\bibinfo {title} {Focused ion beam induced deposition of
  superconducting thin films},}\ }\href
  {https://doi.org/10.1016/j.physc.2005.02.151} {\bibfield  {journal} {\bibinfo
   {journal} {Physica C}\ }\textbf {\bibinfo {volume} {426}},\ \bibinfo {pages}
  {1547--1551} (\bibinfo {year} {2005})}\BibitemShut {NoStop}%
\bibitem [{\citenamefont {Spoddig}\ \emph {et~al.}(2007)\citenamefont
  {Spoddig}, \citenamefont {Schindler}, \citenamefont {R{\"{o}}diger},
  \citenamefont {Barzola-Quiquia}, \citenamefont {Fritsch}, \citenamefont
  {Mulders},\ and\ \citenamefont {Esquinazi}}]{Spoddig_2007}%
  \BibitemOpen
  \bibfield  {author} {\bibinfo {author} {\bibfnamefont {D.}~\bibnamefont
  {Spoddig}}, \bibinfo {author} {\bibfnamefont {K.}~\bibnamefont {Schindler}},
  \bibinfo {author} {\bibfnamefont {P.}~\bibnamefont {R{\"{o}}diger}}, \bibinfo
  {author} {\bibfnamefont {J.}~\bibnamefont {Barzola-Quiquia}}, \bibinfo
  {author} {\bibfnamefont {K.}~\bibnamefont {Fritsch}}, \bibinfo {author}
  {\bibfnamefont {H.}~\bibnamefont {Mulders}},\ and\ \bibinfo {author}
  {\bibfnamefont {P.}~\bibnamefont {Esquinazi}},\ }\bibfield  {title} {\enquote
  {\bibinfo {title} {Transport properties and growth parameters of {PdC} and
  {WC} nanowires prepared in a dual-beam microscope},}\ }\href
  {https://doi.org/10.1088/0957-4484/18/49/495202} {\bibfield  {journal}
  {\bibinfo  {journal} {Nanotechnology}\ }\textbf {\bibinfo {volume} {18}},\
  \bibinfo {pages} {495202} (\bibinfo {year} {2007})}\BibitemShut {NoStop}%
\bibitem [{\citenamefont {Luxmoore}\ \emph {et~al.}(2007)\citenamefont
  {Luxmoore}, \citenamefont {Ross}, \citenamefont {Cullis}, \citenamefont
  {Fry}, \citenamefont {Orr}, \citenamefont {Buckle},\ and\ \citenamefont
  {Jefferson}}]{Luxmoore_2007}%
  \BibitemOpen
  \bibfield  {author} {\bibinfo {author} {\bibfnamefont {I.~J.}\ \bibnamefont
  {Luxmoore}}, \bibinfo {author} {\bibfnamefont {I.~M.}\ \bibnamefont {Ross}},
  \bibinfo {author} {\bibfnamefont {A.~G.}\ \bibnamefont {Cullis}}, \bibinfo
  {author} {\bibfnamefont {P.~W.}\ \bibnamefont {Fry}}, \bibinfo {author}
  {\bibfnamefont {J.}~\bibnamefont {Orr}}, \bibinfo {author} {\bibfnamefont
  {P.~D.}\ \bibnamefont {Buckle}},\ and\ \bibinfo {author} {\bibfnamefont
  {J.~H.}\ \bibnamefont {Jefferson}},\ }\bibfield  {title} {{\selectlanguage
  {english}\enquote {\bibinfo {title} {Low temperature electrical
  characterisation of tungsten nano-wires fabricated by electron and ion beam
  induced chemical vapour deposition},}\ }}\href
  {https://doi.org/10.1016/j.tsf.2007.02.029} {\bibfield  {journal} {\bibinfo
  {journal} {Thin Solid Films}\ }\textbf {\bibinfo {volume} {515}},\ \bibinfo
  {pages} {6791--6797} (\bibinfo {year} {2007})}\BibitemShut {NoStop}%
\bibitem [{\citenamefont {Guillam{\'{o}}n}\ \emph {et~al.}(2008)\citenamefont
  {Guillam{\'{o}}n}, \citenamefont {Suderow}, \citenamefont {Vieira},
  \citenamefont {Fern{\'{a}}ndez-Pacheco}, \citenamefont {Ses{\'{e}}},
  \citenamefont {C{\'{o}}rdoba}, \citenamefont {De~Teresa},\ and\ \citenamefont
  {Ibarra}}]{Guillamon_2008}%
  \BibitemOpen
  \bibfield  {author} {\bibinfo {author} {\bibfnamefont {I.}~\bibnamefont
  {Guillam{\'{o}}n}}, \bibinfo {author} {\bibfnamefont {H.}~\bibnamefont
  {Suderow}}, \bibinfo {author} {\bibfnamefont {S.}~\bibnamefont {Vieira}},
  \bibinfo {author} {\bibfnamefont {A.}~\bibnamefont
  {Fern{\'{a}}ndez-Pacheco}}, \bibinfo {author} {\bibfnamefont
  {J.}~\bibnamefont {Ses{\'{e}}}}, \bibinfo {author} {\bibfnamefont
  {R.}~\bibnamefont {C{\'{o}}rdoba}}, \bibinfo {author} {\bibfnamefont {J.~M.}\
  \bibnamefont {De~Teresa}},\ and\ \bibinfo {author} {\bibfnamefont {M.~R.}\
  \bibnamefont {Ibarra}},\ }\bibfield  {title} {\enquote {\bibinfo {title}
  {Nanoscale superconducting properties of amorphous w-based deposits grown
  with a focused-ion-beam},}\ }\href
  {https://doi.org/10.1088/1367-2630/10/9/093005} {\bibfield  {journal}
  {\bibinfo  {journal} {New J. Phys.}\ }\textbf {\bibinfo {volume} {10}},\
  \bibinfo {pages} {093005} (\bibinfo {year} {2008})}\BibitemShut {NoStop}%
\bibitem [{\citenamefont {Li}\ \emph {et~al.}(2008)\citenamefont {Li},
  \citenamefont {Fenton}, \citenamefont {Wang}, \citenamefont {McComb},\ and\
  \citenamefont {Warburton}}]{Li_2008}%
  \BibitemOpen
  \bibfield  {author} {\bibinfo {author} {\bibfnamefont {W.}~\bibnamefont
  {Li}}, \bibinfo {author} {\bibfnamefont {J.~C.}\ \bibnamefont {Fenton}},
  \bibinfo {author} {\bibfnamefont {Y.}~\bibnamefont {Wang}}, \bibinfo {author}
  {\bibfnamefont {D.~W.}\ \bibnamefont {McComb}},\ and\ \bibinfo {author}
  {\bibfnamefont {P.~A.}\ \bibnamefont {Warburton}},\ }\bibfield  {title}
  {\enquote {\bibinfo {title} {Tunability of the superconductivity of tungsten
  films grown by focused-ion-beam direct writing},}\ }\href
  {https://doi.org/10.1063/1.3013444} {\bibfield  {journal} {\bibinfo
  {journal} {J. Appl. Phys.}\ }\textbf {\bibinfo {volume} {104}},\ \bibinfo
  {pages} {093913} (\bibinfo {year} {2008})}\BibitemShut {NoStop}%
\bibitem [{\citenamefont {Sun}\ \emph {et~al.}(2013)\citenamefont {Sun},
  \citenamefont {Wang}, \citenamefont {Zhao}, \citenamefont {Tian},
  \citenamefont {Singh},\ and\ \citenamefont {Chan}}]{Sun_2013}%
  \BibitemOpen
  \bibfield  {author} {\bibinfo {author} {\bibfnamefont {Y.}~\bibnamefont
  {Sun}}, \bibinfo {author} {\bibfnamefont {J.}~\bibnamefont {Wang}}, \bibinfo
  {author} {\bibfnamefont {W.}~\bibnamefont {Zhao}}, \bibinfo {author}
  {\bibfnamefont {M.}~\bibnamefont {Tian}}, \bibinfo {author} {\bibfnamefont
  {M.}~\bibnamefont {Singh}},\ and\ \bibinfo {author} {\bibfnamefont
  {M.~H.~W.}\ \bibnamefont {Chan}},\ }\bibfield  {title} {\enquote {\bibinfo
  {title} {Voltage-current properties of superconducting amorphous tungsten
  nanostrips},}\ }\href {https://doi.org/10.1038/srep02307} {\bibfield
  {journal} {\bibinfo  {journal} {Sci. Rep.}\ }\textbf {\bibinfo {volume}
  {3}},\ \bibinfo {pages} {1--7} (\bibinfo {year} {2013})}\BibitemShut
  {NoStop}%
\bibitem [{\citenamefont {C{\'{o}}rdoba}\ \emph {et~al.}(2013)\citenamefont
  {C{\'{o}}rdoba}, \citenamefont {Baturina}, \citenamefont {Ses{\'{e}}},
  \citenamefont {Yu~Mironov}, \citenamefont {De~Teresa}, \citenamefont
  {Ibarra}, \citenamefont {Nasimov}, \citenamefont {Gutakovskii}, \citenamefont
  {Latyshev}, \citenamefont {Guillam{\'{o}}n}, \citenamefont {Suderow},
  \citenamefont {Vieira}, \citenamefont {Baklanov}, \citenamefont {Palacios},\
  and\ \citenamefont {Vinokur}}]{Cordoba_2013}%
  \BibitemOpen
  \bibfield  {author} {\bibinfo {author} {\bibfnamefont {R.}~\bibnamefont
  {C{\'{o}}rdoba}}, \bibinfo {author} {\bibfnamefont {T.~I.}\ \bibnamefont
  {Baturina}}, \bibinfo {author} {\bibfnamefont {J.}~\bibnamefont
  {Ses{\'{e}}}}, \bibinfo {author} {\bibfnamefont {A.}~\bibnamefont
  {Yu~Mironov}}, \bibinfo {author} {\bibfnamefont {J.~M.}\ \bibnamefont
  {De~Teresa}}, \bibinfo {author} {\bibfnamefont {M.~R.}\ \bibnamefont
  {Ibarra}}, \bibinfo {author} {\bibfnamefont {D.~A.}\ \bibnamefont {Nasimov}},
  \bibinfo {author} {\bibfnamefont {A.~K.}\ \bibnamefont {Gutakovskii}},
  \bibinfo {author} {\bibfnamefont {V.~A.}\ \bibnamefont {Latyshev}}, \bibinfo
  {author} {\bibfnamefont {I.}~\bibnamefont {Guillam{\'{o}}n}}, \bibinfo
  {author} {\bibfnamefont {H.}~\bibnamefont {Suderow}}, \bibinfo {author}
  {\bibfnamefont {S.}~\bibnamefont {Vieira}}, \bibinfo {author} {\bibfnamefont
  {M.~R.}\ \bibnamefont {Baklanov}}, \bibinfo {author} {\bibfnamefont {J.~J.}\
  \bibnamefont {Palacios}},\ and\ \bibinfo {author} {\bibfnamefont {V.~M.}\
  \bibnamefont {Vinokur}},\ }\bibfield  {title} {{\selectlanguage
  {english}\enquote {\bibinfo {title} {Magnetic field-induced dissipation-free
  state in superconducting nanostructures},}\ }}\href
  {https://doi.org/10.1038/ncomms2437} {\bibfield  {journal} {\bibinfo
  {journal} {Nat. Commun.}\ }\textbf {\bibinfo {volume} {4}},\ \bibinfo {pages}
  {1437} (\bibinfo {year} {2013})}\BibitemShut {NoStop}%
\bibitem [{\citenamefont
  {C{\'{o}}rdoba~Castillo}(2014)}]{CordobaCastillo_2014}%
  \BibitemOpen
  \bibfield  {author} {\bibinfo {author} {\bibfnamefont {R.}~\bibnamefont
  {C{\'{o}}rdoba~Castillo}},\ }\emph {\bibinfo {title} {Functional
  nanostructures fabricated by focused electron/ion beam induced deposition}},\
  \href {https://doi.org/10.1007/978-3-319-02081-5} {\bibinfo {type}
  {phdthesis}},\ \bibinfo  {school} {University of Zaragoza} (\bibinfo {year}
  {2014})\BibitemShut {NoStop}%
\bibitem [{\citenamefont {C{\'{o}}rdoba}\ \emph
  {et~al.}(2019{\natexlab{b}})\citenamefont {C{\'{o}}rdoba}, \citenamefont
  {Or{\'{u}}s}, \citenamefont {Strohauer}, \citenamefont {Torres},\ and\
  \citenamefont {De~Teresa}}]{Cordoba_2019b}%
  \BibitemOpen
  \bibfield  {author} {\bibinfo {author} {\bibfnamefont {R.}~\bibnamefont
  {C{\'{o}}rdoba}}, \bibinfo {author} {\bibfnamefont {P.}~\bibnamefont
  {Or{\'{u}}s}}, \bibinfo {author} {\bibfnamefont {S.}~\bibnamefont
  {Strohauer}}, \bibinfo {author} {\bibfnamefont {T.~E.}\ \bibnamefont
  {Torres}},\ and\ \bibinfo {author} {\bibfnamefont {J.~M.}\ \bibnamefont
  {De~Teresa}},\ }\bibfield  {title} {\enquote {\bibinfo {title} {Ultra-fast
  direct growth of metallic micro- and nano-structures by focused ion beam
  irradiation},}\ }\href {https://doi.org/10.1038/s41598-019-50411-w}
  {\bibfield  {journal} {\bibinfo  {journal} {Sci. Rep.}\ }\textbf {\bibinfo
  {volume} {9}},\ \bibinfo {pages} {14076} (\bibinfo {year}
  {2019}{\natexlab{b}})}\BibitemShut {NoStop}%
\bibitem [{\citenamefont {Or{\'{u}}s}\ \emph
  {et~al.}(2021{\natexlab{b}})\citenamefont {Or{\'{u}}s}, \citenamefont
  {Fomin}, \citenamefont {De~Teresa},\ and\ \citenamefont
  {C{\'{o}}rdoba}}]{Orus_2021}%
  \BibitemOpen
  \bibfield  {author} {\bibinfo {author} {\bibfnamefont {P.}~\bibnamefont
  {Or{\'{u}}s}}, \bibinfo {author} {\bibfnamefont {V.~M.}\ \bibnamefont
  {Fomin}}, \bibinfo {author} {\bibfnamefont {J.~M.}\ \bibnamefont
  {De~Teresa}},\ and\ \bibinfo {author} {\bibfnamefont {R.}~\bibnamefont
  {C{\'{o}}rdoba}},\ }\bibfield  {title} {\enquote {\bibinfo {title} {Critical
  current modulation induced by an electric field in superconducting
  tungsten-carbon nanowires},}\ }\href
  {https://doi.org/10.1038/s41598-021-97075-z} {\bibfield  {journal} {\bibinfo
  {journal} {Sci. Rep.}\ }\textbf {\bibinfo {volume} {11}},\ \bibinfo {pages}
  {17698} (\bibinfo {year} {2021}{\natexlab{b}})}\BibitemShut {NoStop}%
\bibitem [{\citenamefont {Porrati}\ \emph {et~al.}(2019)\citenamefont
  {Porrati}, \citenamefont {Barth}, \citenamefont {Sachser}, \citenamefont
  {Dobrovolskiy}, \citenamefont {Seybert}, \citenamefont {Frangakis},\ and\
  \citenamefont {Huth}}]{Porrati_2019}%
  \BibitemOpen
  \bibfield  {author} {\bibinfo {author} {\bibfnamefont {F.}~\bibnamefont
  {Porrati}}, \bibinfo {author} {\bibfnamefont {S.}~\bibnamefont {Barth}},
  \bibinfo {author} {\bibfnamefont {R.}~\bibnamefont {Sachser}}, \bibinfo
  {author} {\bibfnamefont {V.~O.}\ \bibnamefont {Dobrovolskiy}}, \bibinfo
  {author} {\bibfnamefont {A.}~\bibnamefont {Seybert}}, \bibinfo {author}
  {\bibfnamefont {A.~S.}\ \bibnamefont {Frangakis}},\ and\ \bibinfo {author}
  {\bibfnamefont {M.}~\bibnamefont {Huth}},\ }\bibfield  {title} {\enquote
  {\bibinfo {title} {Crystalline niobium carbide superconducting nanowires
  prepared by focused ion beam direct writing},}\ }\href
  {https://doi.org/10.1021/acsnano.9b00059} {\bibfield  {journal} {\bibinfo
  {journal} {ACS Nano}\ }\textbf {\bibinfo {volume} {13}},\ \bibinfo {pages}
  {6287--6296} (\bibinfo {year} {2019})}\BibitemShut {NoStop}%
\bibitem [{\citenamefont {Dobrovolskiy}\ \emph {et~al.}(2020)\citenamefont
  {Dobrovolskiy}, \citenamefont {Vodolazov}, \citenamefont {Porrati},
  \citenamefont {Sachser}, \citenamefont {Bevz}, \citenamefont {Mikhailov},
  \citenamefont {Chumak},\ and\ \citenamefont {Huth}}]{Dobrovolskiy_2020}%
  \BibitemOpen
  \bibfield  {author} {\bibinfo {author} {\bibfnamefont {O.~V.}\ \bibnamefont
  {Dobrovolskiy}}, \bibinfo {author} {\bibfnamefont {D.~Y.}\ \bibnamefont
  {Vodolazov}}, \bibinfo {author} {\bibfnamefont {F.}~\bibnamefont {Porrati}},
  \bibinfo {author} {\bibfnamefont {R.}~\bibnamefont {Sachser}}, \bibinfo
  {author} {\bibfnamefont {V.~M.}\ \bibnamefont {Bevz}}, \bibinfo {author}
  {\bibfnamefont {M.~Y.}\ \bibnamefont {Mikhailov}}, \bibinfo {author}
  {\bibfnamefont {A.~V.}\ \bibnamefont {Chumak}},\ and\ \bibinfo {author}
  {\bibfnamefont {M.}~\bibnamefont {Huth}},\ }\bibfield  {title} {\enquote
  {\bibinfo {title} {Ultra-fast vortex motion in a direct-write {Nb-C}
  superconductor},}\ }\href {https://doi.org/10.1038/s41467-020-16987-y}
  {\bibfield  {journal} {\bibinfo  {journal} {Nat. Commun.}\ }\textbf {\bibinfo
  {volume} {11}},\ \bibinfo {pages} {3291} (\bibinfo {year}
  {2020})}\BibitemShut {NoStop}%
\bibitem [{\citenamefont {Porrati}\ \emph {et~al.}(2022)\citenamefont
  {Porrati}, \citenamefont {Jungwirth}, \citenamefont {Barth}, \citenamefont
  {Gazzadi}, \citenamefont {Frabboni}, \citenamefont {Dobrovolskiy},\ and\
  \citenamefont {Huth}}]{Porrati_2022}%
  \BibitemOpen
  \bibfield  {author} {\bibinfo {author} {\bibfnamefont {F.}~\bibnamefont
  {Porrati}}, \bibinfo {author} {\bibfnamefont {F.}~\bibnamefont {Jungwirth}},
  \bibinfo {author} {\bibfnamefont {S.}~\bibnamefont {Barth}}, \bibinfo
  {author} {\bibfnamefont {G.~C.}\ \bibnamefont {Gazzadi}}, \bibinfo {author}
  {\bibfnamefont {S.}~\bibnamefont {Frabboni}}, \bibinfo {author}
  {\bibfnamefont {O.~V.}\ \bibnamefont {Dobrovolskiy}},\ and\ \bibinfo {author}
  {\bibfnamefont {M.}~\bibnamefont {Huth}},\ }\bibfield  {title} {\enquote
  {\bibinfo {title} {Highly-packed proximity-coupled {DC-Josephson} junction
  arrays by a direct-write approach},}\ }\href
  {https://doi.org/10.1002/adfm.202203889} {\bibfield  {journal} {\bibinfo
  {journal} {Adv. Func. Mater.}\ }\textbf {\bibinfo {volume} {n/a}},\ \bibinfo
  {pages} {2203889} (\bibinfo {year} {2022})}\BibitemShut {NoStop}%
\bibitem [{\citenamefont {Romans}\ \emph {et~al.}(2010)\citenamefont {Romans},
  \citenamefont {Osley}, \citenamefont {Young}, \citenamefont {Warburton},\
  and\ \citenamefont {Li}}]{Romans_2010}%
  \BibitemOpen
  \bibfield  {author} {\bibinfo {author} {\bibfnamefont {E.~J.}\ \bibnamefont
  {Romans}}, \bibinfo {author} {\bibfnamefont {E.~J.}\ \bibnamefont {Osley}},
  \bibinfo {author} {\bibfnamefont {L.}~\bibnamefont {Young}}, \bibinfo
  {author} {\bibfnamefont {P.~A.}\ \bibnamefont {Warburton}},\ and\ \bibinfo
  {author} {\bibfnamefont {W.}~\bibnamefont {Li}},\ }\bibfield  {title}
  {\enquote {\bibinfo {title} {Three-dimensional nanoscale superconducting
  quantum interference device pickup loops},}\ }\href
  {https://doi.org/10.1063/1.3521262} {\bibfield  {journal} {\bibinfo
  {journal} {Appl. Phys. Lett.}\ }\textbf {\bibinfo {volume} {97}},\ \bibinfo
  {pages} {222506} (\bibinfo {year} {2010})}\BibitemShut {NoStop}%
\bibitem [{\citenamefont {Sigloch}\ \emph {et~al.}(2022)\citenamefont
  {Sigloch}, \citenamefont {Sangiao}, \citenamefont {Or{\'{u}}s},\ and\
  \citenamefont {De~Teresa}}]{Sigloch_2022}%
  \BibitemOpen
  \bibfield  {author} {\bibinfo {author} {\bibfnamefont {F.}~\bibnamefont
  {Sigloch}}, \bibinfo {author} {\bibfnamefont {S.}~\bibnamefont {Sangiao}},
  \bibinfo {author} {\bibfnamefont {P.}~\bibnamefont {Or{\'{u}}s}},\ and\
  \bibinfo {author} {\bibfnamefont {J.~M.}\ \bibnamefont {De~Teresa}},\
  }\bibfield  {title} {\enquote {\bibinfo {title} {Direct-write of
  tungsten-carbide {nanoSQUIDs} based on focused ion beam induced
  deposition},}\ }\href {https://doi.org/10.1039/d2na00602b} {\bibfield
  {journal} {\bibinfo  {journal} {Nanoscale Adv.}\ }\textbf {\bibinfo {volume}
  {4}},\ \bibinfo {pages} {4628--4634} (\bibinfo {year} {2022})}\BibitemShut
  {NoStop}%
\bibitem [{\citenamefont {Lapicki}, \citenamefont {Ahmad},\ and\ \citenamefont
  {Suzuki}(2002)}]{Lapicki_2002}%
  \BibitemOpen
  \bibfield  {author} {\bibinfo {author} {\bibfnamefont {A.}~\bibnamefont
  {Lapicki}}, \bibinfo {author} {\bibfnamefont {E.}~\bibnamefont {Ahmad}},\
  and\ \bibinfo {author} {\bibfnamefont {T.}~\bibnamefont {Suzuki}},\
  }\bibfield  {title} {{\selectlanguage {english}\enquote {\bibinfo {title}
  {Ion beam induced chemical vapor deposition ({IBICVD}) of cobalt
  particles},}\ }}\href {https://doi.org/10.1016/S0304-8853(01)00725-9}
  {\bibfield  {journal} {\bibinfo  {journal} {J. Magn. Magn. Mater.}\ }\textbf
  {\bibinfo {volume} {240}},\ \bibinfo {pages} {47--49} (\bibinfo {year}
  {2002})}\BibitemShut {NoStop}%
\bibitem [{\citenamefont {Sanz-Mart{\'{\i}}n}, \citenamefont {Mag{\'{e}}n},\
  and\ \citenamefont {De~Teresa}(2019)}]{SanzMartin_2019}%
  \BibitemOpen
  \bibfield  {author} {\bibinfo {author} {\bibfnamefont {C.}~\bibnamefont
  {Sanz-Mart{\'{\i}}n}}, \bibinfo {author} {\bibfnamefont {C.}~\bibnamefont
  {Mag{\'{e}}n}},\ and\ \bibinfo {author} {\bibfnamefont {J.~M.}\ \bibnamefont
  {De~Teresa}},\ }\bibfield  {title} {\enquote {\bibinfo {title} {High
  volume-per-dose and low resistivity of cobalt nanowires grown by {G}a$^+$
  focused ion beam induced deposition},}\ }\href
  {https://doi.org/10.3390/nano9121715} {\bibfield  {journal} {\bibinfo
  {journal} {Nanomaterials}\ }\textbf {\bibinfo {volume} {9}},\ \bibinfo
  {pages} {1715} (\bibinfo {year} {2019})}\BibitemShut {NoStop}%
\bibitem [{\citenamefont {Khizroev}, \citenamefont {Bain},\ and\ \citenamefont
  {Litvinov}(2002)}]{Khizroev_2002}%
  \BibitemOpen
  \bibfield  {author} {\bibinfo {author} {\bibfnamefont {S.}~\bibnamefont
  {Khizroev}}, \bibinfo {author} {\bibfnamefont {J.~A.}\ \bibnamefont {Bain}},\
  and\ \bibinfo {author} {\bibfnamefont {D.}~\bibnamefont {Litvinov}},\
  }\bibfield  {title} {\enquote {\bibinfo {title} {Fabrication of nanomagnetic
  probes via focused ion beam etching and deposition},}\ }\href
  {https://doi.org/10.1088/0957-4484/13/5/315} {\bibfield  {journal} {\bibinfo
  {journal} {Nanotechnology}\ }\textbf {\bibinfo {volume} {13}},\ \bibinfo
  {pages} {619--622} (\bibinfo {year} {2002})}\BibitemShut {NoStop}%
\bibitem [{\citenamefont {Hyun}, \citenamefont {Lee},\ and\ \citenamefont
  {de~Lozanne}(2006)}]{Hyun_2006a}%
  \BibitemOpen
  \bibfield  {author} {\bibinfo {author} {\bibfnamefont {C.}~\bibnamefont
  {Hyun}}, \bibinfo {author} {\bibfnamefont {A.~K.~H.}\ \bibnamefont {Lee}},\
  and\ \bibinfo {author} {\bibfnamefont {A.}~\bibnamefont {de~Lozanne}},\
  }\bibfield  {title} {\enquote {\bibinfo {title} {Focused ion beam deposition
  of {C}o$_{71}${C}r$_{17}${P}t$_{12}$ and {N}i$_{80}${F}e$_{20}$ on tips for
  magnetic force microscopy},}\ }\href
  {https://doi.org/10.1088/0957-4484/17/3/051} {\bibfield  {journal} {\bibinfo
  {journal} {Nanotechnology}\ }\textbf {\bibinfo {volume} {17}},\ \bibinfo
  {pages} {921--925} (\bibinfo {year} {2006})}\BibitemShut {NoStop}%
\bibitem [{\citenamefont {Gabureac}\ \emph {et~al.}(2010)\citenamefont
  {Gabureac}, \citenamefont {Bernau}, \citenamefont {Utke},\ and\ \citenamefont
  {Boero}}]{Gabureac_2010a}%
  \BibitemOpen
  \bibfield  {author} {\bibinfo {author} {\bibfnamefont {M.}~\bibnamefont
  {Gabureac}}, \bibinfo {author} {\bibfnamefont {L.}~\bibnamefont {Bernau}},
  \bibinfo {author} {\bibfnamefont {I.}~\bibnamefont {Utke}},\ and\ \bibinfo
  {author} {\bibfnamefont {G.}~\bibnamefont {Boero}},\ }\bibfield  {title}
  {\enquote {\bibinfo {title} {Granular {C}o-{C} nano-hall sensors by
  focused-beam-induced deposition},}\ }\href
  {https://doi.org/10.1088/0957-4484/21/11/115503} {\bibfield  {journal}
  {\bibinfo  {journal} {Nanotechnology}\ }\textbf {\bibinfo {volume} {21}},\
  \bibinfo {pages} {115503} (\bibinfo {year} {2010})}\BibitemShut {NoStop}%
\bibitem [{\citenamefont {Vogelaar}\ \emph {et~al.}(2001)\citenamefont
  {Vogelaar}, \citenamefont {Nijdam}, \citenamefont {van Wolferen},
  \citenamefont {de~Ridder}, \citenamefont {Segerink}, \citenamefont
  {Fl{\"{u}}ck}, \citenamefont {Kuipers},\ and\ \citenamefont {van
  Hulst}}]{Vogelaar_2001}%
  \BibitemOpen
  \bibfield  {author} {\bibinfo {author} {\bibfnamefont {L.}~\bibnamefont
  {Vogelaar}}, \bibinfo {author} {\bibfnamefont {W.}~\bibnamefont {Nijdam}},
  \bibinfo {author} {\bibfnamefont {H.~A. G.~M.}\ \bibnamefont {van Wolferen}},
  \bibinfo {author} {\bibfnamefont {R.~M.}\ \bibnamefont {de~Ridder}}, \bibinfo
  {author} {\bibfnamefont {F.~B.}\ \bibnamefont {Segerink}}, \bibinfo {author}
  {\bibfnamefont {E.}~\bibnamefont {Fl{\"{u}}ck}}, \bibinfo {author}
  {\bibfnamefont {L.}~\bibnamefont {Kuipers}},\ and\ \bibinfo {author}
  {\bibfnamefont {N.~F.}\ \bibnamefont {van Hulst}},\ }\bibfield  {title}
  {\enquote {\bibinfo {title} {Large area photonic crystal slabs for visible
  light with waveguiding defect structures: fabrication with focused ion beam
  assisted laser interference lithography},}\ }\href
  {https://doi.org/10.1002/1521-4095(200110)13:20<1551::aid-adma1551>3.0.co;2-v}
  {\bibfield  {journal} {\bibinfo  {journal} {Adv. Mater.}\ }\textbf {\bibinfo
  {volume} {13}},\ \bibinfo {pages} {1551} (\bibinfo {year}
  {2001})}\BibitemShut {NoStop}%
\bibitem [{\citenamefont {Esposito}\ \emph {et~al.}(2014)\citenamefont
  {Esposito}, \citenamefont {Tasco}, \citenamefont {Cuscun\`{a}}, \citenamefont
  {Todisco}, \citenamefont {Benedetti}, \citenamefont {Tarantini},
  \citenamefont {Giorgi}, \citenamefont {Sanvitto},\ and\ \citenamefont
  {Passaseo}}]{Esposito_2014a}%
  \BibitemOpen
  \bibfield  {author} {\bibinfo {author} {\bibfnamefont {M.}~\bibnamefont
  {Esposito}}, \bibinfo {author} {\bibfnamefont {V.}~\bibnamefont {Tasco}},
  \bibinfo {author} {\bibfnamefont {M.}~\bibnamefont {Cuscun\`{a}}}, \bibinfo
  {author} {\bibfnamefont {F.}~\bibnamefont {Todisco}}, \bibinfo {author}
  {\bibfnamefont {A.}~\bibnamefont {Benedetti}}, \bibinfo {author}
  {\bibfnamefont {I.}~\bibnamefont {Tarantini}}, \bibinfo {author}
  {\bibfnamefont {M.~D.}\ \bibnamefont {Giorgi}}, \bibinfo {author}
  {\bibfnamefont {D.}~\bibnamefont {Sanvitto}},\ and\ \bibinfo {author}
  {\bibfnamefont {A.}~\bibnamefont {Passaseo}},\ }\bibfield  {title} {\enquote
  {\bibinfo {title} {Nanoscale 3{D} chiral plasmonic helices with circular
  dichroism at visible frequencies},}\ }\href
  {https://doi.org/10.1021/ph500318p} {\bibfield  {journal} {\bibinfo
  {journal} {{ACS} Photonics}\ }\textbf {\bibinfo {volume} {2}},\ \bibinfo
  {pages} {105--114} (\bibinfo {year} {2014})}\BibitemShut {NoStop}%
\bibitem [{\citenamefont {Manoccio}\ \emph {et~al.}(2020)\citenamefont
  {Manoccio}, \citenamefont {Esposito}, \citenamefont {Passaseo}, \citenamefont
  {Cuscun{\`{a}}},\ and\ \citenamefont {Tasco}}]{Manoccio_2020}%
  \BibitemOpen
  \bibfield  {author} {\bibinfo {author} {\bibfnamefont {M.}~\bibnamefont
  {Manoccio}}, \bibinfo {author} {\bibfnamefont {M.}~\bibnamefont {Esposito}},
  \bibinfo {author} {\bibfnamefont {A.}~\bibnamefont {Passaseo}}, \bibinfo
  {author} {\bibfnamefont {M.}~\bibnamefont {Cuscun{\`{a}}}},\ and\ \bibinfo
  {author} {\bibfnamefont {V.}~\bibnamefont {Tasco}},\ }\bibfield  {title}
  {\enquote {\bibinfo {title} {Focused ion beam processing for 3{D} chiral
  photonics nanostructures},}\ }\href {https://doi.org/10.3390/mi12010006}
  {\bibfield  {journal} {\bibinfo  {journal} {Micromachines}\ }\textbf
  {\bibinfo {volume} {12}},\ \bibinfo {pages} {6} (\bibinfo {year}
  {2020})}\BibitemShut {NoStop}%
\bibitem [{\citenamefont {Esposito}\ \emph {et~al.}(2015)\citenamefont
  {Esposito}, \citenamefont {Tasco}, \citenamefont {Todisco}, \citenamefont
  {Cuscun{\`{a}}}, \citenamefont {Benedetti}, \citenamefont {Sanvitto},\ and\
  \citenamefont {Passaseo}}]{Esposito_2015}%
  \BibitemOpen
  \bibfield  {author} {\bibinfo {author} {\bibfnamefont {M.}~\bibnamefont
  {Esposito}}, \bibinfo {author} {\bibfnamefont {V.}~\bibnamefont {Tasco}},
  \bibinfo {author} {\bibfnamefont {F.}~\bibnamefont {Todisco}}, \bibinfo
  {author} {\bibfnamefont {M.}~\bibnamefont {Cuscun{\`{a}}}}, \bibinfo {author}
  {\bibfnamefont {A.}~\bibnamefont {Benedetti}}, \bibinfo {author}
  {\bibfnamefont {D.}~\bibnamefont {Sanvitto}},\ and\ \bibinfo {author}
  {\bibfnamefont {A.}~\bibnamefont {Passaseo}},\ }\bibfield  {title} {\enquote
  {\bibinfo {title} {Triple-helical nanowires by tomographic rotatory growth
  for chiral photonics},}\ }\href {https://doi.org/10.1038/ncomms7484}
  {\bibfield  {journal} {\bibinfo  {journal} {Nat. Commun.}\ }\textbf {\bibinfo
  {volume} {6}},\ \bibinfo {pages} {6484} (\bibinfo {year} {2015})}\BibitemShut
  {NoStop}%
\bibitem [{\citenamefont {Esposito}\ \emph {et~al.}(2016)\citenamefont
  {Esposito}, \citenamefont {Tasco}, \citenamefont {Todisco}, \citenamefont
  {Cuscun{\`{a}}}, \citenamefont {Benedetti}, \citenamefont {Scuderi},
  \citenamefont {Nicotra},\ and\ \citenamefont {Passaseo}}]{Esposito_2016a}%
  \BibitemOpen
  \bibfield  {author} {\bibinfo {author} {\bibfnamefont {M.}~\bibnamefont
  {Esposito}}, \bibinfo {author} {\bibfnamefont {V.}~\bibnamefont {Tasco}},
  \bibinfo {author} {\bibfnamefont {F.}~\bibnamefont {Todisco}}, \bibinfo
  {author} {\bibfnamefont {M.}~\bibnamefont {Cuscun{\`{a}}}}, \bibinfo {author}
  {\bibfnamefont {A.}~\bibnamefont {Benedetti}}, \bibinfo {author}
  {\bibfnamefont {M.}~\bibnamefont {Scuderi}}, \bibinfo {author} {\bibfnamefont
  {G.}~\bibnamefont {Nicotra}},\ and\ \bibinfo {author} {\bibfnamefont
  {A.}~\bibnamefont {Passaseo}},\ }\bibfield  {title} {\enquote {\bibinfo
  {title} {Programmable extreme chirality in the visible by helix-shaped
  metamaterial platform},}\ }\href
  {https://doi.org/10.1021/acs.nanolett.6b02583} {\bibfield  {journal}
  {\bibinfo  {journal} {Nano Lett.}\ }\textbf {\bibinfo {volume} {16}},\
  \bibinfo {pages} {5823--5828} (\bibinfo {year} {2016})}\BibitemShut {NoStop}%
\bibitem [{\citenamefont {Wang}\ \emph
  {et~al.}(2019{\natexlab{b}})\citenamefont {Wang}, \citenamefont {Salut},
  \citenamefont {Suarez}, \citenamefont {Martin},\ and\ \citenamefont
  {Grosjean}}]{Wang_2019a}%
  \BibitemOpen
  \bibfield  {author} {\bibinfo {author} {\bibfnamefont {M.}~\bibnamefont
  {Wang}}, \bibinfo {author} {\bibfnamefont {R.}~\bibnamefont {Salut}},
  \bibinfo {author} {\bibfnamefont {M.~A.}\ \bibnamefont {Suarez}}, \bibinfo
  {author} {\bibfnamefont {N.}~\bibnamefont {Martin}},\ and\ \bibinfo {author}
  {\bibfnamefont {T.}~\bibnamefont {Grosjean}},\ }\bibfield  {title} {\enquote
  {\bibinfo {title} {Chiroptical transmission through a plasmonic helical
  traveling-wave nanoantenna, towards on-tip chiroptical probes},}\ }\href
  {https://doi.org/10.1364/OL.44.004861} {\bibfield  {journal} {\bibinfo
  {journal} {Opt. Lett.}\ }\textbf {\bibinfo {volume} {44}},\ \bibinfo {pages}
  {4861} (\bibinfo {year} {2019}{\natexlab{b}})}\BibitemShut {NoStop}%
\bibitem [{\citenamefont {Wang}\ \emph
  {et~al.}(2019{\natexlab{c}})\citenamefont {Wang}, \citenamefont {Salut},
  \citenamefont {Lu}, \citenamefont {Suarez}, \citenamefont {Martin},\ and\
  \citenamefont {Grosjean}}]{Wang_2019c}%
  \BibitemOpen
  \bibfield  {author} {\bibinfo {author} {\bibfnamefont {M.}~\bibnamefont
  {Wang}}, \bibinfo {author} {\bibfnamefont {R.}~\bibnamefont {Salut}},
  \bibinfo {author} {\bibfnamefont {H.}~\bibnamefont {Lu}}, \bibinfo {author}
  {\bibfnamefont {M.-A.}\ \bibnamefont {Suarez}}, \bibinfo {author}
  {\bibfnamefont {N.}~\bibnamefont {Martin}},\ and\ \bibinfo {author}
  {\bibfnamefont {T.}~\bibnamefont {Grosjean}},\ }\bibfield  {title} {\enquote
  {\bibinfo {title} {Subwavelength polarization optics via individual and
  coupled helical traveling-wave nanoantennas},}\ }\href
  {https://doi.org/10.1038/s41377-019-0186-2} {\bibfield  {journal} {\bibinfo
  {journal} {Light Sci. Appl.}\ }\textbf {\bibinfo {volume} {8}},\ \bibinfo
  {pages} {76} (\bibinfo {year} {2019}{\natexlab{c}})}\BibitemShut {NoStop}%
\bibitem [{\citenamefont {Wang}\ \emph
  {et~al.}(2021{\natexlab{c}})\citenamefont {Wang}, \citenamefont {Huang},
  \citenamefont {Salut}, \citenamefont {Suarez}, \citenamefont {Lu},
  \citenamefont {Martin},\ and\ \citenamefont {Grosjean}}]{Wang_2021a}%
  \BibitemOpen
  \bibfield  {author} {\bibinfo {author} {\bibfnamefont {M.}~\bibnamefont
  {Wang}}, \bibinfo {author} {\bibfnamefont {Z.}~\bibnamefont {Huang}},
  \bibinfo {author} {\bibfnamefont {R.}~\bibnamefont {Salut}}, \bibinfo
  {author} {\bibfnamefont {M.~A.}\ \bibnamefont {Suarez}}, \bibinfo {author}
  {\bibfnamefont {H.}~\bibnamefont {Lu}}, \bibinfo {author} {\bibfnamefont
  {N.}~\bibnamefont {Martin}},\ and\ \bibinfo {author} {\bibfnamefont
  {T.}~\bibnamefont {Grosjean}},\ }\bibfield  {title} {\enquote {\bibinfo
  {title} {Plasmonic helical nanoantenna as a converter between longitudinal
  fields and circularly polarized waves},}\ }\href
  {https://doi.org/10.1021/acs.nanolett.0c04948} {\bibfield  {journal}
  {\bibinfo  {journal} {Nano Lett.}\ }\textbf {\bibinfo {volume} {21}},\
  \bibinfo {pages} {3410--3417} (\bibinfo {year}
  {2021}{\natexlab{c}})}\BibitemShut {NoStop}%
\bibitem [{\citenamefont {Manoccio}\ \emph {et~al.}(2021)\citenamefont
  {Manoccio}, \citenamefont {Esposito}, \citenamefont {Primiceri},
  \citenamefont {Leo}, \citenamefont {Tasco}, \citenamefont {Cuscun{\`{a}}},
  \citenamefont {Zuev}, \citenamefont {Sun}, \citenamefont {Maruccio},
  \citenamefont {Romano}, \citenamefont {Quattrini}, \citenamefont {Gigli},\
  and\ \citenamefont {Passaseo}}]{Manoccio_2021}%
  \BibitemOpen
  \bibfield  {author} {\bibinfo {author} {\bibfnamefont {M.}~\bibnamefont
  {Manoccio}}, \bibinfo {author} {\bibfnamefont {M.}~\bibnamefont {Esposito}},
  \bibinfo {author} {\bibfnamefont {E.}~\bibnamefont {Primiceri}}, \bibinfo
  {author} {\bibfnamefont {A.}~\bibnamefont {Leo}}, \bibinfo {author}
  {\bibfnamefont {V.}~\bibnamefont {Tasco}}, \bibinfo {author} {\bibfnamefont
  {M.}~\bibnamefont {Cuscun{\`{a}}}}, \bibinfo {author} {\bibfnamefont
  {D.}~\bibnamefont {Zuev}}, \bibinfo {author} {\bibfnamefont {Y.}~\bibnamefont
  {Sun}}, \bibinfo {author} {\bibfnamefont {G.}~\bibnamefont {Maruccio}},
  \bibinfo {author} {\bibfnamefont {A.}~\bibnamefont {Romano}}, \bibinfo
  {author} {\bibfnamefont {A.}~\bibnamefont {Quattrini}}, \bibinfo {author}
  {\bibfnamefont {G.}~\bibnamefont {Gigli}},\ and\ \bibinfo {author}
  {\bibfnamefont {A.}~\bibnamefont {Passaseo}},\ }\bibfield  {title} {\enquote
  {\bibinfo {title} {Femtomolar biodetection by a compact
  core{\textendash}shell 3{D} chiral metamaterial},}\ }\href
  {https://doi.org/10.1021/acs.nanolett.1c01791} {\bibfield  {journal}
  {\bibinfo  {journal} {Nano Lett.}\ }\textbf {\bibinfo {volume} {21}},\
  \bibinfo {pages} {6179--6187} (\bibinfo {year} {2021})}\BibitemShut {NoStop}%
\bibitem [{\citenamefont {Nanda}\ \emph {et~al.}(2015)\citenamefont {Nanda},
  \citenamefont {van Veldhoven}, \citenamefont {Maas}, \citenamefont
  {Sadeghian},\ and\ \citenamefont {Alkemade}}]{Nanda2015}%
  \BibitemOpen
  \bibfield  {author} {\bibinfo {author} {\bibfnamefont {G.}~\bibnamefont
  {Nanda}}, \bibinfo {author} {\bibfnamefont {E.}~\bibnamefont {van
  Veldhoven}}, \bibinfo {author} {\bibfnamefont {D.}~\bibnamefont {Maas}},
  \bibinfo {author} {\bibfnamefont {H.}~\bibnamefont {Sadeghian}},\ and\
  \bibinfo {author} {\bibfnamefont {P.~F.~A.}\ \bibnamefont {Alkemade}},\
  }\bibfield  {title} {\enquote {\bibinfo {title} {Helium ion beam induced
  growth of hammerhead {AFM} probes},}\ }\href
  {https://doi.org/10.1116/1.4936068} {\bibfield  {journal} {\bibinfo
  {journal} {J. Vac. Sci. Technol. B}\ }\textbf {\bibinfo {volume} {33}},\
  \bibinfo {pages} {06F503} (\bibinfo {year} {2015})}\BibitemShut {NoStop}%
\bibitem [{\citenamefont {Fujita}\ \emph {et~al.}(2007)\citenamefont {Fujita},
  \citenamefont {Okada}, \citenamefont {Ueki}, \citenamefont {Ishida},
  \citenamefont {Kaito},\ and\ \citenamefont {Matsui}}]{Fujita_2007}%
  \BibitemOpen
  \bibfield  {author} {\bibinfo {author} {\bibfnamefont {J.-i.}\ \bibnamefont
  {Fujita}}, \bibinfo {author} {\bibfnamefont {S.}~\bibnamefont {Okada}},
  \bibinfo {author} {\bibfnamefont {R.}~\bibnamefont {Ueki}}, \bibinfo {author}
  {\bibfnamefont {M.}~\bibnamefont {Ishida}}, \bibinfo {author} {\bibfnamefont
  {T.}~\bibnamefont {Kaito}},\ and\ \bibinfo {author} {\bibfnamefont
  {S.}~\bibnamefont {Matsui}},\ }\bibfield  {title} {\enquote {\bibinfo {title}
  {Elastic double structure of amorphous carbon pillar grown by
  focused-ion-beam chemical vapor deposition},}\ }\href
  {https://doi.org/10.1143/JJAP.46.6286} {\bibfield  {journal} {\bibinfo
  {journal} {Jpn. J. Appl. Phys.}\ }\textbf {\bibinfo {volume} {46}},\ \bibinfo
  {pages} {6286--6289} (\bibinfo {year} {2007})}\BibitemShut {NoStop}%
\bibitem [{\citenamefont {Nakamatsu}\ \emph {et~al.}(2009)\citenamefont
  {Nakamatsu}, \citenamefont {Ichihashi}, \citenamefont {Kanda}, \citenamefont
  {Haruyama}, \citenamefont {Kaito},\ and\ \citenamefont
  {Matsui}}]{Nakamatsu_2009}%
  \BibitemOpen
  \bibfield  {author} {\bibinfo {author} {\bibfnamefont {K.-i.}\ \bibnamefont
  {Nakamatsu}}, \bibinfo {author} {\bibfnamefont {T.}~\bibnamefont
  {Ichihashi}}, \bibinfo {author} {\bibfnamefont {K.}~\bibnamefont {Kanda}},
  \bibinfo {author} {\bibfnamefont {Y.}~\bibnamefont {Haruyama}}, \bibinfo
  {author} {\bibfnamefont {T.}~\bibnamefont {Kaito}},\ and\ \bibinfo {author}
  {\bibfnamefont {S.}~\bibnamefont {Matsui}},\ }\bibfield  {title} {\enquote
  {\bibinfo {title} {Nanostructure analysis of nanosprings fabricated by
  focused-ion-beam chemical vapor deposition},}\ }\href
  {https://doi.org/10.1143/JJAP.48.105001} {\bibfield  {journal} {\bibinfo
  {journal} {Jpn. J. Appl. Phys.}\ }\textbf {\bibinfo {volume} {48}},\ \bibinfo
  {pages} {105001} (\bibinfo {year} {2009})}\BibitemShut {NoStop}%
\bibitem [{\citenamefont {Nakai}\ \emph {et~al.}(2010)\citenamefont {Nakai},
  \citenamefont {Kang}, \citenamefont {Okada}, \citenamefont {Haruyama},
  \citenamefont {Kanda}, \citenamefont {Ichihashi},\ and\ \citenamefont
  {Matsui}}]{Nakai_2010}%
  \BibitemOpen
  \bibfield  {author} {\bibinfo {author} {\bibfnamefont {Y.}~\bibnamefont
  {Nakai}}, \bibinfo {author} {\bibfnamefont {Y.}~\bibnamefont {Kang}},
  \bibinfo {author} {\bibfnamefont {M.}~\bibnamefont {Okada}}, \bibinfo
  {author} {\bibfnamefont {Y.}~\bibnamefont {Haruyama}}, \bibinfo {author}
  {\bibfnamefont {K.}~\bibnamefont {Kanda}}, \bibinfo {author} {\bibfnamefont
  {T.}~\bibnamefont {Ichihashi}},\ and\ \bibinfo {author} {\bibfnamefont
  {S.}~\bibnamefont {Matsui}},\ }\bibfield  {title} {\enquote {\bibinfo {title}
  {Mechanical characteristics of nanosprings fabricated by focused-ion-beam
  chemical vapor deposition using ferrocene source gas},}\ }\href
  {https://doi.org/10.1143/JJAP.49.06GH07} {\bibfield  {journal} {\bibinfo
  {journal} {Jpn. J. Appl. Phys.}\ }\textbf {\bibinfo {volume} {49}},\ \bibinfo
  {pages} {06GH07} (\bibinfo {year} {2010})}\BibitemShut {NoStop}%
\bibitem [{\citenamefont {Reiser}\ \emph {et~al.}(2020)\citenamefont {Reiser},
  \citenamefont {Koch}, \citenamefont {Dunn}, \citenamefont {Matsuura},
  \citenamefont {Iwata}, \citenamefont {Fogel}, \citenamefont {Kotler},
  \citenamefont {Zhou}, \citenamefont {Charipar}, \citenamefont {Piqu{\'{e}}},
  \citenamefont {Rohner}, \citenamefont {Poulikakos}, \citenamefont {Lee},
  \citenamefont {Seol}, \citenamefont {Utke}, \citenamefont {Nisselroy},
  \citenamefont {Zambelli}, \citenamefont {Wheeler},\ and\ \citenamefont
  {Spolenak}}]{Reiser_2020}%
  \BibitemOpen
  \bibfield  {author} {\bibinfo {author} {\bibfnamefont {A.}~\bibnamefont
  {Reiser}}, \bibinfo {author} {\bibfnamefont {L.}~\bibnamefont {Koch}},
  \bibinfo {author} {\bibfnamefont {K.~A.}\ \bibnamefont {Dunn}}, \bibinfo
  {author} {\bibfnamefont {T.}~\bibnamefont {Matsuura}}, \bibinfo {author}
  {\bibfnamefont {F.}~\bibnamefont {Iwata}}, \bibinfo {author} {\bibfnamefont
  {O.}~\bibnamefont {Fogel}}, \bibinfo {author} {\bibfnamefont
  {Z.}~\bibnamefont {Kotler}}, \bibinfo {author} {\bibfnamefont
  {N.}~\bibnamefont {Zhou}}, \bibinfo {author} {\bibfnamefont {K.}~\bibnamefont
  {Charipar}}, \bibinfo {author} {\bibfnamefont {A.}~\bibnamefont
  {Piqu{\'{e}}}}, \bibinfo {author} {\bibfnamefont {P.}~\bibnamefont {Rohner}},
  \bibinfo {author} {\bibfnamefont {D.}~\bibnamefont {Poulikakos}}, \bibinfo
  {author} {\bibfnamefont {S.}~\bibnamefont {Lee}}, \bibinfo {author}
  {\bibfnamefont {S.~K.}\ \bibnamefont {Seol}}, \bibinfo {author}
  {\bibfnamefont {I.}~\bibnamefont {Utke}}, \bibinfo {author} {\bibfnamefont
  {C.}~\bibnamefont {Nisselroy}}, \bibinfo {author} {\bibfnamefont
  {T.}~\bibnamefont {Zambelli}}, \bibinfo {author} {\bibfnamefont {J.~M.}\
  \bibnamefont {Wheeler}},\ and\ \bibinfo {author} {\bibfnamefont
  {R.}~\bibnamefont {Spolenak}},\ }\bibfield  {title} {\enquote {\bibinfo
  {title} {Metals by micro‐scale additive manufacturing: comparison of
  microstructure and mechanical properties},}\ }\href
  {https://doi.org/10.1002/adfm.201910491} {\bibfield  {journal} {\bibinfo
  {journal} {Adv. Funct. Mater.}\ }\textbf {\bibinfo {volume} {30}},\ \bibinfo
  {pages} {1910491} (\bibinfo {year} {2020})}\BibitemShut {NoStop}%
\bibitem [{\citenamefont {C{\'{o}}rdoba}\ \emph {et~al.}(2017)\citenamefont
  {C{\'{o}}rdoba}, \citenamefont {Lorenzoni}, \citenamefont {Pablo-Navarro},
  \citenamefont {Mag{\'{e}}n}, \citenamefont {P{\'{e}}rez-Murano},\ and\
  \citenamefont {De~Teresa}}]{Cordoba_2017}%
  \BibitemOpen
  \bibfield  {author} {\bibinfo {author} {\bibfnamefont {R.}~\bibnamefont
  {C{\'{o}}rdoba}}, \bibinfo {author} {\bibfnamefont {M.}~\bibnamefont
  {Lorenzoni}}, \bibinfo {author} {\bibfnamefont {J.}~\bibnamefont
  {Pablo-Navarro}}, \bibinfo {author} {\bibfnamefont {C.}~\bibnamefont
  {Mag{\'{e}}n}}, \bibinfo {author} {\bibfnamefont {F.}~\bibnamefont
  {P{\'{e}}rez-Murano}},\ and\ \bibinfo {author} {\bibfnamefont {J.~M.}\
  \bibnamefont {De~Teresa}},\ }\bibfield  {title} {\enquote {\bibinfo {title}
  {Suspended tungsten-based nanowires with enhanced mechanical properties grown
  by focused ion beam induced deposition},}\ }\href
  {https://doi.org/10.1088/1361-6528/aa873c} {\bibfield  {journal} {\bibinfo
  {journal} {Nanotechnology}\ }\textbf {\bibinfo {volume} {28}},\ \bibinfo
  {pages} {445301} (\bibinfo {year} {2017})}\BibitemShut {NoStop}%
\bibitem [{\citenamefont {Reyntjens}\ and\ \citenamefont
  {Puers}(2000)}]{Reyntjens_2000}%
  \BibitemOpen
  \bibfield  {author} {\bibinfo {author} {\bibfnamefont {S.}~\bibnamefont
  {Reyntjens}}\ and\ \bibinfo {author} {\bibfnamefont {R.}~\bibnamefont
  {Puers}},\ }\bibfield  {title} {{\selectlanguage {english}\enquote {\bibinfo
  {title} {Focused ion beam induced deposition: fabrication of
  three-dimensional microstructures and young's modulus of the deposited
  material},}\ }}\href {https://doi.org/10.1088/0960-1317/10/2/314} {\bibfield
  {journal} {\bibinfo  {journal} {J. Micromech. Microeng.}\ }\textbf {\bibinfo
  {volume} {10}},\ \bibinfo {pages} {181--188} (\bibinfo {year}
  {2000})}\BibitemShut {NoStop}%
\bibitem [{\citenamefont {Yu}\ \emph {et~al.}(2021)\citenamefont {Yu},
  \citenamefont {Abdel-Rahman}, \citenamefont {Fairbrother},\ and\
  \citenamefont {McElwee-White}}]{Yu_2021}%
  \BibitemOpen
  \bibfield  {author} {\bibinfo {author} {\bibfnamefont {J.-C.}\ \bibnamefont
  {Yu}}, \bibinfo {author} {\bibfnamefont {M.~K.}\ \bibnamefont
  {Abdel-Rahman}}, \bibinfo {author} {\bibfnamefont {D.~H.}\ \bibnamefont
  {Fairbrother}},\ and\ \bibinfo {author} {\bibfnamefont {L.}~\bibnamefont
  {McElwee-White}},\ }\bibfield  {title} {\enquote {\bibinfo {title} {Charged
  particle-induced surface reactions of organometallic complexes as a guide to
  precursor design for electron- and ion-induced deposition of
  nanostructures},}\ }\href {https://doi.org/10.1021/acsami.1c12327} {\bibfield
   {journal} {\bibinfo  {journal} {ACS Appl. Mater. Interfaces}\ }\textbf
  {\bibinfo {volume} {13}},\ \bibinfo {pages} {48333--48348} (\bibinfo {year}
  {2021})}\BibitemShut {NoStop}%
\bibitem [{\citenamefont {Thorman}\ \emph
  {et~al.}(2020{\natexlab{a}})\citenamefont {Thorman}, \citenamefont {Matsuda},
  \citenamefont {McElwee-White},\ and\ \citenamefont
  {Fairbrother}}]{Thorman_2020a}%
  \BibitemOpen
  \bibfield  {author} {\bibinfo {author} {\bibfnamefont {R.~M.}\ \bibnamefont
  {Thorman}}, \bibinfo {author} {\bibfnamefont {S.~J.}\ \bibnamefont
  {Matsuda}}, \bibinfo {author} {\bibfnamefont {L.}~\bibnamefont
  {McElwee-White}},\ and\ \bibinfo {author} {\bibfnamefont {D.~H.}\
  \bibnamefont {Fairbrother}},\ }\bibfield  {title} {\enquote {\bibinfo {title}
  {Identifying and rationalizing the differing surface reactions of low-energy
  electrons and ions with an organometallic precursor},}\ }\href
  {https://doi.org/10.1021/acs.jpclett.0c00061} {\bibfield  {journal} {\bibinfo
   {journal} {J. Phys. Chem. Lett.}\ }\textbf {\bibinfo {volume} {11}},\
  \bibinfo {pages} {2006--2013} (\bibinfo {year}
  {2020}{\natexlab{a}})}\BibitemShut {NoStop}%
\bibitem [{\citenamefont {Carden}\ \emph {et~al.}(2018)\citenamefont {Carden},
  \citenamefont {Lu}, \citenamefont {Spencer}, \citenamefont {Fairbrother},\
  and\ \citenamefont {McElwee-White}}]{Carden_2018}%
  \BibitemOpen
  \bibfield  {author} {\bibinfo {author} {\bibfnamefont {W.~G.}\ \bibnamefont
  {Carden}}, \bibinfo {author} {\bibfnamefont {H.}~\bibnamefont {Lu}}, \bibinfo
  {author} {\bibfnamefont {J.~A.}\ \bibnamefont {Spencer}}, \bibinfo {author}
  {\bibfnamefont {D.~H.}\ \bibnamefont {Fairbrother}},\ and\ \bibinfo {author}
  {\bibfnamefont {L.}~\bibnamefont {McElwee-White}},\ }\bibfield  {title}
  {\enquote {\bibinfo {title} {Mechanism-based design of precursors for focused
  electron beam-induced deposition},}\ }\href
  {https://doi.org/10.1557/mrc.2018.77} {\bibfield  {journal} {\bibinfo
  {journal} {{MRS} Communications}\ }\textbf {\bibinfo {volume} {8}},\ \bibinfo
  {pages} {343--357} (\bibinfo {year} {2018})}\BibitemShut {NoStop}%
\bibitem [{\citenamefont {Indrajith}\ \emph {et~al.}(2019)\citenamefont
  {Indrajith}, \citenamefont {Rousseau}, \citenamefont {Huber}, \citenamefont
  {Nicolafrancesco}, \citenamefont {Domaracka}, \citenamefont {Grygoryeva},
  \citenamefont {Nag}, \citenamefont {Sedmidubsk{\'{a}}}, \citenamefont
  {Fedor},\ and\ \citenamefont {Ko{\v{c}}i{\v{s}}ek}}]{Indrajith_2019}%
  \BibitemOpen
  \bibfield  {author} {\bibinfo {author} {\bibfnamefont {S.}~\bibnamefont
  {Indrajith}}, \bibinfo {author} {\bibfnamefont {P.}~\bibnamefont {Rousseau}},
  \bibinfo {author} {\bibfnamefont {B.~A.}\ \bibnamefont {Huber}}, \bibinfo
  {author} {\bibfnamefont {C.}~\bibnamefont {Nicolafrancesco}}, \bibinfo
  {author} {\bibfnamefont {A.}~\bibnamefont {Domaracka}}, \bibinfo {author}
  {\bibfnamefont {K.}~\bibnamefont {Grygoryeva}}, \bibinfo {author}
  {\bibfnamefont {P.}~\bibnamefont {Nag}}, \bibinfo {author} {\bibfnamefont
  {B.}~\bibnamefont {Sedmidubsk{\'{a}}}}, \bibinfo {author} {\bibfnamefont
  {J.}~\bibnamefont {Fedor}},\ and\ \bibinfo {author} {\bibfnamefont
  {J.}~\bibnamefont {Ko{\v{c}}i{\v{s}}ek}},\ }\bibfield  {title} {\enquote
  {\bibinfo {title} {Decomposition of iron pentacarbonyl induced by singly and
  multiply charged ions and implications for focused ion beam-induced
  deposition},}\ }\href {https://doi.org/10.1021/acs.jpcc.9b00289} {\bibfield
  {journal} {\bibinfo  {journal} {J. Phys. Chem. C}\ }\textbf {\bibinfo
  {volume} {123}},\ \bibinfo {pages} {10639--10645} (\bibinfo {year}
  {2019})}\BibitemShut {NoStop}%
\bibitem [{\citenamefont {Thorman}\ \emph
  {et~al.}(2020{\natexlab{b}})\citenamefont {Thorman}, \citenamefont {Jensen},
  \citenamefont {Yu}, \citenamefont {Matsuda}, \citenamefont {McElwee-White},
  \citenamefont {Ing{\'{o}}lfsson},\ and\ \citenamefont
  {Fairbrother}}]{Thorman_2020}%
  \BibitemOpen
  \bibfield  {author} {\bibinfo {author} {\bibfnamefont {R.~M.}\ \bibnamefont
  {Thorman}}, \bibinfo {author} {\bibfnamefont {P.~A.}\ \bibnamefont {Jensen}},
  \bibinfo {author} {\bibfnamefont {J.-C.}\ \bibnamefont {Yu}}, \bibinfo
  {author} {\bibfnamefont {S.~J.}\ \bibnamefont {Matsuda}}, \bibinfo {author}
  {\bibfnamefont {L.}~\bibnamefont {McElwee-White}}, \bibinfo {author}
  {\bibfnamefont {O.}~\bibnamefont {Ing{\'{o}}lfsson}},\ and\ \bibinfo {author}
  {\bibfnamefont {D.~H.}\ \bibnamefont {Fairbrother}},\ }\bibfield  {title}
  {\enquote {\bibinfo {title} {Electron-induced reactions of
  {R}u({CO})$_4${I}$_2$: gas phase, surface, and electron beam-induced
  deposition},}\ }\href {https://doi.org/10.1021/acs.jpcc.0c01801} {\bibfield
  {journal} {\bibinfo  {journal} {J. Phys. Chem. C}\ }\textbf {\bibinfo
  {volume} {124}},\ \bibinfo {pages} {10593--10604} (\bibinfo {year}
  {2020}{\natexlab{b}})}\BibitemShut {NoStop}%
\bibitem [{\citenamefont {Ro}, \citenamefont {Thompson},\ and\ \citenamefont
  {Melngailis}(1994)}]{Ro_1994}%
  \BibitemOpen
  \bibfield  {author} {\bibinfo {author} {\bibfnamefont {J.~S.}\ \bibnamefont
  {Ro}}, \bibinfo {author} {\bibfnamefont {C.~V.}\ \bibnamefont {Thompson}},\
  and\ \bibinfo {author} {\bibfnamefont {J.}~\bibnamefont {Melngailis}},\
  }\bibfield  {title} {\enquote {\bibinfo {title} {Mechanism of ion beam
  induced deposition of gold},}\ }\href {https://doi.org/10.1116/1.587111}
  {\bibfield  {journal} {\bibinfo  {journal} {J. Vac. Sci. Technol. B}\
  }\textbf {\bibinfo {volume} {12}},\ \bibinfo {pages} {73} (\bibinfo {year}
  {1994})}\BibitemShut {NoStop}%
\bibitem [{\citenamefont {Joy}(2013)}]{Joy_2013}%
  \BibitemOpen
  \bibfield  {author} {\bibinfo {author} {\bibfnamefont {D.~C.}\ \bibnamefont
  {Joy}},\ }\href {https://doi.org/10.1007/978-1-4614-8660-2} {\emph {\bibinfo
  {title} {{Helium Ion Microscopy}}}},\ SpringerBriefs in Materials\ (\bibinfo
  {publisher} {Springer New York},\ \bibinfo {address} {New York, NY},\
  \bibinfo {year} {2013})\ p.~\bibinfo {pages} {64}\BibitemShut {NoStop}%
\bibitem [{\citenamefont {Lapicki}\ and\ \citenamefont
  {Suzuki}(2002)}]{Lapicki_2002a}%
  \BibitemOpen
  \bibfield  {author} {\bibinfo {author} {\bibfnamefont {A.}~\bibnamefont
  {Lapicki}}\ and\ \bibinfo {author} {\bibfnamefont {T.}~\bibnamefont
  {Suzuki}},\ }\bibfield  {title} {\enquote {\bibinfo {title} {C dot arrays by
  ion beam induced chemical vapor deposition ({IBICVD})},}\ }in\ \href
  {https://doi.org/10.1109/INTMAG.2002.1001262} {\emph {\bibinfo {booktitle}
  {2002 IEEE International Magnetics Conference (INTERMAG)}}},\ Vol.~\bibinfo
  {volume} {38}\ (\bibinfo  {publisher} {{IEEE}},\ \bibinfo {year} {2002})\
  pp.\ \bibinfo {pages} {377--377}\BibitemShut {NoStop}%
\bibitem [{\citenamefont {Utke}\ \emph {et~al.}(2020)\citenamefont {Utke},
  \citenamefont {Michler}, \citenamefont {Winkler},\ and\ \citenamefont
  {Plank}}]{Utke_2020}%
  \BibitemOpen
  \bibfield  {author} {\bibinfo {author} {\bibfnamefont {I.}~\bibnamefont
  {Utke}}, \bibinfo {author} {\bibfnamefont {J.}~\bibnamefont {Michler}},
  \bibinfo {author} {\bibfnamefont {R.}~\bibnamefont {Winkler}},\ and\ \bibinfo
  {author} {\bibfnamefont {H.}~\bibnamefont {Plank}},\ }\bibfield  {title}
  {\enquote {\bibinfo {title} {Mechanical properties of 3{D} nanostructures
  obtained by focused electron/ion beam-induced deposition: a review},}\ }\href
  {https://doi.org/10.3390/mi11040397} {\bibfield  {journal} {\bibinfo
  {journal} {Micromachines}\ }\textbf {\bibinfo {volume} {11}},\ \bibinfo
  {pages} {397} (\bibinfo {year} {2020})}\BibitemShut {NoStop}%
\bibitem [{\citenamefont {Plank}\ \emph {et~al.}(2019)\citenamefont {Plank},
  \citenamefont {Winkler}, \citenamefont {Schwalb}, \citenamefont
  {H{\"{u}}tner}, \citenamefont {Fowlkes}, \citenamefont {Rack}, \citenamefont
  {Utke},\ and\ \citenamefont {Huth}}]{Plank_2019}%
  \BibitemOpen
  \bibfield  {author} {\bibinfo {author} {\bibfnamefont {H.}~\bibnamefont
  {Plank}}, \bibinfo {author} {\bibfnamefont {R.}~\bibnamefont {Winkler}},
  \bibinfo {author} {\bibfnamefont {C.~H.}\ \bibnamefont {Schwalb}}, \bibinfo
  {author} {\bibfnamefont {J.}~\bibnamefont {H{\"{u}}tner}}, \bibinfo {author}
  {\bibfnamefont {J.~D.}\ \bibnamefont {Fowlkes}}, \bibinfo {author}
  {\bibfnamefont {P.~D.}\ \bibnamefont {Rack}}, \bibinfo {author}
  {\bibfnamefont {I.}~\bibnamefont {Utke}},\ and\ \bibinfo {author}
  {\bibfnamefont {M.}~\bibnamefont {Huth}},\ }\bibfield  {title} {\enquote
  {\bibinfo {title} {Focused electron beam-based 3{D} nanoprinting for scanning
  probe microscopy: a review},}\ }\href {https://doi.org/10.3390/mi11010048}
  {\bibfield  {journal} {\bibinfo  {journal} {Micromachines}\ }\textbf
  {\bibinfo {volume} {11}},\ \bibinfo {pages} {48} (\bibinfo {year}
  {2019})}\BibitemShut {NoStop}%
\bibitem [{\citenamefont {Dukic}\ \emph {et~al.}(2016)\citenamefont {Dukic},
  \citenamefont {Winhold}, \citenamefont {Schwalb}, \citenamefont {Adams},
  \citenamefont {Stavrov}, \citenamefont {Huth},\ and\ \citenamefont
  {Fantner}}]{Dukic_2016}%
  \BibitemOpen
  \bibfield  {author} {\bibinfo {author} {\bibfnamefont {M.}~\bibnamefont
  {Dukic}}, \bibinfo {author} {\bibfnamefont {M.}~\bibnamefont {Winhold}},
  \bibinfo {author} {\bibfnamefont {C.~H.}\ \bibnamefont {Schwalb}}, \bibinfo
  {author} {\bibfnamefont {J.~D.}\ \bibnamefont {Adams}}, \bibinfo {author}
  {\bibfnamefont {V.}~\bibnamefont {Stavrov}}, \bibinfo {author} {\bibfnamefont
  {M.}~\bibnamefont {Huth}},\ and\ \bibinfo {author} {\bibfnamefont {G.~E.}\
  \bibnamefont {Fantner}},\ }\bibfield  {title} {\enquote {\bibinfo {title}
  {Direct-write nanoscale printing of nanogranular tunnelling strain sensors
  for sub-micrometre cantilevers},}\ }\href
  {https://doi.org/10.1038/ncomms12487} {\bibfield  {journal} {\bibinfo
  {journal} {Nat. Commun.}\ }\textbf {\bibinfo {volume} {7}},\ \bibinfo {pages}
  {12487} (\bibinfo {year} {2016})}\BibitemShut {NoStop}%
\bibitem [{\citenamefont {Matsui}(2007)}]{Matsui_2007}%
  \BibitemOpen
  \bibfield  {author} {\bibinfo {author} {\bibfnamefont {S.}~\bibnamefont
  {Matsui}},\ }\bibfield  {title} {\enquote {\bibinfo {title} {Focused-ion-beam
  deposition for 3-d nanostructure fabrication},}\ }\href
  {https://doi.org/10.1016/j.nimb.2007.01.077} {\bibfield  {journal} {\bibinfo
  {journal} {Nucl. Instrum. Meth. Phys. Res. B}\ }\textbf {\bibinfo {volume}
  {257}},\ \bibinfo {pages} {758--764} (\bibinfo {year} {2007})}\BibitemShut
  {NoStop}%
\bibitem [{\citenamefont {De~Ridder}, \citenamefont {Hopman},\ and\
  \citenamefont {Ay}(2007)}]{DeRidder_2007}%
  \BibitemOpen
  \bibfield  {author} {\bibinfo {author} {\bibfnamefont {R.~M.}\ \bibnamefont
  {De~Ridder}}, \bibinfo {author} {\bibfnamefont {W.~C.~L.}\ \bibnamefont
  {Hopman}},\ and\ \bibinfo {author} {\bibfnamefont {F.}~\bibnamefont {Ay}},\
  }\bibfield  {title} {\enquote {\bibinfo {title} {Focused-ion-beam processing
  for photonics},}\ }in\ \href {https://doi.org/10.1109/ICTON.2007.4296183}
  {\emph {\bibinfo {booktitle} {Proceedings of 2007 9th International
  Conference on Transparent Optical Networks, ICTON 2007}}},\ Vol.~\bibinfo
  {volume} {2}\ (\bibinfo {year} {2007})\ pp.\ \bibinfo {pages}
  {212--215}\BibitemShut {NoStop}%
\bibitem [{\citenamefont {Peinado}, \citenamefont {Sangiao},\ and\
  \citenamefont {De~Teresa}(2015)}]{Peinado_2015}%
  \BibitemOpen
  \bibfield  {author} {\bibinfo {author} {\bibfnamefont {P.}~\bibnamefont
  {Peinado}}, \bibinfo {author} {\bibfnamefont {S.}~\bibnamefont {Sangiao}},\
  and\ \bibinfo {author} {\bibfnamefont {J.~M.}\ \bibnamefont {De~Teresa}},\
  }\bibfield  {title} {\enquote {\bibinfo {title} {Focused electron and ion
  beam induced deposition on flexible and transparent polycarbonate
  substrates},}\ }\href {https://doi.org/10.1021/acsnano.5b01383} {\bibfield
  {journal} {\bibinfo  {journal} {ACS Nano}\ }\textbf {\bibinfo {volume} {9}},\
  \bibinfo {pages} {6139--6146} (\bibinfo {year} {2015})}\BibitemShut {NoStop}%
\bibitem [{\citenamefont {Ro}, \citenamefont {Thompson},\ and\ \citenamefont
  {Melngailis}(1995)}]{Ro_1995}%
  \BibitemOpen
  \bibfield  {author} {\bibinfo {author} {\bibfnamefont {J.~S.}\ \bibnamefont
  {Ro}}, \bibinfo {author} {\bibfnamefont {C.~V.}\ \bibnamefont {Thompson}},\
  and\ \bibinfo {author} {\bibfnamefont {J.}~\bibnamefont {Melngailis}},\
  }\bibfield  {title} {{\selectlanguage {english}\enquote {\bibinfo {title}
  {Microstructure of gold grown by ion-induced deposition},}\ }}\href
  {https://doi.org/10.1016/0040-6090(94)06399-0} {\bibfield  {journal}
  {\bibinfo  {journal} {Thin Solid Films}\ }\textbf {\bibinfo {volume} {258}},\
  \bibinfo {pages} {333--335} (\bibinfo {year} {1995})}\BibitemShut {NoStop}%
\bibitem [{\citenamefont {Matsui}(2012)}]{Matsui_2012}%
  \BibitemOpen
  \bibfield  {author} {\bibinfo {author} {\bibfnamefont {S.}~\bibnamefont
  {Matsui}},\ }\bibfield  {title} {\enquote {\bibinfo {title} {Focused-ion-beam
  chemical-vapor-deposition ({FIB}-{CVD})},}\ \ }(\bibinfo  {publisher}
  {Springer Netherlands},\ \bibinfo {address} {Dordrecht},\ \bibinfo {year}
  {2012})\ pp.\ \bibinfo {pages} {866--876}\BibitemShut {NoStop}%
\bibitem [{\citenamefont {Perez-Roldan}, \citenamefont {Mulders},\ and\
  \citenamefont {Trompenaars}(2017)}]{PerezRoldan_2017}%
  \BibitemOpen
  \bibfield  {author} {\bibinfo {author} {\bibfnamefont {M.~J.}\ \bibnamefont
  {Perez-Roldan}}, \bibinfo {author} {\bibfnamefont {J.~J.~L.}\ \bibnamefont
  {Mulders}},\ and\ \bibinfo {author} {\bibfnamefont {P.~H.~F.}\ \bibnamefont
  {Trompenaars}},\ }\bibfield  {title} {\enquote {\bibinfo {title}
  {Oxygen-assisted purification of platinum structures deposited by ion and
  electron beam induced processes},}\ }\href
  {https://doi.org/10.1088/1361-6463/aa69e2} {\bibfield  {journal} {\bibinfo
  {journal} {J. Phys. D: Appl. Phys.}\ }\textbf {\bibinfo {volume} {50}},\
  \bibinfo {pages} {205307} (\bibinfo {year} {2017})}\BibitemShut {NoStop}%
\bibitem [{\citenamefont {Shawrav}\ \emph {et~al.}(2016)\citenamefont
  {Shawrav}, \citenamefont {Taus}, \citenamefont {Wanzenboeck}, \citenamefont
  {Schinnerl}, \citenamefont {St{\"{o}}ger-Pollach}, \citenamefont {Schwarz},
  \citenamefont {Steiger-Thirsfeld},\ and\ \citenamefont
  {Bertagnolli}}]{Shawrav_2016}%
  \BibitemOpen
  \bibfield  {author} {\bibinfo {author} {\bibfnamefont {M.~M.}\ \bibnamefont
  {Shawrav}}, \bibinfo {author} {\bibfnamefont {P.}~\bibnamefont {Taus}},
  \bibinfo {author} {\bibfnamefont {H.~D.}\ \bibnamefont {Wanzenboeck}},
  \bibinfo {author} {\bibfnamefont {M.}~\bibnamefont {Schinnerl}}, \bibinfo
  {author} {\bibfnamefont {M.}~\bibnamefont {St{\"{o}}ger-Pollach}}, \bibinfo
  {author} {\bibfnamefont {S.}~\bibnamefont {Schwarz}}, \bibinfo {author}
  {\bibfnamefont {A.}~\bibnamefont {Steiger-Thirsfeld}},\ and\ \bibinfo
  {author} {\bibfnamefont {E.}~\bibnamefont {Bertagnolli}},\ }\bibfield
  {title} {\enquote {\bibinfo {title} {Highly conductive and pure gold
  nanostructures grown by electron beam induced deposition},}\ }\href
  {https://doi.org/10.1038/srep34003} {\bibfield  {journal} {\bibinfo
  {journal} {Sci. Rep.}\ }\textbf {\bibinfo {volume} {6}},\ \bibinfo {pages}
  {1--10} (\bibinfo {year} {2016})}\BibitemShut {NoStop}%
\bibitem [{\citenamefont {Botman}, \citenamefont {Mulders},\ and\ \citenamefont
  {Hagen}(2009)}]{Botman_2009}%
  \BibitemOpen
  \bibfield  {author} {\bibinfo {author} {\bibfnamefont {A.}~\bibnamefont
  {Botman}}, \bibinfo {author} {\bibfnamefont {J.~J.~L.}\ \bibnamefont
  {Mulders}},\ and\ \bibinfo {author} {\bibfnamefont {C.~W.}\ \bibnamefont
  {Hagen}},\ }\bibfield  {title} {\enquote {\bibinfo {title} {Creating pure
  nanostructures from electron-beam-induced deposition using purification
  techniques: a technology perspective},}\ }\href
  {https://doi.org/10.1088/0957-4484/20/37/372001} {\bibfield  {journal}
  {\bibinfo  {journal} {Nanotechnology}\ }\textbf {\bibinfo {volume} {20}},\
  \bibinfo {pages} {372001} (\bibinfo {year} {2009})}\BibitemShut {NoStop}%
\bibitem [{\citenamefont {H{\"{o}}flich}\ \emph {et~al.}(2011)\citenamefont
  {H{\"{o}}flich}, \citenamefont {Yang}, \citenamefont {Berger}, \citenamefont
  {Leuchs},\ and\ \citenamefont {Christiansen}}]{Hoeflich_2011}%
  \BibitemOpen
  \bibfield  {author} {\bibinfo {author} {\bibfnamefont {K.}~\bibnamefont
  {H{\"{o}}flich}}, \bibinfo {author} {\bibfnamefont {B.~R.}\ \bibnamefont
  {Yang}}, \bibinfo {author} {\bibfnamefont {A.}~\bibnamefont {Berger}},
  \bibinfo {author} {\bibfnamefont {G.}~\bibnamefont {Leuchs}},\ and\ \bibinfo
  {author} {\bibfnamefont {S.}~\bibnamefont {Christiansen}},\ }\bibfield
  {title} {\enquote {\bibinfo {title} {The direct writing of plasmonic gold
  nanostructures by electron-beam- induced deposition},}\ }\href
  {https://doi.org/10.1002/adma.201004114} {\bibfield  {journal} {\bibinfo
  {journal} {Adv. Mater.}\ }\textbf {\bibinfo {volume} {23}},\ \bibinfo {pages}
  {2657--2661} (\bibinfo {year} {2011})}\BibitemShut {NoStop}%
\bibitem [{\citenamefont {Cremers}, \citenamefont {Puurunen},\ and\
  \citenamefont {Dendooven}(2019)}]{Cremers_2019}%
  \BibitemOpen
  \bibfield  {author} {\bibinfo {author} {\bibfnamefont {V.}~\bibnamefont
  {Cremers}}, \bibinfo {author} {\bibfnamefont {R.~L.}\ \bibnamefont
  {Puurunen}},\ and\ \bibinfo {author} {\bibfnamefont {J.}~\bibnamefont
  {Dendooven}},\ }\bibfield  {title} {\enquote {\bibinfo {title} {Conformality
  in atomic layer deposition: current status overview of analysis and
  modelling},}\ }\href {https://doi.org/10.1063/1.5060967} {\bibfield
  {journal} {\bibinfo  {journal} {Appl. Phys. Rev.}\ }\textbf {\bibinfo
  {volume} {6}},\ \bibinfo {pages} {21302} (\bibinfo {year}
  {2019})}\BibitemShut {NoStop}%
\bibitem [{\citenamefont {Cuscun{\`{a}}}\ \emph {et~al.}(2021)\citenamefont
  {Cuscun{\`{a}}}, \citenamefont {Manoccio}, \citenamefont {Esposito},
  \citenamefont {Scuderi}, \citenamefont {Nicotra}, \citenamefont {Tarantini},
  \citenamefont {Melcarne}, \citenamefont {Tasco}, \citenamefont {Losurdo},\
  and\ \citenamefont {Passaseo}}]{Cuscuna_2021}%
  \BibitemOpen
  \bibfield  {author} {\bibinfo {author} {\bibfnamefont {M.}~\bibnamefont
  {Cuscun{\`{a}}}}, \bibinfo {author} {\bibfnamefont {M.}~\bibnamefont
  {Manoccio}}, \bibinfo {author} {\bibfnamefont {M.}~\bibnamefont {Esposito}},
  \bibinfo {author} {\bibfnamefont {M.}~\bibnamefont {Scuderi}}, \bibinfo
  {author} {\bibfnamefont {G.}~\bibnamefont {Nicotra}}, \bibinfo {author}
  {\bibfnamefont {I.}~\bibnamefont {Tarantini}}, \bibinfo {author}
  {\bibfnamefont {A.}~\bibnamefont {Melcarne}}, \bibinfo {author}
  {\bibfnamefont {V.}~\bibnamefont {Tasco}}, \bibinfo {author} {\bibfnamefont
  {M.}~\bibnamefont {Losurdo}},\ and\ \bibinfo {author} {\bibfnamefont
  {A.}~\bibnamefont {Passaseo}},\ }\bibfield  {title} {\enquote {\bibinfo
  {title} {Gallium chiral nanoshaping for circular polarization handling},}\
  }\href {https://doi.org/10.1039/D0MH01078B} {\bibfield  {journal} {\bibinfo
  {journal} {Mater. Horiz.}\ }\textbf {\bibinfo {volume} {8}},\ \bibinfo
  {pages} {187--196} (\bibinfo {year} {2021})}\BibitemShut {NoStop}%
\bibitem [{\citenamefont {Kometani}, \citenamefont {Warisawa},\ and\
  \citenamefont {Ishihara}(2010)}]{Kometani_2010}%
  \BibitemOpen
  \bibfield  {author} {\bibinfo {author} {\bibfnamefont {R.}~\bibnamefont
  {Kometani}}, \bibinfo {author} {\bibfnamefont {S.}~\bibnamefont {Warisawa}},\
  and\ \bibinfo {author} {\bibfnamefont {S.}~\bibnamefont {Ishihara}},\
  }\bibfield  {title} {\enquote {\bibinfo {title} {Non-core--shell
  nanostructure deposition on focused-ion-beam chemical vapor deposition},}\
  }\href {https://doi.org/10.1143/jjap.49.06ge03} {\bibfield  {journal}
  {\bibinfo  {journal} {Jpn. J. Appl. Phys.}\ }\textbf {\bibinfo {volume}
  {49}},\ \bibinfo {pages} {06GE03} (\bibinfo {year} {2010})}\BibitemShut
  {NoStop}%
\bibitem [{\citenamefont {Hentschel}, \citenamefont {Karst},\ and\
  \citenamefont {Giessen}(2020)}]{Hentschel_2020}%
  \BibitemOpen
  \bibfield  {author} {\bibinfo {author} {\bibfnamefont {M.}~\bibnamefont
  {Hentschel}}, \bibinfo {author} {\bibfnamefont {J.}~\bibnamefont {Karst}},\
  and\ \bibinfo {author} {\bibfnamefont {H.}~\bibnamefont {Giessen}},\
  }\bibfield  {title} {\enquote {\bibinfo {title} {Tailored optical
  functionality by combining electron-beam and focused gold-ion beam
  lithography for solid and inverse coupled plasmonic nanostructures},}\ }\href
  {https://doi.org/10.1002/adom.202000879} {\bibfield  {journal} {\bibinfo
  {journal} {Adv. Opt. Mater.}\ }\textbf {\bibinfo {volume} {8}},\ \bibinfo
  {pages} {2000879} (\bibinfo {year} {2020})}\BibitemShut {NoStop}%
\bibitem [{\citenamefont {De~Teresa~Nogueras}(2020)}]{DeTeresaNogueras_2020}%
  \BibitemOpen
  \bibfield  {author} {\bibinfo {author} {\bibfnamefont {J.~M.}\ \bibnamefont
  {De~Teresa~Nogueras}},\ }\href {https://doi.org/10.1088/978-0-7503-2608-7}
  {\emph {\bibinfo {title} {Nanofabrication}}}\ (\bibinfo  {publisher} {{IOP}
  Publishing Ltd},\ \bibinfo {year} {2020})\BibitemShut {NoStop}%
\bibitem [{\citenamefont {Atkinson}(1992)}]{Atkinson_1992a}%
  \BibitemOpen
  \bibfield  {author} {\bibinfo {author} {\bibfnamefont {G.~M.}\ \bibnamefont
  {Atkinson}},\ }\bibfield  {title} {\enquote {\bibinfo {title} {30 nm
  resolution zero proximity lithography on high-z substrates},}\ }\href
  {https://doi.org/10.1116/1.585937} {\bibfield  {journal} {\bibinfo  {journal}
  {J. Vac. Sci. Technol. B}\ }\textbf {\bibinfo {volume} {10}},\ \bibinfo
  {pages} {3104} (\bibinfo {year} {1992})}\BibitemShut {NoStop}%
\bibitem [{\citenamefont {Hartney}(1991)}]{Hartney_1991}%
  \BibitemOpen
  \bibfield  {author} {\bibinfo {author} {\bibfnamefont {M.~A.}\ \bibnamefont
  {Hartney}},\ }\bibfield  {title} {\enquote {\bibinfo {title} {Surface imaging
  of focused ion-beam exposed resists},}\ }\href
  {https://doi.org/10.1116/1.585817} {\bibfield  {journal} {\bibinfo  {journal}
  {J. Vac. Sci. Technol. B}\ }\textbf {\bibinfo {volume} {9}},\ \bibinfo
  {pages} {3432} (\bibinfo {year} {1991})}\BibitemShut {NoStop}%
\bibitem [{\citenamefont {Stratton}(1998)}]{Stratton_1998}%
  \BibitemOpen
  \bibfield  {author} {\bibinfo {author} {\bibfnamefont {F.~P.}\ \bibnamefont
  {Stratton}},\ }\bibfield  {title} {\enquote {\bibinfo {title}
  {Microelectromechanical tunneling sensor fabrication and post-processing
  characterization using focused ion beams},}\ }\href
  {https://doi.org/10.1116/1.590189} {\bibfield  {journal} {\bibinfo  {journal}
  {J. Vac. Sci. Technol. B}\ }\textbf {\bibinfo {volume} {16}},\ \bibinfo
  {pages} {2449} (\bibinfo {year} {1998})}\BibitemShut {NoStop}%
\bibitem [{\citenamefont {Kiran}\ \emph {et~al.}(2020)\citenamefont {Kiran},
  \citenamefont {Chauhan}, \citenamefont {Sharma}, \citenamefont {Ghosh},\ and\
  \citenamefont {Gonsalves}}]{Kiran_2020}%
  \BibitemOpen
  \bibfield  {author} {\bibinfo {author} {\bibfnamefont {N.~R.}\ \bibnamefont
  {Kiran}}, \bibinfo {author} {\bibfnamefont {M.}~\bibnamefont {Chauhan}},
  \bibinfo {author} {\bibfnamefont {S.~K.}\ \bibnamefont {Sharma}}, \bibinfo
  {author} {\bibfnamefont {S.}~\bibnamefont {Ghosh}},\ and\ \bibinfo {author}
  {\bibfnamefont {K.~E.}\ \bibnamefont {Gonsalves}},\ }\bibfield  {title}
  {\enquote {\bibinfo {title} {Resists for helium ion beam lithography: recent
  advances},}\ }\href {https://doi.org/10.1021/acsaelm.0c00627} {\bibfield
  {journal} {\bibinfo  {journal} {{ACS} Applied Electronic Materials}\ }\textbf
  {\bibinfo {volume} {2}},\ \bibinfo {pages} {3805--3817} (\bibinfo {year}
  {2020})}\BibitemShut {NoStop}%
\bibitem [{\citenamefont {Winston}\ \emph {et~al.}(2009)\citenamefont
  {Winston}, \citenamefont {Cord}, \citenamefont {Ming}, \citenamefont {Bell},
  \citenamefont {DiNatale}, \citenamefont {Stern}, \citenamefont {Vladar},
  \citenamefont {Postek}, \citenamefont {Mondol}, \citenamefont {Yang},\ and\
  \citenamefont {Berggren}}]{Winston_2009a}%
  \BibitemOpen
  \bibfield  {author} {\bibinfo {author} {\bibfnamefont {D.}~\bibnamefont
  {Winston}}, \bibinfo {author} {\bibfnamefont {B.~M.}\ \bibnamefont {Cord}},
  \bibinfo {author} {\bibfnamefont {B.}~\bibnamefont {Ming}}, \bibinfo {author}
  {\bibfnamefont {D.~C.}\ \bibnamefont {Bell}}, \bibinfo {author}
  {\bibfnamefont {W.~F.}\ \bibnamefont {DiNatale}}, \bibinfo {author}
  {\bibfnamefont {L.~A.}\ \bibnamefont {Stern}}, \bibinfo {author}
  {\bibfnamefont {A.~E.}\ \bibnamefont {Vladar}}, \bibinfo {author}
  {\bibfnamefont {M.~T.}\ \bibnamefont {Postek}}, \bibinfo {author}
  {\bibfnamefont {M.~K.}\ \bibnamefont {Mondol}}, \bibinfo {author}
  {\bibfnamefont {J.~K.~W.}\ \bibnamefont {Yang}},\ and\ \bibinfo {author}
  {\bibfnamefont {K.~K.}\ \bibnamefont {Berggren}},\ }\bibfield  {title}
  {\enquote {\bibinfo {title} {Scanning-helium-ion-beam lithography with
  hydrogen silsesquioxane resist},}\ }\href {https://doi.org/10.1116/1.3250204}
  {\bibfield  {journal} {\bibinfo  {journal} {J. Vac. Sci. Technol. B}\
  }\textbf {\bibinfo {volume} {27}},\ \bibinfo {pages} {2702} (\bibinfo {year}
  {2009})}\BibitemShut {NoStop}%
\bibitem [{\citenamefont {Sidorkin}\ \emph {et~al.}(2009)\citenamefont
  {Sidorkin}, \citenamefont {van Veldhoven}, \citenamefont {van~der Drift},
  \citenamefont {Alkemade}, \citenamefont {Salemink},\ and\ \citenamefont
  {Maas}}]{Sidorkin_2009b}%
  \BibitemOpen
  \bibfield  {author} {\bibinfo {author} {\bibfnamefont {V.}~\bibnamefont
  {Sidorkin}}, \bibinfo {author} {\bibfnamefont {E.}~\bibnamefont {van
  Veldhoven}}, \bibinfo {author} {\bibfnamefont {E.}~\bibnamefont {van~der
  Drift}}, \bibinfo {author} {\bibfnamefont {P.}~\bibnamefont {Alkemade}},
  \bibinfo {author} {\bibfnamefont {H.}~\bibnamefont {Salemink}},\ and\
  \bibinfo {author} {\bibfnamefont {D.}~\bibnamefont {Maas}},\ }\bibfield
  {title} {\enquote {\bibinfo {title} {Sub-10-nm nanolithography with a
  scanning helium beam},}\ }\href {https://doi.org/10.1116/1.3182742}
  {\bibfield  {journal} {\bibinfo  {journal} {J. Vac. Sci. Technol. B}\
  }\textbf {\bibinfo {volume} {27}},\ \bibinfo {pages} {L18--L20} (\bibinfo
  {year} {2009})}\BibitemShut {NoStop}%
\bibitem [{\citenamefont {Chang}\ \emph {et~al.}(2021)\citenamefont {Chang},
  \citenamefont {Wang}, \citenamefont {Wang}, \citenamefont {Huang},
  \citenamefont {Jiang}, \citenamefont {Chou}, \citenamefont {Kao},\ and\
  \citenamefont {Chen}}]{Chang_2021a}%
  \BibitemOpen
  \bibfield  {author} {\bibinfo {author} {\bibfnamefont {T.-J.}\ \bibnamefont
  {Chang}}, \bibinfo {author} {\bibfnamefont {T.-Y.}\ \bibnamefont {Wang}},
  \bibinfo {author} {\bibfnamefont {C.-I.}\ \bibnamefont {Wang}}, \bibinfo
  {author} {\bibfnamefont {Z.-d.}\ \bibnamefont {Huang}}, \bibinfo {author}
  {\bibfnamefont {Y.-S.}\ \bibnamefont {Jiang}}, \bibinfo {author}
  {\bibfnamefont {C.-Y.}\ \bibnamefont {Chou}}, \bibinfo {author}
  {\bibfnamefont {W.-C.}\ \bibnamefont {Kao}},\ and\ \bibinfo {author}
  {\bibfnamefont {M.-J.}\ \bibnamefont {Chen}},\ }\bibfield  {title} {\enquote
  {\bibinfo {title} {Suppression of short channel effects in ferroelectric si
  junctionless transistors with a sub-10 nm gate length defined by helium ion
  beam lithography},}\ }\href {https://doi.org/10.1039/d1tc00431j} {\bibfield
  {journal} {\bibinfo  {journal} {J. Mater. Chem. C}\ }\textbf {\bibinfo
  {volume} {9}},\ \bibinfo {pages} {8285--8293} (\bibinfo {year}
  {2021})}\BibitemShut {NoStop}%
\bibitem [{\citenamefont {Li}, \citenamefont {Wu},\ and\ \citenamefont
  {Stanley~Williams}(2012)}]{Li_2012}%
  \BibitemOpen
  \bibfield  {author} {\bibinfo {author} {\bibfnamefont {W.-D.}\ \bibnamefont
  {Li}}, \bibinfo {author} {\bibfnamefont {W.}~\bibnamefont {Wu}},\ and\
  \bibinfo {author} {\bibfnamefont {R.}~\bibnamefont {Stanley~Williams}},\
  }\bibfield  {title} {\enquote {\bibinfo {title} {Combined helium ion beam and
  nanoimprint lithography attains 4 nm half-pitch dense patterns},}\ }\href
  {https://doi.org/10.1116/1.4758768} {\bibfield  {journal} {\bibinfo
  {journal} {J. Vac. Sci. Technol. B}\ }\textbf {\bibinfo {volume} {30}},\
  \bibinfo {pages} {06F304} (\bibinfo {year} {2012})}\BibitemShut {NoStop}%
\bibitem [{\citenamefont {Cai}\ \emph {et~al.}(2018)\citenamefont {Cai},
  \citenamefont {Zhu}, \citenamefont {Alkemade}, \citenamefont {van Veldhoven},
  \citenamefont {Wang}, \citenamefont {Ge}, \citenamefont {Rodrigues},
  \citenamefont {Cai},\ and\ \citenamefont {Li}}]{Cai_2018a}%
  \BibitemOpen
  \bibfield  {author} {\bibinfo {author} {\bibfnamefont {J.}~\bibnamefont
  {Cai}}, \bibinfo {author} {\bibfnamefont {Z.}~\bibnamefont {Zhu}}, \bibinfo
  {author} {\bibfnamefont {P.~F.~A.}\ \bibnamefont {Alkemade}}, \bibinfo
  {author} {\bibfnamefont {E.}~\bibnamefont {van Veldhoven}}, \bibinfo {author}
  {\bibfnamefont {Q.}~\bibnamefont {Wang}}, \bibinfo {author} {\bibfnamefont
  {H.}~\bibnamefont {Ge}}, \bibinfo {author} {\bibfnamefont {S.~P.}\
  \bibnamefont {Rodrigues}}, \bibinfo {author} {\bibfnamefont {W.}~\bibnamefont
  {Cai}},\ and\ \bibinfo {author} {\bibfnamefont {W.-D.}\ \bibnamefont {Li}},\
  }\bibfield  {title} {\enquote {\bibinfo {title} {3{D} volumetric energy
  deposition of focused helium ion beam lithography: visualization, modeling,
  and applications in nanofabrication},}\ }\href
  {https://doi.org/10.1002/admi.201800203} {\bibfield  {journal} {\bibinfo
  {journal} {Adv. Mater. Interfaces}\ }\textbf {\bibinfo {volume} {5}},\
  \bibinfo {pages} {1800203} (\bibinfo {year} {2018})}\BibitemShut {NoStop}%
\bibitem [{\citenamefont {Flatab{\o}}\ \emph {et~al.}(2018)\citenamefont
  {Flatab{\o}}, \citenamefont {Agarwal}, \citenamefont {Hobbs}, \citenamefont
  {Greve}, \citenamefont {Holst},\ and\ \citenamefont
  {Berggren}}]{Flataboe_2018}%
  \BibitemOpen
  \bibfield  {author} {\bibinfo {author} {\bibfnamefont {R.}~\bibnamefont
  {Flatab{\o}}}, \bibinfo {author} {\bibfnamefont {A.}~\bibnamefont {Agarwal}},
  \bibinfo {author} {\bibfnamefont {R.}~\bibnamefont {Hobbs}}, \bibinfo
  {author} {\bibfnamefont {M.~M.}\ \bibnamefont {Greve}}, \bibinfo {author}
  {\bibfnamefont {B.}~\bibnamefont {Holst}},\ and\ \bibinfo {author}
  {\bibfnamefont {K.~K.}\ \bibnamefont {Berggren}},\ }\bibfield  {title}
  {\enquote {\bibinfo {title} {Exploring proximity effects and large depth of
  field in helium ion beam lithography: large-area dense patterns and tilted
  surface exposure},}\ }\href {https://doi.org/10.1088/1361-6528/aabe22}
  {\bibfield  {journal} {\bibinfo  {journal} {Nanotechnology}\ }\textbf
  {\bibinfo {volume} {29}},\ \bibinfo {pages} {275301} (\bibinfo {year}
  {2018})}\BibitemShut {NoStop}%
\bibitem [{\citenamefont {Winston}\ \emph {et~al.}(2011)\citenamefont
  {Winston}, \citenamefont {Manfrinato}, \citenamefont {Nicaise}, \citenamefont
  {Cheong}, \citenamefont {Duan}, \citenamefont {Ferranti}, \citenamefont
  {Marshman}, \citenamefont {McVey}, \citenamefont {Stern}, \citenamefont
  {Notte},\ and\ \citenamefont {Berggren}}]{Winston_2011a}%
  \BibitemOpen
  \bibfield  {author} {\bibinfo {author} {\bibfnamefont {D.}~\bibnamefont
  {Winston}}, \bibinfo {author} {\bibfnamefont {V.~R.}\ \bibnamefont
  {Manfrinato}}, \bibinfo {author} {\bibfnamefont {S.~M.}\ \bibnamefont
  {Nicaise}}, \bibinfo {author} {\bibfnamefont {L.~L.}\ \bibnamefont {Cheong}},
  \bibinfo {author} {\bibfnamefont {H.}~\bibnamefont {Duan}}, \bibinfo {author}
  {\bibfnamefont {D.}~\bibnamefont {Ferranti}}, \bibinfo {author}
  {\bibfnamefont {J.}~\bibnamefont {Marshman}}, \bibinfo {author}
  {\bibfnamefont {S.}~\bibnamefont {McVey}}, \bibinfo {author} {\bibfnamefont
  {L.}~\bibnamefont {Stern}}, \bibinfo {author} {\bibfnamefont
  {J.}~\bibnamefont {Notte}},\ and\ \bibinfo {author} {\bibfnamefont {K.~K.}\
  \bibnamefont {Berggren}},\ }\bibfield  {title} {\enquote {\bibinfo {title}
  {Neon ion beam lithography ({NIBL})},}\ }\href
  {https://doi.org/10.1021/nl202447n} {\bibfield  {journal} {\bibinfo
  {journal} {Nano Lett.}\ }\textbf {\bibinfo {volume} {11}},\ \bibinfo {pages}
  {4343--4347} (\bibinfo {year} {2011})}\BibitemShut {NoStop}%
\bibitem [{\citenamefont {Shi}\ and\ \citenamefont {Boden}(2016)}]{Shi_2016a}%
  \BibitemOpen
  \bibfield  {author} {\bibinfo {author} {\bibfnamefont {X.}~\bibnamefont
  {Shi}}\ and\ \bibinfo {author} {\bibfnamefont {S.~A.}\ \bibnamefont
  {Boden}},\ }\bibfield  {title} {\enquote {\bibinfo {title} {Scanning helium
  ion beam lithography},}\ }in\ \href
  {https://doi.org/10.1016/b978-0-08-100354-1.00017-x} {\emph {\bibinfo
  {booktitle} {Materials and Processes for Next Generation Lithography}}},\
  \bibinfo {editor} {edited by\ \bibinfo {editor} {\bibfnamefont
  {A.}~\bibnamefont {Robinson}}}\ (\bibinfo  {publisher} {Elsevier},\ \bibinfo
  {year} {2016})\ pp.\ \bibinfo {pages} {563--594}\BibitemShut {NoStop}%
\bibitem [{\citenamefont {Cattoni}\ \emph {et~al.}(2018)\citenamefont
  {Cattoni}, \citenamefont {Mailly}, \citenamefont {Dalstein}, \citenamefont
  {Faustini}, \citenamefont {Seniutinas}, \citenamefont {R{\"{o}}sner},\ and\
  \citenamefont {David}}]{Cattoni_2018a}%
  \BibitemOpen
  \bibfield  {author} {\bibinfo {author} {\bibfnamefont {A.}~\bibnamefont
  {Cattoni}}, \bibinfo {author} {\bibfnamefont {D.}~\bibnamefont {Mailly}},
  \bibinfo {author} {\bibfnamefont {O.}~\bibnamefont {Dalstein}}, \bibinfo
  {author} {\bibfnamefont {M.}~\bibnamefont {Faustini}}, \bibinfo {author}
  {\bibfnamefont {G.}~\bibnamefont {Seniutinas}}, \bibinfo {author}
  {\bibfnamefont {B.}~\bibnamefont {R{\"{o}}sner}},\ and\ \bibinfo {author}
  {\bibfnamefont {C.}~\bibnamefont {David}},\ }\bibfield  {title} {\enquote
  {\bibinfo {title} {Sub-10 nm electron and helium ion beam lithography using a
  recently developed alumina resist},}\ }\href
  {https://doi.org/10.1016/j.mee.2018.02.015} {\bibfield  {journal} {\bibinfo
  {journal} {Microelectron. Eng.}\ }\textbf {\bibinfo {volume} {193}},\
  \bibinfo {pages} {18--22} (\bibinfo {year} {2018})}\BibitemShut {NoStop}%
\bibitem [{\citenamefont {Lewis}\ \emph {et~al.}(2019)\citenamefont {Lewis},
  \citenamefont {Hunt}, \citenamefont {DeRose}, \citenamefont {Alty},
  \citenamefont {Li}, \citenamefont {Wertheim}, \citenamefont {Rose},
  \citenamefont {Timco}, \citenamefont {Scherer}, \citenamefont {Yeates},\ and\
  \citenamefont {Winpenny}}]{Lewis_2019a}%
  \BibitemOpen
  \bibfield  {author} {\bibinfo {author} {\bibfnamefont {S.~M.}\ \bibnamefont
  {Lewis}}, \bibinfo {author} {\bibfnamefont {M.~S.}\ \bibnamefont {Hunt}},
  \bibinfo {author} {\bibfnamefont {G.~A.}\ \bibnamefont {DeRose}}, \bibinfo
  {author} {\bibfnamefont {H.~R.}\ \bibnamefont {Alty}}, \bibinfo {author}
  {\bibfnamefont {J.}~\bibnamefont {Li}}, \bibinfo {author} {\bibfnamefont
  {A.}~\bibnamefont {Wertheim}}, \bibinfo {author} {\bibfnamefont {L.~D.}\
  \bibnamefont {Rose}}, \bibinfo {author} {\bibfnamefont {G.~A.}\ \bibnamefont
  {Timco}}, \bibinfo {author} {\bibfnamefont {A.}~\bibnamefont {Scherer}},
  \bibinfo {author} {\bibfnamefont {S.~G.}\ \bibnamefont {Yeates}},\ and\
  \bibinfo {author} {\bibfnamefont {R.~E.~P.}\ \bibnamefont {Winpenny}},\
  }\bibfield  {title} {\enquote {\bibinfo {title} {Plasma-etched pattern
  transfer of sub-10 nm structures using a metal--organic resist and helium ion
  beam lithography},}\ }\href {https://doi.org/10.1021/acs.nanolett.9b01911}
  {\bibfield  {journal} {\bibinfo  {journal} {Nano Lett.}\ }\textbf {\bibinfo
  {volume} {19}},\ \bibinfo {pages} {6043--6048} (\bibinfo {year}
  {2019})}\BibitemShut {NoStop}%
\bibitem [{\citenamefont {Kumar}\ \emph {et~al.}(2020)\citenamefont {Kumar},
  \citenamefont {Chauhan}, \citenamefont {Moinuddin}, \citenamefont {Sharma},\
  and\ \citenamefont {Gonsalves}}]{Kumar_2020a}%
  \BibitemOpen
  \bibfield  {author} {\bibinfo {author} {\bibfnamefont {R.}~\bibnamefont
  {Kumar}}, \bibinfo {author} {\bibfnamefont {M.}~\bibnamefont {Chauhan}},
  \bibinfo {author} {\bibfnamefont {M.~G.}\ \bibnamefont {Moinuddin}}, \bibinfo
  {author} {\bibfnamefont {S.~K.}\ \bibnamefont {Sharma}},\ and\ \bibinfo
  {author} {\bibfnamefont {K.~E.}\ \bibnamefont {Gonsalves}},\ }\bibfield
  {title} {\enquote {\bibinfo {title} {Development of nickel-based negative
  tone metal oxide cluster resists for sub-10 nm electron beam and helium ion
  beam lithography},}\ }\href {https://doi.org/10.1021/acsami.9b21414}
  {\bibfield  {journal} {\bibinfo  {journal} {ACS Appl. Mater. Interfaces}\
  }\textbf {\bibinfo {volume} {12}},\ \bibinfo {pages} {19616--19624} (\bibinfo
  {year} {2020})}\BibitemShut {NoStop}%
\bibitem [{\citenamefont {Yogesh}\ \emph {et~al.}(2021)\citenamefont {Yogesh},
  \citenamefont {Moinuddin}, \citenamefont {Chauhan}, \citenamefont {Sharma},
  \citenamefont {Ghosh},\ and\ \citenamefont {Gonsalves}}]{Yogesh_2021a}%
  \BibitemOpen
  \bibfield  {author} {\bibinfo {author} {\bibfnamefont {M.}~\bibnamefont
  {Yogesh}}, \bibinfo {author} {\bibfnamefont {M.~G.}\ \bibnamefont
  {Moinuddin}}, \bibinfo {author} {\bibfnamefont {M.}~\bibnamefont {Chauhan}},
  \bibinfo {author} {\bibfnamefont {S.~K.}\ \bibnamefont {Sharma}}, \bibinfo
  {author} {\bibfnamefont {S.}~\bibnamefont {Ghosh}},\ and\ \bibinfo {author}
  {\bibfnamefont {K.~E.}\ \bibnamefont {Gonsalves}},\ }\bibfield  {title}
  {\enquote {\bibinfo {title} {Organoiodine functionality bearing resists for
  electron-beam and helium ion beam lithography: complex and sub-16 nm
  patterning},}\ }\href {https://doi.org/10.1021/acsaelm.0c01120} {\bibfield
  {journal} {\bibinfo  {journal} {{ACS} Applied Electronic Materials}\ }\textbf
  {\bibinfo {volume} {3}},\ \bibinfo {pages} {1996--2004} (\bibinfo {year}
  {2021})}\BibitemShut {NoStop}%
\bibitem [{\citenamefont {Zhang}\ \emph {et~al.}(2020)\citenamefont {Zhang},
  \citenamefont {Thomas}, \citenamefont {Guan}, \citenamefont {Heinig},\ and\
  \citenamefont {Leung}}]{Zhang_2020a}%
  \BibitemOpen
  \bibfield  {author} {\bibinfo {author} {\bibfnamefont {L.}~\bibnamefont
  {Zhang}}, \bibinfo {author} {\bibfnamefont {J.~P.}\ \bibnamefont {Thomas}},
  \bibinfo {author} {\bibfnamefont {X.}~\bibnamefont {Guan}}, \bibinfo {author}
  {\bibfnamefont {N.~F.}\ \bibnamefont {Heinig}},\ and\ \bibinfo {author}
  {\bibfnamefont {K.~T.}\ \bibnamefont {Leung}},\ }\bibfield  {title} {\enquote
  {\bibinfo {title} {High-energy ion ({H}e$^+$, {S}i$^{++}$, {G}a$^+$,
  {A}u$^{++}$) interactions with {PMMA} in ion beam lithography},}\ }\href
  {https://doi.org/10.1088/1361-6528/ab8d69} {\bibfield  {journal} {\bibinfo
  {journal} {Nanotechnology}\ }\textbf {\bibinfo {volume} {31}},\ \bibinfo
  {pages} {325301} (\bibinfo {year} {2020})}\BibitemShut {NoStop}%
\bibitem [{\citenamefont {Rashid}\ \emph {et~al.}(2021)\citenamefont {Rashid},
  \citenamefont {Walia}, \citenamefont {Northfield}, \citenamefont {Hahn},
  \citenamefont {Olivieri}, \citenamefont {Lesina}, \citenamefont {Variola},
  \citenamefont {Weck}, \citenamefont {Ramunno},\ and\ \citenamefont
  {Berini}}]{Rashid_2021a}%
  \BibitemOpen
  \bibfield  {author} {\bibinfo {author} {\bibfnamefont {S.}~\bibnamefont
  {Rashid}}, \bibinfo {author} {\bibfnamefont {J.}~\bibnamefont {Walia}},
  \bibinfo {author} {\bibfnamefont {H.}~\bibnamefont {Northfield}}, \bibinfo
  {author} {\bibfnamefont {C.}~\bibnamefont {Hahn}}, \bibinfo {author}
  {\bibfnamefont {A.}~\bibnamefont {Olivieri}}, \bibinfo {author}
  {\bibfnamefont {A.~C.}\ \bibnamefont {Lesina}}, \bibinfo {author}
  {\bibfnamefont {F.}~\bibnamefont {Variola}}, \bibinfo {author} {\bibfnamefont
  {A.}~\bibnamefont {Weck}}, \bibinfo {author} {\bibfnamefont {L.}~\bibnamefont
  {Ramunno}},\ and\ \bibinfo {author} {\bibfnamefont {P.}~\bibnamefont
  {Berini}},\ }\bibfield  {title} {\enquote {\bibinfo {title} {Helium ion beam
  lithography and liftoff},}\ }\href {https://doi.org/10.1088/2399-1984/abfd98}
  {\bibfield  {journal} {\bibinfo  {journal} {Nano Futur.}\ }\textbf {\bibinfo
  {volume} {5}},\ \bibinfo {pages} {025003} (\bibinfo {year}
  {2021})}\BibitemShut {NoStop}%
\bibitem [{\citenamefont {Xia}\ \emph {et~al.}(2020)\citenamefont {Xia},
  \citenamefont {Zhu}, \citenamefont {Khanom},\ and\ \citenamefont
  {Runt}}]{Xia_2020a}%
  \BibitemOpen
  \bibfield  {author} {\bibinfo {author} {\bibfnamefont {D.}~\bibnamefont
  {Xia}}, \bibinfo {author} {\bibfnamefont {X.}~\bibnamefont {Zhu}}, \bibinfo
  {author} {\bibfnamefont {F.}~\bibnamefont {Khanom}},\ and\ \bibinfo {author}
  {\bibfnamefont {D.}~\bibnamefont {Runt}},\ }\bibfield  {title} {\enquote
  {\bibinfo {title} {Neon and helium focused ion beam etching of resist
  patterns},}\ }\href {https://doi.org/10.1088/1361-6528/abafd6} {\bibfield
  {journal} {\bibinfo  {journal} {Nanotechnology}\ }\textbf {\bibinfo {volume}
  {31}},\ \bibinfo {pages} {475301} (\bibinfo {year} {2020})}\BibitemShut
  {NoStop}%
\bibitem [{\citenamefont {Maas}\ \emph {et~al.}(2014)\citenamefont {Maas},
  \citenamefont {van Veldhoven}, \citenamefont {van
  Langen{\textendash}Suurling}, \citenamefont {Alkemade}, \citenamefont
  {Wuister}, \citenamefont {Hoefnagels}, \citenamefont {Verspaget},
  \citenamefont {Meessen},\ and\ \citenamefont {Fliervoet}}]{Maas_2014}%
  \BibitemOpen
  \bibfield  {author} {\bibinfo {author} {\bibfnamefont {D.}~\bibnamefont
  {Maas}}, \bibinfo {author} {\bibfnamefont {E.}~\bibnamefont {van Veldhoven}},
  \bibinfo {author} {\bibfnamefont {A.}~\bibnamefont {van
  Langen{\textendash}Suurling}}, \bibinfo {author} {\bibfnamefont {P.~F.~A.}\
  \bibnamefont {Alkemade}}, \bibinfo {author} {\bibfnamefont {S.}~\bibnamefont
  {Wuister}}, \bibinfo {author} {\bibfnamefont {R.}~\bibnamefont {Hoefnagels}},
  \bibinfo {author} {\bibfnamefont {C.}~\bibnamefont {Verspaget}}, \bibinfo
  {author} {\bibfnamefont {J.}~\bibnamefont {Meessen}},\ and\ \bibinfo {author}
  {\bibfnamefont {T.}~\bibnamefont {Fliervoet}},\ }\bibfield  {title} {\enquote
  {\bibinfo {title} {Evaluation of {EUV} resist performance below 20nm {CD}
  using helium ion lithography},}\ }in\ \href
  {https://doi.org/10.1117/12.2046917} {\emph {\bibinfo {booktitle} {{SPIE}
  Proceedings}}},\ \bibinfo {editor} {edited by\ \bibinfo {editor}
  {\bibfnamefont {O.~R.}\ \bibnamefont {Wood}}\ and\ \bibinfo {editor}
  {\bibfnamefont {E.~M.}\ \bibnamefont {Panning}}}\ (\bibinfo  {publisher}
  {{SPIE}},\ \bibinfo {year} {2014})\BibitemShut {NoStop}%
\bibitem [{\citenamefont {Kalhor}\ and\ \citenamefont
  {Alkemade}(2016)}]{Kalhor_2016}%
  \BibitemOpen
  \bibfield  {author} {\bibinfo {author} {\bibfnamefont {N.}~\bibnamefont
  {Kalhor}}\ and\ \bibinfo {author} {\bibfnamefont {P.~F.~A.}\ \bibnamefont
  {Alkemade}},\ }\bibfield  {title} {\enquote {\bibinfo {title} {{Resist
  Assisted Patterning}},}\ }in\  \cite{Hlawacek_2016},\ Chap.~\bibinfo
  {chapter} {16}, pp.\ \bibinfo {pages} {395--414}\BibitemShut {NoStop}%
\bibitem [{\citenamefont {Funsten}\ \emph {et~al.}(1992)\citenamefont
  {Funsten}, \citenamefont {Boring}, \citenamefont {Johnson},\ and\
  \citenamefont {Brown}}]{Funsten_1992a}%
  \BibitemOpen
  \bibfield  {author} {\bibinfo {author} {\bibfnamefont {H.~O.}\ \bibnamefont
  {Funsten}}, \bibinfo {author} {\bibfnamefont {J.~W.}\ \bibnamefont {Boring}},
  \bibinfo {author} {\bibfnamefont {R.~E.}\ \bibnamefont {Johnson}},\ and\
  \bibinfo {author} {\bibfnamefont {W.~L.}\ \bibnamefont {Brown}},\ }\bibfield
  {title} {\enquote {\bibinfo {title} {Low-temperature beam-induced deposition
  of thin tin films},}\ }\href {https://doi.org/10.1063/1.351241} {\bibfield
  {journal} {\bibinfo  {journal} {J. Appl. Phys.}\ }\textbf {\bibinfo {volume}
  {71}},\ \bibinfo {pages} {1475--1484} (\bibinfo {year} {1992})}\BibitemShut
  {NoStop}%
\bibitem [{\citenamefont {Or{\'{u}}s}\ \emph
  {et~al.}(2021{\natexlab{c}})\citenamefont {Or{\'{u}}s}, \citenamefont
  {Sigloch}, \citenamefont {Sangiao},\ and\ \citenamefont
  {De~Teresa}}]{Orus_2021b}%
  \BibitemOpen
  \bibfield  {author} {\bibinfo {author} {\bibfnamefont {P.}~\bibnamefont
  {Or{\'{u}}s}}, \bibinfo {author} {\bibfnamefont {F.}~\bibnamefont {Sigloch}},
  \bibinfo {author} {\bibfnamefont {S.}~\bibnamefont {Sangiao}},\ and\ \bibinfo
  {author} {\bibfnamefont {J.~M.}\ \bibnamefont {De~Teresa}},\ }\bibfield
  {title} {\enquote {\bibinfo {title} {Cryo-focused ion beam-induced deposition
  of tungsten{\textendash}carbon nanostructures using a thermoelectric
  plate},}\ }\href {https://doi.org/10.3390/app112110123} {\bibfield  {journal}
  {\bibinfo  {journal} {Applied Sciences}\ }\textbf {\bibinfo {volume} {11}},\
  \bibinfo {pages} {10123} (\bibinfo {year} {2021}{\natexlab{c}})}\BibitemShut
  {NoStop}%
\bibitem [{\citenamefont {Bilgilisoy}\ \emph {et~al.}(2021)\citenamefont
  {Bilgilisoy}, \citenamefont {Thorman}, \citenamefont {Barclay}, \citenamefont
  {Marbach},\ and\ \citenamefont {Fairbrother}}]{Bilgilisoy_2021a}%
  \BibitemOpen
  \bibfield  {author} {\bibinfo {author} {\bibfnamefont {E.}~\bibnamefont
  {Bilgilisoy}}, \bibinfo {author} {\bibfnamefont {R.~M.}\ \bibnamefont
  {Thorman}}, \bibinfo {author} {\bibfnamefont {M.~S.}\ \bibnamefont
  {Barclay}}, \bibinfo {author} {\bibfnamefont {H.}~\bibnamefont {Marbach}},\
  and\ \bibinfo {author} {\bibfnamefont {D.~H.}\ \bibnamefont {Fairbrother}},\
  }\bibfield  {title} {\enquote {\bibinfo {title} {Low energy electron- and
  ion-induced surface reactions of {F}e({CO})$_5$ thin films},}\ }\href
  {https://doi.org/10.1021/acs.jpcc.1c05826} {\bibfield  {journal} {\bibinfo
  {journal} {J. Phys. Chem. C}\ }\textbf {\bibinfo {volume} {125}},\ \bibinfo
  {pages} {17749--17760} (\bibinfo {year} {2021})}\BibitemShut {NoStop}%
\bibitem [{\citenamefont {Gross}\ \emph {et~al.}(1989)\citenamefont {Gross},
  \citenamefont {Brown}, \citenamefont {Harriott}, \citenamefont {Cummings},
  \citenamefont {Linnros},\ and\ \citenamefont {Funsten}}]{Gross_1989a}%
  \BibitemOpen
  \bibfield  {author} {\bibinfo {author} {\bibfnamefont {M.~E.}\ \bibnamefont
  {Gross}}, \bibinfo {author} {\bibfnamefont {W.~L.}\ \bibnamefont {Brown}},
  \bibinfo {author} {\bibfnamefont {L.~R.}\ \bibnamefont {Harriott}}, \bibinfo
  {author} {\bibfnamefont {K.~D.}\ \bibnamefont {Cummings}}, \bibinfo {author}
  {\bibfnamefont {J.}~\bibnamefont {Linnros}},\ and\ \bibinfo {author}
  {\bibfnamefont {H.}~\bibnamefont {Funsten}},\ }\bibfield  {title} {\enquote
  {\bibinfo {title} {Ion-beam direct-write mechanisms in palladium acetate
  films},}\ }\href {https://doi.org/10.1063/1.344444} {\bibfield  {journal}
  {\bibinfo  {journal} {J. Appl. Phys.}\ }\textbf {\bibinfo {volume} {66}},\
  \bibinfo {pages} {1403--1410} (\bibinfo {year} {1989})}\BibitemShut {NoStop}%
\bibitem [{\citenamefont {Hoffmann}(1991)}]{Hoffmann_1991a}%
  \BibitemOpen
  \bibfield  {author} {\bibinfo {author} {\bibfnamefont {P.}~\bibnamefont
  {Hoffmann}},\ }\bibfield  {title} {\enquote {\bibinfo {title} {Direct writing
  of iridium lines with a focused ion beam},}\ }\href
  {https://doi.org/10.1116/1.585828} {\bibfield  {journal} {\bibinfo  {journal}
  {J. Vac. Sci. Technol. B}\ }\textbf {\bibinfo {volume} {9}},\ \bibinfo
  {pages} {3483} (\bibinfo {year} {1991})}\BibitemShut {NoStop}%
\bibitem [{\citenamefont {Hoffmann}\ \emph {et~al.}(1993)\citenamefont
  {Hoffmann}, \citenamefont {Assayag}, \citenamefont {Gierak}, \citenamefont
  {Flicstein}, \citenamefont {Maar-Stumm},\ and\ \citenamefont {van~den
  Bergh}}]{Hoffmann_1993a}%
  \BibitemOpen
  \bibfield  {author} {\bibinfo {author} {\bibfnamefont {P.}~\bibnamefont
  {Hoffmann}}, \bibinfo {author} {\bibfnamefont {G.~B.}\ \bibnamefont
  {Assayag}}, \bibinfo {author} {\bibfnamefont {J.}~\bibnamefont {Gierak}},
  \bibinfo {author} {\bibfnamefont {J.}~\bibnamefont {Flicstein}}, \bibinfo
  {author} {\bibfnamefont {M.}~\bibnamefont {Maar-Stumm}},\ and\ \bibinfo
  {author} {\bibfnamefont {H.}~\bibnamefont {van~den Bergh}},\ }\bibfield
  {title} {\enquote {\bibinfo {title} {Direct writing of gold nanostructures
  using a gold-cluster compound and a focused-ion beam},}\ }\href
  {https://doi.org/10.1063/1.354985} {\bibfield  {journal} {\bibinfo  {journal}
  {J. Appl. Phys.}\ }\textbf {\bibinfo {volume} {74}},\ \bibinfo {pages}
  {7588--7591} (\bibinfo {year} {1993})}\BibitemShut {NoStop}%
\bibitem [{\citenamefont {Bhuvana}\ and\ \citenamefont
  {Kulkarni}(2008)}]{Bhuvana_2008a}%
  \BibitemOpen
  \bibfield  {author} {\bibinfo {author} {\bibfnamefont {T.}~\bibnamefont
  {Bhuvana}}\ and\ \bibinfo {author} {\bibfnamefont {G.~U.}\ \bibnamefont
  {Kulkarni}},\ }\bibfield  {title} {\enquote {\bibinfo {title} {Highly
  conducting patterned pd nanowires by direct-write electron beam
  lithography},}\ }\href {https://doi.org/10.1021/nn700372h} {\bibfield
  {journal} {\bibinfo  {journal} {{ACS} Nano}\ }\textbf {\bibinfo {volume}
  {2}},\ \bibinfo {pages} {457--462} (\bibinfo {year} {2008})}\BibitemShut
  {NoStop}%
\bibitem [{\citenamefont {Salvador-Porroche}\ \emph
  {et~al.}(2022{\natexlab{a}})\citenamefont {Salvador-Porroche}, \citenamefont
  {Herrer}, \citenamefont {Sangiao}, \citenamefont {De~Teresa},\ and\
  \citenamefont {Cea}}]{SalvadorPorroche_2022}%
  \BibitemOpen
  \bibfield  {author} {\bibinfo {author} {\bibfnamefont {A.}~\bibnamefont
  {Salvador-Porroche}}, \bibinfo {author} {\bibfnamefont {L.}~\bibnamefont
  {Herrer}}, \bibinfo {author} {\bibfnamefont {S.}~\bibnamefont {Sangiao}},
  \bibinfo {author} {\bibfnamefont {J.~M.}\ \bibnamefont {De~Teresa}},\ and\
  \bibinfo {author} {\bibfnamefont {P.}~\bibnamefont {Cea}},\ }\bibfield
  {title} {\enquote {\bibinfo {title} {Low-resistivity {P}d nanopatterns
  created by a direct electron beam irradiation process free of post-treatment
  steps},}\ }\href {https://doi.org/10.1088/1361-6528/ac47cf} {\bibfield
  {journal} {\bibinfo  {journal} {Nanotechnology}\ }\textbf {\bibinfo {volume}
  {33}},\ \bibinfo {pages} {405302} (\bibinfo {year}
  {2022}{\natexlab{a}})}\BibitemShut {NoStop}%
\bibitem [{\citenamefont {Salvador-Porroche}\ \emph
  {et~al.}(2022{\natexlab{b}})\citenamefont {Salvador-Porroche}, \citenamefont
  {Herrer}, \citenamefont {Sangiao}, \citenamefont {Philipp}, \citenamefont
  {Cea},\ and\ \citenamefont {De~Teresa}}]{SalvadorPorroche_2022a}%
  \BibitemOpen
  \bibfield  {author} {\bibinfo {author} {\bibfnamefont {A.}~\bibnamefont
  {Salvador-Porroche}}, \bibinfo {author} {\bibfnamefont {L.}~\bibnamefont
  {Herrer}}, \bibinfo {author} {\bibfnamefont {S.}~\bibnamefont {Sangiao}},
  \bibinfo {author} {\bibfnamefont {P.}~\bibnamefont {Philipp}}, \bibinfo
  {author} {\bibfnamefont {P.}~\bibnamefont {Cea}},\ and\ \bibinfo {author}
  {\bibfnamefont {J.~M.}\ \bibnamefont {De~Teresa}},\ }\bibfield  {title}
  {\enquote {\bibinfo {title} {High-throughput direct writing of metallic
  micro- and nano-structures by focused ga$^+$ beam irradiation of palladium
  acetate films},}\ }\href {https://doi.org/10.1021/acsami.2c05218} {\bibfield
  {journal} {\bibinfo  {journal} {ACS Appl. Mater. Interfaces}\ }\textbf
  {\bibinfo {volume} {14}},\ \bibinfo {pages} {28211--28220} (\bibinfo {year}
  {2022}{\natexlab{b}})}\BibitemShut {NoStop}%
\bibitem [{\citenamefont {Aramesh}(2017)}]{Aramesh2017}%
  \BibitemOpen
  \bibfield  {author} {\bibinfo {author} {\bibfnamefont {M.}~\bibnamefont
  {Aramesh}},\ }\bibfield  {title} {\enquote {\bibinfo {title} {Ion-beam
  sculpting of nanowires},}\ }\href {https://doi.org/10.1002/pssr.201700333}
  {\bibfield  {journal} {\bibinfo  {journal} {Phys. Status Solidi RRL}\
  }\textbf {\bibinfo {volume} {12}},\ \bibinfo {pages} {1700333} (\bibinfo
  {year} {2017})}\BibitemShut {NoStop}%
\bibitem [{\citenamefont {Keller}\ and\ \citenamefont
  {Facsko}(2010)}]{Keller2010}%
  \BibitemOpen
  \bibfield  {author} {\bibinfo {author} {\bibfnamefont {A.}~\bibnamefont
  {Keller}}\ and\ \bibinfo {author} {\bibfnamefont {S.}~\bibnamefont
  {Facsko}},\ }\bibfield  {title} {\enquote {\bibinfo {title} {Ion-induced
  nanoscale ripple patterns on {S}i surfaces: {T}heory and experiment},}\
  }\href {https://doi.org/10.3390/ma3104811} {\bibfield  {journal} {\bibinfo
  {journal} {Materials}\ }\textbf {\bibinfo {volume} {3}},\ \bibinfo {pages}
  {4811--4841} (\bibinfo {year} {2010})}\BibitemShut {NoStop}%
\bibitem [{\citenamefont {Ou}\ \emph {et~al.}(2013)\citenamefont {Ou},
  \citenamefont {Keller}, \citenamefont {Helm}, \citenamefont {Fassbender},\
  and\ \citenamefont {Facsko}}]{Ou2013}%
  \BibitemOpen
  \bibfield  {author} {\bibinfo {author} {\bibfnamefont {X.}~\bibnamefont
  {Ou}}, \bibinfo {author} {\bibfnamefont {A.}~\bibnamefont {Keller}}, \bibinfo
  {author} {\bibfnamefont {M.}~\bibnamefont {Helm}}, \bibinfo {author}
  {\bibfnamefont {J.}~\bibnamefont {Fassbender}},\ and\ \bibinfo {author}
  {\bibfnamefont {S.}~\bibnamefont {Facsko}},\ }\bibfield  {title} {\enquote
  {\bibinfo {title} {Reverse epitaxy of {G}e: {O}rdered and faceted surface
  patterns},}\ }\href {https://doi.org/10.1103/physrevlett.111.016101}
  {\bibfield  {journal} {\bibinfo  {journal} {Phys. Rev. Lett.}\ }\textbf
  {\bibinfo {volume} {111}},\ \bibinfo {pages} {016101} (\bibinfo {year}
  {2013})}\BibitemShut {NoStop}%
\bibitem [{\citenamefont {Hofs{\"{a}}ss}(2013)}]{Hofsaess2013}%
  \BibitemOpen
  \bibfield  {author} {\bibinfo {author} {\bibfnamefont {H.}~\bibnamefont
  {Hofs{\"{a}}ss}},\ }\bibfield  {title} {\enquote {\bibinfo {title} {Surface
  instability and pattern formation by ion-induced erosion and mass
  redistribution},}\ }\href {https://doi.org/10.1007/s00339-013-8170-9}
  {\bibfield  {journal} {\bibinfo  {journal} {Appl. Phys. A}\ }\textbf
  {\bibinfo {volume} {114}},\ \bibinfo {pages} {401--422} (\bibinfo {year}
  {2013})}\BibitemShut {NoStop}%
\bibitem [{\citenamefont {Datta}, \citenamefont {Wu},\ and\ \citenamefont
  {Wang}(2001)}]{Datta2001}%
  \BibitemOpen
  \bibfield  {author} {\bibinfo {author} {\bibfnamefont {A.}~\bibnamefont
  {Datta}}, \bibinfo {author} {\bibfnamefont {Y.-R.}\ \bibnamefont {Wu}},\ and\
  \bibinfo {author} {\bibfnamefont {Y.~L.}\ \bibnamefont {Wang}},\ }\bibfield
  {title} {\enquote {\bibinfo {title} {Real-time observation of ripple
  structure formation on a diamond surface under focused ion-beam
  bombardment},}\ }\href {https://doi.org/10.1103/PhysRevB.63.125407}
  {\bibfield  {journal} {\bibinfo  {journal} {Phys. Rev. B}\ }\textbf {\bibinfo
  {volume} {63}},\ \bibinfo {pages} {125407} (\bibinfo {year}
  {2001})}\BibitemShut {NoStop}%
\bibitem [{\citenamefont {Habenicht}\ \emph {et~al.}(2002)\citenamefont
  {Habenicht}, \citenamefont {Lieb}, \citenamefont {Koch},\ and\ \citenamefont
  {Wieck}}]{Habenicht2002}%
  \BibitemOpen
  \bibfield  {author} {\bibinfo {author} {\bibfnamefont {S.}~\bibnamefont
  {Habenicht}}, \bibinfo {author} {\bibfnamefont {K.~P.}\ \bibnamefont {Lieb}},
  \bibinfo {author} {\bibfnamefont {J.}~\bibnamefont {Koch}},\ and\ \bibinfo
  {author} {\bibfnamefont {A.~D.}\ \bibnamefont {Wieck}},\ }\bibfield  {title}
  {\enquote {\bibinfo {title} {Ripple propagation and velocity dispersion on
  ion-beam-eroded silicon surfaces},}\ }\href
  {https://doi.org/10.1103/PhysRevB.65.115327} {\bibfield  {journal} {\bibinfo
  {journal} {Phys. Rev. B}\ }\textbf {\bibinfo {volume} {65}},\ \bibinfo
  {pages} {115327} (\bibinfo {year} {2002})}\BibitemShut {NoStop}%
\bibitem [{\citenamefont {Gnaser}, \citenamefont {Reuscher},\ and\
  \citenamefont {Zeuner}(2012)}]{Gnaser2012}%
  \BibitemOpen
  \bibfield  {author} {\bibinfo {author} {\bibfnamefont {H.}~\bibnamefont
  {Gnaser}}, \bibinfo {author} {\bibfnamefont {B.}~\bibnamefont {Reuscher}},\
  and\ \bibinfo {author} {\bibfnamefont {A.}~\bibnamefont {Zeuner}},\
  }\bibfield  {title} {\enquote {\bibinfo {title} {Propagation of nanoscale
  ripples on ion-irradiated surfaces},}\ }\href
  {https://doi.org/10.1016/j.nimb.2012.05.028} {\bibfield  {journal} {\bibinfo
  {journal} {Nucl. Instrum. Meth. Phys. Res. B}\ }\textbf {\bibinfo {volume}
  {285}},\ \bibinfo {pages} {142--147} (\bibinfo {year} {2012})}\BibitemShut
  {NoStop}%
\bibitem [{\citenamefont {Bischoff}\ \emph {et~al.}(2012)\citenamefont
  {Bischoff}, \citenamefont {Heinig}, \citenamefont {Schmidt}, \citenamefont
  {Facsko},\ and\ \citenamefont {Pilz}}]{Bischoff2012}%
  \BibitemOpen
  \bibfield  {author} {\bibinfo {author} {\bibfnamefont {L.}~\bibnamefont
  {Bischoff}}, \bibinfo {author} {\bibfnamefont {K.-H.}\ \bibnamefont
  {Heinig}}, \bibinfo {author} {\bibfnamefont {B.}~\bibnamefont {Schmidt}},
  \bibinfo {author} {\bibfnamefont {S.}~\bibnamefont {Facsko}},\ and\ \bibinfo
  {author} {\bibfnamefont {W.}~\bibnamefont {Pilz}},\ }\bibfield  {title}
  {\enquote {\bibinfo {title} {Self-organization of {Ge} nanopattern under
  erosion with heavy {Bi} monomer and cluster ions},}\ }\href
  {https://doi.org/10.1016/j.nimb.2011.01.064} {\bibfield  {journal} {\bibinfo
  {journal} {Nucl. Instrum. Meth. Phys. Res. B.}\ }\textbf {\bibinfo {volume}
  {272}},\ \bibinfo {pages} {198--201} (\bibinfo {year} {2012})}\BibitemShut
  {NoStop}%
\bibitem [{\citenamefont {B{\"{o}}ttger}\ \emph {et~al.}(2012)\citenamefont
  {B{\"{o}}ttger}, \citenamefont {Bischoff}, \citenamefont {Facsko},\ and\
  \citenamefont {Schmidt}}]{Boettger2012}%
  \BibitemOpen
  \bibfield  {author} {\bibinfo {author} {\bibfnamefont {R.}~\bibnamefont
  {B{\"{o}}ttger}}, \bibinfo {author} {\bibfnamefont {L.}~\bibnamefont
  {Bischoff}}, \bibinfo {author} {\bibfnamefont {S.}~\bibnamefont {Facsko}},\
  and\ \bibinfo {author} {\bibfnamefont {B.}~\bibnamefont {Schmidt}},\
  }\bibfield  {title} {\enquote {\bibinfo {title} {Quantitative analysis of the
  order of {Bi} ion induced dot patterns on {Ge}},}\ }\href
  {https://doi.org/10.1209/0295-5075/98/16009} {\bibfield  {journal} {\bibinfo
  {journal} {Europhys. Lett.}\ }\textbf {\bibinfo {volume} {98}},\ \bibinfo
  {pages} {Article No. 16009} (\bibinfo {year} {2012})}\BibitemShut {NoStop}%
\bibitem [{\citenamefont {B{\"{o}}ttger}\ \emph {et~al.}(2013)\citenamefont
  {B{\"{o}}ttger}, \citenamefont {Heinig}, \citenamefont {Bischoff},
  \citenamefont {Liedke}, \citenamefont {H{\"{u}}bner},\ and\ \citenamefont
  {Pilz}}]{Boettger2013}%
  \BibitemOpen
  \bibfield  {author} {\bibinfo {author} {\bibfnamefont {R.}~\bibnamefont
  {B{\"{o}}ttger}}, \bibinfo {author} {\bibfnamefont {K.-H.}\ \bibnamefont
  {Heinig}}, \bibinfo {author} {\bibfnamefont {L.}~\bibnamefont {Bischoff}},
  \bibinfo {author} {\bibfnamefont {B.}~\bibnamefont {Liedke}}, \bibinfo
  {author} {\bibfnamefont {R.}~\bibnamefont {H{\"{u}}bner}},\ and\ \bibinfo
  {author} {\bibfnamefont {W.}~\bibnamefont {Pilz}},\ }\bibfield  {title}
  {\enquote {\bibinfo {title} {Silicon nanodot formation and self-ordering
  under bombardment with heavy {Bi}$_3$ ions},}\ }\href
  {https://doi.org/10.1002/pssr.201307127} {\bibfield  {journal} {\bibinfo
  {journal} {Phys. Status Solidi RRL}\ }\textbf {\bibinfo {volume} {7}},\
  \bibinfo {pages} {501--505} (\bibinfo {year} {2013})}\BibitemShut {NoStop}%
\bibitem [{\citenamefont {Bischoff}\ \emph {et~al.}(2014)\citenamefont
  {Bischoff}, \citenamefont {B{\"{o}}ttger}, \citenamefont {Heinig},
  \citenamefont {Facsko},\ and\ \citenamefont {Pilz}}]{Bischoff2014}%
  \BibitemOpen
  \bibfield  {author} {\bibinfo {author} {\bibfnamefont {L.}~\bibnamefont
  {Bischoff}}, \bibinfo {author} {\bibfnamefont {R.}~\bibnamefont
  {B{\"{o}}ttger}}, \bibinfo {author} {\bibfnamefont {K.-H.}\ \bibnamefont
  {Heinig}}, \bibinfo {author} {\bibfnamefont {S.}~\bibnamefont {Facsko}},\
  and\ \bibinfo {author} {\bibfnamefont {W.}~\bibnamefont {Pilz}},\ }\bibfield
  {title} {\enquote {\bibinfo {title} {Surface patterning of {GaAs} under
  irradiation with very heavy polyatomic {Au} ions},}\ }\href
  {https://doi.org/10.1016/j.apsusc.2014.03.166} {\bibfield  {journal}
  {\bibinfo  {journal} {Appl. Surf. Sci.}\ }\textbf {\bibinfo {volume} {310}},\
  \bibinfo {pages} {154--157} (\bibinfo {year} {2014})}\BibitemShut {NoStop}%
\bibitem [{\citenamefont {Smirnova}\ \emph {et~al.}(2022)\citenamefont
  {Smirnova}, \citenamefont {Bachurin}, \citenamefont {Lebedev}, \citenamefont
  {Mazaletsky}, \citenamefont {Pukhov}, \citenamefont {Churilov},\ and\
  \citenamefont {Rudy}}]{Smirnova2022}%
  \BibitemOpen
  \bibfield  {author} {\bibinfo {author} {\bibfnamefont {M.~A.}\ \bibnamefont
  {Smirnova}}, \bibinfo {author} {\bibfnamefont {V.~I.}\ \bibnamefont
  {Bachurin}}, \bibinfo {author} {\bibfnamefont {M.~E.}\ \bibnamefont
  {Lebedev}}, \bibinfo {author} {\bibfnamefont {L.~A.}\ \bibnamefont
  {Mazaletsky}}, \bibinfo {author} {\bibfnamefont {D.~E.}\ \bibnamefont
  {Pukhov}}, \bibinfo {author} {\bibfnamefont {A.~B.}\ \bibnamefont
  {Churilov}},\ and\ \bibinfo {author} {\bibfnamefont {A.~S.}\ \bibnamefont
  {Rudy}},\ }\bibfield  {title} {{\selectlanguage {en}\enquote {\bibinfo
  {title} {Study of the topography of silicon surface evolution under
  irradiation by a gallium ion beam},}\ }}\href
  {https://doi.org/10.1016/j.vacuum.2022.111238} {\bibfield  {journal}
  {\bibinfo  {journal} {Vacuum}\ }\textbf {\bibinfo {volume} {203}},\ \bibinfo
  {pages} {111238} (\bibinfo {year} {2022})}\BibitemShut {NoStop}%
\bibitem [{\citenamefont {Ichim}\ and\ \citenamefont {Aziz}(2005)}]{Ichim2005}%
  \BibitemOpen
  \bibfield  {author} {\bibinfo {author} {\bibfnamefont {S.}~\bibnamefont
  {Ichim}}\ and\ \bibinfo {author} {\bibfnamefont {M.~J.}\ \bibnamefont
  {Aziz}},\ }\bibfield  {title} {\enquote {\bibinfo {title} {Lateral templating
  of self-organized ripple morphologies during focused ion beam milling of
  {Ge}},}\ }\href {https://doi.org/10.1116/1.1897711} {\bibfield  {journal}
  {\bibinfo  {journal} {J. Vac. Sci. Technol. B}\ }\textbf {\bibinfo {volume}
  {23}},\ \bibinfo {pages} {1068--1071} (\bibinfo {year} {2005})}\BibitemShut
  {NoStop}%
\bibitem [{\citenamefont {Kramczynski}, \citenamefont {Reuscher},\ and\
  \citenamefont {Gnaser}(2014)}]{Kramczynski2014}%
  \BibitemOpen
  \bibfield  {author} {\bibinfo {author} {\bibfnamefont {D.}~\bibnamefont
  {Kramczynski}}, \bibinfo {author} {\bibfnamefont {B.}~\bibnamefont
  {Reuscher}},\ and\ \bibinfo {author} {\bibfnamefont {H.}~\bibnamefont
  {Gnaser}},\ }\bibfield  {title} {\enquote {\bibinfo {title}
  {Wavelength-dependent ripple propagation on ion-irradiated prepatterned
  surfaces driven by viscous flow corroborates two-field continuum model},}\
  }\href {https://doi.org/10.1103/PhysRevB.89.205422} {\bibfield  {journal}
  {\bibinfo  {journal} {Phys. Rev. B}\ }\textbf {\bibinfo {volume} {89}},\
  \bibinfo {pages} {205422} (\bibinfo {year} {2014})}\BibitemShut {NoStop}%
\bibitem [{\citenamefont {Esmek}\ \emph {et~al.}(2019)\citenamefont {Esmek},
  \citenamefont {Bayat}, \citenamefont {P{\'e}rez-Willard}, \citenamefont
  {Volkenandt}, \citenamefont {Blick},\ and\ \citenamefont
  {Fernandez-Cuesta}}]{Esmek_2019}%
  \BibitemOpen
  \bibfield  {author} {\bibinfo {author} {\bibfnamefont {F.~M.}\ \bibnamefont
  {Esmek}}, \bibinfo {author} {\bibfnamefont {P.}~\bibnamefont {Bayat}},
  \bibinfo {author} {\bibfnamefont {F.}~\bibnamefont {P{\'e}rez-Willard}},
  \bibinfo {author} {\bibfnamefont {T.}~\bibnamefont {Volkenandt}}, \bibinfo
  {author} {\bibfnamefont {R.~H.}\ \bibnamefont {Blick}},\ and\ \bibinfo
  {author} {\bibfnamefont {I.}~\bibnamefont {Fernandez-Cuesta}},\ }\bibfield
  {title} {{\selectlanguage {english}\enquote {\bibinfo {title} {Sculpturing
  wafer-scale nanofluidic devices for {DNA} single molecule analysis},}\
  }}\href {https://doi.org/10.1039/c9nr02979f} {\bibfield  {journal} {\bibinfo
  {journal} {Nanoscale}\ }\textbf {\bibinfo {volume} {11}},\ \bibinfo {pages}
  {13620--13631} (\bibinfo {year} {2019})}\BibitemShut {NoStop}%
\bibitem [{\citenamefont {Lebedev}\ \emph {et~al.}(2021)\citenamefont
  {Lebedev}, \citenamefont {Malyshev}, \citenamefont {Ryzhkov}, \citenamefont
  {Mozharov}, \citenamefont {Shugurov}, \citenamefont {Sharov}, \citenamefont
  {Panov}, \citenamefont {Tumkin}, \citenamefont {Afonicheva}, \citenamefont
  {Evstrapov}, \citenamefont {Bukatin},\ and\ \citenamefont
  {Mukhin}}]{Lebedev_2021}%
  \BibitemOpen
  \bibfield  {author} {\bibinfo {author} {\bibfnamefont {D.}~\bibnamefont
  {Lebedev}}, \bibinfo {author} {\bibfnamefont {G.}~\bibnamefont {Malyshev}},
  \bibinfo {author} {\bibfnamefont {I.}~\bibnamefont {Ryzhkov}}, \bibinfo
  {author} {\bibfnamefont {A.}~\bibnamefont {Mozharov}}, \bibinfo {author}
  {\bibfnamefont {K.}~\bibnamefont {Shugurov}}, \bibinfo {author}
  {\bibfnamefont {V.}~\bibnamefont {Sharov}}, \bibinfo {author} {\bibfnamefont
  {M.}~\bibnamefont {Panov}}, \bibinfo {author} {\bibfnamefont
  {I.}~\bibnamefont {Tumkin}}, \bibinfo {author} {\bibfnamefont
  {P.}~\bibnamefont {Afonicheva}}, \bibinfo {author} {\bibfnamefont
  {A.}~\bibnamefont {Evstrapov}}, \bibinfo {author} {\bibfnamefont
  {A.}~\bibnamefont {Bukatin}},\ and\ \bibinfo {author} {\bibfnamefont
  {I.}~\bibnamefont {Mukhin}},\ }\bibfield  {title} {\enquote {\bibinfo {title}
  {Focused ion beam milling based formation of nanochannels in silicon-glass
  microfluidic chips for the study of ion transport},}\ }\href
  {https://doi.org/10.1007/s10404-021-02450-x} {\bibfield  {journal} {\bibinfo
  {journal} {Microfluid. Nanofluidics}\ }\textbf {\bibinfo {volume} {25}},\
  \bibinfo {pages} {51} (\bibinfo {year} {2021})}\BibitemShut {NoStop}%
\bibitem [{\citenamefont {Postigo}\ \emph {et~al.}(2016)\citenamefont
  {Postigo}, \citenamefont {Alvaro}, \citenamefont {Juarros},\ and\
  \citenamefont {Merino}}]{Postigo_2016}%
  \BibitemOpen
  \bibfield  {author} {\bibinfo {author} {\bibfnamefont {P.~A.}\ \bibnamefont
  {Postigo}}, \bibinfo {author} {\bibfnamefont {R.}~\bibnamefont {Alvaro}},
  \bibinfo {author} {\bibfnamefont {A.}~\bibnamefont {Juarros}},\ and\ \bibinfo
  {author} {\bibfnamefont {S.}~\bibnamefont {Merino}},\ }\bibfield  {title}
  {{\selectlanguage {en}\enquote {\bibinfo {title} {Optofluidic chips with
  nanochannels for dynamic molecular detection using enhanced fluorescence},}\
  }}\href {https://doi.org/10.1364/BOE.7.003289} {\bibfield  {journal}
  {\bibinfo  {journal} {Biomed. Opt. Express}\ }\textbf {\bibinfo {volume}
  {7}},\ \bibinfo {pages} {3289--3298} (\bibinfo {year} {2016})}\BibitemShut
  {NoStop}%
\bibitem [{\citenamefont {Liao}\ \emph {et~al.}(2017)\citenamefont {Liao},
  \citenamefont {Schumacher}, \citenamefont {Lezec},\ and\ \citenamefont
  {Stavis}}]{Liao_2017}%
  \BibitemOpen
  \bibfield  {author} {\bibinfo {author} {\bibfnamefont {K.-T.}\ \bibnamefont
  {Liao}}, \bibinfo {author} {\bibfnamefont {J.}~\bibnamefont {Schumacher}},
  \bibinfo {author} {\bibfnamefont {H.~J.}\ \bibnamefont {Lezec}},\ and\
  \bibinfo {author} {\bibfnamefont {S.~M.}\ \bibnamefont {Stavis}},\ }\bibfield
   {title} {{\selectlanguage {en}\enquote {\bibinfo {title} {Subnanometer
  structure and function from ion beams through complex fluidics to fluorescent
  particles},}\ }}\href {https://doi.org/10.1039/c7lc01047h} {\bibfield
  {journal} {\bibinfo  {journal} {Lab Chip}\ }\textbf {\bibinfo {volume}
  {18}},\ \bibinfo {pages} {139--152} (\bibinfo {year} {2017})}\BibitemShut
  {NoStop}%
\bibitem [{\citenamefont {Wang}\ \emph {et~al.}(2005)\citenamefont {Wang},
  \citenamefont {Wang}, \citenamefont {Li}, \citenamefont {Li}, \citenamefont
  {Xu}, \citenamefont {Luo}, \citenamefont {Jin}, \citenamefont {Yang},\ and\
  \citenamefont {Gu}}]{Wang_2005}%
  \BibitemOpen
  \bibfield  {author} {\bibinfo {author} {\bibfnamefont {Z.~L.}\ \bibnamefont
  {Wang}}, \bibinfo {author} {\bibfnamefont {Q.}~\bibnamefont {Wang}}, \bibinfo
  {author} {\bibfnamefont {H.~J.}\ \bibnamefont {Li}}, \bibinfo {author}
  {\bibfnamefont {J.~J.}\ \bibnamefont {Li}}, \bibinfo {author} {\bibfnamefont
  {P.}~\bibnamefont {Xu}}, \bibinfo {author} {\bibfnamefont {Q.}~\bibnamefont
  {Luo}}, \bibinfo {author} {\bibfnamefont {A.~Z.}\ \bibnamefont {Jin}},
  \bibinfo {author} {\bibfnamefont {H.~F.}\ \bibnamefont {Yang}},\ and\
  \bibinfo {author} {\bibfnamefont {C.~Z.}\ \bibnamefont {Gu}},\ }\bibfield
  {title} {\enquote {\bibinfo {title} {The field emission properties of high
  aspect ratio diamond nanocone arrays fabricated by focused ion beam
  milling},}\ }\href {https://doi.org/10.1016/j.stam.2005.06.018} {\bibfield
  {journal} {\bibinfo  {journal} {Sci. Technol. Adv. Mat.}\ }\textbf {\bibinfo
  {volume} {6}},\ \bibinfo {pages} {799--803} (\bibinfo {year}
  {2005})}\BibitemShut {NoStop}%
\bibitem [{\citenamefont {Waid}\ \emph {et~al.}(2014)\citenamefont {Waid},
  \citenamefont {Wanzenboeck}, \citenamefont {Muehlberger}, \citenamefont
  {Gavagnin},\ and\ \citenamefont {Bertagnolli}}]{Waid_2014}%
  \BibitemOpen
  \bibfield  {author} {\bibinfo {author} {\bibfnamefont {S.}~\bibnamefont
  {Waid}}, \bibinfo {author} {\bibfnamefont {H.~D.}\ \bibnamefont
  {Wanzenboeck}}, \bibinfo {author} {\bibfnamefont {M.}~\bibnamefont
  {Muehlberger}}, \bibinfo {author} {\bibfnamefont {M.}~\bibnamefont
  {Gavagnin}},\ and\ \bibinfo {author} {\bibfnamefont {E.}~\bibnamefont
  {Bertagnolli}},\ }\bibfield  {title} {\enquote {\bibinfo {title} {Focused ion
  beam direct patterning of hardmask layers},}\ }\href
  {https://doi.org/10.1116/1.4884777} {\bibfield  {journal} {\bibinfo
  {journal} {J. Vac. Sci. Technol. B}\ }\textbf {\bibinfo {volume} {32}},\
  \bibinfo {pages} {041602} (\bibinfo {year} {2014})}\BibitemShut {NoStop}%
\bibitem [{\citenamefont {Rius}\ \emph
  {et~al.}(2009{\natexlab{a}})\citenamefont {Rius}, \citenamefont {Llobet},
  \citenamefont {Borris{\'e}}, \citenamefont {Mestres}, \citenamefont
  {Retolaza}, \citenamefont {Merino},\ and\ \citenamefont
  {Perez-Murano}}]{Rius_2009}%
  \BibitemOpen
  \bibfield  {author} {\bibinfo {author} {\bibfnamefont {G.}~\bibnamefont
  {Rius}}, \bibinfo {author} {\bibfnamefont {J.}~\bibnamefont {Llobet}},
  \bibinfo {author} {\bibfnamefont {X.}~\bibnamefont {Borris{\'e}}}, \bibinfo
  {author} {\bibfnamefont {N.}~\bibnamefont {Mestres}}, \bibinfo {author}
  {\bibfnamefont {A.}~\bibnamefont {Retolaza}}, \bibinfo {author}
  {\bibfnamefont {S.}~\bibnamefont {Merino}},\ and\ \bibinfo {author}
  {\bibfnamefont {F.}~\bibnamefont {Perez-Murano}},\ }\bibfield  {title}
  {\enquote {\bibinfo {title} {Fabrication of complementary
  metal-oxide-semiconductor integrated nanomechanical devices by ion beam
  patterning},}\ }\href {https://doi.org/10.1116/1.3253550} {\bibfield
  {journal} {\bibinfo  {journal} {J. Vac. Sci. Technol. B}\ }\textbf {\bibinfo
  {volume} {27}},\ \bibinfo {pages} {2691--2697} (\bibinfo {year}
  {2009}{\natexlab{a}})}\BibitemShut {NoStop}%
\bibitem [{\citenamefont {Llobet}\ \emph {et~al.}(2018)\citenamefont {Llobet},
  \citenamefont {Rius}, \citenamefont {Chuquitarqui}, \citenamefont
  {Borris{\'e}}, \citenamefont {Koops}, \citenamefont {van Veghel},\ and\
  \citenamefont {Perez-Murano}}]{Llobet_2018}%
  \BibitemOpen
  \bibfield  {author} {\bibinfo {author} {\bibfnamefont {J.}~\bibnamefont
  {Llobet}}, \bibinfo {author} {\bibfnamefont {G.}~\bibnamefont {Rius}},
  \bibinfo {author} {\bibfnamefont {A.}~\bibnamefont {Chuquitarqui}}, \bibinfo
  {author} {\bibfnamefont {X.}~\bibnamefont {Borris{\'e}}}, \bibinfo {author}
  {\bibfnamefont {R.}~\bibnamefont {Koops}}, \bibinfo {author} {\bibfnamefont
  {M.}~\bibnamefont {van Veghel}},\ and\ \bibinfo {author} {\bibfnamefont
  {F.}~\bibnamefont {Perez-Murano}},\ }\bibfield  {title} {{\selectlanguage
  {english}\enquote {\bibinfo {title} {Arrays of suspended silicon nanowires
  defined by ion beam implantation: mechanical coupling and combination with
  {CMOS} technology},}\ }}\href {https://doi.org/10.1088/1361-6528/aaac67}
  {\bibfield  {journal} {\bibinfo  {journal} {Nanotechnology}\ }\textbf
  {\bibinfo {volume} {29}},\ \bibinfo {pages} {155303} (\bibinfo {year}
  {2018})}\BibitemShut {NoStop}%
\bibitem [{\citenamefont {Rius}\ \emph {et~al.}(2007)\citenamefont {Rius},
  \citenamefont {Mart{\'{\i}}n}, \citenamefont {Godignon}, \citenamefont
  {Bachtold}, \citenamefont {Bausells}, \citenamefont {Lora-Tamayo},\ and\
  \citenamefont {P{\'{e}}rez-Murano}}]{Rius_2007}%
  \BibitemOpen
  \bibfield  {author} {\bibinfo {author} {\bibfnamefont {G.}~\bibnamefont
  {Rius}}, \bibinfo {author} {\bibfnamefont {I.}~\bibnamefont {Mart{\'{\i}}n}},
  \bibinfo {author} {\bibfnamefont {P.}~\bibnamefont {Godignon}}, \bibinfo
  {author} {\bibfnamefont {A.}~\bibnamefont {Bachtold}}, \bibinfo {author}
  {\bibfnamefont {J.}~\bibnamefont {Bausells}}, \bibinfo {author}
  {\bibfnamefont {E.}~\bibnamefont {Lora-Tamayo}},\ and\ \bibinfo {author}
  {\bibfnamefont {F.}~\bibnamefont {P{\'{e}}rez-Murano}},\ }\bibfield  {title}
  {\enquote {\bibinfo {title} {Response of carbon nanotube transistors to
  electron beam exposure},}\ }\href {https://doi.org/10.1016/j.mee.2007.01.237}
  {\bibfield  {journal} {\bibinfo  {journal} {Microelectron. Eng.}\ }\textbf
  {\bibinfo {volume} {84}},\ \bibinfo {pages} {1596--1600} (\bibinfo {year}
  {2007})}\BibitemShut {NoStop}%
\bibitem [{\citenamefont {Rius}\ \emph
  {et~al.}(2009{\natexlab{b}})\citenamefont {Rius}, \citenamefont {Llobet},
  \citenamefont {Esplandiu}, \citenamefont {Sol{\'{e}}}, \citenamefont
  {Borris{\'{e}}},\ and\ \citenamefont {P{\'{e}}rez-Murano}}]{Rius_2009a}%
  \BibitemOpen
  \bibfield  {author} {\bibinfo {author} {\bibfnamefont {G.}~\bibnamefont
  {Rius}}, \bibinfo {author} {\bibfnamefont {J.}~\bibnamefont {Llobet}},
  \bibinfo {author} {\bibfnamefont {M.~J.}\ \bibnamefont {Esplandiu}}, \bibinfo
  {author} {\bibfnamefont {L.}~\bibnamefont {Sol{\'{e}}}}, \bibinfo {author}
  {\bibfnamefont {X.}~\bibnamefont {Borris{\'{e}}}},\ and\ \bibinfo {author}
  {\bibfnamefont {F.}~\bibnamefont {P{\'{e}}rez-Murano}},\ }\bibfield  {title}
  {\enquote {\bibinfo {title} {Using electron and ion beams on carbon
  nanotube-based devices. effects and considerations for nanofabrication},}\
  }\href {https://doi.org/10.1016/j.mee.2008.12.007} {\bibfield  {journal}
  {\bibinfo  {journal} {Microelectron. Eng.}\ }\textbf {\bibinfo {volume}
  {86}},\ \bibinfo {pages} {892--894} (\bibinfo {year}
  {2009}{\natexlab{b}})}\BibitemShut {NoStop}%
\bibitem [{\citenamefont {Matsubara}\ \emph {et~al.}(2016)\citenamefont
  {Matsubara}, \citenamefont {Shichi}, \citenamefont {Kawanami},\ and\
  \citenamefont {Hashizume}}]{Matsubara_2016}%
  \BibitemOpen
  \bibfield  {author} {\bibinfo {author} {\bibfnamefont {S.}~\bibnamefont
  {Matsubara}}, \bibinfo {author} {\bibfnamefont {H.}~\bibnamefont {Shichi}},
  \bibinfo {author} {\bibfnamefont {Y.}~\bibnamefont {Kawanami}},\ and\
  \bibinfo {author} {\bibfnamefont {T.}~\bibnamefont {Hashizume}},\ }\bibfield
  {title} {\enquote {\bibinfo {title} {Novel scanning ion microscope with
  {H}$_3^+$ gas field ionization source},}\ }\href
  {https://doi.org/10.1017/s1431927616003925} {\bibfield  {journal} {\bibinfo
  {journal} {Microsc. Microanal.}\ }\textbf {\bibinfo {volume} {22}},\ \bibinfo
  {pages} {614--615} (\bibinfo {year} {2016})}\BibitemShut {NoStop}%
\bibitem [{\citenamefont {Bischoff}\ and\ \citenamefont
  {Pilz}(2010)}]{Bischoff_2010a}%
  \BibitemOpen
  \bibfield  {author} {\bibinfo {author} {\bibfnamefont {L.}~\bibnamefont
  {Bischoff}}\ and\ \bibinfo {author} {\bibfnamefont {W.}~\bibnamefont
  {Pilz}},\ }\href {https://patents.google.com/patent/DE102007027097B4}
  {{\selectlanguage {German}\enquote {\bibinfo {title}
  {Fl{\"{u}}ssigmetall-{I}onenquelle zur {E}rzeugung von
  {L}ithium-{I}onenstrahlen},}\ }}\bibinfo {type} {patent}\ \bibinfo {number}
  {{DE}102007027097B4}\ (\bibinfo  {institution} {Forschungszentrum Dresden -
  Rossendorf e.V.},\ \bibinfo {year} {2010})\BibitemShut {NoStop}%
\bibitem [{\citenamefont {Viteau}\ \emph {et~al.}(2016)\citenamefont {Viteau},
  \citenamefont {Reveillard}, \citenamefont {Kime}, \citenamefont {Rasser},
  \citenamefont {Sudraud}, \citenamefont {Bruneau}, \citenamefont {Khalili},
  \citenamefont {Pillet}, \citenamefont {Comparat}, \citenamefont {Guerri},
  \citenamefont {Fioretti}, \citenamefont {Ciampini}, \citenamefont
  {Allegrini},\ and\ \citenamefont {Fuso}}]{Viteau_2016}%
  \BibitemOpen
  \bibfield  {author} {\bibinfo {author} {\bibfnamefont {M.}~\bibnamefont
  {Viteau}}, \bibinfo {author} {\bibfnamefont {M.}~\bibnamefont {Reveillard}},
  \bibinfo {author} {\bibfnamefont {L.}~\bibnamefont {Kime}}, \bibinfo {author}
  {\bibfnamefont {B.}~\bibnamefont {Rasser}}, \bibinfo {author} {\bibfnamefont
  {P.}~\bibnamefont {Sudraud}}, \bibinfo {author} {\bibfnamefont
  {Y.}~\bibnamefont {Bruneau}}, \bibinfo {author} {\bibfnamefont
  {G.}~\bibnamefont {Khalili}}, \bibinfo {author} {\bibfnamefont
  {P.}~\bibnamefont {Pillet}}, \bibinfo {author} {\bibfnamefont
  {D.}~\bibnamefont {Comparat}}, \bibinfo {author} {\bibfnamefont
  {I.}~\bibnamefont {Guerri}}, \bibinfo {author} {\bibfnamefont
  {A.}~\bibnamefont {Fioretti}}, \bibinfo {author} {\bibfnamefont
  {D.}~\bibnamefont {Ciampini}}, \bibinfo {author} {\bibfnamefont
  {M.}~\bibnamefont {Allegrini}},\ and\ \bibinfo {author} {\bibfnamefont
  {F.}~\bibnamefont {Fuso}},\ }\bibfield  {title} {\enquote {\bibinfo {title}
  {Ion microscopy based on laser-cooled cesium atoms},}\ }\href
  {https://doi.org/10.1016/j.ultramic.2015.12.007} {\bibfield  {journal}
  {\bibinfo  {journal} {Ultramicroscopy}\ }\textbf {\bibinfo {volume} {164}},\
  \bibinfo {pages} {70--77} (\bibinfo {year} {2016})},\ \Eprint
  {https://arxiv.org/abs/1601.01446} {arXiv:1601.01446} \BibitemShut {NoStop}%
\bibitem [{\citenamefont {RAITH}(2023)}]{RAITH2023}%
  \BibitemOpen
  \bibfield  {author} {\bibinfo {author} {\bibnamefont {RAITH}},\ }\href
  {https://raith.com/ion-microscopy/} {\enquote {\bibinfo {title} {Ion
  microscopy with gabili for fib imaging, 3d analysis, and nanoscale
  tomography},}\ }\bibinfo {howpublished} {Online} (\bibinfo {year} {2023}),\
  \bibinfo {note} {/url{https://raith.com/ion-microscopy/}}\BibitemShut
  {NoStop}%
\bibitem [{\citenamefont {Laurencich}\ \emph {et~al.}(2022)\citenamefont
  {Laurencich}, \citenamefont {Lapena}, \citenamefont {Lagaize}, \citenamefont
  {Hou{\"{e}}l}, \citenamefont {Delobbe},\ and\ \citenamefont
  {Salan{\c{c}}on}}]{Laurencich_2022}%
  \BibitemOpen
  \bibfield  {author} {\bibinfo {author} {\bibfnamefont {M.}~\bibnamefont
  {Laurencich}}, \bibinfo {author} {\bibfnamefont {L.}~\bibnamefont {Lapena}},
  \bibinfo {author} {\bibfnamefont {M.}~\bibnamefont {Lagaize}}, \bibinfo
  {author} {\bibfnamefont {A.}~\bibnamefont {Hou{\"{e}}l}}, \bibinfo {author}
  {\bibfnamefont {A.}~\bibnamefont {Delobbe}},\ and\ \bibinfo {author}
  {\bibfnamefont {E.}~\bibnamefont {Salan{\c{c}}on}},\ }\bibfield  {title}
  {\enquote {\bibinfo {title} {Insulator on conductor ion source: first results
  and measurements},}\ }in\ \href
  {https://fit4nano.eu/wp-content/uploads/2022/07/FIT4NANO_BoA.pdf} {\emph
  {\bibinfo {booktitle} {FIT4NANO workshop}}},\ \bibinfo {editor} {edited by\
  \bibinfo {editor} {\bibfnamefont {G.}~\bibnamefont {Hlawacek}}, \bibinfo
  {editor} {\bibfnamefont {G.}~\bibnamefont {Hobler}}, \bibinfo {editor}
  {\bibfnamefont {K.}~\bibnamefont {H{\"{o}}flich}}, \bibinfo {editor}
  {\bibfnamefont {T.}~\bibnamefont {Wirtz}},\ and\ \bibinfo {editor}
  {\bibfnamefont {M.}~\bibnamefont {Marszalek}}}\ (\bibinfo {organization}
  {FIT4NANO},\ \bibinfo {address} {Krakow},\ \bibinfo {year} {2022})\
  p.~\bibinfo {pages} {40}\BibitemShut {NoStop}%
\bibitem [{\citenamefont {Hawkes}\ and\ \citenamefont
  {Krivanek}(2019)}]{Hawkes_2019}%
  \BibitemOpen
  \bibfield  {author} {\bibinfo {author} {\bibfnamefont {P.~W.}\ \bibnamefont
  {Hawkes}}\ and\ \bibinfo {author} {\bibfnamefont {O.~L.}\ \bibnamefont
  {Krivanek}},\ }\bibfield  {title} {\enquote {\bibinfo {title} {Aberration
  correctors, monochromators, spectrometers},}\ }in\ \href
  {https://doi.org/10.1007/978-3-030-00069-1_13} {\emph {\bibinfo {booktitle}
  {Springer Handbook of Microscopy}}},\ \bibinfo {series and number} {Springer
  Handbooks},\ \bibinfo {editor} {edited by\ \bibinfo {editor} {\bibfnamefont
  {P.~W.}\ \bibnamefont {Hawkes}}\ and\ \bibinfo {editor} {\bibfnamefont
  {J.~C.~H.}\ \bibnamefont {Spence}}}\ (\bibinfo  {publisher} {Springer
  International Publishing},\ \bibinfo {address} {Cham},\ \bibinfo {year}
  {2019})\ Chap.~\bibinfo {chapter} {13}, pp.\ \bibinfo {pages}
  {625--675}\BibitemShut {NoStop}%
\bibitem [{\citenamefont {Klingner}(2016)}]{Klingner2016}%
  \BibitemOpen
  \bibfield  {author} {\bibinfo {author} {\bibfnamefont {N.}~\bibnamefont
  {Klingner}},\ }\emph {\bibinfo {title} {{Ionenstrahlanalytik im
  Helium-Ionen-Mikroskop}}},\ \href
  {https://katalog.slub-dresden.de/en/id/0-1655323466} {\bibinfo {type}
  {phdthesis}},\ \bibinfo  {school} {Technical University of Dresden} (\bibinfo
  {year} {2016})\BibitemShut {NoStop}%
\bibitem [{\citenamefont {Belianinov}\ \emph {et~al.}(2018)\citenamefont
  {Belianinov}, \citenamefont {Pawlicki}, \citenamefont {Burch}, \citenamefont
  {Kim}, \citenamefont {Ievlev}, \citenamefont {Fowler},\ and\ \citenamefont
  {Ovchinnikova}}]{Belianinov2018}%
  \BibitemOpen
  \bibfield  {author} {\bibinfo {author} {\bibfnamefont {A.}~\bibnamefont
  {Belianinov}}, \bibinfo {author} {\bibfnamefont {A.}~\bibnamefont
  {Pawlicki}}, \bibinfo {author} {\bibfnamefont {M.}~\bibnamefont {Burch}},
  \bibinfo {author} {\bibfnamefont {S.}~\bibnamefont {Kim}}, \bibinfo {author}
  {\bibfnamefont {A.}~\bibnamefont {Ievlev}}, \bibinfo {author} {\bibfnamefont
  {J.}~\bibnamefont {Fowler}},\ and\ \bibinfo {author} {\bibfnamefont
  {O.}~\bibnamefont {Ovchinnikova}},\ }\bibfield  {title} {\enquote {\bibinfo
  {title} {In-situ liquid cell crystallization and imaging of thiamethoxam by
  helium ion microscopy},}\ }\href {https://doi.org/10.1116/1.5040849}
  {\bibfield  {journal} {\bibinfo  {journal} {J. Vac. Sci. Technol. B}\
  }\textbf {\bibinfo {volume} {36}} (\bibinfo {year} {2018}),\
  10.1116/1.5040849}\BibitemShut {NoStop}%
\bibitem [{\citenamefont {Stanford}\ \emph
  {et~al.}(2016{\natexlab{b}})\citenamefont {Stanford}, \citenamefont {Lewis},
  \citenamefont {Iberi}, \citenamefont {Fowlkes}, \citenamefont {Tan},
  \citenamefont {Livengood},\ and\ \citenamefont {Rack}}]{Stanford_2016}%
  \BibitemOpen
  \bibfield  {author} {\bibinfo {author} {\bibfnamefont {M.~G.}\ \bibnamefont
  {Stanford}}, \bibinfo {author} {\bibfnamefont {B.~B.}\ \bibnamefont {Lewis}},
  \bibinfo {author} {\bibfnamefont {V.}~\bibnamefont {Iberi}}, \bibinfo
  {author} {\bibfnamefont {J.~D.}\ \bibnamefont {Fowlkes}}, \bibinfo {author}
  {\bibfnamefont {S.}~\bibnamefont {Tan}}, \bibinfo {author} {\bibfnamefont
  {R.~H.}\ \bibnamefont {Livengood}},\ and\ \bibinfo {author} {\bibfnamefont
  {P.~D.}\ \bibnamefont {Rack}},\ }\bibfield  {title} {\enquote {\bibinfo
  {title} {{In Situ Mitigation of Subsurface and Peripheral Focused Ion Beam
  Damage via Simultaneous Pulsed Laser Heating}},}\ }\href
  {https://doi.org/10.1002/smll.201503680} {\bibfield  {journal} {\bibinfo
  {journal} {Small}\ }\textbf {\bibinfo {volume} {12}},\ \bibinfo {pages}
  {1779--1787} (\bibinfo {year} {2016}{\natexlab{b}})}\BibitemShut {NoStop}%
\bibitem [{\citenamefont {Mele}\ \emph {et~al.}(2016)\citenamefont {Mele},
  \citenamefont {Konings}, \citenamefont {Dona}, \citenamefont {Evertz},
  \citenamefont {Mitterbauer}, \citenamefont {Faber}, \citenamefont
  {Schampers},\ and\ \citenamefont {Jinschek}}]{Mele_2016}%
  \BibitemOpen
  \bibfield  {author} {\bibinfo {author} {\bibfnamefont {L.}~\bibnamefont
  {Mele}}, \bibinfo {author} {\bibfnamefont {S.}~\bibnamefont {Konings}},
  \bibinfo {author} {\bibfnamefont {P.}~\bibnamefont {Dona}}, \bibinfo {author}
  {\bibfnamefont {F.}~\bibnamefont {Evertz}}, \bibinfo {author} {\bibfnamefont
  {C.}~\bibnamefont {Mitterbauer}}, \bibinfo {author} {\bibfnamefont
  {P.}~\bibnamefont {Faber}}, \bibinfo {author} {\bibfnamefont
  {R.}~\bibnamefont {Schampers}},\ and\ \bibinfo {author} {\bibfnamefont
  {J.~R.}\ \bibnamefont {Jinschek}},\ }\bibfield  {title} {\enquote {\bibinfo
  {title} {A {MEMS}-based heating holder for the direct imaging of simultaneous
  in-situ heating and biasing experiments in scanning/transmission electron
  microscopes},}\ }\href {https://doi.org/10.1002/jemt.22623} {\bibfield
  {journal} {\bibinfo  {journal} {Microsc. Res. Tech.}\ }\textbf {\bibinfo
  {volume} {79}},\ \bibinfo {pages} {239--250} (\bibinfo {year}
  {2016})}\BibitemShut {NoStop}%
\bibitem [{\citenamefont {Joy}\ \emph {et~al.}(2007)\citenamefont {Joy},
  \citenamefont {Meyer}, \citenamefont {Bolorizadeh}, \citenamefont {Lin},\
  and\ \citenamefont {Newbury}}]{Joy_2007}%
  \BibitemOpen
  \bibfield  {author} {\bibinfo {author} {\bibfnamefont {D.~C.}\ \bibnamefont
  {Joy}}, \bibinfo {author} {\bibfnamefont {H.~M.}\ \bibnamefont {Meyer}},
  \bibinfo {author} {\bibfnamefont {M.}~\bibnamefont {Bolorizadeh}}, \bibinfo
  {author} {\bibfnamefont {Y.}~\bibnamefont {Lin}},\ and\ \bibinfo {author}
  {\bibfnamefont {D.}~\bibnamefont {Newbury}},\ }\bibfield  {title} {\enquote
  {\bibinfo {title} {On the production of x-rays by low energy ion beams},}\
  }\href {https://doi.org/10.1002/sca.20002} {\bibfield  {journal} {\bibinfo
  {journal} {Scanning}\ }\textbf {\bibinfo {volume} {29}},\ \bibinfo {pages}
  {1--4} (\bibinfo {year} {2007})}\BibitemShut {NoStop}%
\bibitem [{\citenamefont {Sugioka}\ and\ \citenamefont
  {Cheng}(2014)}]{Sugioka_2014}%
  \BibitemOpen
  \bibfield  {author} {\bibinfo {author} {\bibfnamefont {K.}~\bibnamefont
  {Sugioka}}\ and\ \bibinfo {author} {\bibfnamefont {Y.}~\bibnamefont
  {Cheng}},\ }\bibfield  {title} {\enquote {\bibinfo {title} {Ultrafast
  lasers---reliable tools for advanced materials processing},}\ }\href
  {https://doi.org/10.1038/lsa.2014.30} {\bibfield  {journal} {\bibinfo
  {journal} {Light Sci. Appl.}\ }\textbf {\bibinfo {volume} {3}},\ \bibinfo
  {pages} {e149--e149} (\bibinfo {year} {2014})}\BibitemShut {NoStop}%
\bibitem [{\citenamefont {{European Research Executive
  Agency}}(2023)}]{EREA2023}%
  \BibitemOpen
  \bibfield  {author} {\bibinfo {author} {\bibnamefont {{European Research
  Executive Agency}}},\ }\href {https://rea.ec.europa.eu/open-science_en}
  {\enquote {\bibinfo {title} {Open science},}\ }\bibinfo {howpublished}
  {Online} (\bibinfo {year} {2023}),\ \bibinfo {note}
  {\url{https://rea.ec.europa.eu/open-science_en}}\BibitemShut {NoStop}%
\bibitem [{\citenamefont {Settles}(2009)}]{Settles_2009}%
  \BibitemOpen
  \bibfield  {author} {\bibinfo {author} {\bibfnamefont {B.}~\bibnamefont
  {Settles}},\ }\href {http://digital.library.wisc.edu/1793/60660} {\enquote
  {\bibinfo {title} {Active learning literature survey},}\ }\bibinfo {type}
  {techreport}\ (\bibinfo  {institution} {University of Wisconsin-Madison
  Department of Computer Sciences},\ \bibinfo {year} {2009})\BibitemShut
  {NoStop}%
\bibitem [{\citenamefont {Wilkinson}\ \emph {et~al.}(2016)\citenamefont
  {Wilkinson}, \citenamefont {Dumontier}, \citenamefont {Aalbersberg},
  \citenamefont {Appleton}, \citenamefont {Axton}, \citenamefont {Baak},
  \citenamefont {Blomberg}, \citenamefont {Boiten}, \citenamefont
  {da~Silva~Santos}, \citenamefont {Bourne}, \citenamefont {Bouwman},
  \citenamefont {Brookes}, \citenamefont {Clark}, \citenamefont {Crosas},
  \citenamefont {Dillo}, \citenamefont {Dumon}, \citenamefont {Edmunds},
  \citenamefont {Evelo}, \citenamefont {Finkers}, \citenamefont
  {Gonzalez-Beltran}, \citenamefont {Gray}, \citenamefont {Groth},
  \citenamefont {Goble}, \citenamefont {Grethe}, \citenamefont {Heringa},
  \citenamefont {'t~Hoen}, \citenamefont {Hooft}, \citenamefont {Kuhn},
  \citenamefont {Kok}, \citenamefont {Kok}, \citenamefont {Lusher},
  \citenamefont {Martone}, \citenamefont {Mons}, \citenamefont {Packer},
  \citenamefont {Persson}, \citenamefont {Rocca-Serra}, \citenamefont {Roos},
  \citenamefont {van Schaik}, \citenamefont {Sansone}, \citenamefont
  {Schultes}, \citenamefont {Sengstag}, \citenamefont {Slater}, \citenamefont
  {Strawn}, \citenamefont {Swertz}, \citenamefont {Thompson}, \citenamefont
  {van~der Lei}, \citenamefont {van Mulligen}, \citenamefont {Velterop},
  \citenamefont {Waagmeester}, \citenamefont {Wittenburg}, \citenamefont
  {Wolstencroft}, \citenamefont {Zhao},\ and\ \citenamefont
  {Mons}}]{Wilkinson2016}%
  \BibitemOpen
  \bibfield  {author} {\bibinfo {author} {\bibfnamefont {M.~D.}\ \bibnamefont
  {Wilkinson}}, \bibinfo {author} {\bibfnamefont {M.}~\bibnamefont
  {Dumontier}}, \bibinfo {author} {\bibfnamefont {I.~J.}\ \bibnamefont
  {Aalbersberg}}, \bibinfo {author} {\bibfnamefont {G.}~\bibnamefont
  {Appleton}}, \bibinfo {author} {\bibfnamefont {M.}~\bibnamefont {Axton}},
  \bibinfo {author} {\bibfnamefont {A.}~\bibnamefont {Baak}}, \bibinfo {author}
  {\bibfnamefont {N.}~\bibnamefont {Blomberg}}, \bibinfo {author}
  {\bibfnamefont {J.-W.}\ \bibnamefont {Boiten}}, \bibinfo {author}
  {\bibfnamefont {L.~B.}\ \bibnamefont {da~Silva~Santos}}, \bibinfo {author}
  {\bibfnamefont {P.~E.}\ \bibnamefont {Bourne}}, \bibinfo {author}
  {\bibfnamefont {J.}~\bibnamefont {Bouwman}}, \bibinfo {author} {\bibfnamefont
  {A.~J.}\ \bibnamefont {Brookes}}, \bibinfo {author} {\bibfnamefont
  {T.}~\bibnamefont {Clark}}, \bibinfo {author} {\bibfnamefont
  {M.}~\bibnamefont {Crosas}}, \bibinfo {author} {\bibfnamefont
  {I.}~\bibnamefont {Dillo}}, \bibinfo {author} {\bibfnamefont
  {O.}~\bibnamefont {Dumon}}, \bibinfo {author} {\bibfnamefont
  {S.}~\bibnamefont {Edmunds}}, \bibinfo {author} {\bibfnamefont {C.~T.}\
  \bibnamefont {Evelo}}, \bibinfo {author} {\bibfnamefont {R.}~\bibnamefont
  {Finkers}}, \bibinfo {author} {\bibfnamefont {A.}~\bibnamefont
  {Gonzalez-Beltran}}, \bibinfo {author} {\bibfnamefont {A.~J.~G.}\
  \bibnamefont {Gray}}, \bibinfo {author} {\bibfnamefont {P.}~\bibnamefont
  {Groth}}, \bibinfo {author} {\bibfnamefont {C.}~\bibnamefont {Goble}},
  \bibinfo {author} {\bibfnamefont {J.~S.}\ \bibnamefont {Grethe}}, \bibinfo
  {author} {\bibfnamefont {J.}~\bibnamefont {Heringa}}, \bibinfo {author}
  {\bibfnamefont {P.~A.~C.}\ \bibnamefont {'t~Hoen}}, \bibinfo {author}
  {\bibfnamefont {R.}~\bibnamefont {Hooft}}, \bibinfo {author} {\bibfnamefont
  {T.}~\bibnamefont {Kuhn}}, \bibinfo {author} {\bibfnamefont {R.}~\bibnamefont
  {Kok}}, \bibinfo {author} {\bibfnamefont {J.}~\bibnamefont {Kok}}, \bibinfo
  {author} {\bibfnamefont {S.~J.}\ \bibnamefont {Lusher}}, \bibinfo {author}
  {\bibfnamefont {M.~E.}\ \bibnamefont {Martone}}, \bibinfo {author}
  {\bibfnamefont {A.}~\bibnamefont {Mons}}, \bibinfo {author} {\bibfnamefont
  {A.~L.}\ \bibnamefont {Packer}}, \bibinfo {author} {\bibfnamefont
  {B.}~\bibnamefont {Persson}}, \bibinfo {author} {\bibfnamefont
  {P.}~\bibnamefont {Rocca-Serra}}, \bibinfo {author} {\bibfnamefont
  {M.}~\bibnamefont {Roos}}, \bibinfo {author} {\bibfnamefont {R.}~\bibnamefont
  {van Schaik}}, \bibinfo {author} {\bibfnamefont {S.-A.}\ \bibnamefont
  {Sansone}}, \bibinfo {author} {\bibfnamefont {E.}~\bibnamefont {Schultes}},
  \bibinfo {author} {\bibfnamefont {T.}~\bibnamefont {Sengstag}}, \bibinfo
  {author} {\bibfnamefont {T.}~\bibnamefont {Slater}}, \bibinfo {author}
  {\bibfnamefont {G.}~\bibnamefont {Strawn}}, \bibinfo {author} {\bibfnamefont
  {M.~A.}\ \bibnamefont {Swertz}}, \bibinfo {author} {\bibfnamefont
  {M.}~\bibnamefont {Thompson}}, \bibinfo {author} {\bibfnamefont
  {J.}~\bibnamefont {van~der Lei}}, \bibinfo {author} {\bibfnamefont
  {E.}~\bibnamefont {van Mulligen}}, \bibinfo {author} {\bibfnamefont
  {J.}~\bibnamefont {Velterop}}, \bibinfo {author} {\bibfnamefont
  {A.}~\bibnamefont {Waagmeester}}, \bibinfo {author} {\bibfnamefont
  {P.}~\bibnamefont {Wittenburg}}, \bibinfo {author} {\bibfnamefont
  {K.}~\bibnamefont {Wolstencroft}}, \bibinfo {author} {\bibfnamefont
  {J.}~\bibnamefont {Zhao}},\ and\ \bibinfo {author} {\bibfnamefont
  {B.}~\bibnamefont {Mons}},\ }\bibfield  {title} {\enquote {\bibinfo {title}
  {The {FAIR} guiding principles for scientific data management and
  stewardship},}\ }\href {https://doi.org/10.1038/sdata.2016.18} {\bibfield
  {journal} {\bibinfo  {journal} {Sci. Data}\ }\textbf {\bibinfo {volume} {3}}
  (\bibinfo {year} {2016}),\ 10.1038/sdata.2016.18}\BibitemShut {NoStop}%
\bibitem [{\citenamefont {Wittenburg}\ \emph {et~al.}(2020)\citenamefont
  {Wittenburg}, \citenamefont {Lautenschlager}, \citenamefont {Thiemann},
  \citenamefont {Baldauf},\ and\ \citenamefont {Trilsbeek}}]{Wittenburg2020}%
  \BibitemOpen
  \bibfield  {author} {\bibinfo {author} {\bibfnamefont {P.}~\bibnamefont
  {Wittenburg}}, \bibinfo {author} {\bibfnamefont {M.}~\bibnamefont
  {Lautenschlager}}, \bibinfo {author} {\bibfnamefont {H.}~\bibnamefont
  {Thiemann}}, \bibinfo {author} {\bibfnamefont {C.}~\bibnamefont {Baldauf}},\
  and\ \bibinfo {author} {\bibfnamefont {P.}~\bibnamefont {Trilsbeek}},\
  }\bibfield  {title} {\enquote {\bibinfo {title} {{FAIR} practices in
  europe},}\ }\href {https://doi.org/10.1162/dint_a_00048} {\bibfield
  {journal} {\bibinfo  {journal} {Data Intelligence}\ }\textbf {\bibinfo
  {volume} {2}},\ \bibinfo {pages} {257--263} (\bibinfo {year}
  {2020})}\BibitemShut {NoStop}%
\bibitem [{\citenamefont {{WorldFAIR~project}}(2024)}]{WorldFAIRproject2024}%
  \BibitemOpen
  \bibfield  {author} {\bibinfo {author} {\bibnamefont {{WorldFAIR~project}}},\
  }\href {https://worldfair-project.eu/} {\enquote {\bibinfo {title} {The
  {WorldFAIR} project},}\ }\bibinfo {howpublished} {Online} (\bibinfo {year}
  {2024}),\ \bibinfo {note} {\url{https://worldfair-project.eu/}}\BibitemShut
  {NoStop}%
\bibitem [{\citenamefont {{Directorate-General for
  Communication}}(2023)}]{DGC2023}%
  \BibitemOpen
  \bibfield  {author} {\bibinfo {author} {\bibnamefont {{Directorate-General
  for Communication}}},\ }\href
  {https://commission.europa.eu/strategy-and-policy/priorities-2019-2024/europe-fit-digital-age/european-data-strategy_en}
  {\enquote {\bibinfo {title} {European data strategy},}\ }\bibinfo
  {howpublished} {Online} (\bibinfo {year} {2023}),\ \bibinfo {note}
  {\url{https://commission.europa.eu/strategy-and-policy/priorities-2019-2024/europe-fit-digital-age/european-data-strategy_en}}\BibitemShut
  {NoStop}%
\bibitem [{\citenamefont {{Nationale Forschungsdateninfrastruktur (NFDI)
  e.V.}}(2023)}]{NationaleForschungsdateninfrastruktur2023}%
  \BibitemOpen
  \bibfield  {author} {\bibinfo {author} {\bibnamefont {{Nationale
  Forschungsdateninfrastruktur (NFDI) e.V.}}},\ }\href
  {https://www.nfdi.de/?lang=en} {\enquote {\bibinfo {title} {German national
  research data infrastructure},}\ }\bibinfo {howpublished} {Online} (\bibinfo
  {year} {2023}),\ \bibinfo {note}
  {\url{https://www.nfdi.de/?lang=en}}\BibitemShut {NoStop}%
\bibitem [{\citenamefont {{NOMAD}}(2023)}]{NOMAD2023}%
  \BibitemOpen
  \bibfield  {author} {\bibinfo {author} {\bibnamefont {{NOMAD}}},\ }\href
  {https://nomad-lab.eu} {\enquote {\bibinfo {title} {Novel materials discovery
  ({NOMAD}) laboratory},}\ }\bibinfo {howpublished} {Online} (\bibinfo {year}
  {2023}),\ \bibinfo {note} {\url{https://nomad-lab.eu}}\BibitemShut {NoStop}%
\bibitem [{\citenamefont {Draxl}\ and\ \citenamefont
  {Scheffler}(2019)}]{Draxl2019}%
  \BibitemOpen
  \bibfield  {author} {\bibinfo {author} {\bibfnamefont {C.}~\bibnamefont
  {Draxl}}\ and\ \bibinfo {author} {\bibfnamefont {M.}~\bibnamefont
  {Scheffler}},\ }\bibfield  {title} {\enquote {\bibinfo {title} {The {NOMAD}
  laboratory: from data sharing to artificial intelligence},}\ }\href
  {https://doi.org/10.1088/2515-7639/ab13bb} {\bibfield  {journal} {\bibinfo
  {journal} {J. Phys. Mater.}\ }\textbf {\bibinfo {volume} {2}},\ \bibinfo
  {pages} {036001} (\bibinfo {year} {2019})}\BibitemShut {NoStop}%
\bibitem [{\citenamefont {{HDF group}}(2023)}]{HDF}%
  \BibitemOpen
  \bibfield  {author} {\bibinfo {author} {\bibnamefont {{HDF group}}},\ }\href
  {https://www.hdfgroup.org/solutions/} {\enquote {\bibinfo {title} {Open
  source data management and storage suite},}\ }\bibinfo {howpublished}
  {Online} (\bibinfo {year} {2023}),\ \bibinfo {note}
  {\url{https://www.hdfgroup.org/solutions/}}\BibitemShut {NoStop}%
\bibitem [{\citenamefont {K{\"{o}}nnecke}\ \emph {et~al.}(2015)\citenamefont
  {K{\"{o}}nnecke}, \citenamefont {Akeroyd}, \citenamefont {Bernstein},
  \citenamefont {Brewster}, \citenamefont {Campbell}, \citenamefont {Clausen},
  \citenamefont {Cottrell}, \citenamefont {Hoffmann}, \citenamefont {Jemian},
  \citenamefont {M{\"{a}}nnicke}, \citenamefont {Osborn}, \citenamefont
  {Peterson}, \citenamefont {Richter}, \citenamefont {Suzuki}, \citenamefont
  {Watts}, \citenamefont {Wintersberger},\ and\ \citenamefont
  {Wuttke}}]{K_nnecke_2015}%
  \BibitemOpen
  \bibfield  {author} {\bibinfo {author} {\bibfnamefont {M.}~\bibnamefont
  {K{\"{o}}nnecke}}, \bibinfo {author} {\bibfnamefont {F.~A.}\ \bibnamefont
  {Akeroyd}}, \bibinfo {author} {\bibfnamefont {H.~J.}\ \bibnamefont
  {Bernstein}}, \bibinfo {author} {\bibfnamefont {A.~S.}\ \bibnamefont
  {Brewster}}, \bibinfo {author} {\bibfnamefont {S.~I.}\ \bibnamefont
  {Campbell}}, \bibinfo {author} {\bibfnamefont {B.}~\bibnamefont {Clausen}},
  \bibinfo {author} {\bibfnamefont {S.}~\bibnamefont {Cottrell}}, \bibinfo
  {author} {\bibfnamefont {J.~U.}\ \bibnamefont {Hoffmann}}, \bibinfo {author}
  {\bibfnamefont {P.~R.}\ \bibnamefont {Jemian}}, \bibinfo {author}
  {\bibfnamefont {D.}~\bibnamefont {M{\"{a}}nnicke}}, \bibinfo {author}
  {\bibfnamefont {R.}~\bibnamefont {Osborn}}, \bibinfo {author} {\bibfnamefont
  {P.~F.}\ \bibnamefont {Peterson}}, \bibinfo {author} {\bibfnamefont
  {T.}~\bibnamefont {Richter}}, \bibinfo {author} {\bibfnamefont
  {J.}~\bibnamefont {Suzuki}}, \bibinfo {author} {\bibfnamefont
  {B.}~\bibnamefont {Watts}}, \bibinfo {author} {\bibfnamefont
  {E.}~\bibnamefont {Wintersberger}},\ and\ \bibinfo {author} {\bibfnamefont
  {J.}~\bibnamefont {Wuttke}},\ }\bibfield  {title} {\enquote {\bibinfo {title}
  {The {NeXus} data format},}\ }\href
  {https://doi.org/10.1107/S1600576714027575} {\bibfield  {journal} {\bibinfo
  {journal} {J. Appl. Crystallogr.}\ }\textbf {\bibinfo {volume} {48}},\
  \bibinfo {pages} {301--305} (\bibinfo {year} {2015})},\ \bibinfo {note}
  {\url{https://www.nexusformat.org/}}\BibitemShut {NoStop}%
\bibitem [{\citenamefont {Rasband}(1997)}]{Rasband_1997}%
  \BibitemOpen
  \bibfield  {author} {\bibinfo {author} {\bibfnamefont {W.}~\bibnamefont
  {Rasband}},\ }\href {https://imagej.net/} {\enquote {\bibinfo {title}
  {{ImageJ}},}\ }\bibinfo {type} {software}\ (\bibinfo  {institution} {National
  Institute of Mental Health},\ \bibinfo {address} {Bethesda, Maryland, USA},\
  \bibinfo {year} {1997})\BibitemShut {NoStop}%
\bibitem [{\citenamefont {Schneider}, \citenamefont {Rasband},\ and\
  \citenamefont {Eliceiri}(2012)}]{Schneider_2012}%
  \BibitemOpen
  \bibfield  {author} {\bibinfo {author} {\bibfnamefont {C.}~\bibnamefont
  {Schneider}}, \bibinfo {author} {\bibfnamefont {W.}~\bibnamefont {Rasband}},\
  and\ \bibinfo {author} {\bibfnamefont {K.}~\bibnamefont {Eliceiri}},\
  }\bibfield  {title} {\enquote {\bibinfo {title} {{NIH} {I}mage to {ImageJ}:
  25 years of image analysis},}\ }\href {https://doi.org/10.1038/nmeth.2089}
  {\bibfield  {journal} {\bibinfo  {journal} {Nat. Methods}\ }\textbf {\bibinfo
  {volume} {9}},\ \bibinfo {pages} {671--675} (\bibinfo {year}
  {2012})}\BibitemShut {NoStop}%
\bibitem [{\citenamefont {Pedregosa}\ \emph {et~al.}(2011)\citenamefont
  {Pedregosa}, \citenamefont {Varoquaux}, \citenamefont {Gramfort},
  \citenamefont {Michel}, \citenamefont {Thirion}, \citenamefont {Grisel},
  \citenamefont {Blondel}, \citenamefont {Prettenhofer}, \citenamefont {Weiss},
  \citenamefont {Dubourg}, \citenamefont {Vanderplas}, \citenamefont {Passos},
  \citenamefont {Cournapeau}, \citenamefont {Brucher}, \citenamefont {Perrot},\
  and\ \citenamefont {Duchesnay}}]{Pedregosa_2011}%
  \BibitemOpen
  \bibfield  {author} {\bibinfo {author} {\bibfnamefont {F.}~\bibnamefont
  {Pedregosa}}, \bibinfo {author} {\bibfnamefont {G.}~\bibnamefont
  {Varoquaux}}, \bibinfo {author} {\bibfnamefont {A.}~\bibnamefont {Gramfort}},
  \bibinfo {author} {\bibfnamefont {V.}~\bibnamefont {Michel}}, \bibinfo
  {author} {\bibfnamefont {B.}~\bibnamefont {Thirion}}, \bibinfo {author}
  {\bibfnamefont {O.}~\bibnamefont {Grisel}}, \bibinfo {author} {\bibfnamefont
  {M.}~\bibnamefont {Blondel}}, \bibinfo {author} {\bibfnamefont
  {P.}~\bibnamefont {Prettenhofer}}, \bibinfo {author} {\bibfnamefont
  {R.}~\bibnamefont {Weiss}}, \bibinfo {author} {\bibfnamefont
  {V.}~\bibnamefont {Dubourg}}, \bibinfo {author} {\bibfnamefont
  {J.}~\bibnamefont {Vanderplas}}, \bibinfo {author} {\bibfnamefont
  {A.}~\bibnamefont {Passos}}, \bibinfo {author} {\bibfnamefont
  {D.}~\bibnamefont {Cournapeau}}, \bibinfo {author} {\bibfnamefont
  {M.}~\bibnamefont {Brucher}}, \bibinfo {author} {\bibfnamefont
  {M.}~\bibnamefont {Perrot}},\ and\ \bibinfo {author} {\bibfnamefont
  {{\'E}.}~\bibnamefont {Duchesnay}},\ }\bibfield  {title} {\enquote {\bibinfo
  {title} {Scikit-learn: machine learning in python},}\ }\href
  {http://jmlr.org/papers/v12/pedregosa11a.html} {\bibfield  {journal}
  {\bibinfo  {journal} {J. Mach. Learn. Res.}\ }\textbf {\bibinfo {volume}
  {12}},\ \bibinfo {pages} {2825--2830} (\bibinfo {year} {2011})}\BibitemShut
  {NoStop}%
\bibitem [{\citenamefont {Abadi}\ \emph {et~al.}(2016)\citenamefont {Abadi},
  \citenamefont {Barham}, \citenamefont {Chen}, \citenamefont {Chen},
  \citenamefont {Davis}, \citenamefont {Dean}, \citenamefont {Devin},
  \citenamefont {Ghemawat}, \citenamefont {Irving}, \citenamefont {Isard} \emph
  {et~al.}}]{abadi2016tensorflow}%
  \BibitemOpen
  \bibfield  {author} {\bibinfo {author} {\bibfnamefont {M.}~\bibnamefont
  {Abadi}}, \bibinfo {author} {\bibfnamefont {P.}~\bibnamefont {Barham}},
  \bibinfo {author} {\bibfnamefont {J.}~\bibnamefont {Chen}}, \bibinfo {author}
  {\bibfnamefont {Z.}~\bibnamefont {Chen}}, \bibinfo {author} {\bibfnamefont
  {A.}~\bibnamefont {Davis}}, \bibinfo {author} {\bibfnamefont
  {J.}~\bibnamefont {Dean}}, \bibinfo {author} {\bibfnamefont {M.}~\bibnamefont
  {Devin}}, \bibinfo {author} {\bibfnamefont {S.}~\bibnamefont {Ghemawat}},
  \bibinfo {author} {\bibfnamefont {G.}~\bibnamefont {Irving}}, \bibinfo
  {author} {\bibfnamefont {M.}~\bibnamefont {Isard}}, \emph {et~al.},\
  }\bibfield  {title} {\enquote {\bibinfo {title} {Tensorflow: a system for
  large-scale machine learning},}\ }in\ \href@noop {} {\emph {\bibinfo
  {booktitle} {OSDI'16: Proceedings of the 12th USENIX Conference on Operating
  Systems Design and Implementation}}}\ (\bibinfo  {publisher} {USENIX
  Association},\ \bibinfo {address} {USA},\ \bibinfo {year} {2016})\ pp.\
  \bibinfo {pages} {265--283}\BibitemShut {NoStop}%
\bibitem [{\citenamefont {Paszke}\ \emph {et~al.}(2019)\citenamefont {Paszke},
  \citenamefont {Gross}, \citenamefont {Massa}, \citenamefont {Lerer},
  \citenamefont {Bradbury}, \citenamefont {Chanan}, \citenamefont {Killeen},
  \citenamefont {Lin}, \citenamefont {Gimelshein}, \citenamefont {Antiga} \emph
  {et~al.}}]{paszke2019pytorch}%
  \BibitemOpen
  \bibfield  {author} {\bibinfo {author} {\bibfnamefont {A.}~\bibnamefont
  {Paszke}}, \bibinfo {author} {\bibfnamefont {S.}~\bibnamefont {Gross}},
  \bibinfo {author} {\bibfnamefont {F.}~\bibnamefont {Massa}}, \bibinfo
  {author} {\bibfnamefont {A.}~\bibnamefont {Lerer}}, \bibinfo {author}
  {\bibfnamefont {J.}~\bibnamefont {Bradbury}}, \bibinfo {author}
  {\bibfnamefont {G.}~\bibnamefont {Chanan}}, \bibinfo {author} {\bibfnamefont
  {T.}~\bibnamefont {Killeen}}, \bibinfo {author} {\bibfnamefont
  {Z.}~\bibnamefont {Lin}}, \bibinfo {author} {\bibfnamefont {N.}~\bibnamefont
  {Gimelshein}}, \bibinfo {author} {\bibfnamefont {L.}~\bibnamefont {Antiga}},
  \emph {et~al.},\ }\bibfield  {title} {\enquote {\bibinfo {title} {Pytorch: an
  imperative style, high-performance deep learning library},}\ }\href@noop {}
  {\bibfield  {journal} {\bibinfo  {journal} {Advances in neural information
  processing systems}\ }\textbf {\bibinfo {volume} {32}} (\bibinfo {year}
  {2019})}\BibitemShut {NoStop}%
\bibitem [{\citenamefont {{Comet Technologies Canada Inc.}}(2023)}]{ORSI_2020}%
  \BibitemOpen
  \bibfield  {author} {\bibinfo {author} {\bibnamefont {{Comet Technologies
  Canada Inc.}}},\ }\href {http://www.theobjects.com/dragonfly} {\enquote
  {\bibinfo {title} {Dragonfly 2020.2},}\ }\bibinfo {type} {software}\
  (\bibinfo  {institution} {Comet Technologies Canada Inc.},\ \bibinfo {year}
  {2023})\ \bibinfo {note}
  {\url{http://www.theobjects.com/dragonfly}}\BibitemShut {NoStop}%
\bibitem [{\citenamefont {Cea}\ \emph {et~al.}(2013)\citenamefont {Cea},
  \citenamefont {Botelho}, \citenamefont {Chaudhry}, \citenamefont
  {Fleischmann}, \citenamefont {Giles}, \citenamefont {Grigoriev},
  \citenamefont {Kaushik}, \citenamefont {Keys}, \citenamefont {Kennel},
  \citenamefont {Lilak}, \citenamefont {Mehandru}, \citenamefont {Stettler},
  \citenamefont {Voinov}, \citenamefont {Voynich}, \citenamefont {Weber},\ and\
  \citenamefont {Zhavoronok}}]{cea_process_2013}%
  \BibitemOpen
  \bibfield  {author} {\bibinfo {author} {\bibfnamefont {S.~M.}\ \bibnamefont
  {Cea}}, \bibinfo {author} {\bibfnamefont {S.}~\bibnamefont {Botelho}},
  \bibinfo {author} {\bibfnamefont {A.}~\bibnamefont {Chaudhry}}, \bibinfo
  {author} {\bibfnamefont {P.}~\bibnamefont {Fleischmann}}, \bibinfo {author}
  {\bibfnamefont {M.~D.}\ \bibnamefont {Giles}}, \bibinfo {author}
  {\bibfnamefont {A.}~\bibnamefont {Grigoriev}}, \bibinfo {author}
  {\bibfnamefont {A.}~\bibnamefont {Kaushik}}, \bibinfo {author} {\bibfnamefont
  {P.~H.}\ \bibnamefont {Keys}}, \bibinfo {author} {\bibfnamefont {H.~W.}\
  \bibnamefont {Kennel}}, \bibinfo {author} {\bibfnamefont {A.~D.}\
  \bibnamefont {Lilak}}, \bibinfo {author} {\bibfnamefont {R.}~\bibnamefont
  {Mehandru}}, \bibinfo {author} {\bibfnamefont {M.}~\bibnamefont {Stettler}},
  \bibinfo {author} {\bibfnamefont {B.}~\bibnamefont {Voinov}}, \bibinfo
  {author} {\bibfnamefont {N.}~\bibnamefont {Voynich}}, \bibinfo {author}
  {\bibfnamefont {C.}~\bibnamefont {Weber}},\ and\ \bibinfo {author}
  {\bibfnamefont {N.}~\bibnamefont {Zhavoronok}},\ }\bibfield  {title}
  {{\selectlanguage {english}\enquote {\bibinfo {title} {Process modeling for
  advanced device technologies},}\ }}\href
  {https://doi.org/10.1007/s10825-013-0491-6} {\bibfield  {journal} {\bibinfo
  {journal} {J. Comput. Electron.}\ }\textbf {\bibinfo {volume} {13}},\
  \bibinfo {pages} {18--32} (\bibinfo {year} {2013})}\BibitemShut {NoStop}%
\bibitem [{\citenamefont {Lorenz}\ \emph {et~al.}(2013)\citenamefont {Lorenz},
  \citenamefont {Baer}, \citenamefont {Burenkov}, \citenamefont {Erdmann},
  \citenamefont {Evanschitzky},\ and\ \citenamefont
  {Pichler}}]{lorenz_challenges_2013}%
  \BibitemOpen
  \bibfield  {author} {\bibinfo {author} {\bibfnamefont {J.~K.}\ \bibnamefont
  {Lorenz}}, \bibinfo {author} {\bibfnamefont {E.}~\bibnamefont {Baer}},
  \bibinfo {author} {\bibfnamefont {A.}~\bibnamefont {Burenkov}}, \bibinfo
  {author} {\bibfnamefont {A.}~\bibnamefont {Erdmann}}, \bibinfo {author}
  {\bibfnamefont {P.}~\bibnamefont {Evanschitzky}},\ and\ \bibinfo {author}
  {\bibfnamefont {P.}~\bibnamefont {Pichler}},\ }\bibfield  {title}
  {{\selectlanguage {english}\enquote {\bibinfo {title} {Challenges and
  opportunities for process modeling in the nanotechnology era},}\ }}\href
  {https://doi.org/10.1007/s10825-013-0477-4} {\bibfield  {journal} {\bibinfo
  {journal} {J. Comput. Electron.}\ }\textbf {\bibinfo {volume} {13}},\
  \bibinfo {pages} {3--17} (\bibinfo {year} {2013})}\BibitemShut {NoStop}%
\bibitem [{\citenamefont {Balestra}\ \emph {et~al.}(2022)\citenamefont
  {Balestra}, \citenamefont {Graef}, \citenamefont {Hayashi}, \citenamefont
  {Takaura}, \citenamefont {Ishiuchi}, \citenamefont {Gargini},\ and\
  \citenamefont {Conte}}]{irds_2022}%
  \BibitemOpen
  \bibfield  {author} {\bibinfo {author} {\bibfnamefont {F.}~\bibnamefont
  {Balestra}}, \bibinfo {author} {\bibfnamefont {M.}~\bibnamefont {Graef}},
  \bibinfo {author} {\bibfnamefont {Y.}~\bibnamefont {Hayashi}}, \bibinfo
  {author} {\bibfnamefont {N.}~\bibnamefont {Takaura}}, \bibinfo {author}
  {\bibfnamefont {H.}~\bibnamefont {Ishiuchi}}, \bibinfo {author}
  {\bibfnamefont {P.}~\bibnamefont {Gargini}},\ and\ \bibinfo {author}
  {\bibfnamefont {T.}~\bibnamefont {Conte}},\ }\href
  {https://irds.ieee.org/editions/2022/executive-summary} {\enquote {\bibinfo
  {title} {International roadmap for devices and systems 2022: {Executive}
  {Summary} - {IEEE} {IRDS}\texttrademark},}\ }\bibinfo {type} {techreport}\
  (\bibinfo  {institution} {IEEE},\ \bibinfo {year} {2022})\BibitemShut
  {NoStop}%
\bibitem [{\citenamefont {Szabo}\ \emph {et~al.}(2022)\citenamefont {Szabo},
  \citenamefont {Weichselbaum}, \citenamefont {Biber}, \citenamefont {Cupak},
  \citenamefont {Mutzke}, \citenamefont {Wilhelm},\ and\ \citenamefont
  {Aumayr}}]{szabo_graphical_2022}%
  \BibitemOpen
  \bibfield  {author} {\bibinfo {author} {\bibfnamefont {P.~S.}\ \bibnamefont
  {Szabo}}, \bibinfo {author} {\bibfnamefont {D.}~\bibnamefont {Weichselbaum}},
  \bibinfo {author} {\bibfnamefont {H.}~\bibnamefont {Biber}}, \bibinfo
  {author} {\bibfnamefont {C.}~\bibnamefont {Cupak}}, \bibinfo {author}
  {\bibfnamefont {A.}~\bibnamefont {Mutzke}}, \bibinfo {author} {\bibfnamefont
  {R.~A.}\ \bibnamefont {Wilhelm}},\ and\ \bibinfo {author} {\bibfnamefont
  {F.}~\bibnamefont {Aumayr}},\ }\bibfield  {title} {{\selectlanguage
  {english}\enquote {\bibinfo {title} {Graphical user interface for {SDTrimSP}
  to simulate sputtering, ion implantation and the dynamic effects of ion
  irradiation},}\ }}\href {https://doi.org/10.1016/j.nimb.2022.04.008}
  {\bibfield  {journal} {\bibinfo  {journal} {Nucl. Instrum. Meth. Phys. Res.
  B}\ }\textbf {\bibinfo {volume} {522}},\ \bibinfo {pages} {47--53} (\bibinfo
  {year} {2022})}\BibitemShut {NoStop}%
\bibitem [{\citenamefont {Buchnev}\ \emph {et~al.}(2022)\citenamefont
  {Buchnev}, \citenamefont {Grant-Jacob}, \citenamefont {Eason}, \citenamefont
  {Zheludev}, \citenamefont {Mills},\ and\ \citenamefont
  {MacDonald}}]{buchnev_deep-learning-assisted_2022}%
  \BibitemOpen
  \bibfield  {author} {\bibinfo {author} {\bibfnamefont {O.}~\bibnamefont
  {Buchnev}}, \bibinfo {author} {\bibfnamefont {J.~A.}\ \bibnamefont
  {Grant-Jacob}}, \bibinfo {author} {\bibfnamefont {R.~W.}\ \bibnamefont
  {Eason}}, \bibinfo {author} {\bibfnamefont {N.~I.}\ \bibnamefont {Zheludev}},
  \bibinfo {author} {\bibfnamefont {B.}~\bibnamefont {Mills}},\ and\ \bibinfo
  {author} {\bibfnamefont {K.~F.}\ \bibnamefont {MacDonald}},\ }\bibfield
  {title} {\enquote {\bibinfo {title} {Deep-learning-assisted focused ion beam
  nanofabrication},}\ }\href {https://doi.org/10.1021/acs.nanolett.1c04604}
  {\bibfield  {journal} {\bibinfo  {journal} {Nano Lett.}\ } (\bibinfo {year}
  {2022}),\ 10.1021/acs.nanolett.1c04604},\ \bibinfo {note} {publisher:
  American Chemical Society}\BibitemShut {NoStop}%
\bibitem [{\citenamefont {Jain}\ \emph {et~al.}(2022)\citenamefont {Jain},
  \citenamefont {Kretschmer}, \citenamefont {H{\"{o}}flich}, \citenamefont
  {Lopes},\ and\ \citenamefont {Krasheninnikov}}]{Jain2022}%
  \BibitemOpen
  \bibfield  {author} {\bibinfo {author} {\bibfnamefont {M.}~\bibnamefont
  {Jain}}, \bibinfo {author} {\bibfnamefont {S.}~\bibnamefont {Kretschmer}},
  \bibinfo {author} {\bibfnamefont {K.}~\bibnamefont {H{\"{o}}flich}}, \bibinfo
  {author} {\bibfnamefont {J.~M.~J.}\ \bibnamefont {Lopes}},\ and\ \bibinfo
  {author} {\bibfnamefont {A.~V.}\ \bibnamefont {Krasheninnikov}},\ }\bibfield
  {title} {\enquote {\bibinfo {title} {{Atomistic Simulations of Defects
  Production under Ion Irradiation in Epitaxial Graphene on SiC}},}\ }\href
  {https://doi.org/10.1002/pssr.202200292} {\bibfield  {journal} {\bibinfo
  {journal} {Phys. Status Solidi RRL}\ }\textbf {\bibinfo {volume} {2200292}},\
  \bibinfo {pages} {2200292} (\bibinfo {year} {2022})}\BibitemShut {NoStop}%
\bibitem [{\citenamefont {Refaely-Abramson}\ \emph {et~al.}(2018)\citenamefont
  {Refaely-Abramson}, \citenamefont {Qiu}, \citenamefont {Louie},\ and\
  \citenamefont {Neaton}}]{Refaely-Abramson2018}%
  \BibitemOpen
  \bibfield  {author} {\bibinfo {author} {\bibfnamefont {S.}~\bibnamefont
  {Refaely-Abramson}}, \bibinfo {author} {\bibfnamefont {D.~Y.}\ \bibnamefont
  {Qiu}}, \bibinfo {author} {\bibfnamefont {S.}~\bibnamefont {Louie}},\ and\
  \bibinfo {author} {\bibfnamefont {J.~B.}\ \bibnamefont {Neaton}},\ }\bibfield
   {title} {\enquote {\bibinfo {title} {{Defect-Induced Modification of
  Low-Lying Excitons and Valley Selectivity in Monolayer Transition Metal
  Dichalcogenides}},}\ }\href {https://doi.org/10.1103/PhysRevLett.121.167402}
  {\bibfield  {journal} {\bibinfo  {journal} {Phys. Rev. Lett.}\ }\textbf
  {\bibinfo {volume} {121}},\ \bibinfo {pages} {167402} (\bibinfo {year}
  {2018})},\ \Eprint {https://arxiv.org/abs/1804.05719} {1804.05719}
  \BibitemShut {NoStop}%
\bibitem [{\citenamefont {Bertoldo}\ \emph {et~al.}(2022)\citenamefont
  {Bertoldo}, \citenamefont {Ali}, \citenamefont {Manti},\ and\ \citenamefont
  {Thygesen}}]{Bertoldo2022}%
  \BibitemOpen
  \bibfield  {author} {\bibinfo {author} {\bibfnamefont {F.}~\bibnamefont
  {Bertoldo}}, \bibinfo {author} {\bibfnamefont {S.}~\bibnamefont {Ali}},
  \bibinfo {author} {\bibfnamefont {S.}~\bibnamefont {Manti}},\ and\ \bibinfo
  {author} {\bibfnamefont {K.~S.}\ \bibnamefont {Thygesen}},\ }\bibfield
  {title} {\enquote {\bibinfo {title} {{Quantum point defects in 2D materials -
  the QPOD database}},}\ }\href {https://doi.org/10.1038/s41524-022-00730-w}
  {\bibfield  {journal} {\bibinfo  {journal} {npj Comput. Mater.}\ }\textbf
  {\bibinfo {volume} {8}},\ \bibinfo {pages} {56} (\bibinfo {year} {2022})},\
  \Eprint {https://arxiv.org/abs/2110.01961} {2110.01961} \BibitemShut
  {NoStop}%
\bibitem [{\citenamefont {Mendelson}\ \emph {et~al.}(2021)\citenamefont
  {Mendelson}, \citenamefont {Chugh}, \citenamefont {Reimers}, \citenamefont
  {Cheng}, \citenamefont {Gottscholl}, \citenamefont {Long}, \citenamefont
  {Mellor}, \citenamefont {Zettl}, \citenamefont {Dyakonov}, \citenamefont
  {Beton}, \citenamefont {Novikov}, \citenamefont {Jagadish}, \citenamefont
  {Tan}, \citenamefont {Ford}, \citenamefont {Toth}, \citenamefont {Bradac},\
  and\ \citenamefont {Aharonovich}}]{Mendelson2021}%
  \BibitemOpen
  \bibfield  {author} {\bibinfo {author} {\bibfnamefont {N.}~\bibnamefont
  {Mendelson}}, \bibinfo {author} {\bibfnamefont {D.}~\bibnamefont {Chugh}},
  \bibinfo {author} {\bibfnamefont {J.~R.}\ \bibnamefont {Reimers}}, \bibinfo
  {author} {\bibfnamefont {T.~S.}\ \bibnamefont {Cheng}}, \bibinfo {author}
  {\bibfnamefont {A.}~\bibnamefont {Gottscholl}}, \bibinfo {author}
  {\bibfnamefont {H.}~\bibnamefont {Long}}, \bibinfo {author} {\bibfnamefont
  {C.~J.}\ \bibnamefont {Mellor}}, \bibinfo {author} {\bibfnamefont
  {A.}~\bibnamefont {Zettl}}, \bibinfo {author} {\bibfnamefont
  {V.}~\bibnamefont {Dyakonov}}, \bibinfo {author} {\bibfnamefont {P.~H.}\
  \bibnamefont {Beton}}, \bibinfo {author} {\bibfnamefont {S.~V.}\ \bibnamefont
  {Novikov}}, \bibinfo {author} {\bibfnamefont {C.}~\bibnamefont {Jagadish}},
  \bibinfo {author} {\bibfnamefont {H.~H.}\ \bibnamefont {Tan}}, \bibinfo
  {author} {\bibfnamefont {M.~J.}\ \bibnamefont {Ford}}, \bibinfo {author}
  {\bibfnamefont {M.}~\bibnamefont {Toth}}, \bibinfo {author} {\bibfnamefont
  {C.}~\bibnamefont {Bradac}},\ and\ \bibinfo {author} {\bibfnamefont
  {I.}~\bibnamefont {Aharonovich}},\ }\bibfield  {title} {\enquote {\bibinfo
  {title} {{Identifying carbon as the source of visible single-photon emission
  from hexagonal boron nitride}},}\ }\href
  {https://doi.org/10.1038/s41563-020-00850-y} {\bibfield  {journal} {\bibinfo
  {journal} {Nat. Mater.}\ }\textbf {\bibinfo {volume} {20}},\ \bibinfo {pages}
  {321--328} (\bibinfo {year} {2021})}\BibitemShut {NoStop}%
\bibitem [{\citenamefont {Tran}\ \emph {et~al.}(2016)\citenamefont {Tran},
  \citenamefont {Bray}, \citenamefont {Ford}, \citenamefont {Toth},\ and\
  \citenamefont {Aharonovich}}]{tran2016quantum}%
  \BibitemOpen
  \bibfield  {author} {\bibinfo {author} {\bibfnamefont {T.~T.}\ \bibnamefont
  {Tran}}, \bibinfo {author} {\bibfnamefont {K.}~\bibnamefont {Bray}}, \bibinfo
  {author} {\bibfnamefont {M.~J.}\ \bibnamefont {Ford}}, \bibinfo {author}
  {\bibfnamefont {M.}~\bibnamefont {Toth}},\ and\ \bibinfo {author}
  {\bibfnamefont {I.}~\bibnamefont {Aharonovich}},\ }\bibfield  {title}
  {\enquote {\bibinfo {title} {Quantum emission from hexagonal boron nitride
  monolayers},}\ }in\ \href {https://doi.org/10.1364/cleo_qels.2016.ftu4d.1}
  {\emph {\bibinfo {booktitle} {Conference on Lasers and Electro-Optics}}}\
  (\bibinfo  {publisher} {{OSA}},\ \bibinfo {year} {2016})\BibitemShut
  {NoStop}%
\bibitem [{\citenamefont {Shang}\ \emph {et~al.}(2020)\citenamefont {Shang},
  \citenamefont {Hashemi}, \citenamefont {Berenc{\'{e}}n}, \citenamefont
  {Komsa}, \citenamefont {Erhart}, \citenamefont {Zhou}, \citenamefont {Helm},
  \citenamefont {Krasheninnikov},\ and\ \citenamefont {Astakhov}}]{Shang2020}%
  \BibitemOpen
  \bibfield  {author} {\bibinfo {author} {\bibfnamefont {Z.}~\bibnamefont
  {Shang}}, \bibinfo {author} {\bibfnamefont {A.}~\bibnamefont {Hashemi}},
  \bibinfo {author} {\bibfnamefont {Y.}~\bibnamefont {Berenc{\'{e}}n}},
  \bibinfo {author} {\bibfnamefont {H.-P.}\ \bibnamefont {Komsa}}, \bibinfo
  {author} {\bibfnamefont {P.}~\bibnamefont {Erhart}}, \bibinfo {author}
  {\bibfnamefont {S.}~\bibnamefont {Zhou}}, \bibinfo {author} {\bibfnamefont
  {M.}~\bibnamefont {Helm}}, \bibinfo {author} {\bibfnamefont {A.~V.}\
  \bibnamefont {Krasheninnikov}},\ and\ \bibinfo {author} {\bibfnamefont
  {G.~V.}\ \bibnamefont {Astakhov}},\ }\bibfield  {title} {\enquote {\bibinfo
  {title} {{Local vibrational modes of Si vacancy spin qubits in SiC}},}\
  }\href {https://doi.org/10.1103/PhysRevB.101.144109} {\bibfield  {journal}
  {\bibinfo  {journal} {Phys. Rev. B}\ }\textbf {\bibinfo {volume} {101}},\
  \bibinfo {pages} {144109} (\bibinfo {year} {2020})},\ \Eprint
  {https://arxiv.org/abs/2002.00067} {2002.00067} \BibitemShut {NoStop}%
\end{thebibliography}%

\end{document}